%% file: arXiv-submit.tex
\documentclass[aps,prb,reprint,superscriptaddress,nobibnotes]{revtex4-2}

\usepackage{amsmath}
\usepackage{amssymb}
\usepackage{amsfonts}
\usepackage{amsthm}
\usepackage{mathtools}
\usepackage{hyperref}
\usepackage{graphicx}
\usepackage{bm}
\usepackage{dsfont}
\usepackage{xcolor}
\usepackage{soul}

\newcommand{\ket}[1]{| #1 \rangle}
\newcommand{\bra}[1]{\langle #1 |}
\newcommand{\braket}[2]{\langle #1 | #2 \rangle}

\newtheorem{theorem}{Theorem\textsuperscript{ph}}
\newtheorem{conjecture}{Conjecture}

\newcommand{\irep}[1]{}

\begin{document}

\title{Most {two-dimensional bosonic} topological orders forbid sign-problem-free quantum Monte Carlo: Nonpositive Gauss sum as an indicator}

\author{Donghae Seo}
\affiliation{Department of Physics, Pohang University of Science and Technology, Pohang 37673, Republic of Korea}
\affiliation{Center for Artificial Low Dimensional Electronic Systems, Institute for Basic Science, Pohang 37673, Republic of Korea}
\author{Minyoung You}
\affiliation{POSTECH Basic Science Research Institute, Pohang 37673, Republic of Korea}
\affiliation{Yukawa Institute for Theoretical Physics, Kyoto University, Kyoto 606-8502, Japan}
\author{Hee-Cheol Kim}
\email{heecheol@postech.ac.kr}
\affiliation{Department of Physics, Pohang University of Science and Technology, Pohang 37673, Republic of Korea}
\affiliation{Asia-Pacific Center for Theoretical Physics, Pohang 37673, Republic of Korea}
\author{Gil Young Cho}
\email{gilyoungcho@kaist.ac.kr}
\affiliation{Department of Physics, Korea Advanced Institute of Science and Technology, Daejeon 34141, Republic of Korea}
\affiliation{Center for Artificial Low Dimensional Electronic Systems, Institute for Basic Science, Pohang 37673, Republic of Korea}
\affiliation{Asia-Pacific Center for Theoretical Physics, Pohang 37673, Republic of Korea}

\date{\today} 

\begin{abstract}
Quantum Monte Carlo is a powerful tool for studying quantum many-body physics, yet its efficacy is often curtailed by the notorious sign problem. In this Letter, we introduce a novel criterion for the ``intrinsic" sign problem in two-dimensional bosonic topological orders, which cannot be resolved by local basis transformations or adiabatic deformations of the Hamiltonian. Specifically, we {find that the positivity of higher Gauss sums is a necessary condition for a two-dimensional bosonic topological order to be realized by a stoquastic Hamiltonian, and hence sign-problem-free. Equivalently,} a nonpositive higher Gauss sum for a given topological order indicates the presence of an intrinsic sign problem. This condition not only aligns with prior findings but significantly broadens their scope. Using this new criterion, we examine the Gauss sums of all 405 bosonic topological orders classified up to rank 12, and strikingly find that 398 of them exhibit intrinsic sign problems. We also uncover intriguing links between the intrinsic sign problem, gappability of boundary theories, and time-reversal symmetry, suggesting that sign-problem-free quantum Monte Carlo may fundamentally rely on both time-reversal symmetry and gapped boundaries. These results highlight the deep connection between the intrinsic sign problem and fundamental properties of topological phases, offering valuable insights into their classical simulability.
\end{abstract}
 
\maketitle

Studying quantum many-body systems poses a significant challenge due to the exponential growth of their Hilbert space dimension, making only a few special models tractable. Quantum Monte Carlo (QMC) methods~\cite{sandvik1991quantum}, which employ random sampling, offer a scalable way to analyze such systems, frequently providing more accurate estimates of physical observables than density functional theory or mean-field theory~\cite{foulkes2001quantum,jones2015density,altland2010condensed}. However, these algorithms become much less efficient when the sign problem arises~\cite{pan2024sign}, namely when the partition function cannot be expressed solely as a sum of classical probability weights. Although the sign problem can occasionally be remedied by a suitable basis transformation, finding a universal approach is widely believed to be computationally intractable~\cite{marvian2019computational}. This naturally leads to the question of whether certain systems possess a sign problem so robust that it cannot be alleviated by any finite-depth local unitary circuit (FDLU). {Such finite-depth circuits, composed of local unitary gates, can be interpreted as local basis transformations. A sign problem that cannot be eliminated by FDLU is referred to as} the `intrinsic sign problem'~\cite{hastings2015how}. Intriguingly, several topological orders are argued to exhibit this intractable obstacle~\cite{ringel2017quantized,smith2020intrinsic,golan2020intrinsic,kim2022chiral}.

\begin{figure}
    \centering
    \includegraphics[width=\linewidth]{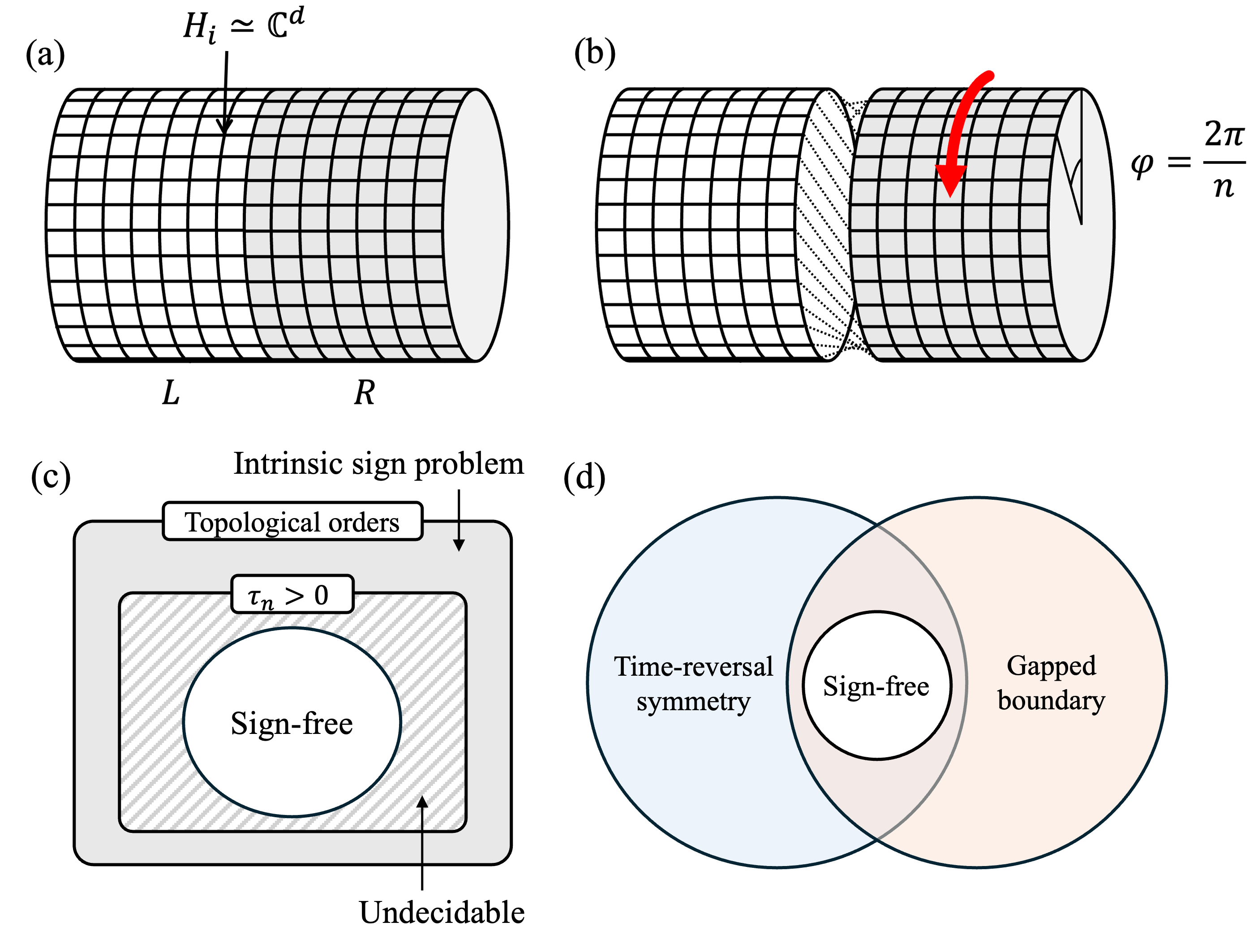}
    \caption{(a) Lattice system on a cylinder. The linear system size is sufficiently longer than the correlation length. Each lattice site is associated with a local Hilbert space $\mathcal{H}_i \simeq \mathbb{C}^d$. The cylinder is partitioned into two subregions, $L$ and $R$. (b) Partial rotation by an angle $\varphi = 2\pi / n$ on the subregion $R$. {(c) Venn diagram illustrating the relationship between our sign-free condition $\tau_n > 0$ and the intrinsic sign problem of two-dimensional bosonic topological orders. The grey and white regions represent topological orders exhibiting the intrinsic sign problem and sign-free topological orders, respectively. Sign-free topological orders necessarily satisfy our sign-free condition $\tau_n > 0$. {The set of topological orders whose intrinsic sign problems are undecidable is represented by the shaded region. Among 405 topological orders up to rank 12~\cite{ng2023classification}, we find 398 cases suffering the intrinsic sign problem. Only 3 topological orders up to rank 12 are sign-free. The intrinsic sign problems of the other 4 are undecidable.} (d) Venn diagram illustrating the relationship between time-reversal symmetry, gapped boundaries, and sign-free Abelian topological orders. Every sign-free Abelian topological order is time-reversal symmetric and admits a gapped boundary. For non-Abelian topological orders, we derive similar, though slightly
    weaker, conclusions. Please refer to the main text for details.}}
    \label{fig:partial_rotation}
\end{figure}

In this Letter, we introduce a rigorous and efficient criterion for detecting the intrinsic sign problem in two-dimensional bosonic topological orders. More precisely, we prove that all higher Gauss sums must be positive for a topological order to be free of an intrinsic sign problem:
\begin{align} \label{eq:sign-free_condition} 
\tau_n \equiv \sum_{a \in \mathcal{A}} d_a^2 \theta_a^n > 0 \quad \forall n \in \mathbb{N}.
\end{align}
Otherwise, the topological order exhibits the intrinsic sign problem, irrespective of detailed forms of microscopic Hamiltonians [Fig.~\ref{fig:partial_rotation}(c)]. Here, $\mathcal{A}$ is the finite set of anyons in the topological order, {which are particle-like excitations obeying fractional statistics (We present a brief review on anyon and its data in ~\cite{supp}).} $d_a$ and $\theta_a$ are the quantum dimension and the topological spin of the corresponding anyon $a \in \mathcal{A}$. Our results not only align with previous findings \cite{hastings2015how,ringel2017quantized,smith2020intrinsic,golan2020intrinsic,kim2022chiral} but also significantly extend their scope. To highlight the power of this condition, we analyze the Gauss sums of 405 different topological orders classified up to rank 12~\cite{ng2023classification}, which also include the complete classification up to rank 11, and find that 398 suffer from the intrinsic sign problem. Additionally, we uncover intriguing connections between the intrinsic sign problem, gappability of the boundary theory, and time-reversal symmetry {[Fig.~\ref{fig:partial_rotation}(d)]}. 

QMC methods and their sign problem have a venerable history in the study of quantum many-body systems, attracting sustained interest in the literature~\cite{foulkes2001quantum,carlson2015quantum,li2015solving,li2016majorana,li2022asymptotic,grossman2023robust}. Other popular approaches, such as exact diagonalization and tensor network methods, face inherent limitations when dealing with large or highly entangled systems: the former is constrained by the exponential growth of the Hilbert space, while the latter becomes less effective for states with substantial entanglement entropy and is particularly challenging to implement in two-dimensional settings~\cite{stoudenmire2012studying}. Although QMC often provides a promising alternative, determining whether a given many-body problem can be efficiently simulated—i.e., is free of the sign problem \footnote{Note that being free of the sign problem does not guarantee an efficient simulation. Computing partition functions in certain classical spin glass models, which are expressible as sums of positive weights, is an NP-complete task~\cite{barahona1982computational}. Hence, no polynomial-time algorithm is believed to exist in these cases.}—remains a nontrivial challenge. Moreover, the precise physical criteria dictating the presence or absence of the sign problem are not well understood, underscoring its significance not only in condensed matter physics but also in the broader domain of quantum complexity theory~\cite{troyer2005computational,bravyi2015monte,marvian2019computational,ellison2021symmetry}.

Two-dimensional bosonic topological order is an ideal system for advancing our understanding of the sign problem. Despite the common belief that bosonic systems do not suffer from the sign problem, bosonic topological orders surprisingly can exhibit the sign problem~\cite{hastings2015how,ringel2017quantized,smith2020intrinsic,golan2020intrinsic,kim2022chiral}. Furthermore, the physics of two-dimensional topological orders is relatively well understood, with known classifying data represented by a small set of discrete numbers such as modular data, fusion coefficients, and central charge \cite{kitaev2006anyons,wen2015theory}, {which we briefly review in \cite{supp}}. This makes it easier to pinpoint the sources of the sign problem compared to more general quantum systems. Additionally, these states are not only of mathematical interest but are also actively studied in quantum simulations and condensed matter physics \cite{georgescu2014quantum,wen2017colloquium}. Indeed, there has been substantial effort in identifying the origins of the unavoidable sign problem, known as the intrinsic sign problem, in bosonic topological orders \cite{hastings2015how,ringel2017quantized,smith2020intrinsic,golan2020intrinsic,kim2022chiral}. Building upon these results, we present a new indicator of the intrinsic sign problem, i.e., a nonpositive Gauss sum Eq.~\eqref{eq:sign-free_condition}. 

\textbf{1. Backgrounds.} We briefly introduce several key concepts such as QMC, the sign problem, and stoquastic Hamiltonians~\cite{sandvik1991quantum,pan2024sign,bravyi2008complexity}, which we will use extensively.  

QMC is a family of algorithms designed to simulate quantum many-body systems by sampling the configuration space of a given model \cite{sandvik1991quantum}. Efficient implementation of a QMC algorithm requires decomposing the quantum partition function into a sum of terms interpretable as as probability weights, which must be nonnegative real numbers. If such a decomposition is not known, the system is said to suffer from the sign problem; otherwise, it is said to be sign-problem-free, or sign-free in short \cite{pan2024sign}. Crucially, a system is sign-free if the Hamiltonian is stoquastic in some basis \cite{bravyi2015monte}. A Hamiltonian $H$ is defined as stoquastic with respect to a basis $\mathcal{B}$ if and only if the matrix representation of $H$ in $\mathcal{B}$ has off-diagonal elements that are all nonpositive real numbers \cite{bravyi2008complexity}. 

{It is straightforward to see that QMC indeed works without the sign problem when the Hamiltonian is stoquastic. At a finite temperature $T = 1 / \beta$, the partition function $\mathcal{Z} = \operatorname{Tr} e^{- \beta H}$ can be expanded as 
\begin{equation*}
    \mathcal{Z} = \sum_{\{\bm{k}\}} \sum_{n=0}^\infty \frac{(-\beta)^n}{n!} \bra{\bm{k}_1} H \ket{\bm{k}_2} \dots \bra{\bm{k}_{n}} H \ket{\bm{k}_1}
\end{equation*}
where $\ket{\bm{k}_i}$ is a basis state. When $H$ is stoquastic in this basis, the minus signs from $\bra{\bm{k}_i} H \ket{\bm{k}_{i+1}}$ cancel those from $-\beta$ so that $\mathcal{Z}$ is a sum of nonnegative real weights. This allows us to efficiently estimate the expectation values of observables by taking the weights as the classical probability distribution. Moreover, the Frobenius-Perron theorem dictates that the ground state has nonnegative real amplitudes, which can also be sampled efficiently.}

Being stoquastic is a basis-dependent property. In some cases, nonstoquastic Hamiltonians can be transformed into stoquastic ones through an appropriate basis transformation~\cite{marvian2019computational}. Consequently, being nonstoquastic in a given basis does not necessarily indicate the presence of sign problem. However, certain nonstoquastic Hamiltonians cannot be transformed into stoquastic ones~\cite{hastings2015how}, and such systems are said to exhibit an intrinsic sign problem, which is the central focus of our work. 

\textbf{2. Derivation.} We will demonstrate that {all the higher Gauss sums must be positive for a topological order to be sign-free. This then implies, as a contrapositive, that} a nonpositive Gauss sum cannot be compatible with stoquastic Hamiltonians and therefore results in an intrinsic sign problem. Our derivation closely follows and generalizes the approach of \cite{golan2020intrinsic}. {To this end, we compute the expectation value of the partial rotation operator $T_R(\varphi)$ for a topological order on a cylinder [Fig.~\ref{fig:partial_rotation}(a,b)]. Importantly, we show that if the Hamiltonian realizing the topological order is stoquastic, this expectation value must be positive. We then compare this result with the corresponding calculation from topological field theory, which expresses the expectation value in terms of the higher Gauss sum. This comparison leads directly to our proposed sign-free condition.}

{We begin by outlining a few important details of the setup.} We consider a quantum lattice system on the cylinder, which is much longer than the correlation length. At each lattice site, a qudit degree of freedom is located, as dipicted in Fig.~\ref{fig:partial_rotation}(a). The Hilbert space is given by $\mathcal{H} = \bigotimes_{i \in \Lambda} \mathcal{H}_i$ where $\Lambda$ is the set of lattice sites and $\mathcal{H}_i \simeq \mathbb{C}^d$ for all $i \in \Lambda$ with a fixed positive integer $d$. We assume that the Hamiltonian $H$, whose ground state exhibits the target topological order, is stoquastic with respect to the onsite basis $\mathcal{B} = \left\{\ket{\bm{k}} = \bigotimes_{i \in \Lambda} \ket{k_i} | k_i = 0, \cdots, d -1 \right\}$. We denote the ground state by $\ket{G_a}$ where $a$ labels the superselection sector of the topological order (anyon label)~\cite{supp}. 


{We now calculate the expectation value of} $T_R(2\pi/n)$ for a positive integer $n$ {that rotates} the right half of the system [Fig.~\ref{fig:partial_rotation}{(a,b)}]. {In terms of the basis $\mathcal{B}$, the expectation value can be written as 
\begin{equation} \label{eq:expectation_value_expansion}
    \begin{aligned}
        \mathcal{T}_{a,n} &\equiv \bra{G_a} T_R(2\pi/n) \ket{G_a} \\
        &= \sum_{\bm{k},\bm{k}'} \braket{G_a}{\bm{k}} \bra{\bm{k}} T_R(2\pi/n) \ket{\bm{k}'} \braket{\bm{k}'}{G_a}
    \end{aligned}
\end{equation}
where $\bm{k}$ and $\bm{k}'$ label the consite basis states in $\mathcal{B}$.

We first show that $\mathcal{T}_{a, n}$ is positive if $H$ is stoquastic in $\mathcal{B}$. The fact that a Hamiltonian is stoquastic also determines some structure of its ground state wavefunciton. Precisely speaking, the Frobenius-Perron theorem for nonnegative matrices implies that the ground state of a stoquastic Hamiltonian $H$ has only nonnegative real amplitudes in the basis in which the Hamiltonian is stoquastic. Hence, all the coefficients $\langle \bm{k}|G_a\rangle$ in the onsite basis $|\bm{k} \rangle \in \mathcal{B}$ are real and nonnegative. Furthermore, $T_R$ in the onsite basis is represented as a permutation matrix, which has only nonnegative elements. Since every summand on the second line of Eq.~\eqref{eq:expectation_value_expansion} is given by a product of nonnegative real numbers, $\mathcal{T}_{a, n}$ is real and nonnegative. In fact, we can further show that $\mathcal{T}_{a, n}$ is nonzero \cite{supp}, leading to the conclusion 
\begin{equation} \label{eq:partial_rotation_positivity}
    \mathcal{T}_{a,n} > 0.
\end{equation}}

{Notably, $\mathcal{T}_{a,n}$ can also be computed within the framework of topological field theory.} Ref.~\cite{kobayashi2024extracting} {showed} that $\mathcal{T}_{a,n}$ is given by the (twisted) $n$-th higher Gauss sum $\tau_{a,n} = \sum_{b \in \mathcal{A}} S_{ab} d_b \theta_b^n$, where $S_{ab}$ is the modular $S$-matrix of the topological order \cite{wen2015theory}. More precisely, $\mathcal{T}_{a,n}$ is 
\begin{equation} \label{eq:extracting_gauss_sum}
    \mathcal{T}_{a, n} \propto e^{\frac{2\pi i}{n} h_a - 2\pi i (\frac{2}{n} + n) \frac{c_-}{24}} \tau_{a,n}  
\end{equation}
where $\propto$ means being proportional up to a positive real number, $h_a$ is the conformal weight of a state in the conformal family associated to the superselection sector $a \in \mathcal{A}$, and $c_{-}$ is the chiral central charge of the topological order. This formula has been used to numerically extract the higher central charges of bosonic topological orders, e.g., the $\mathbb{Z}_2$ toric code topological order~\cite{kobayashi2024extracting}.

Combining Eqs.~\eqref{eq:partial_rotation_positivity} and \eqref{eq:extracting_gauss_sum}, we find 
\begin{equation} \label{sign-free-condition}
    e^{\frac{2\pi i}{n} h_a - 2\pi i (\frac{2}{n} + n) \frac{c_-}{24}} \tau_{a,n} > 0 
\end{equation}
for some superselection sector $a\in \mathcal{A}$ and for all $n$. This results in several nontrivial consequences on each of the topological data {appearing} in the above equation.  

First, the anyon label $a$ should be the {vacuum} superselection sector $\mathds{1}$. {(One should not confuse the vacuum superselection sector with the trivial topological order. The ground state in the vacuum superselection sector still exhibits the nontrivial topological order~\cite{supp}.)} To see this, we note that $\tau_{a,n}$ vanishes for $n = N_{FS}$ unless $a = \mathds{1}$, where $N_{FS}$ is the Frobenius-Schur exponent \cite{ng2010congruence}, since $\tau_{a,N_{FS}} = D \sum_{b \in \mathcal{A}} S_{ab} S_{b\mathds{1}}^* = D \delta_{a,\mathds{1}}$. Thus, if a topological order is realized through a stoquastic local Hamiltonian on a cylinder, the ground state is in the {vacuum} sector. Further, we assume $c_- = 0$ since the nonzero chiral central charge is a known source of the intrinsic sign problem \cite{ringel2017quantized,kim2022chiral}. With these, the sign-free condition Eq.~\eqref{sign-free-condition} is reduced to 
\begin{equation} \label{reduced_condition}
    e^{\frac{2\pi i}{n} h_{\mathds{1}}} \tau_n > 0.
\end{equation}
From this, we further derive $h_{\mathds{1}} = 0$ and $\tau_n > 0$. To see this, note that $\tau_n$ lies in the cyclotomic field $\mathbb{Q}(e^{2 \pi i / N_{FS}})$ \cite{ng2010congruence}. Meanwhile, the phase factor $e^{\frac{2\pi i}{n} h_{\mathds{1}}}$ will not be in the same cyclotomic field for sufficiently large $n$, unless $h_{\mathds{1}} = 0$ (Introduction to the cyclotomic field and some details can be found in \cite{supp}). Hence, Eq.~\eqref{reduced_condition} requires $h_\mathds{1} = 0$. If we plug this back to Eq.~\eqref{reduced_condition}, we finally arrive at our sign-free condition, i.e., $\tau_n > 0$ for all $n \in \mathbb{N}$.

A few clarifying notes are in order. First, although $\tau_n$ is defined for any integer $n$, it exhibits periodicity with respect to the Frobenius-Schur exponent $N_{FS}$, satisfying $\tau_n = \tau_{n+N_{FS}}$. Consequently, for a given topological order, determining the presence of the intrinsic sign problem requires computing only a finite number of $\tau_n$ values. Second, while our derivation assumes that the stoquastic Hamiltonian produces a ground state within a single superselection sector, the results can be readily generalized to a superposition of multiple superselection sectors. For details, please refer to Supplementary Information \cite{supp}. Third, while we have focused on the case where the basis $\mathcal{B}$ corresponds to the tensor product of the Hilbert space at each microscopic lattice site, it is straightforward to generalize this to include a local but non-onsite basis by clustering multiple sites into a superlattice site. Similarly, if the dimensions of Hilbert spaces per each lattice site differ between sites, one can apply a generalized local unitary transformation \cite{chen2010local}, which preserves the topological order, to ensure that all lattice sites share the same local Hilbert space dimension. Since the sign-free condition depends solely on the topological data, such transformations do not affect the conclusion. 

\textbf{3. Consistency with previous results.} Our sign-free condition, $\tau_n > 0$, demonstrates nontrivial consistency with the previous results on the intrinsic sign problem \cite{hastings2015how,ringel2017quantized,smith2020intrinsic,golan2020intrinsic,kim2022chiral}, providing strong support for the validity.  

First, our criterion correctly identifies the intrinsic sign problem of the double semion topological order~\cite{hastings2015how} as the topological order has a nonpositive higher Gauss sum $\tau_2 = 0$. Second, {our sign-free condition is consistent with the results of Refs.~\cite{ringel2017quantized,kim2022chiral}, which showed that a nonvanshining chiral central charge signals intrinsic sign problem. Without assuming \textit{a priori} that $c_{-}=0$, our derivation allows us to prove that every sign-free topological order must satisfy $c_{-}=0$ mod $12$, which includes $c_{-}=0$ \cite{supp}.} Third, our classification of the intrinsic sign problem of topological orders \cite{supp} is consistent with that of Ref.~\cite{golan2020intrinsic}, by identifying the intrinsic sign problem of bosonic Laughlin states, Kitaev's spin liquid, and the Fibonacci anyon model. {Lastly, the necessary condition for a bosonic Abelian topological order to be sign-free~\cite{smith2020intrinsic} is also consistent with our sign-free condition~\cite{supp}.}

\textbf{4. Physical consequences of intrinsic sign problem.} The fact that our sign-free condition is expressed entirely in terms of topological data suggests that the intrinsic sign problem is indeed a fundamental property of topological phases of matter, being consistent with, for example, Ref.~\cite{hastings2015how}. Recall that the intrinsic sign problem refers to the sign problem that cannot be cured by any local basis transformation or, equivalently, by any finite-depth local unitary circuit. This immediately implies that the intrinsic sign problem is preserved throughout the entire region of the corresponding topological phase in the phase diagram, as any two gapped Hamiltonians within the same phase are related by a finite-depth local quantum circuit. Consequently, it is natural to expect that the intrinsic sign problem in a topological phase should be formulated solely in terms of topological data, which remain invariant under finite-depth local unitary circuits. Our result aligns well with this expectation. 
 
Interestingly, our sign-free condition Eq.~\eqref{eq:sign-free_condition} has direct implications on the compatibility {of a topological order} with gapped boundary conditions \cite{kaidi2022higher} and time-reversal symmetries \cite{geiko2023when}, {as illustrated in Fig.~\ref{fig:partial_rotation}(d)}. 

{First, every sign-free bosonic Abelian topological order must admit a gapped boundary. In general, topological orders do not necessarily admit gapped boundaries, owing to the constraints imposed by their anyonic data~\cite{levin2013protected}. The compatibility with a gapped boundary can be examined by computing their higher Gauss sums: an Abelian topological order admits a gapped boundary if and only if certain higher Gauss sums are positive \cite{kaidi2022higher}. Since our sign-free condition requires \emph{all} higher Gauss sums to be positive, the gapped boundary immediately follows~\cite{supp}. In addition, non-Abelian topological orders satisfying our sign-free condition fulfill a necessary condition for the existence of a gapped boundary gapped boundary~\cite{kaidi2022higher}.}

{Second, sign-free bosonic Abelian topological orders are time-reversal symmetric. A topological orders is said to be time-reversal symmetric if its anyonic data---such as braiding statistics and topological spins---are invariant under time-reversal conjugation, up to relabeling. For example, the toric code topological order has topological spins $\{\theta_\mathds{1}, \theta_e, \theta_m, \theta_f\} = \{1, 1, 1, -1\}$, which are invariant under the time-reversal conjugation $\theta \to \theta^*$. Recently, it was proven for Abelian cases and conjectured for non-Abelian cases that a topological order is time-reversal symmetric if and only if all its higher Gauss sums are real \cite{geiko2023when}. Because our sign-free condition enforces positivity of all higher Gauss sums, the time-reversal symmetry follows for Abelian topological orders, and holds for non-Abelian topological orders under the conjecture of Ref.~\cite{geiko2023when}. We refer the reader to \cite{supp} for further details.}

\textbf{5. Classification of topological orders with intrinsic sign problems.} Our sign-free condition is powerful enough to identify the intrinsic sign problems of almost all bosonic topological orders up to rank 12~\cite{ng2023classification}, which includes the complete list of topological orders up to rank 11. There are in total 405 distinct bosonic topological orders up to rank 12. Among these, we find that the majority, 398 out of 405, fails to satisfy our sign-free condition, indicating that they exhibit the intrinsic sign problem. Of the remaining seven, only 3 topological orders, i.e., the toric code, the $S_3$ gauge theory, and the $\mathbb{Z}_3$ gauge theory, can be shown to be sign-free, as their quantum double realizations~\cite{hu2013twisted} yield stoquastic Hamiltonians. Notably, this aligns with the findings of Ref.~\cite{isakov2011topological}, where the toric code topological order was successfully analyzed using QMC. {In \cite{supp}, we present several microscopic models that realize the sign-free topological orders, explicitly demonstrating the stoquasticity of their Hamiltonians.} For the other four topological orders---namely, the double Fibonacci, the double Ising, the twisted double Ising, and the Drinfeld center of $\tilde{R}_{\mathrm{PSU}(2)_5}$---we cannot diagnose their intrinsic sign problem, as their stoquastic realizations are unknown and they do not violate our sign-free condition simultaneously. The data of full classification of topological orders with the intrinsic sign problem is avaliable in Supplementary Information \cite{supp}. 

\textbf{6. Conclusions.} We present a novel condition for identifying the intrinsic sign problems in bosonic topological orders Eq.~\eqref{eq:sign-free_condition}, formulated in terms of the higher Gauss sums{, which have been extensively studied both in pure mathematics and physics \cite{ng2019higher,kaidi2022higher,ng2022higher,geiko2023when,kobayashi2024extracting,you2024gapped}}. Specifically, we show that a nonpositive higher Gauss sum for a given topological order indicates the presence of the intrinsic sign problem. This seemingly simple condition is not only consistent with previously established criteria~\cite{hastings2015how,ringel2017quantized,smith2020intrinsic,golan2020intrinsic,kim2022chiral}, but also extends their applicability. To demonstrate the power of this condition, we analyzed the classification data of bosonic topological orders up to rank 12~\cite{ng2023classification} and found that an overwhelming majority---398 out of 405---exhibit the intrinsic sign problem. Moreover, we uncover intriguing connections between the intrinsic sign problem, gappability of boundary theories~\cite{kaidi2022higher}, and time-reversal symmetry~\cite{geiko2023when}: sign-free topological orders not only permit gapped boundaries but are also compatible with time-reversal symmetry (for non-Abelian orders, this statement is subject to the conjecture in Ref.~\cite{geiko2023when}). 

Our results open the door to several intriguing future research directions. First, further refinement of our condition for the intrinsic sign problem would be valuable. While we have shown that nearly all bosonic topological orders exhibit intrinsic sign problems, four cases remain undetermined. Developing more precise criteria to resolve these cases would be an interesting direction for future research. 
Second, it is worth exploring whether certain topological orders face fundamental obstructions using other numerical methods, such as tensor network techniques \cite{dubail2015tensor}. If such obstructions exist, identifying the specific topological data that give rise to these limitations would be a valuable contribution. Third, extending our investigation to higher-dimensional topological orders could provide new insights into whether intrinsic sign problems persist or manifest differently in higher dimensions.

\begin{acknowledgments}
    We thank Zi-Yang Meng, Byungmin Kang, Ryohei Kobayashi, Hyukjoon Kwon and Wonjun Lee for helpful discussions. This work is financially supported by Samsung Science and Technology Foundation under Project Number SSTF-BA2002-05 and SSTF-BA2401-03, the NRF of Korea (Grants No.~RS-2023-00208291, No.~2023M3K5A1094810, No.~RS-2023-NR119931, No.~RS-2024-00410027, No.~RS-2024-00444725, No.~2023R1A2C1006542, No.~2021R1A6A1A10042944) funded by the Korean Government (MSIT), the Air Force Office of Scientific Research under Award No.~FA2386-22-1-4061, and No.~FA23862514026, and Institute of Basic Science under project code IBS-R014-D1.
\end{acknowledgments}

\bibliography{ref}

\newpage

\begin{widetext}

\begin{center}
    \Large{\bf Supplementary Information for ``Most two-dimensional bosonic topological orders forbid sign-problem-free quantum Monte Carlo: Nonpositive Gauss sum as an indicator''}
\end{center}

\tableofcontents

{\section{Review on algebraic theory of anyons}

In this section, we briefly review the category-theoretic description of two-dimensional bosonic topological orders. Although this framework is general enough to describe both bosonic and fermionic topological orders and extends to higher dimensions, we do not consider these cases here, as they lie beyond the scope of this work. For further details, we refer the reader to Refs.~\cite{lan2016theory,johnson-freyd_classification_2022,cho2023classification}.

In two dimensions, topologically ordered states can host emergent particlelike excitations with exotic statistics, known as \textit{anyons}. Anyons can even be \textit{non-Abelian}, meaning that the exchange of two identical anyons leads to a nontrivial unitary transformation of the quantum state. Since states with the same topological order share the same set of anyons, two-dimensional topological orders can be characterized by their anyonic content. Unitary modular tensor categories (UMTCs) provide the algebraic framework for describing the anyons in such systems \cite{wen2015theory}. However, two distinct topological orders may share the same set of anyons. To further distinguish them, one must also specify the \textit{chiral central charge} $c_-$, which characterizes the chiral edge theory. It is believed that the pair $(\mathcal{C}, c_-)$, consisting of a UMTC and a chiral central charge, fully characterizes a two-dimensional bosonic topological order \cite{wen2015theory}.

As the rigorous definition of UMTCs is mathematically involved, we review only the key concepts relevant to physics. For comprehensive treatments, we refer the reader to Refs.~\cite{lane_categories_1998,gelaki_tensor_2015,kong_invitation_2022}. A category consists of \textit{objects} and \textit{morphisms}---arrows describing relations between objects. A UMTC is a category equipped with additional structures on both objects and morphisms. Every object in a UMTC can be decomposed as a direct sum of \textit{simple objects}, which, up to equivalence, correspond to the anyon types.

Let $\mathcal{C}$ be a UMTC characterizing a given two-dimensional bosonic topological order $\mathsf{C}$. The distinct anyon types in $\mathsf{C}$ are labeled by the equivalence classes of simple objects $\mathcal{A} = \{\mathds{1}, a, b, \dots\}$ in $\mathcal{C}$. Among them, there is a unique label $\mathds{1} \in \mathcal{A}$ corresponding to local excitations, often referred to as the \textit{vacuum}. Local excitations are those that can be created individually by spatially local operators, whereas nontrivial anyons must be pair-created at the endpoints of a string operator.

As noted earlier, anyons exhibit exotic statistics that differ from those of local bosons or fermions. (We use the adjective \textit{local} to emphasize that the bosons and fermions considered here are local excitations.) The statistics of the anyons are encoded in a pair of unitary matrices $S$ and $T$, collectively known as the \textit{modular data} of the UMTC $\mathcal{C}$. These are $r \times r$ unitary matrices, where $r = |\mathcal{A}|$ is called the \textit{rank} of $\mathcal{C}$. In the \textit{canonical} or \textit{anyon} basis, $S$ is symmetric and $T$ is diagonal. Both rows and columns are labeled by anyon types $\mathds{1}, a, b, \dots$. The mutual braiding statistics between anyons $a$ and $b$ is given by the phase factor
\begin{equation}
    M_{a,b} \equiv e^{i \theta_{a,b}} \coloneqq \frac{S_{ab}^* S_{\mathds{1}\mathds{1}}}{S_{\mathds{1}a} S_{\mathds{1}b}}.
\end{equation}
The topological spin of an anyon $a$, which encodes the phase acquired under a $2\pi$ self-rotation, is given by
\begin{equation}
    \theta_a \equiv e^{2 \pi i s_a} \coloneqq T_{aa}.
\end{equation}
Here, $s_a$ is called the \textit{topological spin} of $a$. (In some references, $\theta_a$ is itself referred to as the topological spin.)

When multiple anyons are present, they can combine to form another anyon or a superposition of anyon types. For example, in the toric code, the fermionic anyon $\epsilon$ results from combining the $e$- and $m$-type excitations: $\epsilon = e \otimes m$. In another example, a pair of non-Abelian Majorana anyons $\psi$ can fuse into either a local fermion or the vacuum: $\psi \otimes \psi = \mathds{1} \oplus f$, where $f$ denotes a local fermionic excitation (which does not appear in bosonic systems considered here). Such properties are described by the \textit{fusion rules}:
\begin{equation}
    a \otimes b = \bigoplus_{c \in \mathcal{A}} N^{ab}_c\, c,
\end{equation}
where the nonnegative integers $N^{ab}_c$ are called \textit{fusion coefficients}. These coefficients can be computed from the $S$ matrix using the \textit{Verlinde formula}:
\begin{equation}
    N^{ab}_c = \sum_{d \in \mathcal{A}} \frac{S_{ad} S_{bd} S_{cd}^*}{S_{\mathds{1}d}}, \quad \forall a, b, c \in \mathcal{A}.
\end{equation}

Whether an anyon $a$ is Abelian or non-Abelian is determined by its \textit{quantum dimension} $d_a \coloneqq D S_{\mathds{1}a}$, where $D \coloneqq \sqrt{\sum_{a \in \mathcal{A}} d_a^2}$ is the \textit{total quantum dimension} of $\mathcal{C}$. An anyon is Abelian if and only if $d_a = 1$. The total quantum dimension $D$ appears in the \textit{topological entanglement entropy} $\gamma$, which characterizes a subleading correction to the area law \cite{kitaev_topological_2006,levin_detecting_2006}:
\begin{equation}
    \gamma = \log_2 D.
\end{equation}

Although the chiral central charge $c_-$ is an independent datum beyond the UMTC, the modular data constrain its possible values. In particular, $c_-$ is determined modulo 8 via the \textit{Gauss-Milgram formula}:
\begin{equation} \label{eq:Gauss-Milgram}
    e^{2 \pi i \frac{c_-}{8}} = \sum_{a \in \mathcal{A}} d_a^2 \theta_a.
\end{equation}
If $c_- \not\equiv 0$ mod 8, then the system necessarily hosts a chiral edge mode, which implies that the boundary cannot be gapped. In this sense, nonzero $c_-$ serves as an obstruction to boundary gappability. However, the converse does not hold: $c_- = 0$ does not guarantee that a gapped boundary exists.

It was recently shown that the \textit{higher Gauss sums} \cite{ng2019higher},
\begin{equation}
    \tau_n = \sum_{a \in \mathcal{A}} d_a^2 \theta_a^n,
\end{equation}
which generalize Eq.~\eqref{eq:Gauss-Milgram}, provide additional constraints on boundary gappability \cite{kaidi2022higher}. More precisely, a two-dimensional Abelian bosonic topological order admits a gapped boundary if and only if $\tau_n > 0$ for all $n$ such that 
\begin{equation}
    \gcd\left(n, \frac{N_{FS}}{\gcd(n, N_{FS})}\right) = 1,
\end{equation}
where $N_{FS}$, the \textit{Frobenius--Schur exponent}, is the smallest positive integer such that $T^{N_{FS}} = 1$ \cite{kaidi2022higher}. A non-Abelian bosonic topological order admits a gapped boundary only if $\tau_n > 0$ for all $n$ with $\gcd(n, N_{FS}) = 1$ \cite{kaidi2022higher}.

Moreover, it was recently proven that, at least for Abelian topological orders, all the higher Gauss sums are real if and only if the topological orders admit time-reversal symmetry \cite{geiko2023when}. For non-Abelian cases, this remains a conjecture \cite{geiko2023when}.}

{\section{Cyclotomic fields}

For the reader’s convenience, we briefly review cyclotomic fields, which will be used in our derivation of the sign-free condition.

For a given positive integer $n$, the $n$-th cyclotomic field is the field generated over $\mathbb{Q}$ by a primitive $n$-th root of unity $e^{2 \pi i / n}$. The $n$-th cyclotomic field is denoted by $\mathbb{Q}[e^{2 \pi i / n}]$.

For example, the first and the second cyclotomic fields are just identical to $\mathbb{Q}$, since both $e^{2 \pi i} = 1$ and $e^{\pi i} = -1$ are rational numbers. The simplest nontrivial example is the third cyclotomic field, whose elements have the form of $a + b e^{2 \pi i / 3}$ with rational coefficients $a$ and $b$. (Note that $e^{- 2 \pi i / 3}$ does not appear in the basis since it is redundant. Indeed, one can see that $e^{- 2 \pi i / 3} = - 1 - e^{2 \pi i / 3}$.)

In the derivation of our sign-free condition, we invoke the fact that $\tau_n$ is an element of $\mathbb{Q}[e^{2 \pi i / N_{FS}}]$ where $N_{FS}$ is the Frobenius-Schur exponent \cite{ng2010congruence}. This means that $\tau_n$ takes the form of 
\begin{equation}
    \tau_n = c_0 + c_1 e^{2 \pi i / N_{FS}} + \cdots + c_{N_{FS} - 1} e^{2 (N_{FS} - 1) \pi i / N_{FS}}, 
\end{equation}
where $c_0, \dots, c_{N_{FS} - 1}$ are rational coefficients. (The present expression is valid though it may be redundant. To be more consice, one can only involve the terms with coefficient $c_k$ where $\gcd(k, N_{FS}) = 1$.)
}

{
\section{Example of Sign-free topological orders}

In this section, we list several sign-free microscopic models that realize topological orders, which are expected to be sign-free from our condition, i.e., the toric code ($\mathbb{Z}_2$ topological order), the $\mathbb{Z}_3$ topological order, or the $S_3$ topological order \footnote{While ``$G$ gauge theory'' and ``$G$ topological order'' were previously used interchangeably for a finite group $G$, we will henceforth adopt the term ``$G$ topological order'' for consistency.}.

\subsection{Toric code model (\texorpdfstring{$\mathbb{Z}_2$}{} topological order)}

The $\mathbb{Z}_2$ topological order is famously realized in the toric code model, which is why it is often referred to as the toric code topological order. While it is trivial, we find it pedagogical to show that the ideal toric code Hamiltonian is stoquastic in the onsite basis, and thus the model is sign-free. The toric code model is defined on a square lattice with periodic boundary conditions, where qubits are placed on the links of the lattice. The Hamiltonian is given by 
\begin{equation} \label{eq:toric_code_Hamiltonian}
    H = - \sum_v A_v - \sum_p B_p
\end{equation}
where $A_v = \prod_{i \in v} \sigma^z_i$ and $B_p = \prod_{i \in p} \sigma^x_i$. Here, $v$ and $p$ label the vertices and the plaquettes of the square lattice, respectively, and $i$ labels the qubits. The notation $i \in v$ means that the qubit $i$ is on a link neighboring to the vertex $v$. Similarly, $i \in p$ means that the qubit $i$ is on a link surrounding the plaquette $p$. In the onsite qubit basis, the Pauli-$Z$ and $X$ operators $\sigma^z_i$ and $\sigma^x_i$ are represented as 
\begin{equation}
    \sigma^z_i = 
    \begin{pmatrix}
        1 & 0 \\
        0 & -1
    \end{pmatrix}, \quad 
    \sigma^x_i = 
    \begin{pmatrix}
        0 & 1 \\
        1 & 0
    \end{pmatrix}.
\end{equation}
Therefore, Eq.~\eqref{eq:toric_code_Hamiltonian} is stoquastic in the onsite qubit basis. We note that the positive diagonal matrix elements due to $\sigma^z$ do not make trouble because we can always add a multiple of identity matrix, which corresponds to the constant energy shift, to make all the diagonal elements nonpositive.

\subsection{Bose-Hubbard model on kagome lattice (\texorpdfstring{$\mathbb{Z}_2$}{} topological order)}

The $\mathbb{Z}_2$ topological order can also be realized in the Bose-Hubbard model defined on the kagome lattice~\cite{isakov2011topological}. In this paragraph, we will show that this model is also sign-free. Indeed, the hard-core Bose-Hubbard model on the kagome lattice was successfully simulated using the SSE QMC method \cite{isakov2011topological}. The Hamiltonian is given as 
\begin{equation} \label{eq:Bose-Hubbard_Hamiltonian}
    H = -t \sum_{\langle ij \rangle} [b_i^\dagger b_j + b_i b_j^\dagger] + V \sum_\mathrm{hex} (n_\mathrm{hex})^2,
\end{equation}
where $\langle ij \rangle$ labels a pair of adjacent lattice sites $i$ and $j$, `$\mathrm{hex}$' labels the hexagonal plaquettes of the kagome lattice, $b_i$ are bosonic annihilation operators, $n_\mathrm{hex} = \sum_{i \in \mathrm{hex}} b_i^\dagger b_i$, and $t$ and $V$ are positive coefficients. In Ref.~\cite{isakov2011topological}, the authors estimated the topological entanglement entropy $\gamma$ to identify the realization of the toric code topological order. \\ 

In the following, we show that Eq.~\eqref{eq:Bose-Hubbard_Hamiltonian} is indeed sign-free by mapping it to the ferromagnetic Hisenberg model, which can be easily shown to be sign-free. We employ the hard-core boson assumption, which is valid in the regime where $V/t$ is sufficiently large---precisely the parameter range in which the toric code topological order emerges~\cite{isakov2011topological}. More precisely, the hopping term can be written as 
\begin{equation}
    \begin{aligned}
        b_i^\dagger b_j + b_i b_j^\dagger &= S_i^+ S_j^- + S_i^- S_j^+ \\
        &= \frac{1}{2} \left(S_i^x S_j^x + S_i^y S_j^y\right) \\
        &= \frac{1}{2} \left(\bm{S}_i \cdot \bm{S}_j - S_i^z S_j^z\right)
    \end{aligned}
\end{equation}
where $S_i$ are the Pauli operators acting on the spin at site $i$. Note that the Heisenberg term $\bm{S}_i \cdot \bm{S}_j$ is represented as 
\begin{equation}
    \bm{S}_i \cdot \bm{S}_j = 
    \begin{pmatrix}
        1 & 0 & 0 & 0 \\
        0 & -1 & 2 & 0 \\
        0 & 2 & -1 & 0 \\
        0 & 0 & 0 & 1
    \end{pmatrix}
\end{equation}
in the onsite qubit basis. On the other hand, the term $S_i^z S_j^z$ is diagonal and can be turned into nonpositive trivially by constant energy shift. Therefore, the ferromagnetic Heisenberg Hamiltonian and the Bose-Hubbard model (realizing the toric code topological order) on a kagome lattice Eq.~\eqref{eq:Bose-Hubbard_Hamiltonian} are stoquastic.

\subsection{Trimer RVB state in Rydberg atom array (\texorpdfstring{$\mathbb{Z}_3$}{} topological order)}

The trimer resonating valence bond (tRVB) state can exhibit the $\mathbb{Z}_3$ topological order~\cite{giudice_trimer_2022}. Intriguingly, this state can be constructed by engineering a Rydberg atom array using the Rydberg blockade~\cite{giudice_trimer_2022}. On a square lattice, the tRVB state is given by~\cite{giudice_trimer_2022}
\begin{equation} \label{eq:tRVB_state}
    \ket{\mathrm{tRVB}(\theta)} = \frac{1}{\mathcal{N}(\theta)} \sum_{\{c\}} \mathcal{W}(c) \ket{c},
\end{equation}
where $\mathcal{W}(c) = (\cos \theta)^{N_\perp(c)} (\sin \theta)^{N_\parallel(c)}$ with $c$ labeling coverings with the number of bent and straight trimers $N_\perp(c)$ and $N_\parallel(c)$. See the figures in Ref.~\cite{giudice_trimer_2022} for the bent and straight trimers. Here, $\theta \in [0,\frac{\pi}{2}]$ and $\mathcal{N}(\theta) = \sqrt{\sum_{\{c\}} [\mathcal{W}(c)]^2}$. For $\theta \in [0, \frac{\pi}{2})$, the state in Eq.~\eqref{eq:tRVB_state} is expected to realize the $\mathbb{Z}_3$ topological order \cite{giudice_trimer_2022}. Since the coefficients $\mathcal{W}(c)$ in Eq.~\eqref{eq:tRVB_state} are real and nonnegative for all such $\theta$, the $\mathbb{Z}_3$ topological order can be realized as a ground state of a stoquastic Hamiltonian. Indeed, Ref.~\cite{giudice_trimer_2022} argued that a model Hamiltonian 
\begin{equation} \label{eq:tRVB_Hamiltonian}
    H = \frac{\Omega}{2} \sum_p (\ket{p;1} \bra{p;0} + h.c.) - \Delta \ket{p;1} \bra{p;1} + R_{\frac{\pi}{2}}
\end{equation}
can realize the tRVB state. Here, $p$ labels the plaquetts of the square lattice, $\ket{p;0}$ and $\ket{p;1}$ are the plaquette states occupied and unoccupied by a bent trimer, respectively. See the figures in Ref.~\cite{giudice_trimer_2022} for details. $R_{\frac{\pi}{2}}$ denotes the terms that can be obtained by 90-degree rotations from the previous terms. Around $\frac{\Delta}{\Omega} \approx 1$, the tRVB ground state similar to Eq.~\eqref{eq:tRVB_state} is expected to emerge, which in turn realizes the $\mathbb{Z}_3$ topological order. Note that the off-diagonal terms in Eq.~\eqref{eq:tRVB_Hamiltonian} are represented as 
\begin{equation}
    \ket{p;1}\bra{p;0} + h.c. = 
    \begin{pmatrix}
        0 & 1 \\
        1 & 0
    \end{pmatrix}
\end{equation}
in the onsite basis, and thus for negative $\Omega$, the Hamiltonian Eq.~\eqref{eq:tRVB_Hamiltonian} is stoquastic.

\subsection{Quantum dimer-pentamer model (\texorpdfstring{$\mathbb{Z}_3$}{} topological order)}

The $\mathbb{Z}_3$ topological order can be realized in the quantum dimer-pentamer model (QDPM)~\cite{myers2017z3}. The QDPM on the square lattice has been studied by Monte Carlo sampling of the ground state \cite{myers2017z3}. The model realizes local $\mathbb{Z}_3$ gauge contraints and thus the $\mathbb{Z}_3$ gauge theory (equivalently, the $\mathbb{Z}_3$ topological order). For each link, a dimer degree of freedom (which can equivalently be thought of as a spin-$\frac{1}{2}$ degree of freedom) is assigned. The Hamiltonian is given as 
\begin{equation} \label{eq:GQDM_Hamiltonian}
    H = - \sum_\nu t_\nu T_\nu + \sum_\nu v_\nu V_\nu,
\end{equation}
where $\nu$ labels sets of connected links, $t_\nu$ and $v_\nu$ are positive coefficients. The kinetic and the potential terms are given by 
\begin{equation}
    T_\nu = \ket{c_\nu} \bra{\bar{c}_\nu} + h.c., \quad V_\nu = \ket{c_\nu} \bra{c_\nu} + \ket{\bar{c}_\nu} \bra{\bar{c}_\nu}.
\end{equation}
where $\ket{c_\nu}$ is an allowed dimer configuration on $\nu$ and $\ket{\bar{c}_\nu}$ is its complement. See figures in Ref.~\cite{myers2017z3} for demonstrating this process. To realize the local $\mathbb{Z}_3$ gauge constraints, each vertex is constrained to be either adjacent to only one dimer or at the center of a pentamer. With the local $\mathbb{Z}_3$ gauge constraints, Eq.~\eqref{eq:GQDM_Hamiltonian} realizes the $\mathbb{Z}_3$ topological order. Note that in the onsite dimer basis, the kinetic term is represented as 
\begin{equation}
    T_\nu = 
    \begin{pmatrix}
        0 & \cdots & 1 & \cdots & 0 \\
        \vdots & & & & \\
        1 & & \ddots & & \\
        \vdots & & & & \\
        0 & & & &
    \end{pmatrix},
\end{equation}
where the $(T_\nu)_{ij} = 1$ if and only if the corresponding dimer configurations are dual to each other (See figures in Ref.~\cite{myers2017z3} for the dual dimer configurations), while the potential term is represented by a diagonal matrix
\begin{equation}
    V_v = 
    \begin{pmatrix}
        1 & 0 & \cdots & 0 \\
        0 & 1 & \cdots & 0 \\
        \vdots & \vdots & \ddots & \vdots \\
        0 & 0 & \cdots & 1
    \end{pmatrix}.
\end{equation}
This implies that the Hamiltonian is stoquastic in that basis. Indeed, the authors of Ref.~\cite{myers2017z3} Monte-Carlo sampled the ground state of Eq.~\eqref{eq:GQDM_Hamiltonian} on a 128 by 128 square lattice, and estimated the expectation values of several observables successfully.

\subsection{Quantum double models (\texorpdfstring{$\mathbb{Z}_2$}{}, \texorpdfstring{$\mathbb{Z}_3$}{}, and \texorpdfstring{$S_3$}{} topological orders)}

While artificial, microscopic Hamiltonians realizing $G$ topological orders for finite groups $G$, including the $\mathbb{Z}_2$, $\mathbb{Z}_3$, and $S_3$ topological orders, can be systematically constructed \cite{kitaev2003fault}. They are called as quantum double models, and we will show that the quantum double models for realizing $\mathbb{Z}_2$, $\mathbb{Z}_3$, and $S_3$ topological orders are indeed stoquatic in the onsite basis below. In fact, the toric code model corresponds to the $\mathbb{Z}_2$ quantum double model.

The quantum double model~\cite{kitaev2003fault} is defined on a square lattice with directed links. A discrete group $G$ is given as an input. For each link, a local Hilbert space $\mathbb{C}[G]$ is assigned, where the canonical basis states are labeled by the group elements of $G$. That is, an arbitrary state in the local onsite Hilbert space can be written as $\sum_{g \in G} c_g \ket{g}$ where $c_g$ are complex coefficients. The Hamiltonian is given by 
\begin{equation} \label{eq:untwisted_quantum_double_Hamiltonian}
    H = - \sum_v A_v - \sum_p B_p
\end{equation}
where $v$ and $p$ are vertices and plaquettes of the lattice. Let $\mathcal{B} = \{\ket{\bm{k}}\}$ be the onsite basis where each basis state $\ket{\bm{k}}$ is given by the tensor product of the local states with definite group element labels, i.e., $\ket{\bm{k}} = \ket{g_1} \otimes \ket{g_2} \otimes \dots$ where $g_1, g_2, \dots \in G$. The plaquette term $B_p$ acts on the basis state as 
\begin{equation}
    B_p \ket{\bm{k}} = 
    \begin{cases}
        \ket{\bm{k}}, \quad &\text{if $g_{i_1} g_{i_2} g_{i_3} g_{i_4} = 1$} \\
        0, \quad &\text{otherwise}
    \end{cases}
\end{equation}
where $i_1$, $i_2$, $i_3$, and $i_4$ label the links around the plaquette $p$. (When $G$ is non-Abelian, we need to fix an ordering of the edges so that the action of $B_p$ to be well-defined.) Thus, the plaquette term $B_p$ is diagonal, which we can always make nonpositive by constant energy shift. The vertex term $A_v$ is written as 
\begin{equation}
    A_v = \sum_{g \in G} A^g_v,
\end{equation}
where $A^g_v$ multiplies the group labels on the edges adjacent to $v$ by $g$ (or by $g^{-1}$, depending on the orientation of the edges). Hence, $A_v$ simply takes $\ket{\bm{k}}$ to a linear combination of other basis states, with no additional phase factor. Thus, $A_v$ is also nonpositive in the onsite basis. For example, let $G = \mathbb{Z}_3 = (\{0,1,2\}, \bullet + \bullet \mod 3)$. The vertex term is then given by 
\begin{equation}
    A_v = A^0_v + A^1_v + A^2_v.
\end{equation}
By definition, the first term $A^0_v$ is the identity operator. The other terms $A^1_v$ and $A^2_v$ are given by tensor products of the generalized Pauli-$X$ operator, which is represented in the onsite basis as
\begin{equation}
    X = 
    \begin{pmatrix}
        0 & 0 & 1 \\
        0 & 1 & 0 \\
        1 & 0 & 0
    \end{pmatrix}.
\end{equation}
In conclusion, both $\bra{\bm{k}} A_v \ket{\bm{k}'}$ and $\bra{\bm{k}} B_p \ket{\bm{k}}$ are real and nonnegative, and thus Eq.~\eqref{eq:untwisted_quantum_double_Hamiltonian} is stoquastic in $\mathcal{B}$. Notably, it is mathematically proven that the $G$ topological orders always have positive higher Gauss sums \cite{ng2019higher}, which strengthens the validity of our result.

\section{Sign Problem in Double Semion Topological Order}

Proving that a model has an intrinsic sign problem---meaning it remains non-stoquastic in any local basis---is generally difficult without invoking known sign-free conditions, such as ours. This is due to the intractability of scanning all local bases in large systems \footnote{The problem of finding a basis on which a given Hamiltonian is stoquastic, if exists, is proven to be NP-complete \cite{marvian2019computational}.}, highlighting the practical importance of sign-free conditions.  

Below, we examplify a model with a known intrinsic sign problem: the twisted $\mathbb{Z}_2$ quantum double model realizing the double semion topological order. We will show that the Hamiltonian is not stoquastic in the onsite basis. This does not prove that the double semion topological order has an intrinsic sign problem because there may be another sign-free model realizing the double semion topological order (in fact, the double semion topological order is already proven to exhibit an intrinsic sign problem and thus the absence of such sign-free model is guaranteed). Still, this model serves as an example of a topologically ordered system with a sign problem.

The twisted $\mathbb{Z}_2$ quantum double model, equivalently the double semion model, is defined on an oriented triangular lattice---or more generally, on a directed planar graph consisting solely of triangular faces---with qubits residing on the edges. To fix the orientation, we impose an ordering on the vertices such that each edge is directed from the latter to the former vertex in the ordering. The Hamiltonian is given by 
\begin{equation} \label{eq:twisted_quantum_double}
    H = - \sum_v A_v - \sum_p B_p
\end{equation}
where $v$ and $p$ label the vertices and the plaquettes of the lattice. Though Eq.~\eqref{eq:twisted_quantum_double} looks similar to the toric code Hamiltonian Eq.~\eqref{eq:toric_code_Hamiltonian}, we will see that the vertex term $A_v$ in Eq.~\eqref{eq:twisted_quantum_double} now has nontrivial phase factors and thus the Hamiltonian is nonstoquastic in the standard onsite basis. The plaquette term is defined as 
\begin{equation}
    B_p \ket{\bm{k}} = 
    \begin{cases}
        \ket{\bm{k}}, \quad &\text{if $g_1 g_2 g_3 = 0$,} \\
        0, \quad &\text{otherwise,}
    \end{cases}
\end{equation}
where $g_1$, $g_2$, and $g_3$ are the group elements of $\mathbb{Z}_2 = (\{0,1\}, \bullet + \bullet \mod 2)$ of the qubits on the edges surrounding $p$ (recall that the qubit states can be represented as the group elements of $\mathbb{Z}_2$), and $\ket{\bm{k}} = \bigotimes_i \ket{g_i}$ are the onsite basis states. Therefore, $B_p$ is diagonal and does not cause the sign problem in this onsite basis. 

However, the vertex term $A_v$ is defined as 
\begin{equation}
    A_v = \sum_{g \in \mathbb{Z}_2} A^g_v.
\end{equation}
Here, the term $A^g_v$ replaces the vertex label $v$ to a new label $v'$ such that $v'$ goes earlier than $v$ but latter than any other labels that goes earlier than $v$ and $[vv'] = g$. Here, we have denoted, and will denote, the edge that goes from $v$ to $v'$ by $[vv']$. Consider a trivalent vertex $v_3$, which is connected to the vertices labeled by $v_1$, $v_2$, and $v_4$. Without loss of generality, assume that the ordering of the vertices is given by $v_1 < v_2 < v_3 < v_4$. For the trivalent vertex $v_3$, the action of $A_{v_3}^g$ is given by 
\begin{multline} \label{eq:vertex_term_action}
    A_{v_3}^g \ket{[v_1v_2],[v_2v_4],[v_1v_4],[v_1v_3],[v_2v_3],[v_3v_4]} = \delta_{[v_3'v_3],g} \alpha([v_1v_2],[v_2v_3'],[v_3'v_3]) \\
    \times \alpha([v_2v_3'],[v_3'v_3],[v_3v_4]) \alpha([v_1v_3'],[v_3'v_3],[v_3v_4])^{-1} \ket{[v_1v_2],[v_2v_4],[v_1v_4],[v_1v_3'],[v_2v_3'],[v_3'v_4]},
\end{multline}
where $\alpha(g_1,g_2,g_3) \in H^3[\mathbb{Z}_2, U(1)]$ is a normalized 3-cocycle. (The group element on the edge from $v'$ to $v$ is denoted by $[vv']$.) For the double semion model, the normalized 3-cocycles are given by 
\begin{equation}
    \alpha(g_1,g_2,g_3) = 
    \begin{cases}
        -1, &\quad g_1 = g_2 = g_3 = 1, \\
        1, &\quad \text{otherwise.}
    \end{cases}
\end{equation}
Explicitly, if we view the matrix representation of $A_{v}$ in the standard basis as a $2 \times 2$ block matrix, the lower-left block is given by 
\begingroup
\footnotesize
\begin{equation}
    [A_{v}]_{21} = \left(
    \begin{array}{cccccccccccccccccccccccccccccccc}
     0 & 0 & 0 & 0 & 0 & 0 & 0 & 0 & 0 & 0 & 0 & 0 & 0 & 0 & 0 & 0 & 0 & 0 & 0 & 0 & 0 & 0 & 0 & 0 & 1 & 0 & 0 & 0 & 0 & 0 & 0 & 0 \\
     0 & 0 & 0 & 0 & 0 & 0 & 0 & 0 & 0 & 0 & 0 & 0 & 0 & 0 & 0 & 0 & 0 & 0 & 0 & 0 & 0 & 0 & 0 & 0 & 0 & -1 & 0 & 0 & 0 & 0 & 0 & 0 \\
     0 & 0 & 0 & 0 & 0 & 0 & 0 & 0 & 0 & 0 & 0 & 0 & 0 & 0 & 0 & 0 & 0 & 0 & 0 & 0 & 0 & 0 & 0 & 0 & 0 & 0 & 1 & 0 & 0 & 0 & 0 & 0 \\
     0 & 0 & 0 & 0 & 0 & 0 & 0 & 0 & 0 & 0 & 0 & 0 & 0 & 0 & 0 & 0 & 0 & 0 & 0 & 0 & 0 & 0 & 0 & 0 & 0 & 0 & 0 & -1 & 0 & 0 & 0 & 0 \\
     0 & 0 & 0 & 0 & 0 & 0 & 0 & 0 & 0 & 0 & 0 & 0 & 0 & 0 & 0 & 0 & 0 & 0 & 0 & 0 & 0 & 0 & 0 & 0 & 0 & 0 & 0 & 0 & 1 & 0 & 0 & 0 \\
     0 & 0 & 0 & 0 & 0 & 0 & 0 & 0 & 0 & 0 & 0 & 0 & 0 & 0 & 0 & 0 & 0 & 0 & 0 & 0 & 0 & 0 & 0 & 0 & 0 & 0 & 0 & 0 & 0 & -1 & 0 & 0 \\
     0 & 0 & 0 & 0 & 0 & 0 & 0 & 0 & 0 & 0 & 0 & 0 & 0 & 0 & 0 & 0 & 0 & 0 & 0 & 0 & 0 & 0 & 0 & 0 & 0 & 0 & 0 & 0 & 0 & 0 & 1 & 0 \\
     0 & 0 & 0 & 0 & 0 & 0 & 0 & 0 & 0 & 0 & 0 & 0 & 0 & 0 & 0 & 0 & 0 & 0 & 0 & 0 & 0 & 0 & 0 & 0 & 0 & 0 & 0 & 0 & 0 & 0 & 0 & -1 \\
     0 & 0 & 0 & 0 & 0 & 0 & 0 & 0 & 0 & 0 & 0 & 0 & 0 & 0 & 0 & 0 & -1 & 0 & 0 & 0 & 0 & 0 & 0 & 0 & 0 & 0 & 0 & 0 & 0 & 0 & 0 & 0 \\
     0 & 0 & 0 & 0 & 0 & 0 & 0 & 0 & 0 & 0 & 0 & 0 & 0 & 0 & 0 & 0 & 0 & 1 & 0 & 0 & 0 & 0 & 0 & 0 & 0 & 0 & 0 & 0 & 0 & 0 & 0 & 0 \\
     0 & 0 & 0 & 0 & 0 & 0 & 0 & 0 & 0 & 0 & 0 & 0 & 0 & 0 & 0 & 0 & 0 & 0 & -1 & 0 & 0 & 0 & 0 & 0 & 0 & 0 & 0 & 0 & 0 & 0 & 0 & 0 \\
     0 & 0 & 0 & 0 & 0 & 0 & 0 & 0 & 0 & 0 & 0 & 0 & 0 & 0 & 0 & 0 & 0 & 0 & 0 & 1 & 0 & 0 & 0 & 0 & 0 & 0 & 0 & 0 & 0 & 0 & 0 & 0 \\
     0 & 0 & 0 & 0 & 0 & 0 & 0 & 0 & 0 & 0 & 0 & 0 & 0 & 0 & 0 & 0 & 0 & 0 & 0 & 0 & -1 & 0 & 0 & 0 & 0 & 0 & 0 & 0 & 0 & 0 & 0 & 0 \\
     0 & 0 & 0 & 0 & 0 & 0 & 0 & 0 & 0 & 0 & 0 & 0 & 0 & 0 & 0 & 0 & 0 & 0 & 0 & 0 & 0 & 1 & 0 & 0 & 0 & 0 & 0 & 0 & 0 & 0 & 0 & 0 \\
     0 & 0 & 0 & 0 & 0 & 0 & 0 & 0 & 0 & 0 & 0 & 0 & 0 & 0 & 0 & 0 & 0 & 0 & 0 & 0 & 0 & 0 & -1 & 0 & 0 & 0 & 0 & 0 & 0 & 0 & 0 & 0 \\
     0 & 0 & 0 & 0 & 0 & 0 & 0 & 0 & 0 & 0 & 0 & 0 & 0 & 0 & 0 & 0 & 0 & 0 & 0 & 0 & 0 & 0 & 0 & 1 & 0 & 0 & 0 & 0 & 0 & 0 & 0 & 0 \\
     0 & 0 & 0 & 0 & 0 & 0 & 0 & 0 & -1 & 0 & 0 & 0 & 0 & 0 & 0 & 0 & 0 & 0 & 0 & 0 & 0 & 0 & 0 & 0 & 0 & 0 & 0 & 0 & 0 & 0 & 0 & 0 \\
     0 & 0 & 0 & 0 & 0 & 0 & 0 & 0 & 0 & -1 & 0 & 0 & 0 & 0 & 0 & 0 & 0 & 0 & 0 & 0 & 0 & 0 & 0 & 0 & 0 & 0 & 0 & 0 & 0 & 0 & 0 & 0 \\
     0 & 0 & 0 & 0 & 0 & 0 & 0 & 0 & 0 & 0 & -1 & 0 & 0 & 0 & 0 & 0 & 0 & 0 & 0 & 0 & 0 & 0 & 0 & 0 & 0 & 0 & 0 & 0 & 0 & 0 & 0 & 0 \\
     0 & 0 & 0 & 0 & 0 & 0 & 0 & 0 & 0 & 0 & 0 & -1 & 0 & 0 & 0 & 0 & 0 & 0 & 0 & 0 & 0 & 0 & 0 & 0 & 0 & 0 & 0 & 0 & 0 & 0 & 0 & 0 \\
     0 & 0 & 0 & 0 & 0 & 0 & 0 & 0 & 0 & 0 & 0 & 0 & -1 & 0 & 0 & 0 & 0 & 0 & 0 & 0 & 0 & 0 & 0 & 0 & 0 & 0 & 0 & 0 & 0 & 0 & 0 & 0 \\
     0 & 0 & 0 & 0 & 0 & 0 & 0 & 0 & 0 & 0 & 0 & 0 & 0 & -1 & 0 & 0 & 0 & 0 & 0 & 0 & 0 & 0 & 0 & 0 & 0 & 0 & 0 & 0 & 0 & 0 & 0 & 0 \\
     0 & 0 & 0 & 0 & 0 & 0 & 0 & 0 & 0 & 0 & 0 & 0 & 0 & 0 & -1 & 0 & 0 & 0 & 0 & 0 & 0 & 0 & 0 & 0 & 0 & 0 & 0 & 0 & 0 & 0 & 0 & 0 \\
     0 & 0 & 0 & 0 & 0 & 0 & 0 & 0 & 0 & 0 & 0 & 0 & 0 & 0 & 0 & -1 & 0 & 0 & 0 & 0 & 0 & 0 & 0 & 0 & 0 & 0 & 0 & 0 & 0 & 0 & 0 & 0 \\
     1 & 0 & 0 & 0 & 0 & 0 & 0 & 0 & 0 & 0 & 0 & 0 & 0 & 0 & 0 & 0 & 0 & 0 & 0 & 0 & 0 & 0 & 0 & 0 & 0 & 0 & 0 & 0 & 0 & 0 & 0 & 0 \\
     0 & 1 & 0 & 0 & 0 & 0 & 0 & 0 & 0 & 0 & 0 & 0 & 0 & 0 & 0 & 0 & 0 & 0 & 0 & 0 & 0 & 0 & 0 & 0 & 0 & 0 & 0 & 0 & 0 & 0 & 0 & 0 \\
     0 & 0 & 1 & 0 & 0 & 0 & 0 & 0 & 0 & 0 & 0 & 0 & 0 & 0 & 0 & 0 & 0 & 0 & 0 & 0 & 0 & 0 & 0 & 0 & 0 & 0 & 0 & 0 & 0 & 0 & 0 & 0 \\
     0 & 0 & 0 & 1 & 0 & 0 & 0 & 0 & 0 & 0 & 0 & 0 & 0 & 0 & 0 & 0 & 0 & 0 & 0 & 0 & 0 & 0 & 0 & 0 & 0 & 0 & 0 & 0 & 0 & 0 & 0 & 0 \\
     0 & 0 & 0 & 0 & 1 & 0 & 0 & 0 & 0 & 0 & 0 & 0 & 0 & 0 & 0 & 0 & 0 & 0 & 0 & 0 & 0 & 0 & 0 & 0 & 0 & 0 & 0 & 0 & 0 & 0 & 0 & 0 \\
     0 & 0 & 0 & 0 & 0 & 1 & 0 & 0 & 0 & 0 & 0 & 0 & 0 & 0 & 0 & 0 & 0 & 0 & 0 & 0 & 0 & 0 & 0 & 0 & 0 & 0 & 0 & 0 & 0 & 0 & 0 & 0 \\
     0 & 0 & 0 & 0 & 0 & 0 & 1 & 0 & 0 & 0 & 0 & 0 & 0 & 0 & 0 & 0 & 0 & 0 & 0 & 0 & 0 & 0 & 0 & 0 & 0 & 0 & 0 & 0 & 0 & 0 & 0 & 0 \\
     0 & 0 & 0 & 0 & 0 & 0 & 0 & 1 & 0 & 0 & 0 & 0 & 0 & 0 & 0 & 0 & 0 & 0 & 0 & 0 & 0 & 0 & 0 & 0 & 0 & 0 & 0 & 0 & 0 & 0 & 0 & 0 \\
    \end{array} \right)
\end{equation}
\endgroup
which is sufficient to show that the matrix representation of $A_v$ contains negative off-diagonal matrix elements. Therefore, the Hamiltonian Eq.~\eqref{eq:twisted_quantum_double} is nonstoquastic in the onsite basis. 
}

{
\section{Relation to gapped boundaries and time-reversal symmetry}

In the main text we stated three claims:

\begin{theorem} \label{thm:gapped_boundary}
    Every sign-free bosonic Abelian topological order admits a gapped boundary.
\end{theorem}

\begin{theorem} \label{thm:time-reversal_abelian}
    Every sign-free bosonic Abelian topological order is compatible with time-reversal symmetry.
\end{theorem}

\begin{conjecture}
\label{conj:time-reversal_nonabelian}
    Every sign-free bosonic non-Abelian topological order is compatible with time-reversal symmetry.
\end{conjecture}

We now prove Theorems~\ref{thm:gapped_boundary} and \ref{thm:time-reversal_abelian} and explain why Conjecture~\ref{conj:time-reversal_nonabelian} follows.

Kaidi \emph{et al.}~\cite{kaidi2022higher} showed that a bosonic Abelian topological order admits a gapped boundary if and only if its $n$-th higher Gauss sum $\tau_n$ obeys
\begin{equation}
    \tau_n>0,
    \qquad
    \gcd\left(n, \frac{N_{\mathrm{FS}}}{\gcd(n,N_{\mathrm{FS}})}\right) = 1,
\end{equation}
where $N_{\mathrm{FS}}$ is the Frobenius–Schur exponent. Our sign-free condition is stronger: it requires $\tau_n>0$ for \emph{all} $n\in\mathbb{N}$. Hence the criterion of Kaidi \emph{et al.}~\cite{kaidi2022higher} is automatically satisfied, proving Theorem~\ref{thm:gapped_boundary}.

Geiko and Moore~\cite{geiko2023when} proved that an Abelian topological order is compatible with time-reversal symmetry if and only if $\tau_n\in\mathbb{R}$ for all $n\in\mathbb{N}$. Because sign-free orders satisfy $\tau_n>0$ for every $n$, the reality condition is immediate; thus Theorem~\ref{thm:time-reversal_abelian} follows.

Geiko and Moore~\cite{geiko2023when} further conjectured that non-Abelian topological orders are compatible with time-reversal symmetry if and only if $\tau_n\in\mathbb{R}$ for all $n$. Applying the same logic as above yields Conjecture~\ref{conj:time-reversal_nonabelian}.
}

\section{Proof for \texorpdfstring{$\mathcal{T}_{a, n} > 0$}{}}

In the main text, we have shown that the ground state expectation value of the partial rotation operator $T_R$ is nonnegative real when the Hamiltonian is stoquastic in the onsite basis. In this section, we will prove that the expectation value $\mathcal{T}_a(2 \pi / n)$ is indeed positive real for all superselection sectors $a$ and for all positive integers $n$.

Note that there always exists a finite-depth local unitary quantum circuit $U$ for a given onsite basis state $\ket{\bm{k}} = \bigotimes_{i \in \Lambda} \ket{k_i}$ such that $U \ket{\bm{k}} = \bigotimes_{i \in \Lambda} \ket{0}_i$. Importantly, the resulting state $U \ket{\bm{k}}$ is invariant under $T_R(\varphi)$ for any $\varphi$, which is consistent with the underlying lattice. Explicitly, such an operator $U$ would be given by $U = \prod_{i \in \Lambda} X_i^{-k_i}$, where $X_i$ is the shift operator acting on the qudit at the site $i$ and $k_i$ labels the local qudit state. Moreover, such $U$ does not introduce any nontrivial phase factor to the amplitudes of $\ket{G_a}$ in the onsite basis. Thus, if we choose an onsite basis state $\ket{\bm{k}'}$ that has a finite overlap with $\ket{G_a}$ and consider finding $U$ for this state, i.e., $U \ket{\bm{k}'} = \bigotimes_{i \in \Lambda} \ket{0}_i$, we have
\begin{equation}
    \bra{G_a} U^\dagger T_R(2\pi / n) U \ket{G_a} = |\alpha_{\bm{k}'}|^2 + \cdots > 0,
\end{equation}
where $\alpha_{\bm{k}'}$ is the amplitudes of $\ket{\bm{k}'}$ in $\ket{G_a}$.

Next, we note that since $U$ is a finite-depth local unitary quantum circuit, $U\ket{G_a}$ and $\ket{G_a}$ are in the same gapped quantum phase. Furthermore, given that $H$ is gapped, $U\ket{G_a}$ is the ground state of $H' = U H U^\dagger$, which is still stoquastic. In consequence, we have
\begin{equation}
    \bra{G_a} U^\dagger T_R(2\pi / n) U \ket{G_a} \propto \bra{G_a} T_R(2 \pi / n) \ket{G_a} \neq 0.
\end{equation}
As we wll explain in Sec.~\ref{sec:superposition}, even when $U$ permutes different superselection sectors, our derivation remains valid for those cases as well.

\section{Proof for \texorpdfstring{$c_- = 0$}{} mod 12}

We prove that every sign-free topological order must have $c_- = 0$ mod $12$ without assuming the results of \cite{ringel2017quantized,kim2022chiral}. Since $c_- = 0$ mod 12 includes $c_- = 0$, this augments the consistency of our sign-free condition with the previous findings. 

As we derived in the main text, any sign-free topological order satisfies
\begin{equation} \label{eq:cond}
    \exp\left\{2 \pi i \left[\frac{h_\mathds{1}}{n} - \frac{c_- / 12}{n} + \frac{n c_-}{24}\right]\right\} \tau_n > 0.
\end{equation}
Since $\tau_n$ has the periodicity $N_{FS}$, i.e., $\tau_n = \tau_{n + N_{FS}}$, and $\tau_{N_{FS}} = \sum_a d_a^2 > 0$, we have $\tau_{m N_{fS}} = \sum_a d_a^2 > 0$ for any positive integer $m$. Here, the Frobenius-Schur exponent $N_{FS}$ is the smallest positive integer such that $T^{N_{FS}} = 1$.  By plugging $n = m N_{FS}$ into Eq.~\eqref{eq:cond}, we get
\begin{equation} \label{eq:h-c+c}
    \exp\left\{2 \pi i \left[\frac{h_\mathds{1}}{m N_{FS}} - \frac{c_- / 12}{m N_{FS}} + \frac{m N_{FS} c_-}{24}\right]\right\} > 0.
\end{equation}
Now, assuming $12 h_\mathds{1} \neq c_-$ leads to a contradiction. To see this, note that we can make $m$ arbitrarily large so that Eq.~\eqref{eq:h-c+c} accompany a nontrivial phase factor so that the expression in Eq.~\eqref{eq:h-c+c} cannot be positive real. Therefore, we must have $12 h_\mathds{1} = c_-$. Plus, note that $h_\mathds{1}$ is a nonnegative integer \cite{difrancesco1997conformal}. This implies that $c_-$ is a nonnegative integer multiple of $12$. Therefore, we conclude that $c_- = 0$ mod $12$.

\section{\label{sec:h=0}Proof for \texorpdfstring{$h_\mathds{1} = 0$}{}}

In the main text, we have argued that for $e^{2 \pi i h_\mathds{1} / n} \tau_n > 0$ to be satisfied for all $n$, we must have $h_\mathds{1} = 0$. In this section, we will prove this by contradiction, utilizing properties of cyclotomic fields.

The cyclotomic field is the field given by adjoining a root of unity to the field of rational numbers $\mathbb{Q}$. When we adjoin the primitive $N$-th root of unity, the resulting cyclotomic field is called the $N$-th cyclotomic field. For example, the third cyclotomic field $\mathbb{Q}(e^{2\pi i / 3})$ has elements of the form $p + q e^{2\pi i / 3} + r e^{4 \pi i / 3}$, where $p$, $q$, and $r$ are rational coefficients.

Suppose that $h_\mathds{1}$ is a positive integer. (Note that the conformal weight of any state in the vacuum sector is always given by a nonnegative integer \cite{difrancesco1997conformal}.) If we choose $n$ to be larger than $N_{FS} h_\mathds{1}$, then $e^{2 \pi i h_\mathds{1} / n}$ cannot belong to the $N_{FS}$-th cyclotomic field. In contrast, it is known that $\tau_n$ lies within the $N_{FS}$-th cyclotomic field \cite{ng2010congruence}. Consequently, the phase factor $e^{2 \pi i h_\mathds{1} / n}$ cannot be canceled by $\tau_n$, and the total expression $e^{2 \pi i h_\mathds{1} / n} \tau_n$ cannot be positive real. Therefore, we must have $h_\mathds{1} = 0$. 

\section{\label{sec:superposition}Superposition of different superselection sectors}

In the main text, we have derived the sign-free condition $\tau_n > 0$ for the cases where the ground state is given by a single superselection sector. In this section, we derive the same result for the cases where the ground state is given by a linear combination of different superselection sectors. Generally, the ground state with nonnegative real amplitudes in a given basis has the form of
\begin{equation}
    \ket{G} = \sum_a \alpha_a \ket{G_a},
    \label{mixed_sectors}
\end{equation}
where $\alpha_a$ are some fixed coefficients. We will see that even for the ground state Eq.~\eqref{mixed_sectors} our sign-free condition can be derived.

{
Before delving into details, let us briefly review what do we mean by ``superselection sector'' throughout this work. Since we consider only the cylindrical geometry in this work, we focus on the superselection sectors on a cylinder. On a cylinder, the degenerate ground states of a topological order are labeled by anyon types. Two ground states are said to belong to different superselection sectors if their anyon labels are distinct. Physically, the ground state labeled by anyon $a$ corresponds to a state in which an anyon flux $a$ is threaded through the cylinder. Such a state can be created by pair-producing anyon $a$ and its antiparticle inside the bulk, and then adiabatically moving them to spatial infinity. See Fig.~\ref{fig:sector} for an illustration. If two ground states have different anyon fluxes, they are said to belong to different superselection sectors.

\begin{figure}[t]
    \centering
    \includegraphics[width=\linewidth]{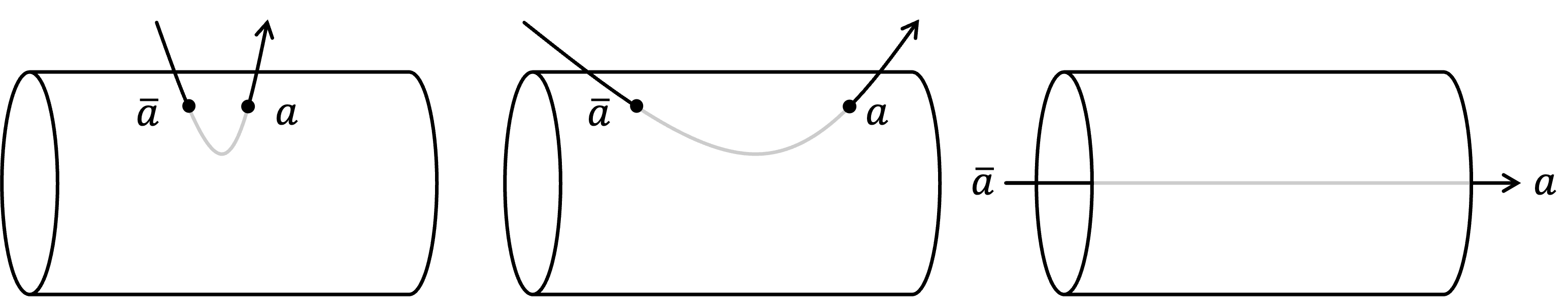}
    \caption{Illustration for the construction of the ground state in the superselection sector labeled by an anyon $a$. Starting with the trivial superselection sector, denoted by $\mathds{1}$, we first create a particle-antiparticle pair in the bulk. The anyon $a$ and its anti-anyon $\bar{a}$ are adiabatically dragged toward each end of the cylinder. When $a$ and $\bar{a}$ move to the spatial infinity, the state is now threaded the anyon flux $a$ through the cylinder and in the superselection sector labeled by $a$.}
    \label{fig:sector}
\end{figure}
}

We first propose that we can assume that $T_R$ does not mix different superselection sectors without loss of generality. Note that a cylinder can be understood as two sub-cylinders glued together. Consider $\ket{G_a}$ defined on a cylinder. If we bipartition the cylinder into $L$ and $R$ by adiabatically turning off the Hamiltonian terms that overlap the cut, the states $\ket{G_a; L}$ and $\ket{G_a; R}$ in the same sector $a$ are obtained for each subregion. Then, $T_R$ acts as the translation operator on $\ket{G_a; R}$. Now, we assume the translation symmetry along the circumference of the cylinder. As a consequence, the state remains a ground state after the translation induced by $T_R$. When the system has an anyon permutation symmetry, the translation may permute the superselection sector of $\ket{G_a; R}$. However, since there are only a finite number of superselection sectors in a topological order, there exists a positive integer $M$ smaller than the rank of the topological order such that the translation of $M$ unit cells leaves the superselection sector invariant after the translation. Consequently, if we consider a sufficiently large system, such an effect can be neglected. Finally, we merge the subsystems $L$ and $R$ by adiabatically restoring the terms that we turned off at the first step. As a result, we get the same state $\ket{G_a}$ up to some phase factor, which gives the topological data. Since $T_R$ does not scramble different superselection sectors under the assumptions above, the expectation value of $T_R(2 \pi / n)$ is given by 
\begin{equation}
    \bra{G} T_R\left(2 \pi / n\right) \ket{G} = \sum_a |\alpha_a|^2 e^{2 \pi i h_a / n} \tau_{a, n} > 0,
    \label{mixed_inequality}
\end{equation}
where we have redefined the positive coefficients $|\alpha_a|^2$ properly. Though we have assumed the translation symmetry to explicitly argue that the partial rotation operator does not scramble different superselection sectors, this is not absolutely necessary. Our argument remains valid as long as the system is well-behaved so that the partial rotation acts as an anyon permutation at most.

Next, we claim that Eq.~\eqref{mixed_inequality} is satisfied for all positive integers $n$ only if $\alpha_\mathds{1} \neq 0$. To see this, note that $\tau_{a, N_{FS}} = 0$ for all $a \neq \mathds{1}$ since $\tau_{a, N_{FS}} = \sum_b S_{ab} d_a \propto \delta_{a, \mathds{1}}$. Consequently, if $\alpha_\mathds{1} = 0$, Eq.~\eqref{mixed_inequality} cannot be satisfied for $n = N_{FS}$. Therefore, we conclude that $\alpha_\mathds{1} \neq 0$. Moreover, since $\tau_{a, n}$ for $a \neq \mathds{1}$ vanish for all $n$ equal to a multiple of $N_{FS}$, we must have $h_\mathds{1} = 0$ for Eq.~\eqref{mixed_inequality} to hold. For details on this claim, we refer to Sec.~\ref{sec:h=0}. Therefore, the inequality in Eq.~\eqref{mixed_inequality} is reduced to
\begin{equation}
    |\alpha_\mathds{1}|^2 \tau_n + \sum_{a: a \neq \mathds{1}} |\alpha_a|^2 e^{2 \pi i h_a / n} \tau_{a, n} > 0.
    \label{mixed_inequality2}
\end{equation}
Lastly, we will derive $\tau_n > 0$ from Eq.~\eqref{mixed_inequality2}. To do so, we note that the first term in Eq.~\eqref{mixed_inequality2} is invariant under $n \to n + k N_{FS}$ for any integer $k$. For a given $n$, consider $\sum_{a: a \neq \mathds{1}} |\alpha_a|^2 e^{2 \pi i h_a / n} \tau_{a, n}$. Note that we can take arbitrarily large $n$ without loss of generality, by adding a sufficiently large integer multiple of $N_{FS}$, since $\tau_{a, n}$ is $N_{FS}$-periodic for any $a$. We will see that $\sum_{a: a \neq \mathds{1}} |\alpha_a|^2 e^{2 \pi i h_a / n} \tau_{a, n}$ and $\sum_{a: a \neq \mathds{1}} |\alpha_a|^2 e^{2 \pi i h_a / (n + N_{FS})} \tau_{a, n}$ have different phase factors for a sufficiently large $n$ or at least one of them vanishes, proving $\tau_n > 0$ for all $n \in \mathbb{N}$ from Eq.~\eqref{mixed_inequality2}. If one of them vanishes, then this directly implies $\tau_n > 0$ for that $n$. If none of them vanishes, it is implied by
\begin{equation}
    e^{2 \pi i h_a / n} = e^{2 \pi i h_a / (n + N_{FS})} e^{2 \pi i h_a N_{FS} / n (n + N_{FS})}
\end{equation}
that $\sum_{a \neq \mathds{1}} |\alpha_a|^2 e^{2 \pi i h_a / n} \tau_{a, n}$ is not in the same cyclotomic field as $\sum_{a \neq \mathds{1}} |\alpha_a|^2 e^{2 \pi i h_a / (n + N_{FS})} \tau_{a, n}$ for a large enough $n$. Therefore, we conclude that Eq.~\eqref{mixed_inequality2} implies $\tau_n > 0$ for $n \in \mathbb{N}$.

\section{Classification of topological orders with intrinsic sign problems}

In this section, we present a comprehensive list of the modular data of topological orders up to rank 12 along with their diagnoses of the intrinsic sign problem, using our sign-free condition $\tau_n > 0$. Modular data, a pair of unitary matrices $S$ and $T$, are key quantities of (2+1)-dimensional bosonic topological orders, since they encode statistics of the anyonic excitations \cite{wen2015theory}. Both $S$ and $T$ are $(r \times r)$-matrices, where $r$ is called the rank and is equal to the number of different types of the anyons $|\mathcal{A}|$, or the superselection sectors. The first row elements of $S$ are given by $S_{\mathds{1}a} = d_a / D$, where $d_a$ is the quantum dimension of the anyon $a \in \mathcal{A}$ and $D = \sqrt{\sum_a d_a^2}$ is the total quantum dimension. The other matrix $T$ only has diagonal elements, which are given by $T_{aa} = \theta_a$ where $\theta_a$ is the topological spin of the anyon $a$. The chiral central charge $c_-$ describes the quantized response of the edge modes associated with the topological order, through the quantized thermal Hall conductance $I = \frac{\pi}{12} c_- T^2$, where $T$ is the temperature and $I$ is the thermal current \cite{kane1997quantized}.

When a given topological order satisfies our sign-free condition, we cannot identify its intrinsic sign problem, unless its stoquastic realization is known. Note that stoquastic realizations of pure $G$-gauge theories for any finite group $G$ are known \cite{hu2013twisted}. In fact, pure $G$-gauge theories are the only topological orders that we can guarantee that they are sign-free. The modular data of the topological orders are retrieved from Ref.~\cite{ng2023classification}, which classified the modular data of two-dimensional bosonic topological orders up to rank 12, including the complete list of modular data up to rank 11, using the congruence representation theory. The rank 12 modular data in \cite{ng2023classification} are incomplete. For each rank $r$, the $k$-th entry in the subsection titled \textit{Rank $r$} correspond to the $k$-th topological order of rank $r$ listed in Appendix D of Ref.~\cite{ng2023classification}. The diagnosis for each topological order is categorized as follows:
\begin{itemize}
    \item \textit{Intrinsic sign problem}: The topological order inherently suffers from the sign problem.
    \item \textit{Sign-problem-free}: The topological order does not exhibit the sign problem.
    \item \textit{Undetermined}: The status of the intrinsic sign problem is not determined.
\end{itemize}

In the following, we use the notations $\zeta^m_n = e^{2 \pi i m / n}$, $c^m_n = \zeta^m_n + \zeta^{-m}_n$, $s^m_n = \zeta^m_n - \zeta^{-m}_n$, $\xi^{m, l}_n = (\zeta^m_{2 n} - \zeta^{-m}_{2 n}) / (\zeta^l_{2 n} - \zeta^{-l}_{2 n})$, $\xi^m_n = \xi^{m, 1}_n$, $\eta^{m, l}_n = (\zeta^m_{2 n} + \zeta^{-m}_{2 n}) / (\zeta^l_{2 n} - \zeta^{-l}_{2 n})$, and $\eta^m_n = \eta^{m, 1}_n$.

\begin{widetext}

\subsection{Sign-free topological orders}

Before presenting the full list, we first present the modular data of topological orders which are guaranteed to be sign-free. They are realized by the quantum double models of $\mathbb{Z}_2$, $S_8$, and $\mathbb{Z}_3$. In particular, the quantum double model of $\mathbb{Z}_2$ is the well-known toric code \cite{kitaev2003fault}.

\small{\input{modular_data/sign-free.tex}}

\subsection{Topological orders with undetermined intrinsic sign problems}

Though our sign-free condition is effective enough that it can identify the intrinsic sign problems of almost all the topological orders up to rank 12, we could not determine the intrinsic sign problems of the following four topological orders. They are the double Fibonacci, the double Ising, the twisted double Ising, and the Drinfeld center of $\tilde{R}_{\mathrm{PSU}(2)_5}$.

\small{\input{modular_data/unknown.tex}}

\subsection{List of intrinsic sign problems}
Here we list the full list of the intrinsic sign problems of topological orders with their modular data and higher Gauss sums. 

\subsubsection{Rank 2}

\small{\input{modular_data/SsL2U_.tex}}

\subsubsection{Rank 3}

\small{\input{modular_data/SsL3U_.tex}}

\subsubsection{Rank 4}

\small{\input{modular_data/SsL4U_.tex}}

\subsubsection{Rank 5}

\small{\input{modular_data/SsL5U_.tex}}

\subsubsection{Rank 6}

\small{\input{modular_data/SsL6U_.tex}}

\subsubsection{Rank 7}

\small{\input{modular_data/SsL7U_.tex}}

\subsubsection{Rank 8}

\small{\input{modular_data/SsL8U_.tex}}

\subsubsection{Rank 9}

\small{\input{modular_data/SsL9U_.tex}}

\subsubsection{Rank 10}

\small{\input{modular_data/SsL10U_.tex}}

\subsubsection{Rank 11}

\small{\input{modular_data/SsL11U_.tex}}

\subsubsection{Rank 12}

\small{\input{modular_data/SsL12U_.tex}}

\end{widetext}

\end{widetext}

\end{document}

%% file: modular_data/sign-free.tex
  \vskip 2ex 

\noindent1. $4_{0,4.}^{2,750}$ \irep{0}:\ \ 
$d_i$ = ($1.0$,
$1.0$,
$1.0$,
$1.0$) 

\vskip 0.7ex
\hangindent=3em \hangafter=1
$D^2= 4.0 = 
4$

\vskip 0.7ex
\hangindent=3em \hangafter=1
$T = ( 0,
0,
0,
\frac{1}{2} )
$,

\vskip 0.7ex
\hangindent=3em \hangafter=1
$S$ = ($ 1$,
$ 1$,
$ 1$,
$ 1$;\ \ 
$ 1$,
$ -1$,
$ -1$;\ \ 
$ 1$,
$ -1$;\ \ 
$ 1$)

\vskip 0.7ex
\hangindent=3em \hangafter=1
$\tau_n$ = ($2.$, $4.$)

\vskip 0.7ex
\hangindent=3em \hangafter=1
\textit{Sign-problem-free}

  \vskip 2ex 

\noindent37. $8_{0,36.}^{6,213}$ \irep{0}:\ \ 
$d_i$ = ($1.0$,
$1.0$,
$2.0$,
$2.0$,
$2.0$,
$2.0$,
$3.0$,
$3.0$) 

\vskip 0.7ex
\hangindent=3em \hangafter=1
$D^2= 36.0 = 
36$

\vskip 0.7ex
\hangindent=3em \hangafter=1
$T = ( 0,
0,
0,
0,
\frac{1}{3},
\frac{2}{3},
0,
\frac{1}{2} )
$,

\vskip 0.7ex
\hangindent=3em \hangafter=1
$S$ = ($ 1$,
$ 1$,
$ 2$,
$ 2$,
$ 2$,
$ 2$,
$ 3$,
$ 3$;\ \ 
$ 1$,
$ 2$,
$ 2$,
$ 2$,
$ 2$,
$ -3$,
$ -3$;\ \ 
$ 4$,
$ -2$,
$ -2$,
$ -2$,
$0$,
$0$;\ \ 
$ 4$,
$ -2$,
$ -2$,
$0$,
$0$;\ \ 
$ -2$,
$ 4$,
$0$,
$0$;\ \ 
$ -2$,
$0$,
$0$;\ \ 
$ 3$,
$ -3$;\ \ 
$ 3$)

\vskip 0.7ex
\hangindent=3em \hangafter=1
$\tau_n$ = ($6.$, $24.$, $18.$, $24.$, $6.$, $36.$)

\vskip 0.7ex
\hangindent=3em \hangafter=1
\textit{Sign-problem-free}

  \vskip 2ex

\noindent1. $9_{0,9.}^{3,113}$ \irep{0}:\ \ 
$d_i$ = ($1.0$,
$1.0$,
$1.0$,
$1.0$,
$1.0$,
$1.0$,
$1.0$,
$1.0$,
$1.0$) 

\vskip 0.7ex
\hangindent=3em \hangafter=1
$D^2= 9.0 = 
9$

\vskip 0.7ex
\hangindent=3em \hangafter=1
$T = ( 0,
0,
0,
0,
0,
\frac{1}{3},
\frac{1}{3},
\frac{2}{3},
\frac{2}{3} )
$,

\vskip 0.7ex
\hangindent=3em \hangafter=1
$S$ = ($ 1$,
$ 1$,
$ 1$,
$ 1$,
$ 1$,
$ 1$,
$ 1$,
$ 1$,
$ 1$;\ \ 
$ 1$,
$ 1$,
$ -\zeta_{6}^{1}$,
$ \zeta_{3}^{1}$,
$ -\zeta_{6}^{1}$,
$ \zeta_{3}^{1}$,
$ -\zeta_{6}^{1}$,
$ \zeta_{3}^{1}$;\ \ 
$ 1$,
$ \zeta_{3}^{1}$,
$ -\zeta_{6}^{1}$,
$ \zeta_{3}^{1}$,
$ -\zeta_{6}^{1}$,
$ \zeta_{3}^{1}$,
$ -\zeta_{6}^{1}$;\ \ 
$ 1$,
$ 1$,
$ -\zeta_{6}^{1}$,
$ \zeta_{3}^{1}$,
$ \zeta_{3}^{1}$,
$ -\zeta_{6}^{1}$;\ \ 
$ 1$,
$ \zeta_{3}^{1}$,
$ -\zeta_{6}^{1}$,
$ -\zeta_{6}^{1}$,
$ \zeta_{3}^{1}$;\ \ 
$ \zeta_{3}^{1}$,
$ -\zeta_{6}^{1}$,
$ 1$,
$ 1$;\ \ 
$ \zeta_{3}^{1}$,
$ 1$,
$ 1$;\ \ 
$ -\zeta_{6}^{1}$,
$ \zeta_{3}^{1}$;\ \ 
$ -\zeta_{6}^{1}$)

Factors = $3_{2,3.}^{3,527}\boxtimes 3_{6,3.}^{3,138}$

\vskip 0.7ex
\hangindent=3em \hangafter=1
$\tau_n$ = ($3.$, $3.$, $9.$)

\vskip 0.7ex
\hangindent=3em \hangafter=1
\textit{Sign-problem-free}

  \vskip 2ex 

%% file: modular_data/unknown.tex
  \vskip 2ex 

\noindent15. $4_{0,13.09}^{5,872}$ \irep{5}:\ \ 
$d_i$ = ($1.0$,
$1.618$,
$1.618$,
$2.618$) 

\vskip 0.7ex
\hangindent=3em \hangafter=1
$D^2= 13.90 = 
\frac{15+5\sqrt{5}}{2}$

\vskip 0.7ex
\hangindent=3em \hangafter=1
$T = ( 0,
\frac{2}{5},
\frac{3}{5},
0 )
$,

\vskip 0.7ex
\hangindent=3em \hangafter=1
$S$ = ($ 1$,
$ \frac{1+\sqrt{5}}{2}$,
$ \frac{1+\sqrt{5}}{2}$,
$ \frac{3+\sqrt{5}}{2}$;\ \ 
$ -1$,
$ \frac{3+\sqrt{5}}{2}$,
$ -\frac{1+\sqrt{5}}{2}$;\ \ 
$ -1$,
$ -\frac{1+\sqrt{5}}{2}$;\ \ 
$ 1$)

Factors = $2_{\frac{14}{5},3.618}^{5,395}\boxtimes 2_{\frac{26}{5},3.618}^{5,720}$

\vskip 0.7ex
\hangindent=3em \hangafter=1
$\tau_n$ = ($3.62$, $9.47$, $9.47$, $3.62$, $13.09$)

\vskip 0.7ex
\hangindent=3em \hangafter=1
\textit{Undetermined}

  \vskip 2ex 

\noindent27. $9_{0,16.}^{16,447}$ \irep{397}:\ \ 
$d_i$ = ($1.0$,
$1.0$,
$1.0$,
$1.0$,
$1.414$,
$1.414$,
$1.414$,
$1.414$,
$2.0$) 

\vskip 0.7ex
\hangindent=3em \hangafter=1
$D^2= 16.0 = 
16$

\vskip 0.7ex
\hangindent=3em \hangafter=1
$T = ( 0,
0,
\frac{1}{2},
\frac{1}{2},
\frac{1}{16},
\frac{7}{16},
\frac{9}{16},
\frac{15}{16},
0 )
$,

\vskip 0.7ex
\hangindent=3em \hangafter=1
$S$ = ($ 1$,
$ 1$,
$ 1$,
$ 1$,
$ \sqrt{2}$,
$ \sqrt{2}$,
$ \sqrt{2}$,
$ \sqrt{2}$,
$ 2$;\ \ 
$ 1$,
$ 1$,
$ 1$,
$ -\sqrt{2}$,
$ -\sqrt{2}$,
$ -\sqrt{2}$,
$ -\sqrt{2}$,
$ 2$;\ \ 
$ 1$,
$ 1$,
$ -\sqrt{2}$,
$ \sqrt{2}$,
$ -\sqrt{2}$,
$ \sqrt{2}$,
$ -2$;\ \ 
$ 1$,
$ \sqrt{2}$,
$ -\sqrt{2}$,
$ \sqrt{2}$,
$ -\sqrt{2}$,
$ -2$;\ \ 
$0$,
$ -2$,
$0$,
$ 2$,
$0$;\ \ 
$0$,
$ 2$,
$0$,
$0$;\ \ 
$0$,
$ -2$,
$0$;\ \ 
$0$,
$0$;\ \ 
$0$)

Factors = $3_{\frac{1}{2},4.}^{16,598}\boxtimes 3_{\frac{15}{2},4.}^{16,639}$

\vskip 0.7ex
\hangindent=3em \hangafter=1
$\tau_n$ = ($4.$, $13.66$, $4.$, $8.$, $4.$, $2.34$, $4.$, $0.$, $4.$, $2.34$, $4.$, $8.$, $4.$, $13.66$, $4.$, $16.$)

\vskip 0.7ex
\hangindent=3em \hangafter=1
\textit{Undetermined}

  \vskip 2ex 

\noindent31. $9_{0,16.}^{16,624}$ \irep{397}:\ \ 
$d_i$ = ($1.0$,
$1.0$,
$1.0$,
$1.0$,
$1.414$,
$1.414$,
$1.414$,
$1.414$,
$2.0$) 

\vskip 0.7ex
\hangindent=3em \hangafter=1
$D^2= 16.0 = 
16$

\vskip 0.7ex
\hangindent=3em \hangafter=1
$T = ( 0,
0,
\frac{1}{2},
\frac{1}{2},
\frac{3}{16},
\frac{5}{16},
\frac{11}{16},
\frac{13}{16},
0 )
$,

\vskip 0.7ex
\hangindent=3em \hangafter=1
$S$ = ($ 1$,
$ 1$,
$ 1$,
$ 1$,
$ \sqrt{2}$,
$ \sqrt{2}$,
$ \sqrt{2}$,
$ \sqrt{2}$,
$ 2$;\ \ 
$ 1$,
$ 1$,
$ 1$,
$ -\sqrt{2}$,
$ -\sqrt{2}$,
$ -\sqrt{2}$,
$ -\sqrt{2}$,
$ 2$;\ \ 
$ 1$,
$ 1$,
$ -\sqrt{2}$,
$ \sqrt{2}$,
$ -\sqrt{2}$,
$ \sqrt{2}$,
$ -2$;\ \ 
$ 1$,
$ \sqrt{2}$,
$ -\sqrt{2}$,
$ \sqrt{2}$,
$ -\sqrt{2}$,
$ -2$;\ \ 
$0$,
$ -2$,
$0$,
$ 2$,
$0$;\ \ 
$0$,
$ 2$,
$0$,
$0$;\ \ 
$0$,
$ -2$,
$0$;\ \ 
$0$,
$0$;\ \ 
$0$)

Factors = $3_{\frac{3}{2},4.}^{16,553}\boxtimes 3_{\frac{13}{2},4.}^{16,330}$

\vskip 0.7ex
\hangindent=3em \hangafter=1
$\tau_n$ = ($4.$, $2.34$, $4.$, $8.$, $4.$, $13.66$, $4.$, $0.$, $4.$, $13.66$, $4.$, $8.$, $4.$, $2.34$, $4.$, $16.$)

\vskip 0.7ex
\hangindent=3em \hangafter=1
\textit{Undetermined}

  \vskip 2ex 

\noindent70. $9_{0,86.41}^{7,161}$ \irep{196}:\ \ 
$d_i$ = ($1.0$,
$1.801$,
$1.801$,
$2.246$,
$2.246$,
$3.246$,
$4.48$,
$4.48$,
$5.48$) 

\vskip 0.7ex
\hangindent=3em \hangafter=1
$D^2= 86.413 = 
49+35c^{1}_{7}
+14c^{2}_{7}
$

\vskip 0.7ex
\hangindent=3em \hangafter=1
$T = ( 0,
\frac{1}{7},
\frac{6}{7},
\frac{2}{7},
\frac{5}{7},
0,
\frac{3}{7},
\frac{4}{7},
0 )
$,

\vskip 0.7ex
\hangindent=3em \hangafter=1
$S$ = ($ 1$,
$ -c_{7}^{3}$,
$ -c_{7}^{3}$,
$ \xi_{7}^{3}$,
$ \xi_{7}^{3}$,
$ 2+c^{1}_{7}
$,
$ 2+2c^{1}_{7}
+c^{2}_{7}
$,
$ 2+2c^{1}_{7}
+c^{2}_{7}
$,
$ 3+2c^{1}_{7}
+c^{2}_{7}
$;\ \ 
$ -\xi_{7}^{3}$,
$ 2+c^{1}_{7}
$,
$ 2+2c^{1}_{7}
+c^{2}_{7}
$,
$ 1$,
$ -2-2  c^{1}_{7}
-c^{2}_{7}
$,
$ -3-2  c^{1}_{7}
-c^{2}_{7}
$,
$ -c_{7}^{3}$,
$ \xi_{7}^{3}$;\ \ 
$ -\xi_{7}^{3}$,
$ 1$,
$ 2+2c^{1}_{7}
+c^{2}_{7}
$,
$ -2-2  c^{1}_{7}
-c^{2}_{7}
$,
$ -c_{7}^{3}$,
$ -3-2  c^{1}_{7}
-c^{2}_{7}
$,
$ \xi_{7}^{3}$;\ \ 
$ c_{7}^{3}$,
$ 3+2c^{1}_{7}
+c^{2}_{7}
$,
$ -c_{7}^{3}$,
$ -2-c^{1}_{7}
$,
$ \xi_{7}^{3}$,
$ -2-2  c^{1}_{7}
-c^{2}_{7}
$;\ \ 
$ c_{7}^{3}$,
$ -c_{7}^{3}$,
$ \xi_{7}^{3}$,
$ -2-c^{1}_{7}
$,
$ -2-2  c^{1}_{7}
-c^{2}_{7}
$;\ \ 
$ 3+2c^{1}_{7}
+c^{2}_{7}
$,
$ -\xi_{7}^{3}$,
$ -\xi_{7}^{3}$,
$ 1$;\ \ 
$ 2+2c^{1}_{7}
+c^{2}_{7}
$,
$ 1$,
$ c_{7}^{3}$;\ \ 
$ 2+2c^{1}_{7}
+c^{2}_{7}
$,
$ c_{7}^{3}$;\ \ 
$ 2+c^{1}_{7}
$)

Factors = $3_{\frac{48}{7},9.295}^{7,790}\boxtimes 3_{\frac{8}{7},9.295}^{7,245}$

\vskip 0.7ex
\hangindent=3em \hangafter=1
$\tau_n$ = ($7.2$, $56.06$, $33.08$, $33.08$, $56.06$, $7.2$, $98.28$)

\vskip 0.7ex
\hangindent=3em \hangafter=1
\textit{Undetermined}

  \vskip 2ex 

%% file: modular_data/SsL2U_.tex
\noindent1. $2_{1,2.}^{4,437}$ \irep{1}:\ \ 
$d_i$ = ($1.0$,
$1.0$) 

\vskip 0.7ex
\hangindent=3em \hangafter=1
$D^2= 2.0 = 
2$

\vskip 0.7ex
\hangindent=3em \hangafter=1
$T = ( 0,
\frac{1}{4} )
$,

\vskip 0.7ex
\hangindent=3em \hangafter=1
$S$ = ($ 1$,
$ 1$;\ \ 
$ -1$)

\vskip 0.7ex
\hangindent=3em \hangafter=1
$\tau_n$ = ($1. + 1. i$, $0.$, $1. - 1. i$, $2.$)

\vskip 0.7ex
\hangindent=3em \hangafter=1
\textit{Intrinsic sign problem}

  \vskip 2ex

\noindent2. $2_{7,2.}^{4,625}$ \irep{1}:\ \ 
$d_i$ = ($1.0$,
$1.0$) 

\vskip 0.7ex
\hangindent=3em \hangafter=1
$D^2= 2.0 = 
2$

\vskip 0.7ex
\hangindent=3em \hangafter=1
$T = ( 0,
\frac{3}{4} )
$,

\vskip 0.7ex
\hangindent=3em \hangafter=1
$S$ = ($ 1$,
$ 1$;\ \ 
$ -1$)

\vskip 0.7ex
\hangindent=3em \hangafter=1
$\tau_n$ = ($1. - 1. i$, $0.$, $1. + 1. i$, $2.$)

\vskip 0.7ex
\hangindent=3em \hangafter=1
\textit{Intrinsic sign problem}

  \vskip 2ex

\noindent3. $2_{\frac{14}{5},3.618}^{5,395}$ \irep{2}:\ \ 
$d_i$ = ($1.0$,
$1.618$) 

\vskip 0.7ex
\hangindent=3em \hangafter=1
$D^2= 3.618 = 
\frac{5+\sqrt{5}}{2}$

\vskip 0.7ex
\hangindent=3em \hangafter=1
$T = ( 0,
\frac{2}{5} )
$,

\vskip 0.7ex
\hangindent=3em \hangafter=1
$S$ = ($ 1$,
$ \frac{1+\sqrt{5}}{2}$;\ \ 
$ -1$)

\vskip 0.7ex
\hangindent=3em \hangafter=1
$\tau_n$ = ($-1.12 + 1.54 i$, $1.81 - 2.49 i$, $1.81 + 2.49 i$, $-1.12 - 1.54 i$, $3.62$)

\vskip 0.7ex
\hangindent=3em \hangafter=1
\textit{Intrinsic sign problem}

  \vskip 2ex

\noindent4. $2_{\frac{26}{5},3.618}^{5,720}$ \irep{2}:\ \ 
$d_i$ = ($1.0$,
$1.618$) 

\vskip 0.7ex
\hangindent=3em \hangafter=1
$D^2= 3.618 = 
\frac{5+\sqrt{5}}{2}$

\vskip 0.7ex
\hangindent=3em \hangafter=1
$T = ( 0,
\frac{3}{5} )
$,

\vskip 0.7ex
\hangindent=3em \hangafter=1
$S$ = ($ 1$,
$ \frac{1+\sqrt{5}}{2}$;\ \ 
$ -1$)

\vskip 0.7ex
\hangindent=3em \hangafter=1
$\tau_n$ = ($-1.12 - 1.54 i$, $1.81 + 2.49 i$, $1.81 - 2.49 i$, $-1.12 + 1.54 i$, $3.62$)

\vskip 0.7ex
\hangindent=3em \hangafter=1
\textit{Intrinsic sign problem}

  \vskip 2ex 

%% file: modular_data/SsL3U_.tex
\noindent1. $3_{2,3.}^{3,527}$ \irep{2}:\ \ 
$d_i$ = ($1.0$,
$1.0$,
$1.0$) 

\vskip 0.7ex
\hangindent=3em \hangafter=1
$D^2= 3.0 = 
3$

\vskip 0.7ex
\hangindent=3em \hangafter=1
$T = ( 0,
\frac{1}{3},
\frac{1}{3} )
$,

\vskip 0.7ex
\hangindent=3em \hangafter=1
$S$ = ($ 1$,
$ 1$,
$ 1$;\ \ 
$ \zeta_{3}^{1}$,
$ -\zeta_{6}^{1}$;\ \ 
$ \zeta_{3}^{1}$)

\vskip 0.7ex
\hangindent=3em \hangafter=1
$\tau_n$ = ($0. + 1.73 i$, $0. - 1.73 i$, $3.$)

\vskip 0.7ex
\hangindent=3em \hangafter=1
\textit{Intrinsic sign problem}

  \vskip 2ex

\noindent2. $3_{6,3.}^{3,138}$ \irep{2}:\ \ 
$d_i$ = ($1.0$,
$1.0$,
$1.0$) 

\vskip 0.7ex
\hangindent=3em \hangafter=1
$D^2= 3.0 = 
3$

\vskip 0.7ex
\hangindent=3em \hangafter=1
$T = ( 0,
\frac{2}{3},
\frac{2}{3} )
$,

\vskip 0.7ex
\hangindent=3em \hangafter=1
$S$ = ($ 1$,
$ 1$,
$ 1$;\ \ 
$ -\zeta_{6}^{1}$,
$ \zeta_{3}^{1}$;\ \ 
$ -\zeta_{6}^{1}$)

\vskip 0.7ex
\hangindent=3em \hangafter=1
$\tau_n$ = ($0. - 1.73 i$, $0. + 1.73 i$, $3.$)

\vskip 0.7ex
\hangindent=3em \hangafter=1
\textit{Intrinsic sign problem}

  \vskip 2ex

\noindent3. $3_{\frac{1}{2},4.}^{16,598}$ \irep{4}:\ \ 
$d_i$ = ($1.0$,
$1.0$,
$1.414$) 

\vskip 0.7ex
\hangindent=3em \hangafter=1
$D^2= 4.0 = 
4$

\vskip 0.7ex
\hangindent=3em \hangafter=1
$T = ( 0,
\frac{1}{2},
\frac{1}{16} )
$,

\vskip 0.7ex
\hangindent=3em \hangafter=1
$S$ = ($ 1$,
$ 1$,
$ \sqrt{2}$;\ \ 
$ 1$,
$ -\sqrt{2}$;\ \ 
$0$)

\vskip 0.7ex
\hangindent=3em \hangafter=1
$\tau_n$ = ($1.85 + 0.77 i$, $3.41 + 1.41 i$, $0.77 + 1.85 i$, $2. + 2. i$, $-0.77 + 1.85 i$, $0.59 + 1.41 i$, $-1.85 + 0.77 i$, $0.$, $-1.85 - 0.77 i$, $0.59 - 1.41 i$, $-0.77 - 1.85 i$, $2. - 2. i$, $0.77 - 1.85 i$, $3.41 - 1.41 i$, $1.85 - 0.77 i$, $4.$)

\vskip 0.7ex
\hangindent=3em \hangafter=1
\textit{Intrinsic sign problem}

  \vskip 2ex

\noindent4. $3_{\frac{3}{2},4.}^{16,553}$ \irep{4}:\ \ 
$d_i$ = ($1.0$,
$1.0$,
$1.414$) 

\vskip 0.7ex
\hangindent=3em \hangafter=1
$D^2= 4.0 = 
4$

\vskip 0.7ex
\hangindent=3em \hangafter=1
$T = ( 0,
\frac{1}{2},
\frac{3}{16} )
$,

\vskip 0.7ex
\hangindent=3em \hangafter=1
$S$ = ($ 1$,
$ 1$,
$ \sqrt{2}$;\ \ 
$ 1$,
$ -\sqrt{2}$;\ \ 
$0$)

\vskip 0.7ex
\hangindent=3em \hangafter=1
$\tau_n$ = ($0.77 + 1.85 i$, $0.59 + 1.41 i$, $-1.85 - 0.77 i$, $2. - 2. i$, $1.85 - 0.77 i$, $3.41 + 1.41 i$, $-0.77 + 1.85 i$, $0.$, $-0.77 - 1.85 i$, $3.41 - 1.41 i$, $1.85 + 0.77 i$, $2. + 2. i$, $-1.85 + 0.77 i$, $0.59 - 1.41 i$, $0.77 - 1.85 i$, $4.$)

\vskip 0.7ex
\hangindent=3em \hangafter=1
\textit{Intrinsic sign problem}

  \vskip 2ex

\noindent5. $3_{\frac{5}{2},4.}^{16,465}$ \irep{4}:\ \ 
$d_i$ = ($1.0$,
$1.0$,
$1.414$) 

\vskip 0.7ex
\hangindent=3em \hangafter=1
$D^2= 4.0 = 
4$

\vskip 0.7ex
\hangindent=3em \hangafter=1
$T = ( 0,
\frac{1}{2},
\frac{5}{16} )
$,

\vskip 0.7ex
\hangindent=3em \hangafter=1
$S$ = ($ 1$,
$ 1$,
$ \sqrt{2}$;\ \ 
$ 1$,
$ -\sqrt{2}$;\ \ 
$0$)

\vskip 0.7ex
\hangindent=3em \hangafter=1
$\tau_n$ = ($-0.77 + 1.85 i$, $0.59 - 1.41 i$, $1.85 - 0.77 i$, $2. + 2. i$, $-1.85 - 0.77 i$, $3.41 - 1.41 i$, $0.77 + 1.85 i$, $0.$, $0.77 - 1.85 i$, $3.41 + 1.41 i$, $-1.85 + 0.77 i$, $2. - 2. i$, $1.85 + 0.77 i$, $0.59 + 1.41 i$, $-0.77 - 1.85 i$, $4.$)

\vskip 0.7ex
\hangindent=3em \hangafter=1
\textit{Intrinsic sign problem}

  \vskip 2ex

\noindent6. $3_{\frac{7}{2},4.}^{16,332}$ \irep{4}:\ \ 
$d_i$ = ($1.0$,
$1.0$,
$1.414$) 

\vskip 0.7ex
\hangindent=3em \hangafter=1
$D^2= 4.0 = 
4$

\vskip 0.7ex
\hangindent=3em \hangafter=1
$T = ( 0,
\frac{1}{2},
\frac{7}{16} )
$,

\vskip 0.7ex
\hangindent=3em \hangafter=1
$S$ = ($ 1$,
$ 1$,
$ \sqrt{2}$;\ \ 
$ 1$,
$ -\sqrt{2}$;\ \ 
$0$)

\vskip 0.7ex
\hangindent=3em \hangafter=1
$\tau_n$ = ($-1.85 + 0.77 i$, $3.41 - 1.41 i$, $-0.77 + 1.85 i$, $2. - 2. i$, $0.77 + 1.85 i$, $0.59 - 1.41 i$, $1.85 + 0.77 i$, $0.$, $1.85 - 0.77 i$, $0.59 + 1.41 i$, $0.77 - 1.85 i$, $2. + 2. i$, $-0.77 - 1.85 i$, $3.41 + 1.41 i$, $-1.85 - 0.77 i$, $4.$)

\vskip 0.7ex
\hangindent=3em \hangafter=1
\textit{Intrinsic sign problem}

  \vskip 2ex

\noindent7. $3_{\frac{9}{2},4.}^{16,156}$ \irep{4}:\ \ 
$d_i$ = ($1.0$,
$1.0$,
$1.414$) 

\vskip 0.7ex
\hangindent=3em \hangafter=1
$D^2= 4.0 = 
4$

\vskip 0.7ex
\hangindent=3em \hangafter=1
$T = ( 0,
\frac{1}{2},
\frac{9}{16} )
$,

\vskip 0.7ex
\hangindent=3em \hangafter=1
$S$ = ($ 1$,
$ 1$,
$ \sqrt{2}$;\ \ 
$ 1$,
$ -\sqrt{2}$;\ \ 
$0$)

\vskip 0.7ex
\hangindent=3em \hangafter=1
$\tau_n$ = ($-1.85 - 0.77 i$, $3.41 + 1.41 i$, $-0.77 - 1.85 i$, $2. + 2. i$, $0.77 - 1.85 i$, $0.59 + 1.41 i$, $1.85 - 0.77 i$, $0.$, $1.85 + 0.77 i$, $0.59 - 1.41 i$, $0.77 + 1.85 i$, $2. - 2. i$, $-0.77 + 1.85 i$, $3.41 - 1.41 i$, $-1.85 + 0.77 i$, $4.$)

\vskip 0.7ex
\hangindent=3em \hangafter=1
\textit{Intrinsic sign problem}

  \vskip 2ex

\noindent8. $3_{\frac{11}{2},4.}^{16,648}$ \irep{4}:\ \ 
$d_i$ = ($1.0$,
$1.0$,
$1.414$) 

\vskip 0.7ex
\hangindent=3em \hangafter=1
$D^2= 4.0 = 
4$

\vskip 0.7ex
\hangindent=3em \hangafter=1
$T = ( 0,
\frac{1}{2},
\frac{11}{16} )
$,

\vskip 0.7ex
\hangindent=3em \hangafter=1
$S$ = ($ 1$,
$ 1$,
$ \sqrt{2}$;\ \ 
$ 1$,
$ -\sqrt{2}$;\ \ 
$0$)

\vskip 0.7ex
\hangindent=3em \hangafter=1
$\tau_n$ = ($-0.77 - 1.85 i$, $0.59 + 1.41 i$, $1.85 + 0.77 i$, $2. - 2. i$, $-1.85 + 0.77 i$, $3.41 + 1.41 i$, $0.77 - 1.85 i$, $0.$, $0.77 + 1.85 i$, $3.41 - 1.41 i$, $-1.85 - 0.77 i$, $2. + 2. i$, $1.85 - 0.77 i$, $0.59 - 1.41 i$, $-0.77 + 1.85 i$, $4.$)

\vskip 0.7ex
\hangindent=3em \hangafter=1
\textit{Intrinsic sign problem}

  \vskip 2ex

\noindent9. $3_{\frac{13}{2},4.}^{16,330}$ \irep{4}:\ \ 
$d_i$ = ($1.0$,
$1.0$,
$1.414$) 

\vskip 0.7ex
\hangindent=3em \hangafter=1
$D^2= 4.0 = 
4$

\vskip 0.7ex
\hangindent=3em \hangafter=1
$T = ( 0,
\frac{1}{2},
\frac{13}{16} )
$,

\vskip 0.7ex
\hangindent=3em \hangafter=1
$S$ = ($ 1$,
$ 1$,
$ \sqrt{2}$;\ \ 
$ 1$,
$ -\sqrt{2}$;\ \ 
$0$)

\vskip 0.7ex
\hangindent=3em \hangafter=1
$\tau_n$ = ($0.77 - 1.85 i$, $0.59 - 1.41 i$, $-1.85 + 0.77 i$, $2. + 2. i$, $1.85 + 0.77 i$, $3.41 - 1.41 i$, $-0.77 - 1.85 i$, $0.$, $-0.77 + 1.85 i$, $3.41 + 1.41 i$, $1.85 - 0.77 i$, $2. - 2. i$, $-1.85 - 0.77 i$, $0.59 + 1.41 i$, $0.77 + 1.85 i$, $4.$)

\vskip 0.7ex
\hangindent=3em \hangafter=1
\textit{Intrinsic sign problem}

  \vskip 2ex

\noindent10. $3_{\frac{15}{2},4.}^{16,639}$ \irep{4}:\ \ 
$d_i$ = ($1.0$,
$1.0$,
$1.414$) 

\vskip 0.7ex
\hangindent=3em \hangafter=1
$D^2= 4.0 = 
4$

\vskip 0.7ex
\hangindent=3em \hangafter=1
$T = ( 0,
\frac{1}{2},
\frac{15}{16} )
$,

\vskip 0.7ex
\hangindent=3em \hangafter=1
$S$ = ($ 1$,
$ 1$,
$ \sqrt{2}$;\ \ 
$ 1$,
$ -\sqrt{2}$;\ \ 
$0$)

\vskip 0.7ex
\hangindent=3em \hangafter=1
$\tau_n$ = ($1.85 - 0.77 i$, $3.41 - 1.41 i$, $0.77 - 1.85 i$, $2. - 2. i$, $-0.77 - 1.85 i$, $0.59 - 1.41 i$, $-1.85 - 0.77 i$, $0.$, $-1.85 + 0.77 i$, $0.59 + 1.41 i$, $-0.77 + 1.85 i$, $2. + 2. i$, $0.77 + 1.85 i$, $3.41 + 1.41 i$, $1.85 + 0.77 i$, $4.$)

\vskip 0.7ex
\hangindent=3em \hangafter=1
\textit{Intrinsic sign problem}

  \vskip 2ex

\noindent11. $3_{\frac{48}{7},9.295}^{7,790}$ \irep{3}:\ \ 
$d_i$ = ($1.0$,
$1.801$,
$2.246$) 

\vskip 0.7ex
\hangindent=3em \hangafter=1
$D^2= 9.295 = 
6+3c^{1}_{7}
+c^{2}_{7}
$

\vskip 0.7ex
\hangindent=3em \hangafter=1
$T = ( 0,
\frac{1}{7},
\frac{5}{7} )
$,

\vskip 0.7ex
\hangindent=3em \hangafter=1
$S$ = ($ 1$,
$ -c_{7}^{3}$,
$ \xi_{7}^{3}$;\ \ 
$ -\xi_{7}^{3}$,
$ 1$;\ \ 
$ c_{7}^{3}$)

\vskip 0.7ex
\hangindent=3em \hangafter=1
$\tau_n$ = ($1.9 - 2.38 i$, $-4.27 + 5.35 i$, $1.22 + 5.35 i$, $1.22 - 5.35 i$, $-4.27 - 5.35 i$, $1.9 + 2.38 i$, $9.29$)

\vskip 0.7ex
\hangindent=3em \hangafter=1
\textit{Intrinsic sign problem}

  \vskip 2ex

\noindent12. $3_{\frac{8}{7},9.295}^{7,245}$ \irep{3}:\ \ 
$d_i$ = ($1.0$,
$1.801$,
$2.246$) 

\vskip 0.7ex
\hangindent=3em \hangafter=1
$D^2= 9.295 = 
6+3c^{1}_{7}
+c^{2}_{7}
$

\vskip 0.7ex
\hangindent=3em \hangafter=1
$T = ( 0,
\frac{6}{7},
\frac{2}{7} )
$,

\vskip 0.7ex
\hangindent=3em \hangafter=1
$S$ = ($ 1$,
$ -c_{7}^{3}$,
$ \xi_{7}^{3}$;\ \ 
$ -\xi_{7}^{3}$,
$ 1$;\ \ 
$ c_{7}^{3}$)

\vskip 0.7ex
\hangindent=3em \hangafter=1
$\tau_n$ = ($1.9 + 2.38 i$, $-4.27 - 5.35 i$, $1.22 - 5.35 i$, $1.22 + 5.35 i$, $-4.27 + 5.35 i$, $1.9 - 2.38 i$, $9.29$)

\vskip 0.7ex
\hangindent=3em \hangafter=1
\textit{Intrinsic sign problem}

  \vskip 2ex 

%% file: modular_data/SsL4U_.tex
\noindent1. $4_{0,4.}^{2,750}$ \irep{0}:\ \ 
$d_i$ = ($1.0$,
$1.0$,
$1.0$,
$1.0$) 

\vskip 0.7ex
\hangindent=3em \hangafter=1
$D^2= 4.0 = 
4$

\vskip 0.7ex
\hangindent=3em \hangafter=1
$T = ( 0,
0,
0,
\frac{1}{2} )
$,

\vskip 0.7ex
\hangindent=3em \hangafter=1
$S$ = ($ 1$,
$ 1$,
$ 1$,
$ 1$;\ \ 
$ 1$,
$ -1$,
$ -1$;\ \ 
$ 1$,
$ -1$;\ \ 
$ 1$)

\vskip 0.7ex
\hangindent=3em \hangafter=1
$\tau_n$ = ($2.$, $4.$)

\vskip 0.7ex
\hangindent=3em \hangafter=1
\textit{Sign-problem-free}

  \vskip 2ex

\noindent2. $4_{4,4.}^{2,250}$ \irep{0}:\ \ 
$d_i$ = ($1.0$,
$1.0$,
$1.0$,
$1.0$) 

\vskip 0.7ex
\hangindent=3em \hangafter=1
$D^2= 4.0 = 
4$

\vskip 0.7ex
\hangindent=3em \hangafter=1
$T = ( 0,
\frac{1}{2},
\frac{1}{2},
\frac{1}{2} )
$,

\vskip 0.7ex
\hangindent=3em \hangafter=1
$S$ = ($ 1$,
$ 1$,
$ 1$,
$ 1$;\ \ 
$ 1$,
$ -1$,
$ -1$;\ \ 
$ 1$,
$ -1$;\ \ 
$ 1$)

\vskip 0.7ex
\hangindent=3em \hangafter=1
$\tau_n$ = ($-2.$, $4.$)

\vskip 0.7ex
\hangindent=3em \hangafter=1
\textit{Intrinsic sign problem}

  \vskip 2ex

\noindent3. $4_{0,4.}^{4,375}$ \irep{0}:\ \ 
$d_i$ = ($1.0$,
$1.0$,
$1.0$,
$1.0$) 

\vskip 0.7ex
\hangindent=3em \hangafter=1
$D^2= 4.0 = 
4$

\vskip 0.7ex
\hangindent=3em \hangafter=1
$T = ( 0,
0,
\frac{1}{4},
\frac{3}{4} )
$,

\vskip 0.7ex
\hangindent=3em \hangafter=1
$S$ = ($ 1$,
$ 1$,
$ 1$,
$ 1$;\ \ 
$ 1$,
$ -1$,
$ -1$;\ \ 
$ -1$,
$ 1$;\ \ 
$ -1$)

Factors = $2_{1,2.}^{4,437}\boxtimes 2_{7,2.}^{4,625}$

\vskip 0.7ex
\hangindent=3em \hangafter=1
$\tau_n$ = ($2.$, $0.$, $2.$, $4.$)

\vskip 0.7ex
\hangindent=3em \hangafter=1
\textit{Intrinsic sign problem}

  \vskip 2ex

\noindent4. $4_{2,4.}^{4,625}$ \irep{0}:\ \ 
$d_i$ = ($1.0$,
$1.0$,
$1.0$,
$1.0$) 

\vskip 0.7ex
\hangindent=3em \hangafter=1
$D^2= 4.0 = 
4$

\vskip 0.7ex
\hangindent=3em \hangafter=1
$T = ( 0,
\frac{1}{2},
\frac{1}{4},
\frac{1}{4} )
$,

\vskip 0.7ex
\hangindent=3em \hangafter=1
$S$ = ($ 1$,
$ 1$,
$ 1$,
$ 1$;\ \ 
$ 1$,
$ -1$,
$ -1$;\ \ 
$ -1$,
$ 1$;\ \ 
$ -1$)

Factors = $2_{1,2.}^{4,437}\boxtimes 2_{1,2.}^{4,437}$

\vskip 0.7ex
\hangindent=3em \hangafter=1
$\tau_n$ = ($0. + 2. i$, $0.$, $0. - 2. i$, $4.$)

\vskip 0.7ex
\hangindent=3em \hangafter=1
\textit{Intrinsic sign problem}

  \vskip 2ex

\noindent5. $4_{6,4.}^{4,375}$ \irep{0}:\ \ 
$d_i$ = ($1.0$,
$1.0$,
$1.0$,
$1.0$) 

\vskip 0.7ex
\hangindent=3em \hangafter=1
$D^2= 4.0 = 
4$

\vskip 0.7ex
\hangindent=3em \hangafter=1
$T = ( 0,
\frac{1}{2},
\frac{3}{4},
\frac{3}{4} )
$,

\vskip 0.7ex
\hangindent=3em \hangafter=1
$S$ = ($ 1$,
$ 1$,
$ 1$,
$ 1$;\ \ 
$ 1$,
$ -1$,
$ -1$;\ \ 
$ -1$,
$ 1$;\ \ 
$ -1$)

Factors = $2_{7,2.}^{4,625}\boxtimes 2_{7,2.}^{4,625}$

\vskip 0.7ex
\hangindent=3em \hangafter=1
$\tau_n$ = ($0. - 2. i$, $0.$, $0. + 2. i$, $4.$)

\vskip 0.7ex
\hangindent=3em \hangafter=1
\textit{Intrinsic sign problem}

  \vskip 2ex

\noindent6. $4_{1,4.}^{8,718}$ \irep{6}:\ \ 
$d_i$ = ($1.0$,
$1.0$,
$1.0$,
$1.0$) 

\vskip 0.7ex
\hangindent=3em \hangafter=1
$D^2= 4.0 = 
4$

\vskip 0.7ex
\hangindent=3em \hangafter=1
$T = ( 0,
\frac{1}{2},
\frac{1}{8},
\frac{1}{8} )
$,

\vskip 0.7ex
\hangindent=3em \hangafter=1
$S$ = ($ 1$,
$ 1$,
$ 1$,
$ 1$;\ \ 
$ 1$,
$ -1$,
$ -1$;\ \ 
$-\mathrm{i}$,
$\mathrm{i}$;\ \ 
$-\mathrm{i}$)

\vskip 0.7ex
\hangindent=3em \hangafter=1
$\tau_n$ = ($1.41 + 1.41 i$, $2. + 2. i$, $-1.41 + 1.41 i$, $0.$, $-1.41 - 1.41 i$, $2. - 2. i$, $1.41 - 1.41 i$, $4.$)

\vskip 0.7ex
\hangindent=3em \hangafter=1
\textit{Intrinsic sign problem}

  \vskip 2ex

\noindent7. $4_{3,4.}^{8,468}$ \irep{6}:\ \ 
$d_i$ = ($1.0$,
$1.0$,
$1.0$,
$1.0$) 

\vskip 0.7ex
\hangindent=3em \hangafter=1
$D^2= 4.0 = 
4$

\vskip 0.7ex
\hangindent=3em \hangafter=1
$T = ( 0,
\frac{1}{2},
\frac{3}{8},
\frac{3}{8} )
$,

\vskip 0.7ex
\hangindent=3em \hangafter=1
$S$ = ($ 1$,
$ 1$,
$ 1$,
$ 1$;\ \ 
$ 1$,
$ -1$,
$ -1$;\ \ 
$\mathrm{i}$,
$-\mathrm{i}$;\ \ 
$\mathrm{i}$)

\vskip 0.7ex
\hangindent=3em \hangafter=1
$\tau_n$ = ($-1.41 + 1.41 i$, $2. - 2. i$, $1.41 + 1.41 i$, $0.$, $1.41 - 1.41 i$, $2. + 2. i$, $-1.41 - 1.41 i$, $4.$)

\vskip 0.7ex
\hangindent=3em \hangafter=1
\textit{Intrinsic sign problem}

  \vskip 2ex

\noindent8. $4_{5,4.}^{8,312}$ \irep{6}:\ \ 
$d_i$ = ($1.0$,
$1.0$,
$1.0$,
$1.0$) 

\vskip 0.7ex
\hangindent=3em \hangafter=1
$D^2= 4.0 = 
4$

\vskip 0.7ex
\hangindent=3em \hangafter=1
$T = ( 0,
\frac{1}{2},
\frac{5}{8},
\frac{5}{8} )
$,

\vskip 0.7ex
\hangindent=3em \hangafter=1
$S$ = ($ 1$,
$ 1$,
$ 1$,
$ 1$;\ \ 
$ 1$,
$ -1$,
$ -1$;\ \ 
$-\mathrm{i}$,
$\mathrm{i}$;\ \ 
$-\mathrm{i}$)

\vskip 0.7ex
\hangindent=3em \hangafter=1
$\tau_n$ = ($-1.41 - 1.41 i$, $2. + 2. i$, $1.41 - 1.41 i$, $0.$, $1.41 + 1.41 i$, $2. - 2. i$, $-1.41 + 1.41 i$, $4.$)

\vskip 0.7ex
\hangindent=3em \hangafter=1
\textit{Intrinsic sign problem}

  \vskip 2ex

\noindent9. $4_{7,4.}^{8,781}$ \irep{6}:\ \ 
$d_i$ = ($1.0$,
$1.0$,
$1.0$,
$1.0$) 

\vskip 0.7ex
\hangindent=3em \hangafter=1
$D^2= 4.0 = 
4$

\vskip 0.7ex
\hangindent=3em \hangafter=1
$T = ( 0,
\frac{1}{2},
\frac{7}{8},
\frac{7}{8} )
$,

\vskip 0.7ex
\hangindent=3em \hangafter=1
$S$ = ($ 1$,
$ 1$,
$ 1$,
$ 1$;\ \ 
$ 1$,
$ -1$,
$ -1$;\ \ 
$\mathrm{i}$,
$-\mathrm{i}$;\ \ 
$\mathrm{i}$)

\vskip 0.7ex
\hangindent=3em \hangafter=1
$\tau_n$ = ($1.41 - 1.41 i$, $2. - 2. i$, $-1.41 - 1.41 i$, $0.$, $-1.41 + 1.41 i$, $2. + 2. i$, $1.41 + 1.41 i$, $4.$)

\vskip 0.7ex
\hangindent=3em \hangafter=1
\textit{Intrinsic sign problem}

  \vskip 2ex

\noindent10. $4_{\frac{19}{5},7.236}^{20,304}$ \irep{8}:\ \ 
$d_i$ = ($1.0$,
$1.0$,
$1.618$,
$1.618$) 

\vskip 0.7ex
\hangindent=3em \hangafter=1
$D^2= 7.236 = 
5+\sqrt{5}$

\vskip 0.7ex
\hangindent=3em \hangafter=1
$T = ( 0,
\frac{1}{4},
\frac{2}{5},
\frac{13}{20} )
$,

\vskip 0.7ex
\hangindent=3em \hangafter=1
$S$ = ($ 1$,
$ 1$,
$ \frac{1+\sqrt{5}}{2}$,
$ \frac{1+\sqrt{5}}{2}$;\ \ 
$ -1$,
$ \frac{1+\sqrt{5}}{2}$,
$ -\frac{1+\sqrt{5}}{2}$;\ \ 
$ -1$,
$ -1$;\ \ 
$ 1$)

Factors = $2_{1,2.}^{4,437}\boxtimes 2_{\frac{14}{5},3.618}^{5,395}$

\vskip 0.7ex
\hangindent=3em \hangafter=1
$\tau_n$ = ($-2.66 + 0.42 i$, $0.$, $4.3 + 0.68 i$, $-2.24 - 3.08 i$, $3.62 + 3.62 i$, $0.$, $-0.68 - 4.3 i$, $3.62 + 4.98 i$, $0.42 - 2.66 i$, $0.$, $0.42 + 2.66 i$, $3.62 - 4.98 i$, $-0.68 + 4.3 i$, $0.$, $3.62 - 3.62 i$, $-2.24 + 3.08 i$, $4.3 - 0.68 i$, $0.$, $-2.66 - 0.42 i$, $7.24$)

\vskip 0.7ex
\hangindent=3em \hangafter=1
\textit{Intrinsic sign problem}

  \vskip 2ex

\noindent11. $4_{\frac{31}{5},7.236}^{20,505}$ \irep{8}:\ \ 
$d_i$ = ($1.0$,
$1.0$,
$1.618$,
$1.618$) 

\vskip 0.7ex
\hangindent=3em \hangafter=1
$D^2= 7.236 = 
5+\sqrt{5}$

\vskip 0.7ex
\hangindent=3em \hangafter=1
$T = ( 0,
\frac{1}{4},
\frac{3}{5},
\frac{17}{20} )
$,

\vskip 0.7ex
\hangindent=3em \hangafter=1
$S$ = ($ 1$,
$ 1$,
$ \frac{1+\sqrt{5}}{2}$,
$ \frac{1+\sqrt{5}}{2}$;\ \ 
$ -1$,
$ \frac{1+\sqrt{5}}{2}$,
$ -\frac{1+\sqrt{5}}{2}$;\ \ 
$ -1$,
$ -1$;\ \ 
$ 1$)

Factors = $2_{1,2.}^{4,437}\boxtimes 2_{\frac{26}{5},3.618}^{5,720}$

\vskip 0.7ex
\hangindent=3em \hangafter=1
$\tau_n$ = ($0.42 - 2.66 i$, $0.$, $-0.68 - 4.3 i$, $-2.24 + 3.08 i$, $3.62 + 3.62 i$, $0.$, $4.3 + 0.68 i$, $3.62 - 4.98 i$, $-2.66 + 0.42 i$, $0.$, $-2.66 - 0.42 i$, $3.62 + 4.98 i$, $4.3 - 0.68 i$, $0.$, $3.62 - 3.62 i$, $-2.24 - 3.08 i$, $-0.68 + 4.3 i$, $0.$, $0.42 + 2.66 i$, $7.24$)

\vskip 0.7ex
\hangindent=3em \hangafter=1
\textit{Intrinsic sign problem}

  \vskip 2ex

\noindent12. $4_{\frac{9}{5},7.236}^{20,451}$ \irep{8}:\ \ 
$d_i$ = ($1.0$,
$1.0$,
$1.618$,
$1.618$) 

\vskip 0.7ex
\hangindent=3em \hangafter=1
$D^2= 7.236 = 
5+\sqrt{5}$

\vskip 0.7ex
\hangindent=3em \hangafter=1
$T = ( 0,
\frac{3}{4},
\frac{2}{5},
\frac{3}{20} )
$,

\vskip 0.7ex
\hangindent=3em \hangafter=1
$S$ = ($ 1$,
$ 1$,
$ \frac{1+\sqrt{5}}{2}$,
$ \frac{1+\sqrt{5}}{2}$;\ \ 
$ -1$,
$ \frac{1+\sqrt{5}}{2}$,
$ -\frac{1+\sqrt{5}}{2}$;\ \ 
$ -1$,
$ -1$;\ \ 
$ 1$)

Factors = $2_{7,2.}^{4,625}\boxtimes 2_{\frac{14}{5},3.618}^{5,395}$

\vskip 0.7ex
\hangindent=3em \hangafter=1
$\tau_n$ = ($0.42 + 2.66 i$, $0.$, $-0.68 + 4.3 i$, $-2.24 - 3.08 i$, $3.62 - 3.62 i$, $0.$, $4.3 - 0.68 i$, $3.62 + 4.98 i$, $-2.66 - 0.42 i$, $0.$, $-2.66 + 0.42 i$, $3.62 - 4.98 i$, $4.3 + 0.68 i$, $0.$, $3.62 + 3.62 i$, $-2.24 + 3.08 i$, $-0.68 - 4.3 i$, $0.$, $0.42 - 2.66 i$, $7.24$)

\vskip 0.7ex
\hangindent=3em \hangafter=1
\textit{Intrinsic sign problem}

  \vskip 2ex

\noindent13. $4_{\frac{21}{5},7.236}^{20,341}$ \irep{8}:\ \ 
$d_i$ = ($1.0$,
$1.0$,
$1.618$,
$1.618$) 

\vskip 0.7ex
\hangindent=3em \hangafter=1
$D^2= 7.236 = 
5+\sqrt{5}$

\vskip 0.7ex
\hangindent=3em \hangafter=1
$T = ( 0,
\frac{3}{4},
\frac{3}{5},
\frac{7}{20} )
$,

\vskip 0.7ex
\hangindent=3em \hangafter=1
$S$ = ($ 1$,
$ 1$,
$ \frac{1+\sqrt{5}}{2}$,
$ \frac{1+\sqrt{5}}{2}$;\ \ 
$ -1$,
$ \frac{1+\sqrt{5}}{2}$,
$ -\frac{1+\sqrt{5}}{2}$;\ \ 
$ -1$,
$ -1$;\ \ 
$ 1$)

Factors = $2_{7,2.}^{4,625}\boxtimes 2_{\frac{26}{5},3.618}^{5,720}$

\vskip 0.7ex
\hangindent=3em \hangafter=1
$\tau_n$ = ($-2.66 - 0.42 i$, $0.$, $4.3 - 0.68 i$, $-2.24 + 3.08 i$, $3.62 - 3.62 i$, $0.$, $-0.68 + 4.3 i$, $3.62 - 4.98 i$, $0.42 + 2.66 i$, $0.$, $0.42 - 2.66 i$, $3.62 + 4.98 i$, $-0.68 - 4.3 i$, $0.$, $3.62 + 3.62 i$, $-2.24 - 3.08 i$, $4.3 + 0.68 i$, $0.$, $-2.66 + 0.42 i$, $7.24$)

\vskip 0.7ex
\hangindent=3em \hangafter=1
\textit{Intrinsic sign problem}

  \vskip 2ex

\noindent14. $4_{\frac{28}{5},13.09}^{5,479}$ \irep{5}:\ \ 
$d_i$ = ($1.0$,
$1.618$,
$1.618$,
$2.618$) 

\vskip 0.7ex
\hangindent=3em \hangafter=1
$D^2= 13.90 = 
\frac{15+5\sqrt{5}}{2}$

\vskip 0.7ex
\hangindent=3em \hangafter=1
$T = ( 0,
\frac{2}{5},
\frac{2}{5},
\frac{4}{5} )
$,

\vskip 0.7ex
\hangindent=3em \hangafter=1
$S$ = ($ 1$,
$ \frac{1+\sqrt{5}}{2}$,
$ \frac{1+\sqrt{5}}{2}$,
$ \frac{3+\sqrt{5}}{2}$;\ \ 
$ -1$,
$ \frac{3+\sqrt{5}}{2}$,
$ -\frac{1+\sqrt{5}}{2}$;\ \ 
$ -1$,
$ -\frac{1+\sqrt{5}}{2}$;\ \ 
$ 1$)

Factors = $2_{\frac{14}{5},3.618}^{5,395}\boxtimes 2_{\frac{14}{5},3.618}^{5,395}$

\vskip 0.7ex
\hangindent=3em \hangafter=1
$\tau_n$ = ($-1.12 - 3.44 i$, $-2.93 - 9.01 i$, $-2.93 + 9.01 i$, $-1.12 + 3.44 i$, $13.09$)

\vskip 0.7ex
\hangindent=3em \hangafter=1
\textit{Intrinsic sign problem}

  \vskip 2ex

\noindent15. $4_{0,13.09}^{5,872}$ \irep{5}:\ \ 
$d_i$ = ($1.0$,
$1.618$,
$1.618$,
$2.618$) 

\vskip 0.7ex
\hangindent=3em \hangafter=1
$D^2= 13.90 = 
\frac{15+5\sqrt{5}}{2}$

\vskip 0.7ex
\hangindent=3em \hangafter=1
$T = ( 0,
\frac{2}{5},
\frac{3}{5},
0 )
$,

\vskip 0.7ex
\hangindent=3em \hangafter=1
$S$ = ($ 1$,
$ \frac{1+\sqrt{5}}{2}$,
$ \frac{1+\sqrt{5}}{2}$,
$ \frac{3+\sqrt{5}}{2}$;\ \ 
$ -1$,
$ \frac{3+\sqrt{5}}{2}$,
$ -\frac{1+\sqrt{5}}{2}$;\ \ 
$ -1$,
$ -\frac{1+\sqrt{5}}{2}$;\ \ 
$ 1$)

Factors = $2_{\frac{14}{5},3.618}^{5,395}\boxtimes 2_{\frac{26}{5},3.618}^{5,720}$

\vskip 0.7ex
\hangindent=3em \hangafter=1
$\tau_n$ = ($3.62$, $9.47$, $9.47$, $3.62$, $13.09$)

\vskip 0.7ex
\hangindent=3em \hangafter=1
\textit{Undetermined}

  \vskip 2ex

\noindent16. $4_{\frac{12}{5},13.09}^{5,443}$ \irep{5}:\ \ 
$d_i$ = ($1.0$,
$1.618$,
$1.618$,
$2.618$) 

\vskip 0.7ex
\hangindent=3em \hangafter=1
$D^2= 13.90 = 
\frac{15+5\sqrt{5}}{2}$

\vskip 0.7ex
\hangindent=3em \hangafter=1
$T = ( 0,
\frac{3}{5},
\frac{3}{5},
\frac{1}{5} )
$,

\vskip 0.7ex
\hangindent=3em \hangafter=1
$S$ = ($ 1$,
$ \frac{1+\sqrt{5}}{2}$,
$ \frac{1+\sqrt{5}}{2}$,
$ \frac{3+\sqrt{5}}{2}$;\ \ 
$ -1$,
$ \frac{3+\sqrt{5}}{2}$,
$ -\frac{1+\sqrt{5}}{2}$;\ \ 
$ -1$,
$ -\frac{1+\sqrt{5}}{2}$;\ \ 
$ 1$)

Factors = $2_{\frac{26}{5},3.618}^{5,720}\boxtimes 2_{\frac{26}{5},3.618}^{5,720}$

\vskip 0.7ex
\hangindent=3em \hangafter=1
$\tau_n$ = ($-1.12 + 3.44 i$, $-2.93 + 9.01 i$, $-2.93 - 9.01 i$, $-1.12 - 3.44 i$, $13.09$)

\vskip 0.7ex
\hangindent=3em \hangafter=1
\textit{Intrinsic sign problem}

  \vskip 2ex

\noindent17. $4_{\frac{10}{3},19.23}^{9,459}$ \irep{7}:\ \ 
$d_i$ = ($1.0$,
$1.879$,
$2.532$,
$2.879$) 

\vskip 0.7ex
\hangindent=3em \hangafter=1
$D^2= 19.234 = 
9+6c^{1}_{9}
+3c^{2}_{9}
$

\vskip 0.7ex
\hangindent=3em \hangafter=1
$T = ( 0,
\frac{1}{3},
\frac{2}{9},
\frac{2}{3} )
$,

\vskip 0.7ex
\hangindent=3em \hangafter=1
$S$ = ($ 1$,
$ -c_{9}^{4}$,
$ \xi_{9}^{3}$,
$ \xi_{9}^{5}$;\ \ 
$ -\xi_{9}^{5}$,
$ \xi_{9}^{3}$,
$ -1$;\ \ 
$0$,
$ -\xi_{9}^{3}$;\ \ 
$ -c_{9}^{4}$)

\vskip 0.7ex
\hangindent=3em \hangafter=1
$\tau_n$ = ($-3.8 + 2.19 i$, $-10.93 + 6.31 i$, $9.61 - 5.55 i$, $0. - 8.24 i$, $0. + 8.24 i$, $9.61 + 5.55 i$, $-10.93 - 6.31 i$, $-3.8 - 2.19 i$, $19.23$)

\vskip 0.7ex
\hangindent=3em \hangafter=1
\textit{Intrinsic sign problem}

  \vskip 2ex

\noindent18. $4_{\frac{14}{3},19.23}^{9,614}$ \irep{7}:\ \ 
$d_i$ = ($1.0$,
$1.879$,
$2.532$,
$2.879$) 

\vskip 0.7ex
\hangindent=3em \hangafter=1
$D^2= 19.234 = 
9+6c^{1}_{9}
+3c^{2}_{9}
$

\vskip 0.7ex
\hangindent=3em \hangafter=1
$T = ( 0,
\frac{2}{3},
\frac{7}{9},
\frac{1}{3} )
$,

\vskip 0.7ex
\hangindent=3em \hangafter=1
$S$ = ($ 1$,
$ -c_{9}^{4}$,
$ \xi_{9}^{3}$,
$ \xi_{9}^{5}$;\ \ 
$ -\xi_{9}^{5}$,
$ \xi_{9}^{3}$,
$ -1$;\ \ 
$0$,
$ -\xi_{9}^{3}$;\ \ 
$ -c_{9}^{4}$)

\vskip 0.7ex
\hangindent=3em \hangafter=1
$\tau_n$ = ($-3.8 - 2.19 i$, $-10.93 - 6.31 i$, $9.61 + 5.55 i$, $0. + 8.24 i$, $0. - 8.24 i$, $9.61 - 5.55 i$, $-10.93 + 6.31 i$, $-3.8 + 2.19 i$, $19.23$)

\vskip 0.7ex
\hangindent=3em \hangafter=1
\textit{Intrinsic sign problem}

  \vskip 2ex 

%% file: modular_data/SsL5U_.tex
\noindent1. $5_{0,5.}^{5,110}$ \irep{9}:\ \ 
$d_i$ = ($1.0$,
$1.0$,
$1.0$,
$1.0$,
$1.0$) 

\vskip 0.7ex
\hangindent=3em \hangafter=1
$D^2= 5.0 = 
5$

\vskip 0.7ex
\hangindent=3em \hangafter=1
$T = ( 0,
\frac{1}{5},
\frac{1}{5},
\frac{4}{5},
\frac{4}{5} )
$,

\vskip 0.7ex
\hangindent=3em \hangafter=1
$S$ = ($ 1$,
$ 1$,
$ 1$,
$ 1$,
$ 1$;\ \ 
$ -\zeta_{10}^{1}$,
$ \zeta_{5}^{2}$,
$ -\zeta_{10}^{3}$,
$ \zeta_{5}^{1}$;\ \ 
$ -\zeta_{10}^{1}$,
$ \zeta_{5}^{1}$,
$ -\zeta_{10}^{3}$;\ \ 
$ \zeta_{5}^{2}$,
$ -\zeta_{10}^{1}$;\ \ 
$ \zeta_{5}^{2}$)

\vskip 0.7ex
\hangindent=3em \hangafter=1
$\tau_n$ = ($2.24$, $-2.24$, $-2.24$, $2.24$, $5.$)

\vskip 0.7ex
\hangindent=3em \hangafter=1
\textit{Intrinsic sign problem}

  \vskip 2ex

\noindent2. $5_{4,5.}^{5,210}$ \irep{9}:\ \ 
$d_i$ = ($1.0$,
$1.0$,
$1.0$,
$1.0$,
$1.0$) 

\vskip 0.7ex
\hangindent=3em \hangafter=1
$D^2= 5.0 = 
5$

\vskip 0.7ex
\hangindent=3em \hangafter=1
$T = ( 0,
\frac{2}{5},
\frac{2}{5},
\frac{3}{5},
\frac{3}{5} )
$,

\vskip 0.7ex
\hangindent=3em \hangafter=1
$S$ = ($ 1$,
$ 1$,
$ 1$,
$ 1$,
$ 1$;\ \ 
$ \zeta_{5}^{1}$,
$ -\zeta_{10}^{3}$,
$ -\zeta_{10}^{1}$,
$ \zeta_{5}^{2}$;\ \ 
$ \zeta_{5}^{1}$,
$ \zeta_{5}^{2}$,
$ -\zeta_{10}^{1}$;\ \ 
$ -\zeta_{10}^{3}$,
$ \zeta_{5}^{1}$;\ \ 
$ -\zeta_{10}^{3}$)

\vskip 0.7ex
\hangindent=3em \hangafter=1
$\tau_n$ = ($-2.24$, $2.24$, $2.24$, $-2.24$, $5.$)

\vskip 0.7ex
\hangindent=3em \hangafter=1
\textit{Intrinsic sign problem}

  \vskip 2ex

\noindent3. $5_{2,12.}^{24,940}$ \irep{17}:\ \ 
$d_i$ = ($1.0$,
$1.0$,
$1.732$,
$1.732$,
$2.0$) 

\vskip 0.7ex
\hangindent=3em \hangafter=1
$D^2= 12.0 = 
12$

\vskip 0.7ex
\hangindent=3em \hangafter=1
$T = ( 0,
0,
\frac{1}{8},
\frac{5}{8},
\frac{1}{3} )
$,

\vskip 0.7ex
\hangindent=3em \hangafter=1
$S$ = ($ 1$,
$ 1$,
$ \sqrt{3}$,
$ \sqrt{3}$,
$ 2$;\ \ 
$ 1$,
$ -\sqrt{3}$,
$ -\sqrt{3}$,
$ 2$;\ \ 
$ \sqrt{3}$,
$ -\sqrt{3}$,
$0$;\ \ 
$ \sqrt{3}$,
$0$;\ \ 
$ -2$)

\vskip 0.7ex
\hangindent=3em \hangafter=1
$\tau_n$ = ($0. + 3.46 i$, $0. + 2.54 i$, $6.$, $-6. + 3.46 i$, $0. - 3.46 i$, $6. - 6. i$, $0. + 3.46 i$, $6. - 3.46 i$, $6.$, $0. + 9.46 i$, $0. - 3.46 i$, $0.$, $0. + 3.46 i$, $0. - 9.46 i$, $6.$, $6. + 3.46 i$, $0. - 3.46 i$, $6. + 6. i$, $0. + 3.46 i$, $-6. - 3.46 i$, $6.$, $0. - 2.54 i$, $0. - 3.46 i$, $12.$)

\vskip 0.7ex
\hangindent=3em \hangafter=1
\textit{Intrinsic sign problem}

  \vskip 2ex

\noindent4. $5_{6,12.}^{24,273}$ \irep{17}:\ \ 
$d_i$ = ($1.0$,
$1.0$,
$1.732$,
$1.732$,
$2.0$) 

\vskip 0.7ex
\hangindent=3em \hangafter=1
$D^2= 12.0 = 
12$

\vskip 0.7ex
\hangindent=3em \hangafter=1
$T = ( 0,
0,
\frac{1}{8},
\frac{5}{8},
\frac{2}{3} )
$,

\vskip 0.7ex
\hangindent=3em \hangafter=1
$S$ = ($ 1$,
$ 1$,
$ \sqrt{3}$,
$ \sqrt{3}$,
$ 2$;\ \ 
$ 1$,
$ -\sqrt{3}$,
$ -\sqrt{3}$,
$ 2$;\ \ 
$ -\sqrt{3}$,
$ \sqrt{3}$,
$0$;\ \ 
$ -\sqrt{3}$,
$0$;\ \ 
$ -2$)

\vskip 0.7ex
\hangindent=3em \hangafter=1
$\tau_n$ = ($0. - 3.46 i$, $0. + 9.46 i$, $6.$, $-6. - 3.46 i$, $0. + 3.46 i$, $6. - 6. i$, $0. - 3.46 i$, $6. + 3.46 i$, $6.$, $0. + 2.54 i$, $0. + 3.46 i$, $0.$, $0. - 3.46 i$, $0. - 2.54 i$, $6.$, $6. - 3.46 i$, $0. + 3.46 i$, $6. + 6. i$, $0. - 3.46 i$, $-6. + 3.46 i$, $6.$, $0. - 9.46 i$, $0. + 3.46 i$, $12.$)

\vskip 0.7ex
\hangindent=3em \hangafter=1
\textit{Intrinsic sign problem}

  \vskip 2ex

\noindent5. $5_{2,12.}^{24,741}$ \irep{17}:\ \ 
$d_i$ = ($1.0$,
$1.0$,
$1.732$,
$1.732$,
$2.0$) 

\vskip 0.7ex
\hangindent=3em \hangafter=1
$D^2= 12.0 = 
12$

\vskip 0.7ex
\hangindent=3em \hangafter=1
$T = ( 0,
0,
\frac{3}{8},
\frac{7}{8},
\frac{1}{3} )
$,

\vskip 0.7ex
\hangindent=3em \hangafter=1
$S$ = ($ 1$,
$ 1$,
$ \sqrt{3}$,
$ \sqrt{3}$,
$ 2$;\ \ 
$ 1$,
$ -\sqrt{3}$,
$ -\sqrt{3}$,
$ 2$;\ \ 
$ -\sqrt{3}$,
$ \sqrt{3}$,
$0$;\ \ 
$ -\sqrt{3}$,
$0$;\ \ 
$ -2$)

\vskip 0.7ex
\hangindent=3em \hangafter=1
$\tau_n$ = ($0. + 3.46 i$, $0. - 9.46 i$, $6.$, $-6. + 3.46 i$, $0. - 3.46 i$, $6. + 6. i$, $0. + 3.46 i$, $6. - 3.46 i$, $6.$, $0. - 2.54 i$, $0. - 3.46 i$, $0.$, $0. + 3.46 i$, $0. + 2.54 i$, $6.$, $6. + 3.46 i$, $0. - 3.46 i$, $6. - 6. i$, $0. + 3.46 i$, $-6. - 3.46 i$, $6.$, $0. + 9.46 i$, $0. - 3.46 i$, $12.$)

\vskip 0.7ex
\hangindent=3em \hangafter=1
\textit{Intrinsic sign problem}

  \vskip 2ex

\noindent6. $5_{6,12.}^{24,592}$ \irep{17}:\ \ 
$d_i$ = ($1.0$,
$1.0$,
$1.732$,
$1.732$,
$2.0$) 

\vskip 0.7ex
\hangindent=3em \hangafter=1
$D^2= 12.0 = 
12$

\vskip 0.7ex
\hangindent=3em \hangafter=1
$T = ( 0,
0,
\frac{3}{8},
\frac{7}{8},
\frac{2}{3} )
$,

\vskip 0.7ex
\hangindent=3em \hangafter=1
$S$ = ($ 1$,
$ 1$,
$ \sqrt{3}$,
$ \sqrt{3}$,
$ 2$;\ \ 
$ 1$,
$ -\sqrt{3}$,
$ -\sqrt{3}$,
$ 2$;\ \ 
$ \sqrt{3}$,
$ -\sqrt{3}$,
$0$;\ \ 
$ \sqrt{3}$,
$0$;\ \ 
$ -2$)

\vskip 0.7ex
\hangindent=3em \hangafter=1
$\tau_n$ = ($0. - 3.46 i$, $0. - 2.54 i$, $6.$, $-6. - 3.46 i$, $0. + 3.46 i$, $6. + 6. i$, $0. - 3.46 i$, $6. + 3.46 i$, $6.$, $0. - 9.46 i$, $0. + 3.46 i$, $0.$, $0. - 3.46 i$, $0. + 9.46 i$, $6.$, $6. - 3.46 i$, $0. + 3.46 i$, $6. - 6. i$, $0. - 3.46 i$, $-6. + 3.46 i$, $6.$, $0. + 2.54 i$, $0. + 3.46 i$, $12.$)

\vskip 0.7ex
\hangindent=3em \hangafter=1
\textit{Intrinsic sign problem}

  \vskip 2ex

\noindent7. $5_{\frac{72}{11},34.64}^{11,216}$ \irep{15}:\ \ 
$d_i$ = ($1.0$,
$1.918$,
$2.682$,
$3.228$,
$3.513$) 

\vskip 0.7ex
\hangindent=3em \hangafter=1
$D^2= 34.646 = 
15+10c^{1}_{11}
+6c^{2}_{11}
+3c^{3}_{11}
+c^{4}_{11}
$

\vskip 0.7ex
\hangindent=3em \hangafter=1
$T = ( 0,
\frac{2}{11},
\frac{9}{11},
\frac{10}{11},
\frac{5}{11} )
$,

\vskip 0.7ex
\hangindent=3em \hangafter=1
$S$ = ($ 1$,
$ -c_{11}^{5}$,
$ \xi_{11}^{3}$,
$ \xi_{11}^{7}$,
$ \xi_{11}^{5}$;\ \ 
$ -\xi_{11}^{7}$,
$ \xi_{11}^{5}$,
$ -\xi_{11}^{3}$,
$ 1$;\ \ 
$ -c_{11}^{5}$,
$ -1$,
$ -\xi_{11}^{7}$;\ \ 
$ \xi_{11}^{5}$,
$ c_{11}^{5}$;\ \ 
$ \xi_{11}^{3}$)

\vskip 0.7ex
\hangindent=3em \hangafter=1
$\tau_n$ = ($2.44 - 5.35 i$, $8.59 - 18.81 i$, $-19.$, $-2.24 - 15.62 i$, $-1.61 + 11.18 i$, $-1.61 - 11.18 i$, $-2.24 + 15.62 i$, $-19.$, $8.59 + 18.81 i$, $2.44 + 5.35 i$, $34.63$)

\vskip 0.7ex
\hangindent=3em \hangafter=1
\textit{Intrinsic sign problem}

  \vskip 2ex

\noindent8. $5_{\frac{16}{11},34.64}^{11,640}$ \irep{15}:\ \ 
$d_i$ = ($1.0$,
$1.918$,
$2.682$,
$3.228$,
$3.513$) 

\vskip 0.7ex
\hangindent=3em \hangafter=1
$D^2= 34.646 = 
15+10c^{1}_{11}
+6c^{2}_{11}
+3c^{3}_{11}
+c^{4}_{11}
$

\vskip 0.7ex
\hangindent=3em \hangafter=1
$T = ( 0,
\frac{9}{11},
\frac{2}{11},
\frac{1}{11},
\frac{6}{11} )
$,

\vskip 0.7ex
\hangindent=3em \hangafter=1
$S$ = ($ 1$,
$ -c_{11}^{5}$,
$ \xi_{11}^{3}$,
$ \xi_{11}^{7}$,
$ \xi_{11}^{5}$;\ \ 
$ -\xi_{11}^{7}$,
$ \xi_{11}^{5}$,
$ -\xi_{11}^{3}$,
$ 1$;\ \ 
$ -c_{11}^{5}$,
$ -1$,
$ -\xi_{11}^{7}$;\ \ 
$ \xi_{11}^{5}$,
$ c_{11}^{5}$;\ \ 
$ \xi_{11}^{3}$)

\vskip 0.7ex
\hangindent=3em \hangafter=1
$\tau_n$ = ($2.44 + 5.35 i$, $8.59 + 18.81 i$, $-19.$, $-2.24 + 15.62 i$, $-1.61 - 11.18 i$, $-1.61 + 11.18 i$, $-2.24 - 15.62 i$, $-19.$, $8.59 - 18.81 i$, $2.44 - 5.35 i$, $34.63$)

\vskip 0.7ex
\hangindent=3em \hangafter=1
\textit{Intrinsic sign problem}

  \vskip 2ex

\noindent9. $5_{\frac{38}{7},35.34}^{7,386}$ \irep{11}:\ \ 
$d_i$ = ($1.0$,
$2.246$,
$2.246$,
$2.801$,
$4.48$) 

\vskip 0.7ex
\hangindent=3em \hangafter=1
$D^2= 35.342 = 
21+14c^{1}_{7}
+7c^{2}_{7}
$

\vskip 0.7ex
\hangindent=3em \hangafter=1
$T = ( 0,
\frac{1}{7},
\frac{1}{7},
\frac{6}{7},
\frac{4}{7} )
$,

\vskip 0.7ex
\hangindent=3em \hangafter=1
$S$ = ($ 1$,
$ \xi_{7}^{3}$,
$ \xi_{7}^{3}$,
$ 2+c^{1}_{7}
+c^{2}_{7}
$,
$ 2+2c^{1}_{7}
+c^{2}_{7}
$;\ \ 
$ s^{1}_{7}
+\zeta^{2}_{7}
+\zeta^{3}_{7}
$,
$ -1-2  \zeta^{1}_{7}
-\zeta^{2}_{7}
-\zeta^{3}_{7}
$,
$ -\xi_{7}^{3}$,
$ \xi_{7}^{3}$;\ \ 
$ s^{1}_{7}
+\zeta^{2}_{7}
+\zeta^{3}_{7}
$,
$ -\xi_{7}^{3}$,
$ \xi_{7}^{3}$;\ \ 
$ 2+2c^{1}_{7}
+c^{2}_{7}
$,
$ -1$;\ \ 
$ -2-c^{1}_{7}
-c^{2}_{7}
$)

\vskip 0.7ex
\hangindent=3em \hangafter=1
$\tau_n$ = ($-5.9 - 6.95 i$, $9.52 + 17.88 i$, $-19.62 - 18.59 i$, $-19.62 + 18.59 i$, $9.52 - 17.88 i$, $-5.9 + 6.95 i$, $39.01$)

\vskip 0.7ex
\hangindent=3em \hangafter=1
\textit{Intrinsic sign problem}

  \vskip 2ex

\noindent10. $5_{\frac{18}{7},35.34}^{7,101}$ \irep{11}:\ \ 
$d_i$ = ($1.0$,
$2.246$,
$2.246$,
$2.801$,
$4.48$) 

\vskip 0.7ex
\hangindent=3em \hangafter=1
$D^2= 35.342 = 
21+14c^{1}_{7}
+7c^{2}_{7}
$

\vskip 0.7ex
\hangindent=3em \hangafter=1
$T = ( 0,
\frac{6}{7},
\frac{6}{7},
\frac{1}{7},
\frac{3}{7} )
$,

\vskip 0.7ex
\hangindent=3em \hangafter=1
$S$ = ($ 1$,
$ \xi_{7}^{3}$,
$ \xi_{7}^{3}$,
$ 2+c^{1}_{7}
+c^{2}_{7}
$,
$ 2+2c^{1}_{7}
+c^{2}_{7}
$;\ \ 
$ -1-2  \zeta^{1}_{7}
-\zeta^{2}_{7}
-\zeta^{3}_{7}
$,
$ s^{1}_{7}
+\zeta^{2}_{7}
+\zeta^{3}_{7}
$,
$ -\xi_{7}^{3}$,
$ \xi_{7}^{3}$;\ \ 
$ -1-2  \zeta^{1}_{7}
-\zeta^{2}_{7}
-\zeta^{3}_{7}
$,
$ -\xi_{7}^{3}$,
$ \xi_{7}^{3}$;\ \ 
$ 2+2c^{1}_{7}
+c^{2}_{7}
$,
$ -1$;\ \ 
$ -2-c^{1}_{7}
-c^{2}_{7}
$)

\vskip 0.7ex
\hangindent=3em \hangafter=1
$\tau_n$ = ($-5.9 + 6.95 i$, $9.52 - 17.88 i$, $-19.62 + 18.59 i$, $-19.62 - 18.59 i$, $9.52 + 17.88 i$, $-5.9 - 6.95 i$, $39.01$)

\vskip 0.7ex
\hangindent=3em \hangafter=1
\textit{Intrinsic sign problem}

  \vskip 2ex 

%% file: modular_data/SsL6U_.tex
\noindent1. $6_{3,6.}^{12,534}$ \irep{34}:\ \ 
$d_i$ = ($1.0$,
$1.0$,
$1.0$,
$1.0$,
$1.0$,
$1.0$) 

\vskip 0.7ex
\hangindent=3em \hangafter=1
$D^2= 6.0 = 
6$

\vskip 0.7ex
\hangindent=3em \hangafter=1
$T = ( 0,
\frac{1}{3},
\frac{1}{3},
\frac{1}{4},
\frac{7}{12},
\frac{7}{12} )
$,

\vskip 0.7ex
\hangindent=3em \hangafter=1
$S$ = ($ 1$,
$ 1$,
$ 1$,
$ 1$,
$ 1$,
$ 1$;\ \ 
$ \zeta_{3}^{1}$,
$ -\zeta_{6}^{1}$,
$ 1$,
$ -\zeta_{6}^{1}$,
$ \zeta_{3}^{1}$;\ \ 
$ \zeta_{3}^{1}$,
$ 1$,
$ \zeta_{3}^{1}$,
$ -\zeta_{6}^{1}$;\ \ 
$ -1$,
$ -1$,
$ -1$;\ \ 
$ -\zeta_{3}^{1}$,
$ \zeta_{6}^{1}$;\ \ 
$ -\zeta_{3}^{1}$)

Factors = $2_{1,2.}^{4,437}\boxtimes 3_{2,3.}^{3,527}$

\vskip 0.7ex
\hangindent=3em \hangafter=1
$\tau_n$ = ($-1.73 + 1.73 i$, $0.$, $3. - 3. i$, $0. + 3.46 i$, $1.73 - 1.73 i$, $0.$, $1.73 + 1.73 i$, $0. - 3.46 i$, $3. + 3. i$, $0.$, $-1.73 - 1.73 i$, $6.$)

\vskip 0.7ex
\hangindent=3em \hangafter=1
\textit{Intrinsic sign problem}

  \vskip 2ex

\noindent2. $6_{1,6.}^{12,701}$ \irep{34}:\ \ 
$d_i$ = ($1.0$,
$1.0$,
$1.0$,
$1.0$,
$1.0$,
$1.0$) 

\vskip 0.7ex
\hangindent=3em \hangafter=1
$D^2= 6.0 = 
6$

\vskip 0.7ex
\hangindent=3em \hangafter=1
$T = ( 0,
\frac{1}{3},
\frac{1}{3},
\frac{3}{4},
\frac{1}{12},
\frac{1}{12} )
$,

\vskip 0.7ex
\hangindent=3em \hangafter=1
$S$ = ($ 1$,
$ 1$,
$ 1$,
$ 1$,
$ 1$,
$ 1$;\ \ 
$ \zeta_{3}^{1}$,
$ -\zeta_{6}^{1}$,
$ 1$,
$ -\zeta_{6}^{1}$,
$ \zeta_{3}^{1}$;\ \ 
$ \zeta_{3}^{1}$,
$ 1$,
$ \zeta_{3}^{1}$,
$ -\zeta_{6}^{1}$;\ \ 
$ -1$,
$ -1$,
$ -1$;\ \ 
$ -\zeta_{3}^{1}$,
$ \zeta_{6}^{1}$;\ \ 
$ -\zeta_{3}^{1}$)

Factors = $2_{7,2.}^{4,625}\boxtimes 3_{2,3.}^{3,527}$

\vskip 0.7ex
\hangindent=3em \hangafter=1
$\tau_n$ = ($1.73 + 1.73 i$, $0.$, $3. + 3. i$, $0. + 3.46 i$, $-1.73 - 1.73 i$, $0.$, $-1.73 + 1.73 i$, $0. - 3.46 i$, $3. - 3. i$, $0.$, $1.73 - 1.73 i$, $6.$)

\vskip 0.7ex
\hangindent=3em \hangafter=1
\textit{Intrinsic sign problem}

  \vskip 2ex

\noindent3. $6_{7,6.}^{12,113}$ \irep{34}:\ \ 
$d_i$ = ($1.0$,
$1.0$,
$1.0$,
$1.0$,
$1.0$,
$1.0$) 

\vskip 0.7ex
\hangindent=3em \hangafter=1
$D^2= 6.0 = 
6$

\vskip 0.7ex
\hangindent=3em \hangafter=1
$T = ( 0,
\frac{2}{3},
\frac{2}{3},
\frac{1}{4},
\frac{11}{12},
\frac{11}{12} )
$,

\vskip 0.7ex
\hangindent=3em \hangafter=1
$S$ = ($ 1$,
$ 1$,
$ 1$,
$ 1$,
$ 1$,
$ 1$;\ \ 
$ -\zeta_{6}^{1}$,
$ \zeta_{3}^{1}$,
$ 1$,
$ -\zeta_{6}^{1}$,
$ \zeta_{3}^{1}$;\ \ 
$ -\zeta_{6}^{1}$,
$ 1$,
$ \zeta_{3}^{1}$,
$ -\zeta_{6}^{1}$;\ \ 
$ -1$,
$ -1$,
$ -1$;\ \ 
$ \zeta_{6}^{1}$,
$ -\zeta_{3}^{1}$;\ \ 
$ \zeta_{6}^{1}$)

Factors = $2_{1,2.}^{4,437}\boxtimes 3_{6,3.}^{3,138}$

\vskip 0.7ex
\hangindent=3em \hangafter=1
$\tau_n$ = ($1.73 - 1.73 i$, $0.$, $3. - 3. i$, $0. - 3.46 i$, $-1.73 + 1.73 i$, $0.$, $-1.73 - 1.73 i$, $0. + 3.46 i$, $3. + 3. i$, $0.$, $1.73 + 1.73 i$, $6.$)

\vskip 0.7ex
\hangindent=3em \hangafter=1
\textit{Intrinsic sign problem}

  \vskip 2ex

\noindent4. $6_{5,6.}^{12,298}$ \irep{34}:\ \ 
$d_i$ = ($1.0$,
$1.0$,
$1.0$,
$1.0$,
$1.0$,
$1.0$) 

\vskip 0.7ex
\hangindent=3em \hangafter=1
$D^2= 6.0 = 
6$

\vskip 0.7ex
\hangindent=3em \hangafter=1
$T = ( 0,
\frac{2}{3},
\frac{2}{3},
\frac{3}{4},
\frac{5}{12},
\frac{5}{12} )
$,

\vskip 0.7ex
\hangindent=3em \hangafter=1
$S$ = ($ 1$,
$ 1$,
$ 1$,
$ 1$,
$ 1$,
$ 1$;\ \ 
$ -\zeta_{6}^{1}$,
$ \zeta_{3}^{1}$,
$ 1$,
$ -\zeta_{6}^{1}$,
$ \zeta_{3}^{1}$;\ \ 
$ -\zeta_{6}^{1}$,
$ 1$,
$ \zeta_{3}^{1}$,
$ -\zeta_{6}^{1}$;\ \ 
$ -1$,
$ -1$,
$ -1$;\ \ 
$ \zeta_{6}^{1}$,
$ -\zeta_{3}^{1}$;\ \ 
$ \zeta_{6}^{1}$)

Factors = $2_{7,2.}^{4,625}\boxtimes 3_{6,3.}^{3,138}$

\vskip 0.7ex
\hangindent=3em \hangafter=1
$\tau_n$ = ($-1.73 - 1.73 i$, $0.$, $3. + 3. i$, $0. - 3.46 i$, $1.73 + 1.73 i$, $0.$, $1.73 - 1.73 i$, $0. + 3.46 i$, $3. - 3. i$, $0.$, $-1.73 + 1.73 i$, $6.$)

\vskip 0.7ex
\hangindent=3em \hangafter=1
\textit{Intrinsic sign problem}

  \vskip 2ex

\noindent5. $6_{\frac{3}{2},8.}^{16,688}$ \irep{39}:\ \ 
$d_i$ = ($1.0$,
$1.0$,
$1.0$,
$1.0$,
$1.414$,
$1.414$) 

\vskip 0.7ex
\hangindent=3em \hangafter=1
$D^2= 8.0 = 
8$

\vskip 0.7ex
\hangindent=3em \hangafter=1
$T = ( 0,
\frac{1}{2},
\frac{1}{4},
\frac{3}{4},
\frac{1}{16},
\frac{5}{16} )
$,

\vskip 0.7ex
\hangindent=3em \hangafter=1
$S$ = ($ 1$,
$ 1$,
$ 1$,
$ 1$,
$ \sqrt{2}$,
$ \sqrt{2}$;\ \ 
$ 1$,
$ 1$,
$ 1$,
$ -\sqrt{2}$,
$ -\sqrt{2}$;\ \ 
$ -1$,
$ -1$,
$ \sqrt{2}$,
$ -\sqrt{2}$;\ \ 
$ -1$,
$ -\sqrt{2}$,
$ \sqrt{2}$;\ \ 
$0$,
$0$;\ \ 
$0$)

Factors = $2_{1,2.}^{4,437}\boxtimes 3_{\frac{1}{2},4.}^{16,598}$

\vskip 0.7ex
\hangindent=3em \hangafter=1
$\tau_n$ = ($1.08 + 2.61 i$, $0.$, $2.61 + 1.08 i$, $4. + 4. i$, $-2.61 + 1.08 i$, $0.$, $-1.08 + 2.61 i$, $0.$, $-1.08 - 2.61 i$, $0.$, $-2.61 - 1.08 i$, $4. - 4. i$, $2.61 - 1.08 i$, $0.$, $1.08 - 2.61 i$, $8.$)

\vskip 0.7ex
\hangindent=3em \hangafter=1
\textit{Intrinsic sign problem}

  \vskip 2ex

\noindent6. $6_{\frac{15}{2},8.}^{16,107}$ \irep{39}:\ \ 
$d_i$ = ($1.0$,
$1.0$,
$1.0$,
$1.0$,
$1.414$,
$1.414$) 

\vskip 0.7ex
\hangindent=3em \hangafter=1
$D^2= 8.0 = 
8$

\vskip 0.7ex
\hangindent=3em \hangafter=1
$T = ( 0,
\frac{1}{2},
\frac{1}{4},
\frac{3}{4},
\frac{1}{16},
\frac{13}{16} )
$,

\vskip 0.7ex
\hangindent=3em \hangafter=1
$S$ = ($ 1$,
$ 1$,
$ 1$,
$ 1$,
$ \sqrt{2}$,
$ \sqrt{2}$;\ \ 
$ 1$,
$ 1$,
$ 1$,
$ -\sqrt{2}$,
$ -\sqrt{2}$;\ \ 
$ -1$,
$ -1$,
$ -\sqrt{2}$,
$ \sqrt{2}$;\ \ 
$ -1$,
$ \sqrt{2}$,
$ -\sqrt{2}$;\ \ 
$0$,
$0$;\ \ 
$0$)

Factors = $2_{7,2.}^{4,625}\boxtimes 3_{\frac{1}{2},4.}^{16,598}$

\vskip 0.7ex
\hangindent=3em \hangafter=1
$\tau_n$ = ($2.61 - 1.08 i$, $0.$, $-1.08 + 2.61 i$, $4. + 4. i$, $1.08 + 2.61 i$, $0.$, $-2.61 - 1.08 i$, $0.$, $-2.61 + 1.08 i$, $0.$, $1.08 - 2.61 i$, $4. - 4. i$, $-1.08 - 2.61 i$, $0.$, $2.61 + 1.08 i$, $8.$)

\vskip 0.7ex
\hangindent=3em \hangafter=1
\textit{Intrinsic sign problem}

  \vskip 2ex

\noindent7. $6_{\frac{5}{2},8.}^{16,511}$ \irep{39}:\ \ 
$d_i$ = ($1.0$,
$1.0$,
$1.0$,
$1.0$,
$1.414$,
$1.414$) 

\vskip 0.7ex
\hangindent=3em \hangafter=1
$D^2= 8.0 = 
8$

\vskip 0.7ex
\hangindent=3em \hangafter=1
$T = ( 0,
\frac{1}{2},
\frac{1}{4},
\frac{3}{4},
\frac{3}{16},
\frac{7}{16} )
$,

\vskip 0.7ex
\hangindent=3em \hangafter=1
$S$ = ($ 1$,
$ 1$,
$ 1$,
$ 1$,
$ \sqrt{2}$,
$ \sqrt{2}$;\ \ 
$ 1$,
$ 1$,
$ 1$,
$ -\sqrt{2}$,
$ -\sqrt{2}$;\ \ 
$ -1$,
$ -1$,
$ \sqrt{2}$,
$ -\sqrt{2}$;\ \ 
$ -1$,
$ -\sqrt{2}$,
$ \sqrt{2}$;\ \ 
$0$,
$0$;\ \ 
$0$)

Factors = $2_{7,2.}^{4,625}\boxtimes 3_{\frac{7}{2},4.}^{16,332}$

\vskip 0.7ex
\hangindent=3em \hangafter=1
$\tau_n$ = ($-1.08 + 2.61 i$, $0.$, $-2.61 + 1.08 i$, $4. - 4. i$, $2.61 + 1.08 i$, $0.$, $1.08 + 2.61 i$, $0.$, $1.08 - 2.61 i$, $0.$, $2.61 - 1.08 i$, $4. + 4. i$, $-2.61 - 1.08 i$, $0.$, $-1.08 - 2.61 i$, $8.$)

\vskip 0.7ex
\hangindent=3em \hangafter=1
\textit{Intrinsic sign problem}

  \vskip 2ex

\noindent8. $6_{\frac{1}{2},8.}^{16,460}$ \irep{39}:\ \ 
$d_i$ = ($1.0$,
$1.0$,
$1.0$,
$1.0$,
$1.414$,
$1.414$) 

\vskip 0.7ex
\hangindent=3em \hangafter=1
$D^2= 8.0 = 
8$

\vskip 0.7ex
\hangindent=3em \hangafter=1
$T = ( 0,
\frac{1}{2},
\frac{1}{4},
\frac{3}{4},
\frac{3}{16},
\frac{15}{16} )
$,

\vskip 0.7ex
\hangindent=3em \hangafter=1
$S$ = ($ 1$,
$ 1$,
$ 1$,
$ 1$,
$ \sqrt{2}$,
$ \sqrt{2}$;\ \ 
$ 1$,
$ 1$,
$ 1$,
$ -\sqrt{2}$,
$ -\sqrt{2}$;\ \ 
$ -1$,
$ -1$,
$ -\sqrt{2}$,
$ \sqrt{2}$;\ \ 
$ -1$,
$ \sqrt{2}$,
$ -\sqrt{2}$;\ \ 
$0$,
$0$;\ \ 
$0$)

Factors = $2_{1,2.}^{4,437}\boxtimes 3_{\frac{15}{2},4.}^{16,639}$

\vskip 0.7ex
\hangindent=3em \hangafter=1
$\tau_n$ = ($2.61 + 1.08 i$, $0.$, $-1.08 - 2.61 i$, $4. - 4. i$, $1.08 - 2.61 i$, $0.$, $-2.61 + 1.08 i$, $0.$, $-2.61 - 1.08 i$, $0.$, $1.08 + 2.61 i$, $4. + 4. i$, $-1.08 + 2.61 i$, $0.$, $2.61 - 1.08 i$, $8.$)

\vskip 0.7ex
\hangindent=3em \hangafter=1
\textit{Intrinsic sign problem}

  \vskip 2ex

\noindent9. $6_{\frac{7}{2},8.}^{16,246}$ \irep{39}:\ \ 
$d_i$ = ($1.0$,
$1.0$,
$1.0$,
$1.0$,
$1.414$,
$1.414$) 

\vskip 0.7ex
\hangindent=3em \hangafter=1
$D^2= 8.0 = 
8$

\vskip 0.7ex
\hangindent=3em \hangafter=1
$T = ( 0,
\frac{1}{2},
\frac{1}{4},
\frac{3}{4},
\frac{5}{16},
\frac{9}{16} )
$,

\vskip 0.7ex
\hangindent=3em \hangafter=1
$S$ = ($ 1$,
$ 1$,
$ 1$,
$ 1$,
$ \sqrt{2}$,
$ \sqrt{2}$;\ \ 
$ 1$,
$ 1$,
$ 1$,
$ -\sqrt{2}$,
$ -\sqrt{2}$;\ \ 
$ -1$,
$ -1$,
$ \sqrt{2}$,
$ -\sqrt{2}$;\ \ 
$ -1$,
$ -\sqrt{2}$,
$ \sqrt{2}$;\ \ 
$0$,
$0$;\ \ 
$0$)

Factors = $2_{7,2.}^{4,625}\boxtimes 3_{\frac{9}{2},4.}^{16,156}$

\vskip 0.7ex
\hangindent=3em \hangafter=1
$\tau_n$ = ($-2.61 + 1.08 i$, $0.$, $1.08 - 2.61 i$, $4. + 4. i$, $-1.08 - 2.61 i$, $0.$, $2.61 + 1.08 i$, $0.$, $2.61 - 1.08 i$, $0.$, $-1.08 + 2.61 i$, $4. - 4. i$, $1.08 + 2.61 i$, $0.$, $-2.61 - 1.08 i$, $8.$)

\vskip 0.7ex
\hangindent=3em \hangafter=1
\textit{Intrinsic sign problem}

  \vskip 2ex

\noindent10. $6_{\frac{9}{2},8.}^{16,107}$ \irep{39}:\ \ 
$d_i$ = ($1.0$,
$1.0$,
$1.0$,
$1.0$,
$1.414$,
$1.414$) 

\vskip 0.7ex
\hangindent=3em \hangafter=1
$D^2= 8.0 = 
8$

\vskip 0.7ex
\hangindent=3em \hangafter=1
$T = ( 0,
\frac{1}{2},
\frac{1}{4},
\frac{3}{4},
\frac{7}{16},
\frac{11}{16} )
$,

\vskip 0.7ex
\hangindent=3em \hangafter=1
$S$ = ($ 1$,
$ 1$,
$ 1$,
$ 1$,
$ \sqrt{2}$,
$ \sqrt{2}$;\ \ 
$ 1$,
$ 1$,
$ 1$,
$ -\sqrt{2}$,
$ -\sqrt{2}$;\ \ 
$ -1$,
$ -1$,
$ \sqrt{2}$,
$ -\sqrt{2}$;\ \ 
$ -1$,
$ -\sqrt{2}$,
$ \sqrt{2}$;\ \ 
$0$,
$0$;\ \ 
$0$)

Factors = $2_{1,2.}^{4,437}\boxtimes 3_{\frac{7}{2},4.}^{16,332}$

\vskip 0.7ex
\hangindent=3em \hangafter=1
$\tau_n$ = ($-2.61 - 1.08 i$, $0.$, $1.08 + 2.61 i$, $4. - 4. i$, $-1.08 + 2.61 i$, $0.$, $2.61 - 1.08 i$, $0.$, $2.61 + 1.08 i$, $0.$, $-1.08 - 2.61 i$, $4. + 4. i$, $1.08 - 2.61 i$, $0.$, $-2.61 + 1.08 i$, $8.$)

\vskip 0.7ex
\hangindent=3em \hangafter=1
\textit{Intrinsic sign problem}

  \vskip 2ex

\noindent11. $6_{\frac{11}{2},8.}^{16,548}$ \irep{39}:\ \ 
$d_i$ = ($1.0$,
$1.0$,
$1.0$,
$1.0$,
$1.414$,
$1.414$) 

\vskip 0.7ex
\hangindent=3em \hangafter=1
$D^2= 8.0 = 
8$

\vskip 0.7ex
\hangindent=3em \hangafter=1
$T = ( 0,
\frac{1}{2},
\frac{1}{4},
\frac{3}{4},
\frac{9}{16},
\frac{13}{16} )
$,

\vskip 0.7ex
\hangindent=3em \hangafter=1
$S$ = ($ 1$,
$ 1$,
$ 1$,
$ 1$,
$ \sqrt{2}$,
$ \sqrt{2}$;\ \ 
$ 1$,
$ 1$,
$ 1$,
$ -\sqrt{2}$,
$ -\sqrt{2}$;\ \ 
$ -1$,
$ -1$,
$ \sqrt{2}$,
$ -\sqrt{2}$;\ \ 
$ -1$,
$ -\sqrt{2}$,
$ \sqrt{2}$;\ \ 
$0$,
$0$;\ \ 
$0$)

Factors = $2_{1,2.}^{4,437}\boxtimes 3_{\frac{9}{2},4.}^{16,156}$

\vskip 0.7ex
\hangindent=3em \hangafter=1
$\tau_n$ = ($-1.08 - 2.61 i$, $0.$, $-2.61 - 1.08 i$, $4. + 4. i$, $2.61 - 1.08 i$, $0.$, $1.08 - 2.61 i$, $0.$, $1.08 + 2.61 i$, $0.$, $2.61 + 1.08 i$, $4. - 4. i$, $-2.61 + 1.08 i$, $0.$, $-1.08 + 2.61 i$, $8.$)

\vskip 0.7ex
\hangindent=3em \hangafter=1
\textit{Intrinsic sign problem}

  \vskip 2ex

\noindent12. $6_{\frac{13}{2},8.}^{16,107}$ \irep{39}:\ \ 
$d_i$ = ($1.0$,
$1.0$,
$1.0$,
$1.0$,
$1.414$,
$1.414$) 

\vskip 0.7ex
\hangindent=3em \hangafter=1
$D^2= 8.0 = 
8$

\vskip 0.7ex
\hangindent=3em \hangafter=1
$T = ( 0,
\frac{1}{2},
\frac{1}{4},
\frac{3}{4},
\frac{11}{16},
\frac{15}{16} )
$,

\vskip 0.7ex
\hangindent=3em \hangafter=1
$S$ = ($ 1$,
$ 1$,
$ 1$,
$ 1$,
$ \sqrt{2}$,
$ \sqrt{2}$;\ \ 
$ 1$,
$ 1$,
$ 1$,
$ -\sqrt{2}$,
$ -\sqrt{2}$;\ \ 
$ -1$,
$ -1$,
$ \sqrt{2}$,
$ -\sqrt{2}$;\ \ 
$ -1$,
$ -\sqrt{2}$,
$ \sqrt{2}$;\ \ 
$0$,
$0$;\ \ 
$0$)

Factors = $2_{7,2.}^{4,625}\boxtimes 3_{\frac{15}{2},4.}^{16,639}$

\vskip 0.7ex
\hangindent=3em \hangafter=1
$\tau_n$ = ($1.08 - 2.61 i$, $0.$, $2.61 - 1.08 i$, $4. - 4. i$, $-2.61 - 1.08 i$, $0.$, $-1.08 - 2.61 i$, $0.$, $-1.08 + 2.61 i$, $0.$, $-2.61 + 1.08 i$, $4. + 4. i$, $2.61 + 1.08 i$, $0.$, $1.08 + 2.61 i$, $8.$)

\vskip 0.7ex
\hangindent=3em \hangafter=1
\textit{Intrinsic sign problem}

  \vskip 2ex

\noindent13. $6_{\frac{24}{5},10.85}^{15,257}$ \irep{38}:\ \ 
$d_i$ = ($1.0$,
$1.0$,
$1.0$,
$1.618$,
$1.618$,
$1.618$) 

\vskip 0.7ex
\hangindent=3em \hangafter=1
$D^2= 10.854 = 
\frac{15+3\sqrt{5}}{2}$

\vskip 0.7ex
\hangindent=3em \hangafter=1
$T = ( 0,
\frac{1}{3},
\frac{1}{3},
\frac{2}{5},
\frac{11}{15},
\frac{11}{15} )
$,

\vskip 0.7ex
\hangindent=3em \hangafter=1
$S$ = ($ 1$,
$ 1$,
$ 1$,
$ \frac{1+\sqrt{5}}{2}$,
$ \frac{1+\sqrt{5}}{2}$,
$ \frac{1+\sqrt{5}}{2}$;\ \ 
$ \zeta_{3}^{1}$,
$ -\zeta_{6}^{1}$,
$ \frac{1+\sqrt{5}}{2}$,
$ -\frac{1+\sqrt{5}}{2}\zeta_{6}^{1}$,
$ \frac{1+\sqrt{5}}{2}\zeta_{3}^{1}$;\ \ 
$ \zeta_{3}^{1}$,
$ \frac{1+\sqrt{5}}{2}$,
$ \frac{1+\sqrt{5}}{2}\zeta_{3}^{1}$,
$ -\frac{1+\sqrt{5}}{2}\zeta_{6}^{1}$;\ \ 
$ -1$,
$ -1$,
$ -1$;\ \ 
$ -\zeta_{3}^{1}$,
$ \zeta_{6}^{1}$;\ \ 
$ -\zeta_{3}^{1}$)

Factors = $2_{\frac{14}{5},3.618}^{5,395}\boxtimes 3_{2,3.}^{3,527}$

\vskip 0.7ex
\hangindent=3em \hangafter=1
$\tau_n$ = ($-2.67 - 1.94 i$, $-4.31 - 3.13 i$, $5.43 + 7.47 i$, $2.67 - 1.94 i$, $0. - 6.27 i$, $-3.35 + 4.62 i$, $4.31 + 3.13 i$, $4.31 - 3.13 i$, $-3.35 - 4.62 i$, $0. + 6.27 i$, $2.67 + 1.94 i$, $5.43 - 7.47 i$, $-4.31 + 3.13 i$, $-2.67 + 1.94 i$, $10.85$)

\vskip 0.7ex
\hangindent=3em \hangafter=1
\textit{Intrinsic sign problem}

  \vskip 2ex

\noindent14. $6_{\frac{36}{5},10.85}^{15,166}$ \irep{38}:\ \ 
$d_i$ = ($1.0$,
$1.0$,
$1.0$,
$1.618$,
$1.618$,
$1.618$) 

\vskip 0.7ex
\hangindent=3em \hangafter=1
$D^2= 10.854 = 
\frac{15+3\sqrt{5}}{2}$

\vskip 0.7ex
\hangindent=3em \hangafter=1
$T = ( 0,
\frac{1}{3},
\frac{1}{3},
\frac{3}{5},
\frac{14}{15},
\frac{14}{15} )
$,

\vskip 0.7ex
\hangindent=3em \hangafter=1
$S$ = ($ 1$,
$ 1$,
$ 1$,
$ \frac{1+\sqrt{5}}{2}$,
$ \frac{1+\sqrt{5}}{2}$,
$ \frac{1+\sqrt{5}}{2}$;\ \ 
$ \zeta_{3}^{1}$,
$ -\zeta_{6}^{1}$,
$ \frac{1+\sqrt{5}}{2}$,
$ -\frac{1+\sqrt{5}}{2}\zeta_{6}^{1}$,
$ \frac{1+\sqrt{5}}{2}\zeta_{3}^{1}$;\ \ 
$ \zeta_{3}^{1}$,
$ \frac{1+\sqrt{5}}{2}$,
$ \frac{1+\sqrt{5}}{2}\zeta_{3}^{1}$,
$ -\frac{1+\sqrt{5}}{2}\zeta_{6}^{1}$;\ \ 
$ -1$,
$ -1$,
$ -1$;\ \ 
$ -\zeta_{3}^{1}$,
$ \zeta_{6}^{1}$;\ \ 
$ -\zeta_{3}^{1}$)

Factors = $2_{\frac{26}{5},3.618}^{5,720}\boxtimes 3_{2,3.}^{3,527}$

\vskip 0.7ex
\hangindent=3em \hangafter=1
$\tau_n$ = ($2.67 - 1.94 i$, $4.31 - 3.13 i$, $5.43 - 7.47 i$, $-2.67 - 1.94 i$, $0. - 6.27 i$, $-3.35 - 4.62 i$, $-4.31 + 3.13 i$, $-4.31 - 3.13 i$, $-3.35 + 4.62 i$, $0. + 6.27 i$, $-2.67 + 1.94 i$, $5.43 + 7.47 i$, $4.31 + 3.13 i$, $2.67 + 1.94 i$, $10.85$)

\vskip 0.7ex
\hangindent=3em \hangafter=1
\textit{Intrinsic sign problem}

  \vskip 2ex

\noindent15. $6_{\frac{4}{5},10.85}^{15,801}$ \irep{38}:\ \ 
$d_i$ = ($1.0$,
$1.0$,
$1.0$,
$1.618$,
$1.618$,
$1.618$) 

\vskip 0.7ex
\hangindent=3em \hangafter=1
$D^2= 10.854 = 
\frac{15+3\sqrt{5}}{2}$

\vskip 0.7ex
\hangindent=3em \hangafter=1
$T = ( 0,
\frac{2}{3},
\frac{2}{3},
\frac{2}{5},
\frac{1}{15},
\frac{1}{15} )
$,

\vskip 0.7ex
\hangindent=3em \hangafter=1
$S$ = ($ 1$,
$ 1$,
$ 1$,
$ \frac{1+\sqrt{5}}{2}$,
$ \frac{1+\sqrt{5}}{2}$,
$ \frac{1+\sqrt{5}}{2}$;\ \ 
$ -\zeta_{6}^{1}$,
$ \zeta_{3}^{1}$,
$ \frac{1+\sqrt{5}}{2}$,
$ -\frac{1+\sqrt{5}}{2}\zeta_{6}^{1}$,
$ \frac{1+\sqrt{5}}{2}\zeta_{3}^{1}$;\ \ 
$ -\zeta_{6}^{1}$,
$ \frac{1+\sqrt{5}}{2}$,
$ \frac{1+\sqrt{5}}{2}\zeta_{3}^{1}$,
$ -\frac{1+\sqrt{5}}{2}\zeta_{6}^{1}$;\ \ 
$ -1$,
$ -1$,
$ -1$;\ \ 
$ \zeta_{6}^{1}$,
$ -\zeta_{3}^{1}$;\ \ 
$ \zeta_{6}^{1}$)

Factors = $2_{\frac{14}{5},3.618}^{5,395}\boxtimes 3_{6,3.}^{3,138}$

\vskip 0.7ex
\hangindent=3em \hangafter=1
$\tau_n$ = ($2.67 + 1.94 i$, $4.31 + 3.13 i$, $5.43 + 7.47 i$, $-2.67 + 1.94 i$, $0. + 6.27 i$, $-3.35 + 4.62 i$, $-4.31 - 3.13 i$, $-4.31 + 3.13 i$, $-3.35 - 4.62 i$, $0. - 6.27 i$, $-2.67 - 1.94 i$, $5.43 - 7.47 i$, $4.31 - 3.13 i$, $2.67 - 1.94 i$, $10.85$)

\vskip 0.7ex
\hangindent=3em \hangafter=1
\textit{Intrinsic sign problem}

  \vskip 2ex

\noindent16. $6_{\frac{16}{5},10.85}^{15,262}$ \irep{38}:\ \ 
$d_i$ = ($1.0$,
$1.0$,
$1.0$,
$1.618$,
$1.618$,
$1.618$) 

\vskip 0.7ex
\hangindent=3em \hangafter=1
$D^2= 10.854 = 
\frac{15+3\sqrt{5}}{2}$

\vskip 0.7ex
\hangindent=3em \hangafter=1
$T = ( 0,
\frac{2}{3},
\frac{2}{3},
\frac{3}{5},
\frac{4}{15},
\frac{4}{15} )
$,

\vskip 0.7ex
\hangindent=3em \hangafter=1
$S$ = ($ 1$,
$ 1$,
$ 1$,
$ \frac{1+\sqrt{5}}{2}$,
$ \frac{1+\sqrt{5}}{2}$,
$ \frac{1+\sqrt{5}}{2}$;\ \ 
$ -\zeta_{6}^{1}$,
$ \zeta_{3}^{1}$,
$ \frac{1+\sqrt{5}}{2}$,
$ -\frac{1+\sqrt{5}}{2}\zeta_{6}^{1}$,
$ \frac{1+\sqrt{5}}{2}\zeta_{3}^{1}$;\ \ 
$ -\zeta_{6}^{1}$,
$ \frac{1+\sqrt{5}}{2}$,
$ \frac{1+\sqrt{5}}{2}\zeta_{3}^{1}$,
$ -\frac{1+\sqrt{5}}{2}\zeta_{6}^{1}$;\ \ 
$ -1$,
$ -1$,
$ -1$;\ \ 
$ \zeta_{6}^{1}$,
$ -\zeta_{3}^{1}$;\ \ 
$ \zeta_{6}^{1}$)

Factors = $2_{\frac{26}{5},3.618}^{5,720}\boxtimes 3_{6,3.}^{3,138}$

\vskip 0.7ex
\hangindent=3em \hangafter=1
$\tau_n$ = ($-2.67 + 1.94 i$, $-4.31 + 3.13 i$, $5.43 - 7.47 i$, $2.67 + 1.94 i$, $0. + 6.27 i$, $-3.35 - 4.62 i$, $4.31 - 3.13 i$, $4.31 + 3.13 i$, $-3.35 + 4.62 i$, $0. - 6.27 i$, $2.67 - 1.94 i$, $5.43 + 7.47 i$, $-4.31 - 3.13 i$, $-2.67 - 1.94 i$, $10.85$)

\vskip 0.7ex
\hangindent=3em \hangafter=1
\textit{Intrinsic sign problem}

  \vskip 2ex

\noindent17. $6_{\frac{33}{10},14.47}^{80,798}$ \irep{48}:\ \ 
$d_i$ = ($1.0$,
$1.0$,
$1.414$,
$1.618$,
$1.618$,
$2.288$) 

\vskip 0.7ex
\hangindent=3em \hangafter=1
$D^2= 14.472 = 
10+2\sqrt{5}$

\vskip 0.7ex
\hangindent=3em \hangafter=1
$T = ( 0,
\frac{1}{2},
\frac{1}{16},
\frac{2}{5},
\frac{9}{10},
\frac{37}{80} )
$,

\vskip 0.7ex
\hangindent=3em \hangafter=1
$S$ = ($ 1$,
$ 1$,
$ \sqrt{2}$,
$ \frac{1+\sqrt{5}}{2}$,
$ \frac{1+\sqrt{5}}{2}$,
$ \frac{5+\sqrt{5}}{\sqrt{10}}$;\ \ 
$ 1$,
$ -\sqrt{2}$,
$ \frac{1+\sqrt{5}}{2}$,
$ \frac{1+\sqrt{5}}{2}$,
$ \frac{-5-\sqrt{5}}{\sqrt{10}}$;\ \ 
$0$,
$ \frac{5+\sqrt{5}}{\sqrt{10}}$,
$ \frac{-5-\sqrt{5}}{\sqrt{10}}$,
$0$;\ \ 
$ -1$,
$ -1$,
$ -\sqrt{2}$;\ \ 
$ -1$,
$ \sqrt{2}$;\ \ 
$0$)

Factors = $2_{\frac{14}{5},3.618}^{5,395}\boxtimes 3_{\frac{1}{2},4.}^{16,598}$

\vskip 0.7ex
\hangindent=3em \hangafter=1
$\tau_n$ = ($-3.24 + 1.99 i$, $9.7 - 5.94 i$, $-3.22 + 5.25 i$, $0.84 - 5.31 i$, $-2.77 + 6.68 i$, $-2.83 - 0.68 i$, $-1.44 + 5.98 i$, $0.$, $0.89 + 3.7 i$, $2.12 - 5.12 i$, $3.7 + 0.89 i$, $-1.36 - 8.6 i$, $5.98 - 1.44 i$, $-5.99 - 3.67 i$, $6.68 - 2.77 i$, $-4.47 + 6.15 i$, $5.25 - 3.22 i$, $2.66 + 11.06 i$, $1.99 - 3.24 i$, $7.24 + 7.23 i$, $-1.99 - 3.24 i$, $4.58 + 1.1 i$, $-5.25 - 3.22 i$, $0.$, $-6.68 - 2.77 i$, $1.52 + 2.48 i$, $-5.98 - 1.44 i$, $8.6 + 1.36 i$, $-3.7 + 0.89 i$, $12.35 - 5.12 i$, $-0.89 + 3.7 i$, $7.24 - 9.96 i$, $1.44 + 5.98 i$, $-1.64 - 6.83 i$, $2.77 + 6.68 i$, $-5.31 + 0.84 i$, $3.22 + 5.25 i$, $-2.46 + 4.02 i$, $3.24 + 1.99 i$, $0.$, $3.24 - 1.99 i$, $-2.46 - 4.02 i$, $3.22 - 5.25 i$, $-5.31 - 0.84 i$, $2.77 - 6.68 i$, $-1.64 + 6.83 i$, $1.44 - 5.98 i$, $7.24 + 9.96 i$, $-0.89 - 3.7 i$, $12.35 + 5.12 i$, $-3.7 - 0.89 i$, $8.6 - 1.36 i$, $-5.98 + 1.44 i$, $1.52 - 2.48 i$, $-6.68 + 2.77 i$, $0.$, $-5.25 + 3.22 i$, $4.58 - 1.1 i$, $-1.99 + 3.24 i$, $7.24 - 7.23 i$, $1.99 + 3.24 i$, $2.66 - 11.06 i$, $5.25 + 3.22 i$, $-4.47 - 6.15 i$, $6.68 + 2.77 i$, $-5.99 + 3.67 i$, $5.98 + 1.44 i$, $-1.36 + 8.6 i$, $3.7 - 0.89 i$, $2.12 + 5.12 i$, $0.89 - 3.7 i$, $0.$, $-1.44 - 5.98 i$, $-2.83 + 0.68 i$, $-2.77 - 6.68 i$, $0.84 + 5.31 i$, $-3.22 - 5.25 i$, $9.7 + 5.94 i$, $-3.24 - 1.99 i$, $14.47$)

\vskip 0.7ex
\hangindent=3em \hangafter=1
\textit{Intrinsic sign problem}

  \vskip 2ex

\noindent18. $6_{\frac{57}{10},14.47}^{80,376}$ \irep{48}:\ \ 
$d_i$ = ($1.0$,
$1.0$,
$1.414$,
$1.618$,
$1.618$,
$2.288$) 

\vskip 0.7ex
\hangindent=3em \hangafter=1
$D^2= 14.472 = 
10+2\sqrt{5}$

\vskip 0.7ex
\hangindent=3em \hangafter=1
$T = ( 0,
\frac{1}{2},
\frac{1}{16},
\frac{3}{5},
\frac{1}{10},
\frac{53}{80} )
$,

\vskip 0.7ex
\hangindent=3em \hangafter=1
$S$ = ($ 1$,
$ 1$,
$ \sqrt{2}$,
$ \frac{1+\sqrt{5}}{2}$,
$ \frac{1+\sqrt{5}}{2}$,
$ \frac{5+\sqrt{5}}{\sqrt{10}}$;\ \ 
$ 1$,
$ -\sqrt{2}$,
$ \frac{1+\sqrt{5}}{2}$,
$ \frac{1+\sqrt{5}}{2}$,
$ \frac{-5-\sqrt{5}}{\sqrt{10}}$;\ \ 
$0$,
$ \frac{5+\sqrt{5}}{\sqrt{10}}$,
$ \frac{-5-\sqrt{5}}{\sqrt{10}}$,
$0$;\ \ 
$ -1$,
$ -1$,
$ -\sqrt{2}$;\ \ 
$ -1$,
$ \sqrt{2}$;\ \ 
$0$)

Factors = $2_{\frac{26}{5},3.618}^{5,720}\boxtimes 3_{\frac{1}{2},4.}^{16,598}$

\vskip 0.7ex
\hangindent=3em \hangafter=1
$\tau_n$ = ($-0.89 - 3.7 i$, $2.66 + 11.06 i$, $5.98 + 1.44 i$, $-5.31 + 0.84 i$, $-2.77 + 6.68 i$, $1.52 - 2.48 i$, $-5.25 - 3.22 i$, $0.$, $3.24 - 1.99 i$, $2.12 - 5.12 i$, $-1.99 + 3.24 i$, $8.6 + 1.36 i$, $-3.22 - 5.25 i$, $-1.64 + 6.83 i$, $6.68 - 2.77 i$, $-4.47 - 6.15 i$, $1.44 + 5.98 i$, $9.7 - 5.94 i$, $-3.7 - 0.89 i$, $7.24 + 7.23 i$, $3.7 - 0.89 i$, $-2.46 + 4.02 i$, $-1.44 + 5.98 i$, $0.$, $-6.68 - 2.77 i$, $-2.83 + 0.68 i$, $3.22 - 5.25 i$, $-1.36 - 8.6 i$, $1.99 + 3.24 i$, $12.35 - 5.12 i$, $-3.24 - 1.99 i$, $7.24 + 9.96 i$, $5.25 - 3.22 i$, $-5.99 + 3.67 i$, $2.77 + 6.68 i$, $0.84 - 5.31 i$, $-5.98 + 1.44 i$, $4.58 + 1.1 i$, $0.89 - 3.7 i$, $0.$, $0.89 + 3.7 i$, $4.58 - 1.1 i$, $-5.98 - 1.44 i$, $0.84 + 5.31 i$, $2.77 - 6.68 i$, $-5.99 - 3.67 i$, $5.25 + 3.22 i$, $7.24 - 9.96 i$, $-3.24 + 1.99 i$, $12.35 + 5.12 i$, $1.99 - 3.24 i$, $-1.36 + 8.6 i$, $3.22 + 5.25 i$, $-2.83 - 0.68 i$, $-6.68 + 2.77 i$, $0.$, $-1.44 - 5.98 i$, $-2.46 - 4.02 i$, $3.7 + 0.89 i$, $7.24 - 7.23 i$, $-3.7 + 0.89 i$, $9.7 + 5.94 i$, $1.44 - 5.98 i$, $-4.47 + 6.15 i$, $6.68 + 2.77 i$, $-1.64 - 6.83 i$, $-3.22 + 5.25 i$, $8.6 - 1.36 i$, $-1.99 - 3.24 i$, $2.12 + 5.12 i$, $3.24 + 1.99 i$, $0.$, $-5.25 + 3.22 i$, $1.52 + 2.48 i$, $-2.77 - 6.68 i$, $-5.31 - 0.84 i$, $5.98 - 1.44 i$, $2.66 - 11.06 i$, $-0.89 + 3.7 i$, $14.47$)

\vskip 0.7ex
\hangindent=3em \hangafter=1
\textit{Intrinsic sign problem}

  \vskip 2ex

\noindent19. $6_{\frac{43}{10},14.47}^{80,424}$ \irep{48}:\ \ 
$d_i$ = ($1.0$,
$1.0$,
$1.414$,
$1.618$,
$1.618$,
$2.288$) 

\vskip 0.7ex
\hangindent=3em \hangafter=1
$D^2= 14.472 = 
10+2\sqrt{5}$

\vskip 0.7ex
\hangindent=3em \hangafter=1
$T = ( 0,
\frac{1}{2},
\frac{3}{16},
\frac{2}{5},
\frac{9}{10},
\frac{47}{80} )
$,

\vskip 0.7ex
\hangindent=3em \hangafter=1
$S$ = ($ 1$,
$ 1$,
$ \sqrt{2}$,
$ \frac{1+\sqrt{5}}{2}$,
$ \frac{1+\sqrt{5}}{2}$,
$ \frac{5+\sqrt{5}}{\sqrt{10}}$;\ \ 
$ 1$,
$ -\sqrt{2}$,
$ \frac{1+\sqrt{5}}{2}$,
$ \frac{1+\sqrt{5}}{2}$,
$ \frac{-5-\sqrt{5}}{\sqrt{10}}$;\ \ 
$0$,
$ \frac{5+\sqrt{5}}{\sqrt{10}}$,
$ \frac{-5-\sqrt{5}}{\sqrt{10}}$,
$0$;\ \ 
$ -1$,
$ -1$,
$ -\sqrt{2}$;\ \ 
$ -1$,
$ \sqrt{2}$;\ \ 
$0$)

Factors = $2_{\frac{14}{5},3.618}^{5,395}\boxtimes 3_{\frac{3}{2},4.}^{16,553}$

\vskip 0.7ex
\hangindent=3em \hangafter=1
$\tau_n$ = ($-3.7 - 0.89 i$, $4.58 + 1.1 i$, $-1.44 - 5.98 i$, $-5.31 - 0.84 i$, $6.68 - 2.77 i$, $-5.99 + 3.67 i$, $3.22 + 5.25 i$, $0.$, $-1.99 + 3.24 i$, $12.35 - 5.12 i$, $-3.24 + 1.99 i$, $8.6 - 1.36 i$, $-5.25 - 3.22 i$, $-2.83 + 0.68 i$, $2.77 - 6.68 i$, $-4.47 + 6.15 i$, $5.98 + 1.44 i$, $-2.46 + 4.02 i$, $0.89 + 3.7 i$, $7.24 - 7.23 i$, $-0.89 + 3.7 i$, $9.7 - 5.94 i$, $-5.98 + 1.44 i$, $0.$, $-2.77 - 6.68 i$, $-1.64 + 6.83 i$, $5.25 - 3.22 i$, $-1.36 + 8.6 i$, $3.24 + 1.99 i$, $2.12 - 5.12 i$, $1.99 + 3.24 i$, $7.24 - 9.96 i$, $-3.22 + 5.25 i$, $1.52 - 2.48 i$, $-6.68 - 2.77 i$, $0.84 + 5.31 i$, $1.44 - 5.98 i$, $2.66 + 11.06 i$, $3.7 - 0.89 i$, $0.$, $3.7 + 0.89 i$, $2.66 - 11.06 i$, $1.44 + 5.98 i$, $0.84 - 5.31 i$, $-6.68 + 2.77 i$, $1.52 + 2.48 i$, $-3.22 - 5.25 i$, $7.24 + 9.96 i$, $1.99 - 3.24 i$, $2.12 + 5.12 i$, $3.24 - 1.99 i$, $-1.36 - 8.6 i$, $5.25 + 3.22 i$, $-1.64 - 6.83 i$, $-2.77 + 6.68 i$, $0.$, $-5.98 - 1.44 i$, $9.7 + 5.94 i$, $-0.89 - 3.7 i$, $7.24 + 7.23 i$, $0.89 - 3.7 i$, $-2.46 - 4.02 i$, $5.98 - 1.44 i$, $-4.47 - 6.15 i$, $2.77 + 6.68 i$, $-2.83 - 0.68 i$, $-5.25 + 3.22 i$, $8.6 + 1.36 i$, $-3.24 - 1.99 i$, $12.35 + 5.12 i$, $-1.99 - 3.24 i$, $0.$, $3.22 - 5.25 i$, $-5.99 - 3.67 i$, $6.68 + 2.77 i$, $-5.31 + 0.84 i$, $-1.44 + 5.98 i$, $4.58 - 1.1 i$, $-3.7 + 0.89 i$, $14.47$)

\vskip 0.7ex
\hangindent=3em \hangafter=1
\textit{Intrinsic sign problem}

  \vskip 2ex

\noindent20. $6_{\frac{67}{10},14.47}^{80,828}$ \irep{48}:\ \ 
$d_i$ = ($1.0$,
$1.0$,
$1.414$,
$1.618$,
$1.618$,
$2.288$) 

\vskip 0.7ex
\hangindent=3em \hangafter=1
$D^2= 14.472 = 
10+2\sqrt{5}$

\vskip 0.7ex
\hangindent=3em \hangafter=1
$T = ( 0,
\frac{1}{2},
\frac{3}{16},
\frac{3}{5},
\frac{1}{10},
\frac{63}{80} )
$,

\vskip 0.7ex
\hangindent=3em \hangafter=1
$S$ = ($ 1$,
$ 1$,
$ \sqrt{2}$,
$ \frac{1+\sqrt{5}}{2}$,
$ \frac{1+\sqrt{5}}{2}$,
$ \frac{5+\sqrt{5}}{\sqrt{10}}$;\ \ 
$ 1$,
$ -\sqrt{2}$,
$ \frac{1+\sqrt{5}}{2}$,
$ \frac{1+\sqrt{5}}{2}$,
$ \frac{-5-\sqrt{5}}{\sqrt{10}}$;\ \ 
$0$,
$ \frac{5+\sqrt{5}}{\sqrt{10}}$,
$ \frac{-5-\sqrt{5}}{\sqrt{10}}$,
$0$;\ \ 
$ -1$,
$ -1$,
$ -\sqrt{2}$;\ \ 
$ -1$,
$ \sqrt{2}$;\ \ 
$0$)

Factors = $2_{\frac{26}{5},3.618}^{5,720}\boxtimes 3_{\frac{3}{2},4.}^{16,553}$

\vskip 0.7ex
\hangindent=3em \hangafter=1
$\tau_n$ = ($1.99 - 3.24 i$, $-2.46 + 4.02 i$, $-5.25 + 3.22 i$, $0.84 + 5.31 i$, $6.68 - 2.77 i$, $-1.64 - 6.83 i$, $-5.98 + 1.44 i$, $0.$, $3.7 + 0.89 i$, $12.35 - 5.12 i$, $-0.89 - 3.7 i$, $-1.36 + 8.6 i$, $-1.44 + 5.98 i$, $1.52 + 2.48 i$, $2.77 - 6.68 i$, $-4.47 - 6.15 i$, $-3.22 + 5.25 i$, $4.58 + 1.1 i$, $3.24 - 1.99 i$, $7.24 - 7.23 i$, $-3.24 - 1.99 i$, $2.66 + 11.06 i$, $3.22 + 5.25 i$, $0.$, $-2.77 - 6.68 i$, $-5.99 - 3.67 i$, $1.44 + 5.98 i$, $8.6 - 1.36 i$, $0.89 - 3.7 i$, $2.12 - 5.12 i$, $-3.7 + 0.89 i$, $7.24 + 9.96 i$, $5.98 + 1.44 i$, $-2.83 - 0.68 i$, $-6.68 - 2.77 i$, $-5.31 - 0.84 i$, $5.25 + 3.22 i$, $9.7 - 5.94 i$, $-1.99 - 3.24 i$, $0.$, $-1.99 + 3.24 i$, $9.7 + 5.94 i$, $5.25 - 3.22 i$, $-5.31 + 0.84 i$, $-6.68 + 2.77 i$, $-2.83 + 0.68 i$, $5.98 - 1.44 i$, $7.24 - 9.96 i$, $-3.7 - 0.89 i$, $2.12 + 5.12 i$, $0.89 + 3.7 i$, $8.6 + 1.36 i$, $1.44 - 5.98 i$, $-5.99 + 3.67 i$, $-2.77 + 6.68 i$, $0.$, $3.22 - 5.25 i$, $2.66 - 11.06 i$, $-3.24 + 1.99 i$, $7.24 + 7.23 i$, $3.24 + 1.99 i$, $4.58 - 1.1 i$, $-3.22 - 5.25 i$, $-4.47 + 6.15 i$, $2.77 + 6.68 i$, $1.52 - 2.48 i$, $-1.44 - 5.98 i$, $-1.36 - 8.6 i$, $-0.89 + 3.7 i$, $12.35 + 5.12 i$, $3.7 - 0.89 i$, $0.$, $-5.98 - 1.44 i$, $-1.64 + 6.83 i$, $6.68 + 2.77 i$, $0.84 - 5.31 i$, $-5.25 - 3.22 i$, $-2.46 - 4.02 i$, $1.99 + 3.24 i$, $14.47$)

\vskip 0.7ex
\hangindent=3em \hangafter=1
\textit{Intrinsic sign problem}

  \vskip 2ex

\noindent21. $6_{\frac{53}{10},14.47}^{80,884}$ \irep{48}:\ \ 
$d_i$ = ($1.0$,
$1.0$,
$1.414$,
$1.618$,
$1.618$,
$2.288$) 

\vskip 0.7ex
\hangindent=3em \hangafter=1
$D^2= 14.472 = 
10+2\sqrt{5}$

\vskip 0.7ex
\hangindent=3em \hangafter=1
$T = ( 0,
\frac{1}{2},
\frac{5}{16},
\frac{2}{5},
\frac{9}{10},
\frac{57}{80} )
$,

\vskip 0.7ex
\hangindent=3em \hangafter=1
$S$ = ($ 1$,
$ 1$,
$ \sqrt{2}$,
$ \frac{1+\sqrt{5}}{2}$,
$ \frac{1+\sqrt{5}}{2}$,
$ \frac{5+\sqrt{5}}{\sqrt{10}}$;\ \ 
$ 1$,
$ -\sqrt{2}$,
$ \frac{1+\sqrt{5}}{2}$,
$ \frac{1+\sqrt{5}}{2}$,
$ \frac{-5-\sqrt{5}}{\sqrt{10}}$;\ \ 
$0$,
$ \frac{5+\sqrt{5}}{\sqrt{10}}$,
$ \frac{-5-\sqrt{5}}{\sqrt{10}}$,
$0$;\ \ 
$ -1$,
$ -1$,
$ -\sqrt{2}$;\ \ 
$ -1$,
$ \sqrt{2}$;\ \ 
$0$)

Factors = $2_{\frac{14}{5},3.618}^{5,395}\boxtimes 3_{\frac{5}{2},4.}^{16,465}$

\vskip 0.7ex
\hangindent=3em \hangafter=1
$\tau_n$ = ($-1.99 - 3.24 i$, $-2.46 - 4.02 i$, $5.25 + 3.22 i$, $0.84 - 5.31 i$, $-6.68 - 2.77 i$, $-1.64 + 6.83 i$, $5.98 + 1.44 i$, $0.$, $-3.7 + 0.89 i$, $12.35 + 5.12 i$, $0.89 - 3.7 i$, $-1.36 - 8.6 i$, $1.44 + 5.98 i$, $1.52 - 2.48 i$, $-2.77 - 6.68 i$, $-4.47 + 6.15 i$, $3.22 + 5.25 i$, $4.58 - 1.1 i$, $-3.24 - 1.99 i$, $7.24 + 7.23 i$, $3.24 - 1.99 i$, $2.66 - 11.06 i$, $-3.22 + 5.25 i$, $0.$, $2.77 - 6.68 i$, $-5.99 + 3.67 i$, $-1.44 + 5.98 i$, $8.6 + 1.36 i$, $-0.89 - 3.7 i$, $2.12 + 5.12 i$, $3.7 + 0.89 i$, $7.24 - 9.96 i$, $-5.98 + 1.44 i$, $-2.83 + 0.68 i$, $6.68 - 2.77 i$, $-5.31 + 0.84 i$, $-5.25 + 3.22 i$, $9.7 + 5.94 i$, $1.99 - 3.24 i$, $0.$, $1.99 + 3.24 i$, $9.7 - 5.94 i$, $-5.25 - 3.22 i$, $-5.31 - 0.84 i$, $6.68 + 2.77 i$, $-2.83 - 0.68 i$, $-5.98 - 1.44 i$, $7.24 + 9.96 i$, $3.7 - 0.89 i$, $2.12 - 5.12 i$, $-0.89 + 3.7 i$, $8.6 - 1.36 i$, $-1.44 - 5.98 i$, $-5.99 - 3.67 i$, $2.77 + 6.68 i$, $0.$, $-3.22 - 5.25 i$, $2.66 + 11.06 i$, $3.24 + 1.99 i$, $7.24 - 7.23 i$, $-3.24 + 1.99 i$, $4.58 + 1.1 i$, $3.22 - 5.25 i$, $-4.47 - 6.15 i$, $-2.77 + 6.68 i$, $1.52 + 2.48 i$, $1.44 - 5.98 i$, $-1.36 + 8.6 i$, $0.89 + 3.7 i$, $12.35 - 5.12 i$, $-3.7 - 0.89 i$, $0.$, $5.98 - 1.44 i$, $-1.64 - 6.83 i$, $-6.68 + 2.77 i$, $0.84 + 5.31 i$, $5.25 - 3.22 i$, $-2.46 + 4.02 i$, $-1.99 + 3.24 i$, $14.47$)

\vskip 0.7ex
\hangindent=3em \hangafter=1
\textit{Intrinsic sign problem}

  \vskip 2ex

\noindent22. $6_{\frac{77}{10},14.47}^{80,657}$ \irep{48}:\ \ 
$d_i$ = ($1.0$,
$1.0$,
$1.414$,
$1.618$,
$1.618$,
$2.288$) 

\vskip 0.7ex
\hangindent=3em \hangafter=1
$D^2= 14.472 = 
10+2\sqrt{5}$

\vskip 0.7ex
\hangindent=3em \hangafter=1
$T = ( 0,
\frac{1}{2},
\frac{5}{16},
\frac{3}{5},
\frac{1}{10},
\frac{73}{80} )
$,

\vskip 0.7ex
\hangindent=3em \hangafter=1
$S$ = ($ 1$,
$ 1$,
$ \sqrt{2}$,
$ \frac{1+\sqrt{5}}{2}$,
$ \frac{1+\sqrt{5}}{2}$,
$ \frac{5+\sqrt{5}}{\sqrt{10}}$;\ \ 
$ 1$,
$ -\sqrt{2}$,
$ \frac{1+\sqrt{5}}{2}$,
$ \frac{1+\sqrt{5}}{2}$,
$ \frac{-5-\sqrt{5}}{\sqrt{10}}$;\ \ 
$0$,
$ \frac{5+\sqrt{5}}{\sqrt{10}}$,
$ \frac{-5-\sqrt{5}}{\sqrt{10}}$,
$0$;\ \ 
$ -1$,
$ -1$,
$ -\sqrt{2}$;\ \ 
$ -1$,
$ \sqrt{2}$;\ \ 
$0$)

Factors = $2_{\frac{26}{5},3.618}^{5,720}\boxtimes 3_{\frac{5}{2},4.}^{16,465}$

\vskip 0.7ex
\hangindent=3em \hangafter=1
$\tau_n$ = ($3.7 - 0.89 i$, $4.58 - 1.1 i$, $1.44 - 5.98 i$, $-5.31 + 0.84 i$, $-6.68 - 2.77 i$, $-5.99 - 3.67 i$, $-3.22 + 5.25 i$, $0.$, $1.99 + 3.24 i$, $12.35 + 5.12 i$, $3.24 + 1.99 i$, $8.6 + 1.36 i$, $5.25 - 3.22 i$, $-2.83 - 0.68 i$, $-2.77 - 6.68 i$, $-4.47 - 6.15 i$, $-5.98 + 1.44 i$, $-2.46 - 4.02 i$, $-0.89 + 3.7 i$, $7.24 + 7.23 i$, $0.89 + 3.7 i$, $9.7 + 5.94 i$, $5.98 + 1.44 i$, $0.$, $2.77 - 6.68 i$, $-1.64 - 6.83 i$, $-5.25 - 3.22 i$, $-1.36 - 8.6 i$, $-3.24 + 1.99 i$, $2.12 + 5.12 i$, $-1.99 + 3.24 i$, $7.24 + 9.96 i$, $3.22 + 5.25 i$, $1.52 + 2.48 i$, $6.68 - 2.77 i$, $0.84 - 5.31 i$, $-1.44 - 5.98 i$, $2.66 - 11.06 i$, $-3.7 - 0.89 i$, $0.$, $-3.7 + 0.89 i$, $2.66 + 11.06 i$, $-1.44 + 5.98 i$, $0.84 + 5.31 i$, $6.68 + 2.77 i$, $1.52 - 2.48 i$, $3.22 - 5.25 i$, $7.24 - 9.96 i$, $-1.99 - 3.24 i$, $2.12 - 5.12 i$, $-3.24 - 1.99 i$, $-1.36 + 8.6 i$, $-5.25 + 3.22 i$, $-1.64 + 6.83 i$, $2.77 + 6.68 i$, $0.$, $5.98 - 1.44 i$, $9.7 - 5.94 i$, $0.89 - 3.7 i$, $7.24 - 7.23 i$, $-0.89 - 3.7 i$, $-2.46 + 4.02 i$, $-5.98 - 1.44 i$, $-4.47 + 6.15 i$, $-2.77 + 6.68 i$, $-2.83 + 0.68 i$, $5.25 + 3.22 i$, $8.6 - 1.36 i$, $3.24 - 1.99 i$, $12.35 - 5.12 i$, $1.99 - 3.24 i$, $0.$, $-3.22 - 5.25 i$, $-5.99 + 3.67 i$, $-6.68 + 2.77 i$, $-5.31 - 0.84 i$, $1.44 + 5.98 i$, $4.58 + 1.1 i$, $3.7 + 0.89 i$, $14.47$)

\vskip 0.7ex
\hangindent=3em \hangafter=1
\textit{Intrinsic sign problem}

  \vskip 2ex

\noindent23. $6_{\frac{63}{10},14.47}^{80,146}$ \irep{48}:\ \ 
$d_i$ = ($1.0$,
$1.0$,
$1.414$,
$1.618$,
$1.618$,
$2.288$) 

\vskip 0.7ex
\hangindent=3em \hangafter=1
$D^2= 14.472 = 
10+2\sqrt{5}$

\vskip 0.7ex
\hangindent=3em \hangafter=1
$T = ( 0,
\frac{1}{2},
\frac{7}{16},
\frac{2}{5},
\frac{9}{10},
\frac{67}{80} )
$,

\vskip 0.7ex
\hangindent=3em \hangafter=1
$S$ = ($ 1$,
$ 1$,
$ \sqrt{2}$,
$ \frac{1+\sqrt{5}}{2}$,
$ \frac{1+\sqrt{5}}{2}$,
$ \frac{5+\sqrt{5}}{\sqrt{10}}$;\ \ 
$ 1$,
$ -\sqrt{2}$,
$ \frac{1+\sqrt{5}}{2}$,
$ \frac{1+\sqrt{5}}{2}$,
$ \frac{-5-\sqrt{5}}{\sqrt{10}}$;\ \ 
$0$,
$ \frac{5+\sqrt{5}}{\sqrt{10}}$,
$ \frac{-5-\sqrt{5}}{\sqrt{10}}$,
$0$;\ \ 
$ -1$,
$ -1$,
$ -\sqrt{2}$;\ \ 
$ -1$,
$ \sqrt{2}$;\ \ 
$0$)

Factors = $2_{\frac{14}{5},3.618}^{5,395}\boxtimes 3_{\frac{7}{2},4.}^{16,332}$

\vskip 0.7ex
\hangindent=3em \hangafter=1
$\tau_n$ = ($0.89 - 3.7 i$, $2.66 - 11.06 i$, $-5.98 + 1.44 i$, $-5.31 - 0.84 i$, $2.77 + 6.68 i$, $1.52 + 2.48 i$, $5.25 - 3.22 i$, $0.$, $-3.24 - 1.99 i$, $2.12 + 5.12 i$, $1.99 + 3.24 i$, $8.6 - 1.36 i$, $3.22 - 5.25 i$, $-1.64 - 6.83 i$, $-6.68 - 2.77 i$, $-4.47 + 6.15 i$, $-1.44 + 5.98 i$, $9.7 + 5.94 i$, $3.7 - 0.89 i$, $7.24 - 7.23 i$, $-3.7 - 0.89 i$, $-2.46 - 4.02 i$, $1.44 + 5.98 i$, $0.$, $6.68 - 2.77 i$, $-2.83 - 0.68 i$, $-3.22 - 5.25 i$, $-1.36 + 8.6 i$, $-1.99 + 3.24 i$, $12.35 + 5.12 i$, $3.24 - 1.99 i$, $7.24 - 9.96 i$, $-5.25 - 3.22 i$, $-5.99 - 3.67 i$, $-2.77 + 6.68 i$, $0.84 + 5.31 i$, $5.98 + 1.44 i$, $4.58 - 1.1 i$, $-0.89 - 3.7 i$, $0.$, $-0.89 + 3.7 i$, $4.58 + 1.1 i$, $5.98 - 1.44 i$, $0.84 - 5.31 i$, $-2.77 - 6.68 i$, $-5.99 + 3.67 i$, $-5.25 + 3.22 i$, $7.24 + 9.96 i$, $3.24 + 1.99 i$, $12.35 - 5.12 i$, $-1.99 - 3.24 i$, $-1.36 - 8.6 i$, $-3.22 + 5.25 i$, $-2.83 + 0.68 i$, $6.68 + 2.77 i$, $0.$, $1.44 - 5.98 i$, $-2.46 + 4.02 i$, $-3.7 + 0.89 i$, $7.24 + 7.23 i$, $3.7 + 0.89 i$, $9.7 - 5.94 i$, $-1.44 - 5.98 i$, $-4.47 - 6.15 i$, $-6.68 + 2.77 i$, $-1.64 + 6.83 i$, $3.22 + 5.25 i$, $8.6 + 1.36 i$, $1.99 - 3.24 i$, $2.12 - 5.12 i$, $-3.24 + 1.99 i$, $0.$, $5.25 + 3.22 i$, $1.52 - 2.48 i$, $2.77 - 6.68 i$, $-5.31 + 0.84 i$, $-5.98 - 1.44 i$, $2.66 + 11.06 i$, $0.89 + 3.7 i$, $14.47$)

\vskip 0.7ex
\hangindent=3em \hangafter=1
\textit{Intrinsic sign problem}

  \vskip 2ex

\noindent24. $6_{\frac{7}{10},14.47}^{80,111}$ \irep{48}:\ \ 
$d_i$ = ($1.0$,
$1.0$,
$1.414$,
$1.618$,
$1.618$,
$2.288$) 

\vskip 0.7ex
\hangindent=3em \hangafter=1
$D^2= 14.472 = 
10+2\sqrt{5}$

\vskip 0.7ex
\hangindent=3em \hangafter=1
$T = ( 0,
\frac{1}{2},
\frac{7}{16},
\frac{3}{5},
\frac{1}{10},
\frac{3}{80} )
$,

\vskip 0.7ex
\hangindent=3em \hangafter=1
$S$ = ($ 1$,
$ 1$,
$ \sqrt{2}$,
$ \frac{1+\sqrt{5}}{2}$,
$ \frac{1+\sqrt{5}}{2}$,
$ \frac{5+\sqrt{5}}{\sqrt{10}}$;\ \ 
$ 1$,
$ -\sqrt{2}$,
$ \frac{1+\sqrt{5}}{2}$,
$ \frac{1+\sqrt{5}}{2}$,
$ \frac{-5-\sqrt{5}}{\sqrt{10}}$;\ \ 
$0$,
$ \frac{5+\sqrt{5}}{\sqrt{10}}$,
$ \frac{-5-\sqrt{5}}{\sqrt{10}}$,
$0$;\ \ 
$ -1$,
$ -1$,
$ -\sqrt{2}$;\ \ 
$ -1$,
$ \sqrt{2}$;\ \ 
$0$)

Factors = $2_{\frac{26}{5},3.618}^{5,720}\boxtimes 3_{\frac{7}{2},4.}^{16,332}$

\vskip 0.7ex
\hangindent=3em \hangafter=1
$\tau_n$ = ($3.24 + 1.99 i$, $9.7 + 5.94 i$, $3.22 + 5.25 i$, $0.84 + 5.31 i$, $2.77 + 6.68 i$, $-2.83 + 0.68 i$, $1.44 + 5.98 i$, $0.$, $-0.89 + 3.7 i$, $2.12 + 5.12 i$, $-3.7 + 0.89 i$, $-1.36 + 8.6 i$, $-5.98 - 1.44 i$, $-5.99 + 3.67 i$, $-6.68 - 2.77 i$, $-4.47 - 6.15 i$, $-5.25 - 3.22 i$, $2.66 - 11.06 i$, $-1.99 - 3.24 i$, $7.24 - 7.23 i$, $1.99 - 3.24 i$, $4.58 - 1.1 i$, $5.25 - 3.22 i$, $0.$, $6.68 - 2.77 i$, $1.52 - 2.48 i$, $5.98 - 1.44 i$, $8.6 - 1.36 i$, $3.7 + 0.89 i$, $12.35 + 5.12 i$, $0.89 + 3.7 i$, $7.24 + 9.96 i$, $-1.44 + 5.98 i$, $-1.64 + 6.83 i$, $-2.77 + 6.68 i$, $-5.31 - 0.84 i$, $-3.22 + 5.25 i$, $-2.46 - 4.02 i$, $-3.24 + 1.99 i$, $0.$, $-3.24 - 1.99 i$, $-2.46 + 4.02 i$, $-3.22 - 5.25 i$, $-5.31 + 0.84 i$, $-2.77 - 6.68 i$, $-1.64 - 6.83 i$, $-1.44 - 5.98 i$, $7.24 - 9.96 i$, $0.89 - 3.7 i$, $12.35 - 5.12 i$, $3.7 - 0.89 i$, $8.6 + 1.36 i$, $5.98 + 1.44 i$, $1.52 + 2.48 i$, $6.68 + 2.77 i$, $0.$, $5.25 + 3.22 i$, $4.58 + 1.1 i$, $1.99 + 3.24 i$, $7.24 + 7.23 i$, $-1.99 + 3.24 i$, $2.66 + 11.06 i$, $-5.25 + 3.22 i$, $-4.47 + 6.15 i$, $-6.68 + 2.77 i$, $-5.99 - 3.67 i$, $-5.98 + 1.44 i$, $-1.36 - 8.6 i$, $-3.7 - 0.89 i$, $2.12 - 5.12 i$, $-0.89 - 3.7 i$, $0.$, $1.44 - 5.98 i$, $-2.83 - 0.68 i$, $2.77 - 6.68 i$, $0.84 - 5.31 i$, $3.22 - 5.25 i$, $9.7 - 5.94 i$, $3.24 - 1.99 i$, $14.47$)

\vskip 0.7ex
\hangindent=3em \hangafter=1
\textit{Intrinsic sign problem}

  \vskip 2ex

\noindent25. $6_{\frac{73}{10},14.47}^{80,215}$ \irep{48}:\ \ 
$d_i$ = ($1.0$,
$1.0$,
$1.414$,
$1.618$,
$1.618$,
$2.288$) 

\vskip 0.7ex
\hangindent=3em \hangafter=1
$D^2= 14.472 = 
10+2\sqrt{5}$

\vskip 0.7ex
\hangindent=3em \hangafter=1
$T = ( 0,
\frac{1}{2},
\frac{9}{16},
\frac{2}{5},
\frac{9}{10},
\frac{77}{80} )
$,

\vskip 0.7ex
\hangindent=3em \hangafter=1
$S$ = ($ 1$,
$ 1$,
$ \sqrt{2}$,
$ \frac{1+\sqrt{5}}{2}$,
$ \frac{1+\sqrt{5}}{2}$,
$ \frac{5+\sqrt{5}}{\sqrt{10}}$;\ \ 
$ 1$,
$ -\sqrt{2}$,
$ \frac{1+\sqrt{5}}{2}$,
$ \frac{1+\sqrt{5}}{2}$,
$ \frac{-5-\sqrt{5}}{\sqrt{10}}$;\ \ 
$0$,
$ \frac{5+\sqrt{5}}{\sqrt{10}}$,
$ \frac{-5-\sqrt{5}}{\sqrt{10}}$,
$0$;\ \ 
$ -1$,
$ -1$,
$ -\sqrt{2}$;\ \ 
$ -1$,
$ \sqrt{2}$;\ \ 
$0$)

Factors = $2_{\frac{14}{5},3.618}^{5,395}\boxtimes 3_{\frac{9}{2},4.}^{16,156}$

\vskip 0.7ex
\hangindent=3em \hangafter=1
$\tau_n$ = ($3.24 - 1.99 i$, $9.7 - 5.94 i$, $3.22 - 5.25 i$, $0.84 - 5.31 i$, $2.77 - 6.68 i$, $-2.83 - 0.68 i$, $1.44 - 5.98 i$, $0.$, $-0.89 - 3.7 i$, $2.12 - 5.12 i$, $-3.7 - 0.89 i$, $-1.36 - 8.6 i$, $-5.98 + 1.44 i$, $-5.99 - 3.67 i$, $-6.68 + 2.77 i$, $-4.47 + 6.15 i$, $-5.25 + 3.22 i$, $2.66 + 11.06 i$, $-1.99 + 3.24 i$, $7.24 + 7.23 i$, $1.99 + 3.24 i$, $4.58 + 1.1 i$, $5.25 + 3.22 i$, $0.$, $6.68 + 2.77 i$, $1.52 + 2.48 i$, $5.98 + 1.44 i$, $8.6 + 1.36 i$, $3.7 - 0.89 i$, $12.35 - 5.12 i$, $0.89 - 3.7 i$, $7.24 - 9.96 i$, $-1.44 - 5.98 i$, $-1.64 - 6.83 i$, $-2.77 - 6.68 i$, $-5.31 + 0.84 i$, $-3.22 - 5.25 i$, $-2.46 + 4.02 i$, $-3.24 - 1.99 i$, $0.$, $-3.24 + 1.99 i$, $-2.46 - 4.02 i$, $-3.22 + 5.25 i$, $-5.31 - 0.84 i$, $-2.77 + 6.68 i$, $-1.64 + 6.83 i$, $-1.44 + 5.98 i$, $7.24 + 9.96 i$, $0.89 + 3.7 i$, $12.35 + 5.12 i$, $3.7 + 0.89 i$, $8.6 - 1.36 i$, $5.98 - 1.44 i$, $1.52 - 2.48 i$, $6.68 - 2.77 i$, $0.$, $5.25 - 3.22 i$, $4.58 - 1.1 i$, $1.99 - 3.24 i$, $7.24 - 7.23 i$, $-1.99 - 3.24 i$, $2.66 - 11.06 i$, $-5.25 - 3.22 i$, $-4.47 - 6.15 i$, $-6.68 - 2.77 i$, $-5.99 + 3.67 i$, $-5.98 - 1.44 i$, $-1.36 + 8.6 i$, $-3.7 + 0.89 i$, $2.12 + 5.12 i$, $-0.89 + 3.7 i$, $0.$, $1.44 + 5.98 i$, $-2.83 + 0.68 i$, $2.77 + 6.68 i$, $0.84 + 5.31 i$, $3.22 + 5.25 i$, $9.7 + 5.94 i$, $3.24 + 1.99 i$, $14.47$)

\vskip 0.7ex
\hangindent=3em \hangafter=1
\textit{Intrinsic sign problem}

  \vskip 2ex

\noindent26. $6_{\frac{17}{10},14.47}^{80,878}$ \irep{48}:\ \ 
$d_i$ = ($1.0$,
$1.0$,
$1.414$,
$1.618$,
$1.618$,
$2.288$) 

\vskip 0.7ex
\hangindent=3em \hangafter=1
$D^2= 14.472 = 
10+2\sqrt{5}$

\vskip 0.7ex
\hangindent=3em \hangafter=1
$T = ( 0,
\frac{1}{2},
\frac{9}{16},
\frac{3}{5},
\frac{1}{10},
\frac{13}{80} )
$,

\vskip 0.7ex
\hangindent=3em \hangafter=1
$S$ = ($ 1$,
$ 1$,
$ \sqrt{2}$,
$ \frac{1+\sqrt{5}}{2}$,
$ \frac{1+\sqrt{5}}{2}$,
$ \frac{5+\sqrt{5}}{\sqrt{10}}$;\ \ 
$ 1$,
$ -\sqrt{2}$,
$ \frac{1+\sqrt{5}}{2}$,
$ \frac{1+\sqrt{5}}{2}$,
$ \frac{-5-\sqrt{5}}{\sqrt{10}}$;\ \ 
$0$,
$ \frac{5+\sqrt{5}}{\sqrt{10}}$,
$ \frac{-5-\sqrt{5}}{\sqrt{10}}$,
$0$;\ \ 
$ -1$,
$ -1$,
$ -\sqrt{2}$;\ \ 
$ -1$,
$ \sqrt{2}$;\ \ 
$0$)

Factors = $2_{\frac{26}{5},3.618}^{5,720}\boxtimes 3_{\frac{9}{2},4.}^{16,156}$

\vskip 0.7ex
\hangindent=3em \hangafter=1
$\tau_n$ = ($0.89 + 3.7 i$, $2.66 + 11.06 i$, $-5.98 - 1.44 i$, $-5.31 + 0.84 i$, $2.77 - 6.68 i$, $1.52 - 2.48 i$, $5.25 + 3.22 i$, $0.$, $-3.24 + 1.99 i$, $2.12 - 5.12 i$, $1.99 - 3.24 i$, $8.6 + 1.36 i$, $3.22 + 5.25 i$, $-1.64 + 6.83 i$, $-6.68 + 2.77 i$, $-4.47 - 6.15 i$, $-1.44 - 5.98 i$, $9.7 - 5.94 i$, $3.7 + 0.89 i$, $7.24 + 7.23 i$, $-3.7 + 0.89 i$, $-2.46 + 4.02 i$, $1.44 - 5.98 i$, $0.$, $6.68 + 2.77 i$, $-2.83 + 0.68 i$, $-3.22 + 5.25 i$, $-1.36 - 8.6 i$, $-1.99 - 3.24 i$, $12.35 - 5.12 i$, $3.24 + 1.99 i$, $7.24 + 9.96 i$, $-5.25 + 3.22 i$, $-5.99 + 3.67 i$, $-2.77 - 6.68 i$, $0.84 - 5.31 i$, $5.98 - 1.44 i$, $4.58 + 1.1 i$, $-0.89 + 3.7 i$, $0.$, $-0.89 - 3.7 i$, $4.58 - 1.1 i$, $5.98 + 1.44 i$, $0.84 + 5.31 i$, $-2.77 + 6.68 i$, $-5.99 - 3.67 i$, $-5.25 - 3.22 i$, $7.24 - 9.96 i$, $3.24 - 1.99 i$, $12.35 + 5.12 i$, $-1.99 + 3.24 i$, $-1.36 + 8.6 i$, $-3.22 - 5.25 i$, $-2.83 - 0.68 i$, $6.68 - 2.77 i$, $0.$, $1.44 + 5.98 i$, $-2.46 - 4.02 i$, $-3.7 - 0.89 i$, $7.24 - 7.23 i$, $3.7 - 0.89 i$, $9.7 + 5.94 i$, $-1.44 + 5.98 i$, $-4.47 + 6.15 i$, $-6.68 - 2.77 i$, $-1.64 - 6.83 i$, $3.22 - 5.25 i$, $8.6 - 1.36 i$, $1.99 + 3.24 i$, $2.12 + 5.12 i$, $-3.24 - 1.99 i$, $0.$, $5.25 - 3.22 i$, $1.52 + 2.48 i$, $2.77 + 6.68 i$, $-5.31 - 0.84 i$, $-5.98 + 1.44 i$, $2.66 - 11.06 i$, $0.89 - 3.7 i$, $14.47$)

\vskip 0.7ex
\hangindent=3em \hangafter=1
\textit{Intrinsic sign problem}

  \vskip 2ex

\noindent27. $6_{\frac{3}{10},14.47}^{80,270}$ \irep{48}:\ \ 
$d_i$ = ($1.0$,
$1.0$,
$1.414$,
$1.618$,
$1.618$,
$2.288$) 

\vskip 0.7ex
\hangindent=3em \hangafter=1
$D^2= 14.472 = 
10+2\sqrt{5}$

\vskip 0.7ex
\hangindent=3em \hangafter=1
$T = ( 0,
\frac{1}{2},
\frac{11}{16},
\frac{2}{5},
\frac{9}{10},
\frac{7}{80} )
$,

\vskip 0.7ex
\hangindent=3em \hangafter=1
$S$ = ($ 1$,
$ 1$,
$ \sqrt{2}$,
$ \frac{1+\sqrt{5}}{2}$,
$ \frac{1+\sqrt{5}}{2}$,
$ \frac{5+\sqrt{5}}{\sqrt{10}}$;\ \ 
$ 1$,
$ -\sqrt{2}$,
$ \frac{1+\sqrt{5}}{2}$,
$ \frac{1+\sqrt{5}}{2}$,
$ \frac{-5-\sqrt{5}}{\sqrt{10}}$;\ \ 
$0$,
$ \frac{5+\sqrt{5}}{\sqrt{10}}$,
$ \frac{-5-\sqrt{5}}{\sqrt{10}}$,
$0$;\ \ 
$ -1$,
$ -1$,
$ -\sqrt{2}$;\ \ 
$ -1$,
$ \sqrt{2}$;\ \ 
$0$)

Factors = $2_{\frac{14}{5},3.618}^{5,395}\boxtimes 3_{\frac{11}{2},4.}^{16,648}$

\vskip 0.7ex
\hangindent=3em \hangafter=1
$\tau_n$ = ($3.7 + 0.89 i$, $4.58 + 1.1 i$, $1.44 + 5.98 i$, $-5.31 - 0.84 i$, $-6.68 + 2.77 i$, $-5.99 + 3.67 i$, $-3.22 - 5.25 i$, $0.$, $1.99 - 3.24 i$, $12.35 - 5.12 i$, $3.24 - 1.99 i$, $8.6 - 1.36 i$, $5.25 + 3.22 i$, $-2.83 + 0.68 i$, $-2.77 + 6.68 i$, $-4.47 + 6.15 i$, $-5.98 - 1.44 i$, $-2.46 + 4.02 i$, $-0.89 - 3.7 i$, $7.24 - 7.23 i$, $0.89 - 3.7 i$, $9.7 - 5.94 i$, $5.98 - 1.44 i$, $0.$, $2.77 + 6.68 i$, $-1.64 + 6.83 i$, $-5.25 + 3.22 i$, $-1.36 + 8.6 i$, $-3.24 - 1.99 i$, $2.12 - 5.12 i$, $-1.99 - 3.24 i$, $7.24 - 9.96 i$, $3.22 - 5.25 i$, $1.52 - 2.48 i$, $6.68 + 2.77 i$, $0.84 + 5.31 i$, $-1.44 + 5.98 i$, $2.66 + 11.06 i$, $-3.7 + 0.89 i$, $0.$, $-3.7 - 0.89 i$, $2.66 - 11.06 i$, $-1.44 - 5.98 i$, $0.84 - 5.31 i$, $6.68 - 2.77 i$, $1.52 + 2.48 i$, $3.22 + 5.25 i$, $7.24 + 9.96 i$, $-1.99 + 3.24 i$, $2.12 + 5.12 i$, $-3.24 + 1.99 i$, $-1.36 - 8.6 i$, $-5.25 - 3.22 i$, $-1.64 - 6.83 i$, $2.77 - 6.68 i$, $0.$, $5.98 + 1.44 i$, $9.7 + 5.94 i$, $0.89 + 3.7 i$, $7.24 + 7.23 i$, $-0.89 + 3.7 i$, $-2.46 - 4.02 i$, $-5.98 + 1.44 i$, $-4.47 - 6.15 i$, $-2.77 - 6.68 i$, $-2.83 - 0.68 i$, $5.25 - 3.22 i$, $8.6 + 1.36 i$, $3.24 + 1.99 i$, $12.35 + 5.12 i$, $1.99 + 3.24 i$, $0.$, $-3.22 + 5.25 i$, $-5.99 - 3.67 i$, $-6.68 - 2.77 i$, $-5.31 + 0.84 i$, $1.44 - 5.98 i$, $4.58 - 1.1 i$, $3.7 - 0.89 i$, $14.47$)

\vskip 0.7ex
\hangindent=3em \hangafter=1
\textit{Intrinsic sign problem}

  \vskip 2ex

\noindent28. $6_{\frac{27}{10},14.47}^{80,528}$ \irep{48}:\ \ 
$d_i$ = ($1.0$,
$1.0$,
$1.414$,
$1.618$,
$1.618$,
$2.288$) 

\vskip 0.7ex
\hangindent=3em \hangafter=1
$D^2= 14.472 = 
10+2\sqrt{5}$

\vskip 0.7ex
\hangindent=3em \hangafter=1
$T = ( 0,
\frac{1}{2},
\frac{11}{16},
\frac{3}{5},
\frac{1}{10},
\frac{23}{80} )
$,

\vskip 0.7ex
\hangindent=3em \hangafter=1
$S$ = ($ 1$,
$ 1$,
$ \sqrt{2}$,
$ \frac{1+\sqrt{5}}{2}$,
$ \frac{1+\sqrt{5}}{2}$,
$ \frac{5+\sqrt{5}}{\sqrt{10}}$;\ \ 
$ 1$,
$ -\sqrt{2}$,
$ \frac{1+\sqrt{5}}{2}$,
$ \frac{1+\sqrt{5}}{2}$,
$ \frac{-5-\sqrt{5}}{\sqrt{10}}$;\ \ 
$0$,
$ \frac{5+\sqrt{5}}{\sqrt{10}}$,
$ \frac{-5-\sqrt{5}}{\sqrt{10}}$,
$0$;\ \ 
$ -1$,
$ -1$,
$ -\sqrt{2}$;\ \ 
$ -1$,
$ \sqrt{2}$;\ \ 
$0$)

Factors = $2_{\frac{26}{5},3.618}^{5,720}\boxtimes 3_{\frac{11}{2},4.}^{16,648}$

\vskip 0.7ex
\hangindent=3em \hangafter=1
$\tau_n$ = ($-1.99 + 3.24 i$, $-2.46 + 4.02 i$, $5.25 - 3.22 i$, $0.84 + 5.31 i$, $-6.68 + 2.77 i$, $-1.64 - 6.83 i$, $5.98 - 1.44 i$, $0.$, $-3.7 - 0.89 i$, $12.35 - 5.12 i$, $0.89 + 3.7 i$, $-1.36 + 8.6 i$, $1.44 - 5.98 i$, $1.52 + 2.48 i$, $-2.77 + 6.68 i$, $-4.47 - 6.15 i$, $3.22 - 5.25 i$, $4.58 + 1.1 i$, $-3.24 + 1.99 i$, $7.24 - 7.23 i$, $3.24 + 1.99 i$, $2.66 + 11.06 i$, $-3.22 - 5.25 i$, $0.$, $2.77 + 6.68 i$, $-5.99 - 3.67 i$, $-1.44 - 5.98 i$, $8.6 - 1.36 i$, $-0.89 + 3.7 i$, $2.12 - 5.12 i$, $3.7 - 0.89 i$, $7.24 + 9.96 i$, $-5.98 - 1.44 i$, $-2.83 - 0.68 i$, $6.68 + 2.77 i$, $-5.31 - 0.84 i$, $-5.25 - 3.22 i$, $9.7 - 5.94 i$, $1.99 + 3.24 i$, $0.$, $1.99 - 3.24 i$, $9.7 + 5.94 i$, $-5.25 + 3.22 i$, $-5.31 + 0.84 i$, $6.68 - 2.77 i$, $-2.83 + 0.68 i$, $-5.98 + 1.44 i$, $7.24 - 9.96 i$, $3.7 + 0.89 i$, $2.12 + 5.12 i$, $-0.89 - 3.7 i$, $8.6 + 1.36 i$, $-1.44 + 5.98 i$, $-5.99 + 3.67 i$, $2.77 - 6.68 i$, $0.$, $-3.22 + 5.25 i$, $2.66 - 11.06 i$, $3.24 - 1.99 i$, $7.24 + 7.23 i$, $-3.24 - 1.99 i$, $4.58 - 1.1 i$, $3.22 + 5.25 i$, $-4.47 + 6.15 i$, $-2.77 - 6.68 i$, $1.52 - 2.48 i$, $1.44 + 5.98 i$, $-1.36 - 8.6 i$, $0.89 - 3.7 i$, $12.35 + 5.12 i$, $-3.7 + 0.89 i$, $0.$, $5.98 + 1.44 i$, $-1.64 + 6.83 i$, $-6.68 - 2.77 i$, $0.84 - 5.31 i$, $5.25 + 3.22 i$, $-2.46 - 4.02 i$, $-1.99 - 3.24 i$, $14.47$)

\vskip 0.7ex
\hangindent=3em \hangafter=1
\textit{Intrinsic sign problem}

  \vskip 2ex

\noindent29. $6_{\frac{13}{10},14.47}^{80,621}$ \irep{48}:\ \ 
$d_i$ = ($1.0$,
$1.0$,
$1.414$,
$1.618$,
$1.618$,
$2.288$) 

\vskip 0.7ex
\hangindent=3em \hangafter=1
$D^2= 14.472 = 
10+2\sqrt{5}$

\vskip 0.7ex
\hangindent=3em \hangafter=1
$T = ( 0,
\frac{1}{2},
\frac{13}{16},
\frac{2}{5},
\frac{9}{10},
\frac{17}{80} )
$,

\vskip 0.7ex
\hangindent=3em \hangafter=1
$S$ = ($ 1$,
$ 1$,
$ \sqrt{2}$,
$ \frac{1+\sqrt{5}}{2}$,
$ \frac{1+\sqrt{5}}{2}$,
$ \frac{5+\sqrt{5}}{\sqrt{10}}$;\ \ 
$ 1$,
$ -\sqrt{2}$,
$ \frac{1+\sqrt{5}}{2}$,
$ \frac{1+\sqrt{5}}{2}$,
$ \frac{-5-\sqrt{5}}{\sqrt{10}}$;\ \ 
$0$,
$ \frac{5+\sqrt{5}}{\sqrt{10}}$,
$ \frac{-5-\sqrt{5}}{\sqrt{10}}$,
$0$;\ \ 
$ -1$,
$ -1$,
$ -\sqrt{2}$;\ \ 
$ -1$,
$ \sqrt{2}$;\ \ 
$0$)

Factors = $2_{\frac{14}{5},3.618}^{5,395}\boxtimes 3_{\frac{13}{2},4.}^{16,330}$

\vskip 0.7ex
\hangindent=3em \hangafter=1
$\tau_n$ = ($1.99 + 3.24 i$, $-2.46 - 4.02 i$, $-5.25 - 3.22 i$, $0.84 - 5.31 i$, $6.68 + 2.77 i$, $-1.64 + 6.83 i$, $-5.98 - 1.44 i$, $0.$, $3.7 - 0.89 i$, $12.35 + 5.12 i$, $-0.89 + 3.7 i$, $-1.36 - 8.6 i$, $-1.44 - 5.98 i$, $1.52 - 2.48 i$, $2.77 + 6.68 i$, $-4.47 + 6.15 i$, $-3.22 - 5.25 i$, $4.58 - 1.1 i$, $3.24 + 1.99 i$, $7.24 + 7.23 i$, $-3.24 + 1.99 i$, $2.66 - 11.06 i$, $3.22 - 5.25 i$, $0.$, $-2.77 + 6.68 i$, $-5.99 + 3.67 i$, $1.44 - 5.98 i$, $8.6 + 1.36 i$, $0.89 + 3.7 i$, $2.12 + 5.12 i$, $-3.7 - 0.89 i$, $7.24 - 9.96 i$, $5.98 - 1.44 i$, $-2.83 + 0.68 i$, $-6.68 + 2.77 i$, $-5.31 + 0.84 i$, $5.25 - 3.22 i$, $9.7 + 5.94 i$, $-1.99 + 3.24 i$, $0.$, $-1.99 - 3.24 i$, $9.7 - 5.94 i$, $5.25 + 3.22 i$, $-5.31 - 0.84 i$, $-6.68 - 2.77 i$, $-2.83 - 0.68 i$, $5.98 + 1.44 i$, $7.24 + 9.96 i$, $-3.7 + 0.89 i$, $2.12 - 5.12 i$, $0.89 - 3.7 i$, $8.6 - 1.36 i$, $1.44 + 5.98 i$, $-5.99 - 3.67 i$, $-2.77 - 6.68 i$, $0.$, $3.22 + 5.25 i$, $2.66 + 11.06 i$, $-3.24 - 1.99 i$, $7.24 - 7.23 i$, $3.24 - 1.99 i$, $4.58 + 1.1 i$, $-3.22 + 5.25 i$, $-4.47 - 6.15 i$, $2.77 - 6.68 i$, $1.52 + 2.48 i$, $-1.44 + 5.98 i$, $-1.36 + 8.6 i$, $-0.89 - 3.7 i$, $12.35 - 5.12 i$, $3.7 + 0.89 i$, $0.$, $-5.98 + 1.44 i$, $-1.64 - 6.83 i$, $6.68 - 2.77 i$, $0.84 + 5.31 i$, $-5.25 + 3.22 i$, $-2.46 + 4.02 i$, $1.99 - 3.24 i$, $14.47$)

\vskip 0.7ex
\hangindent=3em \hangafter=1
\textit{Intrinsic sign problem}

  \vskip 2ex

\noindent30. $6_{\frac{37}{10},14.47}^{80,629}$ \irep{48}:\ \ 
$d_i$ = ($1.0$,
$1.0$,
$1.414$,
$1.618$,
$1.618$,
$2.288$) 

\vskip 0.7ex
\hangindent=3em \hangafter=1
$D^2= 14.472 = 
10+2\sqrt{5}$

\vskip 0.7ex
\hangindent=3em \hangafter=1
$T = ( 0,
\frac{1}{2},
\frac{13}{16},
\frac{3}{5},
\frac{1}{10},
\frac{33}{80} )
$,

\vskip 0.7ex
\hangindent=3em \hangafter=1
$S$ = ($ 1$,
$ 1$,
$ \sqrt{2}$,
$ \frac{1+\sqrt{5}}{2}$,
$ \frac{1+\sqrt{5}}{2}$,
$ \frac{5+\sqrt{5}}{\sqrt{10}}$;\ \ 
$ 1$,
$ -\sqrt{2}$,
$ \frac{1+\sqrt{5}}{2}$,
$ \frac{1+\sqrt{5}}{2}$,
$ \frac{-5-\sqrt{5}}{\sqrt{10}}$;\ \ 
$0$,
$ \frac{5+\sqrt{5}}{\sqrt{10}}$,
$ \frac{-5-\sqrt{5}}{\sqrt{10}}$,
$0$;\ \ 
$ -1$,
$ -1$,
$ -\sqrt{2}$;\ \ 
$ -1$,
$ \sqrt{2}$;\ \ 
$0$)

Factors = $2_{\frac{26}{5},3.618}^{5,720}\boxtimes 3_{\frac{13}{2},4.}^{16,330}$

\vskip 0.7ex
\hangindent=3em \hangafter=1
$\tau_n$ = ($-3.7 + 0.89 i$, $4.58 - 1.1 i$, $-1.44 + 5.98 i$, $-5.31 + 0.84 i$, $6.68 + 2.77 i$, $-5.99 - 3.67 i$, $3.22 - 5.25 i$, $0.$, $-1.99 - 3.24 i$, $12.35 + 5.12 i$, $-3.24 - 1.99 i$, $8.6 + 1.36 i$, $-5.25 + 3.22 i$, $-2.83 - 0.68 i$, $2.77 + 6.68 i$, $-4.47 - 6.15 i$, $5.98 - 1.44 i$, $-2.46 - 4.02 i$, $0.89 - 3.7 i$, $7.24 + 7.23 i$, $-0.89 - 3.7 i$, $9.7 + 5.94 i$, $-5.98 - 1.44 i$, $0.$, $-2.77 + 6.68 i$, $-1.64 - 6.83 i$, $5.25 + 3.22 i$, $-1.36 - 8.6 i$, $3.24 - 1.99 i$, $2.12 + 5.12 i$, $1.99 - 3.24 i$, $7.24 + 9.96 i$, $-3.22 - 5.25 i$, $1.52 + 2.48 i$, $-6.68 + 2.77 i$, $0.84 - 5.31 i$, $1.44 + 5.98 i$, $2.66 - 11.06 i$, $3.7 + 0.89 i$, $0.$, $3.7 - 0.89 i$, $2.66 + 11.06 i$, $1.44 - 5.98 i$, $0.84 + 5.31 i$, $-6.68 - 2.77 i$, $1.52 - 2.48 i$, $-3.22 + 5.25 i$, $7.24 - 9.96 i$, $1.99 + 3.24 i$, $2.12 - 5.12 i$, $3.24 + 1.99 i$, $-1.36 + 8.6 i$, $5.25 - 3.22 i$, $-1.64 + 6.83 i$, $-2.77 - 6.68 i$, $0.$, $-5.98 + 1.44 i$, $9.7 - 5.94 i$, $-0.89 + 3.7 i$, $7.24 - 7.23 i$, $0.89 + 3.7 i$, $-2.46 + 4.02 i$, $5.98 + 1.44 i$, $-4.47 + 6.15 i$, $2.77 - 6.68 i$, $-2.83 + 0.68 i$, $-5.25 - 3.22 i$, $8.6 - 1.36 i$, $-3.24 + 1.99 i$, $12.35 - 5.12 i$, $-1.99 + 3.24 i$, $0.$, $3.22 + 5.25 i$, $-5.99 + 3.67 i$, $6.68 - 2.77 i$, $-5.31 - 0.84 i$, $-1.44 - 5.98 i$, $4.58 + 1.1 i$, $-3.7 - 0.89 i$, $14.47$)

\vskip 0.7ex
\hangindent=3em \hangafter=1
\textit{Intrinsic sign problem}

  \vskip 2ex

\noindent31. $6_{\frac{23}{10},14.47}^{80,108}$ \irep{48}:\ \ 
$d_i$ = ($1.0$,
$1.0$,
$1.414$,
$1.618$,
$1.618$,
$2.288$) 

\vskip 0.7ex
\hangindent=3em \hangafter=1
$D^2= 14.472 = 
10+2\sqrt{5}$

\vskip 0.7ex
\hangindent=3em \hangafter=1
$T = ( 0,
\frac{1}{2},
\frac{15}{16},
\frac{2}{5},
\frac{9}{10},
\frac{27}{80} )
$,

\vskip 0.7ex
\hangindent=3em \hangafter=1
$S$ = ($ 1$,
$ 1$,
$ \sqrt{2}$,
$ \frac{1+\sqrt{5}}{2}$,
$ \frac{1+\sqrt{5}}{2}$,
$ \frac{5+\sqrt{5}}{\sqrt{10}}$;\ \ 
$ 1$,
$ -\sqrt{2}$,
$ \frac{1+\sqrt{5}}{2}$,
$ \frac{1+\sqrt{5}}{2}$,
$ \frac{-5-\sqrt{5}}{\sqrt{10}}$;\ \ 
$0$,
$ \frac{5+\sqrt{5}}{\sqrt{10}}$,
$ \frac{-5-\sqrt{5}}{\sqrt{10}}$,
$0$;\ \ 
$ -1$,
$ -1$,
$ -\sqrt{2}$;\ \ 
$ -1$,
$ \sqrt{2}$;\ \ 
$0$)

Factors = $2_{\frac{14}{5},3.618}^{5,395}\boxtimes 3_{\frac{15}{2},4.}^{16,639}$

\vskip 0.7ex
\hangindent=3em \hangafter=1
$\tau_n$ = ($-0.89 + 3.7 i$, $2.66 - 11.06 i$, $5.98 - 1.44 i$, $-5.31 - 0.84 i$, $-2.77 - 6.68 i$, $1.52 + 2.48 i$, $-5.25 + 3.22 i$, $0.$, $3.24 + 1.99 i$, $2.12 + 5.12 i$, $-1.99 - 3.24 i$, $8.6 - 1.36 i$, $-3.22 + 5.25 i$, $-1.64 - 6.83 i$, $6.68 + 2.77 i$, $-4.47 + 6.15 i$, $1.44 - 5.98 i$, $9.7 + 5.94 i$, $-3.7 + 0.89 i$, $7.24 - 7.23 i$, $3.7 + 0.89 i$, $-2.46 - 4.02 i$, $-1.44 - 5.98 i$, $0.$, $-6.68 + 2.77 i$, $-2.83 - 0.68 i$, $3.22 + 5.25 i$, $-1.36 + 8.6 i$, $1.99 - 3.24 i$, $12.35 + 5.12 i$, $-3.24 + 1.99 i$, $7.24 - 9.96 i$, $5.25 + 3.22 i$, $-5.99 - 3.67 i$, $2.77 - 6.68 i$, $0.84 + 5.31 i$, $-5.98 - 1.44 i$, $4.58 - 1.1 i$, $0.89 + 3.7 i$, $0.$, $0.89 - 3.7 i$, $4.58 + 1.1 i$, $-5.98 + 1.44 i$, $0.84 - 5.31 i$, $2.77 + 6.68 i$, $-5.99 + 3.67 i$, $5.25 - 3.22 i$, $7.24 + 9.96 i$, $-3.24 - 1.99 i$, $12.35 - 5.12 i$, $1.99 + 3.24 i$, $-1.36 - 8.6 i$, $3.22 - 5.25 i$, $-2.83 + 0.68 i$, $-6.68 - 2.77 i$, $0.$, $-1.44 + 5.98 i$, $-2.46 + 4.02 i$, $3.7 - 0.89 i$, $7.24 + 7.23 i$, $-3.7 - 0.89 i$, $9.7 - 5.94 i$, $1.44 + 5.98 i$, $-4.47 - 6.15 i$, $6.68 - 2.77 i$, $-1.64 + 6.83 i$, $-3.22 - 5.25 i$, $8.6 + 1.36 i$, $-1.99 + 3.24 i$, $2.12 - 5.12 i$, $3.24 - 1.99 i$, $0.$, $-5.25 - 3.22 i$, $1.52 - 2.48 i$, $-2.77 + 6.68 i$, $-5.31 + 0.84 i$, $5.98 + 1.44 i$, $2.66 + 11.06 i$, $-0.89 - 3.7 i$, $14.47$)

\vskip 0.7ex
\hangindent=3em \hangafter=1
\textit{Intrinsic sign problem}

  \vskip 2ex

\noindent32. $6_{\frac{47}{10},14.47}^{80,518}$ \irep{48}:\ \ 
$d_i$ = ($1.0$,
$1.0$,
$1.414$,
$1.618$,
$1.618$,
$2.288$) 

\vskip 0.7ex
\hangindent=3em \hangafter=1
$D^2= 14.472 = 
10+2\sqrt{5}$

\vskip 0.7ex
\hangindent=3em \hangafter=1
$T = ( 0,
\frac{1}{2},
\frac{15}{16},
\frac{3}{5},
\frac{1}{10},
\frac{43}{80} )
$,

\vskip 0.7ex
\hangindent=3em \hangafter=1
$S$ = ($ 1$,
$ 1$,
$ \sqrt{2}$,
$ \frac{1+\sqrt{5}}{2}$,
$ \frac{1+\sqrt{5}}{2}$,
$ \frac{5+\sqrt{5}}{\sqrt{10}}$;\ \ 
$ 1$,
$ -\sqrt{2}$,
$ \frac{1+\sqrt{5}}{2}$,
$ \frac{1+\sqrt{5}}{2}$,
$ \frac{-5-\sqrt{5}}{\sqrt{10}}$;\ \ 
$0$,
$ \frac{5+\sqrt{5}}{\sqrt{10}}$,
$ \frac{-5-\sqrt{5}}{\sqrt{10}}$,
$0$;\ \ 
$ -1$,
$ -1$,
$ -\sqrt{2}$;\ \ 
$ -1$,
$ \sqrt{2}$;\ \ 
$0$)

Factors = $2_{\frac{26}{5},3.618}^{5,720}\boxtimes 3_{\frac{15}{2},4.}^{16,639}$

\vskip 0.7ex
\hangindent=3em \hangafter=1
$\tau_n$ = ($-3.24 - 1.99 i$, $9.7 + 5.94 i$, $-3.22 - 5.25 i$, $0.84 + 5.31 i$, $-2.77 - 6.68 i$, $-2.83 + 0.68 i$, $-1.44 - 5.98 i$, $0.$, $0.89 - 3.7 i$, $2.12 + 5.12 i$, $3.7 - 0.89 i$, $-1.36 + 8.6 i$, $5.98 + 1.44 i$, $-5.99 + 3.67 i$, $6.68 + 2.77 i$, $-4.47 - 6.15 i$, $5.25 + 3.22 i$, $2.66 - 11.06 i$, $1.99 + 3.24 i$, $7.24 - 7.23 i$, $-1.99 + 3.24 i$, $4.58 - 1.1 i$, $-5.25 + 3.22 i$, $0.$, $-6.68 + 2.77 i$, $1.52 - 2.48 i$, $-5.98 + 1.44 i$, $8.6 - 1.36 i$, $-3.7 - 0.89 i$, $12.35 + 5.12 i$, $-0.89 - 3.7 i$, $7.24 + 9.96 i$, $1.44 - 5.98 i$, $-1.64 + 6.83 i$, $2.77 - 6.68 i$, $-5.31 - 0.84 i$, $3.22 - 5.25 i$, $-2.46 - 4.02 i$, $3.24 - 1.99 i$, $0.$, $3.24 + 1.99 i$, $-2.46 + 4.02 i$, $3.22 + 5.25 i$, $-5.31 + 0.84 i$, $2.77 + 6.68 i$, $-1.64 - 6.83 i$, $1.44 + 5.98 i$, $7.24 - 9.96 i$, $-0.89 + 3.7 i$, $12.35 - 5.12 i$, $-3.7 + 0.89 i$, $8.6 + 1.36 i$, $-5.98 - 1.44 i$, $1.52 + 2.48 i$, $-6.68 - 2.77 i$, $0.$, $-5.25 - 3.22 i$, $4.58 + 1.1 i$, $-1.99 - 3.24 i$, $7.24 + 7.23 i$, $1.99 - 3.24 i$, $2.66 + 11.06 i$, $5.25 - 3.22 i$, $-4.47 + 6.15 i$, $6.68 - 2.77 i$, $-5.99 - 3.67 i$, $5.98 - 1.44 i$, $-1.36 - 8.6 i$, $3.7 + 0.89 i$, $2.12 - 5.12 i$, $0.89 + 3.7 i$, $0.$, $-1.44 + 5.98 i$, $-2.83 - 0.68 i$, $-2.77 + 6.68 i$, $0.84 - 5.31 i$, $-3.22 + 5.25 i$, $9.7 - 5.94 i$, $-3.24 + 1.99 i$, $14.47$)

\vskip 0.7ex
\hangindent=3em \hangafter=1
\textit{Intrinsic sign problem}

  \vskip 2ex

\noindent33. $6_{\frac{55}{7},18.59}^{28,108}$ \irep{46}:\ \ 
$d_i$ = ($1.0$,
$1.0$,
$1.801$,
$1.801$,
$2.246$,
$2.246$) 

\vskip 0.7ex
\hangindent=3em \hangafter=1
$D^2= 18.591 = 
12+6c^{1}_{7}
+2c^{2}_{7}
$

\vskip 0.7ex
\hangindent=3em \hangafter=1
$T = ( 0,
\frac{1}{4},
\frac{1}{7},
\frac{11}{28},
\frac{5}{7},
\frac{27}{28} )
$,

\vskip 0.7ex
\hangindent=3em \hangafter=1
$S$ = ($ 1$,
$ 1$,
$ -c_{7}^{3}$,
$ -c_{7}^{3}$,
$ \xi_{7}^{3}$,
$ \xi_{7}^{3}$;\ \ 
$ -1$,
$ -c_{7}^{3}$,
$ c_{7}^{3}$,
$ \xi_{7}^{3}$,
$ -\xi_{7}^{3}$;\ \ 
$ -\xi_{7}^{3}$,
$ -\xi_{7}^{3}$,
$ 1$,
$ 1$;\ \ 
$ \xi_{7}^{3}$,
$ 1$,
$ -1$;\ \ 
$ c_{7}^{3}$,
$ c_{7}^{3}$;\ \ 
$ -c_{7}^{3}$)

Factors = $2_{1,2.}^{4,437}\boxtimes 3_{\frac{48}{7},9.295}^{7,790}$

\vskip 0.7ex
\hangindent=3em \hangafter=1
$\tau_n$ = ($4.28 - 0.48 i$, $0.$, $6.57 + 4.13 i$, $2.45 - 10.7 i$, $1.08 - 9.62 i$, $0.$, $9.29 - 9.29 i$, $3.8 - 4.76 i$, $-9.62 + 1.08 i$, $0.$, $-4.13 - 6.57 i$, $-8.53 - 10.7 i$, $-0.48 + 4.28 i$, $0.$, $-0.48 - 4.28 i$, $-8.53 + 10.7 i$, $-4.13 + 6.57 i$, $0.$, $-9.62 - 1.08 i$, $3.8 + 4.76 i$, $9.29 + 9.29 i$, $0.$, $1.08 + 9.62 i$, $2.45 + 10.7 i$, $6.57 - 4.13 i$, $0.$, $4.28 + 0.48 i$, $18.58$)

\vskip 0.7ex
\hangindent=3em \hangafter=1
\textit{Intrinsic sign problem}

  \vskip 2ex

\noindent34. $6_{\frac{15}{7},18.59}^{28,289}$ \irep{46}:\ \ 
$d_i$ = ($1.0$,
$1.0$,
$1.801$,
$1.801$,
$2.246$,
$2.246$) 

\vskip 0.7ex
\hangindent=3em \hangafter=1
$D^2= 18.591 = 
12+6c^{1}_{7}
+2c^{2}_{7}
$

\vskip 0.7ex
\hangindent=3em \hangafter=1
$T = ( 0,
\frac{1}{4},
\frac{6}{7},
\frac{3}{28},
\frac{2}{7},
\frac{15}{28} )
$,

\vskip 0.7ex
\hangindent=3em \hangafter=1
$S$ = ($ 1$,
$ 1$,
$ -c_{7}^{3}$,
$ -c_{7}^{3}$,
$ \xi_{7}^{3}$,
$ \xi_{7}^{3}$;\ \ 
$ -1$,
$ -c_{7}^{3}$,
$ c_{7}^{3}$,
$ \xi_{7}^{3}$,
$ -\xi_{7}^{3}$;\ \ 
$ -\xi_{7}^{3}$,
$ -\xi_{7}^{3}$,
$ 1$,
$ 1$;\ \ 
$ \xi_{7}^{3}$,
$ 1$,
$ -1$;\ \ 
$ c_{7}^{3}$,
$ c_{7}^{3}$;\ \ 
$ -c_{7}^{3}$)

Factors = $2_{1,2.}^{4,437}\boxtimes 3_{\frac{8}{7},9.295}^{7,245}$

\vskip 0.7ex
\hangindent=3em \hangafter=1
$\tau_n$ = ($-0.48 + 4.28 i$, $0.$, $-4.13 - 6.57 i$, $2.45 + 10.7 i$, $-9.62 + 1.08 i$, $0.$, $9.29 - 9.29 i$, $3.8 + 4.76 i$, $1.08 - 9.62 i$, $0.$, $6.57 + 4.13 i$, $-8.53 + 10.7 i$, $4.28 - 0.48 i$, $0.$, $4.28 + 0.48 i$, $-8.53 - 10.7 i$, $6.57 - 4.13 i$, $0.$, $1.08 + 9.62 i$, $3.8 - 4.76 i$, $9.29 + 9.29 i$, $0.$, $-9.62 - 1.08 i$, $2.45 - 10.7 i$, $-4.13 + 6.57 i$, $0.$, $-0.48 - 4.28 i$, $18.58$)

\vskip 0.7ex
\hangindent=3em \hangafter=1
\textit{Intrinsic sign problem}

  \vskip 2ex

\noindent35. $6_{\frac{41}{7},18.59}^{28,114}$ \irep{46}:\ \ 
$d_i$ = ($1.0$,
$1.0$,
$1.801$,
$1.801$,
$2.246$,
$2.246$) 

\vskip 0.7ex
\hangindent=3em \hangafter=1
$D^2= 18.591 = 
12+6c^{1}_{7}
+2c^{2}_{7}
$

\vskip 0.7ex
\hangindent=3em \hangafter=1
$T = ( 0,
\frac{3}{4},
\frac{1}{7},
\frac{25}{28},
\frac{5}{7},
\frac{13}{28} )
$,

\vskip 0.7ex
\hangindent=3em \hangafter=1
$S$ = ($ 1$,
$ 1$,
$ -c_{7}^{3}$,
$ -c_{7}^{3}$,
$ \xi_{7}^{3}$,
$ \xi_{7}^{3}$;\ \ 
$ -1$,
$ -c_{7}^{3}$,
$ c_{7}^{3}$,
$ \xi_{7}^{3}$,
$ -\xi_{7}^{3}$;\ \ 
$ -\xi_{7}^{3}$,
$ -\xi_{7}^{3}$,
$ 1$,
$ 1$;\ \ 
$ \xi_{7}^{3}$,
$ 1$,
$ -1$;\ \ 
$ c_{7}^{3}$,
$ c_{7}^{3}$;\ \ 
$ -c_{7}^{3}$)

Factors = $2_{7,2.}^{4,625}\boxtimes 3_{\frac{48}{7},9.295}^{7,790}$

\vskip 0.7ex
\hangindent=3em \hangafter=1
$\tau_n$ = ($-0.48 - 4.28 i$, $0.$, $-4.13 + 6.57 i$, $2.45 - 10.7 i$, $-9.62 - 1.08 i$, $0.$, $9.29 + 9.29 i$, $3.8 - 4.76 i$, $1.08 + 9.62 i$, $0.$, $6.57 - 4.13 i$, $-8.53 - 10.7 i$, $4.28 + 0.48 i$, $0.$, $4.28 - 0.48 i$, $-8.53 + 10.7 i$, $6.57 + 4.13 i$, $0.$, $1.08 - 9.62 i$, $3.8 + 4.76 i$, $9.29 - 9.29 i$, $0.$, $-9.62 + 1.08 i$, $2.45 + 10.7 i$, $-4.13 - 6.57 i$, $0.$, $-0.48 + 4.28 i$, $18.58$)

\vskip 0.7ex
\hangindent=3em \hangafter=1
\textit{Intrinsic sign problem}

  \vskip 2ex

\noindent36. $6_{\frac{1}{7},18.59}^{28,212}$ \irep{46}:\ \ 
$d_i$ = ($1.0$,
$1.0$,
$1.801$,
$1.801$,
$2.246$,
$2.246$) 

\vskip 0.7ex
\hangindent=3em \hangafter=1
$D^2= 18.591 = 
12+6c^{1}_{7}
+2c^{2}_{7}
$

\vskip 0.7ex
\hangindent=3em \hangafter=1
$T = ( 0,
\frac{3}{4},
\frac{6}{7},
\frac{17}{28},
\frac{2}{7},
\frac{1}{28} )
$,

\vskip 0.7ex
\hangindent=3em \hangafter=1
$S$ = ($ 1$,
$ 1$,
$ -c_{7}^{3}$,
$ -c_{7}^{3}$,
$ \xi_{7}^{3}$,
$ \xi_{7}^{3}$;\ \ 
$ -1$,
$ -c_{7}^{3}$,
$ c_{7}^{3}$,
$ \xi_{7}^{3}$,
$ -\xi_{7}^{3}$;\ \ 
$ -\xi_{7}^{3}$,
$ -\xi_{7}^{3}$,
$ 1$,
$ 1$;\ \ 
$ \xi_{7}^{3}$,
$ 1$,
$ -1$;\ \ 
$ c_{7}^{3}$,
$ c_{7}^{3}$;\ \ 
$ -c_{7}^{3}$)

Factors = $2_{7,2.}^{4,625}\boxtimes 3_{\frac{8}{7},9.295}^{7,245}$

\vskip 0.7ex
\hangindent=3em \hangafter=1
$\tau_n$ = ($4.28 + 0.48 i$, $0.$, $6.57 - 4.13 i$, $2.45 + 10.7 i$, $1.08 + 9.62 i$, $0.$, $9.29 + 9.29 i$, $3.8 + 4.76 i$, $-9.62 - 1.08 i$, $0.$, $-4.13 + 6.57 i$, $-8.53 + 10.7 i$, $-0.48 - 4.28 i$, $0.$, $-0.48 + 4.28 i$, $-8.53 - 10.7 i$, $-4.13 - 6.57 i$, $0.$, $-9.62 + 1.08 i$, $3.8 - 4.76 i$, $9.29 - 9.29 i$, $0.$, $1.08 - 9.62 i$, $2.45 - 10.7 i$, $6.57 + 4.13 i$, $0.$, $4.28 - 0.48 i$, $18.58$)

\vskip 0.7ex
\hangindent=3em \hangafter=1
\textit{Intrinsic sign problem}

  \vskip 2ex

\noindent37. $6_{0,20.}^{10,699}$ \irep{27}:\ \ 
$d_i$ = ($1.0$,
$1.0$,
$2.0$,
$2.0$,
$2.236$,
$2.236$) 

\vskip 0.7ex
\hangindent=3em \hangafter=1
$D^2= 20.0 = 
20$

\vskip 0.7ex
\hangindent=3em \hangafter=1
$T = ( 0,
0,
\frac{1}{5},
\frac{4}{5},
0,
\frac{1}{2} )
$,

\vskip 0.7ex
\hangindent=3em \hangafter=1
$S$ = ($ 1$,
$ 1$,
$ 2$,
$ 2$,
$ \sqrt{5}$,
$ \sqrt{5}$;\ \ 
$ 1$,
$ 2$,
$ 2$,
$ -\sqrt{5}$,
$ -\sqrt{5}$;\ \ 
$ -1-\sqrt{5}$,
$ -1+\sqrt{5}$,
$0$,
$0$;\ \ 
$ -1-\sqrt{5}$,
$0$,
$0$;\ \ 
$ \sqrt{5}$,
$ -\sqrt{5}$;\ \ 
$ \sqrt{5}$)

\vskip 0.7ex
\hangindent=3em \hangafter=1
$\tau_n$ = ($4.47$, $5.53$, $-4.47$, $14.47$, $10.$, $14.47$, $-4.47$, $5.53$, $4.47$, $20.$)

\vskip 0.7ex
\hangindent=3em \hangafter=1
\textit{Intrinsic sign problem}

  \vskip 2ex

\noindent38. $6_{4,20.}^{10,101}$ \irep{27}:\ \ 
$d_i$ = ($1.0$,
$1.0$,
$2.0$,
$2.0$,
$2.236$,
$2.236$) 

\vskip 0.7ex
\hangindent=3em \hangafter=1
$D^2= 20.0 = 
20$

\vskip 0.7ex
\hangindent=3em \hangafter=1
$T = ( 0,
0,
\frac{2}{5},
\frac{3}{5},
0,
\frac{1}{2} )
$,

\vskip 0.7ex
\hangindent=3em \hangafter=1
$S$ = ($ 1$,
$ 1$,
$ 2$,
$ 2$,
$ \sqrt{5}$,
$ \sqrt{5}$;\ \ 
$ 1$,
$ 2$,
$ 2$,
$ -\sqrt{5}$,
$ -\sqrt{5}$;\ \ 
$ -1+\sqrt{5}$,
$ -1-\sqrt{5}$,
$0$,
$0$;\ \ 
$ -1+\sqrt{5}$,
$0$,
$0$;\ \ 
$ -\sqrt{5}$,
$ \sqrt{5}$;\ \ 
$ -\sqrt{5}$)

\vskip 0.7ex
\hangindent=3em \hangafter=1
$\tau_n$ = ($-4.47$, $14.47$, $4.47$, $5.53$, $10.$, $5.53$, $4.47$, $14.47$, $-4.47$, $20.$)

\vskip 0.7ex
\hangindent=3em \hangafter=1
\textit{Intrinsic sign problem}

  \vskip 2ex

\noindent39. $6_{0,20.}^{20,139}$ \irep{42}:\ \ 
$d_i$ = ($1.0$,
$1.0$,
$2.0$,
$2.0$,
$2.236$,
$2.236$) 

\vskip 0.7ex
\hangindent=3em \hangafter=1
$D^2= 20.0 = 
20$

\vskip 0.7ex
\hangindent=3em \hangafter=1
$T = ( 0,
0,
\frac{1}{5},
\frac{4}{5},
\frac{1}{4},
\frac{3}{4} )
$,

\vskip 0.7ex
\hangindent=3em \hangafter=1
$S$ = ($ 1$,
$ 1$,
$ 2$,
$ 2$,
$ \sqrt{5}$,
$ \sqrt{5}$;\ \ 
$ 1$,
$ 2$,
$ 2$,
$ -\sqrt{5}$,
$ -\sqrt{5}$;\ \ 
$ -1-\sqrt{5}$,
$ -1+\sqrt{5}$,
$0$,
$0$;\ \ 
$ -1-\sqrt{5}$,
$0$,
$0$;\ \ 
$ -\sqrt{5}$,
$ \sqrt{5}$;\ \ 
$ -\sqrt{5}$)

\vskip 0.7ex
\hangindent=3em \hangafter=1
$\tau_n$ = ($4.47$, $-14.47$, $-4.47$, $14.47$, $10.$, $-5.53$, $-4.47$, $5.53$, $4.47$, $0.$, $4.47$, $5.53$, $-4.47$, $-5.53$, $10.$, $14.47$, $-4.47$, $-14.47$, $4.47$, $20.$)

\vskip 0.7ex
\hangindent=3em \hangafter=1
\textit{Intrinsic sign problem}

  \vskip 2ex

\noindent40. $6_{4,20.}^{20,180}$ \irep{42}:\ \ 
$d_i$ = ($1.0$,
$1.0$,
$2.0$,
$2.0$,
$2.236$,
$2.236$) 

\vskip 0.7ex
\hangindent=3em \hangafter=1
$D^2= 20.0 = 
20$

\vskip 0.7ex
\hangindent=3em \hangafter=1
$T = ( 0,
0,
\frac{2}{5},
\frac{3}{5},
\frac{1}{4},
\frac{3}{4} )
$,

\vskip 0.7ex
\hangindent=3em \hangafter=1
$S$ = ($ 1$,
$ 1$,
$ 2$,
$ 2$,
$ \sqrt{5}$,
$ \sqrt{5}$;\ \ 
$ 1$,
$ 2$,
$ 2$,
$ -\sqrt{5}$,
$ -\sqrt{5}$;\ \ 
$ -1+\sqrt{5}$,
$ -1-\sqrt{5}$,
$0$,
$0$;\ \ 
$ -1+\sqrt{5}$,
$0$,
$0$;\ \ 
$ \sqrt{5}$,
$ -\sqrt{5}$;\ \ 
$ \sqrt{5}$)

\vskip 0.7ex
\hangindent=3em \hangafter=1
$\tau_n$ = ($-4.47$, $-5.53$, $4.47$, $5.53$, $10.$, $-14.47$, $4.47$, $14.47$, $-4.47$, $0.$, $-4.47$, $14.47$, $4.47$, $-14.47$, $10.$, $5.53$, $4.47$, $-5.53$, $-4.47$, $20.$)

\vskip 0.7ex
\hangindent=3em \hangafter=1
\textit{Intrinsic sign problem}

  \vskip 2ex

\noindent41. $6_{\frac{58}{35},33.63}^{35,955}$ \irep{47}:\ \ 
$d_i$ = ($1.0$,
$1.618$,
$1.801$,
$2.246$,
$2.915$,
$3.635$) 

\vskip 0.7ex
\hangindent=3em \hangafter=1
$D^2= 33.632 = 
15+3c^{1}_{35}
+2c^{4}_{35}
+6c^{5}_{35}
+3c^{6}_{35}
+3c^{7}_{35}
+2c^{10}_{35}
+2c^{11}_{35}
$

\vskip 0.7ex
\hangindent=3em \hangafter=1
$T = ( 0,
\frac{2}{5},
\frac{1}{7},
\frac{5}{7},
\frac{19}{35},
\frac{4}{35} )
$,

\vskip 0.7ex
\hangindent=3em \hangafter=1
$S$ = ($ 1$,
$ \frac{1+\sqrt{5}}{2}$,
$ -c_{7}^{3}$,
$ \xi_{7}^{3}$,
$ c^{1}_{35}
+c^{6}_{35}
$,
$ c^{1}_{35}
+c^{4}_{35}
+c^{6}_{35}
+c^{11}_{35}
$;\ \ 
$ -1$,
$ c^{1}_{35}
+c^{6}_{35}
$,
$ c^{1}_{35}
+c^{4}_{35}
+c^{6}_{35}
+c^{11}_{35}
$,
$ c_{7}^{3}$,
$ -\xi_{7}^{3}$;\ \ 
$ -\xi_{7}^{3}$,
$ 1$,
$ -c^{1}_{35}
-c^{4}_{35}
-c^{6}_{35}
-c^{11}_{35}
$,
$ \frac{1+\sqrt{5}}{2}$;\ \ 
$ c_{7}^{3}$,
$ \frac{1+\sqrt{5}}{2}$,
$ -c^{1}_{35}
-c^{6}_{35}
$;\ \ 
$ \xi_{7}^{3}$,
$ -1$;\ \ 
$ -c_{7}^{3}$)

Factors = $2_{\frac{14}{5},3.618}^{5,395}\boxtimes 3_{\frac{48}{7},9.295}^{7,790}$

\vskip 0.7ex
\hangindent=3em \hangafter=1
$\tau_n$ = ($1.54 + 5.59 i$, $5.61 + 20.31 i$, $-11.12 + 12.73 i$, $-9.61 + 4.11 i$, $-15.44 - 19.37 i$, $-5.79 + 0.26 i$, $16.81 - 23.14 i$, $9.37 + 0.42 i$, $13.02 + 0.58 i$, $4.42 + 19.37 i$, $6.87 + 7.87 i$, $-21.05 + 0.95 i$, $-2.5 + 9.04 i$, $-10.39 - 14.3 i$, $6.88 - 8.62 i$, $-3.46 - 12.56 i$, $15.54 + 6.64 i$, $15.54 - 6.64 i$, $-3.46 + 12.56 i$, $6.88 + 8.62 i$, $-10.39 + 14.3 i$, $-2.5 - 9.04 i$, $-21.05 - 0.95 i$, $6.87 - 7.87 i$, $4.42 - 19.37 i$, $13.02 - 0.58 i$, $9.37 - 0.42 i$, $16.81 + 23.14 i$, $-5.79 - 0.26 i$, $-15.44 + 19.37 i$, $-9.61 - 4.11 i$, $-11.12 - 12.73 i$, $5.61 - 20.31 i$, $1.54 - 5.59 i$, $33.62$)

\vskip 0.7ex
\hangindent=3em \hangafter=1
\textit{Intrinsic sign problem}

  \vskip 2ex

\noindent42. $6_{\frac{138}{35},33.63}^{35,363}$ \irep{47}:\ \ 
$d_i$ = ($1.0$,
$1.618$,
$1.801$,
$2.246$,
$2.915$,
$3.635$) 

\vskip 0.7ex
\hangindent=3em \hangafter=1
$D^2= 33.632 = 
15+3c^{1}_{35}
+2c^{4}_{35}
+6c^{5}_{35}
+3c^{6}_{35}
+3c^{7}_{35}
+2c^{10}_{35}
+2c^{11}_{35}
$

\vskip 0.7ex
\hangindent=3em \hangafter=1
$T = ( 0,
\frac{2}{5},
\frac{6}{7},
\frac{2}{7},
\frac{9}{35},
\frac{24}{35} )
$,

\vskip 0.7ex
\hangindent=3em \hangafter=1
$S$ = ($ 1$,
$ \frac{1+\sqrt{5}}{2}$,
$ -c_{7}^{3}$,
$ \xi_{7}^{3}$,
$ c^{1}_{35}
+c^{6}_{35}
$,
$ c^{1}_{35}
+c^{4}_{35}
+c^{6}_{35}
+c^{11}_{35}
$;\ \ 
$ -1$,
$ c^{1}_{35}
+c^{6}_{35}
$,
$ c^{1}_{35}
+c^{4}_{35}
+c^{6}_{35}
+c^{11}_{35}
$,
$ c_{7}^{3}$,
$ -\xi_{7}^{3}$;\ \ 
$ -\xi_{7}^{3}$,
$ 1$,
$ -c^{1}_{35}
-c^{4}_{35}
-c^{6}_{35}
-c^{11}_{35}
$,
$ \frac{1+\sqrt{5}}{2}$;\ \ 
$ c_{7}^{3}$,
$ \frac{1+\sqrt{5}}{2}$,
$ -c^{1}_{35}
-c^{6}_{35}
$;\ \ 
$ \xi_{7}^{3}$,
$ -1$;\ \ 
$ -c_{7}^{3}$)

Factors = $2_{\frac{14}{5},3.618}^{5,395}\boxtimes 3_{\frac{8}{7},9.295}^{7,245}$

\vskip 0.7ex
\hangindent=3em \hangafter=1
$\tau_n$ = ($-5.79 + 0.26 i$, $-21.05 + 0.95 i$, $15.54 - 6.64 i$, $6.87 - 7.87 i$, $-15.44 + 19.37 i$, $1.54 + 5.59 i$, $16.81 - 23.14 i$, $-2.5 + 9.04 i$, $-3.46 + 12.56 i$, $4.42 - 19.37 i$, $-9.61 - 4.11 i$, $5.61 + 20.31 i$, $9.37 + 0.42 i$, $-10.39 - 14.3 i$, $6.88 + 8.62 i$, $13.02 - 0.58 i$, $-11.12 - 12.73 i$, $-11.12 + 12.73 i$, $13.02 + 0.58 i$, $6.88 - 8.62 i$, $-10.39 + 14.3 i$, $9.37 - 0.42 i$, $5.61 - 20.31 i$, $-9.61 + 4.11 i$, $4.42 + 19.37 i$, $-3.46 - 12.56 i$, $-2.5 - 9.04 i$, $16.81 + 23.14 i$, $1.54 - 5.59 i$, $-15.44 - 19.37 i$, $6.87 + 7.87 i$, $15.54 + 6.64 i$, $-21.05 - 0.95 i$, $-5.79 - 0.26 i$, $33.62$)

\vskip 0.7ex
\hangindent=3em \hangafter=1
\textit{Intrinsic sign problem}

  \vskip 2ex

\noindent43. $6_{\frac{142}{35},33.63}^{35,429}$ \irep{47}:\ \ 
$d_i$ = ($1.0$,
$1.618$,
$1.801$,
$2.246$,
$2.915$,
$3.635$) 

\vskip 0.7ex
\hangindent=3em \hangafter=1
$D^2= 33.632 = 
15+3c^{1}_{35}
+2c^{4}_{35}
+6c^{5}_{35}
+3c^{6}_{35}
+3c^{7}_{35}
+2c^{10}_{35}
+2c^{11}_{35}
$

\vskip 0.7ex
\hangindent=3em \hangafter=1
$T = ( 0,
\frac{3}{5},
\frac{1}{7},
\frac{5}{7},
\frac{26}{35},
\frac{11}{35} )
$,

\vskip 0.7ex
\hangindent=3em \hangafter=1
$S$ = ($ 1$,
$ \frac{1+\sqrt{5}}{2}$,
$ -c_{7}^{3}$,
$ \xi_{7}^{3}$,
$ c^{1}_{35}
+c^{6}_{35}
$,
$ c^{1}_{35}
+c^{4}_{35}
+c^{6}_{35}
+c^{11}_{35}
$;\ \ 
$ -1$,
$ c^{1}_{35}
+c^{6}_{35}
$,
$ c^{1}_{35}
+c^{4}_{35}
+c^{6}_{35}
+c^{11}_{35}
$,
$ c_{7}^{3}$,
$ -\xi_{7}^{3}$;\ \ 
$ -\xi_{7}^{3}$,
$ 1$,
$ -c^{1}_{35}
-c^{4}_{35}
-c^{6}_{35}
-c^{11}_{35}
$,
$ \frac{1+\sqrt{5}}{2}$;\ \ 
$ c_{7}^{3}$,
$ \frac{1+\sqrt{5}}{2}$,
$ -c^{1}_{35}
-c^{6}_{35}
$;\ \ 
$ \xi_{7}^{3}$,
$ -1$;\ \ 
$ -c_{7}^{3}$)

Factors = $2_{\frac{26}{5},3.618}^{5,720}\boxtimes 3_{\frac{48}{7},9.295}^{7,790}$

\vskip 0.7ex
\hangindent=3em \hangafter=1
$\tau_n$ = ($-5.79 - 0.26 i$, $-21.05 - 0.95 i$, $15.54 + 6.64 i$, $6.87 + 7.87 i$, $-15.44 - 19.37 i$, $1.54 - 5.59 i$, $16.81 + 23.14 i$, $-2.5 - 9.04 i$, $-3.46 - 12.56 i$, $4.42 + 19.37 i$, $-9.61 + 4.11 i$, $5.61 - 20.31 i$, $9.37 - 0.42 i$, $-10.39 + 14.3 i$, $6.88 - 8.62 i$, $13.02 + 0.58 i$, $-11.12 + 12.73 i$, $-11.12 - 12.73 i$, $13.02 - 0.58 i$, $6.88 + 8.62 i$, $-10.39 - 14.3 i$, $9.37 + 0.42 i$, $5.61 + 20.31 i$, $-9.61 - 4.11 i$, $4.42 - 19.37 i$, $-3.46 + 12.56 i$, $-2.5 + 9.04 i$, $16.81 - 23.14 i$, $1.54 + 5.59 i$, $-15.44 + 19.37 i$, $6.87 - 7.87 i$, $15.54 - 6.64 i$, $-21.05 + 0.95 i$, $-5.79 + 0.26 i$, $33.62$)

\vskip 0.7ex
\hangindent=3em \hangafter=1
\textit{Intrinsic sign problem}

  \vskip 2ex

\noindent44. $6_{\frac{222}{35},33.63}^{35,224}$ \irep{47}:\ \ 
$d_i$ = ($1.0$,
$1.618$,
$1.801$,
$2.246$,
$2.915$,
$3.635$) 

\vskip 0.7ex
\hangindent=3em \hangafter=1
$D^2= 33.632 = 
15+3c^{1}_{35}
+2c^{4}_{35}
+6c^{5}_{35}
+3c^{6}_{35}
+3c^{7}_{35}
+2c^{10}_{35}
+2c^{11}_{35}
$

\vskip 0.7ex
\hangindent=3em \hangafter=1
$T = ( 0,
\frac{3}{5},
\frac{6}{7},
\frac{2}{7},
\frac{16}{35},
\frac{31}{35} )
$,

\vskip 0.7ex
\hangindent=3em \hangafter=1
$S$ = ($ 1$,
$ \frac{1+\sqrt{5}}{2}$,
$ -c_{7}^{3}$,
$ \xi_{7}^{3}$,
$ c^{1}_{35}
+c^{6}_{35}
$,
$ c^{1}_{35}
+c^{4}_{35}
+c^{6}_{35}
+c^{11}_{35}
$;\ \ 
$ -1$,
$ c^{1}_{35}
+c^{6}_{35}
$,
$ c^{1}_{35}
+c^{4}_{35}
+c^{6}_{35}
+c^{11}_{35}
$,
$ c_{7}^{3}$,
$ -\xi_{7}^{3}$;\ \ 
$ -\xi_{7}^{3}$,
$ 1$,
$ -c^{1}_{35}
-c^{4}_{35}
-c^{6}_{35}
-c^{11}_{35}
$,
$ \frac{1+\sqrt{5}}{2}$;\ \ 
$ c_{7}^{3}$,
$ \frac{1+\sqrt{5}}{2}$,
$ -c^{1}_{35}
-c^{6}_{35}
$;\ \ 
$ \xi_{7}^{3}$,
$ -1$;\ \ 
$ -c_{7}^{3}$)

Factors = $2_{\frac{26}{5},3.618}^{5,720}\boxtimes 3_{\frac{8}{7},9.295}^{7,245}$

\vskip 0.7ex
\hangindent=3em \hangafter=1
$\tau_n$ = ($1.54 - 5.59 i$, $5.61 - 20.31 i$, $-11.12 - 12.73 i$, $-9.61 - 4.11 i$, $-15.44 + 19.37 i$, $-5.79 - 0.26 i$, $16.81 + 23.14 i$, $9.37 - 0.42 i$, $13.02 - 0.58 i$, $4.42 - 19.37 i$, $6.87 - 7.87 i$, $-21.05 - 0.95 i$, $-2.5 - 9.04 i$, $-10.39 + 14.3 i$, $6.88 + 8.62 i$, $-3.46 + 12.56 i$, $15.54 - 6.64 i$, $15.54 + 6.64 i$, $-3.46 - 12.56 i$, $6.88 - 8.62 i$, $-10.39 - 14.3 i$, $-2.5 + 9.04 i$, $-21.05 + 0.95 i$, $6.87 + 7.87 i$, $4.42 + 19.37 i$, $13.02 + 0.58 i$, $9.37 + 0.42 i$, $16.81 - 23.14 i$, $-5.79 + 0.26 i$, $-15.44 - 19.37 i$, $-9.61 + 4.11 i$, $-11.12 + 12.73 i$, $5.61 + 20.31 i$, $1.54 + 5.59 i$, $33.62$)

\vskip 0.7ex
\hangindent=3em \hangafter=1
\textit{Intrinsic sign problem}

  \vskip 2ex

\noindent45. $6_{\frac{46}{13},56.74}^{13,131}$ \irep{35}:\ \ 
$d_i$ = ($1.0$,
$1.941$,
$2.770$,
$3.438$,
$3.907$,
$4.148$) 

\vskip 0.7ex
\hangindent=3em \hangafter=1
$D^2= 56.746 = 
21+15c^{1}_{13}
+10c^{2}_{13}
+6c^{3}_{13}
+3c^{4}_{13}
+c^{5}_{13}
$

\vskip 0.7ex
\hangindent=3em \hangafter=1
$T = ( 0,
\frac{4}{13},
\frac{2}{13},
\frac{7}{13},
\frac{6}{13},
\frac{12}{13} )
$,

\vskip 0.7ex
\hangindent=3em \hangafter=1
$S$ = ($ 1$,
$ -c_{13}^{6}$,
$ \xi_{13}^{3}$,
$ \xi_{13}^{9}$,
$ \xi_{13}^{5}$,
$ \xi_{13}^{7}$;\ \ 
$ -\xi_{13}^{9}$,
$ \xi_{13}^{7}$,
$ -\xi_{13}^{5}$,
$ \xi_{13}^{3}$,
$ -1$;\ \ 
$ \xi_{13}^{9}$,
$ 1$,
$ c_{13}^{6}$,
$ -\xi_{13}^{5}$;\ \ 
$ \xi_{13}^{3}$,
$ -\xi_{13}^{7}$,
$ -c_{13}^{6}$;\ \ 
$ -1$,
$ \xi_{13}^{9}$;\ \ 
$ -\xi_{13}^{3}$)

\vskip 0.7ex
\hangindent=3em \hangafter=1
$\tau_n$ = ($-7.04 + 2.67 i$, $29.22 - 11.09 i$, $-21.31 - 14.71 i$, $5. - 20.27 i$, $-24.22 - 16.71 i$, $-3.51 - 14.2 i$, $-3.51 + 14.2 i$, $-24.22 + 16.71 i$, $5. + 20.27 i$, $-21.31 + 14.71 i$, $29.22 + 11.09 i$, $-7.04 - 2.67 i$, $56.73$)

\vskip 0.7ex
\hangindent=3em \hangafter=1
\textit{Intrinsic sign problem}

  \vskip 2ex

\noindent46. $6_{\frac{58}{13},56.74}^{13,502}$ \irep{35}:\ \ 
$d_i$ = ($1.0$,
$1.941$,
$2.770$,
$3.438$,
$3.907$,
$4.148$) 

\vskip 0.7ex
\hangindent=3em \hangafter=1
$D^2= 56.746 = 
21+15c^{1}_{13}
+10c^{2}_{13}
+6c^{3}_{13}
+3c^{4}_{13}
+c^{5}_{13}
$

\vskip 0.7ex
\hangindent=3em \hangafter=1
$T = ( 0,
\frac{9}{13},
\frac{11}{13},
\frac{6}{13},
\frac{7}{13},
\frac{1}{13} )
$,

\vskip 0.7ex
\hangindent=3em \hangafter=1
$S$ = ($ 1$,
$ -c_{13}^{6}$,
$ \xi_{13}^{3}$,
$ \xi_{13}^{9}$,
$ \xi_{13}^{5}$,
$ \xi_{13}^{7}$;\ \ 
$ -\xi_{13}^{9}$,
$ \xi_{13}^{7}$,
$ -\xi_{13}^{5}$,
$ \xi_{13}^{3}$,
$ -1$;\ \ 
$ \xi_{13}^{9}$,
$ 1$,
$ c_{13}^{6}$,
$ -\xi_{13}^{5}$;\ \ 
$ \xi_{13}^{3}$,
$ -\xi_{13}^{7}$,
$ -c_{13}^{6}$;\ \ 
$ -1$,
$ \xi_{13}^{9}$;\ \ 
$ -\xi_{13}^{3}$)

\vskip 0.7ex
\hangindent=3em \hangafter=1
$\tau_n$ = ($-7.04 - 2.67 i$, $29.22 + 11.09 i$, $-21.31 + 14.71 i$, $5. + 20.27 i$, $-24.22 + 16.71 i$, $-3.51 + 14.2 i$, $-3.51 - 14.2 i$, $-24.22 - 16.71 i$, $5. - 20.27 i$, $-21.31 - 14.71 i$, $29.22 - 11.09 i$, $-7.04 + 2.67 i$, $56.73$)

\vskip 0.7ex
\hangindent=3em \hangafter=1
\textit{Intrinsic sign problem}

  \vskip 2ex

\noindent47. $6_{\frac{8}{3},74.61}^{9,186}$ \irep{26}:\ \ 
$d_i$ = ($1.0$,
$2.879$,
$2.879$,
$2.879$,
$4.411$,
$5.411$) 

\vskip 0.7ex
\hangindent=3em \hangafter=1
$D^2= 74.617 = 
27+27c^{1}_{9}
+18c^{2}_{9}
$

\vskip 0.7ex
\hangindent=3em \hangafter=1
$T = ( 0,
\frac{1}{9},
\frac{1}{9},
\frac{1}{9},
\frac{1}{3},
\frac{2}{3} )
$,

\vskip 0.7ex
\hangindent=3em \hangafter=1
$S$ = ($ 1$,
$ \xi_{9}^{5}$,
$ \xi_{9}^{5}$,
$ \xi_{9}^{5}$,
$ 1+2c^{1}_{9}
+c^{2}_{9}
$,
$ 2+2c^{1}_{9}
+c^{2}_{9}
$;\ \ 
$ 2\xi_{9}^{5}$,
$ -\xi_{9}^{5}$,
$ -\xi_{9}^{5}$,
$ \xi_{9}^{5}$,
$ -\xi_{9}^{5}$;\ \ 
$ 2\xi_{9}^{5}$,
$ -\xi_{9}^{5}$,
$ \xi_{9}^{5}$,
$ -\xi_{9}^{5}$;\ \ 
$ 2\xi_{9}^{5}$,
$ \xi_{9}^{5}$,
$ -\xi_{9}^{5}$;\ \ 
$ -2-2  c^{1}_{9}
-c^{2}_{9}
$,
$ -1$;\ \ 
$ 1+2c^{1}_{9}
+c^{2}_{9}
$)

\vskip 0.7ex
\hangindent=3em \hangafter=1
$\tau_n$ = ($-4.32 + 7.48 i$, $-19.05 + 32.99 i$, $37.3 + 21.53 i$, $-46.73$, $-46.73$, $37.3 - 21.53 i$, $-19.05 - 32.99 i$, $-4.32 - 7.48 i$, $74.6$)

\vskip 0.7ex
\hangindent=3em \hangafter=1
\textit{Intrinsic sign problem}

  \vskip 2ex

\noindent48. $6_{\frac{16}{3},74.61}^{9,452}$ \irep{26}:\ \ 
$d_i$ = ($1.0$,
$2.879$,
$2.879$,
$2.879$,
$4.411$,
$5.411$) 

\vskip 0.7ex
\hangindent=3em \hangafter=1
$D^2= 74.617 = 
27+27c^{1}_{9}
+18c^{2}_{9}
$

\vskip 0.7ex
\hangindent=3em \hangafter=1
$T = ( 0,
\frac{8}{9},
\frac{8}{9},
\frac{8}{9},
\frac{2}{3},
\frac{1}{3} )
$,

\vskip 0.7ex
\hangindent=3em \hangafter=1
$S$ = ($ 1$,
$ \xi_{9}^{5}$,
$ \xi_{9}^{5}$,
$ \xi_{9}^{5}$,
$ 1+2c^{1}_{9}
+c^{2}_{9}
$,
$ 2+2c^{1}_{9}
+c^{2}_{9}
$;\ \ 
$ 2\xi_{9}^{5}$,
$ -\xi_{9}^{5}$,
$ -\xi_{9}^{5}$,
$ \xi_{9}^{5}$,
$ -\xi_{9}^{5}$;\ \ 
$ 2\xi_{9}^{5}$,
$ -\xi_{9}^{5}$,
$ \xi_{9}^{5}$,
$ -\xi_{9}^{5}$;\ \ 
$ 2\xi_{9}^{5}$,
$ \xi_{9}^{5}$,
$ -\xi_{9}^{5}$;\ \ 
$ -2-2  c^{1}_{9}
-c^{2}_{9}
$,
$ -1$;\ \ 
$ 1+2c^{1}_{9}
+c^{2}_{9}
$)

\vskip 0.7ex
\hangindent=3em \hangafter=1
$\tau_n$ = ($-4.32 - 7.48 i$, $-19.05 - 32.99 i$, $37.3 - 21.53 i$, $-46.73$, $-46.73$, $37.3 + 21.53 i$, $-19.05 + 32.99 i$, $-4.32 + 7.48 i$, $74.6$)

\vskip 0.7ex
\hangindent=3em \hangafter=1
\textit{Intrinsic sign problem}

  \vskip 2ex

\noindent49. $6_{6,100.6}^{21,154}$ \irep{43}:\ \ 
$d_i$ = ($1.0$,
$3.791$,
$3.791$,
$3.791$,
$4.791$,
$5.791$) 

\vskip 0.7ex
\hangindent=3em \hangafter=1
$D^2= 100.617 = 
\frac{105+21\sqrt{21}}{2}$

\vskip 0.7ex
\hangindent=3em \hangafter=1
$T = ( 0,
\frac{1}{7},
\frac{2}{7},
\frac{4}{7},
0,
\frac{2}{3} )
$,

\vskip 0.7ex
\hangindent=3em \hangafter=1
$S$ = ($ 1$,
$ \frac{3+\sqrt{21}}{2}$,
$ \frac{3+\sqrt{21}}{2}$,
$ \frac{3+\sqrt{21}}{2}$,
$ \frac{5+\sqrt{21}}{2}$,
$ \frac{7+\sqrt{21}}{2}$;\ \ 
$ 2-c^{1}_{21}
-2  c^{2}_{21}
+3c^{3}_{21}
+2c^{4}_{21}
-2  c^{5}_{21}
$,
$ -c^{2}_{21}
-2  c^{3}_{21}
-c^{4}_{21}
+c^{5}_{21}
$,
$ -1+2c^{1}_{21}
+3c^{2}_{21}
-c^{3}_{21}
+2c^{5}_{21}
$,
$ -\frac{3+\sqrt{21}}{2}$,
$0$;\ \ 
$ -1+2c^{1}_{21}
+3c^{2}_{21}
-c^{3}_{21}
+2c^{5}_{21}
$,
$ 2-c^{1}_{21}
-2  c^{2}_{21}
+3c^{3}_{21}
+2c^{4}_{21}
-2  c^{5}_{21}
$,
$ -\frac{3+\sqrt{21}}{2}$,
$0$;\ \ 
$ -c^{2}_{21}
-2  c^{3}_{21}
-c^{4}_{21}
+c^{5}_{21}
$,
$ -\frac{3+\sqrt{21}}{2}$,
$0$;\ \ 
$ 1$,
$ \frac{7+\sqrt{21}}{2}$;\ \ 
$ -\frac{7+\sqrt{21}}{2}$)

\vskip 0.7ex
\hangindent=3em \hangafter=1
$\tau_n$ = ($0. - 10.03 i$, $0. + 48.05 i$, $50.3 - 19.01 i$, $0. - 10.03 i$, $0. + 10.03 i$, $50.3 - 19.01 i$, $50.3 - 29.04 i$, $0. + 48.05 i$, $50.3 + 19.01 i$, $0. - 48.05 i$, $0. + 48.05 i$, $50.3 - 19.01 i$, $0. - 48.05 i$, $50.3 + 29.04 i$, $50.3 + 19.01 i$, $0. - 10.03 i$, $0. + 10.03 i$, $50.3 + 19.01 i$, $0. - 48.05 i$, $0. + 10.03 i$, $100.6$)

\vskip 0.7ex
\hangindent=3em \hangafter=1
\textit{Intrinsic sign problem}

  \vskip 2ex

\noindent50. $6_{2,100.6}^{21,320}$ \irep{43}:\ \ 
$d_i$ = ($1.0$,
$3.791$,
$3.791$,
$3.791$,
$4.791$,
$5.791$) 

\vskip 0.7ex
\hangindent=3em \hangafter=1
$D^2= 100.617 = 
\frac{105+21\sqrt{21}}{2}$

\vskip 0.7ex
\hangindent=3em \hangafter=1
$T = ( 0,
\frac{3}{7},
\frac{5}{7},
\frac{6}{7},
0,
\frac{1}{3} )
$,

\vskip 0.7ex
\hangindent=3em \hangafter=1
$S$ = ($ 1$,
$ \frac{3+\sqrt{21}}{2}$,
$ \frac{3+\sqrt{21}}{2}$,
$ \frac{3+\sqrt{21}}{2}$,
$ \frac{5+\sqrt{21}}{2}$,
$ \frac{7+\sqrt{21}}{2}$;\ \ 
$ -c^{2}_{21}
-2  c^{3}_{21}
-c^{4}_{21}
+c^{5}_{21}
$,
$ 2-c^{1}_{21}
-2  c^{2}_{21}
+3c^{3}_{21}
+2c^{4}_{21}
-2  c^{5}_{21}
$,
$ -1+2c^{1}_{21}
+3c^{2}_{21}
-c^{3}_{21}
+2c^{5}_{21}
$,
$ -\frac{3+\sqrt{21}}{2}$,
$0$;\ \ 
$ -1+2c^{1}_{21}
+3c^{2}_{21}
-c^{3}_{21}
+2c^{5}_{21}
$,
$ -c^{2}_{21}
-2  c^{3}_{21}
-c^{4}_{21}
+c^{5}_{21}
$,
$ -\frac{3+\sqrt{21}}{2}$,
$0$;\ \ 
$ 2-c^{1}_{21}
-2  c^{2}_{21}
+3c^{3}_{21}
+2c^{4}_{21}
-2  c^{5}_{21}
$,
$ -\frac{3+\sqrt{21}}{2}$,
$0$;\ \ 
$ 1$,
$ \frac{7+\sqrt{21}}{2}$;\ \ 
$ -\frac{7+\sqrt{21}}{2}$)

\vskip 0.7ex
\hangindent=3em \hangafter=1
$\tau_n$ = ($0. + 10.03 i$, $0. - 48.05 i$, $50.3 + 19.01 i$, $0. + 10.03 i$, $0. - 10.03 i$, $50.3 + 19.01 i$, $50.3 + 29.04 i$, $0. - 48.05 i$, $50.3 - 19.01 i$, $0. + 48.05 i$, $0. - 48.05 i$, $50.3 + 19.01 i$, $0. + 48.05 i$, $50.3 - 29.04 i$, $50.3 - 19.01 i$, $0. + 10.03 i$, $0. - 10.03 i$, $50.3 - 19.01 i$, $0. + 48.05 i$, $0. - 10.03 i$, $100.6$)

\vskip 0.7ex
\hangindent=3em \hangafter=1
\textit{Intrinsic sign problem}

  \vskip 2ex 

%% file: modular_data/SsL7U_.tex
\noindent1. $7_{2,7.}^{7,892}$ \irep{48}:\ \ 
$d_i$ = ($1.0$,
$1.0$,
$1.0$,
$1.0$,
$1.0$,
$1.0$,
$1.0$) 

\vskip 0.7ex
\hangindent=3em \hangafter=1
$D^2= 7.0 = 
7$

\vskip 0.7ex
\hangindent=3em \hangafter=1
$T = ( 0,
\frac{1}{7},
\frac{1}{7},
\frac{2}{7},
\frac{2}{7},
\frac{4}{7},
\frac{4}{7} )
$,

\vskip 0.7ex
\hangindent=3em \hangafter=1
$S$ = ($ 1$,
$ 1$,
$ 1$,
$ 1$,
$ 1$,
$ 1$,
$ 1$;\ \ 
$ -\zeta_{14}^{3}$,
$ \zeta_{7}^{2}$,
$ -\zeta_{14}^{5}$,
$ \zeta_{7}^{1}$,
$ -\zeta_{14}^{1}$,
$ \zeta_{7}^{3}$;\ \ 
$ -\zeta_{14}^{3}$,
$ \zeta_{7}^{1}$,
$ -\zeta_{14}^{5}$,
$ \zeta_{7}^{3}$,
$ -\zeta_{14}^{1}$;\ \ 
$ \zeta_{7}^{3}$,
$ -\zeta_{14}^{1}$,
$ \zeta_{7}^{2}$,
$ -\zeta_{14}^{3}$;\ \ 
$ \zeta_{7}^{3}$,
$ -\zeta_{14}^{3}$,
$ \zeta_{7}^{2}$;\ \ 
$ -\zeta_{14}^{5}$,
$ \zeta_{7}^{1}$;\ \ 
$ -\zeta_{14}^{5}$)

\vskip 0.7ex
\hangindent=3em \hangafter=1
$\tau_n$ = ($0. + 2.65 i$, $0. + 2.65 i$, $0. - 2.65 i$, $0. + 2.65 i$, $0. - 2.65 i$, $0. - 2.65 i$, $7.$)

\vskip 0.7ex
\hangindent=3em \hangafter=1
\textit{Intrinsic sign problem}

  \vskip 2ex

\noindent2. $7_{6,7.}^{7,110}$ \irep{48}:\ \ 
$d_i$ = ($1.0$,
$1.0$,
$1.0$,
$1.0$,
$1.0$,
$1.0$,
$1.0$) 

\vskip 0.7ex
\hangindent=3em \hangafter=1
$D^2= 7.0 = 
7$

\vskip 0.7ex
\hangindent=3em \hangafter=1
$T = ( 0,
\frac{3}{7},
\frac{3}{7},
\frac{5}{7},
\frac{5}{7},
\frac{6}{7},
\frac{6}{7} )
$,

\vskip 0.7ex
\hangindent=3em \hangafter=1
$S$ = ($ 1$,
$ 1$,
$ 1$,
$ 1$,
$ 1$,
$ 1$,
$ 1$;\ \ 
$ \zeta_{7}^{1}$,
$ -\zeta_{14}^{5}$,
$ -\zeta_{14}^{3}$,
$ \zeta_{7}^{2}$,
$ -\zeta_{14}^{1}$,
$ \zeta_{7}^{3}$;\ \ 
$ \zeta_{7}^{1}$,
$ \zeta_{7}^{2}$,
$ -\zeta_{14}^{3}$,
$ \zeta_{7}^{3}$,
$ -\zeta_{14}^{1}$;\ \ 
$ -\zeta_{14}^{1}$,
$ \zeta_{7}^{3}$,
$ -\zeta_{14}^{5}$,
$ \zeta_{7}^{1}$;\ \ 
$ -\zeta_{14}^{1}$,
$ \zeta_{7}^{1}$,
$ -\zeta_{14}^{5}$;\ \ 
$ \zeta_{7}^{2}$,
$ -\zeta_{14}^{3}$;\ \ 
$ \zeta_{7}^{2}$)

\vskip 0.7ex
\hangindent=3em \hangafter=1
$\tau_n$ = ($0. - 2.65 i$, $0. - 2.65 i$, $0. + 2.65 i$, $0. - 2.65 i$, $0. + 2.65 i$, $0. + 2.65 i$, $7.$)

\vskip 0.7ex
\hangindent=3em \hangafter=1
\textit{Intrinsic sign problem}

  \vskip 2ex

\noindent3. $7_{\frac{27}{4},27.31}^{32,396}$ \irep{96}:\ \ 
$d_i$ = ($1.0$,
$1.0$,
$1.847$,
$1.847$,
$2.414$,
$2.414$,
$2.613$) 

\vskip 0.7ex
\hangindent=3em \hangafter=1
$D^2= 27.313 = 
16+8\sqrt{2}$

\vskip 0.7ex
\hangindent=3em \hangafter=1
$T = ( 0,
\frac{1}{2},
\frac{1}{32},
\frac{1}{32},
\frac{1}{4},
\frac{3}{4},
\frac{21}{32} )
$,

\vskip 0.7ex
\hangindent=3em \hangafter=1
$S$ = ($ 1$,
$ 1$,
$ c_{16}^{1}$,
$ c_{16}^{1}$,
$ 1+\sqrt{2}$,
$ 1+\sqrt{2}$,
$ c^{1}_{16}
+c^{3}_{16}
$;\ \ 
$ 1$,
$ -c_{16}^{1}$,
$ -c_{16}^{1}$,
$ 1+\sqrt{2}$,
$ 1+\sqrt{2}$,
$ -c^{1}_{16}
-c^{3}_{16}
$;\ \ 
$(-c^{1}_{16}
-c^{3}_{16}
)\mathrm{i}$,
$(c^{1}_{16}
+c^{3}_{16}
)\mathrm{i}$,
$ -c_{16}^{1}$,
$ c_{16}^{1}$,
$0$;\ \ 
$(-c^{1}_{16}
-c^{3}_{16}
)\mathrm{i}$,
$ -c_{16}^{1}$,
$ c_{16}^{1}$,
$0$;\ \ 
$ -1$,
$ -1$,
$ c^{1}_{16}
+c^{3}_{16}
$;\ \ 
$ -1$,
$ -c^{1}_{16}
-c^{3}_{16}
$;\ \ 
$0$)

\vskip 0.7ex
\hangindent=3em \hangafter=1
$\tau_n$ = ($2.9 - 4.35 i$, $-5.96 + 8.92 i$, $12.37 + 2.46 i$, $13.65$, $2.46 + 12.37 i$, $-0.74 + 3.69 i$, $-4.35 + 2.9 i$, $13.65 + 13.65 i$, $4.35 + 2.9 i$, $-18.57 + 3.69 i$, $-2.46 + 12.37 i$, $13.66$, $-12.37 + 2.46 i$, $-13.35 + 8.92 i$, $-2.9 - 4.35 i$, $0.$, $-2.9 + 4.35 i$, $-13.35 - 8.92 i$, $-12.37 - 2.46 i$, $13.66$, $-2.46 - 12.37 i$, $-18.57 - 3.69 i$, $4.35 - 2.9 i$, $13.65 - 13.65 i$, $-4.35 - 2.9 i$, $-0.74 - 3.69 i$, $2.46 - 12.37 i$, $13.65$, $12.37 - 2.46 i$, $-5.96 - 8.92 i$, $2.9 + 4.35 i$, $27.31$)

\vskip 0.7ex
\hangindent=3em \hangafter=1
\textit{Intrinsic sign problem}

  \vskip 2ex

\noindent4. $7_{\frac{9}{4},27.31}^{32,918}$ \irep{97}:\ \ 
$d_i$ = ($1.0$,
$1.0$,
$1.847$,
$1.847$,
$2.414$,
$2.414$,
$2.613$) 

\vskip 0.7ex
\hangindent=3em \hangafter=1
$D^2= 27.313 = 
16+8\sqrt{2}$

\vskip 0.7ex
\hangindent=3em \hangafter=1
$T = ( 0,
\frac{1}{2},
\frac{3}{32},
\frac{3}{32},
\frac{1}{4},
\frac{3}{4},
\frac{15}{32} )
$,

\vskip 0.7ex
\hangindent=3em \hangafter=1
$S$ = ($ 1$,
$ 1$,
$ c_{16}^{1}$,
$ c_{16}^{1}$,
$ 1+\sqrt{2}$,
$ 1+\sqrt{2}$,
$ c^{1}_{16}
+c^{3}_{16}
$;\ \ 
$ 1$,
$ -c_{16}^{1}$,
$ -c_{16}^{1}$,
$ 1+\sqrt{2}$,
$ 1+\sqrt{2}$,
$ -c^{1}_{16}
-c^{3}_{16}
$;\ \ 
$ c^{1}_{16}
+c^{3}_{16}
$,
$ -c^{1}_{16}
-c^{3}_{16}
$,
$ c_{16}^{1}$,
$ -c_{16}^{1}$,
$0$;\ \ 
$ c^{1}_{16}
+c^{3}_{16}
$,
$ c_{16}^{1}$,
$ -c_{16}^{1}$,
$0$;\ \ 
$ -1$,
$ -1$,
$ -c^{1}_{16}
-c^{3}_{16}
$;\ \ 
$ -1$,
$ c^{1}_{16}
+c^{3}_{16}
$;\ \ 
$0$)

\vskip 0.7ex
\hangindent=3em \hangafter=1
$\tau_n$ = ($-1.02 + 5.12 i$, $-0.74 + 3.69 i$, $-7.01 + 10.49 i$, $13.66$, $-10.49 + 7.01 i$, $-13.35 - 8.92 i$, $-5.12 + 1.02 i$, $13.65 - 13.65 i$, $5.12 + 1.02 i$, $-5.96 - 8.92 i$, $10.49 + 7.01 i$, $13.65$, $7.01 + 10.49 i$, $-18.57 + 3.69 i$, $1.02 + 5.12 i$, $0.$, $1.02 - 5.12 i$, $-18.57 - 3.69 i$, $7.01 - 10.49 i$, $13.65$, $10.49 - 7.01 i$, $-5.96 + 8.92 i$, $5.12 - 1.02 i$, $13.65 + 13.65 i$, $-5.12 - 1.02 i$, $-13.35 + 8.92 i$, $-10.49 - 7.01 i$, $13.66$, $-7.01 - 10.49 i$, $-0.74 - 3.69 i$, $-1.02 - 5.12 i$, $27.31$)

\vskip 0.7ex
\hangindent=3em \hangafter=1
\textit{Intrinsic sign problem}

  \vskip 2ex

\noindent5. $7_{\frac{31}{4},27.31}^{32,159}$ \irep{97}:\ \ 
$d_i$ = ($1.0$,
$1.0$,
$1.847$,
$1.847$,
$2.414$,
$2.414$,
$2.613$) 

\vskip 0.7ex
\hangindent=3em \hangafter=1
$D^2= 27.313 = 
16+8\sqrt{2}$

\vskip 0.7ex
\hangindent=3em \hangafter=1
$T = ( 0,
\frac{1}{2},
\frac{5}{32},
\frac{5}{32},
\frac{1}{4},
\frac{3}{4},
\frac{25}{32} )
$,

\vskip 0.7ex
\hangindent=3em \hangafter=1
$S$ = ($ 1$,
$ 1$,
$ c_{16}^{1}$,
$ c_{16}^{1}$,
$ 1+\sqrt{2}$,
$ 1+\sqrt{2}$,
$ c^{1}_{16}
+c^{3}_{16}
$;\ \ 
$ 1$,
$ -c_{16}^{1}$,
$ -c_{16}^{1}$,
$ 1+\sqrt{2}$,
$ 1+\sqrt{2}$,
$ -c^{1}_{16}
-c^{3}_{16}
$;\ \ 
$ -c^{1}_{16}
-c^{3}_{16}
$,
$ c^{1}_{16}
+c^{3}_{16}
$,
$ -c_{16}^{1}$,
$ c_{16}^{1}$,
$0$;\ \ 
$ -c^{1}_{16}
-c^{3}_{16}
$,
$ -c_{16}^{1}$,
$ c_{16}^{1}$,
$0$;\ \ 
$ -1$,
$ -1$,
$ c^{1}_{16}
+c^{3}_{16}
$;\ \ 
$ -1$,
$ -c^{1}_{16}
-c^{3}_{16}
$;\ \ 
$0$)

\vskip 0.7ex
\hangindent=3em \hangafter=1
$\tau_n$ = ($5.12 - 1.02 i$, $-18.57 + 3.69 i$, $-10.49 + 7.01 i$, $13.66$, $7.01 - 10.49 i$, $-5.96 - 8.92 i$, $-1.02 + 5.12 i$, $13.65 + 13.65 i$, $1.02 + 5.12 i$, $-13.35 - 8.92 i$, $-7.01 - 10.49 i$, $13.65$, $10.49 + 7.01 i$, $-0.74 + 3.69 i$, $-5.12 - 1.02 i$, $0.$, $-5.12 + 1.02 i$, $-0.74 - 3.69 i$, $10.49 - 7.01 i$, $13.65$, $-7.01 + 10.49 i$, $-13.35 + 8.92 i$, $1.02 - 5.12 i$, $13.65 - 13.65 i$, $-1.02 - 5.12 i$, $-5.96 + 8.92 i$, $7.01 + 10.49 i$, $13.66$, $-10.49 - 7.01 i$, $-18.57 - 3.69 i$, $5.12 + 1.02 i$, $27.31$)

\vskip 0.7ex
\hangindent=3em \hangafter=1
\textit{Intrinsic sign problem}

  \vskip 2ex

\noindent6. $7_{\frac{13}{4},27.31}^{32,427}$ \irep{96}:\ \ 
$d_i$ = ($1.0$,
$1.0$,
$1.847$,
$1.847$,
$2.414$,
$2.414$,
$2.613$) 

\vskip 0.7ex
\hangindent=3em \hangafter=1
$D^2= 27.313 = 
16+8\sqrt{2}$

\vskip 0.7ex
\hangindent=3em \hangafter=1
$T = ( 0,
\frac{1}{2},
\frac{7}{32},
\frac{7}{32},
\frac{1}{4},
\frac{3}{4},
\frac{19}{32} )
$,

\vskip 0.7ex
\hangindent=3em \hangafter=1
$S$ = ($ 1$,
$ 1$,
$ c_{16}^{1}$,
$ c_{16}^{1}$,
$ 1+\sqrt{2}$,
$ 1+\sqrt{2}$,
$ c^{1}_{16}
+c^{3}_{16}
$;\ \ 
$ 1$,
$ -c_{16}^{1}$,
$ -c_{16}^{1}$,
$ 1+\sqrt{2}$,
$ 1+\sqrt{2}$,
$ -c^{1}_{16}
-c^{3}_{16}
$;\ \ 
$(-c^{1}_{16}
-c^{3}_{16}
)\mathrm{i}$,
$(c^{1}_{16}
+c^{3}_{16}
)\mathrm{i}$,
$ c_{16}^{1}$,
$ -c_{16}^{1}$,
$0$;\ \ 
$(-c^{1}_{16}
-c^{3}_{16}
)\mathrm{i}$,
$ c_{16}^{1}$,
$ -c_{16}^{1}$,
$0$;\ \ 
$ -1$,
$ -1$,
$ -c^{1}_{16}
-c^{3}_{16}
$;\ \ 
$ -1$,
$ c^{1}_{16}
+c^{3}_{16}
$;\ \ 
$0$)

\vskip 0.7ex
\hangindent=3em \hangafter=1
$\tau_n$ = ($-4.35 + 2.9 i$, $-13.35 + 8.92 i$, $-2.46 - 12.37 i$, $13.65$, $12.37 + 2.46 i$, $-18.57 + 3.69 i$, $-2.9 + 4.35 i$, $13.65 - 13.65 i$, $2.9 + 4.35 i$, $-0.74 + 3.69 i$, $-12.37 + 2.46 i$, $13.66$, $2.46 - 12.37 i$, $-5.96 + 8.92 i$, $4.35 + 2.9 i$, $0.$, $4.35 - 2.9 i$, $-5.96 - 8.92 i$, $2.46 + 12.37 i$, $13.66$, $-12.37 - 2.46 i$, $-0.74 - 3.69 i$, $2.9 - 4.35 i$, $13.65 + 13.65 i$, $-2.9 - 4.35 i$, $-18.57 - 3.69 i$, $12.37 - 2.46 i$, $13.65$, $-2.46 + 12.37 i$, $-13.35 - 8.92 i$, $-4.35 - 2.9 i$, $27.31$)

\vskip 0.7ex
\hangindent=3em \hangafter=1
\textit{Intrinsic sign problem}

  \vskip 2ex

\noindent7. $7_{\frac{3}{4},27.31}^{32,913}$ \irep{96}:\ \ 
$d_i$ = ($1.0$,
$1.0$,
$1.847$,
$1.847$,
$2.414$,
$2.414$,
$2.613$) 

\vskip 0.7ex
\hangindent=3em \hangafter=1
$D^2= 27.313 = 
16+8\sqrt{2}$

\vskip 0.7ex
\hangindent=3em \hangafter=1
$T = ( 0,
\frac{1}{2},
\frac{9}{32},
\frac{9}{32},
\frac{1}{4},
\frac{3}{4},
\frac{29}{32} )
$,

\vskip 0.7ex
\hangindent=3em \hangafter=1
$S$ = ($ 1$,
$ 1$,
$ c_{16}^{1}$,
$ c_{16}^{1}$,
$ 1+\sqrt{2}$,
$ 1+\sqrt{2}$,
$ c^{1}_{16}
+c^{3}_{16}
$;\ \ 
$ 1$,
$ -c_{16}^{1}$,
$ -c_{16}^{1}$,
$ 1+\sqrt{2}$,
$ 1+\sqrt{2}$,
$ -c^{1}_{16}
-c^{3}_{16}
$;\ \ 
$(c^{1}_{16}
+c^{3}_{16}
)\mathrm{i}$,
$(-c^{1}_{16}
-c^{3}_{16}
)\mathrm{i}$,
$ -c_{16}^{1}$,
$ c_{16}^{1}$,
$0$;\ \ 
$(c^{1}_{16}
+c^{3}_{16}
)\mathrm{i}$,
$ -c_{16}^{1}$,
$ c_{16}^{1}$,
$0$;\ \ 
$ -1$,
$ -1$,
$ c^{1}_{16}
+c^{3}_{16}
$;\ \ 
$ -1$,
$ -c^{1}_{16}
-c^{3}_{16}
$;\ \ 
$0$)

\vskip 0.7ex
\hangindent=3em \hangafter=1
$\tau_n$ = ($4.35 + 2.9 i$, $-13.35 - 8.92 i$, $2.46 - 12.37 i$, $13.65$, $-12.37 + 2.46 i$, $-18.57 - 3.69 i$, $2.9 + 4.35 i$, $13.65 + 13.65 i$, $-2.9 + 4.35 i$, $-0.74 - 3.69 i$, $12.37 + 2.46 i$, $13.66$, $-2.46 - 12.37 i$, $-5.96 - 8.92 i$, $-4.35 + 2.9 i$, $0.$, $-4.35 - 2.9 i$, $-5.96 + 8.92 i$, $-2.46 + 12.37 i$, $13.66$, $12.37 - 2.46 i$, $-0.74 + 3.69 i$, $-2.9 - 4.35 i$, $13.65 - 13.65 i$, $2.9 - 4.35 i$, $-18.57 + 3.69 i$, $-12.37 - 2.46 i$, $13.65$, $2.46 + 12.37 i$, $-13.35 + 8.92 i$, $4.35 - 2.9 i$, $27.31$)

\vskip 0.7ex
\hangindent=3em \hangafter=1
\textit{Intrinsic sign problem}

  \vskip 2ex

\noindent8. $7_{\frac{17}{4},27.31}^{32,261}$ \irep{97}:\ \ 
$d_i$ = ($1.0$,
$1.0$,
$1.847$,
$1.847$,
$2.414$,
$2.414$,
$2.613$) 

\vskip 0.7ex
\hangindent=3em \hangafter=1
$D^2= 27.313 = 
16+8\sqrt{2}$

\vskip 0.7ex
\hangindent=3em \hangafter=1
$T = ( 0,
\frac{1}{2},
\frac{11}{32},
\frac{11}{32},
\frac{1}{4},
\frac{3}{4},
\frac{23}{32} )
$,

\vskip 0.7ex
\hangindent=3em \hangafter=1
$S$ = ($ 1$,
$ 1$,
$ c_{16}^{1}$,
$ c_{16}^{1}$,
$ 1+\sqrt{2}$,
$ 1+\sqrt{2}$,
$ c^{1}_{16}
+c^{3}_{16}
$;\ \ 
$ 1$,
$ -c_{16}^{1}$,
$ -c_{16}^{1}$,
$ 1+\sqrt{2}$,
$ 1+\sqrt{2}$,
$ -c^{1}_{16}
-c^{3}_{16}
$;\ \ 
$ -c^{1}_{16}
-c^{3}_{16}
$,
$ c^{1}_{16}
+c^{3}_{16}
$,
$ c_{16}^{1}$,
$ -c_{16}^{1}$,
$0$;\ \ 
$ -c^{1}_{16}
-c^{3}_{16}
$,
$ c_{16}^{1}$,
$ -c_{16}^{1}$,
$0$;\ \ 
$ -1$,
$ -1$,
$ -c^{1}_{16}
-c^{3}_{16}
$;\ \ 
$ -1$,
$ c^{1}_{16}
+c^{3}_{16}
$;\ \ 
$0$)

\vskip 0.7ex
\hangindent=3em \hangafter=1
$\tau_n$ = ($-5.12 - 1.02 i$, $-18.57 - 3.69 i$, $10.49 + 7.01 i$, $13.66$, $-7.01 - 10.49 i$, $-5.96 + 8.92 i$, $1.02 + 5.12 i$, $13.65 - 13.65 i$, $-1.02 + 5.12 i$, $-13.35 + 8.92 i$, $7.01 - 10.49 i$, $13.65$, $-10.49 + 7.01 i$, $-0.74 - 3.69 i$, $5.12 - 1.02 i$, $0.$, $5.12 + 1.02 i$, $-0.74 + 3.69 i$, $-10.49 - 7.01 i$, $13.65$, $7.01 + 10.49 i$, $-13.35 - 8.92 i$, $-1.02 - 5.12 i$, $13.65 + 13.65 i$, $1.02 - 5.12 i$, $-5.96 - 8.92 i$, $-7.01 + 10.49 i$, $13.66$, $10.49 - 7.01 i$, $-18.57 + 3.69 i$, $-5.12 + 1.02 i$, $27.31$)

\vskip 0.7ex
\hangindent=3em \hangafter=1
\textit{Intrinsic sign problem}

  \vskip 2ex

\noindent9. $7_{\frac{7}{4},27.31}^{32,912}$ \irep{97}:\ \ 
$d_i$ = ($1.0$,
$1.0$,
$1.847$,
$1.847$,
$2.414$,
$2.414$,
$2.613$) 

\vskip 0.7ex
\hangindent=3em \hangafter=1
$D^2= 27.313 = 
16+8\sqrt{2}$

\vskip 0.7ex
\hangindent=3em \hangafter=1
$T = ( 0,
\frac{1}{2},
\frac{13}{32},
\frac{13}{32},
\frac{1}{4},
\frac{3}{4},
\frac{1}{32} )
$,

\vskip 0.7ex
\hangindent=3em \hangafter=1
$S$ = ($ 1$,
$ 1$,
$ c_{16}^{1}$,
$ c_{16}^{1}$,
$ 1+\sqrt{2}$,
$ 1+\sqrt{2}$,
$ c^{1}_{16}
+c^{3}_{16}
$;\ \ 
$ 1$,
$ -c_{16}^{1}$,
$ -c_{16}^{1}$,
$ 1+\sqrt{2}$,
$ 1+\sqrt{2}$,
$ -c^{1}_{16}
-c^{3}_{16}
$;\ \ 
$ c^{1}_{16}
+c^{3}_{16}
$,
$ -c^{1}_{16}
-c^{3}_{16}
$,
$ -c_{16}^{1}$,
$ c_{16}^{1}$,
$0$;\ \ 
$ c^{1}_{16}
+c^{3}_{16}
$,
$ -c_{16}^{1}$,
$ c_{16}^{1}$,
$0$;\ \ 
$ -1$,
$ -1$,
$ c^{1}_{16}
+c^{3}_{16}
$;\ \ 
$ -1$,
$ -c^{1}_{16}
-c^{3}_{16}
$;\ \ 
$0$)

\vskip 0.7ex
\hangindent=3em \hangafter=1
$\tau_n$ = ($1.02 + 5.12 i$, $-0.74 - 3.69 i$, $7.01 + 10.49 i$, $13.66$, $10.49 + 7.01 i$, $-13.35 + 8.92 i$, $5.12 + 1.02 i$, $13.65 + 13.65 i$, $-5.12 + 1.02 i$, $-5.96 + 8.92 i$, $-10.49 + 7.01 i$, $13.65$, $-7.01 + 10.49 i$, $-18.57 - 3.69 i$, $-1.02 + 5.12 i$, $0.$, $-1.02 - 5.12 i$, $-18.57 + 3.69 i$, $-7.01 - 10.49 i$, $13.65$, $-10.49 - 7.01 i$, $-5.96 - 8.92 i$, $-5.12 - 1.02 i$, $13.65 - 13.65 i$, $5.12 - 1.02 i$, $-13.35 - 8.92 i$, $10.49 - 7.01 i$, $13.66$, $7.01 - 10.49 i$, $-0.74 + 3.69 i$, $1.02 - 5.12 i$, $27.31$)

\vskip 0.7ex
\hangindent=3em \hangafter=1
\textit{Intrinsic sign problem}

  \vskip 2ex

\noindent10. $7_{\frac{21}{4},27.31}^{32,114}$ \irep{96}:\ \ 
$d_i$ = ($1.0$,
$1.0$,
$1.847$,
$1.847$,
$2.414$,
$2.414$,
$2.613$) 

\vskip 0.7ex
\hangindent=3em \hangafter=1
$D^2= 27.313 = 
16+8\sqrt{2}$

\vskip 0.7ex
\hangindent=3em \hangafter=1
$T = ( 0,
\frac{1}{2},
\frac{15}{32},
\frac{15}{32},
\frac{1}{4},
\frac{3}{4},
\frac{27}{32} )
$,

\vskip 0.7ex
\hangindent=3em \hangafter=1
$S$ = ($ 1$,
$ 1$,
$ c_{16}^{1}$,
$ c_{16}^{1}$,
$ 1+\sqrt{2}$,
$ 1+\sqrt{2}$,
$ c^{1}_{16}
+c^{3}_{16}
$;\ \ 
$ 1$,
$ -c_{16}^{1}$,
$ -c_{16}^{1}$,
$ 1+\sqrt{2}$,
$ 1+\sqrt{2}$,
$ -c^{1}_{16}
-c^{3}_{16}
$;\ \ 
$(c^{1}_{16}
+c^{3}_{16}
)\mathrm{i}$,
$(-c^{1}_{16}
-c^{3}_{16}
)\mathrm{i}$,
$ c_{16}^{1}$,
$ -c_{16}^{1}$,
$0$;\ \ 
$(c^{1}_{16}
+c^{3}_{16}
)\mathrm{i}$,
$ c_{16}^{1}$,
$ -c_{16}^{1}$,
$0$;\ \ 
$ -1$,
$ -1$,
$ -c^{1}_{16}
-c^{3}_{16}
$;\ \ 
$ -1$,
$ c^{1}_{16}
+c^{3}_{16}
$;\ \ 
$0$)

\vskip 0.7ex
\hangindent=3em \hangafter=1
$\tau_n$ = ($-2.9 - 4.35 i$, $-5.96 - 8.92 i$, $-12.37 + 2.46 i$, $13.65$, $-2.46 + 12.37 i$, $-0.74 - 3.69 i$, $4.35 + 2.9 i$, $13.65 - 13.65 i$, $-4.35 + 2.9 i$, $-18.57 - 3.69 i$, $2.46 + 12.37 i$, $13.66$, $12.37 + 2.46 i$, $-13.35 - 8.92 i$, $2.9 - 4.35 i$, $0.$, $2.9 + 4.35 i$, $-13.35 + 8.92 i$, $12.37 - 2.46 i$, $13.66$, $2.46 - 12.37 i$, $-18.57 + 3.69 i$, $-4.35 - 2.9 i$, $13.65 + 13.65 i$, $4.35 - 2.9 i$, $-0.74 + 3.69 i$, $-2.46 - 12.37 i$, $13.65$, $-12.37 - 2.46 i$, $-5.96 + 8.92 i$, $-2.9 + 4.35 i$, $27.31$)

\vskip 0.7ex
\hangindent=3em \hangafter=1
\textit{Intrinsic sign problem}

  \vskip 2ex

\noindent11. $7_{\frac{11}{4},27.31}^{32,418}$ \irep{96}:\ \ 
$d_i$ = ($1.0$,
$1.0$,
$1.847$,
$1.847$,
$2.414$,
$2.414$,
$2.613$) 

\vskip 0.7ex
\hangindent=3em \hangafter=1
$D^2= 27.313 = 
16+8\sqrt{2}$

\vskip 0.7ex
\hangindent=3em \hangafter=1
$T = ( 0,
\frac{1}{2},
\frac{17}{32},
\frac{17}{32},
\frac{1}{4},
\frac{3}{4},
\frac{5}{32} )
$,

\vskip 0.7ex
\hangindent=3em \hangafter=1
$S$ = ($ 1$,
$ 1$,
$ c_{16}^{1}$,
$ c_{16}^{1}$,
$ 1+\sqrt{2}$,
$ 1+\sqrt{2}$,
$ c^{1}_{16}
+c^{3}_{16}
$;\ \ 
$ 1$,
$ -c_{16}^{1}$,
$ -c_{16}^{1}$,
$ 1+\sqrt{2}$,
$ 1+\sqrt{2}$,
$ -c^{1}_{16}
-c^{3}_{16}
$;\ \ 
$(-c^{1}_{16}
-c^{3}_{16}
)\mathrm{i}$,
$(c^{1}_{16}
+c^{3}_{16}
)\mathrm{i}$,
$ -c_{16}^{1}$,
$ c_{16}^{1}$,
$0$;\ \ 
$(-c^{1}_{16}
-c^{3}_{16}
)\mathrm{i}$,
$ -c_{16}^{1}$,
$ c_{16}^{1}$,
$0$;\ \ 
$ -1$,
$ -1$,
$ c^{1}_{16}
+c^{3}_{16}
$;\ \ 
$ -1$,
$ -c^{1}_{16}
-c^{3}_{16}
$;\ \ 
$0$)

\vskip 0.7ex
\hangindent=3em \hangafter=1
$\tau_n$ = ($-2.9 + 4.35 i$, $-5.96 + 8.92 i$, $-12.37 - 2.46 i$, $13.65$, $-2.46 - 12.37 i$, $-0.74 + 3.69 i$, $4.35 - 2.9 i$, $13.65 + 13.65 i$, $-4.35 - 2.9 i$, $-18.57 + 3.69 i$, $2.46 - 12.37 i$, $13.66$, $12.37 - 2.46 i$, $-13.35 + 8.92 i$, $2.9 + 4.35 i$, $0.$, $2.9 - 4.35 i$, $-13.35 - 8.92 i$, $12.37 + 2.46 i$, $13.66$, $2.46 + 12.37 i$, $-18.57 - 3.69 i$, $-4.35 + 2.9 i$, $13.65 - 13.65 i$, $4.35 + 2.9 i$, $-0.74 - 3.69 i$, $-2.46 + 12.37 i$, $13.65$, $-12.37 + 2.46 i$, $-5.96 - 8.92 i$, $-2.9 - 4.35 i$, $27.31$)

\vskip 0.7ex
\hangindent=3em \hangafter=1
\textit{Intrinsic sign problem}

  \vskip 2ex

\noindent12. $7_{\frac{25}{4},27.31}^{32,222}$ \irep{97}:\ \ 
$d_i$ = ($1.0$,
$1.0$,
$1.847$,
$1.847$,
$2.414$,
$2.414$,
$2.613$) 

\vskip 0.7ex
\hangindent=3em \hangafter=1
$D^2= 27.313 = 
16+8\sqrt{2}$

\vskip 0.7ex
\hangindent=3em \hangafter=1
$T = ( 0,
\frac{1}{2},
\frac{19}{32},
\frac{19}{32},
\frac{1}{4},
\frac{3}{4},
\frac{31}{32} )
$,

\vskip 0.7ex
\hangindent=3em \hangafter=1
$S$ = ($ 1$,
$ 1$,
$ c_{16}^{1}$,
$ c_{16}^{1}$,
$ 1+\sqrt{2}$,
$ 1+\sqrt{2}$,
$ c^{1}_{16}
+c^{3}_{16}
$;\ \ 
$ 1$,
$ -c_{16}^{1}$,
$ -c_{16}^{1}$,
$ 1+\sqrt{2}$,
$ 1+\sqrt{2}$,
$ -c^{1}_{16}
-c^{3}_{16}
$;\ \ 
$ c^{1}_{16}
+c^{3}_{16}
$,
$ -c^{1}_{16}
-c^{3}_{16}
$,
$ c_{16}^{1}$,
$ -c_{16}^{1}$,
$0$;\ \ 
$ c^{1}_{16}
+c^{3}_{16}
$,
$ c_{16}^{1}$,
$ -c_{16}^{1}$,
$0$;\ \ 
$ -1$,
$ -1$,
$ -c^{1}_{16}
-c^{3}_{16}
$;\ \ 
$ -1$,
$ c^{1}_{16}
+c^{3}_{16}
$;\ \ 
$0$)

\vskip 0.7ex
\hangindent=3em \hangafter=1
$\tau_n$ = ($1.02 - 5.12 i$, $-0.74 + 3.69 i$, $7.01 - 10.49 i$, $13.66$, $10.49 - 7.01 i$, $-13.35 - 8.92 i$, $5.12 - 1.02 i$, $13.65 - 13.65 i$, $-5.12 - 1.02 i$, $-5.96 - 8.92 i$, $-10.49 - 7.01 i$, $13.65$, $-7.01 - 10.49 i$, $-18.57 + 3.69 i$, $-1.02 - 5.12 i$, $0.$, $-1.02 + 5.12 i$, $-18.57 - 3.69 i$, $-7.01 + 10.49 i$, $13.65$, $-10.49 + 7.01 i$, $-5.96 + 8.92 i$, $-5.12 + 1.02 i$, $13.65 + 13.65 i$, $5.12 + 1.02 i$, $-13.35 + 8.92 i$, $10.49 + 7.01 i$, $13.66$, $7.01 + 10.49 i$, $-0.74 - 3.69 i$, $1.02 + 5.12 i$, $27.31$)

\vskip 0.7ex
\hangindent=3em \hangafter=1
\textit{Intrinsic sign problem}

  \vskip 2ex

\noindent13. $7_{\frac{15}{4},27.31}^{32,272}$ \irep{97}:\ \ 
$d_i$ = ($1.0$,
$1.0$,
$1.847$,
$1.847$,
$2.414$,
$2.414$,
$2.613$) 

\vskip 0.7ex
\hangindent=3em \hangafter=1
$D^2= 27.313 = 
16+8\sqrt{2}$

\vskip 0.7ex
\hangindent=3em \hangafter=1
$T = ( 0,
\frac{1}{2},
\frac{21}{32},
\frac{21}{32},
\frac{1}{4},
\frac{3}{4},
\frac{9}{32} )
$,

\vskip 0.7ex
\hangindent=3em \hangafter=1
$S$ = ($ 1$,
$ 1$,
$ c_{16}^{1}$,
$ c_{16}^{1}$,
$ 1+\sqrt{2}$,
$ 1+\sqrt{2}$,
$ c^{1}_{16}
+c^{3}_{16}
$;\ \ 
$ 1$,
$ -c_{16}^{1}$,
$ -c_{16}^{1}$,
$ 1+\sqrt{2}$,
$ 1+\sqrt{2}$,
$ -c^{1}_{16}
-c^{3}_{16}
$;\ \ 
$ -c^{1}_{16}
-c^{3}_{16}
$,
$ c^{1}_{16}
+c^{3}_{16}
$,
$ -c_{16}^{1}$,
$ c_{16}^{1}$,
$0$;\ \ 
$ -c^{1}_{16}
-c^{3}_{16}
$,
$ -c_{16}^{1}$,
$ c_{16}^{1}$,
$0$;\ \ 
$ -1$,
$ -1$,
$ c^{1}_{16}
+c^{3}_{16}
$;\ \ 
$ -1$,
$ -c^{1}_{16}
-c^{3}_{16}
$;\ \ 
$0$)

\vskip 0.7ex
\hangindent=3em \hangafter=1
$\tau_n$ = ($-5.12 + 1.02 i$, $-18.57 + 3.69 i$, $10.49 - 7.01 i$, $13.66$, $-7.01 + 10.49 i$, $-5.96 - 8.92 i$, $1.02 - 5.12 i$, $13.65 + 13.65 i$, $-1.02 - 5.12 i$, $-13.35 - 8.92 i$, $7.01 + 10.49 i$, $13.65$, $-10.49 - 7.01 i$, $-0.74 + 3.69 i$, $5.12 + 1.02 i$, $0.$, $5.12 - 1.02 i$, $-0.74 - 3.69 i$, $-10.49 + 7.01 i$, $13.65$, $7.01 - 10.49 i$, $-13.35 + 8.92 i$, $-1.02 + 5.12 i$, $13.65 - 13.65 i$, $1.02 + 5.12 i$, $-5.96 + 8.92 i$, $-7.01 - 10.49 i$, $13.66$, $10.49 + 7.01 i$, $-18.57 - 3.69 i$, $-5.12 - 1.02 i$, $27.31$)

\vskip 0.7ex
\hangindent=3em \hangafter=1
\textit{Intrinsic sign problem}

  \vskip 2ex

\noindent14. $7_{\frac{29}{4},27.31}^{32,406}$ \irep{96}:\ \ 
$d_i$ = ($1.0$,
$1.0$,
$1.847$,
$1.847$,
$2.414$,
$2.414$,
$2.613$) 

\vskip 0.7ex
\hangindent=3em \hangafter=1
$D^2= 27.313 = 
16+8\sqrt{2}$

\vskip 0.7ex
\hangindent=3em \hangafter=1
$T = ( 0,
\frac{1}{2},
\frac{23}{32},
\frac{23}{32},
\frac{1}{4},
\frac{3}{4},
\frac{3}{32} )
$,

\vskip 0.7ex
\hangindent=3em \hangafter=1
$S$ = ($ 1$,
$ 1$,
$ c_{16}^{1}$,
$ c_{16}^{1}$,
$ 1+\sqrt{2}$,
$ 1+\sqrt{2}$,
$ c^{1}_{16}
+c^{3}_{16}
$;\ \ 
$ 1$,
$ -c_{16}^{1}$,
$ -c_{16}^{1}$,
$ 1+\sqrt{2}$,
$ 1+\sqrt{2}$,
$ -c^{1}_{16}
-c^{3}_{16}
$;\ \ 
$(-c^{1}_{16}
-c^{3}_{16}
)\mathrm{i}$,
$(c^{1}_{16}
+c^{3}_{16}
)\mathrm{i}$,
$ c_{16}^{1}$,
$ -c_{16}^{1}$,
$0$;\ \ 
$(-c^{1}_{16}
-c^{3}_{16}
)\mathrm{i}$,
$ c_{16}^{1}$,
$ -c_{16}^{1}$,
$0$;\ \ 
$ -1$,
$ -1$,
$ -c^{1}_{16}
-c^{3}_{16}
$;\ \ 
$ -1$,
$ c^{1}_{16}
+c^{3}_{16}
$;\ \ 
$0$)

\vskip 0.7ex
\hangindent=3em \hangafter=1
$\tau_n$ = ($4.35 - 2.9 i$, $-13.35 + 8.92 i$, $2.46 + 12.37 i$, $13.65$, $-12.37 - 2.46 i$, $-18.57 + 3.69 i$, $2.9 - 4.35 i$, $13.65 - 13.65 i$, $-2.9 - 4.35 i$, $-0.74 + 3.69 i$, $12.37 - 2.46 i$, $13.66$, $-2.46 + 12.37 i$, $-5.96 + 8.92 i$, $-4.35 - 2.9 i$, $0.$, $-4.35 + 2.9 i$, $-5.96 - 8.92 i$, $-2.46 - 12.37 i$, $13.66$, $12.37 + 2.46 i$, $-0.74 - 3.69 i$, $-2.9 + 4.35 i$, $13.65 + 13.65 i$, $2.9 + 4.35 i$, $-18.57 - 3.69 i$, $-12.37 + 2.46 i$, $13.65$, $2.46 - 12.37 i$, $-13.35 - 8.92 i$, $4.35 + 2.9 i$, $27.31$)

\vskip 0.7ex
\hangindent=3em \hangafter=1
\textit{Intrinsic sign problem}

  \vskip 2ex

\noindent15. $7_{\frac{19}{4},27.31}^{32,116}$ \irep{96}:\ \ 
$d_i$ = ($1.0$,
$1.0$,
$1.847$,
$1.847$,
$2.414$,
$2.414$,
$2.613$) 

\vskip 0.7ex
\hangindent=3em \hangafter=1
$D^2= 27.313 = 
16+8\sqrt{2}$

\vskip 0.7ex
\hangindent=3em \hangafter=1
$T = ( 0,
\frac{1}{2},
\frac{25}{32},
\frac{25}{32},
\frac{1}{4},
\frac{3}{4},
\frac{13}{32} )
$,

\vskip 0.7ex
\hangindent=3em \hangafter=1
$S$ = ($ 1$,
$ 1$,
$ c_{16}^{1}$,
$ c_{16}^{1}$,
$ 1+\sqrt{2}$,
$ 1+\sqrt{2}$,
$ c^{1}_{16}
+c^{3}_{16}
$;\ \ 
$ 1$,
$ -c_{16}^{1}$,
$ -c_{16}^{1}$,
$ 1+\sqrt{2}$,
$ 1+\sqrt{2}$,
$ -c^{1}_{16}
-c^{3}_{16}
$;\ \ 
$(c^{1}_{16}
+c^{3}_{16}
)\mathrm{i}$,
$(-c^{1}_{16}
-c^{3}_{16}
)\mathrm{i}$,
$ -c_{16}^{1}$,
$ c_{16}^{1}$,
$0$;\ \ 
$(c^{1}_{16}
+c^{3}_{16}
)\mathrm{i}$,
$ -c_{16}^{1}$,
$ c_{16}^{1}$,
$0$;\ \ 
$ -1$,
$ -1$,
$ c^{1}_{16}
+c^{3}_{16}
$;\ \ 
$ -1$,
$ -c^{1}_{16}
-c^{3}_{16}
$;\ \ 
$0$)

\vskip 0.7ex
\hangindent=3em \hangafter=1
$\tau_n$ = ($-4.35 - 2.9 i$, $-13.35 - 8.92 i$, $-2.46 + 12.37 i$, $13.65$, $12.37 - 2.46 i$, $-18.57 - 3.69 i$, $-2.9 - 4.35 i$, $13.65 + 13.65 i$, $2.9 - 4.35 i$, $-0.74 - 3.69 i$, $-12.37 - 2.46 i$, $13.66$, $2.46 + 12.37 i$, $-5.96 - 8.92 i$, $4.35 - 2.9 i$, $0.$, $4.35 + 2.9 i$, $-5.96 + 8.92 i$, $2.46 - 12.37 i$, $13.66$, $-12.37 + 2.46 i$, $-0.74 + 3.69 i$, $2.9 + 4.35 i$, $13.65 - 13.65 i$, $-2.9 + 4.35 i$, $-18.57 + 3.69 i$, $12.37 + 2.46 i$, $13.65$, $-2.46 - 12.37 i$, $-13.35 + 8.92 i$, $-4.35 + 2.9 i$, $27.31$)

\vskip 0.7ex
\hangindent=3em \hangafter=1
\textit{Intrinsic sign problem}

  \vskip 2ex

\noindent16. $7_{\frac{1}{4},27.31}^{32,123}$ \irep{97}:\ \ 
$d_i$ = ($1.0$,
$1.0$,
$1.847$,
$1.847$,
$2.414$,
$2.414$,
$2.613$) 

\vskip 0.7ex
\hangindent=3em \hangafter=1
$D^2= 27.313 = 
16+8\sqrt{2}$

\vskip 0.7ex
\hangindent=3em \hangafter=1
$T = ( 0,
\frac{1}{2},
\frac{27}{32},
\frac{27}{32},
\frac{1}{4},
\frac{3}{4},
\frac{7}{32} )
$,

\vskip 0.7ex
\hangindent=3em \hangafter=1
$S$ = ($ 1$,
$ 1$,
$ c_{16}^{1}$,
$ c_{16}^{1}$,
$ 1+\sqrt{2}$,
$ 1+\sqrt{2}$,
$ c^{1}_{16}
+c^{3}_{16}
$;\ \ 
$ 1$,
$ -c_{16}^{1}$,
$ -c_{16}^{1}$,
$ 1+\sqrt{2}$,
$ 1+\sqrt{2}$,
$ -c^{1}_{16}
-c^{3}_{16}
$;\ \ 
$ -c^{1}_{16}
-c^{3}_{16}
$,
$ c^{1}_{16}
+c^{3}_{16}
$,
$ c_{16}^{1}$,
$ -c_{16}^{1}$,
$0$;\ \ 
$ -c^{1}_{16}
-c^{3}_{16}
$,
$ c_{16}^{1}$,
$ -c_{16}^{1}$,
$0$;\ \ 
$ -1$,
$ -1$,
$ -c^{1}_{16}
-c^{3}_{16}
$;\ \ 
$ -1$,
$ c^{1}_{16}
+c^{3}_{16}
$;\ \ 
$0$)

\vskip 0.7ex
\hangindent=3em \hangafter=1
$\tau_n$ = ($5.12 + 1.02 i$, $-18.57 - 3.69 i$, $-10.49 - 7.01 i$, $13.66$, $7.01 + 10.49 i$, $-5.96 + 8.92 i$, $-1.02 - 5.12 i$, $13.65 - 13.65 i$, $1.02 - 5.12 i$, $-13.35 + 8.92 i$, $-7.01 + 10.49 i$, $13.65$, $10.49 - 7.01 i$, $-0.74 - 3.69 i$, $-5.12 + 1.02 i$, $0.$, $-5.12 - 1.02 i$, $-0.74 + 3.69 i$, $10.49 + 7.01 i$, $13.65$, $-7.01 - 10.49 i$, $-13.35 - 8.92 i$, $1.02 + 5.12 i$, $13.65 + 13.65 i$, $-1.02 + 5.12 i$, $-5.96 - 8.92 i$, $7.01 - 10.49 i$, $13.66$, $-10.49 + 7.01 i$, $-18.57 + 3.69 i$, $5.12 - 1.02 i$, $27.31$)

\vskip 0.7ex
\hangindent=3em \hangafter=1
\textit{Intrinsic sign problem}

  \vskip 2ex

\noindent17. $7_{\frac{23}{4},27.31}^{32,224}$ \irep{97}:\ \ 
$d_i$ = ($1.0$,
$1.0$,
$1.847$,
$1.847$,
$2.414$,
$2.414$,
$2.613$) 

\vskip 0.7ex
\hangindent=3em \hangafter=1
$D^2= 27.313 = 
16+8\sqrt{2}$

\vskip 0.7ex
\hangindent=3em \hangafter=1
$T = ( 0,
\frac{1}{2},
\frac{29}{32},
\frac{29}{32},
\frac{1}{4},
\frac{3}{4},
\frac{17}{32} )
$,

\vskip 0.7ex
\hangindent=3em \hangafter=1
$S$ = ($ 1$,
$ 1$,
$ c_{16}^{1}$,
$ c_{16}^{1}$,
$ 1+\sqrt{2}$,
$ 1+\sqrt{2}$,
$ c^{1}_{16}
+c^{3}_{16}
$;\ \ 
$ 1$,
$ -c_{16}^{1}$,
$ -c_{16}^{1}$,
$ 1+\sqrt{2}$,
$ 1+\sqrt{2}$,
$ -c^{1}_{16}
-c^{3}_{16}
$;\ \ 
$ c^{1}_{16}
+c^{3}_{16}
$,
$ -c^{1}_{16}
-c^{3}_{16}
$,
$ -c_{16}^{1}$,
$ c_{16}^{1}$,
$0$;\ \ 
$ c^{1}_{16}
+c^{3}_{16}
$,
$ -c_{16}^{1}$,
$ c_{16}^{1}$,
$0$;\ \ 
$ -1$,
$ -1$,
$ c^{1}_{16}
+c^{3}_{16}
$;\ \ 
$ -1$,
$ -c^{1}_{16}
-c^{3}_{16}
$;\ \ 
$0$)

\vskip 0.7ex
\hangindent=3em \hangafter=1
$\tau_n$ = ($-1.02 - 5.12 i$, $-0.74 - 3.69 i$, $-7.01 - 10.49 i$, $13.66$, $-10.49 - 7.01 i$, $-13.35 + 8.92 i$, $-5.12 - 1.02 i$, $13.65 + 13.65 i$, $5.12 - 1.02 i$, $-5.96 + 8.92 i$, $10.49 - 7.01 i$, $13.65$, $7.01 - 10.49 i$, $-18.57 - 3.69 i$, $1.02 - 5.12 i$, $0.$, $1.02 + 5.12 i$, $-18.57 + 3.69 i$, $7.01 + 10.49 i$, $13.65$, $10.49 + 7.01 i$, $-5.96 - 8.92 i$, $5.12 + 1.02 i$, $13.65 - 13.65 i$, $-5.12 + 1.02 i$, $-13.35 - 8.92 i$, $-10.49 + 7.01 i$, $13.66$, $-7.01 + 10.49 i$, $-0.74 + 3.69 i$, $-1.02 + 5.12 i$, $27.31$)

\vskip 0.7ex
\hangindent=3em \hangafter=1
\textit{Intrinsic sign problem}

  \vskip 2ex

\noindent18. $7_{\frac{5}{4},27.31}^{32,225}$ \irep{96}:\ \ 
$d_i$ = ($1.0$,
$1.0$,
$1.847$,
$1.847$,
$2.414$,
$2.414$,
$2.613$) 

\vskip 0.7ex
\hangindent=3em \hangafter=1
$D^2= 27.313 = 
16+8\sqrt{2}$

\vskip 0.7ex
\hangindent=3em \hangafter=1
$T = ( 0,
\frac{1}{2},
\frac{31}{32},
\frac{31}{32},
\frac{1}{4},
\frac{3}{4},
\frac{11}{32} )
$,

\vskip 0.7ex
\hangindent=3em \hangafter=1
$S$ = ($ 1$,
$ 1$,
$ c_{16}^{1}$,
$ c_{16}^{1}$,
$ 1+\sqrt{2}$,
$ 1+\sqrt{2}$,
$ c^{1}_{16}
+c^{3}_{16}
$;\ \ 
$ 1$,
$ -c_{16}^{1}$,
$ -c_{16}^{1}$,
$ 1+\sqrt{2}$,
$ 1+\sqrt{2}$,
$ -c^{1}_{16}
-c^{3}_{16}
$;\ \ 
$(c^{1}_{16}
+c^{3}_{16}
)\mathrm{i}$,
$(-c^{1}_{16}
-c^{3}_{16}
)\mathrm{i}$,
$ c_{16}^{1}$,
$ -c_{16}^{1}$,
$0$;\ \ 
$(c^{1}_{16}
+c^{3}_{16}
)\mathrm{i}$,
$ c_{16}^{1}$,
$ -c_{16}^{1}$,
$0$;\ \ 
$ -1$,
$ -1$,
$ -c^{1}_{16}
-c^{3}_{16}
$;\ \ 
$ -1$,
$ c^{1}_{16}
+c^{3}_{16}
$;\ \ 
$0$)

\vskip 0.7ex
\hangindent=3em \hangafter=1
$\tau_n$ = ($2.9 + 4.35 i$, $-5.96 - 8.92 i$, $12.37 - 2.46 i$, $13.65$, $2.46 - 12.37 i$, $-0.74 - 3.69 i$, $-4.35 - 2.9 i$, $13.65 - 13.65 i$, $4.35 - 2.9 i$, $-18.57 - 3.69 i$, $-2.46 - 12.37 i$, $13.66$, $-12.37 - 2.46 i$, $-13.35 - 8.92 i$, $-2.9 + 4.35 i$, $0.$, $-2.9 - 4.35 i$, $-13.35 + 8.92 i$, $-12.37 + 2.46 i$, $13.66$, $-2.46 + 12.37 i$, $-18.57 + 3.69 i$, $4.35 + 2.9 i$, $13.65 + 13.65 i$, $-4.35 + 2.9 i$, $-0.74 + 3.69 i$, $2.46 + 12.37 i$, $13.65$, $12.37 + 2.46 i$, $-5.96 + 8.92 i$, $2.9 - 4.35 i$, $27.31$)

\vskip 0.7ex
\hangindent=3em \hangafter=1
\textit{Intrinsic sign problem}

  \vskip 2ex

\noindent19. $7_{2,28.}^{56,139}$ \irep{99}:\ \ 
$d_i$ = ($1.0$,
$1.0$,
$2.0$,
$2.0$,
$2.0$,
$2.645$,
$2.645$) 

\vskip 0.7ex
\hangindent=3em \hangafter=1
$D^2= 28.0 = 
28$

\vskip 0.7ex
\hangindent=3em \hangafter=1
$T = ( 0,
0,
\frac{1}{7},
\frac{2}{7},
\frac{4}{7},
\frac{1}{8},
\frac{5}{8} )
$,

\vskip 0.7ex
\hangindent=3em \hangafter=1
$S$ = ($ 1$,
$ 1$,
$ 2$,
$ 2$,
$ 2$,
$ \sqrt{7}$,
$ \sqrt{7}$;\ \ 
$ 1$,
$ 2$,
$ 2$,
$ 2$,
$ -\sqrt{7}$,
$ -\sqrt{7}$;\ \ 
$ 2c_{7}^{2}$,
$ 2c_{7}^{1}$,
$ 2c_{7}^{3}$,
$0$,
$0$;\ \ 
$ 2c_{7}^{3}$,
$ 2c_{7}^{2}$,
$0$,
$0$;\ \ 
$ 2c_{7}^{1}$,
$0$,
$0$;\ \ 
$ \sqrt{7}$,
$ -\sqrt{7}$;\ \ 
$ \sqrt{7}$)

\vskip 0.7ex
\hangindent=3em \hangafter=1
$\tau_n$ = ($0. + 5.29 i$, $0. + 19.28 i$, $0. - 5.29 i$, $-13.99 + 5.29 i$, $0. - 5.29 i$, $0. - 19.28 i$, $14.$, $13.99 + 5.29 i$, $0. + 5.29 i$, $0. + 8.7 i$, $0. + 5.29 i$, $-13.99 - 5.29 i$, $0. - 5.29 i$, $14. - 13.99 i$, $0. + 5.29 i$, $13.99 + 5.29 i$, $0. - 5.29 i$, $0. + 19.28 i$, $0. - 5.29 i$, $-13.99 - 5.29 i$, $14.$, $0. - 8.7 i$, $0. + 5.29 i$, $13.99 - 5.29 i$, $0. + 5.29 i$, $0. + 8.7 i$, $0. - 5.29 i$, $0.01$, $0. + 5.29 i$, $0. - 8.7 i$, $0. - 5.29 i$, $13.99 + 5.29 i$, $0. - 5.29 i$, $0. + 8.7 i$, $14.$, $-13.99 + 5.29 i$, $0. + 5.29 i$, $0. - 19.28 i$, $0. + 5.29 i$, $13.99 - 5.29 i$, $0. - 5.29 i$, $14. + 13.99 i$, $0. + 5.29 i$, $-13.99 + 5.29 i$, $0. - 5.29 i$, $0. - 8.7 i$, $0. - 5.29 i$, $13.99 - 5.29 i$, $14.$, $0. + 19.28 i$, $0. + 5.29 i$, $-13.99 - 5.29 i$, $0. + 5.29 i$, $0. - 19.28 i$, $0. - 5.29 i$, $27.99$)

\vskip 0.7ex
\hangindent=3em \hangafter=1
\textit{Intrinsic sign problem}

  \vskip 2ex

\noindent20. $7_{2,28.}^{56,680}$ \irep{99}:\ \ 
$d_i$ = ($1.0$,
$1.0$,
$2.0$,
$2.0$,
$2.0$,
$2.645$,
$2.645$) 

\vskip 0.7ex
\hangindent=3em \hangafter=1
$D^2= 28.0 = 
28$

\vskip 0.7ex
\hangindent=3em \hangafter=1
$T = ( 0,
0,
\frac{1}{7},
\frac{2}{7},
\frac{4}{7},
\frac{3}{8},
\frac{7}{8} )
$,

\vskip 0.7ex
\hangindent=3em \hangafter=1
$S$ = ($ 1$,
$ 1$,
$ 2$,
$ 2$,
$ 2$,
$ \sqrt{7}$,
$ \sqrt{7}$;\ \ 
$ 1$,
$ 2$,
$ 2$,
$ 2$,
$ -\sqrt{7}$,
$ -\sqrt{7}$;\ \ 
$ 2c_{7}^{2}$,
$ 2c_{7}^{1}$,
$ 2c_{7}^{3}$,
$0$,
$0$;\ \ 
$ 2c_{7}^{3}$,
$ 2c_{7}^{2}$,
$0$,
$0$;\ \ 
$ 2c_{7}^{1}$,
$0$,
$0$;\ \ 
$ -\sqrt{7}$,
$ \sqrt{7}$;\ \ 
$ -\sqrt{7}$)

\vskip 0.7ex
\hangindent=3em \hangafter=1
$\tau_n$ = ($0. + 5.29 i$, $0. - 8.7 i$, $0. - 5.29 i$, $-13.99 + 5.29 i$, $0. - 5.29 i$, $0. + 8.7 i$, $14.$, $13.99 + 5.29 i$, $0. + 5.29 i$, $0. - 19.28 i$, $0. + 5.29 i$, $-13.99 - 5.29 i$, $0. - 5.29 i$, $14. + 13.99 i$, $0. + 5.29 i$, $13.99 + 5.29 i$, $0. - 5.29 i$, $0. - 8.7 i$, $0. - 5.29 i$, $-13.99 - 5.29 i$, $14.$, $0. + 19.28 i$, $0. + 5.29 i$, $13.99 - 5.29 i$, $0. + 5.29 i$, $0. - 19.28 i$, $0. - 5.29 i$, $0.01$, $0. + 5.29 i$, $0. + 19.28 i$, $0. - 5.29 i$, $13.99 + 5.29 i$, $0. - 5.29 i$, $0. - 19.28 i$, $14.$, $-13.99 + 5.29 i$, $0. + 5.29 i$, $0. + 8.7 i$, $0. + 5.29 i$, $13.99 - 5.29 i$, $0. - 5.29 i$, $14. - 13.99 i$, $0. + 5.29 i$, $-13.99 + 5.29 i$, $0. - 5.29 i$, $0. + 19.28 i$, $0. - 5.29 i$, $13.99 - 5.29 i$, $14.$, $0. - 8.7 i$, $0. + 5.29 i$, $-13.99 - 5.29 i$, $0. + 5.29 i$, $0. + 8.7 i$, $0. - 5.29 i$, $27.99$)

\vskip 0.7ex
\hangindent=3em \hangafter=1
\textit{Intrinsic sign problem}

  \vskip 2ex

\noindent21. $7_{6,28.}^{56,609}$ \irep{99}:\ \ 
$d_i$ = ($1.0$,
$1.0$,
$2.0$,
$2.0$,
$2.0$,
$2.645$,
$2.645$) 

\vskip 0.7ex
\hangindent=3em \hangafter=1
$D^2= 28.0 = 
28$

\vskip 0.7ex
\hangindent=3em \hangafter=1
$T = ( 0,
0,
\frac{3}{7},
\frac{5}{7},
\frac{6}{7},
\frac{1}{8},
\frac{5}{8} )
$,

\vskip 0.7ex
\hangindent=3em \hangafter=1
$S$ = ($ 1$,
$ 1$,
$ 2$,
$ 2$,
$ 2$,
$ \sqrt{7}$,
$ \sqrt{7}$;\ \ 
$ 1$,
$ 2$,
$ 2$,
$ 2$,
$ -\sqrt{7}$,
$ -\sqrt{7}$;\ \ 
$ 2c_{7}^{1}$,
$ 2c_{7}^{2}$,
$ 2c_{7}^{3}$,
$0$,
$0$;\ \ 
$ 2c_{7}^{3}$,
$ 2c_{7}^{1}$,
$0$,
$0$;\ \ 
$ 2c_{7}^{2}$,
$0$,
$0$;\ \ 
$ -\sqrt{7}$,
$ \sqrt{7}$;\ \ 
$ -\sqrt{7}$)

\vskip 0.7ex
\hangindent=3em \hangafter=1
$\tau_n$ = ($0. - 5.29 i$, $0. + 8.7 i$, $0. + 5.29 i$, $-13.99 - 5.29 i$, $0. + 5.29 i$, $0. - 8.7 i$, $14.$, $13.99 - 5.29 i$, $0. - 5.29 i$, $0. + 19.28 i$, $0. - 5.29 i$, $-13.99 + 5.29 i$, $0. + 5.29 i$, $14. - 13.99 i$, $0. - 5.29 i$, $13.99 - 5.29 i$, $0. + 5.29 i$, $0. + 8.7 i$, $0. + 5.29 i$, $-13.99 + 5.29 i$, $14.$, $0. - 19.28 i$, $0. - 5.29 i$, $13.99 + 5.29 i$, $0. - 5.29 i$, $0. + 19.28 i$, $0. + 5.29 i$, $0.01$, $0. - 5.29 i$, $0. - 19.28 i$, $0. + 5.29 i$, $13.99 - 5.29 i$, $0. + 5.29 i$, $0. + 19.28 i$, $14.$, $-13.99 - 5.29 i$, $0. - 5.29 i$, $0. - 8.7 i$, $0. - 5.29 i$, $13.99 + 5.29 i$, $0. + 5.29 i$, $14. + 13.99 i$, $0. - 5.29 i$, $-13.99 - 5.29 i$, $0. + 5.29 i$, $0. - 19.28 i$, $0. + 5.29 i$, $13.99 + 5.29 i$, $14.$, $0. + 8.7 i$, $0. - 5.29 i$, $-13.99 + 5.29 i$, $0. - 5.29 i$, $0. - 8.7 i$, $0. + 5.29 i$, $27.99$)

\vskip 0.7ex
\hangindent=3em \hangafter=1
\textit{Intrinsic sign problem}

  \vskip 2ex

\noindent22. $7_{6,28.}^{56,193}$ \irep{99}:\ \ 
$d_i$ = ($1.0$,
$1.0$,
$2.0$,
$2.0$,
$2.0$,
$2.645$,
$2.645$) 

\vskip 0.7ex
\hangindent=3em \hangafter=1
$D^2= 28.0 = 
28$

\vskip 0.7ex
\hangindent=3em \hangafter=1
$T = ( 0,
0,
\frac{3}{7},
\frac{5}{7},
\frac{6}{7},
\frac{3}{8},
\frac{7}{8} )
$,

\vskip 0.7ex
\hangindent=3em \hangafter=1
$S$ = ($ 1$,
$ 1$,
$ 2$,
$ 2$,
$ 2$,
$ \sqrt{7}$,
$ \sqrt{7}$;\ \ 
$ 1$,
$ 2$,
$ 2$,
$ 2$,
$ -\sqrt{7}$,
$ -\sqrt{7}$;\ \ 
$ 2c_{7}^{1}$,
$ 2c_{7}^{2}$,
$ 2c_{7}^{3}$,
$0$,
$0$;\ \ 
$ 2c_{7}^{3}$,
$ 2c_{7}^{1}$,
$0$,
$0$;\ \ 
$ 2c_{7}^{2}$,
$0$,
$0$;\ \ 
$ \sqrt{7}$,
$ -\sqrt{7}$;\ \ 
$ \sqrt{7}$)

\vskip 0.7ex
\hangindent=3em \hangafter=1
$\tau_n$ = ($0. - 5.29 i$, $0. - 19.28 i$, $0. + 5.29 i$, $-13.99 - 5.29 i$, $0. + 5.29 i$, $0. + 19.28 i$, $14.$, $13.99 - 5.29 i$, $0. - 5.29 i$, $0. - 8.7 i$, $0. - 5.29 i$, $-13.99 + 5.29 i$, $0. + 5.29 i$, $14. + 13.99 i$, $0. - 5.29 i$, $13.99 - 5.29 i$, $0. + 5.29 i$, $0. - 19.28 i$, $0. + 5.29 i$, $-13.99 + 5.29 i$, $14.$, $0. + 8.7 i$, $0. - 5.29 i$, $13.99 + 5.29 i$, $0. - 5.29 i$, $0. - 8.7 i$, $0. + 5.29 i$, $0.01$, $0. - 5.29 i$, $0. + 8.7 i$, $0. + 5.29 i$, $13.99 - 5.29 i$, $0. + 5.29 i$, $0. - 8.7 i$, $14.$, $-13.99 - 5.29 i$, $0. - 5.29 i$, $0. + 19.28 i$, $0. - 5.29 i$, $13.99 + 5.29 i$, $0. + 5.29 i$, $14. - 13.99 i$, $0. - 5.29 i$, $-13.99 - 5.29 i$, $0. + 5.29 i$, $0. + 8.7 i$, $0. + 5.29 i$, $13.99 + 5.29 i$, $14.$, $0. - 19.28 i$, $0. - 5.29 i$, $-13.99 + 5.29 i$, $0. - 5.29 i$, $0. + 19.28 i$, $0. + 5.29 i$, $27.99$)

\vskip 0.7ex
\hangindent=3em \hangafter=1
\textit{Intrinsic sign problem}

  \vskip 2ex

\noindent23. $7_{\frac{32}{5},86.75}^{15,205}$ \irep{80}:\ \ 
$d_i$ = ($1.0$,
$1.956$,
$2.827$,
$3.574$,
$4.165$,
$4.574$,
$4.783$) 

\vskip 0.7ex
\hangindent=3em \hangafter=1
$D^2= 86.750 = 
30+15c^{1}_{15}
+15c^{2}_{15}
+15c^{3}_{15}
$

\vskip 0.7ex
\hangindent=3em \hangafter=1
$T = ( 0,
\frac{1}{5},
\frac{13}{15},
0,
\frac{3}{5},
\frac{2}{3},
\frac{1}{5} )
$,

\vskip 0.7ex
\hangindent=3em \hangafter=1
$S$ = ($ 1$,
$ -c_{15}^{7}$,
$ \xi_{15}^{3}$,
$ \xi_{15}^{11}$,
$ \xi_{15}^{5}$,
$ \xi_{15}^{9}$,
$ \xi_{15}^{7}$;\ \ 
$ -\xi_{15}^{11}$,
$ \xi_{15}^{9}$,
$ -\xi_{15}^{7}$,
$ \xi_{15}^{5}$,
$ -\xi_{15}^{3}$,
$ 1$;\ \ 
$ \xi_{15}^{9}$,
$ \xi_{15}^{3}$,
$0$,
$ -\xi_{15}^{3}$,
$ -\xi_{15}^{9}$;\ \ 
$ 1$,
$ -\xi_{15}^{5}$,
$ \xi_{15}^{9}$,
$ c_{15}^{7}$;\ \ 
$ -\xi_{15}^{5}$,
$0$,
$ \xi_{15}^{5}$;\ \ 
$ -\xi_{15}^{9}$,
$ \xi_{15}^{3}$;\ \ 
$ -\xi_{15}^{11}$)

\vskip 0.7ex
\hangindent=3em \hangafter=1
$\tau_n$ = ($2.88 - 8.86 i$, $-13.77 + 42.36 i$, $11.99 - 36.89 i$, $-10.29 - 31.66 i$, $43.37 + 25.04 i$, $31.38 + 22.8 i$, $-5.63 + 17.33 i$, $-5.63 - 17.33 i$, $31.38 - 22.8 i$, $43.37 - 25.04 i$, $-10.29 + 31.66 i$, $11.99 + 36.89 i$, $-13.77 - 42.36 i$, $2.88 + 8.86 i$, $86.74$)
$\tau_n$ = ()

\vskip 0.7ex
\hangindent=3em \hangafter=1
\textit{Intrinsic sign problem}

  \vskip 2ex

\noindent24. $7_{\frac{8}{5},86.75}^{15,181}$ \irep{80}:\ \ 
$d_i$ = ($1.0$,
$1.956$,
$2.827$,
$3.574$,
$4.165$,
$4.574$,
$4.783$) 

\vskip 0.7ex
\hangindent=3em \hangafter=1
$D^2= 86.750 = 
30+15c^{1}_{15}
+15c^{2}_{15}
+15c^{3}_{15}
$

\vskip 0.7ex
\hangindent=3em \hangafter=1
$T = ( 0,
\frac{4}{5},
\frac{2}{15},
0,
\frac{2}{5},
\frac{1}{3},
\frac{4}{5} )
$,

\vskip 0.7ex
\hangindent=3em \hangafter=1
$S$ = ($ 1$,
$ -c_{15}^{7}$,
$ \xi_{15}^{3}$,
$ \xi_{15}^{11}$,
$ \xi_{15}^{5}$,
$ \xi_{15}^{9}$,
$ \xi_{15}^{7}$;\ \ 
$ -\xi_{15}^{11}$,
$ \xi_{15}^{9}$,
$ -\xi_{15}^{7}$,
$ \xi_{15}^{5}$,
$ -\xi_{15}^{3}$,
$ 1$;\ \ 
$ \xi_{15}^{9}$,
$ \xi_{15}^{3}$,
$0$,
$ -\xi_{15}^{3}$,
$ -\xi_{15}^{9}$;\ \ 
$ 1$,
$ -\xi_{15}^{5}$,
$ \xi_{15}^{9}$,
$ c_{15}^{7}$;\ \ 
$ -\xi_{15}^{5}$,
$0$,
$ \xi_{15}^{5}$;\ \ 
$ -\xi_{15}^{9}$,
$ \xi_{15}^{3}$;\ \ 
$ -\xi_{15}^{11}$)

\vskip 0.7ex
\hangindent=3em \hangafter=1
$\tau_n$ = ($2.88 + 8.86 i$, $-13.77 - 42.36 i$, $11.99 + 36.89 i$, $-10.29 + 31.66 i$, $43.37 - 25.04 i$, $31.38 - 22.8 i$, $-5.63 - 17.33 i$, $-5.63 + 17.33 i$, $31.38 + 22.8 i$, $43.37 + 25.04 i$, $-10.29 - 31.66 i$, $11.99 - 36.89 i$, $-13.77 + 42.36 i$, $2.88 - 8.86 i$, $86.74$)

\vskip 0.7ex
\hangindent=3em \hangafter=1
\textit{Intrinsic sign problem}

  \vskip 2ex

\noindent25. $7_{1,93.25}^{8,230}$ \irep{54}:\ \ 
$d_i$ = ($1.0$,
$2.414$,
$2.414$,
$3.414$,
$3.414$,
$4.828$,
$5.828$) 

\vskip 0.7ex
\hangindent=3em \hangafter=1
$D^2= 93.254 = 
48+32\sqrt{2}$

\vskip 0.7ex
\hangindent=3em \hangafter=1
$T = ( 0,
\frac{1}{2},
\frac{1}{2},
\frac{1}{4},
\frac{1}{4},
\frac{5}{8},
0 )
$,

\vskip 0.7ex
\hangindent=3em \hangafter=1
$S$ = ($ 1$,
$ 1+\sqrt{2}$,
$ 1+\sqrt{2}$,
$ 2+\sqrt{2}$,
$ 2+\sqrt{2}$,
$ 2+2\sqrt{2}$,
$ 3+2\sqrt{2}$;\ \ 
$ -1-2  \zeta^{1}_{8}
-2  \zeta^{2}_{8}
$,
$ -1-2  \zeta^{-1}_{8}
+2\zeta^{2}_{8}
$,
$(-2-\sqrt{2})\mathrm{i}$,
$(2+\sqrt{2})\mathrm{i}$,
$ 2+2\sqrt{2}$,
$ -1-\sqrt{2}$;\ \ 
$ -1-2  \zeta^{1}_{8}
-2  \zeta^{2}_{8}
$,
$(2+\sqrt{2})\mathrm{i}$,
$(-2-\sqrt{2})\mathrm{i}$,
$ 2+2\sqrt{2}$,
$ -1-\sqrt{2}$;\ \ 
$ (2+2\sqrt{2})\zeta_{8}^{3}$,
$ (-2-2\sqrt{2})\zeta_{8}^{1}$,
$0$,
$ 2+\sqrt{2}$;\ \ 
$ (2+2\sqrt{2})\zeta_{8}^{3}$,
$0$,
$ 2+\sqrt{2}$;\ \ 
$0$,
$ -2-2\sqrt{2}$;\ \ 
$ 1$)

\vskip 0.7ex
\hangindent=3em \hangafter=1
$\tau_n$ = ($6.83 + 6.83 i$, $23.31 + 23.31 i$, $39.79 - 39.79 i$, $46.62$, $39.79 + 39.79 i$, $23.31 - 23.31 i$, $6.83 - 6.83 i$, $93.24$)

\vskip 0.7ex
\hangindent=3em \hangafter=1
\textit{Intrinsic sign problem}

  \vskip 2ex

\noindent26. $7_{7,93.25}^{8,101}$ \irep{54}:\ \ 
$d_i$ = ($1.0$,
$2.414$,
$2.414$,
$3.414$,
$3.414$,
$4.828$,
$5.828$) 

\vskip 0.7ex
\hangindent=3em \hangafter=1
$D^2= 93.254 = 
48+32\sqrt{2}$

\vskip 0.7ex
\hangindent=3em \hangafter=1
$T = ( 0,
\frac{1}{2},
\frac{1}{2},
\frac{3}{4},
\frac{3}{4},
\frac{3}{8},
0 )
$,

\vskip 0.7ex
\hangindent=3em \hangafter=1
$S$ = ($ 1$,
$ 1+\sqrt{2}$,
$ 1+\sqrt{2}$,
$ 2+\sqrt{2}$,
$ 2+\sqrt{2}$,
$ 2+2\sqrt{2}$,
$ 3+2\sqrt{2}$;\ \ 
$ -1-2  \zeta^{-1}_{8}
+2\zeta^{2}_{8}
$,
$ -1-2  \zeta^{1}_{8}
-2  \zeta^{2}_{8}
$,
$(-2-\sqrt{2})\mathrm{i}$,
$(2+\sqrt{2})\mathrm{i}$,
$ 2+2\sqrt{2}$,
$ -1-\sqrt{2}$;\ \ 
$ -1-2  \zeta^{-1}_{8}
+2\zeta^{2}_{8}
$,
$(2+\sqrt{2})\mathrm{i}$,
$(-2-\sqrt{2})\mathrm{i}$,
$ 2+2\sqrt{2}$,
$ -1-\sqrt{2}$;\ \ 
$ (-2-2\sqrt{2})\zeta_{8}^{1}$,
$ (2+2\sqrt{2})\zeta_{8}^{3}$,
$0$,
$ 2+\sqrt{2}$;\ \ 
$ (-2-2\sqrt{2})\zeta_{8}^{1}$,
$0$,
$ 2+\sqrt{2}$;\ \ 
$0$,
$ -2-2\sqrt{2}$;\ \ 
$ 1$)

\vskip 0.7ex
\hangindent=3em \hangafter=1
$\tau_n$ = ($6.83 - 6.83 i$, $23.31 - 23.31 i$, $39.79 + 39.79 i$, $46.62$, $39.79 - 39.79 i$, $23.31 + 23.31 i$, $6.83 + 6.83 i$, $93.24$)

\vskip 0.7ex
\hangindent=3em \hangafter=1
\textit{Intrinsic sign problem}

  \vskip 2ex

\noindent27. $7_{\frac{30}{11},135.7}^{11,157}$ \irep{66}:\ \ 
$d_i$ = ($1.0$,
$2.918$,
$3.513$,
$3.513$,
$4.601$,
$5.911$,
$6.742$) 

\vskip 0.7ex
\hangindent=3em \hangafter=1
$D^2= 135.778 = 
55+44c^{1}_{11}
+33c^{2}_{11}
+22c^{3}_{11}
+11c^{4}_{11}
$

\vskip 0.7ex
\hangindent=3em \hangafter=1
$T = ( 0,
\frac{1}{11},
\frac{4}{11},
\frac{4}{11},
\frac{3}{11},
\frac{6}{11},
\frac{10}{11} )
$,

\vskip 0.7ex
\hangindent=3em \hangafter=1
$S$ = ($ 1$,
$ 2+c^{1}_{11}
+c^{2}_{11}
+c^{3}_{11}
+c^{4}_{11}
$,
$ \xi_{11}^{5}$,
$ \xi_{11}^{5}$,
$ 2+2c^{1}_{11}
+c^{2}_{11}
+c^{3}_{11}
+c^{4}_{11}
$,
$ 2+2c^{1}_{11}
+c^{2}_{11}
+c^{3}_{11}
$,
$ 2+2c^{1}_{11}
+2c^{2}_{11}
+c^{3}_{11}
$;\ \ 
$ 2+2c^{1}_{11}
+2c^{2}_{11}
+c^{3}_{11}
$,
$ -\xi_{11}^{5}$,
$ -\xi_{11}^{5}$,
$ 2+2c^{1}_{11}
+c^{2}_{11}
+c^{3}_{11}
$,
$ 1$,
$ -2-2  c^{1}_{11}
-c^{2}_{11}
-c^{3}_{11}
-c^{4}_{11}
$;\ \ 
$ s^{2}_{11}
+2\zeta^{3}_{11}
-\zeta^{-3}_{11}
+\zeta^{4}_{11}
+\zeta^{5}_{11}
$,
$ -1-c^{1}_{11}
-2  \zeta^{2}_{11}
-2  \zeta^{3}_{11}
+\zeta^{-3}_{11}
-\zeta^{4}_{11}
-\zeta^{5}_{11}
$,
$ \xi_{11}^{5}$,
$ -\xi_{11}^{5}$,
$ \xi_{11}^{5}$;\ \ 
$ s^{2}_{11}
+2\zeta^{3}_{11}
-\zeta^{-3}_{11}
+\zeta^{4}_{11}
+\zeta^{5}_{11}
$,
$ \xi_{11}^{5}$,
$ -\xi_{11}^{5}$,
$ \xi_{11}^{5}$;\ \ 
$ -2-c^{1}_{11}
-c^{2}_{11}
-c^{3}_{11}
-c^{4}_{11}
$,
$ -2-2  c^{1}_{11}
-2  c^{2}_{11}
-c^{3}_{11}
$,
$ 1$;\ \ 
$ 2+2c^{1}_{11}
+c^{2}_{11}
+c^{3}_{11}
+c^{4}_{11}
$,
$ 2+c^{1}_{11}
+c^{2}_{11}
+c^{3}_{11}
+c^{4}_{11}
$;\ \ 
$ -2-2  c^{1}_{11}
-c^{2}_{11}
-c^{3}_{11}
$)

\vskip 0.7ex
\hangindent=3em \hangafter=1
$\tau_n$ = ($-6.3 + 9.79 i$, $28.99 - 45.11 i$, $0. - 68.88 i$, $-25.7 + 22.26 i$, $-59.37 - 51.44 i$, $-59.37 + 51.44 i$, $-25.7 - 22.26 i$, $0. + 68.88 i$, $28.99 + 45.11 i$, $-6.3 - 9.79 i$, $135.76$)

\vskip 0.7ex
\hangindent=3em \hangafter=1
\textit{Intrinsic sign problem}

  \vskip 2ex

\noindent28. $7_{\frac{58}{11},135.7}^{11,191}$ \irep{66}:\ \ 
$d_i$ = ($1.0$,
$2.918$,
$3.513$,
$3.513$,
$4.601$,
$5.911$,
$6.742$) 

\vskip 0.7ex
\hangindent=3em \hangafter=1
$D^2= 135.778 = 
55+44c^{1}_{11}
+33c^{2}_{11}
+22c^{3}_{11}
+11c^{4}_{11}
$

\vskip 0.7ex
\hangindent=3em \hangafter=1
$T = ( 0,
\frac{10}{11},
\frac{7}{11},
\frac{7}{11},
\frac{8}{11},
\frac{5}{11},
\frac{1}{11} )
$,

\vskip 0.7ex
\hangindent=3em \hangafter=1
$S$ = ($ 1$,
$ 2+c^{1}_{11}
+c^{2}_{11}
+c^{3}_{11}
+c^{4}_{11}
$,
$ \xi_{11}^{5}$,
$ \xi_{11}^{5}$,
$ 2+2c^{1}_{11}
+c^{2}_{11}
+c^{3}_{11}
+c^{4}_{11}
$,
$ 2+2c^{1}_{11}
+c^{2}_{11}
+c^{3}_{11}
$,
$ 2+2c^{1}_{11}
+2c^{2}_{11}
+c^{3}_{11}
$;\ \ 
$ 2+2c^{1}_{11}
+2c^{2}_{11}
+c^{3}_{11}
$,
$ -\xi_{11}^{5}$,
$ -\xi_{11}^{5}$,
$ 2+2c^{1}_{11}
+c^{2}_{11}
+c^{3}_{11}
$,
$ 1$,
$ -2-2  c^{1}_{11}
-c^{2}_{11}
-c^{3}_{11}
-c^{4}_{11}
$;\ \ 
$ -1-c^{1}_{11}
-2  \zeta^{2}_{11}
-2  \zeta^{3}_{11}
+\zeta^{-3}_{11}
-\zeta^{4}_{11}
-\zeta^{5}_{11}
$,
$ s^{2}_{11}
+2\zeta^{3}_{11}
-\zeta^{-3}_{11}
+\zeta^{4}_{11}
+\zeta^{5}_{11}
$,
$ \xi_{11}^{5}$,
$ -\xi_{11}^{5}$,
$ \xi_{11}^{5}$;\ \ 
$ -1-c^{1}_{11}
-2  \zeta^{2}_{11}
-2  \zeta^{3}_{11}
+\zeta^{-3}_{11}
-\zeta^{4}_{11}
-\zeta^{5}_{11}
$,
$ \xi_{11}^{5}$,
$ -\xi_{11}^{5}$,
$ \xi_{11}^{5}$;\ \ 
$ -2-c^{1}_{11}
-c^{2}_{11}
-c^{3}_{11}
-c^{4}_{11}
$,
$ -2-2  c^{1}_{11}
-2  c^{2}_{11}
-c^{3}_{11}
$,
$ 1$;\ \ 
$ 2+2c^{1}_{11}
+c^{2}_{11}
+c^{3}_{11}
+c^{4}_{11}
$,
$ 2+c^{1}_{11}
+c^{2}_{11}
+c^{3}_{11}
+c^{4}_{11}
$;\ \ 
$ -2-2  c^{1}_{11}
-c^{2}_{11}
-c^{3}_{11}
$)

\vskip 0.7ex
\hangindent=3em \hangafter=1
$\tau_n$ = ($-6.3 - 9.79 i$, $28.99 + 45.11 i$, $0. + 68.88 i$, $-25.7 - 22.26 i$, $-59.37 + 51.44 i$, $-59.37 - 51.44 i$, $-25.7 + 22.26 i$, $0. - 68.88 i$, $28.99 - 45.11 i$, $-6.3 + 9.79 i$, $135.76$)

\vskip 0.7ex
\hangindent=3em \hangafter=1
\textit{Intrinsic sign problem}

  \vskip 2ex 

%% file: modular_data/SsL8U_.tex
\noindent1. $8_{1,8.}^{4,100}$ \irep{0}:\ \ 
$d_i$ = ($1.0$,
$1.0$,
$1.0$,
$1.0$,
$1.0$,
$1.0$,
$1.0$,
$1.0$) 

\vskip 0.7ex
\hangindent=3em \hangafter=1
$D^2= 8.0 = 
8$

\vskip 0.7ex
\hangindent=3em \hangafter=1
$T = ( 0,
0,
0,
\frac{1}{2},
\frac{1}{4},
\frac{1}{4},
\frac{1}{4},
\frac{3}{4} )
$,

\vskip 0.7ex
\hangindent=3em \hangafter=1
$S$ = ($ 1$,
$ 1$,
$ 1$,
$ 1$,
$ 1$,
$ 1$,
$ 1$,
$ 1$;\ \ 
$ 1$,
$ -1$,
$ -1$,
$ -1$,
$ 1$,
$ 1$,
$ -1$;\ \ 
$ 1$,
$ -1$,
$ 1$,
$ 1$,
$ -1$,
$ -1$;\ \ 
$ 1$,
$ -1$,
$ 1$,
$ -1$,
$ 1$;\ \ 
$ -1$,
$ -1$,
$ 1$,
$ 1$;\ \ 
$ -1$,
$ -1$,
$ -1$;\ \ 
$ -1$,
$ 1$;\ \ 
$ -1$)

Factors = $2_{1,2.}^{4,437}\boxtimes 4_{0,4.}^{2,750}$

\vskip 0.7ex
\hangindent=3em \hangafter=1
$\tau_n$ = ($2. + 2. i$, $0.$, $2. - 2. i$, $8.$)

\vskip 0.7ex
\hangindent=3em \hangafter=1
\textit{Intrinsic sign problem}

  \vskip 2ex

\noindent2. $8_{7,8.}^{4,000}$ \irep{0}:\ \ 
$d_i$ = ($1.0$,
$1.0$,
$1.0$,
$1.0$,
$1.0$,
$1.0$,
$1.0$,
$1.0$) 

\vskip 0.7ex
\hangindent=3em \hangafter=1
$D^2= 8.0 = 
8$

\vskip 0.7ex
\hangindent=3em \hangafter=1
$T = ( 0,
0,
0,
\frac{1}{2},
\frac{1}{4},
\frac{3}{4},
\frac{3}{4},
\frac{3}{4} )
$,

\vskip 0.7ex
\hangindent=3em \hangafter=1
$S$ = ($ 1$,
$ 1$,
$ 1$,
$ 1$,
$ 1$,
$ 1$,
$ 1$,
$ 1$;\ \ 
$ 1$,
$ -1$,
$ -1$,
$ -1$,
$ 1$,
$ -1$,
$ 1$;\ \ 
$ 1$,
$ -1$,
$ -1$,
$ -1$,
$ 1$,
$ 1$;\ \ 
$ 1$,
$ 1$,
$ -1$,
$ -1$,
$ 1$;\ \ 
$ -1$,
$ 1$,
$ 1$,
$ -1$;\ \ 
$ -1$,
$ 1$,
$ -1$;\ \ 
$ -1$,
$ -1$;\ \ 
$ -1$)

Factors = $2_{7,2.}^{4,625}\boxtimes 4_{0,4.}^{2,750}$

\vskip 0.7ex
\hangindent=3em \hangafter=1
$\tau_n$ = ($2. - 2. i$, $0.$, $2. + 2. i$, $8.$)

\vskip 0.7ex
\hangindent=3em \hangafter=1
\textit{Intrinsic sign problem}

  \vskip 2ex

\noindent3. $8_{3,8.}^{4,500}$ \irep{0}:\ \ 
$d_i$ = ($1.0$,
$1.0$,
$1.0$,
$1.0$,
$1.0$,
$1.0$,
$1.0$,
$1.0$) 

\vskip 0.7ex
\hangindent=3em \hangafter=1
$D^2= 8.0 = 
8$

\vskip 0.7ex
\hangindent=3em \hangafter=1
$T = ( 0,
\frac{1}{2},
\frac{1}{2},
\frac{1}{2},
\frac{1}{4},
\frac{1}{4},
\frac{1}{4},
\frac{3}{4} )
$,

\vskip 0.7ex
\hangindent=3em \hangafter=1
$S$ = ($ 1$,
$ 1$,
$ 1$,
$ 1$,
$ 1$,
$ 1$,
$ 1$,
$ 1$;\ \ 
$ 1$,
$ -1$,
$ -1$,
$ -1$,
$ -1$,
$ 1$,
$ 1$;\ \ 
$ 1$,
$ -1$,
$ -1$,
$ 1$,
$ -1$,
$ 1$;\ \ 
$ 1$,
$ 1$,
$ -1$,
$ -1$,
$ 1$;\ \ 
$ -1$,
$ 1$,
$ 1$,
$ -1$;\ \ 
$ -1$,
$ 1$,
$ -1$;\ \ 
$ -1$,
$ -1$;\ \ 
$ -1$)

Factors = $2_{7,2.}^{4,625}\boxtimes 4_{4,4.}^{2,250}$

\vskip 0.7ex
\hangindent=3em \hangafter=1
$\tau_n$ = ($-2. + 2. i$, $0.$, $-2. - 2. i$, $8.$)

\vskip 0.7ex
\hangindent=3em \hangafter=1
\textit{Intrinsic sign problem}

  \vskip 2ex

\noindent4. $8_{5,8.}^{4,500}$ \irep{0}:\ \ 
$d_i$ = ($1.0$,
$1.0$,
$1.0$,
$1.0$,
$1.0$,
$1.0$,
$1.0$,
$1.0$) 

\vskip 0.7ex
\hangindent=3em \hangafter=1
$D^2= 8.0 = 
8$

\vskip 0.7ex
\hangindent=3em \hangafter=1
$T = ( 0,
\frac{1}{2},
\frac{1}{2},
\frac{1}{2},
\frac{1}{4},
\frac{3}{4},
\frac{3}{4},
\frac{3}{4} )
$,

\vskip 0.7ex
\hangindent=3em \hangafter=1
$S$ = ($ 1$,
$ 1$,
$ 1$,
$ 1$,
$ 1$,
$ 1$,
$ 1$,
$ 1$;\ \ 
$ 1$,
$ -1$,
$ -1$,
$ 1$,
$ -1$,
$ -1$,
$ 1$;\ \ 
$ 1$,
$ -1$,
$ 1$,
$ -1$,
$ 1$,
$ -1$;\ \ 
$ 1$,
$ 1$,
$ 1$,
$ -1$,
$ -1$;\ \ 
$ -1$,
$ -1$,
$ -1$,
$ -1$;\ \ 
$ -1$,
$ 1$,
$ 1$;\ \ 
$ -1$,
$ 1$;\ \ 
$ -1$)

Factors = $2_{1,2.}^{4,437}\boxtimes 4_{4,4.}^{2,250}$

\vskip 0.7ex
\hangindent=3em \hangafter=1
$\tau_n$ = ($-2. - 2. i$, $0.$, $-2. + 2. i$, $8.$)

\vskip 0.7ex
\hangindent=3em \hangafter=1
\textit{Intrinsic sign problem}

  \vskip 2ex

\noindent5. $8_{2,8.}^{8,812}$ \irep{103}:\ \ 
$d_i$ = ($1.0$,
$1.0$,
$1.0$,
$1.0$,
$1.0$,
$1.0$,
$1.0$,
$1.0$) 

\vskip 0.7ex
\hangindent=3em \hangafter=1
$D^2= 8.0 = 
8$

\vskip 0.7ex
\hangindent=3em \hangafter=1
$T = ( 0,
\frac{1}{2},
\frac{1}{4},
\frac{3}{4},
\frac{1}{8},
\frac{1}{8},
\frac{3}{8},
\frac{3}{8} )
$,

\vskip 0.7ex
\hangindent=3em \hangafter=1
$S$ = ($ 1$,
$ 1$,
$ 1$,
$ 1$,
$ 1$,
$ 1$,
$ 1$,
$ 1$;\ \ 
$ 1$,
$ 1$,
$ 1$,
$ -1$,
$ -1$,
$ -1$,
$ -1$;\ \ 
$ -1$,
$ -1$,
$ 1$,
$ 1$,
$ -1$,
$ -1$;\ \ 
$ -1$,
$ -1$,
$ -1$,
$ 1$,
$ 1$;\ \ 
$-\mathrm{i}$,
$\mathrm{i}$,
$-\mathrm{i}$,
$\mathrm{i}$;\ \ 
$-\mathrm{i}$,
$\mathrm{i}$,
$-\mathrm{i}$;\ \ 
$\mathrm{i}$,
$-\mathrm{i}$;\ \ 
$\mathrm{i}$)

Factors = $2_{1,2.}^{4,437}\boxtimes 4_{1,4.}^{8,718}$

\vskip 0.7ex
\hangindent=3em \hangafter=1
$\tau_n$ = ($0. + 2.83 i$, $0.$, $0. + 2.83 i$, $0.$, $0. - 2.83 i$, $0.$, $0. - 2.83 i$, $8.$)

\vskip 0.7ex
\hangindent=3em \hangafter=1
\textit{Intrinsic sign problem}

  \vskip 2ex

\noindent6. $8_{0,8.}^{8,437}$ \irep{103}:\ \ 
$d_i$ = ($1.0$,
$1.0$,
$1.0$,
$1.0$,
$1.0$,
$1.0$,
$1.0$,
$1.0$) 

\vskip 0.7ex
\hangindent=3em \hangafter=1
$D^2= 8.0 = 
8$

\vskip 0.7ex
\hangindent=3em \hangafter=1
$T = ( 0,
\frac{1}{2},
\frac{1}{4},
\frac{3}{4},
\frac{1}{8},
\frac{1}{8},
\frac{7}{8},
\frac{7}{8} )
$,

\vskip 0.7ex
\hangindent=3em \hangafter=1
$S$ = ($ 1$,
$ 1$,
$ 1$,
$ 1$,
$ 1$,
$ 1$,
$ 1$,
$ 1$;\ \ 
$ 1$,
$ 1$,
$ 1$,
$ -1$,
$ -1$,
$ -1$,
$ -1$;\ \ 
$ -1$,
$ -1$,
$ -1$,
$ -1$,
$ 1$,
$ 1$;\ \ 
$ -1$,
$ 1$,
$ 1$,
$ -1$,
$ -1$;\ \ 
$-\mathrm{i}$,
$\mathrm{i}$,
$-\mathrm{i}$,
$\mathrm{i}$;\ \ 
$-\mathrm{i}$,
$\mathrm{i}$,
$-\mathrm{i}$;\ \ 
$\mathrm{i}$,
$-\mathrm{i}$;\ \ 
$\mathrm{i}$)

Factors = $2_{1,2.}^{4,437}\boxtimes 4_{7,4.}^{8,781}$

\vskip 0.7ex
\hangindent=3em \hangafter=1
$\tau_n$ = ($2.83$, $0.$, $-2.83$, $0.$, $-2.83$, $0.$, $2.83$, $8.$)

\vskip 0.7ex
\hangindent=3em \hangafter=1
\textit{Intrinsic sign problem}

  \vskip 2ex

\noindent7. $8_{4,8.}^{8,625}$ \irep{103}:\ \ 
$d_i$ = ($1.0$,
$1.0$,
$1.0$,
$1.0$,
$1.0$,
$1.0$,
$1.0$,
$1.0$) 

\vskip 0.7ex
\hangindent=3em \hangafter=1
$D^2= 8.0 = 
8$

\vskip 0.7ex
\hangindent=3em \hangafter=1
$T = ( 0,
\frac{1}{2},
\frac{1}{4},
\frac{3}{4},
\frac{3}{8},
\frac{3}{8},
\frac{5}{8},
\frac{5}{8} )
$,

\vskip 0.7ex
\hangindent=3em \hangafter=1
$S$ = ($ 1$,
$ 1$,
$ 1$,
$ 1$,
$ 1$,
$ 1$,
$ 1$,
$ 1$;\ \ 
$ 1$,
$ 1$,
$ 1$,
$ -1$,
$ -1$,
$ -1$,
$ -1$;\ \ 
$ -1$,
$ -1$,
$ 1$,
$ 1$,
$ -1$,
$ -1$;\ \ 
$ -1$,
$ -1$,
$ -1$,
$ 1$,
$ 1$;\ \ 
$\mathrm{i}$,
$-\mathrm{i}$,
$-\mathrm{i}$,
$\mathrm{i}$;\ \ 
$\mathrm{i}$,
$\mathrm{i}$,
$-\mathrm{i}$;\ \ 
$-\mathrm{i}$,
$\mathrm{i}$;\ \ 
$-\mathrm{i}$)

Factors = $2_{1,2.}^{4,437}\boxtimes 4_{3,4.}^{8,468}$

\vskip 0.7ex
\hangindent=3em \hangafter=1
$\tau_n$ = ($-2.83$, $0.$, $2.83$, $0.$, $2.83$, $0.$, $-2.83$, $8.$)

\vskip 0.7ex
\hangindent=3em \hangafter=1
\textit{Intrinsic sign problem}

  \vskip 2ex

\noindent8. $8_{6,8.}^{8,118}$ \irep{103}:\ \ 
$d_i$ = ($1.0$,
$1.0$,
$1.0$,
$1.0$,
$1.0$,
$1.0$,
$1.0$,
$1.0$) 

\vskip 0.7ex
\hangindent=3em \hangafter=1
$D^2= 8.0 = 
8$

\vskip 0.7ex
\hangindent=3em \hangafter=1
$T = ( 0,
\frac{1}{2},
\frac{1}{4},
\frac{3}{4},
\frac{5}{8},
\frac{5}{8},
\frac{7}{8},
\frac{7}{8} )
$,

\vskip 0.7ex
\hangindent=3em \hangafter=1
$S$ = ($ 1$,
$ 1$,
$ 1$,
$ 1$,
$ 1$,
$ 1$,
$ 1$,
$ 1$;\ \ 
$ 1$,
$ 1$,
$ 1$,
$ -1$,
$ -1$,
$ -1$,
$ -1$;\ \ 
$ -1$,
$ -1$,
$ 1$,
$ 1$,
$ -1$,
$ -1$;\ \ 
$ -1$,
$ -1$,
$ -1$,
$ 1$,
$ 1$;\ \ 
$-\mathrm{i}$,
$\mathrm{i}$,
$-\mathrm{i}$,
$\mathrm{i}$;\ \ 
$-\mathrm{i}$,
$\mathrm{i}$,
$-\mathrm{i}$;\ \ 
$\mathrm{i}$,
$-\mathrm{i}$;\ \ 
$\mathrm{i}$)

Factors = $2_{1,2.}^{4,437}\boxtimes 4_{5,4.}^{8,312}$

\vskip 0.7ex
\hangindent=3em \hangafter=1
$\tau_n$ = ($0. - 2.83 i$, $0.$, $0. - 2.83 i$, $0.$, $0. + 2.83 i$, $0.$, $0. + 2.83 i$, $8.$)

\vskip 0.7ex
\hangindent=3em \hangafter=1
\textit{Intrinsic sign problem}

  \vskip 2ex

\noindent9. $8_{1,8.}^{16,123}$ \irep{0}:\ \ 
$d_i$ = ($1.0$,
$1.0$,
$1.0$,
$1.0$,
$1.0$,
$1.0$,
$1.0$,
$1.0$) 

\vskip 0.7ex
\hangindent=3em \hangafter=1
$D^2= 8.0 = 
8$

\vskip 0.7ex
\hangindent=3em \hangafter=1
$T = ( 0,
0,
\frac{1}{4},
\frac{1}{4},
\frac{1}{16},
\frac{1}{16},
\frac{9}{16},
\frac{9}{16} )
$,

\vskip 0.7ex
\hangindent=3em \hangafter=1
$S$ = ($ 1$,
$ 1$,
$ 1$,
$ 1$,
$ 1$,
$ 1$,
$ 1$,
$ 1$;\ \ 
$ 1$,
$ 1$,
$ 1$,
$ -1$,
$ -1$,
$ -1$,
$ -1$;\ \ 
$ -1$,
$ -1$,
$-\mathrm{i}$,
$\mathrm{i}$,
$-\mathrm{i}$,
$\mathrm{i}$;\ \ 
$ -1$,
$\mathrm{i}$,
$-\mathrm{i}$,
$\mathrm{i}$,
$-\mathrm{i}$;\ \ 
$ -\zeta_{8}^{3}$,
$ \zeta_{8}^{1}$,
$ \zeta_{8}^{3}$,
$ -\zeta_{8}^{1}$;\ \ 
$ -\zeta_{8}^{3}$,
$ -\zeta_{8}^{1}$,
$ \zeta_{8}^{3}$;\ \ 
$ -\zeta_{8}^{3}$,
$ \zeta_{8}^{1}$;\ \ 
$ -\zeta_{8}^{3}$)

\vskip 0.7ex
\hangindent=3em \hangafter=1
$\tau_n$ = ($2. + 2. i$, $2.83 + 2.83 i$, $2. - 2. i$, $4. + 4. i$, $2. + 2. i$, $-2.83 + 2.83 i$, $2. - 2. i$, $0.$, $2. + 2. i$, $-2.83 - 2.83 i$, $2. - 2. i$, $4. - 4. i$, $2. + 2. i$, $2.83 - 2.83 i$, $2. - 2. i$, $8.$)

\vskip 0.7ex
\hangindent=3em \hangafter=1
\textit{Intrinsic sign problem}

  \vskip 2ex

\noindent10. $8_{1,8.}^{16,359}$ \irep{0}:\ \ 
$d_i$ = ($1.0$,
$1.0$,
$1.0$,
$1.0$,
$1.0$,
$1.0$,
$1.0$,
$1.0$) 

\vskip 0.7ex
\hangindent=3em \hangafter=1
$D^2= 8.0 = 
8$

\vskip 0.7ex
\hangindent=3em \hangafter=1
$T = ( 0,
0,
\frac{1}{4},
\frac{1}{4},
\frac{5}{16},
\frac{5}{16},
\frac{13}{16},
\frac{13}{16} )
$,

\vskip 0.7ex
\hangindent=3em \hangafter=1
$S$ = ($ 1$,
$ 1$,
$ 1$,
$ 1$,
$ 1$,
$ 1$,
$ 1$,
$ 1$;\ \ 
$ 1$,
$ 1$,
$ 1$,
$ -1$,
$ -1$,
$ -1$,
$ -1$;\ \ 
$ -1$,
$ -1$,
$-\mathrm{i}$,
$\mathrm{i}$,
$-\mathrm{i}$,
$\mathrm{i}$;\ \ 
$ -1$,
$\mathrm{i}$,
$-\mathrm{i}$,
$\mathrm{i}$,
$-\mathrm{i}$;\ \ 
$ \zeta_{8}^{3}$,
$ -\zeta_{8}^{1}$,
$ -\zeta_{8}^{3}$,
$ \zeta_{8}^{1}$;\ \ 
$ \zeta_{8}^{3}$,
$ \zeta_{8}^{1}$,
$ -\zeta_{8}^{3}$;\ \ 
$ \zeta_{8}^{3}$,
$ -\zeta_{8}^{1}$;\ \ 
$ \zeta_{8}^{3}$)

\vskip 0.7ex
\hangindent=3em \hangafter=1
$\tau_n$ = ($2. + 2. i$, $-2.83 - 2.83 i$, $2. - 2. i$, $4. + 4. i$, $2. + 2. i$, $2.83 - 2.83 i$, $2. - 2. i$, $0.$, $2. + 2. i$, $2.83 + 2.83 i$, $2. - 2. i$, $4. - 4. i$, $2. + 2. i$, $-2.83 + 2.83 i$, $2. - 2. i$, $8.$)

\vskip 0.7ex
\hangindent=3em \hangafter=1
\textit{Intrinsic sign problem}

  \vskip 2ex

\noindent11. $8_{7,8.}^{16,140}$ \irep{0}:\ \ 
$d_i$ = ($1.0$,
$1.0$,
$1.0$,
$1.0$,
$1.0$,
$1.0$,
$1.0$,
$1.0$) 

\vskip 0.7ex
\hangindent=3em \hangafter=1
$D^2= 8.0 = 
8$

\vskip 0.7ex
\hangindent=3em \hangafter=1
$T = ( 0,
0,
\frac{3}{4},
\frac{3}{4},
\frac{3}{16},
\frac{3}{16},
\frac{11}{16},
\frac{11}{16} )
$,

\vskip 0.7ex
\hangindent=3em \hangafter=1
$S$ = ($ 1$,
$ 1$,
$ 1$,
$ 1$,
$ 1$,
$ 1$,
$ 1$,
$ 1$;\ \ 
$ 1$,
$ 1$,
$ 1$,
$ -1$,
$ -1$,
$ -1$,
$ -1$;\ \ 
$ -1$,
$ -1$,
$-\mathrm{i}$,
$\mathrm{i}$,
$-\mathrm{i}$,
$\mathrm{i}$;\ \ 
$ -1$,
$\mathrm{i}$,
$-\mathrm{i}$,
$\mathrm{i}$,
$-\mathrm{i}$;\ \ 
$ -\zeta_{8}^{1}$,
$ \zeta_{8}^{3}$,
$ \zeta_{8}^{1}$,
$ -\zeta_{8}^{3}$;\ \ 
$ -\zeta_{8}^{1}$,
$ -\zeta_{8}^{3}$,
$ \zeta_{8}^{1}$;\ \ 
$ -\zeta_{8}^{1}$,
$ \zeta_{8}^{3}$;\ \ 
$ -\zeta_{8}^{1}$)

\vskip 0.7ex
\hangindent=3em \hangafter=1
$\tau_n$ = ($2. - 2. i$, $-2.83 + 2.83 i$, $2. + 2. i$, $4. - 4. i$, $2. - 2. i$, $2.83 + 2.83 i$, $2. + 2. i$, $0.$, $2. - 2. i$, $2.83 - 2.83 i$, $2. + 2. i$, $4. + 4. i$, $2. - 2. i$, $-2.83 - 2.83 i$, $2. + 2. i$, $8.$)

\vskip 0.7ex
\hangindent=3em \hangafter=1
\textit{Intrinsic sign problem}

  \vskip 2ex

\noindent12. $8_{7,8.}^{16,126}$ \irep{0}:\ \ 
$d_i$ = ($1.0$,
$1.0$,
$1.0$,
$1.0$,
$1.0$,
$1.0$,
$1.0$,
$1.0$) 

\vskip 0.7ex
\hangindent=3em \hangafter=1
$D^2= 8.0 = 
8$

\vskip 0.7ex
\hangindent=3em \hangafter=1
$T = ( 0,
0,
\frac{3}{4},
\frac{3}{4},
\frac{7}{16},
\frac{7}{16},
\frac{15}{16},
\frac{15}{16} )
$,

\vskip 0.7ex
\hangindent=3em \hangafter=1
$S$ = ($ 1$,
$ 1$,
$ 1$,
$ 1$,
$ 1$,
$ 1$,
$ 1$,
$ 1$;\ \ 
$ 1$,
$ 1$,
$ 1$,
$ -1$,
$ -1$,
$ -1$,
$ -1$;\ \ 
$ -1$,
$ -1$,
$-\mathrm{i}$,
$\mathrm{i}$,
$-\mathrm{i}$,
$\mathrm{i}$;\ \ 
$ -1$,
$\mathrm{i}$,
$-\mathrm{i}$,
$\mathrm{i}$,
$-\mathrm{i}$;\ \ 
$ \zeta_{8}^{1}$,
$ -\zeta_{8}^{3}$,
$ -\zeta_{8}^{1}$,
$ \zeta_{8}^{3}$;\ \ 
$ \zeta_{8}^{1}$,
$ \zeta_{8}^{3}$,
$ -\zeta_{8}^{1}$;\ \ 
$ \zeta_{8}^{1}$,
$ -\zeta_{8}^{3}$;\ \ 
$ \zeta_{8}^{1}$)

\vskip 0.7ex
\hangindent=3em \hangafter=1
$\tau_n$ = ($2. - 2. i$, $2.83 - 2.83 i$, $2. + 2. i$, $4. - 4. i$, $2. - 2. i$, $-2.83 - 2.83 i$, $2. + 2. i$, $0.$, $2. - 2. i$, $-2.83 + 2.83 i$, $2. + 2. i$, $4. + 4. i$, $2. - 2. i$, $2.83 + 2.83 i$, $2. + 2. i$, $8.$)

\vskip 0.7ex
\hangindent=3em \hangafter=1
\textit{Intrinsic sign problem}

  \vskip 2ex

\noindent13. $8_{\frac{14}{5},14.47}^{10,280}$ \irep{138}:\ \ 
$d_i$ = ($1.0$,
$1.0$,
$1.0$,
$1.0$,
$1.618$,
$1.618$,
$1.618$,
$1.618$) 

\vskip 0.7ex
\hangindent=3em \hangafter=1
$D^2= 14.472 = 
10+2\sqrt{5}$

\vskip 0.7ex
\hangindent=3em \hangafter=1
$T = ( 0,
0,
0,
\frac{1}{2},
\frac{2}{5},
\frac{2}{5},
\frac{2}{5},
\frac{9}{10} )
$,

\vskip 0.7ex
\hangindent=3em \hangafter=1
$S$ = ($ 1$,
$ 1$,
$ 1$,
$ 1$,
$ \frac{1+\sqrt{5}}{2}$,
$ \frac{1+\sqrt{5}}{2}$,
$ \frac{1+\sqrt{5}}{2}$,
$ \frac{1+\sqrt{5}}{2}$;\ \ 
$ 1$,
$ -1$,
$ -1$,
$ \frac{1+\sqrt{5}}{2}$,
$ \frac{1+\sqrt{5}}{2}$,
$ -\frac{1+\sqrt{5}}{2}$,
$ -\frac{1+\sqrt{5}}{2}$;\ \ 
$ 1$,
$ -1$,
$ \frac{1+\sqrt{5}}{2}$,
$ -\frac{1+\sqrt{5}}{2}$,
$ \frac{1+\sqrt{5}}{2}$,
$ -\frac{1+\sqrt{5}}{2}$;\ \ 
$ 1$,
$ \frac{1+\sqrt{5}}{2}$,
$ -\frac{1+\sqrt{5}}{2}$,
$ -\frac{1+\sqrt{5}}{2}$,
$ \frac{1+\sqrt{5}}{2}$;\ \ 
$ -1$,
$ -1$,
$ -1$,
$ -1$;\ \ 
$ -1$,
$ 1$,
$ 1$;\ \ 
$ -1$,
$ 1$;\ \ 
$ -1$)

Factors = $2_{\frac{14}{5},3.618}^{5,395}\boxtimes 4_{0,4.}^{2,750}$

\vskip 0.7ex
\hangindent=3em \hangafter=1
$\tau_n$ = ($-2.24 + 3.08 i$, $7.24 - 9.96 i$, $3.62 + 4.98 i$, $-4.47 - 6.16 i$, $7.24$, $-4.47 + 6.16 i$, $3.62 - 4.98 i$, $7.24 + 9.96 i$, $-2.24 - 3.08 i$, $14.47$)

\vskip 0.7ex
\hangindent=3em \hangafter=1
\textit{Intrinsic sign problem}

  \vskip 2ex

\noindent14. $8_{\frac{26}{5},14.47}^{10,604}$ \irep{138}:\ \ 
$d_i$ = ($1.0$,
$1.0$,
$1.0$,
$1.0$,
$1.618$,
$1.618$,
$1.618$,
$1.618$) 

\vskip 0.7ex
\hangindent=3em \hangafter=1
$D^2= 14.472 = 
10+2\sqrt{5}$

\vskip 0.7ex
\hangindent=3em \hangafter=1
$T = ( 0,
0,
0,
\frac{1}{2},
\frac{3}{5},
\frac{3}{5},
\frac{3}{5},
\frac{1}{10} )
$,

\vskip 0.7ex
\hangindent=3em \hangafter=1
$S$ = ($ 1$,
$ 1$,
$ 1$,
$ 1$,
$ \frac{1+\sqrt{5}}{2}$,
$ \frac{1+\sqrt{5}}{2}$,
$ \frac{1+\sqrt{5}}{2}$,
$ \frac{1+\sqrt{5}}{2}$;\ \ 
$ 1$,
$ -1$,
$ -1$,
$ \frac{1+\sqrt{5}}{2}$,
$ \frac{1+\sqrt{5}}{2}$,
$ -\frac{1+\sqrt{5}}{2}$,
$ -\frac{1+\sqrt{5}}{2}$;\ \ 
$ 1$,
$ -1$,
$ \frac{1+\sqrt{5}}{2}$,
$ -\frac{1+\sqrt{5}}{2}$,
$ \frac{1+\sqrt{5}}{2}$,
$ -\frac{1+\sqrt{5}}{2}$;\ \ 
$ 1$,
$ \frac{1+\sqrt{5}}{2}$,
$ -\frac{1+\sqrt{5}}{2}$,
$ -\frac{1+\sqrt{5}}{2}$,
$ \frac{1+\sqrt{5}}{2}$;\ \ 
$ -1$,
$ -1$,
$ -1$,
$ -1$;\ \ 
$ -1$,
$ 1$,
$ 1$;\ \ 
$ -1$,
$ 1$;\ \ 
$ -1$)

Factors = $2_{\frac{26}{5},3.618}^{5,720}\boxtimes 4_{0,4.}^{2,750}$

\vskip 0.7ex
\hangindent=3em \hangafter=1
$\tau_n$ = ($-2.24 - 3.08 i$, $7.24 + 9.96 i$, $3.62 - 4.98 i$, $-4.47 + 6.16 i$, $7.24$, $-4.47 - 6.16 i$, $3.62 + 4.98 i$, $7.24 - 9.96 i$, $-2.24 + 3.08 i$, $14.47$)

\vskip 0.7ex
\hangindent=3em \hangafter=1
\textit{Intrinsic sign problem}

  \vskip 2ex

\noindent15. $8_{\frac{34}{5},14.47}^{10,232}$ \irep{138}:\ \ 
$d_i$ = ($1.0$,
$1.0$,
$1.0$,
$1.0$,
$1.618$,
$1.618$,
$1.618$,
$1.618$) 

\vskip 0.7ex
\hangindent=3em \hangafter=1
$D^2= 14.472 = 
10+2\sqrt{5}$

\vskip 0.7ex
\hangindent=3em \hangafter=1
$T = ( 0,
\frac{1}{2},
\frac{1}{2},
\frac{1}{2},
\frac{2}{5},
\frac{9}{10},
\frac{9}{10},
\frac{9}{10} )
$,

\vskip 0.7ex
\hangindent=3em \hangafter=1
$S$ = ($ 1$,
$ 1$,
$ 1$,
$ 1$,
$ \frac{1+\sqrt{5}}{2}$,
$ \frac{1+\sqrt{5}}{2}$,
$ \frac{1+\sqrt{5}}{2}$,
$ \frac{1+\sqrt{5}}{2}$;\ \ 
$ 1$,
$ -1$,
$ -1$,
$ \frac{1+\sqrt{5}}{2}$,
$ \frac{1+\sqrt{5}}{2}$,
$ -\frac{1+\sqrt{5}}{2}$,
$ -\frac{1+\sqrt{5}}{2}$;\ \ 
$ 1$,
$ -1$,
$ \frac{1+\sqrt{5}}{2}$,
$ -\frac{1+\sqrt{5}}{2}$,
$ \frac{1+\sqrt{5}}{2}$,
$ -\frac{1+\sqrt{5}}{2}$;\ \ 
$ 1$,
$ \frac{1+\sqrt{5}}{2}$,
$ -\frac{1+\sqrt{5}}{2}$,
$ -\frac{1+\sqrt{5}}{2}$,
$ \frac{1+\sqrt{5}}{2}$;\ \ 
$ -1$,
$ -1$,
$ -1$,
$ -1$;\ \ 
$ -1$,
$ 1$,
$ 1$;\ \ 
$ -1$,
$ 1$;\ \ 
$ -1$)

Factors = $2_{\frac{14}{5},3.618}^{5,395}\boxtimes 4_{4,4.}^{2,250}$

\vskip 0.7ex
\hangindent=3em \hangafter=1
$\tau_n$ = ($2.24 - 3.08 i$, $7.24 - 9.96 i$, $-3.62 - 4.98 i$, $-4.47 - 6.16 i$, $-7.24$, $-4.47 + 6.16 i$, $-3.62 + 4.98 i$, $7.24 + 9.96 i$, $2.24 + 3.08 i$, $14.47$)

\vskip 0.7ex
\hangindent=3em \hangafter=1
\textit{Intrinsic sign problem}

  \vskip 2ex

\noindent16. $8_{\frac{6}{5},14.47}^{10,123}$ \irep{138}:\ \ 
$d_i$ = ($1.0$,
$1.0$,
$1.0$,
$1.0$,
$1.618$,
$1.618$,
$1.618$,
$1.618$) 

\vskip 0.7ex
\hangindent=3em \hangafter=1
$D^2= 14.472 = 
10+2\sqrt{5}$

\vskip 0.7ex
\hangindent=3em \hangafter=1
$T = ( 0,
\frac{1}{2},
\frac{1}{2},
\frac{1}{2},
\frac{3}{5},
\frac{1}{10},
\frac{1}{10},
\frac{1}{10} )
$,

\vskip 0.7ex
\hangindent=3em \hangafter=1
$S$ = ($ 1$,
$ 1$,
$ 1$,
$ 1$,
$ \frac{1+\sqrt{5}}{2}$,
$ \frac{1+\sqrt{5}}{2}$,
$ \frac{1+\sqrt{5}}{2}$,
$ \frac{1+\sqrt{5}}{2}$;\ \ 
$ 1$,
$ -1$,
$ -1$,
$ \frac{1+\sqrt{5}}{2}$,
$ \frac{1+\sqrt{5}}{2}$,
$ -\frac{1+\sqrt{5}}{2}$,
$ -\frac{1+\sqrt{5}}{2}$;\ \ 
$ 1$,
$ -1$,
$ \frac{1+\sqrt{5}}{2}$,
$ -\frac{1+\sqrt{5}}{2}$,
$ \frac{1+\sqrt{5}}{2}$,
$ -\frac{1+\sqrt{5}}{2}$;\ \ 
$ 1$,
$ \frac{1+\sqrt{5}}{2}$,
$ -\frac{1+\sqrt{5}}{2}$,
$ -\frac{1+\sqrt{5}}{2}$,
$ \frac{1+\sqrt{5}}{2}$;\ \ 
$ -1$,
$ -1$,
$ -1$,
$ -1$;\ \ 
$ -1$,
$ 1$,
$ 1$;\ \ 
$ -1$,
$ 1$;\ \ 
$ -1$)

Factors = $2_{\frac{26}{5},3.618}^{5,720}\boxtimes 4_{4,4.}^{2,250}$

\vskip 0.7ex
\hangindent=3em \hangafter=1
$\tau_n$ = ($2.24 + 3.08 i$, $7.24 + 9.96 i$, $-3.62 + 4.98 i$, $-4.47 + 6.16 i$, $-7.24$, $-4.47 - 6.16 i$, $-3.62 - 4.98 i$, $7.24 - 9.96 i$, $2.24 - 3.08 i$, $14.47$)

\vskip 0.7ex
\hangindent=3em \hangafter=1
\textit{Intrinsic sign problem}

  \vskip 2ex

\noindent17. $8_{\frac{14}{5},14.47}^{20,755}$ \irep{207}:\ \ 
$d_i$ = ($1.0$,
$1.0$,
$1.0$,
$1.0$,
$1.618$,
$1.618$,
$1.618$,
$1.618$) 

\vskip 0.7ex
\hangindent=3em \hangafter=1
$D^2= 14.472 = 
10+2\sqrt{5}$

\vskip 0.7ex
\hangindent=3em \hangafter=1
$T = ( 0,
0,
\frac{1}{4},
\frac{3}{4},
\frac{2}{5},
\frac{2}{5},
\frac{3}{20},
\frac{13}{20} )
$,

\vskip 0.7ex
\hangindent=3em \hangafter=1
$S$ = ($ 1$,
$ 1$,
$ 1$,
$ 1$,
$ \frac{1+\sqrt{5}}{2}$,
$ \frac{1+\sqrt{5}}{2}$,
$ \frac{1+\sqrt{5}}{2}$,
$ \frac{1+\sqrt{5}}{2}$;\ \ 
$ 1$,
$ -1$,
$ -1$,
$ \frac{1+\sqrt{5}}{2}$,
$ \frac{1+\sqrt{5}}{2}$,
$ -\frac{1+\sqrt{5}}{2}$,
$ -\frac{1+\sqrt{5}}{2}$;\ \ 
$ -1$,
$ 1$,
$ \frac{1+\sqrt{5}}{2}$,
$ -\frac{1+\sqrt{5}}{2}$,
$ \frac{1+\sqrt{5}}{2}$,
$ -\frac{1+\sqrt{5}}{2}$;\ \ 
$ -1$,
$ \frac{1+\sqrt{5}}{2}$,
$ -\frac{1+\sqrt{5}}{2}$,
$ -\frac{1+\sqrt{5}}{2}$,
$ \frac{1+\sqrt{5}}{2}$;\ \ 
$ -1$,
$ -1$,
$ -1$,
$ -1$;\ \ 
$ -1$,
$ 1$,
$ 1$;\ \ 
$ 1$,
$ -1$;\ \ 
$ 1$)

Factors = $2_{1,2.}^{4,437}\boxtimes 4_{\frac{9}{5},7.236}^{20,451}$

\vskip 0.7ex
\hangindent=3em \hangafter=1
$\tau_n$ = ($-2.24 + 3.08 i$, $0.$, $3.62 + 4.98 i$, $-4.47 - 6.16 i$, $7.24$, $0.$, $3.62 - 4.98 i$, $7.24 + 9.96 i$, $-2.24 - 3.08 i$, $0.$, $-2.24 + 3.08 i$, $7.24 - 9.96 i$, $3.62 + 4.98 i$, $0.$, $7.24$, $-4.47 + 6.16 i$, $3.62 - 4.98 i$, $0.$, $-2.24 - 3.08 i$, $14.47$)

\vskip 0.7ex
\hangindent=3em \hangafter=1
\textit{Intrinsic sign problem}

  \vskip 2ex

\noindent18. $8_{\frac{26}{5},14.47}^{20,539}$ \irep{207}:\ \ 
$d_i$ = ($1.0$,
$1.0$,
$1.0$,
$1.0$,
$1.618$,
$1.618$,
$1.618$,
$1.618$) 

\vskip 0.7ex
\hangindent=3em \hangafter=1
$D^2= 14.472 = 
10+2\sqrt{5}$

\vskip 0.7ex
\hangindent=3em \hangafter=1
$T = ( 0,
0,
\frac{1}{4},
\frac{3}{4},
\frac{3}{5},
\frac{3}{5},
\frac{7}{20},
\frac{17}{20} )
$,

\vskip 0.7ex
\hangindent=3em \hangafter=1
$S$ = ($ 1$,
$ 1$,
$ 1$,
$ 1$,
$ \frac{1+\sqrt{5}}{2}$,
$ \frac{1+\sqrt{5}}{2}$,
$ \frac{1+\sqrt{5}}{2}$,
$ \frac{1+\sqrt{5}}{2}$;\ \ 
$ 1$,
$ -1$,
$ -1$,
$ \frac{1+\sqrt{5}}{2}$,
$ \frac{1+\sqrt{5}}{2}$,
$ -\frac{1+\sqrt{5}}{2}$,
$ -\frac{1+\sqrt{5}}{2}$;\ \ 
$ -1$,
$ 1$,
$ \frac{1+\sqrt{5}}{2}$,
$ -\frac{1+\sqrt{5}}{2}$,
$ \frac{1+\sqrt{5}}{2}$,
$ -\frac{1+\sqrt{5}}{2}$;\ \ 
$ -1$,
$ \frac{1+\sqrt{5}}{2}$,
$ -\frac{1+\sqrt{5}}{2}$,
$ -\frac{1+\sqrt{5}}{2}$,
$ \frac{1+\sqrt{5}}{2}$;\ \ 
$ -1$,
$ -1$,
$ -1$,
$ -1$;\ \ 
$ -1$,
$ 1$,
$ 1$;\ \ 
$ 1$,
$ -1$;\ \ 
$ 1$)

Factors = $2_{1,2.}^{4,437}\boxtimes 4_{\frac{21}{5},7.236}^{20,341}$

\vskip 0.7ex
\hangindent=3em \hangafter=1
$\tau_n$ = ($-2.24 - 3.08 i$, $0.$, $3.62 - 4.98 i$, $-4.47 + 6.16 i$, $7.24$, $0.$, $3.62 + 4.98 i$, $7.24 - 9.96 i$, $-2.24 + 3.08 i$, $0.$, $-2.24 - 3.08 i$, $7.24 + 9.96 i$, $3.62 - 4.98 i$, $0.$, $7.24$, $-4.47 - 6.16 i$, $3.62 + 4.98 i$, $0.$, $-2.24 + 3.08 i$, $14.47$)

\vskip 0.7ex
\hangindent=3em \hangafter=1
\textit{Intrinsic sign problem}

  \vskip 2ex

\noindent19. $8_{\frac{24}{5},14.47}^{20,693}$ \irep{207}:\ \ 
$d_i$ = ($1.0$,
$1.0$,
$1.0$,
$1.0$,
$1.618$,
$1.618$,
$1.618$,
$1.618$) 

\vskip 0.7ex
\hangindent=3em \hangafter=1
$D^2= 14.472 = 
10+2\sqrt{5}$

\vskip 0.7ex
\hangindent=3em \hangafter=1
$T = ( 0,
\frac{1}{2},
\frac{1}{4},
\frac{1}{4},
\frac{2}{5},
\frac{9}{10},
\frac{13}{20},
\frac{13}{20} )
$,

\vskip 0.7ex
\hangindent=3em \hangafter=1
$S$ = ($ 1$,
$ 1$,
$ 1$,
$ 1$,
$ \frac{1+\sqrt{5}}{2}$,
$ \frac{1+\sqrt{5}}{2}$,
$ \frac{1+\sqrt{5}}{2}$,
$ \frac{1+\sqrt{5}}{2}$;\ \ 
$ 1$,
$ -1$,
$ -1$,
$ \frac{1+\sqrt{5}}{2}$,
$ \frac{1+\sqrt{5}}{2}$,
$ -\frac{1+\sqrt{5}}{2}$,
$ -\frac{1+\sqrt{5}}{2}$;\ \ 
$ -1$,
$ 1$,
$ \frac{1+\sqrt{5}}{2}$,
$ -\frac{1+\sqrt{5}}{2}$,
$ \frac{1+\sqrt{5}}{2}$,
$ -\frac{1+\sqrt{5}}{2}$;\ \ 
$ -1$,
$ \frac{1+\sqrt{5}}{2}$,
$ -\frac{1+\sqrt{5}}{2}$,
$ -\frac{1+\sqrt{5}}{2}$,
$ \frac{1+\sqrt{5}}{2}$;\ \ 
$ -1$,
$ -1$,
$ -1$,
$ -1$;\ \ 
$ -1$,
$ 1$,
$ 1$;\ \ 
$ 1$,
$ -1$;\ \ 
$ 1$)

Factors = $2_{1,2.}^{4,437}\boxtimes 4_{\frac{19}{5},7.236}^{20,304}$

\vskip 0.7ex
\hangindent=3em \hangafter=1
$\tau_n$ = ($-3.08 - 2.24 i$, $0.$, $4.98 - 3.62 i$, $-4.47 - 6.16 i$, $0. + 7.24 i$, $0.$, $-4.98 - 3.62 i$, $7.24 + 9.96 i$, $3.08 - 2.24 i$, $0.$, $3.08 + 2.24 i$, $7.24 - 9.96 i$, $-4.98 + 3.62 i$, $0.$, $0. - 7.24 i$, $-4.47 + 6.16 i$, $4.98 + 3.62 i$, $0.$, $-3.08 + 2.24 i$, $14.47$)

\vskip 0.7ex
\hangindent=3em \hangafter=1
\textit{Intrinsic sign problem}

  \vskip 2ex

\noindent20. $8_{\frac{36}{5},14.47}^{20,693}$ \irep{207}:\ \ 
$d_i$ = ($1.0$,
$1.0$,
$1.0$,
$1.0$,
$1.618$,
$1.618$,
$1.618$,
$1.618$) 

\vskip 0.7ex
\hangindent=3em \hangafter=1
$D^2= 14.472 = 
10+2\sqrt{5}$

\vskip 0.7ex
\hangindent=3em \hangafter=1
$T = ( 0,
\frac{1}{2},
\frac{1}{4},
\frac{1}{4},
\frac{3}{5},
\frac{1}{10},
\frac{17}{20},
\frac{17}{20} )
$,

\vskip 0.7ex
\hangindent=3em \hangafter=1
$S$ = ($ 1$,
$ 1$,
$ 1$,
$ 1$,
$ \frac{1+\sqrt{5}}{2}$,
$ \frac{1+\sqrt{5}}{2}$,
$ \frac{1+\sqrt{5}}{2}$,
$ \frac{1+\sqrt{5}}{2}$;\ \ 
$ 1$,
$ -1$,
$ -1$,
$ \frac{1+\sqrt{5}}{2}$,
$ \frac{1+\sqrt{5}}{2}$,
$ -\frac{1+\sqrt{5}}{2}$,
$ -\frac{1+\sqrt{5}}{2}$;\ \ 
$ -1$,
$ 1$,
$ \frac{1+\sqrt{5}}{2}$,
$ -\frac{1+\sqrt{5}}{2}$,
$ \frac{1+\sqrt{5}}{2}$,
$ -\frac{1+\sqrt{5}}{2}$;\ \ 
$ -1$,
$ \frac{1+\sqrt{5}}{2}$,
$ -\frac{1+\sqrt{5}}{2}$,
$ -\frac{1+\sqrt{5}}{2}$,
$ \frac{1+\sqrt{5}}{2}$;\ \ 
$ -1$,
$ -1$,
$ -1$,
$ -1$;\ \ 
$ -1$,
$ 1$,
$ 1$;\ \ 
$ 1$,
$ -1$;\ \ 
$ 1$)

Factors = $2_{1,2.}^{4,437}\boxtimes 4_{\frac{31}{5},7.236}^{20,505}$

\vskip 0.7ex
\hangindent=3em \hangafter=1
$\tau_n$ = ($3.08 - 2.24 i$, $0.$, $-4.98 - 3.62 i$, $-4.47 + 6.16 i$, $0. + 7.24 i$, $0.$, $4.98 - 3.62 i$, $7.24 - 9.96 i$, $-3.08 - 2.24 i$, $0.$, $-3.08 + 2.24 i$, $7.24 + 9.96 i$, $4.98 + 3.62 i$, $0.$, $0. - 7.24 i$, $-4.47 - 6.16 i$, $-4.98 + 3.62 i$, $0.$, $3.08 + 2.24 i$, $14.47$)

\vskip 0.7ex
\hangindent=3em \hangafter=1
\textit{Intrinsic sign problem}

  \vskip 2ex

\noindent21. $8_{\frac{4}{5},14.47}^{20,399}$ \irep{207}:\ \ 
$d_i$ = ($1.0$,
$1.0$,
$1.0$,
$1.0$,
$1.618$,
$1.618$,
$1.618$,
$1.618$) 

\vskip 0.7ex
\hangindent=3em \hangafter=1
$D^2= 14.472 = 
10+2\sqrt{5}$

\vskip 0.7ex
\hangindent=3em \hangafter=1
$T = ( 0,
\frac{1}{2},
\frac{3}{4},
\frac{3}{4},
\frac{2}{5},
\frac{9}{10},
\frac{3}{20},
\frac{3}{20} )
$,

\vskip 0.7ex
\hangindent=3em \hangafter=1
$S$ = ($ 1$,
$ 1$,
$ 1$,
$ 1$,
$ \frac{1+\sqrt{5}}{2}$,
$ \frac{1+\sqrt{5}}{2}$,
$ \frac{1+\sqrt{5}}{2}$,
$ \frac{1+\sqrt{5}}{2}$;\ \ 
$ 1$,
$ -1$,
$ -1$,
$ \frac{1+\sqrt{5}}{2}$,
$ \frac{1+\sqrt{5}}{2}$,
$ -\frac{1+\sqrt{5}}{2}$,
$ -\frac{1+\sqrt{5}}{2}$;\ \ 
$ -1$,
$ 1$,
$ \frac{1+\sqrt{5}}{2}$,
$ -\frac{1+\sqrt{5}}{2}$,
$ \frac{1+\sqrt{5}}{2}$,
$ -\frac{1+\sqrt{5}}{2}$;\ \ 
$ -1$,
$ \frac{1+\sqrt{5}}{2}$,
$ -\frac{1+\sqrt{5}}{2}$,
$ -\frac{1+\sqrt{5}}{2}$,
$ \frac{1+\sqrt{5}}{2}$;\ \ 
$ -1$,
$ -1$,
$ -1$,
$ -1$;\ \ 
$ -1$,
$ 1$,
$ 1$;\ \ 
$ 1$,
$ -1$;\ \ 
$ 1$)

Factors = $2_{7,2.}^{4,625}\boxtimes 4_{\frac{9}{5},7.236}^{20,451}$

\vskip 0.7ex
\hangindent=3em \hangafter=1
$\tau_n$ = ($3.08 + 2.24 i$, $0.$, $-4.98 + 3.62 i$, $-4.47 - 6.16 i$, $0. - 7.24 i$, $0.$, $4.98 + 3.62 i$, $7.24 + 9.96 i$, $-3.08 + 2.24 i$, $0.$, $-3.08 - 2.24 i$, $7.24 - 9.96 i$, $4.98 - 3.62 i$, $0.$, $0. + 7.24 i$, $-4.47 + 6.16 i$, $-4.98 - 3.62 i$, $0.$, $3.08 - 2.24 i$, $14.47$)

\vskip 0.7ex
\hangindent=3em \hangafter=1
\textit{Intrinsic sign problem}

  \vskip 2ex

\noindent22. $8_{\frac{16}{5},14.47}^{20,247}$ \irep{207}:\ \ 
$d_i$ = ($1.0$,
$1.0$,
$1.0$,
$1.0$,
$1.618$,
$1.618$,
$1.618$,
$1.618$) 

\vskip 0.7ex
\hangindent=3em \hangafter=1
$D^2= 14.472 = 
10+2\sqrt{5}$

\vskip 0.7ex
\hangindent=3em \hangafter=1
$T = ( 0,
\frac{1}{2},
\frac{3}{4},
\frac{3}{4},
\frac{3}{5},
\frac{1}{10},
\frac{7}{20},
\frac{7}{20} )
$,

\vskip 0.7ex
\hangindent=3em \hangafter=1
$S$ = ($ 1$,
$ 1$,
$ 1$,
$ 1$,
$ \frac{1+\sqrt{5}}{2}$,
$ \frac{1+\sqrt{5}}{2}$,
$ \frac{1+\sqrt{5}}{2}$,
$ \frac{1+\sqrt{5}}{2}$;\ \ 
$ 1$,
$ -1$,
$ -1$,
$ \frac{1+\sqrt{5}}{2}$,
$ \frac{1+\sqrt{5}}{2}$,
$ -\frac{1+\sqrt{5}}{2}$,
$ -\frac{1+\sqrt{5}}{2}$;\ \ 
$ -1$,
$ 1$,
$ \frac{1+\sqrt{5}}{2}$,
$ -\frac{1+\sqrt{5}}{2}$,
$ \frac{1+\sqrt{5}}{2}$,
$ -\frac{1+\sqrt{5}}{2}$;\ \ 
$ -1$,
$ \frac{1+\sqrt{5}}{2}$,
$ -\frac{1+\sqrt{5}}{2}$,
$ -\frac{1+\sqrt{5}}{2}$,
$ \frac{1+\sqrt{5}}{2}$;\ \ 
$ -1$,
$ -1$,
$ -1$,
$ -1$;\ \ 
$ -1$,
$ 1$,
$ 1$;\ \ 
$ 1$,
$ -1$;\ \ 
$ 1$)

Factors = $2_{7,2.}^{4,625}\boxtimes 4_{\frac{21}{5},7.236}^{20,341}$

\vskip 0.7ex
\hangindent=3em \hangafter=1
$\tau_n$ = ($-3.08 + 2.24 i$, $0.$, $4.98 + 3.62 i$, $-4.47 + 6.16 i$, $0. - 7.24 i$, $0.$, $-4.98 + 3.62 i$, $7.24 - 9.96 i$, $3.08 + 2.24 i$, $0.$, $3.08 - 2.24 i$, $7.24 + 9.96 i$, $-4.98 - 3.62 i$, $0.$, $0. + 7.24 i$, $-4.47 - 6.16 i$, $4.98 - 3.62 i$, $0.$, $-3.08 - 2.24 i$, $14.47$)

\vskip 0.7ex
\hangindent=3em \hangafter=1
\textit{Intrinsic sign problem}

  \vskip 2ex

\noindent23. $8_{\frac{19}{5},14.47}^{40,124}$ \irep{233}:\ \ 
$d_i$ = ($1.0$,
$1.0$,
$1.0$,
$1.0$,
$1.618$,
$1.618$,
$1.618$,
$1.618$) 

\vskip 0.7ex
\hangindent=3em \hangafter=1
$D^2= 14.472 = 
10+2\sqrt{5}$

\vskip 0.7ex
\hangindent=3em \hangafter=1
$T = ( 0,
\frac{1}{2},
\frac{1}{8},
\frac{1}{8},
\frac{2}{5},
\frac{9}{10},
\frac{21}{40},
\frac{21}{40} )
$,

\vskip 0.7ex
\hangindent=3em \hangafter=1
$S$ = ($ 1$,
$ 1$,
$ 1$,
$ 1$,
$ \frac{1+\sqrt{5}}{2}$,
$ \frac{1+\sqrt{5}}{2}$,
$ \frac{1+\sqrt{5}}{2}$,
$ \frac{1+\sqrt{5}}{2}$;\ \ 
$ 1$,
$ -1$,
$ -1$,
$ \frac{1+\sqrt{5}}{2}$,
$ \frac{1+\sqrt{5}}{2}$,
$ -\frac{1+\sqrt{5}}{2}$,
$ -\frac{1+\sqrt{5}}{2}$;\ \ 
$-\mathrm{i}$,
$\mathrm{i}$,
$ \frac{1+\sqrt{5}}{2}$,
$ -\frac{1+\sqrt{5}}{2}$,
$(\frac{1+\sqrt{5}}{2})\mathrm{i}$,
$(-\frac{1+\sqrt{5}}{2})\mathrm{i}$;\ \ 
$-\mathrm{i}$,
$ \frac{1+\sqrt{5}}{2}$,
$ -\frac{1+\sqrt{5}}{2}$,
$(-\frac{1+\sqrt{5}}{2})\mathrm{i}$,
$(\frac{1+\sqrt{5}}{2})\mathrm{i}$;\ \ 
$ -1$,
$ -1$,
$ -1$,
$ -1$;\ \ 
$ -1$,
$ 1$,
$ 1$;\ \ 
$\mathrm{i}$,
$-\mathrm{i}$;\ \ 
$\mathrm{i}$)

Factors = $2_{\frac{14}{5},3.618}^{5,395}\boxtimes 4_{1,4.}^{8,718}$

\vskip 0.7ex
\hangindent=3em \hangafter=1
$\tau_n$ = ($-3.76 + 0.6 i$, $8.6 - 1.36 i$, $-6.08 - 0.96 i$, $0.$, $-5.12 - 5.12 i$, $0.84 + 5.31 i$, $-0.96 - 6.08 i$, $7.24 + 9.96 i$, $0.6 - 3.76 i$, $7.24 + 7.24 i$, $-0.6 - 3.76 i$, $0.$, $0.96 - 6.08 i$, $-5.31 - 0.84 i$, $5.12 - 5.12 i$, $-4.47 + 6.16 i$, $6.08 - 0.96 i$, $-1.36 + 8.6 i$, $3.76 + 0.6 i$, $0.$, $3.76 - 0.6 i$, $-1.36 - 8.6 i$, $6.08 + 0.96 i$, $-4.47 - 6.16 i$, $5.12 + 5.12 i$, $-5.31 + 0.84 i$, $0.96 + 6.08 i$, $0.$, $-0.6 + 3.76 i$, $7.24 - 7.24 i$, $0.6 + 3.76 i$, $7.24 - 9.96 i$, $-0.96 + 6.08 i$, $0.84 - 5.31 i$, $-5.12 + 5.12 i$, $0.$, $-6.08 + 0.96 i$, $8.6 + 1.36 i$, $-3.76 - 0.6 i$, $14.47$)

\vskip 0.7ex
\hangindent=3em \hangafter=1
\textit{Intrinsic sign problem}

  \vskip 2ex

\noindent24. $8_{\frac{31}{5},14.47}^{40,371}$ \irep{233}:\ \ 
$d_i$ = ($1.0$,
$1.0$,
$1.0$,
$1.0$,
$1.618$,
$1.618$,
$1.618$,
$1.618$) 

\vskip 0.7ex
\hangindent=3em \hangafter=1
$D^2= 14.472 = 
10+2\sqrt{5}$

\vskip 0.7ex
\hangindent=3em \hangafter=1
$T = ( 0,
\frac{1}{2},
\frac{1}{8},
\frac{1}{8},
\frac{3}{5},
\frac{1}{10},
\frac{29}{40},
\frac{29}{40} )
$,

\vskip 0.7ex
\hangindent=3em \hangafter=1
$S$ = ($ 1$,
$ 1$,
$ 1$,
$ 1$,
$ \frac{1+\sqrt{5}}{2}$,
$ \frac{1+\sqrt{5}}{2}$,
$ \frac{1+\sqrt{5}}{2}$,
$ \frac{1+\sqrt{5}}{2}$;\ \ 
$ 1$,
$ -1$,
$ -1$,
$ \frac{1+\sqrt{5}}{2}$,
$ \frac{1+\sqrt{5}}{2}$,
$ -\frac{1+\sqrt{5}}{2}$,
$ -\frac{1+\sqrt{5}}{2}$;\ \ 
$-\mathrm{i}$,
$\mathrm{i}$,
$ \frac{1+\sqrt{5}}{2}$,
$ -\frac{1+\sqrt{5}}{2}$,
$(\frac{1+\sqrt{5}}{2})\mathrm{i}$,
$(-\frac{1+\sqrt{5}}{2})\mathrm{i}$;\ \ 
$-\mathrm{i}$,
$ \frac{1+\sqrt{5}}{2}$,
$ -\frac{1+\sqrt{5}}{2}$,
$(-\frac{1+\sqrt{5}}{2})\mathrm{i}$,
$(\frac{1+\sqrt{5}}{2})\mathrm{i}$;\ \ 
$ -1$,
$ -1$,
$ -1$,
$ -1$;\ \ 
$ -1$,
$ 1$,
$ 1$;\ \ 
$\mathrm{i}$,
$-\mathrm{i}$;\ \ 
$\mathrm{i}$)

Factors = $2_{\frac{26}{5},3.618}^{5,720}\boxtimes 4_{1,4.}^{8,718}$

\vskip 0.7ex
\hangindent=3em \hangafter=1
$\tau_n$ = ($0.6 - 3.76 i$, $-1.36 + 8.6 i$, $0.96 + 6.08 i$, $0.$, $-5.12 - 5.12 i$, $-5.31 - 0.84 i$, $6.08 + 0.96 i$, $7.24 - 9.96 i$, $-3.76 + 0.6 i$, $7.24 + 7.24 i$, $3.76 + 0.6 i$, $0.$, $-6.08 + 0.96 i$, $0.84 + 5.31 i$, $5.12 - 5.12 i$, $-4.47 - 6.16 i$, $-0.96 + 6.08 i$, $8.6 - 1.36 i$, $-0.6 - 3.76 i$, $0.$, $-0.6 + 3.76 i$, $8.6 + 1.36 i$, $-0.96 - 6.08 i$, $-4.47 + 6.16 i$, $5.12 + 5.12 i$, $0.84 - 5.31 i$, $-6.08 - 0.96 i$, $0.$, $3.76 - 0.6 i$, $7.24 - 7.24 i$, $-3.76 - 0.6 i$, $7.24 + 9.96 i$, $6.08 - 0.96 i$, $-5.31 + 0.84 i$, $-5.12 + 5.12 i$, $0.$, $0.96 - 6.08 i$, $-1.36 - 8.6 i$, $0.6 + 3.76 i$, $14.47$)

\vskip 0.7ex
\hangindent=3em \hangafter=1
\textit{Intrinsic sign problem}

  \vskip 2ex

\noindent25. $8_{\frac{29}{5},14.47}^{40,142}$ \irep{233}:\ \ 
$d_i$ = ($1.0$,
$1.0$,
$1.0$,
$1.0$,
$1.618$,
$1.618$,
$1.618$,
$1.618$) 

\vskip 0.7ex
\hangindent=3em \hangafter=1
$D^2= 14.472 = 
10+2\sqrt{5}$

\vskip 0.7ex
\hangindent=3em \hangafter=1
$T = ( 0,
\frac{1}{2},
\frac{3}{8},
\frac{3}{8},
\frac{2}{5},
\frac{9}{10},
\frac{31}{40},
\frac{31}{40} )
$,

\vskip 0.7ex
\hangindent=3em \hangafter=1
$S$ = ($ 1$,
$ 1$,
$ 1$,
$ 1$,
$ \frac{1+\sqrt{5}}{2}$,
$ \frac{1+\sqrt{5}}{2}$,
$ \frac{1+\sqrt{5}}{2}$,
$ \frac{1+\sqrt{5}}{2}$;\ \ 
$ 1$,
$ -1$,
$ -1$,
$ \frac{1+\sqrt{5}}{2}$,
$ \frac{1+\sqrt{5}}{2}$,
$ -\frac{1+\sqrt{5}}{2}$,
$ -\frac{1+\sqrt{5}}{2}$;\ \ 
$\mathrm{i}$,
$-\mathrm{i}$,
$ \frac{1+\sqrt{5}}{2}$,
$ -\frac{1+\sqrt{5}}{2}$,
$(\frac{1+\sqrt{5}}{2})\mathrm{i}$,
$(-\frac{1+\sqrt{5}}{2})\mathrm{i}$;\ \ 
$\mathrm{i}$,
$ \frac{1+\sqrt{5}}{2}$,
$ -\frac{1+\sqrt{5}}{2}$,
$(-\frac{1+\sqrt{5}}{2})\mathrm{i}$,
$(\frac{1+\sqrt{5}}{2})\mathrm{i}$;\ \ 
$ -1$,
$ -1$,
$ -1$,
$ -1$;\ \ 
$ -1$,
$ 1$,
$ 1$;\ \ 
$-\mathrm{i}$,
$\mathrm{i}$;\ \ 
$-\mathrm{i}$)

Factors = $2_{\frac{14}{5},3.618}^{5,395}\boxtimes 4_{3,4.}^{8,468}$

\vskip 0.7ex
\hangindent=3em \hangafter=1
$\tau_n$ = ($-0.6 - 3.76 i$, $-1.36 - 8.6 i$, $-0.96 + 6.08 i$, $0.$, $5.12 - 5.12 i$, $-5.31 + 0.84 i$, $-6.08 + 0.96 i$, $7.24 + 9.96 i$, $3.76 + 0.6 i$, $7.24 - 7.24 i$, $-3.76 + 0.6 i$, $0.$, $6.08 + 0.96 i$, $0.84 - 5.31 i$, $-5.12 - 5.12 i$, $-4.47 + 6.16 i$, $0.96 + 6.08 i$, $8.6 + 1.36 i$, $0.6 - 3.76 i$, $0.$, $0.6 + 3.76 i$, $8.6 - 1.36 i$, $0.96 - 6.08 i$, $-4.47 - 6.16 i$, $-5.12 + 5.12 i$, $0.84 + 5.31 i$, $6.08 - 0.96 i$, $0.$, $-3.76 - 0.6 i$, $7.24 + 7.24 i$, $3.76 - 0.6 i$, $7.24 - 9.96 i$, $-6.08 - 0.96 i$, $-5.31 - 0.84 i$, $5.12 + 5.12 i$, $0.$, $-0.96 - 6.08 i$, $-1.36 + 8.6 i$, $-0.6 + 3.76 i$, $14.47$)

\vskip 0.7ex
\hangindent=3em \hangafter=1
\textit{Intrinsic sign problem}

  \vskip 2ex

\noindent26. $8_{\frac{1}{5},14.47}^{40,158}$ \irep{233}:\ \ 
$d_i$ = ($1.0$,
$1.0$,
$1.0$,
$1.0$,
$1.618$,
$1.618$,
$1.618$,
$1.618$) 

\vskip 0.7ex
\hangindent=3em \hangafter=1
$D^2= 14.472 = 
10+2\sqrt{5}$

\vskip 0.7ex
\hangindent=3em \hangafter=1
$T = ( 0,
\frac{1}{2},
\frac{3}{8},
\frac{3}{8},
\frac{3}{5},
\frac{1}{10},
\frac{39}{40},
\frac{39}{40} )
$,

\vskip 0.7ex
\hangindent=3em \hangafter=1
$S$ = ($ 1$,
$ 1$,
$ 1$,
$ 1$,
$ \frac{1+\sqrt{5}}{2}$,
$ \frac{1+\sqrt{5}}{2}$,
$ \frac{1+\sqrt{5}}{2}$,
$ \frac{1+\sqrt{5}}{2}$;\ \ 
$ 1$,
$ -1$,
$ -1$,
$ \frac{1+\sqrt{5}}{2}$,
$ \frac{1+\sqrt{5}}{2}$,
$ -\frac{1+\sqrt{5}}{2}$,
$ -\frac{1+\sqrt{5}}{2}$;\ \ 
$\mathrm{i}$,
$-\mathrm{i}$,
$ \frac{1+\sqrt{5}}{2}$,
$ -\frac{1+\sqrt{5}}{2}$,
$(\frac{1+\sqrt{5}}{2})\mathrm{i}$,
$(-\frac{1+\sqrt{5}}{2})\mathrm{i}$;\ \ 
$\mathrm{i}$,
$ \frac{1+\sqrt{5}}{2}$,
$ -\frac{1+\sqrt{5}}{2}$,
$(-\frac{1+\sqrt{5}}{2})\mathrm{i}$,
$(\frac{1+\sqrt{5}}{2})\mathrm{i}$;\ \ 
$ -1$,
$ -1$,
$ -1$,
$ -1$;\ \ 
$ -1$,
$ 1$,
$ 1$;\ \ 
$-\mathrm{i}$,
$\mathrm{i}$;\ \ 
$-\mathrm{i}$)

Factors = $2_{\frac{26}{5},3.618}^{5,720}\boxtimes 4_{3,4.}^{8,468}$

\vskip 0.7ex
\hangindent=3em \hangafter=1
$\tau_n$ = ($3.76 + 0.6 i$, $8.6 + 1.36 i$, $6.08 - 0.96 i$, $0.$, $5.12 - 5.12 i$, $0.84 - 5.31 i$, $0.96 - 6.08 i$, $7.24 - 9.96 i$, $-0.6 - 3.76 i$, $7.24 - 7.24 i$, $0.6 - 3.76 i$, $0.$, $-0.96 - 6.08 i$, $-5.31 + 0.84 i$, $-5.12 - 5.12 i$, $-4.47 - 6.16 i$, $-6.08 - 0.96 i$, $-1.36 - 8.6 i$, $-3.76 + 0.6 i$, $0.$, $-3.76 - 0.6 i$, $-1.36 + 8.6 i$, $-6.08 + 0.96 i$, $-4.47 + 6.16 i$, $-5.12 + 5.12 i$, $-5.31 - 0.84 i$, $-0.96 + 6.08 i$, $0.$, $0.6 + 3.76 i$, $7.24 + 7.24 i$, $-0.6 + 3.76 i$, $7.24 + 9.96 i$, $0.96 + 6.08 i$, $0.84 + 5.31 i$, $5.12 + 5.12 i$, $0.$, $6.08 + 0.96 i$, $8.6 - 1.36 i$, $3.76 - 0.6 i$, $14.47$)

\vskip 0.7ex
\hangindent=3em \hangafter=1
\textit{Intrinsic sign problem}

  \vskip 2ex

\noindent27. $8_{\frac{39}{5},14.47}^{40,152}$ \irep{233}:\ \ 
$d_i$ = ($1.0$,
$1.0$,
$1.0$,
$1.0$,
$1.618$,
$1.618$,
$1.618$,
$1.618$) 

\vskip 0.7ex
\hangindent=3em \hangafter=1
$D^2= 14.472 = 
10+2\sqrt{5}$

\vskip 0.7ex
\hangindent=3em \hangafter=1
$T = ( 0,
\frac{1}{2},
\frac{5}{8},
\frac{5}{8},
\frac{2}{5},
\frac{9}{10},
\frac{1}{40},
\frac{1}{40} )
$,

\vskip 0.7ex
\hangindent=3em \hangafter=1
$S$ = ($ 1$,
$ 1$,
$ 1$,
$ 1$,
$ \frac{1+\sqrt{5}}{2}$,
$ \frac{1+\sqrt{5}}{2}$,
$ \frac{1+\sqrt{5}}{2}$,
$ \frac{1+\sqrt{5}}{2}$;\ \ 
$ 1$,
$ -1$,
$ -1$,
$ \frac{1+\sqrt{5}}{2}$,
$ \frac{1+\sqrt{5}}{2}$,
$ -\frac{1+\sqrt{5}}{2}$,
$ -\frac{1+\sqrt{5}}{2}$;\ \ 
$-\mathrm{i}$,
$\mathrm{i}$,
$ \frac{1+\sqrt{5}}{2}$,
$ -\frac{1+\sqrt{5}}{2}$,
$(\frac{1+\sqrt{5}}{2})\mathrm{i}$,
$(-\frac{1+\sqrt{5}}{2})\mathrm{i}$;\ \ 
$-\mathrm{i}$,
$ \frac{1+\sqrt{5}}{2}$,
$ -\frac{1+\sqrt{5}}{2}$,
$(-\frac{1+\sqrt{5}}{2})\mathrm{i}$,
$(\frac{1+\sqrt{5}}{2})\mathrm{i}$;\ \ 
$ -1$,
$ -1$,
$ -1$,
$ -1$;\ \ 
$ -1$,
$ 1$,
$ 1$;\ \ 
$\mathrm{i}$,
$-\mathrm{i}$;\ \ 
$\mathrm{i}$)

Factors = $2_{\frac{14}{5},3.618}^{5,395}\boxtimes 4_{5,4.}^{8,312}$

\vskip 0.7ex
\hangindent=3em \hangafter=1
$\tau_n$ = ($3.76 - 0.6 i$, $8.6 - 1.36 i$, $6.08 + 0.96 i$, $0.$, $5.12 + 5.12 i$, $0.84 + 5.31 i$, $0.96 + 6.08 i$, $7.24 + 9.96 i$, $-0.6 + 3.76 i$, $7.24 + 7.24 i$, $0.6 + 3.76 i$, $0.$, $-0.96 + 6.08 i$, $-5.31 - 0.84 i$, $-5.12 + 5.12 i$, $-4.47 + 6.16 i$, $-6.08 + 0.96 i$, $-1.36 + 8.6 i$, $-3.76 - 0.6 i$, $0.$, $-3.76 + 0.6 i$, $-1.36 - 8.6 i$, $-6.08 - 0.96 i$, $-4.47 - 6.16 i$, $-5.12 - 5.12 i$, $-5.31 + 0.84 i$, $-0.96 - 6.08 i$, $0.$, $0.6 - 3.76 i$, $7.24 - 7.24 i$, $-0.6 - 3.76 i$, $7.24 - 9.96 i$, $0.96 - 6.08 i$, $0.84 - 5.31 i$, $5.12 - 5.12 i$, $0.$, $6.08 - 0.96 i$, $8.6 + 1.36 i$, $3.76 + 0.6 i$, $14.47$)

\vskip 0.7ex
\hangindent=3em \hangafter=1
\textit{Intrinsic sign problem}

  \vskip 2ex

\noindent28. $8_{\frac{11}{5},14.47}^{40,824}$ \irep{233}:\ \ 
$d_i$ = ($1.0$,
$1.0$,
$1.0$,
$1.0$,
$1.618$,
$1.618$,
$1.618$,
$1.618$) 

\vskip 0.7ex
\hangindent=3em \hangafter=1
$D^2= 14.472 = 
10+2\sqrt{5}$

\vskip 0.7ex
\hangindent=3em \hangafter=1
$T = ( 0,
\frac{1}{2},
\frac{5}{8},
\frac{5}{8},
\frac{3}{5},
\frac{1}{10},
\frac{9}{40},
\frac{9}{40} )
$,

\vskip 0.7ex
\hangindent=3em \hangafter=1
$S$ = ($ 1$,
$ 1$,
$ 1$,
$ 1$,
$ \frac{1+\sqrt{5}}{2}$,
$ \frac{1+\sqrt{5}}{2}$,
$ \frac{1+\sqrt{5}}{2}$,
$ \frac{1+\sqrt{5}}{2}$;\ \ 
$ 1$,
$ -1$,
$ -1$,
$ \frac{1+\sqrt{5}}{2}$,
$ \frac{1+\sqrt{5}}{2}$,
$ -\frac{1+\sqrt{5}}{2}$,
$ -\frac{1+\sqrt{5}}{2}$;\ \ 
$-\mathrm{i}$,
$\mathrm{i}$,
$ \frac{1+\sqrt{5}}{2}$,
$ -\frac{1+\sqrt{5}}{2}$,
$(\frac{1+\sqrt{5}}{2})\mathrm{i}$,
$(-\frac{1+\sqrt{5}}{2})\mathrm{i}$;\ \ 
$-\mathrm{i}$,
$ \frac{1+\sqrt{5}}{2}$,
$ -\frac{1+\sqrt{5}}{2}$,
$(-\frac{1+\sqrt{5}}{2})\mathrm{i}$,
$(\frac{1+\sqrt{5}}{2})\mathrm{i}$;\ \ 
$ -1$,
$ -1$,
$ -1$,
$ -1$;\ \ 
$ -1$,
$ 1$,
$ 1$;\ \ 
$\mathrm{i}$,
$-\mathrm{i}$;\ \ 
$\mathrm{i}$)

Factors = $2_{\frac{26}{5},3.618}^{5,720}\boxtimes 4_{5,4.}^{8,312}$

\vskip 0.7ex
\hangindent=3em \hangafter=1
$\tau_n$ = ($-0.6 + 3.76 i$, $-1.36 + 8.6 i$, $-0.96 - 6.08 i$, $0.$, $5.12 + 5.12 i$, $-5.31 - 0.84 i$, $-6.08 - 0.96 i$, $7.24 - 9.96 i$, $3.76 - 0.6 i$, $7.24 + 7.24 i$, $-3.76 - 0.6 i$, $0.$, $6.08 - 0.96 i$, $0.84 + 5.31 i$, $-5.12 + 5.12 i$, $-4.47 - 6.16 i$, $0.96 - 6.08 i$, $8.6 - 1.36 i$, $0.6 + 3.76 i$, $0.$, $0.6 - 3.76 i$, $8.6 + 1.36 i$, $0.96 + 6.08 i$, $-4.47 + 6.16 i$, $-5.12 - 5.12 i$, $0.84 - 5.31 i$, $6.08 + 0.96 i$, $0.$, $-3.76 + 0.6 i$, $7.24 - 7.24 i$, $3.76 + 0.6 i$, $7.24 + 9.96 i$, $-6.08 + 0.96 i$, $-5.31 + 0.84 i$, $5.12 - 5.12 i$, $0.$, $-0.96 + 6.08 i$, $-1.36 - 8.6 i$, $-0.6 - 3.76 i$, $14.47$)

\vskip 0.7ex
\hangindent=3em \hangafter=1
\textit{Intrinsic sign problem}

  \vskip 2ex

\noindent29. $8_{\frac{9}{5},14.47}^{40,977}$ \irep{233}:\ \ 
$d_i$ = ($1.0$,
$1.0$,
$1.0$,
$1.0$,
$1.618$,
$1.618$,
$1.618$,
$1.618$) 

\vskip 0.7ex
\hangindent=3em \hangafter=1
$D^2= 14.472 = 
10+2\sqrt{5}$

\vskip 0.7ex
\hangindent=3em \hangafter=1
$T = ( 0,
\frac{1}{2},
\frac{7}{8},
\frac{7}{8},
\frac{2}{5},
\frac{9}{10},
\frac{11}{40},
\frac{11}{40} )
$,

\vskip 0.7ex
\hangindent=3em \hangafter=1
$S$ = ($ 1$,
$ 1$,
$ 1$,
$ 1$,
$ \frac{1+\sqrt{5}}{2}$,
$ \frac{1+\sqrt{5}}{2}$,
$ \frac{1+\sqrt{5}}{2}$,
$ \frac{1+\sqrt{5}}{2}$;\ \ 
$ 1$,
$ -1$,
$ -1$,
$ \frac{1+\sqrt{5}}{2}$,
$ \frac{1+\sqrt{5}}{2}$,
$ -\frac{1+\sqrt{5}}{2}$,
$ -\frac{1+\sqrt{5}}{2}$;\ \ 
$\mathrm{i}$,
$-\mathrm{i}$,
$ \frac{1+\sqrt{5}}{2}$,
$ -\frac{1+\sqrt{5}}{2}$,
$(\frac{1+\sqrt{5}}{2})\mathrm{i}$,
$(-\frac{1+\sqrt{5}}{2})\mathrm{i}$;\ \ 
$\mathrm{i}$,
$ \frac{1+\sqrt{5}}{2}$,
$ -\frac{1+\sqrt{5}}{2}$,
$(-\frac{1+\sqrt{5}}{2})\mathrm{i}$,
$(\frac{1+\sqrt{5}}{2})\mathrm{i}$;\ \ 
$ -1$,
$ -1$,
$ -1$,
$ -1$;\ \ 
$ -1$,
$ 1$,
$ 1$;\ \ 
$-\mathrm{i}$,
$\mathrm{i}$;\ \ 
$-\mathrm{i}$)

Factors = $2_{\frac{14}{5},3.618}^{5,395}\boxtimes 4_{7,4.}^{8,781}$

\vskip 0.7ex
\hangindent=3em \hangafter=1
$\tau_n$ = ($0.6 + 3.76 i$, $-1.36 - 8.6 i$, $0.96 - 6.08 i$, $0.$, $-5.12 + 5.12 i$, $-5.31 + 0.84 i$, $6.08 - 0.96 i$, $7.24 + 9.96 i$, $-3.76 - 0.6 i$, $7.24 - 7.24 i$, $3.76 - 0.6 i$, $0.$, $-6.08 - 0.96 i$, $0.84 - 5.31 i$, $5.12 + 5.12 i$, $-4.47 + 6.16 i$, $-0.96 - 6.08 i$, $8.6 + 1.36 i$, $-0.6 + 3.76 i$, $0.$, $-0.6 - 3.76 i$, $8.6 - 1.36 i$, $-0.96 + 6.08 i$, $-4.47 - 6.16 i$, $5.12 - 5.12 i$, $0.84 + 5.31 i$, $-6.08 + 0.96 i$, $0.$, $3.76 + 0.6 i$, $7.24 + 7.24 i$, $-3.76 + 0.6 i$, $7.24 - 9.96 i$, $6.08 + 0.96 i$, $-5.31 - 0.84 i$, $-5.12 - 5.12 i$, $0.$, $0.96 + 6.08 i$, $-1.36 + 8.6 i$, $0.6 - 3.76 i$, $14.47$)

\vskip 0.7ex
\hangindent=3em \hangafter=1
\textit{Intrinsic sign problem}

  \vskip 2ex

\noindent30. $8_{\frac{21}{5},14.47}^{40,492}$ \irep{233}:\ \ 
$d_i$ = ($1.0$,
$1.0$,
$1.0$,
$1.0$,
$1.618$,
$1.618$,
$1.618$,
$1.618$) 

\vskip 0.7ex
\hangindent=3em \hangafter=1
$D^2= 14.472 = 
10+2\sqrt{5}$

\vskip 0.7ex
\hangindent=3em \hangafter=1
$T = ( 0,
\frac{1}{2},
\frac{7}{8},
\frac{7}{8},
\frac{3}{5},
\frac{1}{10},
\frac{19}{40},
\frac{19}{40} )
$,

\vskip 0.7ex
\hangindent=3em \hangafter=1
$S$ = ($ 1$,
$ 1$,
$ 1$,
$ 1$,
$ \frac{1+\sqrt{5}}{2}$,
$ \frac{1+\sqrt{5}}{2}$,
$ \frac{1+\sqrt{5}}{2}$,
$ \frac{1+\sqrt{5}}{2}$;\ \ 
$ 1$,
$ -1$,
$ -1$,
$ \frac{1+\sqrt{5}}{2}$,
$ \frac{1+\sqrt{5}}{2}$,
$ -\frac{1+\sqrt{5}}{2}$,
$ -\frac{1+\sqrt{5}}{2}$;\ \ 
$\mathrm{i}$,
$-\mathrm{i}$,
$ \frac{1+\sqrt{5}}{2}$,
$ -\frac{1+\sqrt{5}}{2}$,
$(\frac{1+\sqrt{5}}{2})\mathrm{i}$,
$(-\frac{1+\sqrt{5}}{2})\mathrm{i}$;\ \ 
$\mathrm{i}$,
$ \frac{1+\sqrt{5}}{2}$,
$ -\frac{1+\sqrt{5}}{2}$,
$(-\frac{1+\sqrt{5}}{2})\mathrm{i}$,
$(\frac{1+\sqrt{5}}{2})\mathrm{i}$;\ \ 
$ -1$,
$ -1$,
$ -1$,
$ -1$;\ \ 
$ -1$,
$ 1$,
$ 1$;\ \ 
$-\mathrm{i}$,
$\mathrm{i}$;\ \ 
$-\mathrm{i}$)

Factors = $2_{\frac{26}{5},3.618}^{5,720}\boxtimes 4_{7,4.}^{8,781}$

\vskip 0.7ex
\hangindent=3em \hangafter=1
$\tau_n$ = ($-3.76 - 0.6 i$, $8.6 + 1.36 i$, $-6.08 + 0.96 i$, $0.$, $-5.12 + 5.12 i$, $0.84 - 5.31 i$, $-0.96 + 6.08 i$, $7.24 - 9.96 i$, $0.6 + 3.76 i$, $7.24 - 7.24 i$, $-0.6 + 3.76 i$, $0.$, $0.96 + 6.08 i$, $-5.31 + 0.84 i$, $5.12 + 5.12 i$, $-4.47 - 6.16 i$, $6.08 + 0.96 i$, $-1.36 - 8.6 i$, $3.76 - 0.6 i$, $0.$, $3.76 + 0.6 i$, $-1.36 + 8.6 i$, $6.08 - 0.96 i$, $-4.47 + 6.16 i$, $5.12 - 5.12 i$, $-5.31 - 0.84 i$, $0.96 - 6.08 i$, $0.$, $-0.6 - 3.76 i$, $7.24 + 7.24 i$, $0.6 - 3.76 i$, $7.24 + 9.96 i$, $-0.96 - 6.08 i$, $0.84 + 5.31 i$, $-5.12 - 5.12 i$, $0.$, $-6.08 - 0.96 i$, $8.6 - 1.36 i$, $-3.76 + 0.6 i$, $14.47$)

\vskip 0.7ex
\hangindent=3em \hangafter=1
\textit{Intrinsic sign problem}

  \vskip 2ex

\noindent31. $8_{\frac{33}{5},26.18}^{20,202}$ \irep{208}:\ \ 
$d_i$ = ($1.0$,
$1.0$,
$1.618$,
$1.618$,
$1.618$,
$1.618$,
$2.618$,
$2.618$) 

\vskip 0.7ex
\hangindent=3em \hangafter=1
$D^2= 26.180 = 
15+5\sqrt{5}$

\vskip 0.7ex
\hangindent=3em \hangafter=1
$T = ( 0,
\frac{1}{4},
\frac{2}{5},
\frac{2}{5},
\frac{13}{20},
\frac{13}{20},
\frac{4}{5},
\frac{1}{20} )
$,

\vskip 0.7ex
\hangindent=3em \hangafter=1
$S$ = ($ 1$,
$ 1$,
$ \frac{1+\sqrt{5}}{2}$,
$ \frac{1+\sqrt{5}}{2}$,
$ \frac{1+\sqrt{5}}{2}$,
$ \frac{1+\sqrt{5}}{2}$,
$ \frac{3+\sqrt{5}}{2}$,
$ \frac{3+\sqrt{5}}{2}$;\ \ 
$ -1$,
$ \frac{1+\sqrt{5}}{2}$,
$ \frac{1+\sqrt{5}}{2}$,
$ -\frac{1+\sqrt{5}}{2}$,
$ -\frac{1+\sqrt{5}}{2}$,
$ \frac{3+\sqrt{5}}{2}$,
$ -\frac{3+\sqrt{5}}{2}$;\ \ 
$ -1$,
$ \frac{3+\sqrt{5}}{2}$,
$ -1$,
$ \frac{3+\sqrt{5}}{2}$,
$ -\frac{1+\sqrt{5}}{2}$,
$ -\frac{1+\sqrt{5}}{2}$;\ \ 
$ -1$,
$ \frac{3+\sqrt{5}}{2}$,
$ -1$,
$ -\frac{1+\sqrt{5}}{2}$,
$ -\frac{1+\sqrt{5}}{2}$;\ \ 
$ 1$,
$ -\frac{3+\sqrt{5}}{2}$,
$ -\frac{1+\sqrt{5}}{2}$,
$ \frac{1+\sqrt{5}}{2}$;\ \ 
$ 1$,
$ -\frac{1+\sqrt{5}}{2}$,
$ \frac{1+\sqrt{5}}{2}$;\ \ 
$ 1$,
$ 1$;\ \ 
$ -1$)

Factors = $2_{1,2.}^{4,437}\boxtimes 4_{\frac{28}{5},13.09}^{5,479}$

\vskip 0.7ex
\hangindent=3em \hangafter=1
$\tau_n$ = ($2.32 - 4.56 i$, $0.$, $6.08 + 11.94 i$, $-2.24 + 6.88 i$, $13.09 + 13.09 i$, $0.$, $-11.94 - 6.08 i$, $-5.85 + 18.02 i$, $-4.56 + 2.32 i$, $0.$, $-4.56 - 2.32 i$, $-5.85 - 18.02 i$, $-11.94 + 6.08 i$, $0.$, $13.09 - 13.09 i$, $-2.24 - 6.88 i$, $6.08 - 11.94 i$, $0.$, $2.32 + 4.56 i$, $26.18$)

\vskip 0.7ex
\hangindent=3em \hangafter=1
\textit{Intrinsic sign problem}

  \vskip 2ex

\noindent32. $8_{1,26.18}^{20,506}$ \irep{208}:\ \ 
$d_i$ = ($1.0$,
$1.0$,
$1.618$,
$1.618$,
$1.618$,
$1.618$,
$2.618$,
$2.618$) 

\vskip 0.7ex
\hangindent=3em \hangafter=1
$D^2= 26.180 = 
15+5\sqrt{5}$

\vskip 0.7ex
\hangindent=3em \hangafter=1
$T = ( 0,
\frac{1}{4},
\frac{2}{5},
\frac{3}{5},
\frac{13}{20},
\frac{17}{20},
0,
\frac{1}{4} )
$,

\vskip 0.7ex
\hangindent=3em \hangafter=1
$S$ = ($ 1$,
$ 1$,
$ \frac{1+\sqrt{5}}{2}$,
$ \frac{1+\sqrt{5}}{2}$,
$ \frac{1+\sqrt{5}}{2}$,
$ \frac{1+\sqrt{5}}{2}$,
$ \frac{3+\sqrt{5}}{2}$,
$ \frac{3+\sqrt{5}}{2}$;\ \ 
$ -1$,
$ \frac{1+\sqrt{5}}{2}$,
$ \frac{1+\sqrt{5}}{2}$,
$ -\frac{1+\sqrt{5}}{2}$,
$ -\frac{1+\sqrt{5}}{2}$,
$ \frac{3+\sqrt{5}}{2}$,
$ -\frac{3+\sqrt{5}}{2}$;\ \ 
$ -1$,
$ \frac{3+\sqrt{5}}{2}$,
$ -1$,
$ \frac{3+\sqrt{5}}{2}$,
$ -\frac{1+\sqrt{5}}{2}$,
$ -\frac{1+\sqrt{5}}{2}$;\ \ 
$ -1$,
$ \frac{3+\sqrt{5}}{2}$,
$ -1$,
$ -\frac{1+\sqrt{5}}{2}$,
$ -\frac{1+\sqrt{5}}{2}$;\ \ 
$ 1$,
$ -\frac{3+\sqrt{5}}{2}$,
$ -\frac{1+\sqrt{5}}{2}$,
$ \frac{1+\sqrt{5}}{2}$;\ \ 
$ 1$,
$ -\frac{1+\sqrt{5}}{2}$,
$ \frac{1+\sqrt{5}}{2}$;\ \ 
$ 1$,
$ 1$;\ \ 
$ -1$)

Factors = $2_{1,2.}^{4,437}\boxtimes 4_{0,13.09}^{5,872}$

\vskip 0.7ex
\hangindent=3em \hangafter=1
$\tau_n$ = ($3.62 + 3.62 i$, $0.$, $9.47 - 9.47 i$, $7.24$, $13.09 + 13.09 i$, $0.$, $9.47 - 9.47 i$, $18.94$, $3.62 + 3.62 i$, $0.$, $3.62 - 3.62 i$, $18.94$, $9.47 + 9.47 i$, $0.$, $13.09 - 13.09 i$, $7.24$, $9.47 + 9.47 i$, $0.$, $3.62 - 3.62 i$, $26.18$)

\vskip 0.7ex
\hangindent=3em \hangafter=1
\textit{Intrinsic sign problem}

  \vskip 2ex

\noindent33. $8_{\frac{17}{5},26.18}^{20,773}$ \irep{208}:\ \ 
$d_i$ = ($1.0$,
$1.0$,
$1.618$,
$1.618$,
$1.618$,
$1.618$,
$2.618$,
$2.618$) 

\vskip 0.7ex
\hangindent=3em \hangafter=1
$D^2= 26.180 = 
15+5\sqrt{5}$

\vskip 0.7ex
\hangindent=3em \hangafter=1
$T = ( 0,
\frac{1}{4},
\frac{3}{5},
\frac{3}{5},
\frac{17}{20},
\frac{17}{20},
\frac{1}{5},
\frac{9}{20} )
$,

\vskip 0.7ex
\hangindent=3em \hangafter=1
$S$ = ($ 1$,
$ 1$,
$ \frac{1+\sqrt{5}}{2}$,
$ \frac{1+\sqrt{5}}{2}$,
$ \frac{1+\sqrt{5}}{2}$,
$ \frac{1+\sqrt{5}}{2}$,
$ \frac{3+\sqrt{5}}{2}$,
$ \frac{3+\sqrt{5}}{2}$;\ \ 
$ -1$,
$ \frac{1+\sqrt{5}}{2}$,
$ \frac{1+\sqrt{5}}{2}$,
$ -\frac{1+\sqrt{5}}{2}$,
$ -\frac{1+\sqrt{5}}{2}$,
$ \frac{3+\sqrt{5}}{2}$,
$ -\frac{3+\sqrt{5}}{2}$;\ \ 
$ -1$,
$ \frac{3+\sqrt{5}}{2}$,
$ -1$,
$ \frac{3+\sqrt{5}}{2}$,
$ -\frac{1+\sqrt{5}}{2}$,
$ -\frac{1+\sqrt{5}}{2}$;\ \ 
$ -1$,
$ \frac{3+\sqrt{5}}{2}$,
$ -1$,
$ -\frac{1+\sqrt{5}}{2}$,
$ -\frac{1+\sqrt{5}}{2}$;\ \ 
$ 1$,
$ -\frac{3+\sqrt{5}}{2}$,
$ -\frac{1+\sqrt{5}}{2}$,
$ \frac{1+\sqrt{5}}{2}$;\ \ 
$ 1$,
$ -\frac{1+\sqrt{5}}{2}$,
$ \frac{1+\sqrt{5}}{2}$;\ \ 
$ 1$,
$ 1$;\ \ 
$ -1$)

Factors = $2_{1,2.}^{4,437}\boxtimes 4_{\frac{12}{5},13.09}^{5,443}$

\vskip 0.7ex
\hangindent=3em \hangafter=1
$\tau_n$ = ($-4.56 + 2.32 i$, $0.$, $-11.94 - 6.08 i$, $-2.24 - 6.88 i$, $13.09 + 13.09 i$, $0.$, $6.08 + 11.94 i$, $-5.85 - 18.02 i$, $2.32 - 4.56 i$, $0.$, $2.32 + 4.56 i$, $-5.85 + 18.02 i$, $6.08 - 11.94 i$, $0.$, $13.09 - 13.09 i$, $-2.24 + 6.88 i$, $-11.94 + 6.08 i$, $0.$, $-4.56 - 2.32 i$, $26.18$)

\vskip 0.7ex
\hangindent=3em \hangafter=1
\textit{Intrinsic sign problem}

  \vskip 2ex

\noindent34. $8_{\frac{23}{5},26.18}^{20,193}$ \irep{208}:\ \ 
$d_i$ = ($1.0$,
$1.0$,
$1.618$,
$1.618$,
$1.618$,
$1.618$,
$2.618$,
$2.618$) 

\vskip 0.7ex
\hangindent=3em \hangafter=1
$D^2= 26.180 = 
15+5\sqrt{5}$

\vskip 0.7ex
\hangindent=3em \hangafter=1
$T = ( 0,
\frac{3}{4},
\frac{2}{5},
\frac{2}{5},
\frac{3}{20},
\frac{3}{20},
\frac{4}{5},
\frac{11}{20} )
$,

\vskip 0.7ex
\hangindent=3em \hangafter=1
$S$ = ($ 1$,
$ 1$,
$ \frac{1+\sqrt{5}}{2}$,
$ \frac{1+\sqrt{5}}{2}$,
$ \frac{1+\sqrt{5}}{2}$,
$ \frac{1+\sqrt{5}}{2}$,
$ \frac{3+\sqrt{5}}{2}$,
$ \frac{3+\sqrt{5}}{2}$;\ \ 
$ -1$,
$ \frac{1+\sqrt{5}}{2}$,
$ \frac{1+\sqrt{5}}{2}$,
$ -\frac{1+\sqrt{5}}{2}$,
$ -\frac{1+\sqrt{5}}{2}$,
$ \frac{3+\sqrt{5}}{2}$,
$ -\frac{3+\sqrt{5}}{2}$;\ \ 
$ -1$,
$ \frac{3+\sqrt{5}}{2}$,
$ -1$,
$ \frac{3+\sqrt{5}}{2}$,
$ -\frac{1+\sqrt{5}}{2}$,
$ -\frac{1+\sqrt{5}}{2}$;\ \ 
$ -1$,
$ \frac{3+\sqrt{5}}{2}$,
$ -1$,
$ -\frac{1+\sqrt{5}}{2}$,
$ -\frac{1+\sqrt{5}}{2}$;\ \ 
$ 1$,
$ -\frac{3+\sqrt{5}}{2}$,
$ -\frac{1+\sqrt{5}}{2}$,
$ \frac{1+\sqrt{5}}{2}$;\ \ 
$ 1$,
$ -\frac{1+\sqrt{5}}{2}$,
$ \frac{1+\sqrt{5}}{2}$;\ \ 
$ 1$,
$ 1$;\ \ 
$ -1$)

Factors = $2_{7,2.}^{4,625}\boxtimes 4_{\frac{28}{5},13.09}^{5,479}$

\vskip 0.7ex
\hangindent=3em \hangafter=1
$\tau_n$ = ($-4.56 - 2.32 i$, $0.$, $-11.94 + 6.08 i$, $-2.24 + 6.88 i$, $13.09 - 13.09 i$, $0.$, $6.08 - 11.94 i$, $-5.85 + 18.02 i$, $2.32 + 4.56 i$, $0.$, $2.32 - 4.56 i$, $-5.85 - 18.02 i$, $6.08 + 11.94 i$, $0.$, $13.09 + 13.09 i$, $-2.24 - 6.88 i$, $-11.94 - 6.08 i$, $0.$, $-4.56 + 2.32 i$, $26.18$)

\vskip 0.7ex
\hangindent=3em \hangafter=1
\textit{Intrinsic sign problem}

  \vskip 2ex

\noindent35. $8_{7,26.18}^{20,315}$ \irep{208}:\ \ 
$d_i$ = ($1.0$,
$1.0$,
$1.618$,
$1.618$,
$1.618$,
$1.618$,
$2.618$,
$2.618$) 

\vskip 0.7ex
\hangindent=3em \hangafter=1
$D^2= 26.180 = 
15+5\sqrt{5}$

\vskip 0.7ex
\hangindent=3em \hangafter=1
$T = ( 0,
\frac{3}{4},
\frac{2}{5},
\frac{3}{5},
\frac{3}{20},
\frac{7}{20},
0,
\frac{3}{4} )
$,

\vskip 0.7ex
\hangindent=3em \hangafter=1
$S$ = ($ 1$,
$ 1$,
$ \frac{1+\sqrt{5}}{2}$,
$ \frac{1+\sqrt{5}}{2}$,
$ \frac{1+\sqrt{5}}{2}$,
$ \frac{1+\sqrt{5}}{2}$,
$ \frac{3+\sqrt{5}}{2}$,
$ \frac{3+\sqrt{5}}{2}$;\ \ 
$ -1$,
$ \frac{1+\sqrt{5}}{2}$,
$ \frac{1+\sqrt{5}}{2}$,
$ -\frac{1+\sqrt{5}}{2}$,
$ -\frac{1+\sqrt{5}}{2}$,
$ \frac{3+\sqrt{5}}{2}$,
$ -\frac{3+\sqrt{5}}{2}$;\ \ 
$ -1$,
$ \frac{3+\sqrt{5}}{2}$,
$ -1$,
$ \frac{3+\sqrt{5}}{2}$,
$ -\frac{1+\sqrt{5}}{2}$,
$ -\frac{1+\sqrt{5}}{2}$;\ \ 
$ -1$,
$ \frac{3+\sqrt{5}}{2}$,
$ -1$,
$ -\frac{1+\sqrt{5}}{2}$,
$ -\frac{1+\sqrt{5}}{2}$;\ \ 
$ 1$,
$ -\frac{3+\sqrt{5}}{2}$,
$ -\frac{1+\sqrt{5}}{2}$,
$ \frac{1+\sqrt{5}}{2}$;\ \ 
$ 1$,
$ -\frac{1+\sqrt{5}}{2}$,
$ \frac{1+\sqrt{5}}{2}$;\ \ 
$ 1$,
$ 1$;\ \ 
$ -1$)

Factors = $2_{7,2.}^{4,625}\boxtimes 4_{0,13.09}^{5,872}$

\vskip 0.7ex
\hangindent=3em \hangafter=1
$\tau_n$ = ($3.62 - 3.62 i$, $0.$, $9.47 + 9.47 i$, $7.24$, $13.09 - 13.09 i$, $0.$, $9.47 + 9.47 i$, $18.94$, $3.62 - 3.62 i$, $0.$, $3.62 + 3.62 i$, $18.94$, $9.47 - 9.47 i$, $0.$, $13.09 + 13.09 i$, $7.24$, $9.47 - 9.47 i$, $0.$, $3.62 + 3.62 i$, $26.18$)

\vskip 0.7ex
\hangindent=3em \hangafter=1
\textit{Intrinsic sign problem}

  \vskip 2ex

\noindent36. $8_{\frac{7}{5},26.18}^{20,116}$ \irep{208}:\ \ 
$d_i$ = ($1.0$,
$1.0$,
$1.618$,
$1.618$,
$1.618$,
$1.618$,
$2.618$,
$2.618$) 

\vskip 0.7ex
\hangindent=3em \hangafter=1
$D^2= 26.180 = 
15+5\sqrt{5}$

\vskip 0.7ex
\hangindent=3em \hangafter=1
$T = ( 0,
\frac{3}{4},
\frac{3}{5},
\frac{3}{5},
\frac{7}{20},
\frac{7}{20},
\frac{1}{5},
\frac{19}{20} )
$,

\vskip 0.7ex
\hangindent=3em \hangafter=1
$S$ = ($ 1$,
$ 1$,
$ \frac{1+\sqrt{5}}{2}$,
$ \frac{1+\sqrt{5}}{2}$,
$ \frac{1+\sqrt{5}}{2}$,
$ \frac{1+\sqrt{5}}{2}$,
$ \frac{3+\sqrt{5}}{2}$,
$ \frac{3+\sqrt{5}}{2}$;\ \ 
$ -1$,
$ \frac{1+\sqrt{5}}{2}$,
$ \frac{1+\sqrt{5}}{2}$,
$ -\frac{1+\sqrt{5}}{2}$,
$ -\frac{1+\sqrt{5}}{2}$,
$ \frac{3+\sqrt{5}}{2}$,
$ -\frac{3+\sqrt{5}}{2}$;\ \ 
$ -1$,
$ \frac{3+\sqrt{5}}{2}$,
$ -1$,
$ \frac{3+\sqrt{5}}{2}$,
$ -\frac{1+\sqrt{5}}{2}$,
$ -\frac{1+\sqrt{5}}{2}$;\ \ 
$ -1$,
$ \frac{3+\sqrt{5}}{2}$,
$ -1$,
$ -\frac{1+\sqrt{5}}{2}$,
$ -\frac{1+\sqrt{5}}{2}$;\ \ 
$ 1$,
$ -\frac{3+\sqrt{5}}{2}$,
$ -\frac{1+\sqrt{5}}{2}$,
$ \frac{1+\sqrt{5}}{2}$;\ \ 
$ 1$,
$ -\frac{1+\sqrt{5}}{2}$,
$ \frac{1+\sqrt{5}}{2}$;\ \ 
$ 1$,
$ 1$;\ \ 
$ -1$)

Factors = $2_{7,2.}^{4,625}\boxtimes 4_{\frac{12}{5},13.09}^{5,443}$

\vskip 0.7ex
\hangindent=3em \hangafter=1
$\tau_n$ = ($2.32 + 4.56 i$, $0.$, $6.08 - 11.94 i$, $-2.24 - 6.88 i$, $13.09 - 13.09 i$, $0.$, $-11.94 + 6.08 i$, $-5.85 - 18.02 i$, $-4.56 - 2.32 i$, $0.$, $-4.56 + 2.32 i$, $-5.85 + 18.02 i$, $-11.94 - 6.08 i$, $0.$, $13.09 + 13.09 i$, $-2.24 + 6.88 i$, $6.08 + 11.94 i$, $0.$, $2.32 - 4.56 i$, $26.18$)

\vskip 0.7ex
\hangindent=3em \hangafter=1
\textit{Intrinsic sign problem}

  \vskip 2ex

\noindent37. $8_{0,36.}^{6,213}$ \irep{0}:\ \ 
$d_i$ = ($1.0$,
$1.0$,
$2.0$,
$2.0$,
$2.0$,
$2.0$,
$3.0$,
$3.0$) 

\vskip 0.7ex
\hangindent=3em \hangafter=1
$D^2= 36.0 = 
36$

\vskip 0.7ex
\hangindent=3em \hangafter=1
$T = ( 0,
0,
0,
0,
\frac{1}{3},
\frac{2}{3},
0,
\frac{1}{2} )
$,

\vskip 0.7ex
\hangindent=3em \hangafter=1
$S$ = ($ 1$,
$ 1$,
$ 2$,
$ 2$,
$ 2$,
$ 2$,
$ 3$,
$ 3$;\ \ 
$ 1$,
$ 2$,
$ 2$,
$ 2$,
$ 2$,
$ -3$,
$ -3$;\ \ 
$ 4$,
$ -2$,
$ -2$,
$ -2$,
$0$,
$0$;\ \ 
$ 4$,
$ -2$,
$ -2$,
$0$,
$0$;\ \ 
$ -2$,
$ 4$,
$0$,
$0$;\ \ 
$ -2$,
$0$,
$0$;\ \ 
$ 3$,
$ -3$;\ \ 
$ 3$)

\vskip 0.7ex
\hangindent=3em \hangafter=1
$\tau_n$ = ($6.$, $24.$, $18.$, $24.$, $6.$, $36.$)

\vskip 0.7ex
\hangindent=3em \hangafter=1
\textit{Sign-problem-free}

  \vskip 2ex

\noindent38. $8_{4,36.}^{6,102}$ \irep{0}:\ \ 
$d_i$ = ($1.0$,
$1.0$,
$2.0$,
$2.0$,
$2.0$,
$2.0$,
$3.0$,
$3.0$) 

\vskip 0.7ex
\hangindent=3em \hangafter=1
$D^2= 36.0 = 
36$

\vskip 0.7ex
\hangindent=3em \hangafter=1
$T = ( 0,
0,
\frac{1}{3},
\frac{1}{3},
\frac{2}{3},
\frac{2}{3},
0,
\frac{1}{2} )
$,

\vskip 0.7ex
\hangindent=3em \hangafter=1
$S$ = ($ 1$,
$ 1$,
$ 2$,
$ 2$,
$ 2$,
$ 2$,
$ 3$,
$ 3$;\ \ 
$ 1$,
$ 2$,
$ 2$,
$ 2$,
$ 2$,
$ -3$,
$ -3$;\ \ 
$ -2$,
$ 4$,
$ -2$,
$ -2$,
$0$,
$0$;\ \ 
$ -2$,
$ -2$,
$ -2$,
$0$,
$0$;\ \ 
$ -2$,
$ 4$,
$0$,
$0$;\ \ 
$ -2$,
$0$,
$0$;\ \ 
$ -3$,
$ 3$;\ \ 
$ -3$)

\vskip 0.7ex
\hangindent=3em \hangafter=1
$\tau_n$ = ($-6.$, $12.$, $18.$, $12.$, $-6.$, $36.$)

\vskip 0.7ex
\hangindent=3em \hangafter=1
\textit{Intrinsic sign problem}

  \vskip 2ex

\noindent39. $8_{0,36.}^{12,101}$ \irep{0}:\ \ 
$d_i$ = ($1.0$,
$1.0$,
$2.0$,
$2.0$,
$2.0$,
$2.0$,
$3.0$,
$3.0$) 

\vskip 0.7ex
\hangindent=3em \hangafter=1
$D^2= 36.0 = 
36$

\vskip 0.7ex
\hangindent=3em \hangafter=1
$T = ( 0,
0,
0,
0,
\frac{1}{3},
\frac{2}{3},
\frac{1}{4},
\frac{3}{4} )
$,

\vskip 0.7ex
\hangindent=3em \hangafter=1
$S$ = ($ 1$,
$ 1$,
$ 2$,
$ 2$,
$ 2$,
$ 2$,
$ 3$,
$ 3$;\ \ 
$ 1$,
$ 2$,
$ 2$,
$ 2$,
$ 2$,
$ -3$,
$ -3$;\ \ 
$ 4$,
$ -2$,
$ -2$,
$ -2$,
$0$,
$0$;\ \ 
$ 4$,
$ -2$,
$ -2$,
$0$,
$0$;\ \ 
$ -2$,
$ 4$,
$0$,
$0$;\ \ 
$ -2$,
$0$,
$0$;\ \ 
$ -3$,
$ 3$;\ \ 
$ -3$)

\vskip 0.7ex
\hangindent=3em \hangafter=1
$\tau_n$ = ($6.$, $-12.$, $18.$, $24.$, $6.$, $0.$, $6.$, $24.$, $18.$, $-12.$, $6.$, $36.$)

\vskip 0.7ex
\hangindent=3em \hangafter=1
\textit{Intrinsic sign problem}

  \vskip 2ex

\noindent40. $8_{4,36.}^{12,972}$ \irep{0}:\ \ 
$d_i$ = ($1.0$,
$1.0$,
$2.0$,
$2.0$,
$2.0$,
$2.0$,
$3.0$,
$3.0$) 

\vskip 0.7ex
\hangindent=3em \hangafter=1
$D^2= 36.0 = 
36$

\vskip 0.7ex
\hangindent=3em \hangafter=1
$T = ( 0,
0,
\frac{1}{3},
\frac{1}{3},
\frac{2}{3},
\frac{2}{3},
\frac{1}{4},
\frac{3}{4} )
$,

\vskip 0.7ex
\hangindent=3em \hangafter=1
$S$ = ($ 1$,
$ 1$,
$ 2$,
$ 2$,
$ 2$,
$ 2$,
$ 3$,
$ 3$;\ \ 
$ 1$,
$ 2$,
$ 2$,
$ 2$,
$ 2$,
$ -3$,
$ -3$;\ \ 
$ -2$,
$ 4$,
$ -2$,
$ -2$,
$0$,
$0$;\ \ 
$ -2$,
$ -2$,
$ -2$,
$0$,
$0$;\ \ 
$ -2$,
$ 4$,
$0$,
$0$;\ \ 
$ -2$,
$0$,
$0$;\ \ 
$ 3$,
$ -3$;\ \ 
$ 3$)

\vskip 0.7ex
\hangindent=3em \hangafter=1
$\tau_n$ = ($-6.$, $-24.$, $18.$, $12.$, $-6.$, $0.$, $-6.$, $12.$, $18.$, $-24.$, $-6.$, $36.$)

\vskip 0.7ex
\hangindent=3em \hangafter=1
\textit{Intrinsic sign problem}

  \vskip 2ex

\noindent41. $8_{0,36.}^{18,162}$ \irep{0}:\ \ 
$d_i$ = ($1.0$,
$1.0$,
$2.0$,
$2.0$,
$2.0$,
$2.0$,
$3.0$,
$3.0$) 

\vskip 0.7ex
\hangindent=3em \hangafter=1
$D^2= 36.0 = 
36$

\vskip 0.7ex
\hangindent=3em \hangafter=1
$T = ( 0,
0,
0,
\frac{1}{9},
\frac{4}{9},
\frac{7}{9},
0,
\frac{1}{2} )
$,

\vskip 0.7ex
\hangindent=3em \hangafter=1
$S$ = ($ 1$,
$ 1$,
$ 2$,
$ 2$,
$ 2$,
$ 2$,
$ 3$,
$ 3$;\ \ 
$ 1$,
$ 2$,
$ 2$,
$ 2$,
$ 2$,
$ -3$,
$ -3$;\ \ 
$ 4$,
$ -2$,
$ -2$,
$ -2$,
$0$,
$0$;\ \ 
$ 2c_{9}^{2}$,
$ 2c_{9}^{4}$,
$ 2c_{9}^{1}$,
$0$,
$0$;\ \ 
$ 2c_{9}^{1}$,
$ 2c_{9}^{2}$,
$0$,
$0$;\ \ 
$ 2c_{9}^{4}$,
$0$,
$0$;\ \ 
$ 3$,
$ -3$;\ \ 
$ 3$)

\vskip 0.7ex
\hangindent=3em \hangafter=1
$\tau_n$ = ($6.$, $24.$, $0. + 10.39 i$, $24.$, $6.$, $18. - 10.39 i$, $6.$, $24.$, $18.$, $24.$, $6.$, $18. + 10.39 i$, $6.$, $24.$, $0. - 10.39 i$, $24.$, $6.$, $36.$)

\vskip 0.7ex
\hangindent=3em \hangafter=1
\textit{Intrinsic sign problem}

  \vskip 2ex

\noindent42. $8_{0,36.}^{18,953}$ \irep{0}:\ \ 
$d_i$ = ($1.0$,
$1.0$,
$2.0$,
$2.0$,
$2.0$,
$2.0$,
$3.0$,
$3.0$) 

\vskip 0.7ex
\hangindent=3em \hangafter=1
$D^2= 36.0 = 
36$

\vskip 0.7ex
\hangindent=3em \hangafter=1
$T = ( 0,
0,
0,
\frac{2}{9},
\frac{5}{9},
\frac{8}{9},
0,
\frac{1}{2} )
$,

\vskip 0.7ex
\hangindent=3em \hangafter=1
$S$ = ($ 1$,
$ 1$,
$ 2$,
$ 2$,
$ 2$,
$ 2$,
$ 3$,
$ 3$;\ \ 
$ 1$,
$ 2$,
$ 2$,
$ 2$,
$ 2$,
$ -3$,
$ -3$;\ \ 
$ 4$,
$ -2$,
$ -2$,
$ -2$,
$0$,
$0$;\ \ 
$ 2c_{9}^{4}$,
$ 2c_{9}^{2}$,
$ 2c_{9}^{1}$,
$0$,
$0$;\ \ 
$ 2c_{9}^{1}$,
$ 2c_{9}^{4}$,
$0$,
$0$;\ \ 
$ 2c_{9}^{2}$,
$0$,
$0$;\ \ 
$ 3$,
$ -3$;\ \ 
$ 3$)

\vskip 0.7ex
\hangindent=3em \hangafter=1
$\tau_n$ = ($6.$, $24.$, $0. - 10.39 i$, $24.$, $6.$, $18. + 10.39 i$, $6.$, $24.$, $18.$, $24.$, $6.$, $18. - 10.39 i$, $6.$, $24.$, $0. + 10.39 i$, $24.$, $6.$, $36.$)

\vskip 0.7ex
\hangindent=3em \hangafter=1
\textit{Intrinsic sign problem}

  \vskip 2ex

\noindent43. $8_{0,36.}^{36,495}$ \irep{0}:\ \ 
$d_i$ = ($1.0$,
$1.0$,
$2.0$,
$2.0$,
$2.0$,
$2.0$,
$3.0$,
$3.0$) 

\vskip 0.7ex
\hangindent=3em \hangafter=1
$D^2= 36.0 = 
36$

\vskip 0.7ex
\hangindent=3em \hangafter=1
$T = ( 0,
0,
0,
\frac{1}{9},
\frac{4}{9},
\frac{7}{9},
\frac{1}{4},
\frac{3}{4} )
$,

\vskip 0.7ex
\hangindent=3em \hangafter=1
$S$ = ($ 1$,
$ 1$,
$ 2$,
$ 2$,
$ 2$,
$ 2$,
$ 3$,
$ 3$;\ \ 
$ 1$,
$ 2$,
$ 2$,
$ 2$,
$ 2$,
$ -3$,
$ -3$;\ \ 
$ 4$,
$ -2$,
$ -2$,
$ -2$,
$0$,
$0$;\ \ 
$ 2c_{9}^{2}$,
$ 2c_{9}^{4}$,
$ 2c_{9}^{1}$,
$0$,
$0$;\ \ 
$ 2c_{9}^{1}$,
$ 2c_{9}^{2}$,
$0$,
$0$;\ \ 
$ 2c_{9}^{4}$,
$0$,
$0$;\ \ 
$ -3$,
$ 3$;\ \ 
$ -3$)

\vskip 0.7ex
\hangindent=3em \hangafter=1
$\tau_n$ = ($6.$, $-12.$, $0. + 10.39 i$, $24.$, $6.$, $-18. - 10.39 i$, $6.$, $24.$, $18.$, $-12.$, $6.$, $18. + 10.39 i$, $6.$, $-12.$, $0. - 10.39 i$, $24.$, $6.$, $0.$, $6.$, $24.$, $0. + 10.39 i$, $-12.$, $6.$, $18. - 10.39 i$, $6.$, $-12.$, $18.$, $24.$, $6.$, $-18. + 10.39 i$, $6.$, $24.$, $0. - 10.39 i$, $-12.$, $6.$, $36.$)

\vskip 0.7ex
\hangindent=3em \hangafter=1
\textit{Intrinsic sign problem}

  \vskip 2ex

\noindent44. $8_{0,36.}^{36,171}$ \irep{0}:\ \ 
$d_i$ = ($1.0$,
$1.0$,
$2.0$,
$2.0$,
$2.0$,
$2.0$,
$3.0$,
$3.0$) 

\vskip 0.7ex
\hangindent=3em \hangafter=1
$D^2= 36.0 = 
36$

\vskip 0.7ex
\hangindent=3em \hangafter=1
$T = ( 0,
0,
0,
\frac{2}{9},
\frac{5}{9},
\frac{8}{9},
\frac{1}{4},
\frac{3}{4} )
$,

\vskip 0.7ex
\hangindent=3em \hangafter=1
$S$ = ($ 1$,
$ 1$,
$ 2$,
$ 2$,
$ 2$,
$ 2$,
$ 3$,
$ 3$;\ \ 
$ 1$,
$ 2$,
$ 2$,
$ 2$,
$ 2$,
$ -3$,
$ -3$;\ \ 
$ 4$,
$ -2$,
$ -2$,
$ -2$,
$0$,
$0$;\ \ 
$ 2c_{9}^{4}$,
$ 2c_{9}^{2}$,
$ 2c_{9}^{1}$,
$0$,
$0$;\ \ 
$ 2c_{9}^{1}$,
$ 2c_{9}^{4}$,
$0$,
$0$;\ \ 
$ 2c_{9}^{2}$,
$0$,
$0$;\ \ 
$ -3$,
$ 3$;\ \ 
$ -3$)

\vskip 0.7ex
\hangindent=3em \hangafter=1
$\tau_n$ = ($6.$, $-12.$, $0. - 10.39 i$, $24.$, $6.$, $-18. + 10.39 i$, $6.$, $24.$, $18.$, $-12.$, $6.$, $18. - 10.39 i$, $6.$, $-12.$, $0. + 10.39 i$, $24.$, $6.$, $0.$, $6.$, $24.$, $0. - 10.39 i$, $-12.$, $6.$, $18. + 10.39 i$, $6.$, $-12.$, $18.$, $24.$, $6.$, $-18. - 10.39 i$, $6.$, $24.$, $0. + 10.39 i$, $-12.$, $6.$, $36.$)

\vskip 0.7ex
\hangindent=3em \hangafter=1
\textit{Intrinsic sign problem}

  \vskip 2ex

\noindent45. $8_{\frac{13}{3},38.46}^{36,115}$ \irep{229}:\ \ 
$d_i$ = ($1.0$,
$1.0$,
$1.879$,
$1.879$,
$2.532$,
$2.532$,
$2.879$,
$2.879$) 

\vskip 0.7ex
\hangindent=3em \hangafter=1
$D^2= 38.468 = 
18+12c^{1}_{9}
+6c^{2}_{9}
$

\vskip 0.7ex
\hangindent=3em \hangafter=1
$T = ( 0,
\frac{1}{4},
\frac{1}{3},
\frac{7}{12},
\frac{2}{9},
\frac{17}{36},
\frac{2}{3},
\frac{11}{12} )
$,

\vskip 0.7ex
\hangindent=3em \hangafter=1
$S$ = ($ 1$,
$ 1$,
$ -c_{9}^{4}$,
$ -c_{9}^{4}$,
$ \xi_{9}^{3}$,
$ \xi_{9}^{3}$,
$ \xi_{9}^{5}$,
$ \xi_{9}^{5}$;\ \ 
$ -1$,
$ -c_{9}^{4}$,
$ c_{9}^{4}$,
$ \xi_{9}^{3}$,
$ -\xi_{9}^{3}$,
$ \xi_{9}^{5}$,
$ -\xi_{9}^{5}$;\ \ 
$ -\xi_{9}^{5}$,
$ -\xi_{9}^{5}$,
$ \xi_{9}^{3}$,
$ \xi_{9}^{3}$,
$ -1$,
$ -1$;\ \ 
$ \xi_{9}^{5}$,
$ \xi_{9}^{3}$,
$ -\xi_{9}^{3}$,
$ -1$,
$ 1$;\ \ 
$0$,
$0$,
$ -\xi_{9}^{3}$,
$ -\xi_{9}^{3}$;\ \ 
$0$,
$ -\xi_{9}^{3}$,
$ \xi_{9}^{3}$;\ \ 
$ -c_{9}^{4}$,
$ -c_{9}^{4}$;\ \ 
$ c_{9}^{4}$)

Factors = $2_{1,2.}^{4,437}\boxtimes 4_{\frac{10}{3},19.23}^{9,459}$

\vskip 0.7ex
\hangindent=3em \hangafter=1
$\tau_n$ = ($-5.99 - 1.6 i$, $0.$, $4.06 - 15.17 i$, $0. - 16.48 i$, $-8.24 + 8.24 i$, $0.$, $-17.25 + 4.62 i$, $-7.59 - 4.39 i$, $19.23 + 19.23 i$, $0.$, $-4.62 + 17.25 i$, $19.23 - 11.1 i$, $8.24 - 8.24 i$, $0.$, $15.17 - 4.06 i$, $-21.87 - 12.63 i$, $-1.6 - 5.99 i$, $0.$, $-1.6 + 5.99 i$, $-21.87 + 12.63 i$, $15.17 + 4.06 i$, $0.$, $8.24 + 8.24 i$, $19.23 + 11.1 i$, $-4.62 - 17.25 i$, $0.$, $19.23 - 19.23 i$, $-7.59 + 4.39 i$, $-17.25 - 4.62 i$, $0.$, $-8.24 - 8.24 i$, $0. + 16.48 i$, $4.06 + 15.17 i$, $0.$, $-5.99 + 1.6 i$, $38.46$)

\vskip 0.7ex
\hangindent=3em \hangafter=1
\textit{Intrinsic sign problem}

  \vskip 2ex

\noindent46. $8_{\frac{17}{3},38.46}^{36,116}$ \irep{229}:\ \ 
$d_i$ = ($1.0$,
$1.0$,
$1.879$,
$1.879$,
$2.532$,
$2.532$,
$2.879$,
$2.879$) 

\vskip 0.7ex
\hangindent=3em \hangafter=1
$D^2= 38.468 = 
18+12c^{1}_{9}
+6c^{2}_{9}
$

\vskip 0.7ex
\hangindent=3em \hangafter=1
$T = ( 0,
\frac{1}{4},
\frac{2}{3},
\frac{11}{12},
\frac{7}{9},
\frac{1}{36},
\frac{1}{3},
\frac{7}{12} )
$,

\vskip 0.7ex
\hangindent=3em \hangafter=1
$S$ = ($ 1$,
$ 1$,
$ -c_{9}^{4}$,
$ -c_{9}^{4}$,
$ \xi_{9}^{3}$,
$ \xi_{9}^{3}$,
$ \xi_{9}^{5}$,
$ \xi_{9}^{5}$;\ \ 
$ -1$,
$ -c_{9}^{4}$,
$ c_{9}^{4}$,
$ \xi_{9}^{3}$,
$ -\xi_{9}^{3}$,
$ \xi_{9}^{5}$,
$ -\xi_{9}^{5}$;\ \ 
$ -\xi_{9}^{5}$,
$ -\xi_{9}^{5}$,
$ \xi_{9}^{3}$,
$ \xi_{9}^{3}$,
$ -1$,
$ -1$;\ \ 
$ \xi_{9}^{5}$,
$ \xi_{9}^{3}$,
$ -\xi_{9}^{3}$,
$ -1$,
$ 1$;\ \ 
$0$,
$0$,
$ -\xi_{9}^{3}$,
$ -\xi_{9}^{3}$;\ \ 
$0$,
$ -\xi_{9}^{3}$,
$ \xi_{9}^{3}$;\ \ 
$ -c_{9}^{4}$,
$ -c_{9}^{4}$;\ \ 
$ c_{9}^{4}$)

Factors = $2_{1,2.}^{4,437}\boxtimes 4_{\frac{14}{3},19.23}^{9,614}$

\vskip 0.7ex
\hangindent=3em \hangafter=1
$\tau_n$ = ($-1.6 - 5.99 i$, $0.$, $15.17 - 4.06 i$, $0. + 16.48 i$, $8.24 - 8.24 i$, $0.$, $-4.62 + 17.25 i$, $-7.59 + 4.39 i$, $19.23 + 19.23 i$, $0.$, $-17.25 + 4.62 i$, $19.23 + 11.1 i$, $-8.24 + 8.24 i$, $0.$, $4.06 - 15.17 i$, $-21.87 + 12.63 i$, $-5.99 - 1.6 i$, $0.$, $-5.99 + 1.6 i$, $-21.87 - 12.63 i$, $4.06 + 15.17 i$, $0.$, $-8.24 - 8.24 i$, $19.23 - 11.1 i$, $-17.25 - 4.62 i$, $0.$, $19.23 - 19.23 i$, $-7.59 - 4.39 i$, $-4.62 - 17.25 i$, $0.$, $8.24 + 8.24 i$, $0. - 16.48 i$, $15.17 + 4.06 i$, $0.$, $-1.6 + 5.99 i$, $38.46$)

\vskip 0.7ex
\hangindent=3em \hangafter=1
\textit{Intrinsic sign problem}

  \vskip 2ex

\noindent47. $8_{\frac{7}{3},38.46}^{36,936}$ \irep{229}:\ \ 
$d_i$ = ($1.0$,
$1.0$,
$1.879$,
$1.879$,
$2.532$,
$2.532$,
$2.879$,
$2.879$) 

\vskip 0.7ex
\hangindent=3em \hangafter=1
$D^2= 38.468 = 
18+12c^{1}_{9}
+6c^{2}_{9}
$

\vskip 0.7ex
\hangindent=3em \hangafter=1
$T = ( 0,
\frac{3}{4},
\frac{1}{3},
\frac{1}{12},
\frac{2}{9},
\frac{35}{36},
\frac{2}{3},
\frac{5}{12} )
$,

\vskip 0.7ex
\hangindent=3em \hangafter=1
$S$ = ($ 1$,
$ 1$,
$ -c_{9}^{4}$,
$ -c_{9}^{4}$,
$ \xi_{9}^{3}$,
$ \xi_{9}^{3}$,
$ \xi_{9}^{5}$,
$ \xi_{9}^{5}$;\ \ 
$ -1$,
$ -c_{9}^{4}$,
$ c_{9}^{4}$,
$ \xi_{9}^{3}$,
$ -\xi_{9}^{3}$,
$ \xi_{9}^{5}$,
$ -\xi_{9}^{5}$;\ \ 
$ -\xi_{9}^{5}$,
$ -\xi_{9}^{5}$,
$ \xi_{9}^{3}$,
$ \xi_{9}^{3}$,
$ -1$,
$ -1$;\ \ 
$ \xi_{9}^{5}$,
$ \xi_{9}^{3}$,
$ -\xi_{9}^{3}$,
$ -1$,
$ 1$;\ \ 
$0$,
$0$,
$ -\xi_{9}^{3}$,
$ -\xi_{9}^{3}$;\ \ 
$0$,
$ -\xi_{9}^{3}$,
$ \xi_{9}^{3}$;\ \ 
$ -c_{9}^{4}$,
$ -c_{9}^{4}$;\ \ 
$ c_{9}^{4}$)

Factors = $2_{7,2.}^{4,625}\boxtimes 4_{\frac{10}{3},19.23}^{9,459}$

\vskip 0.7ex
\hangindent=3em \hangafter=1
$\tau_n$ = ($-1.6 + 5.99 i$, $0.$, $15.17 + 4.06 i$, $0. - 16.48 i$, $8.24 + 8.24 i$, $0.$, $-4.62 - 17.25 i$, $-7.59 - 4.39 i$, $19.23 - 19.23 i$, $0.$, $-17.25 - 4.62 i$, $19.23 - 11.1 i$, $-8.24 - 8.24 i$, $0.$, $4.06 + 15.17 i$, $-21.87 - 12.63 i$, $-5.99 + 1.6 i$, $0.$, $-5.99 - 1.6 i$, $-21.87 + 12.63 i$, $4.06 - 15.17 i$, $0.$, $-8.24 + 8.24 i$, $19.23 + 11.1 i$, $-17.25 + 4.62 i$, $0.$, $19.23 + 19.23 i$, $-7.59 + 4.39 i$, $-4.62 + 17.25 i$, $0.$, $8.24 - 8.24 i$, $0. + 16.48 i$, $15.17 - 4.06 i$, $0.$, $-1.6 - 5.99 i$, $38.46$)

\vskip 0.7ex
\hangindent=3em \hangafter=1
\textit{Intrinsic sign problem}

  \vskip 2ex

\noindent48. $8_{\frac{11}{3},38.46}^{36,155}$ \irep{229}:\ \ 
$d_i$ = ($1.0$,
$1.0$,
$1.879$,
$1.879$,
$2.532$,
$2.532$,
$2.879$,
$2.879$) 

\vskip 0.7ex
\hangindent=3em \hangafter=1
$D^2= 38.468 = 
18+12c^{1}_{9}
+6c^{2}_{9}
$

\vskip 0.7ex
\hangindent=3em \hangafter=1
$T = ( 0,
\frac{3}{4},
\frac{2}{3},
\frac{5}{12},
\frac{7}{9},
\frac{19}{36},
\frac{1}{3},
\frac{1}{12} )
$,

\vskip 0.7ex
\hangindent=3em \hangafter=1
$S$ = ($ 1$,
$ 1$,
$ -c_{9}^{4}$,
$ -c_{9}^{4}$,
$ \xi_{9}^{3}$,
$ \xi_{9}^{3}$,
$ \xi_{9}^{5}$,
$ \xi_{9}^{5}$;\ \ 
$ -1$,
$ -c_{9}^{4}$,
$ c_{9}^{4}$,
$ \xi_{9}^{3}$,
$ -\xi_{9}^{3}$,
$ \xi_{9}^{5}$,
$ -\xi_{9}^{5}$;\ \ 
$ -\xi_{9}^{5}$,
$ -\xi_{9}^{5}$,
$ \xi_{9}^{3}$,
$ \xi_{9}^{3}$,
$ -1$,
$ -1$;\ \ 
$ \xi_{9}^{5}$,
$ \xi_{9}^{3}$,
$ -\xi_{9}^{3}$,
$ -1$,
$ 1$;\ \ 
$0$,
$0$,
$ -\xi_{9}^{3}$,
$ -\xi_{9}^{3}$;\ \ 
$0$,
$ -\xi_{9}^{3}$,
$ \xi_{9}^{3}$;\ \ 
$ -c_{9}^{4}$,
$ -c_{9}^{4}$;\ \ 
$ c_{9}^{4}$)

Factors = $2_{7,2.}^{4,625}\boxtimes 4_{\frac{14}{3},19.23}^{9,614}$

\vskip 0.7ex
\hangindent=3em \hangafter=1
$\tau_n$ = ($-5.99 + 1.6 i$, $0.$, $4.06 + 15.17 i$, $0. + 16.48 i$, $-8.24 - 8.24 i$, $0.$, $-17.25 - 4.62 i$, $-7.59 + 4.39 i$, $19.23 - 19.23 i$, $0.$, $-4.62 - 17.25 i$, $19.23 + 11.1 i$, $8.24 + 8.24 i$, $0.$, $15.17 + 4.06 i$, $-21.87 + 12.63 i$, $-1.6 + 5.99 i$, $0.$, $-1.6 - 5.99 i$, $-21.87 - 12.63 i$, $15.17 - 4.06 i$, $0.$, $8.24 - 8.24 i$, $19.23 - 11.1 i$, $-4.62 + 17.25 i$, $0.$, $19.23 + 19.23 i$, $-7.59 - 4.39 i$, $-17.25 + 4.62 i$, $0.$, $-8.24 + 8.24 i$, $0. - 16.48 i$, $4.06 - 15.17 i$, $0.$, $-5.99 - 1.6 i$, $38.46$)

\vskip 0.7ex
\hangindent=3em \hangafter=1
\textit{Intrinsic sign problem}

  \vskip 2ex

\noindent49. $8_{\frac{2}{5},47.36}^{5,148}$ \irep{52}:\ \ 
$d_i$ = ($1.0$,
$1.618$,
$1.618$,
$1.618$,
$2.618$,
$2.618$,
$2.618$,
$4.236$) 

\vskip 0.7ex
\hangindent=3em \hangafter=1
$D^2= 47.360 = 
25+10\sqrt{5}$

\vskip 0.7ex
\hangindent=3em \hangafter=1
$T = ( 0,
\frac{2}{5},
\frac{2}{5},
\frac{2}{5},
\frac{4}{5},
\frac{4}{5},
\frac{4}{5},
\frac{1}{5} )
$,

\vskip 0.7ex
\hangindent=3em \hangafter=1
$S$ = ($ 1$,
$ \frac{1+\sqrt{5}}{2}$,
$ \frac{1+\sqrt{5}}{2}$,
$ \frac{1+\sqrt{5}}{2}$,
$ \frac{3+\sqrt{5}}{2}$,
$ \frac{3+\sqrt{5}}{2}$,
$ \frac{3+\sqrt{5}}{2}$,
$ 2+\sqrt{5}$;\ \ 
$ -1$,
$ \frac{3+\sqrt{5}}{2}$,
$ \frac{3+\sqrt{5}}{2}$,
$ 2+\sqrt{5}$,
$ -\frac{1+\sqrt{5}}{2}$,
$ -\frac{1+\sqrt{5}}{2}$,
$ -\frac{3+\sqrt{5}}{2}$;\ \ 
$ -1$,
$ \frac{3+\sqrt{5}}{2}$,
$ -\frac{1+\sqrt{5}}{2}$,
$ 2+\sqrt{5}$,
$ -\frac{1+\sqrt{5}}{2}$,
$ -\frac{3+\sqrt{5}}{2}$;\ \ 
$ -1$,
$ -\frac{1+\sqrt{5}}{2}$,
$ -\frac{1+\sqrt{5}}{2}$,
$ 2+\sqrt{5}$,
$ -\frac{3+\sqrt{5}}{2}$;\ \ 
$ 1$,
$ -\frac{3+\sqrt{5}}{2}$,
$ -\frac{3+\sqrt{5}}{2}$,
$ \frac{1+\sqrt{5}}{2}$;\ \ 
$ 1$,
$ -\frac{3+\sqrt{5}}{2}$,
$ \frac{1+\sqrt{5}}{2}$;\ \ 
$ 1$,
$ \frac{1+\sqrt{5}}{2}$;\ \ 
$ -1$)

Factors = $2_{\frac{14}{5},3.618}^{5,395}\boxtimes 4_{\frac{28}{5},13.09}^{5,479}$

\vskip 0.7ex
\hangindent=3em \hangafter=1
$\tau_n$ = ($6.55 + 2.13 i$, $-27.72 - 9.01 i$, $-27.72 + 9.01 i$, $6.55 - 2.13 i$, $47.36$)

\vskip 0.7ex
\hangindent=3em \hangafter=1
\textit{Intrinsic sign problem}

  \vskip 2ex

\noindent50. $8_{\frac{14}{5},47.36}^{5,103}$ \irep{52}:\ \ 
$d_i$ = ($1.0$,
$1.618$,
$1.618$,
$1.618$,
$2.618$,
$2.618$,
$2.618$,
$4.236$) 

\vskip 0.7ex
\hangindent=3em \hangafter=1
$D^2= 47.360 = 
25+10\sqrt{5}$

\vskip 0.7ex
\hangindent=3em \hangafter=1
$T = ( 0,
\frac{2}{5},
\frac{2}{5},
\frac{3}{5},
0,
0,
\frac{4}{5},
\frac{2}{5} )
$,

\vskip 0.7ex
\hangindent=3em \hangafter=1
$S$ = ($ 1$,
$ \frac{1+\sqrt{5}}{2}$,
$ \frac{1+\sqrt{5}}{2}$,
$ \frac{1+\sqrt{5}}{2}$,
$ \frac{3+\sqrt{5}}{2}$,
$ \frac{3+\sqrt{5}}{2}$,
$ \frac{3+\sqrt{5}}{2}$,
$ 2+\sqrt{5}$;\ \ 
$ -1$,
$ \frac{3+\sqrt{5}}{2}$,
$ \frac{3+\sqrt{5}}{2}$,
$ 2+\sqrt{5}$,
$ -\frac{1+\sqrt{5}}{2}$,
$ -\frac{1+\sqrt{5}}{2}$,
$ -\frac{3+\sqrt{5}}{2}$;\ \ 
$ -1$,
$ \frac{3+\sqrt{5}}{2}$,
$ -\frac{1+\sqrt{5}}{2}$,
$ 2+\sqrt{5}$,
$ -\frac{1+\sqrt{5}}{2}$,
$ -\frac{3+\sqrt{5}}{2}$;\ \ 
$ -1$,
$ -\frac{1+\sqrt{5}}{2}$,
$ -\frac{1+\sqrt{5}}{2}$,
$ 2+\sqrt{5}$,
$ -\frac{3+\sqrt{5}}{2}$;\ \ 
$ 1$,
$ -\frac{3+\sqrt{5}}{2}$,
$ -\frac{3+\sqrt{5}}{2}$,
$ \frac{1+\sqrt{5}}{2}$;\ \ 
$ 1$,
$ -\frac{3+\sqrt{5}}{2}$,
$ \frac{1+\sqrt{5}}{2}$;\ \ 
$ 1$,
$ \frac{1+\sqrt{5}}{2}$;\ \ 
$ -1$)

Factors = $2_{\frac{26}{5},3.618}^{5,720}\boxtimes 4_{\frac{28}{5},13.09}^{5,479}$

\vskip 0.7ex
\hangindent=3em \hangafter=1
$\tau_n$ = ($-4.04 + 5.57 i$, $17.13 - 23.58 i$, $17.13 + 23.58 i$, $-4.04 - 5.57 i$, $47.36$)

\vskip 0.7ex
\hangindent=3em \hangafter=1
\textit{Intrinsic sign problem}

  \vskip 2ex

\noindent51. $8_{\frac{26}{5},47.36}^{5,143}$ \irep{52}:\ \ 
$d_i$ = ($1.0$,
$1.618$,
$1.618$,
$1.618$,
$2.618$,
$2.618$,
$2.618$,
$4.236$) 

\vskip 0.7ex
\hangindent=3em \hangafter=1
$D^2= 47.360 = 
25+10\sqrt{5}$

\vskip 0.7ex
\hangindent=3em \hangafter=1
$T = ( 0,
\frac{2}{5},
\frac{3}{5},
\frac{3}{5},
0,
0,
\frac{1}{5},
\frac{3}{5} )
$,

\vskip 0.7ex
\hangindent=3em \hangafter=1
$S$ = ($ 1$,
$ \frac{1+\sqrt{5}}{2}$,
$ \frac{1+\sqrt{5}}{2}$,
$ \frac{1+\sqrt{5}}{2}$,
$ \frac{3+\sqrt{5}}{2}$,
$ \frac{3+\sqrt{5}}{2}$,
$ \frac{3+\sqrt{5}}{2}$,
$ 2+\sqrt{5}$;\ \ 
$ -1$,
$ \frac{3+\sqrt{5}}{2}$,
$ \frac{3+\sqrt{5}}{2}$,
$ -\frac{1+\sqrt{5}}{2}$,
$ -\frac{1+\sqrt{5}}{2}$,
$ 2+\sqrt{5}$,
$ -\frac{3+\sqrt{5}}{2}$;\ \ 
$ -1$,
$ \frac{3+\sqrt{5}}{2}$,
$ 2+\sqrt{5}$,
$ -\frac{1+\sqrt{5}}{2}$,
$ -\frac{1+\sqrt{5}}{2}$,
$ -\frac{3+\sqrt{5}}{2}$;\ \ 
$ -1$,
$ -\frac{1+\sqrt{5}}{2}$,
$ 2+\sqrt{5}$,
$ -\frac{1+\sqrt{5}}{2}$,
$ -\frac{3+\sqrt{5}}{2}$;\ \ 
$ 1$,
$ -\frac{3+\sqrt{5}}{2}$,
$ -\frac{3+\sqrt{5}}{2}$,
$ \frac{1+\sqrt{5}}{2}$;\ \ 
$ 1$,
$ -\frac{3+\sqrt{5}}{2}$,
$ \frac{1+\sqrt{5}}{2}$;\ \ 
$ 1$,
$ \frac{1+\sqrt{5}}{2}$;\ \ 
$ -1$)

Factors = $2_{\frac{14}{5},3.618}^{5,395}\boxtimes 4_{\frac{12}{5},13.09}^{5,443}$

\vskip 0.7ex
\hangindent=3em \hangafter=1
$\tau_n$ = ($-4.04 - 5.57 i$, $17.13 + 23.58 i$, $17.13 - 23.58 i$, $-4.04 + 5.57 i$, $47.36$)

\vskip 0.7ex
\hangindent=3em \hangafter=1
\textit{Intrinsic sign problem}

  \vskip 2ex

\noindent52. $8_{\frac{38}{5},47.36}^{5,286}$ \irep{52}:\ \ 
$d_i$ = ($1.0$,
$1.618$,
$1.618$,
$1.618$,
$2.618$,
$2.618$,
$2.618$,
$4.236$) 

\vskip 0.7ex
\hangindent=3em \hangafter=1
$D^2= 47.360 = 
25+10\sqrt{5}$

\vskip 0.7ex
\hangindent=3em \hangafter=1
$T = ( 0,
\frac{3}{5},
\frac{3}{5},
\frac{3}{5},
\frac{1}{5},
\frac{1}{5},
\frac{1}{5},
\frac{4}{5} )
$,

\vskip 0.7ex
\hangindent=3em \hangafter=1
$S$ = ($ 1$,
$ \frac{1+\sqrt{5}}{2}$,
$ \frac{1+\sqrt{5}}{2}$,
$ \frac{1+\sqrt{5}}{2}$,
$ \frac{3+\sqrt{5}}{2}$,
$ \frac{3+\sqrt{5}}{2}$,
$ \frac{3+\sqrt{5}}{2}$,
$ 2+\sqrt{5}$;\ \ 
$ -1$,
$ \frac{3+\sqrt{5}}{2}$,
$ \frac{3+\sqrt{5}}{2}$,
$ 2+\sqrt{5}$,
$ -\frac{1+\sqrt{5}}{2}$,
$ -\frac{1+\sqrt{5}}{2}$,
$ -\frac{3+\sqrt{5}}{2}$;\ \ 
$ -1$,
$ \frac{3+\sqrt{5}}{2}$,
$ -\frac{1+\sqrt{5}}{2}$,
$ 2+\sqrt{5}$,
$ -\frac{1+\sqrt{5}}{2}$,
$ -\frac{3+\sqrt{5}}{2}$;\ \ 
$ -1$,
$ -\frac{1+\sqrt{5}}{2}$,
$ -\frac{1+\sqrt{5}}{2}$,
$ 2+\sqrt{5}$,
$ -\frac{3+\sqrt{5}}{2}$;\ \ 
$ 1$,
$ -\frac{3+\sqrt{5}}{2}$,
$ -\frac{3+\sqrt{5}}{2}$,
$ \frac{1+\sqrt{5}}{2}$;\ \ 
$ 1$,
$ -\frac{3+\sqrt{5}}{2}$,
$ \frac{1+\sqrt{5}}{2}$;\ \ 
$ 1$,
$ \frac{1+\sqrt{5}}{2}$;\ \ 
$ -1$)

Factors = $2_{\frac{26}{5},3.618}^{5,720}\boxtimes 4_{\frac{12}{5},13.09}^{5,443}$

\vskip 0.7ex
\hangindent=3em \hangafter=1
$\tau_n$ = ($6.55 - 2.13 i$, $-27.72 + 9.01 i$, $-27.72 - 9.01 i$, $6.55 + 2.13 i$, $47.36$)

\vskip 0.7ex
\hangindent=3em \hangafter=1
\textit{Intrinsic sign problem}

  \vskip 2ex

\noindent53. $8_{\frac{92}{15},69.59}^{45,311}$ \irep{234}:\ \ 
$d_i$ = ($1.0$,
$1.618$,
$1.879$,
$2.532$,
$2.879$,
$3.40$,
$4.97$,
$4.658$) 

\vskip 0.7ex
\hangindent=3em \hangafter=1
$D^2= 69.590 = 
27-3  c^{1}_{45}
+6c^{2}_{45}
+3c^{4}_{45}
+12c^{5}_{45}
+6c^{7}_{45}
+9c^{9}_{45}
+3c^{10}_{45}
+3c^{11}_{45}
$

\vskip 0.7ex
\hangindent=3em \hangafter=1
$T = ( 0,
\frac{2}{5},
\frac{1}{3},
\frac{2}{9},
\frac{2}{3},
\frac{11}{15},
\frac{28}{45},
\frac{1}{15} )
$,

\vskip 0.7ex
\hangindent=3em \hangafter=1
$S$ = ($ 1$,
$ \frac{1+\sqrt{5}}{2}$,
$ -c_{9}^{4}$,
$ \xi_{9}^{3}$,
$ \xi_{9}^{5}$,
$ c^{2}_{45}
+c^{7}_{45}
$,
$ 1-c^{1}_{45}
+c^{2}_{45}
+c^{4}_{45}
+c^{7}_{45}
+c^{9}_{45}
-c^{10}_{45}
+c^{11}_{45}
$,
$ 1+c^{2}_{45}
+c^{7}_{45}
+c^{9}_{45}
$;\ \ 
$ -1$,
$ c^{2}_{45}
+c^{7}_{45}
$,
$ 1-c^{1}_{45}
+c^{2}_{45}
+c^{4}_{45}
+c^{7}_{45}
+c^{9}_{45}
-c^{10}_{45}
+c^{11}_{45}
$,
$ 1+c^{2}_{45}
+c^{7}_{45}
+c^{9}_{45}
$,
$ c_{9}^{4}$,
$ -\xi_{9}^{3}$,
$ -\xi_{9}^{5}$;\ \ 
$ -\xi_{9}^{5}$,
$ \xi_{9}^{3}$,
$ -1$,
$ -1-c^{2}_{45}
-c^{7}_{45}
-c^{9}_{45}
$,
$ 1-c^{1}_{45}
+c^{2}_{45}
+c^{4}_{45}
+c^{7}_{45}
+c^{9}_{45}
-c^{10}_{45}
+c^{11}_{45}
$,
$ -\frac{1+\sqrt{5}}{2}$;\ \ 
$0$,
$ -\xi_{9}^{3}$,
$ 1-c^{1}_{45}
+c^{2}_{45}
+c^{4}_{45}
+c^{7}_{45}
+c^{9}_{45}
-c^{10}_{45}
+c^{11}_{45}
$,
$0$,
$ -1+c^{1}_{45}
-c^{2}_{45}
-c^{4}_{45}
-c^{7}_{45}
-c^{9}_{45}
+c^{10}_{45}
-c^{11}_{45}
$;\ \ 
$ -c_{9}^{4}$,
$ -\frac{1+\sqrt{5}}{2}$,
$ -1+c^{1}_{45}
-c^{2}_{45}
-c^{4}_{45}
-c^{7}_{45}
-c^{9}_{45}
+c^{10}_{45}
-c^{11}_{45}
$,
$ c^{2}_{45}
+c^{7}_{45}
$;\ \ 
$ \xi_{9}^{5}$,
$ -\xi_{9}^{3}$,
$ 1$;\ \ 
$0$,
$ \xi_{9}^{3}$;\ \ 
$ c_{9}^{4}$)

Factors = $2_{\frac{14}{5},3.618}^{5,395}\boxtimes 4_{\frac{10}{3},19.23}^{9,459}$

\vskip 0.7ex
\hangindent=3em \hangafter=1
$\tau_n$ = ($-5.07 - 16.1 i$, $-6.05 + 47.04 i$, $37.23 + 10.21 i$, $-18.46 + 8.82 i$, $4.91 + 32.9 i$, $-21.99 + 2.07 i$, $-38.82 + 23.76 i$, $7.99 - 16.24 i$, $-29.78 - 35.61 i$, $-13.52 + 17.74 i$, $9.05 - 29.5 i$, $-3.46 - 34.54 i$, $24.97 - 11.19 i$, $10.52 - 14.59 i$, $33.14 + 26.94 i$, $29.3 - 14.25 i$, $-21.59 + 2.25 i$, $37.95 + 57.61 i$, $13.21 - 4.67 i$, $-48.16 + 23.54 i$, $3.16 + 25.58 i$, $-21.93 - 20.53 i$, $-21.93 + 20.53 i$, $3.16 - 25.58 i$, $-48.16 - 23.54 i$, $13.21 + 4.67 i$, $37.95 - 57.61 i$, $-21.59 - 2.25 i$, $29.3 + 14.25 i$, $33.14 - 26.94 i$, $10.52 + 14.59 i$, $24.97 + 11.19 i$, $-3.46 + 34.54 i$, $9.05 + 29.5 i$, $-13.52 - 17.74 i$, $-29.78 + 35.61 i$, $7.99 + 16.24 i$, $-38.82 - 23.76 i$, $-21.99 - 2.07 i$, $4.91 - 32.9 i$, $-18.46 - 8.82 i$, $37.23 - 10.21 i$, $-6.05 - 47.04 i$, $-5.07 + 16.1 i$, $79.81$)

\vskip 0.7ex
\hangindent=3em \hangafter=1
\textit{Intrinsic sign problem}

  \vskip 2ex

\noindent54. $8_{\frac{112}{15},69.59}^{45,167}$ \irep{234}:\ \ 
$d_i$ = ($1.0$,
$1.618$,
$1.879$,
$2.532$,
$2.879$,
$3.40$,
$4.97$,
$4.658$) 

\vskip 0.7ex
\hangindent=3em \hangafter=1
$D^2= 69.590 = 
27-3  c^{1}_{45}
+6c^{2}_{45}
+3c^{4}_{45}
+12c^{5}_{45}
+6c^{7}_{45}
+9c^{9}_{45}
+3c^{10}_{45}
+3c^{11}_{45}
$

\vskip 0.7ex
\hangindent=3em \hangafter=1
$T = ( 0,
\frac{2}{5},
\frac{2}{3},
\frac{7}{9},
\frac{1}{3},
\frac{1}{15},
\frac{8}{45},
\frac{11}{15} )
$,

\vskip 0.7ex
\hangindent=3em \hangafter=1
$S$ = ($ 1$,
$ \frac{1+\sqrt{5}}{2}$,
$ -c_{9}^{4}$,
$ \xi_{9}^{3}$,
$ \xi_{9}^{5}$,
$ c^{2}_{45}
+c^{7}_{45}
$,
$ 1-c^{1}_{45}
+c^{2}_{45}
+c^{4}_{45}
+c^{7}_{45}
+c^{9}_{45}
-c^{10}_{45}
+c^{11}_{45}
$,
$ 1+c^{2}_{45}
+c^{7}_{45}
+c^{9}_{45}
$;\ \ 
$ -1$,
$ c^{2}_{45}
+c^{7}_{45}
$,
$ 1-c^{1}_{45}
+c^{2}_{45}
+c^{4}_{45}
+c^{7}_{45}
+c^{9}_{45}
-c^{10}_{45}
+c^{11}_{45}
$,
$ 1+c^{2}_{45}
+c^{7}_{45}
+c^{9}_{45}
$,
$ c_{9}^{4}$,
$ -\xi_{9}^{3}$,
$ -\xi_{9}^{5}$;\ \ 
$ -\xi_{9}^{5}$,
$ \xi_{9}^{3}$,
$ -1$,
$ -1-c^{2}_{45}
-c^{7}_{45}
-c^{9}_{45}
$,
$ 1-c^{1}_{45}
+c^{2}_{45}
+c^{4}_{45}
+c^{7}_{45}
+c^{9}_{45}
-c^{10}_{45}
+c^{11}_{45}
$,
$ -\frac{1+\sqrt{5}}{2}$;\ \ 
$0$,
$ -\xi_{9}^{3}$,
$ 1-c^{1}_{45}
+c^{2}_{45}
+c^{4}_{45}
+c^{7}_{45}
+c^{9}_{45}
-c^{10}_{45}
+c^{11}_{45}
$,
$0$,
$ -1+c^{1}_{45}
-c^{2}_{45}
-c^{4}_{45}
-c^{7}_{45}
-c^{9}_{45}
+c^{10}_{45}
-c^{11}_{45}
$;\ \ 
$ -c_{9}^{4}$,
$ -\frac{1+\sqrt{5}}{2}$,
$ -1+c^{1}_{45}
-c^{2}_{45}
-c^{4}_{45}
-c^{7}_{45}
-c^{9}_{45}
+c^{10}_{45}
-c^{11}_{45}
$,
$ c^{2}_{45}
+c^{7}_{45}
$;\ \ 
$ \xi_{9}^{5}$,
$ -\xi_{9}^{3}$,
$ 1$;\ \ 
$0$,
$ \xi_{9}^{3}$;\ \ 
$ c_{9}^{4}$)

Factors = $2_{\frac{14}{5},3.618}^{5,395}\boxtimes 4_{\frac{14}{3},19.23}^{9,614}$

\vskip 0.7ex
\hangindent=3em \hangafter=1
$\tau_n$ = ($13.21 + 4.67 i$, $-38.82 + 23.76 i$, $-3.46 + 34.54 i$, $10.52 - 14.59 i$, $4.91 - 32.9 i$, $3.16 + 25.58 i$, $-6.05 + 47.04 i$, $-21.59 - 2.25 i$, $-29.78 - 35.61 i$, $-13.52 - 17.74 i$, $29.3 - 14.25 i$, $37.23 - 10.21 i$, $-21.93 + 20.53 i$, $-18.46 + 8.82 i$, $33.14 - 26.94 i$, $9.05 - 29.5 i$, $7.99 + 16.24 i$, $37.95 + 57.61 i$, $-5.07 + 16.1 i$, $-48.16 - 23.54 i$, $-21.99 + 2.07 i$, $24.97 + 11.19 i$, $24.97 - 11.19 i$, $-21.99 - 2.07 i$, $-48.16 + 23.54 i$, $-5.07 - 16.1 i$, $37.95 - 57.61 i$, $7.99 - 16.24 i$, $9.05 + 29.5 i$, $33.14 + 26.94 i$, $-18.46 - 8.82 i$, $-21.93 - 20.53 i$, $37.23 + 10.21 i$, $29.3 + 14.25 i$, $-13.52 + 17.74 i$, $-29.78 + 35.61 i$, $-21.59 + 2.25 i$, $-6.05 - 47.04 i$, $3.16 - 25.58 i$, $4.91 + 32.9 i$, $10.52 + 14.59 i$, $-3.46 - 34.54 i$, $-38.82 - 23.76 i$, $13.21 - 4.67 i$, $79.81$)

\vskip 0.7ex
\hangindent=3em \hangafter=1
\textit{Intrinsic sign problem}

  \vskip 2ex

\noindent55. $8_{\frac{8}{15},69.59}^{45,251}$ \irep{234}:\ \ 
$d_i$ = ($1.0$,
$1.618$,
$1.879$,
$2.532$,
$2.879$,
$3.40$,
$4.97$,
$4.658$) 

\vskip 0.7ex
\hangindent=3em \hangafter=1
$D^2= 69.590 = 
27-3  c^{1}_{45}
+6c^{2}_{45}
+3c^{4}_{45}
+12c^{5}_{45}
+6c^{7}_{45}
+9c^{9}_{45}
+3c^{10}_{45}
+3c^{11}_{45}
$

\vskip 0.7ex
\hangindent=3em \hangafter=1
$T = ( 0,
\frac{3}{5},
\frac{1}{3},
\frac{2}{9},
\frac{2}{3},
\frac{14}{15},
\frac{37}{45},
\frac{4}{15} )
$,

\vskip 0.7ex
\hangindent=3em \hangafter=1
$S$ = ($ 1$,
$ \frac{1+\sqrt{5}}{2}$,
$ -c_{9}^{4}$,
$ \xi_{9}^{3}$,
$ \xi_{9}^{5}$,
$ c^{2}_{45}
+c^{7}_{45}
$,
$ 1-c^{1}_{45}
+c^{2}_{45}
+c^{4}_{45}
+c^{7}_{45}
+c^{9}_{45}
-c^{10}_{45}
+c^{11}_{45}
$,
$ 1+c^{2}_{45}
+c^{7}_{45}
+c^{9}_{45}
$;\ \ 
$ -1$,
$ c^{2}_{45}
+c^{7}_{45}
$,
$ 1-c^{1}_{45}
+c^{2}_{45}
+c^{4}_{45}
+c^{7}_{45}
+c^{9}_{45}
-c^{10}_{45}
+c^{11}_{45}
$,
$ 1+c^{2}_{45}
+c^{7}_{45}
+c^{9}_{45}
$,
$ c_{9}^{4}$,
$ -\xi_{9}^{3}$,
$ -\xi_{9}^{5}$;\ \ 
$ -\xi_{9}^{5}$,
$ \xi_{9}^{3}$,
$ -1$,
$ -1-c^{2}_{45}
-c^{7}_{45}
-c^{9}_{45}
$,
$ 1-c^{1}_{45}
+c^{2}_{45}
+c^{4}_{45}
+c^{7}_{45}
+c^{9}_{45}
-c^{10}_{45}
+c^{11}_{45}
$,
$ -\frac{1+\sqrt{5}}{2}$;\ \ 
$0$,
$ -\xi_{9}^{3}$,
$ 1-c^{1}_{45}
+c^{2}_{45}
+c^{4}_{45}
+c^{7}_{45}
+c^{9}_{45}
-c^{10}_{45}
+c^{11}_{45}
$,
$0$,
$ -1+c^{1}_{45}
-c^{2}_{45}
-c^{4}_{45}
-c^{7}_{45}
-c^{9}_{45}
+c^{10}_{45}
-c^{11}_{45}
$;\ \ 
$ -c_{9}^{4}$,
$ -\frac{1+\sqrt{5}}{2}$,
$ -1+c^{1}_{45}
-c^{2}_{45}
-c^{4}_{45}
-c^{7}_{45}
-c^{9}_{45}
+c^{10}_{45}
-c^{11}_{45}
$,
$ c^{2}_{45}
+c^{7}_{45}
$;\ \ 
$ \xi_{9}^{5}$,
$ -\xi_{9}^{3}$,
$ 1$;\ \ 
$0$,
$ \xi_{9}^{3}$;\ \ 
$ c_{9}^{4}$)

Factors = $2_{\frac{26}{5},3.618}^{5,720}\boxtimes 4_{\frac{10}{3},19.23}^{9,459}$

\vskip 0.7ex
\hangindent=3em \hangafter=1
$\tau_n$ = ($13.21 - 4.67 i$, $-38.82 - 23.76 i$, $-3.46 - 34.54 i$, $10.52 + 14.59 i$, $4.91 + 32.9 i$, $3.16 - 25.58 i$, $-6.05 - 47.04 i$, $-21.59 + 2.25 i$, $-29.78 + 35.61 i$, $-13.52 + 17.74 i$, $29.3 + 14.25 i$, $37.23 + 10.21 i$, $-21.93 - 20.53 i$, $-18.46 - 8.82 i$, $33.14 + 26.94 i$, $9.05 + 29.5 i$, $7.99 - 16.24 i$, $37.95 - 57.61 i$, $-5.07 - 16.1 i$, $-48.16 + 23.54 i$, $-21.99 - 2.07 i$, $24.97 - 11.19 i$, $24.97 + 11.19 i$, $-21.99 + 2.07 i$, $-48.16 - 23.54 i$, $-5.07 + 16.1 i$, $37.95 + 57.61 i$, $7.99 + 16.24 i$, $9.05 - 29.5 i$, $33.14 - 26.94 i$, $-18.46 + 8.82 i$, $-21.93 + 20.53 i$, $37.23 - 10.21 i$, $29.3 - 14.25 i$, $-13.52 - 17.74 i$, $-29.78 - 35.61 i$, $-21.59 - 2.25 i$, $-6.05 + 47.04 i$, $3.16 + 25.58 i$, $4.91 - 32.9 i$, $10.52 - 14.59 i$, $-3.46 + 34.54 i$, $-38.82 + 23.76 i$, $13.21 + 4.67 i$, $79.81$)

\vskip 0.7ex
\hangindent=3em \hangafter=1
\textit{Intrinsic sign problem}

  \vskip 2ex

\noindent56. $8_{\frac{28}{15},69.59}^{45,270}$ \irep{234}:\ \ 
$d_i$ = ($1.0$,
$1.618$,
$1.879$,
$2.532$,
$2.879$,
$3.40$,
$4.97$,
$4.658$) 

\vskip 0.7ex
\hangindent=3em \hangafter=1
$D^2= 69.590 = 
27-3  c^{1}_{45}
+6c^{2}_{45}
+3c^{4}_{45}
+12c^{5}_{45}
+6c^{7}_{45}
+9c^{9}_{45}
+3c^{10}_{45}
+3c^{11}_{45}
$

\vskip 0.7ex
\hangindent=3em \hangafter=1
$T = ( 0,
\frac{3}{5},
\frac{2}{3},
\frac{7}{9},
\frac{1}{3},
\frac{4}{15},
\frac{17}{45},
\frac{14}{15} )
$,

\vskip 0.7ex
\hangindent=3em \hangafter=1
$S$ = ($ 1$,
$ \frac{1+\sqrt{5}}{2}$,
$ -c_{9}^{4}$,
$ \xi_{9}^{3}$,
$ \xi_{9}^{5}$,
$ c^{2}_{45}
+c^{7}_{45}
$,
$ 1-c^{1}_{45}
+c^{2}_{45}
+c^{4}_{45}
+c^{7}_{45}
+c^{9}_{45}
-c^{10}_{45}
+c^{11}_{45}
$,
$ 1+c^{2}_{45}
+c^{7}_{45}
+c^{9}_{45}
$;\ \ 
$ -1$,
$ c^{2}_{45}
+c^{7}_{45}
$,
$ 1-c^{1}_{45}
+c^{2}_{45}
+c^{4}_{45}
+c^{7}_{45}
+c^{9}_{45}
-c^{10}_{45}
+c^{11}_{45}
$,
$ 1+c^{2}_{45}
+c^{7}_{45}
+c^{9}_{45}
$,
$ c_{9}^{4}$,
$ -\xi_{9}^{3}$,
$ -\xi_{9}^{5}$;\ \ 
$ -\xi_{9}^{5}$,
$ \xi_{9}^{3}$,
$ -1$,
$ -1-c^{2}_{45}
-c^{7}_{45}
-c^{9}_{45}
$,
$ 1-c^{1}_{45}
+c^{2}_{45}
+c^{4}_{45}
+c^{7}_{45}
+c^{9}_{45}
-c^{10}_{45}
+c^{11}_{45}
$,
$ -\frac{1+\sqrt{5}}{2}$;\ \ 
$0$,
$ -\xi_{9}^{3}$,
$ 1-c^{1}_{45}
+c^{2}_{45}
+c^{4}_{45}
+c^{7}_{45}
+c^{9}_{45}
-c^{10}_{45}
+c^{11}_{45}
$,
$0$,
$ -1+c^{1}_{45}
-c^{2}_{45}
-c^{4}_{45}
-c^{7}_{45}
-c^{9}_{45}
+c^{10}_{45}
-c^{11}_{45}
$;\ \ 
$ -c_{9}^{4}$,
$ -\frac{1+\sqrt{5}}{2}$,
$ -1+c^{1}_{45}
-c^{2}_{45}
-c^{4}_{45}
-c^{7}_{45}
-c^{9}_{45}
+c^{10}_{45}
-c^{11}_{45}
$,
$ c^{2}_{45}
+c^{7}_{45}
$;\ \ 
$ \xi_{9}^{5}$,
$ -\xi_{9}^{3}$,
$ 1$;\ \ 
$0$,
$ \xi_{9}^{3}$;\ \ 
$ c_{9}^{4}$)

Factors = $2_{\frac{26}{5},3.618}^{5,720}\boxtimes 4_{\frac{14}{3},19.23}^{9,614}$

\vskip 0.7ex
\hangindent=3em \hangafter=1
$\tau_n$ = ($-5.07 + 16.1 i$, $-6.05 - 47.04 i$, $37.23 - 10.21 i$, $-18.46 - 8.82 i$, $4.91 - 32.9 i$, $-21.99 - 2.07 i$, $-38.82 - 23.76 i$, $7.99 + 16.24 i$, $-29.78 + 35.61 i$, $-13.52 - 17.74 i$, $9.05 + 29.5 i$, $-3.46 + 34.54 i$, $24.97 + 11.19 i$, $10.52 + 14.59 i$, $33.14 - 26.94 i$, $29.3 + 14.25 i$, $-21.59 - 2.25 i$, $37.95 - 57.61 i$, $13.21 + 4.67 i$, $-48.16 - 23.54 i$, $3.16 - 25.58 i$, $-21.93 + 20.53 i$, $-21.93 - 20.53 i$, $3.16 + 25.58 i$, $-48.16 + 23.54 i$, $13.21 - 4.67 i$, $37.95 + 57.61 i$, $-21.59 + 2.25 i$, $29.3 - 14.25 i$, $33.14 + 26.94 i$, $10.52 - 14.59 i$, $24.97 - 11.19 i$, $-3.46 - 34.54 i$, $9.05 - 29.5 i$, $-13.52 + 17.74 i$, $-29.78 - 35.61 i$, $7.99 - 16.24 i$, $-38.82 + 23.76 i$, $-21.99 + 2.07 i$, $4.91 + 32.9 i$, $-18.46 + 8.82 i$, $37.23 + 10.21 i$, $-6.05 + 47.04 i$, $-5.07 - 16.1 i$, $79.81$)

\vskip 0.7ex
\hangindent=3em \hangafter=1
\textit{Intrinsic sign problem}

  \vskip 2ex

\noindent57. $8_{\frac{62}{17},125.8}^{17,152}$ \irep{199}:\ \ 
$d_i$ = ($1.0$,
$1.965$,
$2.864$,
$3.666$,
$4.342$,
$4.871$,
$5.234$,
$5.418$) 

\vskip 0.7ex
\hangindent=3em \hangafter=1
$D^2= 125.874 = 
36+28c^{1}_{17}
+21c^{2}_{17}
+15c^{3}_{17}
+10c^{4}_{17}
+6c^{5}_{17}
+3c^{6}_{17}
+c^{7}_{17}
$

\vskip 0.7ex
\hangindent=3em \hangafter=1
$T = ( 0,
\frac{5}{17},
\frac{2}{17},
\frac{8}{17},
\frac{6}{17},
\frac{13}{17},
\frac{12}{17},
\frac{3}{17} )
$,

\vskip 0.7ex
\hangindent=3em \hangafter=1
$S$ = ($ 1$,
$ -c_{17}^{8}$,
$ \xi_{17}^{3}$,
$ \xi_{17}^{13}$,
$ \xi_{17}^{5}$,
$ \xi_{17}^{11}$,
$ \xi_{17}^{7}$,
$ \xi_{17}^{9}$;\ \ 
$ -\xi_{17}^{13}$,
$ \xi_{17}^{11}$,
$ -\xi_{17}^{9}$,
$ \xi_{17}^{7}$,
$ -\xi_{17}^{5}$,
$ \xi_{17}^{3}$,
$ -1$;\ \ 
$ \xi_{17}^{9}$,
$ \xi_{17}^{5}$,
$ -c_{17}^{8}$,
$ -1$,
$ -\xi_{17}^{13}$,
$ -\xi_{17}^{7}$;\ \ 
$ -1$,
$ -\xi_{17}^{3}$,
$ \xi_{17}^{7}$,
$ -\xi_{17}^{11}$,
$ -c_{17}^{8}$;\ \ 
$ -\xi_{17}^{9}$,
$ -\xi_{17}^{13}$,
$ 1$,
$ \xi_{17}^{11}$;\ \ 
$ c_{17}^{8}$,
$ \xi_{17}^{9}$,
$ -\xi_{17}^{3}$;\ \ 
$ -c_{17}^{8}$,
$ -\xi_{17}^{5}$;\ \ 
$ \xi_{17}^{13}$)

\vskip 0.7ex
\hangindent=3em \hangafter=1
$\tau_n$ = ($-10.79 + 3.06 i$, $-58.46 + 16.63 i$, $-10.04 + 53.71 i$, $14.86 - 38.35 i$, $-10.79 - 57.71 i$, $28.76 + 14.33 i$, $0. + 48.71 i$, $-7.97 - 20.56 i$, $-7.97 + 20.56 i$, $0. - 48.71 i$, $28.76 - 14.33 i$, $-10.79 + 57.71 i$, $14.86 + 38.35 i$, $-10.04 - 53.71 i$, $-58.46 - 16.63 i$, $-10.79 - 3.06 i$, $125.83$)

\vskip 0.7ex
\hangindent=3em \hangafter=1
\textit{Intrinsic sign problem}

  \vskip 2ex

\noindent58. $8_{\frac{74}{17},125.8}^{17,311}$ \irep{199}:\ \ 
$d_i$ = ($1.0$,
$1.965$,
$2.864$,
$3.666$,
$4.342$,
$4.871$,
$5.234$,
$5.418$) 

\vskip 0.7ex
\hangindent=3em \hangafter=1
$D^2= 125.874 = 
36+28c^{1}_{17}
+21c^{2}_{17}
+15c^{3}_{17}
+10c^{4}_{17}
+6c^{5}_{17}
+3c^{6}_{17}
+c^{7}_{17}
$

\vskip 0.7ex
\hangindent=3em \hangafter=1
$T = ( 0,
\frac{12}{17},
\frac{15}{17},
\frac{9}{17},
\frac{11}{17},
\frac{4}{17},
\frac{5}{17},
\frac{14}{17} )
$,

\vskip 0.7ex
\hangindent=3em \hangafter=1
$S$ = ($ 1$,
$ -c_{17}^{8}$,
$ \xi_{17}^{3}$,
$ \xi_{17}^{13}$,
$ \xi_{17}^{5}$,
$ \xi_{17}^{11}$,
$ \xi_{17}^{7}$,
$ \xi_{17}^{9}$;\ \ 
$ -\xi_{17}^{13}$,
$ \xi_{17}^{11}$,
$ -\xi_{17}^{9}$,
$ \xi_{17}^{7}$,
$ -\xi_{17}^{5}$,
$ \xi_{17}^{3}$,
$ -1$;\ \ 
$ \xi_{17}^{9}$,
$ \xi_{17}^{5}$,
$ -c_{17}^{8}$,
$ -1$,
$ -\xi_{17}^{13}$,
$ -\xi_{17}^{7}$;\ \ 
$ -1$,
$ -\xi_{17}^{3}$,
$ \xi_{17}^{7}$,
$ -\xi_{17}^{11}$,
$ -c_{17}^{8}$;\ \ 
$ -\xi_{17}^{9}$,
$ -\xi_{17}^{13}$,
$ 1$,
$ \xi_{17}^{11}$;\ \ 
$ c_{17}^{8}$,
$ \xi_{17}^{9}$,
$ -\xi_{17}^{3}$;\ \ 
$ -c_{17}^{8}$,
$ -\xi_{17}^{5}$;\ \ 
$ \xi_{17}^{13}$)

\vskip 0.7ex
\hangindent=3em \hangafter=1
$\tau_n$ = ($-10.79 - 3.06 i$, $-58.46 - 16.63 i$, $-10.04 - 53.71 i$, $14.86 + 38.35 i$, $-10.79 + 57.71 i$, $28.76 - 14.33 i$, $0. - 48.71 i$, $-7.97 + 20.56 i$, $-7.97 - 20.56 i$, $0. + 48.71 i$, $28.76 + 14.33 i$, $-10.79 - 57.71 i$, $14.86 - 38.35 i$, $-10.04 + 53.71 i$, $-58.46 + 16.63 i$, $-10.79 + 3.06 i$, $125.83$)

\vskip 0.7ex
\hangindent=3em \hangafter=1
\textit{Intrinsic sign problem}

  \vskip 2ex

\noindent59. $8_{\frac{36}{13},223.6}^{13,370}$ \irep{179}:\ \ 
$d_i$ = ($1.0$,
$2.941$,
$4.148$,
$4.148$,
$4.712$,
$6.209$,
$7.345$,
$8.55$) 

\vskip 0.7ex
\hangindent=3em \hangafter=1
$D^2= 223.689 = 
78+65c^{1}_{13}
+52c^{2}_{13}
+39c^{3}_{13}
+26c^{4}_{13}
+13c^{5}_{13}
$

\vskip 0.7ex
\hangindent=3em \hangafter=1
$T = ( 0,
\frac{1}{13},
\frac{8}{13},
\frac{8}{13},
\frac{3}{13},
\frac{6}{13},
\frac{10}{13},
\frac{2}{13} )
$,

\vskip 0.7ex
\hangindent=3em \hangafter=1
$S$ = ($ 1$,
$ 2+c^{1}_{13}
+c^{2}_{13}
+c^{3}_{13}
+c^{4}_{13}
+c^{5}_{13}
$,
$ \xi_{13}^{7}$,
$ \xi_{13}^{7}$,
$ 2+2c^{1}_{13}
+c^{2}_{13}
+c^{3}_{13}
+c^{4}_{13}
+c^{5}_{13}
$,
$ 2+2c^{1}_{13}
+c^{2}_{13}
+c^{3}_{13}
+c^{4}_{13}
$,
$ 2+2c^{1}_{13}
+2c^{2}_{13}
+c^{3}_{13}
+c^{4}_{13}
$,
$ 2+2c^{1}_{13}
+2c^{2}_{13}
+c^{3}_{13}
$;\ \ 
$ 2+2c^{1}_{13}
+2c^{2}_{13}
+c^{3}_{13}
+c^{4}_{13}
$,
$ -\xi_{13}^{7}$,
$ -\xi_{13}^{7}$,
$ 2+2c^{1}_{13}
+2c^{2}_{13}
+c^{3}_{13}
$,
$ 2+2c^{1}_{13}
+c^{2}_{13}
+c^{3}_{13}
+c^{4}_{13}
+c^{5}_{13}
$,
$ -1$,
$ -2-2  c^{1}_{13}
-c^{2}_{13}
-c^{3}_{13}
-c^{4}_{13}
$;\ \ 
$ -1-c^{1}_{13}
-c^{2}_{13}
+c^{5}_{13}
$,
$ 2+2c^{1}_{13}
+2c^{2}_{13}
+c^{3}_{13}
-c^{5}_{13}
$,
$ \xi_{13}^{7}$,
$ -\xi_{13}^{7}$,
$ \xi_{13}^{7}$,
$ -\xi_{13}^{7}$;\ \ 
$ -1-c^{1}_{13}
-c^{2}_{13}
+c^{5}_{13}
$,
$ \xi_{13}^{7}$,
$ -\xi_{13}^{7}$,
$ \xi_{13}^{7}$,
$ -\xi_{13}^{7}$;\ \ 
$ 1$,
$ -2-2  c^{1}_{13}
-2  c^{2}_{13}
-c^{3}_{13}
-c^{4}_{13}
$,
$ -2-2  c^{1}_{13}
-c^{2}_{13}
-c^{3}_{13}
-c^{4}_{13}
$,
$ 2+c^{1}_{13}
+c^{2}_{13}
+c^{3}_{13}
+c^{4}_{13}
+c^{5}_{13}
$;\ \ 
$ -2-c^{1}_{13}
-c^{2}_{13}
-c^{3}_{13}
-c^{4}_{13}
-c^{5}_{13}
$,
$ 2+2c^{1}_{13}
+2c^{2}_{13}
+c^{3}_{13}
$,
$ 1$;\ \ 
$ -2-c^{1}_{13}
-c^{2}_{13}
-c^{3}_{13}
-c^{4}_{13}
-c^{5}_{13}
$,
$ -2-2  c^{1}_{13}
-c^{2}_{13}
-c^{3}_{13}
-c^{4}_{13}
-c^{5}_{13}
$;\ \ 
$ 2+2c^{1}_{13}
+2c^{2}_{13}
+c^{3}_{13}
+c^{4}_{13}
$)

\vskip 0.7ex
\hangindent=3em \hangafter=1
$\tau_n$ = ($-3.82 + 19.07 i$, $-55.66 + 84.12 i$, $-105.25 + 53.01 i$, $-0.87 - 49.13 i$, $63.4 - 40.92 i$, $-7.23 - 123.4 i$, $-7.23 + 123.4 i$, $63.4 + 40.92 i$, $-0.87 + 49.13 i$, $-105.25 - 53.01 i$, $-55.66 - 84.12 i$, $-3.82 - 19.07 i$, $231.87$)

\vskip 0.7ex
\hangindent=3em \hangafter=1
\textit{Intrinsic sign problem}

  \vskip 2ex

\noindent60. $8_{\frac{68}{13},223.6}^{13,484}$ \irep{179}:\ \ 
$d_i$ = ($1.0$,
$2.941$,
$4.148$,
$4.148$,
$4.712$,
$6.209$,
$7.345$,
$8.55$) 

\vskip 0.7ex
\hangindent=3em \hangafter=1
$D^2= 223.689 = 
78+65c^{1}_{13}
+52c^{2}_{13}
+39c^{3}_{13}
+26c^{4}_{13}
+13c^{5}_{13}
$

\vskip 0.7ex
\hangindent=3em \hangafter=1
$T = ( 0,
\frac{12}{13},
\frac{5}{13},
\frac{5}{13},
\frac{10}{13},
\frac{7}{13},
\frac{3}{13},
\frac{11}{13} )
$,

\vskip 0.7ex
\hangindent=3em \hangafter=1
$S$ = ($ 1$,
$ 2+c^{1}_{13}
+c^{2}_{13}
+c^{3}_{13}
+c^{4}_{13}
+c^{5}_{13}
$,
$ \xi_{13}^{7}$,
$ \xi_{13}^{7}$,
$ 2+2c^{1}_{13}
+c^{2}_{13}
+c^{3}_{13}
+c^{4}_{13}
+c^{5}_{13}
$,
$ 2+2c^{1}_{13}
+c^{2}_{13}
+c^{3}_{13}
+c^{4}_{13}
$,
$ 2+2c^{1}_{13}
+2c^{2}_{13}
+c^{3}_{13}
+c^{4}_{13}
$,
$ 2+2c^{1}_{13}
+2c^{2}_{13}
+c^{3}_{13}
$;\ \ 
$ 2+2c^{1}_{13}
+2c^{2}_{13}
+c^{3}_{13}
+c^{4}_{13}
$,
$ -\xi_{13}^{7}$,
$ -\xi_{13}^{7}$,
$ 2+2c^{1}_{13}
+2c^{2}_{13}
+c^{3}_{13}
$,
$ 2+2c^{1}_{13}
+c^{2}_{13}
+c^{3}_{13}
+c^{4}_{13}
+c^{5}_{13}
$,
$ -1$,
$ -2-2  c^{1}_{13}
-c^{2}_{13}
-c^{3}_{13}
-c^{4}_{13}
$;\ \ 
$ -1-c^{1}_{13}
-c^{2}_{13}
+c^{5}_{13}
$,
$ 2+2c^{1}_{13}
+2c^{2}_{13}
+c^{3}_{13}
-c^{5}_{13}
$,
$ \xi_{13}^{7}$,
$ -\xi_{13}^{7}$,
$ \xi_{13}^{7}$,
$ -\xi_{13}^{7}$;\ \ 
$ -1-c^{1}_{13}
-c^{2}_{13}
+c^{5}_{13}
$,
$ \xi_{13}^{7}$,
$ -\xi_{13}^{7}$,
$ \xi_{13}^{7}$,
$ -\xi_{13}^{7}$;\ \ 
$ 1$,
$ -2-2  c^{1}_{13}
-2  c^{2}_{13}
-c^{3}_{13}
-c^{4}_{13}
$,
$ -2-2  c^{1}_{13}
-c^{2}_{13}
-c^{3}_{13}
-c^{4}_{13}
$,
$ 2+c^{1}_{13}
+c^{2}_{13}
+c^{3}_{13}
+c^{4}_{13}
+c^{5}_{13}
$;\ \ 
$ -2-c^{1}_{13}
-c^{2}_{13}
-c^{3}_{13}
-c^{4}_{13}
-c^{5}_{13}
$,
$ 2+2c^{1}_{13}
+2c^{2}_{13}
+c^{3}_{13}
$,
$ 1$;\ \ 
$ -2-c^{1}_{13}
-c^{2}_{13}
-c^{3}_{13}
-c^{4}_{13}
-c^{5}_{13}
$,
$ -2-2  c^{1}_{13}
-c^{2}_{13}
-c^{3}_{13}
-c^{4}_{13}
-c^{5}_{13}
$;\ \ 
$ 2+2c^{1}_{13}
+2c^{2}_{13}
+c^{3}_{13}
+c^{4}_{13}
$)

\vskip 0.7ex
\hangindent=3em \hangafter=1
$\tau_n$ = ($-3.82 - 19.07 i$, $-55.66 - 84.12 i$, $-105.25 - 53.01 i$, $-0.87 + 49.13 i$, $63.4 + 40.92 i$, $-7.23 + 123.4 i$, $-7.23 - 123.4 i$, $63.4 - 40.92 i$, $-0.87 - 49.13 i$, $-105.25 + 53.01 i$, $-55.66 + 84.12 i$, $-3.82 + 19.07 i$, $231.87$)

\vskip 0.7ex
\hangindent=3em \hangafter=1
\textit{Intrinsic sign problem}

  \vskip 2ex

\noindent61. $8_{4,308.4}^{15,440}$ \irep{183}:\ \ 
$d_i$ = ($1.0$,
$5.854$,
$5.854$,
$5.854$,
$5.854$,
$6.854$,
$7.854$,
$7.854$) 

\vskip 0.7ex
\hangindent=3em \hangafter=1
$D^2= 308.434 = 
\frac{315+135\sqrt{5}}{2}$

\vskip 0.7ex
\hangindent=3em \hangafter=1
$T = ( 0,
0,
0,
\frac{1}{3},
\frac{2}{3},
0,
\frac{2}{5},
\frac{3}{5} )
$,

\vskip 0.7ex
\hangindent=3em \hangafter=1
$S$ = ($ 1$,
$ \frac{5+3\sqrt{5}}{2}$,
$ \frac{5+3\sqrt{5}}{2}$,
$ \frac{5+3\sqrt{5}}{2}$,
$ \frac{5+3\sqrt{5}}{2}$,
$ \frac{7+3\sqrt{5}}{2}$,
$ \frac{9+3\sqrt{5}}{2}$,
$ \frac{9+3\sqrt{5}}{2}$;\ \ 
$ -5-3\sqrt{5}$,
$ \frac{5+3\sqrt{5}}{2}$,
$ \frac{5+3\sqrt{5}}{2}$,
$ \frac{5+3\sqrt{5}}{2}$,
$ -\frac{5+3\sqrt{5}}{2}$,
$0$,
$0$;\ \ 
$ -5-3\sqrt{5}$,
$ \frac{5+3\sqrt{5}}{2}$,
$ \frac{5+3\sqrt{5}}{2}$,
$ -\frac{5+3\sqrt{5}}{2}$,
$0$,
$0$;\ \ 
$ \frac{5+3\sqrt{5}}{2}$,
$ -5-3\sqrt{5}$,
$ -\frac{5+3\sqrt{5}}{2}$,
$0$,
$0$;\ \ 
$ \frac{5+3\sqrt{5}}{2}$,
$ -\frac{5+3\sqrt{5}}{2}$,
$0$,
$0$;\ \ 
$ 1$,
$ \frac{9+3\sqrt{5}}{2}$,
$ \frac{9+3\sqrt{5}}{2}$;\ \ 
$ \frac{3+3\sqrt{5}}{2}$,
$ -6-3\sqrt{5}$;\ \ 
$ \frac{3+3\sqrt{5}}{2}$)

\vskip 0.7ex
\hangindent=3em \hangafter=1
$\tau_n$ = ($-17.56$, $120.37$, $223.18$, $-17.56$, $205.62$, $85.25$, $120.37$, $120.37$, $85.25$, $205.62$, $-17.56$, $223.18$, $120.37$, $-17.56$, $308.43$)

\vskip 0.7ex
\hangindent=3em \hangafter=1
\textit{Intrinsic sign problem}

  \vskip 2ex

\noindent62. $8_{0,308.4}^{15,100}$ \irep{188}:\ \ 
$d_i$ = ($1.0$,
$5.854$,
$5.854$,
$5.854$,
$5.854$,
$6.854$,
$7.854$,
$7.854$) 

\vskip 0.7ex
\hangindent=3em \hangafter=1
$D^2= 308.434 = 
\frac{315+135\sqrt{5}}{2}$

\vskip 0.7ex
\hangindent=3em \hangafter=1
$T = ( 0,
\frac{1}{3},
\frac{1}{3},
\frac{2}{3},
\frac{2}{3},
0,
\frac{1}{5},
\frac{4}{5} )
$,

\vskip 0.7ex
\hangindent=3em \hangafter=1
$S$ = ($ 1$,
$ \frac{5+3\sqrt{5}}{2}$,
$ \frac{5+3\sqrt{5}}{2}$,
$ \frac{5+3\sqrt{5}}{2}$,
$ \frac{5+3\sqrt{5}}{2}$,
$ \frac{7+3\sqrt{5}}{2}$,
$ \frac{9+3\sqrt{5}}{2}$,
$ \frac{9+3\sqrt{5}}{2}$;\ \ 
$ \frac{5+3\sqrt{5}}{2}$,
$ -5-3\sqrt{5}$,
$ \frac{5+3\sqrt{5}}{2}$,
$ \frac{5+3\sqrt{5}}{2}$,
$ -\frac{5+3\sqrt{5}}{2}$,
$0$,
$0$;\ \ 
$ \frac{5+3\sqrt{5}}{2}$,
$ \frac{5+3\sqrt{5}}{2}$,
$ \frac{5+3\sqrt{5}}{2}$,
$ -\frac{5+3\sqrt{5}}{2}$,
$0$,
$0$;\ \ 
$ \frac{5+3\sqrt{5}}{2}$,
$ -5-3\sqrt{5}$,
$ -\frac{5+3\sqrt{5}}{2}$,
$0$,
$0$;\ \ 
$ \frac{5+3\sqrt{5}}{2}$,
$ -\frac{5+3\sqrt{5}}{2}$,
$0$,
$0$;\ \ 
$ 1$,
$ \frac{9+3\sqrt{5}}{2}$,
$ \frac{9+3\sqrt{5}}{2}$;\ \ 
$ -6-3\sqrt{5}$,
$ \frac{3+3\sqrt{5}}{2}$;\ \ 
$ -6-3\sqrt{5}$)

\vskip 0.7ex
\hangindent=3em \hangafter=1
$\tau_n$ = ($17.56$, $-120.37$, $85.25$, $17.56$, $102.81$, $223.18$, $-120.37$, $-120.37$, $223.18$, $102.81$, $17.56$, $85.25$, $-120.37$, $17.56$, $308.43$)

\vskip 0.7ex
\hangindent=3em \hangafter=1
\textit{Intrinsic sign problem}

  \vskip 2ex

\noindent63. $8_{4,308.4}^{45,289}$ \irep{235}:\ \ 
$d_i$ = ($1.0$,
$5.854$,
$5.854$,
$5.854$,
$5.854$,
$6.854$,
$7.854$,
$7.854$) 

\vskip 0.7ex
\hangindent=3em \hangafter=1
$D^2= 308.434 = 
\frac{315+135\sqrt{5}}{2}$

\vskip 0.7ex
\hangindent=3em \hangafter=1
$T = ( 0,
0,
\frac{1}{9},
\frac{4}{9},
\frac{7}{9},
0,
\frac{2}{5},
\frac{3}{5} )
$,

\vskip 0.7ex
\hangindent=3em \hangafter=1
$S$ = ($ 1$,
$ \frac{5+3\sqrt{5}}{2}$,
$ \frac{5+3\sqrt{5}}{2}$,
$ \frac{5+3\sqrt{5}}{2}$,
$ \frac{5+3\sqrt{5}}{2}$,
$ \frac{7+3\sqrt{5}}{2}$,
$ \frac{9+3\sqrt{5}}{2}$,
$ \frac{9+3\sqrt{5}}{2}$;\ \ 
$ -5-3\sqrt{5}$,
$ \frac{5+3\sqrt{5}}{2}$,
$ \frac{5+3\sqrt{5}}{2}$,
$ \frac{5+3\sqrt{5}}{2}$,
$ -\frac{5+3\sqrt{5}}{2}$,
$0$,
$0$;\ \ 
$ -3  c^{1}_{45}
+3c^{4}_{45}
-4  c^{10}_{45}
+3c^{11}_{45}
$,
$ 3c^{2}_{45}
+c^{5}_{45}
+3c^{7}_{45}
+c^{10}_{45}
$,
$ 3c^{1}_{45}
-3  c^{2}_{45}
-3  c^{4}_{45}
-c^{5}_{45}
-3  c^{7}_{45}
+3c^{10}_{45}
-3  c^{11}_{45}
$,
$ -\frac{5+3\sqrt{5}}{2}$,
$0$,
$0$;\ \ 
$ 3c^{1}_{45}
-3  c^{2}_{45}
-3  c^{4}_{45}
-c^{5}_{45}
-3  c^{7}_{45}
+3c^{10}_{45}
-3  c^{11}_{45}
$,
$ -3  c^{1}_{45}
+3c^{4}_{45}
-4  c^{10}_{45}
+3c^{11}_{45}
$,
$ -\frac{5+3\sqrt{5}}{2}$,
$0$,
$0$;\ \ 
$ 3c^{2}_{45}
+c^{5}_{45}
+3c^{7}_{45}
+c^{10}_{45}
$,
$ -\frac{5+3\sqrt{5}}{2}$,
$0$,
$0$;\ \ 
$ 1$,
$ \frac{9+3\sqrt{5}}{2}$,
$ \frac{9+3\sqrt{5}}{2}$;\ \ 
$ \frac{3+3\sqrt{5}}{2}$,
$ -6-3\sqrt{5}$;\ \ 
$ \frac{3+3\sqrt{5}}{2}$)

\vskip 0.7ex
\hangindent=3em \hangafter=1
$\tau_n$ = ($-17.56$, $120.37$, $68.97 + 89.03 i$, $-17.56$, $205.62$, $-68.97 - 89.03 i$, $120.37$, $120.37$, $85.25$, $205.62$, $-17.56$, $68.97 + 89.03 i$, $120.37$, $-17.56$, $154.21 - 89.03 i$, $-17.56$, $120.37$, $223.18$, $-17.56$, $205.62$, $-68.97 + 89.03 i$, $120.37$, $120.37$, $-68.97 - 89.03 i$, $205.62$, $-17.56$, $223.18$, $120.37$, $-17.56$, $154.21 + 89.03 i$, $-17.56$, $120.37$, $68.97 - 89.03 i$, $-17.56$, $205.62$, $85.25$, $120.37$, $120.37$, $-68.97 + 89.03 i$, $205.62$, $-17.56$, $68.97 - 89.03 i$, $120.37$, $-17.56$, $308.43$)

\vskip 0.7ex
\hangindent=3em \hangafter=1
\textit{Intrinsic sign problem}

  \vskip 2ex

\noindent64. $8_{4,308.4}^{45,939}$ \irep{235}:\ \ 
$d_i$ = ($1.0$,
$5.854$,
$5.854$,
$5.854$,
$5.854$,
$6.854$,
$7.854$,
$7.854$) 

\vskip 0.7ex
\hangindent=3em \hangafter=1
$D^2= 308.434 = 
\frac{315+135\sqrt{5}}{2}$

\vskip 0.7ex
\hangindent=3em \hangafter=1
$T = ( 0,
0,
\frac{2}{9},
\frac{5}{9},
\frac{8}{9},
0,
\frac{2}{5},
\frac{3}{5} )
$,

\vskip 0.7ex
\hangindent=3em \hangafter=1
$S$ = ($ 1$,
$ \frac{5+3\sqrt{5}}{2}$,
$ \frac{5+3\sqrt{5}}{2}$,
$ \frac{5+3\sqrt{5}}{2}$,
$ \frac{5+3\sqrt{5}}{2}$,
$ \frac{7+3\sqrt{5}}{2}$,
$ \frac{9+3\sqrt{5}}{2}$,
$ \frac{9+3\sqrt{5}}{2}$;\ \ 
$ -5-3\sqrt{5}$,
$ \frac{5+3\sqrt{5}}{2}$,
$ \frac{5+3\sqrt{5}}{2}$,
$ \frac{5+3\sqrt{5}}{2}$,
$ -\frac{5+3\sqrt{5}}{2}$,
$0$,
$0$;\ \ 
$ 3c^{2}_{45}
+c^{5}_{45}
+3c^{7}_{45}
+c^{10}_{45}
$,
$ -3  c^{1}_{45}
+3c^{4}_{45}
-4  c^{10}_{45}
+3c^{11}_{45}
$,
$ 3c^{1}_{45}
-3  c^{2}_{45}
-3  c^{4}_{45}
-c^{5}_{45}
-3  c^{7}_{45}
+3c^{10}_{45}
-3  c^{11}_{45}
$,
$ -\frac{5+3\sqrt{5}}{2}$,
$0$,
$0$;\ \ 
$ 3c^{1}_{45}
-3  c^{2}_{45}
-3  c^{4}_{45}
-c^{5}_{45}
-3  c^{7}_{45}
+3c^{10}_{45}
-3  c^{11}_{45}
$,
$ 3c^{2}_{45}
+c^{5}_{45}
+3c^{7}_{45}
+c^{10}_{45}
$,
$ -\frac{5+3\sqrt{5}}{2}$,
$0$,
$0$;\ \ 
$ -3  c^{1}_{45}
+3c^{4}_{45}
-4  c^{10}_{45}
+3c^{11}_{45}
$,
$ -\frac{5+3\sqrt{5}}{2}$,
$0$,
$0$;\ \ 
$ 1$,
$ \frac{9+3\sqrt{5}}{2}$,
$ \frac{9+3\sqrt{5}}{2}$;\ \ 
$ \frac{3+3\sqrt{5}}{2}$,
$ -6-3\sqrt{5}$;\ \ 
$ \frac{3+3\sqrt{5}}{2}$)

\vskip 0.7ex
\hangindent=3em \hangafter=1
$\tau_n$ = ($-17.56$, $120.37$, $68.97 - 89.03 i$, $-17.56$, $205.62$, $-68.97 + 89.03 i$, $120.37$, $120.37$, $85.25$, $205.62$, $-17.56$, $68.97 - 89.03 i$, $120.37$, $-17.56$, $154.21 + 89.03 i$, $-17.56$, $120.37$, $223.18$, $-17.56$, $205.62$, $-68.97 - 89.03 i$, $120.37$, $120.37$, $-68.97 + 89.03 i$, $205.62$, $-17.56$, $223.18$, $120.37$, $-17.56$, $154.21 - 89.03 i$, $-17.56$, $120.37$, $68.97 + 89.03 i$, $-17.56$, $205.62$, $85.25$, $120.37$, $120.37$, $-68.97 - 89.03 i$, $205.62$, $-17.56$, $68.97 + 89.03 i$, $120.37$, $-17.56$, $308.43$)

\vskip 0.7ex
\hangindent=3em \hangafter=1
\textit{Intrinsic sign problem}

  \vskip 2ex 

%% file: modular_data/SsL9U_.tex
\noindent1. $9_{0,9.}^{3,113}$ \irep{0}:\ \ 
$d_i$ = ($1.0$,
$1.0$,
$1.0$,
$1.0$,
$1.0$,
$1.0$,
$1.0$,
$1.0$,
$1.0$) 

\vskip 0.7ex
\hangindent=3em \hangafter=1
$D^2= 9.0 = 
9$

\vskip 0.7ex
\hangindent=3em \hangafter=1
$T = ( 0,
0,
0,
0,
0,
\frac{1}{3},
\frac{1}{3},
\frac{2}{3},
\frac{2}{3} )
$,

\vskip 0.7ex
\hangindent=3em \hangafter=1
$S$ = ($ 1$,
$ 1$,
$ 1$,
$ 1$,
$ 1$,
$ 1$,
$ 1$,
$ 1$,
$ 1$;\ \ 
$ 1$,
$ 1$,
$ -\zeta_{6}^{1}$,
$ \zeta_{3}^{1}$,
$ -\zeta_{6}^{1}$,
$ \zeta_{3}^{1}$,
$ -\zeta_{6}^{1}$,
$ \zeta_{3}^{1}$;\ \ 
$ 1$,
$ \zeta_{3}^{1}$,
$ -\zeta_{6}^{1}$,
$ \zeta_{3}^{1}$,
$ -\zeta_{6}^{1}$,
$ \zeta_{3}^{1}$,
$ -\zeta_{6}^{1}$;\ \ 
$ 1$,
$ 1$,
$ -\zeta_{6}^{1}$,
$ \zeta_{3}^{1}$,
$ \zeta_{3}^{1}$,
$ -\zeta_{6}^{1}$;\ \ 
$ 1$,
$ \zeta_{3}^{1}$,
$ -\zeta_{6}^{1}$,
$ -\zeta_{6}^{1}$,
$ \zeta_{3}^{1}$;\ \ 
$ \zeta_{3}^{1}$,
$ -\zeta_{6}^{1}$,
$ 1$,
$ 1$;\ \ 
$ \zeta_{3}^{1}$,
$ 1$,
$ 1$;\ \ 
$ -\zeta_{6}^{1}$,
$ \zeta_{3}^{1}$;\ \ 
$ -\zeta_{6}^{1}$)

Factors = $3_{2,3.}^{3,527}\boxtimes 3_{6,3.}^{3,138}$

\vskip 0.7ex
\hangindent=3em \hangafter=1
$\tau_n$ = ($3.$, $3.$, $9.$)

\vskip 0.7ex
\hangindent=3em \hangafter=1
\textit{Sign-problem-free}

  \vskip 2ex

\noindent2. $9_{4,9.}^{3,277}$ \irep{0}:\ \ 
$d_i$ = ($1.0$,
$1.0$,
$1.0$,
$1.0$,
$1.0$,
$1.0$,
$1.0$,
$1.0$,
$1.0$) 

\vskip 0.7ex
\hangindent=3em \hangafter=1
$D^2= 9.0 = 
9$

\vskip 0.7ex
\hangindent=3em \hangafter=1
$T = ( 0,
\frac{1}{3},
\frac{1}{3},
\frac{1}{3},
\frac{1}{3},
\frac{2}{3},
\frac{2}{3},
\frac{2}{3},
\frac{2}{3} )
$,

\vskip 0.7ex
\hangindent=3em \hangafter=1
$S$ = ($ 1$,
$ 1$,
$ 1$,
$ 1$,
$ 1$,
$ 1$,
$ 1$,
$ 1$,
$ 1$;\ \ 
$ \zeta_{3}^{1}$,
$ 1$,
$ 1$,
$ -\zeta_{6}^{1}$,
$ -\zeta_{6}^{1}$,
$ \zeta_{3}^{1}$,
$ -\zeta_{6}^{1}$,
$ \zeta_{3}^{1}$;\ \ 
$ \zeta_{3}^{1}$,
$ -\zeta_{6}^{1}$,
$ 1$,
$ \zeta_{3}^{1}$,
$ -\zeta_{6}^{1}$,
$ -\zeta_{6}^{1}$,
$ \zeta_{3}^{1}$;\ \ 
$ \zeta_{3}^{1}$,
$ 1$,
$ -\zeta_{6}^{1}$,
$ \zeta_{3}^{1}$,
$ \zeta_{3}^{1}$,
$ -\zeta_{6}^{1}$;\ \ 
$ \zeta_{3}^{1}$,
$ \zeta_{3}^{1}$,
$ -\zeta_{6}^{1}$,
$ \zeta_{3}^{1}$,
$ -\zeta_{6}^{1}$;\ \ 
$ -\zeta_{6}^{1}$,
$ \zeta_{3}^{1}$,
$ 1$,
$ 1$;\ \ 
$ -\zeta_{6}^{1}$,
$ 1$,
$ 1$;\ \ 
$ -\zeta_{6}^{1}$,
$ \zeta_{3}^{1}$;\ \ 
$ -\zeta_{6}^{1}$)

Factors = $3_{2,3.}^{3,527}\boxtimes 3_{2,3.}^{3,527}$

\vskip 0.7ex
\hangindent=3em \hangafter=1
$\tau_n$ = ($-3.$, $-3.$, $9.$)

\vskip 0.7ex
\hangindent=3em \hangafter=1
\textit{Intrinsic sign problem}

  \vskip 2ex

\noindent3. $9_{0,9.}^{9,620}$ \irep{0}:\ \ 
$d_i$ = ($1.0$,
$1.0$,
$1.0$,
$1.0$,
$1.0$,
$1.0$,
$1.0$,
$1.0$,
$1.0$) 

\vskip 0.7ex
\hangindent=3em \hangafter=1
$D^2= 9.0 = 
9$

\vskip 0.7ex
\hangindent=3em \hangafter=1
$T = ( 0,
0,
0,
\frac{1}{9},
\frac{1}{9},
\frac{4}{9},
\frac{4}{9},
\frac{7}{9},
\frac{7}{9} )
$,

\vskip 0.7ex
\hangindent=3em \hangafter=1
$S$ = ($ 1$,
$ 1$,
$ 1$,
$ 1$,
$ 1$,
$ 1$,
$ 1$,
$ 1$,
$ 1$;\ \ 
$ 1$,
$ 1$,
$ -\zeta_{6}^{1}$,
$ \zeta_{3}^{1}$,
$ -\zeta_{6}^{1}$,
$ \zeta_{3}^{1}$,
$ -\zeta_{6}^{1}$,
$ \zeta_{3}^{1}$;\ \ 
$ 1$,
$ \zeta_{3}^{1}$,
$ -\zeta_{6}^{1}$,
$ \zeta_{3}^{1}$,
$ -\zeta_{6}^{1}$,
$ \zeta_{3}^{1}$,
$ -\zeta_{6}^{1}$;\ \ 
$ -\zeta_{18}^{5}$,
$ \zeta_{9}^{2}$,
$ \zeta_{9}^{4}$,
$ -\zeta_{18}^{1}$,
$ \zeta_{9}^{1}$,
$ -\zeta_{18}^{7}$;\ \ 
$ -\zeta_{18}^{5}$,
$ -\zeta_{18}^{1}$,
$ \zeta_{9}^{4}$,
$ -\zeta_{18}^{7}$,
$ \zeta_{9}^{1}$;\ \ 
$ \zeta_{9}^{1}$,
$ -\zeta_{18}^{7}$,
$ -\zeta_{18}^{5}$,
$ \zeta_{9}^{2}$;\ \ 
$ \zeta_{9}^{1}$,
$ \zeta_{9}^{2}$,
$ -\zeta_{18}^{5}$;\ \ 
$ \zeta_{9}^{4}$,
$ -\zeta_{18}^{1}$;\ \ 
$ \zeta_{9}^{4}$)

\vskip 0.7ex
\hangindent=3em \hangafter=1
$\tau_n$ = ($3.$, $3.$, $0. + 5.2 i$, $3.$, $3.$, $0. - 5.2 i$, $3.$, $3.$, $9.$)

\vskip 0.7ex
\hangindent=3em \hangafter=1
\textit{Intrinsic sign problem}

  \vskip 2ex

\noindent4. $9_{0,9.}^{9,462}$ \irep{0}:\ \ 
$d_i$ = ($1.0$,
$1.0$,
$1.0$,
$1.0$,
$1.0$,
$1.0$,
$1.0$,
$1.0$,
$1.0$) 

\vskip 0.7ex
\hangindent=3em \hangafter=1
$D^2= 9.0 = 
9$

\vskip 0.7ex
\hangindent=3em \hangafter=1
$T = ( 0,
0,
0,
\frac{2}{9},
\frac{2}{9},
\frac{5}{9},
\frac{5}{9},
\frac{8}{9},
\frac{8}{9} )
$,

\vskip 0.7ex
\hangindent=3em \hangafter=1
$S$ = ($ 1$,
$ 1$,
$ 1$,
$ 1$,
$ 1$,
$ 1$,
$ 1$,
$ 1$,
$ 1$;\ \ 
$ 1$,
$ 1$,
$ -\zeta_{6}^{1}$,
$ \zeta_{3}^{1}$,
$ -\zeta_{6}^{1}$,
$ \zeta_{3}^{1}$,
$ -\zeta_{6}^{1}$,
$ \zeta_{3}^{1}$;\ \ 
$ 1$,
$ \zeta_{3}^{1}$,
$ -\zeta_{6}^{1}$,
$ \zeta_{3}^{1}$,
$ -\zeta_{6}^{1}$,
$ \zeta_{3}^{1}$,
$ -\zeta_{6}^{1}$;\ \ 
$ -\zeta_{18}^{1}$,
$ \zeta_{9}^{4}$,
$ \zeta_{9}^{2}$,
$ -\zeta_{18}^{5}$,
$ -\zeta_{18}^{7}$,
$ \zeta_{9}^{1}$;\ \ 
$ -\zeta_{18}^{1}$,
$ -\zeta_{18}^{5}$,
$ \zeta_{9}^{2}$,
$ \zeta_{9}^{1}$,
$ -\zeta_{18}^{7}$;\ \ 
$ -\zeta_{18}^{7}$,
$ \zeta_{9}^{1}$,
$ -\zeta_{18}^{1}$,
$ \zeta_{9}^{4}$;\ \ 
$ -\zeta_{18}^{7}$,
$ \zeta_{9}^{4}$,
$ -\zeta_{18}^{1}$;\ \ 
$ \zeta_{9}^{2}$,
$ -\zeta_{18}^{5}$;\ \ 
$ \zeta_{9}^{2}$)

\vskip 0.7ex
\hangindent=3em \hangafter=1
$\tau_n$ = ($3.$, $3.$, $0. - 5.2 i$, $3.$, $3.$, $0. + 5.2 i$, $3.$, $3.$, $9.$)

\vskip 0.7ex
\hangindent=3em \hangafter=1
\textit{Intrinsic sign problem}

  \vskip 2ex

\noindent5. $9_{\frac{5}{2},12.}^{48,250}$ \irep{490}:\ \ 
$d_i$ = ($1.0$,
$1.0$,
$1.0$,
$1.0$,
$1.0$,
$1.0$,
$1.414$,
$1.414$,
$1.414$) 

\vskip 0.7ex
\hangindent=3em \hangafter=1
$D^2= 12.0 = 
12$

\vskip 0.7ex
\hangindent=3em \hangafter=1
$T = ( 0,
\frac{1}{2},
\frac{1}{3},
\frac{1}{3},
\frac{5}{6},
\frac{5}{6},
\frac{1}{16},
\frac{19}{48},
\frac{19}{48} )
$,

\vskip 0.7ex
\hangindent=3em \hangafter=1
$S$ = ($ 1$,
$ 1$,
$ 1$,
$ 1$,
$ 1$,
$ 1$,
$ \sqrt{2}$,
$ \sqrt{2}$,
$ \sqrt{2}$;\ \ 
$ 1$,
$ 1$,
$ 1$,
$ 1$,
$ 1$,
$ -\sqrt{2}$,
$ -\sqrt{2}$,
$ -\sqrt{2}$;\ \ 
$ \zeta_{3}^{1}$,
$ -\zeta_{6}^{1}$,
$ -\zeta_{6}^{1}$,
$ \zeta_{3}^{1}$,
$ \sqrt{2}$,
$ -\sqrt{2}\zeta_{6}^{1}$,
$ \sqrt{2}\zeta_{3}^{1}$;\ \ 
$ \zeta_{3}^{1}$,
$ \zeta_{3}^{1}$,
$ -\zeta_{6}^{1}$,
$ \sqrt{2}$,
$ \sqrt{2}\zeta_{3}^{1}$,
$ -\sqrt{2}\zeta_{6}^{1}$;\ \ 
$ \zeta_{3}^{1}$,
$ -\zeta_{6}^{1}$,
$ -\sqrt{2}$,
$ -\sqrt{2}\zeta_{3}^{1}$,
$ \sqrt{2}\zeta_{6}^{1}$;\ \ 
$ \zeta_{3}^{1}$,
$ -\sqrt{2}$,
$ \sqrt{2}\zeta_{6}^{1}$,
$ -\sqrt{2}\zeta_{3}^{1}$;\ \ 
$0$,
$0$,
$0$;\ \ 
$0$,
$0$;\ \ 
$0$)

Factors = $3_{2,3.}^{3,527}\boxtimes 3_{\frac{1}{2},4.}^{16,598}$

\vskip 0.7ex
\hangindent=3em \hangafter=1
$\tau_n$ = ($-1.33 + 3.2 i$, $2.45 - 5.91 i$, $2.3 + 5.54 i$, $-3.46 + 3.46 i$, $3.2 + 1.33 i$, $1.76 + 4.24 i$, $-1.33 - 3.2 i$, $0.$, $-5.54 - 2.3 i$, $2.45 + 1.02 i$, $-3.2 + 1.33 i$, $6. - 6. i$, $3.2 + 1.33 i$, $-2.45 - 5.91 i$, $5.54 - 2.3 i$, $0. + 6.93 i$, $1.33 - 3.2 i$, $10.24 + 4.24 i$, $-3.2 + 1.33 i$, $3.46 - 3.46 i$, $-2.3 + 5.54 i$, $-2.45 + 1.02 i$, $1.33 + 3.2 i$, $0.$, $1.33 - 3.2 i$, $-2.45 - 1.02 i$, $-2.3 - 5.54 i$, $3.46 + 3.46 i$, $-3.2 - 1.33 i$, $10.24 - 4.24 i$, $1.33 + 3.2 i$, $0. - 6.93 i$, $5.54 + 2.3 i$, $-2.45 + 5.91 i$, $3.2 - 1.33 i$, $6. + 6. i$, $-3.2 - 1.33 i$, $2.45 - 1.02 i$, $-5.54 + 2.3 i$, $0.$, $-1.33 + 3.2 i$, $1.76 - 4.24 i$, $3.2 - 1.33 i$, $-3.46 - 3.46 i$, $2.3 - 5.54 i$, $2.45 + 5.91 i$, $-1.33 - 3.2 i$, $12.$)

\vskip 0.7ex
\hangindent=3em \hangafter=1
\textit{Intrinsic sign problem}

  \vskip 2ex

\noindent6. $9_{\frac{7}{2},12.}^{48,117}$ \irep{490}:\ \ 
$d_i$ = ($1.0$,
$1.0$,
$1.0$,
$1.0$,
$1.0$,
$1.0$,
$1.414$,
$1.414$,
$1.414$) 

\vskip 0.7ex
\hangindent=3em \hangafter=1
$D^2= 12.0 = 
12$

\vskip 0.7ex
\hangindent=3em \hangafter=1
$T = ( 0,
\frac{1}{2},
\frac{1}{3},
\frac{1}{3},
\frac{5}{6},
\frac{5}{6},
\frac{3}{16},
\frac{25}{48},
\frac{25}{48} )
$,

\vskip 0.7ex
\hangindent=3em \hangafter=1
$S$ = ($ 1$,
$ 1$,
$ 1$,
$ 1$,
$ 1$,
$ 1$,
$ \sqrt{2}$,
$ \sqrt{2}$,
$ \sqrt{2}$;\ \ 
$ 1$,
$ 1$,
$ 1$,
$ 1$,
$ 1$,
$ -\sqrt{2}$,
$ -\sqrt{2}$,
$ -\sqrt{2}$;\ \ 
$ \zeta_{3}^{1}$,
$ -\zeta_{6}^{1}$,
$ -\zeta_{6}^{1}$,
$ \zeta_{3}^{1}$,
$ \sqrt{2}$,
$ -\sqrt{2}\zeta_{6}^{1}$,
$ \sqrt{2}\zeta_{3}^{1}$;\ \ 
$ \zeta_{3}^{1}$,
$ \zeta_{3}^{1}$,
$ -\zeta_{6}^{1}$,
$ \sqrt{2}$,
$ \sqrt{2}\zeta_{3}^{1}$,
$ -\sqrt{2}\zeta_{6}^{1}$;\ \ 
$ \zeta_{3}^{1}$,
$ -\zeta_{6}^{1}$,
$ -\sqrt{2}$,
$ -\sqrt{2}\zeta_{3}^{1}$,
$ \sqrt{2}\zeta_{6}^{1}$;\ \ 
$ \zeta_{3}^{1}$,
$ -\sqrt{2}$,
$ \sqrt{2}\zeta_{6}^{1}$,
$ -\sqrt{2}\zeta_{3}^{1}$;\ \ 
$0$,
$0$,
$0$;\ \ 
$0$,
$0$;\ \ 
$0$)

Factors = $3_{2,3.}^{3,527}\boxtimes 3_{\frac{3}{2},4.}^{16,553}$

\vskip 0.7ex
\hangindent=3em \hangafter=1
$\tau_n$ = ($-3.2 + 1.33 i$, $2.45 - 1.02 i$, $-5.54 - 2.3 i$, $3.46 + 3.46 i$, $-1.33 - 3.2 i$, $10.24 + 4.24 i$, $-3.2 - 1.33 i$, $0.$, $-2.3 - 5.54 i$, $2.45 + 5.91 i$, $1.33 - 3.2 i$, $6. + 6. i$, $-1.33 - 3.2 i$, $-2.45 - 1.02 i$, $2.3 - 5.54 i$, $0. + 6.93 i$, $3.2 - 1.33 i$, $1.76 + 4.24 i$, $1.33 - 3.2 i$, $-3.46 - 3.46 i$, $5.54 - 2.3 i$, $-2.45 + 5.91 i$, $3.2 + 1.33 i$, $0.$, $3.2 - 1.33 i$, $-2.45 - 5.91 i$, $5.54 + 2.3 i$, $-3.46 + 3.46 i$, $1.33 + 3.2 i$, $1.76 - 4.24 i$, $3.2 + 1.33 i$, $0. - 6.93 i$, $2.3 + 5.54 i$, $-2.45 + 1.02 i$, $-1.33 + 3.2 i$, $6. - 6. i$, $1.33 + 3.2 i$, $2.45 - 5.91 i$, $-2.3 + 5.54 i$, $0.$, $-3.2 + 1.33 i$, $10.24 - 4.24 i$, $-1.33 + 3.2 i$, $3.46 - 3.46 i$, $-5.54 + 2.3 i$, $2.45 + 1.02 i$, $-3.2 - 1.33 i$, $12.$)

\vskip 0.7ex
\hangindent=3em \hangafter=1
\textit{Intrinsic sign problem}

  \vskip 2ex

\noindent7. $9_{\frac{9}{2},12.}^{48,618}$ \irep{490}:\ \ 
$d_i$ = ($1.0$,
$1.0$,
$1.0$,
$1.0$,
$1.0$,
$1.0$,
$1.414$,
$1.414$,
$1.414$) 

\vskip 0.7ex
\hangindent=3em \hangafter=1
$D^2= 12.0 = 
12$

\vskip 0.7ex
\hangindent=3em \hangafter=1
$T = ( 0,
\frac{1}{2},
\frac{1}{3},
\frac{1}{3},
\frac{5}{6},
\frac{5}{6},
\frac{5}{16},
\frac{31}{48},
\frac{31}{48} )
$,

\vskip 0.7ex
\hangindent=3em \hangafter=1
$S$ = ($ 1$,
$ 1$,
$ 1$,
$ 1$,
$ 1$,
$ 1$,
$ \sqrt{2}$,
$ \sqrt{2}$,
$ \sqrt{2}$;\ \ 
$ 1$,
$ 1$,
$ 1$,
$ 1$,
$ 1$,
$ -\sqrt{2}$,
$ -\sqrt{2}$,
$ -\sqrt{2}$;\ \ 
$ \zeta_{3}^{1}$,
$ -\zeta_{6}^{1}$,
$ -\zeta_{6}^{1}$,
$ \zeta_{3}^{1}$,
$ \sqrt{2}$,
$ -\sqrt{2}\zeta_{6}^{1}$,
$ \sqrt{2}\zeta_{3}^{1}$;\ \ 
$ \zeta_{3}^{1}$,
$ \zeta_{3}^{1}$,
$ -\zeta_{6}^{1}$,
$ \sqrt{2}$,
$ \sqrt{2}\zeta_{3}^{1}$,
$ -\sqrt{2}\zeta_{6}^{1}$;\ \ 
$ \zeta_{3}^{1}$,
$ -\zeta_{6}^{1}$,
$ -\sqrt{2}$,
$ -\sqrt{2}\zeta_{3}^{1}$,
$ \sqrt{2}\zeta_{6}^{1}$;\ \ 
$ \zeta_{3}^{1}$,
$ -\sqrt{2}$,
$ \sqrt{2}\zeta_{6}^{1}$,
$ -\sqrt{2}\zeta_{3}^{1}$;\ \ 
$0$,
$0$,
$0$;\ \ 
$0$,
$0$;\ \ 
$0$)

Factors = $3_{2,3.}^{3,527}\boxtimes 3_{\frac{5}{2},4.}^{16,465}$

\vskip 0.7ex
\hangindent=3em \hangafter=1
$\tau_n$ = ($-3.2 - 1.33 i$, $-2.45 - 1.02 i$, $5.54 - 2.3 i$, $-3.46 + 3.46 i$, $-1.33 + 3.2 i$, $10.24 - 4.24 i$, $-3.2 + 1.33 i$, $0.$, $2.3 - 5.54 i$, $-2.45 + 5.91 i$, $1.33 + 3.2 i$, $6. - 6. i$, $-1.33 + 3.2 i$, $2.45 - 1.02 i$, $-2.3 - 5.54 i$, $0. + 6.93 i$, $3.2 + 1.33 i$, $1.76 - 4.24 i$, $1.33 + 3.2 i$, $3.46 - 3.46 i$, $-5.54 - 2.3 i$, $2.45 + 5.91 i$, $3.2 - 1.33 i$, $0.$, $3.2 + 1.33 i$, $2.45 - 5.91 i$, $-5.54 + 2.3 i$, $3.46 + 3.46 i$, $1.33 - 3.2 i$, $1.76 + 4.24 i$, $3.2 - 1.33 i$, $0. - 6.93 i$, $-2.3 + 5.54 i$, $2.45 + 1.02 i$, $-1.33 - 3.2 i$, $6. + 6. i$, $1.33 - 3.2 i$, $-2.45 - 5.91 i$, $2.3 + 5.54 i$, $0.$, $-3.2 - 1.33 i$, $10.24 + 4.24 i$, $-1.33 - 3.2 i$, $-3.46 - 3.46 i$, $5.54 + 2.3 i$, $-2.45 + 1.02 i$, $-3.2 + 1.33 i$, $12.$)

\vskip 0.7ex
\hangindent=3em \hangafter=1
\textit{Intrinsic sign problem}

  \vskip 2ex

\noindent8. $9_{\frac{11}{2},12.}^{48,125}$ \irep{490}:\ \ 
$d_i$ = ($1.0$,
$1.0$,
$1.0$,
$1.0$,
$1.0$,
$1.0$,
$1.414$,
$1.414$,
$1.414$) 

\vskip 0.7ex
\hangindent=3em \hangafter=1
$D^2= 12.0 = 
12$

\vskip 0.7ex
\hangindent=3em \hangafter=1
$T = ( 0,
\frac{1}{2},
\frac{1}{3},
\frac{1}{3},
\frac{5}{6},
\frac{5}{6},
\frac{7}{16},
\frac{37}{48},
\frac{37}{48} )
$,

\vskip 0.7ex
\hangindent=3em \hangafter=1
$S$ = ($ 1$,
$ 1$,
$ 1$,
$ 1$,
$ 1$,
$ 1$,
$ \sqrt{2}$,
$ \sqrt{2}$,
$ \sqrt{2}$;\ \ 
$ 1$,
$ 1$,
$ 1$,
$ 1$,
$ 1$,
$ -\sqrt{2}$,
$ -\sqrt{2}$,
$ -\sqrt{2}$;\ \ 
$ \zeta_{3}^{1}$,
$ -\zeta_{6}^{1}$,
$ -\zeta_{6}^{1}$,
$ \zeta_{3}^{1}$,
$ \sqrt{2}$,
$ -\sqrt{2}\zeta_{6}^{1}$,
$ \sqrt{2}\zeta_{3}^{1}$;\ \ 
$ \zeta_{3}^{1}$,
$ \zeta_{3}^{1}$,
$ -\zeta_{6}^{1}$,
$ \sqrt{2}$,
$ \sqrt{2}\zeta_{3}^{1}$,
$ -\sqrt{2}\zeta_{6}^{1}$;\ \ 
$ \zeta_{3}^{1}$,
$ -\zeta_{6}^{1}$,
$ -\sqrt{2}$,
$ -\sqrt{2}\zeta_{3}^{1}$,
$ \sqrt{2}\zeta_{6}^{1}$;\ \ 
$ \zeta_{3}^{1}$,
$ -\sqrt{2}$,
$ \sqrt{2}\zeta_{6}^{1}$,
$ -\sqrt{2}\zeta_{3}^{1}$;\ \ 
$0$,
$0$,
$0$;\ \ 
$0$,
$0$;\ \ 
$0$)

Factors = $3_{2,3.}^{3,527}\boxtimes 3_{\frac{7}{2},4.}^{16,332}$

\vskip 0.7ex
\hangindent=3em \hangafter=1
$\tau_n$ = ($-1.33 - 3.2 i$, $-2.45 - 5.91 i$, $-2.3 + 5.54 i$, $3.46 + 3.46 i$, $3.2 - 1.33 i$, $1.76 - 4.24 i$, $-1.33 + 3.2 i$, $0.$, $5.54 - 2.3 i$, $-2.45 + 1.02 i$, $-3.2 - 1.33 i$, $6. + 6. i$, $3.2 - 1.33 i$, $2.45 - 5.91 i$, $-5.54 - 2.3 i$, $0. + 6.93 i$, $1.33 + 3.2 i$, $10.24 - 4.24 i$, $-3.2 - 1.33 i$, $-3.46 - 3.46 i$, $2.3 + 5.54 i$, $2.45 + 1.02 i$, $1.33 - 3.2 i$, $0.$, $1.33 + 3.2 i$, $2.45 - 1.02 i$, $2.3 - 5.54 i$, $-3.46 + 3.46 i$, $-3.2 + 1.33 i$, $10.24 + 4.24 i$, $1.33 - 3.2 i$, $0. - 6.93 i$, $-5.54 + 2.3 i$, $2.45 + 5.91 i$, $3.2 + 1.33 i$, $6. - 6. i$, $-3.2 + 1.33 i$, $-2.45 - 1.02 i$, $5.54 + 2.3 i$, $0.$, $-1.33 - 3.2 i$, $1.76 + 4.24 i$, $3.2 + 1.33 i$, $3.46 - 3.46 i$, $-2.3 - 5.54 i$, $-2.45 + 5.91 i$, $-1.33 + 3.2 i$, $12.$)

\vskip 0.7ex
\hangindent=3em \hangafter=1
\textit{Intrinsic sign problem}

  \vskip 2ex

\noindent9. $9_{\frac{13}{2},12.}^{48,201}$ \irep{490}:\ \ 
$d_i$ = ($1.0$,
$1.0$,
$1.0$,
$1.0$,
$1.0$,
$1.0$,
$1.414$,
$1.414$,
$1.414$) 

\vskip 0.7ex
\hangindent=3em \hangafter=1
$D^2= 12.0 = 
12$

\vskip 0.7ex
\hangindent=3em \hangafter=1
$T = ( 0,
\frac{1}{2},
\frac{1}{3},
\frac{1}{3},
\frac{5}{6},
\frac{5}{6},
\frac{9}{16},
\frac{43}{48},
\frac{43}{48} )
$,

\vskip 0.7ex
\hangindent=3em \hangafter=1
$S$ = ($ 1$,
$ 1$,
$ 1$,
$ 1$,
$ 1$,
$ 1$,
$ \sqrt{2}$,
$ \sqrt{2}$,
$ \sqrt{2}$;\ \ 
$ 1$,
$ 1$,
$ 1$,
$ 1$,
$ 1$,
$ -\sqrt{2}$,
$ -\sqrt{2}$,
$ -\sqrt{2}$;\ \ 
$ \zeta_{3}^{1}$,
$ -\zeta_{6}^{1}$,
$ -\zeta_{6}^{1}$,
$ \zeta_{3}^{1}$,
$ \sqrt{2}$,
$ -\sqrt{2}\zeta_{6}^{1}$,
$ \sqrt{2}\zeta_{3}^{1}$;\ \ 
$ \zeta_{3}^{1}$,
$ \zeta_{3}^{1}$,
$ -\zeta_{6}^{1}$,
$ \sqrt{2}$,
$ \sqrt{2}\zeta_{3}^{1}$,
$ -\sqrt{2}\zeta_{6}^{1}$;\ \ 
$ \zeta_{3}^{1}$,
$ -\zeta_{6}^{1}$,
$ -\sqrt{2}$,
$ -\sqrt{2}\zeta_{3}^{1}$,
$ \sqrt{2}\zeta_{6}^{1}$;\ \ 
$ \zeta_{3}^{1}$,
$ -\sqrt{2}$,
$ \sqrt{2}\zeta_{6}^{1}$,
$ -\sqrt{2}\zeta_{3}^{1}$;\ \ 
$0$,
$0$,
$0$;\ \ 
$0$,
$0$;\ \ 
$0$)

Factors = $3_{2,3.}^{3,527}\boxtimes 3_{\frac{9}{2},4.}^{16,156}$

\vskip 0.7ex
\hangindent=3em \hangafter=1
$\tau_n$ = ($1.33 - 3.2 i$, $2.45 - 5.91 i$, $-2.3 - 5.54 i$, $-3.46 + 3.46 i$, $-3.2 - 1.33 i$, $1.76 + 4.24 i$, $1.33 + 3.2 i$, $0.$, $5.54 + 2.3 i$, $2.45 + 1.02 i$, $3.2 - 1.33 i$, $6. - 6. i$, $-3.2 - 1.33 i$, $-2.45 - 5.91 i$, $-5.54 + 2.3 i$, $0. + 6.93 i$, $-1.33 + 3.2 i$, $10.24 + 4.24 i$, $3.2 - 1.33 i$, $3.46 - 3.46 i$, $2.3 - 5.54 i$, $-2.45 + 1.02 i$, $-1.33 - 3.2 i$, $0.$, $-1.33 + 3.2 i$, $-2.45 - 1.02 i$, $2.3 + 5.54 i$, $3.46 + 3.46 i$, $3.2 + 1.33 i$, $10.24 - 4.24 i$, $-1.33 - 3.2 i$, $0. - 6.93 i$, $-5.54 - 2.3 i$, $-2.45 + 5.91 i$, $-3.2 + 1.33 i$, $6. + 6. i$, $3.2 + 1.33 i$, $2.45 - 1.02 i$, $5.54 - 2.3 i$, $0.$, $1.33 - 3.2 i$, $1.76 - 4.24 i$, $-3.2 + 1.33 i$, $-3.46 - 3.46 i$, $-2.3 + 5.54 i$, $2.45 + 5.91 i$, $1.33 + 3.2 i$, $12.$)

\vskip 0.7ex
\hangindent=3em \hangafter=1
\textit{Intrinsic sign problem}

  \vskip 2ex

\noindent10. $9_{\frac{15}{2},12.}^{48,298}$ \irep{490}:\ \ 
$d_i$ = ($1.0$,
$1.0$,
$1.0$,
$1.0$,
$1.0$,
$1.0$,
$1.414$,
$1.414$,
$1.414$) 

\vskip 0.7ex
\hangindent=3em \hangafter=1
$D^2= 12.0 = 
12$

\vskip 0.7ex
\hangindent=3em \hangafter=1
$T = ( 0,
\frac{1}{2},
\frac{1}{3},
\frac{1}{3},
\frac{5}{6},
\frac{5}{6},
\frac{11}{16},
\frac{1}{48},
\frac{1}{48} )
$,

\vskip 0.7ex
\hangindent=3em \hangafter=1
$S$ = ($ 1$,
$ 1$,
$ 1$,
$ 1$,
$ 1$,
$ 1$,
$ \sqrt{2}$,
$ \sqrt{2}$,
$ \sqrt{2}$;\ \ 
$ 1$,
$ 1$,
$ 1$,
$ 1$,
$ 1$,
$ -\sqrt{2}$,
$ -\sqrt{2}$,
$ -\sqrt{2}$;\ \ 
$ \zeta_{3}^{1}$,
$ -\zeta_{6}^{1}$,
$ -\zeta_{6}^{1}$,
$ \zeta_{3}^{1}$,
$ \sqrt{2}$,
$ -\sqrt{2}\zeta_{6}^{1}$,
$ \sqrt{2}\zeta_{3}^{1}$;\ \ 
$ \zeta_{3}^{1}$,
$ \zeta_{3}^{1}$,
$ -\zeta_{6}^{1}$,
$ \sqrt{2}$,
$ \sqrt{2}\zeta_{3}^{1}$,
$ -\sqrt{2}\zeta_{6}^{1}$;\ \ 
$ \zeta_{3}^{1}$,
$ -\zeta_{6}^{1}$,
$ -\sqrt{2}$,
$ -\sqrt{2}\zeta_{3}^{1}$,
$ \sqrt{2}\zeta_{6}^{1}$;\ \ 
$ \zeta_{3}^{1}$,
$ -\sqrt{2}$,
$ \sqrt{2}\zeta_{6}^{1}$,
$ -\sqrt{2}\zeta_{3}^{1}$;\ \ 
$0$,
$0$,
$0$;\ \ 
$0$,
$0$;\ \ 
$0$)

Factors = $3_{2,3.}^{3,527}\boxtimes 3_{\frac{11}{2},4.}^{16,648}$

\vskip 0.7ex
\hangindent=3em \hangafter=1
$\tau_n$ = ($3.2 - 1.33 i$, $2.45 - 1.02 i$, $5.54 + 2.3 i$, $3.46 + 3.46 i$, $1.33 + 3.2 i$, $10.24 + 4.24 i$, $3.2 + 1.33 i$, $0.$, $2.3 + 5.54 i$, $2.45 + 5.91 i$, $-1.33 + 3.2 i$, $6. + 6. i$, $1.33 + 3.2 i$, $-2.45 - 1.02 i$, $-2.3 + 5.54 i$, $0. + 6.93 i$, $-3.2 + 1.33 i$, $1.76 + 4.24 i$, $-1.33 + 3.2 i$, $-3.46 - 3.46 i$, $-5.54 + 2.3 i$, $-2.45 + 5.91 i$, $-3.2 - 1.33 i$, $0.$, $-3.2 + 1.33 i$, $-2.45 - 5.91 i$, $-5.54 - 2.3 i$, $-3.46 + 3.46 i$, $-1.33 - 3.2 i$, $1.76 - 4.24 i$, $-3.2 - 1.33 i$, $0. - 6.93 i$, $-2.3 - 5.54 i$, $-2.45 + 1.02 i$, $1.33 - 3.2 i$, $6. - 6. i$, $-1.33 - 3.2 i$, $2.45 - 5.91 i$, $2.3 - 5.54 i$, $0.$, $3.2 - 1.33 i$, $10.24 - 4.24 i$, $1.33 - 3.2 i$, $3.46 - 3.46 i$, $5.54 - 2.3 i$, $2.45 + 1.02 i$, $3.2 + 1.33 i$, $12.$)

\vskip 0.7ex
\hangindent=3em \hangafter=1
\textit{Intrinsic sign problem}

  \vskip 2ex

\noindent11. $9_{\frac{1}{2},12.}^{48,294}$ \irep{490}:\ \ 
$d_i$ = ($1.0$,
$1.0$,
$1.0$,
$1.0$,
$1.0$,
$1.0$,
$1.414$,
$1.414$,
$1.414$) 

\vskip 0.7ex
\hangindent=3em \hangafter=1
$D^2= 12.0 = 
12$

\vskip 0.7ex
\hangindent=3em \hangafter=1
$T = ( 0,
\frac{1}{2},
\frac{1}{3},
\frac{1}{3},
\frac{5}{6},
\frac{5}{6},
\frac{13}{16},
\frac{7}{48},
\frac{7}{48} )
$,

\vskip 0.7ex
\hangindent=3em \hangafter=1
$S$ = ($ 1$,
$ 1$,
$ 1$,
$ 1$,
$ 1$,
$ 1$,
$ \sqrt{2}$,
$ \sqrt{2}$,
$ \sqrt{2}$;\ \ 
$ 1$,
$ 1$,
$ 1$,
$ 1$,
$ 1$,
$ -\sqrt{2}$,
$ -\sqrt{2}$,
$ -\sqrt{2}$;\ \ 
$ \zeta_{3}^{1}$,
$ -\zeta_{6}^{1}$,
$ -\zeta_{6}^{1}$,
$ \zeta_{3}^{1}$,
$ \sqrt{2}$,
$ -\sqrt{2}\zeta_{6}^{1}$,
$ \sqrt{2}\zeta_{3}^{1}$;\ \ 
$ \zeta_{3}^{1}$,
$ \zeta_{3}^{1}$,
$ -\zeta_{6}^{1}$,
$ \sqrt{2}$,
$ \sqrt{2}\zeta_{3}^{1}$,
$ -\sqrt{2}\zeta_{6}^{1}$;\ \ 
$ \zeta_{3}^{1}$,
$ -\zeta_{6}^{1}$,
$ -\sqrt{2}$,
$ -\sqrt{2}\zeta_{3}^{1}$,
$ \sqrt{2}\zeta_{6}^{1}$;\ \ 
$ \zeta_{3}^{1}$,
$ -\sqrt{2}$,
$ \sqrt{2}\zeta_{6}^{1}$,
$ -\sqrt{2}\zeta_{3}^{1}$;\ \ 
$0$,
$0$,
$0$;\ \ 
$0$,
$0$;\ \ 
$0$)

Factors = $3_{2,3.}^{3,527}\boxtimes 3_{\frac{13}{2},4.}^{16,330}$

\vskip 0.7ex
\hangindent=3em \hangafter=1
$\tau_n$ = ($3.2 + 1.33 i$, $-2.45 - 1.02 i$, $-5.54 + 2.3 i$, $-3.46 + 3.46 i$, $1.33 - 3.2 i$, $10.24 - 4.24 i$, $3.2 - 1.33 i$, $0.$, $-2.3 + 5.54 i$, $-2.45 + 5.91 i$, $-1.33 - 3.2 i$, $6. - 6. i$, $1.33 - 3.2 i$, $2.45 - 1.02 i$, $2.3 + 5.54 i$, $0. + 6.93 i$, $-3.2 - 1.33 i$, $1.76 - 4.24 i$, $-1.33 - 3.2 i$, $3.46 - 3.46 i$, $5.54 + 2.3 i$, $2.45 + 5.91 i$, $-3.2 + 1.33 i$, $0.$, $-3.2 - 1.33 i$, $2.45 - 5.91 i$, $5.54 - 2.3 i$, $3.46 + 3.46 i$, $-1.33 + 3.2 i$, $1.76 + 4.24 i$, $-3.2 + 1.33 i$, $0. - 6.93 i$, $2.3 - 5.54 i$, $2.45 + 1.02 i$, $1.33 + 3.2 i$, $6. + 6. i$, $-1.33 + 3.2 i$, $-2.45 - 5.91 i$, $-2.3 - 5.54 i$, $0.$, $3.2 + 1.33 i$, $10.24 + 4.24 i$, $1.33 + 3.2 i$, $-3.46 - 3.46 i$, $-5.54 - 2.3 i$, $-2.45 + 1.02 i$, $3.2 - 1.33 i$, $12.$)

\vskip 0.7ex
\hangindent=3em \hangafter=1
\textit{Intrinsic sign problem}

  \vskip 2ex

\noindent12. $9_{\frac{3}{2},12.}^{48,750}$ \irep{490}:\ \ 
$d_i$ = ($1.0$,
$1.0$,
$1.0$,
$1.0$,
$1.0$,
$1.0$,
$1.414$,
$1.414$,
$1.414$) 

\vskip 0.7ex
\hangindent=3em \hangafter=1
$D^2= 12.0 = 
12$

\vskip 0.7ex
\hangindent=3em \hangafter=1
$T = ( 0,
\frac{1}{2},
\frac{1}{3},
\frac{1}{3},
\frac{5}{6},
\frac{5}{6},
\frac{15}{16},
\frac{13}{48},
\frac{13}{48} )
$,

\vskip 0.7ex
\hangindent=3em \hangafter=1
$S$ = ($ 1$,
$ 1$,
$ 1$,
$ 1$,
$ 1$,
$ 1$,
$ \sqrt{2}$,
$ \sqrt{2}$,
$ \sqrt{2}$;\ \ 
$ 1$,
$ 1$,
$ 1$,
$ 1$,
$ 1$,
$ -\sqrt{2}$,
$ -\sqrt{2}$,
$ -\sqrt{2}$;\ \ 
$ \zeta_{3}^{1}$,
$ -\zeta_{6}^{1}$,
$ -\zeta_{6}^{1}$,
$ \zeta_{3}^{1}$,
$ \sqrt{2}$,
$ -\sqrt{2}\zeta_{6}^{1}$,
$ \sqrt{2}\zeta_{3}^{1}$;\ \ 
$ \zeta_{3}^{1}$,
$ \zeta_{3}^{1}$,
$ -\zeta_{6}^{1}$,
$ \sqrt{2}$,
$ \sqrt{2}\zeta_{3}^{1}$,
$ -\sqrt{2}\zeta_{6}^{1}$;\ \ 
$ \zeta_{3}^{1}$,
$ -\zeta_{6}^{1}$,
$ -\sqrt{2}$,
$ -\sqrt{2}\zeta_{3}^{1}$,
$ \sqrt{2}\zeta_{6}^{1}$;\ \ 
$ \zeta_{3}^{1}$,
$ -\sqrt{2}$,
$ \sqrt{2}\zeta_{6}^{1}$,
$ -\sqrt{2}\zeta_{3}^{1}$;\ \ 
$0$,
$0$,
$0$;\ \ 
$0$,
$0$;\ \ 
$0$)

Factors = $3_{2,3.}^{3,527}\boxtimes 3_{\frac{15}{2},4.}^{16,639}$

\vskip 0.7ex
\hangindent=3em \hangafter=1
$\tau_n$ = ($1.33 + 3.2 i$, $-2.45 - 5.91 i$, $2.3 - 5.54 i$, $3.46 + 3.46 i$, $-3.2 + 1.33 i$, $1.76 - 4.24 i$, $1.33 - 3.2 i$, $0.$, $-5.54 + 2.3 i$, $-2.45 + 1.02 i$, $3.2 + 1.33 i$, $6. + 6. i$, $-3.2 + 1.33 i$, $2.45 - 5.91 i$, $5.54 + 2.3 i$, $0. + 6.93 i$, $-1.33 - 3.2 i$, $10.24 - 4.24 i$, $3.2 + 1.33 i$, $-3.46 - 3.46 i$, $-2.3 - 5.54 i$, $2.45 + 1.02 i$, $-1.33 + 3.2 i$, $0.$, $-1.33 - 3.2 i$, $2.45 - 1.02 i$, $-2.3 + 5.54 i$, $-3.46 + 3.46 i$, $3.2 - 1.33 i$, $10.24 + 4.24 i$, $-1.33 + 3.2 i$, $0. - 6.93 i$, $5.54 - 2.3 i$, $2.45 + 5.91 i$, $-3.2 - 1.33 i$, $6. - 6. i$, $3.2 - 1.33 i$, $-2.45 - 1.02 i$, $-5.54 - 2.3 i$, $0.$, $1.33 + 3.2 i$, $1.76 + 4.24 i$, $-3.2 - 1.33 i$, $3.46 - 3.46 i$, $2.3 + 5.54 i$, $-2.45 + 5.91 i$, $1.33 - 3.2 i$, $12.$)

\vskip 0.7ex
\hangindent=3em \hangafter=1
\textit{Intrinsic sign problem}

  \vskip 2ex

\noindent13. $9_{\frac{13}{2},12.}^{48,143}$ \irep{490}:\ \ 
$d_i$ = ($1.0$,
$1.0$,
$1.0$,
$1.0$,
$1.0$,
$1.0$,
$1.414$,
$1.414$,
$1.414$) 

\vskip 0.7ex
\hangindent=3em \hangafter=1
$D^2= 12.0 = 
12$

\vskip 0.7ex
\hangindent=3em \hangafter=1
$T = ( 0,
\frac{1}{2},
\frac{2}{3},
\frac{2}{3},
\frac{1}{6},
\frac{1}{6},
\frac{1}{16},
\frac{35}{48},
\frac{35}{48} )
$,

\vskip 0.7ex
\hangindent=3em \hangafter=1
$S$ = ($ 1$,
$ 1$,
$ 1$,
$ 1$,
$ 1$,
$ 1$,
$ \sqrt{2}$,
$ \sqrt{2}$,
$ \sqrt{2}$;\ \ 
$ 1$,
$ 1$,
$ 1$,
$ 1$,
$ 1$,
$ -\sqrt{2}$,
$ -\sqrt{2}$,
$ -\sqrt{2}$;\ \ 
$ -\zeta_{6}^{1}$,
$ \zeta_{3}^{1}$,
$ -\zeta_{6}^{1}$,
$ \zeta_{3}^{1}$,
$ \sqrt{2}$,
$ -\sqrt{2}\zeta_{6}^{1}$,
$ \sqrt{2}\zeta_{3}^{1}$;\ \ 
$ -\zeta_{6}^{1}$,
$ \zeta_{3}^{1}$,
$ -\zeta_{6}^{1}$,
$ \sqrt{2}$,
$ \sqrt{2}\zeta_{3}^{1}$,
$ -\sqrt{2}\zeta_{6}^{1}$;\ \ 
$ -\zeta_{6}^{1}$,
$ \zeta_{3}^{1}$,
$ -\sqrt{2}$,
$ \sqrt{2}\zeta_{6}^{1}$,
$ -\sqrt{2}\zeta_{3}^{1}$;\ \ 
$ -\zeta_{6}^{1}$,
$ -\sqrt{2}$,
$ -\sqrt{2}\zeta_{3}^{1}$,
$ \sqrt{2}\zeta_{6}^{1}$;\ \ 
$0$,
$0$,
$0$;\ \ 
$0$,
$0$;\ \ 
$0$)

Factors = $3_{6,3.}^{3,138}\boxtimes 3_{\frac{1}{2},4.}^{16,598}$

\vskip 0.7ex
\hangindent=3em \hangafter=1
$\tau_n$ = ($1.33 - 3.2 i$, $-2.45 + 5.91 i$, $2.3 + 5.54 i$, $3.46 - 3.46 i$, $-3.2 - 1.33 i$, $1.76 + 4.24 i$, $1.33 + 3.2 i$, $0.$, $-5.54 - 2.3 i$, $-2.45 - 1.02 i$, $3.2 - 1.33 i$, $6. - 6. i$, $-3.2 - 1.33 i$, $2.45 + 5.91 i$, $5.54 - 2.3 i$, $0. - 6.93 i$, $-1.33 + 3.2 i$, $10.24 + 4.24 i$, $3.2 - 1.33 i$, $-3.46 + 3.46 i$, $-2.3 + 5.54 i$, $2.45 - 1.02 i$, $-1.33 - 3.2 i$, $0.$, $-1.33 + 3.2 i$, $2.45 + 1.02 i$, $-2.3 - 5.54 i$, $-3.46 - 3.46 i$, $3.2 + 1.33 i$, $10.24 - 4.24 i$, $-1.33 - 3.2 i$, $0. + 6.93 i$, $5.54 + 2.3 i$, $2.45 - 5.91 i$, $-3.2 + 1.33 i$, $6. + 6. i$, $3.2 + 1.33 i$, $-2.45 + 1.02 i$, $-5.54 + 2.3 i$, $0.$, $1.33 - 3.2 i$, $1.76 - 4.24 i$, $-3.2 + 1.33 i$, $3.46 + 3.46 i$, $2.3 - 5.54 i$, $-2.45 - 5.91 i$, $1.33 + 3.2 i$, $12.$)

\vskip 0.7ex
\hangindent=3em \hangafter=1
\textit{Intrinsic sign problem}

  \vskip 2ex

\noindent14. $9_{\frac{15}{2},12.}^{48,747}$ \irep{490}:\ \ 
$d_i$ = ($1.0$,
$1.0$,
$1.0$,
$1.0$,
$1.0$,
$1.0$,
$1.414$,
$1.414$,
$1.414$) 

\vskip 0.7ex
\hangindent=3em \hangafter=1
$D^2= 12.0 = 
12$

\vskip 0.7ex
\hangindent=3em \hangafter=1
$T = ( 0,
\frac{1}{2},
\frac{2}{3},
\frac{2}{3},
\frac{1}{6},
\frac{1}{6},
\frac{3}{16},
\frac{41}{48},
\frac{41}{48} )
$,

\vskip 0.7ex
\hangindent=3em \hangafter=1
$S$ = ($ 1$,
$ 1$,
$ 1$,
$ 1$,
$ 1$,
$ 1$,
$ \sqrt{2}$,
$ \sqrt{2}$,
$ \sqrt{2}$;\ \ 
$ 1$,
$ 1$,
$ 1$,
$ 1$,
$ 1$,
$ -\sqrt{2}$,
$ -\sqrt{2}$,
$ -\sqrt{2}$;\ \ 
$ -\zeta_{6}^{1}$,
$ \zeta_{3}^{1}$,
$ -\zeta_{6}^{1}$,
$ \zeta_{3}^{1}$,
$ \sqrt{2}$,
$ -\sqrt{2}\zeta_{6}^{1}$,
$ \sqrt{2}\zeta_{3}^{1}$;\ \ 
$ -\zeta_{6}^{1}$,
$ \zeta_{3}^{1}$,
$ -\zeta_{6}^{1}$,
$ \sqrt{2}$,
$ \sqrt{2}\zeta_{3}^{1}$,
$ -\sqrt{2}\zeta_{6}^{1}$;\ \ 
$ -\zeta_{6}^{1}$,
$ \zeta_{3}^{1}$,
$ -\sqrt{2}$,
$ \sqrt{2}\zeta_{6}^{1}$,
$ -\sqrt{2}\zeta_{3}^{1}$;\ \ 
$ -\zeta_{6}^{1}$,
$ -\sqrt{2}$,
$ -\sqrt{2}\zeta_{3}^{1}$,
$ \sqrt{2}\zeta_{6}^{1}$;\ \ 
$0$,
$0$,
$0$;\ \ 
$0$,
$0$;\ \ 
$0$)

Factors = $3_{6,3.}^{3,138}\boxtimes 3_{\frac{3}{2},4.}^{16,553}$

\vskip 0.7ex
\hangindent=3em \hangafter=1
$\tau_n$ = ($3.2 - 1.33 i$, $-2.45 + 1.02 i$, $-5.54 - 2.3 i$, $-3.46 - 3.46 i$, $1.33 + 3.2 i$, $10.24 + 4.24 i$, $3.2 + 1.33 i$, $0.$, $-2.3 - 5.54 i$, $-2.45 - 5.91 i$, $-1.33 + 3.2 i$, $6. + 6. i$, $1.33 + 3.2 i$, $2.45 + 1.02 i$, $2.3 - 5.54 i$, $0. - 6.93 i$, $-3.2 + 1.33 i$, $1.76 + 4.24 i$, $-1.33 + 3.2 i$, $3.46 + 3.46 i$, $5.54 - 2.3 i$, $2.45 - 5.91 i$, $-3.2 - 1.33 i$, $0.$, $-3.2 + 1.33 i$, $2.45 + 5.91 i$, $5.54 + 2.3 i$, $3.46 - 3.46 i$, $-1.33 - 3.2 i$, $1.76 - 4.24 i$, $-3.2 - 1.33 i$, $0. + 6.93 i$, $2.3 + 5.54 i$, $2.45 - 1.02 i$, $1.33 - 3.2 i$, $6. - 6. i$, $-1.33 - 3.2 i$, $-2.45 + 5.91 i$, $-2.3 + 5.54 i$, $0.$, $3.2 - 1.33 i$, $10.24 - 4.24 i$, $1.33 - 3.2 i$, $-3.46 + 3.46 i$, $-5.54 + 2.3 i$, $-2.45 - 1.02 i$, $3.2 + 1.33 i$, $12.$)

\vskip 0.7ex
\hangindent=3em \hangafter=1
\textit{Intrinsic sign problem}

  \vskip 2ex

\noindent15. $9_{\frac{1}{2},12.}^{48,148}$ \irep{490}:\ \ 
$d_i$ = ($1.0$,
$1.0$,
$1.0$,
$1.0$,
$1.0$,
$1.0$,
$1.414$,
$1.414$,
$1.414$) 

\vskip 0.7ex
\hangindent=3em \hangafter=1
$D^2= 12.0 = 
12$

\vskip 0.7ex
\hangindent=3em \hangafter=1
$T = ( 0,
\frac{1}{2},
\frac{2}{3},
\frac{2}{3},
\frac{1}{6},
\frac{1}{6},
\frac{5}{16},
\frac{47}{48},
\frac{47}{48} )
$,

\vskip 0.7ex
\hangindent=3em \hangafter=1
$S$ = ($ 1$,
$ 1$,
$ 1$,
$ 1$,
$ 1$,
$ 1$,
$ \sqrt{2}$,
$ \sqrt{2}$,
$ \sqrt{2}$;\ \ 
$ 1$,
$ 1$,
$ 1$,
$ 1$,
$ 1$,
$ -\sqrt{2}$,
$ -\sqrt{2}$,
$ -\sqrt{2}$;\ \ 
$ -\zeta_{6}^{1}$,
$ \zeta_{3}^{1}$,
$ -\zeta_{6}^{1}$,
$ \zeta_{3}^{1}$,
$ \sqrt{2}$,
$ -\sqrt{2}\zeta_{6}^{1}$,
$ \sqrt{2}\zeta_{3}^{1}$;\ \ 
$ -\zeta_{6}^{1}$,
$ \zeta_{3}^{1}$,
$ -\zeta_{6}^{1}$,
$ \sqrt{2}$,
$ \sqrt{2}\zeta_{3}^{1}$,
$ -\sqrt{2}\zeta_{6}^{1}$;\ \ 
$ -\zeta_{6}^{1}$,
$ \zeta_{3}^{1}$,
$ -\sqrt{2}$,
$ \sqrt{2}\zeta_{6}^{1}$,
$ -\sqrt{2}\zeta_{3}^{1}$;\ \ 
$ -\zeta_{6}^{1}$,
$ -\sqrt{2}$,
$ -\sqrt{2}\zeta_{3}^{1}$,
$ \sqrt{2}\zeta_{6}^{1}$;\ \ 
$0$,
$0$,
$0$;\ \ 
$0$,
$0$;\ \ 
$0$)

Factors = $3_{6,3.}^{3,138}\boxtimes 3_{\frac{5}{2},4.}^{16,465}$

\vskip 0.7ex
\hangindent=3em \hangafter=1
$\tau_n$ = ($3.2 + 1.33 i$, $2.45 + 1.02 i$, $5.54 - 2.3 i$, $3.46 - 3.46 i$, $1.33 - 3.2 i$, $10.24 - 4.24 i$, $3.2 - 1.33 i$, $0.$, $2.3 - 5.54 i$, $2.45 - 5.91 i$, $-1.33 - 3.2 i$, $6. - 6. i$, $1.33 - 3.2 i$, $-2.45 + 1.02 i$, $-2.3 - 5.54 i$, $0. - 6.93 i$, $-3.2 - 1.33 i$, $1.76 - 4.24 i$, $-1.33 - 3.2 i$, $-3.46 + 3.46 i$, $-5.54 - 2.3 i$, $-2.45 - 5.91 i$, $-3.2 + 1.33 i$, $0.$, $-3.2 - 1.33 i$, $-2.45 + 5.91 i$, $-5.54 + 2.3 i$, $-3.46 - 3.46 i$, $-1.33 + 3.2 i$, $1.76 + 4.24 i$, $-3.2 + 1.33 i$, $0. + 6.93 i$, $-2.3 + 5.54 i$, $-2.45 - 1.02 i$, $1.33 + 3.2 i$, $6. + 6. i$, $-1.33 + 3.2 i$, $2.45 + 5.91 i$, $2.3 + 5.54 i$, $0.$, $3.2 + 1.33 i$, $10.24 + 4.24 i$, $1.33 + 3.2 i$, $3.46 + 3.46 i$, $5.54 + 2.3 i$, $2.45 - 1.02 i$, $3.2 - 1.33 i$, $12.$)

\vskip 0.7ex
\hangindent=3em \hangafter=1
\textit{Intrinsic sign problem}

  \vskip 2ex

\noindent16. $9_{\frac{3}{2},12.}^{48,106}$ \irep{490}:\ \ 
$d_i$ = ($1.0$,
$1.0$,
$1.0$,
$1.0$,
$1.0$,
$1.0$,
$1.414$,
$1.414$,
$1.414$) 

\vskip 0.7ex
\hangindent=3em \hangafter=1
$D^2= 12.0 = 
12$

\vskip 0.7ex
\hangindent=3em \hangafter=1
$T = ( 0,
\frac{1}{2},
\frac{2}{3},
\frac{2}{3},
\frac{1}{6},
\frac{1}{6},
\frac{7}{16},
\frac{5}{48},
\frac{5}{48} )
$,

\vskip 0.7ex
\hangindent=3em \hangafter=1
$S$ = ($ 1$,
$ 1$,
$ 1$,
$ 1$,
$ 1$,
$ 1$,
$ \sqrt{2}$,
$ \sqrt{2}$,
$ \sqrt{2}$;\ \ 
$ 1$,
$ 1$,
$ 1$,
$ 1$,
$ 1$,
$ -\sqrt{2}$,
$ -\sqrt{2}$,
$ -\sqrt{2}$;\ \ 
$ -\zeta_{6}^{1}$,
$ \zeta_{3}^{1}$,
$ -\zeta_{6}^{1}$,
$ \zeta_{3}^{1}$,
$ \sqrt{2}$,
$ -\sqrt{2}\zeta_{6}^{1}$,
$ \sqrt{2}\zeta_{3}^{1}$;\ \ 
$ -\zeta_{6}^{1}$,
$ \zeta_{3}^{1}$,
$ -\zeta_{6}^{1}$,
$ \sqrt{2}$,
$ \sqrt{2}\zeta_{3}^{1}$,
$ -\sqrt{2}\zeta_{6}^{1}$;\ \ 
$ -\zeta_{6}^{1}$,
$ \zeta_{3}^{1}$,
$ -\sqrt{2}$,
$ \sqrt{2}\zeta_{6}^{1}$,
$ -\sqrt{2}\zeta_{3}^{1}$;\ \ 
$ -\zeta_{6}^{1}$,
$ -\sqrt{2}$,
$ -\sqrt{2}\zeta_{3}^{1}$,
$ \sqrt{2}\zeta_{6}^{1}$;\ \ 
$0$,
$0$,
$0$;\ \ 
$0$,
$0$;\ \ 
$0$)

Factors = $3_{6,3.}^{3,138}\boxtimes 3_{\frac{7}{2},4.}^{16,332}$

\vskip 0.7ex
\hangindent=3em \hangafter=1
$\tau_n$ = ($1.33 + 3.2 i$, $2.45 + 5.91 i$, $-2.3 + 5.54 i$, $-3.46 - 3.46 i$, $-3.2 + 1.33 i$, $1.76 - 4.24 i$, $1.33 - 3.2 i$, $0.$, $5.54 - 2.3 i$, $2.45 - 1.02 i$, $3.2 + 1.33 i$, $6. + 6. i$, $-3.2 + 1.33 i$, $-2.45 + 5.91 i$, $-5.54 - 2.3 i$, $0. - 6.93 i$, $-1.33 - 3.2 i$, $10.24 - 4.24 i$, $3.2 + 1.33 i$, $3.46 + 3.46 i$, $2.3 + 5.54 i$, $-2.45 - 1.02 i$, $-1.33 + 3.2 i$, $0.$, $-1.33 - 3.2 i$, $-2.45 + 1.02 i$, $2.3 - 5.54 i$, $3.46 - 3.46 i$, $3.2 - 1.33 i$, $10.24 + 4.24 i$, $-1.33 + 3.2 i$, $0. + 6.93 i$, $-5.54 + 2.3 i$, $-2.45 - 5.91 i$, $-3.2 - 1.33 i$, $6. - 6. i$, $3.2 - 1.33 i$, $2.45 + 1.02 i$, $5.54 + 2.3 i$, $0.$, $1.33 + 3.2 i$, $1.76 + 4.24 i$, $-3.2 - 1.33 i$, $-3.46 + 3.46 i$, $-2.3 - 5.54 i$, $2.45 - 5.91 i$, $1.33 - 3.2 i$, $12.$)

\vskip 0.7ex
\hangindent=3em \hangafter=1
\textit{Intrinsic sign problem}

  \vskip 2ex

\noindent17. $9_{\frac{5}{2},12.}^{48,770}$ \irep{490}:\ \ 
$d_i$ = ($1.0$,
$1.0$,
$1.0$,
$1.0$,
$1.0$,
$1.0$,
$1.414$,
$1.414$,
$1.414$) 

\vskip 0.7ex
\hangindent=3em \hangafter=1
$D^2= 12.0 = 
12$

\vskip 0.7ex
\hangindent=3em \hangafter=1
$T = ( 0,
\frac{1}{2},
\frac{2}{3},
\frac{2}{3},
\frac{1}{6},
\frac{1}{6},
\frac{9}{16},
\frac{11}{48},
\frac{11}{48} )
$,

\vskip 0.7ex
\hangindent=3em \hangafter=1
$S$ = ($ 1$,
$ 1$,
$ 1$,
$ 1$,
$ 1$,
$ 1$,
$ \sqrt{2}$,
$ \sqrt{2}$,
$ \sqrt{2}$;\ \ 
$ 1$,
$ 1$,
$ 1$,
$ 1$,
$ 1$,
$ -\sqrt{2}$,
$ -\sqrt{2}$,
$ -\sqrt{2}$;\ \ 
$ -\zeta_{6}^{1}$,
$ \zeta_{3}^{1}$,
$ -\zeta_{6}^{1}$,
$ \zeta_{3}^{1}$,
$ \sqrt{2}$,
$ -\sqrt{2}\zeta_{6}^{1}$,
$ \sqrt{2}\zeta_{3}^{1}$;\ \ 
$ -\zeta_{6}^{1}$,
$ \zeta_{3}^{1}$,
$ -\zeta_{6}^{1}$,
$ \sqrt{2}$,
$ \sqrt{2}\zeta_{3}^{1}$,
$ -\sqrt{2}\zeta_{6}^{1}$;\ \ 
$ -\zeta_{6}^{1}$,
$ \zeta_{3}^{1}$,
$ -\sqrt{2}$,
$ \sqrt{2}\zeta_{6}^{1}$,
$ -\sqrt{2}\zeta_{3}^{1}$;\ \ 
$ -\zeta_{6}^{1}$,
$ -\sqrt{2}$,
$ -\sqrt{2}\zeta_{3}^{1}$,
$ \sqrt{2}\zeta_{6}^{1}$;\ \ 
$0$,
$0$,
$0$;\ \ 
$0$,
$0$;\ \ 
$0$)

Factors = $3_{6,3.}^{3,138}\boxtimes 3_{\frac{9}{2},4.}^{16,156}$

\vskip 0.7ex
\hangindent=3em \hangafter=1
$\tau_n$ = ($-1.33 + 3.2 i$, $-2.45 + 5.91 i$, $-2.3 - 5.54 i$, $3.46 - 3.46 i$, $3.2 + 1.33 i$, $1.76 + 4.24 i$, $-1.33 - 3.2 i$, $0.$, $5.54 + 2.3 i$, $-2.45 - 1.02 i$, $-3.2 + 1.33 i$, $6. - 6. i$, $3.2 + 1.33 i$, $2.45 + 5.91 i$, $-5.54 + 2.3 i$, $0. - 6.93 i$, $1.33 - 3.2 i$, $10.24 + 4.24 i$, $-3.2 + 1.33 i$, $-3.46 + 3.46 i$, $2.3 - 5.54 i$, $2.45 - 1.02 i$, $1.33 + 3.2 i$, $0.$, $1.33 - 3.2 i$, $2.45 + 1.02 i$, $2.3 + 5.54 i$, $-3.46 - 3.46 i$, $-3.2 - 1.33 i$, $10.24 - 4.24 i$, $1.33 + 3.2 i$, $0. + 6.93 i$, $-5.54 - 2.3 i$, $2.45 - 5.91 i$, $3.2 - 1.33 i$, $6. + 6. i$, $-3.2 - 1.33 i$, $-2.45 + 1.02 i$, $5.54 - 2.3 i$, $0.$, $-1.33 + 3.2 i$, $1.76 - 4.24 i$, $3.2 - 1.33 i$, $3.46 + 3.46 i$, $-2.3 + 5.54 i$, $-2.45 - 5.91 i$, $-1.33 - 3.2 i$, $12.$)

\vskip 0.7ex
\hangindent=3em \hangafter=1
\textit{Intrinsic sign problem}

  \vskip 2ex

\noindent18. $9_{\frac{7}{2},12.}^{48,342}$ \irep{490}:\ \ 
$d_i$ = ($1.0$,
$1.0$,
$1.0$,
$1.0$,
$1.0$,
$1.0$,
$1.414$,
$1.414$,
$1.414$) 

\vskip 0.7ex
\hangindent=3em \hangafter=1
$D^2= 12.0 = 
12$

\vskip 0.7ex
\hangindent=3em \hangafter=1
$T = ( 0,
\frac{1}{2},
\frac{2}{3},
\frac{2}{3},
\frac{1}{6},
\frac{1}{6},
\frac{11}{16},
\frac{17}{48},
\frac{17}{48} )
$,

\vskip 0.7ex
\hangindent=3em \hangafter=1
$S$ = ($ 1$,
$ 1$,
$ 1$,
$ 1$,
$ 1$,
$ 1$,
$ \sqrt{2}$,
$ \sqrt{2}$,
$ \sqrt{2}$;\ \ 
$ 1$,
$ 1$,
$ 1$,
$ 1$,
$ 1$,
$ -\sqrt{2}$,
$ -\sqrt{2}$,
$ -\sqrt{2}$;\ \ 
$ -\zeta_{6}^{1}$,
$ \zeta_{3}^{1}$,
$ -\zeta_{6}^{1}$,
$ \zeta_{3}^{1}$,
$ \sqrt{2}$,
$ -\sqrt{2}\zeta_{6}^{1}$,
$ \sqrt{2}\zeta_{3}^{1}$;\ \ 
$ -\zeta_{6}^{1}$,
$ \zeta_{3}^{1}$,
$ -\zeta_{6}^{1}$,
$ \sqrt{2}$,
$ \sqrt{2}\zeta_{3}^{1}$,
$ -\sqrt{2}\zeta_{6}^{1}$;\ \ 
$ -\zeta_{6}^{1}$,
$ \zeta_{3}^{1}$,
$ -\sqrt{2}$,
$ \sqrt{2}\zeta_{6}^{1}$,
$ -\sqrt{2}\zeta_{3}^{1}$;\ \ 
$ -\zeta_{6}^{1}$,
$ -\sqrt{2}$,
$ -\sqrt{2}\zeta_{3}^{1}$,
$ \sqrt{2}\zeta_{6}^{1}$;\ \ 
$0$,
$0$,
$0$;\ \ 
$0$,
$0$;\ \ 
$0$)

Factors = $3_{6,3.}^{3,138}\boxtimes 3_{\frac{11}{2},4.}^{16,648}$

\vskip 0.7ex
\hangindent=3em \hangafter=1
$\tau_n$ = ($-3.2 + 1.33 i$, $-2.45 + 1.02 i$, $5.54 + 2.3 i$, $-3.46 - 3.46 i$, $-1.33 - 3.2 i$, $10.24 + 4.24 i$, $-3.2 - 1.33 i$, $0.$, $2.3 + 5.54 i$, $-2.45 - 5.91 i$, $1.33 - 3.2 i$, $6. + 6. i$, $-1.33 - 3.2 i$, $2.45 + 1.02 i$, $-2.3 + 5.54 i$, $0. - 6.93 i$, $3.2 - 1.33 i$, $1.76 + 4.24 i$, $1.33 - 3.2 i$, $3.46 + 3.46 i$, $-5.54 + 2.3 i$, $2.45 - 5.91 i$, $3.2 + 1.33 i$, $0.$, $3.2 - 1.33 i$, $2.45 + 5.91 i$, $-5.54 - 2.3 i$, $3.46 - 3.46 i$, $1.33 + 3.2 i$, $1.76 - 4.24 i$, $3.2 + 1.33 i$, $0. + 6.93 i$, $-2.3 - 5.54 i$, $2.45 - 1.02 i$, $-1.33 + 3.2 i$, $6. - 6. i$, $1.33 + 3.2 i$, $-2.45 + 5.91 i$, $2.3 - 5.54 i$, $0.$, $-3.2 + 1.33 i$, $10.24 - 4.24 i$, $-1.33 + 3.2 i$, $-3.46 + 3.46 i$, $5.54 - 2.3 i$, $-2.45 - 1.02 i$, $-3.2 - 1.33 i$, $12.$)

\vskip 0.7ex
\hangindent=3em \hangafter=1
\textit{Intrinsic sign problem}

  \vskip 2ex

\noindent19. $9_{\frac{9}{2},12.}^{48,216}$ \irep{490}:\ \ 
$d_i$ = ($1.0$,
$1.0$,
$1.0$,
$1.0$,
$1.0$,
$1.0$,
$1.414$,
$1.414$,
$1.414$) 

\vskip 0.7ex
\hangindent=3em \hangafter=1
$D^2= 12.0 = 
12$

\vskip 0.7ex
\hangindent=3em \hangafter=1
$T = ( 0,
\frac{1}{2},
\frac{2}{3},
\frac{2}{3},
\frac{1}{6},
\frac{1}{6},
\frac{13}{16},
\frac{23}{48},
\frac{23}{48} )
$,

\vskip 0.7ex
\hangindent=3em \hangafter=1
$S$ = ($ 1$,
$ 1$,
$ 1$,
$ 1$,
$ 1$,
$ 1$,
$ \sqrt{2}$,
$ \sqrt{2}$,
$ \sqrt{2}$;\ \ 
$ 1$,
$ 1$,
$ 1$,
$ 1$,
$ 1$,
$ -\sqrt{2}$,
$ -\sqrt{2}$,
$ -\sqrt{2}$;\ \ 
$ -\zeta_{6}^{1}$,
$ \zeta_{3}^{1}$,
$ -\zeta_{6}^{1}$,
$ \zeta_{3}^{1}$,
$ \sqrt{2}$,
$ -\sqrt{2}\zeta_{6}^{1}$,
$ \sqrt{2}\zeta_{3}^{1}$;\ \ 
$ -\zeta_{6}^{1}$,
$ \zeta_{3}^{1}$,
$ -\zeta_{6}^{1}$,
$ \sqrt{2}$,
$ \sqrt{2}\zeta_{3}^{1}$,
$ -\sqrt{2}\zeta_{6}^{1}$;\ \ 
$ -\zeta_{6}^{1}$,
$ \zeta_{3}^{1}$,
$ -\sqrt{2}$,
$ \sqrt{2}\zeta_{6}^{1}$,
$ -\sqrt{2}\zeta_{3}^{1}$;\ \ 
$ -\zeta_{6}^{1}$,
$ -\sqrt{2}$,
$ -\sqrt{2}\zeta_{3}^{1}$,
$ \sqrt{2}\zeta_{6}^{1}$;\ \ 
$0$,
$0$,
$0$;\ \ 
$0$,
$0$;\ \ 
$0$)

Factors = $3_{6,3.}^{3,138}\boxtimes 3_{\frac{13}{2},4.}^{16,330}$

\vskip 0.7ex
\hangindent=3em \hangafter=1
$\tau_n$ = ($-3.2 - 1.33 i$, $2.45 + 1.02 i$, $-5.54 + 2.3 i$, $3.46 - 3.46 i$, $-1.33 + 3.2 i$, $10.24 - 4.24 i$, $-3.2 + 1.33 i$, $0.$, $-2.3 + 5.54 i$, $2.45 - 5.91 i$, $1.33 + 3.2 i$, $6. - 6. i$, $-1.33 + 3.2 i$, $-2.45 + 1.02 i$, $2.3 + 5.54 i$, $0. - 6.93 i$, $3.2 + 1.33 i$, $1.76 - 4.24 i$, $1.33 + 3.2 i$, $-3.46 + 3.46 i$, $5.54 + 2.3 i$, $-2.45 - 5.91 i$, $3.2 - 1.33 i$, $0.$, $3.2 + 1.33 i$, $-2.45 + 5.91 i$, $5.54 - 2.3 i$, $-3.46 - 3.46 i$, $1.33 - 3.2 i$, $1.76 + 4.24 i$, $3.2 - 1.33 i$, $0. + 6.93 i$, $2.3 - 5.54 i$, $-2.45 - 1.02 i$, $-1.33 - 3.2 i$, $6. + 6. i$, $1.33 - 3.2 i$, $2.45 + 5.91 i$, $-2.3 - 5.54 i$, $0.$, $-3.2 - 1.33 i$, $10.24 + 4.24 i$, $-1.33 - 3.2 i$, $3.46 + 3.46 i$, $-5.54 - 2.3 i$, $2.45 - 1.02 i$, $-3.2 + 1.33 i$, $12.$)

\vskip 0.7ex
\hangindent=3em \hangafter=1
\textit{Intrinsic sign problem}

  \vskip 2ex

\noindent20. $9_{\frac{11}{2},12.}^{48,909}$ \irep{490}:\ \ 
$d_i$ = ($1.0$,
$1.0$,
$1.0$,
$1.0$,
$1.0$,
$1.0$,
$1.414$,
$1.414$,
$1.414$) 

\vskip 0.7ex
\hangindent=3em \hangafter=1
$D^2= 12.0 = 
12$

\vskip 0.7ex
\hangindent=3em \hangafter=1
$T = ( 0,
\frac{1}{2},
\frac{2}{3},
\frac{2}{3},
\frac{1}{6},
\frac{1}{6},
\frac{15}{16},
\frac{29}{48},
\frac{29}{48} )
$,

\vskip 0.7ex
\hangindent=3em \hangafter=1
$S$ = ($ 1$,
$ 1$,
$ 1$,
$ 1$,
$ 1$,
$ 1$,
$ \sqrt{2}$,
$ \sqrt{2}$,
$ \sqrt{2}$;\ \ 
$ 1$,
$ 1$,
$ 1$,
$ 1$,
$ 1$,
$ -\sqrt{2}$,
$ -\sqrt{2}$,
$ -\sqrt{2}$;\ \ 
$ -\zeta_{6}^{1}$,
$ \zeta_{3}^{1}$,
$ -\zeta_{6}^{1}$,
$ \zeta_{3}^{1}$,
$ \sqrt{2}$,
$ -\sqrt{2}\zeta_{6}^{1}$,
$ \sqrt{2}\zeta_{3}^{1}$;\ \ 
$ -\zeta_{6}^{1}$,
$ \zeta_{3}^{1}$,
$ -\zeta_{6}^{1}$,
$ \sqrt{2}$,
$ \sqrt{2}\zeta_{3}^{1}$,
$ -\sqrt{2}\zeta_{6}^{1}$;\ \ 
$ -\zeta_{6}^{1}$,
$ \zeta_{3}^{1}$,
$ -\sqrt{2}$,
$ \sqrt{2}\zeta_{6}^{1}$,
$ -\sqrt{2}\zeta_{3}^{1}$;\ \ 
$ -\zeta_{6}^{1}$,
$ -\sqrt{2}$,
$ -\sqrt{2}\zeta_{3}^{1}$,
$ \sqrt{2}\zeta_{6}^{1}$;\ \ 
$0$,
$0$,
$0$;\ \ 
$0$,
$0$;\ \ 
$0$)

Factors = $3_{6,3.}^{3,138}\boxtimes 3_{\frac{15}{2},4.}^{16,639}$

\vskip 0.7ex
\hangindent=3em \hangafter=1
$\tau_n$ = ($-1.33 - 3.2 i$, $2.45 + 5.91 i$, $2.3 - 5.54 i$, $-3.46 - 3.46 i$, $3.2 - 1.33 i$, $1.76 - 4.24 i$, $-1.33 + 3.2 i$, $0.$, $-5.54 + 2.3 i$, $2.45 - 1.02 i$, $-3.2 - 1.33 i$, $6. + 6. i$, $3.2 - 1.33 i$, $-2.45 + 5.91 i$, $5.54 + 2.3 i$, $0. - 6.93 i$, $1.33 + 3.2 i$, $10.24 - 4.24 i$, $-3.2 - 1.33 i$, $3.46 + 3.46 i$, $-2.3 - 5.54 i$, $-2.45 - 1.02 i$, $1.33 - 3.2 i$, $0.$, $1.33 + 3.2 i$, $-2.45 + 1.02 i$, $-2.3 + 5.54 i$, $3.46 - 3.46 i$, $-3.2 + 1.33 i$, $10.24 + 4.24 i$, $1.33 - 3.2 i$, $0. + 6.93 i$, $5.54 - 2.3 i$, $-2.45 - 5.91 i$, $3.2 + 1.33 i$, $6. - 6. i$, $-3.2 + 1.33 i$, $2.45 + 1.02 i$, $-5.54 - 2.3 i$, $0.$, $-1.33 - 3.2 i$, $1.76 + 4.24 i$, $3.2 + 1.33 i$, $-3.46 + 3.46 i$, $2.3 + 5.54 i$, $2.45 - 5.91 i$, $-1.33 + 3.2 i$, $12.$)

\vskip 0.7ex
\hangindent=3em \hangafter=1
\textit{Intrinsic sign problem}

  \vskip 2ex

\noindent21. $9_{1,16.}^{16,147}$ \irep{403}:\ \ 
$d_i$ = ($1.0$,
$1.0$,
$1.0$,
$1.0$,
$1.414$,
$1.414$,
$1.414$,
$1.414$,
$2.0$) 

\vskip 0.7ex
\hangindent=3em \hangafter=1
$D^2= 16.0 = 
16$

\vskip 0.7ex
\hangindent=3em \hangafter=1
$T = ( 0,
0,
\frac{1}{2},
\frac{1}{2},
\frac{1}{16},
\frac{1}{16},
\frac{9}{16},
\frac{9}{16},
\frac{1}{8} )
$,

\vskip 0.7ex
\hangindent=3em \hangafter=1
$S$ = ($ 1$,
$ 1$,
$ 1$,
$ 1$,
$ \sqrt{2}$,
$ \sqrt{2}$,
$ \sqrt{2}$,
$ \sqrt{2}$,
$ 2$;\ \ 
$ 1$,
$ 1$,
$ 1$,
$ -\sqrt{2}$,
$ -\sqrt{2}$,
$ -\sqrt{2}$,
$ -\sqrt{2}$,
$ 2$;\ \ 
$ 1$,
$ 1$,
$ -\sqrt{2}$,
$ \sqrt{2}$,
$ -\sqrt{2}$,
$ \sqrt{2}$,
$ -2$;\ \ 
$ 1$,
$ \sqrt{2}$,
$ -\sqrt{2}$,
$ \sqrt{2}$,
$ -\sqrt{2}$,
$ -2$;\ \ 
$0$,
$ 2$,
$0$,
$ -2$,
$0$;\ \ 
$0$,
$ -2$,
$0$,
$0$;\ \ 
$0$,
$ 2$,
$0$;\ \ 
$0$,
$0$;\ \ 
$0$)

Factors = $3_{\frac{1}{2},4.}^{16,598}\boxtimes 3_{\frac{1}{2},4.}^{16,598}$

\vskip 0.7ex
\hangindent=3em \hangafter=1
$\tau_n$ = ($2.83 + 2.83 i$, $9.66 + 9.66 i$, $-2.83 + 2.83 i$, $0. + 8. i$, $-2.83 - 2.83 i$, $-1.66 + 1.66 i$, $2.83 - 2.83 i$, $0.$, $2.83 + 2.83 i$, $-1.66 - 1.66 i$, $-2.83 + 2.83 i$, $0. - 8. i$, $-2.83 - 2.83 i$, $9.66 - 9.66 i$, $2.83 - 2.83 i$, $16.$)

\vskip 0.7ex
\hangindent=3em \hangafter=1
\textit{Intrinsic sign problem}

  \vskip 2ex

\noindent22. $9_{5,16.}^{16,726}$ \irep{403}:\ \ 
$d_i$ = ($1.0$,
$1.0$,
$1.0$,
$1.0$,
$1.414$,
$1.414$,
$1.414$,
$1.414$,
$2.0$) 

\vskip 0.7ex
\hangindent=3em \hangafter=1
$D^2= 16.0 = 
16$

\vskip 0.7ex
\hangindent=3em \hangafter=1
$T = ( 0,
0,
\frac{1}{2},
\frac{1}{2},
\frac{1}{16},
\frac{1}{16},
\frac{9}{16},
\frac{9}{16},
\frac{5}{8} )
$,

\vskip 0.7ex
\hangindent=3em \hangafter=1
$S$ = ($ 1$,
$ 1$,
$ 1$,
$ 1$,
$ \sqrt{2}$,
$ \sqrt{2}$,
$ \sqrt{2}$,
$ \sqrt{2}$,
$ 2$;\ \ 
$ 1$,
$ 1$,
$ 1$,
$ -\sqrt{2}$,
$ -\sqrt{2}$,
$ -\sqrt{2}$,
$ -\sqrt{2}$,
$ 2$;\ \ 
$ 1$,
$ 1$,
$ -\sqrt{2}$,
$ \sqrt{2}$,
$ -\sqrt{2}$,
$ \sqrt{2}$,
$ -2$;\ \ 
$ 1$,
$ \sqrt{2}$,
$ -\sqrt{2}$,
$ \sqrt{2}$,
$ -\sqrt{2}$,
$ -2$;\ \ 
$0$,
$ -2$,
$0$,
$ 2$,
$0$;\ \ 
$0$,
$ 2$,
$0$,
$0$;\ \ 
$0$,
$ -2$,
$0$;\ \ 
$0$,
$0$;\ \ 
$0$)

Factors = $3_{\frac{1}{2},4.}^{16,598}\boxtimes 3_{\frac{9}{2},4.}^{16,156}$

\vskip 0.7ex
\hangindent=3em \hangafter=1
$\tau_n$ = ($-2.83 - 2.83 i$, $9.66 + 9.66 i$, $2.83 - 2.83 i$, $0. + 8. i$, $2.83 + 2.83 i$, $-1.66 + 1.66 i$, $-2.83 + 2.83 i$, $0.$, $-2.83 - 2.83 i$, $-1.66 - 1.66 i$, $2.83 - 2.83 i$, $0. - 8. i$, $2.83 + 2.83 i$, $9.66 - 9.66 i$, $-2.83 + 2.83 i$, $16.$)

\vskip 0.7ex
\hangindent=3em \hangafter=1
\textit{Intrinsic sign problem}

  \vskip 2ex

\noindent23. $9_{2,16.}^{16,111}$ \irep{389}:\ \ 
$d_i$ = ($1.0$,
$1.0$,
$1.0$,
$1.0$,
$1.414$,
$1.414$,
$1.414$,
$1.414$,
$2.0$) 

\vskip 0.7ex
\hangindent=3em \hangafter=1
$D^2= 16.0 = 
16$

\vskip 0.7ex
\hangindent=3em \hangafter=1
$T = ( 0,
0,
\frac{1}{2},
\frac{1}{2},
\frac{1}{16},
\frac{3}{16},
\frac{9}{16},
\frac{11}{16},
\frac{1}{4} )
$,

\vskip 0.7ex
\hangindent=3em \hangafter=1
$S$ = ($ 1$,
$ 1$,
$ 1$,
$ 1$,
$ \sqrt{2}$,
$ \sqrt{2}$,
$ \sqrt{2}$,
$ \sqrt{2}$,
$ 2$;\ \ 
$ 1$,
$ 1$,
$ 1$,
$ -\sqrt{2}$,
$ -\sqrt{2}$,
$ -\sqrt{2}$,
$ -\sqrt{2}$,
$ 2$;\ \ 
$ 1$,
$ 1$,
$ -\sqrt{2}$,
$ \sqrt{2}$,
$ -\sqrt{2}$,
$ \sqrt{2}$,
$ -2$;\ \ 
$ 1$,
$ \sqrt{2}$,
$ -\sqrt{2}$,
$ \sqrt{2}$,
$ -\sqrt{2}$,
$ -2$;\ \ 
$0$,
$ 2$,
$0$,
$ -2$,
$0$;\ \ 
$0$,
$ -2$,
$0$,
$0$;\ \ 
$0$,
$ 2$,
$0$;\ \ 
$0$,
$0$;\ \ 
$0$)

Factors = $3_{\frac{1}{2},4.}^{16,598}\boxtimes 3_{\frac{3}{2},4.}^{16,553}$

\vskip 0.7ex
\hangindent=3em \hangafter=1
$\tau_n$ = ($0. + 4. i$, $0. + 5.66 i$, $0. - 4. i$, $8.$, $0. + 4. i$, $0. + 5.66 i$, $0. - 4. i$, $0.$, $0. + 4. i$, $0. - 5.66 i$, $0. - 4. i$, $8.$, $0. + 4. i$, $0. - 5.66 i$, $0. - 4. i$, $16.$)

\vskip 0.7ex
\hangindent=3em \hangafter=1
\textit{Intrinsic sign problem}

  \vskip 2ex

\noindent24. $9_{6,16.}^{16,118}$ \irep{389}:\ \ 
$d_i$ = ($1.0$,
$1.0$,
$1.0$,
$1.0$,
$1.414$,
$1.414$,
$1.414$,
$1.414$,
$2.0$) 

\vskip 0.7ex
\hangindent=3em \hangafter=1
$D^2= 16.0 = 
16$

\vskip 0.7ex
\hangindent=3em \hangafter=1
$T = ( 0,
0,
\frac{1}{2},
\frac{1}{2},
\frac{1}{16},
\frac{3}{16},
\frac{9}{16},
\frac{11}{16},
\frac{3}{4} )
$,

\vskip 0.7ex
\hangindent=3em \hangafter=1
$S$ = ($ 1$,
$ 1$,
$ 1$,
$ 1$,
$ \sqrt{2}$,
$ \sqrt{2}$,
$ \sqrt{2}$,
$ \sqrt{2}$,
$ 2$;\ \ 
$ 1$,
$ 1$,
$ 1$,
$ -\sqrt{2}$,
$ -\sqrt{2}$,
$ -\sqrt{2}$,
$ -\sqrt{2}$,
$ 2$;\ \ 
$ 1$,
$ 1$,
$ -\sqrt{2}$,
$ \sqrt{2}$,
$ -\sqrt{2}$,
$ \sqrt{2}$,
$ -2$;\ \ 
$ 1$,
$ \sqrt{2}$,
$ -\sqrt{2}$,
$ \sqrt{2}$,
$ -\sqrt{2}$,
$ -2$;\ \ 
$0$,
$ -2$,
$0$,
$ 2$,
$0$;\ \ 
$0$,
$ 2$,
$0$,
$0$;\ \ 
$0$,
$ -2$,
$0$;\ \ 
$0$,
$0$;\ \ 
$0$)

Factors = $3_{\frac{1}{2},4.}^{16,598}\boxtimes 3_{\frac{11}{2},4.}^{16,648}$

\vskip 0.7ex
\hangindent=3em \hangafter=1
$\tau_n$ = ($0. - 4. i$, $0. + 5.66 i$, $0. + 4. i$, $8.$, $0. - 4. i$, $0. + 5.66 i$, $0. + 4. i$, $0.$, $0. - 4. i$, $0. - 5.66 i$, $0. + 4. i$, $8.$, $0. - 4. i$, $0. - 5.66 i$, $0. + 4. i$, $16.$)

\vskip 0.7ex
\hangindent=3em \hangafter=1
\textit{Intrinsic sign problem}

  \vskip 2ex

\noindent25. $9_{3,16.}^{16,608}$ \irep{388}:\ \ 
$d_i$ = ($1.0$,
$1.0$,
$1.0$,
$1.0$,
$1.414$,
$1.414$,
$1.414$,
$1.414$,
$2.0$) 

\vskip 0.7ex
\hangindent=3em \hangafter=1
$D^2= 16.0 = 
16$

\vskip 0.7ex
\hangindent=3em \hangafter=1
$T = ( 0,
0,
\frac{1}{2},
\frac{1}{2},
\frac{1}{16},
\frac{5}{16},
\frac{9}{16},
\frac{13}{16},
\frac{3}{8} )
$,

\vskip 0.7ex
\hangindent=3em \hangafter=1
$S$ = ($ 1$,
$ 1$,
$ 1$,
$ 1$,
$ \sqrt{2}$,
$ \sqrt{2}$,
$ \sqrt{2}$,
$ \sqrt{2}$,
$ 2$;\ \ 
$ 1$,
$ 1$,
$ 1$,
$ -\sqrt{2}$,
$ -\sqrt{2}$,
$ -\sqrt{2}$,
$ -\sqrt{2}$,
$ 2$;\ \ 
$ 1$,
$ 1$,
$ -\sqrt{2}$,
$ \sqrt{2}$,
$ -\sqrt{2}$,
$ \sqrt{2}$,
$ -2$;\ \ 
$ 1$,
$ \sqrt{2}$,
$ -\sqrt{2}$,
$ \sqrt{2}$,
$ -\sqrt{2}$,
$ -2$;\ \ 
$0$,
$ 2$,
$0$,
$ -2$,
$0$;\ \ 
$0$,
$ -2$,
$0$,
$0$;\ \ 
$0$,
$ 2$,
$0$;\ \ 
$0$,
$0$;\ \ 
$0$)

Factors = $3_{\frac{1}{2},4.}^{16,598}\boxtimes 3_{\frac{5}{2},4.}^{16,465}$

\vskip 0.7ex
\hangindent=3em \hangafter=1
$\tau_n$ = ($-2.83 + 2.83 i$, $4. - 4. i$, $2.83 + 2.83 i$, $0. + 8. i$, $2.83 - 2.83 i$, $4. + 4. i$, $-2.83 - 2.83 i$, $0.$, $-2.83 + 2.83 i$, $4. - 4. i$, $2.83 + 2.83 i$, $0. - 8. i$, $2.83 - 2.83 i$, $4. + 4. i$, $-2.83 - 2.83 i$, $16.$)

\vskip 0.7ex
\hangindent=3em \hangafter=1
\textit{Intrinsic sign problem}

  \vskip 2ex

\noindent26. $9_{7,16.}^{16,641}$ \irep{388}:\ \ 
$d_i$ = ($1.0$,
$1.0$,
$1.0$,
$1.0$,
$1.414$,
$1.414$,
$1.414$,
$1.414$,
$2.0$) 

\vskip 0.7ex
\hangindent=3em \hangafter=1
$D^2= 16.0 = 
16$

\vskip 0.7ex
\hangindent=3em \hangafter=1
$T = ( 0,
0,
\frac{1}{2},
\frac{1}{2},
\frac{1}{16},
\frac{5}{16},
\frac{9}{16},
\frac{13}{16},
\frac{7}{8} )
$,

\vskip 0.7ex
\hangindent=3em \hangafter=1
$S$ = ($ 1$,
$ 1$,
$ 1$,
$ 1$,
$ \sqrt{2}$,
$ \sqrt{2}$,
$ \sqrt{2}$,
$ \sqrt{2}$,
$ 2$;\ \ 
$ 1$,
$ 1$,
$ 1$,
$ -\sqrt{2}$,
$ -\sqrt{2}$,
$ -\sqrt{2}$,
$ -\sqrt{2}$,
$ 2$;\ \ 
$ 1$,
$ 1$,
$ -\sqrt{2}$,
$ \sqrt{2}$,
$ -\sqrt{2}$,
$ \sqrt{2}$,
$ -2$;\ \ 
$ 1$,
$ \sqrt{2}$,
$ -\sqrt{2}$,
$ \sqrt{2}$,
$ -\sqrt{2}$,
$ -2$;\ \ 
$0$,
$ -2$,
$0$,
$ 2$,
$0$;\ \ 
$0$,
$ 2$,
$0$,
$0$;\ \ 
$0$,
$ -2$,
$0$;\ \ 
$0$,
$0$;\ \ 
$0$)

Factors = $3_{\frac{1}{2},4.}^{16,598}\boxtimes 3_{\frac{13}{2},4.}^{16,330}$

\vskip 0.7ex
\hangindent=3em \hangafter=1
$\tau_n$ = ($2.83 - 2.83 i$, $4. - 4. i$, $-2.83 - 2.83 i$, $0. + 8. i$, $-2.83 + 2.83 i$, $4. + 4. i$, $2.83 + 2.83 i$, $0.$, $2.83 - 2.83 i$, $4. - 4. i$, $-2.83 - 2.83 i$, $0. - 8. i$, $-2.83 + 2.83 i$, $4. + 4. i$, $2.83 + 2.83 i$, $16.$)

\vskip 0.7ex
\hangindent=3em \hangafter=1
\textit{Intrinsic sign problem}

  \vskip 2ex

\noindent27. $9_{0,16.}^{16,447}$ \irep{397}:\ \ 
$d_i$ = ($1.0$,
$1.0$,
$1.0$,
$1.0$,
$1.414$,
$1.414$,
$1.414$,
$1.414$,
$2.0$) 

\vskip 0.7ex
\hangindent=3em \hangafter=1
$D^2= 16.0 = 
16$

\vskip 0.7ex
\hangindent=3em \hangafter=1
$T = ( 0,
0,
\frac{1}{2},
\frac{1}{2},
\frac{1}{16},
\frac{7}{16},
\frac{9}{16},
\frac{15}{16},
0 )
$,

\vskip 0.7ex
\hangindent=3em \hangafter=1
$S$ = ($ 1$,
$ 1$,
$ 1$,
$ 1$,
$ \sqrt{2}$,
$ \sqrt{2}$,
$ \sqrt{2}$,
$ \sqrt{2}$,
$ 2$;\ \ 
$ 1$,
$ 1$,
$ 1$,
$ -\sqrt{2}$,
$ -\sqrt{2}$,
$ -\sqrt{2}$,
$ -\sqrt{2}$,
$ 2$;\ \ 
$ 1$,
$ 1$,
$ -\sqrt{2}$,
$ \sqrt{2}$,
$ -\sqrt{2}$,
$ \sqrt{2}$,
$ -2$;\ \ 
$ 1$,
$ \sqrt{2}$,
$ -\sqrt{2}$,
$ \sqrt{2}$,
$ -\sqrt{2}$,
$ -2$;\ \ 
$0$,
$ -2$,
$0$,
$ 2$,
$0$;\ \ 
$0$,
$ 2$,
$0$,
$0$;\ \ 
$0$,
$ -2$,
$0$;\ \ 
$0$,
$0$;\ \ 
$0$)

Factors = $3_{\frac{1}{2},4.}^{16,598}\boxtimes 3_{\frac{15}{2},4.}^{16,639}$

\vskip 0.7ex
\hangindent=3em \hangafter=1
$\tau_n$ = ($4.$, $13.66$, $4.$, $8.$, $4.$, $2.34$, $4.$, $0.$, $4.$, $2.34$, $4.$, $8.$, $4.$, $13.66$, $4.$, $16.$)

\vskip 0.7ex
\hangindent=3em \hangafter=1
\textit{Undetermined}

  \vskip 2ex

\noindent28. $9_{4,16.}^{16,524}$ \irep{397}:\ \ 
$d_i$ = ($1.0$,
$1.0$,
$1.0$,
$1.0$,
$1.414$,
$1.414$,
$1.414$,
$1.414$,
$2.0$) 

\vskip 0.7ex
\hangindent=3em \hangafter=1
$D^2= 16.0 = 
16$

\vskip 0.7ex
\hangindent=3em \hangafter=1
$T = ( 0,
0,
\frac{1}{2},
\frac{1}{2},
\frac{1}{16},
\frac{7}{16},
\frac{9}{16},
\frac{15}{16},
\frac{1}{2} )
$,

\vskip 0.7ex
\hangindent=3em \hangafter=1
$S$ = ($ 1$,
$ 1$,
$ 1$,
$ 1$,
$ \sqrt{2}$,
$ \sqrt{2}$,
$ \sqrt{2}$,
$ \sqrt{2}$,
$ 2$;\ \ 
$ 1$,
$ 1$,
$ 1$,
$ -\sqrt{2}$,
$ -\sqrt{2}$,
$ -\sqrt{2}$,
$ -\sqrt{2}$,
$ 2$;\ \ 
$ 1$,
$ 1$,
$ -\sqrt{2}$,
$ \sqrt{2}$,
$ -\sqrt{2}$,
$ \sqrt{2}$,
$ -2$;\ \ 
$ 1$,
$ \sqrt{2}$,
$ -\sqrt{2}$,
$ \sqrt{2}$,
$ -\sqrt{2}$,
$ -2$;\ \ 
$0$,
$ 2$,
$0$,
$ -2$,
$0$;\ \ 
$0$,
$ -2$,
$0$,
$0$;\ \ 
$0$,
$ 2$,
$0$;\ \ 
$0$,
$0$;\ \ 
$0$)

Factors = $3_{\frac{1}{2},4.}^{16,598}\boxtimes 3_{\frac{7}{2},4.}^{16,332}$

\vskip 0.7ex
\hangindent=3em \hangafter=1
$\tau_n$ = ($-4.$, $13.66$, $-4.$, $8.$, $-4.$, $2.34$, $-4.$, $0.$, $-4.$, $2.34$, $-4.$, $8.$, $-4.$, $13.66$, $-4.$, $16.$)

\vskip 0.7ex
\hangindent=3em \hangafter=1
\textit{Intrinsic sign problem}

  \vskip 2ex

\noindent29. $9_{3,16.}^{16,696}$ \irep{403}:\ \ 
$d_i$ = ($1.0$,
$1.0$,
$1.0$,
$1.0$,
$1.414$,
$1.414$,
$1.414$,
$1.414$,
$2.0$) 

\vskip 0.7ex
\hangindent=3em \hangafter=1
$D^2= 16.0 = 
16$

\vskip 0.7ex
\hangindent=3em \hangafter=1
$T = ( 0,
0,
\frac{1}{2},
\frac{1}{2},
\frac{3}{16},
\frac{3}{16},
\frac{11}{16},
\frac{11}{16},
\frac{3}{8} )
$,

\vskip 0.7ex
\hangindent=3em \hangafter=1
$S$ = ($ 1$,
$ 1$,
$ 1$,
$ 1$,
$ \sqrt{2}$,
$ \sqrt{2}$,
$ \sqrt{2}$,
$ \sqrt{2}$,
$ 2$;\ \ 
$ 1$,
$ 1$,
$ 1$,
$ -\sqrt{2}$,
$ -\sqrt{2}$,
$ -\sqrt{2}$,
$ -\sqrt{2}$,
$ 2$;\ \ 
$ 1$,
$ 1$,
$ -\sqrt{2}$,
$ \sqrt{2}$,
$ -\sqrt{2}$,
$ \sqrt{2}$,
$ -2$;\ \ 
$ 1$,
$ \sqrt{2}$,
$ -\sqrt{2}$,
$ \sqrt{2}$,
$ -\sqrt{2}$,
$ -2$;\ \ 
$0$,
$ 2$,
$0$,
$ -2$,
$0$;\ \ 
$0$,
$ -2$,
$0$,
$0$;\ \ 
$0$,
$ 2$,
$0$;\ \ 
$0$,
$0$;\ \ 
$0$)

Factors = $3_{\frac{3}{2},4.}^{16,553}\boxtimes 3_{\frac{3}{2},4.}^{16,553}$

\vskip 0.7ex
\hangindent=3em \hangafter=1
$\tau_n$ = ($-2.83 + 2.83 i$, $-1.66 + 1.66 i$, $2.83 + 2.83 i$, $0. - 8. i$, $2.83 - 2.83 i$, $9.66 + 9.66 i$, $-2.83 - 2.83 i$, $0.$, $-2.83 + 2.83 i$, $9.66 - 9.66 i$, $2.83 + 2.83 i$, $0. + 8. i$, $2.83 - 2.83 i$, $-1.66 - 1.66 i$, $-2.83 - 2.83 i$, $16.$)

\vskip 0.7ex
\hangindent=3em \hangafter=1
\textit{Intrinsic sign problem}

  \vskip 2ex

\noindent30. $9_{7,16.}^{16,553}$ \irep{403}:\ \ 
$d_i$ = ($1.0$,
$1.0$,
$1.0$,
$1.0$,
$1.414$,
$1.414$,
$1.414$,
$1.414$,
$2.0$) 

\vskip 0.7ex
\hangindent=3em \hangafter=1
$D^2= 16.0 = 
16$

\vskip 0.7ex
\hangindent=3em \hangafter=1
$T = ( 0,
0,
\frac{1}{2},
\frac{1}{2},
\frac{3}{16},
\frac{3}{16},
\frac{11}{16},
\frac{11}{16},
\frac{7}{8} )
$,

\vskip 0.7ex
\hangindent=3em \hangafter=1
$S$ = ($ 1$,
$ 1$,
$ 1$,
$ 1$,
$ \sqrt{2}$,
$ \sqrt{2}$,
$ \sqrt{2}$,
$ \sqrt{2}$,
$ 2$;\ \ 
$ 1$,
$ 1$,
$ 1$,
$ -\sqrt{2}$,
$ -\sqrt{2}$,
$ -\sqrt{2}$,
$ -\sqrt{2}$,
$ 2$;\ \ 
$ 1$,
$ 1$,
$ -\sqrt{2}$,
$ \sqrt{2}$,
$ -\sqrt{2}$,
$ \sqrt{2}$,
$ -2$;\ \ 
$ 1$,
$ \sqrt{2}$,
$ -\sqrt{2}$,
$ \sqrt{2}$,
$ -\sqrt{2}$,
$ -2$;\ \ 
$0$,
$ -2$,
$0$,
$ 2$,
$0$;\ \ 
$0$,
$ 2$,
$0$,
$0$;\ \ 
$0$,
$ -2$,
$0$;\ \ 
$0$,
$0$;\ \ 
$0$)

Factors = $3_{\frac{3}{2},4.}^{16,553}\boxtimes 3_{\frac{11}{2},4.}^{16,648}$

\vskip 0.7ex
\hangindent=3em \hangafter=1
$\tau_n$ = ($2.83 - 2.83 i$, $-1.66 + 1.66 i$, $-2.83 - 2.83 i$, $0. - 8. i$, $-2.83 + 2.83 i$, $9.66 + 9.66 i$, $2.83 + 2.83 i$, $0.$, $2.83 - 2.83 i$, $9.66 - 9.66 i$, $-2.83 - 2.83 i$, $0. + 8. i$, $-2.83 + 2.83 i$, $-1.66 - 1.66 i$, $2.83 + 2.83 i$, $16.$)

\vskip 0.7ex
\hangindent=3em \hangafter=1
\textit{Intrinsic sign problem}

  \vskip 2ex

\noindent31. $9_{0,16.}^{16,624}$ \irep{397}:\ \ 
$d_i$ = ($1.0$,
$1.0$,
$1.0$,
$1.0$,
$1.414$,
$1.414$,
$1.414$,
$1.414$,
$2.0$) 

\vskip 0.7ex
\hangindent=3em \hangafter=1
$D^2= 16.0 = 
16$

\vskip 0.7ex
\hangindent=3em \hangafter=1
$T = ( 0,
0,
\frac{1}{2},
\frac{1}{2},
\frac{3}{16},
\frac{5}{16},
\frac{11}{16},
\frac{13}{16},
0 )
$,

\vskip 0.7ex
\hangindent=3em \hangafter=1
$S$ = ($ 1$,
$ 1$,
$ 1$,
$ 1$,
$ \sqrt{2}$,
$ \sqrt{2}$,
$ \sqrt{2}$,
$ \sqrt{2}$,
$ 2$;\ \ 
$ 1$,
$ 1$,
$ 1$,
$ -\sqrt{2}$,
$ -\sqrt{2}$,
$ -\sqrt{2}$,
$ -\sqrt{2}$,
$ 2$;\ \ 
$ 1$,
$ 1$,
$ -\sqrt{2}$,
$ \sqrt{2}$,
$ -\sqrt{2}$,
$ \sqrt{2}$,
$ -2$;\ \ 
$ 1$,
$ \sqrt{2}$,
$ -\sqrt{2}$,
$ \sqrt{2}$,
$ -\sqrt{2}$,
$ -2$;\ \ 
$0$,
$ -2$,
$0$,
$ 2$,
$0$;\ \ 
$0$,
$ 2$,
$0$,
$0$;\ \ 
$0$,
$ -2$,
$0$;\ \ 
$0$,
$0$;\ \ 
$0$)

Factors = $3_{\frac{3}{2},4.}^{16,553}\boxtimes 3_{\frac{13}{2},4.}^{16,330}$

\vskip 0.7ex
\hangindent=3em \hangafter=1
$\tau_n$ = ($4.$, $2.34$, $4.$, $8.$, $4.$, $13.66$, $4.$, $0.$, $4.$, $13.66$, $4.$, $8.$, $4.$, $2.34$, $4.$, $16.$)

\vskip 0.7ex
\hangindent=3em \hangafter=1
\textit{Undetermined}

  \vskip 2ex

\noindent32. $9_{4,16.}^{16,124}$ \irep{397}:\ \ 
$d_i$ = ($1.0$,
$1.0$,
$1.0$,
$1.0$,
$1.414$,
$1.414$,
$1.414$,
$1.414$,
$2.0$) 

\vskip 0.7ex
\hangindent=3em \hangafter=1
$D^2= 16.0 = 
16$

\vskip 0.7ex
\hangindent=3em \hangafter=1
$T = ( 0,
0,
\frac{1}{2},
\frac{1}{2},
\frac{3}{16},
\frac{5}{16},
\frac{11}{16},
\frac{13}{16},
\frac{1}{2} )
$,

\vskip 0.7ex
\hangindent=3em \hangafter=1
$S$ = ($ 1$,
$ 1$,
$ 1$,
$ 1$,
$ \sqrt{2}$,
$ \sqrt{2}$,
$ \sqrt{2}$,
$ \sqrt{2}$,
$ 2$;\ \ 
$ 1$,
$ 1$,
$ 1$,
$ -\sqrt{2}$,
$ -\sqrt{2}$,
$ -\sqrt{2}$,
$ -\sqrt{2}$,
$ 2$;\ \ 
$ 1$,
$ 1$,
$ -\sqrt{2}$,
$ \sqrt{2}$,
$ -\sqrt{2}$,
$ \sqrt{2}$,
$ -2$;\ \ 
$ 1$,
$ \sqrt{2}$,
$ -\sqrt{2}$,
$ \sqrt{2}$,
$ -\sqrt{2}$,
$ -2$;\ \ 
$0$,
$ 2$,
$0$,
$ -2$,
$0$;\ \ 
$0$,
$ -2$,
$0$,
$0$;\ \ 
$0$,
$ 2$,
$0$;\ \ 
$0$,
$0$;\ \ 
$0$)

Factors = $3_{\frac{3}{2},4.}^{16,553}\boxtimes 3_{\frac{5}{2},4.}^{16,465}$

\vskip 0.7ex
\hangindent=3em \hangafter=1
$\tau_n$ = ($-4.$, $2.34$, $-4.$, $8.$, $-4.$, $13.66$, $-4.$, $0.$, $-4.$, $13.66$, $-4.$, $8.$, $-4.$, $2.34$, $-4.$, $16.$)

\vskip 0.7ex
\hangindent=3em \hangafter=1
\textit{Intrinsic sign problem}

  \vskip 2ex

\noindent33. $9_{1,16.}^{16,151}$ \irep{388}:\ \ 
$d_i$ = ($1.0$,
$1.0$,
$1.0$,
$1.0$,
$1.414$,
$1.414$,
$1.414$,
$1.414$,
$2.0$) 

\vskip 0.7ex
\hangindent=3em \hangafter=1
$D^2= 16.0 = 
16$

\vskip 0.7ex
\hangindent=3em \hangafter=1
$T = ( 0,
0,
\frac{1}{2},
\frac{1}{2},
\frac{3}{16},
\frac{7}{16},
\frac{11}{16},
\frac{15}{16},
\frac{1}{8} )
$,

\vskip 0.7ex
\hangindent=3em \hangafter=1
$S$ = ($ 1$,
$ 1$,
$ 1$,
$ 1$,
$ \sqrt{2}$,
$ \sqrt{2}$,
$ \sqrt{2}$,
$ \sqrt{2}$,
$ 2$;\ \ 
$ 1$,
$ 1$,
$ 1$,
$ -\sqrt{2}$,
$ -\sqrt{2}$,
$ -\sqrt{2}$,
$ -\sqrt{2}$,
$ 2$;\ \ 
$ 1$,
$ 1$,
$ -\sqrt{2}$,
$ \sqrt{2}$,
$ -\sqrt{2}$,
$ \sqrt{2}$,
$ -2$;\ \ 
$ 1$,
$ \sqrt{2}$,
$ -\sqrt{2}$,
$ \sqrt{2}$,
$ -\sqrt{2}$,
$ -2$;\ \ 
$0$,
$ -2$,
$0$,
$ 2$,
$0$;\ \ 
$0$,
$ 2$,
$0$,
$0$;\ \ 
$0$,
$ -2$,
$0$;\ \ 
$0$,
$0$;\ \ 
$0$)

Factors = $3_{\frac{7}{2},4.}^{16,332}\boxtimes 3_{\frac{11}{2},4.}^{16,648}$

\vskip 0.7ex
\hangindent=3em \hangafter=1
$\tau_n$ = ($2.83 + 2.83 i$, $4. + 4. i$, $-2.83 + 2.83 i$, $0. - 8. i$, $-2.83 - 2.83 i$, $4. - 4. i$, $2.83 - 2.83 i$, $0.$, $2.83 + 2.83 i$, $4. + 4. i$, $-2.83 + 2.83 i$, $0. + 8. i$, $-2.83 - 2.83 i$, $4. - 4. i$, $2.83 - 2.83 i$, $16.$)

\vskip 0.7ex
\hangindent=3em \hangafter=1
\textit{Intrinsic sign problem}

  \vskip 2ex

\noindent34. $9_{5,16.}^{16,598}$ \irep{388}:\ \ 
$d_i$ = ($1.0$,
$1.0$,
$1.0$,
$1.0$,
$1.414$,
$1.414$,
$1.414$,
$1.414$,
$2.0$) 

\vskip 0.7ex
\hangindent=3em \hangafter=1
$D^2= 16.0 = 
16$

\vskip 0.7ex
\hangindent=3em \hangafter=1
$T = ( 0,
0,
\frac{1}{2},
\frac{1}{2},
\frac{3}{16},
\frac{7}{16},
\frac{11}{16},
\frac{15}{16},
\frac{5}{8} )
$,

\vskip 0.7ex
\hangindent=3em \hangafter=1
$S$ = ($ 1$,
$ 1$,
$ 1$,
$ 1$,
$ \sqrt{2}$,
$ \sqrt{2}$,
$ \sqrt{2}$,
$ \sqrt{2}$,
$ 2$;\ \ 
$ 1$,
$ 1$,
$ 1$,
$ -\sqrt{2}$,
$ -\sqrt{2}$,
$ -\sqrt{2}$,
$ -\sqrt{2}$,
$ 2$;\ \ 
$ 1$,
$ 1$,
$ -\sqrt{2}$,
$ \sqrt{2}$,
$ -\sqrt{2}$,
$ \sqrt{2}$,
$ -2$;\ \ 
$ 1$,
$ \sqrt{2}$,
$ -\sqrt{2}$,
$ \sqrt{2}$,
$ -\sqrt{2}$,
$ -2$;\ \ 
$0$,
$ 2$,
$0$,
$ -2$,
$0$;\ \ 
$0$,
$ -2$,
$0$,
$0$;\ \ 
$0$,
$ 2$,
$0$;\ \ 
$0$,
$0$;\ \ 
$0$)

Factors = $3_{\frac{7}{2},4.}^{16,332}\boxtimes 3_{\frac{3}{2},4.}^{16,553}$

\vskip 0.7ex
\hangindent=3em \hangafter=1
$\tau_n$ = ($-2.83 - 2.83 i$, $4. + 4. i$, $2.83 - 2.83 i$, $0. - 8. i$, $2.83 + 2.83 i$, $4. - 4. i$, $-2.83 + 2.83 i$, $0.$, $-2.83 - 2.83 i$, $4. + 4. i$, $2.83 - 2.83 i$, $0. + 8. i$, $2.83 + 2.83 i$, $4. - 4. i$, $-2.83 + 2.83 i$, $16.$)

\vskip 0.7ex
\hangindent=3em \hangafter=1
\textit{Intrinsic sign problem}

  \vskip 2ex

\noindent35. $9_{1,16.}^{16,239}$ \irep{403}:\ \ 
$d_i$ = ($1.0$,
$1.0$,
$1.0$,
$1.0$,
$1.414$,
$1.414$,
$1.414$,
$1.414$,
$2.0$) 

\vskip 0.7ex
\hangindent=3em \hangafter=1
$D^2= 16.0 = 
16$

\vskip 0.7ex
\hangindent=3em \hangafter=1
$T = ( 0,
0,
\frac{1}{2},
\frac{1}{2},
\frac{5}{16},
\frac{5}{16},
\frac{13}{16},
\frac{13}{16},
\frac{1}{8} )
$,

\vskip 0.7ex
\hangindent=3em \hangafter=1
$S$ = ($ 1$,
$ 1$,
$ 1$,
$ 1$,
$ \sqrt{2}$,
$ \sqrt{2}$,
$ \sqrt{2}$,
$ \sqrt{2}$,
$ 2$;\ \ 
$ 1$,
$ 1$,
$ 1$,
$ -\sqrt{2}$,
$ -\sqrt{2}$,
$ -\sqrt{2}$,
$ -\sqrt{2}$,
$ 2$;\ \ 
$ 1$,
$ 1$,
$ -\sqrt{2}$,
$ \sqrt{2}$,
$ -\sqrt{2}$,
$ \sqrt{2}$,
$ -2$;\ \ 
$ 1$,
$ \sqrt{2}$,
$ -\sqrt{2}$,
$ \sqrt{2}$,
$ -\sqrt{2}$,
$ -2$;\ \ 
$0$,
$ -2$,
$0$,
$ 2$,
$0$;\ \ 
$0$,
$ 2$,
$0$,
$0$;\ \ 
$0$,
$ -2$,
$0$;\ \ 
$0$,
$0$;\ \ 
$0$)

Factors = $3_{\frac{5}{2},4.}^{16,465}\boxtimes 3_{\frac{13}{2},4.}^{16,330}$

\vskip 0.7ex
\hangindent=3em \hangafter=1
$\tau_n$ = ($2.83 + 2.83 i$, $-1.66 - 1.66 i$, $-2.83 + 2.83 i$, $0. + 8. i$, $-2.83 - 2.83 i$, $9.66 - 9.66 i$, $2.83 - 2.83 i$, $0.$, $2.83 + 2.83 i$, $9.66 + 9.66 i$, $-2.83 + 2.83 i$, $0. - 8. i$, $-2.83 - 2.83 i$, $-1.66 + 1.66 i$, $2.83 - 2.83 i$, $16.$)

\vskip 0.7ex
\hangindent=3em \hangafter=1
\textit{Intrinsic sign problem}

  \vskip 2ex

\noindent36. $9_{5,16.}^{16,510}$ \irep{403}:\ \ 
$d_i$ = ($1.0$,
$1.0$,
$1.0$,
$1.0$,
$1.414$,
$1.414$,
$1.414$,
$1.414$,
$2.0$) 

\vskip 0.7ex
\hangindent=3em \hangafter=1
$D^2= 16.0 = 
16$

\vskip 0.7ex
\hangindent=3em \hangafter=1
$T = ( 0,
0,
\frac{1}{2},
\frac{1}{2},
\frac{5}{16},
\frac{5}{16},
\frac{13}{16},
\frac{13}{16},
\frac{5}{8} )
$,

\vskip 0.7ex
\hangindent=3em \hangafter=1
$S$ = ($ 1$,
$ 1$,
$ 1$,
$ 1$,
$ \sqrt{2}$,
$ \sqrt{2}$,
$ \sqrt{2}$,
$ \sqrt{2}$,
$ 2$;\ \ 
$ 1$,
$ 1$,
$ 1$,
$ -\sqrt{2}$,
$ -\sqrt{2}$,
$ -\sqrt{2}$,
$ -\sqrt{2}$,
$ 2$;\ \ 
$ 1$,
$ 1$,
$ -\sqrt{2}$,
$ \sqrt{2}$,
$ -\sqrt{2}$,
$ \sqrt{2}$,
$ -2$;\ \ 
$ 1$,
$ \sqrt{2}$,
$ -\sqrt{2}$,
$ \sqrt{2}$,
$ -\sqrt{2}$,
$ -2$;\ \ 
$0$,
$ 2$,
$0$,
$ -2$,
$0$;\ \ 
$0$,
$ -2$,
$0$,
$0$;\ \ 
$0$,
$ 2$,
$0$;\ \ 
$0$,
$0$;\ \ 
$0$)

Factors = $3_{\frac{5}{2},4.}^{16,465}\boxtimes 3_{\frac{5}{2},4.}^{16,465}$

\vskip 0.7ex
\hangindent=3em \hangafter=1
$\tau_n$ = ($-2.83 - 2.83 i$, $-1.66 - 1.66 i$, $2.83 - 2.83 i$, $0. + 8. i$, $2.83 + 2.83 i$, $9.66 - 9.66 i$, $-2.83 + 2.83 i$, $0.$, $-2.83 - 2.83 i$, $9.66 + 9.66 i$, $2.83 - 2.83 i$, $0. - 8. i$, $2.83 + 2.83 i$, $-1.66 + 1.66 i$, $-2.83 + 2.83 i$, $16.$)

\vskip 0.7ex
\hangindent=3em \hangafter=1
\textit{Intrinsic sign problem}

  \vskip 2ex

\noindent37. $9_{2,16.}^{16,296}$ \irep{389}:\ \ 
$d_i$ = ($1.0$,
$1.0$,
$1.0$,
$1.0$,
$1.414$,
$1.414$,
$1.414$,
$1.414$,
$2.0$) 

\vskip 0.7ex
\hangindent=3em \hangafter=1
$D^2= 16.0 = 
16$

\vskip 0.7ex
\hangindent=3em \hangafter=1
$T = ( 0,
0,
\frac{1}{2},
\frac{1}{2},
\frac{5}{16},
\frac{7}{16},
\frac{13}{16},
\frac{15}{16},
\frac{1}{4} )
$,

\vskip 0.7ex
\hangindent=3em \hangafter=1
$S$ = ($ 1$,
$ 1$,
$ 1$,
$ 1$,
$ \sqrt{2}$,
$ \sqrt{2}$,
$ \sqrt{2}$,
$ \sqrt{2}$,
$ 2$;\ \ 
$ 1$,
$ 1$,
$ 1$,
$ -\sqrt{2}$,
$ -\sqrt{2}$,
$ -\sqrt{2}$,
$ -\sqrt{2}$,
$ 2$;\ \ 
$ 1$,
$ 1$,
$ -\sqrt{2}$,
$ \sqrt{2}$,
$ -\sqrt{2}$,
$ \sqrt{2}$,
$ -2$;\ \ 
$ 1$,
$ \sqrt{2}$,
$ -\sqrt{2}$,
$ \sqrt{2}$,
$ -\sqrt{2}$,
$ -2$;\ \ 
$0$,
$ -2$,
$0$,
$ 2$,
$0$;\ \ 
$0$,
$ 2$,
$0$,
$0$;\ \ 
$0$,
$ -2$,
$0$;\ \ 
$0$,
$0$;\ \ 
$0$)

Factors = $3_{\frac{7}{2},4.}^{16,332}\boxtimes 3_{\frac{13}{2},4.}^{16,330}$

\vskip 0.7ex
\hangindent=3em \hangafter=1
$\tau_n$ = ($0. + 4. i$, $0. - 5.66 i$, $0. - 4. i$, $8.$, $0. + 4. i$, $0. - 5.66 i$, $0. - 4. i$, $0.$, $0. + 4. i$, $0. + 5.66 i$, $0. - 4. i$, $8.$, $0. + 4. i$, $0. + 5.66 i$, $0. - 4. i$, $16.$)

\vskip 0.7ex
\hangindent=3em \hangafter=1
\textit{Intrinsic sign problem}

  \vskip 2ex

\noindent38. $9_{6,16.}^{16,129}$ \irep{389}:\ \ 
$d_i$ = ($1.0$,
$1.0$,
$1.0$,
$1.0$,
$1.414$,
$1.414$,
$1.414$,
$1.414$,
$2.0$) 

\vskip 0.7ex
\hangindent=3em \hangafter=1
$D^2= 16.0 = 
16$

\vskip 0.7ex
\hangindent=3em \hangafter=1
$T = ( 0,
0,
\frac{1}{2},
\frac{1}{2},
\frac{5}{16},
\frac{7}{16},
\frac{13}{16},
\frac{15}{16},
\frac{3}{4} )
$,

\vskip 0.7ex
\hangindent=3em \hangafter=1
$S$ = ($ 1$,
$ 1$,
$ 1$,
$ 1$,
$ \sqrt{2}$,
$ \sqrt{2}$,
$ \sqrt{2}$,
$ \sqrt{2}$,
$ 2$;\ \ 
$ 1$,
$ 1$,
$ 1$,
$ -\sqrt{2}$,
$ -\sqrt{2}$,
$ -\sqrt{2}$,
$ -\sqrt{2}$,
$ 2$;\ \ 
$ 1$,
$ 1$,
$ -\sqrt{2}$,
$ \sqrt{2}$,
$ -\sqrt{2}$,
$ \sqrt{2}$,
$ -2$;\ \ 
$ 1$,
$ \sqrt{2}$,
$ -\sqrt{2}$,
$ \sqrt{2}$,
$ -\sqrt{2}$,
$ -2$;\ \ 
$0$,
$ 2$,
$0$,
$ -2$,
$0$;\ \ 
$0$,
$ -2$,
$0$,
$0$;\ \ 
$0$,
$ 2$,
$0$;\ \ 
$0$,
$0$;\ \ 
$0$)

Factors = $3_{\frac{7}{2},4.}^{16,332}\boxtimes 3_{\frac{5}{2},4.}^{16,465}$

\vskip 0.7ex
\hangindent=3em \hangafter=1
$\tau_n$ = ($0. - 4. i$, $0. - 5.66 i$, $0. + 4. i$, $8.$, $0. - 4. i$, $0. - 5.66 i$, $0. + 4. i$, $0.$, $0. - 4. i$, $0. + 5.66 i$, $0. + 4. i$, $8.$, $0. - 4. i$, $0. + 5.66 i$, $0. + 4. i$, $16.$)

\vskip 0.7ex
\hangindent=3em \hangafter=1
\textit{Intrinsic sign problem}

  \vskip 2ex

\noindent39. $9_{3,16.}^{16,894}$ \irep{403}:\ \ 
$d_i$ = ($1.0$,
$1.0$,
$1.0$,
$1.0$,
$1.414$,
$1.414$,
$1.414$,
$1.414$,
$2.0$) 

\vskip 0.7ex
\hangindent=3em \hangafter=1
$D^2= 16.0 = 
16$

\vskip 0.7ex
\hangindent=3em \hangafter=1
$T = ( 0,
0,
\frac{1}{2},
\frac{1}{2},
\frac{7}{16},
\frac{7}{16},
\frac{15}{16},
\frac{15}{16},
\frac{3}{8} )
$,

\vskip 0.7ex
\hangindent=3em \hangafter=1
$S$ = ($ 1$,
$ 1$,
$ 1$,
$ 1$,
$ \sqrt{2}$,
$ \sqrt{2}$,
$ \sqrt{2}$,
$ \sqrt{2}$,
$ 2$;\ \ 
$ 1$,
$ 1$,
$ 1$,
$ -\sqrt{2}$,
$ -\sqrt{2}$,
$ -\sqrt{2}$,
$ -\sqrt{2}$,
$ 2$;\ \ 
$ 1$,
$ 1$,
$ -\sqrt{2}$,
$ \sqrt{2}$,
$ -\sqrt{2}$,
$ \sqrt{2}$,
$ -2$;\ \ 
$ 1$,
$ \sqrt{2}$,
$ -\sqrt{2}$,
$ \sqrt{2}$,
$ -\sqrt{2}$,
$ -2$;\ \ 
$0$,
$ -2$,
$0$,
$ 2$,
$0$;\ \ 
$0$,
$ 2$,
$0$,
$0$;\ \ 
$0$,
$ -2$,
$0$;\ \ 
$0$,
$0$;\ \ 
$0$)

Factors = $3_{\frac{7}{2},4.}^{16,332}\boxtimes 3_{\frac{15}{2},4.}^{16,639}$

\vskip 0.7ex
\hangindent=3em \hangafter=1
$\tau_n$ = ($-2.83 + 2.83 i$, $9.66 - 9.66 i$, $2.83 + 2.83 i$, $0. - 8. i$, $2.83 - 2.83 i$, $-1.66 - 1.66 i$, $-2.83 - 2.83 i$, $0.$, $-2.83 + 2.83 i$, $-1.66 + 1.66 i$, $2.83 + 2.83 i$, $0. + 8. i$, $2.83 - 2.83 i$, $9.66 + 9.66 i$, $-2.83 - 2.83 i$, $16.$)

\vskip 0.7ex
\hangindent=3em \hangafter=1
\textit{Intrinsic sign problem}

  \vskip 2ex

\noindent40. $9_{7,16.}^{16,214}$ \irep{403}:\ \ 
$d_i$ = ($1.0$,
$1.0$,
$1.0$,
$1.0$,
$1.414$,
$1.414$,
$1.414$,
$1.414$,
$2.0$) 

\vskip 0.7ex
\hangindent=3em \hangafter=1
$D^2= 16.0 = 
16$

\vskip 0.7ex
\hangindent=3em \hangafter=1
$T = ( 0,
0,
\frac{1}{2},
\frac{1}{2},
\frac{7}{16},
\frac{7}{16},
\frac{15}{16},
\frac{15}{16},
\frac{7}{8} )
$,

\vskip 0.7ex
\hangindent=3em \hangafter=1
$S$ = ($ 1$,
$ 1$,
$ 1$,
$ 1$,
$ \sqrt{2}$,
$ \sqrt{2}$,
$ \sqrt{2}$,
$ \sqrt{2}$,
$ 2$;\ \ 
$ 1$,
$ 1$,
$ 1$,
$ -\sqrt{2}$,
$ -\sqrt{2}$,
$ -\sqrt{2}$,
$ -\sqrt{2}$,
$ 2$;\ \ 
$ 1$,
$ 1$,
$ -\sqrt{2}$,
$ \sqrt{2}$,
$ -\sqrt{2}$,
$ \sqrt{2}$,
$ -2$;\ \ 
$ 1$,
$ \sqrt{2}$,
$ -\sqrt{2}$,
$ \sqrt{2}$,
$ -\sqrt{2}$,
$ -2$;\ \ 
$0$,
$ 2$,
$0$,
$ -2$,
$0$;\ \ 
$0$,
$ -2$,
$0$,
$0$;\ \ 
$0$,
$ 2$,
$0$;\ \ 
$0$,
$0$;\ \ 
$0$)

Factors = $3_{\frac{7}{2},4.}^{16,332}\boxtimes 3_{\frac{7}{2},4.}^{16,332}$

\vskip 0.7ex
\hangindent=3em \hangafter=1
$\tau_n$ = ($2.83 - 2.83 i$, $9.66 - 9.66 i$, $-2.83 - 2.83 i$, $0. - 8. i$, $-2.83 + 2.83 i$, $-1.66 - 1.66 i$, $2.83 + 2.83 i$, $0.$, $2.83 - 2.83 i$, $-1.66 + 1.66 i$, $-2.83 - 2.83 i$, $0. + 8. i$, $-2.83 + 2.83 i$, $9.66 + 9.66 i$, $2.83 + 2.83 i$, $16.$)

\vskip 0.7ex
\hangindent=3em \hangafter=1
\textit{Intrinsic sign problem}

  \vskip 2ex

\noindent41. $9_{\frac{6}{7},27.88}^{21,155}$ \irep{432}:\ \ 
$d_i$ = ($1.0$,
$1.0$,
$1.0$,
$1.801$,
$1.801$,
$1.801$,
$2.246$,
$2.246$,
$2.246$) 

\vskip 0.7ex
\hangindent=3em \hangafter=1
$D^2= 27.887 = 
18+9c^{1}_{7}
+3c^{2}_{7}
$

\vskip 0.7ex
\hangindent=3em \hangafter=1
$T = ( 0,
\frac{1}{3},
\frac{1}{3},
\frac{1}{7},
\frac{10}{21},
\frac{10}{21},
\frac{5}{7},
\frac{1}{21},
\frac{1}{21} )
$,

\vskip 0.7ex
\hangindent=3em \hangafter=1
$S$ = ($ 1$,
$ 1$,
$ 1$,
$ -c_{7}^{3}$,
$ -c_{7}^{3}$,
$ -c_{7}^{3}$,
$ \xi_{7}^{3}$,
$ \xi_{7}^{3}$,
$ \xi_{7}^{3}$;\ \ 
$ \zeta_{3}^{1}$,
$ -\zeta_{6}^{1}$,
$ -c_{7}^{3}$,
$ c_{7}^{3}\zeta_{6}^{1}$,
$ -c_{7}^{3}\zeta_{3}^{1}$,
$ \xi_{7}^{3}$,
$ \xi_{7}^{3}\zeta_{3}^{1}$,
$ -\xi_{7}^{3}\zeta_{6}^{1}$;\ \ 
$ \zeta_{3}^{1}$,
$ -c_{7}^{3}$,
$ -c_{7}^{3}\zeta_{3}^{1}$,
$ c_{7}^{3}\zeta_{6}^{1}$,
$ \xi_{7}^{3}$,
$ -\xi_{7}^{3}\zeta_{6}^{1}$,
$ \xi_{7}^{3}\zeta_{3}^{1}$;\ \ 
$ -\xi_{7}^{3}$,
$ -\xi_{7}^{3}$,
$ -\xi_{7}^{3}$,
$ 1$,
$ 1$,
$ 1$;\ \ 
$ -\xi_{7}^{3}\zeta_{3}^{1}$,
$ \xi_{7}^{3}\zeta_{6}^{1}$,
$ 1$,
$ -\zeta_{6}^{1}$,
$ \zeta_{3}^{1}$;\ \ 
$ -\xi_{7}^{3}\zeta_{3}^{1}$,
$ 1$,
$ \zeta_{3}^{1}$,
$ -\zeta_{6}^{1}$;\ \ 
$ c_{7}^{3}$,
$ c_{7}^{3}$,
$ c_{7}^{3}$;\ \ 
$ c_{7}^{3}\zeta_{3}^{1}$,
$ -c_{7}^{3}\zeta_{6}^{1}$;\ \ 
$ c_{7}^{3}\zeta_{3}^{1}$)

Factors = $3_{2,3.}^{3,527}\boxtimes 3_{\frac{48}{7},9.295}^{7,790}$

\vskip 0.7ex
\hangindent=3em \hangafter=1
$\tau_n$ = ($4.13 + 3.29 i$, $9.27 + 7.39 i$, $3.67 + 16.05 i$, $9.27 + 2.12 i$, $-9.27 + 7.39 i$, $5.7 + 7.15 i$, $0. + 16.09 i$, $-4.13 - 3.29 i$, $-12.8 + 16.05 i$, $-9.27 + 2.12 i$, $-9.27 - 2.12 i$, $-12.8 - 16.05 i$, $-4.13 + 3.29 i$, $0. - 16.09 i$, $5.7 - 7.15 i$, $-9.27 - 7.39 i$, $9.27 - 2.12 i$, $3.67 - 16.05 i$, $9.27 - 7.39 i$, $4.13 - 3.29 i$, $27.86$)

\vskip 0.7ex
\hangindent=3em \hangafter=1
\textit{Intrinsic sign problem}

  \vskip 2ex

\noindent42. $9_{\frac{22}{7},27.88}^{21,204}$ \irep{432}:\ \ 
$d_i$ = ($1.0$,
$1.0$,
$1.0$,
$1.801$,
$1.801$,
$1.801$,
$2.246$,
$2.246$,
$2.246$) 

\vskip 0.7ex
\hangindent=3em \hangafter=1
$D^2= 27.887 = 
18+9c^{1}_{7}
+3c^{2}_{7}
$

\vskip 0.7ex
\hangindent=3em \hangafter=1
$T = ( 0,
\frac{1}{3},
\frac{1}{3},
\frac{6}{7},
\frac{4}{21},
\frac{4}{21},
\frac{2}{7},
\frac{13}{21},
\frac{13}{21} )
$,

\vskip 0.7ex
\hangindent=3em \hangafter=1
$S$ = ($ 1$,
$ 1$,
$ 1$,
$ -c_{7}^{3}$,
$ -c_{7}^{3}$,
$ -c_{7}^{3}$,
$ \xi_{7}^{3}$,
$ \xi_{7}^{3}$,
$ \xi_{7}^{3}$;\ \ 
$ \zeta_{3}^{1}$,
$ -\zeta_{6}^{1}$,
$ -c_{7}^{3}$,
$ c_{7}^{3}\zeta_{6}^{1}$,
$ -c_{7}^{3}\zeta_{3}^{1}$,
$ \xi_{7}^{3}$,
$ \xi_{7}^{3}\zeta_{3}^{1}$,
$ -\xi_{7}^{3}\zeta_{6}^{1}$;\ \ 
$ \zeta_{3}^{1}$,
$ -c_{7}^{3}$,
$ -c_{7}^{3}\zeta_{3}^{1}$,
$ c_{7}^{3}\zeta_{6}^{1}$,
$ \xi_{7}^{3}$,
$ -\xi_{7}^{3}\zeta_{6}^{1}$,
$ \xi_{7}^{3}\zeta_{3}^{1}$;\ \ 
$ -\xi_{7}^{3}$,
$ -\xi_{7}^{3}$,
$ -\xi_{7}^{3}$,
$ 1$,
$ 1$,
$ 1$;\ \ 
$ -\xi_{7}^{3}\zeta_{3}^{1}$,
$ \xi_{7}^{3}\zeta_{6}^{1}$,
$ 1$,
$ -\zeta_{6}^{1}$,
$ \zeta_{3}^{1}$;\ \ 
$ -\xi_{7}^{3}\zeta_{3}^{1}$,
$ 1$,
$ \zeta_{3}^{1}$,
$ -\zeta_{6}^{1}$;\ \ 
$ c_{7}^{3}$,
$ c_{7}^{3}$,
$ c_{7}^{3}$;\ \ 
$ c_{7}^{3}\zeta_{3}^{1}$,
$ -c_{7}^{3}\zeta_{6}^{1}$;\ \ 
$ c_{7}^{3}\zeta_{3}^{1}$)

Factors = $3_{2,3.}^{3,527}\boxtimes 3_{\frac{8}{7},9.295}^{7,245}$

\vskip 0.7ex
\hangindent=3em \hangafter=1
$\tau_n$ = ($-4.13 + 3.29 i$, $-9.27 + 7.39 i$, $3.67 - 16.05 i$, $-9.27 + 2.12 i$, $9.27 + 7.39 i$, $5.7 - 7.15 i$, $0. + 16.09 i$, $4.13 - 3.29 i$, $-12.8 - 16.05 i$, $9.27 + 2.12 i$, $9.27 - 2.12 i$, $-12.8 + 16.05 i$, $4.13 + 3.29 i$, $0. - 16.09 i$, $5.7 + 7.15 i$, $9.27 - 7.39 i$, $-9.27 - 2.12 i$, $3.67 + 16.05 i$, $-9.27 - 7.39 i$, $-4.13 - 3.29 i$, $27.86$)

\vskip 0.7ex
\hangindent=3em \hangafter=1
\textit{Intrinsic sign problem}

  \vskip 2ex

\noindent43. $9_{\frac{34}{7},27.88}^{21,129}$ \irep{432}:\ \ 
$d_i$ = ($1.0$,
$1.0$,
$1.0$,
$1.801$,
$1.801$,
$1.801$,
$2.246$,
$2.246$,
$2.246$) 

\vskip 0.7ex
\hangindent=3em \hangafter=1
$D^2= 27.887 = 
18+9c^{1}_{7}
+3c^{2}_{7}
$

\vskip 0.7ex
\hangindent=3em \hangafter=1
$T = ( 0,
\frac{2}{3},
\frac{2}{3},
\frac{1}{7},
\frac{17}{21},
\frac{17}{21},
\frac{5}{7},
\frac{8}{21},
\frac{8}{21} )
$,

\vskip 0.7ex
\hangindent=3em \hangafter=1
$S$ = ($ 1$,
$ 1$,
$ 1$,
$ -c_{7}^{3}$,
$ -c_{7}^{3}$,
$ -c_{7}^{3}$,
$ \xi_{7}^{3}$,
$ \xi_{7}^{3}$,
$ \xi_{7}^{3}$;\ \ 
$ -\zeta_{6}^{1}$,
$ \zeta_{3}^{1}$,
$ -c_{7}^{3}$,
$ c_{7}^{3}\zeta_{6}^{1}$,
$ -c_{7}^{3}\zeta_{3}^{1}$,
$ \xi_{7}^{3}$,
$ \xi_{7}^{3}\zeta_{3}^{1}$,
$ -\xi_{7}^{3}\zeta_{6}^{1}$;\ \ 
$ -\zeta_{6}^{1}$,
$ -c_{7}^{3}$,
$ -c_{7}^{3}\zeta_{3}^{1}$,
$ c_{7}^{3}\zeta_{6}^{1}$,
$ \xi_{7}^{3}$,
$ -\xi_{7}^{3}\zeta_{6}^{1}$,
$ \xi_{7}^{3}\zeta_{3}^{1}$;\ \ 
$ -\xi_{7}^{3}$,
$ -\xi_{7}^{3}$,
$ -\xi_{7}^{3}$,
$ 1$,
$ 1$,
$ 1$;\ \ 
$ \xi_{7}^{3}\zeta_{6}^{1}$,
$ -\xi_{7}^{3}\zeta_{3}^{1}$,
$ 1$,
$ \zeta_{3}^{1}$,
$ -\zeta_{6}^{1}$;\ \ 
$ \xi_{7}^{3}\zeta_{6}^{1}$,
$ 1$,
$ -\zeta_{6}^{1}$,
$ \zeta_{3}^{1}$;\ \ 
$ c_{7}^{3}$,
$ c_{7}^{3}$,
$ c_{7}^{3}$;\ \ 
$ -c_{7}^{3}\zeta_{6}^{1}$,
$ c_{7}^{3}\zeta_{3}^{1}$;\ \ 
$ -c_{7}^{3}\zeta_{6}^{1}$)

Factors = $3_{6,3.}^{3,138}\boxtimes 3_{\frac{48}{7},9.295}^{7,790}$

\vskip 0.7ex
\hangindent=3em \hangafter=1
$\tau_n$ = ($-4.13 - 3.29 i$, $-9.27 - 7.39 i$, $3.67 + 16.05 i$, $-9.27 - 2.12 i$, $9.27 - 7.39 i$, $5.7 + 7.15 i$, $0. - 16.09 i$, $4.13 + 3.29 i$, $-12.8 + 16.05 i$, $9.27 - 2.12 i$, $9.27 + 2.12 i$, $-12.8 - 16.05 i$, $4.13 - 3.29 i$, $0. + 16.09 i$, $5.7 - 7.15 i$, $9.27 + 7.39 i$, $-9.27 + 2.12 i$, $3.67 - 16.05 i$, $-9.27 + 7.39 i$, $-4.13 + 3.29 i$, $27.86$)

\vskip 0.7ex
\hangindent=3em \hangafter=1
\textit{Intrinsic sign problem}

  \vskip 2ex

\noindent44. $9_{\frac{50}{7},27.88}^{21,367}$ \irep{432}:\ \ 
$d_i$ = ($1.0$,
$1.0$,
$1.0$,
$1.801$,
$1.801$,
$1.801$,
$2.246$,
$2.246$,
$2.246$) 

\vskip 0.7ex
\hangindent=3em \hangafter=1
$D^2= 27.887 = 
18+9c^{1}_{7}
+3c^{2}_{7}
$

\vskip 0.7ex
\hangindent=3em \hangafter=1
$T = ( 0,
\frac{2}{3},
\frac{2}{3},
\frac{6}{7},
\frac{11}{21},
\frac{11}{21},
\frac{2}{7},
\frac{20}{21},
\frac{20}{21} )
$,

\vskip 0.7ex
\hangindent=3em \hangafter=1
$S$ = ($ 1$,
$ 1$,
$ 1$,
$ -c_{7}^{3}$,
$ -c_{7}^{3}$,
$ -c_{7}^{3}$,
$ \xi_{7}^{3}$,
$ \xi_{7}^{3}$,
$ \xi_{7}^{3}$;\ \ 
$ -\zeta_{6}^{1}$,
$ \zeta_{3}^{1}$,
$ -c_{7}^{3}$,
$ c_{7}^{3}\zeta_{6}^{1}$,
$ -c_{7}^{3}\zeta_{3}^{1}$,
$ \xi_{7}^{3}$,
$ \xi_{7}^{3}\zeta_{3}^{1}$,
$ -\xi_{7}^{3}\zeta_{6}^{1}$;\ \ 
$ -\zeta_{6}^{1}$,
$ -c_{7}^{3}$,
$ -c_{7}^{3}\zeta_{3}^{1}$,
$ c_{7}^{3}\zeta_{6}^{1}$,
$ \xi_{7}^{3}$,
$ -\xi_{7}^{3}\zeta_{6}^{1}$,
$ \xi_{7}^{3}\zeta_{3}^{1}$;\ \ 
$ -\xi_{7}^{3}$,
$ -\xi_{7}^{3}$,
$ -\xi_{7}^{3}$,
$ 1$,
$ 1$,
$ 1$;\ \ 
$ \xi_{7}^{3}\zeta_{6}^{1}$,
$ -\xi_{7}^{3}\zeta_{3}^{1}$,
$ 1$,
$ \zeta_{3}^{1}$,
$ -\zeta_{6}^{1}$;\ \ 
$ \xi_{7}^{3}\zeta_{6}^{1}$,
$ 1$,
$ -\zeta_{6}^{1}$,
$ \zeta_{3}^{1}$;\ \ 
$ c_{7}^{3}$,
$ c_{7}^{3}$,
$ c_{7}^{3}$;\ \ 
$ -c_{7}^{3}\zeta_{6}^{1}$,
$ c_{7}^{3}\zeta_{3}^{1}$;\ \ 
$ -c_{7}^{3}\zeta_{6}^{1}$)

Factors = $3_{6,3.}^{3,138}\boxtimes 3_{\frac{8}{7},9.295}^{7,245}$

\vskip 0.7ex
\hangindent=3em \hangafter=1
$\tau_n$ = ($4.13 - 3.29 i$, $9.27 - 7.39 i$, $3.67 - 16.05 i$, $9.27 - 2.12 i$, $-9.27 - 7.39 i$, $5.7 - 7.15 i$, $0. - 16.09 i$, $-4.13 + 3.29 i$, $-12.8 - 16.05 i$, $-9.27 - 2.12 i$, $-9.27 + 2.12 i$, $-12.8 + 16.05 i$, $-4.13 - 3.29 i$, $0. + 16.09 i$, $5.7 + 7.15 i$, $-9.27 + 7.39 i$, $9.27 + 2.12 i$, $3.67 + 16.05 i$, $9.27 + 7.39 i$, $4.13 + 3.29 i$, $27.86$)

\vskip 0.7ex
\hangindent=3em \hangafter=1
\textit{Intrinsic sign problem}

  \vskip 2ex

\noindent45. $9_{\frac{103}{14},37.18}^{112,193}$ \irep{498}:\ \ 
$d_i$ = ($1.0$,
$1.0$,
$1.414$,
$1.801$,
$1.801$,
$2.246$,
$2.246$,
$2.548$,
$3.177$) 

\vskip 0.7ex
\hangindent=3em \hangafter=1
$D^2= 37.183 = 
24+12c^{1}_{7}
+4c^{2}_{7}
$

\vskip 0.7ex
\hangindent=3em \hangafter=1
$T = ( 0,
\frac{1}{2},
\frac{1}{16},
\frac{1}{7},
\frac{9}{14},
\frac{5}{7},
\frac{3}{14},
\frac{23}{112},
\frac{87}{112} )
$,

\vskip 0.7ex
\hangindent=3em \hangafter=1
$S$ = ($ 1$,
$ 1$,
$ \sqrt{2}$,
$ -c_{7}^{3}$,
$ -c_{7}^{3}$,
$ \xi_{7}^{3}$,
$ \xi_{7}^{3}$,
$ c^{3}_{56}
+c^{11}_{56}
$,
$ c^{3}_{56}
+c^{5}_{56}
-c^{9}_{56}
+c^{11}_{56}
$;\ \ 
$ 1$,
$ -\sqrt{2}$,
$ -c_{7}^{3}$,
$ -c_{7}^{3}$,
$ \xi_{7}^{3}$,
$ \xi_{7}^{3}$,
$ -c^{3}_{56}
-c^{11}_{56}
$,
$ -c^{3}_{56}
-c^{5}_{56}
+c^{9}_{56}
-c^{11}_{56}
$;\ \ 
$0$,
$ c^{3}_{56}
+c^{11}_{56}
$,
$ -c^{3}_{56}
-c^{11}_{56}
$,
$ c^{3}_{56}
+c^{5}_{56}
-c^{9}_{56}
+c^{11}_{56}
$,
$ -c^{3}_{56}
-c^{5}_{56}
+c^{9}_{56}
-c^{11}_{56}
$,
$0$,
$0$;\ \ 
$ -\xi_{7}^{3}$,
$ -\xi_{7}^{3}$,
$ 1$,
$ 1$,
$ -c^{3}_{56}
-c^{5}_{56}
+c^{9}_{56}
-c^{11}_{56}
$,
$ \sqrt{2}$;\ \ 
$ -\xi_{7}^{3}$,
$ 1$,
$ 1$,
$ c^{3}_{56}
+c^{5}_{56}
-c^{9}_{56}
+c^{11}_{56}
$,
$ -\sqrt{2}$;\ \ 
$ c_{7}^{3}$,
$ c_{7}^{3}$,
$ \sqrt{2}$,
$ -c^{3}_{56}
-c^{11}_{56}
$;\ \ 
$ c_{7}^{3}$,
$ -\sqrt{2}$,
$ c^{3}_{56}
+c^{11}_{56}
$;\ \ 
$0$,
$0$;\ \ 
$0$)

Factors = $3_{\frac{1}{2},4.}^{16,598}\boxtimes 3_{\frac{48}{7},9.295}^{7,790}$

\vskip 0.7ex
\hangindent=3em \hangafter=1
$\tau_n$ = ($5.34 - 2.95 i$, $-22.14 + 12.24 i$, $-8.96 + 6.35 i$, $13.15 - 8.26 i$, $13.16 - 3.79 i$, $-2.26 + 4.08 i$, $-17.17 + 7.11 i$, $0.$, $11.99 - 6.63 i$, $8.29 + 1.4 i$, $-10.83 + 1.84 i$, $-19.24 - 2.16 i$, $5.86 - 1.69 i$, $31.72 - 13.14 i$, $1.69 - 5.86 i$, $-17.07 + 21.41 i$, $-1.84 + 10.83 i$, $11.75 - 16.55 i$, $6.63 - 11.99 i$, $-0.96 + 8.57 i$, $-7.11 + 17.17 i$, $4.48 + 1.29 i$, $3.79 - 13.16 i$, $0. - 0.01 i$, $-6.35 + 8.96 i$, $-10.07 + 2.91 i$, $2.95 - 5.34 i$, $18.58 - 18.59 i$, $-2.95 - 5.34 i$, $-7. + 24.31 i$, $6.35 + 8.96 i$, $4.89 - 21.41 i$, $-3.79 - 13.16 i$, $3.12 + 10.82 i$, $7.11 + 17.17 i$, $8.56 - 0.96 i$, $-6.63 - 11.99 i$, $-6.85 + 4.86 i$, $1.84 + 10.83 i$, $0.01 + 0.01 i$, $-1.69 - 5.86 i$, $5.43 - 13.14 i$, $-5.86 - 1.69 i$, $2.18 + 19.24 i$, $10.83 + 1.84 i$, $-3.4 - 20. i$, $-11.99 - 6.63 i$, $7.6 + 9.53 i$, $17.17 + 7.11 i$, $9.86 - 5.45 i$, $-13.16 - 3.79 i$, $-8.26 + 13.15 i$, $8.96 + 6.35 i$, $5.08 - 9.17 i$, $-5.34 - 2.95 i$, $-0.01$, $-5.34 + 2.95 i$, $5.08 + 9.17 i$, $8.96 - 6.35 i$, $-8.26 - 13.15 i$, $-13.16 + 3.79 i$, $9.86 + 5.45 i$, $17.17 - 7.11 i$, $7.6 - 9.53 i$, $-11.99 + 6.63 i$, $-3.4 + 20. i$, $10.83 - 1.84 i$, $2.18 - 19.24 i$, $-5.86 + 1.69 i$, $5.43 + 13.14 i$, $-1.69 + 5.86 i$, $0.01 - 0.01 i$, $1.84 - 10.83 i$, $-6.85 - 4.86 i$, $-6.63 + 11.99 i$, $8.56 + 0.96 i$, $7.11 - 17.17 i$, $3.12 - 10.82 i$, $-3.79 + 13.16 i$, $4.89 + 21.41 i$, $6.35 - 8.96 i$, $-7. - 24.31 i$, $-2.95 + 5.34 i$, $18.58 + 18.59 i$, $2.95 + 5.34 i$, $-10.07 - 2.91 i$, $-6.35 - 8.96 i$, $0. + 0.01 i$, $3.79 + 13.16 i$, $4.48 - 1.29 i$, $-7.11 - 17.17 i$, $-0.96 - 8.57 i$, $6.63 + 11.99 i$, $11.75 + 16.55 i$, $-1.84 - 10.83 i$, $-17.07 - 21.41 i$, $1.69 + 5.86 i$, $31.72 + 13.14 i$, $5.86 + 1.69 i$, $-19.24 + 2.16 i$, $-10.83 - 1.84 i$, $8.29 - 1.4 i$, $11.99 + 6.63 i$, $0.$, $-17.17 - 7.11 i$, $-2.26 - 4.08 i$, $13.16 + 3.79 i$, $13.15 + 8.26 i$, $-8.96 - 6.35 i$, $-22.14 - 12.24 i$, $5.34 + 2.95 i$, $37.16$)

\vskip 0.7ex
\hangindent=3em \hangafter=1
\textit{Intrinsic sign problem}

  \vskip 2ex

\noindent46. $9_{\frac{23}{14},37.18}^{112,161}$ \irep{498}:\ \ 
$d_i$ = ($1.0$,
$1.0$,
$1.414$,
$1.801$,
$1.801$,
$2.246$,
$2.246$,
$2.548$,
$3.177$) 

\vskip 0.7ex
\hangindent=3em \hangafter=1
$D^2= 37.183 = 
24+12c^{1}_{7}
+4c^{2}_{7}
$

\vskip 0.7ex
\hangindent=3em \hangafter=1
$T = ( 0,
\frac{1}{2},
\frac{1}{16},
\frac{6}{7},
\frac{5}{14},
\frac{2}{7},
\frac{11}{14},
\frac{103}{112},
\frac{39}{112} )
$,

\vskip 0.7ex
\hangindent=3em \hangafter=1
$S$ = ($ 1$,
$ 1$,
$ \sqrt{2}$,
$ -c_{7}^{3}$,
$ -c_{7}^{3}$,
$ \xi_{7}^{3}$,
$ \xi_{7}^{3}$,
$ c^{3}_{56}
+c^{11}_{56}
$,
$ c^{3}_{56}
+c^{5}_{56}
-c^{9}_{56}
+c^{11}_{56}
$;\ \ 
$ 1$,
$ -\sqrt{2}$,
$ -c_{7}^{3}$,
$ -c_{7}^{3}$,
$ \xi_{7}^{3}$,
$ \xi_{7}^{3}$,
$ -c^{3}_{56}
-c^{11}_{56}
$,
$ -c^{3}_{56}
-c^{5}_{56}
+c^{9}_{56}
-c^{11}_{56}
$;\ \ 
$0$,
$ c^{3}_{56}
+c^{11}_{56}
$,
$ -c^{3}_{56}
-c^{11}_{56}
$,
$ c^{3}_{56}
+c^{5}_{56}
-c^{9}_{56}
+c^{11}_{56}
$,
$ -c^{3}_{56}
-c^{5}_{56}
+c^{9}_{56}
-c^{11}_{56}
$,
$0$,
$0$;\ \ 
$ -\xi_{7}^{3}$,
$ -\xi_{7}^{3}$,
$ 1$,
$ 1$,
$ -c^{3}_{56}
-c^{5}_{56}
+c^{9}_{56}
-c^{11}_{56}
$,
$ \sqrt{2}$;\ \ 
$ -\xi_{7}^{3}$,
$ 1$,
$ 1$,
$ c^{3}_{56}
+c^{5}_{56}
-c^{9}_{56}
+c^{11}_{56}
$,
$ -\sqrt{2}$;\ \ 
$ c_{7}^{3}$,
$ c_{7}^{3}$,
$ \sqrt{2}$,
$ -c^{3}_{56}
-c^{11}_{56}
$;\ \ 
$ c_{7}^{3}$,
$ -\sqrt{2}$,
$ c^{3}_{56}
+c^{11}_{56}
$;\ \ 
$0$,
$0$;\ \ 
$0$)

Factors = $3_{\frac{1}{2},4.}^{16,598}\boxtimes 3_{\frac{8}{7},9.295}^{7,245}$

\vskip 0.7ex
\hangindent=3em \hangafter=1
$\tau_n$ = ($1.69 + 5.86 i$, $-7. - 24.31 i$, $10.83 - 1.84 i$, $-8.26 + 13.15 i$, $-6.63 - 11.99 i$, $4.48 + 1.29 i$, $-17.17 + 7.11 i$, $0.$, $3.79 + 13.16 i$, $-6.85 - 4.86 i$, $8.96 - 6.35 i$, $2.18 + 19.24 i$, $-2.95 - 5.34 i$, $31.72 - 13.14 i$, $5.34 + 2.95 i$, $-17.07 - 21.41 i$, $6.35 - 8.96 i$, $-3.4 + 20. i$, $-13.16 - 3.79 i$, $8.56 - 0.96 i$, $-7.11 + 17.17 i$, $-2.26 + 4.08 i$, $11.99 + 6.63 i$, $0. + 0.01 i$, $1.84 - 10.83 i$, $5.08 + 9.17 i$, $-5.86 - 1.69 i$, $18.58 - 18.59 i$, $5.86 - 1.69 i$, $-22.14 - 12.24 i$, $-1.84 - 10.83 i$, $4.89 + 21.41 i$, $-11.99 + 6.63 i$, $9.86 - 5.45 i$, $7.11 + 17.17 i$, $-0.96 + 8.57 i$, $13.16 - 3.79 i$, $8.29 - 1.4 i$, $-6.35 - 8.96 i$, $0.01 - 0.01 i$, $-5.34 + 2.95 i$, $5.43 - 13.14 i$, $2.95 - 5.34 i$, $-19.24 - 2.16 i$, $-8.96 - 6.35 i$, $11.75 + 16.55 i$, $-3.79 + 13.16 i$, $7.6 - 9.53 i$, $17.17 + 7.11 i$, $3.12 + 10.82 i$, $6.63 - 11.99 i$, $13.15 - 8.26 i$, $-10.83 - 1.84 i$, $-10.07 - 2.91 i$, $-1.69 + 5.86 i$, $-0.01$, $-1.69 - 5.86 i$, $-10.07 + 2.91 i$, $-10.83 + 1.84 i$, $13.15 + 8.26 i$, $6.63 + 11.99 i$, $3.12 - 10.82 i$, $17.17 - 7.11 i$, $7.6 + 9.53 i$, $-3.79 - 13.16 i$, $11.75 - 16.55 i$, $-8.96 + 6.35 i$, $-19.24 + 2.16 i$, $2.95 + 5.34 i$, $5.43 + 13.14 i$, $-5.34 - 2.95 i$, $0.01 + 0.01 i$, $-6.35 + 8.96 i$, $8.29 + 1.4 i$, $13.16 + 3.79 i$, $-0.96 - 8.57 i$, $7.11 - 17.17 i$, $9.86 + 5.45 i$, $-11.99 - 6.63 i$, $4.89 - 21.41 i$, $-1.84 + 10.83 i$, $-22.14 + 12.24 i$, $5.86 + 1.69 i$, $18.58 + 18.59 i$, $-5.86 + 1.69 i$, $5.08 - 9.17 i$, $1.84 + 10.83 i$, $0. - 0.01 i$, $11.99 - 6.63 i$, $-2.26 - 4.08 i$, $-7.11 - 17.17 i$, $8.56 + 0.96 i$, $-13.16 + 3.79 i$, $-3.4 - 20. i$, $6.35 + 8.96 i$, $-17.07 + 21.41 i$, $5.34 - 2.95 i$, $31.72 + 13.14 i$, $-2.95 + 5.34 i$, $2.18 - 19.24 i$, $8.96 + 6.35 i$, $-6.85 + 4.86 i$, $3.79 - 13.16 i$, $0.$, $-17.17 - 7.11 i$, $4.48 - 1.29 i$, $-6.63 + 11.99 i$, $-8.26 - 13.15 i$, $10.83 + 1.84 i$, $-7. + 24.31 i$, $1.69 - 5.86 i$, $37.16$)

\vskip 0.7ex
\hangindent=3em \hangafter=1
\textit{Intrinsic sign problem}

  \vskip 2ex

\noindent47. $9_{\frac{5}{14},37.18}^{112,883}$ \irep{498}:\ \ 
$d_i$ = ($1.0$,
$1.0$,
$1.414$,
$1.801$,
$1.801$,
$2.246$,
$2.246$,
$2.548$,
$3.177$) 

\vskip 0.7ex
\hangindent=3em \hangafter=1
$D^2= 37.183 = 
24+12c^{1}_{7}
+4c^{2}_{7}
$

\vskip 0.7ex
\hangindent=3em \hangafter=1
$T = ( 0,
\frac{1}{2},
\frac{3}{16},
\frac{1}{7},
\frac{9}{14},
\frac{5}{7},
\frac{3}{14},
\frac{37}{112},
\frac{101}{112} )
$,

\vskip 0.7ex
\hangindent=3em \hangafter=1
$S$ = ($ 1$,
$ 1$,
$ \sqrt{2}$,
$ -c_{7}^{3}$,
$ -c_{7}^{3}$,
$ \xi_{7}^{3}$,
$ \xi_{7}^{3}$,
$ c^{3}_{56}
+c^{11}_{56}
$,
$ c^{3}_{56}
+c^{5}_{56}
-c^{9}_{56}
+c^{11}_{56}
$;\ \ 
$ 1$,
$ -\sqrt{2}$,
$ -c_{7}^{3}$,
$ -c_{7}^{3}$,
$ \xi_{7}^{3}$,
$ \xi_{7}^{3}$,
$ -c^{3}_{56}
-c^{11}_{56}
$,
$ -c^{3}_{56}
-c^{5}_{56}
+c^{9}_{56}
-c^{11}_{56}
$;\ \ 
$0$,
$ c^{3}_{56}
+c^{11}_{56}
$,
$ -c^{3}_{56}
-c^{11}_{56}
$,
$ c^{3}_{56}
+c^{5}_{56}
-c^{9}_{56}
+c^{11}_{56}
$,
$ -c^{3}_{56}
-c^{5}_{56}
+c^{9}_{56}
-c^{11}_{56}
$,
$0$,
$0$;\ \ 
$ -\xi_{7}^{3}$,
$ -\xi_{7}^{3}$,
$ 1$,
$ 1$,
$ -c^{3}_{56}
-c^{5}_{56}
+c^{9}_{56}
-c^{11}_{56}
$,
$ \sqrt{2}$;\ \ 
$ -\xi_{7}^{3}$,
$ 1$,
$ 1$,
$ c^{3}_{56}
+c^{5}_{56}
-c^{9}_{56}
+c^{11}_{56}
$,
$ -\sqrt{2}$;\ \ 
$ c_{7}^{3}$,
$ c_{7}^{3}$,
$ \sqrt{2}$,
$ -c^{3}_{56}
-c^{11}_{56}
$;\ \ 
$ c_{7}^{3}$,
$ -\sqrt{2}$,
$ c^{3}_{56}
+c^{11}_{56}
$;\ \ 
$0$,
$0$;\ \ 
$0$)

Factors = $3_{\frac{3}{2},4.}^{16,553}\boxtimes 3_{\frac{48}{7},9.295}^{7,790}$

\vskip 0.7ex
\hangindent=3em \hangafter=1
$\tau_n$ = ($5.86 + 1.69 i$, $-10.07 - 2.91 i$, $1.84 - 10.83 i$, $-8.26 - 13.15 i$, $-11.99 - 6.63 i$, $3.12 + 10.82 i$, $-7.11 + 17.17 i$, $0.$, $13.16 + 3.79 i$, $11.75 + 16.55 i$, $6.35 - 8.96 i$, $2.18 - 19.24 i$, $-5.34 - 2.95 i$, $5.43 - 13.14 i$, $-2.95 - 5.34 i$, $-17.07 + 21.41 i$, $-8.96 + 6.35 i$, $8.29 - 1.4 i$, $3.79 + 13.16 i$, $8.56 + 0.96 i$, $17.17 - 7.11 i$, $9.86 - 5.45 i$, $-6.63 - 11.99 i$, $0. - 0.01 i$, $-10.83 + 1.84 i$, $-22.14 - 12.24 i$, $1.69 + 5.86 i$, $18.58 + 18.59 i$, $-1.69 + 5.86 i$, $5.08 + 9.17 i$, $10.83 + 1.84 i$, $4.89 - 21.41 i$, $6.63 - 11.99 i$, $-2.26 + 4.08 i$, $-17.17 - 7.11 i$, $-0.96 - 8.57 i$, $-3.79 + 13.16 i$, $-3.4 + 20. i$, $8.96 + 6.35 i$, $0.01 + 0.01 i$, $2.95 - 5.34 i$, $31.72 - 13.14 i$, $5.34 - 2.95 i$, $-19.24 + 2.16 i$, $-6.35 - 8.96 i$, $-6.85 - 4.86 i$, $-13.16 + 3.79 i$, $7.6 + 9.53 i$, $7.11 + 17.17 i$, $4.48 + 1.29 i$, $11.99 - 6.63 i$, $13.15 + 8.26 i$, $-1.84 - 10.83 i$, $-7. - 24.31 i$, $-5.86 + 1.69 i$, $-0.01$, $-5.86 - 1.69 i$, $-7. + 24.31 i$, $-1.84 + 10.83 i$, $13.15 - 8.26 i$, $11.99 + 6.63 i$, $4.48 - 1.29 i$, $7.11 - 17.17 i$, $7.6 - 9.53 i$, $-13.16 - 3.79 i$, $-6.85 + 4.86 i$, $-6.35 + 8.96 i$, $-19.24 - 2.16 i$, $5.34 + 2.95 i$, $31.72 + 13.14 i$, $2.95 + 5.34 i$, $0.01 - 0.01 i$, $8.96 - 6.35 i$, $-3.4 - 20. i$, $-3.79 - 13.16 i$, $-0.96 + 8.57 i$, $-17.17 + 7.11 i$, $-2.26 - 4.08 i$, $6.63 + 11.99 i$, $4.89 + 21.41 i$, $10.83 - 1.84 i$, $5.08 - 9.17 i$, $-1.69 - 5.86 i$, $18.58 - 18.59 i$, $1.69 - 5.86 i$, $-22.14 + 12.24 i$, $-10.83 - 1.84 i$, $0. + 0.01 i$, $-6.63 + 11.99 i$, $9.86 + 5.45 i$, $17.17 + 7.11 i$, $8.56 - 0.96 i$, $3.79 - 13.16 i$, $8.29 + 1.4 i$, $-8.96 - 6.35 i$, $-17.07 - 21.41 i$, $-2.95 + 5.34 i$, $5.43 + 13.14 i$, $-5.34 + 2.95 i$, $2.18 + 19.24 i$, $6.35 + 8.96 i$, $11.75 - 16.55 i$, $13.16 - 3.79 i$, $0.$, $-7.11 - 17.17 i$, $3.12 - 10.82 i$, $-11.99 + 6.63 i$, $-8.26 + 13.15 i$, $1.84 + 10.83 i$, $-10.07 + 2.91 i$, $5.86 - 1.69 i$, $37.16$)

\vskip 0.7ex
\hangindent=3em \hangafter=1
\textit{Intrinsic sign problem}

  \vskip 2ex

\noindent48. $9_{\frac{37}{14},37.18}^{112,168}$ \irep{498}:\ \ 
$d_i$ = ($1.0$,
$1.0$,
$1.414$,
$1.801$,
$1.801$,
$2.246$,
$2.246$,
$2.548$,
$3.177$) 

\vskip 0.7ex
\hangindent=3em \hangafter=1
$D^2= 37.183 = 
24+12c^{1}_{7}
+4c^{2}_{7}
$

\vskip 0.7ex
\hangindent=3em \hangafter=1
$T = ( 0,
\frac{1}{2},
\frac{3}{16},
\frac{6}{7},
\frac{5}{14},
\frac{2}{7},
\frac{11}{14},
\frac{5}{112},
\frac{53}{112} )
$,

\vskip 0.7ex
\hangindent=3em \hangafter=1
$S$ = ($ 1$,
$ 1$,
$ \sqrt{2}$,
$ -c_{7}^{3}$,
$ -c_{7}^{3}$,
$ \xi_{7}^{3}$,
$ \xi_{7}^{3}$,
$ c^{3}_{56}
+c^{11}_{56}
$,
$ c^{3}_{56}
+c^{5}_{56}
-c^{9}_{56}
+c^{11}_{56}
$;\ \ 
$ 1$,
$ -\sqrt{2}$,
$ -c_{7}^{3}$,
$ -c_{7}^{3}$,
$ \xi_{7}^{3}$,
$ \xi_{7}^{3}$,
$ -c^{3}_{56}
-c^{11}_{56}
$,
$ -c^{3}_{56}
-c^{5}_{56}
+c^{9}_{56}
-c^{11}_{56}
$;\ \ 
$0$,
$ c^{3}_{56}
+c^{11}_{56}
$,
$ -c^{3}_{56}
-c^{11}_{56}
$,
$ c^{3}_{56}
+c^{5}_{56}
-c^{9}_{56}
+c^{11}_{56}
$,
$ -c^{3}_{56}
-c^{5}_{56}
+c^{9}_{56}
-c^{11}_{56}
$,
$0$,
$0$;\ \ 
$ -\xi_{7}^{3}$,
$ -\xi_{7}^{3}$,
$ 1$,
$ 1$,
$ -c^{3}_{56}
-c^{5}_{56}
+c^{9}_{56}
-c^{11}_{56}
$,
$ \sqrt{2}$;\ \ 
$ -\xi_{7}^{3}$,
$ 1$,
$ 1$,
$ c^{3}_{56}
+c^{5}_{56}
-c^{9}_{56}
+c^{11}_{56}
$,
$ -\sqrt{2}$;\ \ 
$ c_{7}^{3}$,
$ c_{7}^{3}$,
$ \sqrt{2}$,
$ -c^{3}_{56}
-c^{11}_{56}
$;\ \ 
$ c_{7}^{3}$,
$ -\sqrt{2}$,
$ c^{3}_{56}
+c^{11}_{56}
$;\ \ 
$0$,
$0$;\ \ 
$0$)

Factors = $3_{\frac{3}{2},4.}^{16,553}\boxtimes 3_{\frac{8}{7},9.295}^{7,245}$

\vskip 0.7ex
\hangindent=3em \hangafter=1
$\tau_n$ = ($-2.95 + 5.34 i$, $5.08 - 9.17 i$, $-6.35 + 8.96 i$, $13.15 + 8.26 i$, $-3.79 + 13.16 i$, $9.86 - 5.45 i$, $-7.11 + 17.17 i$, $0.$, $-6.63 + 11.99 i$, $-3.4 - 20. i$, $-1.84 + 10.83 i$, $-19.24 + 2.16 i$, $-1.69 + 5.86 i$, $5.43 - 13.14 i$, $5.86 - 1.69 i$, $-17.07 - 21.41 i$, $10.83 - 1.84 i$, $-6.85 + 4.86 i$, $11.99 - 6.63 i$, $-0.96 - 8.57 i$, $17.17 - 7.11 i$, $3.12 + 10.82 i$, $13.16 - 3.79 i$, $0. + 0.01 i$, $8.96 - 6.35 i$, $-7. + 24.31 i$, $5.34 - 2.95 i$, $18.58 + 18.59 i$, $-5.34 - 2.95 i$, $-10.07 + 2.91 i$, $-8.96 - 6.35 i$, $4.89 + 21.41 i$, $-13.16 - 3.79 i$, $4.48 + 1.29 i$, $-17.17 - 7.11 i$, $8.56 + 0.96 i$, $-11.99 - 6.63 i$, $11.75 - 16.55 i$, $-10.83 - 1.84 i$, $0.01 - 0.01 i$, $-5.86 - 1.69 i$, $31.72 - 13.14 i$, $1.69 + 5.86 i$, $2.18 - 19.24 i$, $1.84 + 10.83 i$, $8.29 + 1.4 i$, $6.63 + 11.99 i$, $7.6 - 9.53 i$, $7.11 + 17.17 i$, $-2.26 + 4.08 i$, $3.79 + 13.16 i$, $-8.26 - 13.15 i$, $6.35 + 8.96 i$, $-22.14 + 12.24 i$, $2.95 + 5.34 i$, $-0.01$, $2.95 - 5.34 i$, $-22.14 - 12.24 i$, $6.35 - 8.96 i$, $-8.26 + 13.15 i$, $3.79 - 13.16 i$, $-2.26 - 4.08 i$, $7.11 - 17.17 i$, $7.6 + 9.53 i$, $6.63 - 11.99 i$, $8.29 - 1.4 i$, $1.84 - 10.83 i$, $2.18 + 19.24 i$, $1.69 - 5.86 i$, $31.72 + 13.14 i$, $-5.86 + 1.69 i$, $0.01 + 0.01 i$, $-10.83 + 1.84 i$, $11.75 + 16.55 i$, $-11.99 + 6.63 i$, $8.56 - 0.96 i$, $-17.17 + 7.11 i$, $4.48 - 1.29 i$, $-13.16 + 3.79 i$, $4.89 - 21.41 i$, $-8.96 + 6.35 i$, $-10.07 - 2.91 i$, $-5.34 + 2.95 i$, $18.58 - 18.59 i$, $5.34 + 2.95 i$, $-7. - 24.31 i$, $8.96 + 6.35 i$, $0. - 0.01 i$, $13.16 + 3.79 i$, $3.12 - 10.82 i$, $17.17 + 7.11 i$, $-0.96 + 8.57 i$, $11.99 + 6.63 i$, $-6.85 - 4.86 i$, $10.83 + 1.84 i$, $-17.07 + 21.41 i$, $5.86 + 1.69 i$, $5.43 + 13.14 i$, $-1.69 - 5.86 i$, $-19.24 - 2.16 i$, $-1.84 - 10.83 i$, $-3.4 + 20. i$, $-6.63 - 11.99 i$, $0.$, $-7.11 - 17.17 i$, $9.86 + 5.45 i$, $-3.79 - 13.16 i$, $13.15 - 8.26 i$, $-6.35 - 8.96 i$, $5.08 + 9.17 i$, $-2.95 - 5.34 i$, $37.16$)

\vskip 0.7ex
\hangindent=3em \hangafter=1
\textit{Intrinsic sign problem}

  \vskip 2ex

\noindent49. $9_{\frac{19}{14},37.18}^{112,135}$ \irep{498}:\ \ 
$d_i$ = ($1.0$,
$1.0$,
$1.414$,
$1.801$,
$1.801$,
$2.246$,
$2.246$,
$2.548$,
$3.177$) 

\vskip 0.7ex
\hangindent=3em \hangafter=1
$D^2= 37.183 = 
24+12c^{1}_{7}
+4c^{2}_{7}
$

\vskip 0.7ex
\hangindent=3em \hangafter=1
$T = ( 0,
\frac{1}{2},
\frac{5}{16},
\frac{1}{7},
\frac{9}{14},
\frac{5}{7},
\frac{3}{14},
\frac{51}{112},
\frac{3}{112} )
$,

\vskip 0.7ex
\hangindent=3em \hangafter=1
$S$ = ($ 1$,
$ 1$,
$ \sqrt{2}$,
$ -c_{7}^{3}$,
$ -c_{7}^{3}$,
$ \xi_{7}^{3}$,
$ \xi_{7}^{3}$,
$ c^{3}_{56}
+c^{11}_{56}
$,
$ c^{3}_{56}
+c^{5}_{56}
-c^{9}_{56}
+c^{11}_{56}
$;\ \ 
$ 1$,
$ -\sqrt{2}$,
$ -c_{7}^{3}$,
$ -c_{7}^{3}$,
$ \xi_{7}^{3}$,
$ \xi_{7}^{3}$,
$ -c^{3}_{56}
-c^{11}_{56}
$,
$ -c^{3}_{56}
-c^{5}_{56}
+c^{9}_{56}
-c^{11}_{56}
$;\ \ 
$0$,
$ c^{3}_{56}
+c^{11}_{56}
$,
$ -c^{3}_{56}
-c^{11}_{56}
$,
$ c^{3}_{56}
+c^{5}_{56}
-c^{9}_{56}
+c^{11}_{56}
$,
$ -c^{3}_{56}
-c^{5}_{56}
+c^{9}_{56}
-c^{11}_{56}
$,
$0$,
$0$;\ \ 
$ -\xi_{7}^{3}$,
$ -\xi_{7}^{3}$,
$ 1$,
$ 1$,
$ -c^{3}_{56}
-c^{5}_{56}
+c^{9}_{56}
-c^{11}_{56}
$,
$ \sqrt{2}$;\ \ 
$ -\xi_{7}^{3}$,
$ 1$,
$ 1$,
$ c^{3}_{56}
+c^{5}_{56}
-c^{9}_{56}
+c^{11}_{56}
$,
$ -\sqrt{2}$;\ \ 
$ c_{7}^{3}$,
$ c_{7}^{3}$,
$ \sqrt{2}$,
$ -c^{3}_{56}
-c^{11}_{56}
$;\ \ 
$ c_{7}^{3}$,
$ -\sqrt{2}$,
$ c^{3}_{56}
+c^{11}_{56}
$;\ \ 
$0$,
$0$;\ \ 
$0$)

Factors = $3_{\frac{5}{2},4.}^{16,465}\boxtimes 3_{\frac{48}{7},9.295}^{7,790}$

\vskip 0.7ex
\hangindent=3em \hangafter=1
$\tau_n$ = ($2.95 + 5.34 i$, $5.08 + 9.17 i$, $6.35 + 8.96 i$, $13.15 - 8.26 i$, $3.79 + 13.16 i$, $9.86 + 5.45 i$, $7.11 + 17.17 i$, $0.$, $6.63 + 11.99 i$, $-3.4 + 20. i$, $1.84 + 10.83 i$, $-19.24 - 2.16 i$, $1.69 + 5.86 i$, $5.43 + 13.14 i$, $-5.86 - 1.69 i$, $-17.07 + 21.41 i$, $-10.83 - 1.84 i$, $-6.85 - 4.86 i$, $-11.99 - 6.63 i$, $-0.96 + 8.57 i$, $-17.17 - 7.11 i$, $3.12 - 10.82 i$, $-13.16 - 3.79 i$, $0. - 0.01 i$, $-8.96 - 6.35 i$, $-7. - 24.31 i$, $-5.34 - 2.95 i$, $18.58 - 18.59 i$, $5.34 - 2.95 i$, $-10.07 - 2.91 i$, $8.96 - 6.35 i$, $4.89 - 21.41 i$, $13.16 - 3.79 i$, $4.48 - 1.29 i$, $17.17 - 7.11 i$, $8.56 - 0.96 i$, $11.99 - 6.63 i$, $11.75 + 16.55 i$, $10.83 - 1.84 i$, $0.01 + 0.01 i$, $5.86 - 1.69 i$, $31.72 + 13.14 i$, $-1.69 + 5.86 i$, $2.18 + 19.24 i$, $-1.84 + 10.83 i$, $8.29 - 1.4 i$, $-6.63 + 11.99 i$, $7.6 + 9.53 i$, $-7.11 + 17.17 i$, $-2.26 - 4.08 i$, $-3.79 + 13.16 i$, $-8.26 + 13.15 i$, $-6.35 + 8.96 i$, $-22.14 - 12.24 i$, $-2.95 + 5.34 i$, $-0.01$, $-2.95 - 5.34 i$, $-22.14 + 12.24 i$, $-6.35 - 8.96 i$, $-8.26 - 13.15 i$, $-3.79 - 13.16 i$, $-2.26 + 4.08 i$, $-7.11 - 17.17 i$, $7.6 - 9.53 i$, $-6.63 - 11.99 i$, $8.29 + 1.4 i$, $-1.84 - 10.83 i$, $2.18 - 19.24 i$, $-1.69 - 5.86 i$, $31.72 - 13.14 i$, $5.86 + 1.69 i$, $0.01 - 0.01 i$, $10.83 + 1.84 i$, $11.75 - 16.55 i$, $11.99 + 6.63 i$, $8.56 + 0.96 i$, $17.17 + 7.11 i$, $4.48 + 1.29 i$, $13.16 + 3.79 i$, $4.89 + 21.41 i$, $8.96 + 6.35 i$, $-10.07 + 2.91 i$, $5.34 + 2.95 i$, $18.58 + 18.59 i$, $-5.34 + 2.95 i$, $-7. + 24.31 i$, $-8.96 + 6.35 i$, $0. + 0.01 i$, $-13.16 + 3.79 i$, $3.12 + 10.82 i$, $-17.17 + 7.11 i$, $-0.96 - 8.57 i$, $-11.99 + 6.63 i$, $-6.85 + 4.86 i$, $-10.83 + 1.84 i$, $-17.07 - 21.41 i$, $-5.86 + 1.69 i$, $5.43 - 13.14 i$, $1.69 - 5.86 i$, $-19.24 + 2.16 i$, $1.84 - 10.83 i$, $-3.4 - 20. i$, $6.63 - 11.99 i$, $0.$, $7.11 - 17.17 i$, $9.86 - 5.45 i$, $3.79 - 13.16 i$, $13.15 + 8.26 i$, $6.35 - 8.96 i$, $5.08 - 9.17 i$, $2.95 - 5.34 i$, $37.16$)

\vskip 0.7ex
\hangindent=3em \hangafter=1
\textit{Intrinsic sign problem}

  \vskip 2ex

\noindent50. $9_{\frac{51}{14},37.18}^{112,413}$ \irep{498}:\ \ 
$d_i$ = ($1.0$,
$1.0$,
$1.414$,
$1.801$,
$1.801$,
$2.246$,
$2.246$,
$2.548$,
$3.177$) 

\vskip 0.7ex
\hangindent=3em \hangafter=1
$D^2= 37.183 = 
24+12c^{1}_{7}
+4c^{2}_{7}
$

\vskip 0.7ex
\hangindent=3em \hangafter=1
$T = ( 0,
\frac{1}{2},
\frac{5}{16},
\frac{6}{7},
\frac{5}{14},
\frac{2}{7},
\frac{11}{14},
\frac{19}{112},
\frac{67}{112} )
$,

\vskip 0.7ex
\hangindent=3em \hangafter=1
$S$ = ($ 1$,
$ 1$,
$ \sqrt{2}$,
$ -c_{7}^{3}$,
$ -c_{7}^{3}$,
$ \xi_{7}^{3}$,
$ \xi_{7}^{3}$,
$ c^{3}_{56}
+c^{11}_{56}
$,
$ c^{3}_{56}
+c^{5}_{56}
-c^{9}_{56}
+c^{11}_{56}
$;\ \ 
$ 1$,
$ -\sqrt{2}$,
$ -c_{7}^{3}$,
$ -c_{7}^{3}$,
$ \xi_{7}^{3}$,
$ \xi_{7}^{3}$,
$ -c^{3}_{56}
-c^{11}_{56}
$,
$ -c^{3}_{56}
-c^{5}_{56}
+c^{9}_{56}
-c^{11}_{56}
$;\ \ 
$0$,
$ c^{3}_{56}
+c^{11}_{56}
$,
$ -c^{3}_{56}
-c^{11}_{56}
$,
$ c^{3}_{56}
+c^{5}_{56}
-c^{9}_{56}
+c^{11}_{56}
$,
$ -c^{3}_{56}
-c^{5}_{56}
+c^{9}_{56}
-c^{11}_{56}
$,
$0$,
$0$;\ \ 
$ -\xi_{7}^{3}$,
$ -\xi_{7}^{3}$,
$ 1$,
$ 1$,
$ -c^{3}_{56}
-c^{5}_{56}
+c^{9}_{56}
-c^{11}_{56}
$,
$ \sqrt{2}$;\ \ 
$ -\xi_{7}^{3}$,
$ 1$,
$ 1$,
$ c^{3}_{56}
+c^{5}_{56}
-c^{9}_{56}
+c^{11}_{56}
$,
$ -\sqrt{2}$;\ \ 
$ c_{7}^{3}$,
$ c_{7}^{3}$,
$ \sqrt{2}$,
$ -c^{3}_{56}
-c^{11}_{56}
$;\ \ 
$ c_{7}^{3}$,
$ -\sqrt{2}$,
$ c^{3}_{56}
+c^{11}_{56}
$;\ \ 
$0$,
$0$;\ \ 
$0$)

Factors = $3_{\frac{5}{2},4.}^{16,465}\boxtimes 3_{\frac{8}{7},9.295}^{7,245}$

\vskip 0.7ex
\hangindent=3em \hangafter=1
$\tau_n$ = ($-5.86 + 1.69 i$, $-10.07 + 2.91 i$, $-1.84 - 10.83 i$, $-8.26 + 13.15 i$, $11.99 - 6.63 i$, $3.12 - 10.82 i$, $7.11 + 17.17 i$, $0.$, $-13.16 + 3.79 i$, $11.75 - 16.55 i$, $-6.35 - 8.96 i$, $2.18 + 19.24 i$, $5.34 - 2.95 i$, $5.43 + 13.14 i$, $2.95 - 5.34 i$, $-17.07 - 21.41 i$, $8.96 + 6.35 i$, $8.29 + 1.4 i$, $-3.79 + 13.16 i$, $8.56 - 0.96 i$, $-17.17 - 7.11 i$, $9.86 + 5.45 i$, $6.63 - 11.99 i$, $0. + 0.01 i$, $10.83 + 1.84 i$, $-22.14 + 12.24 i$, $-1.69 + 5.86 i$, $18.58 - 18.59 i$, $1.69 + 5.86 i$, $5.08 - 9.17 i$, $-10.83 + 1.84 i$, $4.89 + 21.41 i$, $-6.63 - 11.99 i$, $-2.26 - 4.08 i$, $17.17 - 7.11 i$, $-0.96 + 8.57 i$, $3.79 + 13.16 i$, $-3.4 - 20. i$, $-8.96 + 6.35 i$, $0.01 - 0.01 i$, $-2.95 - 5.34 i$, $31.72 + 13.14 i$, $-5.34 - 2.95 i$, $-19.24 - 2.16 i$, $6.35 - 8.96 i$, $-6.85 + 4.86 i$, $13.16 + 3.79 i$, $7.6 - 9.53 i$, $-7.11 + 17.17 i$, $4.48 - 1.29 i$, $-11.99 - 6.63 i$, $13.15 - 8.26 i$, $1.84 - 10.83 i$, $-7. + 24.31 i$, $5.86 + 1.69 i$, $-0.01$, $5.86 - 1.69 i$, $-7. - 24.31 i$, $1.84 + 10.83 i$, $13.15 + 8.26 i$, $-11.99 + 6.63 i$, $4.48 + 1.29 i$, $-7.11 - 17.17 i$, $7.6 + 9.53 i$, $13.16 - 3.79 i$, $-6.85 - 4.86 i$, $6.35 + 8.96 i$, $-19.24 + 2.16 i$, $-5.34 + 2.95 i$, $31.72 - 13.14 i$, $-2.95 + 5.34 i$, $0.01 + 0.01 i$, $-8.96 - 6.35 i$, $-3.4 + 20. i$, $3.79 - 13.16 i$, $-0.96 - 8.57 i$, $17.17 + 7.11 i$, $-2.26 + 4.08 i$, $-6.63 + 11.99 i$, $4.89 - 21.41 i$, $-10.83 - 1.84 i$, $5.08 + 9.17 i$, $1.69 - 5.86 i$, $18.58 + 18.59 i$, $-1.69 - 5.86 i$, $-22.14 - 12.24 i$, $10.83 - 1.84 i$, $0. - 0.01 i$, $6.63 + 11.99 i$, $9.86 - 5.45 i$, $-17.17 + 7.11 i$, $8.56 + 0.96 i$, $-3.79 - 13.16 i$, $8.29 - 1.4 i$, $8.96 - 6.35 i$, $-17.07 + 21.41 i$, $2.95 + 5.34 i$, $5.43 - 13.14 i$, $5.34 + 2.95 i$, $2.18 - 19.24 i$, $-6.35 + 8.96 i$, $11.75 + 16.55 i$, $-13.16 - 3.79 i$, $0.$, $7.11 - 17.17 i$, $3.12 + 10.82 i$, $11.99 + 6.63 i$, $-8.26 - 13.15 i$, $-1.84 + 10.83 i$, $-10.07 - 2.91 i$, $-5.86 - 1.69 i$, $37.16$)

\vskip 0.7ex
\hangindent=3em \hangafter=1
\textit{Intrinsic sign problem}

  \vskip 2ex

\noindent51. $9_{\frac{33}{14},37.18}^{112,826}$ \irep{498}:\ \ 
$d_i$ = ($1.0$,
$1.0$,
$1.414$,
$1.801$,
$1.801$,
$2.246$,
$2.246$,
$2.548$,
$3.177$) 

\vskip 0.7ex
\hangindent=3em \hangafter=1
$D^2= 37.183 = 
24+12c^{1}_{7}
+4c^{2}_{7}
$

\vskip 0.7ex
\hangindent=3em \hangafter=1
$T = ( 0,
\frac{1}{2},
\frac{7}{16},
\frac{1}{7},
\frac{9}{14},
\frac{5}{7},
\frac{3}{14},
\frac{65}{112},
\frac{17}{112} )
$,

\vskip 0.7ex
\hangindent=3em \hangafter=1
$S$ = ($ 1$,
$ 1$,
$ \sqrt{2}$,
$ -c_{7}^{3}$,
$ -c_{7}^{3}$,
$ \xi_{7}^{3}$,
$ \xi_{7}^{3}$,
$ c^{3}_{56}
+c^{11}_{56}
$,
$ c^{3}_{56}
+c^{5}_{56}
-c^{9}_{56}
+c^{11}_{56}
$;\ \ 
$ 1$,
$ -\sqrt{2}$,
$ -c_{7}^{3}$,
$ -c_{7}^{3}$,
$ \xi_{7}^{3}$,
$ \xi_{7}^{3}$,
$ -c^{3}_{56}
-c^{11}_{56}
$,
$ -c^{3}_{56}
-c^{5}_{56}
+c^{9}_{56}
-c^{11}_{56}
$;\ \ 
$0$,
$ c^{3}_{56}
+c^{11}_{56}
$,
$ -c^{3}_{56}
-c^{11}_{56}
$,
$ c^{3}_{56}
+c^{5}_{56}
-c^{9}_{56}
+c^{11}_{56}
$,
$ -c^{3}_{56}
-c^{5}_{56}
+c^{9}_{56}
-c^{11}_{56}
$,
$0$,
$0$;\ \ 
$ -\xi_{7}^{3}$,
$ -\xi_{7}^{3}$,
$ 1$,
$ 1$,
$ -c^{3}_{56}
-c^{5}_{56}
+c^{9}_{56}
-c^{11}_{56}
$,
$ \sqrt{2}$;\ \ 
$ -\xi_{7}^{3}$,
$ 1$,
$ 1$,
$ c^{3}_{56}
+c^{5}_{56}
-c^{9}_{56}
+c^{11}_{56}
$,
$ -\sqrt{2}$;\ \ 
$ c_{7}^{3}$,
$ c_{7}^{3}$,
$ \sqrt{2}$,
$ -c^{3}_{56}
-c^{11}_{56}
$;\ \ 
$ c_{7}^{3}$,
$ -\sqrt{2}$,
$ c^{3}_{56}
+c^{11}_{56}
$;\ \ 
$0$,
$0$;\ \ 
$0$)

Factors = $3_{\frac{7}{2},4.}^{16,332}\boxtimes 3_{\frac{48}{7},9.295}^{7,790}$

\vskip 0.7ex
\hangindent=3em \hangafter=1
$\tau_n$ = ($-1.69 + 5.86 i$, $-7. + 24.31 i$, $-10.83 - 1.84 i$, $-8.26 - 13.15 i$, $6.63 - 11.99 i$, $4.48 - 1.29 i$, $17.17 + 7.11 i$, $0.$, $-3.79 + 13.16 i$, $-6.85 + 4.86 i$, $-8.96 - 6.35 i$, $2.18 - 19.24 i$, $2.95 - 5.34 i$, $31.72 + 13.14 i$, $-5.34 + 2.95 i$, $-17.07 + 21.41 i$, $-6.35 - 8.96 i$, $-3.4 - 20. i$, $13.16 - 3.79 i$, $8.56 + 0.96 i$, $7.11 + 17.17 i$, $-2.26 - 4.08 i$, $-11.99 + 6.63 i$, $0. - 0.01 i$, $-1.84 - 10.83 i$, $5.08 - 9.17 i$, $5.86 - 1.69 i$, $18.58 + 18.59 i$, $-5.86 - 1.69 i$, $-22.14 + 12.24 i$, $1.84 - 10.83 i$, $4.89 - 21.41 i$, $11.99 + 6.63 i$, $9.86 + 5.45 i$, $-7.11 + 17.17 i$, $-0.96 - 8.57 i$, $-13.16 - 3.79 i$, $8.29 + 1.4 i$, $6.35 - 8.96 i$, $0.01 + 0.01 i$, $5.34 + 2.95 i$, $5.43 + 13.14 i$, $-2.95 - 5.34 i$, $-19.24 + 2.16 i$, $8.96 - 6.35 i$, $11.75 - 16.55 i$, $3.79 + 13.16 i$, $7.6 + 9.53 i$, $-17.17 + 7.11 i$, $3.12 - 10.82 i$, $-6.63 - 11.99 i$, $13.15 + 8.26 i$, $10.83 - 1.84 i$, $-10.07 + 2.91 i$, $1.69 + 5.86 i$, $-0.01$, $1.69 - 5.86 i$, $-10.07 - 2.91 i$, $10.83 + 1.84 i$, $13.15 - 8.26 i$, $-6.63 + 11.99 i$, $3.12 + 10.82 i$, $-17.17 - 7.11 i$, $7.6 - 9.53 i$, $3.79 - 13.16 i$, $11.75 + 16.55 i$, $8.96 + 6.35 i$, $-19.24 - 2.16 i$, $-2.95 + 5.34 i$, $5.43 - 13.14 i$, $5.34 - 2.95 i$, $0.01 - 0.01 i$, $6.35 + 8.96 i$, $8.29 - 1.4 i$, $-13.16 + 3.79 i$, $-0.96 + 8.57 i$, $-7.11 - 17.17 i$, $9.86 - 5.45 i$, $11.99 - 6.63 i$, $4.89 + 21.41 i$, $1.84 + 10.83 i$, $-22.14 - 12.24 i$, $-5.86 + 1.69 i$, $18.58 - 18.59 i$, $5.86 + 1.69 i$, $5.08 + 9.17 i$, $-1.84 + 10.83 i$, $0. + 0.01 i$, $-11.99 - 6.63 i$, $-2.26 + 4.08 i$, $7.11 - 17.17 i$, $8.56 - 0.96 i$, $13.16 + 3.79 i$, $-3.4 + 20. i$, $-6.35 + 8.96 i$, $-17.07 - 21.41 i$, $-5.34 - 2.95 i$, $31.72 - 13.14 i$, $2.95 + 5.34 i$, $2.18 + 19.24 i$, $-8.96 + 6.35 i$, $-6.85 - 4.86 i$, $-3.79 - 13.16 i$, $0.$, $17.17 - 7.11 i$, $4.48 + 1.29 i$, $6.63 + 11.99 i$, $-8.26 + 13.15 i$, $-10.83 + 1.84 i$, $-7. - 24.31 i$, $-1.69 - 5.86 i$, $37.16$)

\vskip 0.7ex
\hangindent=3em \hangafter=1
\textit{Intrinsic sign problem}

  \vskip 2ex

\noindent52. $9_{\frac{65}{14},37.18}^{112,121}$ \irep{498}:\ \ 
$d_i$ = ($1.0$,
$1.0$,
$1.414$,
$1.801$,
$1.801$,
$2.246$,
$2.246$,
$2.548$,
$3.177$) 

\vskip 0.7ex
\hangindent=3em \hangafter=1
$D^2= 37.183 = 
24+12c^{1}_{7}
+4c^{2}_{7}
$

\vskip 0.7ex
\hangindent=3em \hangafter=1
$T = ( 0,
\frac{1}{2},
\frac{7}{16},
\frac{6}{7},
\frac{5}{14},
\frac{2}{7},
\frac{11}{14},
\frac{33}{112},
\frac{81}{112} )
$,

\vskip 0.7ex
\hangindent=3em \hangafter=1
$S$ = ($ 1$,
$ 1$,
$ \sqrt{2}$,
$ -c_{7}^{3}$,
$ -c_{7}^{3}$,
$ \xi_{7}^{3}$,
$ \xi_{7}^{3}$,
$ c^{3}_{56}
+c^{11}_{56}
$,
$ c^{3}_{56}
+c^{5}_{56}
-c^{9}_{56}
+c^{11}_{56}
$;\ \ 
$ 1$,
$ -\sqrt{2}$,
$ -c_{7}^{3}$,
$ -c_{7}^{3}$,
$ \xi_{7}^{3}$,
$ \xi_{7}^{3}$,
$ -c^{3}_{56}
-c^{11}_{56}
$,
$ -c^{3}_{56}
-c^{5}_{56}
+c^{9}_{56}
-c^{11}_{56}
$;\ \ 
$0$,
$ c^{3}_{56}
+c^{11}_{56}
$,
$ -c^{3}_{56}
-c^{11}_{56}
$,
$ c^{3}_{56}
+c^{5}_{56}
-c^{9}_{56}
+c^{11}_{56}
$,
$ -c^{3}_{56}
-c^{5}_{56}
+c^{9}_{56}
-c^{11}_{56}
$,
$0$,
$0$;\ \ 
$ -\xi_{7}^{3}$,
$ -\xi_{7}^{3}$,
$ 1$,
$ 1$,
$ -c^{3}_{56}
-c^{5}_{56}
+c^{9}_{56}
-c^{11}_{56}
$,
$ \sqrt{2}$;\ \ 
$ -\xi_{7}^{3}$,
$ 1$,
$ 1$,
$ c^{3}_{56}
+c^{5}_{56}
-c^{9}_{56}
+c^{11}_{56}
$,
$ -\sqrt{2}$;\ \ 
$ c_{7}^{3}$,
$ c_{7}^{3}$,
$ \sqrt{2}$,
$ -c^{3}_{56}
-c^{11}_{56}
$;\ \ 
$ c_{7}^{3}$,
$ -\sqrt{2}$,
$ c^{3}_{56}
+c^{11}_{56}
$;\ \ 
$0$,
$0$;\ \ 
$0$)

Factors = $3_{\frac{7}{2},4.}^{16,332}\boxtimes 3_{\frac{8}{7},9.295}^{7,245}$

\vskip 0.7ex
\hangindent=3em \hangafter=1
$\tau_n$ = ($-5.34 - 2.95 i$, $-22.14 - 12.24 i$, $8.96 + 6.35 i$, $13.15 + 8.26 i$, $-13.16 - 3.79 i$, $-2.26 - 4.08 i$, $17.17 + 7.11 i$, $0.$, $-11.99 - 6.63 i$, $8.29 - 1.4 i$, $10.83 + 1.84 i$, $-19.24 + 2.16 i$, $-5.86 - 1.69 i$, $31.72 + 13.14 i$, $-1.69 - 5.86 i$, $-17.07 - 21.41 i$, $1.84 + 10.83 i$, $11.75 + 16.55 i$, $-6.63 - 11.99 i$, $-0.96 - 8.57 i$, $7.11 + 17.17 i$, $4.48 - 1.29 i$, $-3.79 - 13.16 i$, $0. + 0.01 i$, $6.35 + 8.96 i$, $-10.07 - 2.91 i$, $-2.95 - 5.34 i$, $18.58 + 18.59 i$, $2.95 - 5.34 i$, $-7. - 24.31 i$, $-6.35 + 8.96 i$, $4.89 + 21.41 i$, $3.79 - 13.16 i$, $3.12 - 10.82 i$, $-7.11 + 17.17 i$, $8.56 + 0.96 i$, $6.63 - 11.99 i$, $-6.85 - 4.86 i$, $-1.84 + 10.83 i$, $0.01 - 0.01 i$, $1.69 - 5.86 i$, $5.43 + 13.14 i$, $5.86 - 1.69 i$, $2.18 - 19.24 i$, $-10.83 + 1.84 i$, $-3.4 + 20. i$, $11.99 - 6.63 i$, $7.6 - 9.53 i$, $-17.17 + 7.11 i$, $9.86 + 5.45 i$, $13.16 - 3.79 i$, $-8.26 - 13.15 i$, $-8.96 + 6.35 i$, $5.08 + 9.17 i$, $5.34 - 2.95 i$, $-0.01$, $5.34 + 2.95 i$, $5.08 - 9.17 i$, $-8.96 - 6.35 i$, $-8.26 + 13.15 i$, $13.16 + 3.79 i$, $9.86 - 5.45 i$, $-17.17 - 7.11 i$, $7.6 + 9.53 i$, $11.99 + 6.63 i$, $-3.4 - 20. i$, $-10.83 - 1.84 i$, $2.18 + 19.24 i$, $5.86 + 1.69 i$, $5.43 - 13.14 i$, $1.69 + 5.86 i$, $0.01 + 0.01 i$, $-1.84 - 10.83 i$, $-6.85 + 4.86 i$, $6.63 + 11.99 i$, $8.56 - 0.96 i$, $-7.11 - 17.17 i$, $3.12 + 10.82 i$, $3.79 + 13.16 i$, $4.89 - 21.41 i$, $-6.35 - 8.96 i$, $-7. + 24.31 i$, $2.95 + 5.34 i$, $18.58 - 18.59 i$, $-2.95 + 5.34 i$, $-10.07 + 2.91 i$, $6.35 - 8.96 i$, $0. - 0.01 i$, $-3.79 + 13.16 i$, $4.48 + 1.29 i$, $7.11 - 17.17 i$, $-0.96 + 8.57 i$, $-6.63 + 11.99 i$, $11.75 - 16.55 i$, $1.84 - 10.83 i$, $-17.07 + 21.41 i$, $-1.69 + 5.86 i$, $31.72 - 13.14 i$, $-5.86 + 1.69 i$, $-19.24 - 2.16 i$, $10.83 - 1.84 i$, $8.29 + 1.4 i$, $-11.99 + 6.63 i$, $0.$, $17.17 - 7.11 i$, $-2.26 + 4.08 i$, $-13.16 + 3.79 i$, $13.15 - 8.26 i$, $8.96 - 6.35 i$, $-22.14 + 12.24 i$, $-5.34 + 2.95 i$, $37.16$)

\vskip 0.7ex
\hangindent=3em \hangafter=1
\textit{Intrinsic sign problem}

  \vskip 2ex

\noindent53. $9_{\frac{47}{14},37.18}^{112,696}$ \irep{498}:\ \ 
$d_i$ = ($1.0$,
$1.0$,
$1.414$,
$1.801$,
$1.801$,
$2.246$,
$2.246$,
$2.548$,
$3.177$) 

\vskip 0.7ex
\hangindent=3em \hangafter=1
$D^2= 37.183 = 
24+12c^{1}_{7}
+4c^{2}_{7}
$

\vskip 0.7ex
\hangindent=3em \hangafter=1
$T = ( 0,
\frac{1}{2},
\frac{9}{16},
\frac{1}{7},
\frac{9}{14},
\frac{5}{7},
\frac{3}{14},
\frac{79}{112},
\frac{31}{112} )
$,

\vskip 0.7ex
\hangindent=3em \hangafter=1
$S$ = ($ 1$,
$ 1$,
$ \sqrt{2}$,
$ -c_{7}^{3}$,
$ -c_{7}^{3}$,
$ \xi_{7}^{3}$,
$ \xi_{7}^{3}$,
$ c^{3}_{56}
+c^{11}_{56}
$,
$ c^{3}_{56}
+c^{5}_{56}
-c^{9}_{56}
+c^{11}_{56}
$;\ \ 
$ 1$,
$ -\sqrt{2}$,
$ -c_{7}^{3}$,
$ -c_{7}^{3}$,
$ \xi_{7}^{3}$,
$ \xi_{7}^{3}$,
$ -c^{3}_{56}
-c^{11}_{56}
$,
$ -c^{3}_{56}
-c^{5}_{56}
+c^{9}_{56}
-c^{11}_{56}
$;\ \ 
$0$,
$ c^{3}_{56}
+c^{11}_{56}
$,
$ -c^{3}_{56}
-c^{11}_{56}
$,
$ c^{3}_{56}
+c^{5}_{56}
-c^{9}_{56}
+c^{11}_{56}
$,
$ -c^{3}_{56}
-c^{5}_{56}
+c^{9}_{56}
-c^{11}_{56}
$,
$0$,
$0$;\ \ 
$ -\xi_{7}^{3}$,
$ -\xi_{7}^{3}$,
$ 1$,
$ 1$,
$ -c^{3}_{56}
-c^{5}_{56}
+c^{9}_{56}
-c^{11}_{56}
$,
$ \sqrt{2}$;\ \ 
$ -\xi_{7}^{3}$,
$ 1$,
$ 1$,
$ c^{3}_{56}
+c^{5}_{56}
-c^{9}_{56}
+c^{11}_{56}
$,
$ -\sqrt{2}$;\ \ 
$ c_{7}^{3}$,
$ c_{7}^{3}$,
$ \sqrt{2}$,
$ -c^{3}_{56}
-c^{11}_{56}
$;\ \ 
$ c_{7}^{3}$,
$ -\sqrt{2}$,
$ c^{3}_{56}
+c^{11}_{56}
$;\ \ 
$0$,
$0$;\ \ 
$0$)

Factors = $3_{\frac{9}{2},4.}^{16,156}\boxtimes 3_{\frac{48}{7},9.295}^{7,790}$

\vskip 0.7ex
\hangindent=3em \hangafter=1
$\tau_n$ = ($-5.34 + 2.95 i$, $-22.14 + 12.24 i$, $8.96 - 6.35 i$, $13.15 - 8.26 i$, $-13.16 + 3.79 i$, $-2.26 + 4.08 i$, $17.17 - 7.11 i$, $0.$, $-11.99 + 6.63 i$, $8.29 + 1.4 i$, $10.83 - 1.84 i$, $-19.24 - 2.16 i$, $-5.86 + 1.69 i$, $31.72 - 13.14 i$, $-1.69 + 5.86 i$, $-17.07 + 21.41 i$, $1.84 - 10.83 i$, $11.75 - 16.55 i$, $-6.63 + 11.99 i$, $-0.96 + 8.57 i$, $7.11 - 17.17 i$, $4.48 + 1.29 i$, $-3.79 + 13.16 i$, $0. - 0.01 i$, $6.35 - 8.96 i$, $-10.07 + 2.91 i$, $-2.95 + 5.34 i$, $18.58 - 18.59 i$, $2.95 + 5.34 i$, $-7. + 24.31 i$, $-6.35 - 8.96 i$, $4.89 - 21.41 i$, $3.79 + 13.16 i$, $3.12 + 10.82 i$, $-7.11 - 17.17 i$, $8.56 - 0.96 i$, $6.63 + 11.99 i$, $-6.85 + 4.86 i$, $-1.84 - 10.83 i$, $0.01 + 0.01 i$, $1.69 + 5.86 i$, $5.43 - 13.14 i$, $5.86 + 1.69 i$, $2.18 + 19.24 i$, $-10.83 - 1.84 i$, $-3.4 - 20. i$, $11.99 + 6.63 i$, $7.6 + 9.53 i$, $-17.17 - 7.11 i$, $9.86 - 5.45 i$, $13.16 + 3.79 i$, $-8.26 + 13.15 i$, $-8.96 - 6.35 i$, $5.08 - 9.17 i$, $5.34 + 2.95 i$, $-0.01$, $5.34 - 2.95 i$, $5.08 + 9.17 i$, $-8.96 + 6.35 i$, $-8.26 - 13.15 i$, $13.16 - 3.79 i$, $9.86 + 5.45 i$, $-17.17 + 7.11 i$, $7.6 - 9.53 i$, $11.99 - 6.63 i$, $-3.4 + 20. i$, $-10.83 + 1.84 i$, $2.18 - 19.24 i$, $5.86 - 1.69 i$, $5.43 + 13.14 i$, $1.69 - 5.86 i$, $0.01 - 0.01 i$, $-1.84 + 10.83 i$, $-6.85 - 4.86 i$, $6.63 - 11.99 i$, $8.56 + 0.96 i$, $-7.11 + 17.17 i$, $3.12 - 10.82 i$, $3.79 - 13.16 i$, $4.89 + 21.41 i$, $-6.35 + 8.96 i$, $-7. - 24.31 i$, $2.95 - 5.34 i$, $18.58 + 18.59 i$, $-2.95 - 5.34 i$, $-10.07 - 2.91 i$, $6.35 + 8.96 i$, $0. + 0.01 i$, $-3.79 - 13.16 i$, $4.48 - 1.29 i$, $7.11 + 17.17 i$, $-0.96 - 8.57 i$, $-6.63 - 11.99 i$, $11.75 + 16.55 i$, $1.84 + 10.83 i$, $-17.07 - 21.41 i$, $-1.69 - 5.86 i$, $31.72 + 13.14 i$, $-5.86 - 1.69 i$, $-19.24 + 2.16 i$, $10.83 + 1.84 i$, $8.29 - 1.4 i$, $-11.99 - 6.63 i$, $0.$, $17.17 + 7.11 i$, $-2.26 - 4.08 i$, $-13.16 - 3.79 i$, $13.15 + 8.26 i$, $8.96 + 6.35 i$, $-22.14 - 12.24 i$, $-5.34 - 2.95 i$, $37.16$)

\vskip 0.7ex
\hangindent=3em \hangafter=1
\textit{Intrinsic sign problem}

  \vskip 2ex

\noindent54. $9_{\frac{79}{14},37.18}^{112,224}$ \irep{498}:\ \ 
$d_i$ = ($1.0$,
$1.0$,
$1.414$,
$1.801$,
$1.801$,
$2.246$,
$2.246$,
$2.548$,
$3.177$) 

\vskip 0.7ex
\hangindent=3em \hangafter=1
$D^2= 37.183 = 
24+12c^{1}_{7}
+4c^{2}_{7}
$

\vskip 0.7ex
\hangindent=3em \hangafter=1
$T = ( 0,
\frac{1}{2},
\frac{9}{16},
\frac{6}{7},
\frac{5}{14},
\frac{2}{7},
\frac{11}{14},
\frac{47}{112},
\frac{95}{112} )
$,

\vskip 0.7ex
\hangindent=3em \hangafter=1
$S$ = ($ 1$,
$ 1$,
$ \sqrt{2}$,
$ -c_{7}^{3}$,
$ -c_{7}^{3}$,
$ \xi_{7}^{3}$,
$ \xi_{7}^{3}$,
$ c^{3}_{56}
+c^{11}_{56}
$,
$ c^{3}_{56}
+c^{5}_{56}
-c^{9}_{56}
+c^{11}_{56}
$;\ \ 
$ 1$,
$ -\sqrt{2}$,
$ -c_{7}^{3}$,
$ -c_{7}^{3}$,
$ \xi_{7}^{3}$,
$ \xi_{7}^{3}$,
$ -c^{3}_{56}
-c^{11}_{56}
$,
$ -c^{3}_{56}
-c^{5}_{56}
+c^{9}_{56}
-c^{11}_{56}
$;\ \ 
$0$,
$ c^{3}_{56}
+c^{11}_{56}
$,
$ -c^{3}_{56}
-c^{11}_{56}
$,
$ c^{3}_{56}
+c^{5}_{56}
-c^{9}_{56}
+c^{11}_{56}
$,
$ -c^{3}_{56}
-c^{5}_{56}
+c^{9}_{56}
-c^{11}_{56}
$,
$0$,
$0$;\ \ 
$ -\xi_{7}^{3}$,
$ -\xi_{7}^{3}$,
$ 1$,
$ 1$,
$ -c^{3}_{56}
-c^{5}_{56}
+c^{9}_{56}
-c^{11}_{56}
$,
$ \sqrt{2}$;\ \ 
$ -\xi_{7}^{3}$,
$ 1$,
$ 1$,
$ c^{3}_{56}
+c^{5}_{56}
-c^{9}_{56}
+c^{11}_{56}
$,
$ -\sqrt{2}$;\ \ 
$ c_{7}^{3}$,
$ c_{7}^{3}$,
$ \sqrt{2}$,
$ -c^{3}_{56}
-c^{11}_{56}
$;\ \ 
$ c_{7}^{3}$,
$ -\sqrt{2}$,
$ c^{3}_{56}
+c^{11}_{56}
$;\ \ 
$0$,
$0$;\ \ 
$0$)

Factors = $3_{\frac{9}{2},4.}^{16,156}\boxtimes 3_{\frac{8}{7},9.295}^{7,245}$

\vskip 0.7ex
\hangindent=3em \hangafter=1
$\tau_n$ = ($-1.69 - 5.86 i$, $-7. - 24.31 i$, $-10.83 + 1.84 i$, $-8.26 + 13.15 i$, $6.63 + 11.99 i$, $4.48 + 1.29 i$, $17.17 - 7.11 i$, $0.$, $-3.79 - 13.16 i$, $-6.85 - 4.86 i$, $-8.96 + 6.35 i$, $2.18 + 19.24 i$, $2.95 + 5.34 i$, $31.72 - 13.14 i$, $-5.34 - 2.95 i$, $-17.07 - 21.41 i$, $-6.35 + 8.96 i$, $-3.4 + 20. i$, $13.16 + 3.79 i$, $8.56 - 0.96 i$, $7.11 - 17.17 i$, $-2.26 + 4.08 i$, $-11.99 - 6.63 i$, $0. + 0.01 i$, $-1.84 + 10.83 i$, $5.08 + 9.17 i$, $5.86 + 1.69 i$, $18.58 - 18.59 i$, $-5.86 + 1.69 i$, $-22.14 - 12.24 i$, $1.84 + 10.83 i$, $4.89 + 21.41 i$, $11.99 - 6.63 i$, $9.86 - 5.45 i$, $-7.11 - 17.17 i$, $-0.96 + 8.57 i$, $-13.16 + 3.79 i$, $8.29 - 1.4 i$, $6.35 + 8.96 i$, $0.01 - 0.01 i$, $5.34 - 2.95 i$, $5.43 - 13.14 i$, $-2.95 + 5.34 i$, $-19.24 - 2.16 i$, $8.96 + 6.35 i$, $11.75 + 16.55 i$, $3.79 - 13.16 i$, $7.6 - 9.53 i$, $-17.17 - 7.11 i$, $3.12 + 10.82 i$, $-6.63 + 11.99 i$, $13.15 - 8.26 i$, $10.83 + 1.84 i$, $-10.07 - 2.91 i$, $1.69 - 5.86 i$, $-0.01$, $1.69 + 5.86 i$, $-10.07 + 2.91 i$, $10.83 - 1.84 i$, $13.15 + 8.26 i$, $-6.63 - 11.99 i$, $3.12 - 10.82 i$, $-17.17 + 7.11 i$, $7.6 + 9.53 i$, $3.79 + 13.16 i$, $11.75 - 16.55 i$, $8.96 - 6.35 i$, $-19.24 + 2.16 i$, $-2.95 - 5.34 i$, $5.43 + 13.14 i$, $5.34 + 2.95 i$, $0.01 + 0.01 i$, $6.35 - 8.96 i$, $8.29 + 1.4 i$, $-13.16 - 3.79 i$, $-0.96 - 8.57 i$, $-7.11 + 17.17 i$, $9.86 + 5.45 i$, $11.99 + 6.63 i$, $4.89 - 21.41 i$, $1.84 - 10.83 i$, $-22.14 + 12.24 i$, $-5.86 - 1.69 i$, $18.58 + 18.59 i$, $5.86 - 1.69 i$, $5.08 - 9.17 i$, $-1.84 - 10.83 i$, $0. - 0.01 i$, $-11.99 + 6.63 i$, $-2.26 - 4.08 i$, $7.11 + 17.17 i$, $8.56 + 0.96 i$, $13.16 - 3.79 i$, $-3.4 - 20. i$, $-6.35 - 8.96 i$, $-17.07 + 21.41 i$, $-5.34 + 2.95 i$, $31.72 + 13.14 i$, $2.95 - 5.34 i$, $2.18 - 19.24 i$, $-8.96 - 6.35 i$, $-6.85 + 4.86 i$, $-3.79 + 13.16 i$, $0.$, $17.17 + 7.11 i$, $4.48 - 1.29 i$, $6.63 - 11.99 i$, $-8.26 - 13.15 i$, $-10.83 - 1.84 i$, $-7. + 24.31 i$, $-1.69 + 5.86 i$, $37.16$)

\vskip 0.7ex
\hangindent=3em \hangafter=1
\textit{Intrinsic sign problem}

  \vskip 2ex

\noindent55. $9_{\frac{61}{14},37.18}^{112,909}$ \irep{498}:\ \ 
$d_i$ = ($1.0$,
$1.0$,
$1.414$,
$1.801$,
$1.801$,
$2.246$,
$2.246$,
$2.548$,
$3.177$) 

\vskip 0.7ex
\hangindent=3em \hangafter=1
$D^2= 37.183 = 
24+12c^{1}_{7}
+4c^{2}_{7}
$

\vskip 0.7ex
\hangindent=3em \hangafter=1
$T = ( 0,
\frac{1}{2},
\frac{11}{16},
\frac{1}{7},
\frac{9}{14},
\frac{5}{7},
\frac{3}{14},
\frac{93}{112},
\frac{45}{112} )
$,

\vskip 0.7ex
\hangindent=3em \hangafter=1
$S$ = ($ 1$,
$ 1$,
$ \sqrt{2}$,
$ -c_{7}^{3}$,
$ -c_{7}^{3}$,
$ \xi_{7}^{3}$,
$ \xi_{7}^{3}$,
$ c^{3}_{56}
+c^{11}_{56}
$,
$ c^{3}_{56}
+c^{5}_{56}
-c^{9}_{56}
+c^{11}_{56}
$;\ \ 
$ 1$,
$ -\sqrt{2}$,
$ -c_{7}^{3}$,
$ -c_{7}^{3}$,
$ \xi_{7}^{3}$,
$ \xi_{7}^{3}$,
$ -c^{3}_{56}
-c^{11}_{56}
$,
$ -c^{3}_{56}
-c^{5}_{56}
+c^{9}_{56}
-c^{11}_{56}
$;\ \ 
$0$,
$ c^{3}_{56}
+c^{11}_{56}
$,
$ -c^{3}_{56}
-c^{11}_{56}
$,
$ c^{3}_{56}
+c^{5}_{56}
-c^{9}_{56}
+c^{11}_{56}
$,
$ -c^{3}_{56}
-c^{5}_{56}
+c^{9}_{56}
-c^{11}_{56}
$,
$0$,
$0$;\ \ 
$ -\xi_{7}^{3}$,
$ -\xi_{7}^{3}$,
$ 1$,
$ 1$,
$ -c^{3}_{56}
-c^{5}_{56}
+c^{9}_{56}
-c^{11}_{56}
$,
$ \sqrt{2}$;\ \ 
$ -\xi_{7}^{3}$,
$ 1$,
$ 1$,
$ c^{3}_{56}
+c^{5}_{56}
-c^{9}_{56}
+c^{11}_{56}
$,
$ -\sqrt{2}$;\ \ 
$ c_{7}^{3}$,
$ c_{7}^{3}$,
$ \sqrt{2}$,
$ -c^{3}_{56}
-c^{11}_{56}
$;\ \ 
$ c_{7}^{3}$,
$ -\sqrt{2}$,
$ c^{3}_{56}
+c^{11}_{56}
$;\ \ 
$0$,
$0$;\ \ 
$0$)

Factors = $3_{\frac{11}{2},4.}^{16,648}\boxtimes 3_{\frac{48}{7},9.295}^{7,790}$

\vskip 0.7ex
\hangindent=3em \hangafter=1
$\tau_n$ = ($-5.86 - 1.69 i$, $-10.07 - 2.91 i$, $-1.84 + 10.83 i$, $-8.26 - 13.15 i$, $11.99 + 6.63 i$, $3.12 + 10.82 i$, $7.11 - 17.17 i$, $0.$, $-13.16 - 3.79 i$, $11.75 + 16.55 i$, $-6.35 + 8.96 i$, $2.18 - 19.24 i$, $5.34 + 2.95 i$, $5.43 - 13.14 i$, $2.95 + 5.34 i$, $-17.07 + 21.41 i$, $8.96 - 6.35 i$, $8.29 - 1.4 i$, $-3.79 - 13.16 i$, $8.56 + 0.96 i$, $-17.17 + 7.11 i$, $9.86 - 5.45 i$, $6.63 + 11.99 i$, $0. - 0.01 i$, $10.83 - 1.84 i$, $-22.14 - 12.24 i$, $-1.69 - 5.86 i$, $18.58 + 18.59 i$, $1.69 - 5.86 i$, $5.08 + 9.17 i$, $-10.83 - 1.84 i$, $4.89 - 21.41 i$, $-6.63 + 11.99 i$, $-2.26 + 4.08 i$, $17.17 + 7.11 i$, $-0.96 - 8.57 i$, $3.79 - 13.16 i$, $-3.4 + 20. i$, $-8.96 - 6.35 i$, $0.01 + 0.01 i$, $-2.95 + 5.34 i$, $31.72 - 13.14 i$, $-5.34 + 2.95 i$, $-19.24 + 2.16 i$, $6.35 + 8.96 i$, $-6.85 - 4.86 i$, $13.16 - 3.79 i$, $7.6 + 9.53 i$, $-7.11 - 17.17 i$, $4.48 + 1.29 i$, $-11.99 + 6.63 i$, $13.15 + 8.26 i$, $1.84 + 10.83 i$, $-7. - 24.31 i$, $5.86 - 1.69 i$, $-0.01$, $5.86 + 1.69 i$, $-7. + 24.31 i$, $1.84 - 10.83 i$, $13.15 - 8.26 i$, $-11.99 - 6.63 i$, $4.48 - 1.29 i$, $-7.11 + 17.17 i$, $7.6 - 9.53 i$, $13.16 + 3.79 i$, $-6.85 + 4.86 i$, $6.35 - 8.96 i$, $-19.24 - 2.16 i$, $-5.34 - 2.95 i$, $31.72 + 13.14 i$, $-2.95 - 5.34 i$, $0.01 - 0.01 i$, $-8.96 + 6.35 i$, $-3.4 - 20. i$, $3.79 + 13.16 i$, $-0.96 + 8.57 i$, $17.17 - 7.11 i$, $-2.26 - 4.08 i$, $-6.63 - 11.99 i$, $4.89 + 21.41 i$, $-10.83 + 1.84 i$, $5.08 - 9.17 i$, $1.69 + 5.86 i$, $18.58 - 18.59 i$, $-1.69 + 5.86 i$, $-22.14 + 12.24 i$, $10.83 + 1.84 i$, $0. + 0.01 i$, $6.63 - 11.99 i$, $9.86 + 5.45 i$, $-17.17 - 7.11 i$, $8.56 - 0.96 i$, $-3.79 + 13.16 i$, $8.29 + 1.4 i$, $8.96 + 6.35 i$, $-17.07 - 21.41 i$, $2.95 - 5.34 i$, $5.43 + 13.14 i$, $5.34 - 2.95 i$, $2.18 + 19.24 i$, $-6.35 - 8.96 i$, $11.75 - 16.55 i$, $-13.16 + 3.79 i$, $0.$, $7.11 + 17.17 i$, $3.12 - 10.82 i$, $11.99 - 6.63 i$, $-8.26 + 13.15 i$, $-1.84 - 10.83 i$, $-10.07 + 2.91 i$, $-5.86 + 1.69 i$, $37.16$)

\vskip 0.7ex
\hangindent=3em \hangafter=1
\textit{Intrinsic sign problem}

  \vskip 2ex

\noindent56. $9_{\frac{93}{14},37.18}^{112,349}$ \irep{498}:\ \ 
$d_i$ = ($1.0$,
$1.0$,
$1.414$,
$1.801$,
$1.801$,
$2.246$,
$2.246$,
$2.548$,
$3.177$) 

\vskip 0.7ex
\hangindent=3em \hangafter=1
$D^2= 37.183 = 
24+12c^{1}_{7}
+4c^{2}_{7}
$

\vskip 0.7ex
\hangindent=3em \hangafter=1
$T = ( 0,
\frac{1}{2},
\frac{11}{16},
\frac{6}{7},
\frac{5}{14},
\frac{2}{7},
\frac{11}{14},
\frac{61}{112},
\frac{109}{112} )
$,

\vskip 0.7ex
\hangindent=3em \hangafter=1
$S$ = ($ 1$,
$ 1$,
$ \sqrt{2}$,
$ -c_{7}^{3}$,
$ -c_{7}^{3}$,
$ \xi_{7}^{3}$,
$ \xi_{7}^{3}$,
$ c^{3}_{56}
+c^{11}_{56}
$,
$ c^{3}_{56}
+c^{5}_{56}
-c^{9}_{56}
+c^{11}_{56}
$;\ \ 
$ 1$,
$ -\sqrt{2}$,
$ -c_{7}^{3}$,
$ -c_{7}^{3}$,
$ \xi_{7}^{3}$,
$ \xi_{7}^{3}$,
$ -c^{3}_{56}
-c^{11}_{56}
$,
$ -c^{3}_{56}
-c^{5}_{56}
+c^{9}_{56}
-c^{11}_{56}
$;\ \ 
$0$,
$ c^{3}_{56}
+c^{11}_{56}
$,
$ -c^{3}_{56}
-c^{11}_{56}
$,
$ c^{3}_{56}
+c^{5}_{56}
-c^{9}_{56}
+c^{11}_{56}
$,
$ -c^{3}_{56}
-c^{5}_{56}
+c^{9}_{56}
-c^{11}_{56}
$,
$0$,
$0$;\ \ 
$ -\xi_{7}^{3}$,
$ -\xi_{7}^{3}$,
$ 1$,
$ 1$,
$ -c^{3}_{56}
-c^{5}_{56}
+c^{9}_{56}
-c^{11}_{56}
$,
$ \sqrt{2}$;\ \ 
$ -\xi_{7}^{3}$,
$ 1$,
$ 1$,
$ c^{3}_{56}
+c^{5}_{56}
-c^{9}_{56}
+c^{11}_{56}
$,
$ -\sqrt{2}$;\ \ 
$ c_{7}^{3}$,
$ c_{7}^{3}$,
$ \sqrt{2}$,
$ -c^{3}_{56}
-c^{11}_{56}
$;\ \ 
$ c_{7}^{3}$,
$ -\sqrt{2}$,
$ c^{3}_{56}
+c^{11}_{56}
$;\ \ 
$0$,
$0$;\ \ 
$0$)

Factors = $3_{\frac{11}{2},4.}^{16,648}\boxtimes 3_{\frac{8}{7},9.295}^{7,245}$

\vskip 0.7ex
\hangindent=3em \hangafter=1
$\tau_n$ = ($2.95 - 5.34 i$, $5.08 - 9.17 i$, $6.35 - 8.96 i$, $13.15 + 8.26 i$, $3.79 - 13.16 i$, $9.86 - 5.45 i$, $7.11 - 17.17 i$, $0.$, $6.63 - 11.99 i$, $-3.4 - 20. i$, $1.84 - 10.83 i$, $-19.24 + 2.16 i$, $1.69 - 5.86 i$, $5.43 - 13.14 i$, $-5.86 + 1.69 i$, $-17.07 - 21.41 i$, $-10.83 + 1.84 i$, $-6.85 + 4.86 i$, $-11.99 + 6.63 i$, $-0.96 - 8.57 i$, $-17.17 + 7.11 i$, $3.12 + 10.82 i$, $-13.16 + 3.79 i$, $0. + 0.01 i$, $-8.96 + 6.35 i$, $-7. + 24.31 i$, $-5.34 + 2.95 i$, $18.58 + 18.59 i$, $5.34 + 2.95 i$, $-10.07 + 2.91 i$, $8.96 + 6.35 i$, $4.89 + 21.41 i$, $13.16 + 3.79 i$, $4.48 + 1.29 i$, $17.17 + 7.11 i$, $8.56 + 0.96 i$, $11.99 + 6.63 i$, $11.75 - 16.55 i$, $10.83 + 1.84 i$, $0.01 - 0.01 i$, $5.86 + 1.69 i$, $31.72 - 13.14 i$, $-1.69 - 5.86 i$, $2.18 - 19.24 i$, $-1.84 - 10.83 i$, $8.29 + 1.4 i$, $-6.63 - 11.99 i$, $7.6 - 9.53 i$, $-7.11 - 17.17 i$, $-2.26 + 4.08 i$, $-3.79 - 13.16 i$, $-8.26 - 13.15 i$, $-6.35 - 8.96 i$, $-22.14 + 12.24 i$, $-2.95 - 5.34 i$, $-0.01$, $-2.95 + 5.34 i$, $-22.14 - 12.24 i$, $-6.35 + 8.96 i$, $-8.26 + 13.15 i$, $-3.79 + 13.16 i$, $-2.26 - 4.08 i$, $-7.11 + 17.17 i$, $7.6 + 9.53 i$, $-6.63 + 11.99 i$, $8.29 - 1.4 i$, $-1.84 + 10.83 i$, $2.18 + 19.24 i$, $-1.69 + 5.86 i$, $31.72 + 13.14 i$, $5.86 - 1.69 i$, $0.01 + 0.01 i$, $10.83 - 1.84 i$, $11.75 + 16.55 i$, $11.99 - 6.63 i$, $8.56 - 0.96 i$, $17.17 - 7.11 i$, $4.48 - 1.29 i$, $13.16 - 3.79 i$, $4.89 - 21.41 i$, $8.96 - 6.35 i$, $-10.07 - 2.91 i$, $5.34 - 2.95 i$, $18.58 - 18.59 i$, $-5.34 - 2.95 i$, $-7. - 24.31 i$, $-8.96 - 6.35 i$, $0. - 0.01 i$, $-13.16 - 3.79 i$, $3.12 - 10.82 i$, $-17.17 - 7.11 i$, $-0.96 + 8.57 i$, $-11.99 - 6.63 i$, $-6.85 - 4.86 i$, $-10.83 - 1.84 i$, $-17.07 + 21.41 i$, $-5.86 - 1.69 i$, $5.43 + 13.14 i$, $1.69 + 5.86 i$, $-19.24 - 2.16 i$, $1.84 + 10.83 i$, $-3.4 + 20. i$, $6.63 + 11.99 i$, $0.$, $7.11 + 17.17 i$, $9.86 + 5.45 i$, $3.79 + 13.16 i$, $13.15 - 8.26 i$, $6.35 + 8.96 i$, $5.08 + 9.17 i$, $2.95 + 5.34 i$, $37.16$)

\vskip 0.7ex
\hangindent=3em \hangafter=1
\textit{Intrinsic sign problem}

  \vskip 2ex

\noindent57. $9_{\frac{75}{14},37.18}^{112,211}$ \irep{498}:\ \ 
$d_i$ = ($1.0$,
$1.0$,
$1.414$,
$1.801$,
$1.801$,
$2.246$,
$2.246$,
$2.548$,
$3.177$) 

\vskip 0.7ex
\hangindent=3em \hangafter=1
$D^2= 37.183 = 
24+12c^{1}_{7}
+4c^{2}_{7}
$

\vskip 0.7ex
\hangindent=3em \hangafter=1
$T = ( 0,
\frac{1}{2},
\frac{13}{16},
\frac{1}{7},
\frac{9}{14},
\frac{5}{7},
\frac{3}{14},
\frac{107}{112},
\frac{59}{112} )
$,

\vskip 0.7ex
\hangindent=3em \hangafter=1
$S$ = ($ 1$,
$ 1$,
$ \sqrt{2}$,
$ -c_{7}^{3}$,
$ -c_{7}^{3}$,
$ \xi_{7}^{3}$,
$ \xi_{7}^{3}$,
$ c^{3}_{56}
+c^{11}_{56}
$,
$ c^{3}_{56}
+c^{5}_{56}
-c^{9}_{56}
+c^{11}_{56}
$;\ \ 
$ 1$,
$ -\sqrt{2}$,
$ -c_{7}^{3}$,
$ -c_{7}^{3}$,
$ \xi_{7}^{3}$,
$ \xi_{7}^{3}$,
$ -c^{3}_{56}
-c^{11}_{56}
$,
$ -c^{3}_{56}
-c^{5}_{56}
+c^{9}_{56}
-c^{11}_{56}
$;\ \ 
$0$,
$ c^{3}_{56}
+c^{11}_{56}
$,
$ -c^{3}_{56}
-c^{11}_{56}
$,
$ c^{3}_{56}
+c^{5}_{56}
-c^{9}_{56}
+c^{11}_{56}
$,
$ -c^{3}_{56}
-c^{5}_{56}
+c^{9}_{56}
-c^{11}_{56}
$,
$0$,
$0$;\ \ 
$ -\xi_{7}^{3}$,
$ -\xi_{7}^{3}$,
$ 1$,
$ 1$,
$ -c^{3}_{56}
-c^{5}_{56}
+c^{9}_{56}
-c^{11}_{56}
$,
$ \sqrt{2}$;\ \ 
$ -\xi_{7}^{3}$,
$ 1$,
$ 1$,
$ c^{3}_{56}
+c^{5}_{56}
-c^{9}_{56}
+c^{11}_{56}
$,
$ -\sqrt{2}$;\ \ 
$ c_{7}^{3}$,
$ c_{7}^{3}$,
$ \sqrt{2}$,
$ -c^{3}_{56}
-c^{11}_{56}
$;\ \ 
$ c_{7}^{3}$,
$ -\sqrt{2}$,
$ c^{3}_{56}
+c^{11}_{56}
$;\ \ 
$0$,
$0$;\ \ 
$0$)

Factors = $3_{\frac{13}{2},4.}^{16,330}\boxtimes 3_{\frac{48}{7},9.295}^{7,790}$

\vskip 0.7ex
\hangindent=3em \hangafter=1
$\tau_n$ = ($-2.95 - 5.34 i$, $5.08 + 9.17 i$, $-6.35 - 8.96 i$, $13.15 - 8.26 i$, $-3.79 - 13.16 i$, $9.86 + 5.45 i$, $-7.11 - 17.17 i$, $0.$, $-6.63 - 11.99 i$, $-3.4 + 20. i$, $-1.84 - 10.83 i$, $-19.24 - 2.16 i$, $-1.69 - 5.86 i$, $5.43 + 13.14 i$, $5.86 + 1.69 i$, $-17.07 + 21.41 i$, $10.83 + 1.84 i$, $-6.85 - 4.86 i$, $11.99 + 6.63 i$, $-0.96 + 8.57 i$, $17.17 + 7.11 i$, $3.12 - 10.82 i$, $13.16 + 3.79 i$, $0. - 0.01 i$, $8.96 + 6.35 i$, $-7. - 24.31 i$, $5.34 + 2.95 i$, $18.58 - 18.59 i$, $-5.34 + 2.95 i$, $-10.07 - 2.91 i$, $-8.96 + 6.35 i$, $4.89 - 21.41 i$, $-13.16 + 3.79 i$, $4.48 - 1.29 i$, $-17.17 + 7.11 i$, $8.56 - 0.96 i$, $-11.99 + 6.63 i$, $11.75 + 16.55 i$, $-10.83 + 1.84 i$, $0.01 + 0.01 i$, $-5.86 + 1.69 i$, $31.72 + 13.14 i$, $1.69 - 5.86 i$, $2.18 + 19.24 i$, $1.84 - 10.83 i$, $8.29 - 1.4 i$, $6.63 - 11.99 i$, $7.6 + 9.53 i$, $7.11 - 17.17 i$, $-2.26 - 4.08 i$, $3.79 - 13.16 i$, $-8.26 + 13.15 i$, $6.35 - 8.96 i$, $-22.14 - 12.24 i$, $2.95 - 5.34 i$, $-0.01$, $2.95 + 5.34 i$, $-22.14 + 12.24 i$, $6.35 + 8.96 i$, $-8.26 - 13.15 i$, $3.79 + 13.16 i$, $-2.26 + 4.08 i$, $7.11 + 17.17 i$, $7.6 - 9.53 i$, $6.63 + 11.99 i$, $8.29 + 1.4 i$, $1.84 + 10.83 i$, $2.18 - 19.24 i$, $1.69 + 5.86 i$, $31.72 - 13.14 i$, $-5.86 - 1.69 i$, $0.01 - 0.01 i$, $-10.83 - 1.84 i$, $11.75 - 16.55 i$, $-11.99 - 6.63 i$, $8.56 + 0.96 i$, $-17.17 - 7.11 i$, $4.48 + 1.29 i$, $-13.16 - 3.79 i$, $4.89 + 21.41 i$, $-8.96 - 6.35 i$, $-10.07 + 2.91 i$, $-5.34 - 2.95 i$, $18.58 + 18.59 i$, $5.34 - 2.95 i$, $-7. + 24.31 i$, $8.96 - 6.35 i$, $0. + 0.01 i$, $13.16 - 3.79 i$, $3.12 + 10.82 i$, $17.17 - 7.11 i$, $-0.96 - 8.57 i$, $11.99 - 6.63 i$, $-6.85 + 4.86 i$, $10.83 - 1.84 i$, $-17.07 - 21.41 i$, $5.86 - 1.69 i$, $5.43 - 13.14 i$, $-1.69 + 5.86 i$, $-19.24 + 2.16 i$, $-1.84 + 10.83 i$, $-3.4 - 20. i$, $-6.63 + 11.99 i$, $0.$, $-7.11 + 17.17 i$, $9.86 - 5.45 i$, $-3.79 + 13.16 i$, $13.15 + 8.26 i$, $-6.35 + 8.96 i$, $5.08 - 9.17 i$, $-2.95 + 5.34 i$, $37.16$)

\vskip 0.7ex
\hangindent=3em \hangafter=1
\textit{Intrinsic sign problem}

  \vskip 2ex

\noindent58. $9_{\frac{107}{14},37.18}^{112,117}$ \irep{498}:\ \ 
$d_i$ = ($1.0$,
$1.0$,
$1.414$,
$1.801$,
$1.801$,
$2.246$,
$2.246$,
$2.548$,
$3.177$) 

\vskip 0.7ex
\hangindent=3em \hangafter=1
$D^2= 37.183 = 
24+12c^{1}_{7}
+4c^{2}_{7}
$

\vskip 0.7ex
\hangindent=3em \hangafter=1
$T = ( 0,
\frac{1}{2},
\frac{13}{16},
\frac{6}{7},
\frac{5}{14},
\frac{2}{7},
\frac{11}{14},
\frac{75}{112},
\frac{11}{112} )
$,

\vskip 0.7ex
\hangindent=3em \hangafter=1
$S$ = ($ 1$,
$ 1$,
$ \sqrt{2}$,
$ -c_{7}^{3}$,
$ -c_{7}^{3}$,
$ \xi_{7}^{3}$,
$ \xi_{7}^{3}$,
$ c^{3}_{56}
+c^{11}_{56}
$,
$ c^{3}_{56}
+c^{5}_{56}
-c^{9}_{56}
+c^{11}_{56}
$;\ \ 
$ 1$,
$ -\sqrt{2}$,
$ -c_{7}^{3}$,
$ -c_{7}^{3}$,
$ \xi_{7}^{3}$,
$ \xi_{7}^{3}$,
$ -c^{3}_{56}
-c^{11}_{56}
$,
$ -c^{3}_{56}
-c^{5}_{56}
+c^{9}_{56}
-c^{11}_{56}
$;\ \ 
$0$,
$ c^{3}_{56}
+c^{11}_{56}
$,
$ -c^{3}_{56}
-c^{11}_{56}
$,
$ c^{3}_{56}
+c^{5}_{56}
-c^{9}_{56}
+c^{11}_{56}
$,
$ -c^{3}_{56}
-c^{5}_{56}
+c^{9}_{56}
-c^{11}_{56}
$,
$0$,
$0$;\ \ 
$ -\xi_{7}^{3}$,
$ -\xi_{7}^{3}$,
$ 1$,
$ 1$,
$ -c^{3}_{56}
-c^{5}_{56}
+c^{9}_{56}
-c^{11}_{56}
$,
$ \sqrt{2}$;\ \ 
$ -\xi_{7}^{3}$,
$ 1$,
$ 1$,
$ c^{3}_{56}
+c^{5}_{56}
-c^{9}_{56}
+c^{11}_{56}
$,
$ -\sqrt{2}$;\ \ 
$ c_{7}^{3}$,
$ c_{7}^{3}$,
$ \sqrt{2}$,
$ -c^{3}_{56}
-c^{11}_{56}
$;\ \ 
$ c_{7}^{3}$,
$ -\sqrt{2}$,
$ c^{3}_{56}
+c^{11}_{56}
$;\ \ 
$0$,
$0$;\ \ 
$0$)

Factors = $3_{\frac{13}{2},4.}^{16,330}\boxtimes 3_{\frac{8}{7},9.295}^{7,245}$

\vskip 0.7ex
\hangindent=3em \hangafter=1
$\tau_n$ = ($5.86 - 1.69 i$, $-10.07 + 2.91 i$, $1.84 + 10.83 i$, $-8.26 + 13.15 i$, $-11.99 + 6.63 i$, $3.12 - 10.82 i$, $-7.11 - 17.17 i$, $0.$, $13.16 - 3.79 i$, $11.75 - 16.55 i$, $6.35 + 8.96 i$, $2.18 + 19.24 i$, $-5.34 + 2.95 i$, $5.43 + 13.14 i$, $-2.95 + 5.34 i$, $-17.07 - 21.41 i$, $-8.96 - 6.35 i$, $8.29 + 1.4 i$, $3.79 - 13.16 i$, $8.56 - 0.96 i$, $17.17 + 7.11 i$, $9.86 + 5.45 i$, $-6.63 + 11.99 i$, $0. + 0.01 i$, $-10.83 - 1.84 i$, $-22.14 + 12.24 i$, $1.69 - 5.86 i$, $18.58 - 18.59 i$, $-1.69 - 5.86 i$, $5.08 - 9.17 i$, $10.83 - 1.84 i$, $4.89 + 21.41 i$, $6.63 + 11.99 i$, $-2.26 - 4.08 i$, $-17.17 + 7.11 i$, $-0.96 + 8.57 i$, $-3.79 - 13.16 i$, $-3.4 - 20. i$, $8.96 - 6.35 i$, $0.01 - 0.01 i$, $2.95 + 5.34 i$, $31.72 + 13.14 i$, $5.34 + 2.95 i$, $-19.24 - 2.16 i$, $-6.35 + 8.96 i$, $-6.85 + 4.86 i$, $-13.16 - 3.79 i$, $7.6 - 9.53 i$, $7.11 - 17.17 i$, $4.48 - 1.29 i$, $11.99 + 6.63 i$, $13.15 - 8.26 i$, $-1.84 + 10.83 i$, $-7. + 24.31 i$, $-5.86 - 1.69 i$, $-0.01$, $-5.86 + 1.69 i$, $-7. - 24.31 i$, $-1.84 - 10.83 i$, $13.15 + 8.26 i$, $11.99 - 6.63 i$, $4.48 + 1.29 i$, $7.11 + 17.17 i$, $7.6 + 9.53 i$, $-13.16 + 3.79 i$, $-6.85 - 4.86 i$, $-6.35 - 8.96 i$, $-19.24 + 2.16 i$, $5.34 - 2.95 i$, $31.72 - 13.14 i$, $2.95 - 5.34 i$, $0.01 + 0.01 i$, $8.96 + 6.35 i$, $-3.4 + 20. i$, $-3.79 + 13.16 i$, $-0.96 - 8.57 i$, $-17.17 - 7.11 i$, $-2.26 + 4.08 i$, $6.63 - 11.99 i$, $4.89 - 21.41 i$, $10.83 + 1.84 i$, $5.08 + 9.17 i$, $-1.69 + 5.86 i$, $18.58 + 18.59 i$, $1.69 + 5.86 i$, $-22.14 - 12.24 i$, $-10.83 + 1.84 i$, $0. - 0.01 i$, $-6.63 - 11.99 i$, $9.86 - 5.45 i$, $17.17 - 7.11 i$, $8.56 + 0.96 i$, $3.79 + 13.16 i$, $8.29 - 1.4 i$, $-8.96 + 6.35 i$, $-17.07 + 21.41 i$, $-2.95 - 5.34 i$, $5.43 - 13.14 i$, $-5.34 - 2.95 i$, $2.18 - 19.24 i$, $6.35 - 8.96 i$, $11.75 + 16.55 i$, $13.16 + 3.79 i$, $0.$, $-7.11 + 17.17 i$, $3.12 + 10.82 i$, $-11.99 - 6.63 i$, $-8.26 - 13.15 i$, $1.84 - 10.83 i$, $-10.07 - 2.91 i$, $5.86 + 1.69 i$, $37.16$)

\vskip 0.7ex
\hangindent=3em \hangafter=1
\textit{Intrinsic sign problem}

  \vskip 2ex

\noindent59. $9_{\frac{89}{14},37.18}^{112,580}$ \irep{498}:\ \ 
$d_i$ = ($1.0$,
$1.0$,
$1.414$,
$1.801$,
$1.801$,
$2.246$,
$2.246$,
$2.548$,
$3.177$) 

\vskip 0.7ex
\hangindent=3em \hangafter=1
$D^2= 37.183 = 
24+12c^{1}_{7}
+4c^{2}_{7}
$

\vskip 0.7ex
\hangindent=3em \hangafter=1
$T = ( 0,
\frac{1}{2},
\frac{15}{16},
\frac{1}{7},
\frac{9}{14},
\frac{5}{7},
\frac{3}{14},
\frac{9}{112},
\frac{73}{112} )
$,

\vskip 0.7ex
\hangindent=3em \hangafter=1
$S$ = ($ 1$,
$ 1$,
$ \sqrt{2}$,
$ -c_{7}^{3}$,
$ -c_{7}^{3}$,
$ \xi_{7}^{3}$,
$ \xi_{7}^{3}$,
$ c^{3}_{56}
+c^{11}_{56}
$,
$ c^{3}_{56}
+c^{5}_{56}
-c^{9}_{56}
+c^{11}_{56}
$;\ \ 
$ 1$,
$ -\sqrt{2}$,
$ -c_{7}^{3}$,
$ -c_{7}^{3}$,
$ \xi_{7}^{3}$,
$ \xi_{7}^{3}$,
$ -c^{3}_{56}
-c^{11}_{56}
$,
$ -c^{3}_{56}
-c^{5}_{56}
+c^{9}_{56}
-c^{11}_{56}
$;\ \ 
$0$,
$ c^{3}_{56}
+c^{11}_{56}
$,
$ -c^{3}_{56}
-c^{11}_{56}
$,
$ c^{3}_{56}
+c^{5}_{56}
-c^{9}_{56}
+c^{11}_{56}
$,
$ -c^{3}_{56}
-c^{5}_{56}
+c^{9}_{56}
-c^{11}_{56}
$,
$0$,
$0$;\ \ 
$ -\xi_{7}^{3}$,
$ -\xi_{7}^{3}$,
$ 1$,
$ 1$,
$ -c^{3}_{56}
-c^{5}_{56}
+c^{9}_{56}
-c^{11}_{56}
$,
$ \sqrt{2}$;\ \ 
$ -\xi_{7}^{3}$,
$ 1$,
$ 1$,
$ c^{3}_{56}
+c^{5}_{56}
-c^{9}_{56}
+c^{11}_{56}
$,
$ -\sqrt{2}$;\ \ 
$ c_{7}^{3}$,
$ c_{7}^{3}$,
$ \sqrt{2}$,
$ -c^{3}_{56}
-c^{11}_{56}
$;\ \ 
$ c_{7}^{3}$,
$ -\sqrt{2}$,
$ c^{3}_{56}
+c^{11}_{56}
$;\ \ 
$0$,
$0$;\ \ 
$0$)

Factors = $3_{\frac{15}{2},4.}^{16,639}\boxtimes 3_{\frac{48}{7},9.295}^{7,790}$

\vskip 0.7ex
\hangindent=3em \hangafter=1
$\tau_n$ = ($1.69 - 5.86 i$, $-7. + 24.31 i$, $10.83 + 1.84 i$, $-8.26 - 13.15 i$, $-6.63 + 11.99 i$, $4.48 - 1.29 i$, $-17.17 - 7.11 i$, $0.$, $3.79 - 13.16 i$, $-6.85 + 4.86 i$, $8.96 + 6.35 i$, $2.18 - 19.24 i$, $-2.95 + 5.34 i$, $31.72 + 13.14 i$, $5.34 - 2.95 i$, $-17.07 + 21.41 i$, $6.35 + 8.96 i$, $-3.4 - 20. i$, $-13.16 + 3.79 i$, $8.56 + 0.96 i$, $-7.11 - 17.17 i$, $-2.26 - 4.08 i$, $11.99 - 6.63 i$, $0. - 0.01 i$, $1.84 + 10.83 i$, $5.08 - 9.17 i$, $-5.86 + 1.69 i$, $18.58 + 18.59 i$, $5.86 + 1.69 i$, $-22.14 + 12.24 i$, $-1.84 + 10.83 i$, $4.89 - 21.41 i$, $-11.99 - 6.63 i$, $9.86 + 5.45 i$, $7.11 - 17.17 i$, $-0.96 - 8.57 i$, $13.16 + 3.79 i$, $8.29 + 1.4 i$, $-6.35 + 8.96 i$, $0.01 + 0.01 i$, $-5.34 - 2.95 i$, $5.43 + 13.14 i$, $2.95 + 5.34 i$, $-19.24 + 2.16 i$, $-8.96 + 6.35 i$, $11.75 - 16.55 i$, $-3.79 - 13.16 i$, $7.6 + 9.53 i$, $17.17 - 7.11 i$, $3.12 - 10.82 i$, $6.63 + 11.99 i$, $13.15 + 8.26 i$, $-10.83 + 1.84 i$, $-10.07 + 2.91 i$, $-1.69 - 5.86 i$, $-0.01$, $-1.69 + 5.86 i$, $-10.07 - 2.91 i$, $-10.83 - 1.84 i$, $13.15 - 8.26 i$, $6.63 - 11.99 i$, $3.12 + 10.82 i$, $17.17 + 7.11 i$, $7.6 - 9.53 i$, $-3.79 + 13.16 i$, $11.75 + 16.55 i$, $-8.96 - 6.35 i$, $-19.24 - 2.16 i$, $2.95 - 5.34 i$, $5.43 - 13.14 i$, $-5.34 + 2.95 i$, $0.01 - 0.01 i$, $-6.35 - 8.96 i$, $8.29 - 1.4 i$, $13.16 - 3.79 i$, $-0.96 + 8.57 i$, $7.11 + 17.17 i$, $9.86 - 5.45 i$, $-11.99 + 6.63 i$, $4.89 + 21.41 i$, $-1.84 - 10.83 i$, $-22.14 - 12.24 i$, $5.86 - 1.69 i$, $18.58 - 18.59 i$, $-5.86 - 1.69 i$, $5.08 + 9.17 i$, $1.84 - 10.83 i$, $0. + 0.01 i$, $11.99 + 6.63 i$, $-2.26 + 4.08 i$, $-7.11 + 17.17 i$, $8.56 - 0.96 i$, $-13.16 - 3.79 i$, $-3.4 + 20. i$, $6.35 - 8.96 i$, $-17.07 - 21.41 i$, $5.34 + 2.95 i$, $31.72 - 13.14 i$, $-2.95 - 5.34 i$, $2.18 + 19.24 i$, $8.96 - 6.35 i$, $-6.85 - 4.86 i$, $3.79 + 13.16 i$, $0.$, $-17.17 + 7.11 i$, $4.48 + 1.29 i$, $-6.63 - 11.99 i$, $-8.26 + 13.15 i$, $10.83 - 1.84 i$, $-7. - 24.31 i$, $1.69 + 5.86 i$, $37.16$)

\vskip 0.7ex
\hangindent=3em \hangafter=1
\textit{Intrinsic sign problem}

  \vskip 2ex

\noindent60. $9_{\frac{9}{14},37.18}^{112,207}$ \irep{498}:\ \ 
$d_i$ = ($1.0$,
$1.0$,
$1.414$,
$1.801$,
$1.801$,
$2.246$,
$2.246$,
$2.548$,
$3.177$) 

\vskip 0.7ex
\hangindent=3em \hangafter=1
$D^2= 37.183 = 
24+12c^{1}_{7}
+4c^{2}_{7}
$

\vskip 0.7ex
\hangindent=3em \hangafter=1
$T = ( 0,
\frac{1}{2},
\frac{15}{16},
\frac{6}{7},
\frac{5}{14},
\frac{2}{7},
\frac{11}{14},
\frac{89}{112},
\frac{25}{112} )
$,

\vskip 0.7ex
\hangindent=3em \hangafter=1
$S$ = ($ 1$,
$ 1$,
$ \sqrt{2}$,
$ -c_{7}^{3}$,
$ -c_{7}^{3}$,
$ \xi_{7}^{3}$,
$ \xi_{7}^{3}$,
$ c^{3}_{56}
+c^{11}_{56}
$,
$ c^{3}_{56}
+c^{5}_{56}
-c^{9}_{56}
+c^{11}_{56}
$;\ \ 
$ 1$,
$ -\sqrt{2}$,
$ -c_{7}^{3}$,
$ -c_{7}^{3}$,
$ \xi_{7}^{3}$,
$ \xi_{7}^{3}$,
$ -c^{3}_{56}
-c^{11}_{56}
$,
$ -c^{3}_{56}
-c^{5}_{56}
+c^{9}_{56}
-c^{11}_{56}
$;\ \ 
$0$,
$ c^{3}_{56}
+c^{11}_{56}
$,
$ -c^{3}_{56}
-c^{11}_{56}
$,
$ c^{3}_{56}
+c^{5}_{56}
-c^{9}_{56}
+c^{11}_{56}
$,
$ -c^{3}_{56}
-c^{5}_{56}
+c^{9}_{56}
-c^{11}_{56}
$,
$0$,
$0$;\ \ 
$ -\xi_{7}^{3}$,
$ -\xi_{7}^{3}$,
$ 1$,
$ 1$,
$ -c^{3}_{56}
-c^{5}_{56}
+c^{9}_{56}
-c^{11}_{56}
$,
$ \sqrt{2}$;\ \ 
$ -\xi_{7}^{3}$,
$ 1$,
$ 1$,
$ c^{3}_{56}
+c^{5}_{56}
-c^{9}_{56}
+c^{11}_{56}
$,
$ -\sqrt{2}$;\ \ 
$ c_{7}^{3}$,
$ c_{7}^{3}$,
$ \sqrt{2}$,
$ -c^{3}_{56}
-c^{11}_{56}
$;\ \ 
$ c_{7}^{3}$,
$ -\sqrt{2}$,
$ c^{3}_{56}
+c^{11}_{56}
$;\ \ 
$0$,
$0$;\ \ 
$0$)

Factors = $3_{\frac{15}{2},4.}^{16,639}\boxtimes 3_{\frac{8}{7},9.295}^{7,245}$

\vskip 0.7ex
\hangindent=3em \hangafter=1
$\tau_n$ = ($5.34 + 2.95 i$, $-22.14 - 12.24 i$, $-8.96 - 6.35 i$, $13.15 + 8.26 i$, $13.16 + 3.79 i$, $-2.26 - 4.08 i$, $-17.17 - 7.11 i$, $0.$, $11.99 + 6.63 i$, $8.29 - 1.4 i$, $-10.83 - 1.84 i$, $-19.24 + 2.16 i$, $5.86 + 1.69 i$, $31.72 + 13.14 i$, $1.69 + 5.86 i$, $-17.07 - 21.41 i$, $-1.84 - 10.83 i$, $11.75 + 16.55 i$, $6.63 + 11.99 i$, $-0.96 - 8.57 i$, $-7.11 - 17.17 i$, $4.48 - 1.29 i$, $3.79 + 13.16 i$, $0. + 0.01 i$, $-6.35 - 8.96 i$, $-10.07 - 2.91 i$, $2.95 + 5.34 i$, $18.58 + 18.59 i$, $-2.95 + 5.34 i$, $-7. - 24.31 i$, $6.35 - 8.96 i$, $4.89 + 21.41 i$, $-3.79 + 13.16 i$, $3.12 - 10.82 i$, $7.11 - 17.17 i$, $8.56 + 0.96 i$, $-6.63 + 11.99 i$, $-6.85 - 4.86 i$, $1.84 - 10.83 i$, $0.01 - 0.01 i$, $-1.69 + 5.86 i$, $5.43 + 13.14 i$, $-5.86 + 1.69 i$, $2.18 - 19.24 i$, $10.83 - 1.84 i$, $-3.4 + 20. i$, $-11.99 + 6.63 i$, $7.6 - 9.53 i$, $17.17 - 7.11 i$, $9.86 + 5.45 i$, $-13.16 + 3.79 i$, $-8.26 - 13.15 i$, $8.96 - 6.35 i$, $5.08 + 9.17 i$, $-5.34 + 2.95 i$, $-0.01$, $-5.34 - 2.95 i$, $5.08 - 9.17 i$, $8.96 + 6.35 i$, $-8.26 + 13.15 i$, $-13.16 - 3.79 i$, $9.86 - 5.45 i$, $17.17 + 7.11 i$, $7.6 + 9.53 i$, $-11.99 - 6.63 i$, $-3.4 - 20. i$, $10.83 + 1.84 i$, $2.18 + 19.24 i$, $-5.86 - 1.69 i$, $5.43 - 13.14 i$, $-1.69 - 5.86 i$, $0.01 + 0.01 i$, $1.84 + 10.83 i$, $-6.85 + 4.86 i$, $-6.63 - 11.99 i$, $8.56 - 0.96 i$, $7.11 + 17.17 i$, $3.12 + 10.82 i$, $-3.79 - 13.16 i$, $4.89 - 21.41 i$, $6.35 + 8.96 i$, $-7. + 24.31 i$, $-2.95 - 5.34 i$, $18.58 - 18.59 i$, $2.95 - 5.34 i$, $-10.07 + 2.91 i$, $-6.35 + 8.96 i$, $0. - 0.01 i$, $3.79 - 13.16 i$, $4.48 + 1.29 i$, $-7.11 + 17.17 i$, $-0.96 + 8.57 i$, $6.63 - 11.99 i$, $11.75 - 16.55 i$, $-1.84 + 10.83 i$, $-17.07 + 21.41 i$, $1.69 - 5.86 i$, $31.72 - 13.14 i$, $5.86 - 1.69 i$, $-19.24 - 2.16 i$, $-10.83 + 1.84 i$, $8.29 + 1.4 i$, $11.99 - 6.63 i$, $0.$, $-17.17 + 7.11 i$, $-2.26 + 4.08 i$, $13.16 - 3.79 i$, $13.15 - 8.26 i$, $-8.96 + 6.35 i$, $-22.14 + 12.24 i$, $5.34 - 2.95 i$, $37.16$)

\vskip 0.7ex
\hangindent=3em \hangafter=1
\textit{Intrinsic sign problem}

  \vskip 2ex

\noindent61. $9_{2,44.}^{88,112}$ \irep{496}:\ \ 
$d_i$ = ($1.0$,
$1.0$,
$2.0$,
$2.0$,
$2.0$,
$2.0$,
$2.0$,
$3.316$,
$3.316$) 

\vskip 0.7ex
\hangindent=3em \hangafter=1
$D^2= 44.0 = 
44$

\vskip 0.7ex
\hangindent=3em \hangafter=1
$T = ( 0,
0,
\frac{1}{11},
\frac{3}{11},
\frac{4}{11},
\frac{5}{11},
\frac{9}{11},
\frac{1}{8},
\frac{5}{8} )
$,

\vskip 0.7ex
\hangindent=3em \hangafter=1
$S$ = ($ 1$,
$ 1$,
$ 2$,
$ 2$,
$ 2$,
$ 2$,
$ 2$,
$ \sqrt{11}$,
$ \sqrt{11}$;\ \ 
$ 1$,
$ 2$,
$ 2$,
$ 2$,
$ 2$,
$ 2$,
$ -\sqrt{11}$,
$ -\sqrt{11}$;\ \ 
$ 2c_{11}^{2}$,
$ 2c_{11}^{1}$,
$ 2c_{11}^{4}$,
$ 2c_{11}^{3}$,
$ 2c_{11}^{5}$,
$0$,
$0$;\ \ 
$ 2c_{11}^{5}$,
$ 2c_{11}^{2}$,
$ 2c_{11}^{4}$,
$ 2c_{11}^{3}$,
$0$,
$0$;\ \ 
$ 2c_{11}^{3}$,
$ 2c_{11}^{5}$,
$ 2c_{11}^{1}$,
$0$,
$0$;\ \ 
$ 2c_{11}^{1}$,
$ 2c_{11}^{2}$,
$0$,
$0$;\ \ 
$ 2c_{11}^{4}$,
$0$,
$0$;\ \ 
$ \sqrt{11}$,
$ -\sqrt{11}$;\ \ 
$ \sqrt{11}$)

\vskip 0.7ex
\hangindent=3em \hangafter=1
$\tau_n$ = ($0. + 6.63 i$, $0. + 15.36 i$, $0. + 6.63 i$, $-21.99 + 6.63 i$, $0. + 6.63 i$, $0. - 28.62 i$, $0. - 6.63 i$, $21.99 - 6.63 i$, $0. + 6.63 i$, $0. + 15.36 i$, $22.$, $-21.99 + 6.63 i$, $0. - 6.63 i$, $0. - 15.36 i$, $0. + 6.63 i$, $21.99 + 6.63 i$, $0. - 6.63 i$, $0. + 15.36 i$, $0. - 6.63 i$, $-21.99 + 6.63 i$, $0. - 6.63 i$, $22. - 21.99 i$, $0. + 6.63 i$, $21.99 - 6.63 i$, $0. + 6.63 i$, $0. + 28.62 i$, $0. + 6.63 i$, $-21.99 - 6.63 i$, $0. - 6.63 i$, $0. - 28.62 i$, $0. + 6.63 i$, $21.99 - 6.63 i$, $22.$, $0. + 28.62 i$, $0. - 6.63 i$, $-21.99 + 6.63 i$, $0. + 6.63 i$, $0. - 15.36 i$, $0. - 6.63 i$, $21.99 - 6.63 i$, $0. - 6.63 i$, $0. + 28.62 i$, $0. - 6.63 i$, $0.01$, $0. + 6.63 i$, $0. - 28.62 i$, $0. + 6.63 i$, $21.99 + 6.63 i$, $0. + 6.63 i$, $0. + 15.36 i$, $0. - 6.63 i$, $-21.99 - 6.63 i$, $0. + 6.63 i$, $0. - 28.62 i$, $22.$, $21.99 + 6.63 i$, $0. - 6.63 i$, $0. + 28.62 i$, $0. + 6.63 i$, $-21.99 + 6.63 i$, $0. - 6.63 i$, $0. - 28.62 i$, $0. - 6.63 i$, $21.99 + 6.63 i$, $0. - 6.63 i$, $22. + 21.99 i$, $0. + 6.63 i$, $-21.99 - 6.63 i$, $0. + 6.63 i$, $0. - 15.36 i$, $0. + 6.63 i$, $21.99 - 6.63 i$, $0. - 6.63 i$, $0. + 15.36 i$, $0. + 6.63 i$, $-21.99 - 6.63 i$, $22.$, $0. - 15.36 i$, $0. - 6.63 i$, $21.99 + 6.63 i$, $0. + 6.63 i$, $0. + 28.62 i$, $0. - 6.63 i$, $-21.99 - 6.63 i$, $0. - 6.63 i$, $0. - 15.36 i$, $0. - 6.63 i$, $43.99$)

\vskip 0.7ex
\hangindent=3em \hangafter=1
\textit{Intrinsic sign problem}

  \vskip 2ex

\noindent62. $9_{2,44.}^{88,529}$ \irep{496}:\ \ 
$d_i$ = ($1.0$,
$1.0$,
$2.0$,
$2.0$,
$2.0$,
$2.0$,
$2.0$,
$3.316$,
$3.316$) 

\vskip 0.7ex
\hangindent=3em \hangafter=1
$D^2= 44.0 = 
44$

\vskip 0.7ex
\hangindent=3em \hangafter=1
$T = ( 0,
0,
\frac{1}{11},
\frac{3}{11},
\frac{4}{11},
\frac{5}{11},
\frac{9}{11},
\frac{3}{8},
\frac{7}{8} )
$,

\vskip 0.7ex
\hangindent=3em \hangafter=1
$S$ = ($ 1$,
$ 1$,
$ 2$,
$ 2$,
$ 2$,
$ 2$,
$ 2$,
$ \sqrt{11}$,
$ \sqrt{11}$;\ \ 
$ 1$,
$ 2$,
$ 2$,
$ 2$,
$ 2$,
$ 2$,
$ -\sqrt{11}$,
$ -\sqrt{11}$;\ \ 
$ 2c_{11}^{2}$,
$ 2c_{11}^{1}$,
$ 2c_{11}^{4}$,
$ 2c_{11}^{3}$,
$ 2c_{11}^{5}$,
$0$,
$0$;\ \ 
$ 2c_{11}^{5}$,
$ 2c_{11}^{2}$,
$ 2c_{11}^{4}$,
$ 2c_{11}^{3}$,
$0$,
$0$;\ \ 
$ 2c_{11}^{3}$,
$ 2c_{11}^{5}$,
$ 2c_{11}^{1}$,
$0$,
$0$;\ \ 
$ 2c_{11}^{1}$,
$ 2c_{11}^{2}$,
$0$,
$0$;\ \ 
$ 2c_{11}^{4}$,
$0$,
$0$;\ \ 
$ -\sqrt{11}$,
$ \sqrt{11}$;\ \ 
$ -\sqrt{11}$)

\vskip 0.7ex
\hangindent=3em \hangafter=1
$\tau_n$ = ($0. + 6.63 i$, $0. - 28.62 i$, $0. + 6.63 i$, $-21.99 + 6.63 i$, $0. + 6.63 i$, $0. + 15.36 i$, $0. - 6.63 i$, $21.99 - 6.63 i$, $0. + 6.63 i$, $0. - 28.62 i$, $22.$, $-21.99 + 6.63 i$, $0. - 6.63 i$, $0. + 28.62 i$, $0. + 6.63 i$, $21.99 + 6.63 i$, $0. - 6.63 i$, $0. - 28.62 i$, $0. - 6.63 i$, $-21.99 + 6.63 i$, $0. - 6.63 i$, $22. + 21.99 i$, $0. + 6.63 i$, $21.99 - 6.63 i$, $0. + 6.63 i$, $0. - 15.36 i$, $0. + 6.63 i$, $-21.99 - 6.63 i$, $0. - 6.63 i$, $0. + 15.36 i$, $0. + 6.63 i$, $21.99 - 6.63 i$, $22.$, $0. - 15.36 i$, $0. - 6.63 i$, $-21.99 + 6.63 i$, $0. + 6.63 i$, $0. + 28.62 i$, $0. - 6.63 i$, $21.99 - 6.63 i$, $0. - 6.63 i$, $0. - 15.36 i$, $0. - 6.63 i$, $0.01$, $0. + 6.63 i$, $0. + 15.36 i$, $0. + 6.63 i$, $21.99 + 6.63 i$, $0. + 6.63 i$, $0. - 28.62 i$, $0. - 6.63 i$, $-21.99 - 6.63 i$, $0. + 6.63 i$, $0. + 15.36 i$, $22.$, $21.99 + 6.63 i$, $0. - 6.63 i$, $0. - 15.36 i$, $0. + 6.63 i$, $-21.99 + 6.63 i$, $0. - 6.63 i$, $0. + 15.36 i$, $0. - 6.63 i$, $21.99 + 6.63 i$, $0. - 6.63 i$, $22. - 21.99 i$, $0. + 6.63 i$, $-21.99 - 6.63 i$, $0. + 6.63 i$, $0. + 28.62 i$, $0. + 6.63 i$, $21.99 - 6.63 i$, $0. - 6.63 i$, $0. - 28.62 i$, $0. + 6.63 i$, $-21.99 - 6.63 i$, $22.$, $0. + 28.62 i$, $0. - 6.63 i$, $21.99 + 6.63 i$, $0. + 6.63 i$, $0. - 15.36 i$, $0. - 6.63 i$, $-21.99 - 6.63 i$, $0. - 6.63 i$, $0. + 28.62 i$, $0. - 6.63 i$, $43.99$)

\vskip 0.7ex
\hangindent=3em \hangafter=1
\textit{Intrinsic sign problem}

  \vskip 2ex

\noindent63. $9_{6,44.}^{88,870}$ \irep{496}:\ \ 
$d_i$ = ($1.0$,
$1.0$,
$2.0$,
$2.0$,
$2.0$,
$2.0$,
$2.0$,
$3.316$,
$3.316$) 

\vskip 0.7ex
\hangindent=3em \hangafter=1
$D^2= 44.0 = 
44$

\vskip 0.7ex
\hangindent=3em \hangafter=1
$T = ( 0,
0,
\frac{2}{11},
\frac{6}{11},
\frac{7}{11},
\frac{8}{11},
\frac{10}{11},
\frac{1}{8},
\frac{5}{8} )
$,

\vskip 0.7ex
\hangindent=3em \hangafter=1
$S$ = ($ 1$,
$ 1$,
$ 2$,
$ 2$,
$ 2$,
$ 2$,
$ 2$,
$ \sqrt{11}$,
$ \sqrt{11}$;\ \ 
$ 1$,
$ 2$,
$ 2$,
$ 2$,
$ 2$,
$ 2$,
$ -\sqrt{11}$,
$ -\sqrt{11}$;\ \ 
$ 2c_{11}^{4}$,
$ 2c_{11}^{2}$,
$ 2c_{11}^{1}$,
$ 2c_{11}^{3}$,
$ 2c_{11}^{5}$,
$0$,
$0$;\ \ 
$ 2c_{11}^{1}$,
$ 2c_{11}^{5}$,
$ 2c_{11}^{4}$,
$ 2c_{11}^{3}$,
$0$,
$0$;\ \ 
$ 2c_{11}^{3}$,
$ 2c_{11}^{2}$,
$ 2c_{11}^{4}$,
$0$,
$0$;\ \ 
$ 2c_{11}^{5}$,
$ 2c_{11}^{1}$,
$0$,
$0$;\ \ 
$ 2c_{11}^{2}$,
$0$,
$0$;\ \ 
$ -\sqrt{11}$,
$ \sqrt{11}$;\ \ 
$ -\sqrt{11}$)

\vskip 0.7ex
\hangindent=3em \hangafter=1
$\tau_n$ = ($0. - 6.63 i$, $0. + 28.62 i$, $0. - 6.63 i$, $-21.99 - 6.63 i$, $0. - 6.63 i$, $0. - 15.36 i$, $0. + 6.63 i$, $21.99 + 6.63 i$, $0. - 6.63 i$, $0. + 28.62 i$, $22.$, $-21.99 - 6.63 i$, $0. + 6.63 i$, $0. - 28.62 i$, $0. - 6.63 i$, $21.99 - 6.63 i$, $0. + 6.63 i$, $0. + 28.62 i$, $0. + 6.63 i$, $-21.99 - 6.63 i$, $0. + 6.63 i$, $22. - 21.99 i$, $0. - 6.63 i$, $21.99 + 6.63 i$, $0. - 6.63 i$, $0. + 15.36 i$, $0. - 6.63 i$, $-21.99 + 6.63 i$, $0. + 6.63 i$, $0. - 15.36 i$, $0. - 6.63 i$, $21.99 + 6.63 i$, $22.$, $0. + 15.36 i$, $0. + 6.63 i$, $-21.99 - 6.63 i$, $0. - 6.63 i$, $0. - 28.62 i$, $0. + 6.63 i$, $21.99 + 6.63 i$, $0. + 6.63 i$, $0. + 15.36 i$, $0. + 6.63 i$, $0.01$, $0. - 6.63 i$, $0. - 15.36 i$, $0. - 6.63 i$, $21.99 - 6.63 i$, $0. - 6.63 i$, $0. + 28.62 i$, $0. + 6.63 i$, $-21.99 + 6.63 i$, $0. - 6.63 i$, $0. - 15.36 i$, $22.$, $21.99 - 6.63 i$, $0. + 6.63 i$, $0. + 15.36 i$, $0. - 6.63 i$, $-21.99 - 6.63 i$, $0. + 6.63 i$, $0. - 15.36 i$, $0. + 6.63 i$, $21.99 - 6.63 i$, $0. + 6.63 i$, $22. + 21.99 i$, $0. - 6.63 i$, $-21.99 + 6.63 i$, $0. - 6.63 i$, $0. - 28.62 i$, $0. - 6.63 i$, $21.99 + 6.63 i$, $0. + 6.63 i$, $0. + 28.62 i$, $0. - 6.63 i$, $-21.99 + 6.63 i$, $22.$, $0. - 28.62 i$, $0. + 6.63 i$, $21.99 - 6.63 i$, $0. - 6.63 i$, $0. + 15.36 i$, $0. + 6.63 i$, $-21.99 + 6.63 i$, $0. + 6.63 i$, $0. - 28.62 i$, $0. + 6.63 i$, $43.99$)

\vskip 0.7ex
\hangindent=3em \hangafter=1
\textit{Intrinsic sign problem}

  \vskip 2ex

\noindent64. $9_{6,44.}^{88,252}$ \irep{496}:\ \ 
$d_i$ = ($1.0$,
$1.0$,
$2.0$,
$2.0$,
$2.0$,
$2.0$,
$2.0$,
$3.316$,
$3.316$) 

\vskip 0.7ex
\hangindent=3em \hangafter=1
$D^2= 44.0 = 
44$

\vskip 0.7ex
\hangindent=3em \hangafter=1
$T = ( 0,
0,
\frac{2}{11},
\frac{6}{11},
\frac{7}{11},
\frac{8}{11},
\frac{10}{11},
\frac{3}{8},
\frac{7}{8} )
$,

\vskip 0.7ex
\hangindent=3em \hangafter=1
$S$ = ($ 1$,
$ 1$,
$ 2$,
$ 2$,
$ 2$,
$ 2$,
$ 2$,
$ \sqrt{11}$,
$ \sqrt{11}$;\ \ 
$ 1$,
$ 2$,
$ 2$,
$ 2$,
$ 2$,
$ 2$,
$ -\sqrt{11}$,
$ -\sqrt{11}$;\ \ 
$ 2c_{11}^{4}$,
$ 2c_{11}^{2}$,
$ 2c_{11}^{1}$,
$ 2c_{11}^{3}$,
$ 2c_{11}^{5}$,
$0$,
$0$;\ \ 
$ 2c_{11}^{1}$,
$ 2c_{11}^{5}$,
$ 2c_{11}^{4}$,
$ 2c_{11}^{3}$,
$0$,
$0$;\ \ 
$ 2c_{11}^{3}$,
$ 2c_{11}^{2}$,
$ 2c_{11}^{4}$,
$0$,
$0$;\ \ 
$ 2c_{11}^{5}$,
$ 2c_{11}^{1}$,
$0$,
$0$;\ \ 
$ 2c_{11}^{2}$,
$0$,
$0$;\ \ 
$ \sqrt{11}$,
$ -\sqrt{11}$;\ \ 
$ \sqrt{11}$)

\vskip 0.7ex
\hangindent=3em \hangafter=1
$\tau_n$ = ($0. - 6.63 i$, $0. - 15.36 i$, $0. - 6.63 i$, $-21.99 - 6.63 i$, $0. - 6.63 i$, $0. + 28.62 i$, $0. + 6.63 i$, $21.99 + 6.63 i$, $0. - 6.63 i$, $0. - 15.36 i$, $22.$, $-21.99 - 6.63 i$, $0. + 6.63 i$, $0. + 15.36 i$, $0. - 6.63 i$, $21.99 - 6.63 i$, $0. + 6.63 i$, $0. - 15.36 i$, $0. + 6.63 i$, $-21.99 - 6.63 i$, $0. + 6.63 i$, $22. + 21.99 i$, $0. - 6.63 i$, $21.99 + 6.63 i$, $0. - 6.63 i$, $0. - 28.62 i$, $0. - 6.63 i$, $-21.99 + 6.63 i$, $0. + 6.63 i$, $0. + 28.62 i$, $0. - 6.63 i$, $21.99 + 6.63 i$, $22.$, $0. - 28.62 i$, $0. + 6.63 i$, $-21.99 - 6.63 i$, $0. - 6.63 i$, $0. + 15.36 i$, $0. + 6.63 i$, $21.99 + 6.63 i$, $0. + 6.63 i$, $0. - 28.62 i$, $0. + 6.63 i$, $0.01$, $0. - 6.63 i$, $0. + 28.62 i$, $0. - 6.63 i$, $21.99 - 6.63 i$, $0. - 6.63 i$, $0. - 15.36 i$, $0. + 6.63 i$, $-21.99 + 6.63 i$, $0. - 6.63 i$, $0. + 28.62 i$, $22.$, $21.99 - 6.63 i$, $0. + 6.63 i$, $0. - 28.62 i$, $0. - 6.63 i$, $-21.99 - 6.63 i$, $0. + 6.63 i$, $0. + 28.62 i$, $0. + 6.63 i$, $21.99 - 6.63 i$, $0. + 6.63 i$, $22. - 21.99 i$, $0. - 6.63 i$, $-21.99 + 6.63 i$, $0. - 6.63 i$, $0. + 15.36 i$, $0. - 6.63 i$, $21.99 + 6.63 i$, $0. + 6.63 i$, $0. - 15.36 i$, $0. - 6.63 i$, $-21.99 + 6.63 i$, $22.$, $0. + 15.36 i$, $0. + 6.63 i$, $21.99 - 6.63 i$, $0. - 6.63 i$, $0. - 28.62 i$, $0. + 6.63 i$, $-21.99 + 6.63 i$, $0. + 6.63 i$, $0. + 15.36 i$, $0. + 6.63 i$, $43.99$)

\vskip 0.7ex
\hangindent=3em \hangafter=1
\textit{Intrinsic sign problem}

  \vskip 2ex

\noindent65. $9_{\frac{12}{5},52.36}^{40,304}$ \irep{483}:\ \ 
$d_i$ = ($1.0$,
$1.0$,
$1.902$,
$1.902$,
$2.618$,
$2.618$,
$3.77$,
$3.77$,
$3.236$) 

\vskip 0.7ex
\hangindent=3em \hangafter=1
$D^2= 52.360 = 
30+10\sqrt{5}$

\vskip 0.7ex
\hangindent=3em \hangafter=1
$T = ( 0,
0,
\frac{3}{40},
\frac{23}{40},
\frac{1}{5},
\frac{1}{5},
\frac{3}{8},
\frac{7}{8},
\frac{3}{5} )
$,

\vskip 0.7ex
\hangindent=3em \hangafter=1
$S$ = ($ 1$,
$ 1$,
$ c_{20}^{1}$,
$ c_{20}^{1}$,
$ \frac{3+\sqrt{5}}{2}$,
$ \frac{3+\sqrt{5}}{2}$,
$ c^{1}_{20}
+c^{3}_{20}
$,
$ c^{1}_{20}
+c^{3}_{20}
$,
$ 1+\sqrt{5}$;\ \ 
$ 1$,
$ -c_{20}^{1}$,
$ -c_{20}^{1}$,
$ \frac{3+\sqrt{5}}{2}$,
$ \frac{3+\sqrt{5}}{2}$,
$ -c^{1}_{20}
-c^{3}_{20}
$,
$ -c^{1}_{20}
-c^{3}_{20}
$,
$ 1+\sqrt{5}$;\ \ 
$ c^{1}_{20}
+c^{3}_{20}
$,
$ -c^{1}_{20}
-c^{3}_{20}
$,
$ -c^{1}_{20}
-c^{3}_{20}
$,
$ c^{1}_{20}
+c^{3}_{20}
$,
$ c_{20}^{1}$,
$ -c_{20}^{1}$,
$0$;\ \ 
$ c^{1}_{20}
+c^{3}_{20}
$,
$ -c^{1}_{20}
-c^{3}_{20}
$,
$ c^{1}_{20}
+c^{3}_{20}
$,
$ -c_{20}^{1}$,
$ c_{20}^{1}$,
$0$;\ \ 
$ 1$,
$ 1$,
$ c_{20}^{1}$,
$ c_{20}^{1}$,
$ -1-\sqrt{5}$;\ \ 
$ 1$,
$ -c_{20}^{1}$,
$ -c_{20}^{1}$,
$ -1-\sqrt{5}$;\ \ 
$ -c^{1}_{20}
-c^{3}_{20}
$,
$ c^{1}_{20}
+c^{3}_{20}
$,
$0$;\ \ 
$ -c^{1}_{20}
-c^{3}_{20}
$,
$0$;\ \ 
$ 1+\sqrt{5}$)

\vskip 0.7ex
\hangindent=3em \hangafter=1
$\tau_n$ = ($-2.24 + 6.88 i$, $-1.6 - 4.56 i$, $-5.85 - 18.02 i$, $-32.9$, $26.18$, $-9.12 + 37.54 i$, $-5.85 + 18.02 i$, $16.72 - 22.27 i$, $-2.24 - 6.88 i$, $26.18 - 35.66 i$, $-2.24 + 6.88 i$, $-28.43 + 13.76 i$, $-5.85 - 18.02 i$, $4.65 + 23.78 i$, $26.18$, $28.43 + 13.76 i$, $-5.85 + 18.02 i$, $-10.11 - 40.59 i$, $-2.24 - 6.88 i$, $-9.48$, $-2.24 + 6.88 i$, $-10.11 + 40.59 i$, $-5.85 - 18.02 i$, $28.43 - 13.76 i$, $26.18$, $4.65 - 23.78 i$, $-5.85 + 18.02 i$, $-28.43 - 13.76 i$, $-2.24 - 6.88 i$, $26.18 + 35.66 i$, $-2.24 + 6.88 i$, $16.72 + 22.27 i$, $-5.85 - 18.02 i$, $-9.12 - 37.54 i$, $26.18$, $-32.9$, $-5.85 + 18.02 i$, $-1.6 + 4.56 i$, $-2.24 - 6.88 i$, $61.84$)

\vskip 0.7ex
\hangindent=3em \hangafter=1
\textit{Intrinsic sign problem}

  \vskip 2ex

\noindent66. $9_{\frac{28}{5},52.36}^{40,247}$ \irep{483}:\ \ 
$d_i$ = ($1.0$,
$1.0$,
$1.902$,
$1.902$,
$2.618$,
$2.618$,
$3.77$,
$3.77$,
$3.236$) 

\vskip 0.7ex
\hangindent=3em \hangafter=1
$D^2= 52.360 = 
30+10\sqrt{5}$

\vskip 0.7ex
\hangindent=3em \hangafter=1
$T = ( 0,
0,
\frac{7}{40},
\frac{27}{40},
\frac{4}{5},
\frac{4}{5},
\frac{3}{8},
\frac{7}{8},
\frac{2}{5} )
$,

\vskip 0.7ex
\hangindent=3em \hangafter=1
$S$ = ($ 1$,
$ 1$,
$ c_{20}^{1}$,
$ c_{20}^{1}$,
$ \frac{3+\sqrt{5}}{2}$,
$ \frac{3+\sqrt{5}}{2}$,
$ c^{1}_{20}
+c^{3}_{20}
$,
$ c^{1}_{20}
+c^{3}_{20}
$,
$ 1+\sqrt{5}$;\ \ 
$ 1$,
$ -c_{20}^{1}$,
$ -c_{20}^{1}$,
$ \frac{3+\sqrt{5}}{2}$,
$ \frac{3+\sqrt{5}}{2}$,
$ -c^{1}_{20}
-c^{3}_{20}
$,
$ -c^{1}_{20}
-c^{3}_{20}
$,
$ 1+\sqrt{5}$;\ \ 
$ -c^{1}_{20}
-c^{3}_{20}
$,
$ c^{1}_{20}
+c^{3}_{20}
$,
$ -c^{1}_{20}
-c^{3}_{20}
$,
$ c^{1}_{20}
+c^{3}_{20}
$,
$ c_{20}^{1}$,
$ -c_{20}^{1}$,
$0$;\ \ 
$ -c^{1}_{20}
-c^{3}_{20}
$,
$ -c^{1}_{20}
-c^{3}_{20}
$,
$ c^{1}_{20}
+c^{3}_{20}
$,
$ -c_{20}^{1}$,
$ c_{20}^{1}$,
$0$;\ \ 
$ 1$,
$ 1$,
$ c_{20}^{1}$,
$ c_{20}^{1}$,
$ -1-\sqrt{5}$;\ \ 
$ 1$,
$ -c_{20}^{1}$,
$ -c_{20}^{1}$,
$ -1-\sqrt{5}$;\ \ 
$ c^{1}_{20}
+c^{3}_{20}
$,
$ -c^{1}_{20}
-c^{3}_{20}
$,
$0$;\ \ 
$ c^{1}_{20}
+c^{3}_{20}
$,
$0$;\ \ 
$ 1+\sqrt{5}$)

\vskip 0.7ex
\hangindent=3em \hangafter=1
$\tau_n$ = ($-2.24 - 6.88 i$, $-10.11 - 40.59 i$, $-5.85 + 18.02 i$, $-32.9$, $26.18$, $4.65 + 23.78 i$, $-5.85 - 18.02 i$, $16.72 + 22.27 i$, $-2.24 + 6.88 i$, $26.18 - 35.66 i$, $-2.24 - 6.88 i$, $-28.43 - 13.76 i$, $-5.85 + 18.02 i$, $-9.12 + 37.54 i$, $26.18$, $28.43 - 13.76 i$, $-5.85 - 18.02 i$, $-1.6 - 4.56 i$, $-2.24 + 6.88 i$, $-9.48$, $-2.24 - 6.88 i$, $-1.6 + 4.56 i$, $-5.85 + 18.02 i$, $28.43 + 13.76 i$, $26.18$, $-9.12 - 37.54 i$, $-5.85 - 18.02 i$, $-28.43 + 13.76 i$, $-2.24 + 6.88 i$, $26.18 + 35.66 i$, $-2.24 - 6.88 i$, $16.72 - 22.27 i$, $-5.85 + 18.02 i$, $4.65 - 23.78 i$, $26.18$, $-32.9$, $-5.85 - 18.02 i$, $-10.11 + 40.59 i$, $-2.24 + 6.88 i$, $61.84$)

\vskip 0.7ex
\hangindent=3em \hangafter=1
\textit{Intrinsic sign problem}

  \vskip 2ex

\noindent67. $9_{\frac{12}{5},52.36}^{40,987}$ \irep{483}:\ \ 
$d_i$ = ($1.0$,
$1.0$,
$1.902$,
$1.902$,
$2.618$,
$2.618$,
$3.77$,
$3.77$,
$3.236$) 

\vskip 0.7ex
\hangindent=3em \hangafter=1
$D^2= 52.360 = 
30+10\sqrt{5}$

\vskip 0.7ex
\hangindent=3em \hangafter=1
$T = ( 0,
0,
\frac{13}{40},
\frac{33}{40},
\frac{1}{5},
\frac{1}{5},
\frac{1}{8},
\frac{5}{8},
\frac{3}{5} )
$,

\vskip 0.7ex
\hangindent=3em \hangafter=1
$S$ = ($ 1$,
$ 1$,
$ c_{20}^{1}$,
$ c_{20}^{1}$,
$ \frac{3+\sqrt{5}}{2}$,
$ \frac{3+\sqrt{5}}{2}$,
$ c^{1}_{20}
+c^{3}_{20}
$,
$ c^{1}_{20}
+c^{3}_{20}
$,
$ 1+\sqrt{5}$;\ \ 
$ 1$,
$ -c_{20}^{1}$,
$ -c_{20}^{1}$,
$ \frac{3+\sqrt{5}}{2}$,
$ \frac{3+\sqrt{5}}{2}$,
$ -c^{1}_{20}
-c^{3}_{20}
$,
$ -c^{1}_{20}
-c^{3}_{20}
$,
$ 1+\sqrt{5}$;\ \ 
$ -c^{1}_{20}
-c^{3}_{20}
$,
$ c^{1}_{20}
+c^{3}_{20}
$,
$ -c^{1}_{20}
-c^{3}_{20}
$,
$ c^{1}_{20}
+c^{3}_{20}
$,
$ c_{20}^{1}$,
$ -c_{20}^{1}$,
$0$;\ \ 
$ -c^{1}_{20}
-c^{3}_{20}
$,
$ -c^{1}_{20}
-c^{3}_{20}
$,
$ c^{1}_{20}
+c^{3}_{20}
$,
$ -c_{20}^{1}$,
$ c_{20}^{1}$,
$0$;\ \ 
$ 1$,
$ 1$,
$ c_{20}^{1}$,
$ c_{20}^{1}$,
$ -1-\sqrt{5}$;\ \ 
$ 1$,
$ -c_{20}^{1}$,
$ -c_{20}^{1}$,
$ -1-\sqrt{5}$;\ \ 
$ c^{1}_{20}
+c^{3}_{20}
$,
$ -c^{1}_{20}
-c^{3}_{20}
$,
$0$;\ \ 
$ c^{1}_{20}
+c^{3}_{20}
$,
$0$;\ \ 
$ 1+\sqrt{5}$)

\vskip 0.7ex
\hangindent=3em \hangafter=1
$\tau_n$ = ($-2.24 + 6.88 i$, $-10.11 + 40.59 i$, $-5.85 - 18.02 i$, $-32.9$, $26.18$, $4.65 - 23.78 i$, $-5.85 + 18.02 i$, $16.72 - 22.27 i$, $-2.24 - 6.88 i$, $26.18 + 35.66 i$, $-2.24 + 6.88 i$, $-28.43 + 13.76 i$, $-5.85 - 18.02 i$, $-9.12 - 37.54 i$, $26.18$, $28.43 + 13.76 i$, $-5.85 + 18.02 i$, $-1.6 + 4.56 i$, $-2.24 - 6.88 i$, $-9.48$, $-2.24 + 6.88 i$, $-1.6 - 4.56 i$, $-5.85 - 18.02 i$, $28.43 - 13.76 i$, $26.18$, $-9.12 + 37.54 i$, $-5.85 + 18.02 i$, $-28.43 - 13.76 i$, $-2.24 - 6.88 i$, $26.18 - 35.66 i$, $-2.24 + 6.88 i$, $16.72 + 22.27 i$, $-5.85 - 18.02 i$, $4.65 + 23.78 i$, $26.18$, $-32.9$, $-5.85 + 18.02 i$, $-10.11 - 40.59 i$, $-2.24 - 6.88 i$, $61.84$)

\vskip 0.7ex
\hangindent=3em \hangafter=1
\textit{Intrinsic sign problem}

  \vskip 2ex

\noindent68. $9_{\frac{28}{5},52.36}^{40,198}$ \irep{483}:\ \ 
$d_i$ = ($1.0$,
$1.0$,
$1.902$,
$1.902$,
$2.618$,
$2.618$,
$3.77$,
$3.77$,
$3.236$) 

\vskip 0.7ex
\hangindent=3em \hangafter=1
$D^2= 52.360 = 
30+10\sqrt{5}$

\vskip 0.7ex
\hangindent=3em \hangafter=1
$T = ( 0,
0,
\frac{17}{40},
\frac{37}{40},
\frac{4}{5},
\frac{4}{5},
\frac{1}{8},
\frac{5}{8},
\frac{2}{5} )
$,

\vskip 0.7ex
\hangindent=3em \hangafter=1
$S$ = ($ 1$,
$ 1$,
$ c_{20}^{1}$,
$ c_{20}^{1}$,
$ \frac{3+\sqrt{5}}{2}$,
$ \frac{3+\sqrt{5}}{2}$,
$ c^{1}_{20}
+c^{3}_{20}
$,
$ c^{1}_{20}
+c^{3}_{20}
$,
$ 1+\sqrt{5}$;\ \ 
$ 1$,
$ -c_{20}^{1}$,
$ -c_{20}^{1}$,
$ \frac{3+\sqrt{5}}{2}$,
$ \frac{3+\sqrt{5}}{2}$,
$ -c^{1}_{20}
-c^{3}_{20}
$,
$ -c^{1}_{20}
-c^{3}_{20}
$,
$ 1+\sqrt{5}$;\ \ 
$ c^{1}_{20}
+c^{3}_{20}
$,
$ -c^{1}_{20}
-c^{3}_{20}
$,
$ -c^{1}_{20}
-c^{3}_{20}
$,
$ c^{1}_{20}
+c^{3}_{20}
$,
$ c_{20}^{1}$,
$ -c_{20}^{1}$,
$0$;\ \ 
$ c^{1}_{20}
+c^{3}_{20}
$,
$ -c^{1}_{20}
-c^{3}_{20}
$,
$ c^{1}_{20}
+c^{3}_{20}
$,
$ -c_{20}^{1}$,
$ c_{20}^{1}$,
$0$;\ \ 
$ 1$,
$ 1$,
$ c_{20}^{1}$,
$ c_{20}^{1}$,
$ -1-\sqrt{5}$;\ \ 
$ 1$,
$ -c_{20}^{1}$,
$ -c_{20}^{1}$,
$ -1-\sqrt{5}$;\ \ 
$ -c^{1}_{20}
-c^{3}_{20}
$,
$ c^{1}_{20}
+c^{3}_{20}
$,
$0$;\ \ 
$ -c^{1}_{20}
-c^{3}_{20}
$,
$0$;\ \ 
$ 1+\sqrt{5}$)

\vskip 0.7ex
\hangindent=3em \hangafter=1
$\tau_n$ = ($-2.24 - 6.88 i$, $-1.6 + 4.56 i$, $-5.85 + 18.02 i$, $-32.9$, $26.18$, $-9.12 - 37.54 i$, $-5.85 - 18.02 i$, $16.72 + 22.27 i$, $-2.24 + 6.88 i$, $26.18 + 35.66 i$, $-2.24 - 6.88 i$, $-28.43 - 13.76 i$, $-5.85 + 18.02 i$, $4.65 - 23.78 i$, $26.18$, $28.43 - 13.76 i$, $-5.85 - 18.02 i$, $-10.11 + 40.59 i$, $-2.24 + 6.88 i$, $-9.48$, $-2.24 - 6.88 i$, $-10.11 - 40.59 i$, $-5.85 + 18.02 i$, $28.43 + 13.76 i$, $26.18$, $4.65 + 23.78 i$, $-5.85 - 18.02 i$, $-28.43 + 13.76 i$, $-2.24 + 6.88 i$, $26.18 - 35.66 i$, $-2.24 - 6.88 i$, $16.72 - 22.27 i$, $-5.85 + 18.02 i$, $-9.12 + 37.54 i$, $26.18$, $-32.9$, $-5.85 - 18.02 i$, $-1.6 - 4.56 i$, $-2.24 + 6.88 i$, $61.84$)

\vskip 0.7ex
\hangindent=3em \hangafter=1
\textit{Intrinsic sign problem}

  \vskip 2ex

\noindent69. $9_{\frac{40}{7},86.41}^{7,313}$ \irep{197}:\ \ 
$d_i$ = ($1.0$,
$1.801$,
$1.801$,
$2.246$,
$2.246$,
$3.246$,
$4.48$,
$4.48$,
$5.48$) 

\vskip 0.7ex
\hangindent=3em \hangafter=1
$D^2= 86.413 = 
49+35c^{1}_{7}
+14c^{2}_{7}
$

\vskip 0.7ex
\hangindent=3em \hangafter=1
$T = ( 0,
\frac{1}{7},
\frac{1}{7},
\frac{5}{7},
\frac{5}{7},
\frac{2}{7},
\frac{6}{7},
\frac{6}{7},
\frac{3}{7} )
$,

\vskip 0.7ex
\hangindent=3em \hangafter=1
$S$ = ($ 1$,
$ -c_{7}^{3}$,
$ -c_{7}^{3}$,
$ \xi_{7}^{3}$,
$ \xi_{7}^{3}$,
$ 2+c^{1}_{7}
$,
$ 2+2c^{1}_{7}
+c^{2}_{7}
$,
$ 2+2c^{1}_{7}
+c^{2}_{7}
$,
$ 3+2c^{1}_{7}
+c^{2}_{7}
$;\ \ 
$ -\xi_{7}^{3}$,
$ 2+c^{1}_{7}
$,
$ 2+2c^{1}_{7}
+c^{2}_{7}
$,
$ 1$,
$ -2-2  c^{1}_{7}
-c^{2}_{7}
$,
$ -3-2  c^{1}_{7}
-c^{2}_{7}
$,
$ -c_{7}^{3}$,
$ \xi_{7}^{3}$;\ \ 
$ -\xi_{7}^{3}$,
$ 1$,
$ 2+2c^{1}_{7}
+c^{2}_{7}
$,
$ -2-2  c^{1}_{7}
-c^{2}_{7}
$,
$ -c_{7}^{3}$,
$ -3-2  c^{1}_{7}
-c^{2}_{7}
$,
$ \xi_{7}^{3}$;\ \ 
$ c_{7}^{3}$,
$ 3+2c^{1}_{7}
+c^{2}_{7}
$,
$ -c_{7}^{3}$,
$ -2-c^{1}_{7}
$,
$ \xi_{7}^{3}$,
$ -2-2  c^{1}_{7}
-c^{2}_{7}
$;\ \ 
$ c_{7}^{3}$,
$ -c_{7}^{3}$,
$ \xi_{7}^{3}$,
$ -2-c^{1}_{7}
$,
$ -2-2  c^{1}_{7}
-c^{2}_{7}
$;\ \ 
$ 3+2c^{1}_{7}
+c^{2}_{7}
$,
$ -\xi_{7}^{3}$,
$ -\xi_{7}^{3}$,
$ 1$;\ \ 
$ 2+2c^{1}_{7}
+c^{2}_{7}
$,
$ 1$,
$ c_{7}^{3}$;\ \ 
$ 2+2c^{1}_{7}
+c^{2}_{7}
$,
$ c_{7}^{3}$;\ \ 
$ 2+c^{1}_{7}
$)

Factors = $3_{\frac{48}{7},9.295}^{7,790}\boxtimes 3_{\frac{48}{7},9.295}^{7,790}$

\vskip 0.7ex
\hangindent=3em \hangafter=1
$\tau_n$ = ($-1.57 - 12.85 i$, $-9.24 - 56.48 i$, $-34.83 + 14.33 i$, $-34.83 - 14.33 i$, $-9.24 + 56.48 i$, $-1.57 + 12.85 i$, $98.28$)

\vskip 0.7ex
\hangindent=3em \hangafter=1
\textit{Intrinsic sign problem}

  \vskip 2ex

\noindent70. $9_{0,86.41}^{7,161}$ \irep{196}:\ \ 
$d_i$ = ($1.0$,
$1.801$,
$1.801$,
$2.246$,
$2.246$,
$3.246$,
$4.48$,
$4.48$,
$5.48$) 

\vskip 0.7ex
\hangindent=3em \hangafter=1
$D^2= 86.413 = 
49+35c^{1}_{7}
+14c^{2}_{7}
$

\vskip 0.7ex
\hangindent=3em \hangafter=1
$T = ( 0,
\frac{1}{7},
\frac{6}{7},
\frac{2}{7},
\frac{5}{7},
0,
\frac{3}{7},
\frac{4}{7},
0 )
$,

\vskip 0.7ex
\hangindent=3em \hangafter=1
$S$ = ($ 1$,
$ -c_{7}^{3}$,
$ -c_{7}^{3}$,
$ \xi_{7}^{3}$,
$ \xi_{7}^{3}$,
$ 2+c^{1}_{7}
$,
$ 2+2c^{1}_{7}
+c^{2}_{7}
$,
$ 2+2c^{1}_{7}
+c^{2}_{7}
$,
$ 3+2c^{1}_{7}
+c^{2}_{7}
$;\ \ 
$ -\xi_{7}^{3}$,
$ 2+c^{1}_{7}
$,
$ 2+2c^{1}_{7}
+c^{2}_{7}
$,
$ 1$,
$ -2-2  c^{1}_{7}
-c^{2}_{7}
$,
$ -3-2  c^{1}_{7}
-c^{2}_{7}
$,
$ -c_{7}^{3}$,
$ \xi_{7}^{3}$;\ \ 
$ -\xi_{7}^{3}$,
$ 1$,
$ 2+2c^{1}_{7}
+c^{2}_{7}
$,
$ -2-2  c^{1}_{7}
-c^{2}_{7}
$,
$ -c_{7}^{3}$,
$ -3-2  c^{1}_{7}
-c^{2}_{7}
$,
$ \xi_{7}^{3}$;\ \ 
$ c_{7}^{3}$,
$ 3+2c^{1}_{7}
+c^{2}_{7}
$,
$ -c_{7}^{3}$,
$ -2-c^{1}_{7}
$,
$ \xi_{7}^{3}$,
$ -2-2  c^{1}_{7}
-c^{2}_{7}
$;\ \ 
$ c_{7}^{3}$,
$ -c_{7}^{3}$,
$ \xi_{7}^{3}$,
$ -2-c^{1}_{7}
$,
$ -2-2  c^{1}_{7}
-c^{2}_{7}
$;\ \ 
$ 3+2c^{1}_{7}
+c^{2}_{7}
$,
$ -\xi_{7}^{3}$,
$ -\xi_{7}^{3}$,
$ 1$;\ \ 
$ 2+2c^{1}_{7}
+c^{2}_{7}
$,
$ 1$,
$ c_{7}^{3}$;\ \ 
$ 2+2c^{1}_{7}
+c^{2}_{7}
$,
$ c_{7}^{3}$;\ \ 
$ 2+c^{1}_{7}
$)

Factors = $3_{\frac{48}{7},9.295}^{7,790}\boxtimes 3_{\frac{8}{7},9.295}^{7,245}$

\vskip 0.7ex
\hangindent=3em \hangafter=1
$\tau_n$ = ($7.2$, $56.06$, $33.08$, $33.08$, $56.06$, $7.2$, $98.28$)

\vskip 0.7ex
\hangindent=3em \hangafter=1
\textit{Undetermined}

  \vskip 2ex

\noindent71. $9_{\frac{16}{7},86.41}^{7,112}$ \irep{197}:\ \ 
$d_i$ = ($1.0$,
$1.801$,
$1.801$,
$2.246$,
$2.246$,
$3.246$,
$4.48$,
$4.48$,
$5.48$) 

\vskip 0.7ex
\hangindent=3em \hangafter=1
$D^2= 86.413 = 
49+35c^{1}_{7}
+14c^{2}_{7}
$

\vskip 0.7ex
\hangindent=3em \hangafter=1
$T = ( 0,
\frac{6}{7},
\frac{6}{7},
\frac{2}{7},
\frac{2}{7},
\frac{5}{7},
\frac{1}{7},
\frac{1}{7},
\frac{4}{7} )
$,

\vskip 0.7ex
\hangindent=3em \hangafter=1
$S$ = ($ 1$,
$ -c_{7}^{3}$,
$ -c_{7}^{3}$,
$ \xi_{7}^{3}$,
$ \xi_{7}^{3}$,
$ 2+c^{1}_{7}
$,
$ 2+2c^{1}_{7}
+c^{2}_{7}
$,
$ 2+2c^{1}_{7}
+c^{2}_{7}
$,
$ 3+2c^{1}_{7}
+c^{2}_{7}
$;\ \ 
$ -\xi_{7}^{3}$,
$ 2+c^{1}_{7}
$,
$ 2+2c^{1}_{7}
+c^{2}_{7}
$,
$ 1$,
$ -2-2  c^{1}_{7}
-c^{2}_{7}
$,
$ -3-2  c^{1}_{7}
-c^{2}_{7}
$,
$ -c_{7}^{3}$,
$ \xi_{7}^{3}$;\ \ 
$ -\xi_{7}^{3}$,
$ 1$,
$ 2+2c^{1}_{7}
+c^{2}_{7}
$,
$ -2-2  c^{1}_{7}
-c^{2}_{7}
$,
$ -c_{7}^{3}$,
$ -3-2  c^{1}_{7}
-c^{2}_{7}
$,
$ \xi_{7}^{3}$;\ \ 
$ c_{7}^{3}$,
$ 3+2c^{1}_{7}
+c^{2}_{7}
$,
$ -c_{7}^{3}$,
$ -2-c^{1}_{7}
$,
$ \xi_{7}^{3}$,
$ -2-2  c^{1}_{7}
-c^{2}_{7}
$;\ \ 
$ c_{7}^{3}$,
$ -c_{7}^{3}$,
$ \xi_{7}^{3}$,
$ -2-c^{1}_{7}
$,
$ -2-2  c^{1}_{7}
-c^{2}_{7}
$;\ \ 
$ 3+2c^{1}_{7}
+c^{2}_{7}
$,
$ -\xi_{7}^{3}$,
$ -\xi_{7}^{3}$,
$ 1$;\ \ 
$ 2+2c^{1}_{7}
+c^{2}_{7}
$,
$ 1$,
$ c_{7}^{3}$;\ \ 
$ 2+2c^{1}_{7}
+c^{2}_{7}
$,
$ c_{7}^{3}$;\ \ 
$ 2+c^{1}_{7}
$)

Factors = $3_{\frac{8}{7},9.295}^{7,245}\boxtimes 3_{\frac{8}{7},9.295}^{7,245}$

\vskip 0.7ex
\hangindent=3em \hangafter=1
$\tau_n$ = ($-1.57 + 12.85 i$, $-9.24 + 56.48 i$, $-34.83 - 14.33 i$, $-34.83 + 14.33 i$, $-9.24 - 56.48 i$, $-1.57 - 12.85 i$, $98.28$)

\vskip 0.7ex
\hangindent=3em \hangafter=1
\textit{Intrinsic sign problem}

  \vskip 2ex

\noindent72. $9_{\frac{120}{19},175.3}^{19,574}$ \irep{414}:\ \ 
$d_i$ = ($1.0$,
$1.972$,
$2.891$,
$3.731$,
$4.469$,
$5.86$,
$5.563$,
$5.889$,
$6.54$) 

\vskip 0.7ex
\hangindent=3em \hangafter=1
$D^2= 175.332 = 
45+36c^{1}_{19}
+28c^{2}_{19}
+21c^{3}_{19}
+15c^{4}_{19}
+10c^{5}_{19}
+6c^{6}_{19}
+3c^{7}_{19}
+c^{8}_{19}
$

\vskip 0.7ex
\hangindent=3em \hangafter=1
$T = ( 0,
\frac{4}{19},
\frac{17}{19},
\frac{1}{19},
\frac{13}{19},
\frac{15}{19},
\frac{7}{19},
\frac{8}{19},
\frac{18}{19} )
$,

\vskip 0.7ex
\hangindent=3em \hangafter=1
$S$ = ($ 1$,
$ -c_{19}^{9}$,
$ \xi_{19}^{3}$,
$ \xi_{19}^{15}$,
$ \xi_{19}^{5}$,
$ \xi_{19}^{13}$,
$ \xi_{19}^{7}$,
$ \xi_{19}^{11}$,
$ \xi_{19}^{9}$;\ \ 
$ -\xi_{19}^{15}$,
$ \xi_{19}^{13}$,
$ -\xi_{19}^{11}$,
$ \xi_{19}^{9}$,
$ -\xi_{19}^{7}$,
$ \xi_{19}^{5}$,
$ -\xi_{19}^{3}$,
$ 1$;\ \ 
$ \xi_{19}^{9}$,
$ \xi_{19}^{7}$,
$ \xi_{19}^{15}$,
$ 1$,
$ c_{19}^{9}$,
$ -\xi_{19}^{5}$,
$ -\xi_{19}^{11}$;\ \ 
$ -\xi_{19}^{3}$,
$ -1$,
$ \xi_{19}^{5}$,
$ -\xi_{19}^{9}$,
$ \xi_{19}^{13}$,
$ c_{19}^{9}$;\ \ 
$ -\xi_{19}^{13}$,
$ -\xi_{19}^{11}$,
$ -\xi_{19}^{3}$,
$ -c_{19}^{9}$,
$ \xi_{19}^{7}$;\ \ 
$ -c_{19}^{9}$,
$ \xi_{19}^{15}$,
$ -\xi_{19}^{9}$,
$ \xi_{19}^{3}$;\ \ 
$ \xi_{19}^{11}$,
$ 1$,
$ -\xi_{19}^{13}$;\ \ 
$ \xi_{19}^{7}$,
$ -\xi_{19}^{15}$;\ \ 
$ \xi_{19}^{5}$)

\vskip 0.7ex
\hangindent=3em \hangafter=1
$\tau_n$ = ($11.12 - 23.04 i$, $17.05 - 85.5 i$, $43.21 + 50.65 i$, $-17.63 - 53.02 i$, $50.96 - 32.37 i$, $-41.44 - 14.06 i$, $-65.31 - 63.26 i$, $-73.56 + 39.91 i$, $-9.84 + 28.11 i$, $-9.84 - 28.11 i$, $-73.56 - 39.91 i$, $-65.31 + 63.26 i$, $-41.44 + 14.06 i$, $50.96 + 32.37 i$, $-17.63 + 53.02 i$, $43.21 - 50.65 i$, $17.05 + 85.5 i$, $11.12 + 23.04 i$, $189.88$)

\vskip 0.7ex
\hangindent=3em \hangafter=1
\textit{Intrinsic sign problem}

  \vskip 2ex

\noindent73. $9_{\frac{32}{19},175.3}^{19,327}$ \irep{414}:\ \ 
$d_i$ = ($1.0$,
$1.972$,
$2.891$,
$3.731$,
$4.469$,
$5.86$,
$5.563$,
$5.889$,
$6.54$) 

\vskip 0.7ex
\hangindent=3em \hangafter=1
$D^2= 175.332 = 
45+36c^{1}_{19}
+28c^{2}_{19}
+21c^{3}_{19}
+15c^{4}_{19}
+10c^{5}_{19}
+6c^{6}_{19}
+3c^{7}_{19}
+c^{8}_{19}
$

\vskip 0.7ex
\hangindent=3em \hangafter=1
$T = ( 0,
\frac{15}{19},
\frac{2}{19},
\frac{18}{19},
\frac{6}{19},
\frac{4}{19},
\frac{12}{19},
\frac{11}{19},
\frac{1}{19} )
$,

\vskip 0.7ex
\hangindent=3em \hangafter=1
$S$ = ($ 1$,
$ -c_{19}^{9}$,
$ \xi_{19}^{3}$,
$ \xi_{19}^{15}$,
$ \xi_{19}^{5}$,
$ \xi_{19}^{13}$,
$ \xi_{19}^{7}$,
$ \xi_{19}^{11}$,
$ \xi_{19}^{9}$;\ \ 
$ -\xi_{19}^{15}$,
$ \xi_{19}^{13}$,
$ -\xi_{19}^{11}$,
$ \xi_{19}^{9}$,
$ -\xi_{19}^{7}$,
$ \xi_{19}^{5}$,
$ -\xi_{19}^{3}$,
$ 1$;\ \ 
$ \xi_{19}^{9}$,
$ \xi_{19}^{7}$,
$ \xi_{19}^{15}$,
$ 1$,
$ c_{19}^{9}$,
$ -\xi_{19}^{5}$,
$ -\xi_{19}^{11}$;\ \ 
$ -\xi_{19}^{3}$,
$ -1$,
$ \xi_{19}^{5}$,
$ -\xi_{19}^{9}$,
$ \xi_{19}^{13}$,
$ c_{19}^{9}$;\ \ 
$ -\xi_{19}^{13}$,
$ -\xi_{19}^{11}$,
$ -\xi_{19}^{3}$,
$ -c_{19}^{9}$,
$ \xi_{19}^{7}$;\ \ 
$ -c_{19}^{9}$,
$ \xi_{19}^{15}$,
$ -\xi_{19}^{9}$,
$ \xi_{19}^{3}$;\ \ 
$ \xi_{19}^{11}$,
$ 1$,
$ -\xi_{19}^{13}$;\ \ 
$ \xi_{19}^{7}$,
$ -\xi_{19}^{15}$;\ \ 
$ \xi_{19}^{5}$)

\vskip 0.7ex
\hangindent=3em \hangafter=1
$\tau_n$ = ($11.12 + 23.04 i$, $17.05 + 85.5 i$, $43.21 - 50.65 i$, $-17.63 + 53.02 i$, $50.96 + 32.37 i$, $-41.44 + 14.06 i$, $-65.31 + 63.26 i$, $-73.56 - 39.91 i$, $-9.84 - 28.11 i$, $-9.84 + 28.11 i$, $-73.56 + 39.91 i$, $-65.31 - 63.26 i$, $-41.44 - 14.06 i$, $50.96 - 32.37 i$, $-17.63 - 53.02 i$, $43.21 + 50.65 i$, $17.05 - 85.5 i$, $11.12 - 23.04 i$, $189.88$)

\vskip 0.7ex
\hangindent=3em \hangafter=1
\textit{Intrinsic sign problem}

  \vskip 2ex

\noindent74. $9_{\frac{14}{5},343.2}^{15,715}$ \irep{361}:\ \ 
$d_i$ = ($1.0$,
$2.956$,
$4.783$,
$4.783$,
$4.783$,
$6.401$,
$7.739$,
$8.739$,
$9.357$) 

\vskip 0.7ex
\hangindent=3em \hangafter=1
$D^2= 343.211 = 
105+45c^{1}_{15}
+75c^{2}_{15}
+90c^{3}_{15}
$

\vskip 0.7ex
\hangindent=3em \hangafter=1
$T = ( 0,
\frac{1}{15},
\frac{1}{5},
\frac{13}{15},
\frac{13}{15},
\frac{2}{5},
\frac{2}{3},
0,
\frac{2}{5} )
$,

\vskip 0.7ex
\hangindent=3em \hangafter=1
$S$ = ($ 1$,
$ 1+c^{2}_{15}
+c^{3}_{15}
$,
$ \xi_{15}^{7}$,
$ \xi_{15}^{7}$,
$ \xi_{15}^{7}$,
$ 2+c^{1}_{15}
+c^{2}_{15}
+2c^{3}_{15}
$,
$ 2+c^{1}_{15}
+2c^{2}_{15}
+2c^{3}_{15}
$,
$ 3+c^{1}_{15}
+2c^{2}_{15}
+2c^{3}_{15}
$,
$ 3+c^{1}_{15}
+2c^{2}_{15}
+3c^{3}_{15}
$;\ \ 
$ 2+c^{1}_{15}
+2c^{2}_{15}
+2c^{3}_{15}
$,
$ 2\xi_{15}^{7}$,
$ -\xi_{15}^{7}$,
$ -\xi_{15}^{7}$,
$ 2+c^{1}_{15}
+2c^{2}_{15}
+2c^{3}_{15}
$,
$ 1+c^{2}_{15}
+c^{3}_{15}
$,
$ -1-c^{2}_{15}
-c^{3}_{15}
$,
$ -2-c^{1}_{15}
-2  c^{2}_{15}
-2  c^{3}_{15}
$;\ \ 
$ \xi_{15}^{7}$,
$ \xi_{15}^{7}$,
$ \xi_{15}^{7}$,
$ -\xi_{15}^{7}$,
$ -2\xi_{15}^{7}$,
$ -\xi_{15}^{7}$,
$ \xi_{15}^{7}$;\ \ 
$ 2-4  \zeta^{1}_{15}
-2  \zeta^{-1}_{15}
+\zeta^{2}_{15}
-2  \zeta^{-2}_{15}
-\zeta^{3}_{15}
+5\zeta^{-3}_{15}
-5  \zeta^{4}_{15}
$,
$ -3+3\zeta^{1}_{15}
+\zeta^{-1}_{15}
-2  \zeta^{2}_{15}
+\zeta^{-2}_{15}
-6  \zeta^{-3}_{15}
+5\zeta^{4}_{15}
$,
$ -\xi_{15}^{7}$,
$ \xi_{15}^{7}$,
$ -\xi_{15}^{7}$,
$ \xi_{15}^{7}$;\ \ 
$ 2-4  \zeta^{1}_{15}
-2  \zeta^{-1}_{15}
+\zeta^{2}_{15}
-2  \zeta^{-2}_{15}
-\zeta^{3}_{15}
+5\zeta^{-3}_{15}
-5  \zeta^{4}_{15}
$,
$ -\xi_{15}^{7}$,
$ \xi_{15}^{7}$,
$ -\xi_{15}^{7}$,
$ \xi_{15}^{7}$;\ \ 
$ -3-c^{1}_{15}
-2  c^{2}_{15}
-2  c^{3}_{15}
$,
$ 1+c^{2}_{15}
+c^{3}_{15}
$,
$ 3+c^{1}_{15}
+2c^{2}_{15}
+3c^{3}_{15}
$,
$ -1$;\ \ 
$ 2+c^{1}_{15}
+2c^{2}_{15}
+2c^{3}_{15}
$,
$ -2-c^{1}_{15}
-2  c^{2}_{15}
-2  c^{3}_{15}
$,
$ -1-c^{2}_{15}
-c^{3}_{15}
$;\ \ 
$ 1$,
$ 2+c^{1}_{15}
+c^{2}_{15}
+2c^{3}_{15}
$;\ \ 
$ -3-c^{1}_{15}
-2  c^{2}_{15}
-2  c^{3}_{15}
$)

\vskip 0.7ex
\hangindent=3em \hangafter=1
$\tau_n$ = ($-10.89 + 14.99 i$, $69.7 - 95.93 i$, $124.16 + 90.21 i$, $-95.15 - 130.97 i$, $171.58 + 99.06 i$, $47.42 + 145.95 i$, $101.88 - 140.23 i$, $101.88 + 140.23 i$, $47.42 - 145.95 i$, $171.58 - 99.06 i$, $-95.15 + 130.97 i$, $124.16 - 90.21 i$, $69.7 + 95.93 i$, $-10.89 - 14.99 i$, $343.16$)

\vskip 0.7ex
\hangindent=3em \hangafter=1
\textit{Intrinsic sign problem}

  \vskip 2ex

\noindent75. $9_{\frac{26}{5},343.2}^{15,296}$ \irep{361}:\ \ 
$d_i$ = ($1.0$,
$2.956$,
$4.783$,
$4.783$,
$4.783$,
$6.401$,
$7.739$,
$8.739$,
$9.357$) 

\vskip 0.7ex
\hangindent=3em \hangafter=1
$D^2= 343.211 = 
105+45c^{1}_{15}
+75c^{2}_{15}
+90c^{3}_{15}
$

\vskip 0.7ex
\hangindent=3em \hangafter=1
$T = ( 0,
\frac{14}{15},
\frac{4}{5},
\frac{2}{15},
\frac{2}{15},
\frac{3}{5},
\frac{1}{3},
0,
\frac{3}{5} )
$,

\vskip 0.7ex
\hangindent=3em \hangafter=1
$S$ = ($ 1$,
$ 1+c^{2}_{15}
+c^{3}_{15}
$,
$ \xi_{15}^{7}$,
$ \xi_{15}^{7}$,
$ \xi_{15}^{7}$,
$ 2+c^{1}_{15}
+c^{2}_{15}
+2c^{3}_{15}
$,
$ 2+c^{1}_{15}
+2c^{2}_{15}
+2c^{3}_{15}
$,
$ 3+c^{1}_{15}
+2c^{2}_{15}
+2c^{3}_{15}
$,
$ 3+c^{1}_{15}
+2c^{2}_{15}
+3c^{3}_{15}
$;\ \ 
$ 2+c^{1}_{15}
+2c^{2}_{15}
+2c^{3}_{15}
$,
$ 2\xi_{15}^{7}$,
$ -\xi_{15}^{7}$,
$ -\xi_{15}^{7}$,
$ 2+c^{1}_{15}
+2c^{2}_{15}
+2c^{3}_{15}
$,
$ 1+c^{2}_{15}
+c^{3}_{15}
$,
$ -1-c^{2}_{15}
-c^{3}_{15}
$,
$ -2-c^{1}_{15}
-2  c^{2}_{15}
-2  c^{3}_{15}
$;\ \ 
$ \xi_{15}^{7}$,
$ \xi_{15}^{7}$,
$ \xi_{15}^{7}$,
$ -\xi_{15}^{7}$,
$ -2\xi_{15}^{7}$,
$ -\xi_{15}^{7}$,
$ \xi_{15}^{7}$;\ \ 
$ -3+3\zeta^{1}_{15}
+\zeta^{-1}_{15}
-2  \zeta^{2}_{15}
+\zeta^{-2}_{15}
-6  \zeta^{-3}_{15}
+5\zeta^{4}_{15}
$,
$ 2-4  \zeta^{1}_{15}
-2  \zeta^{-1}_{15}
+\zeta^{2}_{15}
-2  \zeta^{-2}_{15}
-\zeta^{3}_{15}
+5\zeta^{-3}_{15}
-5  \zeta^{4}_{15}
$,
$ -\xi_{15}^{7}$,
$ \xi_{15}^{7}$,
$ -\xi_{15}^{7}$,
$ \xi_{15}^{7}$;\ \ 
$ -3+3\zeta^{1}_{15}
+\zeta^{-1}_{15}
-2  \zeta^{2}_{15}
+\zeta^{-2}_{15}
-6  \zeta^{-3}_{15}
+5\zeta^{4}_{15}
$,
$ -\xi_{15}^{7}$,
$ \xi_{15}^{7}$,
$ -\xi_{15}^{7}$,
$ \xi_{15}^{7}$;\ \ 
$ -3-c^{1}_{15}
-2  c^{2}_{15}
-2  c^{3}_{15}
$,
$ 1+c^{2}_{15}
+c^{3}_{15}
$,
$ 3+c^{1}_{15}
+2c^{2}_{15}
+3c^{3}_{15}
$,
$ -1$;\ \ 
$ 2+c^{1}_{15}
+2c^{2}_{15}
+2c^{3}_{15}
$,
$ -2-c^{1}_{15}
-2  c^{2}_{15}
-2  c^{3}_{15}
$,
$ -1-c^{2}_{15}
-c^{3}_{15}
$;\ \ 
$ 1$,
$ 2+c^{1}_{15}
+c^{2}_{15}
+2c^{3}_{15}
$;\ \ 
$ -3-c^{1}_{15}
-2  c^{2}_{15}
-2  c^{3}_{15}
$)

\vskip 0.7ex
\hangindent=3em \hangafter=1
$\tau_n$ = ($-10.89 - 14.99 i$, $69.7 + 95.93 i$, $124.16 - 90.21 i$, $-95.15 + 130.97 i$, $171.58 - 99.06 i$, $47.42 - 145.95 i$, $101.88 + 140.23 i$, $101.88 - 140.23 i$, $47.42 + 145.95 i$, $171.58 + 99.06 i$, $-95.15 - 130.97 i$, $124.16 + 90.21 i$, $69.7 - 95.93 i$, $-10.89 + 14.99 i$, $343.16$)

\vskip 0.7ex
\hangindent=3em \hangafter=1
\textit{Intrinsic sign problem}

  \vskip 2ex

\noindent76. $9_{7,475.1}^{24,793}$ \irep{448}:\ \ 
$d_i$ = ($1.0$,
$4.449$,
$4.449$,
$5.449$,
$5.449$,
$8.898$,
$8.898$,
$9.898$,
$10.898$) 

\vskip 0.7ex
\hangindent=3em \hangafter=1
$D^2= 475.151 = 
240+96\sqrt{6}$

\vskip 0.7ex
\hangindent=3em \hangafter=1
$T = ( 0,
\frac{1}{4},
\frac{1}{4},
\frac{1}{2},
\frac{1}{2},
\frac{1}{3},
\frac{7}{12},
0,
\frac{7}{8} )
$,

\vskip 0.7ex
\hangindent=3em \hangafter=1
$S$ = ($ 1$,
$ 2+\sqrt{6}$,
$ 2+\sqrt{6}$,
$ 3+\sqrt{6}$,
$ 3+\sqrt{6}$,
$ 4+2\sqrt{6}$,
$ 4+2\sqrt{6}$,
$ 5+2\sqrt{6}$,
$ 6+2\sqrt{6}$;\ \ 
$ 2\xi_{24}^{7}$,
$ 2+2c^{1}_{24}
-2  c^{2}_{24}
-4  c^{3}_{24}
$,
$ -2  c^{2}_{24}
-3  c^{3}_{24}
$,
$ 2c^{2}_{24}
+3c^{3}_{24}
$,
$ -4-2\sqrt{6}$,
$ 4+2\sqrt{6}$,
$ -2-\sqrt{6}$,
$0$;\ \ 
$ 2\xi_{24}^{7}$,
$ 2c^{2}_{24}
+3c^{3}_{24}
$,
$ -2  c^{2}_{24}
-3  c^{3}_{24}
$,
$ -4-2\sqrt{6}$,
$ 4+2\sqrt{6}$,
$ -2-\sqrt{6}$,
$0$;\ \ 
$ 3+2c^{1}_{24}
-2  c^{2}_{24}
-4  c^{3}_{24}
$,
$ 3+2c^{1}_{24}
+2c^{2}_{24}
+2c^{3}_{24}
$,
$0$,
$0$,
$ 3+\sqrt{6}$,
$ -6-2\sqrt{6}$;\ \ 
$ 3+2c^{1}_{24}
-2  c^{2}_{24}
-4  c^{3}_{24}
$,
$0$,
$0$,
$ 3+\sqrt{6}$,
$ -6-2\sqrt{6}$;\ \ 
$ 4+2\sqrt{6}$,
$ 4+2\sqrt{6}$,
$ -4-2\sqrt{6}$,
$0$;\ \ 
$ -4-2\sqrt{6}$,
$ -4-2\sqrt{6}$,
$0$;\ \ 
$ 1$,
$ 6+2\sqrt{6}$;\ \ 
$0$)

\vskip 0.7ex
\hangindent=3em \hangafter=1
$\tau_n$ = ($15.41 - 15.41 i$, $118.77 - 118.77 i$, $34.78 - 202.74 i$, $0. + 137.13 i$, $-15.41 + 15.41 i$, $118.77 + 118.77 i$, $152.55 + 152.55 i$, $237.53 - 137.13 i$, $202.74 + 34.78 i$, $118.77 - 118.77 i$, $-152.55 - 152.55 i$, $237.52$, $-152.55 + 152.55 i$, $118.77 + 118.77 i$, $202.74 - 34.78 i$, $237.53 + 137.13 i$, $152.55 - 152.55 i$, $118.77 - 118.77 i$, $-15.41 - 15.41 i$, $0. - 137.13 i$, $34.78 + 202.74 i$, $118.77 + 118.77 i$, $15.41 + 15.41 i$, $475.06$)

\vskip 0.7ex
\hangindent=3em \hangafter=1
\textit{Intrinsic sign problem}

  \vskip 2ex

\noindent77. $9_{3,475.1}^{24,139}$ \irep{448}:\ \ 
$d_i$ = ($1.0$,
$4.449$,
$4.449$,
$5.449$,
$5.449$,
$8.898$,
$8.898$,
$9.898$,
$10.898$) 

\vskip 0.7ex
\hangindent=3em \hangafter=1
$D^2= 475.151 = 
240+96\sqrt{6}$

\vskip 0.7ex
\hangindent=3em \hangafter=1
$T = ( 0,
\frac{1}{4},
\frac{1}{4},
\frac{1}{2},
\frac{1}{2},
\frac{2}{3},
\frac{11}{12},
0,
\frac{3}{8} )
$,

\vskip 0.7ex
\hangindent=3em \hangafter=1
$S$ = ($ 1$,
$ 2+\sqrt{6}$,
$ 2+\sqrt{6}$,
$ 3+\sqrt{6}$,
$ 3+\sqrt{6}$,
$ 4+2\sqrt{6}$,
$ 4+2\sqrt{6}$,
$ 5+2\sqrt{6}$,
$ 6+2\sqrt{6}$;\ \ 
$ 2+2c^{1}_{24}
-2  c^{2}_{24}
-4  c^{3}_{24}
$,
$ 2\xi_{24}^{7}$,
$ -2  c^{2}_{24}
-3  c^{3}_{24}
$,
$ 2c^{2}_{24}
+3c^{3}_{24}
$,
$ -4-2\sqrt{6}$,
$ 4+2\sqrt{6}$,
$ -2-\sqrt{6}$,
$0$;\ \ 
$ 2+2c^{1}_{24}
-2  c^{2}_{24}
-4  c^{3}_{24}
$,
$ 2c^{2}_{24}
+3c^{3}_{24}
$,
$ -2  c^{2}_{24}
-3  c^{3}_{24}
$,
$ -4-2\sqrt{6}$,
$ 4+2\sqrt{6}$,
$ -2-\sqrt{6}$,
$0$;\ \ 
$ 3+2c^{1}_{24}
+2c^{2}_{24}
+2c^{3}_{24}
$,
$ 3+2c^{1}_{24}
-2  c^{2}_{24}
-4  c^{3}_{24}
$,
$0$,
$0$,
$ 3+\sqrt{6}$,
$ -6-2\sqrt{6}$;\ \ 
$ 3+2c^{1}_{24}
+2c^{2}_{24}
+2c^{3}_{24}
$,
$0$,
$0$,
$ 3+\sqrt{6}$,
$ -6-2\sqrt{6}$;\ \ 
$ 4+2\sqrt{6}$,
$ 4+2\sqrt{6}$,
$ -4-2\sqrt{6}$,
$0$;\ \ 
$ -4-2\sqrt{6}$,
$ -4-2\sqrt{6}$,
$0$;\ \ 
$ 1$,
$ 6+2\sqrt{6}$;\ \ 
$0$)

\vskip 0.7ex
\hangindent=3em \hangafter=1
$\tau_n$ = ($-15.41 + 15.41 i$, $118.77 - 118.77 i$, $202.74 - 34.78 i$, $0. - 137.13 i$, $15.41 - 15.41 i$, $118.77 + 118.77 i$, $-152.55 - 152.55 i$, $237.53 + 137.13 i$, $34.78 + 202.74 i$, $118.77 - 118.77 i$, $152.55 + 152.55 i$, $237.52$, $152.55 - 152.55 i$, $118.77 + 118.77 i$, $34.78 - 202.74 i$, $237.53 - 137.13 i$, $-152.55 + 152.55 i$, $118.77 - 118.77 i$, $15.41 + 15.41 i$, $0. + 137.13 i$, $202.74 + 34.78 i$, $118.77 + 118.77 i$, $-15.41 - 15.41 i$, $475.06$)

\vskip 0.7ex
\hangindent=3em \hangafter=1
\textit{Intrinsic sign problem}

  \vskip 2ex

\noindent78. $9_{5,475.1}^{24,181}$ \irep{448}:\ \ 
$d_i$ = ($1.0$,
$4.449$,
$4.449$,
$5.449$,
$5.449$,
$8.898$,
$8.898$,
$9.898$,
$10.898$) 

\vskip 0.7ex
\hangindent=3em \hangafter=1
$D^2= 475.151 = 
240+96\sqrt{6}$

\vskip 0.7ex
\hangindent=3em \hangafter=1
$T = ( 0,
\frac{3}{4},
\frac{3}{4},
\frac{1}{2},
\frac{1}{2},
\frac{1}{3},
\frac{1}{12},
0,
\frac{5}{8} )
$,

\vskip 0.7ex
\hangindent=3em \hangafter=1
$S$ = ($ 1$,
$ 2+\sqrt{6}$,
$ 2+\sqrt{6}$,
$ 3+\sqrt{6}$,
$ 3+\sqrt{6}$,
$ 4+2\sqrt{6}$,
$ 4+2\sqrt{6}$,
$ 5+2\sqrt{6}$,
$ 6+2\sqrt{6}$;\ \ 
$ 2+2c^{1}_{24}
-2  c^{2}_{24}
-4  c^{3}_{24}
$,
$ 2\xi_{24}^{7}$,
$ -2  c^{2}_{24}
-3  c^{3}_{24}
$,
$ 2c^{2}_{24}
+3c^{3}_{24}
$,
$ -4-2\sqrt{6}$,
$ 4+2\sqrt{6}$,
$ -2-\sqrt{6}$,
$0$;\ \ 
$ 2+2c^{1}_{24}
-2  c^{2}_{24}
-4  c^{3}_{24}
$,
$ 2c^{2}_{24}
+3c^{3}_{24}
$,
$ -2  c^{2}_{24}
-3  c^{3}_{24}
$,
$ -4-2\sqrt{6}$,
$ 4+2\sqrt{6}$,
$ -2-\sqrt{6}$,
$0$;\ \ 
$ 3+2c^{1}_{24}
+2c^{2}_{24}
+2c^{3}_{24}
$,
$ 3+2c^{1}_{24}
-2  c^{2}_{24}
-4  c^{3}_{24}
$,
$0$,
$0$,
$ 3+\sqrt{6}$,
$ -6-2\sqrt{6}$;\ \ 
$ 3+2c^{1}_{24}
+2c^{2}_{24}
+2c^{3}_{24}
$,
$0$,
$0$,
$ 3+\sqrt{6}$,
$ -6-2\sqrt{6}$;\ \ 
$ 4+2\sqrt{6}$,
$ 4+2\sqrt{6}$,
$ -4-2\sqrt{6}$,
$0$;\ \ 
$ -4-2\sqrt{6}$,
$ -4-2\sqrt{6}$,
$0$;\ \ 
$ 1$,
$ 6+2\sqrt{6}$;\ \ 
$0$)

\vskip 0.7ex
\hangindent=3em \hangafter=1
$\tau_n$ = ($-15.41 - 15.41 i$, $118.77 + 118.77 i$, $202.74 + 34.78 i$, $0. + 137.13 i$, $15.41 + 15.41 i$, $118.77 - 118.77 i$, $-152.55 + 152.55 i$, $237.53 - 137.13 i$, $34.78 - 202.74 i$, $118.77 + 118.77 i$, $152.55 - 152.55 i$, $237.52$, $152.55 + 152.55 i$, $118.77 - 118.77 i$, $34.78 + 202.74 i$, $237.53 + 137.13 i$, $-152.55 - 152.55 i$, $118.77 + 118.77 i$, $15.41 - 15.41 i$, $0. - 137.13 i$, $202.74 - 34.78 i$, $118.77 - 118.77 i$, $-15.41 + 15.41 i$, $475.06$)

\vskip 0.7ex
\hangindent=3em \hangafter=1
\textit{Intrinsic sign problem}

  \vskip 2ex

\noindent79. $9_{1,475.1}^{24,144}$ \irep{448}:\ \ 
$d_i$ = ($1.0$,
$4.449$,
$4.449$,
$5.449$,
$5.449$,
$8.898$,
$8.898$,
$9.898$,
$10.898$) 

\vskip 0.7ex
\hangindent=3em \hangafter=1
$D^2= 475.151 = 
240+96\sqrt{6}$

\vskip 0.7ex
\hangindent=3em \hangafter=1
$T = ( 0,
\frac{3}{4},
\frac{3}{4},
\frac{1}{2},
\frac{1}{2},
\frac{2}{3},
\frac{5}{12},
0,
\frac{1}{8} )
$,

\vskip 0.7ex
\hangindent=3em \hangafter=1
$S$ = ($ 1$,
$ 2+\sqrt{6}$,
$ 2+\sqrt{6}$,
$ 3+\sqrt{6}$,
$ 3+\sqrt{6}$,
$ 4+2\sqrt{6}$,
$ 4+2\sqrt{6}$,
$ 5+2\sqrt{6}$,
$ 6+2\sqrt{6}$;\ \ 
$ 2\xi_{24}^{7}$,
$ 2+2c^{1}_{24}
-2  c^{2}_{24}
-4  c^{3}_{24}
$,
$ -2  c^{2}_{24}
-3  c^{3}_{24}
$,
$ 2c^{2}_{24}
+3c^{3}_{24}
$,
$ -4-2\sqrt{6}$,
$ 4+2\sqrt{6}$,
$ -2-\sqrt{6}$,
$0$;\ \ 
$ 2\xi_{24}^{7}$,
$ 2c^{2}_{24}
+3c^{3}_{24}
$,
$ -2  c^{2}_{24}
-3  c^{3}_{24}
$,
$ -4-2\sqrt{6}$,
$ 4+2\sqrt{6}$,
$ -2-\sqrt{6}$,
$0$;\ \ 
$ 3+2c^{1}_{24}
-2  c^{2}_{24}
-4  c^{3}_{24}
$,
$ 3+2c^{1}_{24}
+2c^{2}_{24}
+2c^{3}_{24}
$,
$0$,
$0$,
$ 3+\sqrt{6}$,
$ -6-2\sqrt{6}$;\ \ 
$ 3+2c^{1}_{24}
-2  c^{2}_{24}
-4  c^{3}_{24}
$,
$0$,
$0$,
$ 3+\sqrt{6}$,
$ -6-2\sqrt{6}$;\ \ 
$ 4+2\sqrt{6}$,
$ 4+2\sqrt{6}$,
$ -4-2\sqrt{6}$,
$0$;\ \ 
$ -4-2\sqrt{6}$,
$ -4-2\sqrt{6}$,
$0$;\ \ 
$ 1$,
$ 6+2\sqrt{6}$;\ \ 
$0$)

\vskip 0.7ex
\hangindent=3em \hangafter=1
$\tau_n$ = ($15.41 + 15.41 i$, $118.77 + 118.77 i$, $34.78 + 202.74 i$, $0. - 137.13 i$, $-15.41 - 15.41 i$, $118.77 - 118.77 i$, $152.55 - 152.55 i$, $237.53 + 137.13 i$, $202.74 - 34.78 i$, $118.77 + 118.77 i$, $-152.55 + 152.55 i$, $237.52$, $-152.55 - 152.55 i$, $118.77 - 118.77 i$, $202.74 + 34.78 i$, $237.53 - 137.13 i$, $152.55 + 152.55 i$, $118.77 + 118.77 i$, $-15.41 + 15.41 i$, $0. + 137.13 i$, $34.78 - 202.74 i$, $118.77 - 118.77 i$, $15.41 - 15.41 i$, $475.06$)

\vskip 0.7ex
\hangindent=3em \hangafter=1
\textit{Intrinsic sign problem}

  \vskip 2ex

\noindent80. $9_{6,668.5}^{12,567}$ \irep{330}:\ \ 
$d_i$ = ($1.0$,
$6.464$,
$6.464$,
$6.464$,
$6.464$,
$6.464$,
$6.464$,
$13.928$,
$14.928$) 

\vskip 0.7ex
\hangindent=3em \hangafter=1
$D^2= 668.553 = 
336+192\sqrt{3}$

\vskip 0.7ex
\hangindent=3em \hangafter=1
$T = ( 0,
\frac{1}{2},
\frac{1}{2},
\frac{1}{4},
\frac{1}{4},
\frac{1}{4},
\frac{1}{4},
0,
\frac{2}{3} )
$,

\vskip 0.7ex
\hangindent=3em \hangafter=1
$S$ = ($ 1$,
$ 3+2\sqrt{3}$,
$ 3+2\sqrt{3}$,
$ 3+2\sqrt{3}$,
$ 3+2\sqrt{3}$,
$ 3+2\sqrt{3}$,
$ 3+2\sqrt{3}$,
$ 7+4\sqrt{3}$,
$ 8+4\sqrt{3}$;\ \ 
$ -7+4\zeta^{1}_{12}
-8  \zeta^{-1}_{12}
+8\zeta^{2}_{12}
$,
$ 1-8  \zeta^{1}_{12}
+4\zeta^{-1}_{12}
-8  \zeta^{2}_{12}
$,
$ 3+2\sqrt{3}$,
$ 3+2\sqrt{3}$,
$ 3+2\sqrt{3}$,
$ 3+2\sqrt{3}$,
$ -3-2\sqrt{3}$,
$0$;\ \ 
$ -7+4\zeta^{1}_{12}
-8  \zeta^{-1}_{12}
+8\zeta^{2}_{12}
$,
$ 3+2\sqrt{3}$,
$ 3+2\sqrt{3}$,
$ 3+2\sqrt{3}$,
$ 3+2\sqrt{3}$,
$ -3-2\sqrt{3}$,
$0$;\ \ 
$ 9+6\sqrt{3}$,
$ -3-2\sqrt{3}$,
$ -3-2\sqrt{3}$,
$ -3-2\sqrt{3}$,
$ -3-2\sqrt{3}$,
$0$;\ \ 
$ 9+6\sqrt{3}$,
$ -3-2\sqrt{3}$,
$ -3-2\sqrt{3}$,
$ -3-2\sqrt{3}$,
$0$;\ \ 
$ 9+6\sqrt{3}$,
$ -3-2\sqrt{3}$,
$ -3-2\sqrt{3}$,
$0$;\ \ 
$ 9+6\sqrt{3}$,
$ -3-2\sqrt{3}$,
$0$;\ \ 
$ 1$,
$ 8+4\sqrt{3}$;\ \ 
$ -8-4\sqrt{3}$)

\vskip 0.7ex
\hangindent=3em \hangafter=1
$\tau_n$ = ($0. - 25.86 i$, $0. + 192.99 i$, $334.27 - 167.13 i$, $334.27 - 192.99 i$, $0. + 360.12 i$, $334.27$, $0. - 360.12 i$, $334.27 + 192.99 i$, $334.27 + 167.13 i$, $0. - 192.99 i$, $0. + 25.86 i$, $668.53$)

\vskip 0.7ex
\hangindent=3em \hangafter=1
\textit{Intrinsic sign problem}

  \vskip 2ex

\noindent81. $9_{2,668.5}^{12,227}$ \irep{330}:\ \ 
$d_i$ = ($1.0$,
$6.464$,
$6.464$,
$6.464$,
$6.464$,
$6.464$,
$6.464$,
$13.928$,
$14.928$) 

\vskip 0.7ex
\hangindent=3em \hangafter=1
$D^2= 668.553 = 
336+192\sqrt{3}$

\vskip 0.7ex
\hangindent=3em \hangafter=1
$T = ( 0,
\frac{1}{2},
\frac{1}{2},
\frac{3}{4},
\frac{3}{4},
\frac{3}{4},
\frac{3}{4},
0,
\frac{1}{3} )
$,

\vskip 0.7ex
\hangindent=3em \hangafter=1
$S$ = ($ 1$,
$ 3+2\sqrt{3}$,
$ 3+2\sqrt{3}$,
$ 3+2\sqrt{3}$,
$ 3+2\sqrt{3}$,
$ 3+2\sqrt{3}$,
$ 3+2\sqrt{3}$,
$ 7+4\sqrt{3}$,
$ 8+4\sqrt{3}$;\ \ 
$ 1-8  \zeta^{1}_{12}
+4\zeta^{-1}_{12}
-8  \zeta^{2}_{12}
$,
$ -7+4\zeta^{1}_{12}
-8  \zeta^{-1}_{12}
+8\zeta^{2}_{12}
$,
$ 3+2\sqrt{3}$,
$ 3+2\sqrt{3}$,
$ 3+2\sqrt{3}$,
$ 3+2\sqrt{3}$,
$ -3-2\sqrt{3}$,
$0$;\ \ 
$ 1-8  \zeta^{1}_{12}
+4\zeta^{-1}_{12}
-8  \zeta^{2}_{12}
$,
$ 3+2\sqrt{3}$,
$ 3+2\sqrt{3}$,
$ 3+2\sqrt{3}$,
$ 3+2\sqrt{3}$,
$ -3-2\sqrt{3}$,
$0$;\ \ 
$ 9+6\sqrt{3}$,
$ -3-2\sqrt{3}$,
$ -3-2\sqrt{3}$,
$ -3-2\sqrt{3}$,
$ -3-2\sqrt{3}$,
$0$;\ \ 
$ 9+6\sqrt{3}$,
$ -3-2\sqrt{3}$,
$ -3-2\sqrt{3}$,
$ -3-2\sqrt{3}$,
$0$;\ \ 
$ 9+6\sqrt{3}$,
$ -3-2\sqrt{3}$,
$ -3-2\sqrt{3}$,
$0$;\ \ 
$ 9+6\sqrt{3}$,
$ -3-2\sqrt{3}$,
$0$;\ \ 
$ 1$,
$ 8+4\sqrt{3}$;\ \ 
$ -8-4\sqrt{3}$)

\vskip 0.7ex
\hangindent=3em \hangafter=1
$\tau_n$ = ($0. + 25.86 i$, $0. - 192.99 i$, $334.27 + 167.13 i$, $334.27 + 192.99 i$, $0. - 360.12 i$, $334.27$, $0. + 360.12 i$, $334.27 - 192.99 i$, $334.27 - 167.13 i$, $0. + 192.99 i$, $0. - 25.86 i$, $668.53$)

\vskip 0.7ex
\hangindent=3em \hangafter=1
\textit{Intrinsic sign problem}

  \vskip 2ex 

%% file: modular_data/SsL10U_.tex
\noindent1. $10_{1,10.}^{20,667}$ \irep{947}:\ \ 
$d_i$ = ($1.0$,
$1.0$,
$1.0$,
$1.0$,
$1.0$,
$1.0$,
$1.0$,
$1.0$,
$1.0$,
$1.0$) 

\vskip 0.7ex
\hangindent=3em \hangafter=1
$D^2= 10.0 = 
10$

\vskip 0.7ex
\hangindent=3em \hangafter=1
$T = ( 0,
\frac{1}{4},
\frac{1}{5},
\frac{1}{5},
\frac{4}{5},
\frac{4}{5},
\frac{1}{20},
\frac{1}{20},
\frac{9}{20},
\frac{9}{20} )
$,

\vskip 0.7ex
\hangindent=3em \hangafter=1
$S$ = ($ 1$,
$ 1$,
$ 1$,
$ 1$,
$ 1$,
$ 1$,
$ 1$,
$ 1$,
$ 1$,
$ 1$;\ \ 
$ -1$,
$ 1$,
$ 1$,
$ 1$,
$ 1$,
$ -1$,
$ -1$,
$ -1$,
$ -1$;\ \ 
$ -\zeta_{10}^{1}$,
$ \zeta_{5}^{2}$,
$ -\zeta_{10}^{3}$,
$ \zeta_{5}^{1}$,
$ -\zeta_{10}^{3}$,
$ \zeta_{5}^{1}$,
$ -\zeta_{10}^{1}$,
$ \zeta_{5}^{2}$;\ \ 
$ -\zeta_{10}^{1}$,
$ \zeta_{5}^{1}$,
$ -\zeta_{10}^{3}$,
$ \zeta_{5}^{1}$,
$ -\zeta_{10}^{3}$,
$ \zeta_{5}^{2}$,
$ -\zeta_{10}^{1}$;\ \ 
$ \zeta_{5}^{2}$,
$ -\zeta_{10}^{1}$,
$ \zeta_{5}^{2}$,
$ -\zeta_{10}^{1}$,
$ -\zeta_{10}^{3}$,
$ \zeta_{5}^{1}$;\ \ 
$ \zeta_{5}^{2}$,
$ -\zeta_{10}^{1}$,
$ \zeta_{5}^{2}$,
$ \zeta_{5}^{1}$,
$ -\zeta_{10}^{3}$;\ \ 
$ -\zeta_{5}^{2}$,
$ \zeta_{10}^{1}$,
$ \zeta_{10}^{3}$,
$ -\zeta_{5}^{1}$;\ \ 
$ -\zeta_{5}^{2}$,
$ -\zeta_{5}^{1}$,
$ \zeta_{10}^{3}$;\ \ 
$ \zeta_{10}^{1}$,
$ -\zeta_{5}^{2}$;\ \ 
$ \zeta_{10}^{1}$)

Factors = $2_{1,2.}^{4,437}\boxtimes 5_{0,5.}^{5,110}$

\vskip 0.7ex
\hangindent=3em \hangafter=1
$\tau_n$ = ($2.24 + 2.24 i$, $0.$, $-2.24 + 2.24 i$, $4.47$, $5. + 5. i$, $0.$, $-2.24 + 2.24 i$, $-4.47$, $2.24 + 2.24 i$, $0.$, $2.24 - 2.24 i$, $-4.47$, $-2.24 - 2.24 i$, $0.$, $5. - 5. i$, $4.47$, $-2.24 - 2.24 i$, $0.$, $2.24 - 2.24 i$, $10.$)

\vskip 0.7ex
\hangindent=3em \hangafter=1
\textit{Intrinsic sign problem}

  \vskip 2ex

\noindent2. $10_{5,10.}^{20,892}$ \irep{947}:\ \ 
$d_i$ = ($1.0$,
$1.0$,
$1.0$,
$1.0$,
$1.0$,
$1.0$,
$1.0$,
$1.0$,
$1.0$,
$1.0$) 

\vskip 0.7ex
\hangindent=3em \hangafter=1
$D^2= 10.0 = 
10$

\vskip 0.7ex
\hangindent=3em \hangafter=1
$T = ( 0,
\frac{1}{4},
\frac{2}{5},
\frac{2}{5},
\frac{3}{5},
\frac{3}{5},
\frac{13}{20},
\frac{13}{20},
\frac{17}{20},
\frac{17}{20} )
$,

\vskip 0.7ex
\hangindent=3em \hangafter=1
$S$ = ($ 1$,
$ 1$,
$ 1$,
$ 1$,
$ 1$,
$ 1$,
$ 1$,
$ 1$,
$ 1$,
$ 1$;\ \ 
$ -1$,
$ 1$,
$ 1$,
$ 1$,
$ 1$,
$ -1$,
$ -1$,
$ -1$,
$ -1$;\ \ 
$ \zeta_{5}^{1}$,
$ -\zeta_{10}^{3}$,
$ -\zeta_{10}^{1}$,
$ \zeta_{5}^{2}$,
$ -\zeta_{10}^{3}$,
$ \zeta_{5}^{1}$,
$ \zeta_{5}^{2}$,
$ -\zeta_{10}^{1}$;\ \ 
$ \zeta_{5}^{1}$,
$ \zeta_{5}^{2}$,
$ -\zeta_{10}^{1}$,
$ \zeta_{5}^{1}$,
$ -\zeta_{10}^{3}$,
$ -\zeta_{10}^{1}$,
$ \zeta_{5}^{2}$;\ \ 
$ -\zeta_{10}^{3}$,
$ \zeta_{5}^{1}$,
$ \zeta_{5}^{2}$,
$ -\zeta_{10}^{1}$,
$ \zeta_{5}^{1}$,
$ -\zeta_{10}^{3}$;\ \ 
$ -\zeta_{10}^{3}$,
$ -\zeta_{10}^{1}$,
$ \zeta_{5}^{2}$,
$ -\zeta_{10}^{3}$,
$ \zeta_{5}^{1}$;\ \ 
$ -\zeta_{5}^{1}$,
$ \zeta_{10}^{3}$,
$ \zeta_{10}^{1}$,
$ -\zeta_{5}^{2}$;\ \ 
$ -\zeta_{5}^{1}$,
$ -\zeta_{5}^{2}$,
$ \zeta_{10}^{1}$;\ \ 
$ \zeta_{10}^{3}$,
$ -\zeta_{5}^{1}$;\ \ 
$ \zeta_{10}^{3}$)

Factors = $2_{1,2.}^{4,437}\boxtimes 5_{4,5.}^{5,210}$

\vskip 0.7ex
\hangindent=3em \hangafter=1
$\tau_n$ = ($-2.24 - 2.24 i$, $0.$, $2.24 - 2.24 i$, $-4.47$, $5. + 5. i$, $0.$, $2.24 - 2.24 i$, $4.47$, $-2.24 - 2.24 i$, $0.$, $-2.24 + 2.24 i$, $4.47$, $2.24 + 2.24 i$, $0.$, $5. - 5. i$, $-4.47$, $2.24 + 2.24 i$, $0.$, $-2.24 + 2.24 i$, $10.$)

\vskip 0.7ex
\hangindent=3em \hangafter=1
\textit{Intrinsic sign problem}

  \vskip 2ex

\noindent3. $10_{7,10.}^{20,183}$ \irep{947}:\ \ 
$d_i$ = ($1.0$,
$1.0$,
$1.0$,
$1.0$,
$1.0$,
$1.0$,
$1.0$,
$1.0$,
$1.0$,
$1.0$) 

\vskip 0.7ex
\hangindent=3em \hangafter=1
$D^2= 10.0 = 
10$

\vskip 0.7ex
\hangindent=3em \hangafter=1
$T = ( 0,
\frac{3}{4},
\frac{1}{5},
\frac{1}{5},
\frac{4}{5},
\frac{4}{5},
\frac{11}{20},
\frac{11}{20},
\frac{19}{20},
\frac{19}{20} )
$,

\vskip 0.7ex
\hangindent=3em \hangafter=1
$S$ = ($ 1$,
$ 1$,
$ 1$,
$ 1$,
$ 1$,
$ 1$,
$ 1$,
$ 1$,
$ 1$,
$ 1$;\ \ 
$ -1$,
$ 1$,
$ 1$,
$ 1$,
$ 1$,
$ -1$,
$ -1$,
$ -1$,
$ -1$;\ \ 
$ -\zeta_{10}^{1}$,
$ \zeta_{5}^{2}$,
$ -\zeta_{10}^{3}$,
$ \zeta_{5}^{1}$,
$ -\zeta_{10}^{3}$,
$ \zeta_{5}^{1}$,
$ -\zeta_{10}^{1}$,
$ \zeta_{5}^{2}$;\ \ 
$ -\zeta_{10}^{1}$,
$ \zeta_{5}^{1}$,
$ -\zeta_{10}^{3}$,
$ \zeta_{5}^{1}$,
$ -\zeta_{10}^{3}$,
$ \zeta_{5}^{2}$,
$ -\zeta_{10}^{1}$;\ \ 
$ \zeta_{5}^{2}$,
$ -\zeta_{10}^{1}$,
$ \zeta_{5}^{2}$,
$ -\zeta_{10}^{1}$,
$ -\zeta_{10}^{3}$,
$ \zeta_{5}^{1}$;\ \ 
$ \zeta_{5}^{2}$,
$ -\zeta_{10}^{1}$,
$ \zeta_{5}^{2}$,
$ \zeta_{5}^{1}$,
$ -\zeta_{10}^{3}$;\ \ 
$ -\zeta_{5}^{2}$,
$ \zeta_{10}^{1}$,
$ \zeta_{10}^{3}$,
$ -\zeta_{5}^{1}$;\ \ 
$ -\zeta_{5}^{2}$,
$ -\zeta_{5}^{1}$,
$ \zeta_{10}^{3}$;\ \ 
$ \zeta_{10}^{1}$,
$ -\zeta_{5}^{2}$;\ \ 
$ \zeta_{10}^{1}$)

Factors = $2_{7,2.}^{4,625}\boxtimes 5_{0,5.}^{5,110}$

\vskip 0.7ex
\hangindent=3em \hangafter=1
$\tau_n$ = ($2.24 - 2.24 i$, $0.$, $-2.24 - 2.24 i$, $4.47$, $5. - 5. i$, $0.$, $-2.24 - 2.24 i$, $-4.47$, $2.24 - 2.24 i$, $0.$, $2.24 + 2.24 i$, $-4.47$, $-2.24 + 2.24 i$, $0.$, $5. + 5. i$, $4.47$, $-2.24 + 2.24 i$, $0.$, $2.24 + 2.24 i$, $10.$)

\vskip 0.7ex
\hangindent=3em \hangafter=1
\textit{Intrinsic sign problem}

  \vskip 2ex

\noindent4. $10_{3,10.}^{20,607}$ \irep{947}:\ \ 
$d_i$ = ($1.0$,
$1.0$,
$1.0$,
$1.0$,
$1.0$,
$1.0$,
$1.0$,
$1.0$,
$1.0$,
$1.0$) 

\vskip 0.7ex
\hangindent=3em \hangafter=1
$D^2= 10.0 = 
10$

\vskip 0.7ex
\hangindent=3em \hangafter=1
$T = ( 0,
\frac{3}{4},
\frac{2}{5},
\frac{2}{5},
\frac{3}{5},
\frac{3}{5},
\frac{3}{20},
\frac{3}{20},
\frac{7}{20},
\frac{7}{20} )
$,

\vskip 0.7ex
\hangindent=3em \hangafter=1
$S$ = ($ 1$,
$ 1$,
$ 1$,
$ 1$,
$ 1$,
$ 1$,
$ 1$,
$ 1$,
$ 1$,
$ 1$;\ \ 
$ -1$,
$ 1$,
$ 1$,
$ 1$,
$ 1$,
$ -1$,
$ -1$,
$ -1$,
$ -1$;\ \ 
$ \zeta_{5}^{1}$,
$ -\zeta_{10}^{3}$,
$ -\zeta_{10}^{1}$,
$ \zeta_{5}^{2}$,
$ -\zeta_{10}^{3}$,
$ \zeta_{5}^{1}$,
$ \zeta_{5}^{2}$,
$ -\zeta_{10}^{1}$;\ \ 
$ \zeta_{5}^{1}$,
$ \zeta_{5}^{2}$,
$ -\zeta_{10}^{1}$,
$ \zeta_{5}^{1}$,
$ -\zeta_{10}^{3}$,
$ -\zeta_{10}^{1}$,
$ \zeta_{5}^{2}$;\ \ 
$ -\zeta_{10}^{3}$,
$ \zeta_{5}^{1}$,
$ \zeta_{5}^{2}$,
$ -\zeta_{10}^{1}$,
$ \zeta_{5}^{1}$,
$ -\zeta_{10}^{3}$;\ \ 
$ -\zeta_{10}^{3}$,
$ -\zeta_{10}^{1}$,
$ \zeta_{5}^{2}$,
$ -\zeta_{10}^{3}$,
$ \zeta_{5}^{1}$;\ \ 
$ -\zeta_{5}^{1}$,
$ \zeta_{10}^{3}$,
$ \zeta_{10}^{1}$,
$ -\zeta_{5}^{2}$;\ \ 
$ -\zeta_{5}^{1}$,
$ -\zeta_{5}^{2}$,
$ \zeta_{10}^{1}$;\ \ 
$ \zeta_{10}^{3}$,
$ -\zeta_{5}^{1}$;\ \ 
$ \zeta_{10}^{3}$)

Factors = $2_{7,2.}^{4,625}\boxtimes 5_{4,5.}^{5,210}$

\vskip 0.7ex
\hangindent=3em \hangafter=1
$\tau_n$ = ($-2.24 + 2.24 i$, $0.$, $2.24 + 2.24 i$, $-4.47$, $5. - 5. i$, $0.$, $2.24 + 2.24 i$, $4.47$, $-2.24 + 2.24 i$, $0.$, $-2.24 - 2.24 i$, $4.47$, $2.24 - 2.24 i$, $0.$, $5. + 5. i$, $-4.47$, $2.24 - 2.24 i$, $0.$, $-2.24 - 2.24 i$, $10.$)

\vskip 0.7ex
\hangindent=3em \hangafter=1
\textit{Intrinsic sign problem}

  \vskip 2ex

\noindent5. $10_{\frac{14}{5},18.09}^{5,359}$ \irep{125}:\ \ 
$d_i$ = ($1.0$,
$1.0$,
$1.0$,
$1.0$,
$1.0$,
$1.618$,
$1.618$,
$1.618$,
$1.618$,
$1.618$) 

\vskip 0.7ex
\hangindent=3em \hangafter=1
$D^2= 18.90 = 
\frac{25+5\sqrt{5}}{2}$

\vskip 0.7ex
\hangindent=3em \hangafter=1
$T = ( 0,
\frac{1}{5},
\frac{1}{5},
\frac{4}{5},
\frac{4}{5},
\frac{1}{5},
\frac{1}{5},
\frac{2}{5},
\frac{3}{5},
\frac{3}{5} )
$,

\vskip 0.7ex
\hangindent=3em \hangafter=1
$S$ = ($ 1$,
$ 1$,
$ 1$,
$ 1$,
$ 1$,
$ \frac{1+\sqrt{5}}{2}$,
$ \frac{1+\sqrt{5}}{2}$,
$ \frac{1+\sqrt{5}}{2}$,
$ \frac{1+\sqrt{5}}{2}$,
$ \frac{1+\sqrt{5}}{2}$;\ \ 
$ -\zeta_{10}^{1}$,
$ \zeta_{5}^{2}$,
$ -\zeta_{10}^{3}$,
$ \zeta_{5}^{1}$,
$ -\frac{1+\sqrt{5}}{2}\zeta_{10}^{3}$,
$ \frac{1+\sqrt{5}}{2}\zeta_{5}^{1}$,
$ \frac{1+\sqrt{5}}{2}$,
$ -\frac{1+\sqrt{5}}{2}\zeta_{10}^{1}$,
$ \frac{1+\sqrt{5}}{2}\zeta_{5}^{2}$;\ \ 
$ -\zeta_{10}^{1}$,
$ \zeta_{5}^{1}$,
$ -\zeta_{10}^{3}$,
$ \frac{1+\sqrt{5}}{2}\zeta_{5}^{1}$,
$ -\frac{1+\sqrt{5}}{2}\zeta_{10}^{3}$,
$ \frac{1+\sqrt{5}}{2}$,
$ \frac{1+\sqrt{5}}{2}\zeta_{5}^{2}$,
$ -\frac{1+\sqrt{5}}{2}\zeta_{10}^{1}$;\ \ 
$ \zeta_{5}^{2}$,
$ -\zeta_{10}^{1}$,
$ \frac{1+\sqrt{5}}{2}\zeta_{5}^{2}$,
$ -\frac{1+\sqrt{5}}{2}\zeta_{10}^{1}$,
$ \frac{1+\sqrt{5}}{2}$,
$ -\frac{1+\sqrt{5}}{2}\zeta_{10}^{3}$,
$ \frac{1+\sqrt{5}}{2}\zeta_{5}^{1}$;\ \ 
$ \zeta_{5}^{2}$,
$ -\frac{1+\sqrt{5}}{2}\zeta_{10}^{1}$,
$ \frac{1+\sqrt{5}}{2}\zeta_{5}^{2}$,
$ \frac{1+\sqrt{5}}{2}$,
$ \frac{1+\sqrt{5}}{2}\zeta_{5}^{1}$,
$ -\frac{1+\sqrt{5}}{2}\zeta_{10}^{3}$;\ \ 
$ -\zeta_{5}^{2}$,
$ \zeta_{10}^{1}$,
$ -1$,
$ \zeta_{10}^{3}$,
$ -\zeta_{5}^{1}$;\ \ 
$ -\zeta_{5}^{2}$,
$ -1$,
$ -\zeta_{5}^{1}$,
$ \zeta_{10}^{3}$;\ \ 
$ -1$,
$ -1$,
$ -1$;\ \ 
$ \zeta_{10}^{1}$,
$ -\zeta_{5}^{2}$;\ \ 
$ \zeta_{10}^{1}$)

Factors = $2_{\frac{14}{5},3.618}^{5,395}\boxtimes 5_{0,5.}^{5,110}$

\vskip 0.7ex
\hangindent=3em \hangafter=1
$\tau_n$ = ($-2.5 + 3.44 i$, $-4.05 + 5.57 i$, $-4.05 - 5.57 i$, $-2.5 - 3.44 i$, $18.09$)

\vskip 0.7ex
\hangindent=3em \hangafter=1
\textit{Intrinsic sign problem}

  \vskip 2ex

\noindent6. $10_{\frac{26}{5},18.09}^{5,125}$ \irep{125}:\ \ 
$d_i$ = ($1.0$,
$1.0$,
$1.0$,
$1.0$,
$1.0$,
$1.618$,
$1.618$,
$1.618$,
$1.618$,
$1.618$) 

\vskip 0.7ex
\hangindent=3em \hangafter=1
$D^2= 18.90 = 
\frac{25+5\sqrt{5}}{2}$

\vskip 0.7ex
\hangindent=3em \hangafter=1
$T = ( 0,
\frac{1}{5},
\frac{1}{5},
\frac{4}{5},
\frac{4}{5},
\frac{2}{5},
\frac{2}{5},
\frac{3}{5},
\frac{4}{5},
\frac{4}{5} )
$,

\vskip 0.7ex
\hangindent=3em \hangafter=1
$S$ = ($ 1$,
$ 1$,
$ 1$,
$ 1$,
$ 1$,
$ \frac{1+\sqrt{5}}{2}$,
$ \frac{1+\sqrt{5}}{2}$,
$ \frac{1+\sqrt{5}}{2}$,
$ \frac{1+\sqrt{5}}{2}$,
$ \frac{1+\sqrt{5}}{2}$;\ \ 
$ -\zeta_{10}^{1}$,
$ \zeta_{5}^{2}$,
$ -\zeta_{10}^{3}$,
$ \zeta_{5}^{1}$,
$ -\frac{1+\sqrt{5}}{2}\zeta_{10}^{3}$,
$ \frac{1+\sqrt{5}}{2}\zeta_{5}^{1}$,
$ \frac{1+\sqrt{5}}{2}$,
$ -\frac{1+\sqrt{5}}{2}\zeta_{10}^{1}$,
$ \frac{1+\sqrt{5}}{2}\zeta_{5}^{2}$;\ \ 
$ -\zeta_{10}^{1}$,
$ \zeta_{5}^{1}$,
$ -\zeta_{10}^{3}$,
$ \frac{1+\sqrt{5}}{2}\zeta_{5}^{1}$,
$ -\frac{1+\sqrt{5}}{2}\zeta_{10}^{3}$,
$ \frac{1+\sqrt{5}}{2}$,
$ \frac{1+\sqrt{5}}{2}\zeta_{5}^{2}$,
$ -\frac{1+\sqrt{5}}{2}\zeta_{10}^{1}$;\ \ 
$ \zeta_{5}^{2}$,
$ -\zeta_{10}^{1}$,
$ \frac{1+\sqrt{5}}{2}\zeta_{5}^{2}$,
$ -\frac{1+\sqrt{5}}{2}\zeta_{10}^{1}$,
$ \frac{1+\sqrt{5}}{2}$,
$ -\frac{1+\sqrt{5}}{2}\zeta_{10}^{3}$,
$ \frac{1+\sqrt{5}}{2}\zeta_{5}^{1}$;\ \ 
$ \zeta_{5}^{2}$,
$ -\frac{1+\sqrt{5}}{2}\zeta_{10}^{1}$,
$ \frac{1+\sqrt{5}}{2}\zeta_{5}^{2}$,
$ \frac{1+\sqrt{5}}{2}$,
$ \frac{1+\sqrt{5}}{2}\zeta_{5}^{1}$,
$ -\frac{1+\sqrt{5}}{2}\zeta_{10}^{3}$;\ \ 
$ -\zeta_{5}^{2}$,
$ \zeta_{10}^{1}$,
$ -1$,
$ \zeta_{10}^{3}$,
$ -\zeta_{5}^{1}$;\ \ 
$ -\zeta_{5}^{2}$,
$ -1$,
$ -\zeta_{5}^{1}$,
$ \zeta_{10}^{3}$;\ \ 
$ -1$,
$ -1$,
$ -1$;\ \ 
$ \zeta_{10}^{1}$,
$ -\zeta_{5}^{2}$;\ \ 
$ \zeta_{10}^{1}$)

Factors = $2_{\frac{26}{5},3.618}^{5,720}\boxtimes 5_{0,5.}^{5,110}$

\vskip 0.7ex
\hangindent=3em \hangafter=1
$\tau_n$ = ($-2.5 - 3.44 i$, $-4.05 - 5.57 i$, $-4.05 + 5.57 i$, $-2.5 + 3.44 i$, $18.09$)

\vskip 0.7ex
\hangindent=3em \hangafter=1
\textit{Intrinsic sign problem}

  \vskip 2ex

\noindent7. $10_{\frac{6}{5},18.09}^{5,152}$ \irep{145}:\ \ 
$d_i$ = ($1.0$,
$1.0$,
$1.0$,
$1.0$,
$1.0$,
$1.618$,
$1.618$,
$1.618$,
$1.618$,
$1.618$) 

\vskip 0.7ex
\hangindent=3em \hangafter=1
$D^2= 18.90 = 
\frac{25+5\sqrt{5}}{2}$

\vskip 0.7ex
\hangindent=3em \hangafter=1
$T = ( 0,
\frac{2}{5},
\frac{2}{5},
\frac{3}{5},
\frac{3}{5},
0,
0,
\frac{1}{5},
\frac{1}{5},
\frac{3}{5} )
$,

\vskip 0.7ex
\hangindent=3em \hangafter=1
$S$ = ($ 1$,
$ 1$,
$ 1$,
$ 1$,
$ 1$,
$ \frac{1+\sqrt{5}}{2}$,
$ \frac{1+\sqrt{5}}{2}$,
$ \frac{1+\sqrt{5}}{2}$,
$ \frac{1+\sqrt{5}}{2}$,
$ \frac{1+\sqrt{5}}{2}$;\ \ 
$ \zeta_{5}^{1}$,
$ -\zeta_{10}^{3}$,
$ -\zeta_{10}^{1}$,
$ \zeta_{5}^{2}$,
$ -\frac{1+\sqrt{5}}{2}\zeta_{10}^{3}$,
$ \frac{1+\sqrt{5}}{2}\zeta_{5}^{1}$,
$ \frac{1+\sqrt{5}}{2}\zeta_{5}^{2}$,
$ -\frac{1+\sqrt{5}}{2}\zeta_{10}^{1}$,
$ \frac{1+\sqrt{5}}{2}$;\ \ 
$ \zeta_{5}^{1}$,
$ \zeta_{5}^{2}$,
$ -\zeta_{10}^{1}$,
$ \frac{1+\sqrt{5}}{2}\zeta_{5}^{1}$,
$ -\frac{1+\sqrt{5}}{2}\zeta_{10}^{3}$,
$ -\frac{1+\sqrt{5}}{2}\zeta_{10}^{1}$,
$ \frac{1+\sqrt{5}}{2}\zeta_{5}^{2}$,
$ \frac{1+\sqrt{5}}{2}$;\ \ 
$ -\zeta_{10}^{3}$,
$ \zeta_{5}^{1}$,
$ \frac{1+\sqrt{5}}{2}\zeta_{5}^{2}$,
$ -\frac{1+\sqrt{5}}{2}\zeta_{10}^{1}$,
$ \frac{1+\sqrt{5}}{2}\zeta_{5}^{1}$,
$ -\frac{1+\sqrt{5}}{2}\zeta_{10}^{3}$,
$ \frac{1+\sqrt{5}}{2}$;\ \ 
$ -\zeta_{10}^{3}$,
$ -\frac{1+\sqrt{5}}{2}\zeta_{10}^{1}$,
$ \frac{1+\sqrt{5}}{2}\zeta_{5}^{2}$,
$ -\frac{1+\sqrt{5}}{2}\zeta_{10}^{3}$,
$ \frac{1+\sqrt{5}}{2}\zeta_{5}^{1}$,
$ \frac{1+\sqrt{5}}{2}$;\ \ 
$ -\zeta_{5}^{1}$,
$ \zeta_{10}^{3}$,
$ \zeta_{10}^{1}$,
$ -\zeta_{5}^{2}$,
$ -1$;\ \ 
$ -\zeta_{5}^{1}$,
$ -\zeta_{5}^{2}$,
$ \zeta_{10}^{1}$,
$ -1$;\ \ 
$ \zeta_{10}^{3}$,
$ -\zeta_{5}^{1}$,
$ -1$;\ \ 
$ \zeta_{10}^{3}$,
$ -1$;\ \ 
$ -1$)

Factors = $2_{\frac{26}{5},3.618}^{5,720}\boxtimes 5_{4,5.}^{5,210}$

\vskip 0.7ex
\hangindent=3em \hangafter=1
$\tau_n$ = ($2.5 + 3.44 i$, $4.05 + 5.57 i$, $4.05 - 5.57 i$, $2.5 - 3.44 i$, $18.09$)

\vskip 0.7ex
\hangindent=3em \hangafter=1
\textit{Intrinsic sign problem}

  \vskip 2ex

\noindent8. $10_{\frac{34}{5},18.09}^{5,974}$ \irep{145}:\ \ 
$d_i$ = ($1.0$,
$1.0$,
$1.0$,
$1.0$,
$1.0$,
$1.618$,
$1.618$,
$1.618$,
$1.618$,
$1.618$) 

\vskip 0.7ex
\hangindent=3em \hangafter=1
$D^2= 18.90 = 
\frac{25+5\sqrt{5}}{2}$

\vskip 0.7ex
\hangindent=3em \hangafter=1
$T = ( 0,
\frac{2}{5},
\frac{2}{5},
\frac{3}{5},
\frac{3}{5},
0,
0,
\frac{2}{5},
\frac{4}{5},
\frac{4}{5} )
$,

\vskip 0.7ex
\hangindent=3em \hangafter=1
$S$ = ($ 1$,
$ 1$,
$ 1$,
$ 1$,
$ 1$,
$ \frac{1+\sqrt{5}}{2}$,
$ \frac{1+\sqrt{5}}{2}$,
$ \frac{1+\sqrt{5}}{2}$,
$ \frac{1+\sqrt{5}}{2}$,
$ \frac{1+\sqrt{5}}{2}$;\ \ 
$ \zeta_{5}^{1}$,
$ -\zeta_{10}^{3}$,
$ -\zeta_{10}^{1}$,
$ \zeta_{5}^{2}$,
$ -\frac{1+\sqrt{5}}{2}\zeta_{10}^{1}$,
$ \frac{1+\sqrt{5}}{2}\zeta_{5}^{2}$,
$ \frac{1+\sqrt{5}}{2}$,
$ \frac{1+\sqrt{5}}{2}\zeta_{5}^{1}$,
$ -\frac{1+\sqrt{5}}{2}\zeta_{10}^{3}$;\ \ 
$ \zeta_{5}^{1}$,
$ \zeta_{5}^{2}$,
$ -\zeta_{10}^{1}$,
$ \frac{1+\sqrt{5}}{2}\zeta_{5}^{2}$,
$ -\frac{1+\sqrt{5}}{2}\zeta_{10}^{1}$,
$ \frac{1+\sqrt{5}}{2}$,
$ -\frac{1+\sqrt{5}}{2}\zeta_{10}^{3}$,
$ \frac{1+\sqrt{5}}{2}\zeta_{5}^{1}$;\ \ 
$ -\zeta_{10}^{3}$,
$ \zeta_{5}^{1}$,
$ -\frac{1+\sqrt{5}}{2}\zeta_{10}^{3}$,
$ \frac{1+\sqrt{5}}{2}\zeta_{5}^{1}$,
$ \frac{1+\sqrt{5}}{2}$,
$ -\frac{1+\sqrt{5}}{2}\zeta_{10}^{1}$,
$ \frac{1+\sqrt{5}}{2}\zeta_{5}^{2}$;\ \ 
$ -\zeta_{10}^{3}$,
$ \frac{1+\sqrt{5}}{2}\zeta_{5}^{1}$,
$ -\frac{1+\sqrt{5}}{2}\zeta_{10}^{3}$,
$ \frac{1+\sqrt{5}}{2}$,
$ \frac{1+\sqrt{5}}{2}\zeta_{5}^{2}$,
$ -\frac{1+\sqrt{5}}{2}\zeta_{10}^{1}$;\ \ 
$ \zeta_{10}^{3}$,
$ -\zeta_{5}^{1}$,
$ -1$,
$ \zeta_{10}^{1}$,
$ -\zeta_{5}^{2}$;\ \ 
$ \zeta_{10}^{3}$,
$ -1$,
$ -\zeta_{5}^{2}$,
$ \zeta_{10}^{1}$;\ \ 
$ -1$,
$ -1$,
$ -1$;\ \ 
$ -\zeta_{5}^{1}$,
$ \zeta_{10}^{3}$;\ \ 
$ -\zeta_{5}^{1}$)

Factors = $2_{\frac{14}{5},3.618}^{5,395}\boxtimes 5_{4,5.}^{5,210}$

\vskip 0.7ex
\hangindent=3em \hangafter=1
$\tau_n$ = ($2.5 - 3.44 i$, $4.05 - 5.57 i$, $4.05 + 5.57 i$, $2.5 + 3.44 i$, $18.09$)

\vskip 0.7ex
\hangindent=3em \hangafter=1
\textit{Intrinsic sign problem}

  \vskip 2ex

\noindent9. $10_{3,24.}^{24,430}$ \irep{972}:\ \ 
$d_i$ = ($1.0$,
$1.0$,
$1.0$,
$1.0$,
$1.732$,
$1.732$,
$1.732$,
$1.732$,
$2.0$,
$2.0$) 

\vskip 0.7ex
\hangindent=3em \hangafter=1
$D^2= 24.0 = 
24$

\vskip 0.7ex
\hangindent=3em \hangafter=1
$T = ( 0,
0,
\frac{1}{4},
\frac{1}{4},
\frac{1}{8},
\frac{3}{8},
\frac{5}{8},
\frac{7}{8},
\frac{1}{3},
\frac{7}{12} )
$,

\vskip 0.7ex
\hangindent=3em \hangafter=1
$S$ = ($ 1$,
$ 1$,
$ 1$,
$ 1$,
$ \sqrt{3}$,
$ \sqrt{3}$,
$ \sqrt{3}$,
$ \sqrt{3}$,
$ 2$,
$ 2$;\ \ 
$ 1$,
$ 1$,
$ 1$,
$ -\sqrt{3}$,
$ -\sqrt{3}$,
$ -\sqrt{3}$,
$ -\sqrt{3}$,
$ 2$,
$ 2$;\ \ 
$ -1$,
$ -1$,
$ \sqrt{3}$,
$ -\sqrt{3}$,
$ \sqrt{3}$,
$ -\sqrt{3}$,
$ 2$,
$ -2$;\ \ 
$ -1$,
$ -\sqrt{3}$,
$ \sqrt{3}$,
$ -\sqrt{3}$,
$ \sqrt{3}$,
$ 2$,
$ -2$;\ \ 
$ \sqrt{3}$,
$ \sqrt{3}$,
$ -\sqrt{3}$,
$ -\sqrt{3}$,
$0$,
$0$;\ \ 
$ -\sqrt{3}$,
$ -\sqrt{3}$,
$ \sqrt{3}$,
$0$,
$0$;\ \ 
$ \sqrt{3}$,
$ \sqrt{3}$,
$0$,
$0$;\ \ 
$ -\sqrt{3}$,
$0$,
$0$;\ \ 
$ -2$,
$ -2$;\ \ 
$ 2$)

Factors = $2_{1,2.}^{4,437}\boxtimes 5_{2,12.}^{24,940}$

\vskip 0.7ex
\hangindent=3em \hangafter=1
$\tau_n$ = ($-3.46 + 3.46 i$, $0.$, $6. - 6. i$, $-12. + 6.93 i$, $3.46 - 3.46 i$, $0.$, $3.46 + 3.46 i$, $12. - 6.93 i$, $6. + 6. i$, $0.$, $-3.46 - 3.46 i$, $0.$, $-3.46 + 3.46 i$, $0.$, $6. - 6. i$, $12. + 6.93 i$, $3.46 - 3.46 i$, $0.$, $3.46 + 3.46 i$, $-12. - 6.93 i$, $6. + 6. i$, $0.$, $-3.46 - 3.46 i$, $24.$)

\vskip 0.7ex
\hangindent=3em \hangafter=1
\textit{Intrinsic sign problem}

  \vskip 2ex

\noindent10. $10_{7,24.}^{24,123}$ \irep{972}:\ \ 
$d_i$ = ($1.0$,
$1.0$,
$1.0$,
$1.0$,
$1.732$,
$1.732$,
$1.732$,
$1.732$,
$2.0$,
$2.0$) 

\vskip 0.7ex
\hangindent=3em \hangafter=1
$D^2= 24.0 = 
24$

\vskip 0.7ex
\hangindent=3em \hangafter=1
$T = ( 0,
0,
\frac{1}{4},
\frac{1}{4},
\frac{1}{8},
\frac{3}{8},
\frac{5}{8},
\frac{7}{8},
\frac{2}{3},
\frac{11}{12} )
$,

\vskip 0.7ex
\hangindent=3em \hangafter=1
$S$ = ($ 1$,
$ 1$,
$ 1$,
$ 1$,
$ \sqrt{3}$,
$ \sqrt{3}$,
$ \sqrt{3}$,
$ \sqrt{3}$,
$ 2$,
$ 2$;\ \ 
$ 1$,
$ 1$,
$ 1$,
$ -\sqrt{3}$,
$ -\sqrt{3}$,
$ -\sqrt{3}$,
$ -\sqrt{3}$,
$ 2$,
$ 2$;\ \ 
$ -1$,
$ -1$,
$ \sqrt{3}$,
$ -\sqrt{3}$,
$ \sqrt{3}$,
$ -\sqrt{3}$,
$ 2$,
$ -2$;\ \ 
$ -1$,
$ -\sqrt{3}$,
$ \sqrt{3}$,
$ -\sqrt{3}$,
$ \sqrt{3}$,
$ 2$,
$ -2$;\ \ 
$ -\sqrt{3}$,
$ -\sqrt{3}$,
$ \sqrt{3}$,
$ \sqrt{3}$,
$0$,
$0$;\ \ 
$ \sqrt{3}$,
$ \sqrt{3}$,
$ -\sqrt{3}$,
$0$,
$0$;\ \ 
$ -\sqrt{3}$,
$ -\sqrt{3}$,
$0$,
$0$;\ \ 
$ \sqrt{3}$,
$0$,
$0$;\ \ 
$ -2$,
$ -2$;\ \ 
$ 2$)

Factors = $2_{1,2.}^{4,437}\boxtimes 5_{6,12.}^{24,592}$

\vskip 0.7ex
\hangindent=3em \hangafter=1
$\tau_n$ = ($3.46 - 3.46 i$, $0.$, $6. - 6. i$, $-12. - 6.93 i$, $-3.46 + 3.46 i$, $0.$, $-3.46 - 3.46 i$, $12. + 6.93 i$, $6. + 6. i$, $0.$, $3.46 + 3.46 i$, $0.$, $3.46 - 3.46 i$, $0.$, $6. - 6. i$, $12. - 6.93 i$, $-3.46 + 3.46 i$, $0.$, $-3.46 - 3.46 i$, $-12. + 6.93 i$, $6. + 6. i$, $0.$, $3.46 + 3.46 i$, $24.$)

\vskip 0.7ex
\hangindent=3em \hangafter=1
\textit{Intrinsic sign problem}

  \vskip 2ex

\noindent11. $10_{1,24.}^{24,976}$ \irep{972}:\ \ 
$d_i$ = ($1.0$,
$1.0$,
$1.0$,
$1.0$,
$1.732$,
$1.732$,
$1.732$,
$1.732$,
$2.0$,
$2.0$) 

\vskip 0.7ex
\hangindent=3em \hangafter=1
$D^2= 24.0 = 
24$

\vskip 0.7ex
\hangindent=3em \hangafter=1
$T = ( 0,
0,
\frac{3}{4},
\frac{3}{4},
\frac{1}{8},
\frac{3}{8},
\frac{5}{8},
\frac{7}{8},
\frac{1}{3},
\frac{1}{12} )
$,

\vskip 0.7ex
\hangindent=3em \hangafter=1
$S$ = ($ 1$,
$ 1$,
$ 1$,
$ 1$,
$ \sqrt{3}$,
$ \sqrt{3}$,
$ \sqrt{3}$,
$ \sqrt{3}$,
$ 2$,
$ 2$;\ \ 
$ 1$,
$ 1$,
$ 1$,
$ -\sqrt{3}$,
$ -\sqrt{3}$,
$ -\sqrt{3}$,
$ -\sqrt{3}$,
$ 2$,
$ 2$;\ \ 
$ -1$,
$ -1$,
$ \sqrt{3}$,
$ -\sqrt{3}$,
$ \sqrt{3}$,
$ -\sqrt{3}$,
$ 2$,
$ -2$;\ \ 
$ -1$,
$ -\sqrt{3}$,
$ \sqrt{3}$,
$ -\sqrt{3}$,
$ \sqrt{3}$,
$ 2$,
$ -2$;\ \ 
$ \sqrt{3}$,
$ -\sqrt{3}$,
$ -\sqrt{3}$,
$ \sqrt{3}$,
$0$,
$0$;\ \ 
$ -\sqrt{3}$,
$ \sqrt{3}$,
$ \sqrt{3}$,
$0$,
$0$;\ \ 
$ \sqrt{3}$,
$ -\sqrt{3}$,
$0$,
$0$;\ \ 
$ -\sqrt{3}$,
$0$,
$0$;\ \ 
$ -2$,
$ -2$;\ \ 
$ 2$)

Factors = $2_{7,2.}^{4,625}\boxtimes 5_{2,12.}^{24,940}$

\vskip 0.7ex
\hangindent=3em \hangafter=1
$\tau_n$ = ($3.46 + 3.46 i$, $0.$, $6. + 6. i$, $-12. + 6.93 i$, $-3.46 - 3.46 i$, $0.$, $-3.46 + 3.46 i$, $12. - 6.93 i$, $6. - 6. i$, $0.$, $3.46 - 3.46 i$, $0.$, $3.46 + 3.46 i$, $0.$, $6. + 6. i$, $12. + 6.93 i$, $-3.46 - 3.46 i$, $0.$, $-3.46 + 3.46 i$, $-12. - 6.93 i$, $6. - 6. i$, $0.$, $3.46 - 3.46 i$, $24.$)

\vskip 0.7ex
\hangindent=3em \hangafter=1
\textit{Intrinsic sign problem}

  \vskip 2ex

\noindent12. $10_{5,24.}^{24,902}$ \irep{972}:\ \ 
$d_i$ = ($1.0$,
$1.0$,
$1.0$,
$1.0$,
$1.732$,
$1.732$,
$1.732$,
$1.732$,
$2.0$,
$2.0$) 

\vskip 0.7ex
\hangindent=3em \hangafter=1
$D^2= 24.0 = 
24$

\vskip 0.7ex
\hangindent=3em \hangafter=1
$T = ( 0,
0,
\frac{3}{4},
\frac{3}{4},
\frac{1}{8},
\frac{3}{8},
\frac{5}{8},
\frac{7}{8},
\frac{2}{3},
\frac{5}{12} )
$,

\vskip 0.7ex
\hangindent=3em \hangafter=1
$S$ = ($ 1$,
$ 1$,
$ 1$,
$ 1$,
$ \sqrt{3}$,
$ \sqrt{3}$,
$ \sqrt{3}$,
$ \sqrt{3}$,
$ 2$,
$ 2$;\ \ 
$ 1$,
$ 1$,
$ 1$,
$ -\sqrt{3}$,
$ -\sqrt{3}$,
$ -\sqrt{3}$,
$ -\sqrt{3}$,
$ 2$,
$ 2$;\ \ 
$ -1$,
$ -1$,
$ \sqrt{3}$,
$ -\sqrt{3}$,
$ \sqrt{3}$,
$ -\sqrt{3}$,
$ 2$,
$ -2$;\ \ 
$ -1$,
$ -\sqrt{3}$,
$ \sqrt{3}$,
$ -\sqrt{3}$,
$ \sqrt{3}$,
$ 2$,
$ -2$;\ \ 
$ -\sqrt{3}$,
$ \sqrt{3}$,
$ \sqrt{3}$,
$ -\sqrt{3}$,
$0$,
$0$;\ \ 
$ \sqrt{3}$,
$ -\sqrt{3}$,
$ -\sqrt{3}$,
$0$,
$0$;\ \ 
$ -\sqrt{3}$,
$ \sqrt{3}$,
$0$,
$0$;\ \ 
$ \sqrt{3}$,
$0$,
$0$;\ \ 
$ -2$,
$ -2$;\ \ 
$ 2$)

Factors = $2_{7,2.}^{4,625}\boxtimes 5_{6,12.}^{24,592}$

\vskip 0.7ex
\hangindent=3em \hangafter=1
$\tau_n$ = ($-3.46 - 3.46 i$, $0.$, $6. + 6. i$, $-12. - 6.93 i$, $3.46 + 3.46 i$, $0.$, $3.46 - 3.46 i$, $12. + 6.93 i$, $6. - 6. i$, $0.$, $-3.46 + 3.46 i$, $0.$, $-3.46 - 3.46 i$, $0.$, $6. + 6. i$, $12. - 6.93 i$, $3.46 + 3.46 i$, $0.$, $3.46 - 3.46 i$, $-12. + 6.93 i$, $6. - 6. i$, $0.$, $-3.46 + 3.46 i$, $24.$)

\vskip 0.7ex
\hangindent=3em \hangafter=1
\textit{Intrinsic sign problem}

  \vskip 2ex

\noindent13. $10_{3,24.}^{48,945}$ \irep{1119}:\ \ 
$d_i$ = ($1.0$,
$1.0$,
$1.0$,
$1.0$,
$1.732$,
$1.732$,
$1.732$,
$1.732$,
$2.0$,
$2.0$) 

\vskip 0.7ex
\hangindent=3em \hangafter=1
$D^2= 24.0 = 
24$

\vskip 0.7ex
\hangindent=3em \hangafter=1
$T = ( 0,
0,
\frac{1}{4},
\frac{1}{4},
\frac{3}{16},
\frac{3}{16},
\frac{11}{16},
\frac{11}{16},
\frac{1}{3},
\frac{7}{12} )
$,

\vskip 0.7ex
\hangindent=3em \hangafter=1
$S$ = ($ 1$,
$ 1$,
$ 1$,
$ 1$,
$ \sqrt{3}$,
$ \sqrt{3}$,
$ \sqrt{3}$,
$ \sqrt{3}$,
$ 2$,
$ 2$;\ \ 
$ 1$,
$ 1$,
$ 1$,
$ -\sqrt{3}$,
$ -\sqrt{3}$,
$ -\sqrt{3}$,
$ -\sqrt{3}$,
$ 2$,
$ 2$;\ \ 
$ -1$,
$ -1$,
$(-\sqrt{3})\mathrm{i}$,
$(\sqrt{3})\mathrm{i}$,
$(-\sqrt{3})\mathrm{i}$,
$(\sqrt{3})\mathrm{i}$,
$ 2$,
$ -2$;\ \ 
$ -1$,
$(\sqrt{3})\mathrm{i}$,
$(-\sqrt{3})\mathrm{i}$,
$(\sqrt{3})\mathrm{i}$,
$(-\sqrt{3})\mathrm{i}$,
$ 2$,
$ -2$;\ \ 
$ -\sqrt{3}\zeta_{8}^{3}$,
$ \sqrt{3}\zeta_{8}^{1}$,
$ \sqrt{3}\zeta_{8}^{3}$,
$ -\sqrt{3}\zeta_{8}^{1}$,
$0$,
$0$;\ \ 
$ -\sqrt{3}\zeta_{8}^{3}$,
$ -\sqrt{3}\zeta_{8}^{1}$,
$ \sqrt{3}\zeta_{8}^{3}$,
$0$,
$0$;\ \ 
$ -\sqrt{3}\zeta_{8}^{3}$,
$ \sqrt{3}\zeta_{8}^{1}$,
$0$,
$0$;\ \ 
$ -\sqrt{3}\zeta_{8}^{3}$,
$0$,
$0$;\ \ 
$ -2$,
$ -2$;\ \ 
$ 2$)

\vskip 0.7ex
\hangindent=3em \hangafter=1
$\tau_n$ = ($-3.46 + 3.46 i$, $-8.48 + 8.48 i$, $6. - 6. i$, $0. - 5.07 i$, $3.46 - 3.46 i$, $8.48 + 8.48 i$, $3.46 + 3.46 i$, $-12. - 6.93 i$, $6. + 6. i$, $8.48 - 8.48 i$, $-3.46 - 3.46 i$, $12. + 12. i$, $-3.46 + 3.46 i$, $-8.48 - 8.48 i$, $6. - 6. i$, $12. + 6.93 i$, $3.46 - 3.46 i$, $-8.48 + 8.48 i$, $3.46 + 3.46 i$, $0. - 18.93 i$, $6. + 6. i$, $8.48 + 8.48 i$, $-3.46 - 3.46 i$, $0.$, $-3.46 + 3.46 i$, $8.48 - 8.48 i$, $6. - 6. i$, $0. + 18.93 i$, $3.46 - 3.46 i$, $-8.48 - 8.48 i$, $3.46 + 3.46 i$, $12. - 6.93 i$, $6. + 6. i$, $-8.48 + 8.48 i$, $-3.46 - 3.46 i$, $12. - 12. i$, $-3.46 + 3.46 i$, $8.48 + 8.48 i$, $6. - 6. i$, $-12. + 6.93 i$, $3.46 - 3.46 i$, $8.48 - 8.48 i$, $3.46 + 3.46 i$, $0. + 5.07 i$, $6. + 6. i$, $-8.48 - 8.48 i$, $-3.46 - 3.46 i$, $24.$)

\vskip 0.7ex
\hangindent=3em \hangafter=1
\textit{Intrinsic sign problem}

  \vskip 2ex

\noindent14. $10_{7,24.}^{48,721}$ \irep{1119}:\ \ 
$d_i$ = ($1.0$,
$1.0$,
$1.0$,
$1.0$,
$1.732$,
$1.732$,
$1.732$,
$1.732$,
$2.0$,
$2.0$) 

\vskip 0.7ex
\hangindent=3em \hangafter=1
$D^2= 24.0 = 
24$

\vskip 0.7ex
\hangindent=3em \hangafter=1
$T = ( 0,
0,
\frac{1}{4},
\frac{1}{4},
\frac{3}{16},
\frac{3}{16},
\frac{11}{16},
\frac{11}{16},
\frac{2}{3},
\frac{11}{12} )
$,

\vskip 0.7ex
\hangindent=3em \hangafter=1
$S$ = ($ 1$,
$ 1$,
$ 1$,
$ 1$,
$ \sqrt{3}$,
$ \sqrt{3}$,
$ \sqrt{3}$,
$ \sqrt{3}$,
$ 2$,
$ 2$;\ \ 
$ 1$,
$ 1$,
$ 1$,
$ -\sqrt{3}$,
$ -\sqrt{3}$,
$ -\sqrt{3}$,
$ -\sqrt{3}$,
$ 2$,
$ 2$;\ \ 
$ -1$,
$ -1$,
$(-\sqrt{3})\mathrm{i}$,
$(\sqrt{3})\mathrm{i}$,
$(-\sqrt{3})\mathrm{i}$,
$(\sqrt{3})\mathrm{i}$,
$ 2$,
$ -2$;\ \ 
$ -1$,
$(\sqrt{3})\mathrm{i}$,
$(-\sqrt{3})\mathrm{i}$,
$(\sqrt{3})\mathrm{i}$,
$(-\sqrt{3})\mathrm{i}$,
$ 2$,
$ -2$;\ \ 
$ \sqrt{3}\zeta_{8}^{3}$,
$ -\sqrt{3}\zeta_{8}^{1}$,
$ -\sqrt{3}\zeta_{8}^{3}$,
$ \sqrt{3}\zeta_{8}^{1}$,
$0$,
$0$;\ \ 
$ \sqrt{3}\zeta_{8}^{3}$,
$ \sqrt{3}\zeta_{8}^{1}$,
$ -\sqrt{3}\zeta_{8}^{3}$,
$0$,
$0$;\ \ 
$ \sqrt{3}\zeta_{8}^{3}$,
$ -\sqrt{3}\zeta_{8}^{1}$,
$0$,
$0$;\ \ 
$ \sqrt{3}\zeta_{8}^{3}$,
$0$,
$0$;\ \ 
$ -2$,
$ -2$;\ \ 
$ 2$)

\vskip 0.7ex
\hangindent=3em \hangafter=1
$\tau_n$ = ($3.46 - 3.46 i$, $-8.48 + 8.48 i$, $6. - 6. i$, $0. - 18.93 i$, $-3.46 + 3.46 i$, $8.48 + 8.48 i$, $-3.46 - 3.46 i$, $-12. + 6.93 i$, $6. + 6. i$, $8.48 - 8.48 i$, $3.46 + 3.46 i$, $12. + 12. i$, $3.46 - 3.46 i$, $-8.48 - 8.48 i$, $6. - 6. i$, $12. - 6.93 i$, $-3.46 + 3.46 i$, $-8.48 + 8.48 i$, $-3.46 - 3.46 i$, $0. - 5.07 i$, $6. + 6. i$, $8.48 + 8.48 i$, $3.46 + 3.46 i$, $0.$, $3.46 - 3.46 i$, $8.48 - 8.48 i$, $6. - 6. i$, $0. + 5.07 i$, $-3.46 + 3.46 i$, $-8.48 - 8.48 i$, $-3.46 - 3.46 i$, $12. + 6.93 i$, $6. + 6. i$, $-8.48 + 8.48 i$, $3.46 + 3.46 i$, $12. - 12. i$, $3.46 - 3.46 i$, $8.48 + 8.48 i$, $6. - 6. i$, $-12. - 6.93 i$, $-3.46 + 3.46 i$, $8.48 - 8.48 i$, $-3.46 - 3.46 i$, $0. + 18.93 i$, $6. + 6. i$, $-8.48 - 8.48 i$, $3.46 + 3.46 i$, $24.$)

\vskip 0.7ex
\hangindent=3em \hangafter=1
\textit{Intrinsic sign problem}

  \vskip 2ex

\noindent15. $10_{3,24.}^{48,100}$ \irep{1119}:\ \ 
$d_i$ = ($1.0$,
$1.0$,
$1.0$,
$1.0$,
$1.732$,
$1.732$,
$1.732$,
$1.732$,
$2.0$,
$2.0$) 

\vskip 0.7ex
\hangindent=3em \hangafter=1
$D^2= 24.0 = 
24$

\vskip 0.7ex
\hangindent=3em \hangafter=1
$T = ( 0,
0,
\frac{1}{4},
\frac{1}{4},
\frac{7}{16},
\frac{7}{16},
\frac{15}{16},
\frac{15}{16},
\frac{1}{3},
\frac{7}{12} )
$,

\vskip 0.7ex
\hangindent=3em \hangafter=1
$S$ = ($ 1$,
$ 1$,
$ 1$,
$ 1$,
$ \sqrt{3}$,
$ \sqrt{3}$,
$ \sqrt{3}$,
$ \sqrt{3}$,
$ 2$,
$ 2$;\ \ 
$ 1$,
$ 1$,
$ 1$,
$ -\sqrt{3}$,
$ -\sqrt{3}$,
$ -\sqrt{3}$,
$ -\sqrt{3}$,
$ 2$,
$ 2$;\ \ 
$ -1$,
$ -1$,
$(-\sqrt{3})\mathrm{i}$,
$(\sqrt{3})\mathrm{i}$,
$(-\sqrt{3})\mathrm{i}$,
$(\sqrt{3})\mathrm{i}$,
$ 2$,
$ -2$;\ \ 
$ -1$,
$(\sqrt{3})\mathrm{i}$,
$(-\sqrt{3})\mathrm{i}$,
$(\sqrt{3})\mathrm{i}$,
$(-\sqrt{3})\mathrm{i}$,
$ 2$,
$ -2$;\ \ 
$ \sqrt{3}\zeta_{8}^{3}$,
$ -\sqrt{3}\zeta_{8}^{1}$,
$ -\sqrt{3}\zeta_{8}^{3}$,
$ \sqrt{3}\zeta_{8}^{1}$,
$0$,
$0$;\ \ 
$ \sqrt{3}\zeta_{8}^{3}$,
$ \sqrt{3}\zeta_{8}^{1}$,
$ -\sqrt{3}\zeta_{8}^{3}$,
$0$,
$0$;\ \ 
$ \sqrt{3}\zeta_{8}^{3}$,
$ -\sqrt{3}\zeta_{8}^{1}$,
$0$,
$0$;\ \ 
$ \sqrt{3}\zeta_{8}^{3}$,
$0$,
$0$;\ \ 
$ -2$,
$ -2$;\ \ 
$ 2$)

\vskip 0.7ex
\hangindent=3em \hangafter=1
$\tau_n$ = ($-3.46 + 3.46 i$, $8.48 - 8.48 i$, $6. - 6. i$, $0. - 5.07 i$, $3.46 - 3.46 i$, $-8.48 - 8.48 i$, $3.46 + 3.46 i$, $-12. - 6.93 i$, $6. + 6. i$, $-8.48 + 8.48 i$, $-3.46 - 3.46 i$, $12. + 12. i$, $-3.46 + 3.46 i$, $8.48 + 8.48 i$, $6. - 6. i$, $12. + 6.93 i$, $3.46 - 3.46 i$, $8.48 - 8.48 i$, $3.46 + 3.46 i$, $0. - 18.93 i$, $6. + 6. i$, $-8.48 - 8.48 i$, $-3.46 - 3.46 i$, $0.$, $-3.46 + 3.46 i$, $-8.48 + 8.48 i$, $6. - 6. i$, $0. + 18.93 i$, $3.46 - 3.46 i$, $8.48 + 8.48 i$, $3.46 + 3.46 i$, $12. - 6.93 i$, $6. + 6. i$, $8.48 - 8.48 i$, $-3.46 - 3.46 i$, $12. - 12. i$, $-3.46 + 3.46 i$, $-8.48 - 8.48 i$, $6. - 6. i$, $-12. + 6.93 i$, $3.46 - 3.46 i$, $-8.48 + 8.48 i$, $3.46 + 3.46 i$, $0. + 5.07 i$, $6. + 6. i$, $8.48 + 8.48 i$, $-3.46 - 3.46 i$, $24.$)

\vskip 0.7ex
\hangindent=3em \hangafter=1
\textit{Intrinsic sign problem}

  \vskip 2ex

\noindent16. $10_{7,24.}^{48,267}$ \irep{1119}:\ \ 
$d_i$ = ($1.0$,
$1.0$,
$1.0$,
$1.0$,
$1.732$,
$1.732$,
$1.732$,
$1.732$,
$2.0$,
$2.0$) 

\vskip 0.7ex
\hangindent=3em \hangafter=1
$D^2= 24.0 = 
24$

\vskip 0.7ex
\hangindent=3em \hangafter=1
$T = ( 0,
0,
\frac{1}{4},
\frac{1}{4},
\frac{7}{16},
\frac{7}{16},
\frac{15}{16},
\frac{15}{16},
\frac{2}{3},
\frac{11}{12} )
$,

\vskip 0.7ex
\hangindent=3em \hangafter=1
$S$ = ($ 1$,
$ 1$,
$ 1$,
$ 1$,
$ \sqrt{3}$,
$ \sqrt{3}$,
$ \sqrt{3}$,
$ \sqrt{3}$,
$ 2$,
$ 2$;\ \ 
$ 1$,
$ 1$,
$ 1$,
$ -\sqrt{3}$,
$ -\sqrt{3}$,
$ -\sqrt{3}$,
$ -\sqrt{3}$,
$ 2$,
$ 2$;\ \ 
$ -1$,
$ -1$,
$(-\sqrt{3})\mathrm{i}$,
$(\sqrt{3})\mathrm{i}$,
$(-\sqrt{3})\mathrm{i}$,
$(\sqrt{3})\mathrm{i}$,
$ 2$,
$ -2$;\ \ 
$ -1$,
$(\sqrt{3})\mathrm{i}$,
$(-\sqrt{3})\mathrm{i}$,
$(\sqrt{3})\mathrm{i}$,
$(-\sqrt{3})\mathrm{i}$,
$ 2$,
$ -2$;\ \ 
$ -\sqrt{3}\zeta_{8}^{3}$,
$ \sqrt{3}\zeta_{8}^{1}$,
$ \sqrt{3}\zeta_{8}^{3}$,
$ -\sqrt{3}\zeta_{8}^{1}$,
$0$,
$0$;\ \ 
$ -\sqrt{3}\zeta_{8}^{3}$,
$ -\sqrt{3}\zeta_{8}^{1}$,
$ \sqrt{3}\zeta_{8}^{3}$,
$0$,
$0$;\ \ 
$ -\sqrt{3}\zeta_{8}^{3}$,
$ \sqrt{3}\zeta_{8}^{1}$,
$0$,
$0$;\ \ 
$ -\sqrt{3}\zeta_{8}^{3}$,
$0$,
$0$;\ \ 
$ -2$,
$ -2$;\ \ 
$ 2$)

\vskip 0.7ex
\hangindent=3em \hangafter=1
$\tau_n$ = ($3.46 - 3.46 i$, $8.48 - 8.48 i$, $6. - 6. i$, $0. - 18.93 i$, $-3.46 + 3.46 i$, $-8.48 - 8.48 i$, $-3.46 - 3.46 i$, $-12. + 6.93 i$, $6. + 6. i$, $-8.48 + 8.48 i$, $3.46 + 3.46 i$, $12. + 12. i$, $3.46 - 3.46 i$, $8.48 + 8.48 i$, $6. - 6. i$, $12. - 6.93 i$, $-3.46 + 3.46 i$, $8.48 - 8.48 i$, $-3.46 - 3.46 i$, $0. - 5.07 i$, $6. + 6. i$, $-8.48 - 8.48 i$, $3.46 + 3.46 i$, $0.$, $3.46 - 3.46 i$, $-8.48 + 8.48 i$, $6. - 6. i$, $0. + 5.07 i$, $-3.46 + 3.46 i$, $8.48 + 8.48 i$, $-3.46 - 3.46 i$, $12. + 6.93 i$, $6. + 6. i$, $8.48 - 8.48 i$, $3.46 + 3.46 i$, $12. - 12. i$, $3.46 - 3.46 i$, $-8.48 - 8.48 i$, $6. - 6. i$, $-12. - 6.93 i$, $-3.46 + 3.46 i$, $-8.48 + 8.48 i$, $-3.46 - 3.46 i$, $0. + 18.93 i$, $6. + 6. i$, $8.48 + 8.48 i$, $3.46 + 3.46 i$, $24.$)

\vskip 0.7ex
\hangindent=3em \hangafter=1
\textit{Intrinsic sign problem}

  \vskip 2ex

\noindent17. $10_{1,24.}^{48,126}$ \irep{1119}:\ \ 
$d_i$ = ($1.0$,
$1.0$,
$1.0$,
$1.0$,
$1.732$,
$1.732$,
$1.732$,
$1.732$,
$2.0$,
$2.0$) 

\vskip 0.7ex
\hangindent=3em \hangafter=1
$D^2= 24.0 = 
24$

\vskip 0.7ex
\hangindent=3em \hangafter=1
$T = ( 0,
0,
\frac{3}{4},
\frac{3}{4},
\frac{1}{16},
\frac{1}{16},
\frac{9}{16},
\frac{9}{16},
\frac{1}{3},
\frac{1}{12} )
$,

\vskip 0.7ex
\hangindent=3em \hangafter=1
$S$ = ($ 1$,
$ 1$,
$ 1$,
$ 1$,
$ \sqrt{3}$,
$ \sqrt{3}$,
$ \sqrt{3}$,
$ \sqrt{3}$,
$ 2$,
$ 2$;\ \ 
$ 1$,
$ 1$,
$ 1$,
$ -\sqrt{3}$,
$ -\sqrt{3}$,
$ -\sqrt{3}$,
$ -\sqrt{3}$,
$ 2$,
$ 2$;\ \ 
$ -1$,
$ -1$,
$(-\sqrt{3})\mathrm{i}$,
$(\sqrt{3})\mathrm{i}$,
$(-\sqrt{3})\mathrm{i}$,
$(\sqrt{3})\mathrm{i}$,
$ 2$,
$ -2$;\ \ 
$ -1$,
$(\sqrt{3})\mathrm{i}$,
$(-\sqrt{3})\mathrm{i}$,
$(\sqrt{3})\mathrm{i}$,
$(-\sqrt{3})\mathrm{i}$,
$ 2$,
$ -2$;\ \ 
$ \sqrt{3}\zeta_{8}^{1}$,
$ -\sqrt{3}\zeta_{8}^{3}$,
$ -\sqrt{3}\zeta_{8}^{1}$,
$ \sqrt{3}\zeta_{8}^{3}$,
$0$,
$0$;\ \ 
$ \sqrt{3}\zeta_{8}^{1}$,
$ \sqrt{3}\zeta_{8}^{3}$,
$ -\sqrt{3}\zeta_{8}^{1}$,
$0$,
$0$;\ \ 
$ \sqrt{3}\zeta_{8}^{1}$,
$ -\sqrt{3}\zeta_{8}^{3}$,
$0$,
$0$;\ \ 
$ \sqrt{3}\zeta_{8}^{1}$,
$0$,
$0$;\ \ 
$ -2$,
$ -2$;\ \ 
$ 2$)

\vskip 0.7ex
\hangindent=3em \hangafter=1
$\tau_n$ = ($3.46 + 3.46 i$, $8.48 + 8.48 i$, $6. + 6. i$, $0. + 18.93 i$, $-3.46 - 3.46 i$, $-8.48 + 8.48 i$, $-3.46 + 3.46 i$, $-12. - 6.93 i$, $6. - 6. i$, $-8.48 - 8.48 i$, $3.46 - 3.46 i$, $12. - 12. i$, $3.46 + 3.46 i$, $8.48 - 8.48 i$, $6. + 6. i$, $12. + 6.93 i$, $-3.46 - 3.46 i$, $8.48 + 8.48 i$, $-3.46 + 3.46 i$, $0. + 5.07 i$, $6. - 6. i$, $-8.48 + 8.48 i$, $3.46 - 3.46 i$, $0.$, $3.46 + 3.46 i$, $-8.48 - 8.48 i$, $6. + 6. i$, $0. - 5.07 i$, $-3.46 - 3.46 i$, $8.48 - 8.48 i$, $-3.46 + 3.46 i$, $12. - 6.93 i$, $6. - 6. i$, $8.48 + 8.48 i$, $3.46 - 3.46 i$, $12. + 12. i$, $3.46 + 3.46 i$, $-8.48 + 8.48 i$, $6. + 6. i$, $-12. + 6.93 i$, $-3.46 - 3.46 i$, $-8.48 - 8.48 i$, $-3.46 + 3.46 i$, $0. - 18.93 i$, $6. - 6. i$, $8.48 - 8.48 i$, $3.46 - 3.46 i$, $24.$)

\vskip 0.7ex
\hangindent=3em \hangafter=1
\textit{Intrinsic sign problem}

  \vskip 2ex

\noindent18. $10_{5,24.}^{48,261}$ \irep{1119}:\ \ 
$d_i$ = ($1.0$,
$1.0$,
$1.0$,
$1.0$,
$1.732$,
$1.732$,
$1.732$,
$1.732$,
$2.0$,
$2.0$) 

\vskip 0.7ex
\hangindent=3em \hangafter=1
$D^2= 24.0 = 
24$

\vskip 0.7ex
\hangindent=3em \hangafter=1
$T = ( 0,
0,
\frac{3}{4},
\frac{3}{4},
\frac{1}{16},
\frac{1}{16},
\frac{9}{16},
\frac{9}{16},
\frac{2}{3},
\frac{5}{12} )
$,

\vskip 0.7ex
\hangindent=3em \hangafter=1
$S$ = ($ 1$,
$ 1$,
$ 1$,
$ 1$,
$ \sqrt{3}$,
$ \sqrt{3}$,
$ \sqrt{3}$,
$ \sqrt{3}$,
$ 2$,
$ 2$;\ \ 
$ 1$,
$ 1$,
$ 1$,
$ -\sqrt{3}$,
$ -\sqrt{3}$,
$ -\sqrt{3}$,
$ -\sqrt{3}$,
$ 2$,
$ 2$;\ \ 
$ -1$,
$ -1$,
$(-\sqrt{3})\mathrm{i}$,
$(\sqrt{3})\mathrm{i}$,
$(-\sqrt{3})\mathrm{i}$,
$(\sqrt{3})\mathrm{i}$,
$ 2$,
$ -2$;\ \ 
$ -1$,
$(\sqrt{3})\mathrm{i}$,
$(-\sqrt{3})\mathrm{i}$,
$(\sqrt{3})\mathrm{i}$,
$(-\sqrt{3})\mathrm{i}$,
$ 2$,
$ -2$;\ \ 
$ -\sqrt{3}\zeta_{8}^{1}$,
$ \sqrt{3}\zeta_{8}^{3}$,
$ \sqrt{3}\zeta_{8}^{1}$,
$ -\sqrt{3}\zeta_{8}^{3}$,
$0$,
$0$;\ \ 
$ -\sqrt{3}\zeta_{8}^{1}$,
$ -\sqrt{3}\zeta_{8}^{3}$,
$ \sqrt{3}\zeta_{8}^{1}$,
$0$,
$0$;\ \ 
$ -\sqrt{3}\zeta_{8}^{1}$,
$ \sqrt{3}\zeta_{8}^{3}$,
$0$,
$0$;\ \ 
$ -\sqrt{3}\zeta_{8}^{1}$,
$0$,
$0$;\ \ 
$ -2$,
$ -2$;\ \ 
$ 2$)

\vskip 0.7ex
\hangindent=3em \hangafter=1
$\tau_n$ = ($-3.46 - 3.46 i$, $8.48 + 8.48 i$, $6. + 6. i$, $0. + 5.07 i$, $3.46 + 3.46 i$, $-8.48 + 8.48 i$, $3.46 - 3.46 i$, $-12. + 6.93 i$, $6. - 6. i$, $-8.48 - 8.48 i$, $-3.46 + 3.46 i$, $12. - 12. i$, $-3.46 - 3.46 i$, $8.48 - 8.48 i$, $6. + 6. i$, $12. - 6.93 i$, $3.46 + 3.46 i$, $8.48 + 8.48 i$, $3.46 - 3.46 i$, $0. + 18.93 i$, $6. - 6. i$, $-8.48 + 8.48 i$, $-3.46 + 3.46 i$, $0.$, $-3.46 - 3.46 i$, $-8.48 - 8.48 i$, $6. + 6. i$, $0. - 18.93 i$, $3.46 + 3.46 i$, $8.48 - 8.48 i$, $3.46 - 3.46 i$, $12. + 6.93 i$, $6. - 6. i$, $8.48 + 8.48 i$, $-3.46 + 3.46 i$, $12. + 12. i$, $-3.46 - 3.46 i$, $-8.48 + 8.48 i$, $6. + 6. i$, $-12. - 6.93 i$, $3.46 + 3.46 i$, $-8.48 - 8.48 i$, $3.46 - 3.46 i$, $0. - 5.07 i$, $6. - 6. i$, $8.48 - 8.48 i$, $-3.46 + 3.46 i$, $24.$)

\vskip 0.7ex
\hangindent=3em \hangafter=1
\textit{Intrinsic sign problem}

  \vskip 2ex

\noindent19. $10_{1,24.}^{48,254}$ \irep{1119}:\ \ 
$d_i$ = ($1.0$,
$1.0$,
$1.0$,
$1.0$,
$1.732$,
$1.732$,
$1.732$,
$1.732$,
$2.0$,
$2.0$) 

\vskip 0.7ex
\hangindent=3em \hangafter=1
$D^2= 24.0 = 
24$

\vskip 0.7ex
\hangindent=3em \hangafter=1
$T = ( 0,
0,
\frac{3}{4},
\frac{3}{4},
\frac{5}{16},
\frac{5}{16},
\frac{13}{16},
\frac{13}{16},
\frac{1}{3},
\frac{1}{12} )
$,

\vskip 0.7ex
\hangindent=3em \hangafter=1
$S$ = ($ 1$,
$ 1$,
$ 1$,
$ 1$,
$ \sqrt{3}$,
$ \sqrt{3}$,
$ \sqrt{3}$,
$ \sqrt{3}$,
$ 2$,
$ 2$;\ \ 
$ 1$,
$ 1$,
$ 1$,
$ -\sqrt{3}$,
$ -\sqrt{3}$,
$ -\sqrt{3}$,
$ -\sqrt{3}$,
$ 2$,
$ 2$;\ \ 
$ -1$,
$ -1$,
$(-\sqrt{3})\mathrm{i}$,
$(\sqrt{3})\mathrm{i}$,
$(-\sqrt{3})\mathrm{i}$,
$(\sqrt{3})\mathrm{i}$,
$ 2$,
$ -2$;\ \ 
$ -1$,
$(\sqrt{3})\mathrm{i}$,
$(-\sqrt{3})\mathrm{i}$,
$(\sqrt{3})\mathrm{i}$,
$(-\sqrt{3})\mathrm{i}$,
$ 2$,
$ -2$;\ \ 
$ -\sqrt{3}\zeta_{8}^{1}$,
$ \sqrt{3}\zeta_{8}^{3}$,
$ \sqrt{3}\zeta_{8}^{1}$,
$ -\sqrt{3}\zeta_{8}^{3}$,
$0$,
$0$;\ \ 
$ -\sqrt{3}\zeta_{8}^{1}$,
$ -\sqrt{3}\zeta_{8}^{3}$,
$ \sqrt{3}\zeta_{8}^{1}$,
$0$,
$0$;\ \ 
$ -\sqrt{3}\zeta_{8}^{1}$,
$ \sqrt{3}\zeta_{8}^{3}$,
$0$,
$0$;\ \ 
$ -\sqrt{3}\zeta_{8}^{1}$,
$0$,
$0$;\ \ 
$ -2$,
$ -2$;\ \ 
$ 2$)

\vskip 0.7ex
\hangindent=3em \hangafter=1
$\tau_n$ = ($3.46 + 3.46 i$, $-8.48 - 8.48 i$, $6. + 6. i$, $0. + 18.93 i$, $-3.46 - 3.46 i$, $8.48 - 8.48 i$, $-3.46 + 3.46 i$, $-12. - 6.93 i$, $6. - 6. i$, $8.48 + 8.48 i$, $3.46 - 3.46 i$, $12. - 12. i$, $3.46 + 3.46 i$, $-8.48 + 8.48 i$, $6. + 6. i$, $12. + 6.93 i$, $-3.46 - 3.46 i$, $-8.48 - 8.48 i$, $-3.46 + 3.46 i$, $0. + 5.07 i$, $6. - 6. i$, $8.48 - 8.48 i$, $3.46 - 3.46 i$, $0.$, $3.46 + 3.46 i$, $8.48 + 8.48 i$, $6. + 6. i$, $0. - 5.07 i$, $-3.46 - 3.46 i$, $-8.48 + 8.48 i$, $-3.46 + 3.46 i$, $12. - 6.93 i$, $6. - 6. i$, $-8.48 - 8.48 i$, $3.46 - 3.46 i$, $12. + 12. i$, $3.46 + 3.46 i$, $8.48 - 8.48 i$, $6. + 6. i$, $-12. + 6.93 i$, $-3.46 - 3.46 i$, $8.48 + 8.48 i$, $-3.46 + 3.46 i$, $0. - 18.93 i$, $6. - 6. i$, $-8.48 + 8.48 i$, $3.46 - 3.46 i$, $24.$)

\vskip 0.7ex
\hangindent=3em \hangafter=1
\textit{Intrinsic sign problem}

  \vskip 2ex

\noindent20. $10_{5,24.}^{48,125}$ \irep{1119}:\ \ 
$d_i$ = ($1.0$,
$1.0$,
$1.0$,
$1.0$,
$1.732$,
$1.732$,
$1.732$,
$1.732$,
$2.0$,
$2.0$) 

\vskip 0.7ex
\hangindent=3em \hangafter=1
$D^2= 24.0 = 
24$

\vskip 0.7ex
\hangindent=3em \hangafter=1
$T = ( 0,
0,
\frac{3}{4},
\frac{3}{4},
\frac{5}{16},
\frac{5}{16},
\frac{13}{16},
\frac{13}{16},
\frac{2}{3},
\frac{5}{12} )
$,

\vskip 0.7ex
\hangindent=3em \hangafter=1
$S$ = ($ 1$,
$ 1$,
$ 1$,
$ 1$,
$ \sqrt{3}$,
$ \sqrt{3}$,
$ \sqrt{3}$,
$ \sqrt{3}$,
$ 2$,
$ 2$;\ \ 
$ 1$,
$ 1$,
$ 1$,
$ -\sqrt{3}$,
$ -\sqrt{3}$,
$ -\sqrt{3}$,
$ -\sqrt{3}$,
$ 2$,
$ 2$;\ \ 
$ -1$,
$ -1$,
$(-\sqrt{3})\mathrm{i}$,
$(\sqrt{3})\mathrm{i}$,
$(-\sqrt{3})\mathrm{i}$,
$(\sqrt{3})\mathrm{i}$,
$ 2$,
$ -2$;\ \ 
$ -1$,
$(\sqrt{3})\mathrm{i}$,
$(-\sqrt{3})\mathrm{i}$,
$(\sqrt{3})\mathrm{i}$,
$(-\sqrt{3})\mathrm{i}$,
$ 2$,
$ -2$;\ \ 
$ \sqrt{3}\zeta_{8}^{1}$,
$ -\sqrt{3}\zeta_{8}^{3}$,
$ -\sqrt{3}\zeta_{8}^{1}$,
$ \sqrt{3}\zeta_{8}^{3}$,
$0$,
$0$;\ \ 
$ \sqrt{3}\zeta_{8}^{1}$,
$ \sqrt{3}\zeta_{8}^{3}$,
$ -\sqrt{3}\zeta_{8}^{1}$,
$0$,
$0$;\ \ 
$ \sqrt{3}\zeta_{8}^{1}$,
$ -\sqrt{3}\zeta_{8}^{3}$,
$0$,
$0$;\ \ 
$ \sqrt{3}\zeta_{8}^{1}$,
$0$,
$0$;\ \ 
$ -2$,
$ -2$;\ \ 
$ 2$)

\vskip 0.7ex
\hangindent=3em \hangafter=1
$\tau_n$ = ($-3.46 - 3.46 i$, $-8.48 - 8.48 i$, $6. + 6. i$, $0. + 5.07 i$, $3.46 + 3.46 i$, $8.48 - 8.48 i$, $3.46 - 3.46 i$, $-12. + 6.93 i$, $6. - 6. i$, $8.48 + 8.48 i$, $-3.46 + 3.46 i$, $12. - 12. i$, $-3.46 - 3.46 i$, $-8.48 + 8.48 i$, $6. + 6. i$, $12. - 6.93 i$, $3.46 + 3.46 i$, $-8.48 - 8.48 i$, $3.46 - 3.46 i$, $0. + 18.93 i$, $6. - 6. i$, $8.48 - 8.48 i$, $-3.46 + 3.46 i$, $0.$, $-3.46 - 3.46 i$, $8.48 + 8.48 i$, $6. + 6. i$, $0. - 18.93 i$, $3.46 + 3.46 i$, $-8.48 + 8.48 i$, $3.46 - 3.46 i$, $12. + 6.93 i$, $6. - 6. i$, $-8.48 - 8.48 i$, $-3.46 + 3.46 i$, $12. + 12. i$, $-3.46 - 3.46 i$, $8.48 - 8.48 i$, $6. + 6. i$, $-12. - 6.93 i$, $3.46 + 3.46 i$, $8.48 + 8.48 i$, $3.46 - 3.46 i$, $0. - 5.07 i$, $6. - 6. i$, $-8.48 + 8.48 i$, $-3.46 + 3.46 i$, $24.$)

\vskip 0.7ex
\hangindent=3em \hangafter=1
\textit{Intrinsic sign problem}

  \vskip 2ex

\noindent21. $10_{4,36.}^{6,152}$ \irep{0}:\ \ 
$d_i$ = ($1.0$,
$1.0$,
$1.0$,
$2.0$,
$2.0$,
$2.0$,
$2.0$,
$2.0$,
$2.0$,
$3.0$) 

\vskip 0.7ex
\hangindent=3em \hangafter=1
$D^2= 36.0 = 
36$

\vskip 0.7ex
\hangindent=3em \hangafter=1
$T = ( 0,
0,
0,
0,
0,
\frac{1}{3},
\frac{1}{3},
\frac{2}{3},
\frac{2}{3},
\frac{1}{2} )
$,

\vskip 0.7ex
\hangindent=3em \hangafter=1
$S$ = ($ 1$,
$ 1$,
$ 1$,
$ 2$,
$ 2$,
$ 2$,
$ 2$,
$ 2$,
$ 2$,
$ 3$;\ \ 
$ 1$,
$ 1$,
$ -2\zeta_{6}^{1}$,
$ 2\zeta_{3}^{1}$,
$ -2\zeta_{6}^{1}$,
$ 2\zeta_{3}^{1}$,
$ -2\zeta_{6}^{1}$,
$ 2\zeta_{3}^{1}$,
$ 3$;\ \ 
$ 1$,
$ 2\zeta_{3}^{1}$,
$ -2\zeta_{6}^{1}$,
$ 2\zeta_{3}^{1}$,
$ -2\zeta_{6}^{1}$,
$ 2\zeta_{3}^{1}$,
$ -2\zeta_{6}^{1}$,
$ 3$;\ \ 
$ -2$,
$ -2$,
$ 2\zeta_{6}^{1}$,
$ -2\zeta_{3}^{1}$,
$ -2\zeta_{3}^{1}$,
$ 2\zeta_{6}^{1}$,
$0$;\ \ 
$ -2$,
$ -2\zeta_{3}^{1}$,
$ 2\zeta_{6}^{1}$,
$ 2\zeta_{6}^{1}$,
$ -2\zeta_{3}^{1}$,
$0$;\ \ 
$ -2\zeta_{3}^{1}$,
$ 2\zeta_{6}^{1}$,
$ -2$,
$ -2$,
$0$;\ \ 
$ -2\zeta_{3}^{1}$,
$ -2$,
$ -2$,
$0$;\ \ 
$ 2\zeta_{6}^{1}$,
$ -2\zeta_{3}^{1}$,
$0$;\ \ 
$ 2\zeta_{6}^{1}$,
$0$;\ \ 
$ -3$)

\vskip 0.7ex
\hangindent=3em \hangafter=1
$\tau_n$ = ($-6.$, $12.$, $18.$, $12.$, $-6.$, $36.$)

\vskip 0.7ex
\hangindent=3em \hangafter=1
\textit{Intrinsic sign problem}

  \vskip 2ex

\noindent22. $10_{4,36.}^{18,490}$ \irep{0}:\ \ 
$d_i$ = ($1.0$,
$1.0$,
$1.0$,
$2.0$,
$2.0$,
$2.0$,
$2.0$,
$2.0$,
$2.0$,
$3.0$) 

\vskip 0.7ex
\hangindent=3em \hangafter=1
$D^2= 36.0 = 
36$

\vskip 0.7ex
\hangindent=3em \hangafter=1
$T = ( 0,
0,
0,
\frac{1}{9},
\frac{1}{9},
\frac{4}{9},
\frac{4}{9},
\frac{7}{9},
\frac{7}{9},
\frac{1}{2} )
$,

\vskip 0.7ex
\hangindent=3em \hangafter=1
$S$ = ($ 1$,
$ 1$,
$ 1$,
$ 2$,
$ 2$,
$ 2$,
$ 2$,
$ 2$,
$ 2$,
$ 3$;\ \ 
$ 1$,
$ 1$,
$ -2\zeta_{6}^{1}$,
$ 2\zeta_{3}^{1}$,
$ -2\zeta_{6}^{1}$,
$ 2\zeta_{3}^{1}$,
$ -2\zeta_{6}^{1}$,
$ 2\zeta_{3}^{1}$,
$ 3$;\ \ 
$ 1$,
$ 2\zeta_{3}^{1}$,
$ -2\zeta_{6}^{1}$,
$ 2\zeta_{3}^{1}$,
$ -2\zeta_{6}^{1}$,
$ 2\zeta_{3}^{1}$,
$ -2\zeta_{6}^{1}$,
$ 3$;\ \ 
$ 2\zeta_{18}^{5}$,
$ -2\zeta_{9}^{2}$,
$ -2\zeta_{9}^{4}$,
$ 2\zeta_{18}^{1}$,
$ -2\zeta_{9}^{1}$,
$ 2\zeta_{18}^{7}$,
$0$;\ \ 
$ 2\zeta_{18}^{5}$,
$ 2\zeta_{18}^{1}$,
$ -2\zeta_{9}^{4}$,
$ 2\zeta_{18}^{7}$,
$ -2\zeta_{9}^{1}$,
$0$;\ \ 
$ -2\zeta_{9}^{1}$,
$ 2\zeta_{18}^{7}$,
$ 2\zeta_{18}^{5}$,
$ -2\zeta_{9}^{2}$,
$0$;\ \ 
$ -2\zeta_{9}^{1}$,
$ -2\zeta_{9}^{2}$,
$ 2\zeta_{18}^{5}$,
$0$;\ \ 
$ -2\zeta_{9}^{4}$,
$ 2\zeta_{18}^{1}$,
$0$;\ \ 
$ -2\zeta_{9}^{4}$,
$0$;\ \ 
$ -3$)

\vskip 0.7ex
\hangindent=3em \hangafter=1
$\tau_n$ = ($-6.$, $12.$, $-18. + 20.78 i$, $12.$, $-6.$, $0. - 20.78 i$, $-6.$, $12.$, $18.$, $12.$, $-6.$, $0. + 20.78 i$, $-6.$, $12.$, $-18. - 20.78 i$, $12.$, $-6.$, $36.$)

\vskip 0.7ex
\hangindent=3em \hangafter=1
\textit{Intrinsic sign problem}

  \vskip 2ex

\noindent23. $10_{4,36.}^{18,842}$ \irep{0}:\ \ 
$d_i$ = ($1.0$,
$1.0$,
$1.0$,
$2.0$,
$2.0$,
$2.0$,
$2.0$,
$2.0$,
$2.0$,
$3.0$) 

\vskip 0.7ex
\hangindent=3em \hangafter=1
$D^2= 36.0 = 
36$

\vskip 0.7ex
\hangindent=3em \hangafter=1
$T = ( 0,
0,
0,
\frac{2}{9},
\frac{2}{9},
\frac{5}{9},
\frac{5}{9},
\frac{8}{9},
\frac{8}{9},
\frac{1}{2} )
$,

\vskip 0.7ex
\hangindent=3em \hangafter=1
$S$ = ($ 1$,
$ 1$,
$ 1$,
$ 2$,
$ 2$,
$ 2$,
$ 2$,
$ 2$,
$ 2$,
$ 3$;\ \ 
$ 1$,
$ 1$,
$ -2\zeta_{6}^{1}$,
$ 2\zeta_{3}^{1}$,
$ -2\zeta_{6}^{1}$,
$ 2\zeta_{3}^{1}$,
$ -2\zeta_{6}^{1}$,
$ 2\zeta_{3}^{1}$,
$ 3$;\ \ 
$ 1$,
$ 2\zeta_{3}^{1}$,
$ -2\zeta_{6}^{1}$,
$ 2\zeta_{3}^{1}$,
$ -2\zeta_{6}^{1}$,
$ 2\zeta_{3}^{1}$,
$ -2\zeta_{6}^{1}$,
$ 3$;\ \ 
$ 2\zeta_{18}^{1}$,
$ -2\zeta_{9}^{4}$,
$ -2\zeta_{9}^{2}$,
$ 2\zeta_{18}^{5}$,
$ 2\zeta_{18}^{7}$,
$ -2\zeta_{9}^{1}$,
$0$;\ \ 
$ 2\zeta_{18}^{1}$,
$ 2\zeta_{18}^{5}$,
$ -2\zeta_{9}^{2}$,
$ -2\zeta_{9}^{1}$,
$ 2\zeta_{18}^{7}$,
$0$;\ \ 
$ 2\zeta_{18}^{7}$,
$ -2\zeta_{9}^{1}$,
$ 2\zeta_{18}^{1}$,
$ -2\zeta_{9}^{4}$,
$0$;\ \ 
$ 2\zeta_{18}^{7}$,
$ -2\zeta_{9}^{4}$,
$ 2\zeta_{18}^{1}$,
$0$;\ \ 
$ -2\zeta_{9}^{2}$,
$ 2\zeta_{18}^{5}$,
$0$;\ \ 
$ -2\zeta_{9}^{2}$,
$0$;\ \ 
$ -3$)

\vskip 0.7ex
\hangindent=3em \hangafter=1
$\tau_n$ = ($-6.$, $12.$, $-18. - 20.78 i$, $12.$, $-6.$, $0. + 20.78 i$, $-6.$, $12.$, $18.$, $12.$, $-6.$, $0. - 20.78 i$, $-6.$, $12.$, $-18. + 20.78 i$, $12.$, $-6.$, $36.$)

\vskip 0.7ex
\hangindent=3em \hangafter=1
\textit{Intrinsic sign problem}

  \vskip 2ex

\noindent24. $10_{\frac{24}{5},43.41}^{120,927}$ \irep{1146}:\ \ 
$d_i$ = ($1.0$,
$1.0$,
$1.618$,
$1.618$,
$1.732$,
$1.732$,
$2.0$,
$2.802$,
$2.802$,
$3.236$) 

\vskip 0.7ex
\hangindent=3em \hangafter=1
$D^2= 43.416 = 
30+6\sqrt{5}$

\vskip 0.7ex
\hangindent=3em \hangafter=1
$T = ( 0,
0,
\frac{2}{5},
\frac{2}{5},
\frac{1}{8},
\frac{5}{8},
\frac{1}{3},
\frac{1}{40},
\frac{21}{40},
\frac{11}{15} )
$,

\vskip 0.7ex
\hangindent=3em \hangafter=1
$S$ = ($ 1$,
$ 1$,
$ \frac{1+\sqrt{5}}{2}$,
$ \frac{1+\sqrt{5}}{2}$,
$ \sqrt{3}$,
$ \sqrt{3}$,
$ 2$,
$ \frac{15+3\sqrt{5}}{2\sqrt{15}}$,
$ \frac{15+3\sqrt{5}}{2\sqrt{15}}$,
$ 1+\sqrt{5}$;\ \ 
$ 1$,
$ \frac{1+\sqrt{5}}{2}$,
$ \frac{1+\sqrt{5}}{2}$,
$ -\sqrt{3}$,
$ -\sqrt{3}$,
$ 2$,
$ -\frac{15+3\sqrt{5}}{2\sqrt{15}}$,
$ -\frac{15+3\sqrt{5}}{2\sqrt{15}}$,
$ 1+\sqrt{5}$;\ \ 
$ -1$,
$ -1$,
$ -\frac{15+3\sqrt{5}}{2\sqrt{15}}$,
$ -\frac{15+3\sqrt{5}}{2\sqrt{15}}$,
$ 1+\sqrt{5}$,
$ \sqrt{3}$,
$ \sqrt{3}$,
$ -2$;\ \ 
$ -1$,
$ \frac{15+3\sqrt{5}}{2\sqrt{15}}$,
$ \frac{15+3\sqrt{5}}{2\sqrt{15}}$,
$ 1+\sqrt{5}$,
$ -\sqrt{3}$,
$ -\sqrt{3}$,
$ -2$;\ \ 
$ \sqrt{3}$,
$ -\sqrt{3}$,
$0$,
$ -\frac{15+3\sqrt{5}}{2\sqrt{15}}$,
$ \frac{15+3\sqrt{5}}{2\sqrt{15}}$,
$0$;\ \ 
$ \sqrt{3}$,
$0$,
$ \frac{15+3\sqrt{5}}{2\sqrt{15}}$,
$ -\frac{15+3\sqrt{5}}{2\sqrt{15}}$,
$0$;\ \ 
$ -2$,
$0$,
$0$,
$ -1-\sqrt{5}$;\ \ 
$ -\sqrt{3}$,
$ \sqrt{3}$,
$0$;\ \ 
$ -\sqrt{3}$,
$0$;\ \ 
$ 2$)

Factors = $2_{\frac{14}{5},3.618}^{5,395}\boxtimes 5_{2,12.}^{24,940}$

\vskip 0.7ex
\hangindent=3em \hangafter=1
$\tau_n$ = ($-5.33 - 3.87 i$, $6.31 + 4.59 i$, $10.85 + 14.94 i$, $12.03 + 5.36 i$, $0. - 12.53 i$, $2.52 + 15.94 i$, $8.62 + 6.27 i$, $19.48 + 8.67 i$, $-6.71 - 9.23 i$, $0. + 34.23 i$, $5.33 + 3.87 i$, $0.$, $-8.62 + 6.27 i$, $-14.56 + 10.58 i$, $21.71$, $-12.03 + 5.36 i$, $-8.62 - 6.27 i$, $-4.08 + 25.79 i$, $5.33 - 3.87 i$, $-21.7 - 12.53 i$, $-6.71 + 9.23 i$, $-6.31 - 4.59 i$, $8.62 - 6.27 i$, $-13.41 - 18.46 i$, $0. + 12.53 i$, $-3.9 - 2.83 i$, $10.85 - 14.94 i$, $-19.48 - 8.67 i$, $-5.33 + 3.87 i$, $21.71 - 21.7 i$, $-5.33 - 3.87 i$, $2.23 - 21.2 i$, $10.85 + 14.94 i$, $14.56 - 10.58 i$, $0. - 12.53 i$, $0.$, $8.62 + 6.27 i$, $23.56 - 17.12 i$, $-6.71 - 9.23 i$, $21.7 + 12.53 i$, $5.33 + 3.87 i$, $25.79 - 4.09 i$, $-8.62 + 6.27 i$, $1.37 + 13.1 i$, $21.71$, $3.9 + 2.83 i$, $-8.62 - 6.27 i$, $21.71 + 29.87 i$, $5.33 - 3.87 i$, $0. + 9.17 i$, $-6.71 + 9.23 i$, $-2.23 + 21.2 i$, $8.62 - 6.27 i$, $-15.94 - 2.53 i$, $0. + 12.53 i$, $-1.37 + 13.1 i$, $10.85 - 14.94 i$, $-23.56 + 17.12 i$, $-5.33 + 3.87 i$, $0.01$, $-5.33 - 3.87 i$, $-23.56 - 17.12 i$, $10.85 + 14.94 i$, $-1.37 - 13.1 i$, $0. - 12.53 i$, $-15.94 + 2.53 i$, $8.62 + 6.27 i$, $-2.23 - 21.2 i$, $-6.71 - 9.23 i$, $0. - 9.17 i$, $5.33 + 3.87 i$, $21.71 - 29.87 i$, $-8.62 + 6.27 i$, $3.9 - 2.83 i$, $21.71$, $1.37 - 13.1 i$, $-8.62 - 6.27 i$, $25.79 + 4.09 i$, $5.33 - 3.87 i$, $21.7 - 12.53 i$, $-6.71 + 9.23 i$, $23.56 + 17.12 i$, $8.62 - 6.27 i$, $0.$, $0. + 12.53 i$, $14.56 + 10.58 i$, $10.85 - 14.94 i$, $2.23 + 21.2 i$, $-5.33 + 3.87 i$, $21.71 + 21.7 i$, $-5.33 - 3.87 i$, $-19.48 + 8.67 i$, $10.85 + 14.94 i$, $-3.9 + 2.83 i$, $0. - 12.53 i$, $-13.41 + 18.46 i$, $8.62 + 6.27 i$, $-6.31 + 4.59 i$, $-6.71 - 9.23 i$, $-21.7 + 12.53 i$, $5.33 + 3.87 i$, $-4.08 - 25.79 i$, $-8.62 + 6.27 i$, $-12.03 - 5.36 i$, $21.71$, $-14.56 - 10.58 i$, $-8.62 - 6.27 i$, $0.$, $5.33 - 3.87 i$, $0. - 34.23 i$, $-6.71 + 9.23 i$, $19.48 - 8.67 i$, $8.62 - 6.27 i$, $2.52 - 15.94 i$, $0. + 12.53 i$, $12.03 - 5.36 i$, $10.85 - 14.94 i$, $6.31 - 4.59 i$, $-5.33 + 3.87 i$, $43.41$)

\vskip 0.7ex
\hangindent=3em \hangafter=1
\textit{Intrinsic sign problem}

  \vskip 2ex

\noindent25. $10_{\frac{4}{5},43.41}^{120,198}$ \irep{1146}:\ \ 
$d_i$ = ($1.0$,
$1.0$,
$1.618$,
$1.618$,
$1.732$,
$1.732$,
$2.0$,
$2.802$,
$2.802$,
$3.236$) 

\vskip 0.7ex
\hangindent=3em \hangafter=1
$D^2= 43.416 = 
30+6\sqrt{5}$

\vskip 0.7ex
\hangindent=3em \hangafter=1
$T = ( 0,
0,
\frac{2}{5},
\frac{2}{5},
\frac{1}{8},
\frac{5}{8},
\frac{2}{3},
\frac{1}{40},
\frac{21}{40},
\frac{1}{15} )
$,

\vskip 0.7ex
\hangindent=3em \hangafter=1
$S$ = ($ 1$,
$ 1$,
$ \frac{1+\sqrt{5}}{2}$,
$ \frac{1+\sqrt{5}}{2}$,
$ \sqrt{3}$,
$ \sqrt{3}$,
$ 2$,
$ \frac{15+3\sqrt{5}}{2\sqrt{15}}$,
$ \frac{15+3\sqrt{5}}{2\sqrt{15}}$,
$ 1+\sqrt{5}$;\ \ 
$ 1$,
$ \frac{1+\sqrt{5}}{2}$,
$ \frac{1+\sqrt{5}}{2}$,
$ -\sqrt{3}$,
$ -\sqrt{3}$,
$ 2$,
$ -\frac{15+3\sqrt{5}}{2\sqrt{15}}$,
$ -\frac{15+3\sqrt{5}}{2\sqrt{15}}$,
$ 1+\sqrt{5}$;\ \ 
$ -1$,
$ -1$,
$ -\frac{15+3\sqrt{5}}{2\sqrt{15}}$,
$ -\frac{15+3\sqrt{5}}{2\sqrt{15}}$,
$ 1+\sqrt{5}$,
$ \sqrt{3}$,
$ \sqrt{3}$,
$ -2$;\ \ 
$ -1$,
$ \frac{15+3\sqrt{5}}{2\sqrt{15}}$,
$ \frac{15+3\sqrt{5}}{2\sqrt{15}}$,
$ 1+\sqrt{5}$,
$ -\sqrt{3}$,
$ -\sqrt{3}$,
$ -2$;\ \ 
$ -\sqrt{3}$,
$ \sqrt{3}$,
$0$,
$ \frac{15+3\sqrt{5}}{2\sqrt{15}}$,
$ -\frac{15+3\sqrt{5}}{2\sqrt{15}}$,
$0$;\ \ 
$ -\sqrt{3}$,
$0$,
$ -\frac{15+3\sqrt{5}}{2\sqrt{15}}$,
$ \frac{15+3\sqrt{5}}{2\sqrt{15}}$,
$0$;\ \ 
$ -2$,
$0$,
$0$,
$ -1-\sqrt{5}$;\ \ 
$ \sqrt{3}$,
$ -\sqrt{3}$,
$0$;\ \ 
$ \sqrt{3}$,
$0$;\ \ 
$ 2$)

Factors = $2_{\frac{14}{5},3.618}^{5,395}\boxtimes 5_{6,12.}^{24,273}$

\vskip 0.7ex
\hangindent=3em \hangafter=1
$\tau_n$ = ($5.33 + 3.87 i$, $23.56 + 17.12 i$, $10.85 + 14.94 i$, $1.37 + 13.1 i$, $0. + 12.53 i$, $2.52 + 15.94 i$, $-8.62 - 6.27 i$, $2.23 + 21.2 i$, $-6.71 - 9.23 i$, $0. + 9.17 i$, $-5.33 - 3.87 i$, $0.$, $8.62 - 6.27 i$, $-3.9 + 2.83 i$, $21.71$, $-1.37 + 13.1 i$, $8.62 + 6.27 i$, $-4.08 + 25.79 i$, $-5.33 + 3.87 i$, $-21.7 + 12.53 i$, $-6.71 + 9.23 i$, $-23.56 - 17.12 i$, $-8.62 + 6.27 i$, $-13.41 - 18.46 i$, $0. - 12.53 i$, $-14.56 - 10.58 i$, $10.85 - 14.94 i$, $-2.23 - 21.2 i$, $5.33 - 3.87 i$, $21.71 - 21.7 i$, $5.33 + 3.87 i$, $19.48 - 8.67 i$, $10.85 + 14.94 i$, $3.9 - 2.83 i$, $0. + 12.53 i$, $0.$, $-8.62 - 6.27 i$, $6.31 - 4.59 i$, $-6.71 - 9.23 i$, $21.7 - 12.53 i$, $-5.33 - 3.87 i$, $25.79 - 4.09 i$, $8.62 - 6.27 i$, $12.03 + 5.36 i$, $21.71$, $14.56 + 10.58 i$, $8.62 + 6.27 i$, $21.71 + 29.87 i$, $-5.33 + 3.87 i$, $0. + 34.23 i$, $-6.71 + 9.23 i$, $-19.48 + 8.67 i$, $-8.62 + 6.27 i$, $-15.94 - 2.53 i$, $0. - 12.53 i$, $-12.03 + 5.36 i$, $10.85 - 14.94 i$, $-6.31 + 4.59 i$, $5.33 - 3.87 i$, $0.01$, $5.33 + 3.87 i$, $-6.31 - 4.59 i$, $10.85 + 14.94 i$, $-12.03 - 5.36 i$, $0. + 12.53 i$, $-15.94 + 2.53 i$, $-8.62 - 6.27 i$, $-19.48 - 8.67 i$, $-6.71 - 9.23 i$, $0. - 34.23 i$, $-5.33 - 3.87 i$, $21.71 - 29.87 i$, $8.62 - 6.27 i$, $14.56 - 10.58 i$, $21.71$, $12.03 - 5.36 i$, $8.62 + 6.27 i$, $25.79 + 4.09 i$, $-5.33 + 3.87 i$, $21.7 + 12.53 i$, $-6.71 + 9.23 i$, $6.31 + 4.59 i$, $-8.62 + 6.27 i$, $0.$, $0. - 12.53 i$, $3.9 + 2.83 i$, $10.85 - 14.94 i$, $19.48 + 8.67 i$, $5.33 - 3.87 i$, $21.71 + 21.7 i$, $5.33 + 3.87 i$, $-2.23 + 21.2 i$, $10.85 + 14.94 i$, $-14.56 + 10.58 i$, $0. + 12.53 i$, $-13.41 + 18.46 i$, $-8.62 - 6.27 i$, $-23.56 + 17.12 i$, $-6.71 - 9.23 i$, $-21.7 - 12.53 i$, $-5.33 - 3.87 i$, $-4.08 - 25.79 i$, $8.62 - 6.27 i$, $-1.37 - 13.1 i$, $21.71$, $-3.9 - 2.83 i$, $8.62 + 6.27 i$, $0.$, $-5.33 + 3.87 i$, $0. - 9.17 i$, $-6.71 + 9.23 i$, $2.23 - 21.2 i$, $-8.62 + 6.27 i$, $2.52 - 15.94 i$, $0. - 12.53 i$, $1.37 - 13.1 i$, $10.85 - 14.94 i$, $23.56 - 17.12 i$, $5.33 - 3.87 i$, $43.41$)

\vskip 0.7ex
\hangindent=3em \hangafter=1
\textit{Intrinsic sign problem}

  \vskip 2ex

\noindent26. $10_{\frac{24}{5},43.41}^{120,105}$ \irep{1146}:\ \ 
$d_i$ = ($1.0$,
$1.0$,
$1.618$,
$1.618$,
$1.732$,
$1.732$,
$2.0$,
$2.802$,
$2.802$,
$3.236$) 

\vskip 0.7ex
\hangindent=3em \hangafter=1
$D^2= 43.416 = 
30+6\sqrt{5}$

\vskip 0.7ex
\hangindent=3em \hangafter=1
$T = ( 0,
0,
\frac{2}{5},
\frac{2}{5},
\frac{3}{8},
\frac{7}{8},
\frac{1}{3},
\frac{11}{40},
\frac{31}{40},
\frac{11}{15} )
$,

\vskip 0.7ex
\hangindent=3em \hangafter=1
$S$ = ($ 1$,
$ 1$,
$ \frac{1+\sqrt{5}}{2}$,
$ \frac{1+\sqrt{5}}{2}$,
$ \sqrt{3}$,
$ \sqrt{3}$,
$ 2$,
$ \frac{15+3\sqrt{5}}{2\sqrt{15}}$,
$ \frac{15+3\sqrt{5}}{2\sqrt{15}}$,
$ 1+\sqrt{5}$;\ \ 
$ 1$,
$ \frac{1+\sqrt{5}}{2}$,
$ \frac{1+\sqrt{5}}{2}$,
$ -\sqrt{3}$,
$ -\sqrt{3}$,
$ 2$,
$ -\frac{15+3\sqrt{5}}{2\sqrt{15}}$,
$ -\frac{15+3\sqrt{5}}{2\sqrt{15}}$,
$ 1+\sqrt{5}$;\ \ 
$ -1$,
$ -1$,
$ -\frac{15+3\sqrt{5}}{2\sqrt{15}}$,
$ -\frac{15+3\sqrt{5}}{2\sqrt{15}}$,
$ 1+\sqrt{5}$,
$ \sqrt{3}$,
$ \sqrt{3}$,
$ -2$;\ \ 
$ -1$,
$ \frac{15+3\sqrt{5}}{2\sqrt{15}}$,
$ \frac{15+3\sqrt{5}}{2\sqrt{15}}$,
$ 1+\sqrt{5}$,
$ -\sqrt{3}$,
$ -\sqrt{3}$,
$ -2$;\ \ 
$ -\sqrt{3}$,
$ \sqrt{3}$,
$0$,
$ \frac{15+3\sqrt{5}}{2\sqrt{15}}$,
$ -\frac{15+3\sqrt{5}}{2\sqrt{15}}$,
$0$;\ \ 
$ -\sqrt{3}$,
$0$,
$ -\frac{15+3\sqrt{5}}{2\sqrt{15}}$,
$ \frac{15+3\sqrt{5}}{2\sqrt{15}}$,
$0$;\ \ 
$ -2$,
$0$,
$0$,
$ -1-\sqrt{5}$;\ \ 
$ \sqrt{3}$,
$ -\sqrt{3}$,
$0$;\ \ 
$ \sqrt{3}$,
$0$;\ \ 
$ 2$)

Factors = $2_{\frac{14}{5},3.618}^{5,395}\boxtimes 5_{2,12.}^{24,741}$

\vskip 0.7ex
\hangindent=3em \hangafter=1
$\tau_n$ = ($-5.33 - 3.87 i$, $-23.56 - 17.12 i$, $10.85 + 14.94 i$, $12.03 + 5.36 i$, $0. - 12.53 i$, $-15.94 + 2.53 i$, $8.62 + 6.27 i$, $19.48 + 8.67 i$, $-6.71 - 9.23 i$, $0. - 9.17 i$, $5.33 + 3.87 i$, $0.$, $-8.62 + 6.27 i$, $3.9 - 2.83 i$, $21.71$, $-12.03 + 5.36 i$, $-8.62 - 6.27 i$, $25.79 + 4.09 i$, $5.33 - 3.87 i$, $-21.7 - 12.53 i$, $-6.71 + 9.23 i$, $23.56 + 17.12 i$, $8.62 - 6.27 i$, $-13.41 - 18.46 i$, $0. + 12.53 i$, $14.56 + 10.58 i$, $10.85 - 14.94 i$, $-19.48 - 8.67 i$, $-5.33 + 3.87 i$, $21.71 + 21.7 i$, $-5.33 - 3.87 i$, $2.23 - 21.2 i$, $10.85 + 14.94 i$, $-3.9 + 2.83 i$, $0. - 12.53 i$, $0.$, $8.62 + 6.27 i$, $-6.31 + 4.59 i$, $-6.71 - 9.23 i$, $21.7 + 12.53 i$, $5.33 + 3.87 i$, $-4.08 - 25.79 i$, $-8.62 + 6.27 i$, $1.37 + 13.1 i$, $21.71$, $-14.56 - 10.58 i$, $-8.62 - 6.27 i$, $21.71 + 29.87 i$, $5.33 - 3.87 i$, $0. - 34.23 i$, $-6.71 + 9.23 i$, $-2.23 + 21.2 i$, $8.62 - 6.27 i$, $2.52 - 15.94 i$, $0. + 12.53 i$, $-1.37 + 13.1 i$, $10.85 - 14.94 i$, $6.31 - 4.59 i$, $-5.33 + 3.87 i$, $0.01$, $-5.33 - 3.87 i$, $6.31 + 4.59 i$, $10.85 + 14.94 i$, $-1.37 - 13.1 i$, $0. - 12.53 i$, $2.52 + 15.94 i$, $8.62 + 6.27 i$, $-2.23 - 21.2 i$, $-6.71 - 9.23 i$, $0. + 34.23 i$, $5.33 + 3.87 i$, $21.71 - 29.87 i$, $-8.62 + 6.27 i$, $-14.56 + 10.58 i$, $21.71$, $1.37 - 13.1 i$, $-8.62 - 6.27 i$, $-4.08 + 25.79 i$, $5.33 - 3.87 i$, $21.7 - 12.53 i$, $-6.71 + 9.23 i$, $-6.31 - 4.59 i$, $8.62 - 6.27 i$, $0.$, $0. + 12.53 i$, $-3.9 - 2.83 i$, $10.85 - 14.94 i$, $2.23 + 21.2 i$, $-5.33 + 3.87 i$, $21.71 - 21.7 i$, $-5.33 - 3.87 i$, $-19.48 + 8.67 i$, $10.85 + 14.94 i$, $14.56 - 10.58 i$, $0. - 12.53 i$, $-13.41 + 18.46 i$, $8.62 + 6.27 i$, $23.56 - 17.12 i$, $-6.71 - 9.23 i$, $-21.7 + 12.53 i$, $5.33 + 3.87 i$, $25.79 - 4.09 i$, $-8.62 + 6.27 i$, $-12.03 - 5.36 i$, $21.71$, $3.9 + 2.83 i$, $-8.62 - 6.27 i$, $0.$, $5.33 - 3.87 i$, $0. + 9.17 i$, $-6.71 + 9.23 i$, $19.48 - 8.67 i$, $8.62 - 6.27 i$, $-15.94 - 2.53 i$, $0. + 12.53 i$, $12.03 - 5.36 i$, $10.85 - 14.94 i$, $-23.56 + 17.12 i$, $-5.33 + 3.87 i$, $43.41$)

\vskip 0.7ex
\hangindent=3em \hangafter=1
\textit{Intrinsic sign problem}

  \vskip 2ex

\noindent27. $10_{\frac{4}{5},43.41}^{120,597}$ \irep{1146}:\ \ 
$d_i$ = ($1.0$,
$1.0$,
$1.618$,
$1.618$,
$1.732$,
$1.732$,
$2.0$,
$2.802$,
$2.802$,
$3.236$) 

\vskip 0.7ex
\hangindent=3em \hangafter=1
$D^2= 43.416 = 
30+6\sqrt{5}$

\vskip 0.7ex
\hangindent=3em \hangafter=1
$T = ( 0,
0,
\frac{2}{5},
\frac{2}{5},
\frac{3}{8},
\frac{7}{8},
\frac{2}{3},
\frac{11}{40},
\frac{31}{40},
\frac{1}{15} )
$,

\vskip 0.7ex
\hangindent=3em \hangafter=1
$S$ = ($ 1$,
$ 1$,
$ \frac{1+\sqrt{5}}{2}$,
$ \frac{1+\sqrt{5}}{2}$,
$ \sqrt{3}$,
$ \sqrt{3}$,
$ 2$,
$ \frac{15+3\sqrt{5}}{2\sqrt{15}}$,
$ \frac{15+3\sqrt{5}}{2\sqrt{15}}$,
$ 1+\sqrt{5}$;\ \ 
$ 1$,
$ \frac{1+\sqrt{5}}{2}$,
$ \frac{1+\sqrt{5}}{2}$,
$ -\sqrt{3}$,
$ -\sqrt{3}$,
$ 2$,
$ -\frac{15+3\sqrt{5}}{2\sqrt{15}}$,
$ -\frac{15+3\sqrt{5}}{2\sqrt{15}}$,
$ 1+\sqrt{5}$;\ \ 
$ -1$,
$ -1$,
$ -\frac{15+3\sqrt{5}}{2\sqrt{15}}$,
$ -\frac{15+3\sqrt{5}}{2\sqrt{15}}$,
$ 1+\sqrt{5}$,
$ \sqrt{3}$,
$ \sqrt{3}$,
$ -2$;\ \ 
$ -1$,
$ \frac{15+3\sqrt{5}}{2\sqrt{15}}$,
$ \frac{15+3\sqrt{5}}{2\sqrt{15}}$,
$ 1+\sqrt{5}$,
$ -\sqrt{3}$,
$ -\sqrt{3}$,
$ -2$;\ \ 
$ \sqrt{3}$,
$ -\sqrt{3}$,
$0$,
$ -\frac{15+3\sqrt{5}}{2\sqrt{15}}$,
$ \frac{15+3\sqrt{5}}{2\sqrt{15}}$,
$0$;\ \ 
$ \sqrt{3}$,
$0$,
$ \frac{15+3\sqrt{5}}{2\sqrt{15}}$,
$ -\frac{15+3\sqrt{5}}{2\sqrt{15}}$,
$0$;\ \ 
$ -2$,
$0$,
$0$,
$ -1-\sqrt{5}$;\ \ 
$ -\sqrt{3}$,
$ \sqrt{3}$,
$0$;\ \ 
$ -\sqrt{3}$,
$0$;\ \ 
$ 2$)

Factors = $2_{\frac{14}{5},3.618}^{5,395}\boxtimes 5_{6,12.}^{24,592}$

\vskip 0.7ex
\hangindent=3em \hangafter=1
$\tau_n$ = ($5.33 + 3.87 i$, $-6.31 - 4.59 i$, $10.85 + 14.94 i$, $1.37 + 13.1 i$, $0. + 12.53 i$, $-15.94 + 2.53 i$, $-8.62 - 6.27 i$, $2.23 + 21.2 i$, $-6.71 - 9.23 i$, $0. - 34.23 i$, $-5.33 - 3.87 i$, $0.$, $8.62 - 6.27 i$, $14.56 - 10.58 i$, $21.71$, $-1.37 + 13.1 i$, $8.62 + 6.27 i$, $25.79 + 4.09 i$, $-5.33 + 3.87 i$, $-21.7 + 12.53 i$, $-6.71 + 9.23 i$, $6.31 + 4.59 i$, $-8.62 + 6.27 i$, $-13.41 - 18.46 i$, $0. - 12.53 i$, $3.9 + 2.83 i$, $10.85 - 14.94 i$, $-2.23 - 21.2 i$, $5.33 - 3.87 i$, $21.71 + 21.7 i$, $5.33 + 3.87 i$, $19.48 - 8.67 i$, $10.85 + 14.94 i$, $-14.56 + 10.58 i$, $0. + 12.53 i$, $0.$, $-8.62 - 6.27 i$, $-23.56 + 17.12 i$, $-6.71 - 9.23 i$, $21.7 - 12.53 i$, $-5.33 - 3.87 i$, $-4.08 - 25.79 i$, $8.62 - 6.27 i$, $12.03 + 5.36 i$, $21.71$, $-3.9 - 2.83 i$, $8.62 + 6.27 i$, $21.71 + 29.87 i$, $-5.33 + 3.87 i$, $0. - 9.17 i$, $-6.71 + 9.23 i$, $-19.48 + 8.67 i$, $-8.62 + 6.27 i$, $2.52 - 15.94 i$, $0. - 12.53 i$, $-12.03 + 5.36 i$, $10.85 - 14.94 i$, $23.56 - 17.12 i$, $5.33 - 3.87 i$, $0.01$, $5.33 + 3.87 i$, $23.56 + 17.12 i$, $10.85 + 14.94 i$, $-12.03 - 5.36 i$, $0. + 12.53 i$, $2.52 + 15.94 i$, $-8.62 - 6.27 i$, $-19.48 - 8.67 i$, $-6.71 - 9.23 i$, $0. + 9.17 i$, $-5.33 - 3.87 i$, $21.71 - 29.87 i$, $8.62 - 6.27 i$, $-3.9 + 2.83 i$, $21.71$, $12.03 - 5.36 i$, $8.62 + 6.27 i$, $-4.08 + 25.79 i$, $-5.33 + 3.87 i$, $21.7 + 12.53 i$, $-6.71 + 9.23 i$, $-23.56 - 17.12 i$, $-8.62 + 6.27 i$, $0.$, $0. - 12.53 i$, $-14.56 - 10.58 i$, $10.85 - 14.94 i$, $19.48 + 8.67 i$, $5.33 - 3.87 i$, $21.71 - 21.7 i$, $5.33 + 3.87 i$, $-2.23 + 21.2 i$, $10.85 + 14.94 i$, $3.9 - 2.83 i$, $0. + 12.53 i$, $-13.41 + 18.46 i$, $-8.62 - 6.27 i$, $6.31 - 4.59 i$, $-6.71 - 9.23 i$, $-21.7 - 12.53 i$, $-5.33 - 3.87 i$, $25.79 - 4.09 i$, $8.62 - 6.27 i$, $-1.37 - 13.1 i$, $21.71$, $14.56 + 10.58 i$, $8.62 + 6.27 i$, $0.$, $-5.33 + 3.87 i$, $0. + 34.23 i$, $-6.71 + 9.23 i$, $2.23 - 21.2 i$, $-8.62 + 6.27 i$, $-15.94 - 2.53 i$, $0. - 12.53 i$, $1.37 - 13.1 i$, $10.85 - 14.94 i$, $-6.31 + 4.59 i$, $5.33 - 3.87 i$, $43.41$)

\vskip 0.7ex
\hangindent=3em \hangafter=1
\textit{Intrinsic sign problem}

  \vskip 2ex

\noindent28. $10_{\frac{36}{5},43.41}^{120,163}$ \irep{1146}:\ \ 
$d_i$ = ($1.0$,
$1.0$,
$1.618$,
$1.618$,
$1.732$,
$1.732$,
$2.0$,
$2.802$,
$2.802$,
$3.236$) 

\vskip 0.7ex
\hangindent=3em \hangafter=1
$D^2= 43.416 = 
30+6\sqrt{5}$

\vskip 0.7ex
\hangindent=3em \hangafter=1
$T = ( 0,
0,
\frac{3}{5},
\frac{3}{5},
\frac{1}{8},
\frac{5}{8},
\frac{1}{3},
\frac{9}{40},
\frac{29}{40},
\frac{14}{15} )
$,

\vskip 0.7ex
\hangindent=3em \hangafter=1
$S$ = ($ 1$,
$ 1$,
$ \frac{1+\sqrt{5}}{2}$,
$ \frac{1+\sqrt{5}}{2}$,
$ \sqrt{3}$,
$ \sqrt{3}$,
$ 2$,
$ \frac{15+3\sqrt{5}}{2\sqrt{15}}$,
$ \frac{15+3\sqrt{5}}{2\sqrt{15}}$,
$ 1+\sqrt{5}$;\ \ 
$ 1$,
$ \frac{1+\sqrt{5}}{2}$,
$ \frac{1+\sqrt{5}}{2}$,
$ -\sqrt{3}$,
$ -\sqrt{3}$,
$ 2$,
$ -\frac{15+3\sqrt{5}}{2\sqrt{15}}$,
$ -\frac{15+3\sqrt{5}}{2\sqrt{15}}$,
$ 1+\sqrt{5}$;\ \ 
$ -1$,
$ -1$,
$ -\frac{15+3\sqrt{5}}{2\sqrt{15}}$,
$ -\frac{15+3\sqrt{5}}{2\sqrt{15}}$,
$ 1+\sqrt{5}$,
$ \sqrt{3}$,
$ \sqrt{3}$,
$ -2$;\ \ 
$ -1$,
$ \frac{15+3\sqrt{5}}{2\sqrt{15}}$,
$ \frac{15+3\sqrt{5}}{2\sqrt{15}}$,
$ 1+\sqrt{5}$,
$ -\sqrt{3}$,
$ -\sqrt{3}$,
$ -2$;\ \ 
$ \sqrt{3}$,
$ -\sqrt{3}$,
$0$,
$ -\frac{15+3\sqrt{5}}{2\sqrt{15}}$,
$ \frac{15+3\sqrt{5}}{2\sqrt{15}}$,
$0$;\ \ 
$ \sqrt{3}$,
$0$,
$ \frac{15+3\sqrt{5}}{2\sqrt{15}}$,
$ -\frac{15+3\sqrt{5}}{2\sqrt{15}}$,
$0$;\ \ 
$ -2$,
$0$,
$0$,
$ -1-\sqrt{5}$;\ \ 
$ -\sqrt{3}$,
$ \sqrt{3}$,
$0$;\ \ 
$ -\sqrt{3}$,
$0$;\ \ 
$ 2$)

Factors = $2_{\frac{26}{5},3.618}^{5,720}\boxtimes 5_{2,12.}^{24,940}$

\vskip 0.7ex
\hangindent=3em \hangafter=1
$\tau_n$ = ($5.33 - 3.87 i$, $-6.31 + 4.59 i$, $10.85 - 14.94 i$, $1.37 - 13.1 i$, $0. - 12.53 i$, $-15.94 - 2.53 i$, $-8.62 + 6.27 i$, $2.23 - 21.2 i$, $-6.71 + 9.23 i$, $0. + 34.23 i$, $-5.33 + 3.87 i$, $0.$, $8.62 + 6.27 i$, $14.56 + 10.58 i$, $21.71$, $-1.37 - 13.1 i$, $8.62 - 6.27 i$, $25.79 - 4.09 i$, $-5.33 - 3.87 i$, $-21.7 - 12.53 i$, $-6.71 - 9.23 i$, $6.31 - 4.59 i$, $-8.62 - 6.27 i$, $-13.41 + 18.46 i$, $0. + 12.53 i$, $3.9 - 2.83 i$, $10.85 + 14.94 i$, $-2.23 + 21.2 i$, $5.33 + 3.87 i$, $21.71 - 21.7 i$, $5.33 - 3.87 i$, $19.48 + 8.67 i$, $10.85 - 14.94 i$, $-14.56 - 10.58 i$, $0. - 12.53 i$, $0.$, $-8.62 + 6.27 i$, $-23.56 - 17.12 i$, $-6.71 + 9.23 i$, $21.7 + 12.53 i$, $-5.33 + 3.87 i$, $-4.08 + 25.79 i$, $8.62 + 6.27 i$, $12.03 - 5.36 i$, $21.71$, $-3.9 + 2.83 i$, $8.62 - 6.27 i$, $21.71 - 29.87 i$, $-5.33 - 3.87 i$, $0. + 9.17 i$, $-6.71 - 9.23 i$, $-19.48 - 8.67 i$, $-8.62 - 6.27 i$, $2.52 + 15.94 i$, $0. + 12.53 i$, $-12.03 - 5.36 i$, $10.85 + 14.94 i$, $23.56 + 17.12 i$, $5.33 + 3.87 i$, $0.01$, $5.33 - 3.87 i$, $23.56 - 17.12 i$, $10.85 - 14.94 i$, $-12.03 + 5.36 i$, $0. - 12.53 i$, $2.52 - 15.94 i$, $-8.62 + 6.27 i$, $-19.48 + 8.67 i$, $-6.71 + 9.23 i$, $0. - 9.17 i$, $-5.33 + 3.87 i$, $21.71 + 29.87 i$, $8.62 + 6.27 i$, $-3.9 - 2.83 i$, $21.71$, $12.03 + 5.36 i$, $8.62 - 6.27 i$, $-4.08 - 25.79 i$, $-5.33 - 3.87 i$, $21.7 - 12.53 i$, $-6.71 - 9.23 i$, $-23.56 + 17.12 i$, $-8.62 - 6.27 i$, $0.$, $0. + 12.53 i$, $-14.56 + 10.58 i$, $10.85 + 14.94 i$, $19.48 - 8.67 i$, $5.33 + 3.87 i$, $21.71 + 21.7 i$, $5.33 - 3.87 i$, $-2.23 - 21.2 i$, $10.85 - 14.94 i$, $3.9 + 2.83 i$, $0. - 12.53 i$, $-13.41 - 18.46 i$, $-8.62 + 6.27 i$, $6.31 + 4.59 i$, $-6.71 + 9.23 i$, $-21.7 + 12.53 i$, $-5.33 + 3.87 i$, $25.79 + 4.09 i$, $8.62 + 6.27 i$, $-1.37 + 13.1 i$, $21.71$, $14.56 - 10.58 i$, $8.62 - 6.27 i$, $0.$, $-5.33 - 3.87 i$, $0. - 34.23 i$, $-6.71 - 9.23 i$, $2.23 + 21.2 i$, $-8.62 - 6.27 i$, $-15.94 + 2.53 i$, $0. + 12.53 i$, $1.37 + 13.1 i$, $10.85 + 14.94 i$, $-6.31 - 4.59 i$, $5.33 + 3.87 i$, $43.41$)

\vskip 0.7ex
\hangindent=3em \hangafter=1
\textit{Intrinsic sign problem}

  \vskip 2ex

\noindent29. $10_{\frac{16}{5},43.41}^{120,282}$ \irep{1146}:\ \ 
$d_i$ = ($1.0$,
$1.0$,
$1.618$,
$1.618$,
$1.732$,
$1.732$,
$2.0$,
$2.802$,
$2.802$,
$3.236$) 

\vskip 0.7ex
\hangindent=3em \hangafter=1
$D^2= 43.416 = 
30+6\sqrt{5}$

\vskip 0.7ex
\hangindent=3em \hangafter=1
$T = ( 0,
0,
\frac{3}{5},
\frac{3}{5},
\frac{1}{8},
\frac{5}{8},
\frac{2}{3},
\frac{9}{40},
\frac{29}{40},
\frac{4}{15} )
$,

\vskip 0.7ex
\hangindent=3em \hangafter=1
$S$ = ($ 1$,
$ 1$,
$ \frac{1+\sqrt{5}}{2}$,
$ \frac{1+\sqrt{5}}{2}$,
$ \sqrt{3}$,
$ \sqrt{3}$,
$ 2$,
$ \frac{15+3\sqrt{5}}{2\sqrt{15}}$,
$ \frac{15+3\sqrt{5}}{2\sqrt{15}}$,
$ 1+\sqrt{5}$;\ \ 
$ 1$,
$ \frac{1+\sqrt{5}}{2}$,
$ \frac{1+\sqrt{5}}{2}$,
$ -\sqrt{3}$,
$ -\sqrt{3}$,
$ 2$,
$ -\frac{15+3\sqrt{5}}{2\sqrt{15}}$,
$ -\frac{15+3\sqrt{5}}{2\sqrt{15}}$,
$ 1+\sqrt{5}$;\ \ 
$ -1$,
$ -1$,
$ -\frac{15+3\sqrt{5}}{2\sqrt{15}}$,
$ -\frac{15+3\sqrt{5}}{2\sqrt{15}}$,
$ 1+\sqrt{5}$,
$ \sqrt{3}$,
$ \sqrt{3}$,
$ -2$;\ \ 
$ -1$,
$ \frac{15+3\sqrt{5}}{2\sqrt{15}}$,
$ \frac{15+3\sqrt{5}}{2\sqrt{15}}$,
$ 1+\sqrt{5}$,
$ -\sqrt{3}$,
$ -\sqrt{3}$,
$ -2$;\ \ 
$ -\sqrt{3}$,
$ \sqrt{3}$,
$0$,
$ \frac{15+3\sqrt{5}}{2\sqrt{15}}$,
$ -\frac{15+3\sqrt{5}}{2\sqrt{15}}$,
$0$;\ \ 
$ -\sqrt{3}$,
$0$,
$ -\frac{15+3\sqrt{5}}{2\sqrt{15}}$,
$ \frac{15+3\sqrt{5}}{2\sqrt{15}}$,
$0$;\ \ 
$ -2$,
$0$,
$0$,
$ -1-\sqrt{5}$;\ \ 
$ \sqrt{3}$,
$ -\sqrt{3}$,
$0$;\ \ 
$ \sqrt{3}$,
$0$;\ \ 
$ 2$)

Factors = $2_{\frac{26}{5},3.618}^{5,720}\boxtimes 5_{6,12.}^{24,273}$

\vskip 0.7ex
\hangindent=3em \hangafter=1
$\tau_n$ = ($-5.33 + 3.87 i$, $-23.56 + 17.12 i$, $10.85 - 14.94 i$, $12.03 - 5.36 i$, $0. + 12.53 i$, $-15.94 - 2.53 i$, $8.62 - 6.27 i$, $19.48 - 8.67 i$, $-6.71 + 9.23 i$, $0. + 9.17 i$, $5.33 - 3.87 i$, $0.$, $-8.62 - 6.27 i$, $3.9 + 2.83 i$, $21.71$, $-12.03 - 5.36 i$, $-8.62 + 6.27 i$, $25.79 - 4.09 i$, $5.33 + 3.87 i$, $-21.7 + 12.53 i$, $-6.71 - 9.23 i$, $23.56 - 17.12 i$, $8.62 + 6.27 i$, $-13.41 + 18.46 i$, $0. - 12.53 i$, $14.56 - 10.58 i$, $10.85 + 14.94 i$, $-19.48 + 8.67 i$, $-5.33 - 3.87 i$, $21.71 - 21.7 i$, $-5.33 + 3.87 i$, $2.23 + 21.2 i$, $10.85 - 14.94 i$, $-3.9 - 2.83 i$, $0. + 12.53 i$, $0.$, $8.62 - 6.27 i$, $-6.31 - 4.59 i$, $-6.71 + 9.23 i$, $21.7 - 12.53 i$, $5.33 - 3.87 i$, $-4.08 + 25.79 i$, $-8.62 - 6.27 i$, $1.37 - 13.1 i$, $21.71$, $-14.56 + 10.58 i$, $-8.62 + 6.27 i$, $21.71 - 29.87 i$, $5.33 + 3.87 i$, $0. + 34.23 i$, $-6.71 - 9.23 i$, $-2.23 - 21.2 i$, $8.62 + 6.27 i$, $2.52 + 15.94 i$, $0. - 12.53 i$, $-1.37 - 13.1 i$, $10.85 + 14.94 i$, $6.31 + 4.59 i$, $-5.33 - 3.87 i$, $0.01$, $-5.33 + 3.87 i$, $6.31 - 4.59 i$, $10.85 - 14.94 i$, $-1.37 + 13.1 i$, $0. + 12.53 i$, $2.52 - 15.94 i$, $8.62 - 6.27 i$, $-2.23 + 21.2 i$, $-6.71 + 9.23 i$, $0. - 34.23 i$, $5.33 - 3.87 i$, $21.71 + 29.87 i$, $-8.62 - 6.27 i$, $-14.56 - 10.58 i$, $21.71$, $1.37 + 13.1 i$, $-8.62 + 6.27 i$, $-4.08 - 25.79 i$, $5.33 + 3.87 i$, $21.7 + 12.53 i$, $-6.71 - 9.23 i$, $-6.31 + 4.59 i$, $8.62 + 6.27 i$, $0.$, $0. - 12.53 i$, $-3.9 + 2.83 i$, $10.85 + 14.94 i$, $2.23 - 21.2 i$, $-5.33 - 3.87 i$, $21.71 + 21.7 i$, $-5.33 + 3.87 i$, $-19.48 - 8.67 i$, $10.85 - 14.94 i$, $14.56 + 10.58 i$, $0. + 12.53 i$, $-13.41 - 18.46 i$, $8.62 - 6.27 i$, $23.56 + 17.12 i$, $-6.71 + 9.23 i$, $-21.7 - 12.53 i$, $5.33 - 3.87 i$, $25.79 + 4.09 i$, $-8.62 - 6.27 i$, $-12.03 + 5.36 i$, $21.71$, $3.9 - 2.83 i$, $-8.62 + 6.27 i$, $0.$, $5.33 + 3.87 i$, $0. - 9.17 i$, $-6.71 - 9.23 i$, $19.48 + 8.67 i$, $8.62 + 6.27 i$, $-15.94 + 2.53 i$, $0. - 12.53 i$, $12.03 + 5.36 i$, $10.85 + 14.94 i$, $-23.56 - 17.12 i$, $-5.33 - 3.87 i$, $43.41$)

\vskip 0.7ex
\hangindent=3em \hangafter=1
\textit{Intrinsic sign problem}

  \vskip 2ex

\noindent30. $10_{\frac{36}{5},43.41}^{120,418}$ \irep{1146}:\ \ 
$d_i$ = ($1.0$,
$1.0$,
$1.618$,
$1.618$,
$1.732$,
$1.732$,
$2.0$,
$2.802$,
$2.802$,
$3.236$) 

\vskip 0.7ex
\hangindent=3em \hangafter=1
$D^2= 43.416 = 
30+6\sqrt{5}$

\vskip 0.7ex
\hangindent=3em \hangafter=1
$T = ( 0,
0,
\frac{3}{5},
\frac{3}{5},
\frac{3}{8},
\frac{7}{8},
\frac{1}{3},
\frac{19}{40},
\frac{39}{40},
\frac{14}{15} )
$,

\vskip 0.7ex
\hangindent=3em \hangafter=1
$S$ = ($ 1$,
$ 1$,
$ \frac{1+\sqrt{5}}{2}$,
$ \frac{1+\sqrt{5}}{2}$,
$ \sqrt{3}$,
$ \sqrt{3}$,
$ 2$,
$ \frac{15+3\sqrt{5}}{2\sqrt{15}}$,
$ \frac{15+3\sqrt{5}}{2\sqrt{15}}$,
$ 1+\sqrt{5}$;\ \ 
$ 1$,
$ \frac{1+\sqrt{5}}{2}$,
$ \frac{1+\sqrt{5}}{2}$,
$ -\sqrt{3}$,
$ -\sqrt{3}$,
$ 2$,
$ -\frac{15+3\sqrt{5}}{2\sqrt{15}}$,
$ -\frac{15+3\sqrt{5}}{2\sqrt{15}}$,
$ 1+\sqrt{5}$;\ \ 
$ -1$,
$ -1$,
$ -\frac{15+3\sqrt{5}}{2\sqrt{15}}$,
$ -\frac{15+3\sqrt{5}}{2\sqrt{15}}$,
$ 1+\sqrt{5}$,
$ \sqrt{3}$,
$ \sqrt{3}$,
$ -2$;\ \ 
$ -1$,
$ \frac{15+3\sqrt{5}}{2\sqrt{15}}$,
$ \frac{15+3\sqrt{5}}{2\sqrt{15}}$,
$ 1+\sqrt{5}$,
$ -\sqrt{3}$,
$ -\sqrt{3}$,
$ -2$;\ \ 
$ -\sqrt{3}$,
$ \sqrt{3}$,
$0$,
$ \frac{15+3\sqrt{5}}{2\sqrt{15}}$,
$ -\frac{15+3\sqrt{5}}{2\sqrt{15}}$,
$0$;\ \ 
$ -\sqrt{3}$,
$0$,
$ -\frac{15+3\sqrt{5}}{2\sqrt{15}}$,
$ \frac{15+3\sqrt{5}}{2\sqrt{15}}$,
$0$;\ \ 
$ -2$,
$0$,
$0$,
$ -1-\sqrt{5}$;\ \ 
$ \sqrt{3}$,
$ -\sqrt{3}$,
$0$;\ \ 
$ \sqrt{3}$,
$0$;\ \ 
$ 2$)

Factors = $2_{\frac{26}{5},3.618}^{5,720}\boxtimes 5_{2,12.}^{24,741}$

\vskip 0.7ex
\hangindent=3em \hangafter=1
$\tau_n$ = ($5.33 - 3.87 i$, $23.56 - 17.12 i$, $10.85 - 14.94 i$, $1.37 - 13.1 i$, $0. - 12.53 i$, $2.52 - 15.94 i$, $-8.62 + 6.27 i$, $2.23 - 21.2 i$, $-6.71 + 9.23 i$, $0. - 9.17 i$, $-5.33 + 3.87 i$, $0.$, $8.62 + 6.27 i$, $-3.9 - 2.83 i$, $21.71$, $-1.37 - 13.1 i$, $8.62 - 6.27 i$, $-4.08 - 25.79 i$, $-5.33 - 3.87 i$, $-21.7 - 12.53 i$, $-6.71 - 9.23 i$, $-23.56 + 17.12 i$, $-8.62 - 6.27 i$, $-13.41 + 18.46 i$, $0. + 12.53 i$, $-14.56 + 10.58 i$, $10.85 + 14.94 i$, $-2.23 + 21.2 i$, $5.33 + 3.87 i$, $21.71 + 21.7 i$, $5.33 - 3.87 i$, $19.48 + 8.67 i$, $10.85 - 14.94 i$, $3.9 + 2.83 i$, $0. - 12.53 i$, $0.$, $-8.62 + 6.27 i$, $6.31 + 4.59 i$, $-6.71 + 9.23 i$, $21.7 + 12.53 i$, $-5.33 + 3.87 i$, $25.79 + 4.09 i$, $8.62 + 6.27 i$, $12.03 - 5.36 i$, $21.71$, $14.56 - 10.58 i$, $8.62 - 6.27 i$, $21.71 - 29.87 i$, $-5.33 - 3.87 i$, $0. - 34.23 i$, $-6.71 - 9.23 i$, $-19.48 - 8.67 i$, $-8.62 - 6.27 i$, $-15.94 + 2.53 i$, $0. + 12.53 i$, $-12.03 - 5.36 i$, $10.85 + 14.94 i$, $-6.31 - 4.59 i$, $5.33 + 3.87 i$, $0.01$, $5.33 - 3.87 i$, $-6.31 + 4.59 i$, $10.85 - 14.94 i$, $-12.03 + 5.36 i$, $0. - 12.53 i$, $-15.94 - 2.53 i$, $-8.62 + 6.27 i$, $-19.48 + 8.67 i$, $-6.71 + 9.23 i$, $0. + 34.23 i$, $-5.33 + 3.87 i$, $21.71 + 29.87 i$, $8.62 + 6.27 i$, $14.56 + 10.58 i$, $21.71$, $12.03 + 5.36 i$, $8.62 - 6.27 i$, $25.79 - 4.09 i$, $-5.33 - 3.87 i$, $21.7 - 12.53 i$, $-6.71 - 9.23 i$, $6.31 - 4.59 i$, $-8.62 - 6.27 i$, $0.$, $0. + 12.53 i$, $3.9 - 2.83 i$, $10.85 + 14.94 i$, $19.48 - 8.67 i$, $5.33 + 3.87 i$, $21.71 - 21.7 i$, $5.33 - 3.87 i$, $-2.23 - 21.2 i$, $10.85 - 14.94 i$, $-14.56 - 10.58 i$, $0. - 12.53 i$, $-13.41 - 18.46 i$, $-8.62 + 6.27 i$, $-23.56 - 17.12 i$, $-6.71 + 9.23 i$, $-21.7 + 12.53 i$, $-5.33 + 3.87 i$, $-4.08 + 25.79 i$, $8.62 + 6.27 i$, $-1.37 + 13.1 i$, $21.71$, $-3.9 + 2.83 i$, $8.62 - 6.27 i$, $0.$, $-5.33 - 3.87 i$, $0. + 9.17 i$, $-6.71 - 9.23 i$, $2.23 + 21.2 i$, $-8.62 - 6.27 i$, $2.52 + 15.94 i$, $0. + 12.53 i$, $1.37 + 13.1 i$, $10.85 + 14.94 i$, $23.56 + 17.12 i$, $5.33 + 3.87 i$, $43.41$)

\vskip 0.7ex
\hangindent=3em \hangafter=1
\textit{Intrinsic sign problem}

  \vskip 2ex

\noindent31. $10_{\frac{16}{5},43.41}^{120,226}$ \irep{1146}:\ \ 
$d_i$ = ($1.0$,
$1.0$,
$1.618$,
$1.618$,
$1.732$,
$1.732$,
$2.0$,
$2.802$,
$2.802$,
$3.236$) 

\vskip 0.7ex
\hangindent=3em \hangafter=1
$D^2= 43.416 = 
30+6\sqrt{5}$

\vskip 0.7ex
\hangindent=3em \hangafter=1
$T = ( 0,
0,
\frac{3}{5},
\frac{3}{5},
\frac{3}{8},
\frac{7}{8},
\frac{2}{3},
\frac{19}{40},
\frac{39}{40},
\frac{4}{15} )
$,

\vskip 0.7ex
\hangindent=3em \hangafter=1
$S$ = ($ 1$,
$ 1$,
$ \frac{1+\sqrt{5}}{2}$,
$ \frac{1+\sqrt{5}}{2}$,
$ \sqrt{3}$,
$ \sqrt{3}$,
$ 2$,
$ \frac{15+3\sqrt{5}}{2\sqrt{15}}$,
$ \frac{15+3\sqrt{5}}{2\sqrt{15}}$,
$ 1+\sqrt{5}$;\ \ 
$ 1$,
$ \frac{1+\sqrt{5}}{2}$,
$ \frac{1+\sqrt{5}}{2}$,
$ -\sqrt{3}$,
$ -\sqrt{3}$,
$ 2$,
$ -\frac{15+3\sqrt{5}}{2\sqrt{15}}$,
$ -\frac{15+3\sqrt{5}}{2\sqrt{15}}$,
$ 1+\sqrt{5}$;\ \ 
$ -1$,
$ -1$,
$ -\frac{15+3\sqrt{5}}{2\sqrt{15}}$,
$ -\frac{15+3\sqrt{5}}{2\sqrt{15}}$,
$ 1+\sqrt{5}$,
$ \sqrt{3}$,
$ \sqrt{3}$,
$ -2$;\ \ 
$ -1$,
$ \frac{15+3\sqrt{5}}{2\sqrt{15}}$,
$ \frac{15+3\sqrt{5}}{2\sqrt{15}}$,
$ 1+\sqrt{5}$,
$ -\sqrt{3}$,
$ -\sqrt{3}$,
$ -2$;\ \ 
$ \sqrt{3}$,
$ -\sqrt{3}$,
$0$,
$ -\frac{15+3\sqrt{5}}{2\sqrt{15}}$,
$ \frac{15+3\sqrt{5}}{2\sqrt{15}}$,
$0$;\ \ 
$ \sqrt{3}$,
$0$,
$ \frac{15+3\sqrt{5}}{2\sqrt{15}}$,
$ -\frac{15+3\sqrt{5}}{2\sqrt{15}}$,
$0$;\ \ 
$ -2$,
$0$,
$0$,
$ -1-\sqrt{5}$;\ \ 
$ -\sqrt{3}$,
$ \sqrt{3}$,
$0$;\ \ 
$ -\sqrt{3}$,
$0$;\ \ 
$ 2$)

Factors = $2_{\frac{26}{5},3.618}^{5,720}\boxtimes 5_{6,12.}^{24,592}$

\vskip 0.7ex
\hangindent=3em \hangafter=1
$\tau_n$ = ($-5.33 + 3.87 i$, $6.31 - 4.59 i$, $10.85 - 14.94 i$, $12.03 - 5.36 i$, $0. + 12.53 i$, $2.52 - 15.94 i$, $8.62 - 6.27 i$, $19.48 - 8.67 i$, $-6.71 + 9.23 i$, $0. - 34.23 i$, $5.33 - 3.87 i$, $0.$, $-8.62 - 6.27 i$, $-14.56 - 10.58 i$, $21.71$, $-12.03 - 5.36 i$, $-8.62 + 6.27 i$, $-4.08 - 25.79 i$, $5.33 + 3.87 i$, $-21.7 + 12.53 i$, $-6.71 - 9.23 i$, $-6.31 + 4.59 i$, $8.62 + 6.27 i$, $-13.41 + 18.46 i$, $0. - 12.53 i$, $-3.9 + 2.83 i$, $10.85 + 14.94 i$, $-19.48 + 8.67 i$, $-5.33 - 3.87 i$, $21.71 + 21.7 i$, $-5.33 + 3.87 i$, $2.23 + 21.2 i$, $10.85 - 14.94 i$, $14.56 + 10.58 i$, $0. + 12.53 i$, $0.$, $8.62 - 6.27 i$, $23.56 + 17.12 i$, $-6.71 + 9.23 i$, $21.7 - 12.53 i$, $5.33 - 3.87 i$, $25.79 + 4.09 i$, $-8.62 - 6.27 i$, $1.37 - 13.1 i$, $21.71$, $3.9 - 2.83 i$, $-8.62 + 6.27 i$, $21.71 - 29.87 i$, $5.33 + 3.87 i$, $0. - 9.17 i$, $-6.71 - 9.23 i$, $-2.23 - 21.2 i$, $8.62 + 6.27 i$, $-15.94 + 2.53 i$, $0. - 12.53 i$, $-1.37 - 13.1 i$, $10.85 + 14.94 i$, $-23.56 - 17.12 i$, $-5.33 - 3.87 i$, $0.01$, $-5.33 + 3.87 i$, $-23.56 + 17.12 i$, $10.85 - 14.94 i$, $-1.37 + 13.1 i$, $0. + 12.53 i$, $-15.94 - 2.53 i$, $8.62 - 6.27 i$, $-2.23 + 21.2 i$, $-6.71 + 9.23 i$, $0. + 9.17 i$, $5.33 - 3.87 i$, $21.71 + 29.87 i$, $-8.62 - 6.27 i$, $3.9 + 2.83 i$, $21.71$, $1.37 + 13.1 i$, $-8.62 + 6.27 i$, $25.79 - 4.09 i$, $5.33 + 3.87 i$, $21.7 + 12.53 i$, $-6.71 - 9.23 i$, $23.56 - 17.12 i$, $8.62 + 6.27 i$, $0.$, $0. - 12.53 i$, $14.56 - 10.58 i$, $10.85 + 14.94 i$, $2.23 - 21.2 i$, $-5.33 - 3.87 i$, $21.71 - 21.7 i$, $-5.33 + 3.87 i$, $-19.48 - 8.67 i$, $10.85 - 14.94 i$, $-3.9 - 2.83 i$, $0. + 12.53 i$, $-13.41 - 18.46 i$, $8.62 - 6.27 i$, $-6.31 - 4.59 i$, $-6.71 + 9.23 i$, $-21.7 - 12.53 i$, $5.33 - 3.87 i$, $-4.08 + 25.79 i$, $-8.62 - 6.27 i$, $-12.03 + 5.36 i$, $21.71$, $-14.56 + 10.58 i$, $-8.62 + 6.27 i$, $0.$, $5.33 + 3.87 i$, $0. + 34.23 i$, $-6.71 - 9.23 i$, $19.48 + 8.67 i$, $8.62 + 6.27 i$, $2.52 + 15.94 i$, $0. - 12.53 i$, $12.03 + 5.36 i$, $10.85 + 14.94 i$, $6.31 + 4.59 i$, $-5.33 - 3.87 i$, $43.41$)

\vskip 0.7ex
\hangindent=3em \hangafter=1
\textit{Intrinsic sign problem}

  \vskip 2ex

\noindent32. $10_{0,52.}^{26,247}$ \irep{1035}:\ \ 
$d_i$ = ($1.0$,
$1.0$,
$2.0$,
$2.0$,
$2.0$,
$2.0$,
$2.0$,
$2.0$,
$3.605$,
$3.605$) 

\vskip 0.7ex
\hangindent=3em \hangafter=1
$D^2= 52.0 = 
52$

\vskip 0.7ex
\hangindent=3em \hangafter=1
$T = ( 0,
0,
\frac{1}{13},
\frac{3}{13},
\frac{4}{13},
\frac{9}{13},
\frac{10}{13},
\frac{12}{13},
0,
\frac{1}{2} )
$,

\vskip 0.7ex
\hangindent=3em \hangafter=1
$S$ = ($ 1$,
$ 1$,
$ 2$,
$ 2$,
$ 2$,
$ 2$,
$ 2$,
$ 2$,
$ \sqrt{13}$,
$ \sqrt{13}$;\ \ 
$ 1$,
$ 2$,
$ 2$,
$ 2$,
$ 2$,
$ 2$,
$ 2$,
$ -\sqrt{13}$,
$ -\sqrt{13}$;\ \ 
$ 2c_{13}^{2}$,
$ 2c_{13}^{5}$,
$ 2c_{13}^{4}$,
$ 2c_{13}^{6}$,
$ 2c_{13}^{1}$,
$ 2c_{13}^{3}$,
$0$,
$0$;\ \ 
$ 2c_{13}^{6}$,
$ 2c_{13}^{3}$,
$ 2c_{13}^{2}$,
$ 2c_{13}^{4}$,
$ 2c_{13}^{1}$,
$0$,
$0$;\ \ 
$ 2c_{13}^{5}$,
$ 2c_{13}^{1}$,
$ 2c_{13}^{2}$,
$ 2c_{13}^{6}$,
$0$,
$0$;\ \ 
$ 2c_{13}^{5}$,
$ 2c_{13}^{3}$,
$ 2c_{13}^{4}$,
$0$,
$0$;\ \ 
$ 2c_{13}^{6}$,
$ 2c_{13}^{5}$,
$0$,
$0$;\ \ 
$ 2c_{13}^{2}$,
$0$,
$0$;\ \ 
$ \sqrt{13}$,
$ -\sqrt{13}$;\ \ 
$ \sqrt{13}$)

\vskip 0.7ex
\hangindent=3em \hangafter=1
$\tau_n$ = ($7.21$, $18.78$, $7.21$, $33.2$, $-7.21$, $18.78$, $-7.21$, $18.78$, $7.21$, $33.2$, $-7.21$, $33.2$, $26.$, $33.2$, $-7.21$, $33.2$, $7.21$, $18.78$, $-7.21$, $18.78$, $-7.21$, $33.2$, $7.21$, $18.78$, $7.21$, $51.99$)

\vskip 0.7ex
\hangindent=3em \hangafter=1
\textit{Intrinsic sign problem}

  \vskip 2ex

\noindent33. $10_{4,52.}^{26,862}$ \irep{1035}:\ \ 
$d_i$ = ($1.0$,
$1.0$,
$2.0$,
$2.0$,
$2.0$,
$2.0$,
$2.0$,
$2.0$,
$3.605$,
$3.605$) 

\vskip 0.7ex
\hangindent=3em \hangafter=1
$D^2= 52.0 = 
52$

\vskip 0.7ex
\hangindent=3em \hangafter=1
$T = ( 0,
0,
\frac{2}{13},
\frac{5}{13},
\frac{6}{13},
\frac{7}{13},
\frac{8}{13},
\frac{11}{13},
0,
\frac{1}{2} )
$,

\vskip 0.7ex
\hangindent=3em \hangafter=1
$S$ = ($ 1$,
$ 1$,
$ 2$,
$ 2$,
$ 2$,
$ 2$,
$ 2$,
$ 2$,
$ \sqrt{13}$,
$ \sqrt{13}$;\ \ 
$ 1$,
$ 2$,
$ 2$,
$ 2$,
$ 2$,
$ 2$,
$ 2$,
$ -\sqrt{13}$,
$ -\sqrt{13}$;\ \ 
$ 2c_{13}^{4}$,
$ 2c_{13}^{1}$,
$ 2c_{13}^{3}$,
$ 2c_{13}^{2}$,
$ 2c_{13}^{5}$,
$ 2c_{13}^{6}$,
$0$,
$0$;\ \ 
$ 2c_{13}^{3}$,
$ 2c_{13}^{4}$,
$ 2c_{13}^{6}$,
$ 2c_{13}^{2}$,
$ 2c_{13}^{5}$,
$0$,
$0$;\ \ 
$ 2c_{13}^{1}$,
$ 2c_{13}^{5}$,
$ 2c_{13}^{6}$,
$ 2c_{13}^{2}$,
$0$,
$0$;\ \ 
$ 2c_{13}^{1}$,
$ 2c_{13}^{4}$,
$ 2c_{13}^{3}$,
$0$,
$0$;\ \ 
$ 2c_{13}^{3}$,
$ 2c_{13}^{1}$,
$0$,
$0$;\ \ 
$ 2c_{13}^{4}$,
$0$,
$0$;\ \ 
$ -\sqrt{13}$,
$ \sqrt{13}$;\ \ 
$ -\sqrt{13}$)

\vskip 0.7ex
\hangindent=3em \hangafter=1
$\tau_n$ = ($-7.21$, $33.2$, $-7.21$, $18.78$, $7.21$, $33.2$, $7.21$, $33.2$, $-7.21$, $18.78$, $7.21$, $18.78$, $26.$, $18.78$, $7.21$, $18.78$, $-7.21$, $33.2$, $7.21$, $33.2$, $7.21$, $18.78$, $-7.21$, $33.2$, $-7.21$, $51.99$)

\vskip 0.7ex
\hangindent=3em \hangafter=1
\textit{Intrinsic sign problem}

  \vskip 2ex

\noindent34. $10_{0,52.}^{52,110}$ \irep{1120}:\ \ 
$d_i$ = ($1.0$,
$1.0$,
$2.0$,
$2.0$,
$2.0$,
$2.0$,
$2.0$,
$2.0$,
$3.605$,
$3.605$) 

\vskip 0.7ex
\hangindent=3em \hangafter=1
$D^2= 52.0 = 
52$

\vskip 0.7ex
\hangindent=3em \hangafter=1
$T = ( 0,
0,
\frac{1}{13},
\frac{3}{13},
\frac{4}{13},
\frac{9}{13},
\frac{10}{13},
\frac{12}{13},
\frac{1}{4},
\frac{3}{4} )
$,

\vskip 0.7ex
\hangindent=3em \hangafter=1
$S$ = ($ 1$,
$ 1$,
$ 2$,
$ 2$,
$ 2$,
$ 2$,
$ 2$,
$ 2$,
$ \sqrt{13}$,
$ \sqrt{13}$;\ \ 
$ 1$,
$ 2$,
$ 2$,
$ 2$,
$ 2$,
$ 2$,
$ 2$,
$ -\sqrt{13}$,
$ -\sqrt{13}$;\ \ 
$ 2c_{13}^{2}$,
$ 2c_{13}^{5}$,
$ 2c_{13}^{4}$,
$ 2c_{13}^{6}$,
$ 2c_{13}^{1}$,
$ 2c_{13}^{3}$,
$0$,
$0$;\ \ 
$ 2c_{13}^{6}$,
$ 2c_{13}^{3}$,
$ 2c_{13}^{2}$,
$ 2c_{13}^{4}$,
$ 2c_{13}^{1}$,
$0$,
$0$;\ \ 
$ 2c_{13}^{5}$,
$ 2c_{13}^{1}$,
$ 2c_{13}^{2}$,
$ 2c_{13}^{6}$,
$0$,
$0$;\ \ 
$ 2c_{13}^{5}$,
$ 2c_{13}^{3}$,
$ 2c_{13}^{4}$,
$0$,
$0$;\ \ 
$ 2c_{13}^{6}$,
$ 2c_{13}^{5}$,
$0$,
$0$;\ \ 
$ 2c_{13}^{2}$,
$0$,
$0$;\ \ 
$ -\sqrt{13}$,
$ \sqrt{13}$;\ \ 
$ -\sqrt{13}$)

\vskip 0.7ex
\hangindent=3em \hangafter=1
$\tau_n$ = ($7.21$, $-33.2$, $7.21$, $33.2$, $-7.21$, $-33.2$, $-7.21$, $18.78$, $7.21$, $-18.78$, $-7.21$, $33.2$, $26.$, $-18.78$, $-7.21$, $33.2$, $7.21$, $-33.2$, $-7.21$, $18.78$, $-7.21$, $-18.78$, $7.21$, $18.78$, $7.21$, $0.01$, $7.21$, $18.78$, $7.21$, $-18.78$, $-7.21$, $18.78$, $-7.21$, $-33.2$, $7.21$, $33.2$, $-7.21$, $-18.78$, $26.$, $33.2$, $-7.21$, $-18.78$, $7.21$, $18.78$, $-7.21$, $-33.2$, $-7.21$, $33.2$, $7.21$, $-33.2$, $7.21$, $51.99$)

\vskip 0.7ex
\hangindent=3em \hangafter=1
\textit{Intrinsic sign problem}

  \vskip 2ex

\noindent35. $10_{4,52.}^{52,489}$ \irep{1120}:\ \ 
$d_i$ = ($1.0$,
$1.0$,
$2.0$,
$2.0$,
$2.0$,
$2.0$,
$2.0$,
$2.0$,
$3.605$,
$3.605$) 

\vskip 0.7ex
\hangindent=3em \hangafter=1
$D^2= 52.0 = 
52$

\vskip 0.7ex
\hangindent=3em \hangafter=1
$T = ( 0,
0,
\frac{2}{13},
\frac{5}{13},
\frac{6}{13},
\frac{7}{13},
\frac{8}{13},
\frac{11}{13},
\frac{1}{4},
\frac{3}{4} )
$,

\vskip 0.7ex
\hangindent=3em \hangafter=1
$S$ = ($ 1$,
$ 1$,
$ 2$,
$ 2$,
$ 2$,
$ 2$,
$ 2$,
$ 2$,
$ \sqrt{13}$,
$ \sqrt{13}$;\ \ 
$ 1$,
$ 2$,
$ 2$,
$ 2$,
$ 2$,
$ 2$,
$ 2$,
$ -\sqrt{13}$,
$ -\sqrt{13}$;\ \ 
$ 2c_{13}^{4}$,
$ 2c_{13}^{1}$,
$ 2c_{13}^{3}$,
$ 2c_{13}^{2}$,
$ 2c_{13}^{5}$,
$ 2c_{13}^{6}$,
$0$,
$0$;\ \ 
$ 2c_{13}^{3}$,
$ 2c_{13}^{4}$,
$ 2c_{13}^{6}$,
$ 2c_{13}^{2}$,
$ 2c_{13}^{5}$,
$0$,
$0$;\ \ 
$ 2c_{13}^{1}$,
$ 2c_{13}^{5}$,
$ 2c_{13}^{6}$,
$ 2c_{13}^{2}$,
$0$,
$0$;\ \ 
$ 2c_{13}^{1}$,
$ 2c_{13}^{4}$,
$ 2c_{13}^{3}$,
$0$,
$0$;\ \ 
$ 2c_{13}^{3}$,
$ 2c_{13}^{1}$,
$0$,
$0$;\ \ 
$ 2c_{13}^{4}$,
$0$,
$0$;\ \ 
$ \sqrt{13}$,
$ -\sqrt{13}$;\ \ 
$ \sqrt{13}$)

\vskip 0.7ex
\hangindent=3em \hangafter=1
$\tau_n$ = ($-7.21$, $-18.78$, $-7.21$, $18.78$, $7.21$, $-18.78$, $7.21$, $33.2$, $-7.21$, $-33.2$, $7.21$, $18.78$, $26.$, $-33.2$, $7.21$, $18.78$, $-7.21$, $-18.78$, $7.21$, $33.2$, $7.21$, $-33.2$, $-7.21$, $33.2$, $-7.21$, $0.01$, $-7.21$, $33.2$, $-7.21$, $-33.2$, $7.21$, $33.2$, $7.21$, $-18.78$, $-7.21$, $18.78$, $7.21$, $-33.2$, $26.$, $18.78$, $7.21$, $-33.2$, $-7.21$, $33.2$, $7.21$, $-18.78$, $7.21$, $18.78$, $-7.21$, $-18.78$, $-7.21$, $51.99$)

\vskip 0.7ex
\hangindent=3em \hangafter=1
\textit{Intrinsic sign problem}

  \vskip 2ex

\noindent36. $10_{\frac{83}{11},69.29}^{44,134}$ \irep{1106}:\ \ 
$d_i$ = ($1.0$,
$1.0$,
$1.918$,
$1.918$,
$2.682$,
$2.682$,
$3.228$,
$3.228$,
$3.513$,
$3.513$) 

\vskip 0.7ex
\hangindent=3em \hangafter=1
$D^2= 69.292 = 
30+20c^{1}_{11}
+12c^{2}_{11}
+6c^{3}_{11}
+2c^{4}_{11}
$

\vskip 0.7ex
\hangindent=3em \hangafter=1
$T = ( 0,
\frac{1}{4},
\frac{2}{11},
\frac{19}{44},
\frac{9}{11},
\frac{3}{44},
\frac{10}{11},
\frac{7}{44},
\frac{5}{11},
\frac{31}{44} )
$,

\vskip 0.7ex
\hangindent=3em \hangafter=1
$S$ = ($ 1$,
$ 1$,
$ -c_{11}^{5}$,
$ -c_{11}^{5}$,
$ \xi_{11}^{3}$,
$ \xi_{11}^{3}$,
$ \xi_{11}^{7}$,
$ \xi_{11}^{7}$,
$ \xi_{11}^{5}$,
$ \xi_{11}^{5}$;\ \ 
$ -1$,
$ -c_{11}^{5}$,
$ c_{11}^{5}$,
$ \xi_{11}^{3}$,
$ -\xi_{11}^{3}$,
$ \xi_{11}^{7}$,
$ -\xi_{11}^{7}$,
$ \xi_{11}^{5}$,
$ -\xi_{11}^{5}$;\ \ 
$ -\xi_{11}^{7}$,
$ -\xi_{11}^{7}$,
$ \xi_{11}^{5}$,
$ \xi_{11}^{5}$,
$ -\xi_{11}^{3}$,
$ -\xi_{11}^{3}$,
$ 1$,
$ 1$;\ \ 
$ \xi_{11}^{7}$,
$ \xi_{11}^{5}$,
$ -\xi_{11}^{5}$,
$ -\xi_{11}^{3}$,
$ \xi_{11}^{3}$,
$ 1$,
$ -1$;\ \ 
$ -c_{11}^{5}$,
$ -c_{11}^{5}$,
$ -1$,
$ -1$,
$ -\xi_{11}^{7}$,
$ -\xi_{11}^{7}$;\ \ 
$ c_{11}^{5}$,
$ -1$,
$ 1$,
$ -\xi_{11}^{7}$,
$ \xi_{11}^{7}$;\ \ 
$ \xi_{11}^{5}$,
$ \xi_{11}^{5}$,
$ c_{11}^{5}$,
$ c_{11}^{5}$;\ \ 
$ -\xi_{11}^{5}$,
$ c_{11}^{5}$,
$ -c_{11}^{5}$;\ \ 
$ \xi_{11}^{3}$,
$ \xi_{11}^{3}$;\ \ 
$ -\xi_{11}^{3}$)

Factors = $2_{1,2.}^{4,437}\boxtimes 5_{\frac{72}{11},34.64}^{11,216}$

\vskip 0.7ex
\hangindent=3em \hangafter=1
$\tau_n$ = ($7.79 - 2.91 i$, $0.$, $-18.99 + 19. i$, $-4.49 - 31.24 i$, $-12.79 + 9.57 i$, $0.$, $13.38 + 17.87 i$, $-37.99 - 0.01 i$, $-10.22 + 27.4 i$, $0.$, $34.63 - 34.63 i$, $4.88 - 10.71 i$, $27.4 - 10.22 i$, $0.$, $-17.87 - 13.38 i$, $-3.22 + 22.36 i$, $9.57 - 12.79 i$, $0.$, $-19. + 18.99 i$, $17.18 + 37.61 i$, $-2.91 + 7.79 i$, $0.$, $-2.91 - 7.79 i$, $17.18 - 37.61 i$, $-19. - 18.99 i$, $0.$, $9.57 + 12.79 i$, $-3.22 - 22.36 i$, $-17.87 + 13.38 i$, $0.$, $27.4 + 10.22 i$, $4.88 + 10.71 i$, $34.63 + 34.63 i$, $0.$, $-10.22 - 27.4 i$, $-37.99 + 0.01 i$, $13.38 - 17.87 i$, $0.$, $-12.79 - 9.57 i$, $-4.49 + 31.24 i$, $-18.99 - 19. i$, $0.$, $7.79 + 2.91 i$, $69.27$)

\vskip 0.7ex
\hangindent=3em \hangafter=1
\textit{Intrinsic sign problem}

  \vskip 2ex

\noindent37. $10_{\frac{27}{11},69.29}^{44,556}$ \irep{1106}:\ \ 
$d_i$ = ($1.0$,
$1.0$,
$1.918$,
$1.918$,
$2.682$,
$2.682$,
$3.228$,
$3.228$,
$3.513$,
$3.513$) 

\vskip 0.7ex
\hangindent=3em \hangafter=1
$D^2= 69.292 = 
30+20c^{1}_{11}
+12c^{2}_{11}
+6c^{3}_{11}
+2c^{4}_{11}
$

\vskip 0.7ex
\hangindent=3em \hangafter=1
$T = ( 0,
\frac{1}{4},
\frac{9}{11},
\frac{3}{44},
\frac{2}{11},
\frac{19}{44},
\frac{1}{11},
\frac{15}{44},
\frac{6}{11},
\frac{35}{44} )
$,

\vskip 0.7ex
\hangindent=3em \hangafter=1
$S$ = ($ 1$,
$ 1$,
$ -c_{11}^{5}$,
$ -c_{11}^{5}$,
$ \xi_{11}^{3}$,
$ \xi_{11}^{3}$,
$ \xi_{11}^{7}$,
$ \xi_{11}^{7}$,
$ \xi_{11}^{5}$,
$ \xi_{11}^{5}$;\ \ 
$ -1$,
$ -c_{11}^{5}$,
$ c_{11}^{5}$,
$ \xi_{11}^{3}$,
$ -\xi_{11}^{3}$,
$ \xi_{11}^{7}$,
$ -\xi_{11}^{7}$,
$ \xi_{11}^{5}$,
$ -\xi_{11}^{5}$;\ \ 
$ -\xi_{11}^{7}$,
$ -\xi_{11}^{7}$,
$ \xi_{11}^{5}$,
$ \xi_{11}^{5}$,
$ -\xi_{11}^{3}$,
$ -\xi_{11}^{3}$,
$ 1$,
$ 1$;\ \ 
$ \xi_{11}^{7}$,
$ \xi_{11}^{5}$,
$ -\xi_{11}^{5}$,
$ -\xi_{11}^{3}$,
$ \xi_{11}^{3}$,
$ 1$,
$ -1$;\ \ 
$ -c_{11}^{5}$,
$ -c_{11}^{5}$,
$ -1$,
$ -1$,
$ -\xi_{11}^{7}$,
$ -\xi_{11}^{7}$;\ \ 
$ c_{11}^{5}$,
$ -1$,
$ 1$,
$ -\xi_{11}^{7}$,
$ \xi_{11}^{7}$;\ \ 
$ \xi_{11}^{5}$,
$ \xi_{11}^{5}$,
$ c_{11}^{5}$,
$ c_{11}^{5}$;\ \ 
$ -\xi_{11}^{5}$,
$ c_{11}^{5}$,
$ -c_{11}^{5}$;\ \ 
$ \xi_{11}^{3}$,
$ \xi_{11}^{3}$;\ \ 
$ -\xi_{11}^{3}$)

Factors = $2_{1,2.}^{4,437}\boxtimes 5_{\frac{16}{11},34.64}^{11,640}$

\vskip 0.7ex
\hangindent=3em \hangafter=1
$\tau_n$ = ($-2.91 + 7.79 i$, $0.$, $-19. + 18.99 i$, $-4.49 + 31.24 i$, $9.57 - 12.79 i$, $0.$, $-17.87 - 13.38 i$, $-37.99 + 0.01 i$, $27.4 - 10.22 i$, $0.$, $34.63 - 34.63 i$, $4.88 + 10.71 i$, $-10.22 + 27.4 i$, $0.$, $13.38 + 17.87 i$, $-3.22 - 22.36 i$, $-12.79 + 9.57 i$, $0.$, $-18.99 + 19. i$, $17.18 - 37.61 i$, $7.79 - 2.91 i$, $0.$, $7.79 + 2.91 i$, $17.18 + 37.61 i$, $-18.99 - 19. i$, $0.$, $-12.79 - 9.57 i$, $-3.22 + 22.36 i$, $13.38 - 17.87 i$, $0.$, $-10.22 - 27.4 i$, $4.88 - 10.71 i$, $34.63 + 34.63 i$, $0.$, $27.4 + 10.22 i$, $-37.99 - 0.01 i$, $-17.87 + 13.38 i$, $0.$, $9.57 + 12.79 i$, $-4.49 - 31.24 i$, $-19. - 18.99 i$, $0.$, $-2.91 - 7.79 i$, $69.27$)

\vskip 0.7ex
\hangindent=3em \hangafter=1
\textit{Intrinsic sign problem}

  \vskip 2ex

\noindent38. $10_{\frac{61}{11},69.29}^{44,372}$ \irep{1106}:\ \ 
$d_i$ = ($1.0$,
$1.0$,
$1.918$,
$1.918$,
$2.682$,
$2.682$,
$3.228$,
$3.228$,
$3.513$,
$3.513$) 

\vskip 0.7ex
\hangindent=3em \hangafter=1
$D^2= 69.292 = 
30+20c^{1}_{11}
+12c^{2}_{11}
+6c^{3}_{11}
+2c^{4}_{11}
$

\vskip 0.7ex
\hangindent=3em \hangafter=1
$T = ( 0,
\frac{3}{4},
\frac{2}{11},
\frac{41}{44},
\frac{9}{11},
\frac{25}{44},
\frac{10}{11},
\frac{29}{44},
\frac{5}{11},
\frac{9}{44} )
$,

\vskip 0.7ex
\hangindent=3em \hangafter=1
$S$ = ($ 1$,
$ 1$,
$ -c_{11}^{5}$,
$ -c_{11}^{5}$,
$ \xi_{11}^{3}$,
$ \xi_{11}^{3}$,
$ \xi_{11}^{7}$,
$ \xi_{11}^{7}$,
$ \xi_{11}^{5}$,
$ \xi_{11}^{5}$;\ \ 
$ -1$,
$ -c_{11}^{5}$,
$ c_{11}^{5}$,
$ \xi_{11}^{3}$,
$ -\xi_{11}^{3}$,
$ \xi_{11}^{7}$,
$ -\xi_{11}^{7}$,
$ \xi_{11}^{5}$,
$ -\xi_{11}^{5}$;\ \ 
$ -\xi_{11}^{7}$,
$ -\xi_{11}^{7}$,
$ \xi_{11}^{5}$,
$ \xi_{11}^{5}$,
$ -\xi_{11}^{3}$,
$ -\xi_{11}^{3}$,
$ 1$,
$ 1$;\ \ 
$ \xi_{11}^{7}$,
$ \xi_{11}^{5}$,
$ -\xi_{11}^{5}$,
$ -\xi_{11}^{3}$,
$ \xi_{11}^{3}$,
$ 1$,
$ -1$;\ \ 
$ -c_{11}^{5}$,
$ -c_{11}^{5}$,
$ -1$,
$ -1$,
$ -\xi_{11}^{7}$,
$ -\xi_{11}^{7}$;\ \ 
$ c_{11}^{5}$,
$ -1$,
$ 1$,
$ -\xi_{11}^{7}$,
$ \xi_{11}^{7}$;\ \ 
$ \xi_{11}^{5}$,
$ \xi_{11}^{5}$,
$ c_{11}^{5}$,
$ c_{11}^{5}$;\ \ 
$ -\xi_{11}^{5}$,
$ c_{11}^{5}$,
$ -c_{11}^{5}$;\ \ 
$ \xi_{11}^{3}$,
$ \xi_{11}^{3}$;\ \ 
$ -\xi_{11}^{3}$)

Factors = $2_{7,2.}^{4,625}\boxtimes 5_{\frac{72}{11},34.64}^{11,216}$

\vskip 0.7ex
\hangindent=3em \hangafter=1
$\tau_n$ = ($-2.91 - 7.79 i$, $0.$, $-19. - 18.99 i$, $-4.49 - 31.24 i$, $9.57 + 12.79 i$, $0.$, $-17.87 + 13.38 i$, $-37.99 - 0.01 i$, $27.4 + 10.22 i$, $0.$, $34.63 + 34.63 i$, $4.88 - 10.71 i$, $-10.22 - 27.4 i$, $0.$, $13.38 - 17.87 i$, $-3.22 + 22.36 i$, $-12.79 - 9.57 i$, $0.$, $-18.99 - 19. i$, $17.18 + 37.61 i$, $7.79 + 2.91 i$, $0.$, $7.79 - 2.91 i$, $17.18 - 37.61 i$, $-18.99 + 19. i$, $0.$, $-12.79 + 9.57 i$, $-3.22 - 22.36 i$, $13.38 + 17.87 i$, $0.$, $-10.22 + 27.4 i$, $4.88 + 10.71 i$, $34.63 - 34.63 i$, $0.$, $27.4 - 10.22 i$, $-37.99 + 0.01 i$, $-17.87 - 13.38 i$, $0.$, $9.57 - 12.79 i$, $-4.49 + 31.24 i$, $-19. + 18.99 i$, $0.$, $-2.91 + 7.79 i$, $69.27$)

\vskip 0.7ex
\hangindent=3em \hangafter=1
\textit{Intrinsic sign problem}

  \vskip 2ex

\noindent39. $10_{\frac{5}{11},69.29}^{44,237}$ \irep{1106}:\ \ 
$d_i$ = ($1.0$,
$1.0$,
$1.918$,
$1.918$,
$2.682$,
$2.682$,
$3.228$,
$3.228$,
$3.513$,
$3.513$) 

\vskip 0.7ex
\hangindent=3em \hangafter=1
$D^2= 69.292 = 
30+20c^{1}_{11}
+12c^{2}_{11}
+6c^{3}_{11}
+2c^{4}_{11}
$

\vskip 0.7ex
\hangindent=3em \hangafter=1
$T = ( 0,
\frac{3}{4},
\frac{9}{11},
\frac{25}{44},
\frac{2}{11},
\frac{41}{44},
\frac{1}{11},
\frac{37}{44},
\frac{6}{11},
\frac{13}{44} )
$,

\vskip 0.7ex
\hangindent=3em \hangafter=1
$S$ = ($ 1$,
$ 1$,
$ -c_{11}^{5}$,
$ -c_{11}^{5}$,
$ \xi_{11}^{3}$,
$ \xi_{11}^{3}$,
$ \xi_{11}^{7}$,
$ \xi_{11}^{7}$,
$ \xi_{11}^{5}$,
$ \xi_{11}^{5}$;\ \ 
$ -1$,
$ -c_{11}^{5}$,
$ c_{11}^{5}$,
$ \xi_{11}^{3}$,
$ -\xi_{11}^{3}$,
$ \xi_{11}^{7}$,
$ -\xi_{11}^{7}$,
$ \xi_{11}^{5}$,
$ -\xi_{11}^{5}$;\ \ 
$ -\xi_{11}^{7}$,
$ -\xi_{11}^{7}$,
$ \xi_{11}^{5}$,
$ \xi_{11}^{5}$,
$ -\xi_{11}^{3}$,
$ -\xi_{11}^{3}$,
$ 1$,
$ 1$;\ \ 
$ \xi_{11}^{7}$,
$ \xi_{11}^{5}$,
$ -\xi_{11}^{5}$,
$ -\xi_{11}^{3}$,
$ \xi_{11}^{3}$,
$ 1$,
$ -1$;\ \ 
$ -c_{11}^{5}$,
$ -c_{11}^{5}$,
$ -1$,
$ -1$,
$ -\xi_{11}^{7}$,
$ -\xi_{11}^{7}$;\ \ 
$ c_{11}^{5}$,
$ -1$,
$ 1$,
$ -\xi_{11}^{7}$,
$ \xi_{11}^{7}$;\ \ 
$ \xi_{11}^{5}$,
$ \xi_{11}^{5}$,
$ c_{11}^{5}$,
$ c_{11}^{5}$;\ \ 
$ -\xi_{11}^{5}$,
$ c_{11}^{5}$,
$ -c_{11}^{5}$;\ \ 
$ \xi_{11}^{3}$,
$ \xi_{11}^{3}$;\ \ 
$ -\xi_{11}^{3}$)

Factors = $2_{7,2.}^{4,625}\boxtimes 5_{\frac{16}{11},34.64}^{11,640}$

\vskip 0.7ex
\hangindent=3em \hangafter=1
$\tau_n$ = ($7.79 + 2.91 i$, $0.$, $-18.99 - 19. i$, $-4.49 + 31.24 i$, $-12.79 - 9.57 i$, $0.$, $13.38 - 17.87 i$, $-37.99 + 0.01 i$, $-10.22 - 27.4 i$, $0.$, $34.63 + 34.63 i$, $4.88 + 10.71 i$, $27.4 + 10.22 i$, $0.$, $-17.87 + 13.38 i$, $-3.22 - 22.36 i$, $9.57 + 12.79 i$, $0.$, $-19. - 18.99 i$, $17.18 - 37.61 i$, $-2.91 - 7.79 i$, $0.$, $-2.91 + 7.79 i$, $17.18 + 37.61 i$, $-19. + 18.99 i$, $0.$, $9.57 - 12.79 i$, $-3.22 + 22.36 i$, $-17.87 - 13.38 i$, $0.$, $27.4 - 10.22 i$, $4.88 - 10.71 i$, $34.63 - 34.63 i$, $0.$, $-10.22 + 27.4 i$, $-37.99 - 0.01 i$, $13.38 + 17.87 i$, $0.$, $-12.79 + 9.57 i$, $-4.49 - 31.24 i$, $-18.99 + 19. i$, $0.$, $7.79 - 2.91 i$, $69.27$)

\vskip 0.7ex
\hangindent=3em \hangafter=1
\textit{Intrinsic sign problem}

  \vskip 2ex

\noindent40. $10_{\frac{45}{7},70.68}^{28,820}$ \irep{1038}:\ \ 
$d_i$ = ($1.0$,
$1.0$,
$2.246$,
$2.246$,
$2.246$,
$2.246$,
$2.801$,
$2.801$,
$4.48$,
$4.48$) 

\vskip 0.7ex
\hangindent=3em \hangafter=1
$D^2= 70.684 = 
42+28c^{1}_{7}
+14c^{2}_{7}
$

\vskip 0.7ex
\hangindent=3em \hangafter=1
$T = ( 0,
\frac{1}{4},
\frac{1}{7},
\frac{1}{7},
\frac{11}{28},
\frac{11}{28},
\frac{6}{7},
\frac{3}{28},
\frac{4}{7},
\frac{23}{28} )
$,

\vskip 0.7ex
\hangindent=3em \hangafter=1
$S$ = ($ 1$,
$ 1$,
$ \xi_{7}^{3}$,
$ \xi_{7}^{3}$,
$ \xi_{7}^{3}$,
$ \xi_{7}^{3}$,
$ 2+c^{1}_{7}
+c^{2}_{7}
$,
$ 2+c^{1}_{7}
+c^{2}_{7}
$,
$ 2+2c^{1}_{7}
+c^{2}_{7}
$,
$ 2+2c^{1}_{7}
+c^{2}_{7}
$;\ \ 
$ -1$,
$ \xi_{7}^{3}$,
$ \xi_{7}^{3}$,
$ -\xi_{7}^{3}$,
$ -\xi_{7}^{3}$,
$ 2+c^{1}_{7}
+c^{2}_{7}
$,
$ -2-c^{1}_{7}
-c^{2}_{7}
$,
$ 2+2c^{1}_{7}
+c^{2}_{7}
$,
$ -2-2  c^{1}_{7}
-c^{2}_{7}
$;\ \ 
$ s^{1}_{7}
+\zeta^{2}_{7}
+\zeta^{3}_{7}
$,
$ -1-2  \zeta^{1}_{7}
-\zeta^{2}_{7}
-\zeta^{3}_{7}
$,
$ -1-2  \zeta^{1}_{7}
-\zeta^{2}_{7}
-\zeta^{3}_{7}
$,
$ s^{1}_{7}
+\zeta^{2}_{7}
+\zeta^{3}_{7}
$,
$ -\xi_{7}^{3}$,
$ -\xi_{7}^{3}$,
$ \xi_{7}^{3}$,
$ \xi_{7}^{3}$;\ \ 
$ s^{1}_{7}
+\zeta^{2}_{7}
+\zeta^{3}_{7}
$,
$ s^{1}_{7}
+\zeta^{2}_{7}
+\zeta^{3}_{7}
$,
$ -1-2  \zeta^{1}_{7}
-\zeta^{2}_{7}
-\zeta^{3}_{7}
$,
$ -\xi_{7}^{3}$,
$ -\xi_{7}^{3}$,
$ \xi_{7}^{3}$,
$ \xi_{7}^{3}$;\ \ 
$ -s^{1}_{7}
-\zeta^{2}_{7}
-\zeta^{3}_{7}
$,
$ 1+2\zeta^{1}_{7}
+\zeta^{2}_{7}
+\zeta^{3}_{7}
$,
$ -\xi_{7}^{3}$,
$ \xi_{7}^{3}$,
$ \xi_{7}^{3}$,
$ -\xi_{7}^{3}$;\ \ 
$ -s^{1}_{7}
-\zeta^{2}_{7}
-\zeta^{3}_{7}
$,
$ -\xi_{7}^{3}$,
$ \xi_{7}^{3}$,
$ \xi_{7}^{3}$,
$ -\xi_{7}^{3}$;\ \ 
$ 2+2c^{1}_{7}
+c^{2}_{7}
$,
$ 2+2c^{1}_{7}
+c^{2}_{7}
$,
$ -1$,
$ -1$;\ \ 
$ -2-2  c^{1}_{7}
-c^{2}_{7}
$,
$ -1$,
$ 1$;\ \ 
$ -2-c^{1}_{7}
-c^{2}_{7}
$,
$ -2-c^{1}_{7}
-c^{2}_{7}
$;\ \ 
$ 2+c^{1}_{7}
+c^{2}_{7}
$)

Factors = $2_{1,2.}^{4,437}\boxtimes 5_{\frac{38}{7},35.34}^{7,386}$

\vskip 0.7ex
\hangindent=3em \hangafter=1
$\tau_n$ = ($1.05 - 12.85 i$, $0.$, $-38.22 + 1.03 i$, $-39.25 + 37.19 i$, $27.4 - 8.36 i$, $0.$, $39.01 - 39.01 i$, $-11.8 - 13.91 i$, $-8.36 + 27.4 i$, $0.$, $-1.03 + 38.22 i$, $19.05 - 35.76 i$, $-12.85 + 1.05 i$, $0.$, $-12.85 - 1.05 i$, $19.05 + 35.76 i$, $-1.03 - 38.22 i$, $0.$, $-8.36 - 27.4 i$, $-11.8 + 13.91 i$, $39.01 + 39.01 i$, $0.$, $27.4 + 8.36 i$, $-39.25 - 37.19 i$, $-38.22 - 1.03 i$, $0.$, $1.05 + 12.85 i$, $78.01$)

\vskip 0.7ex
\hangindent=3em \hangafter=1
\textit{Intrinsic sign problem}

  \vskip 2ex

\noindent41. $10_{\frac{25}{7},70.68}^{28,341}$ \irep{1038}:\ \ 
$d_i$ = ($1.0$,
$1.0$,
$2.246$,
$2.246$,
$2.246$,
$2.246$,
$2.801$,
$2.801$,
$4.48$,
$4.48$) 

\vskip 0.7ex
\hangindent=3em \hangafter=1
$D^2= 70.684 = 
42+28c^{1}_{7}
+14c^{2}_{7}
$

\vskip 0.7ex
\hangindent=3em \hangafter=1
$T = ( 0,
\frac{1}{4},
\frac{6}{7},
\frac{6}{7},
\frac{3}{28},
\frac{3}{28},
\frac{1}{7},
\frac{11}{28},
\frac{3}{7},
\frac{19}{28} )
$,

\vskip 0.7ex
\hangindent=3em \hangafter=1
$S$ = ($ 1$,
$ 1$,
$ \xi_{7}^{3}$,
$ \xi_{7}^{3}$,
$ \xi_{7}^{3}$,
$ \xi_{7}^{3}$,
$ 2+c^{1}_{7}
+c^{2}_{7}
$,
$ 2+c^{1}_{7}
+c^{2}_{7}
$,
$ 2+2c^{1}_{7}
+c^{2}_{7}
$,
$ 2+2c^{1}_{7}
+c^{2}_{7}
$;\ \ 
$ -1$,
$ \xi_{7}^{3}$,
$ \xi_{7}^{3}$,
$ -\xi_{7}^{3}$,
$ -\xi_{7}^{3}$,
$ 2+c^{1}_{7}
+c^{2}_{7}
$,
$ -2-c^{1}_{7}
-c^{2}_{7}
$,
$ 2+2c^{1}_{7}
+c^{2}_{7}
$,
$ -2-2  c^{1}_{7}
-c^{2}_{7}
$;\ \ 
$ -1-2  \zeta^{1}_{7}
-\zeta^{2}_{7}
-\zeta^{3}_{7}
$,
$ s^{1}_{7}
+\zeta^{2}_{7}
+\zeta^{3}_{7}
$,
$ -1-2  \zeta^{1}_{7}
-\zeta^{2}_{7}
-\zeta^{3}_{7}
$,
$ s^{1}_{7}
+\zeta^{2}_{7}
+\zeta^{3}_{7}
$,
$ -\xi_{7}^{3}$,
$ -\xi_{7}^{3}$,
$ \xi_{7}^{3}$,
$ \xi_{7}^{3}$;\ \ 
$ -1-2  \zeta^{1}_{7}
-\zeta^{2}_{7}
-\zeta^{3}_{7}
$,
$ s^{1}_{7}
+\zeta^{2}_{7}
+\zeta^{3}_{7}
$,
$ -1-2  \zeta^{1}_{7}
-\zeta^{2}_{7}
-\zeta^{3}_{7}
$,
$ -\xi_{7}^{3}$,
$ -\xi_{7}^{3}$,
$ \xi_{7}^{3}$,
$ \xi_{7}^{3}$;\ \ 
$ 1+2\zeta^{1}_{7}
+\zeta^{2}_{7}
+\zeta^{3}_{7}
$,
$ -s^{1}_{7}
-\zeta^{2}_{7}
-\zeta^{3}_{7}
$,
$ -\xi_{7}^{3}$,
$ \xi_{7}^{3}$,
$ \xi_{7}^{3}$,
$ -\xi_{7}^{3}$;\ \ 
$ 1+2\zeta^{1}_{7}
+\zeta^{2}_{7}
+\zeta^{3}_{7}
$,
$ -\xi_{7}^{3}$,
$ \xi_{7}^{3}$,
$ \xi_{7}^{3}$,
$ -\xi_{7}^{3}$;\ \ 
$ 2+2c^{1}_{7}
+c^{2}_{7}
$,
$ 2+2c^{1}_{7}
+c^{2}_{7}
$,
$ -1$,
$ -1$;\ \ 
$ -2-2  c^{1}_{7}
-c^{2}_{7}
$,
$ -1$,
$ 1$;\ \ 
$ -2-c^{1}_{7}
-c^{2}_{7}
$,
$ -2-c^{1}_{7}
-c^{2}_{7}
$;\ \ 
$ 2+c^{1}_{7}
+c^{2}_{7}
$)

Factors = $2_{1,2.}^{4,437}\boxtimes 5_{\frac{18}{7},35.34}^{7,101}$

\vskip 0.7ex
\hangindent=3em \hangafter=1
$\tau_n$ = ($-12.85 + 1.05 i$, $0.$, $-1.03 + 38.22 i$, $-39.25 - 37.19 i$, $-8.36 + 27.4 i$, $0.$, $39.01 - 39.01 i$, $-11.8 + 13.91 i$, $27.4 - 8.36 i$, $0.$, $-38.22 + 1.03 i$, $19.05 + 35.76 i$, $1.05 - 12.85 i$, $0.$, $1.05 + 12.85 i$, $19.05 - 35.76 i$, $-38.22 - 1.03 i$, $0.$, $27.4 + 8.36 i$, $-11.8 - 13.91 i$, $39.01 + 39.01 i$, $0.$, $-8.36 - 27.4 i$, $-39.25 + 37.19 i$, $-1.03 - 38.22 i$, $0.$, $-12.85 - 1.05 i$, $78.01$)

\vskip 0.7ex
\hangindent=3em \hangafter=1
\textit{Intrinsic sign problem}

  \vskip 2ex

\noindent42. $10_{\frac{31}{7},70.68}^{28,289}$ \irep{1038}:\ \ 
$d_i$ = ($1.0$,
$1.0$,
$2.246$,
$2.246$,
$2.246$,
$2.246$,
$2.801$,
$2.801$,
$4.48$,
$4.48$) 

\vskip 0.7ex
\hangindent=3em \hangafter=1
$D^2= 70.684 = 
42+28c^{1}_{7}
+14c^{2}_{7}
$

\vskip 0.7ex
\hangindent=3em \hangafter=1
$T = ( 0,
\frac{3}{4},
\frac{1}{7},
\frac{1}{7},
\frac{25}{28},
\frac{25}{28},
\frac{6}{7},
\frac{17}{28},
\frac{4}{7},
\frac{9}{28} )
$,

\vskip 0.7ex
\hangindent=3em \hangafter=1
$S$ = ($ 1$,
$ 1$,
$ \xi_{7}^{3}$,
$ \xi_{7}^{3}$,
$ \xi_{7}^{3}$,
$ \xi_{7}^{3}$,
$ 2+c^{1}_{7}
+c^{2}_{7}
$,
$ 2+c^{1}_{7}
+c^{2}_{7}
$,
$ 2+2c^{1}_{7}
+c^{2}_{7}
$,
$ 2+2c^{1}_{7}
+c^{2}_{7}
$;\ \ 
$ -1$,
$ \xi_{7}^{3}$,
$ \xi_{7}^{3}$,
$ -\xi_{7}^{3}$,
$ -\xi_{7}^{3}$,
$ 2+c^{1}_{7}
+c^{2}_{7}
$,
$ -2-c^{1}_{7}
-c^{2}_{7}
$,
$ 2+2c^{1}_{7}
+c^{2}_{7}
$,
$ -2-2  c^{1}_{7}
-c^{2}_{7}
$;\ \ 
$ s^{1}_{7}
+\zeta^{2}_{7}
+\zeta^{3}_{7}
$,
$ -1-2  \zeta^{1}_{7}
-\zeta^{2}_{7}
-\zeta^{3}_{7}
$,
$ -1-2  \zeta^{1}_{7}
-\zeta^{2}_{7}
-\zeta^{3}_{7}
$,
$ s^{1}_{7}
+\zeta^{2}_{7}
+\zeta^{3}_{7}
$,
$ -\xi_{7}^{3}$,
$ -\xi_{7}^{3}$,
$ \xi_{7}^{3}$,
$ \xi_{7}^{3}$;\ \ 
$ s^{1}_{7}
+\zeta^{2}_{7}
+\zeta^{3}_{7}
$,
$ s^{1}_{7}
+\zeta^{2}_{7}
+\zeta^{3}_{7}
$,
$ -1-2  \zeta^{1}_{7}
-\zeta^{2}_{7}
-\zeta^{3}_{7}
$,
$ -\xi_{7}^{3}$,
$ -\xi_{7}^{3}$,
$ \xi_{7}^{3}$,
$ \xi_{7}^{3}$;\ \ 
$ -s^{1}_{7}
-\zeta^{2}_{7}
-\zeta^{3}_{7}
$,
$ 1+2\zeta^{1}_{7}
+\zeta^{2}_{7}
+\zeta^{3}_{7}
$,
$ -\xi_{7}^{3}$,
$ \xi_{7}^{3}$,
$ \xi_{7}^{3}$,
$ -\xi_{7}^{3}$;\ \ 
$ -s^{1}_{7}
-\zeta^{2}_{7}
-\zeta^{3}_{7}
$,
$ -\xi_{7}^{3}$,
$ \xi_{7}^{3}$,
$ \xi_{7}^{3}$,
$ -\xi_{7}^{3}$;\ \ 
$ 2+2c^{1}_{7}
+c^{2}_{7}
$,
$ 2+2c^{1}_{7}
+c^{2}_{7}
$,
$ -1$,
$ -1$;\ \ 
$ -2-2  c^{1}_{7}
-c^{2}_{7}
$,
$ -1$,
$ 1$;\ \ 
$ -2-c^{1}_{7}
-c^{2}_{7}
$,
$ -2-c^{1}_{7}
-c^{2}_{7}
$;\ \ 
$ 2+c^{1}_{7}
+c^{2}_{7}
$)

Factors = $2_{7,2.}^{4,625}\boxtimes 5_{\frac{38}{7},35.34}^{7,386}$

\vskip 0.7ex
\hangindent=3em \hangafter=1
$\tau_n$ = ($-12.85 - 1.05 i$, $0.$, $-1.03 - 38.22 i$, $-39.25 + 37.19 i$, $-8.36 - 27.4 i$, $0.$, $39.01 + 39.01 i$, $-11.8 - 13.91 i$, $27.4 + 8.36 i$, $0.$, $-38.22 - 1.03 i$, $19.05 - 35.76 i$, $1.05 + 12.85 i$, $0.$, $1.05 - 12.85 i$, $19.05 + 35.76 i$, $-38.22 + 1.03 i$, $0.$, $27.4 - 8.36 i$, $-11.8 + 13.91 i$, $39.01 - 39.01 i$, $0.$, $-8.36 + 27.4 i$, $-39.25 - 37.19 i$, $-1.03 + 38.22 i$, $0.$, $-12.85 + 1.05 i$, $78.01$)

\vskip 0.7ex
\hangindent=3em \hangafter=1
\textit{Intrinsic sign problem}

  \vskip 2ex

\noindent43. $10_{\frac{11}{7},70.68}^{28,251}$ \irep{1038}:\ \ 
$d_i$ = ($1.0$,
$1.0$,
$2.246$,
$2.246$,
$2.246$,
$2.246$,
$2.801$,
$2.801$,
$4.48$,
$4.48$) 

\vskip 0.7ex
\hangindent=3em \hangafter=1
$D^2= 70.684 = 
42+28c^{1}_{7}
+14c^{2}_{7}
$

\vskip 0.7ex
\hangindent=3em \hangafter=1
$T = ( 0,
\frac{3}{4},
\frac{6}{7},
\frac{6}{7},
\frac{17}{28},
\frac{17}{28},
\frac{1}{7},
\frac{25}{28},
\frac{3}{7},
\frac{5}{28} )
$,

\vskip 0.7ex
\hangindent=3em \hangafter=1
$S$ = ($ 1$,
$ 1$,
$ \xi_{7}^{3}$,
$ \xi_{7}^{3}$,
$ \xi_{7}^{3}$,
$ \xi_{7}^{3}$,
$ 2+c^{1}_{7}
+c^{2}_{7}
$,
$ 2+c^{1}_{7}
+c^{2}_{7}
$,
$ 2+2c^{1}_{7}
+c^{2}_{7}
$,
$ 2+2c^{1}_{7}
+c^{2}_{7}
$;\ \ 
$ -1$,
$ \xi_{7}^{3}$,
$ \xi_{7}^{3}$,
$ -\xi_{7}^{3}$,
$ -\xi_{7}^{3}$,
$ 2+c^{1}_{7}
+c^{2}_{7}
$,
$ -2-c^{1}_{7}
-c^{2}_{7}
$,
$ 2+2c^{1}_{7}
+c^{2}_{7}
$,
$ -2-2  c^{1}_{7}
-c^{2}_{7}
$;\ \ 
$ -1-2  \zeta^{1}_{7}
-\zeta^{2}_{7}
-\zeta^{3}_{7}
$,
$ s^{1}_{7}
+\zeta^{2}_{7}
+\zeta^{3}_{7}
$,
$ -1-2  \zeta^{1}_{7}
-\zeta^{2}_{7}
-\zeta^{3}_{7}
$,
$ s^{1}_{7}
+\zeta^{2}_{7}
+\zeta^{3}_{7}
$,
$ -\xi_{7}^{3}$,
$ -\xi_{7}^{3}$,
$ \xi_{7}^{3}$,
$ \xi_{7}^{3}$;\ \ 
$ -1-2  \zeta^{1}_{7}
-\zeta^{2}_{7}
-\zeta^{3}_{7}
$,
$ s^{1}_{7}
+\zeta^{2}_{7}
+\zeta^{3}_{7}
$,
$ -1-2  \zeta^{1}_{7}
-\zeta^{2}_{7}
-\zeta^{3}_{7}
$,
$ -\xi_{7}^{3}$,
$ -\xi_{7}^{3}$,
$ \xi_{7}^{3}$,
$ \xi_{7}^{3}$;\ \ 
$ 1+2\zeta^{1}_{7}
+\zeta^{2}_{7}
+\zeta^{3}_{7}
$,
$ -s^{1}_{7}
-\zeta^{2}_{7}
-\zeta^{3}_{7}
$,
$ -\xi_{7}^{3}$,
$ \xi_{7}^{3}$,
$ \xi_{7}^{3}$,
$ -\xi_{7}^{3}$;\ \ 
$ 1+2\zeta^{1}_{7}
+\zeta^{2}_{7}
+\zeta^{3}_{7}
$,
$ -\xi_{7}^{3}$,
$ \xi_{7}^{3}$,
$ \xi_{7}^{3}$,
$ -\xi_{7}^{3}$;\ \ 
$ 2+2c^{1}_{7}
+c^{2}_{7}
$,
$ 2+2c^{1}_{7}
+c^{2}_{7}
$,
$ -1$,
$ -1$;\ \ 
$ -2-2  c^{1}_{7}
-c^{2}_{7}
$,
$ -1$,
$ 1$;\ \ 
$ -2-c^{1}_{7}
-c^{2}_{7}
$,
$ -2-c^{1}_{7}
-c^{2}_{7}
$;\ \ 
$ 2+c^{1}_{7}
+c^{2}_{7}
$)

Factors = $2_{7,2.}^{4,625}\boxtimes 5_{\frac{18}{7},35.34}^{7,101}$

\vskip 0.7ex
\hangindent=3em \hangafter=1
$\tau_n$ = ($1.05 + 12.85 i$, $0.$, $-38.22 - 1.03 i$, $-39.25 - 37.19 i$, $27.4 + 8.36 i$, $0.$, $39.01 + 39.01 i$, $-11.8 + 13.91 i$, $-8.36 - 27.4 i$, $0.$, $-1.03 - 38.22 i$, $19.05 + 35.76 i$, $-12.85 - 1.05 i$, $0.$, $-12.85 + 1.05 i$, $19.05 - 35.76 i$, $-1.03 + 38.22 i$, $0.$, $-8.36 + 27.4 i$, $-11.8 - 13.91 i$, $39.01 - 39.01 i$, $0.$, $27.4 - 8.36 i$, $-39.25 + 37.19 i$, $-38.22 + 1.03 i$, $0.$, $1.05 - 12.85 i$, $78.01$)

\vskip 0.7ex
\hangindent=3em \hangafter=1
\textit{Intrinsic sign problem}

  \vskip 2ex

\noindent44. $10_{6,89.56}^{12,311}$ \irep{680}:\ \ 
$d_i$ = ($1.0$,
$1.0$,
$2.732$,
$2.732$,
$2.732$,
$2.732$,
$2.732$,
$3.732$,
$3.732$,
$4.732$) 

\vskip 0.7ex
\hangindent=3em \hangafter=1
$D^2= 89.569 = 
48+24\sqrt{3}$

\vskip 0.7ex
\hangindent=3em \hangafter=1
$T = ( 0,
\frac{1}{2},
0,
\frac{1}{3},
\frac{1}{3},
\frac{1}{3},
\frac{5}{6},
0,
\frac{1}{2},
\frac{3}{4} )
$,

\vskip 0.7ex
\hangindent=3em \hangafter=1
$S$ = ($ 1$,
$ 1$,
$ 1+\sqrt{3}$,
$ 1+\sqrt{3}$,
$ 1+\sqrt{3}$,
$ 1+\sqrt{3}$,
$ 1+\sqrt{3}$,
$ 2+\sqrt{3}$,
$ 2+\sqrt{3}$,
$ 3+\sqrt{3}$;\ \ 
$ 1$,
$ -1-\sqrt{3}$,
$ 1+\sqrt{3}$,
$ -1-\sqrt{3}$,
$ -1-\sqrt{3}$,
$ 1+\sqrt{3}$,
$ 2+\sqrt{3}$,
$ 2+\sqrt{3}$,
$ -3-\sqrt{3}$;\ \ 
$0$,
$ -2-2\sqrt{3}$,
$0$,
$0$,
$ 2+2\sqrt{3}$,
$ -1-\sqrt{3}$,
$ 1+\sqrt{3}$,
$0$;\ \ 
$ 1+\sqrt{3}$,
$ 1+\sqrt{3}$,
$ 1+\sqrt{3}$,
$ 1+\sqrt{3}$,
$ -1-\sqrt{3}$,
$ -1-\sqrt{3}$,
$0$;\ \ 
$(-3-\sqrt{3})\mathrm{i}$,
$(3+\sqrt{3})\mathrm{i}$,
$ -1-\sqrt{3}$,
$ -1-\sqrt{3}$,
$ 1+\sqrt{3}$,
$0$;\ \ 
$(-3-\sqrt{3})\mathrm{i}$,
$ -1-\sqrt{3}$,
$ -1-\sqrt{3}$,
$ 1+\sqrt{3}$,
$0$;\ \ 
$ 1+\sqrt{3}$,
$ -1-\sqrt{3}$,
$ -1-\sqrt{3}$,
$0$;\ \ 
$ 1$,
$ 1$,
$ 3+\sqrt{3}$;\ \ 
$ 1$,
$ -3-\sqrt{3}$;\ \ 
$0$)

\vskip 0.7ex
\hangindent=3em \hangafter=1
$\tau_n$ = ($0. - 9.46 i$, $0. - 25.86 i$, $22.39 + 22.39 i$, $44.78 + 25.86 i$, $0. - 35.32 i$, $44.78$, $0. + 35.32 i$, $44.78 - 25.86 i$, $22.39 - 22.39 i$, $0. + 25.86 i$, $0. + 9.46 i$, $89.57$)

\vskip 0.7ex
\hangindent=3em \hangafter=1
\textit{Intrinsic sign problem}

  \vskip 2ex

\noindent45. $10_{2,89.56}^{12,176}$ \irep{680}:\ \ 
$d_i$ = ($1.0$,
$1.0$,
$2.732$,
$2.732$,
$2.732$,
$2.732$,
$2.732$,
$3.732$,
$3.732$,
$4.732$) 

\vskip 0.7ex
\hangindent=3em \hangafter=1
$D^2= 89.569 = 
48+24\sqrt{3}$

\vskip 0.7ex
\hangindent=3em \hangafter=1
$T = ( 0,
\frac{1}{2},
0,
\frac{2}{3},
\frac{2}{3},
\frac{2}{3},
\frac{1}{6},
0,
\frac{1}{2},
\frac{1}{4} )
$,

\vskip 0.7ex
\hangindent=3em \hangafter=1
$S$ = ($ 1$,
$ 1$,
$ 1+\sqrt{3}$,
$ 1+\sqrt{3}$,
$ 1+\sqrt{3}$,
$ 1+\sqrt{3}$,
$ 1+\sqrt{3}$,
$ 2+\sqrt{3}$,
$ 2+\sqrt{3}$,
$ 3+\sqrt{3}$;\ \ 
$ 1$,
$ -1-\sqrt{3}$,
$ 1+\sqrt{3}$,
$ -1-\sqrt{3}$,
$ -1-\sqrt{3}$,
$ 1+\sqrt{3}$,
$ 2+\sqrt{3}$,
$ 2+\sqrt{3}$,
$ -3-\sqrt{3}$;\ \ 
$0$,
$ -2-2\sqrt{3}$,
$0$,
$0$,
$ 2+2\sqrt{3}$,
$ -1-\sqrt{3}$,
$ 1+\sqrt{3}$,
$0$;\ \ 
$ 1+\sqrt{3}$,
$ 1+\sqrt{3}$,
$ 1+\sqrt{3}$,
$ 1+\sqrt{3}$,
$ -1-\sqrt{3}$,
$ -1-\sqrt{3}$,
$0$;\ \ 
$(3+\sqrt{3})\mathrm{i}$,
$(-3-\sqrt{3})\mathrm{i}$,
$ -1-\sqrt{3}$,
$ -1-\sqrt{3}$,
$ 1+\sqrt{3}$,
$0$;\ \ 
$(3+\sqrt{3})\mathrm{i}$,
$ -1-\sqrt{3}$,
$ -1-\sqrt{3}$,
$ 1+\sqrt{3}$,
$0$;\ \ 
$ 1+\sqrt{3}$,
$ -1-\sqrt{3}$,
$ -1-\sqrt{3}$,
$0$;\ \ 
$ 1$,
$ 1$,
$ 3+\sqrt{3}$;\ \ 
$ 1$,
$ -3-\sqrt{3}$;\ \ 
$0$)

\vskip 0.7ex
\hangindent=3em \hangafter=1
$\tau_n$ = ($0. + 9.46 i$, $0. + 25.86 i$, $22.39 - 22.39 i$, $44.78 - 25.86 i$, $0. + 35.32 i$, $44.78$, $0. - 35.32 i$, $44.78 + 25.86 i$, $22.39 + 22.39 i$, $0. - 25.86 i$, $0. - 9.46 i$, $89.57$)

\vskip 0.7ex
\hangindent=3em \hangafter=1
\textit{Intrinsic sign problem}

  \vskip 2ex

\noindent46. $10_{0,89.56}^{12,155}$ \irep{587}:\ \ 
$d_i$ = ($1.0$,
$1.0$,
$2.732$,
$2.732$,
$2.732$,
$2.732$,
$2.732$,
$3.732$,
$3.732$,
$4.732$) 

\vskip 0.7ex
\hangindent=3em \hangafter=1
$D^2= 89.569 = 
48+24\sqrt{3}$

\vskip 0.7ex
\hangindent=3em \hangafter=1
$T = ( 0,
\frac{1}{2},
\frac{1}{3},
\frac{1}{4},
\frac{5}{6},
\frac{7}{12},
\frac{7}{12},
0,
\frac{1}{2},
0 )
$,

\vskip 0.7ex
\hangindent=3em \hangafter=1
$S$ = ($ 1$,
$ 1$,
$ 1+\sqrt{3}$,
$ 1+\sqrt{3}$,
$ 1+\sqrt{3}$,
$ 1+\sqrt{3}$,
$ 1+\sqrt{3}$,
$ 2+\sqrt{3}$,
$ 2+\sqrt{3}$,
$ 3+\sqrt{3}$;\ \ 
$ 1$,
$ 1+\sqrt{3}$,
$ -1-\sqrt{3}$,
$ 1+\sqrt{3}$,
$ -1-\sqrt{3}$,
$ -1-\sqrt{3}$,
$ 2+\sqrt{3}$,
$ 2+\sqrt{3}$,
$ -3-\sqrt{3}$;\ \ 
$ 1+\sqrt{3}$,
$ -2-2\sqrt{3}$,
$ 1+\sqrt{3}$,
$ 1+\sqrt{3}$,
$ 1+\sqrt{3}$,
$ -1-\sqrt{3}$,
$ -1-\sqrt{3}$,
$0$;\ \ 
$0$,
$ 2+2\sqrt{3}$,
$0$,
$0$,
$ -1-\sqrt{3}$,
$ 1+\sqrt{3}$,
$0$;\ \ 
$ 1+\sqrt{3}$,
$ -1-\sqrt{3}$,
$ -1-\sqrt{3}$,
$ -1-\sqrt{3}$,
$ -1-\sqrt{3}$,
$0$;\ \ 
$(3+\sqrt{3})\mathrm{i}$,
$(-3-\sqrt{3})\mathrm{i}$,
$ -1-\sqrt{3}$,
$ 1+\sqrt{3}$,
$0$;\ \ 
$(3+\sqrt{3})\mathrm{i}$,
$ -1-\sqrt{3}$,
$ 1+\sqrt{3}$,
$0$;\ \ 
$ 1$,
$ 1$,
$ 3+\sqrt{3}$;\ \ 
$ 1$,
$ -3-\sqrt{3}$;\ \ 
$0$)

\vskip 0.7ex
\hangindent=3em \hangafter=1
$\tau_n$ = ($9.46$, $44.78$, $22.39 - 22.39 i$, $44.78 + 25.86 i$, $35.32$, $44.78$, $35.32$, $44.78 - 25.86 i$, $22.39 + 22.39 i$, $44.78$, $9.46$, $89.57$)

\vskip 0.7ex
\hangindent=3em \hangafter=1
\textit{Intrinsic sign problem}

  \vskip 2ex

\noindent47. $10_{4,89.56}^{12,822}$ \irep{587}:\ \ 
$d_i$ = ($1.0$,
$1.0$,
$2.732$,
$2.732$,
$2.732$,
$2.732$,
$2.732$,
$3.732$,
$3.732$,
$4.732$) 

\vskip 0.7ex
\hangindent=3em \hangafter=1
$D^2= 89.569 = 
48+24\sqrt{3}$

\vskip 0.7ex
\hangindent=3em \hangafter=1
$T = ( 0,
\frac{1}{2},
\frac{1}{3},
\frac{3}{4},
\frac{5}{6},
\frac{1}{12},
\frac{1}{12},
0,
\frac{1}{2},
\frac{1}{2} )
$,

\vskip 0.7ex
\hangindent=3em \hangafter=1
$S$ = ($ 1$,
$ 1$,
$ 1+\sqrt{3}$,
$ 1+\sqrt{3}$,
$ 1+\sqrt{3}$,
$ 1+\sqrt{3}$,
$ 1+\sqrt{3}$,
$ 2+\sqrt{3}$,
$ 2+\sqrt{3}$,
$ 3+\sqrt{3}$;\ \ 
$ 1$,
$ 1+\sqrt{3}$,
$ -1-\sqrt{3}$,
$ 1+\sqrt{3}$,
$ -1-\sqrt{3}$,
$ -1-\sqrt{3}$,
$ 2+\sqrt{3}$,
$ 2+\sqrt{3}$,
$ -3-\sqrt{3}$;\ \ 
$ 1+\sqrt{3}$,
$ -2-2\sqrt{3}$,
$ 1+\sqrt{3}$,
$ 1+\sqrt{3}$,
$ 1+\sqrt{3}$,
$ -1-\sqrt{3}$,
$ -1-\sqrt{3}$,
$0$;\ \ 
$0$,
$ 2+2\sqrt{3}$,
$0$,
$0$,
$ -1-\sqrt{3}$,
$ 1+\sqrt{3}$,
$0$;\ \ 
$ 1+\sqrt{3}$,
$ -1-\sqrt{3}$,
$ -1-\sqrt{3}$,
$ -1-\sqrt{3}$,
$ -1-\sqrt{3}$,
$0$;\ \ 
$(3+\sqrt{3})\mathrm{i}$,
$(-3-\sqrt{3})\mathrm{i}$,
$ -1-\sqrt{3}$,
$ 1+\sqrt{3}$,
$0$;\ \ 
$(3+\sqrt{3})\mathrm{i}$,
$ -1-\sqrt{3}$,
$ 1+\sqrt{3}$,
$0$;\ \ 
$ 1$,
$ 1$,
$ 3+\sqrt{3}$;\ \ 
$ 1$,
$ -3-\sqrt{3}$;\ \ 
$0$)

\vskip 0.7ex
\hangindent=3em \hangafter=1
$\tau_n$ = ($-9.46$, $44.78$, $-22.39 + 22.39 i$, $44.78 + 25.86 i$, $-35.32$, $44.78$, $-35.32$, $44.78 - 25.86 i$, $-22.39 - 22.39 i$, $44.78$, $-9.46$, $89.57$)

\vskip 0.7ex
\hangindent=3em \hangafter=1
\textit{Intrinsic sign problem}

  \vskip 2ex

\noindent48. $10_{2,89.56}^{12,119}$ \irep{680}:\ \ 
$d_i$ = ($1.0$,
$1.0$,
$2.732$,
$2.732$,
$2.732$,
$2.732$,
$2.732$,
$3.732$,
$3.732$,
$4.732$) 

\vskip 0.7ex
\hangindent=3em \hangafter=1
$D^2= 89.569 = 
48+24\sqrt{3}$

\vskip 0.7ex
\hangindent=3em \hangafter=1
$T = ( 0,
\frac{1}{2},
\frac{1}{2},
\frac{1}{3},
\frac{5}{6},
\frac{5}{6},
\frac{5}{6},
0,
\frac{1}{2},
\frac{1}{4} )
$,

\vskip 0.7ex
\hangindent=3em \hangafter=1
$S$ = ($ 1$,
$ 1$,
$ 1+\sqrt{3}$,
$ 1+\sqrt{3}$,
$ 1+\sqrt{3}$,
$ 1+\sqrt{3}$,
$ 1+\sqrt{3}$,
$ 2+\sqrt{3}$,
$ 2+\sqrt{3}$,
$ 3+\sqrt{3}$;\ \ 
$ 1$,
$ -1-\sqrt{3}$,
$ 1+\sqrt{3}$,
$ 1+\sqrt{3}$,
$ -1-\sqrt{3}$,
$ -1-\sqrt{3}$,
$ 2+\sqrt{3}$,
$ 2+\sqrt{3}$,
$ -3-\sqrt{3}$;\ \ 
$0$,
$ -2-2\sqrt{3}$,
$ 2+2\sqrt{3}$,
$0$,
$0$,
$ -1-\sqrt{3}$,
$ 1+\sqrt{3}$,
$0$;\ \ 
$ 1+\sqrt{3}$,
$ 1+\sqrt{3}$,
$ 1+\sqrt{3}$,
$ 1+\sqrt{3}$,
$ -1-\sqrt{3}$,
$ -1-\sqrt{3}$,
$0$;\ \ 
$ 1+\sqrt{3}$,
$ -1-\sqrt{3}$,
$ -1-\sqrt{3}$,
$ -1-\sqrt{3}$,
$ -1-\sqrt{3}$,
$0$;\ \ 
$(-3-\sqrt{3})\mathrm{i}$,
$(3+\sqrt{3})\mathrm{i}$,
$ -1-\sqrt{3}$,
$ 1+\sqrt{3}$,
$0$;\ \ 
$(-3-\sqrt{3})\mathrm{i}$,
$ -1-\sqrt{3}$,
$ 1+\sqrt{3}$,
$0$;\ \ 
$ 1$,
$ 1$,
$ 3+\sqrt{3}$;\ \ 
$ 1$,
$ -3-\sqrt{3}$;\ \ 
$0$)

\vskip 0.7ex
\hangindent=3em \hangafter=1
$\tau_n$ = ($0. + 9.46 i$, $0. - 25.86 i$, $-22.39 - 22.39 i$, $44.78 + 25.86 i$, $0. + 35.32 i$, $44.78$, $0. - 35.32 i$, $44.78 - 25.86 i$, $-22.39 + 22.39 i$, $0. + 25.86 i$, $0. - 9.46 i$, $89.57$)

\vskip 0.7ex
\hangindent=3em \hangafter=1
\textit{Intrinsic sign problem}

  \vskip 2ex

\noindent49. $10_{6,89.56}^{12,994}$ \irep{680}:\ \ 
$d_i$ = ($1.0$,
$1.0$,
$2.732$,
$2.732$,
$2.732$,
$2.732$,
$2.732$,
$3.732$,
$3.732$,
$4.732$) 

\vskip 0.7ex
\hangindent=3em \hangafter=1
$D^2= 89.569 = 
48+24\sqrt{3}$

\vskip 0.7ex
\hangindent=3em \hangafter=1
$T = ( 0,
\frac{1}{2},
\frac{1}{2},
\frac{2}{3},
\frac{1}{6},
\frac{1}{6},
\frac{1}{6},
0,
\frac{1}{2},
\frac{3}{4} )
$,

\vskip 0.7ex
\hangindent=3em \hangafter=1
$S$ = ($ 1$,
$ 1$,
$ 1+\sqrt{3}$,
$ 1+\sqrt{3}$,
$ 1+\sqrt{3}$,
$ 1+\sqrt{3}$,
$ 1+\sqrt{3}$,
$ 2+\sqrt{3}$,
$ 2+\sqrt{3}$,
$ 3+\sqrt{3}$;\ \ 
$ 1$,
$ -1-\sqrt{3}$,
$ 1+\sqrt{3}$,
$ 1+\sqrt{3}$,
$ -1-\sqrt{3}$,
$ -1-\sqrt{3}$,
$ 2+\sqrt{3}$,
$ 2+\sqrt{3}$,
$ -3-\sqrt{3}$;\ \ 
$0$,
$ -2-2\sqrt{3}$,
$ 2+2\sqrt{3}$,
$0$,
$0$,
$ -1-\sqrt{3}$,
$ 1+\sqrt{3}$,
$0$;\ \ 
$ 1+\sqrt{3}$,
$ 1+\sqrt{3}$,
$ 1+\sqrt{3}$,
$ 1+\sqrt{3}$,
$ -1-\sqrt{3}$,
$ -1-\sqrt{3}$,
$0$;\ \ 
$ 1+\sqrt{3}$,
$ -1-\sqrt{3}$,
$ -1-\sqrt{3}$,
$ -1-\sqrt{3}$,
$ -1-\sqrt{3}$,
$0$;\ \ 
$(3+\sqrt{3})\mathrm{i}$,
$(-3-\sqrt{3})\mathrm{i}$,
$ -1-\sqrt{3}$,
$ 1+\sqrt{3}$,
$0$;\ \ 
$(3+\sqrt{3})\mathrm{i}$,
$ -1-\sqrt{3}$,
$ 1+\sqrt{3}$,
$0$;\ \ 
$ 1$,
$ 1$,
$ 3+\sqrt{3}$;\ \ 
$ 1$,
$ -3-\sqrt{3}$;\ \ 
$0$)

\vskip 0.7ex
\hangindent=3em \hangafter=1
$\tau_n$ = ($0. - 9.46 i$, $0. + 25.86 i$, $-22.39 + 22.39 i$, $44.78 - 25.86 i$, $0. - 35.32 i$, $44.78$, $0. + 35.32 i$, $44.78 + 25.86 i$, $-22.39 - 22.39 i$, $0. - 25.86 i$, $0. + 9.46 i$, $89.57$)

\vskip 0.7ex
\hangindent=3em \hangafter=1
\textit{Intrinsic sign problem}

  \vskip 2ex

\noindent50. $10_{4,89.56}^{12,145}$ \irep{587}:\ \ 
$d_i$ = ($1.0$,
$1.0$,
$2.732$,
$2.732$,
$2.732$,
$2.732$,
$2.732$,
$3.732$,
$3.732$,
$4.732$) 

\vskip 0.7ex
\hangindent=3em \hangafter=1
$D^2= 89.569 = 
48+24\sqrt{3}$

\vskip 0.7ex
\hangindent=3em \hangafter=1
$T = ( 0,
\frac{1}{2},
\frac{2}{3},
\frac{1}{4},
\frac{1}{6},
\frac{11}{12},
\frac{11}{12},
0,
\frac{1}{2},
\frac{1}{2} )
$,

\vskip 0.7ex
\hangindent=3em \hangafter=1
$S$ = ($ 1$,
$ 1$,
$ 1+\sqrt{3}$,
$ 1+\sqrt{3}$,
$ 1+\sqrt{3}$,
$ 1+\sqrt{3}$,
$ 1+\sqrt{3}$,
$ 2+\sqrt{3}$,
$ 2+\sqrt{3}$,
$ 3+\sqrt{3}$;\ \ 
$ 1$,
$ 1+\sqrt{3}$,
$ -1-\sqrt{3}$,
$ 1+\sqrt{3}$,
$ -1-\sqrt{3}$,
$ -1-\sqrt{3}$,
$ 2+\sqrt{3}$,
$ 2+\sqrt{3}$,
$ -3-\sqrt{3}$;\ \ 
$ 1+\sqrt{3}$,
$ -2-2\sqrt{3}$,
$ 1+\sqrt{3}$,
$ 1+\sqrt{3}$,
$ 1+\sqrt{3}$,
$ -1-\sqrt{3}$,
$ -1-\sqrt{3}$,
$0$;\ \ 
$0$,
$ 2+2\sqrt{3}$,
$0$,
$0$,
$ -1-\sqrt{3}$,
$ 1+\sqrt{3}$,
$0$;\ \ 
$ 1+\sqrt{3}$,
$ -1-\sqrt{3}$,
$ -1-\sqrt{3}$,
$ -1-\sqrt{3}$,
$ -1-\sqrt{3}$,
$0$;\ \ 
$(-3-\sqrt{3})\mathrm{i}$,
$(3+\sqrt{3})\mathrm{i}$,
$ -1-\sqrt{3}$,
$ 1+\sqrt{3}$,
$0$;\ \ 
$(-3-\sqrt{3})\mathrm{i}$,
$ -1-\sqrt{3}$,
$ 1+\sqrt{3}$,
$0$;\ \ 
$ 1$,
$ 1$,
$ 3+\sqrt{3}$;\ \ 
$ 1$,
$ -3-\sqrt{3}$;\ \ 
$0$)

\vskip 0.7ex
\hangindent=3em \hangafter=1
$\tau_n$ = ($-9.46$, $44.78$, $-22.39 - 22.39 i$, $44.78 - 25.86 i$, $-35.32$, $44.78$, $-35.32$, $44.78 + 25.86 i$, $-22.39 + 22.39 i$, $44.78$, $-9.46$, $89.57$)

\vskip 0.7ex
\hangindent=3em \hangafter=1
\textit{Intrinsic sign problem}

  \vskip 2ex

\noindent51. $10_{0,89.56}^{12,200}$ \irep{587}:\ \ 
$d_i$ = ($1.0$,
$1.0$,
$2.732$,
$2.732$,
$2.732$,
$2.732$,
$2.732$,
$3.732$,
$3.732$,
$4.732$) 

\vskip 0.7ex
\hangindent=3em \hangafter=1
$D^2= 89.569 = 
48+24\sqrt{3}$

\vskip 0.7ex
\hangindent=3em \hangafter=1
$T = ( 0,
\frac{1}{2},
\frac{2}{3},
\frac{3}{4},
\frac{1}{6},
\frac{5}{12},
\frac{5}{12},
0,
\frac{1}{2},
0 )
$,

\vskip 0.7ex
\hangindent=3em \hangafter=1
$S$ = ($ 1$,
$ 1$,
$ 1+\sqrt{3}$,
$ 1+\sqrt{3}$,
$ 1+\sqrt{3}$,
$ 1+\sqrt{3}$,
$ 1+\sqrt{3}$,
$ 2+\sqrt{3}$,
$ 2+\sqrt{3}$,
$ 3+\sqrt{3}$;\ \ 
$ 1$,
$ 1+\sqrt{3}$,
$ -1-\sqrt{3}$,
$ 1+\sqrt{3}$,
$ -1-\sqrt{3}$,
$ -1-\sqrt{3}$,
$ 2+\sqrt{3}$,
$ 2+\sqrt{3}$,
$ -3-\sqrt{3}$;\ \ 
$ 1+\sqrt{3}$,
$ -2-2\sqrt{3}$,
$ 1+\sqrt{3}$,
$ 1+\sqrt{3}$,
$ 1+\sqrt{3}$,
$ -1-\sqrt{3}$,
$ -1-\sqrt{3}$,
$0$;\ \ 
$0$,
$ 2+2\sqrt{3}$,
$0$,
$0$,
$ -1-\sqrt{3}$,
$ 1+\sqrt{3}$,
$0$;\ \ 
$ 1+\sqrt{3}$,
$ -1-\sqrt{3}$,
$ -1-\sqrt{3}$,
$ -1-\sqrt{3}$,
$ -1-\sqrt{3}$,
$0$;\ \ 
$(-3-\sqrt{3})\mathrm{i}$,
$(3+\sqrt{3})\mathrm{i}$,
$ -1-\sqrt{3}$,
$ 1+\sqrt{3}$,
$0$;\ \ 
$(-3-\sqrt{3})\mathrm{i}$,
$ -1-\sqrt{3}$,
$ 1+\sqrt{3}$,
$0$;\ \ 
$ 1$,
$ 1$,
$ 3+\sqrt{3}$;\ \ 
$ 1$,
$ -3-\sqrt{3}$;\ \ 
$0$)

\vskip 0.7ex
\hangindent=3em \hangafter=1
$\tau_n$ = ($9.46$, $44.78$, $22.39 + 22.39 i$, $44.78 - 25.86 i$, $35.32$, $44.78$, $35.32$, $44.78 + 25.86 i$, $22.39 - 22.39 i$, $44.78$, $9.46$, $89.57$)

\vskip 0.7ex
\hangindent=3em \hangafter=1
\textit{Intrinsic sign problem}

  \vskip 2ex

\noindent52. $10_{7,89.56}^{24,123}$ \irep{978}:\ \ 
$d_i$ = ($1.0$,
$1.0$,
$2.732$,
$2.732$,
$2.732$,
$2.732$,
$2.732$,
$3.732$,
$3.732$,
$4.732$) 

\vskip 0.7ex
\hangindent=3em \hangafter=1
$D^2= 89.569 = 
48+24\sqrt{3}$

\vskip 0.7ex
\hangindent=3em \hangafter=1
$T = ( 0,
\frac{1}{2},
\frac{1}{3},
\frac{5}{6},
\frac{1}{8},
\frac{11}{24},
\frac{11}{24},
0,
\frac{1}{2},
\frac{7}{8} )
$,

\vskip 0.7ex
\hangindent=3em \hangafter=1
$S$ = ($ 1$,
$ 1$,
$ 1+\sqrt{3}$,
$ 1+\sqrt{3}$,
$ 1+\sqrt{3}$,
$ 1+\sqrt{3}$,
$ 1+\sqrt{3}$,
$ 2+\sqrt{3}$,
$ 2+\sqrt{3}$,
$ 3+\sqrt{3}$;\ \ 
$ 1$,
$ 1+\sqrt{3}$,
$ 1+\sqrt{3}$,
$ -1-\sqrt{3}$,
$ -1-\sqrt{3}$,
$ -1-\sqrt{3}$,
$ 2+\sqrt{3}$,
$ 2+\sqrt{3}$,
$ -3-\sqrt{3}$;\ \ 
$ 1+\sqrt{3}$,
$ 1+\sqrt{3}$,
$ -2-2\sqrt{3}$,
$ 1+\sqrt{3}$,
$ 1+\sqrt{3}$,
$ -1-\sqrt{3}$,
$ -1-\sqrt{3}$,
$0$;\ \ 
$ 1+\sqrt{3}$,
$ 2+2\sqrt{3}$,
$ -1-\sqrt{3}$,
$ -1-\sqrt{3}$,
$ -1-\sqrt{3}$,
$ -1-\sqrt{3}$,
$0$;\ \ 
$0$,
$0$,
$0$,
$ -1-\sqrt{3}$,
$ 1+\sqrt{3}$,
$0$;\ \ 
$ -3-\sqrt{3}$,
$ 3+\sqrt{3}$,
$ -1-\sqrt{3}$,
$ 1+\sqrt{3}$,
$0$;\ \ 
$ -3-\sqrt{3}$,
$ -1-\sqrt{3}$,
$ 1+\sqrt{3}$,
$0$;\ \ 
$ 1$,
$ 1$,
$ 3+\sqrt{3}$;\ \ 
$ 1$,
$ -3-\sqrt{3}$;\ \ 
$0$)

\vskip 0.7ex
\hangindent=3em \hangafter=1
$\tau_n$ = ($6.69 - 6.69 i$, $35.32 - 35.32 i$, $-31.67$, $0.$, $-24.97 + 24.97 i$, $44.78$, $24.97 + 24.97 i$, $44.78 - 25.86 i$, $31.67$, $9.46 - 9.46 i$, $-6.69 - 6.69 i$, $0.$, $-6.69 + 6.69 i$, $9.46 + 9.46 i$, $31.67$, $44.78 + 25.86 i$, $24.97 - 24.97 i$, $44.78$, $-24.97 - 24.97 i$, $0.$, $-31.67$, $35.32 + 35.32 i$, $6.69 + 6.69 i$, $89.57$)

\vskip 0.7ex
\hangindent=3em \hangafter=1
\textit{Intrinsic sign problem}

  \vskip 2ex

\noindent53. $10_{1,89.56}^{24,380}$ \irep{977}:\ \ 
$d_i$ = ($1.0$,
$1.0$,
$2.732$,
$2.732$,
$2.732$,
$2.732$,
$2.732$,
$3.732$,
$3.732$,
$4.732$) 

\vskip 0.7ex
\hangindent=3em \hangafter=1
$D^2= 89.569 = 
48+24\sqrt{3}$

\vskip 0.7ex
\hangindent=3em \hangafter=1
$T = ( 0,
\frac{1}{2},
\frac{1}{3},
\frac{5}{6},
\frac{3}{8},
\frac{17}{24},
\frac{17}{24},
0,
\frac{1}{2},
\frac{1}{8} )
$,

\vskip 0.7ex
\hangindent=3em \hangafter=1
$S$ = ($ 1$,
$ 1$,
$ 1+\sqrt{3}$,
$ 1+\sqrt{3}$,
$ 1+\sqrt{3}$,
$ 1+\sqrt{3}$,
$ 1+\sqrt{3}$,
$ 2+\sqrt{3}$,
$ 2+\sqrt{3}$,
$ 3+\sqrt{3}$;\ \ 
$ 1$,
$ 1+\sqrt{3}$,
$ 1+\sqrt{3}$,
$ -1-\sqrt{3}$,
$ -1-\sqrt{3}$,
$ -1-\sqrt{3}$,
$ 2+\sqrt{3}$,
$ 2+\sqrt{3}$,
$ -3-\sqrt{3}$;\ \ 
$ 1+\sqrt{3}$,
$ 1+\sqrt{3}$,
$ -2-2\sqrt{3}$,
$ 1+\sqrt{3}$,
$ 1+\sqrt{3}$,
$ -1-\sqrt{3}$,
$ -1-\sqrt{3}$,
$0$;\ \ 
$ 1+\sqrt{3}$,
$ 2+2\sqrt{3}$,
$ -1-\sqrt{3}$,
$ -1-\sqrt{3}$,
$ -1-\sqrt{3}$,
$ -1-\sqrt{3}$,
$0$;\ \ 
$0$,
$0$,
$0$,
$ -1-\sqrt{3}$,
$ 1+\sqrt{3}$,
$0$;\ \ 
$ 3+\sqrt{3}$,
$ -3-\sqrt{3}$,
$ -1-\sqrt{3}$,
$ 1+\sqrt{3}$,
$0$;\ \ 
$ 3+\sqrt{3}$,
$ -1-\sqrt{3}$,
$ 1+\sqrt{3}$,
$0$;\ \ 
$ 1$,
$ 1$,
$ 3+\sqrt{3}$;\ \ 
$ 1$,
$ -3-\sqrt{3}$;\ \ 
$0$)

\vskip 0.7ex
\hangindent=3em \hangafter=1
$\tau_n$ = ($6.69 + 6.69 i$, $9.46 + 9.46 i$, $0. + 31.67 i$, $0.$, $-24.97 - 24.97 i$, $44.78$, $24.97 - 24.97 i$, $44.78 - 25.86 i$, $0. + 31.67 i$, $35.32 + 35.32 i$, $-6.69 + 6.69 i$, $0.$, $-6.69 - 6.69 i$, $35.32 - 35.32 i$, $0. - 31.67 i$, $44.78 + 25.86 i$, $24.97 + 24.97 i$, $44.78$, $-24.97 + 24.97 i$, $0.$, $0. - 31.67 i$, $9.46 - 9.46 i$, $6.69 - 6.69 i$, $89.57$)

\vskip 0.7ex
\hangindent=3em \hangafter=1
\textit{Intrinsic sign problem}

  \vskip 2ex

\noindent54. $10_{3,89.56}^{24,317}$ \irep{978}:\ \ 
$d_i$ = ($1.0$,
$1.0$,
$2.732$,
$2.732$,
$2.732$,
$2.732$,
$2.732$,
$3.732$,
$3.732$,
$4.732$) 

\vskip 0.7ex
\hangindent=3em \hangafter=1
$D^2= 89.569 = 
48+24\sqrt{3}$

\vskip 0.7ex
\hangindent=3em \hangafter=1
$T = ( 0,
\frac{1}{2},
\frac{1}{3},
\frac{5}{6},
\frac{5}{8},
\frac{23}{24},
\frac{23}{24},
0,
\frac{1}{2},
\frac{3}{8} )
$,

\vskip 0.7ex
\hangindent=3em \hangafter=1
$S$ = ($ 1$,
$ 1$,
$ 1+\sqrt{3}$,
$ 1+\sqrt{3}$,
$ 1+\sqrt{3}$,
$ 1+\sqrt{3}$,
$ 1+\sqrt{3}$,
$ 2+\sqrt{3}$,
$ 2+\sqrt{3}$,
$ 3+\sqrt{3}$;\ \ 
$ 1$,
$ 1+\sqrt{3}$,
$ 1+\sqrt{3}$,
$ -1-\sqrt{3}$,
$ -1-\sqrt{3}$,
$ -1-\sqrt{3}$,
$ 2+\sqrt{3}$,
$ 2+\sqrt{3}$,
$ -3-\sqrt{3}$;\ \ 
$ 1+\sqrt{3}$,
$ 1+\sqrt{3}$,
$ -2-2\sqrt{3}$,
$ 1+\sqrt{3}$,
$ 1+\sqrt{3}$,
$ -1-\sqrt{3}$,
$ -1-\sqrt{3}$,
$0$;\ \ 
$ 1+\sqrt{3}$,
$ 2+2\sqrt{3}$,
$ -1-\sqrt{3}$,
$ -1-\sqrt{3}$,
$ -1-\sqrt{3}$,
$ -1-\sqrt{3}$,
$0$;\ \ 
$0$,
$0$,
$0$,
$ -1-\sqrt{3}$,
$ 1+\sqrt{3}$,
$0$;\ \ 
$ -3-\sqrt{3}$,
$ 3+\sqrt{3}$,
$ -1-\sqrt{3}$,
$ 1+\sqrt{3}$,
$0$;\ \ 
$ -3-\sqrt{3}$,
$ -1-\sqrt{3}$,
$ 1+\sqrt{3}$,
$0$;\ \ 
$ 1$,
$ 1$,
$ 3+\sqrt{3}$;\ \ 
$ 1$,
$ -3-\sqrt{3}$;\ \ 
$0$)

\vskip 0.7ex
\hangindent=3em \hangafter=1
$\tau_n$ = ($-6.69 + 6.69 i$, $35.32 - 35.32 i$, $31.67$, $0.$, $24.97 - 24.97 i$, $44.78$, $-24.97 - 24.97 i$, $44.78 - 25.86 i$, $-31.67$, $9.46 - 9.46 i$, $6.69 + 6.69 i$, $0.$, $6.69 - 6.69 i$, $9.46 + 9.46 i$, $-31.67$, $44.78 + 25.86 i$, $-24.97 + 24.97 i$, $44.78$, $24.97 + 24.97 i$, $0.$, $31.67$, $35.32 + 35.32 i$, $-6.69 - 6.69 i$, $89.57$)

\vskip 0.7ex
\hangindent=3em \hangafter=1
\textit{Intrinsic sign problem}

  \vskip 2ex

\noindent55. $10_{5,89.56}^{24,597}$ \irep{977}:\ \ 
$d_i$ = ($1.0$,
$1.0$,
$2.732$,
$2.732$,
$2.732$,
$2.732$,
$2.732$,
$3.732$,
$3.732$,
$4.732$) 

\vskip 0.7ex
\hangindent=3em \hangafter=1
$D^2= 89.569 = 
48+24\sqrt{3}$

\vskip 0.7ex
\hangindent=3em \hangafter=1
$T = ( 0,
\frac{1}{2},
\frac{1}{3},
\frac{5}{6},
\frac{7}{8},
\frac{5}{24},
\frac{5}{24},
0,
\frac{1}{2},
\frac{5}{8} )
$,

\vskip 0.7ex
\hangindent=3em \hangafter=1
$S$ = ($ 1$,
$ 1$,
$ 1+\sqrt{3}$,
$ 1+\sqrt{3}$,
$ 1+\sqrt{3}$,
$ 1+\sqrt{3}$,
$ 1+\sqrt{3}$,
$ 2+\sqrt{3}$,
$ 2+\sqrt{3}$,
$ 3+\sqrt{3}$;\ \ 
$ 1$,
$ 1+\sqrt{3}$,
$ 1+\sqrt{3}$,
$ -1-\sqrt{3}$,
$ -1-\sqrt{3}$,
$ -1-\sqrt{3}$,
$ 2+\sqrt{3}$,
$ 2+\sqrt{3}$,
$ -3-\sqrt{3}$;\ \ 
$ 1+\sqrt{3}$,
$ 1+\sqrt{3}$,
$ -2-2\sqrt{3}$,
$ 1+\sqrt{3}$,
$ 1+\sqrt{3}$,
$ -1-\sqrt{3}$,
$ -1-\sqrt{3}$,
$0$;\ \ 
$ 1+\sqrt{3}$,
$ 2+2\sqrt{3}$,
$ -1-\sqrt{3}$,
$ -1-\sqrt{3}$,
$ -1-\sqrt{3}$,
$ -1-\sqrt{3}$,
$0$;\ \ 
$0$,
$0$,
$0$,
$ -1-\sqrt{3}$,
$ 1+\sqrt{3}$,
$0$;\ \ 
$ 3+\sqrt{3}$,
$ -3-\sqrt{3}$,
$ -1-\sqrt{3}$,
$ 1+\sqrt{3}$,
$0$;\ \ 
$ 3+\sqrt{3}$,
$ -1-\sqrt{3}$,
$ 1+\sqrt{3}$,
$0$;\ \ 
$ 1$,
$ 1$,
$ 3+\sqrt{3}$;\ \ 
$ 1$,
$ -3-\sqrt{3}$;\ \ 
$0$)

\vskip 0.7ex
\hangindent=3em \hangafter=1
$\tau_n$ = ($-6.69 - 6.69 i$, $9.46 + 9.46 i$, $0. - 31.67 i$, $0.$, $24.97 + 24.97 i$, $44.78$, $-24.97 + 24.97 i$, $44.78 - 25.86 i$, $0. - 31.67 i$, $35.32 + 35.32 i$, $6.69 - 6.69 i$, $0.$, $6.69 + 6.69 i$, $35.32 - 35.32 i$, $0. + 31.67 i$, $44.78 + 25.86 i$, $-24.97 - 24.97 i$, $44.78$, $24.97 - 24.97 i$, $0.$, $0. + 31.67 i$, $9.46 - 9.46 i$, $-6.69 + 6.69 i$, $89.57$)

\vskip 0.7ex
\hangindent=3em \hangafter=1
\textit{Intrinsic sign problem}

  \vskip 2ex

\noindent56. $10_{3,89.56}^{24,358}$ \irep{977}:\ \ 
$d_i$ = ($1.0$,
$1.0$,
$2.732$,
$2.732$,
$2.732$,
$2.732$,
$2.732$,
$3.732$,
$3.732$,
$4.732$) 

\vskip 0.7ex
\hangindent=3em \hangafter=1
$D^2= 89.569 = 
48+24\sqrt{3}$

\vskip 0.7ex
\hangindent=3em \hangafter=1
$T = ( 0,
\frac{1}{2},
\frac{2}{3},
\frac{1}{6},
\frac{1}{8},
\frac{19}{24},
\frac{19}{24},
0,
\frac{1}{2},
\frac{3}{8} )
$,

\vskip 0.7ex
\hangindent=3em \hangafter=1
$S$ = ($ 1$,
$ 1$,
$ 1+\sqrt{3}$,
$ 1+\sqrt{3}$,
$ 1+\sqrt{3}$,
$ 1+\sqrt{3}$,
$ 1+\sqrt{3}$,
$ 2+\sqrt{3}$,
$ 2+\sqrt{3}$,
$ 3+\sqrt{3}$;\ \ 
$ 1$,
$ 1+\sqrt{3}$,
$ 1+\sqrt{3}$,
$ -1-\sqrt{3}$,
$ -1-\sqrt{3}$,
$ -1-\sqrt{3}$,
$ 2+\sqrt{3}$,
$ 2+\sqrt{3}$,
$ -3-\sqrt{3}$;\ \ 
$ 1+\sqrt{3}$,
$ 1+\sqrt{3}$,
$ -2-2\sqrt{3}$,
$ 1+\sqrt{3}$,
$ 1+\sqrt{3}$,
$ -1-\sqrt{3}$,
$ -1-\sqrt{3}$,
$0$;\ \ 
$ 1+\sqrt{3}$,
$ 2+2\sqrt{3}$,
$ -1-\sqrt{3}$,
$ -1-\sqrt{3}$,
$ -1-\sqrt{3}$,
$ -1-\sqrt{3}$,
$0$;\ \ 
$0$,
$0$,
$0$,
$ -1-\sqrt{3}$,
$ 1+\sqrt{3}$,
$0$;\ \ 
$ 3+\sqrt{3}$,
$ -3-\sqrt{3}$,
$ -1-\sqrt{3}$,
$ 1+\sqrt{3}$,
$0$;\ \ 
$ 3+\sqrt{3}$,
$ -1-\sqrt{3}$,
$ 1+\sqrt{3}$,
$0$;\ \ 
$ 1$,
$ 1$,
$ 3+\sqrt{3}$;\ \ 
$ 1$,
$ -3-\sqrt{3}$;\ \ 
$0$)

\vskip 0.7ex
\hangindent=3em \hangafter=1
$\tau_n$ = ($-6.69 + 6.69 i$, $9.46 - 9.46 i$, $0. + 31.67 i$, $0.$, $24.97 - 24.97 i$, $44.78$, $-24.97 - 24.97 i$, $44.78 + 25.86 i$, $0. + 31.67 i$, $35.32 - 35.32 i$, $6.69 + 6.69 i$, $0.$, $6.69 - 6.69 i$, $35.32 + 35.32 i$, $0. - 31.67 i$, $44.78 - 25.86 i$, $-24.97 + 24.97 i$, $44.78$, $24.97 + 24.97 i$, $0.$, $0. - 31.67 i$, $9.46 + 9.46 i$, $-6.69 - 6.69 i$, $89.57$)

\vskip 0.7ex
\hangindent=3em \hangafter=1
\textit{Intrinsic sign problem}

  \vskip 2ex

\noindent57. $10_{5,89.56}^{24,224}$ \irep{978}:\ \ 
$d_i$ = ($1.0$,
$1.0$,
$2.732$,
$2.732$,
$2.732$,
$2.732$,
$2.732$,
$3.732$,
$3.732$,
$4.732$) 

\vskip 0.7ex
\hangindent=3em \hangafter=1
$D^2= 89.569 = 
48+24\sqrt{3}$

\vskip 0.7ex
\hangindent=3em \hangafter=1
$T = ( 0,
\frac{1}{2},
\frac{2}{3},
\frac{1}{6},
\frac{3}{8},
\frac{1}{24},
\frac{1}{24},
0,
\frac{1}{2},
\frac{5}{8} )
$,

\vskip 0.7ex
\hangindent=3em \hangafter=1
$S$ = ($ 1$,
$ 1$,
$ 1+\sqrt{3}$,
$ 1+\sqrt{3}$,
$ 1+\sqrt{3}$,
$ 1+\sqrt{3}$,
$ 1+\sqrt{3}$,
$ 2+\sqrt{3}$,
$ 2+\sqrt{3}$,
$ 3+\sqrt{3}$;\ \ 
$ 1$,
$ 1+\sqrt{3}$,
$ 1+\sqrt{3}$,
$ -1-\sqrt{3}$,
$ -1-\sqrt{3}$,
$ -1-\sqrt{3}$,
$ 2+\sqrt{3}$,
$ 2+\sqrt{3}$,
$ -3-\sqrt{3}$;\ \ 
$ 1+\sqrt{3}$,
$ 1+\sqrt{3}$,
$ -2-2\sqrt{3}$,
$ 1+\sqrt{3}$,
$ 1+\sqrt{3}$,
$ -1-\sqrt{3}$,
$ -1-\sqrt{3}$,
$0$;\ \ 
$ 1+\sqrt{3}$,
$ 2+2\sqrt{3}$,
$ -1-\sqrt{3}$,
$ -1-\sqrt{3}$,
$ -1-\sqrt{3}$,
$ -1-\sqrt{3}$,
$0$;\ \ 
$0$,
$0$,
$0$,
$ -1-\sqrt{3}$,
$ 1+\sqrt{3}$,
$0$;\ \ 
$ -3-\sqrt{3}$,
$ 3+\sqrt{3}$,
$ -1-\sqrt{3}$,
$ 1+\sqrt{3}$,
$0$;\ \ 
$ -3-\sqrt{3}$,
$ -1-\sqrt{3}$,
$ 1+\sqrt{3}$,
$0$;\ \ 
$ 1$,
$ 1$,
$ 3+\sqrt{3}$;\ \ 
$ 1$,
$ -3-\sqrt{3}$;\ \ 
$0$)

\vskip 0.7ex
\hangindent=3em \hangafter=1
$\tau_n$ = ($-6.69 - 6.69 i$, $35.32 + 35.32 i$, $31.67$, $0.$, $24.97 + 24.97 i$, $44.78$, $-24.97 + 24.97 i$, $44.78 + 25.86 i$, $-31.67$, $9.46 + 9.46 i$, $6.69 - 6.69 i$, $0.$, $6.69 + 6.69 i$, $9.46 - 9.46 i$, $-31.67$, $44.78 - 25.86 i$, $-24.97 - 24.97 i$, $44.78$, $24.97 - 24.97 i$, $0.$, $31.67$, $35.32 - 35.32 i$, $-6.69 + 6.69 i$, $89.57$)

\vskip 0.7ex
\hangindent=3em \hangafter=1
\textit{Intrinsic sign problem}

  \vskip 2ex

\noindent58. $10_{7,89.56}^{24,664}$ \irep{977}:\ \ 
$d_i$ = ($1.0$,
$1.0$,
$2.732$,
$2.732$,
$2.732$,
$2.732$,
$2.732$,
$3.732$,
$3.732$,
$4.732$) 

\vskip 0.7ex
\hangindent=3em \hangafter=1
$D^2= 89.569 = 
48+24\sqrt{3}$

\vskip 0.7ex
\hangindent=3em \hangafter=1
$T = ( 0,
\frac{1}{2},
\frac{2}{3},
\frac{1}{6},
\frac{5}{8},
\frac{7}{24},
\frac{7}{24},
0,
\frac{1}{2},
\frac{7}{8} )
$,

\vskip 0.7ex
\hangindent=3em \hangafter=1
$S$ = ($ 1$,
$ 1$,
$ 1+\sqrt{3}$,
$ 1+\sqrt{3}$,
$ 1+\sqrt{3}$,
$ 1+\sqrt{3}$,
$ 1+\sqrt{3}$,
$ 2+\sqrt{3}$,
$ 2+\sqrt{3}$,
$ 3+\sqrt{3}$;\ \ 
$ 1$,
$ 1+\sqrt{3}$,
$ 1+\sqrt{3}$,
$ -1-\sqrt{3}$,
$ -1-\sqrt{3}$,
$ -1-\sqrt{3}$,
$ 2+\sqrt{3}$,
$ 2+\sqrt{3}$,
$ -3-\sqrt{3}$;\ \ 
$ 1+\sqrt{3}$,
$ 1+\sqrt{3}$,
$ -2-2\sqrt{3}$,
$ 1+\sqrt{3}$,
$ 1+\sqrt{3}$,
$ -1-\sqrt{3}$,
$ -1-\sqrt{3}$,
$0$;\ \ 
$ 1+\sqrt{3}$,
$ 2+2\sqrt{3}$,
$ -1-\sqrt{3}$,
$ -1-\sqrt{3}$,
$ -1-\sqrt{3}$,
$ -1-\sqrt{3}$,
$0$;\ \ 
$0$,
$0$,
$0$,
$ -1-\sqrt{3}$,
$ 1+\sqrt{3}$,
$0$;\ \ 
$ 3+\sqrt{3}$,
$ -3-\sqrt{3}$,
$ -1-\sqrt{3}$,
$ 1+\sqrt{3}$,
$0$;\ \ 
$ 3+\sqrt{3}$,
$ -1-\sqrt{3}$,
$ 1+\sqrt{3}$,
$0$;\ \ 
$ 1$,
$ 1$,
$ 3+\sqrt{3}$;\ \ 
$ 1$,
$ -3-\sqrt{3}$;\ \ 
$0$)

\vskip 0.7ex
\hangindent=3em \hangafter=1
$\tau_n$ = ($6.69 - 6.69 i$, $9.46 - 9.46 i$, $0. - 31.67 i$, $0.$, $-24.97 + 24.97 i$, $44.78$, $24.97 + 24.97 i$, $44.78 + 25.86 i$, $0. - 31.67 i$, $35.32 - 35.32 i$, $-6.69 - 6.69 i$, $0.$, $-6.69 + 6.69 i$, $35.32 + 35.32 i$, $0. + 31.67 i$, $44.78 - 25.86 i$, $24.97 - 24.97 i$, $44.78$, $-24.97 - 24.97 i$, $0.$, $0. + 31.67 i$, $9.46 + 9.46 i$, $6.69 + 6.69 i$, $89.57$)

\vskip 0.7ex
\hangindent=3em \hangafter=1
\textit{Intrinsic sign problem}

  \vskip 2ex

\noindent59. $10_{1,89.56}^{24,722}$ \irep{978}:\ \ 
$d_i$ = ($1.0$,
$1.0$,
$2.732$,
$2.732$,
$2.732$,
$2.732$,
$2.732$,
$3.732$,
$3.732$,
$4.732$) 

\vskip 0.7ex
\hangindent=3em \hangafter=1
$D^2= 89.569 = 
48+24\sqrt{3}$

\vskip 0.7ex
\hangindent=3em \hangafter=1
$T = ( 0,
\frac{1}{2},
\frac{2}{3},
\frac{1}{6},
\frac{7}{8},
\frac{13}{24},
\frac{13}{24},
0,
\frac{1}{2},
\frac{1}{8} )
$,

\vskip 0.7ex
\hangindent=3em \hangafter=1
$S$ = ($ 1$,
$ 1$,
$ 1+\sqrt{3}$,
$ 1+\sqrt{3}$,
$ 1+\sqrt{3}$,
$ 1+\sqrt{3}$,
$ 1+\sqrt{3}$,
$ 2+\sqrt{3}$,
$ 2+\sqrt{3}$,
$ 3+\sqrt{3}$;\ \ 
$ 1$,
$ 1+\sqrt{3}$,
$ 1+\sqrt{3}$,
$ -1-\sqrt{3}$,
$ -1-\sqrt{3}$,
$ -1-\sqrt{3}$,
$ 2+\sqrt{3}$,
$ 2+\sqrt{3}$,
$ -3-\sqrt{3}$;\ \ 
$ 1+\sqrt{3}$,
$ 1+\sqrt{3}$,
$ -2-2\sqrt{3}$,
$ 1+\sqrt{3}$,
$ 1+\sqrt{3}$,
$ -1-\sqrt{3}$,
$ -1-\sqrt{3}$,
$0$;\ \ 
$ 1+\sqrt{3}$,
$ 2+2\sqrt{3}$,
$ -1-\sqrt{3}$,
$ -1-\sqrt{3}$,
$ -1-\sqrt{3}$,
$ -1-\sqrt{3}$,
$0$;\ \ 
$0$,
$0$,
$0$,
$ -1-\sqrt{3}$,
$ 1+\sqrt{3}$,
$0$;\ \ 
$ -3-\sqrt{3}$,
$ 3+\sqrt{3}$,
$ -1-\sqrt{3}$,
$ 1+\sqrt{3}$,
$0$;\ \ 
$ -3-\sqrt{3}$,
$ -1-\sqrt{3}$,
$ 1+\sqrt{3}$,
$0$;\ \ 
$ 1$,
$ 1$,
$ 3+\sqrt{3}$;\ \ 
$ 1$,
$ -3-\sqrt{3}$;\ \ 
$0$)

\vskip 0.7ex
\hangindent=3em \hangafter=1
$\tau_n$ = ($6.69 + 6.69 i$, $35.32 + 35.32 i$, $-31.67$, $0.$, $-24.97 - 24.97 i$, $44.78$, $24.97 - 24.97 i$, $44.78 + 25.86 i$, $31.67$, $9.46 + 9.46 i$, $-6.69 + 6.69 i$, $0.$, $-6.69 - 6.69 i$, $9.46 - 9.46 i$, $31.67$, $44.78 - 25.86 i$, $24.97 + 24.97 i$, $44.78$, $-24.97 + 24.97 i$, $0.$, $-31.67$, $35.32 - 35.32 i$, $6.69 - 6.69 i$, $89.57$)

\vskip 0.7ex
\hangindent=3em \hangafter=1
\textit{Intrinsic sign problem}

  \vskip 2ex

\noindent60. $10_{\frac{74}{55},125.3}^{55,334}$ \irep{1121}:\ \ 
$d_i$ = ($1.0$,
$1.618$,
$1.918$,
$2.682$,
$3.104$,
$3.228$,
$3.513$,
$4.340$,
$5.224$,
$5.684$) 

\vskip 0.7ex
\hangindent=3em \hangafter=1
$D^2= 125.351 = 
37+5c^{2}_{55}
+8c^{3}_{55}
+22c^{5}_{55}
+4c^{6}_{55}
+3c^{7}_{55}
+8c^{8}_{55}
+10c^{10}_{55}
+9c^{11}_{55}
+5c^{13}_{55}
+2c^{14}_{55}
+4c^{15}_{55}
+4c^{16}_{55}
+3c^{18}_{55}
+2c^{19}_{55}
$

\vskip 0.7ex
\hangindent=3em \hangafter=1
$T = ( 0,
\frac{2}{5},
\frac{2}{11},
\frac{9}{11},
\frac{32}{55},
\frac{10}{11},
\frac{5}{11},
\frac{12}{55},
\frac{17}{55},
\frac{47}{55} )
$,

\vskip 0.7ex
\hangindent=3em \hangafter=1
$S$ = ($ 1$,
$ \frac{1+\sqrt{5}}{2}$,
$ -c_{11}^{5}$,
$ \xi_{11}^{3}$,
$ c^{3}_{55}
+c^{8}_{55}
$,
$ \xi_{11}^{7}$,
$ \xi_{11}^{5}$,
$ 1+c^{5}_{55}
+c^{6}_{55}
+c^{11}_{55}
+c^{16}_{55}
$,
$ c^{2}_{55}
+c^{3}_{55}
+c^{8}_{55}
+c^{13}_{55}
$,
$ c^{2}_{55}
+c^{3}_{55}
+c^{7}_{55}
+c^{8}_{55}
+c^{13}_{55}
+c^{18}_{55}
$;\ \ 
$ -1$,
$ c^{3}_{55}
+c^{8}_{55}
$,
$ 1+c^{5}_{55}
+c^{6}_{55}
+c^{11}_{55}
+c^{16}_{55}
$,
$ c_{11}^{5}$,
$ c^{2}_{55}
+c^{3}_{55}
+c^{8}_{55}
+c^{13}_{55}
$,
$ c^{2}_{55}
+c^{3}_{55}
+c^{7}_{55}
+c^{8}_{55}
+c^{13}_{55}
+c^{18}_{55}
$,
$ -\xi_{11}^{3}$,
$ -\xi_{11}^{7}$,
$ -\xi_{11}^{5}$;\ \ 
$ -\xi_{11}^{7}$,
$ \xi_{11}^{5}$,
$ -c^{2}_{55}
-c^{3}_{55}
-c^{8}_{55}
-c^{13}_{55}
$,
$ -\xi_{11}^{3}$,
$ 1$,
$ c^{2}_{55}
+c^{3}_{55}
+c^{7}_{55}
+c^{8}_{55}
+c^{13}_{55}
+c^{18}_{55}
$,
$ -1-c^{5}_{55}
-c^{6}_{55}
-c^{11}_{55}
-c^{16}_{55}
$,
$ \frac{1+\sqrt{5}}{2}$;\ \ 
$ -c_{11}^{5}$,
$ c^{2}_{55}
+c^{3}_{55}
+c^{7}_{55}
+c^{8}_{55}
+c^{13}_{55}
+c^{18}_{55}
$,
$ -1$,
$ -\xi_{11}^{7}$,
$ c^{3}_{55}
+c^{8}_{55}
$,
$ -\frac{1+\sqrt{5}}{2}$,
$ -c^{2}_{55}
-c^{3}_{55}
-c^{8}_{55}
-c^{13}_{55}
$;\ \ 
$ \xi_{11}^{7}$,
$ -1-c^{5}_{55}
-c^{6}_{55}
-c^{11}_{55}
-c^{16}_{55}
$,
$ \frac{1+\sqrt{5}}{2}$,
$ -\xi_{11}^{5}$,
$ \xi_{11}^{3}$,
$ -1$;\ \ 
$ \xi_{11}^{5}$,
$ c_{11}^{5}$,
$ -\frac{1+\sqrt{5}}{2}$,
$ c^{2}_{55}
+c^{3}_{55}
+c^{7}_{55}
+c^{8}_{55}
+c^{13}_{55}
+c^{18}_{55}
$,
$ -c^{3}_{55}
-c^{8}_{55}
$;\ \ 
$ \xi_{11}^{3}$,
$ -c^{2}_{55}
-c^{3}_{55}
-c^{8}_{55}
-c^{13}_{55}
$,
$ -c^{3}_{55}
-c^{8}_{55}
$,
$ 1+c^{5}_{55}
+c^{6}_{55}
+c^{11}_{55}
+c^{16}_{55}
$;\ \ 
$ c_{11}^{5}$,
$ 1$,
$ \xi_{11}^{7}$;\ \ 
$ -\xi_{11}^{5}$,
$ -c_{11}^{5}$;\ \ 
$ -\xi_{11}^{3}$)

Factors = $2_{\frac{14}{5},3.618}^{5,395}\boxtimes 5_{\frac{72}{11},34.64}^{11,216}$

\vskip 0.7ex
\hangindent=3em \hangafter=1
$\tau_n$ = ($5.5 + 9.75 i$, $-31.29 - 55.41 i$, $-34.36 - 47.3 i$, $-21.53 + 20.93 i$, $-5.82 + 40.44 i$, $19. + 10.02 i$, $34.84 + 33.86 i$, $-34.37 - 47.31 i$, $19.34 - 34.25 i$, $8.84 + 19.37 i$, $-38.73 + 53.3 i$, $-8.92 - 15.78 i$, $62.38 - 12.63 i$, $21.24 + 29.24 i$, $-8.13 - 56.53 i$, $-15.4 - 14.97 i$, $-30.74 - 16.22 i$, $-42.97 + 22.67 i$, $21.25 + 29.23 i$, $31.08 + 68.05 i$, $-10.98 - 2.23 i$, $62.66 - 86.25 i$, $17.75 - 3.6 i$, $-38.55 + 7.81 i$, $-68.74$, $26.56 + 14.01 i$, $24.92 + 24.23 i$, $24.92 - 24.23 i$, $26.56 - 14.01 i$, $-68.74$, $-38.55 - 7.81 i$, $17.75 + 3.6 i$, $62.66 + 86.25 i$, $-10.98 + 2.23 i$, $31.08 - 68.05 i$, $21.25 - 29.23 i$, $-42.97 - 22.67 i$, $-30.74 + 16.22 i$, $-15.4 + 14.97 i$, $-8.13 + 56.53 i$, $21.24 - 29.24 i$, $62.38 + 12.63 i$, $-8.92 + 15.78 i$, $-38.73 - 53.3 i$, $8.84 - 19.37 i$, $19.34 + 34.25 i$, $-34.37 + 47.31 i$, $34.84 - 33.86 i$, $19. - 10.02 i$, $-5.82 - 40.44 i$, $-21.53 - 20.93 i$, $-34.36 + 47.3 i$, $-31.29 + 55.41 i$, $5.5 - 9.75 i$, $125.32$)

\vskip 0.7ex
\hangindent=3em \hangafter=1
\textit{Intrinsic sign problem}

  \vskip 2ex

\noindent61. $10_{\frac{234}{55},125.3}^{55,258}$ \irep{1121}:\ \ 
$d_i$ = ($1.0$,
$1.618$,
$1.918$,
$2.682$,
$3.104$,
$3.228$,
$3.513$,
$4.340$,
$5.224$,
$5.684$) 

\vskip 0.7ex
\hangindent=3em \hangafter=1
$D^2= 125.351 = 
37+5c^{2}_{55}
+8c^{3}_{55}
+22c^{5}_{55}
+4c^{6}_{55}
+3c^{7}_{55}
+8c^{8}_{55}
+10c^{10}_{55}
+9c^{11}_{55}
+5c^{13}_{55}
+2c^{14}_{55}
+4c^{15}_{55}
+4c^{16}_{55}
+3c^{18}_{55}
+2c^{19}_{55}
$

\vskip 0.7ex
\hangindent=3em \hangafter=1
$T = ( 0,
\frac{2}{5},
\frac{9}{11},
\frac{2}{11},
\frac{12}{55},
\frac{1}{11},
\frac{6}{11},
\frac{32}{55},
\frac{27}{55},
\frac{52}{55} )
$,

\vskip 0.7ex
\hangindent=3em \hangafter=1
$S$ = ($ 1$,
$ \frac{1+\sqrt{5}}{2}$,
$ -c_{11}^{5}$,
$ \xi_{11}^{3}$,
$ c^{3}_{55}
+c^{8}_{55}
$,
$ \xi_{11}^{7}$,
$ \xi_{11}^{5}$,
$ 1+c^{5}_{55}
+c^{6}_{55}
+c^{11}_{55}
+c^{16}_{55}
$,
$ c^{2}_{55}
+c^{3}_{55}
+c^{8}_{55}
+c^{13}_{55}
$,
$ c^{2}_{55}
+c^{3}_{55}
+c^{7}_{55}
+c^{8}_{55}
+c^{13}_{55}
+c^{18}_{55}
$;\ \ 
$ -1$,
$ c^{3}_{55}
+c^{8}_{55}
$,
$ 1+c^{5}_{55}
+c^{6}_{55}
+c^{11}_{55}
+c^{16}_{55}
$,
$ c_{11}^{5}$,
$ c^{2}_{55}
+c^{3}_{55}
+c^{8}_{55}
+c^{13}_{55}
$,
$ c^{2}_{55}
+c^{3}_{55}
+c^{7}_{55}
+c^{8}_{55}
+c^{13}_{55}
+c^{18}_{55}
$,
$ -\xi_{11}^{3}$,
$ -\xi_{11}^{7}$,
$ -\xi_{11}^{5}$;\ \ 
$ -\xi_{11}^{7}$,
$ \xi_{11}^{5}$,
$ -c^{2}_{55}
-c^{3}_{55}
-c^{8}_{55}
-c^{13}_{55}
$,
$ -\xi_{11}^{3}$,
$ 1$,
$ c^{2}_{55}
+c^{3}_{55}
+c^{7}_{55}
+c^{8}_{55}
+c^{13}_{55}
+c^{18}_{55}
$,
$ -1-c^{5}_{55}
-c^{6}_{55}
-c^{11}_{55}
-c^{16}_{55}
$,
$ \frac{1+\sqrt{5}}{2}$;\ \ 
$ -c_{11}^{5}$,
$ c^{2}_{55}
+c^{3}_{55}
+c^{7}_{55}
+c^{8}_{55}
+c^{13}_{55}
+c^{18}_{55}
$,
$ -1$,
$ -\xi_{11}^{7}$,
$ c^{3}_{55}
+c^{8}_{55}
$,
$ -\frac{1+\sqrt{5}}{2}$,
$ -c^{2}_{55}
-c^{3}_{55}
-c^{8}_{55}
-c^{13}_{55}
$;\ \ 
$ \xi_{11}^{7}$,
$ -1-c^{5}_{55}
-c^{6}_{55}
-c^{11}_{55}
-c^{16}_{55}
$,
$ \frac{1+\sqrt{5}}{2}$,
$ -\xi_{11}^{5}$,
$ \xi_{11}^{3}$,
$ -1$;\ \ 
$ \xi_{11}^{5}$,
$ c_{11}^{5}$,
$ -\frac{1+\sqrt{5}}{2}$,
$ c^{2}_{55}
+c^{3}_{55}
+c^{7}_{55}
+c^{8}_{55}
+c^{13}_{55}
+c^{18}_{55}
$,
$ -c^{3}_{55}
-c^{8}_{55}
$;\ \ 
$ \xi_{11}^{3}$,
$ -c^{2}_{55}
-c^{3}_{55}
-c^{8}_{55}
-c^{13}_{55}
$,
$ -c^{3}_{55}
-c^{8}_{55}
$,
$ 1+c^{5}_{55}
+c^{6}_{55}
+c^{11}_{55}
+c^{16}_{55}
$;\ \ 
$ c_{11}^{5}$,
$ 1$,
$ \xi_{11}^{7}$;\ \ 
$ -\xi_{11}^{5}$,
$ -c_{11}^{5}$;\ \ 
$ -\xi_{11}^{3}$)

Factors = $2_{\frac{14}{5},3.618}^{5,395}\boxtimes 5_{\frac{16}{11},34.64}^{11,640}$

\vskip 0.7ex
\hangindent=3em \hangafter=1
$\tau_n$ = ($-10.98 - 2.23 i$, $62.38 + 12.63 i$, $-34.37 - 47.31 i$, $26.56 - 14.01 i$, $-5.82 - 40.44 i$, $-15.4 - 14.97 i$, $-42.97 - 22.67 i$, $-34.36 - 47.3 i$, $-38.55 + 7.81 i$, $8.84 - 19.37 i$, $-38.73 + 53.3 i$, $17.75 + 3.6 i$, $-31.29 + 55.41 i$, $21.25 + 29.23 i$, $-8.13 + 56.53 i$, $19. + 10.02 i$, $24.92 + 24.23 i$, $34.84 - 33.86 i$, $21.24 + 29.24 i$, $31.08 - 68.05 i$, $5.5 + 9.75 i$, $62.66 - 86.25 i$, $-8.92 + 15.78 i$, $19.34 - 34.25 i$, $-68.74$, $-21.53 - 20.93 i$, $-30.74 - 16.22 i$, $-30.74 + 16.22 i$, $-21.53 + 20.93 i$, $-68.74$, $19.34 + 34.25 i$, $-8.92 - 15.78 i$, $62.66 + 86.25 i$, $5.5 - 9.75 i$, $31.08 + 68.05 i$, $21.24 - 29.24 i$, $34.84 + 33.86 i$, $24.92 - 24.23 i$, $19. - 10.02 i$, $-8.13 - 56.53 i$, $21.25 - 29.23 i$, $-31.29 - 55.41 i$, $17.75 - 3.6 i$, $-38.73 - 53.3 i$, $8.84 + 19.37 i$, $-38.55 - 7.81 i$, $-34.36 + 47.3 i$, $-42.97 + 22.67 i$, $-15.4 + 14.97 i$, $-5.82 + 40.44 i$, $26.56 + 14.01 i$, $-34.37 + 47.31 i$, $62.38 - 12.63 i$, $-10.98 + 2.23 i$, $125.32$)

\vskip 0.7ex
\hangindent=3em \hangafter=1
\textit{Intrinsic sign problem}

  \vskip 2ex

\noindent62. $10_{\frac{206}{55},125.3}^{55,178}$ \irep{1121}:\ \ 
$d_i$ = ($1.0$,
$1.618$,
$1.918$,
$2.682$,
$3.104$,
$3.228$,
$3.513$,
$4.340$,
$5.224$,
$5.684$) 

\vskip 0.7ex
\hangindent=3em \hangafter=1
$D^2= 125.351 = 
37+5c^{2}_{55}
+8c^{3}_{55}
+22c^{5}_{55}
+4c^{6}_{55}
+3c^{7}_{55}
+8c^{8}_{55}
+10c^{10}_{55}
+9c^{11}_{55}
+5c^{13}_{55}
+2c^{14}_{55}
+4c^{15}_{55}
+4c^{16}_{55}
+3c^{18}_{55}
+2c^{19}_{55}
$

\vskip 0.7ex
\hangindent=3em \hangafter=1
$T = ( 0,
\frac{3}{5},
\frac{2}{11},
\frac{9}{11},
\frac{43}{55},
\frac{10}{11},
\frac{5}{11},
\frac{23}{55},
\frac{28}{55},
\frac{3}{55} )
$,

\vskip 0.7ex
\hangindent=3em \hangafter=1
$S$ = ($ 1$,
$ \frac{1+\sqrt{5}}{2}$,
$ -c_{11}^{5}$,
$ \xi_{11}^{3}$,
$ c^{3}_{55}
+c^{8}_{55}
$,
$ \xi_{11}^{7}$,
$ \xi_{11}^{5}$,
$ 1+c^{5}_{55}
+c^{6}_{55}
+c^{11}_{55}
+c^{16}_{55}
$,
$ c^{2}_{55}
+c^{3}_{55}
+c^{8}_{55}
+c^{13}_{55}
$,
$ c^{2}_{55}
+c^{3}_{55}
+c^{7}_{55}
+c^{8}_{55}
+c^{13}_{55}
+c^{18}_{55}
$;\ \ 
$ -1$,
$ c^{3}_{55}
+c^{8}_{55}
$,
$ 1+c^{5}_{55}
+c^{6}_{55}
+c^{11}_{55}
+c^{16}_{55}
$,
$ c_{11}^{5}$,
$ c^{2}_{55}
+c^{3}_{55}
+c^{8}_{55}
+c^{13}_{55}
$,
$ c^{2}_{55}
+c^{3}_{55}
+c^{7}_{55}
+c^{8}_{55}
+c^{13}_{55}
+c^{18}_{55}
$,
$ -\xi_{11}^{3}$,
$ -\xi_{11}^{7}$,
$ -\xi_{11}^{5}$;\ \ 
$ -\xi_{11}^{7}$,
$ \xi_{11}^{5}$,
$ -c^{2}_{55}
-c^{3}_{55}
-c^{8}_{55}
-c^{13}_{55}
$,
$ -\xi_{11}^{3}$,
$ 1$,
$ c^{2}_{55}
+c^{3}_{55}
+c^{7}_{55}
+c^{8}_{55}
+c^{13}_{55}
+c^{18}_{55}
$,
$ -1-c^{5}_{55}
-c^{6}_{55}
-c^{11}_{55}
-c^{16}_{55}
$,
$ \frac{1+\sqrt{5}}{2}$;\ \ 
$ -c_{11}^{5}$,
$ c^{2}_{55}
+c^{3}_{55}
+c^{7}_{55}
+c^{8}_{55}
+c^{13}_{55}
+c^{18}_{55}
$,
$ -1$,
$ -\xi_{11}^{7}$,
$ c^{3}_{55}
+c^{8}_{55}
$,
$ -\frac{1+\sqrt{5}}{2}$,
$ -c^{2}_{55}
-c^{3}_{55}
-c^{8}_{55}
-c^{13}_{55}
$;\ \ 
$ \xi_{11}^{7}$,
$ -1-c^{5}_{55}
-c^{6}_{55}
-c^{11}_{55}
-c^{16}_{55}
$,
$ \frac{1+\sqrt{5}}{2}$,
$ -\xi_{11}^{5}$,
$ \xi_{11}^{3}$,
$ -1$;\ \ 
$ \xi_{11}^{5}$,
$ c_{11}^{5}$,
$ -\frac{1+\sqrt{5}}{2}$,
$ c^{2}_{55}
+c^{3}_{55}
+c^{7}_{55}
+c^{8}_{55}
+c^{13}_{55}
+c^{18}_{55}
$,
$ -c^{3}_{55}
-c^{8}_{55}
$;\ \ 
$ \xi_{11}^{3}$,
$ -c^{2}_{55}
-c^{3}_{55}
-c^{8}_{55}
-c^{13}_{55}
$,
$ -c^{3}_{55}
-c^{8}_{55}
$,
$ 1+c^{5}_{55}
+c^{6}_{55}
+c^{11}_{55}
+c^{16}_{55}
$;\ \ 
$ c_{11}^{5}$,
$ 1$,
$ \xi_{11}^{7}$;\ \ 
$ -\xi_{11}^{5}$,
$ -c_{11}^{5}$;\ \ 
$ -\xi_{11}^{3}$)

Factors = $2_{\frac{26}{5},3.618}^{5,720}\boxtimes 5_{\frac{72}{11},34.64}^{11,216}$

\vskip 0.7ex
\hangindent=3em \hangafter=1
$\tau_n$ = ($-10.98 + 2.23 i$, $62.38 - 12.63 i$, $-34.37 + 47.31 i$, $26.56 + 14.01 i$, $-5.82 + 40.44 i$, $-15.4 + 14.97 i$, $-42.97 + 22.67 i$, $-34.36 + 47.3 i$, $-38.55 - 7.81 i$, $8.84 + 19.37 i$, $-38.73 - 53.3 i$, $17.75 - 3.6 i$, $-31.29 - 55.41 i$, $21.25 - 29.23 i$, $-8.13 - 56.53 i$, $19. - 10.02 i$, $24.92 - 24.23 i$, $34.84 + 33.86 i$, $21.24 - 29.24 i$, $31.08 + 68.05 i$, $5.5 - 9.75 i$, $62.66 + 86.25 i$, $-8.92 - 15.78 i$, $19.34 + 34.25 i$, $-68.74$, $-21.53 + 20.93 i$, $-30.74 + 16.22 i$, $-30.74 - 16.22 i$, $-21.53 - 20.93 i$, $-68.74$, $19.34 - 34.25 i$, $-8.92 + 15.78 i$, $62.66 - 86.25 i$, $5.5 + 9.75 i$, $31.08 - 68.05 i$, $21.24 + 29.24 i$, $34.84 - 33.86 i$, $24.92 + 24.23 i$, $19. + 10.02 i$, $-8.13 + 56.53 i$, $21.25 + 29.23 i$, $-31.29 + 55.41 i$, $17.75 + 3.6 i$, $-38.73 + 53.3 i$, $8.84 - 19.37 i$, $-38.55 + 7.81 i$, $-34.36 - 47.3 i$, $-42.97 - 22.67 i$, $-15.4 - 14.97 i$, $-5.82 - 40.44 i$, $26.56 - 14.01 i$, $-34.37 - 47.31 i$, $62.38 + 12.63 i$, $-10.98 - 2.23 i$, $125.32$)

\vskip 0.7ex
\hangindent=3em \hangafter=1
\textit{Intrinsic sign problem}

  \vskip 2ex

\noindent63. $10_{\frac{366}{55},125.3}^{55,758}$ \irep{1121}:\ \ 
$d_i$ = ($1.0$,
$1.618$,
$1.918$,
$2.682$,
$3.104$,
$3.228$,
$3.513$,
$4.340$,
$5.224$,
$5.684$) 

\vskip 0.7ex
\hangindent=3em \hangafter=1
$D^2= 125.351 = 
37+5c^{2}_{55}
+8c^{3}_{55}
+22c^{5}_{55}
+4c^{6}_{55}
+3c^{7}_{55}
+8c^{8}_{55}
+10c^{10}_{55}
+9c^{11}_{55}
+5c^{13}_{55}
+2c^{14}_{55}
+4c^{15}_{55}
+4c^{16}_{55}
+3c^{18}_{55}
+2c^{19}_{55}
$

\vskip 0.7ex
\hangindent=3em \hangafter=1
$T = ( 0,
\frac{3}{5},
\frac{9}{11},
\frac{2}{11},
\frac{23}{55},
\frac{1}{11},
\frac{6}{11},
\frac{43}{55},
\frac{38}{55},
\frac{8}{55} )
$,

\vskip 0.7ex
\hangindent=3em \hangafter=1
$S$ = ($ 1$,
$ \frac{1+\sqrt{5}}{2}$,
$ -c_{11}^{5}$,
$ \xi_{11}^{3}$,
$ c^{3}_{55}
+c^{8}_{55}
$,
$ \xi_{11}^{7}$,
$ \xi_{11}^{5}$,
$ 1+c^{5}_{55}
+c^{6}_{55}
+c^{11}_{55}
+c^{16}_{55}
$,
$ c^{2}_{55}
+c^{3}_{55}
+c^{8}_{55}
+c^{13}_{55}
$,
$ c^{2}_{55}
+c^{3}_{55}
+c^{7}_{55}
+c^{8}_{55}
+c^{13}_{55}
+c^{18}_{55}
$;\ \ 
$ -1$,
$ c^{3}_{55}
+c^{8}_{55}
$,
$ 1+c^{5}_{55}
+c^{6}_{55}
+c^{11}_{55}
+c^{16}_{55}
$,
$ c_{11}^{5}$,
$ c^{2}_{55}
+c^{3}_{55}
+c^{8}_{55}
+c^{13}_{55}
$,
$ c^{2}_{55}
+c^{3}_{55}
+c^{7}_{55}
+c^{8}_{55}
+c^{13}_{55}
+c^{18}_{55}
$,
$ -\xi_{11}^{3}$,
$ -\xi_{11}^{7}$,
$ -\xi_{11}^{5}$;\ \ 
$ -\xi_{11}^{7}$,
$ \xi_{11}^{5}$,
$ -c^{2}_{55}
-c^{3}_{55}
-c^{8}_{55}
-c^{13}_{55}
$,
$ -\xi_{11}^{3}$,
$ 1$,
$ c^{2}_{55}
+c^{3}_{55}
+c^{7}_{55}
+c^{8}_{55}
+c^{13}_{55}
+c^{18}_{55}
$,
$ -1-c^{5}_{55}
-c^{6}_{55}
-c^{11}_{55}
-c^{16}_{55}
$,
$ \frac{1+\sqrt{5}}{2}$;\ \ 
$ -c_{11}^{5}$,
$ c^{2}_{55}
+c^{3}_{55}
+c^{7}_{55}
+c^{8}_{55}
+c^{13}_{55}
+c^{18}_{55}
$,
$ -1$,
$ -\xi_{11}^{7}$,
$ c^{3}_{55}
+c^{8}_{55}
$,
$ -\frac{1+\sqrt{5}}{2}$,
$ -c^{2}_{55}
-c^{3}_{55}
-c^{8}_{55}
-c^{13}_{55}
$;\ \ 
$ \xi_{11}^{7}$,
$ -1-c^{5}_{55}
-c^{6}_{55}
-c^{11}_{55}
-c^{16}_{55}
$,
$ \frac{1+\sqrt{5}}{2}$,
$ -\xi_{11}^{5}$,
$ \xi_{11}^{3}$,
$ -1$;\ \ 
$ \xi_{11}^{5}$,
$ c_{11}^{5}$,
$ -\frac{1+\sqrt{5}}{2}$,
$ c^{2}_{55}
+c^{3}_{55}
+c^{7}_{55}
+c^{8}_{55}
+c^{13}_{55}
+c^{18}_{55}
$,
$ -c^{3}_{55}
-c^{8}_{55}
$;\ \ 
$ \xi_{11}^{3}$,
$ -c^{2}_{55}
-c^{3}_{55}
-c^{8}_{55}
-c^{13}_{55}
$,
$ -c^{3}_{55}
-c^{8}_{55}
$,
$ 1+c^{5}_{55}
+c^{6}_{55}
+c^{11}_{55}
+c^{16}_{55}
$;\ \ 
$ c_{11}^{5}$,
$ 1$,
$ \xi_{11}^{7}$;\ \ 
$ -\xi_{11}^{5}$,
$ -c_{11}^{5}$;\ \ 
$ -\xi_{11}^{3}$)

Factors = $2_{\frac{26}{5},3.618}^{5,720}\boxtimes 5_{\frac{16}{11},34.64}^{11,640}$

\vskip 0.7ex
\hangindent=3em \hangafter=1
$\tau_n$ = ($5.5 - 9.75 i$, $-31.29 + 55.41 i$, $-34.36 + 47.3 i$, $-21.53 - 20.93 i$, $-5.82 - 40.44 i$, $19. - 10.02 i$, $34.84 - 33.86 i$, $-34.37 + 47.31 i$, $19.34 + 34.25 i$, $8.84 - 19.37 i$, $-38.73 - 53.3 i$, $-8.92 + 15.78 i$, $62.38 + 12.63 i$, $21.24 - 29.24 i$, $-8.13 + 56.53 i$, $-15.4 + 14.97 i$, $-30.74 + 16.22 i$, $-42.97 - 22.67 i$, $21.25 - 29.23 i$, $31.08 - 68.05 i$, $-10.98 + 2.23 i$, $62.66 + 86.25 i$, $17.75 + 3.6 i$, $-38.55 - 7.81 i$, $-68.74$, $26.56 - 14.01 i$, $24.92 - 24.23 i$, $24.92 + 24.23 i$, $26.56 + 14.01 i$, $-68.74$, $-38.55 + 7.81 i$, $17.75 - 3.6 i$, $62.66 - 86.25 i$, $-10.98 - 2.23 i$, $31.08 + 68.05 i$, $21.25 + 29.23 i$, $-42.97 + 22.67 i$, $-30.74 - 16.22 i$, $-15.4 - 14.97 i$, $-8.13 - 56.53 i$, $21.24 + 29.24 i$, $62.38 - 12.63 i$, $-8.92 - 15.78 i$, $-38.73 + 53.3 i$, $8.84 + 19.37 i$, $19.34 - 34.25 i$, $-34.37 - 47.31 i$, $34.84 + 33.86 i$, $19. + 10.02 i$, $-5.82 + 40.44 i$, $-21.53 + 20.93 i$, $-34.36 - 47.3 i$, $-31.29 - 55.41 i$, $5.5 + 9.75 i$, $125.32$)

\vskip 0.7ex
\hangindent=3em \hangafter=1
\textit{Intrinsic sign problem}

  \vskip 2ex

\noindent64. $10_{\frac{8}{35},127.8}^{35,427}$ \irep{1074}:\ \ 
$d_i$ = ($1.0$,
$1.618$,
$2.246$,
$2.246$,
$2.801$,
$3.635$,
$3.635$,
$4.48$,
$4.533$,
$6.551$) 

\vskip 0.7ex
\hangindent=3em \hangafter=1
$D^2= 127.870 = 
49+14c^{1}_{35}
+7c^{4}_{35}
+28c^{5}_{35}
+14c^{6}_{35}
+7c^{7}_{35}
+14c^{10}_{35}
+7c^{11}_{35}
$

\vskip 0.7ex
\hangindent=3em \hangafter=1
$T = ( 0,
\frac{2}{5},
\frac{1}{7},
\frac{1}{7},
\frac{6}{7},
\frac{19}{35},
\frac{19}{35},
\frac{4}{7},
\frac{9}{35},
\frac{34}{35} )
$,

\vskip 0.7ex
\hangindent=3em \hangafter=1
$S$ = ($ 1$,
$ \frac{1+\sqrt{5}}{2}$,
$ \xi_{7}^{3}$,
$ \xi_{7}^{3}$,
$ 2+c^{1}_{7}
+c^{2}_{7}
$,
$ c^{1}_{35}
+c^{4}_{35}
+c^{6}_{35}
+c^{11}_{35}
$,
$ c^{1}_{35}
+c^{4}_{35}
+c^{6}_{35}
+c^{11}_{35}
$,
$ 2+2c^{1}_{7}
+c^{2}_{7}
$,
$ 1+c^{1}_{35}
+c^{6}_{35}
+c^{7}_{35}
$,
$ 2c^{1}_{35}
+c^{4}_{35}
+2c^{6}_{35}
+c^{11}_{35}
$;\ \ 
$ -1$,
$ c^{1}_{35}
+c^{4}_{35}
+c^{6}_{35}
+c^{11}_{35}
$,
$ c^{1}_{35}
+c^{4}_{35}
+c^{6}_{35}
+c^{11}_{35}
$,
$ 1+c^{1}_{35}
+c^{6}_{35}
+c^{7}_{35}
$,
$ -\xi_{7}^{3}$,
$ -\xi_{7}^{3}$,
$ 2c^{1}_{35}
+c^{4}_{35}
+2c^{6}_{35}
+c^{11}_{35}
$,
$ -2-c^{1}_{7}
-c^{2}_{7}
$,
$ -2-2  c^{1}_{7}
-c^{2}_{7}
$;\ \ 
$ s^{1}_{7}
+\zeta^{2}_{7}
+\zeta^{3}_{7}
$,
$ -1-2  \zeta^{1}_{7}
-\zeta^{2}_{7}
-\zeta^{3}_{7}
$,
$ -\xi_{7}^{3}$,
$ 1-2  \zeta^{1}_{35}
-\zeta^{-1}_{35}
+2\zeta^{-2}_{35}
-\zeta^{4}_{35}
-2  \zeta^{-4}_{35}
+2\zeta^{5}_{35}
-\zeta^{6}_{35}
-2  \zeta^{-6}_{35}
+c^{7}_{35}
-\zeta^{11}_{35}
-2  \zeta^{-11}_{35}
+2\zeta^{12}_{35}
$,
$ -1+\zeta^{1}_{35}
-2  \zeta^{-2}_{35}
+\zeta^{-4}_{35}
-2  \zeta^{5}_{35}
+\zeta^{-6}_{35}
-c^{7}_{35}
+\zeta^{-11}_{35}
-2  \zeta^{12}_{35}
$,
$ \xi_{7}^{3}$,
$ -c^{1}_{35}
-c^{4}_{35}
-c^{6}_{35}
-c^{11}_{35}
$,
$ c^{1}_{35}
+c^{4}_{35}
+c^{6}_{35}
+c^{11}_{35}
$;\ \ 
$ s^{1}_{7}
+\zeta^{2}_{7}
+\zeta^{3}_{7}
$,
$ -\xi_{7}^{3}$,
$ -1+\zeta^{1}_{35}
-2  \zeta^{-2}_{35}
+\zeta^{-4}_{35}
-2  \zeta^{5}_{35}
+\zeta^{-6}_{35}
-c^{7}_{35}
+\zeta^{-11}_{35}
-2  \zeta^{12}_{35}
$,
$ 1-2  \zeta^{1}_{35}
-\zeta^{-1}_{35}
+2\zeta^{-2}_{35}
-\zeta^{4}_{35}
-2  \zeta^{-4}_{35}
+2\zeta^{5}_{35}
-\zeta^{6}_{35}
-2  \zeta^{-6}_{35}
+c^{7}_{35}
-\zeta^{11}_{35}
-2  \zeta^{-11}_{35}
+2\zeta^{12}_{35}
$,
$ \xi_{7}^{3}$,
$ -c^{1}_{35}
-c^{4}_{35}
-c^{6}_{35}
-c^{11}_{35}
$,
$ c^{1}_{35}
+c^{4}_{35}
+c^{6}_{35}
+c^{11}_{35}
$;\ \ 
$ 2+2c^{1}_{7}
+c^{2}_{7}
$,
$ -c^{1}_{35}
-c^{4}_{35}
-c^{6}_{35}
-c^{11}_{35}
$,
$ -c^{1}_{35}
-c^{4}_{35}
-c^{6}_{35}
-c^{11}_{35}
$,
$ -1$,
$ 2c^{1}_{35}
+c^{4}_{35}
+2c^{6}_{35}
+c^{11}_{35}
$,
$ -\frac{1+\sqrt{5}}{2}$;\ \ 
$ -s^{1}_{7}
-\zeta^{2}_{7}
-\zeta^{3}_{7}
$,
$ 1+2\zeta^{1}_{7}
+\zeta^{2}_{7}
+\zeta^{3}_{7}
$,
$ c^{1}_{35}
+c^{4}_{35}
+c^{6}_{35}
+c^{11}_{35}
$,
$ \xi_{7}^{3}$,
$ -\xi_{7}^{3}$;\ \ 
$ -s^{1}_{7}
-\zeta^{2}_{7}
-\zeta^{3}_{7}
$,
$ c^{1}_{35}
+c^{4}_{35}
+c^{6}_{35}
+c^{11}_{35}
$,
$ \xi_{7}^{3}$,
$ -\xi_{7}^{3}$;\ \ 
$ -2-c^{1}_{7}
-c^{2}_{7}
$,
$ -\frac{1+\sqrt{5}}{2}$,
$ -1-c^{1}_{35}
-c^{6}_{35}
-c^{7}_{35}
$;\ \ 
$ -2-2  c^{1}_{7}
-c^{2}_{7}
$,
$ 1$;\ \ 
$ 2+c^{1}_{7}
+c^{2}_{7}
$)

Factors = $2_{\frac{14}{5},3.618}^{5,395}\boxtimes 5_{\frac{38}{7},35.34}^{7,386}$

\vskip 0.7ex
\hangindent=3em \hangafter=1
$\tau_n$ = ($7.81 + 0.42 i$, $52.73 + 12.02 i$, $2.52 - 77.58 i$, $43.32 + 15.76 i$, $28.45 - 57.16 i$, $-8.68 - 8.36 i$, $67.59 - 87.98 i$, $5.35 - 17.72 i$, $17.3 - 25.02 i$, $-68.88 - 57.88 i$, $-2.87 - 42.14 i$, $-21.99 - 48.01 i$, $-21.33 + 4.86 i$, $-35.84 - 54.37 i$, $-12.66 - 20.98 i$, $-28.88 - 2.78 i$, $-72.21 + 16.11 i$, $-72.21 - 16.11 i$, $-28.88 + 2.78 i$, $-12.66 + 20.98 i$, $-35.84 + 54.37 i$, $-21.33 - 4.86 i$, $-21.99 + 48.01 i$, $-2.87 + 42.14 i$, $-68.88 + 57.88 i$, $17.3 + 25.02 i$, $5.35 + 17.72 i$, $67.59 + 87.98 i$, $-8.68 + 8.36 i$, $28.45 + 57.16 i$, $43.32 - 15.76 i$, $2.52 + 77.58 i$, $52.73 - 12.02 i$, $7.81 - 0.42 i$, $131.51$)

\vskip 0.7ex
\hangindent=3em \hangafter=1
\textit{Intrinsic sign problem}

  \vskip 2ex

\noindent65. $10_{\frac{188}{35},127.8}^{35,259}$ \irep{1074}:\ \ 
$d_i$ = ($1.0$,
$1.618$,
$2.246$,
$2.246$,
$2.801$,
$3.635$,
$3.635$,
$4.48$,
$4.533$,
$6.551$) 

\vskip 0.7ex
\hangindent=3em \hangafter=1
$D^2= 127.870 = 
49+14c^{1}_{35}
+7c^{4}_{35}
+28c^{5}_{35}
+14c^{6}_{35}
+7c^{7}_{35}
+14c^{10}_{35}
+7c^{11}_{35}
$

\vskip 0.7ex
\hangindent=3em \hangafter=1
$T = ( 0,
\frac{2}{5},
\frac{6}{7},
\frac{6}{7},
\frac{1}{7},
\frac{9}{35},
\frac{9}{35},
\frac{3}{7},
\frac{19}{35},
\frac{29}{35} )
$,

\vskip 0.7ex
\hangindent=3em \hangafter=1
$S$ = ($ 1$,
$ \frac{1+\sqrt{5}}{2}$,
$ \xi_{7}^{3}$,
$ \xi_{7}^{3}$,
$ 2+c^{1}_{7}
+c^{2}_{7}
$,
$ c^{1}_{35}
+c^{4}_{35}
+c^{6}_{35}
+c^{11}_{35}
$,
$ c^{1}_{35}
+c^{4}_{35}
+c^{6}_{35}
+c^{11}_{35}
$,
$ 2+2c^{1}_{7}
+c^{2}_{7}
$,
$ 1+c^{1}_{35}
+c^{6}_{35}
+c^{7}_{35}
$,
$ 2c^{1}_{35}
+c^{4}_{35}
+2c^{6}_{35}
+c^{11}_{35}
$;\ \ 
$ -1$,
$ c^{1}_{35}
+c^{4}_{35}
+c^{6}_{35}
+c^{11}_{35}
$,
$ c^{1}_{35}
+c^{4}_{35}
+c^{6}_{35}
+c^{11}_{35}
$,
$ 1+c^{1}_{35}
+c^{6}_{35}
+c^{7}_{35}
$,
$ -\xi_{7}^{3}$,
$ -\xi_{7}^{3}$,
$ 2c^{1}_{35}
+c^{4}_{35}
+2c^{6}_{35}
+c^{11}_{35}
$,
$ -2-c^{1}_{7}
-c^{2}_{7}
$,
$ -2-2  c^{1}_{7}
-c^{2}_{7}
$;\ \ 
$ -1-2  \zeta^{1}_{7}
-\zeta^{2}_{7}
-\zeta^{3}_{7}
$,
$ s^{1}_{7}
+\zeta^{2}_{7}
+\zeta^{3}_{7}
$,
$ -\xi_{7}^{3}$,
$ 1-2  \zeta^{1}_{35}
-\zeta^{-1}_{35}
+2\zeta^{-2}_{35}
-\zeta^{4}_{35}
-2  \zeta^{-4}_{35}
+2\zeta^{5}_{35}
-\zeta^{6}_{35}
-2  \zeta^{-6}_{35}
+c^{7}_{35}
-\zeta^{11}_{35}
-2  \zeta^{-11}_{35}
+2\zeta^{12}_{35}
$,
$ -1+\zeta^{1}_{35}
-2  \zeta^{-2}_{35}
+\zeta^{-4}_{35}
-2  \zeta^{5}_{35}
+\zeta^{-6}_{35}
-c^{7}_{35}
+\zeta^{-11}_{35}
-2  \zeta^{12}_{35}
$,
$ \xi_{7}^{3}$,
$ -c^{1}_{35}
-c^{4}_{35}
-c^{6}_{35}
-c^{11}_{35}
$,
$ c^{1}_{35}
+c^{4}_{35}
+c^{6}_{35}
+c^{11}_{35}
$;\ \ 
$ -1-2  \zeta^{1}_{7}
-\zeta^{2}_{7}
-\zeta^{3}_{7}
$,
$ -\xi_{7}^{3}$,
$ -1+\zeta^{1}_{35}
-2  \zeta^{-2}_{35}
+\zeta^{-4}_{35}
-2  \zeta^{5}_{35}
+\zeta^{-6}_{35}
-c^{7}_{35}
+\zeta^{-11}_{35}
-2  \zeta^{12}_{35}
$,
$ 1-2  \zeta^{1}_{35}
-\zeta^{-1}_{35}
+2\zeta^{-2}_{35}
-\zeta^{4}_{35}
-2  \zeta^{-4}_{35}
+2\zeta^{5}_{35}
-\zeta^{6}_{35}
-2  \zeta^{-6}_{35}
+c^{7}_{35}
-\zeta^{11}_{35}
-2  \zeta^{-11}_{35}
+2\zeta^{12}_{35}
$,
$ \xi_{7}^{3}$,
$ -c^{1}_{35}
-c^{4}_{35}
-c^{6}_{35}
-c^{11}_{35}
$,
$ c^{1}_{35}
+c^{4}_{35}
+c^{6}_{35}
+c^{11}_{35}
$;\ \ 
$ 2+2c^{1}_{7}
+c^{2}_{7}
$,
$ -c^{1}_{35}
-c^{4}_{35}
-c^{6}_{35}
-c^{11}_{35}
$,
$ -c^{1}_{35}
-c^{4}_{35}
-c^{6}_{35}
-c^{11}_{35}
$,
$ -1$,
$ 2c^{1}_{35}
+c^{4}_{35}
+2c^{6}_{35}
+c^{11}_{35}
$,
$ -\frac{1+\sqrt{5}}{2}$;\ \ 
$ 1+2\zeta^{1}_{7}
+\zeta^{2}_{7}
+\zeta^{3}_{7}
$,
$ -s^{1}_{7}
-\zeta^{2}_{7}
-\zeta^{3}_{7}
$,
$ c^{1}_{35}
+c^{4}_{35}
+c^{6}_{35}
+c^{11}_{35}
$,
$ \xi_{7}^{3}$,
$ -\xi_{7}^{3}$;\ \ 
$ 1+2\zeta^{1}_{7}
+\zeta^{2}_{7}
+\zeta^{3}_{7}
$,
$ c^{1}_{35}
+c^{4}_{35}
+c^{6}_{35}
+c^{11}_{35}
$,
$ \xi_{7}^{3}$,
$ -\xi_{7}^{3}$;\ \ 
$ -2-c^{1}_{7}
-c^{2}_{7}
$,
$ -\frac{1+\sqrt{5}}{2}$,
$ -1-c^{1}_{35}
-c^{6}_{35}
-c^{7}_{35}
$;\ \ 
$ -2-2  c^{1}_{7}
-c^{2}_{7}
$,
$ 1$;\ \ 
$ 2+c^{1}_{7}
+c^{2}_{7}
$)

Factors = $2_{\frac{14}{5},3.618}^{5,395}\boxtimes 5_{\frac{18}{7},35.34}^{7,101}$

\vskip 0.7ex
\hangindent=3em \hangafter=1
$\tau_n$ = ($-8.68 - 8.36 i$, $-21.99 - 48.01 i$, $-72.21 - 16.11 i$, $-2.87 + 42.14 i$, $28.45 + 57.16 i$, $7.81 + 0.42 i$, $67.59 - 87.98 i$, $-21.33 + 4.86 i$, $-28.88 + 2.78 i$, $-68.88 + 57.88 i$, $43.32 - 15.76 i$, $52.73 + 12.02 i$, $5.35 - 17.72 i$, $-35.84 - 54.37 i$, $-12.66 + 20.98 i$, $17.3 + 25.02 i$, $2.52 + 77.58 i$, $2.52 - 77.58 i$, $17.3 - 25.02 i$, $-12.66 - 20.98 i$, $-35.84 + 54.37 i$, $5.35 + 17.72 i$, $52.73 - 12.02 i$, $43.32 + 15.76 i$, $-68.88 - 57.88 i$, $-28.88 - 2.78 i$, $-21.33 - 4.86 i$, $67.59 + 87.98 i$, $7.81 - 0.42 i$, $28.45 - 57.16 i$, $-2.87 - 42.14 i$, $-72.21 + 16.11 i$, $-21.99 + 48.01 i$, $-8.68 + 8.36 i$, $131.51$)

\vskip 0.7ex
\hangindent=3em \hangafter=1
\textit{Intrinsic sign problem}

  \vskip 2ex

\noindent66. $10_{\frac{92}{35},127.8}^{35,112}$ \irep{1074}:\ \ 
$d_i$ = ($1.0$,
$1.618$,
$2.246$,
$2.246$,
$2.801$,
$3.635$,
$3.635$,
$4.48$,
$4.533$,
$6.551$) 

\vskip 0.7ex
\hangindent=3em \hangafter=1
$D^2= 127.870 = 
49+14c^{1}_{35}
+7c^{4}_{35}
+28c^{5}_{35}
+14c^{6}_{35}
+7c^{7}_{35}
+14c^{10}_{35}
+7c^{11}_{35}
$

\vskip 0.7ex
\hangindent=3em \hangafter=1
$T = ( 0,
\frac{3}{5},
\frac{1}{7},
\frac{1}{7},
\frac{6}{7},
\frac{26}{35},
\frac{26}{35},
\frac{4}{7},
\frac{16}{35},
\frac{6}{35} )
$,

\vskip 0.7ex
\hangindent=3em \hangafter=1
$S$ = ($ 1$,
$ \frac{1+\sqrt{5}}{2}$,
$ \xi_{7}^{3}$,
$ \xi_{7}^{3}$,
$ 2+c^{1}_{7}
+c^{2}_{7}
$,
$ c^{1}_{35}
+c^{4}_{35}
+c^{6}_{35}
+c^{11}_{35}
$,
$ c^{1}_{35}
+c^{4}_{35}
+c^{6}_{35}
+c^{11}_{35}
$,
$ 2+2c^{1}_{7}
+c^{2}_{7}
$,
$ 1+c^{1}_{35}
+c^{6}_{35}
+c^{7}_{35}
$,
$ 2c^{1}_{35}
+c^{4}_{35}
+2c^{6}_{35}
+c^{11}_{35}
$;\ \ 
$ -1$,
$ c^{1}_{35}
+c^{4}_{35}
+c^{6}_{35}
+c^{11}_{35}
$,
$ c^{1}_{35}
+c^{4}_{35}
+c^{6}_{35}
+c^{11}_{35}
$,
$ 1+c^{1}_{35}
+c^{6}_{35}
+c^{7}_{35}
$,
$ -\xi_{7}^{3}$,
$ -\xi_{7}^{3}$,
$ 2c^{1}_{35}
+c^{4}_{35}
+2c^{6}_{35}
+c^{11}_{35}
$,
$ -2-c^{1}_{7}
-c^{2}_{7}
$,
$ -2-2  c^{1}_{7}
-c^{2}_{7}
$;\ \ 
$ s^{1}_{7}
+\zeta^{2}_{7}
+\zeta^{3}_{7}
$,
$ -1-2  \zeta^{1}_{7}
-\zeta^{2}_{7}
-\zeta^{3}_{7}
$,
$ -\xi_{7}^{3}$,
$ 1-2  \zeta^{1}_{35}
-\zeta^{-1}_{35}
+2\zeta^{-2}_{35}
-\zeta^{4}_{35}
-2  \zeta^{-4}_{35}
+2\zeta^{5}_{35}
-\zeta^{6}_{35}
-2  \zeta^{-6}_{35}
+c^{7}_{35}
-\zeta^{11}_{35}
-2  \zeta^{-11}_{35}
+2\zeta^{12}_{35}
$,
$ -1+\zeta^{1}_{35}
-2  \zeta^{-2}_{35}
+\zeta^{-4}_{35}
-2  \zeta^{5}_{35}
+\zeta^{-6}_{35}
-c^{7}_{35}
+\zeta^{-11}_{35}
-2  \zeta^{12}_{35}
$,
$ \xi_{7}^{3}$,
$ -c^{1}_{35}
-c^{4}_{35}
-c^{6}_{35}
-c^{11}_{35}
$,
$ c^{1}_{35}
+c^{4}_{35}
+c^{6}_{35}
+c^{11}_{35}
$;\ \ 
$ s^{1}_{7}
+\zeta^{2}_{7}
+\zeta^{3}_{7}
$,
$ -\xi_{7}^{3}$,
$ -1+\zeta^{1}_{35}
-2  \zeta^{-2}_{35}
+\zeta^{-4}_{35}
-2  \zeta^{5}_{35}
+\zeta^{-6}_{35}
-c^{7}_{35}
+\zeta^{-11}_{35}
-2  \zeta^{12}_{35}
$,
$ 1-2  \zeta^{1}_{35}
-\zeta^{-1}_{35}
+2\zeta^{-2}_{35}
-\zeta^{4}_{35}
-2  \zeta^{-4}_{35}
+2\zeta^{5}_{35}
-\zeta^{6}_{35}
-2  \zeta^{-6}_{35}
+c^{7}_{35}
-\zeta^{11}_{35}
-2  \zeta^{-11}_{35}
+2\zeta^{12}_{35}
$,
$ \xi_{7}^{3}$,
$ -c^{1}_{35}
-c^{4}_{35}
-c^{6}_{35}
-c^{11}_{35}
$,
$ c^{1}_{35}
+c^{4}_{35}
+c^{6}_{35}
+c^{11}_{35}
$;\ \ 
$ 2+2c^{1}_{7}
+c^{2}_{7}
$,
$ -c^{1}_{35}
-c^{4}_{35}
-c^{6}_{35}
-c^{11}_{35}
$,
$ -c^{1}_{35}
-c^{4}_{35}
-c^{6}_{35}
-c^{11}_{35}
$,
$ -1$,
$ 2c^{1}_{35}
+c^{4}_{35}
+2c^{6}_{35}
+c^{11}_{35}
$,
$ -\frac{1+\sqrt{5}}{2}$;\ \ 
$ -s^{1}_{7}
-\zeta^{2}_{7}
-\zeta^{3}_{7}
$,
$ 1+2\zeta^{1}_{7}
+\zeta^{2}_{7}
+\zeta^{3}_{7}
$,
$ c^{1}_{35}
+c^{4}_{35}
+c^{6}_{35}
+c^{11}_{35}
$,
$ \xi_{7}^{3}$,
$ -\xi_{7}^{3}$;\ \ 
$ -s^{1}_{7}
-\zeta^{2}_{7}
-\zeta^{3}_{7}
$,
$ c^{1}_{35}
+c^{4}_{35}
+c^{6}_{35}
+c^{11}_{35}
$,
$ \xi_{7}^{3}$,
$ -\xi_{7}^{3}$;\ \ 
$ -2-c^{1}_{7}
-c^{2}_{7}
$,
$ -\frac{1+\sqrt{5}}{2}$,
$ -1-c^{1}_{35}
-c^{6}_{35}
-c^{7}_{35}
$;\ \ 
$ -2-2  c^{1}_{7}
-c^{2}_{7}
$,
$ 1$;\ \ 
$ 2+c^{1}_{7}
+c^{2}_{7}
$)

Factors = $2_{\frac{26}{5},3.618}^{5,720}\boxtimes 5_{\frac{38}{7},35.34}^{7,386}$

\vskip 0.7ex
\hangindent=3em \hangafter=1
$\tau_n$ = ($-8.68 + 8.36 i$, $-21.99 + 48.01 i$, $-72.21 + 16.11 i$, $-2.87 - 42.14 i$, $28.45 - 57.16 i$, $7.81 - 0.42 i$, $67.59 + 87.98 i$, $-21.33 - 4.86 i$, $-28.88 - 2.78 i$, $-68.88 - 57.88 i$, $43.32 + 15.76 i$, $52.73 - 12.02 i$, $5.35 + 17.72 i$, $-35.84 + 54.37 i$, $-12.66 - 20.98 i$, $17.3 - 25.02 i$, $2.52 - 77.58 i$, $2.52 + 77.58 i$, $17.3 + 25.02 i$, $-12.66 + 20.98 i$, $-35.84 - 54.37 i$, $5.35 - 17.72 i$, $52.73 + 12.02 i$, $43.32 - 15.76 i$, $-68.88 + 57.88 i$, $-28.88 + 2.78 i$, $-21.33 + 4.86 i$, $67.59 - 87.98 i$, $7.81 + 0.42 i$, $28.45 + 57.16 i$, $-2.87 + 42.14 i$, $-72.21 - 16.11 i$, $-21.99 - 48.01 i$, $-8.68 - 8.36 i$, $131.51$)

\vskip 0.7ex
\hangindent=3em \hangafter=1
\textit{Intrinsic sign problem}

  \vskip 2ex

\noindent67. $10_{\frac{272}{35},127.8}^{35,631}$ \irep{1074}:\ \ 
$d_i$ = ($1.0$,
$1.618$,
$2.246$,
$2.246$,
$2.801$,
$3.635$,
$3.635$,
$4.48$,
$4.533$,
$6.551$) 

\vskip 0.7ex
\hangindent=3em \hangafter=1
$D^2= 127.870 = 
49+14c^{1}_{35}
+7c^{4}_{35}
+28c^{5}_{35}
+14c^{6}_{35}
+7c^{7}_{35}
+14c^{10}_{35}
+7c^{11}_{35}
$

\vskip 0.7ex
\hangindent=3em \hangafter=1
$T = ( 0,
\frac{3}{5},
\frac{6}{7},
\frac{6}{7},
\frac{1}{7},
\frac{16}{35},
\frac{16}{35},
\frac{3}{7},
\frac{26}{35},
\frac{1}{35} )
$,

\vskip 0.7ex
\hangindent=3em \hangafter=1
$S$ = ($ 1$,
$ \frac{1+\sqrt{5}}{2}$,
$ \xi_{7}^{3}$,
$ \xi_{7}^{3}$,
$ 2+c^{1}_{7}
+c^{2}_{7}
$,
$ c^{1}_{35}
+c^{4}_{35}
+c^{6}_{35}
+c^{11}_{35}
$,
$ c^{1}_{35}
+c^{4}_{35}
+c^{6}_{35}
+c^{11}_{35}
$,
$ 2+2c^{1}_{7}
+c^{2}_{7}
$,
$ 1+c^{1}_{35}
+c^{6}_{35}
+c^{7}_{35}
$,
$ 2c^{1}_{35}
+c^{4}_{35}
+2c^{6}_{35}
+c^{11}_{35}
$;\ \ 
$ -1$,
$ c^{1}_{35}
+c^{4}_{35}
+c^{6}_{35}
+c^{11}_{35}
$,
$ c^{1}_{35}
+c^{4}_{35}
+c^{6}_{35}
+c^{11}_{35}
$,
$ 1+c^{1}_{35}
+c^{6}_{35}
+c^{7}_{35}
$,
$ -\xi_{7}^{3}$,
$ -\xi_{7}^{3}$,
$ 2c^{1}_{35}
+c^{4}_{35}
+2c^{6}_{35}
+c^{11}_{35}
$,
$ -2-c^{1}_{7}
-c^{2}_{7}
$,
$ -2-2  c^{1}_{7}
-c^{2}_{7}
$;\ \ 
$ -1-2  \zeta^{1}_{7}
-\zeta^{2}_{7}
-\zeta^{3}_{7}
$,
$ s^{1}_{7}
+\zeta^{2}_{7}
+\zeta^{3}_{7}
$,
$ -\xi_{7}^{3}$,
$ 1-2  \zeta^{1}_{35}
-\zeta^{-1}_{35}
+2\zeta^{-2}_{35}
-\zeta^{4}_{35}
-2  \zeta^{-4}_{35}
+2\zeta^{5}_{35}
-\zeta^{6}_{35}
-2  \zeta^{-6}_{35}
+c^{7}_{35}
-\zeta^{11}_{35}
-2  \zeta^{-11}_{35}
+2\zeta^{12}_{35}
$,
$ -1+\zeta^{1}_{35}
-2  \zeta^{-2}_{35}
+\zeta^{-4}_{35}
-2  \zeta^{5}_{35}
+\zeta^{-6}_{35}
-c^{7}_{35}
+\zeta^{-11}_{35}
-2  \zeta^{12}_{35}
$,
$ \xi_{7}^{3}$,
$ -c^{1}_{35}
-c^{4}_{35}
-c^{6}_{35}
-c^{11}_{35}
$,
$ c^{1}_{35}
+c^{4}_{35}
+c^{6}_{35}
+c^{11}_{35}
$;\ \ 
$ -1-2  \zeta^{1}_{7}
-\zeta^{2}_{7}
-\zeta^{3}_{7}
$,
$ -\xi_{7}^{3}$,
$ -1+\zeta^{1}_{35}
-2  \zeta^{-2}_{35}
+\zeta^{-4}_{35}
-2  \zeta^{5}_{35}
+\zeta^{-6}_{35}
-c^{7}_{35}
+\zeta^{-11}_{35}
-2  \zeta^{12}_{35}
$,
$ 1-2  \zeta^{1}_{35}
-\zeta^{-1}_{35}
+2\zeta^{-2}_{35}
-\zeta^{4}_{35}
-2  \zeta^{-4}_{35}
+2\zeta^{5}_{35}
-\zeta^{6}_{35}
-2  \zeta^{-6}_{35}
+c^{7}_{35}
-\zeta^{11}_{35}
-2  \zeta^{-11}_{35}
+2\zeta^{12}_{35}
$,
$ \xi_{7}^{3}$,
$ -c^{1}_{35}
-c^{4}_{35}
-c^{6}_{35}
-c^{11}_{35}
$,
$ c^{1}_{35}
+c^{4}_{35}
+c^{6}_{35}
+c^{11}_{35}
$;\ \ 
$ 2+2c^{1}_{7}
+c^{2}_{7}
$,
$ -c^{1}_{35}
-c^{4}_{35}
-c^{6}_{35}
-c^{11}_{35}
$,
$ -c^{1}_{35}
-c^{4}_{35}
-c^{6}_{35}
-c^{11}_{35}
$,
$ -1$,
$ 2c^{1}_{35}
+c^{4}_{35}
+2c^{6}_{35}
+c^{11}_{35}
$,
$ -\frac{1+\sqrt{5}}{2}$;\ \ 
$ 1+2\zeta^{1}_{7}
+\zeta^{2}_{7}
+\zeta^{3}_{7}
$,
$ -s^{1}_{7}
-\zeta^{2}_{7}
-\zeta^{3}_{7}
$,
$ c^{1}_{35}
+c^{4}_{35}
+c^{6}_{35}
+c^{11}_{35}
$,
$ \xi_{7}^{3}$,
$ -\xi_{7}^{3}$;\ \ 
$ 1+2\zeta^{1}_{7}
+\zeta^{2}_{7}
+\zeta^{3}_{7}
$,
$ c^{1}_{35}
+c^{4}_{35}
+c^{6}_{35}
+c^{11}_{35}
$,
$ \xi_{7}^{3}$,
$ -\xi_{7}^{3}$;\ \ 
$ -2-c^{1}_{7}
-c^{2}_{7}
$,
$ -\frac{1+\sqrt{5}}{2}$,
$ -1-c^{1}_{35}
-c^{6}_{35}
-c^{7}_{35}
$;\ \ 
$ -2-2  c^{1}_{7}
-c^{2}_{7}
$,
$ 1$;\ \ 
$ 2+c^{1}_{7}
+c^{2}_{7}
$)

Factors = $2_{\frac{26}{5},3.618}^{5,720}\boxtimes 5_{\frac{18}{7},35.34}^{7,101}$

\vskip 0.7ex
\hangindent=3em \hangafter=1
$\tau_n$ = ($7.81 - 0.42 i$, $52.73 - 12.02 i$, $2.52 + 77.58 i$, $43.32 - 15.76 i$, $28.45 + 57.16 i$, $-8.68 + 8.36 i$, $67.59 + 87.98 i$, $5.35 + 17.72 i$, $17.3 + 25.02 i$, $-68.88 + 57.88 i$, $-2.87 + 42.14 i$, $-21.99 + 48.01 i$, $-21.33 - 4.86 i$, $-35.84 + 54.37 i$, $-12.66 + 20.98 i$, $-28.88 + 2.78 i$, $-72.21 - 16.11 i$, $-72.21 + 16.11 i$, $-28.88 - 2.78 i$, $-12.66 - 20.98 i$, $-35.84 - 54.37 i$, $-21.33 + 4.86 i$, $-21.99 - 48.01 i$, $-2.87 - 42.14 i$, $-68.88 - 57.88 i$, $17.3 - 25.02 i$, $5.35 - 17.72 i$, $67.59 - 87.98 i$, $-8.68 - 8.36 i$, $28.45 - 57.16 i$, $43.32 + 15.76 i$, $2.52 - 77.58 i$, $52.73 + 12.02 i$, $7.81 + 0.42 i$, $131.51$)

\vskip 0.7ex
\hangindent=3em \hangafter=1
\textit{Intrinsic sign problem}

  \vskip 2ex

\noindent68. $10_{\frac{26}{7},236.3}^{21,145}$ \irep{960}:\ \ 
$d_i$ = ($1.0$,
$1.977$,
$2.911$,
$3.779$,
$4.563$,
$5.245$,
$5.810$,
$6.245$,
$6.541$,
$6.690$) 

\vskip 0.7ex
\hangindent=3em \hangafter=1
$D^2= 236.341 = 
42+42c^{1}_{21}
+42c^{2}_{21}
+21c^{3}_{21}
+21c^{4}_{21}
+21c^{5}_{21}
$

\vskip 0.7ex
\hangindent=3em \hangafter=1
$T = ( 0,
\frac{2}{7},
\frac{2}{21},
\frac{3}{7},
\frac{2}{7},
\frac{2}{3},
\frac{4}{7},
0,
\frac{20}{21},
\frac{3}{7} )
$,

\vskip 0.7ex
\hangindent=3em \hangafter=1
$S$ = ($ 1$,
$ -c_{21}^{10}$,
$ \xi_{21}^{3}$,
$ \xi_{21}^{17}$,
$ \xi_{21}^{5}$,
$ \xi_{21}^{15}$,
$ \xi_{21}^{7}$,
$ \xi_{21}^{13}$,
$ \xi_{21}^{9}$,
$ \xi_{21}^{11}$;\ \ 
$ -\xi_{21}^{17}$,
$ \xi_{21}^{15}$,
$ -\xi_{21}^{13}$,
$ \xi_{21}^{11}$,
$ -\xi_{21}^{9}$,
$ \xi_{21}^{7}$,
$ -\xi_{21}^{5}$,
$ \xi_{21}^{3}$,
$ -1$;\ \ 
$ \xi_{21}^{9}$,
$ \xi_{21}^{9}$,
$ \xi_{21}^{15}$,
$ \xi_{21}^{3}$,
$0$,
$ -\xi_{21}^{3}$,
$ -\xi_{21}^{15}$,
$ -\xi_{21}^{9}$;\ \ 
$ -\xi_{21}^{5}$,
$ 1$,
$ \xi_{21}^{3}$,
$ -\xi_{21}^{7}$,
$ \xi_{21}^{11}$,
$ -\xi_{21}^{15}$,
$ -c_{21}^{10}$;\ \ 
$ -\xi_{21}^{17}$,
$ -\xi_{21}^{9}$,
$ -\xi_{21}^{7}$,
$ c_{21}^{10}$,
$ \xi_{21}^{3}$,
$ \xi_{21}^{13}$;\ \ 
$ \xi_{21}^{15}$,
$0$,
$ -\xi_{21}^{15}$,
$ \xi_{21}^{9}$,
$ -\xi_{21}^{3}$;\ \ 
$ \xi_{21}^{7}$,
$ \xi_{21}^{7}$,
$0$,
$ -\xi_{21}^{7}$;\ \ 
$ 1$,
$ -\xi_{21}^{9}$,
$ \xi_{21}^{17}$;\ \ 
$ -\xi_{21}^{3}$,
$ \xi_{21}^{15}$;\ \ 
$ -\xi_{21}^{5}$)

\vskip 0.7ex
\hangindent=3em \hangafter=1
$\tau_n$ = ($-14.98 + 3.42 i$, $100.27 - 22.88 i$, $87.07 - 19.88 i$, $30.43 - 63.2 i$, $56.64 + 12.92 i$, $-38.75 - 80.47 i$, $118.14 - 68.22 i$, $-93.59 + 21.35 i$, $69.82 - 55.68 i$, $-13.19 - 27.39 i$, $-13.19 + 27.39 i$, $69.82 + 55.68 i$, $-93.59 - 21.35 i$, $118.14 + 68.22 i$, $-38.75 + 80.47 i$, $56.64 - 12.92 i$, $30.43 + 63.2 i$, $87.07 + 19.88 i$, $100.27 + 22.88 i$, $-14.98 - 3.42 i$, $236.29$)

\vskip 0.7ex
\hangindent=3em \hangafter=1
\textit{Intrinsic sign problem}

  \vskip 2ex

\noindent69. $10_{\frac{30}{7},236.3}^{21,387}$ \irep{960}:\ \ 
$d_i$ = ($1.0$,
$1.977$,
$2.911$,
$3.779$,
$4.563$,
$5.245$,
$5.810$,
$6.245$,
$6.541$,
$6.690$) 

\vskip 0.7ex
\hangindent=3em \hangafter=1
$D^2= 236.341 = 
42+42c^{1}_{21}
+42c^{2}_{21}
+21c^{3}_{21}
+21c^{4}_{21}
+21c^{5}_{21}
$

\vskip 0.7ex
\hangindent=3em \hangafter=1
$T = ( 0,
\frac{5}{7},
\frac{19}{21},
\frac{4}{7},
\frac{5}{7},
\frac{1}{3},
\frac{3}{7},
0,
\frac{1}{21},
\frac{4}{7} )
$,

\vskip 0.7ex
\hangindent=3em \hangafter=1
$S$ = ($ 1$,
$ -c_{21}^{10}$,
$ \xi_{21}^{3}$,
$ \xi_{21}^{17}$,
$ \xi_{21}^{5}$,
$ \xi_{21}^{15}$,
$ \xi_{21}^{7}$,
$ \xi_{21}^{13}$,
$ \xi_{21}^{9}$,
$ \xi_{21}^{11}$;\ \ 
$ -\xi_{21}^{17}$,
$ \xi_{21}^{15}$,
$ -\xi_{21}^{13}$,
$ \xi_{21}^{11}$,
$ -\xi_{21}^{9}$,
$ \xi_{21}^{7}$,
$ -\xi_{21}^{5}$,
$ \xi_{21}^{3}$,
$ -1$;\ \ 
$ \xi_{21}^{9}$,
$ \xi_{21}^{9}$,
$ \xi_{21}^{15}$,
$ \xi_{21}^{3}$,
$0$,
$ -\xi_{21}^{3}$,
$ -\xi_{21}^{15}$,
$ -\xi_{21}^{9}$;\ \ 
$ -\xi_{21}^{5}$,
$ 1$,
$ \xi_{21}^{3}$,
$ -\xi_{21}^{7}$,
$ \xi_{21}^{11}$,
$ -\xi_{21}^{15}$,
$ -c_{21}^{10}$;\ \ 
$ -\xi_{21}^{17}$,
$ -\xi_{21}^{9}$,
$ -\xi_{21}^{7}$,
$ c_{21}^{10}$,
$ \xi_{21}^{3}$,
$ \xi_{21}^{13}$;\ \ 
$ \xi_{21}^{15}$,
$0$,
$ -\xi_{21}^{15}$,
$ \xi_{21}^{9}$,
$ -\xi_{21}^{3}$;\ \ 
$ \xi_{21}^{7}$,
$ \xi_{21}^{7}$,
$0$,
$ -\xi_{21}^{7}$;\ \ 
$ 1$,
$ -\xi_{21}^{9}$,
$ \xi_{21}^{17}$;\ \ 
$ -\xi_{21}^{3}$,
$ \xi_{21}^{15}$;\ \ 
$ -\xi_{21}^{5}$)

\vskip 0.7ex
\hangindent=3em \hangafter=1
$\tau_n$ = ($-14.98 - 3.42 i$, $100.27 + 22.88 i$, $87.07 + 19.88 i$, $30.43 + 63.2 i$, $56.64 - 12.92 i$, $-38.75 + 80.47 i$, $118.14 + 68.22 i$, $-93.59 - 21.35 i$, $69.82 + 55.68 i$, $-13.19 + 27.39 i$, $-13.19 - 27.39 i$, $69.82 - 55.68 i$, $-93.59 + 21.35 i$, $118.14 - 68.22 i$, $-38.75 - 80.47 i$, $56.64 + 12.92 i$, $30.43 - 63.2 i$, $87.07 - 19.88 i$, $100.27 - 22.88 i$, $-14.98 + 3.42 i$, $236.29$)

\vskip 0.7ex
\hangindent=3em \hangafter=1
\textit{Intrinsic sign problem}

  \vskip 2ex

\noindent70. $10_{\frac{48}{17},499.2}^{17,522}$ \irep{888}:\ \ 
$d_i$ = ($1.0$,
$2.965$,
$4.830$,
$5.418$,
$5.418$,
$6.531$,
$8.9$,
$9.214$,
$10.106$,
$10.653$) 

\vskip 0.7ex
\hangindent=3em \hangafter=1
$D^2= 499.210 = 
136+119c^{1}_{17}
+102c^{2}_{17}
+85c^{3}_{17}
+68c^{4}_{17}
+51c^{5}_{17}
+34c^{6}_{17}
+17c^{7}_{17}
$

\vskip 0.7ex
\hangindent=3em \hangafter=1
$T = ( 0,
\frac{1}{17},
\frac{3}{17},
\frac{2}{17},
\frac{2}{17},
\frac{6}{17},
\frac{10}{17},
\frac{15}{17},
\frac{4}{17},
\frac{11}{17} )
$,

\vskip 0.7ex
\hangindent=3em \hangafter=1
$S$ = ($ 1$,
$ 2+c^{1}_{17}
+c^{2}_{17}
+c^{3}_{17}
+c^{4}_{17}
+c^{5}_{17}
+c^{6}_{17}
+c^{7}_{17}
$,
$ 2+2c^{1}_{17}
+c^{2}_{17}
+c^{3}_{17}
+c^{4}_{17}
+c^{5}_{17}
+c^{6}_{17}
+c^{7}_{17}
$,
$ \xi_{17}^{9}$,
$ \xi_{17}^{9}$,
$ 2+2c^{1}_{17}
+c^{2}_{17}
+c^{3}_{17}
+c^{4}_{17}
+c^{5}_{17}
+c^{6}_{17}
$,
$ 2+2c^{1}_{17}
+2c^{2}_{17}
+c^{3}_{17}
+c^{4}_{17}
+c^{5}_{17}
+c^{6}_{17}
$,
$ 2+2c^{1}_{17}
+2c^{2}_{17}
+c^{3}_{17}
+c^{4}_{17}
+c^{5}_{17}
$,
$ 2+2c^{1}_{17}
+2c^{2}_{17}
+2c^{3}_{17}
+c^{4}_{17}
+c^{5}_{17}
$,
$ 2+2c^{1}_{17}
+2c^{2}_{17}
+2c^{3}_{17}
+c^{4}_{17}
$;\ \ 
$ 2+2c^{1}_{17}
+2c^{2}_{17}
+c^{3}_{17}
+c^{4}_{17}
+c^{5}_{17}
+c^{6}_{17}
$,
$ 2+2c^{1}_{17}
+2c^{2}_{17}
+2c^{3}_{17}
+c^{4}_{17}
$,
$ -\xi_{17}^{9}$,
$ -\xi_{17}^{9}$,
$ 2+2c^{1}_{17}
+2c^{2}_{17}
+2c^{3}_{17}
+c^{4}_{17}
+c^{5}_{17}
$,
$ 2+2c^{1}_{17}
+c^{2}_{17}
+c^{3}_{17}
+c^{4}_{17}
+c^{5}_{17}
+c^{6}_{17}
$,
$ 1$,
$ -2-2  c^{1}_{17}
-c^{2}_{17}
-c^{3}_{17}
-c^{4}_{17}
-c^{5}_{17}
-c^{6}_{17}
-c^{7}_{17}
$,
$ -2-2  c^{1}_{17}
-2  c^{2}_{17}
-c^{3}_{17}
-c^{4}_{17}
-c^{5}_{17}
$;\ \ 
$ 2+2c^{1}_{17}
+2c^{2}_{17}
+c^{3}_{17}
+c^{4}_{17}
+c^{5}_{17}
+c^{6}_{17}
$,
$ \xi_{17}^{9}$,
$ \xi_{17}^{9}$,
$ -1$,
$ -2-2  c^{1}_{17}
-2  c^{2}_{17}
-c^{3}_{17}
-c^{4}_{17}
-c^{5}_{17}
$,
$ -2-2  c^{1}_{17}
-2  c^{2}_{17}
-2  c^{3}_{17}
-c^{4}_{17}
-c^{5}_{17}
$,
$ -2-c^{1}_{17}
-c^{2}_{17}
-c^{3}_{17}
-c^{4}_{17}
-c^{5}_{17}
-c^{6}_{17}
-c^{7}_{17}
$,
$ 2+2c^{1}_{17}
+c^{2}_{17}
+c^{3}_{17}
+c^{4}_{17}
+c^{5}_{17}
+c^{6}_{17}
$;\ \ 
$ 4+3c^{1}_{17}
+3c^{2}_{17}
+2c^{3}_{17}
+2c^{4}_{17}
+2c^{5}_{17}
+c^{6}_{17}
$,
$ -3-2  c^{1}_{17}
-2  c^{2}_{17}
-c^{3}_{17}
-c^{4}_{17}
-2  c^{5}_{17}
-c^{6}_{17}
$,
$ -\xi_{17}^{9}$,
$ \xi_{17}^{9}$,
$ -\xi_{17}^{9}$,
$ \xi_{17}^{9}$,
$ -\xi_{17}^{9}$;\ \ 
$ 4+3c^{1}_{17}
+3c^{2}_{17}
+2c^{3}_{17}
+2c^{4}_{17}
+2c^{5}_{17}
+c^{6}_{17}
$,
$ -\xi_{17}^{9}$,
$ \xi_{17}^{9}$,
$ -\xi_{17}^{9}$,
$ \xi_{17}^{9}$,
$ -\xi_{17}^{9}$;\ \ 
$ -2-2  c^{1}_{17}
-2  c^{2}_{17}
-2  c^{3}_{17}
-c^{4}_{17}
$,
$ -2-2  c^{1}_{17}
-c^{2}_{17}
-c^{3}_{17}
-c^{4}_{17}
-c^{5}_{17}
-c^{6}_{17}
-c^{7}_{17}
$,
$ 2+2c^{1}_{17}
+2c^{2}_{17}
+c^{3}_{17}
+c^{4}_{17}
+c^{5}_{17}
+c^{6}_{17}
$,
$ 2+2c^{1}_{17}
+2c^{2}_{17}
+c^{3}_{17}
+c^{4}_{17}
+c^{5}_{17}
$,
$ -2-c^{1}_{17}
-c^{2}_{17}
-c^{3}_{17}
-c^{4}_{17}
-c^{5}_{17}
-c^{6}_{17}
-c^{7}_{17}
$;\ \ 
$ 2+2c^{1}_{17}
+2c^{2}_{17}
+2c^{3}_{17}
+c^{4}_{17}
+c^{5}_{17}
$,
$ 2+c^{1}_{17}
+c^{2}_{17}
+c^{3}_{17}
+c^{4}_{17}
+c^{5}_{17}
+c^{6}_{17}
+c^{7}_{17}
$,
$ -2-2  c^{1}_{17}
-2  c^{2}_{17}
-2  c^{3}_{17}
-c^{4}_{17}
$,
$ -1$;\ \ 
$ -2-2  c^{1}_{17}
-2  c^{2}_{17}
-2  c^{3}_{17}
-c^{4}_{17}
$,
$ 2+2c^{1}_{17}
+c^{2}_{17}
+c^{3}_{17}
+c^{4}_{17}
+c^{5}_{17}
+c^{6}_{17}
$,
$ 2+2c^{1}_{17}
+c^{2}_{17}
+c^{3}_{17}
+c^{4}_{17}
+c^{5}_{17}
+c^{6}_{17}
+c^{7}_{17}
$;\ \ 
$ 1$,
$ -2-2  c^{1}_{17}
-2  c^{2}_{17}
-c^{3}_{17}
-c^{4}_{17}
-c^{5}_{17}
-c^{6}_{17}
$;\ \ 
$ 2+2c^{1}_{17}
+2c^{2}_{17}
+2c^{3}_{17}
+c^{4}_{17}
+c^{5}_{17}
$)

\vskip 0.7ex
\hangindent=3em \hangafter=1
$\tau_n$ = ($-26.29 + 9.89 i$, $-101.13 + 156.26 i$, $20.4 - 220.01 i$, $-230.99 - 29.47 i$, $27.53 + 139.87 i$, $-71.15 + 32.12 i$, $-96.8 + 10.15 i$, $229.83 - 58.23 i$, $229.83 + 58.23 i$, $-96.8 - 10.15 i$, $-71.15 - 32.12 i$, $27.53 - 139.87 i$, $-230.99 + 29.47 i$, $20.4 + 220.01 i$, $-101.13 - 156.26 i$, $-26.29 - 9.89 i$, $514.21$)

\vskip 0.7ex
\hangindent=3em \hangafter=1
\textit{Intrinsic sign problem}

  \vskip 2ex

\noindent71. $10_{\frac{88}{17},499.2}^{17,976}$ \irep{888}:\ \ 
$d_i$ = ($1.0$,
$2.965$,
$4.830$,
$5.418$,
$5.418$,
$6.531$,
$8.9$,
$9.214$,
$10.106$,
$10.653$) 

\vskip 0.7ex
\hangindent=3em \hangafter=1
$D^2= 499.210 = 
136+119c^{1}_{17}
+102c^{2}_{17}
+85c^{3}_{17}
+68c^{4}_{17}
+51c^{5}_{17}
+34c^{6}_{17}
+17c^{7}_{17}
$

\vskip 0.7ex
\hangindent=3em \hangafter=1
$T = ( 0,
\frac{16}{17},
\frac{14}{17},
\frac{15}{17},
\frac{15}{17},
\frac{11}{17},
\frac{7}{17},
\frac{2}{17},
\frac{13}{17},
\frac{6}{17} )
$,

\vskip 0.7ex
\hangindent=3em \hangafter=1
$S$ = ($ 1$,
$ 2+c^{1}_{17}
+c^{2}_{17}
+c^{3}_{17}
+c^{4}_{17}
+c^{5}_{17}
+c^{6}_{17}
+c^{7}_{17}
$,
$ 2+2c^{1}_{17}
+c^{2}_{17}
+c^{3}_{17}
+c^{4}_{17}
+c^{5}_{17}
+c^{6}_{17}
+c^{7}_{17}
$,
$ \xi_{17}^{9}$,
$ \xi_{17}^{9}$,
$ 2+2c^{1}_{17}
+c^{2}_{17}
+c^{3}_{17}
+c^{4}_{17}
+c^{5}_{17}
+c^{6}_{17}
$,
$ 2+2c^{1}_{17}
+2c^{2}_{17}
+c^{3}_{17}
+c^{4}_{17}
+c^{5}_{17}
+c^{6}_{17}
$,
$ 2+2c^{1}_{17}
+2c^{2}_{17}
+c^{3}_{17}
+c^{4}_{17}
+c^{5}_{17}
$,
$ 2+2c^{1}_{17}
+2c^{2}_{17}
+2c^{3}_{17}
+c^{4}_{17}
+c^{5}_{17}
$,
$ 2+2c^{1}_{17}
+2c^{2}_{17}
+2c^{3}_{17}
+c^{4}_{17}
$;\ \ 
$ 2+2c^{1}_{17}
+2c^{2}_{17}
+c^{3}_{17}
+c^{4}_{17}
+c^{5}_{17}
+c^{6}_{17}
$,
$ 2+2c^{1}_{17}
+2c^{2}_{17}
+2c^{3}_{17}
+c^{4}_{17}
$,
$ -\xi_{17}^{9}$,
$ -\xi_{17}^{9}$,
$ 2+2c^{1}_{17}
+2c^{2}_{17}
+2c^{3}_{17}
+c^{4}_{17}
+c^{5}_{17}
$,
$ 2+2c^{1}_{17}
+c^{2}_{17}
+c^{3}_{17}
+c^{4}_{17}
+c^{5}_{17}
+c^{6}_{17}
$,
$ 1$,
$ -2-2  c^{1}_{17}
-c^{2}_{17}
-c^{3}_{17}
-c^{4}_{17}
-c^{5}_{17}
-c^{6}_{17}
-c^{7}_{17}
$,
$ -2-2  c^{1}_{17}
-2  c^{2}_{17}
-c^{3}_{17}
-c^{4}_{17}
-c^{5}_{17}
$;\ \ 
$ 2+2c^{1}_{17}
+2c^{2}_{17}
+c^{3}_{17}
+c^{4}_{17}
+c^{5}_{17}
+c^{6}_{17}
$,
$ \xi_{17}^{9}$,
$ \xi_{17}^{9}$,
$ -1$,
$ -2-2  c^{1}_{17}
-2  c^{2}_{17}
-c^{3}_{17}
-c^{4}_{17}
-c^{5}_{17}
$,
$ -2-2  c^{1}_{17}
-2  c^{2}_{17}
-2  c^{3}_{17}
-c^{4}_{17}
-c^{5}_{17}
$,
$ -2-c^{1}_{17}
-c^{2}_{17}
-c^{3}_{17}
-c^{4}_{17}
-c^{5}_{17}
-c^{6}_{17}
-c^{7}_{17}
$,
$ 2+2c^{1}_{17}
+c^{2}_{17}
+c^{3}_{17}
+c^{4}_{17}
+c^{5}_{17}
+c^{6}_{17}
$;\ \ 
$ 4+3c^{1}_{17}
+3c^{2}_{17}
+2c^{3}_{17}
+2c^{4}_{17}
+2c^{5}_{17}
+c^{6}_{17}
$,
$ -3-2  c^{1}_{17}
-2  c^{2}_{17}
-c^{3}_{17}
-c^{4}_{17}
-2  c^{5}_{17}
-c^{6}_{17}
$,
$ -\xi_{17}^{9}$,
$ \xi_{17}^{9}$,
$ -\xi_{17}^{9}$,
$ \xi_{17}^{9}$,
$ -\xi_{17}^{9}$;\ \ 
$ 4+3c^{1}_{17}
+3c^{2}_{17}
+2c^{3}_{17}
+2c^{4}_{17}
+2c^{5}_{17}
+c^{6}_{17}
$,
$ -\xi_{17}^{9}$,
$ \xi_{17}^{9}$,
$ -\xi_{17}^{9}$,
$ \xi_{17}^{9}$,
$ -\xi_{17}^{9}$;\ \ 
$ -2-2  c^{1}_{17}
-2  c^{2}_{17}
-2  c^{3}_{17}
-c^{4}_{17}
$,
$ -2-2  c^{1}_{17}
-c^{2}_{17}
-c^{3}_{17}
-c^{4}_{17}
-c^{5}_{17}
-c^{6}_{17}
-c^{7}_{17}
$,
$ 2+2c^{1}_{17}
+2c^{2}_{17}
+c^{3}_{17}
+c^{4}_{17}
+c^{5}_{17}
+c^{6}_{17}
$,
$ 2+2c^{1}_{17}
+2c^{2}_{17}
+c^{3}_{17}
+c^{4}_{17}
+c^{5}_{17}
$,
$ -2-c^{1}_{17}
-c^{2}_{17}
-c^{3}_{17}
-c^{4}_{17}
-c^{5}_{17}
-c^{6}_{17}
-c^{7}_{17}
$;\ \ 
$ 2+2c^{1}_{17}
+2c^{2}_{17}
+2c^{3}_{17}
+c^{4}_{17}
+c^{5}_{17}
$,
$ 2+c^{1}_{17}
+c^{2}_{17}
+c^{3}_{17}
+c^{4}_{17}
+c^{5}_{17}
+c^{6}_{17}
+c^{7}_{17}
$,
$ -2-2  c^{1}_{17}
-2  c^{2}_{17}
-2  c^{3}_{17}
-c^{4}_{17}
$,
$ -1$;\ \ 
$ -2-2  c^{1}_{17}
-2  c^{2}_{17}
-2  c^{3}_{17}
-c^{4}_{17}
$,
$ 2+2c^{1}_{17}
+c^{2}_{17}
+c^{3}_{17}
+c^{4}_{17}
+c^{5}_{17}
+c^{6}_{17}
$,
$ 2+2c^{1}_{17}
+c^{2}_{17}
+c^{3}_{17}
+c^{4}_{17}
+c^{5}_{17}
+c^{6}_{17}
+c^{7}_{17}
$;\ \ 
$ 1$,
$ -2-2  c^{1}_{17}
-2  c^{2}_{17}
-c^{3}_{17}
-c^{4}_{17}
-c^{5}_{17}
-c^{6}_{17}
$;\ \ 
$ 2+2c^{1}_{17}
+2c^{2}_{17}
+2c^{3}_{17}
+c^{4}_{17}
+c^{5}_{17}
$)

\vskip 0.7ex
\hangindent=3em \hangafter=1
$\tau_n$ = ($-26.29 - 9.89 i$, $-101.13 - 156.26 i$, $20.4 + 220.01 i$, $-230.99 + 29.47 i$, $27.53 - 139.87 i$, $-71.15 - 32.12 i$, $-96.8 - 10.15 i$, $229.83 + 58.23 i$, $229.83 - 58.23 i$, $-96.8 + 10.15 i$, $-71.15 + 32.12 i$, $27.53 + 139.87 i$, $-230.99 - 29.47 i$, $20.4 - 220.01 i$, $-101.13 + 156.26 i$, $-26.29 + 9.89 i$, $514.21$)

\vskip 0.7ex
\hangindent=3em \hangafter=1
\textit{Intrinsic sign problem}

  \vskip 2ex

\noindent72. $10_{0,537.4}^{14,352}$ \irep{783}:\ \ 
$d_i$ = ($1.0$,
$3.493$,
$4.493$,
$4.493$,
$5.603$,
$5.603$,
$9.97$,
$10.97$,
$10.97$,
$11.591$) 

\vskip 0.7ex
\hangindent=3em \hangafter=1
$D^2= 537.478 = 
308+224c^{1}_{7}
+112c^{2}_{7}
$

\vskip 0.7ex
\hangindent=3em \hangafter=1
$T = ( 0,
0,
\frac{2}{7},
\frac{5}{7},
\frac{3}{7},
\frac{4}{7},
0,
\frac{1}{7},
\frac{6}{7},
\frac{1}{2} )
$,

\vskip 0.7ex
\hangindent=3em \hangafter=1
$S$ = ($ 1$,
$ 1+2c^{1}_{7}
$,
$ 2\xi_{7}^{3}$,
$ 2\xi_{7}^{3}$,
$ 4+2c^{1}_{7}
+2c^{2}_{7}
$,
$ 4+2c^{1}_{7}
+2c^{2}_{7}
$,
$ 5+4c^{1}_{7}
+2c^{2}_{7}
$,
$ 6+4c^{1}_{7}
+2c^{2}_{7}
$,
$ 6+4c^{1}_{7}
+2c^{2}_{7}
$,
$ 5+6c^{1}_{7}
+2c^{2}_{7}
$;\ \ 
$ -5-4  c^{1}_{7}
-2  c^{2}_{7}
$,
$ -4-2  c^{1}_{7}
-2  c^{2}_{7}
$,
$ -4-2  c^{1}_{7}
-2  c^{2}_{7}
$,
$ 6+4c^{1}_{7}
+2c^{2}_{7}
$,
$ 6+4c^{1}_{7}
+2c^{2}_{7}
$,
$ 1$,
$ 2\xi_{7}^{3}$,
$ 2\xi_{7}^{3}$,
$ -5-6  c^{1}_{7}
-2  c^{2}_{7}
$;\ \ 
$ -2\xi_{7}^{3}$,
$ 6+6c^{1}_{7}
+2c^{2}_{7}
$,
$ 4+4c^{1}_{7}
+2c^{2}_{7}
$,
$ -4-2  c^{1}_{7}
-2  c^{2}_{7}
$,
$ 6+4c^{1}_{7}
+2c^{2}_{7}
$,
$ -6-4  c^{1}_{7}
-2  c^{2}_{7}
$,
$ -2c_{7}^{1}$,
$0$;\ \ 
$ -2\xi_{7}^{3}$,
$ -4-2  c^{1}_{7}
-2  c^{2}_{7}
$,
$ 4+4c^{1}_{7}
+2c^{2}_{7}
$,
$ 6+4c^{1}_{7}
+2c^{2}_{7}
$,
$ -2c_{7}^{1}$,
$ -6-4  c^{1}_{7}
-2  c^{2}_{7}
$,
$0$;\ \ 
$ 6+4c^{1}_{7}
+2c^{2}_{7}
$,
$ 2c_{7}^{1}$,
$ -2\xi_{7}^{3}$,
$ -6-6  c^{1}_{7}
-2  c^{2}_{7}
$,
$ 2\xi_{7}^{3}$,
$0$;\ \ 
$ 6+4c^{1}_{7}
+2c^{2}_{7}
$,
$ -2\xi_{7}^{3}$,
$ 2\xi_{7}^{3}$,
$ -6-6  c^{1}_{7}
-2  c^{2}_{7}
$,
$0$;\ \ 
$ -1-2  c^{1}_{7}
$,
$ 4+2c^{1}_{7}
+2c^{2}_{7}
$,
$ 4+2c^{1}_{7}
+2c^{2}_{7}
$,
$ -5-6  c^{1}_{7}
-2  c^{2}_{7}
$;\ \ 
$ -4-2  c^{1}_{7}
-2  c^{2}_{7}
$,
$ 4+4c^{1}_{7}
+2c^{2}_{7}
$,
$0$;\ \ 
$ -4-2  c^{1}_{7}
-2  c^{2}_{7}
$,
$0$;\ \ 
$ 5+6c^{1}_{7}
+2c^{2}_{7}
$)

\vskip 0.7ex
\hangindent=3em \hangafter=1
$\tau_n$ = ($62.76$, $196.17$, $-227.39$, $41.31$, $-72.53$, $331.46$, $322.09$, $331.46$, $-72.53$, $41.31$, $-227.39$, $196.17$, $62.76$, $590.8$)

\vskip 0.7ex
\hangindent=3em \hangafter=1
\textit{Intrinsic sign problem}

  \vskip 2ex

\noindent73. $10_{6,684.3}^{77,298}$ \irep{1138}:\ \ 
$d_i$ = ($1.0$,
$7.887$,
$7.887$,
$7.887$,
$7.887$,
$7.887$,
$8.887$,
$9.887$,
$9.887$,
$9.887$) 

\vskip 0.7ex
\hangindent=3em \hangafter=1
$D^2= 684.336 = 
\frac{693+77\sqrt{77}}{2}$

\vskip 0.7ex
\hangindent=3em \hangafter=1
$T = ( 0,
\frac{1}{11},
\frac{3}{11},
\frac{4}{11},
\frac{5}{11},
\frac{9}{11},
0,
\frac{3}{7},
\frac{5}{7},
\frac{6}{7} )
$,

\vskip 0.7ex
\hangindent=3em \hangafter=1
$S$ = ($ 1$,
$ \frac{7+\sqrt{77}}{2}$,
$ \frac{7+\sqrt{77}}{2}$,
$ \frac{7+\sqrt{77}}{2}$,
$ \frac{7+\sqrt{77}}{2}$,
$ \frac{7+\sqrt{77}}{2}$,
$ \frac{9+\sqrt{77}}{2}$,
$ \frac{11+\sqrt{77}}{2}$,
$ \frac{11+\sqrt{77}}{2}$,
$ \frac{11+\sqrt{77}}{2}$;\ \ 
$ -1-2  c^{1}_{77}
+c^{2}_{77}
-c^{3}_{77}
-c^{4}_{77}
+c^{5}_{77}
+2c^{6}_{77}
-c^{7}_{77}
-c^{8}_{77}
+c^{9}_{77}
-2  c^{10}_{77}
-c^{11}_{77}
+c^{12}_{77}
-4  c^{14}_{77}
-c^{15}_{77}
+3c^{16}_{77}
+2c^{17}_{77}
-c^{18}_{77}
+c^{19}_{77}
-c^{21}_{77}
-c^{22}_{77}
-c^{23}_{77}
+c^{28}_{77}
-c^{29}_{77}
$,
$ -2+2c^{2}_{77}
-2  c^{3}_{77}
-2  c^{4}_{77}
+2c^{5}_{77}
-2  c^{6}_{77}
-6  c^{7}_{77}
-2  c^{8}_{77}
+2c^{9}_{77}
-2  c^{11}_{77}
+2c^{12}_{77}
-c^{14}_{77}
-2  c^{15}_{77}
-2  c^{17}_{77}
-3  c^{18}_{77}
+2c^{19}_{77}
-2  c^{22}_{77}
+2c^{23}_{77}
-c^{26}_{77}
-c^{28}_{77}
-3  c^{29}_{77}
$,
$ 1+2c^{1}_{77}
+c^{6}_{77}
+c^{7}_{77}
-2  c^{9}_{77}
+2c^{10}_{77}
-2  c^{13}_{77}
+c^{14}_{77}
+c^{16}_{77}
+c^{17}_{77}
+2c^{21}_{77}
+2c^{23}_{77}
-2  c^{24}_{77}
-2  c^{28}_{77}
$,
$ -1+c^{1}_{77}
+2c^{9}_{77}
+c^{10}_{77}
+2c^{13}_{77}
-c^{14}_{77}
+2c^{18}_{77}
-4  c^{21}_{77}
+c^{23}_{77}
+2c^{24}_{77}
+2c^{26}_{77}
-c^{28}_{77}
+2c^{29}_{77}
$,
$ 5-2  c^{2}_{77}
+2c^{3}_{77}
+2c^{4}_{77}
-2  c^{5}_{77}
+4c^{7}_{77}
+2c^{8}_{77}
-c^{9}_{77}
+2c^{11}_{77}
-2  c^{12}_{77}
+c^{13}_{77}
+4c^{14}_{77}
+2c^{15}_{77}
-2  c^{16}_{77}
-2  c^{19}_{77}
+3c^{21}_{77}
+2c^{22}_{77}
-2  c^{23}_{77}
+c^{24}_{77}
-2  c^{26}_{77}
+3c^{28}_{77}
$,
$ -\frac{7+\sqrt{77}}{2}$,
$0$,
$0$,
$0$;\ \ 
$ 5-2  c^{2}_{77}
+2c^{3}_{77}
+2c^{4}_{77}
-2  c^{5}_{77}
+4c^{7}_{77}
+2c^{8}_{77}
-c^{9}_{77}
+2c^{11}_{77}
-2  c^{12}_{77}
+c^{13}_{77}
+4c^{14}_{77}
+2c^{15}_{77}
-2  c^{16}_{77}
-2  c^{19}_{77}
+3c^{21}_{77}
+2c^{22}_{77}
-2  c^{23}_{77}
+c^{24}_{77}
-2  c^{26}_{77}
+3c^{28}_{77}
$,
$ -1-2  c^{1}_{77}
+c^{2}_{77}
-c^{3}_{77}
-c^{4}_{77}
+c^{5}_{77}
+2c^{6}_{77}
-c^{7}_{77}
-c^{8}_{77}
+c^{9}_{77}
-2  c^{10}_{77}
-c^{11}_{77}
+c^{12}_{77}
-4  c^{14}_{77}
-c^{15}_{77}
+3c^{16}_{77}
+2c^{17}_{77}
-c^{18}_{77}
+c^{19}_{77}
-c^{21}_{77}
-c^{22}_{77}
-c^{23}_{77}
+c^{28}_{77}
-c^{29}_{77}
$,
$ 1+2c^{1}_{77}
+c^{6}_{77}
+c^{7}_{77}
-2  c^{9}_{77}
+2c^{10}_{77}
-2  c^{13}_{77}
+c^{14}_{77}
+c^{16}_{77}
+c^{17}_{77}
+2c^{21}_{77}
+2c^{23}_{77}
-2  c^{24}_{77}
-2  c^{28}_{77}
$,
$ -1+c^{1}_{77}
+2c^{9}_{77}
+c^{10}_{77}
+2c^{13}_{77}
-c^{14}_{77}
+2c^{18}_{77}
-4  c^{21}_{77}
+c^{23}_{77}
+2c^{24}_{77}
+2c^{26}_{77}
-c^{28}_{77}
+2c^{29}_{77}
$,
$ -\frac{7+\sqrt{77}}{2}$,
$0$,
$0$,
$0$;\ \ 
$ -1+c^{1}_{77}
+2c^{9}_{77}
+c^{10}_{77}
+2c^{13}_{77}
-c^{14}_{77}
+2c^{18}_{77}
-4  c^{21}_{77}
+c^{23}_{77}
+2c^{24}_{77}
+2c^{26}_{77}
-c^{28}_{77}
+2c^{29}_{77}
$,
$ 5-2  c^{2}_{77}
+2c^{3}_{77}
+2c^{4}_{77}
-2  c^{5}_{77}
+4c^{7}_{77}
+2c^{8}_{77}
-c^{9}_{77}
+2c^{11}_{77}
-2  c^{12}_{77}
+c^{13}_{77}
+4c^{14}_{77}
+2c^{15}_{77}
-2  c^{16}_{77}
-2  c^{19}_{77}
+3c^{21}_{77}
+2c^{22}_{77}
-2  c^{23}_{77}
+c^{24}_{77}
-2  c^{26}_{77}
+3c^{28}_{77}
$,
$ -2+2c^{2}_{77}
-2  c^{3}_{77}
-2  c^{4}_{77}
+2c^{5}_{77}
-2  c^{6}_{77}
-6  c^{7}_{77}
-2  c^{8}_{77}
+2c^{9}_{77}
-2  c^{11}_{77}
+2c^{12}_{77}
-c^{14}_{77}
-2  c^{15}_{77}
-2  c^{17}_{77}
-3  c^{18}_{77}
+2c^{19}_{77}
-2  c^{22}_{77}
+2c^{23}_{77}
-c^{26}_{77}
-c^{28}_{77}
-3  c^{29}_{77}
$,
$ -\frac{7+\sqrt{77}}{2}$,
$0$,
$0$,
$0$;\ \ 
$ -2+2c^{2}_{77}
-2  c^{3}_{77}
-2  c^{4}_{77}
+2c^{5}_{77}
-2  c^{6}_{77}
-6  c^{7}_{77}
-2  c^{8}_{77}
+2c^{9}_{77}
-2  c^{11}_{77}
+2c^{12}_{77}
-c^{14}_{77}
-2  c^{15}_{77}
-2  c^{17}_{77}
-3  c^{18}_{77}
+2c^{19}_{77}
-2  c^{22}_{77}
+2c^{23}_{77}
-c^{26}_{77}
-c^{28}_{77}
-3  c^{29}_{77}
$,
$ -1-2  c^{1}_{77}
+c^{2}_{77}
-c^{3}_{77}
-c^{4}_{77}
+c^{5}_{77}
+2c^{6}_{77}
-c^{7}_{77}
-c^{8}_{77}
+c^{9}_{77}
-2  c^{10}_{77}
-c^{11}_{77}
+c^{12}_{77}
-4  c^{14}_{77}
-c^{15}_{77}
+3c^{16}_{77}
+2c^{17}_{77}
-c^{18}_{77}
+c^{19}_{77}
-c^{21}_{77}
-c^{22}_{77}
-c^{23}_{77}
+c^{28}_{77}
-c^{29}_{77}
$,
$ -\frac{7+\sqrt{77}}{2}$,
$0$,
$0$,
$0$;\ \ 
$ 1+2c^{1}_{77}
+c^{6}_{77}
+c^{7}_{77}
-2  c^{9}_{77}
+2c^{10}_{77}
-2  c^{13}_{77}
+c^{14}_{77}
+c^{16}_{77}
+c^{17}_{77}
+2c^{21}_{77}
+2c^{23}_{77}
-2  c^{24}_{77}
-2  c^{28}_{77}
$,
$ -\frac{7+\sqrt{77}}{2}$,
$0$,
$0$,
$0$;\ \ 
$ 1$,
$ \frac{11+\sqrt{77}}{2}$,
$ \frac{11+\sqrt{77}}{2}$,
$ \frac{11+\sqrt{77}}{2}$;\ \ 
$ 1+3c^{4}_{77}
+2c^{7}_{77}
-2  c^{9}_{77}
+c^{10}_{77}
+7c^{11}_{77}
+2c^{15}_{77}
-2  c^{16}_{77}
+c^{17}_{77}
+2c^{18}_{77}
-2  c^{19}_{77}
+c^{22}_{77}
-2  c^{23}_{77}
+c^{24}_{77}
+c^{25}_{77}
+2c^{26}_{77}
+2c^{29}_{77}
$,
$ -1+c^{1}_{77}
+c^{2}_{77}
-c^{3}_{77}
-3  c^{4}_{77}
+c^{5}_{77}
+c^{6}_{77}
-c^{7}_{77}
-c^{8}_{77}
+c^{9}_{77}
-2  c^{10}_{77}
-2  c^{11}_{77}
+c^{12}_{77}
+c^{13}_{77}
-c^{14}_{77}
+c^{16}_{77}
-2  c^{17}_{77}
-c^{18}_{77}
+5c^{22}_{77}
+c^{23}_{77}
-2  c^{24}_{77}
-3  c^{25}_{77}
-c^{29}_{77}
$,
$ -4-2  c^{1}_{77}
-2  c^{2}_{77}
+2c^{3}_{77}
+c^{4}_{77}
-2  c^{5}_{77}
-2  c^{6}_{77}
+c^{7}_{77}
+2c^{8}_{77}
-c^{9}_{77}
-4  c^{11}_{77}
-2  c^{12}_{77}
-2  c^{13}_{77}
+2c^{14}_{77}
-c^{15}_{77}
-c^{16}_{77}
+c^{18}_{77}
+c^{19}_{77}
-5  c^{22}_{77}
-c^{23}_{77}
+2c^{25}_{77}
-c^{26}_{77}
+c^{29}_{77}
$;\ \ 
$ -4-2  c^{1}_{77}
-2  c^{2}_{77}
+2c^{3}_{77}
+c^{4}_{77}
-2  c^{5}_{77}
-2  c^{6}_{77}
+c^{7}_{77}
+2c^{8}_{77}
-c^{9}_{77}
-4  c^{11}_{77}
-2  c^{12}_{77}
-2  c^{13}_{77}
+2c^{14}_{77}
-c^{15}_{77}
-c^{16}_{77}
+c^{18}_{77}
+c^{19}_{77}
-5  c^{22}_{77}
-c^{23}_{77}
+2c^{25}_{77}
-c^{26}_{77}
+c^{29}_{77}
$,
$ 1+3c^{4}_{77}
+2c^{7}_{77}
-2  c^{9}_{77}
+c^{10}_{77}
+7c^{11}_{77}
+2c^{15}_{77}
-2  c^{16}_{77}
+c^{17}_{77}
+2c^{18}_{77}
-2  c^{19}_{77}
+c^{22}_{77}
-2  c^{23}_{77}
+c^{24}_{77}
+c^{25}_{77}
+2c^{26}_{77}
+2c^{29}_{77}
$;\ \ 
$ -1+c^{1}_{77}
+c^{2}_{77}
-c^{3}_{77}
-3  c^{4}_{77}
+c^{5}_{77}
+c^{6}_{77}
-c^{7}_{77}
-c^{8}_{77}
+c^{9}_{77}
-2  c^{10}_{77}
-2  c^{11}_{77}
+c^{12}_{77}
+c^{13}_{77}
-c^{14}_{77}
+c^{16}_{77}
-2  c^{17}_{77}
-c^{18}_{77}
+5c^{22}_{77}
+c^{23}_{77}
-2  c^{24}_{77}
-3  c^{25}_{77}
-c^{29}_{77}
$)

\vskip 0.7ex
\hangindent=3em \hangafter=1
$\tau_n$ = ($0. - 26.16 i$, $0. - 232.47 i$, $0. + 232.47 i$, $0. - 26.16 i$, $0. + 232.47 i$, $0. + 26.16 i$, $342.13 - 103.15 i$, $0. - 232.47 i$, $0. - 26.16 i$, $0. + 26.16 i$, $342.13 - 129.31 i$, $0. + 232.47 i$, $0. + 26.16 i$, $342.13 + 103.15 i$, $0. - 26.16 i$, $0. - 26.16 i$, $0. + 26.16 i$, $0. - 232.47 i$, $0. + 26.16 i$, $0. + 232.47 i$, $342.13 - 103.15 i$, $342.13 - 129.31 i$, $0. - 26.16 i$, $0. + 26.16 i$, $0. - 26.16 i$, $0. + 232.47 i$, $0. + 232.47 i$, $342.13 - 103.15 i$, $0. - 232.47 i$, $0. - 232.47 i$, $0. + 232.47 i$, $0. - 232.47 i$, $342.13 + 129.31 i$, $0. + 232.47 i$, $342.13 - 103.15 i$, $0. - 26.16 i$, $0. - 26.16 i$, $0. + 232.47 i$, $0. - 232.47 i$, $0. + 26.16 i$, $0. + 26.16 i$, $342.13 + 103.15 i$, $0. - 232.47 i$, $342.13 - 129.31 i$, $0. + 232.47 i$, $0. - 232.47 i$, $0. + 232.47 i$, $0. + 232.47 i$, $342.13 + 103.15 i$, $0. - 232.47 i$, $0. - 232.47 i$, $0. + 26.16 i$, $0. - 26.16 i$, $0. + 26.16 i$, $342.13 + 129.31 i$, $342.13 + 103.15 i$, $0. - 232.47 i$, $0. - 26.16 i$, $0. + 232.47 i$, $0. - 26.16 i$, $0. + 26.16 i$, $0. + 26.16 i$, $342.13 - 103.15 i$, $0. - 26.16 i$, $0. - 232.47 i$, $342.13 + 129.31 i$, $0. - 26.16 i$, $0. + 26.16 i$, $0. + 232.47 i$, $342.13 + 103.15 i$, $0. - 26.16 i$, $0. - 232.47 i$, $0. + 26.16 i$, $0. - 232.47 i$, $0. + 232.47 i$, $0. + 26.16 i$, $684.26$)

\vskip 0.7ex
\hangindent=3em \hangafter=1
\textit{Intrinsic sign problem}

  \vskip 2ex

\noindent74. $10_{2,684.3}^{77,982}$ \irep{1138}:\ \ 
$d_i$ = ($1.0$,
$7.887$,
$7.887$,
$7.887$,
$7.887$,
$7.887$,
$8.887$,
$9.887$,
$9.887$,
$9.887$) 

\vskip 0.7ex
\hangindent=3em \hangafter=1
$D^2= 684.336 = 
\frac{693+77\sqrt{77}}{2}$

\vskip 0.7ex
\hangindent=3em \hangafter=1
$T = ( 0,
\frac{2}{11},
\frac{6}{11},
\frac{7}{11},
\frac{8}{11},
\frac{10}{11},
0,
\frac{1}{7},
\frac{2}{7},
\frac{4}{7} )
$,

\vskip 0.7ex
\hangindent=3em \hangafter=1
$S$ = ($ 1$,
$ \frac{7+\sqrt{77}}{2}$,
$ \frac{7+\sqrt{77}}{2}$,
$ \frac{7+\sqrt{77}}{2}$,
$ \frac{7+\sqrt{77}}{2}$,
$ \frac{7+\sqrt{77}}{2}$,
$ \frac{9+\sqrt{77}}{2}$,
$ \frac{11+\sqrt{77}}{2}$,
$ \frac{11+\sqrt{77}}{2}$,
$ \frac{11+\sqrt{77}}{2}$;\ \ 
$ 1+2c^{1}_{77}
+c^{6}_{77}
+c^{7}_{77}
-2  c^{9}_{77}
+2c^{10}_{77}
-2  c^{13}_{77}
+c^{14}_{77}
+c^{16}_{77}
+c^{17}_{77}
+2c^{21}_{77}
+2c^{23}_{77}
-2  c^{24}_{77}
-2  c^{28}_{77}
$,
$ -1-2  c^{1}_{77}
+c^{2}_{77}
-c^{3}_{77}
-c^{4}_{77}
+c^{5}_{77}
+2c^{6}_{77}
-c^{7}_{77}
-c^{8}_{77}
+c^{9}_{77}
-2  c^{10}_{77}
-c^{11}_{77}
+c^{12}_{77}
-4  c^{14}_{77}
-c^{15}_{77}
+3c^{16}_{77}
+2c^{17}_{77}
-c^{18}_{77}
+c^{19}_{77}
-c^{21}_{77}
-c^{22}_{77}
-c^{23}_{77}
+c^{28}_{77}
-c^{29}_{77}
$,
$ -2+2c^{2}_{77}
-2  c^{3}_{77}
-2  c^{4}_{77}
+2c^{5}_{77}
-2  c^{6}_{77}
-6  c^{7}_{77}
-2  c^{8}_{77}
+2c^{9}_{77}
-2  c^{11}_{77}
+2c^{12}_{77}
-c^{14}_{77}
-2  c^{15}_{77}
-2  c^{17}_{77}
-3  c^{18}_{77}
+2c^{19}_{77}
-2  c^{22}_{77}
+2c^{23}_{77}
-c^{26}_{77}
-c^{28}_{77}
-3  c^{29}_{77}
$,
$ -1+c^{1}_{77}
+2c^{9}_{77}
+c^{10}_{77}
+2c^{13}_{77}
-c^{14}_{77}
+2c^{18}_{77}
-4  c^{21}_{77}
+c^{23}_{77}
+2c^{24}_{77}
+2c^{26}_{77}
-c^{28}_{77}
+2c^{29}_{77}
$,
$ 5-2  c^{2}_{77}
+2c^{3}_{77}
+2c^{4}_{77}
-2  c^{5}_{77}
+4c^{7}_{77}
+2c^{8}_{77}
-c^{9}_{77}
+2c^{11}_{77}
-2  c^{12}_{77}
+c^{13}_{77}
+4c^{14}_{77}
+2c^{15}_{77}
-2  c^{16}_{77}
-2  c^{19}_{77}
+3c^{21}_{77}
+2c^{22}_{77}
-2  c^{23}_{77}
+c^{24}_{77}
-2  c^{26}_{77}
+3c^{28}_{77}
$,
$ -\frac{7+\sqrt{77}}{2}$,
$0$,
$0$,
$0$;\ \ 
$ -2+2c^{2}_{77}
-2  c^{3}_{77}
-2  c^{4}_{77}
+2c^{5}_{77}
-2  c^{6}_{77}
-6  c^{7}_{77}
-2  c^{8}_{77}
+2c^{9}_{77}
-2  c^{11}_{77}
+2c^{12}_{77}
-c^{14}_{77}
-2  c^{15}_{77}
-2  c^{17}_{77}
-3  c^{18}_{77}
+2c^{19}_{77}
-2  c^{22}_{77}
+2c^{23}_{77}
-c^{26}_{77}
-c^{28}_{77}
-3  c^{29}_{77}
$,
$ 5-2  c^{2}_{77}
+2c^{3}_{77}
+2c^{4}_{77}
-2  c^{5}_{77}
+4c^{7}_{77}
+2c^{8}_{77}
-c^{9}_{77}
+2c^{11}_{77}
-2  c^{12}_{77}
+c^{13}_{77}
+4c^{14}_{77}
+2c^{15}_{77}
-2  c^{16}_{77}
-2  c^{19}_{77}
+3c^{21}_{77}
+2c^{22}_{77}
-2  c^{23}_{77}
+c^{24}_{77}
-2  c^{26}_{77}
+3c^{28}_{77}
$,
$ 1+2c^{1}_{77}
+c^{6}_{77}
+c^{7}_{77}
-2  c^{9}_{77}
+2c^{10}_{77}
-2  c^{13}_{77}
+c^{14}_{77}
+c^{16}_{77}
+c^{17}_{77}
+2c^{21}_{77}
+2c^{23}_{77}
-2  c^{24}_{77}
-2  c^{28}_{77}
$,
$ -1+c^{1}_{77}
+2c^{9}_{77}
+c^{10}_{77}
+2c^{13}_{77}
-c^{14}_{77}
+2c^{18}_{77}
-4  c^{21}_{77}
+c^{23}_{77}
+2c^{24}_{77}
+2c^{26}_{77}
-c^{28}_{77}
+2c^{29}_{77}
$,
$ -\frac{7+\sqrt{77}}{2}$,
$0$,
$0$,
$0$;\ \ 
$ -1+c^{1}_{77}
+2c^{9}_{77}
+c^{10}_{77}
+2c^{13}_{77}
-c^{14}_{77}
+2c^{18}_{77}
-4  c^{21}_{77}
+c^{23}_{77}
+2c^{24}_{77}
+2c^{26}_{77}
-c^{28}_{77}
+2c^{29}_{77}
$,
$ -1-2  c^{1}_{77}
+c^{2}_{77}
-c^{3}_{77}
-c^{4}_{77}
+c^{5}_{77}
+2c^{6}_{77}
-c^{7}_{77}
-c^{8}_{77}
+c^{9}_{77}
-2  c^{10}_{77}
-c^{11}_{77}
+c^{12}_{77}
-4  c^{14}_{77}
-c^{15}_{77}
+3c^{16}_{77}
+2c^{17}_{77}
-c^{18}_{77}
+c^{19}_{77}
-c^{21}_{77}
-c^{22}_{77}
-c^{23}_{77}
+c^{28}_{77}
-c^{29}_{77}
$,
$ 1+2c^{1}_{77}
+c^{6}_{77}
+c^{7}_{77}
-2  c^{9}_{77}
+2c^{10}_{77}
-2  c^{13}_{77}
+c^{14}_{77}
+c^{16}_{77}
+c^{17}_{77}
+2c^{21}_{77}
+2c^{23}_{77}
-2  c^{24}_{77}
-2  c^{28}_{77}
$,
$ -\frac{7+\sqrt{77}}{2}$,
$0$,
$0$,
$0$;\ \ 
$ 5-2  c^{2}_{77}
+2c^{3}_{77}
+2c^{4}_{77}
-2  c^{5}_{77}
+4c^{7}_{77}
+2c^{8}_{77}
-c^{9}_{77}
+2c^{11}_{77}
-2  c^{12}_{77}
+c^{13}_{77}
+4c^{14}_{77}
+2c^{15}_{77}
-2  c^{16}_{77}
-2  c^{19}_{77}
+3c^{21}_{77}
+2c^{22}_{77}
-2  c^{23}_{77}
+c^{24}_{77}
-2  c^{26}_{77}
+3c^{28}_{77}
$,
$ -2+2c^{2}_{77}
-2  c^{3}_{77}
-2  c^{4}_{77}
+2c^{5}_{77}
-2  c^{6}_{77}
-6  c^{7}_{77}
-2  c^{8}_{77}
+2c^{9}_{77}
-2  c^{11}_{77}
+2c^{12}_{77}
-c^{14}_{77}
-2  c^{15}_{77}
-2  c^{17}_{77}
-3  c^{18}_{77}
+2c^{19}_{77}
-2  c^{22}_{77}
+2c^{23}_{77}
-c^{26}_{77}
-c^{28}_{77}
-3  c^{29}_{77}
$,
$ -\frac{7+\sqrt{77}}{2}$,
$0$,
$0$,
$0$;\ \ 
$ -1-2  c^{1}_{77}
+c^{2}_{77}
-c^{3}_{77}
-c^{4}_{77}
+c^{5}_{77}
+2c^{6}_{77}
-c^{7}_{77}
-c^{8}_{77}
+c^{9}_{77}
-2  c^{10}_{77}
-c^{11}_{77}
+c^{12}_{77}
-4  c^{14}_{77}
-c^{15}_{77}
+3c^{16}_{77}
+2c^{17}_{77}
-c^{18}_{77}
+c^{19}_{77}
-c^{21}_{77}
-c^{22}_{77}
-c^{23}_{77}
+c^{28}_{77}
-c^{29}_{77}
$,
$ -\frac{7+\sqrt{77}}{2}$,
$0$,
$0$,
$0$;\ \ 
$ 1$,
$ \frac{11+\sqrt{77}}{2}$,
$ \frac{11+\sqrt{77}}{2}$,
$ \frac{11+\sqrt{77}}{2}$;\ \ 
$ -1+c^{1}_{77}
+c^{2}_{77}
-c^{3}_{77}
-3  c^{4}_{77}
+c^{5}_{77}
+c^{6}_{77}
-c^{7}_{77}
-c^{8}_{77}
+c^{9}_{77}
-2  c^{10}_{77}
-2  c^{11}_{77}
+c^{12}_{77}
+c^{13}_{77}
-c^{14}_{77}
+c^{16}_{77}
-2  c^{17}_{77}
-c^{18}_{77}
+5c^{22}_{77}
+c^{23}_{77}
-2  c^{24}_{77}
-3  c^{25}_{77}
-c^{29}_{77}
$,
$ 1+3c^{4}_{77}
+2c^{7}_{77}
-2  c^{9}_{77}
+c^{10}_{77}
+7c^{11}_{77}
+2c^{15}_{77}
-2  c^{16}_{77}
+c^{17}_{77}
+2c^{18}_{77}
-2  c^{19}_{77}
+c^{22}_{77}
-2  c^{23}_{77}
+c^{24}_{77}
+c^{25}_{77}
+2c^{26}_{77}
+2c^{29}_{77}
$,
$ -4-2  c^{1}_{77}
-2  c^{2}_{77}
+2c^{3}_{77}
+c^{4}_{77}
-2  c^{5}_{77}
-2  c^{6}_{77}
+c^{7}_{77}
+2c^{8}_{77}
-c^{9}_{77}
-4  c^{11}_{77}
-2  c^{12}_{77}
-2  c^{13}_{77}
+2c^{14}_{77}
-c^{15}_{77}
-c^{16}_{77}
+c^{18}_{77}
+c^{19}_{77}
-5  c^{22}_{77}
-c^{23}_{77}
+2c^{25}_{77}
-c^{26}_{77}
+c^{29}_{77}
$;\ \ 
$ -4-2  c^{1}_{77}
-2  c^{2}_{77}
+2c^{3}_{77}
+c^{4}_{77}
-2  c^{5}_{77}
-2  c^{6}_{77}
+c^{7}_{77}
+2c^{8}_{77}
-c^{9}_{77}
-4  c^{11}_{77}
-2  c^{12}_{77}
-2  c^{13}_{77}
+2c^{14}_{77}
-c^{15}_{77}
-c^{16}_{77}
+c^{18}_{77}
+c^{19}_{77}
-5  c^{22}_{77}
-c^{23}_{77}
+2c^{25}_{77}
-c^{26}_{77}
+c^{29}_{77}
$,
$ -1+c^{1}_{77}
+c^{2}_{77}
-c^{3}_{77}
-3  c^{4}_{77}
+c^{5}_{77}
+c^{6}_{77}
-c^{7}_{77}
-c^{8}_{77}
+c^{9}_{77}
-2  c^{10}_{77}
-2  c^{11}_{77}
+c^{12}_{77}
+c^{13}_{77}
-c^{14}_{77}
+c^{16}_{77}
-2  c^{17}_{77}
-c^{18}_{77}
+5c^{22}_{77}
+c^{23}_{77}
-2  c^{24}_{77}
-3  c^{25}_{77}
-c^{29}_{77}
$;\ \ 
$ 1+3c^{4}_{77}
+2c^{7}_{77}
-2  c^{9}_{77}
+c^{10}_{77}
+7c^{11}_{77}
+2c^{15}_{77}
-2  c^{16}_{77}
+c^{17}_{77}
+2c^{18}_{77}
-2  c^{19}_{77}
+c^{22}_{77}
-2  c^{23}_{77}
+c^{24}_{77}
+c^{25}_{77}
+2c^{26}_{77}
+2c^{29}_{77}
$)

\vskip 0.7ex
\hangindent=3em \hangafter=1
$\tau_n$ = ($0. + 26.16 i$, $0. + 232.47 i$, $0. - 232.47 i$, $0. + 26.16 i$, $0. - 232.47 i$, $0. - 26.16 i$, $342.13 + 103.15 i$, $0. + 232.47 i$, $0. + 26.16 i$, $0. - 26.16 i$, $342.13 + 129.31 i$, $0. - 232.47 i$, $0. - 26.16 i$, $342.13 - 103.15 i$, $0. + 26.16 i$, $0. + 26.16 i$, $0. - 26.16 i$, $0. + 232.47 i$, $0. - 26.16 i$, $0. - 232.47 i$, $342.13 + 103.15 i$, $342.13 + 129.31 i$, $0. + 26.16 i$, $0. - 26.16 i$, $0. + 26.16 i$, $0. - 232.47 i$, $0. - 232.47 i$, $342.13 + 103.15 i$, $0. + 232.47 i$, $0. + 232.47 i$, $0. - 232.47 i$, $0. + 232.47 i$, $342.13 - 129.31 i$, $0. - 232.47 i$, $342.13 + 103.15 i$, $0. + 26.16 i$, $0. + 26.16 i$, $0. - 232.47 i$, $0. + 232.47 i$, $0. - 26.16 i$, $0. - 26.16 i$, $342.13 - 103.15 i$, $0. + 232.47 i$, $342.13 + 129.31 i$, $0. - 232.47 i$, $0. + 232.47 i$, $0. - 232.47 i$, $0. - 232.47 i$, $342.13 - 103.15 i$, $0. + 232.47 i$, $0. + 232.47 i$, $0. - 26.16 i$, $0. + 26.16 i$, $0. - 26.16 i$, $342.13 - 129.31 i$, $342.13 - 103.15 i$, $0. + 232.47 i$, $0. + 26.16 i$, $0. - 232.47 i$, $0. + 26.16 i$, $0. - 26.16 i$, $0. - 26.16 i$, $342.13 + 103.15 i$, $0. + 26.16 i$, $0. + 232.47 i$, $342.13 - 129.31 i$, $0. + 26.16 i$, $0. - 26.16 i$, $0. - 232.47 i$, $342.13 - 103.15 i$, $0. + 26.16 i$, $0. + 232.47 i$, $0. - 26.16 i$, $0. + 232.47 i$, $0. - 232.47 i$, $0. - 26.16 i$, $684.26$)

\vskip 0.7ex
\hangindent=3em \hangafter=1
\textit{Intrinsic sign problem}

  \vskip 2ex

\noindent75. $10_{4,1435.}^{10,168}$ \irep{538}:\ \ 
$d_i$ = ($1.0$,
$9.472$,
$9.472$,
$9.472$,
$9.472$,
$9.472$,
$9.472$,
$16.944$,
$16.944$,
$17.944$) 

\vskip 0.7ex
\hangindent=3em \hangafter=1
$D^2= 1435.541 = 
720+320\sqrt{5}$

\vskip 0.7ex
\hangindent=3em \hangafter=1
$T = ( 0,
\frac{1}{2},
\frac{1}{2},
\frac{1}{2},
\frac{1}{2},
\frac{1}{2},
\frac{1}{2},
\frac{1}{5},
\frac{4}{5},
0 )
$,

\vskip 0.7ex
\hangindent=3em \hangafter=1
$S$ = ($ 1$,
$ 5+2\sqrt{5}$,
$ 5+2\sqrt{5}$,
$ 5+2\sqrt{5}$,
$ 5+2\sqrt{5}$,
$ 5+2\sqrt{5}$,
$ 5+2\sqrt{5}$,
$ 8+4\sqrt{5}$,
$ 8+4\sqrt{5}$,
$ 9+4\sqrt{5}$;\ \ 
$ 15+6\sqrt{5}$,
$ -5-2\sqrt{5}$,
$ -5-2\sqrt{5}$,
$ -5-2\sqrt{5}$,
$ -5-2\sqrt{5}$,
$ -5-2\sqrt{5}$,
$0$,
$0$,
$ 5+2\sqrt{5}$;\ \ 
$ 15+6\sqrt{5}$,
$ -5-2\sqrt{5}$,
$ -5-2\sqrt{5}$,
$ -5-2\sqrt{5}$,
$ -5-2\sqrt{5}$,
$0$,
$0$,
$ 5+2\sqrt{5}$;\ \ 
$ 15+6\sqrt{5}$,
$ -5-2\sqrt{5}$,
$ -5-2\sqrt{5}$,
$ -5-2\sqrt{5}$,
$0$,
$0$,
$ 5+2\sqrt{5}$;\ \ 
$ 15+6\sqrt{5}$,
$ -5-2\sqrt{5}$,
$ -5-2\sqrt{5}$,
$0$,
$0$,
$ 5+2\sqrt{5}$;\ \ 
$ 15+6\sqrt{5}$,
$ -5-2\sqrt{5}$,
$0$,
$0$,
$ 5+2\sqrt{5}$;\ \ 
$ 15+6\sqrt{5}$,
$0$,
$0$,
$ 5+2\sqrt{5}$;\ \ 
$ 14+6\sqrt{5}$,
$ -6-2\sqrt{5}$,
$ -8-4\sqrt{5}$;\ \ 
$ 14+6\sqrt{5}$,
$ -8-4\sqrt{5}$;\ \ 
$ 1$)

\vskip 0.7ex
\hangindent=3em \hangafter=1
$\tau_n$ = ($-37.89$, $396.76$, $-679.86$, $1038.74$, $358.87$, $1038.74$, $-679.86$, $396.76$, $-37.89$, $1435.5$)

\vskip 0.7ex
\hangindent=3em \hangafter=1
\textit{Intrinsic sign problem}

  \vskip 2ex

\noindent76. $10_{0,1435.}^{20,676}$ \irep{948}:\ \ 
$d_i$ = ($1.0$,
$9.472$,
$9.472$,
$9.472$,
$9.472$,
$9.472$,
$9.472$,
$16.944$,
$16.944$,
$17.944$) 

\vskip 0.7ex
\hangindent=3em \hangafter=1
$D^2= 1435.541 = 
720+320\sqrt{5}$

\vskip 0.7ex
\hangindent=3em \hangafter=1
$T = ( 0,
0,
0,
\frac{1}{4},
\frac{1}{4},
\frac{3}{4},
\frac{3}{4},
\frac{2}{5},
\frac{3}{5},
0 )
$,

\vskip 0.7ex
\hangindent=3em \hangafter=1
$S$ = ($ 1$,
$ 5+2\sqrt{5}$,
$ 5+2\sqrt{5}$,
$ 5+2\sqrt{5}$,
$ 5+2\sqrt{5}$,
$ 5+2\sqrt{5}$,
$ 5+2\sqrt{5}$,
$ 8+4\sqrt{5}$,
$ 8+4\sqrt{5}$,
$ 9+4\sqrt{5}$;\ \ 
$ 15+6\sqrt{5}$,
$ -5-2\sqrt{5}$,
$ -5-2\sqrt{5}$,
$ -5-2\sqrt{5}$,
$ -5-2\sqrt{5}$,
$ -5-2\sqrt{5}$,
$0$,
$0$,
$ 5+2\sqrt{5}$;\ \ 
$ 15+6\sqrt{5}$,
$ -5-2\sqrt{5}$,
$ -5-2\sqrt{5}$,
$ -5-2\sqrt{5}$,
$ -5-2\sqrt{5}$,
$0$,
$0$,
$ 5+2\sqrt{5}$;\ \ 
$ -3-6  s^{1}_{20}
-4  c^{2}_{20}
+14s^{3}_{20}
$,
$ -3+6s^{1}_{20}
-4  c^{2}_{20}
-14  s^{3}_{20}
$,
$ 5+2\sqrt{5}$,
$ 5+2\sqrt{5}$,
$0$,
$0$,
$ 5+2\sqrt{5}$;\ \ 
$ -3-6  s^{1}_{20}
-4  c^{2}_{20}
+14s^{3}_{20}
$,
$ 5+2\sqrt{5}$,
$ 5+2\sqrt{5}$,
$0$,
$0$,
$ 5+2\sqrt{5}$;\ \ 
$ -3+6s^{1}_{20}
-4  c^{2}_{20}
-14  s^{3}_{20}
$,
$ -3-6  s^{1}_{20}
-4  c^{2}_{20}
+14s^{3}_{20}
$,
$0$,
$0$,
$ 5+2\sqrt{5}$;\ \ 
$ -3+6s^{1}_{20}
-4  c^{2}_{20}
-14  s^{3}_{20}
$,
$0$,
$0$,
$ 5+2\sqrt{5}$;\ \ 
$ -6-2\sqrt{5}$,
$ 14+6\sqrt{5}$,
$ -8-4\sqrt{5}$;\ \ 
$ -6-2\sqrt{5}$,
$ -8-4\sqrt{5}$;\ \ 
$ 1$)

\vskip 0.7ex
\hangindent=3em \hangafter=1
$\tau_n$ = ($37.89$, $320.99$, $679.86$, $396.76$, $1076.62$, $-320.99$, $679.86$, $1038.74$, $37.89$, $717.75$, $37.89$, $1038.74$, $679.86$, $-320.99$, $1076.62$, $396.76$, $679.86$, $320.99$, $37.89$, $1435.5$)

\vskip 0.7ex
\hangindent=3em \hangafter=1
\textit{Intrinsic sign problem}

  \vskip 2ex 

%% file: modular_data/SsL11U_.tex
\noindent1. $11_{2,11.}^{11,568}$ \irep{983}:\ \ 
$d_i$ = ($1.0$,
$1.0$,
$1.0$,
$1.0$,
$1.0$,
$1.0$,
$1.0$,
$1.0$,
$1.0$,
$1.0$,
$1.0$) 

\vskip 0.7ex
\hangindent=3em \hangafter=1
$D^2= 11.0 = 
11$

\vskip 0.7ex
\hangindent=3em \hangafter=1
$T = ( 0,
\frac{1}{11},
\frac{1}{11},
\frac{3}{11},
\frac{3}{11},
\frac{4}{11},
\frac{4}{11},
\frac{5}{11},
\frac{5}{11},
\frac{9}{11},
\frac{9}{11} )
$,

\vskip 0.7ex
\hangindent=3em \hangafter=1
$S$ = ($ 1$,
$ 1$,
$ 1$,
$ 1$,
$ 1$,
$ 1$,
$ 1$,
$ 1$,
$ 1$,
$ 1$,
$ 1$;\ \ 
$ -\zeta_{22}^{7}$,
$ \zeta_{11}^{2}$,
$ -\zeta_{22}^{9}$,
$ \zeta_{11}^{1}$,
$ -\zeta_{22}^{3}$,
$ \zeta_{11}^{4}$,
$ -\zeta_{22}^{5}$,
$ \zeta_{11}^{3}$,
$ -\zeta_{22}^{1}$,
$ \zeta_{11}^{5}$;\ \ 
$ -\zeta_{22}^{7}$,
$ \zeta_{11}^{1}$,
$ -\zeta_{22}^{9}$,
$ \zeta_{11}^{4}$,
$ -\zeta_{22}^{3}$,
$ \zeta_{11}^{3}$,
$ -\zeta_{22}^{5}$,
$ \zeta_{11}^{5}$,
$ -\zeta_{22}^{1}$;\ \ 
$ \zeta_{11}^{5}$,
$ -\zeta_{22}^{1}$,
$ -\zeta_{22}^{7}$,
$ \zeta_{11}^{2}$,
$ \zeta_{11}^{4}$,
$ -\zeta_{22}^{3}$,
$ \zeta_{11}^{3}$,
$ -\zeta_{22}^{5}$;\ \ 
$ \zeta_{11}^{5}$,
$ \zeta_{11}^{2}$,
$ -\zeta_{22}^{7}$,
$ -\zeta_{22}^{3}$,
$ \zeta_{11}^{4}$,
$ -\zeta_{22}^{5}$,
$ \zeta_{11}^{3}$;\ \ 
$ \zeta_{11}^{3}$,
$ -\zeta_{22}^{5}$,
$ \zeta_{11}^{5}$,
$ -\zeta_{22}^{1}$,
$ \zeta_{11}^{1}$,
$ -\zeta_{22}^{9}$;\ \ 
$ \zeta_{11}^{3}$,
$ -\zeta_{22}^{1}$,
$ \zeta_{11}^{5}$,
$ -\zeta_{22}^{9}$,
$ \zeta_{11}^{1}$;\ \ 
$ \zeta_{11}^{1}$,
$ -\zeta_{22}^{9}$,
$ -\zeta_{22}^{7}$,
$ \zeta_{11}^{2}$;\ \ 
$ \zeta_{11}^{1}$,
$ \zeta_{11}^{2}$,
$ -\zeta_{22}^{7}$;\ \ 
$ \zeta_{11}^{4}$,
$ -\zeta_{22}^{3}$;\ \ 
$ \zeta_{11}^{4}$)

\vskip 0.7ex
\hangindent=3em \hangafter=1
$\tau_n$ = ($0. + 3.32 i$, $0. - 3.32 i$, $0. + 3.32 i$, $0. + 3.32 i$, $0. + 3.32 i$, $0. - 3.32 i$, $0. - 3.32 i$, $0. - 3.32 i$, $0. + 3.32 i$, $0. - 3.32 i$, $11.$)

\vskip 0.7ex
\hangindent=3em \hangafter=1
\textit{Intrinsic sign problem}

  \vskip 2ex

\noindent2. $11_{6,11.}^{11,143}$ \irep{983}:\ \ 
$d_i$ = ($1.0$,
$1.0$,
$1.0$,
$1.0$,
$1.0$,
$1.0$,
$1.0$,
$1.0$,
$1.0$,
$1.0$,
$1.0$) 

\vskip 0.7ex
\hangindent=3em \hangafter=1
$D^2= 11.0 = 
11$

\vskip 0.7ex
\hangindent=3em \hangafter=1
$T = ( 0,
\frac{2}{11},
\frac{2}{11},
\frac{6}{11},
\frac{6}{11},
\frac{7}{11},
\frac{7}{11},
\frac{8}{11},
\frac{8}{11},
\frac{10}{11},
\frac{10}{11} )
$,

\vskip 0.7ex
\hangindent=3em \hangafter=1
$S$ = ($ 1$,
$ 1$,
$ 1$,
$ 1$,
$ 1$,
$ 1$,
$ 1$,
$ 1$,
$ 1$,
$ 1$,
$ 1$;\ \ 
$ -\zeta_{22}^{3}$,
$ \zeta_{11}^{4}$,
$ -\zeta_{22}^{7}$,
$ \zeta_{11}^{2}$,
$ -\zeta_{22}^{9}$,
$ \zeta_{11}^{1}$,
$ -\zeta_{22}^{5}$,
$ \zeta_{11}^{3}$,
$ -\zeta_{22}^{1}$,
$ \zeta_{11}^{5}$;\ \ 
$ -\zeta_{22}^{3}$,
$ \zeta_{11}^{2}$,
$ -\zeta_{22}^{7}$,
$ \zeta_{11}^{1}$,
$ -\zeta_{22}^{9}$,
$ \zeta_{11}^{3}$,
$ -\zeta_{22}^{5}$,
$ \zeta_{11}^{5}$,
$ -\zeta_{22}^{1}$;\ \ 
$ -\zeta_{22}^{9}$,
$ \zeta_{11}^{1}$,
$ \zeta_{11}^{5}$,
$ -\zeta_{22}^{1}$,
$ \zeta_{11}^{4}$,
$ -\zeta_{22}^{3}$,
$ \zeta_{11}^{3}$,
$ -\zeta_{22}^{5}$;\ \ 
$ -\zeta_{22}^{9}$,
$ -\zeta_{22}^{1}$,
$ \zeta_{11}^{5}$,
$ -\zeta_{22}^{3}$,
$ \zeta_{11}^{4}$,
$ -\zeta_{22}^{5}$,
$ \zeta_{11}^{3}$;\ \ 
$ -\zeta_{22}^{5}$,
$ \zeta_{11}^{3}$,
$ \zeta_{11}^{2}$,
$ -\zeta_{22}^{7}$,
$ -\zeta_{22}^{3}$,
$ \zeta_{11}^{4}$;\ \ 
$ -\zeta_{22}^{5}$,
$ -\zeta_{22}^{7}$,
$ \zeta_{11}^{2}$,
$ \zeta_{11}^{4}$,
$ -\zeta_{22}^{3}$;\ \ 
$ -\zeta_{22}^{1}$,
$ \zeta_{11}^{5}$,
$ -\zeta_{22}^{9}$,
$ \zeta_{11}^{1}$;\ \ 
$ -\zeta_{22}^{1}$,
$ \zeta_{11}^{1}$,
$ -\zeta_{22}^{9}$;\ \ 
$ \zeta_{11}^{2}$,
$ -\zeta_{22}^{7}$;\ \ 
$ \zeta_{11}^{2}$)

\vskip 0.7ex
\hangindent=3em \hangafter=1
$\tau_n$ = ($0. - 3.32 i$, $0. + 3.32 i$, $0. - 3.32 i$, $0. - 3.32 i$, $0. - 3.32 i$, $0. + 3.32 i$, $0. + 3.32 i$, $0. + 3.32 i$, $0. - 3.32 i$, $0. + 3.32 i$, $11.$)

\vskip 0.7ex
\hangindent=3em \hangafter=1
\textit{Intrinsic sign problem}

  \vskip 2ex

\noindent3. $11_{1,32.}^{16,245}$ \irep{0}:\ \ 
$d_i$ = ($1.0$,
$1.0$,
$1.0$,
$1.0$,
$2.0$,
$2.0$,
$2.0$,
$2.0$,
$2.0$,
$2.0$,
$2.0$) 

\vskip 0.7ex
\hangindent=3em \hangafter=1
$D^2= 32.0 = 
32$

\vskip 0.7ex
\hangindent=3em \hangafter=1
$T = ( 0,
0,
0,
0,
\frac{1}{4},
\frac{1}{16},
\frac{1}{16},
\frac{1}{16},
\frac{9}{16},
\frac{9}{16},
\frac{9}{16} )
$,

\vskip 0.7ex
\hangindent=3em \hangafter=1
$S$ = ($ 1$,
$ 1$,
$ 1$,
$ 1$,
$ 2$,
$ 2$,
$ 2$,
$ 2$,
$ 2$,
$ 2$,
$ 2$;\ \ 
$ 1$,
$ 1$,
$ 1$,
$ 2$,
$ -2$,
$ -2$,
$ 2$,
$ -2$,
$ -2$,
$ 2$;\ \ 
$ 1$,
$ 1$,
$ 2$,
$ -2$,
$ 2$,
$ -2$,
$ -2$,
$ 2$,
$ -2$;\ \ 
$ 1$,
$ 2$,
$ 2$,
$ -2$,
$ -2$,
$ 2$,
$ -2$,
$ -2$;\ \ 
$ -4$,
$0$,
$0$,
$0$,
$0$,
$0$,
$0$;\ \ 
$ 2\sqrt{2}$,
$0$,
$0$,
$ -2\sqrt{2}$,
$0$,
$0$;\ \ 
$ 2\sqrt{2}$,
$0$,
$0$,
$ -2\sqrt{2}$,
$0$;\ \ 
$ 2\sqrt{2}$,
$0$,
$0$,
$ -2\sqrt{2}$;\ \ 
$ 2\sqrt{2}$,
$0$,
$0$;\ \ 
$ 2\sqrt{2}$,
$0$;\ \ 
$ 2\sqrt{2}$)

\vskip 0.7ex
\hangindent=3em \hangafter=1
$\tau_n$ = ($4. + 4. i$, $16.97 + 16.97 i$, $4. - 4. i$, $8. + 24. i$, $4. + 4. i$, $-16.97 + 16.97 i$, $4. - 4. i$, $-16.$, $4. + 4. i$, $-16.97 - 16.97 i$, $4. - 4. i$, $8. - 24. i$, $4. + 4. i$, $16.97 - 16.97 i$, $4. - 4. i$, $32.$)

\vskip 0.7ex
\hangindent=3em \hangafter=1
\textit{Intrinsic sign problem}

  \vskip 2ex

\noindent4. $11_{1,32.}^{16,157}$ \irep{0}:\ \ 
$d_i$ = ($1.0$,
$1.0$,
$1.0$,
$1.0$,
$2.0$,
$2.0$,
$2.0$,
$2.0$,
$2.0$,
$2.0$,
$2.0$) 

\vskip 0.7ex
\hangindent=3em \hangafter=1
$D^2= 32.0 = 
32$

\vskip 0.7ex
\hangindent=3em \hangafter=1
$T = ( 0,
0,
0,
0,
\frac{1}{4},
\frac{1}{16},
\frac{1}{16},
\frac{5}{16},
\frac{9}{16},
\frac{9}{16},
\frac{13}{16} )
$,

\vskip 0.7ex
\hangindent=3em \hangafter=1
$S$ = ($ 1$,
$ 1$,
$ 1$,
$ 1$,
$ 2$,
$ 2$,
$ 2$,
$ 2$,
$ 2$,
$ 2$,
$ 2$;\ \ 
$ 1$,
$ 1$,
$ 1$,
$ 2$,
$ -2$,
$ -2$,
$ 2$,
$ -2$,
$ -2$,
$ 2$;\ \ 
$ 1$,
$ 1$,
$ 2$,
$ -2$,
$ 2$,
$ -2$,
$ -2$,
$ 2$,
$ -2$;\ \ 
$ 1$,
$ 2$,
$ 2$,
$ -2$,
$ -2$,
$ 2$,
$ -2$,
$ -2$;\ \ 
$ -4$,
$0$,
$0$,
$0$,
$0$,
$0$,
$0$;\ \ 
$ 2\sqrt{2}$,
$0$,
$0$,
$ -2\sqrt{2}$,
$0$,
$0$;\ \ 
$ 2\sqrt{2}$,
$0$,
$0$,
$ -2\sqrt{2}$,
$0$;\ \ 
$ -2\sqrt{2}$,
$0$,
$0$,
$ 2\sqrt{2}$;\ \ 
$ 2\sqrt{2}$,
$0$,
$0$;\ \ 
$ 2\sqrt{2}$,
$0$;\ \ 
$ -2\sqrt{2}$)

\vskip 0.7ex
\hangindent=3em \hangafter=1
$\tau_n$ = ($4. + 4. i$, $5.66 + 5.66 i$, $4. - 4. i$, $8. + 24. i$, $4. + 4. i$, $-5.66 + 5.66 i$, $4. - 4. i$, $-16.$, $4. + 4. i$, $-5.66 - 5.66 i$, $4. - 4. i$, $8. - 24. i$, $4. + 4. i$, $5.66 - 5.66 i$, $4. - 4. i$, $32.$)

\vskip 0.7ex
\hangindent=3em \hangafter=1
\textit{Intrinsic sign problem}

  \vskip 2ex

\noindent5. $11_{1,32.}^{16,703}$ \irep{0}:\ \ 
$d_i$ = ($1.0$,
$1.0$,
$1.0$,
$1.0$,
$2.0$,
$2.0$,
$2.0$,
$2.0$,
$2.0$,
$2.0$,
$2.0$) 

\vskip 0.7ex
\hangindent=3em \hangafter=1
$D^2= 32.0 = 
32$

\vskip 0.7ex
\hangindent=3em \hangafter=1
$T = ( 0,
0,
0,
0,
\frac{1}{4},
\frac{1}{16},
\frac{5}{16},
\frac{5}{16},
\frac{9}{16},
\frac{13}{16},
\frac{13}{16} )
$,

\vskip 0.7ex
\hangindent=3em \hangafter=1
$S$ = ($ 1$,
$ 1$,
$ 1$,
$ 1$,
$ 2$,
$ 2$,
$ 2$,
$ 2$,
$ 2$,
$ 2$,
$ 2$;\ \ 
$ 1$,
$ 1$,
$ 1$,
$ 2$,
$ -2$,
$ -2$,
$ 2$,
$ -2$,
$ -2$,
$ 2$;\ \ 
$ 1$,
$ 1$,
$ 2$,
$ -2$,
$ 2$,
$ -2$,
$ -2$,
$ 2$,
$ -2$;\ \ 
$ 1$,
$ 2$,
$ 2$,
$ -2$,
$ -2$,
$ 2$,
$ -2$,
$ -2$;\ \ 
$ -4$,
$0$,
$0$,
$0$,
$0$,
$0$,
$0$;\ \ 
$ 2\sqrt{2}$,
$0$,
$0$,
$ -2\sqrt{2}$,
$0$,
$0$;\ \ 
$ -2\sqrt{2}$,
$0$,
$0$,
$ 2\sqrt{2}$,
$0$;\ \ 
$ -2\sqrt{2}$,
$0$,
$0$,
$ 2\sqrt{2}$;\ \ 
$ 2\sqrt{2}$,
$0$,
$0$;\ \ 
$ -2\sqrt{2}$,
$0$;\ \ 
$ -2\sqrt{2}$)

\vskip 0.7ex
\hangindent=3em \hangafter=1
$\tau_n$ = ($4. + 4. i$, $-5.66 - 5.66 i$, $4. - 4. i$, $8. + 24. i$, $4. + 4. i$, $5.66 - 5.66 i$, $4. - 4. i$, $-16.$, $4. + 4. i$, $5.66 + 5.66 i$, $4. - 4. i$, $8. - 24. i$, $4. + 4. i$, $-5.66 + 5.66 i$, $4. - 4. i$, $32.$)

\vskip 0.7ex
\hangindent=3em \hangafter=1
\textit{Intrinsic sign problem}

  \vskip 2ex

\noindent6. $11_{1,32.}^{16,171}$ \irep{0}:\ \ 
$d_i$ = ($1.0$,
$1.0$,
$1.0$,
$1.0$,
$2.0$,
$2.0$,
$2.0$,
$2.0$,
$2.0$,
$2.0$,
$2.0$) 

\vskip 0.7ex
\hangindent=3em \hangafter=1
$D^2= 32.0 = 
32$

\vskip 0.7ex
\hangindent=3em \hangafter=1
$T = ( 0,
0,
0,
0,
\frac{1}{4},
\frac{5}{16},
\frac{5}{16},
\frac{5}{16},
\frac{13}{16},
\frac{13}{16},
\frac{13}{16} )
$,

\vskip 0.7ex
\hangindent=3em \hangafter=1
$S$ = ($ 1$,
$ 1$,
$ 1$,
$ 1$,
$ 2$,
$ 2$,
$ 2$,
$ 2$,
$ 2$,
$ 2$,
$ 2$;\ \ 
$ 1$,
$ 1$,
$ 1$,
$ 2$,
$ -2$,
$ -2$,
$ 2$,
$ -2$,
$ -2$,
$ 2$;\ \ 
$ 1$,
$ 1$,
$ 2$,
$ -2$,
$ 2$,
$ -2$,
$ -2$,
$ 2$,
$ -2$;\ \ 
$ 1$,
$ 2$,
$ 2$,
$ -2$,
$ -2$,
$ 2$,
$ -2$,
$ -2$;\ \ 
$ -4$,
$0$,
$0$,
$0$,
$0$,
$0$,
$0$;\ \ 
$ -2\sqrt{2}$,
$0$,
$0$,
$ 2\sqrt{2}$,
$0$,
$0$;\ \ 
$ -2\sqrt{2}$,
$0$,
$0$,
$ 2\sqrt{2}$,
$0$;\ \ 
$ -2\sqrt{2}$,
$0$,
$0$,
$ 2\sqrt{2}$;\ \ 
$ -2\sqrt{2}$,
$0$,
$0$;\ \ 
$ -2\sqrt{2}$,
$0$;\ \ 
$ -2\sqrt{2}$)

\vskip 0.7ex
\hangindent=3em \hangafter=1
$\tau_n$ = ($4. + 4. i$, $-16.97 - 16.97 i$, $4. - 4. i$, $8. + 24. i$, $4. + 4. i$, $16.97 - 16.97 i$, $4. - 4. i$, $-16.$, $4. + 4. i$, $16.97 + 16.97 i$, $4. - 4. i$, $8. - 24. i$, $4. + 4. i$, $-16.97 + 16.97 i$, $4. - 4. i$, $32.$)

\vskip 0.7ex
\hangindent=3em \hangafter=1
\textit{Intrinsic sign problem}

  \vskip 2ex

\noindent7. $11_{7,32.}^{16,328}$ \irep{0}:\ \ 
$d_i$ = ($1.0$,
$1.0$,
$1.0$,
$1.0$,
$2.0$,
$2.0$,
$2.0$,
$2.0$,
$2.0$,
$2.0$,
$2.0$) 

\vskip 0.7ex
\hangindent=3em \hangafter=1
$D^2= 32.0 = 
32$

\vskip 0.7ex
\hangindent=3em \hangafter=1
$T = ( 0,
0,
0,
0,
\frac{3}{4},
\frac{3}{16},
\frac{3}{16},
\frac{3}{16},
\frac{11}{16},
\frac{11}{16},
\frac{11}{16} )
$,

\vskip 0.7ex
\hangindent=3em \hangafter=1
$S$ = ($ 1$,
$ 1$,
$ 1$,
$ 1$,
$ 2$,
$ 2$,
$ 2$,
$ 2$,
$ 2$,
$ 2$,
$ 2$;\ \ 
$ 1$,
$ 1$,
$ 1$,
$ 2$,
$ -2$,
$ -2$,
$ 2$,
$ -2$,
$ -2$,
$ 2$;\ \ 
$ 1$,
$ 1$,
$ 2$,
$ -2$,
$ 2$,
$ -2$,
$ -2$,
$ 2$,
$ -2$;\ \ 
$ 1$,
$ 2$,
$ 2$,
$ -2$,
$ -2$,
$ 2$,
$ -2$,
$ -2$;\ \ 
$ -4$,
$0$,
$0$,
$0$,
$0$,
$0$,
$0$;\ \ 
$ -2\sqrt{2}$,
$0$,
$0$,
$ 2\sqrt{2}$,
$0$,
$0$;\ \ 
$ -2\sqrt{2}$,
$0$,
$0$,
$ 2\sqrt{2}$,
$0$;\ \ 
$ -2\sqrt{2}$,
$0$,
$0$,
$ 2\sqrt{2}$;\ \ 
$ -2\sqrt{2}$,
$0$,
$0$;\ \ 
$ -2\sqrt{2}$,
$0$;\ \ 
$ -2\sqrt{2}$)

\vskip 0.7ex
\hangindent=3em \hangafter=1
$\tau_n$ = ($4. - 4. i$, $-16.97 + 16.97 i$, $4. + 4. i$, $8. - 24. i$, $4. - 4. i$, $16.97 + 16.97 i$, $4. + 4. i$, $-16.$, $4. - 4. i$, $16.97 - 16.97 i$, $4. + 4. i$, $8. + 24. i$, $4. - 4. i$, $-16.97 - 16.97 i$, $4. + 4. i$, $32.$)

\vskip 0.7ex
\hangindent=3em \hangafter=1
\textit{Intrinsic sign problem}

  \vskip 2ex

\noindent8. $11_{7,32.}^{16,796}$ \irep{0}:\ \ 
$d_i$ = ($1.0$,
$1.0$,
$1.0$,
$1.0$,
$2.0$,
$2.0$,
$2.0$,
$2.0$,
$2.0$,
$2.0$,
$2.0$) 

\vskip 0.7ex
\hangindent=3em \hangafter=1
$D^2= 32.0 = 
32$

\vskip 0.7ex
\hangindent=3em \hangafter=1
$T = ( 0,
0,
0,
0,
\frac{3}{4},
\frac{3}{16},
\frac{3}{16},
\frac{7}{16},
\frac{11}{16},
\frac{11}{16},
\frac{15}{16} )
$,

\vskip 0.7ex
\hangindent=3em \hangafter=1
$S$ = ($ 1$,
$ 1$,
$ 1$,
$ 1$,
$ 2$,
$ 2$,
$ 2$,
$ 2$,
$ 2$,
$ 2$,
$ 2$;\ \ 
$ 1$,
$ 1$,
$ 1$,
$ 2$,
$ -2$,
$ -2$,
$ 2$,
$ -2$,
$ -2$,
$ 2$;\ \ 
$ 1$,
$ 1$,
$ 2$,
$ -2$,
$ 2$,
$ -2$,
$ -2$,
$ 2$,
$ -2$;\ \ 
$ 1$,
$ 2$,
$ 2$,
$ -2$,
$ -2$,
$ 2$,
$ -2$,
$ -2$;\ \ 
$ -4$,
$0$,
$0$,
$0$,
$0$,
$0$,
$0$;\ \ 
$ -2\sqrt{2}$,
$0$,
$0$,
$ 2\sqrt{2}$,
$0$,
$0$;\ \ 
$ -2\sqrt{2}$,
$0$,
$0$,
$ 2\sqrt{2}$,
$0$;\ \ 
$ 2\sqrt{2}$,
$0$,
$0$,
$ -2\sqrt{2}$;\ \ 
$ -2\sqrt{2}$,
$0$,
$0$;\ \ 
$ -2\sqrt{2}$,
$0$;\ \ 
$ 2\sqrt{2}$)

\vskip 0.7ex
\hangindent=3em \hangafter=1
$\tau_n$ = ($4. - 4. i$, $-5.66 + 5.66 i$, $4. + 4. i$, $8. - 24. i$, $4. - 4. i$, $5.66 + 5.66 i$, $4. + 4. i$, $-16.$, $4. - 4. i$, $5.66 - 5.66 i$, $4. + 4. i$, $8. + 24. i$, $4. - 4. i$, $-5.66 - 5.66 i$, $4. + 4. i$, $32.$)

\vskip 0.7ex
\hangindent=3em \hangafter=1
\textit{Intrinsic sign problem}

  \vskip 2ex

\noindent9. $11_{7,32.}^{16,192}$ \irep{0}:\ \ 
$d_i$ = ($1.0$,
$1.0$,
$1.0$,
$1.0$,
$2.0$,
$2.0$,
$2.0$,
$2.0$,
$2.0$,
$2.0$,
$2.0$) 

\vskip 0.7ex
\hangindent=3em \hangafter=1
$D^2= 32.0 = 
32$

\vskip 0.7ex
\hangindent=3em \hangafter=1
$T = ( 0,
0,
0,
0,
\frac{3}{4},
\frac{3}{16},
\frac{7}{16},
\frac{7}{16},
\frac{11}{16},
\frac{15}{16},
\frac{15}{16} )
$,

\vskip 0.7ex
\hangindent=3em \hangafter=1
$S$ = ($ 1$,
$ 1$,
$ 1$,
$ 1$,
$ 2$,
$ 2$,
$ 2$,
$ 2$,
$ 2$,
$ 2$,
$ 2$;\ \ 
$ 1$,
$ 1$,
$ 1$,
$ 2$,
$ -2$,
$ -2$,
$ 2$,
$ -2$,
$ -2$,
$ 2$;\ \ 
$ 1$,
$ 1$,
$ 2$,
$ -2$,
$ 2$,
$ -2$,
$ -2$,
$ 2$,
$ -2$;\ \ 
$ 1$,
$ 2$,
$ 2$,
$ -2$,
$ -2$,
$ 2$,
$ -2$,
$ -2$;\ \ 
$ -4$,
$0$,
$0$,
$0$,
$0$,
$0$,
$0$;\ \ 
$ -2\sqrt{2}$,
$0$,
$0$,
$ 2\sqrt{2}$,
$0$,
$0$;\ \ 
$ 2\sqrt{2}$,
$0$,
$0$,
$ -2\sqrt{2}$,
$0$;\ \ 
$ 2\sqrt{2}$,
$0$,
$0$,
$ -2\sqrt{2}$;\ \ 
$ -2\sqrt{2}$,
$0$,
$0$;\ \ 
$ 2\sqrt{2}$,
$0$;\ \ 
$ 2\sqrt{2}$)

\vskip 0.7ex
\hangindent=3em \hangafter=1
$\tau_n$ = ($4. - 4. i$, $5.66 - 5.66 i$, $4. + 4. i$, $8. - 24. i$, $4. - 4. i$, $-5.66 - 5.66 i$, $4. + 4. i$, $-16.$, $4. - 4. i$, $-5.66 + 5.66 i$, $4. + 4. i$, $8. + 24. i$, $4. - 4. i$, $5.66 + 5.66 i$, $4. + 4. i$, $32.$)

\vskip 0.7ex
\hangindent=3em \hangafter=1
\textit{Intrinsic sign problem}

  \vskip 2ex

\noindent10. $11_{7,32.}^{16,304}$ \irep{0}:\ \ 
$d_i$ = ($1.0$,
$1.0$,
$1.0$,
$1.0$,
$2.0$,
$2.0$,
$2.0$,
$2.0$,
$2.0$,
$2.0$,
$2.0$) 

\vskip 0.7ex
\hangindent=3em \hangafter=1
$D^2= 32.0 = 
32$

\vskip 0.7ex
\hangindent=3em \hangafter=1
$T = ( 0,
0,
0,
0,
\frac{3}{4},
\frac{7}{16},
\frac{7}{16},
\frac{7}{16},
\frac{15}{16},
\frac{15}{16},
\frac{15}{16} )
$,

\vskip 0.7ex
\hangindent=3em \hangafter=1
$S$ = ($ 1$,
$ 1$,
$ 1$,
$ 1$,
$ 2$,
$ 2$,
$ 2$,
$ 2$,
$ 2$,
$ 2$,
$ 2$;\ \ 
$ 1$,
$ 1$,
$ 1$,
$ 2$,
$ -2$,
$ -2$,
$ 2$,
$ -2$,
$ -2$,
$ 2$;\ \ 
$ 1$,
$ 1$,
$ 2$,
$ -2$,
$ 2$,
$ -2$,
$ -2$,
$ 2$,
$ -2$;\ \ 
$ 1$,
$ 2$,
$ 2$,
$ -2$,
$ -2$,
$ 2$,
$ -2$,
$ -2$;\ \ 
$ -4$,
$0$,
$0$,
$0$,
$0$,
$0$,
$0$;\ \ 
$ 2\sqrt{2}$,
$0$,
$0$,
$ -2\sqrt{2}$,
$0$,
$0$;\ \ 
$ 2\sqrt{2}$,
$0$,
$0$,
$ -2\sqrt{2}$,
$0$;\ \ 
$ 2\sqrt{2}$,
$0$,
$0$,
$ -2\sqrt{2}$;\ \ 
$ 2\sqrt{2}$,
$0$,
$0$;\ \ 
$ 2\sqrt{2}$,
$0$;\ \ 
$ 2\sqrt{2}$)

\vskip 0.7ex
\hangindent=3em \hangafter=1
$\tau_n$ = ($4. - 4. i$, $16.97 - 16.97 i$, $4. + 4. i$, $8. - 24. i$, $4. - 4. i$, $-16.97 - 16.97 i$, $4. + 4. i$, $-16.$, $4. - 4. i$, $-16.97 + 16.97 i$, $4. + 4. i$, $8. + 24. i$, $4. - 4. i$, $16.97 + 16.97 i$, $4. + 4. i$, $32.$)

\vskip 0.7ex
\hangindent=3em \hangafter=1
\textit{Intrinsic sign problem}

  \vskip 2ex

\noindent11. $11_{2,60.}^{120,157}$ \irep{2274}:\ \ 
$d_i$ = ($1.0$,
$1.0$,
$2.0$,
$2.0$,
$2.0$,
$2.0$,
$2.0$,
$2.0$,
$2.0$,
$3.872$,
$3.872$) 

\vskip 0.7ex
\hangindent=3em \hangafter=1
$D^2= 60.0 = 
60$

\vskip 0.7ex
\hangindent=3em \hangafter=1
$T = ( 0,
0,
\frac{1}{3},
\frac{1}{5},
\frac{4}{5},
\frac{2}{15},
\frac{2}{15},
\frac{8}{15},
\frac{8}{15},
\frac{1}{8},
\frac{5}{8} )
$,

\vskip 0.7ex
\hangindent=3em \hangafter=1
$S$ = ($ 1$,
$ 1$,
$ 2$,
$ 2$,
$ 2$,
$ 2$,
$ 2$,
$ 2$,
$ 2$,
$ \sqrt{15}$,
$ \sqrt{15}$;\ \ 
$ 1$,
$ 2$,
$ 2$,
$ 2$,
$ 2$,
$ 2$,
$ 2$,
$ 2$,
$ -\sqrt{15}$,
$ -\sqrt{15}$;\ \ 
$ -2$,
$ 4$,
$ 4$,
$ -2$,
$ -2$,
$ -2$,
$ -2$,
$0$,
$0$;\ \ 
$ -1-\sqrt{5}$,
$ -1+\sqrt{5}$,
$ -1+\sqrt{5}$,
$ -1+\sqrt{5}$,
$ -1-\sqrt{5}$,
$ -1-\sqrt{5}$,
$0$,
$0$;\ \ 
$ -1-\sqrt{5}$,
$ -1-\sqrt{5}$,
$ -1-\sqrt{5}$,
$ -1+\sqrt{5}$,
$ -1+\sqrt{5}$,
$0$,
$0$;\ \ 
$ 2c_{15}^{4}$,
$ 2c_{15}^{1}$,
$ 2c_{15}^{7}$,
$ 2c_{15}^{2}$,
$0$,
$0$;\ \ 
$ 2c_{15}^{4}$,
$ 2c_{15}^{2}$,
$ 2c_{15}^{7}$,
$0$,
$0$;\ \ 
$ 2c_{15}^{1}$,
$ 2c_{15}^{4}$,
$0$,
$0$;\ \ 
$ 2c_{15}^{1}$,
$0$,
$0$;\ \ 
$ \sqrt{15}$,
$ -\sqrt{15}$;\ \ 
$ \sqrt{15}$)

\vskip 0.7ex
\hangindent=3em \hangafter=1
$\tau_n$ = ($0. + 7.75 i$, $0. + 37.73 i$, $-13.42$, $-29.98 + 7.75 i$, $0. - 17.32 i$, $13.42 - 29.98 i$, $0. - 7.75 i$, $29.98 + 7.75 i$, $13.42$, $0. + 47.31 i$, $0. - 7.75 i$, $-43.4$, $0. - 7.75 i$, $0. - 37.73 i$, $30.$, $29.98 + 7.75 i$, $0. + 7.75 i$, $-13.42 + 29.98 i$, $0. + 7.75 i$, $-29.98 - 17.32 i$, $13.42$, $0. - 37.73 i$, $0. + 7.75 i$, $43.4$, $0. + 17.32 i$, $0. + 22.24 i$, $-13.42$, $-29.98 - 7.75 i$, $0. - 7.75 i$, $30. - 29.98 i$, $0. + 7.75 i$, $29.98 + 7.75 i$, $-13.42$, $0. + 37.73 i$, $0. - 17.32 i$, $-16.57$, $0. - 7.75 i$, $0. - 22.24 i$, $13.42$, $29.98 + 17.32 i$, $0. - 7.75 i$, $-13.42 + 29.98 i$, $0. - 7.75 i$, $-29.98 - 7.75 i$, $30.$, $0. - 22.24 i$, $0. + 7.75 i$, $16.57$, $0. + 7.75 i$, $0. + 12.66 i$, $13.42$, $-29.98 - 7.75 i$, $0. + 7.75 i$, $13.42 - 29.98 i$, $0. + 17.32 i$, $29.98 - 7.75 i$, $-13.42$, $0. + 22.24 i$, $0. - 7.75 i$, $0.02$, $0. + 7.75 i$, $0. - 22.24 i$, $-13.42$, $29.98 + 7.75 i$, $0. - 17.32 i$, $13.42 + 29.98 i$, $0. - 7.75 i$, $-29.98 + 7.75 i$, $13.42$, $0. - 12.66 i$, $0. - 7.75 i$, $16.57$, $0. - 7.75 i$, $0. + 22.24 i$, $30.$, $-29.98 + 7.75 i$, $0. + 7.75 i$, $-13.42 - 29.98 i$, $0. + 7.75 i$, $29.98 - 17.32 i$, $13.42$, $0. + 22.24 i$, $0. + 7.75 i$, $-16.57$, $0. + 17.32 i$, $0. - 37.73 i$, $-13.42$, $29.98 - 7.75 i$, $0. - 7.75 i$, $30. + 29.98 i$, $0. + 7.75 i$, $-29.98 + 7.75 i$, $-13.42$, $0. - 22.24 i$, $0. - 17.32 i$, $43.4$, $0. - 7.75 i$, $0. + 37.73 i$, $13.42$, $-29.98 + 17.32 i$, $0. - 7.75 i$, $-13.42 - 29.98 i$, $0. - 7.75 i$, $29.98 - 7.75 i$, $30.$, $0. + 37.73 i$, $0. + 7.75 i$, $-43.4$, $0. + 7.75 i$, $0. - 47.31 i$, $13.42$, $29.98 - 7.75 i$, $0. + 7.75 i$, $13.42 + 29.98 i$, $0. + 17.32 i$, $-29.98 - 7.75 i$, $-13.42$, $0. - 37.73 i$, $0. - 7.75 i$, $59.98$)

\vskip 0.7ex
\hangindent=3em \hangafter=1
\textit{Intrinsic sign problem}

  \vskip 2ex

\noindent12. $11_{2,60.}^{120,364}$ \irep{2274}:\ \ 
$d_i$ = ($1.0$,
$1.0$,
$2.0$,
$2.0$,
$2.0$,
$2.0$,
$2.0$,
$2.0$,
$2.0$,
$3.872$,
$3.872$) 

\vskip 0.7ex
\hangindent=3em \hangafter=1
$D^2= 60.0 = 
60$

\vskip 0.7ex
\hangindent=3em \hangafter=1
$T = ( 0,
0,
\frac{1}{3},
\frac{1}{5},
\frac{4}{5},
\frac{2}{15},
\frac{2}{15},
\frac{8}{15},
\frac{8}{15},
\frac{3}{8},
\frac{7}{8} )
$,

\vskip 0.7ex
\hangindent=3em \hangafter=1
$S$ = ($ 1$,
$ 1$,
$ 2$,
$ 2$,
$ 2$,
$ 2$,
$ 2$,
$ 2$,
$ 2$,
$ \sqrt{15}$,
$ \sqrt{15}$;\ \ 
$ 1$,
$ 2$,
$ 2$,
$ 2$,
$ 2$,
$ 2$,
$ 2$,
$ 2$,
$ -\sqrt{15}$,
$ -\sqrt{15}$;\ \ 
$ -2$,
$ 4$,
$ 4$,
$ -2$,
$ -2$,
$ -2$,
$ -2$,
$0$,
$0$;\ \ 
$ -1-\sqrt{5}$,
$ -1+\sqrt{5}$,
$ -1+\sqrt{5}$,
$ -1+\sqrt{5}$,
$ -1-\sqrt{5}$,
$ -1-\sqrt{5}$,
$0$,
$0$;\ \ 
$ -1-\sqrt{5}$,
$ -1-\sqrt{5}$,
$ -1-\sqrt{5}$,
$ -1+\sqrt{5}$,
$ -1+\sqrt{5}$,
$0$,
$0$;\ \ 
$ 2c_{15}^{4}$,
$ 2c_{15}^{1}$,
$ 2c_{15}^{7}$,
$ 2c_{15}^{2}$,
$0$,
$0$;\ \ 
$ 2c_{15}^{4}$,
$ 2c_{15}^{2}$,
$ 2c_{15}^{7}$,
$0$,
$0$;\ \ 
$ 2c_{15}^{1}$,
$ 2c_{15}^{4}$,
$0$,
$0$;\ \ 
$ 2c_{15}^{1}$,
$0$,
$0$;\ \ 
$ -\sqrt{15}$,
$ \sqrt{15}$;\ \ 
$ -\sqrt{15}$)

\vskip 0.7ex
\hangindent=3em \hangafter=1
$\tau_n$ = ($0. + 7.75 i$, $0. - 22.24 i$, $-13.42$, $-29.98 + 7.75 i$, $0. - 17.32 i$, $13.42 + 29.98 i$, $0. - 7.75 i$, $29.98 + 7.75 i$, $13.42$, $0. - 12.66 i$, $0. - 7.75 i$, $-43.4$, $0. - 7.75 i$, $0. + 22.24 i$, $30.$, $29.98 + 7.75 i$, $0. + 7.75 i$, $-13.42 - 29.98 i$, $0. + 7.75 i$, $-29.98 - 17.32 i$, $13.42$, $0. + 22.24 i$, $0. + 7.75 i$, $43.4$, $0. + 17.32 i$, $0. - 37.73 i$, $-13.42$, $-29.98 - 7.75 i$, $0. - 7.75 i$, $30. + 29.98 i$, $0. + 7.75 i$, $29.98 + 7.75 i$, $-13.42$, $0. - 22.24 i$, $0. - 17.32 i$, $-16.57$, $0. - 7.75 i$, $0. + 37.73 i$, $13.42$, $29.98 + 17.32 i$, $0. - 7.75 i$, $-13.42 - 29.98 i$, $0. - 7.75 i$, $-29.98 - 7.75 i$, $30.$, $0. + 37.73 i$, $0. + 7.75 i$, $16.57$, $0. + 7.75 i$, $0. - 47.31 i$, $13.42$, $-29.98 - 7.75 i$, $0. + 7.75 i$, $13.42 + 29.98 i$, $0. + 17.32 i$, $29.98 - 7.75 i$, $-13.42$, $0. - 37.73 i$, $0. - 7.75 i$, $0.02$, $0. + 7.75 i$, $0. + 37.73 i$, $-13.42$, $29.98 + 7.75 i$, $0. - 17.32 i$, $13.42 - 29.98 i$, $0. - 7.75 i$, $-29.98 + 7.75 i$, $13.42$, $0. + 47.31 i$, $0. - 7.75 i$, $16.57$, $0. - 7.75 i$, $0. - 37.73 i$, $30.$, $-29.98 + 7.75 i$, $0. + 7.75 i$, $-13.42 + 29.98 i$, $0. + 7.75 i$, $29.98 - 17.32 i$, $13.42$, $0. - 37.73 i$, $0. + 7.75 i$, $-16.57$, $0. + 17.32 i$, $0. + 22.24 i$, $-13.42$, $29.98 - 7.75 i$, $0. - 7.75 i$, $30. - 29.98 i$, $0. + 7.75 i$, $-29.98 + 7.75 i$, $-13.42$, $0. + 37.73 i$, $0. - 17.32 i$, $43.4$, $0. - 7.75 i$, $0. - 22.24 i$, $13.42$, $-29.98 + 17.32 i$, $0. - 7.75 i$, $-13.42 + 29.98 i$, $0. - 7.75 i$, $29.98 - 7.75 i$, $30.$, $0. - 22.24 i$, $0. + 7.75 i$, $-43.4$, $0. + 7.75 i$, $0. + 12.66 i$, $13.42$, $29.98 - 7.75 i$, $0. + 7.75 i$, $13.42 - 29.98 i$, $0. + 17.32 i$, $-29.98 - 7.75 i$, $-13.42$, $0. + 22.24 i$, $0. - 7.75 i$, $59.98$)

\vskip 0.7ex
\hangindent=3em \hangafter=1
\textit{Intrinsic sign problem}

  \vskip 2ex

\noindent13. $11_{6,60.}^{120,253}$ \irep{2274}:\ \ 
$d_i$ = ($1.0$,
$1.0$,
$2.0$,
$2.0$,
$2.0$,
$2.0$,
$2.0$,
$2.0$,
$2.0$,
$3.872$,
$3.872$) 

\vskip 0.7ex
\hangindent=3em \hangafter=1
$D^2= 60.0 = 
60$

\vskip 0.7ex
\hangindent=3em \hangafter=1
$T = ( 0,
0,
\frac{1}{3},
\frac{2}{5},
\frac{3}{5},
\frac{11}{15},
\frac{11}{15},
\frac{14}{15},
\frac{14}{15},
\frac{1}{8},
\frac{5}{8} )
$,

\vskip 0.7ex
\hangindent=3em \hangafter=1
$S$ = ($ 1$,
$ 1$,
$ 2$,
$ 2$,
$ 2$,
$ 2$,
$ 2$,
$ 2$,
$ 2$,
$ \sqrt{15}$,
$ \sqrt{15}$;\ \ 
$ 1$,
$ 2$,
$ 2$,
$ 2$,
$ 2$,
$ 2$,
$ 2$,
$ 2$,
$ -\sqrt{15}$,
$ -\sqrt{15}$;\ \ 
$ -2$,
$ 4$,
$ 4$,
$ -2$,
$ -2$,
$ -2$,
$ -2$,
$0$,
$0$;\ \ 
$ -1+\sqrt{5}$,
$ -1-\sqrt{5}$,
$ -1+\sqrt{5}$,
$ -1+\sqrt{5}$,
$ -1-\sqrt{5}$,
$ -1-\sqrt{5}$,
$0$,
$0$;\ \ 
$ -1+\sqrt{5}$,
$ -1-\sqrt{5}$,
$ -1-\sqrt{5}$,
$ -1+\sqrt{5}$,
$ -1+\sqrt{5}$,
$0$,
$0$;\ \ 
$ 2c_{15}^{7}$,
$ 2c_{15}^{2}$,
$ 2c_{15}^{4}$,
$ 2c_{15}^{1}$,
$0$,
$0$;\ \ 
$ 2c_{15}^{7}$,
$ 2c_{15}^{1}$,
$ 2c_{15}^{4}$,
$0$,
$0$;\ \ 
$ 2c_{15}^{2}$,
$ 2c_{15}^{7}$,
$0$,
$0$;\ \ 
$ 2c_{15}^{2}$,
$0$,
$0$;\ \ 
$ -\sqrt{15}$,
$ \sqrt{15}$;\ \ 
$ -\sqrt{15}$)

\vskip 0.7ex
\hangindent=3em \hangafter=1
$\tau_n$ = ($0. - 7.75 i$, $0. + 22.24 i$, $13.42$, $-29.98 - 7.75 i$, $0. - 17.32 i$, $-13.42 - 29.98 i$, $0. + 7.75 i$, $29.98 - 7.75 i$, $-13.42$, $0. + 47.31 i$, $0. + 7.75 i$, $-16.57$, $0. + 7.75 i$, $0. - 22.24 i$, $30.$, $29.98 - 7.75 i$, $0. - 7.75 i$, $13.42 + 29.98 i$, $0. - 7.75 i$, $-29.98 - 17.32 i$, $-13.42$, $0. - 22.24 i$, $0. - 7.75 i$, $16.57$, $0. + 17.32 i$, $0. + 37.73 i$, $13.42$, $-29.98 + 7.75 i$, $0. + 7.75 i$, $30. - 29.98 i$, $0. - 7.75 i$, $29.98 - 7.75 i$, $13.42$, $0. + 22.24 i$, $0. - 17.32 i$, $-43.4$, $0. + 7.75 i$, $0. - 37.73 i$, $-13.42$, $29.98 + 17.32 i$, $0. + 7.75 i$, $13.42 + 29.98 i$, $0. + 7.75 i$, $-29.98 + 7.75 i$, $30.$, $0. - 37.73 i$, $0. - 7.75 i$, $43.4$, $0. - 7.75 i$, $0. + 12.66 i$, $-13.42$, $-29.98 + 7.75 i$, $0. - 7.75 i$, $-13.42 - 29.98 i$, $0. + 17.32 i$, $29.98 + 7.75 i$, $13.42$, $0. + 37.73 i$, $0. + 7.75 i$, $0.02$, $0. - 7.75 i$, $0. - 37.73 i$, $13.42$, $29.98 - 7.75 i$, $0. - 17.32 i$, $-13.42 + 29.98 i$, $0. + 7.75 i$, $-29.98 - 7.75 i$, $-13.42$, $0. - 12.66 i$, $0. + 7.75 i$, $43.4$, $0. + 7.75 i$, $0. + 37.73 i$, $30.$, $-29.98 - 7.75 i$, $0. - 7.75 i$, $13.42 - 29.98 i$, $0. - 7.75 i$, $29.98 - 17.32 i$, $-13.42$, $0. + 37.73 i$, $0. - 7.75 i$, $-43.4$, $0. + 17.32 i$, $0. - 22.24 i$, $13.42$, $29.98 + 7.75 i$, $0. + 7.75 i$, $30. + 29.98 i$, $0. - 7.75 i$, $-29.98 - 7.75 i$, $13.42$, $0. - 37.73 i$, $0. - 17.32 i$, $16.57$, $0. + 7.75 i$, $0. + 22.24 i$, $-13.42$, $-29.98 + 17.32 i$, $0. + 7.75 i$, $13.42 - 29.98 i$, $0. + 7.75 i$, $29.98 + 7.75 i$, $30.$, $0. + 22.24 i$, $0. - 7.75 i$, $-16.57$, $0. - 7.75 i$, $0. - 47.31 i$, $-13.42$, $29.98 + 7.75 i$, $0. - 7.75 i$, $-13.42 + 29.98 i$, $0. + 17.32 i$, $-29.98 + 7.75 i$, $13.42$, $0. - 22.24 i$, $0. + 7.75 i$, $59.98$)

\vskip 0.7ex
\hangindent=3em \hangafter=1
\textit{Intrinsic sign problem}

  \vskip 2ex

\noindent14. $11_{6,60.}^{120,447}$ \irep{2274}:\ \ 
$d_i$ = ($1.0$,
$1.0$,
$2.0$,
$2.0$,
$2.0$,
$2.0$,
$2.0$,
$2.0$,
$2.0$,
$3.872$,
$3.872$) 

\vskip 0.7ex
\hangindent=3em \hangafter=1
$D^2= 60.0 = 
60$

\vskip 0.7ex
\hangindent=3em \hangafter=1
$T = ( 0,
0,
\frac{1}{3},
\frac{2}{5},
\frac{3}{5},
\frac{11}{15},
\frac{11}{15},
\frac{14}{15},
\frac{14}{15},
\frac{3}{8},
\frac{7}{8} )
$,

\vskip 0.7ex
\hangindent=3em \hangafter=1
$S$ = ($ 1$,
$ 1$,
$ 2$,
$ 2$,
$ 2$,
$ 2$,
$ 2$,
$ 2$,
$ 2$,
$ \sqrt{15}$,
$ \sqrt{15}$;\ \ 
$ 1$,
$ 2$,
$ 2$,
$ 2$,
$ 2$,
$ 2$,
$ 2$,
$ 2$,
$ -\sqrt{15}$,
$ -\sqrt{15}$;\ \ 
$ -2$,
$ 4$,
$ 4$,
$ -2$,
$ -2$,
$ -2$,
$ -2$,
$0$,
$0$;\ \ 
$ -1+\sqrt{5}$,
$ -1-\sqrt{5}$,
$ -1+\sqrt{5}$,
$ -1+\sqrt{5}$,
$ -1-\sqrt{5}$,
$ -1-\sqrt{5}$,
$0$,
$0$;\ \ 
$ -1+\sqrt{5}$,
$ -1-\sqrt{5}$,
$ -1-\sqrt{5}$,
$ -1+\sqrt{5}$,
$ -1+\sqrt{5}$,
$0$,
$0$;\ \ 
$ 2c_{15}^{7}$,
$ 2c_{15}^{2}$,
$ 2c_{15}^{4}$,
$ 2c_{15}^{1}$,
$0$,
$0$;\ \ 
$ 2c_{15}^{7}$,
$ 2c_{15}^{1}$,
$ 2c_{15}^{4}$,
$0$,
$0$;\ \ 
$ 2c_{15}^{2}$,
$ 2c_{15}^{7}$,
$0$,
$0$;\ \ 
$ 2c_{15}^{2}$,
$0$,
$0$;\ \ 
$ \sqrt{15}$,
$ -\sqrt{15}$;\ \ 
$ \sqrt{15}$)

\vskip 0.7ex
\hangindent=3em \hangafter=1
$\tau_n$ = ($0. - 7.75 i$, $0. - 37.73 i$, $13.42$, $-29.98 - 7.75 i$, $0. - 17.32 i$, $-13.42 + 29.98 i$, $0. + 7.75 i$, $29.98 - 7.75 i$, $-13.42$, $0. - 12.66 i$, $0. + 7.75 i$, $-16.57$, $0. + 7.75 i$, $0. + 37.73 i$, $30.$, $29.98 - 7.75 i$, $0. - 7.75 i$, $13.42 - 29.98 i$, $0. - 7.75 i$, $-29.98 - 17.32 i$, $-13.42$, $0. + 37.73 i$, $0. - 7.75 i$, $16.57$, $0. + 17.32 i$, $0. - 22.24 i$, $13.42$, $-29.98 + 7.75 i$, $0. + 7.75 i$, $30. + 29.98 i$, $0. - 7.75 i$, $29.98 - 7.75 i$, $13.42$, $0. - 37.73 i$, $0. - 17.32 i$, $-43.4$, $0. + 7.75 i$, $0. + 22.24 i$, $-13.42$, $29.98 + 17.32 i$, $0. + 7.75 i$, $13.42 - 29.98 i$, $0. + 7.75 i$, $-29.98 + 7.75 i$, $30.$, $0. + 22.24 i$, $0. - 7.75 i$, $43.4$, $0. - 7.75 i$, $0. - 47.31 i$, $-13.42$, $-29.98 + 7.75 i$, $0. - 7.75 i$, $-13.42 + 29.98 i$, $0. + 17.32 i$, $29.98 + 7.75 i$, $13.42$, $0. - 22.24 i$, $0. + 7.75 i$, $0.02$, $0. - 7.75 i$, $0. + 22.24 i$, $13.42$, $29.98 - 7.75 i$, $0. - 17.32 i$, $-13.42 - 29.98 i$, $0. + 7.75 i$, $-29.98 - 7.75 i$, $-13.42$, $0. + 47.31 i$, $0. + 7.75 i$, $43.4$, $0. + 7.75 i$, $0. - 22.24 i$, $30.$, $-29.98 - 7.75 i$, $0. - 7.75 i$, $13.42 + 29.98 i$, $0. - 7.75 i$, $29.98 - 17.32 i$, $-13.42$, $0. - 22.24 i$, $0. - 7.75 i$, $-43.4$, $0. + 17.32 i$, $0. + 37.73 i$, $13.42$, $29.98 + 7.75 i$, $0. + 7.75 i$, $30. - 29.98 i$, $0. - 7.75 i$, $-29.98 - 7.75 i$, $13.42$, $0. + 22.24 i$, $0. - 17.32 i$, $16.57$, $0. + 7.75 i$, $0. - 37.73 i$, $-13.42$, $-29.98 + 17.32 i$, $0. + 7.75 i$, $13.42 + 29.98 i$, $0. + 7.75 i$, $29.98 + 7.75 i$, $30.$, $0. - 37.73 i$, $0. - 7.75 i$, $-16.57$, $0. - 7.75 i$, $0. + 12.66 i$, $-13.42$, $29.98 + 7.75 i$, $0. - 7.75 i$, $-13.42 - 29.98 i$, $0. + 17.32 i$, $-29.98 + 7.75 i$, $13.42$, $0. + 37.73 i$, $0. + 7.75 i$, $59.98$)

\vskip 0.7ex
\hangindent=3em \hangafter=1
\textit{Intrinsic sign problem}

  \vskip 2ex

\noindent15. $11_{6,60.}^{120,176}$ \irep{2274}:\ \ 
$d_i$ = ($1.0$,
$1.0$,
$2.0$,
$2.0$,
$2.0$,
$2.0$,
$2.0$,
$2.0$,
$2.0$,
$3.872$,
$3.872$) 

\vskip 0.7ex
\hangindent=3em \hangafter=1
$D^2= 60.0 = 
60$

\vskip 0.7ex
\hangindent=3em \hangafter=1
$T = ( 0,
0,
\frac{2}{3},
\frac{1}{5},
\frac{4}{5},
\frac{7}{15},
\frac{7}{15},
\frac{13}{15},
\frac{13}{15},
\frac{1}{8},
\frac{5}{8} )
$,

\vskip 0.7ex
\hangindent=3em \hangafter=1
$S$ = ($ 1$,
$ 1$,
$ 2$,
$ 2$,
$ 2$,
$ 2$,
$ 2$,
$ 2$,
$ 2$,
$ \sqrt{15}$,
$ \sqrt{15}$;\ \ 
$ 1$,
$ 2$,
$ 2$,
$ 2$,
$ 2$,
$ 2$,
$ 2$,
$ 2$,
$ -\sqrt{15}$,
$ -\sqrt{15}$;\ \ 
$ -2$,
$ 4$,
$ 4$,
$ -2$,
$ -2$,
$ -2$,
$ -2$,
$0$,
$0$;\ \ 
$ -1-\sqrt{5}$,
$ -1+\sqrt{5}$,
$ -1+\sqrt{5}$,
$ -1+\sqrt{5}$,
$ -1-\sqrt{5}$,
$ -1-\sqrt{5}$,
$0$,
$0$;\ \ 
$ -1-\sqrt{5}$,
$ -1-\sqrt{5}$,
$ -1-\sqrt{5}$,
$ -1+\sqrt{5}$,
$ -1+\sqrt{5}$,
$0$,
$0$;\ \ 
$ 2c_{15}^{1}$,
$ 2c_{15}^{4}$,
$ 2c_{15}^{7}$,
$ 2c_{15}^{2}$,
$0$,
$0$;\ \ 
$ 2c_{15}^{1}$,
$ 2c_{15}^{2}$,
$ 2c_{15}^{7}$,
$0$,
$0$;\ \ 
$ 2c_{15}^{4}$,
$ 2c_{15}^{1}$,
$0$,
$0$;\ \ 
$ 2c_{15}^{4}$,
$0$,
$0$;\ \ 
$ -\sqrt{15}$,
$ \sqrt{15}$;\ \ 
$ -\sqrt{15}$)

\vskip 0.7ex
\hangindent=3em \hangafter=1
$\tau_n$ = ($0. - 7.75 i$, $0. + 22.24 i$, $-13.42$, $-29.98 - 7.75 i$, $0. + 17.32 i$, $13.42 - 29.98 i$, $0. + 7.75 i$, $29.98 - 7.75 i$, $13.42$, $0. + 12.66 i$, $0. + 7.75 i$, $-43.4$, $0. + 7.75 i$, $0. - 22.24 i$, $30.$, $29.98 - 7.75 i$, $0. - 7.75 i$, $-13.42 + 29.98 i$, $0. - 7.75 i$, $-29.98 + 17.32 i$, $13.42$, $0. - 22.24 i$, $0. - 7.75 i$, $43.4$, $0. - 17.32 i$, $0. + 37.73 i$, $-13.42$, $-29.98 + 7.75 i$, $0. + 7.75 i$, $30. - 29.98 i$, $0. - 7.75 i$, $29.98 - 7.75 i$, $-13.42$, $0. + 22.24 i$, $0. + 17.32 i$, $-16.57$, $0. + 7.75 i$, $0. - 37.73 i$, $13.42$, $29.98 - 17.32 i$, $0. + 7.75 i$, $-13.42 + 29.98 i$, $0. + 7.75 i$, $-29.98 + 7.75 i$, $30.$, $0. - 37.73 i$, $0. - 7.75 i$, $16.57$, $0. - 7.75 i$, $0. + 47.31 i$, $13.42$, $-29.98 + 7.75 i$, $0. - 7.75 i$, $13.42 - 29.98 i$, $0. - 17.32 i$, $29.98 + 7.75 i$, $-13.42$, $0. + 37.73 i$, $0. + 7.75 i$, $0.02$, $0. - 7.75 i$, $0. - 37.73 i$, $-13.42$, $29.98 - 7.75 i$, $0. + 17.32 i$, $13.42 + 29.98 i$, $0. + 7.75 i$, $-29.98 - 7.75 i$, $13.42$, $0. - 47.31 i$, $0. + 7.75 i$, $16.57$, $0. + 7.75 i$, $0. + 37.73 i$, $30.$, $-29.98 - 7.75 i$, $0. - 7.75 i$, $-13.42 - 29.98 i$, $0. - 7.75 i$, $29.98 + 17.32 i$, $13.42$, $0. + 37.73 i$, $0. - 7.75 i$, $-16.57$, $0. - 17.32 i$, $0. - 22.24 i$, $-13.42$, $29.98 + 7.75 i$, $0. + 7.75 i$, $30. + 29.98 i$, $0. - 7.75 i$, $-29.98 - 7.75 i$, $-13.42$, $0. - 37.73 i$, $0. + 17.32 i$, $43.4$, $0. + 7.75 i$, $0. + 22.24 i$, $13.42$, $-29.98 - 17.32 i$, $0. + 7.75 i$, $-13.42 - 29.98 i$, $0. + 7.75 i$, $29.98 + 7.75 i$, $30.$, $0. + 22.24 i$, $0. - 7.75 i$, $-43.4$, $0. - 7.75 i$, $0. - 12.66 i$, $13.42$, $29.98 + 7.75 i$, $0. - 7.75 i$, $13.42 + 29.98 i$, $0. - 17.32 i$, $-29.98 + 7.75 i$, $-13.42$, $0. - 22.24 i$, $0. + 7.75 i$, $59.98$)

\vskip 0.7ex
\hangindent=3em \hangafter=1
\textit{Intrinsic sign problem}

  \vskip 2ex

\noindent16. $11_{6,60.}^{120,369}$ \irep{2274}:\ \ 
$d_i$ = ($1.0$,
$1.0$,
$2.0$,
$2.0$,
$2.0$,
$2.0$,
$2.0$,
$2.0$,
$2.0$,
$3.872$,
$3.872$) 

\vskip 0.7ex
\hangindent=3em \hangafter=1
$D^2= 60.0 = 
60$

\vskip 0.7ex
\hangindent=3em \hangafter=1
$T = ( 0,
0,
\frac{2}{3},
\frac{1}{5},
\frac{4}{5},
\frac{7}{15},
\frac{7}{15},
\frac{13}{15},
\frac{13}{15},
\frac{3}{8},
\frac{7}{8} )
$,

\vskip 0.7ex
\hangindent=3em \hangafter=1
$S$ = ($ 1$,
$ 1$,
$ 2$,
$ 2$,
$ 2$,
$ 2$,
$ 2$,
$ 2$,
$ 2$,
$ \sqrt{15}$,
$ \sqrt{15}$;\ \ 
$ 1$,
$ 2$,
$ 2$,
$ 2$,
$ 2$,
$ 2$,
$ 2$,
$ 2$,
$ -\sqrt{15}$,
$ -\sqrt{15}$;\ \ 
$ -2$,
$ 4$,
$ 4$,
$ -2$,
$ -2$,
$ -2$,
$ -2$,
$0$,
$0$;\ \ 
$ -1-\sqrt{5}$,
$ -1+\sqrt{5}$,
$ -1+\sqrt{5}$,
$ -1+\sqrt{5}$,
$ -1-\sqrt{5}$,
$ -1-\sqrt{5}$,
$0$,
$0$;\ \ 
$ -1-\sqrt{5}$,
$ -1-\sqrt{5}$,
$ -1-\sqrt{5}$,
$ -1+\sqrt{5}$,
$ -1+\sqrt{5}$,
$0$,
$0$;\ \ 
$ 2c_{15}^{1}$,
$ 2c_{15}^{4}$,
$ 2c_{15}^{7}$,
$ 2c_{15}^{2}$,
$0$,
$0$;\ \ 
$ 2c_{15}^{1}$,
$ 2c_{15}^{2}$,
$ 2c_{15}^{7}$,
$0$,
$0$;\ \ 
$ 2c_{15}^{4}$,
$ 2c_{15}^{1}$,
$0$,
$0$;\ \ 
$ 2c_{15}^{4}$,
$0$,
$0$;\ \ 
$ \sqrt{15}$,
$ -\sqrt{15}$;\ \ 
$ \sqrt{15}$)

\vskip 0.7ex
\hangindent=3em \hangafter=1
$\tau_n$ = ($0. - 7.75 i$, $0. - 37.73 i$, $-13.42$, $-29.98 - 7.75 i$, $0. + 17.32 i$, $13.42 + 29.98 i$, $0. + 7.75 i$, $29.98 - 7.75 i$, $13.42$, $0. - 47.31 i$, $0. + 7.75 i$, $-43.4$, $0. + 7.75 i$, $0. + 37.73 i$, $30.$, $29.98 - 7.75 i$, $0. - 7.75 i$, $-13.42 - 29.98 i$, $0. - 7.75 i$, $-29.98 + 17.32 i$, $13.42$, $0. + 37.73 i$, $0. - 7.75 i$, $43.4$, $0. - 17.32 i$, $0. - 22.24 i$, $-13.42$, $-29.98 + 7.75 i$, $0. + 7.75 i$, $30. + 29.98 i$, $0. - 7.75 i$, $29.98 - 7.75 i$, $-13.42$, $0. - 37.73 i$, $0. + 17.32 i$, $-16.57$, $0. + 7.75 i$, $0. + 22.24 i$, $13.42$, $29.98 - 17.32 i$, $0. + 7.75 i$, $-13.42 - 29.98 i$, $0. + 7.75 i$, $-29.98 + 7.75 i$, $30.$, $0. + 22.24 i$, $0. - 7.75 i$, $16.57$, $0. - 7.75 i$, $0. - 12.66 i$, $13.42$, $-29.98 + 7.75 i$, $0. - 7.75 i$, $13.42 + 29.98 i$, $0. - 17.32 i$, $29.98 + 7.75 i$, $-13.42$, $0. - 22.24 i$, $0. + 7.75 i$, $0.02$, $0. - 7.75 i$, $0. + 22.24 i$, $-13.42$, $29.98 - 7.75 i$, $0. + 17.32 i$, $13.42 - 29.98 i$, $0. + 7.75 i$, $-29.98 - 7.75 i$, $13.42$, $0. + 12.66 i$, $0. + 7.75 i$, $16.57$, $0. + 7.75 i$, $0. - 22.24 i$, $30.$, $-29.98 - 7.75 i$, $0. - 7.75 i$, $-13.42 + 29.98 i$, $0. - 7.75 i$, $29.98 + 17.32 i$, $13.42$, $0. - 22.24 i$, $0. - 7.75 i$, $-16.57$, $0. - 17.32 i$, $0. + 37.73 i$, $-13.42$, $29.98 + 7.75 i$, $0. + 7.75 i$, $30. - 29.98 i$, $0. - 7.75 i$, $-29.98 - 7.75 i$, $-13.42$, $0. + 22.24 i$, $0. + 17.32 i$, $43.4$, $0. + 7.75 i$, $0. - 37.73 i$, $13.42$, $-29.98 - 17.32 i$, $0. + 7.75 i$, $-13.42 + 29.98 i$, $0. + 7.75 i$, $29.98 + 7.75 i$, $30.$, $0. - 37.73 i$, $0. - 7.75 i$, $-43.4$, $0. - 7.75 i$, $0. + 47.31 i$, $13.42$, $29.98 + 7.75 i$, $0. - 7.75 i$, $13.42 - 29.98 i$, $0. - 17.32 i$, $-29.98 + 7.75 i$, $-13.42$, $0. + 37.73 i$, $0. + 7.75 i$, $59.98$)

\vskip 0.7ex
\hangindent=3em \hangafter=1
\textit{Intrinsic sign problem}

  \vskip 2ex

\noindent17. $11_{2,60.}^{120,213}$ \irep{2274}:\ \ 
$d_i$ = ($1.0$,
$1.0$,
$2.0$,
$2.0$,
$2.0$,
$2.0$,
$2.0$,
$2.0$,
$2.0$,
$3.872$,
$3.872$) 

\vskip 0.7ex
\hangindent=3em \hangafter=1
$D^2= 60.0 = 
60$

\vskip 0.7ex
\hangindent=3em \hangafter=1
$T = ( 0,
0,
\frac{2}{3},
\frac{2}{5},
\frac{3}{5},
\frac{1}{15},
\frac{1}{15},
\frac{4}{15},
\frac{4}{15},
\frac{1}{8},
\frac{5}{8} )
$,

\vskip 0.7ex
\hangindent=3em \hangafter=1
$S$ = ($ 1$,
$ 1$,
$ 2$,
$ 2$,
$ 2$,
$ 2$,
$ 2$,
$ 2$,
$ 2$,
$ \sqrt{15}$,
$ \sqrt{15}$;\ \ 
$ 1$,
$ 2$,
$ 2$,
$ 2$,
$ 2$,
$ 2$,
$ 2$,
$ 2$,
$ -\sqrt{15}$,
$ -\sqrt{15}$;\ \ 
$ -2$,
$ 4$,
$ 4$,
$ -2$,
$ -2$,
$ -2$,
$ -2$,
$0$,
$0$;\ \ 
$ -1+\sqrt{5}$,
$ -1-\sqrt{5}$,
$ -1+\sqrt{5}$,
$ -1+\sqrt{5}$,
$ -1-\sqrt{5}$,
$ -1-\sqrt{5}$,
$0$,
$0$;\ \ 
$ -1+\sqrt{5}$,
$ -1-\sqrt{5}$,
$ -1-\sqrt{5}$,
$ -1+\sqrt{5}$,
$ -1+\sqrt{5}$,
$0$,
$0$;\ \ 
$ 2c_{15}^{2}$,
$ 2c_{15}^{7}$,
$ 2c_{15}^{4}$,
$ 2c_{15}^{1}$,
$0$,
$0$;\ \ 
$ 2c_{15}^{2}$,
$ 2c_{15}^{1}$,
$ 2c_{15}^{4}$,
$0$,
$0$;\ \ 
$ 2c_{15}^{7}$,
$ 2c_{15}^{2}$,
$0$,
$0$;\ \ 
$ 2c_{15}^{7}$,
$0$,
$0$;\ \ 
$ \sqrt{15}$,
$ -\sqrt{15}$;\ \ 
$ \sqrt{15}$)

\vskip 0.7ex
\hangindent=3em \hangafter=1
$\tau_n$ = ($0. + 7.75 i$, $0. + 37.73 i$, $13.42$, $-29.98 + 7.75 i$, $0. + 17.32 i$, $-13.42 - 29.98 i$, $0. - 7.75 i$, $29.98 + 7.75 i$, $-13.42$, $0. + 12.66 i$, $0. - 7.75 i$, $-16.57$, $0. - 7.75 i$, $0. - 37.73 i$, $30.$, $29.98 + 7.75 i$, $0. + 7.75 i$, $13.42 + 29.98 i$, $0. + 7.75 i$, $-29.98 + 17.32 i$, $-13.42$, $0. - 37.73 i$, $0. + 7.75 i$, $16.57$, $0. - 17.32 i$, $0. + 22.24 i$, $13.42$, $-29.98 - 7.75 i$, $0. - 7.75 i$, $30. - 29.98 i$, $0. + 7.75 i$, $29.98 + 7.75 i$, $13.42$, $0. + 37.73 i$, $0. + 17.32 i$, $-43.4$, $0. - 7.75 i$, $0. - 22.24 i$, $-13.42$, $29.98 - 17.32 i$, $0. - 7.75 i$, $13.42 + 29.98 i$, $0. - 7.75 i$, $-29.98 - 7.75 i$, $30.$, $0. - 22.24 i$, $0. + 7.75 i$, $43.4$, $0. + 7.75 i$, $0. + 47.31 i$, $-13.42$, $-29.98 - 7.75 i$, $0. + 7.75 i$, $-13.42 - 29.98 i$, $0. - 17.32 i$, $29.98 - 7.75 i$, $13.42$, $0. + 22.24 i$, $0. - 7.75 i$, $0.02$, $0. + 7.75 i$, $0. - 22.24 i$, $13.42$, $29.98 + 7.75 i$, $0. + 17.32 i$, $-13.42 + 29.98 i$, $0. - 7.75 i$, $-29.98 + 7.75 i$, $-13.42$, $0. - 47.31 i$, $0. - 7.75 i$, $43.4$, $0. - 7.75 i$, $0. + 22.24 i$, $30.$, $-29.98 + 7.75 i$, $0. + 7.75 i$, $13.42 - 29.98 i$, $0. + 7.75 i$, $29.98 + 17.32 i$, $-13.42$, $0. + 22.24 i$, $0. + 7.75 i$, $-43.4$, $0. - 17.32 i$, $0. - 37.73 i$, $13.42$, $29.98 - 7.75 i$, $0. - 7.75 i$, $30. + 29.98 i$, $0. + 7.75 i$, $-29.98 + 7.75 i$, $13.42$, $0. - 22.24 i$, $0. + 17.32 i$, $16.57$, $0. - 7.75 i$, $0. + 37.73 i$, $-13.42$, $-29.98 - 17.32 i$, $0. - 7.75 i$, $13.42 - 29.98 i$, $0. - 7.75 i$, $29.98 - 7.75 i$, $30.$, $0. + 37.73 i$, $0. + 7.75 i$, $-16.57$, $0. + 7.75 i$, $0. - 12.66 i$, $-13.42$, $29.98 - 7.75 i$, $0. + 7.75 i$, $-13.42 + 29.98 i$, $0. - 17.32 i$, $-29.98 - 7.75 i$, $13.42$, $0. - 37.73 i$, $0. - 7.75 i$, $59.98$)

\vskip 0.7ex
\hangindent=3em \hangafter=1
\textit{Intrinsic sign problem}

  \vskip 2ex

\noindent18. $11_{2,60.}^{120,195}$ \irep{2274}:\ \ 
$d_i$ = ($1.0$,
$1.0$,
$2.0$,
$2.0$,
$2.0$,
$2.0$,
$2.0$,
$2.0$,
$2.0$,
$3.872$,
$3.872$) 

\vskip 0.7ex
\hangindent=3em \hangafter=1
$D^2= 60.0 = 
60$

\vskip 0.7ex
\hangindent=3em \hangafter=1
$T = ( 0,
0,
\frac{2}{3},
\frac{2}{5},
\frac{3}{5},
\frac{1}{15},
\frac{1}{15},
\frac{4}{15},
\frac{4}{15},
\frac{3}{8},
\frac{7}{8} )
$,

\vskip 0.7ex
\hangindent=3em \hangafter=1
$S$ = ($ 1$,
$ 1$,
$ 2$,
$ 2$,
$ 2$,
$ 2$,
$ 2$,
$ 2$,
$ 2$,
$ \sqrt{15}$,
$ \sqrt{15}$;\ \ 
$ 1$,
$ 2$,
$ 2$,
$ 2$,
$ 2$,
$ 2$,
$ 2$,
$ 2$,
$ -\sqrt{15}$,
$ -\sqrt{15}$;\ \ 
$ -2$,
$ 4$,
$ 4$,
$ -2$,
$ -2$,
$ -2$,
$ -2$,
$0$,
$0$;\ \ 
$ -1+\sqrt{5}$,
$ -1-\sqrt{5}$,
$ -1+\sqrt{5}$,
$ -1+\sqrt{5}$,
$ -1-\sqrt{5}$,
$ -1-\sqrt{5}$,
$0$,
$0$;\ \ 
$ -1+\sqrt{5}$,
$ -1-\sqrt{5}$,
$ -1-\sqrt{5}$,
$ -1+\sqrt{5}$,
$ -1+\sqrt{5}$,
$0$,
$0$;\ \ 
$ 2c_{15}^{2}$,
$ 2c_{15}^{7}$,
$ 2c_{15}^{4}$,
$ 2c_{15}^{1}$,
$0$,
$0$;\ \ 
$ 2c_{15}^{2}$,
$ 2c_{15}^{1}$,
$ 2c_{15}^{4}$,
$0$,
$0$;\ \ 
$ 2c_{15}^{7}$,
$ 2c_{15}^{2}$,
$0$,
$0$;\ \ 
$ 2c_{15}^{7}$,
$0$,
$0$;\ \ 
$ -\sqrt{15}$,
$ \sqrt{15}$;\ \ 
$ -\sqrt{15}$)

\vskip 0.7ex
\hangindent=3em \hangafter=1
$\tau_n$ = ($0. + 7.75 i$, $0. - 22.24 i$, $13.42$, $-29.98 + 7.75 i$, $0. + 17.32 i$, $-13.42 + 29.98 i$, $0. - 7.75 i$, $29.98 + 7.75 i$, $-13.42$, $0. - 47.31 i$, $0. - 7.75 i$, $-16.57$, $0. - 7.75 i$, $0. + 22.24 i$, $30.$, $29.98 + 7.75 i$, $0. + 7.75 i$, $13.42 - 29.98 i$, $0. + 7.75 i$, $-29.98 + 17.32 i$, $-13.42$, $0. + 22.24 i$, $0. + 7.75 i$, $16.57$, $0. - 17.32 i$, $0. - 37.73 i$, $13.42$, $-29.98 - 7.75 i$, $0. - 7.75 i$, $30. + 29.98 i$, $0. + 7.75 i$, $29.98 + 7.75 i$, $13.42$, $0. - 22.24 i$, $0. + 17.32 i$, $-43.4$, $0. - 7.75 i$, $0. + 37.73 i$, $-13.42$, $29.98 - 17.32 i$, $0. - 7.75 i$, $13.42 - 29.98 i$, $0. - 7.75 i$, $-29.98 - 7.75 i$, $30.$, $0. + 37.73 i$, $0. + 7.75 i$, $43.4$, $0. + 7.75 i$, $0. - 12.66 i$, $-13.42$, $-29.98 - 7.75 i$, $0. + 7.75 i$, $-13.42 + 29.98 i$, $0. - 17.32 i$, $29.98 - 7.75 i$, $13.42$, $0. - 37.73 i$, $0. - 7.75 i$, $0.02$, $0. + 7.75 i$, $0. + 37.73 i$, $13.42$, $29.98 + 7.75 i$, $0. + 17.32 i$, $-13.42 - 29.98 i$, $0. - 7.75 i$, $-29.98 + 7.75 i$, $-13.42$, $0. + 12.66 i$, $0. - 7.75 i$, $43.4$, $0. - 7.75 i$, $0. - 37.73 i$, $30.$, $-29.98 + 7.75 i$, $0. + 7.75 i$, $13.42 + 29.98 i$, $0. + 7.75 i$, $29.98 + 17.32 i$, $-13.42$, $0. - 37.73 i$, $0. + 7.75 i$, $-43.4$, $0. - 17.32 i$, $0. + 22.24 i$, $13.42$, $29.98 - 7.75 i$, $0. - 7.75 i$, $30. - 29.98 i$, $0. + 7.75 i$, $-29.98 + 7.75 i$, $13.42$, $0. + 37.73 i$, $0. + 17.32 i$, $16.57$, $0. - 7.75 i$, $0. - 22.24 i$, $-13.42$, $-29.98 - 17.32 i$, $0. - 7.75 i$, $13.42 + 29.98 i$, $0. - 7.75 i$, $29.98 - 7.75 i$, $30.$, $0. - 22.24 i$, $0. + 7.75 i$, $-16.57$, $0. + 7.75 i$, $0. + 47.31 i$, $-13.42$, $29.98 - 7.75 i$, $0. + 7.75 i$, $-13.42 - 29.98 i$, $0. - 17.32 i$, $-29.98 - 7.75 i$, $13.42$, $0. + 22.24 i$, $0. - 7.75 i$, $59.98$)

\vskip 0.7ex
\hangindent=3em \hangafter=1
\textit{Intrinsic sign problem}

  \vskip 2ex

\noindent19. $11_{\frac{13}{2},89.56}^{48,108}$ \irep{2191}:\ \ 
$d_i$ = ($1.0$,
$1.0$,
$1.931$,
$1.931$,
$2.732$,
$2.732$,
$3.346$,
$3.346$,
$3.732$,
$3.732$,
$3.863$) 

\vskip 0.7ex
\hangindent=3em \hangafter=1
$D^2= 89.569 = 
48+24\sqrt{3}$

\vskip 0.7ex
\hangindent=3em \hangafter=1
$T = ( 0,
\frac{1}{2},
\frac{1}{16},
\frac{1}{16},
\frac{1}{3},
\frac{5}{6},
\frac{13}{16},
\frac{13}{16},
0,
\frac{1}{2},
\frac{19}{48} )
$,

\vskip 0.7ex
\hangindent=3em \hangafter=1
$S$ = ($ 1$,
$ 1$,
$ c_{24}^{1}$,
$ c_{24}^{1}$,
$ 1+\sqrt{3}$,
$ 1+\sqrt{3}$,
$ \frac{3+3\sqrt{3}}{\sqrt{6}}$,
$ \frac{3+3\sqrt{3}}{\sqrt{6}}$,
$ 2+\sqrt{3}$,
$ 2+\sqrt{3}$,
$ 2c_{24}^{1}$;\ \ 
$ 1$,
$ -c_{24}^{1}$,
$ -c_{24}^{1}$,
$ 1+\sqrt{3}$,
$ 1+\sqrt{3}$,
$ \frac{-3-3\sqrt{3}}{\sqrt{6}}$,
$ \frac{-3-3\sqrt{3}}{\sqrt{6}}$,
$ 2+\sqrt{3}$,
$ 2+\sqrt{3}$,
$ -2c_{24}^{1}$;\ \ 
$(\frac{-3-3\sqrt{3}}{\sqrt{6}})\mathrm{i}$,
$(\frac{3+3\sqrt{3}}{\sqrt{6}})\mathrm{i}$,
$ -2c_{24}^{1}$,
$ 2c_{24}^{1}$,
$(\frac{3+3\sqrt{3}}{\sqrt{6}})\mathrm{i}$,
$(\frac{-3-3\sqrt{3}}{\sqrt{6}})\mathrm{i}$,
$ -c_{24}^{1}$,
$ c_{24}^{1}$,
$0$;\ \ 
$(\frac{-3-3\sqrt{3}}{\sqrt{6}})\mathrm{i}$,
$ -2c_{24}^{1}$,
$ 2c_{24}^{1}$,
$(\frac{-3-3\sqrt{3}}{\sqrt{6}})\mathrm{i}$,
$(\frac{3+3\sqrt{3}}{\sqrt{6}})\mathrm{i}$,
$ -c_{24}^{1}$,
$ c_{24}^{1}$,
$0$;\ \ 
$ 1+\sqrt{3}$,
$ 1+\sqrt{3}$,
$0$,
$0$,
$ -1-\sqrt{3}$,
$ -1-\sqrt{3}$,
$ 2c_{24}^{1}$;\ \ 
$ 1+\sqrt{3}$,
$0$,
$0$,
$ -1-\sqrt{3}$,
$ -1-\sqrt{3}$,
$ -2c_{24}^{1}$;\ \ 
$(\frac{3+3\sqrt{3}}{\sqrt{6}})\mathrm{i}$,
$(\frac{-3-3\sqrt{3}}{\sqrt{6}})\mathrm{i}$,
$ \frac{3+3\sqrt{3}}{\sqrt{6}}$,
$ \frac{-3-3\sqrt{3}}{\sqrt{6}}$,
$0$;\ \ 
$(\frac{3+3\sqrt{3}}{\sqrt{6}})\mathrm{i}$,
$ \frac{3+3\sqrt{3}}{\sqrt{6}}$,
$ \frac{-3-3\sqrt{3}}{\sqrt{6}}$,
$0$;\ \ 
$ 1$,
$ 1$,
$ -2c_{24}^{1}$;\ \ 
$ 1$,
$ 2c_{24}^{1}$;\ \ 
$0$)

\vskip 0.7ex
\hangindent=3em \hangafter=1
$\tau_n$ = ($3.62 - 8.75 i$, $15.69 - 37.9 i$, $-12.12 + 29.25 i$, $9.47 + 35.32 i$, $32.63 + 13.51 i$, $44.79 - 0.01 i$, $-13.51 - 32.63 i$, $0.$, $-29.25 + 12.12 i$, $47.37 + 19.63 i$, $8.75 - 3.62 i$, $44.78 - 44.77 i$, $-8.75 - 3.62 i$, $-2.58 - 6.23 i$, $29.25 + 12.12 i$, $44.78 + 25.85 i$, $13.51 - 32.63 i$, $44.78 - 0.01 i$, $-32.63 + 13.51 i$, $35.32 + 9.46 i$, $12.12 + 29.25 i$, $29.09 - 12.05 i$, $-3.62 - 8.75 i$, $0.01$, $-3.62 + 8.75 i$, $29.09 + 12.05 i$, $12.12 - 29.25 i$, $35.32 - 9.46 i$, $-32.63 - 13.51 i$, $44.78 + 0.01 i$, $13.51 + 32.63 i$, $44.78 - 25.85 i$, $29.25 - 12.12 i$, $-2.58 + 6.23 i$, $-8.75 + 3.62 i$, $44.78 + 44.77 i$, $8.75 + 3.62 i$, $47.37 - 19.63 i$, $-29.25 - 12.12 i$, $0.$, $-13.51 + 32.63 i$, $44.79 + 0.01 i$, $32.63 - 13.51 i$, $9.47 - 35.32 i$, $-12.12 - 29.25 i$, $15.69 + 37.9 i$, $3.62 + 8.75 i$, $89.56$)

\vskip 0.7ex
\hangindent=3em \hangafter=1
\textit{Intrinsic sign problem}

  \vskip 2ex

\noindent20. $11_{\frac{5}{2},89.56}^{48,214}$ \irep{2192}:\ \ 
$d_i$ = ($1.0$,
$1.0$,
$1.931$,
$1.931$,
$2.732$,
$2.732$,
$3.346$,
$3.346$,
$3.732$,
$3.732$,
$3.863$) 

\vskip 0.7ex
\hangindent=3em \hangafter=1
$D^2= 89.569 = 
48+24\sqrt{3}$

\vskip 0.7ex
\hangindent=3em \hangafter=1
$T = ( 0,
\frac{1}{2},
\frac{1}{16},
\frac{1}{16},
\frac{2}{3},
\frac{1}{6},
\frac{5}{16},
\frac{5}{16},
0,
\frac{1}{2},
\frac{35}{48} )
$,

\vskip 0.7ex
\hangindent=3em \hangafter=1
$S$ = ($ 1$,
$ 1$,
$ c_{24}^{1}$,
$ c_{24}^{1}$,
$ 1+\sqrt{3}$,
$ 1+\sqrt{3}$,
$ \frac{3+3\sqrt{3}}{\sqrt{6}}$,
$ \frac{3+3\sqrt{3}}{\sqrt{6}}$,
$ 2+\sqrt{3}$,
$ 2+\sqrt{3}$,
$ 2c_{24}^{1}$;\ \ 
$ 1$,
$ -c_{24}^{1}$,
$ -c_{24}^{1}$,
$ 1+\sqrt{3}$,
$ 1+\sqrt{3}$,
$ \frac{-3-3\sqrt{3}}{\sqrt{6}}$,
$ \frac{-3-3\sqrt{3}}{\sqrt{6}}$,
$ 2+\sqrt{3}$,
$ 2+\sqrt{3}$,
$ -2c_{24}^{1}$;\ \ 
$ \frac{3+3\sqrt{3}}{\sqrt{6}}$,
$ \frac{-3-3\sqrt{3}}{\sqrt{6}}$,
$ -2c_{24}^{1}$,
$ 2c_{24}^{1}$,
$ \frac{-3-3\sqrt{3}}{\sqrt{6}}$,
$ \frac{3+3\sqrt{3}}{\sqrt{6}}$,
$ -c_{24}^{1}$,
$ c_{24}^{1}$,
$0$;\ \ 
$ \frac{3+3\sqrt{3}}{\sqrt{6}}$,
$ -2c_{24}^{1}$,
$ 2c_{24}^{1}$,
$ \frac{3+3\sqrt{3}}{\sqrt{6}}$,
$ \frac{-3-3\sqrt{3}}{\sqrt{6}}$,
$ -c_{24}^{1}$,
$ c_{24}^{1}$,
$0$;\ \ 
$ 1+\sqrt{3}$,
$ 1+\sqrt{3}$,
$0$,
$0$,
$ -1-\sqrt{3}$,
$ -1-\sqrt{3}$,
$ 2c_{24}^{1}$;\ \ 
$ 1+\sqrt{3}$,
$0$,
$0$,
$ -1-\sqrt{3}$,
$ -1-\sqrt{3}$,
$ -2c_{24}^{1}$;\ \ 
$ \frac{-3-3\sqrt{3}}{\sqrt{6}}$,
$ \frac{3+3\sqrt{3}}{\sqrt{6}}$,
$ \frac{3+3\sqrt{3}}{\sqrt{6}}$,
$ \frac{-3-3\sqrt{3}}{\sqrt{6}}$,
$0$;\ \ 
$ \frac{-3-3\sqrt{3}}{\sqrt{6}}$,
$ \frac{3+3\sqrt{3}}{\sqrt{6}}$,
$ \frac{-3-3\sqrt{3}}{\sqrt{6}}$,
$0$;\ \ 
$ 1$,
$ 1$,
$ -2c_{24}^{1}$;\ \ 
$ 1$,
$ 2c_{24}^{1}$;\ \ 
$0$)

\vskip 0.7ex
\hangindent=3em \hangafter=1
$\tau_n$ = ($-3.63 + 8.75 i$, $-2.58 + 6.23 i$, $29.25 + 12.11 i$, $35.32 + 9.46 i$, $-32.63 - 13.52 i$, $44.79 - 0.01 i$, $13.52 + 32.63 i$, $0.$, $-12.11 - 29.25 i$, $29.09 + 12.05 i$, $-8.75 + 3.63 i$, $44.78 - 44.77 i$, $8.75 + 3.63 i$, $15.69 + 37.9 i$, $12.11 - 29.25 i$, $44.78 - 25.85 i$, $-13.52 + 32.63 i$, $44.78 - 0.01 i$, $32.63 - 13.52 i$, $9.47 + 35.32 i$, $-29.25 + 12.11 i$, $47.37 - 19.63 i$, $3.63 + 8.75 i$, $0.01$, $3.63 - 8.75 i$, $47.37 + 19.63 i$, $-29.25 - 12.11 i$, $9.47 - 35.32 i$, $32.63 + 13.52 i$, $44.78 + 0.01 i$, $-13.52 - 32.63 i$, $44.78 + 25.85 i$, $12.11 + 29.25 i$, $15.69 - 37.9 i$, $8.75 - 3.63 i$, $44.78 + 44.77 i$, $-8.75 - 3.63 i$, $29.09 - 12.05 i$, $-12.11 + 29.25 i$, $0.$, $13.52 - 32.63 i$, $44.79 + 0.01 i$, $-32.63 + 13.52 i$, $35.32 - 9.46 i$, $29.25 - 12.11 i$, $-2.58 - 6.23 i$, $-3.63 - 8.75 i$, $89.56$)

\vskip 0.7ex
\hangindent=3em \hangafter=1
\textit{Intrinsic sign problem}

  \vskip 2ex

\noindent21. $11_{\frac{15}{2},89.56}^{48,311}$ \irep{2192}:\ \ 
$d_i$ = ($1.0$,
$1.0$,
$1.931$,
$1.931$,
$2.732$,
$2.732$,
$3.346$,
$3.346$,
$3.732$,
$3.732$,
$3.863$) 

\vskip 0.7ex
\hangindent=3em \hangafter=1
$D^2= 89.569 = 
48+24\sqrt{3}$

\vskip 0.7ex
\hangindent=3em \hangafter=1
$T = ( 0,
\frac{1}{2},
\frac{3}{16},
\frac{3}{16},
\frac{1}{3},
\frac{5}{6},
\frac{15}{16},
\frac{15}{16},
0,
\frac{1}{2},
\frac{25}{48} )
$,

\vskip 0.7ex
\hangindent=3em \hangafter=1
$S$ = ($ 1$,
$ 1$,
$ c_{24}^{1}$,
$ c_{24}^{1}$,
$ 1+\sqrt{3}$,
$ 1+\sqrt{3}$,
$ \frac{3+3\sqrt{3}}{\sqrt{6}}$,
$ \frac{3+3\sqrt{3}}{\sqrt{6}}$,
$ 2+\sqrt{3}$,
$ 2+\sqrt{3}$,
$ 2c_{24}^{1}$;\ \ 
$ 1$,
$ -c_{24}^{1}$,
$ -c_{24}^{1}$,
$ 1+\sqrt{3}$,
$ 1+\sqrt{3}$,
$ \frac{-3-3\sqrt{3}}{\sqrt{6}}$,
$ \frac{-3-3\sqrt{3}}{\sqrt{6}}$,
$ 2+\sqrt{3}$,
$ 2+\sqrt{3}$,
$ -2c_{24}^{1}$;\ \ 
$ \frac{-3-3\sqrt{3}}{\sqrt{6}}$,
$ \frac{3+3\sqrt{3}}{\sqrt{6}}$,
$ -2c_{24}^{1}$,
$ 2c_{24}^{1}$,
$ \frac{-3-3\sqrt{3}}{\sqrt{6}}$,
$ \frac{3+3\sqrt{3}}{\sqrt{6}}$,
$ -c_{24}^{1}$,
$ c_{24}^{1}$,
$0$;\ \ 
$ \frac{-3-3\sqrt{3}}{\sqrt{6}}$,
$ -2c_{24}^{1}$,
$ 2c_{24}^{1}$,
$ \frac{3+3\sqrt{3}}{\sqrt{6}}$,
$ \frac{-3-3\sqrt{3}}{\sqrt{6}}$,
$ -c_{24}^{1}$,
$ c_{24}^{1}$,
$0$;\ \ 
$ 1+\sqrt{3}$,
$ 1+\sqrt{3}$,
$0$,
$0$,
$ -1-\sqrt{3}$,
$ -1-\sqrt{3}$,
$ 2c_{24}^{1}$;\ \ 
$ 1+\sqrt{3}$,
$0$,
$0$,
$ -1-\sqrt{3}$,
$ -1-\sqrt{3}$,
$ -2c_{24}^{1}$;\ \ 
$ \frac{3+3\sqrt{3}}{\sqrt{6}}$,
$ \frac{-3-3\sqrt{3}}{\sqrt{6}}$,
$ \frac{3+3\sqrt{3}}{\sqrt{6}}$,
$ \frac{-3-3\sqrt{3}}{\sqrt{6}}$,
$0$;\ \ 
$ \frac{3+3\sqrt{3}}{\sqrt{6}}$,
$ \frac{3+3\sqrt{3}}{\sqrt{6}}$,
$ \frac{-3-3\sqrt{3}}{\sqrt{6}}$,
$0$;\ \ 
$ 1$,
$ 1$,
$ -2c_{24}^{1}$;\ \ 
$ 1$,
$ 2c_{24}^{1}$;\ \ 
$0$)

\vskip 0.7ex
\hangindent=3em \hangafter=1
$\tau_n$ = ($8.75 - 3.63 i$, $47.37 - 19.63 i$, $-12.11 - 29.25 i$, $35.32 - 9.46 i$, $-13.52 - 32.63 i$, $44.78 - 0.01 i$, $-32.63 - 13.52 i$, $0.$, $-29.25 - 12.11 i$, $15.69 + 37.9 i$, $-3.63 + 8.75 i$, $44.78 + 44.77 i$, $3.63 + 8.75 i$, $29.09 + 12.05 i$, $29.25 - 12.11 i$, $44.78 + 25.85 i$, $32.63 - 13.52 i$, $44.79 - 0.01 i$, $13.52 - 32.63 i$, $9.47 - 35.32 i$, $12.11 - 29.25 i$, $-2.58 + 6.23 i$, $-8.75 - 3.63 i$, $0.01$, $-8.75 + 3.63 i$, $-2.58 - 6.23 i$, $12.11 + 29.25 i$, $9.47 + 35.32 i$, $13.52 + 32.63 i$, $44.79 + 0.01 i$, $32.63 + 13.52 i$, $44.78 - 25.85 i$, $29.25 + 12.11 i$, $29.09 - 12.05 i$, $3.63 - 8.75 i$, $44.78 - 44.77 i$, $-3.63 - 8.75 i$, $15.69 - 37.9 i$, $-29.25 + 12.11 i$, $0.$, $-32.63 + 13.52 i$, $44.78 + 0.01 i$, $-13.52 + 32.63 i$, $35.32 + 9.46 i$, $-12.11 + 29.25 i$, $47.37 + 19.63 i$, $8.75 + 3.63 i$, $89.56$)

\vskip 0.7ex
\hangindent=3em \hangafter=1
\textit{Intrinsic sign problem}

  \vskip 2ex

\noindent22. $11_{\frac{7}{2},89.56}^{48,628}$ \irep{2191}:\ \ 
$d_i$ = ($1.0$,
$1.0$,
$1.931$,
$1.931$,
$2.732$,
$2.732$,
$3.346$,
$3.346$,
$3.732$,
$3.732$,
$3.863$) 

\vskip 0.7ex
\hangindent=3em \hangafter=1
$D^2= 89.569 = 
48+24\sqrt{3}$

\vskip 0.7ex
\hangindent=3em \hangafter=1
$T = ( 0,
\frac{1}{2},
\frac{3}{16},
\frac{3}{16},
\frac{2}{3},
\frac{1}{6},
\frac{7}{16},
\frac{7}{16},
0,
\frac{1}{2},
\frac{41}{48} )
$,

\vskip 0.7ex
\hangindent=3em \hangafter=1
$S$ = ($ 1$,
$ 1$,
$ c_{24}^{1}$,
$ c_{24}^{1}$,
$ 1+\sqrt{3}$,
$ 1+\sqrt{3}$,
$ \frac{3+3\sqrt{3}}{\sqrt{6}}$,
$ \frac{3+3\sqrt{3}}{\sqrt{6}}$,
$ 2+\sqrt{3}$,
$ 2+\sqrt{3}$,
$ 2c_{24}^{1}$;\ \ 
$ 1$,
$ -c_{24}^{1}$,
$ -c_{24}^{1}$,
$ 1+\sqrt{3}$,
$ 1+\sqrt{3}$,
$ \frac{-3-3\sqrt{3}}{\sqrt{6}}$,
$ \frac{-3-3\sqrt{3}}{\sqrt{6}}$,
$ 2+\sqrt{3}$,
$ 2+\sqrt{3}$,
$ -2c_{24}^{1}$;\ \ 
$(\frac{-3-3\sqrt{3}}{\sqrt{6}})\mathrm{i}$,
$(\frac{3+3\sqrt{3}}{\sqrt{6}})\mathrm{i}$,
$ -2c_{24}^{1}$,
$ 2c_{24}^{1}$,
$(\frac{3+3\sqrt{3}}{\sqrt{6}})\mathrm{i}$,
$(\frac{-3-3\sqrt{3}}{\sqrt{6}})\mathrm{i}$,
$ -c_{24}^{1}$,
$ c_{24}^{1}$,
$0$;\ \ 
$(\frac{-3-3\sqrt{3}}{\sqrt{6}})\mathrm{i}$,
$ -2c_{24}^{1}$,
$ 2c_{24}^{1}$,
$(\frac{-3-3\sqrt{3}}{\sqrt{6}})\mathrm{i}$,
$(\frac{3+3\sqrt{3}}{\sqrt{6}})\mathrm{i}$,
$ -c_{24}^{1}$,
$ c_{24}^{1}$,
$0$;\ \ 
$ 1+\sqrt{3}$,
$ 1+\sqrt{3}$,
$0$,
$0$,
$ -1-\sqrt{3}$,
$ -1-\sqrt{3}$,
$ 2c_{24}^{1}$;\ \ 
$ 1+\sqrt{3}$,
$0$,
$0$,
$ -1-\sqrt{3}$,
$ -1-\sqrt{3}$,
$ -2c_{24}^{1}$;\ \ 
$(\frac{3+3\sqrt{3}}{\sqrt{6}})\mathrm{i}$,
$(\frac{-3-3\sqrt{3}}{\sqrt{6}})\mathrm{i}$,
$ \frac{3+3\sqrt{3}}{\sqrt{6}}$,
$ \frac{-3-3\sqrt{3}}{\sqrt{6}}$,
$0$;\ \ 
$(\frac{3+3\sqrt{3}}{\sqrt{6}})\mathrm{i}$,
$ \frac{3+3\sqrt{3}}{\sqrt{6}}$,
$ \frac{-3-3\sqrt{3}}{\sqrt{6}}$,
$0$;\ \ 
$ 1$,
$ 1$,
$ -2c_{24}^{1}$;\ \ 
$ 1$,
$ 2c_{24}^{1}$;\ \ 
$0$)

\vskip 0.7ex
\hangindent=3em \hangafter=1
$\tau_n$ = ($-8.75 + 3.62 i$, $29.09 - 12.05 i$, $-29.25 + 12.12 i$, $9.47 - 35.32 i$, $13.51 + 32.63 i$, $44.78 - 0.01 i$, $32.63 + 13.51 i$, $0.$, $12.12 - 29.25 i$, $-2.58 - 6.23 i$, $3.62 - 8.75 i$, $44.78 + 44.77 i$, $-3.62 - 8.75 i$, $47.37 + 19.63 i$, $-12.12 - 29.25 i$, $44.78 - 25.85 i$, $-32.63 + 13.51 i$, $44.79 - 0.01 i$, $-13.51 + 32.63 i$, $35.32 - 9.46 i$, $29.25 + 12.12 i$, $15.69 - 37.9 i$, $8.75 + 3.62 i$, $0.01$, $8.75 - 3.62 i$, $15.69 + 37.9 i$, $29.25 - 12.12 i$, $35.32 + 9.46 i$, $-13.51 - 32.63 i$, $44.79 + 0.01 i$, $-32.63 - 13.51 i$, $44.78 + 25.85 i$, $-12.12 + 29.25 i$, $47.37 - 19.63 i$, $-3.62 + 8.75 i$, $44.78 - 44.77 i$, $3.62 + 8.75 i$, $-2.58 + 6.23 i$, $12.12 + 29.25 i$, $0.$, $32.63 - 13.51 i$, $44.78 + 0.01 i$, $13.51 - 32.63 i$, $9.47 + 35.32 i$, $-29.25 - 12.12 i$, $29.09 + 12.05 i$, $-8.75 - 3.62 i$, $89.56$)

\vskip 0.7ex
\hangindent=3em \hangafter=1
\textit{Intrinsic sign problem}

  \vskip 2ex

\noindent23. $11_{\frac{1}{2},89.56}^{48,193}$ \irep{2191}:\ \ 
$d_i$ = ($1.0$,
$1.0$,
$1.931$,
$1.931$,
$2.732$,
$2.732$,
$3.346$,
$3.346$,
$3.732$,
$3.732$,
$3.863$) 

\vskip 0.7ex
\hangindent=3em \hangafter=1
$D^2= 89.569 = 
48+24\sqrt{3}$

\vskip 0.7ex
\hangindent=3em \hangafter=1
$T = ( 0,
\frac{1}{2},
\frac{5}{16},
\frac{5}{16},
\frac{1}{3},
\frac{5}{6},
\frac{1}{16},
\frac{1}{16},
0,
\frac{1}{2},
\frac{31}{48} )
$,

\vskip 0.7ex
\hangindent=3em \hangafter=1
$S$ = ($ 1$,
$ 1$,
$ c_{24}^{1}$,
$ c_{24}^{1}$,
$ 1+\sqrt{3}$,
$ 1+\sqrt{3}$,
$ \frac{3+3\sqrt{3}}{\sqrt{6}}$,
$ \frac{3+3\sqrt{3}}{\sqrt{6}}$,
$ 2+\sqrt{3}$,
$ 2+\sqrt{3}$,
$ 2c_{24}^{1}$;\ \ 
$ 1$,
$ -c_{24}^{1}$,
$ -c_{24}^{1}$,
$ 1+\sqrt{3}$,
$ 1+\sqrt{3}$,
$ \frac{-3-3\sqrt{3}}{\sqrt{6}}$,
$ \frac{-3-3\sqrt{3}}{\sqrt{6}}$,
$ 2+\sqrt{3}$,
$ 2+\sqrt{3}$,
$ -2c_{24}^{1}$;\ \ 
$(\frac{3+3\sqrt{3}}{\sqrt{6}})\mathrm{i}$,
$(\frac{-3-3\sqrt{3}}{\sqrt{6}})\mathrm{i}$,
$ -2c_{24}^{1}$,
$ 2c_{24}^{1}$,
$(\frac{3+3\sqrt{3}}{\sqrt{6}})\mathrm{i}$,
$(\frac{-3-3\sqrt{3}}{\sqrt{6}})\mathrm{i}$,
$ -c_{24}^{1}$,
$ c_{24}^{1}$,
$0$;\ \ 
$(\frac{3+3\sqrt{3}}{\sqrt{6}})\mathrm{i}$,
$ -2c_{24}^{1}$,
$ 2c_{24}^{1}$,
$(\frac{-3-3\sqrt{3}}{\sqrt{6}})\mathrm{i}$,
$(\frac{3+3\sqrt{3}}{\sqrt{6}})\mathrm{i}$,
$ -c_{24}^{1}$,
$ c_{24}^{1}$,
$0$;\ \ 
$ 1+\sqrt{3}$,
$ 1+\sqrt{3}$,
$0$,
$0$,
$ -1-\sqrt{3}$,
$ -1-\sqrt{3}$,
$ 2c_{24}^{1}$;\ \ 
$ 1+\sqrt{3}$,
$0$,
$0$,
$ -1-\sqrt{3}$,
$ -1-\sqrt{3}$,
$ -2c_{24}^{1}$;\ \ 
$(\frac{-3-3\sqrt{3}}{\sqrt{6}})\mathrm{i}$,
$(\frac{3+3\sqrt{3}}{\sqrt{6}})\mathrm{i}$,
$ \frac{3+3\sqrt{3}}{\sqrt{6}}$,
$ \frac{-3-3\sqrt{3}}{\sqrt{6}}$,
$0$;\ \ 
$(\frac{-3-3\sqrt{3}}{\sqrt{6}})\mathrm{i}$,
$ \frac{3+3\sqrt{3}}{\sqrt{6}}$,
$ \frac{-3-3\sqrt{3}}{\sqrt{6}}$,
$0$;\ \ 
$ 1$,
$ 1$,
$ -2c_{24}^{1}$;\ \ 
$ 1$,
$ 2c_{24}^{1}$;\ \ 
$0$)

\vskip 0.7ex
\hangindent=3em \hangafter=1
$\tau_n$ = ($8.75 + 3.62 i$, $29.09 + 12.05 i$, $29.25 + 12.12 i$, $9.47 + 35.32 i$, $-13.51 + 32.63 i$, $44.78 + 0.01 i$, $-32.63 + 13.51 i$, $0.$, $-12.12 - 29.25 i$, $-2.58 + 6.23 i$, $-3.62 - 8.75 i$, $44.78 - 44.77 i$, $3.62 - 8.75 i$, $47.37 - 19.63 i$, $12.12 - 29.25 i$, $44.78 + 25.85 i$, $32.63 + 13.51 i$, $44.79 + 0.01 i$, $13.51 + 32.63 i$, $35.32 + 9.46 i$, $-29.25 + 12.12 i$, $15.69 + 37.9 i$, $-8.75 + 3.62 i$, $0.01$, $-8.75 - 3.62 i$, $15.69 - 37.9 i$, $-29.25 - 12.12 i$, $35.32 - 9.46 i$, $13.51 - 32.63 i$, $44.79 - 0.01 i$, $32.63 - 13.51 i$, $44.78 - 25.85 i$, $12.12 + 29.25 i$, $47.37 + 19.63 i$, $3.62 + 8.75 i$, $44.78 + 44.77 i$, $-3.62 + 8.75 i$, $-2.58 - 6.23 i$, $-12.12 + 29.25 i$, $0.$, $-32.63 - 13.51 i$, $44.78 - 0.01 i$, $-13.51 - 32.63 i$, $9.47 - 35.32 i$, $29.25 - 12.12 i$, $29.09 - 12.05 i$, $8.75 - 3.62 i$, $89.56$)

\vskip 0.7ex
\hangindent=3em \hangafter=1
\textit{Intrinsic sign problem}

  \vskip 2ex

\noindent24. $11_{\frac{9}{2},89.56}^{48,133}$ \irep{2192}:\ \ 
$d_i$ = ($1.0$,
$1.0$,
$1.931$,
$1.931$,
$2.732$,
$2.732$,
$3.346$,
$3.346$,
$3.732$,
$3.732$,
$3.863$) 

\vskip 0.7ex
\hangindent=3em \hangafter=1
$D^2= 89.569 = 
48+24\sqrt{3}$

\vskip 0.7ex
\hangindent=3em \hangafter=1
$T = ( 0,
\frac{1}{2},
\frac{5}{16},
\frac{5}{16},
\frac{2}{3},
\frac{1}{6},
\frac{9}{16},
\frac{9}{16},
0,
\frac{1}{2},
\frac{47}{48} )
$,

\vskip 0.7ex
\hangindent=3em \hangafter=1
$S$ = ($ 1$,
$ 1$,
$ c_{24}^{1}$,
$ c_{24}^{1}$,
$ 1+\sqrt{3}$,
$ 1+\sqrt{3}$,
$ \frac{3+3\sqrt{3}}{\sqrt{6}}$,
$ \frac{3+3\sqrt{3}}{\sqrt{6}}$,
$ 2+\sqrt{3}$,
$ 2+\sqrt{3}$,
$ 2c_{24}^{1}$;\ \ 
$ 1$,
$ -c_{24}^{1}$,
$ -c_{24}^{1}$,
$ 1+\sqrt{3}$,
$ 1+\sqrt{3}$,
$ \frac{-3-3\sqrt{3}}{\sqrt{6}}$,
$ \frac{-3-3\sqrt{3}}{\sqrt{6}}$,
$ 2+\sqrt{3}$,
$ 2+\sqrt{3}$,
$ -2c_{24}^{1}$;\ \ 
$ \frac{-3-3\sqrt{3}}{\sqrt{6}}$,
$ \frac{3+3\sqrt{3}}{\sqrt{6}}$,
$ -2c_{24}^{1}$,
$ 2c_{24}^{1}$,
$ \frac{-3-3\sqrt{3}}{\sqrt{6}}$,
$ \frac{3+3\sqrt{3}}{\sqrt{6}}$,
$ -c_{24}^{1}$,
$ c_{24}^{1}$,
$0$;\ \ 
$ \frac{-3-3\sqrt{3}}{\sqrt{6}}$,
$ -2c_{24}^{1}$,
$ 2c_{24}^{1}$,
$ \frac{3+3\sqrt{3}}{\sqrt{6}}$,
$ \frac{-3-3\sqrt{3}}{\sqrt{6}}$,
$ -c_{24}^{1}$,
$ c_{24}^{1}$,
$0$;\ \ 
$ 1+\sqrt{3}$,
$ 1+\sqrt{3}$,
$0$,
$0$,
$ -1-\sqrt{3}$,
$ -1-\sqrt{3}$,
$ 2c_{24}^{1}$;\ \ 
$ 1+\sqrt{3}$,
$0$,
$0$,
$ -1-\sqrt{3}$,
$ -1-\sqrt{3}$,
$ -2c_{24}^{1}$;\ \ 
$ \frac{3+3\sqrt{3}}{\sqrt{6}}$,
$ \frac{-3-3\sqrt{3}}{\sqrt{6}}$,
$ \frac{3+3\sqrt{3}}{\sqrt{6}}$,
$ \frac{-3-3\sqrt{3}}{\sqrt{6}}$,
$0$;\ \ 
$ \frac{3+3\sqrt{3}}{\sqrt{6}}$,
$ \frac{3+3\sqrt{3}}{\sqrt{6}}$,
$ \frac{-3-3\sqrt{3}}{\sqrt{6}}$,
$0$;\ \ 
$ 1$,
$ 1$,
$ -2c_{24}^{1}$;\ \ 
$ 1$,
$ 2c_{24}^{1}$;\ \ 
$0$)

\vskip 0.7ex
\hangindent=3em \hangafter=1
$\tau_n$ = ($-8.75 - 3.63 i$, $47.37 + 19.63 i$, $12.11 - 29.25 i$, $35.32 + 9.46 i$, $13.52 - 32.63 i$, $44.78 + 0.01 i$, $32.63 - 13.52 i$, $0.$, $29.25 - 12.11 i$, $15.69 - 37.9 i$, $3.63 + 8.75 i$, $44.78 - 44.77 i$, $-3.63 + 8.75 i$, $29.09 - 12.05 i$, $-29.25 - 12.11 i$, $44.78 - 25.85 i$, $-32.63 - 13.52 i$, $44.79 + 0.01 i$, $-13.52 - 32.63 i$, $9.47 + 35.32 i$, $-12.11 - 29.25 i$, $-2.58 - 6.23 i$, $8.75 - 3.63 i$, $0.01$, $8.75 + 3.63 i$, $-2.58 + 6.23 i$, $-12.11 + 29.25 i$, $9.47 - 35.32 i$, $-13.52 + 32.63 i$, $44.79 - 0.01 i$, $-32.63 + 13.52 i$, $44.78 + 25.85 i$, $-29.25 + 12.11 i$, $29.09 + 12.05 i$, $-3.63 - 8.75 i$, $44.78 + 44.77 i$, $3.63 - 8.75 i$, $15.69 + 37.9 i$, $29.25 + 12.11 i$, $0.$, $32.63 + 13.52 i$, $44.78 - 0.01 i$, $13.52 + 32.63 i$, $35.32 - 9.46 i$, $12.11 + 29.25 i$, $47.37 - 19.63 i$, $-8.75 + 3.63 i$, $89.56$)

\vskip 0.7ex
\hangindent=3em \hangafter=1
\textit{Intrinsic sign problem}

  \vskip 2ex

\noindent25. $11_{\frac{3}{2},89.56}^{48,682}$ \irep{2192}:\ \ 
$d_i$ = ($1.0$,
$1.0$,
$1.931$,
$1.931$,
$2.732$,
$2.732$,
$3.346$,
$3.346$,
$3.732$,
$3.732$,
$3.863$) 

\vskip 0.7ex
\hangindent=3em \hangafter=1
$D^2= 89.569 = 
48+24\sqrt{3}$

\vskip 0.7ex
\hangindent=3em \hangafter=1
$T = ( 0,
\frac{1}{2},
\frac{7}{16},
\frac{7}{16},
\frac{1}{3},
\frac{5}{6},
\frac{3}{16},
\frac{3}{16},
0,
\frac{1}{2},
\frac{37}{48} )
$,

\vskip 0.7ex
\hangindent=3em \hangafter=1
$S$ = ($ 1$,
$ 1$,
$ c_{24}^{1}$,
$ c_{24}^{1}$,
$ 1+\sqrt{3}$,
$ 1+\sqrt{3}$,
$ \frac{3+3\sqrt{3}}{\sqrt{6}}$,
$ \frac{3+3\sqrt{3}}{\sqrt{6}}$,
$ 2+\sqrt{3}$,
$ 2+\sqrt{3}$,
$ 2c_{24}^{1}$;\ \ 
$ 1$,
$ -c_{24}^{1}$,
$ -c_{24}^{1}$,
$ 1+\sqrt{3}$,
$ 1+\sqrt{3}$,
$ \frac{-3-3\sqrt{3}}{\sqrt{6}}$,
$ \frac{-3-3\sqrt{3}}{\sqrt{6}}$,
$ 2+\sqrt{3}$,
$ 2+\sqrt{3}$,
$ -2c_{24}^{1}$;\ \ 
$ \frac{3+3\sqrt{3}}{\sqrt{6}}$,
$ \frac{-3-3\sqrt{3}}{\sqrt{6}}$,
$ -2c_{24}^{1}$,
$ 2c_{24}^{1}$,
$ \frac{-3-3\sqrt{3}}{\sqrt{6}}$,
$ \frac{3+3\sqrt{3}}{\sqrt{6}}$,
$ -c_{24}^{1}$,
$ c_{24}^{1}$,
$0$;\ \ 
$ \frac{3+3\sqrt{3}}{\sqrt{6}}$,
$ -2c_{24}^{1}$,
$ 2c_{24}^{1}$,
$ \frac{3+3\sqrt{3}}{\sqrt{6}}$,
$ \frac{-3-3\sqrt{3}}{\sqrt{6}}$,
$ -c_{24}^{1}$,
$ c_{24}^{1}$,
$0$;\ \ 
$ 1+\sqrt{3}$,
$ 1+\sqrt{3}$,
$0$,
$0$,
$ -1-\sqrt{3}$,
$ -1-\sqrt{3}$,
$ 2c_{24}^{1}$;\ \ 
$ 1+\sqrt{3}$,
$0$,
$0$,
$ -1-\sqrt{3}$,
$ -1-\sqrt{3}$,
$ -2c_{24}^{1}$;\ \ 
$ \frac{-3-3\sqrt{3}}{\sqrt{6}}$,
$ \frac{3+3\sqrt{3}}{\sqrt{6}}$,
$ \frac{3+3\sqrt{3}}{\sqrt{6}}$,
$ \frac{-3-3\sqrt{3}}{\sqrt{6}}$,
$0$;\ \ 
$ \frac{-3-3\sqrt{3}}{\sqrt{6}}$,
$ \frac{3+3\sqrt{3}}{\sqrt{6}}$,
$ \frac{-3-3\sqrt{3}}{\sqrt{6}}$,
$0$;\ \ 
$ 1$,
$ 1$,
$ -2c_{24}^{1}$;\ \ 
$ 1$,
$ 2c_{24}^{1}$;\ \ 
$0$)

\vskip 0.7ex
\hangindent=3em \hangafter=1
$\tau_n$ = ($3.63 + 8.75 i$, $-2.58 - 6.23 i$, $-29.25 + 12.11 i$, $35.32 - 9.46 i$, $32.63 - 13.52 i$, $44.79 + 0.01 i$, $-13.52 + 32.63 i$, $0.$, $12.11 - 29.25 i$, $29.09 - 12.05 i$, $8.75 + 3.63 i$, $44.78 + 44.77 i$, $-8.75 + 3.63 i$, $15.69 - 37.9 i$, $-12.11 - 29.25 i$, $44.78 + 25.85 i$, $13.52 + 32.63 i$, $44.78 + 0.01 i$, $-32.63 - 13.52 i$, $9.47 - 35.32 i$, $29.25 + 12.11 i$, $47.37 + 19.63 i$, $-3.63 + 8.75 i$, $0.01$, $-3.63 - 8.75 i$, $47.37 - 19.63 i$, $29.25 - 12.11 i$, $9.47 + 35.32 i$, $-32.63 + 13.52 i$, $44.78 - 0.01 i$, $13.52 - 32.63 i$, $44.78 - 25.85 i$, $-12.11 + 29.25 i$, $15.69 + 37.9 i$, $-8.75 - 3.63 i$, $44.78 - 44.77 i$, $8.75 - 3.63 i$, $29.09 + 12.05 i$, $12.11 + 29.25 i$, $0.$, $-13.52 - 32.63 i$, $44.79 - 0.01 i$, $32.63 + 13.52 i$, $35.32 + 9.46 i$, $-29.25 - 12.11 i$, $-2.58 + 6.23 i$, $3.63 - 8.75 i$, $89.56$)

\vskip 0.7ex
\hangindent=3em \hangafter=1
\textit{Intrinsic sign problem}

  \vskip 2ex

\noindent26. $11_{\frac{11}{2},89.56}^{48,919}$ \irep{2191}:\ \ 
$d_i$ = ($1.0$,
$1.0$,
$1.931$,
$1.931$,
$2.732$,
$2.732$,
$3.346$,
$3.346$,
$3.732$,
$3.732$,
$3.863$) 

\vskip 0.7ex
\hangindent=3em \hangafter=1
$D^2= 89.569 = 
48+24\sqrt{3}$

\vskip 0.7ex
\hangindent=3em \hangafter=1
$T = ( 0,
\frac{1}{2},
\frac{7}{16},
\frac{7}{16},
\frac{2}{3},
\frac{1}{6},
\frac{11}{16},
\frac{11}{16},
0,
\frac{1}{2},
\frac{5}{48} )
$,

\vskip 0.7ex
\hangindent=3em \hangafter=1
$S$ = ($ 1$,
$ 1$,
$ c_{24}^{1}$,
$ c_{24}^{1}$,
$ 1+\sqrt{3}$,
$ 1+\sqrt{3}$,
$ \frac{3+3\sqrt{3}}{\sqrt{6}}$,
$ \frac{3+3\sqrt{3}}{\sqrt{6}}$,
$ 2+\sqrt{3}$,
$ 2+\sqrt{3}$,
$ 2c_{24}^{1}$;\ \ 
$ 1$,
$ -c_{24}^{1}$,
$ -c_{24}^{1}$,
$ 1+\sqrt{3}$,
$ 1+\sqrt{3}$,
$ \frac{-3-3\sqrt{3}}{\sqrt{6}}$,
$ \frac{-3-3\sqrt{3}}{\sqrt{6}}$,
$ 2+\sqrt{3}$,
$ 2+\sqrt{3}$,
$ -2c_{24}^{1}$;\ \ 
$(\frac{3+3\sqrt{3}}{\sqrt{6}})\mathrm{i}$,
$(\frac{-3-3\sqrt{3}}{\sqrt{6}})\mathrm{i}$,
$ -2c_{24}^{1}$,
$ 2c_{24}^{1}$,
$(\frac{3+3\sqrt{3}}{\sqrt{6}})\mathrm{i}$,
$(\frac{-3-3\sqrt{3}}{\sqrt{6}})\mathrm{i}$,
$ -c_{24}^{1}$,
$ c_{24}^{1}$,
$0$;\ \ 
$(\frac{3+3\sqrt{3}}{\sqrt{6}})\mathrm{i}$,
$ -2c_{24}^{1}$,
$ 2c_{24}^{1}$,
$(\frac{-3-3\sqrt{3}}{\sqrt{6}})\mathrm{i}$,
$(\frac{3+3\sqrt{3}}{\sqrt{6}})\mathrm{i}$,
$ -c_{24}^{1}$,
$ c_{24}^{1}$,
$0$;\ \ 
$ 1+\sqrt{3}$,
$ 1+\sqrt{3}$,
$0$,
$0$,
$ -1-\sqrt{3}$,
$ -1-\sqrt{3}$,
$ 2c_{24}^{1}$;\ \ 
$ 1+\sqrt{3}$,
$0$,
$0$,
$ -1-\sqrt{3}$,
$ -1-\sqrt{3}$,
$ -2c_{24}^{1}$;\ \ 
$(\frac{-3-3\sqrt{3}}{\sqrt{6}})\mathrm{i}$,
$(\frac{3+3\sqrt{3}}{\sqrt{6}})\mathrm{i}$,
$ \frac{3+3\sqrt{3}}{\sqrt{6}}$,
$ \frac{-3-3\sqrt{3}}{\sqrt{6}}$,
$0$;\ \ 
$(\frac{-3-3\sqrt{3}}{\sqrt{6}})\mathrm{i}$,
$ \frac{3+3\sqrt{3}}{\sqrt{6}}$,
$ \frac{-3-3\sqrt{3}}{\sqrt{6}}$,
$0$;\ \ 
$ 1$,
$ 1$,
$ -2c_{24}^{1}$;\ \ 
$ 1$,
$ 2c_{24}^{1}$;\ \ 
$0$)

\vskip 0.7ex
\hangindent=3em \hangafter=1
$\tau_n$ = ($-3.62 - 8.75 i$, $15.69 + 37.9 i$, $12.12 + 29.25 i$, $9.47 - 35.32 i$, $-32.63 + 13.51 i$, $44.79 + 0.01 i$, $13.51 - 32.63 i$, $0.$, $29.25 + 12.12 i$, $47.37 - 19.63 i$, $-8.75 - 3.62 i$, $44.78 + 44.77 i$, $8.75 - 3.62 i$, $-2.58 + 6.23 i$, $-29.25 + 12.12 i$, $44.78 - 25.85 i$, $-13.51 - 32.63 i$, $44.78 + 0.01 i$, $32.63 + 13.51 i$, $35.32 - 9.46 i$, $-12.12 + 29.25 i$, $29.09 + 12.05 i$, $3.62 - 8.75 i$, $0.01$, $3.62 + 8.75 i$, $29.09 - 12.05 i$, $-12.12 - 29.25 i$, $35.32 + 9.46 i$, $32.63 - 13.51 i$, $44.78 - 0.01 i$, $-13.51 + 32.63 i$, $44.78 + 25.85 i$, $-29.25 - 12.12 i$, $-2.58 - 6.23 i$, $8.75 + 3.62 i$, $44.78 - 44.77 i$, $-8.75 + 3.62 i$, $47.37 + 19.63 i$, $29.25 - 12.12 i$, $0.$, $13.51 + 32.63 i$, $44.79 - 0.01 i$, $-32.63 - 13.51 i$, $9.47 + 35.32 i$, $12.12 - 29.25 i$, $15.69 - 37.9 i$, $-3.62 + 8.75 i$, $89.56$)

\vskip 0.7ex
\hangindent=3em \hangafter=1
\textit{Intrinsic sign problem}

  \vskip 2ex

\noindent27. $11_{\frac{5}{2},89.56}^{48,102}$ \irep{2191}:\ \ 
$d_i$ = ($1.0$,
$1.0$,
$1.931$,
$1.931$,
$2.732$,
$2.732$,
$3.346$,
$3.346$,
$3.732$,
$3.732$,
$3.863$) 

\vskip 0.7ex
\hangindent=3em \hangafter=1
$D^2= 89.569 = 
48+24\sqrt{3}$

\vskip 0.7ex
\hangindent=3em \hangafter=1
$T = ( 0,
\frac{1}{2},
\frac{9}{16},
\frac{9}{16},
\frac{1}{3},
\frac{5}{6},
\frac{5}{16},
\frac{5}{16},
0,
\frac{1}{2},
\frac{43}{48} )
$,

\vskip 0.7ex
\hangindent=3em \hangafter=1
$S$ = ($ 1$,
$ 1$,
$ c_{24}^{1}$,
$ c_{24}^{1}$,
$ 1+\sqrt{3}$,
$ 1+\sqrt{3}$,
$ \frac{3+3\sqrt{3}}{\sqrt{6}}$,
$ \frac{3+3\sqrt{3}}{\sqrt{6}}$,
$ 2+\sqrt{3}$,
$ 2+\sqrt{3}$,
$ 2c_{24}^{1}$;\ \ 
$ 1$,
$ -c_{24}^{1}$,
$ -c_{24}^{1}$,
$ 1+\sqrt{3}$,
$ 1+\sqrt{3}$,
$ \frac{-3-3\sqrt{3}}{\sqrt{6}}$,
$ \frac{-3-3\sqrt{3}}{\sqrt{6}}$,
$ 2+\sqrt{3}$,
$ 2+\sqrt{3}$,
$ -2c_{24}^{1}$;\ \ 
$(\frac{-3-3\sqrt{3}}{\sqrt{6}})\mathrm{i}$,
$(\frac{3+3\sqrt{3}}{\sqrt{6}})\mathrm{i}$,
$ -2c_{24}^{1}$,
$ 2c_{24}^{1}$,
$(\frac{3+3\sqrt{3}}{\sqrt{6}})\mathrm{i}$,
$(\frac{-3-3\sqrt{3}}{\sqrt{6}})\mathrm{i}$,
$ -c_{24}^{1}$,
$ c_{24}^{1}$,
$0$;\ \ 
$(\frac{-3-3\sqrt{3}}{\sqrt{6}})\mathrm{i}$,
$ -2c_{24}^{1}$,
$ 2c_{24}^{1}$,
$(\frac{-3-3\sqrt{3}}{\sqrt{6}})\mathrm{i}$,
$(\frac{3+3\sqrt{3}}{\sqrt{6}})\mathrm{i}$,
$ -c_{24}^{1}$,
$ c_{24}^{1}$,
$0$;\ \ 
$ 1+\sqrt{3}$,
$ 1+\sqrt{3}$,
$0$,
$0$,
$ -1-\sqrt{3}$,
$ -1-\sqrt{3}$,
$ 2c_{24}^{1}$;\ \ 
$ 1+\sqrt{3}$,
$0$,
$0$,
$ -1-\sqrt{3}$,
$ -1-\sqrt{3}$,
$ -2c_{24}^{1}$;\ \ 
$(\frac{3+3\sqrt{3}}{\sqrt{6}})\mathrm{i}$,
$(\frac{-3-3\sqrt{3}}{\sqrt{6}})\mathrm{i}$,
$ \frac{3+3\sqrt{3}}{\sqrt{6}}$,
$ \frac{-3-3\sqrt{3}}{\sqrt{6}}$,
$0$;\ \ 
$(\frac{3+3\sqrt{3}}{\sqrt{6}})\mathrm{i}$,
$ \frac{3+3\sqrt{3}}{\sqrt{6}}$,
$ \frac{-3-3\sqrt{3}}{\sqrt{6}}$,
$0$;\ \ 
$ 1$,
$ 1$,
$ -2c_{24}^{1}$;\ \ 
$ 1$,
$ 2c_{24}^{1}$;\ \ 
$0$)

\vskip 0.7ex
\hangindent=3em \hangafter=1
$\tau_n$ = ($-3.62 + 8.75 i$, $15.69 - 37.9 i$, $12.12 - 29.25 i$, $9.47 + 35.32 i$, $-32.63 - 13.51 i$, $44.79 - 0.01 i$, $13.51 + 32.63 i$, $0.$, $29.25 - 12.12 i$, $47.37 + 19.63 i$, $-8.75 + 3.62 i$, $44.78 - 44.77 i$, $8.75 + 3.62 i$, $-2.58 - 6.23 i$, $-29.25 - 12.12 i$, $44.78 + 25.85 i$, $-13.51 + 32.63 i$, $44.78 - 0.01 i$, $32.63 - 13.51 i$, $35.32 + 9.46 i$, $-12.12 - 29.25 i$, $29.09 - 12.05 i$, $3.62 + 8.75 i$, $0.01$, $3.62 - 8.75 i$, $29.09 + 12.05 i$, $-12.12 + 29.25 i$, $35.32 - 9.46 i$, $32.63 + 13.51 i$, $44.78 + 0.01 i$, $-13.51 - 32.63 i$, $44.78 - 25.85 i$, $-29.25 + 12.12 i$, $-2.58 + 6.23 i$, $8.75 - 3.62 i$, $44.78 + 44.77 i$, $-8.75 - 3.62 i$, $47.37 - 19.63 i$, $29.25 + 12.12 i$, $0.$, $13.51 - 32.63 i$, $44.79 + 0.01 i$, $-32.63 + 13.51 i$, $9.47 - 35.32 i$, $12.12 + 29.25 i$, $15.69 + 37.9 i$, $-3.62 - 8.75 i$, $89.56$)

\vskip 0.7ex
\hangindent=3em \hangafter=1
\textit{Intrinsic sign problem}

  \vskip 2ex

\noindent28. $11_{\frac{13}{2},89.56}^{48,979}$ \irep{2192}:\ \ 
$d_i$ = ($1.0$,
$1.0$,
$1.931$,
$1.931$,
$2.732$,
$2.732$,
$3.346$,
$3.346$,
$3.732$,
$3.732$,
$3.863$) 

\vskip 0.7ex
\hangindent=3em \hangafter=1
$D^2= 89.569 = 
48+24\sqrt{3}$

\vskip 0.7ex
\hangindent=3em \hangafter=1
$T = ( 0,
\frac{1}{2},
\frac{9}{16},
\frac{9}{16},
\frac{2}{3},
\frac{1}{6},
\frac{13}{16},
\frac{13}{16},
0,
\frac{1}{2},
\frac{11}{48} )
$,

\vskip 0.7ex
\hangindent=3em \hangafter=1
$S$ = ($ 1$,
$ 1$,
$ c_{24}^{1}$,
$ c_{24}^{1}$,
$ 1+\sqrt{3}$,
$ 1+\sqrt{3}$,
$ \frac{3+3\sqrt{3}}{\sqrt{6}}$,
$ \frac{3+3\sqrt{3}}{\sqrt{6}}$,
$ 2+\sqrt{3}$,
$ 2+\sqrt{3}$,
$ 2c_{24}^{1}$;\ \ 
$ 1$,
$ -c_{24}^{1}$,
$ -c_{24}^{1}$,
$ 1+\sqrt{3}$,
$ 1+\sqrt{3}$,
$ \frac{-3-3\sqrt{3}}{\sqrt{6}}$,
$ \frac{-3-3\sqrt{3}}{\sqrt{6}}$,
$ 2+\sqrt{3}$,
$ 2+\sqrt{3}$,
$ -2c_{24}^{1}$;\ \ 
$ \frac{3+3\sqrt{3}}{\sqrt{6}}$,
$ \frac{-3-3\sqrt{3}}{\sqrt{6}}$,
$ -2c_{24}^{1}$,
$ 2c_{24}^{1}$,
$ \frac{-3-3\sqrt{3}}{\sqrt{6}}$,
$ \frac{3+3\sqrt{3}}{\sqrt{6}}$,
$ -c_{24}^{1}$,
$ c_{24}^{1}$,
$0$;\ \ 
$ \frac{3+3\sqrt{3}}{\sqrt{6}}$,
$ -2c_{24}^{1}$,
$ 2c_{24}^{1}$,
$ \frac{3+3\sqrt{3}}{\sqrt{6}}$,
$ \frac{-3-3\sqrt{3}}{\sqrt{6}}$,
$ -c_{24}^{1}$,
$ c_{24}^{1}$,
$0$;\ \ 
$ 1+\sqrt{3}$,
$ 1+\sqrt{3}$,
$0$,
$0$,
$ -1-\sqrt{3}$,
$ -1-\sqrt{3}$,
$ 2c_{24}^{1}$;\ \ 
$ 1+\sqrt{3}$,
$0$,
$0$,
$ -1-\sqrt{3}$,
$ -1-\sqrt{3}$,
$ -2c_{24}^{1}$;\ \ 
$ \frac{-3-3\sqrt{3}}{\sqrt{6}}$,
$ \frac{3+3\sqrt{3}}{\sqrt{6}}$,
$ \frac{3+3\sqrt{3}}{\sqrt{6}}$,
$ \frac{-3-3\sqrt{3}}{\sqrt{6}}$,
$0$;\ \ 
$ \frac{-3-3\sqrt{3}}{\sqrt{6}}$,
$ \frac{3+3\sqrt{3}}{\sqrt{6}}$,
$ \frac{-3-3\sqrt{3}}{\sqrt{6}}$,
$0$;\ \ 
$ 1$,
$ 1$,
$ -2c_{24}^{1}$;\ \ 
$ 1$,
$ 2c_{24}^{1}$;\ \ 
$0$)

\vskip 0.7ex
\hangindent=3em \hangafter=1
$\tau_n$ = ($3.63 - 8.75 i$, $-2.58 + 6.23 i$, $-29.25 - 12.11 i$, $35.32 + 9.46 i$, $32.63 + 13.52 i$, $44.79 - 0.01 i$, $-13.52 - 32.63 i$, $0.$, $12.11 + 29.25 i$, $29.09 + 12.05 i$, $8.75 - 3.63 i$, $44.78 - 44.77 i$, $-8.75 - 3.63 i$, $15.69 + 37.9 i$, $-12.11 + 29.25 i$, $44.78 - 25.85 i$, $13.52 - 32.63 i$, $44.78 - 0.01 i$, $-32.63 + 13.52 i$, $9.47 + 35.32 i$, $29.25 - 12.11 i$, $47.37 - 19.63 i$, $-3.63 - 8.75 i$, $0.01$, $-3.63 + 8.75 i$, $47.37 + 19.63 i$, $29.25 + 12.11 i$, $9.47 - 35.32 i$, $-32.63 - 13.52 i$, $44.78 + 0.01 i$, $13.52 + 32.63 i$, $44.78 + 25.85 i$, $-12.11 - 29.25 i$, $15.69 - 37.9 i$, $-8.75 + 3.63 i$, $44.78 + 44.77 i$, $8.75 + 3.63 i$, $29.09 - 12.05 i$, $12.11 - 29.25 i$, $0.$, $-13.52 + 32.63 i$, $44.79 + 0.01 i$, $32.63 - 13.52 i$, $35.32 - 9.46 i$, $-29.25 + 12.11 i$, $-2.58 - 6.23 i$, $3.63 + 8.75 i$, $89.56$)

\vskip 0.7ex
\hangindent=3em \hangafter=1
\textit{Intrinsic sign problem}

  \vskip 2ex

\noindent29. $11_{\frac{7}{2},89.56}^{48,844}$ \irep{2192}:\ \ 
$d_i$ = ($1.0$,
$1.0$,
$1.931$,
$1.931$,
$2.732$,
$2.732$,
$3.346$,
$3.346$,
$3.732$,
$3.732$,
$3.863$) 

\vskip 0.7ex
\hangindent=3em \hangafter=1
$D^2= 89.569 = 
48+24\sqrt{3}$

\vskip 0.7ex
\hangindent=3em \hangafter=1
$T = ( 0,
\frac{1}{2},
\frac{11}{16},
\frac{11}{16},
\frac{1}{3},
\frac{5}{6},
\frac{7}{16},
\frac{7}{16},
0,
\frac{1}{2},
\frac{1}{48} )
$,

\vskip 0.7ex
\hangindent=3em \hangafter=1
$S$ = ($ 1$,
$ 1$,
$ c_{24}^{1}$,
$ c_{24}^{1}$,
$ 1+\sqrt{3}$,
$ 1+\sqrt{3}$,
$ \frac{3+3\sqrt{3}}{\sqrt{6}}$,
$ \frac{3+3\sqrt{3}}{\sqrt{6}}$,
$ 2+\sqrt{3}$,
$ 2+\sqrt{3}$,
$ 2c_{24}^{1}$;\ \ 
$ 1$,
$ -c_{24}^{1}$,
$ -c_{24}^{1}$,
$ 1+\sqrt{3}$,
$ 1+\sqrt{3}$,
$ \frac{-3-3\sqrt{3}}{\sqrt{6}}$,
$ \frac{-3-3\sqrt{3}}{\sqrt{6}}$,
$ 2+\sqrt{3}$,
$ 2+\sqrt{3}$,
$ -2c_{24}^{1}$;\ \ 
$ \frac{-3-3\sqrt{3}}{\sqrt{6}}$,
$ \frac{3+3\sqrt{3}}{\sqrt{6}}$,
$ -2c_{24}^{1}$,
$ 2c_{24}^{1}$,
$ \frac{-3-3\sqrt{3}}{\sqrt{6}}$,
$ \frac{3+3\sqrt{3}}{\sqrt{6}}$,
$ -c_{24}^{1}$,
$ c_{24}^{1}$,
$0$;\ \ 
$ \frac{-3-3\sqrt{3}}{\sqrt{6}}$,
$ -2c_{24}^{1}$,
$ 2c_{24}^{1}$,
$ \frac{3+3\sqrt{3}}{\sqrt{6}}$,
$ \frac{-3-3\sqrt{3}}{\sqrt{6}}$,
$ -c_{24}^{1}$,
$ c_{24}^{1}$,
$0$;\ \ 
$ 1+\sqrt{3}$,
$ 1+\sqrt{3}$,
$0$,
$0$,
$ -1-\sqrt{3}$,
$ -1-\sqrt{3}$,
$ 2c_{24}^{1}$;\ \ 
$ 1+\sqrt{3}$,
$0$,
$0$,
$ -1-\sqrt{3}$,
$ -1-\sqrt{3}$,
$ -2c_{24}^{1}$;\ \ 
$ \frac{3+3\sqrt{3}}{\sqrt{6}}$,
$ \frac{-3-3\sqrt{3}}{\sqrt{6}}$,
$ \frac{3+3\sqrt{3}}{\sqrt{6}}$,
$ \frac{-3-3\sqrt{3}}{\sqrt{6}}$,
$0$;\ \ 
$ \frac{3+3\sqrt{3}}{\sqrt{6}}$,
$ \frac{3+3\sqrt{3}}{\sqrt{6}}$,
$ \frac{-3-3\sqrt{3}}{\sqrt{6}}$,
$0$;\ \ 
$ 1$,
$ 1$,
$ -2c_{24}^{1}$;\ \ 
$ 1$,
$ 2c_{24}^{1}$;\ \ 
$0$)

\vskip 0.7ex
\hangindent=3em \hangafter=1
$\tau_n$ = ($-8.75 + 3.63 i$, $47.37 - 19.63 i$, $12.11 + 29.25 i$, $35.32 - 9.46 i$, $13.52 + 32.63 i$, $44.78 - 0.01 i$, $32.63 + 13.52 i$, $0.$, $29.25 + 12.11 i$, $15.69 + 37.9 i$, $3.63 - 8.75 i$, $44.78 + 44.77 i$, $-3.63 - 8.75 i$, $29.09 + 12.05 i$, $-29.25 + 12.11 i$, $44.78 + 25.85 i$, $-32.63 + 13.52 i$, $44.79 - 0.01 i$, $-13.52 + 32.63 i$, $9.47 - 35.32 i$, $-12.11 + 29.25 i$, $-2.58 + 6.23 i$, $8.75 + 3.63 i$, $0.01$, $8.75 - 3.63 i$, $-2.58 - 6.23 i$, $-12.11 - 29.25 i$, $9.47 + 35.32 i$, $-13.52 - 32.63 i$, $44.79 + 0.01 i$, $-32.63 - 13.52 i$, $44.78 - 25.85 i$, $-29.25 - 12.11 i$, $29.09 - 12.05 i$, $-3.63 + 8.75 i$, $44.78 - 44.77 i$, $3.63 + 8.75 i$, $15.69 - 37.9 i$, $29.25 - 12.11 i$, $0.$, $32.63 - 13.52 i$, $44.78 + 0.01 i$, $13.52 - 32.63 i$, $35.32 + 9.46 i$, $12.11 - 29.25 i$, $47.37 + 19.63 i$, $-8.75 - 3.63 i$, $89.56$)

\vskip 0.7ex
\hangindent=3em \hangafter=1
\textit{Intrinsic sign problem}

  \vskip 2ex

\noindent30. $11_{\frac{15}{2},89.56}^{48,332}$ \irep{2191}:\ \ 
$d_i$ = ($1.0$,
$1.0$,
$1.931$,
$1.931$,
$2.732$,
$2.732$,
$3.346$,
$3.346$,
$3.732$,
$3.732$,
$3.863$) 

\vskip 0.7ex
\hangindent=3em \hangafter=1
$D^2= 89.569 = 
48+24\sqrt{3}$

\vskip 0.7ex
\hangindent=3em \hangafter=1
$T = ( 0,
\frac{1}{2},
\frac{11}{16},
\frac{11}{16},
\frac{2}{3},
\frac{1}{6},
\frac{15}{16},
\frac{15}{16},
0,
\frac{1}{2},
\frac{17}{48} )
$,

\vskip 0.7ex
\hangindent=3em \hangafter=1
$S$ = ($ 1$,
$ 1$,
$ c_{24}^{1}$,
$ c_{24}^{1}$,
$ 1+\sqrt{3}$,
$ 1+\sqrt{3}$,
$ \frac{3+3\sqrt{3}}{\sqrt{6}}$,
$ \frac{3+3\sqrt{3}}{\sqrt{6}}$,
$ 2+\sqrt{3}$,
$ 2+\sqrt{3}$,
$ 2c_{24}^{1}$;\ \ 
$ 1$,
$ -c_{24}^{1}$,
$ -c_{24}^{1}$,
$ 1+\sqrt{3}$,
$ 1+\sqrt{3}$,
$ \frac{-3-3\sqrt{3}}{\sqrt{6}}$,
$ \frac{-3-3\sqrt{3}}{\sqrt{6}}$,
$ 2+\sqrt{3}$,
$ 2+\sqrt{3}$,
$ -2c_{24}^{1}$;\ \ 
$(\frac{-3-3\sqrt{3}}{\sqrt{6}})\mathrm{i}$,
$(\frac{3+3\sqrt{3}}{\sqrt{6}})\mathrm{i}$,
$ -2c_{24}^{1}$,
$ 2c_{24}^{1}$,
$(\frac{3+3\sqrt{3}}{\sqrt{6}})\mathrm{i}$,
$(\frac{-3-3\sqrt{3}}{\sqrt{6}})\mathrm{i}$,
$ -c_{24}^{1}$,
$ c_{24}^{1}$,
$0$;\ \ 
$(\frac{-3-3\sqrt{3}}{\sqrt{6}})\mathrm{i}$,
$ -2c_{24}^{1}$,
$ 2c_{24}^{1}$,
$(\frac{-3-3\sqrt{3}}{\sqrt{6}})\mathrm{i}$,
$(\frac{3+3\sqrt{3}}{\sqrt{6}})\mathrm{i}$,
$ -c_{24}^{1}$,
$ c_{24}^{1}$,
$0$;\ \ 
$ 1+\sqrt{3}$,
$ 1+\sqrt{3}$,
$0$,
$0$,
$ -1-\sqrt{3}$,
$ -1-\sqrt{3}$,
$ 2c_{24}^{1}$;\ \ 
$ 1+\sqrt{3}$,
$0$,
$0$,
$ -1-\sqrt{3}$,
$ -1-\sqrt{3}$,
$ -2c_{24}^{1}$;\ \ 
$(\frac{3+3\sqrt{3}}{\sqrt{6}})\mathrm{i}$,
$(\frac{-3-3\sqrt{3}}{\sqrt{6}})\mathrm{i}$,
$ \frac{3+3\sqrt{3}}{\sqrt{6}}$,
$ \frac{-3-3\sqrt{3}}{\sqrt{6}}$,
$0$;\ \ 
$(\frac{3+3\sqrt{3}}{\sqrt{6}})\mathrm{i}$,
$ \frac{3+3\sqrt{3}}{\sqrt{6}}$,
$ \frac{-3-3\sqrt{3}}{\sqrt{6}}$,
$0$;\ \ 
$ 1$,
$ 1$,
$ -2c_{24}^{1}$;\ \ 
$ 1$,
$ 2c_{24}^{1}$;\ \ 
$0$)

\vskip 0.7ex
\hangindent=3em \hangafter=1
$\tau_n$ = ($8.75 - 3.62 i$, $29.09 - 12.05 i$, $29.25 - 12.12 i$, $9.47 - 35.32 i$, $-13.51 - 32.63 i$, $44.78 - 0.01 i$, $-32.63 - 13.51 i$, $0.$, $-12.12 + 29.25 i$, $-2.58 - 6.23 i$, $-3.62 + 8.75 i$, $44.78 + 44.77 i$, $3.62 + 8.75 i$, $47.37 + 19.63 i$, $12.12 + 29.25 i$, $44.78 - 25.85 i$, $32.63 - 13.51 i$, $44.79 - 0.01 i$, $13.51 - 32.63 i$, $35.32 - 9.46 i$, $-29.25 - 12.12 i$, $15.69 - 37.9 i$, $-8.75 - 3.62 i$, $0.01$, $-8.75 + 3.62 i$, $15.69 + 37.9 i$, $-29.25 + 12.12 i$, $35.32 + 9.46 i$, $13.51 + 32.63 i$, $44.79 + 0.01 i$, $32.63 + 13.51 i$, $44.78 + 25.85 i$, $12.12 - 29.25 i$, $47.37 - 19.63 i$, $3.62 - 8.75 i$, $44.78 - 44.77 i$, $-3.62 - 8.75 i$, $-2.58 + 6.23 i$, $-12.12 - 29.25 i$, $0.$, $-32.63 + 13.51 i$, $44.78 + 0.01 i$, $-13.51 + 32.63 i$, $9.47 + 35.32 i$, $29.25 + 12.12 i$, $29.09 + 12.05 i$, $8.75 + 3.62 i$, $89.56$)

\vskip 0.7ex
\hangindent=3em \hangafter=1
\textit{Intrinsic sign problem}

  \vskip 2ex

\noindent31. $11_{\frac{9}{2},89.56}^{48,797}$ \irep{2191}:\ \ 
$d_i$ = ($1.0$,
$1.0$,
$1.931$,
$1.931$,
$2.732$,
$2.732$,
$3.346$,
$3.346$,
$3.732$,
$3.732$,
$3.863$) 

\vskip 0.7ex
\hangindent=3em \hangafter=1
$D^2= 89.569 = 
48+24\sqrt{3}$

\vskip 0.7ex
\hangindent=3em \hangafter=1
$T = ( 0,
\frac{1}{2},
\frac{13}{16},
\frac{13}{16},
\frac{1}{3},
\frac{5}{6},
\frac{9}{16},
\frac{9}{16},
0,
\frac{1}{2},
\frac{7}{48} )
$,

\vskip 0.7ex
\hangindent=3em \hangafter=1
$S$ = ($ 1$,
$ 1$,
$ c_{24}^{1}$,
$ c_{24}^{1}$,
$ 1+\sqrt{3}$,
$ 1+\sqrt{3}$,
$ \frac{3+3\sqrt{3}}{\sqrt{6}}$,
$ \frac{3+3\sqrt{3}}{\sqrt{6}}$,
$ 2+\sqrt{3}$,
$ 2+\sqrt{3}$,
$ 2c_{24}^{1}$;\ \ 
$ 1$,
$ -c_{24}^{1}$,
$ -c_{24}^{1}$,
$ 1+\sqrt{3}$,
$ 1+\sqrt{3}$,
$ \frac{-3-3\sqrt{3}}{\sqrt{6}}$,
$ \frac{-3-3\sqrt{3}}{\sqrt{6}}$,
$ 2+\sqrt{3}$,
$ 2+\sqrt{3}$,
$ -2c_{24}^{1}$;\ \ 
$(\frac{3+3\sqrt{3}}{\sqrt{6}})\mathrm{i}$,
$(\frac{-3-3\sqrt{3}}{\sqrt{6}})\mathrm{i}$,
$ -2c_{24}^{1}$,
$ 2c_{24}^{1}$,
$(\frac{3+3\sqrt{3}}{\sqrt{6}})\mathrm{i}$,
$(\frac{-3-3\sqrt{3}}{\sqrt{6}})\mathrm{i}$,
$ -c_{24}^{1}$,
$ c_{24}^{1}$,
$0$;\ \ 
$(\frac{3+3\sqrt{3}}{\sqrt{6}})\mathrm{i}$,
$ -2c_{24}^{1}$,
$ 2c_{24}^{1}$,
$(\frac{-3-3\sqrt{3}}{\sqrt{6}})\mathrm{i}$,
$(\frac{3+3\sqrt{3}}{\sqrt{6}})\mathrm{i}$,
$ -c_{24}^{1}$,
$ c_{24}^{1}$,
$0$;\ \ 
$ 1+\sqrt{3}$,
$ 1+\sqrt{3}$,
$0$,
$0$,
$ -1-\sqrt{3}$,
$ -1-\sqrt{3}$,
$ 2c_{24}^{1}$;\ \ 
$ 1+\sqrt{3}$,
$0$,
$0$,
$ -1-\sqrt{3}$,
$ -1-\sqrt{3}$,
$ -2c_{24}^{1}$;\ \ 
$(\frac{-3-3\sqrt{3}}{\sqrt{6}})\mathrm{i}$,
$(\frac{3+3\sqrt{3}}{\sqrt{6}})\mathrm{i}$,
$ \frac{3+3\sqrt{3}}{\sqrt{6}}$,
$ \frac{-3-3\sqrt{3}}{\sqrt{6}}$,
$0$;\ \ 
$(\frac{-3-3\sqrt{3}}{\sqrt{6}})\mathrm{i}$,
$ \frac{3+3\sqrt{3}}{\sqrt{6}}$,
$ \frac{-3-3\sqrt{3}}{\sqrt{6}}$,
$0$;\ \ 
$ 1$,
$ 1$,
$ -2c_{24}^{1}$;\ \ 
$ 1$,
$ 2c_{24}^{1}$;\ \ 
$0$)

\vskip 0.7ex
\hangindent=3em \hangafter=1
$\tau_n$ = ($-8.75 - 3.62 i$, $29.09 + 12.05 i$, $-29.25 - 12.12 i$, $9.47 + 35.32 i$, $13.51 - 32.63 i$, $44.78 + 0.01 i$, $32.63 - 13.51 i$, $0.$, $12.12 + 29.25 i$, $-2.58 + 6.23 i$, $3.62 + 8.75 i$, $44.78 - 44.77 i$, $-3.62 + 8.75 i$, $47.37 - 19.63 i$, $-12.12 + 29.25 i$, $44.78 + 25.85 i$, $-32.63 - 13.51 i$, $44.79 + 0.01 i$, $-13.51 - 32.63 i$, $35.32 + 9.46 i$, $29.25 - 12.12 i$, $15.69 + 37.9 i$, $8.75 - 3.62 i$, $0.01$, $8.75 + 3.62 i$, $15.69 - 37.9 i$, $29.25 + 12.12 i$, $35.32 - 9.46 i$, $-13.51 + 32.63 i$, $44.79 - 0.01 i$, $-32.63 + 13.51 i$, $44.78 - 25.85 i$, $-12.12 - 29.25 i$, $47.37 + 19.63 i$, $-3.62 - 8.75 i$, $44.78 + 44.77 i$, $3.62 - 8.75 i$, $-2.58 - 6.23 i$, $12.12 - 29.25 i$, $0.$, $32.63 + 13.51 i$, $44.78 - 0.01 i$, $13.51 + 32.63 i$, $9.47 - 35.32 i$, $-29.25 + 12.12 i$, $29.09 - 12.05 i$, $-8.75 + 3.62 i$, $89.56$)

\vskip 0.7ex
\hangindent=3em \hangafter=1
\textit{Intrinsic sign problem}

  \vskip 2ex

\noindent32. $11_{\frac{1}{2},89.56}^{48,139}$ \irep{2192}:\ \ 
$d_i$ = ($1.0$,
$1.0$,
$1.931$,
$1.931$,
$2.732$,
$2.732$,
$3.346$,
$3.346$,
$3.732$,
$3.732$,
$3.863$) 

\vskip 0.7ex
\hangindent=3em \hangafter=1
$D^2= 89.569 = 
48+24\sqrt{3}$

\vskip 0.7ex
\hangindent=3em \hangafter=1
$T = ( 0,
\frac{1}{2},
\frac{13}{16},
\frac{13}{16},
\frac{2}{3},
\frac{1}{6},
\frac{1}{16},
\frac{1}{16},
0,
\frac{1}{2},
\frac{23}{48} )
$,

\vskip 0.7ex
\hangindent=3em \hangafter=1
$S$ = ($ 1$,
$ 1$,
$ c_{24}^{1}$,
$ c_{24}^{1}$,
$ 1+\sqrt{3}$,
$ 1+\sqrt{3}$,
$ \frac{3+3\sqrt{3}}{\sqrt{6}}$,
$ \frac{3+3\sqrt{3}}{\sqrt{6}}$,
$ 2+\sqrt{3}$,
$ 2+\sqrt{3}$,
$ 2c_{24}^{1}$;\ \ 
$ 1$,
$ -c_{24}^{1}$,
$ -c_{24}^{1}$,
$ 1+\sqrt{3}$,
$ 1+\sqrt{3}$,
$ \frac{-3-3\sqrt{3}}{\sqrt{6}}$,
$ \frac{-3-3\sqrt{3}}{\sqrt{6}}$,
$ 2+\sqrt{3}$,
$ 2+\sqrt{3}$,
$ -2c_{24}^{1}$;\ \ 
$ \frac{-3-3\sqrt{3}}{\sqrt{6}}$,
$ \frac{3+3\sqrt{3}}{\sqrt{6}}$,
$ -2c_{24}^{1}$,
$ 2c_{24}^{1}$,
$ \frac{-3-3\sqrt{3}}{\sqrt{6}}$,
$ \frac{3+3\sqrt{3}}{\sqrt{6}}$,
$ -c_{24}^{1}$,
$ c_{24}^{1}$,
$0$;\ \ 
$ \frac{-3-3\sqrt{3}}{\sqrt{6}}$,
$ -2c_{24}^{1}$,
$ 2c_{24}^{1}$,
$ \frac{3+3\sqrt{3}}{\sqrt{6}}$,
$ \frac{-3-3\sqrt{3}}{\sqrt{6}}$,
$ -c_{24}^{1}$,
$ c_{24}^{1}$,
$0$;\ \ 
$ 1+\sqrt{3}$,
$ 1+\sqrt{3}$,
$0$,
$0$,
$ -1-\sqrt{3}$,
$ -1-\sqrt{3}$,
$ 2c_{24}^{1}$;\ \ 
$ 1+\sqrt{3}$,
$0$,
$0$,
$ -1-\sqrt{3}$,
$ -1-\sqrt{3}$,
$ -2c_{24}^{1}$;\ \ 
$ \frac{3+3\sqrt{3}}{\sqrt{6}}$,
$ \frac{-3-3\sqrt{3}}{\sqrt{6}}$,
$ \frac{3+3\sqrt{3}}{\sqrt{6}}$,
$ \frac{-3-3\sqrt{3}}{\sqrt{6}}$,
$0$;\ \ 
$ \frac{3+3\sqrt{3}}{\sqrt{6}}$,
$ \frac{3+3\sqrt{3}}{\sqrt{6}}$,
$ \frac{-3-3\sqrt{3}}{\sqrt{6}}$,
$0$;\ \ 
$ 1$,
$ 1$,
$ -2c_{24}^{1}$;\ \ 
$ 1$,
$ 2c_{24}^{1}$;\ \ 
$0$)

\vskip 0.7ex
\hangindent=3em \hangafter=1
$\tau_n$ = ($8.75 + 3.63 i$, $47.37 + 19.63 i$, $-12.11 + 29.25 i$, $35.32 + 9.46 i$, $-13.52 + 32.63 i$, $44.78 + 0.01 i$, $-32.63 + 13.52 i$, $0.$, $-29.25 + 12.11 i$, $15.69 - 37.9 i$, $-3.63 - 8.75 i$, $44.78 - 44.77 i$, $3.63 - 8.75 i$, $29.09 - 12.05 i$, $29.25 + 12.11 i$, $44.78 - 25.85 i$, $32.63 + 13.52 i$, $44.79 + 0.01 i$, $13.52 + 32.63 i$, $9.47 + 35.32 i$, $12.11 + 29.25 i$, $-2.58 - 6.23 i$, $-8.75 + 3.63 i$, $0.01$, $-8.75 - 3.63 i$, $-2.58 + 6.23 i$, $12.11 - 29.25 i$, $9.47 - 35.32 i$, $13.52 - 32.63 i$, $44.79 - 0.01 i$, $32.63 - 13.52 i$, $44.78 + 25.85 i$, $29.25 - 12.11 i$, $29.09 + 12.05 i$, $3.63 + 8.75 i$, $44.78 + 44.77 i$, $-3.63 + 8.75 i$, $15.69 + 37.9 i$, $-29.25 - 12.11 i$, $0.$, $-32.63 - 13.52 i$, $44.78 - 0.01 i$, $-13.52 - 32.63 i$, $35.32 - 9.46 i$, $-12.11 - 29.25 i$, $47.37 - 19.63 i$, $8.75 - 3.63 i$, $89.56$)

\vskip 0.7ex
\hangindent=3em \hangafter=1
\textit{Intrinsic sign problem}

  \vskip 2ex

\noindent33. $11_{\frac{11}{2},89.56}^{48,288}$ \irep{2192}:\ \ 
$d_i$ = ($1.0$,
$1.0$,
$1.931$,
$1.931$,
$2.732$,
$2.732$,
$3.346$,
$3.346$,
$3.732$,
$3.732$,
$3.863$) 

\vskip 0.7ex
\hangindent=3em \hangafter=1
$D^2= 89.569 = 
48+24\sqrt{3}$

\vskip 0.7ex
\hangindent=3em \hangafter=1
$T = ( 0,
\frac{1}{2},
\frac{15}{16},
\frac{15}{16},
\frac{1}{3},
\frac{5}{6},
\frac{11}{16},
\frac{11}{16},
0,
\frac{1}{2},
\frac{13}{48} )
$,

\vskip 0.7ex
\hangindent=3em \hangafter=1
$S$ = ($ 1$,
$ 1$,
$ c_{24}^{1}$,
$ c_{24}^{1}$,
$ 1+\sqrt{3}$,
$ 1+\sqrt{3}$,
$ \frac{3+3\sqrt{3}}{\sqrt{6}}$,
$ \frac{3+3\sqrt{3}}{\sqrt{6}}$,
$ 2+\sqrt{3}$,
$ 2+\sqrt{3}$,
$ 2c_{24}^{1}$;\ \ 
$ 1$,
$ -c_{24}^{1}$,
$ -c_{24}^{1}$,
$ 1+\sqrt{3}$,
$ 1+\sqrt{3}$,
$ \frac{-3-3\sqrt{3}}{\sqrt{6}}$,
$ \frac{-3-3\sqrt{3}}{\sqrt{6}}$,
$ 2+\sqrt{3}$,
$ 2+\sqrt{3}$,
$ -2c_{24}^{1}$;\ \ 
$ \frac{3+3\sqrt{3}}{\sqrt{6}}$,
$ \frac{-3-3\sqrt{3}}{\sqrt{6}}$,
$ -2c_{24}^{1}$,
$ 2c_{24}^{1}$,
$ \frac{-3-3\sqrt{3}}{\sqrt{6}}$,
$ \frac{3+3\sqrt{3}}{\sqrt{6}}$,
$ -c_{24}^{1}$,
$ c_{24}^{1}$,
$0$;\ \ 
$ \frac{3+3\sqrt{3}}{\sqrt{6}}$,
$ -2c_{24}^{1}$,
$ 2c_{24}^{1}$,
$ \frac{3+3\sqrt{3}}{\sqrt{6}}$,
$ \frac{-3-3\sqrt{3}}{\sqrt{6}}$,
$ -c_{24}^{1}$,
$ c_{24}^{1}$,
$0$;\ \ 
$ 1+\sqrt{3}$,
$ 1+\sqrt{3}$,
$0$,
$0$,
$ -1-\sqrt{3}$,
$ -1-\sqrt{3}$,
$ 2c_{24}^{1}$;\ \ 
$ 1+\sqrt{3}$,
$0$,
$0$,
$ -1-\sqrt{3}$,
$ -1-\sqrt{3}$,
$ -2c_{24}^{1}$;\ \ 
$ \frac{-3-3\sqrt{3}}{\sqrt{6}}$,
$ \frac{3+3\sqrt{3}}{\sqrt{6}}$,
$ \frac{3+3\sqrt{3}}{\sqrt{6}}$,
$ \frac{-3-3\sqrt{3}}{\sqrt{6}}$,
$0$;\ \ 
$ \frac{-3-3\sqrt{3}}{\sqrt{6}}$,
$ \frac{3+3\sqrt{3}}{\sqrt{6}}$,
$ \frac{-3-3\sqrt{3}}{\sqrt{6}}$,
$0$;\ \ 
$ 1$,
$ 1$,
$ -2c_{24}^{1}$;\ \ 
$ 1$,
$ 2c_{24}^{1}$;\ \ 
$0$)

\vskip 0.7ex
\hangindent=3em \hangafter=1
$\tau_n$ = ($-3.63 - 8.75 i$, $-2.58 - 6.23 i$, $29.25 - 12.11 i$, $35.32 - 9.46 i$, $-32.63 + 13.52 i$, $44.79 + 0.01 i$, $13.52 - 32.63 i$, $0.$, $-12.11 + 29.25 i$, $29.09 - 12.05 i$, $-8.75 - 3.63 i$, $44.78 + 44.77 i$, $8.75 - 3.63 i$, $15.69 - 37.9 i$, $12.11 + 29.25 i$, $44.78 + 25.85 i$, $-13.52 - 32.63 i$, $44.78 + 0.01 i$, $32.63 + 13.52 i$, $9.47 - 35.32 i$, $-29.25 - 12.11 i$, $47.37 + 19.63 i$, $3.63 - 8.75 i$, $0.01$, $3.63 + 8.75 i$, $47.37 - 19.63 i$, $-29.25 + 12.11 i$, $9.47 + 35.32 i$, $32.63 - 13.52 i$, $44.78 - 0.01 i$, $-13.52 + 32.63 i$, $44.78 - 25.85 i$, $12.11 - 29.25 i$, $15.69 + 37.9 i$, $8.75 + 3.63 i$, $44.78 - 44.77 i$, $-8.75 + 3.63 i$, $29.09 + 12.05 i$, $-12.11 - 29.25 i$, $0.$, $13.52 + 32.63 i$, $44.79 - 0.01 i$, $-32.63 - 13.52 i$, $35.32 + 9.46 i$, $29.25 + 12.11 i$, $-2.58 + 6.23 i$, $-3.63 + 8.75 i$, $89.56$)

\vskip 0.7ex
\hangindent=3em \hangafter=1
\textit{Intrinsic sign problem}

  \vskip 2ex

\noindent34. $11_{\frac{3}{2},89.56}^{48,177}$ \irep{2191}:\ \ 
$d_i$ = ($1.0$,
$1.0$,
$1.931$,
$1.931$,
$2.732$,
$2.732$,
$3.346$,
$3.346$,
$3.732$,
$3.732$,
$3.863$) 

\vskip 0.7ex
\hangindent=3em \hangafter=1
$D^2= 89.569 = 
48+24\sqrt{3}$

\vskip 0.7ex
\hangindent=3em \hangafter=1
$T = ( 0,
\frac{1}{2},
\frac{15}{16},
\frac{15}{16},
\frac{2}{3},
\frac{1}{6},
\frac{3}{16},
\frac{3}{16},
0,
\frac{1}{2},
\frac{29}{48} )
$,

\vskip 0.7ex
\hangindent=3em \hangafter=1
$S$ = ($ 1$,
$ 1$,
$ c_{24}^{1}$,
$ c_{24}^{1}$,
$ 1+\sqrt{3}$,
$ 1+\sqrt{3}$,
$ \frac{3+3\sqrt{3}}{\sqrt{6}}$,
$ \frac{3+3\sqrt{3}}{\sqrt{6}}$,
$ 2+\sqrt{3}$,
$ 2+\sqrt{3}$,
$ 2c_{24}^{1}$;\ \ 
$ 1$,
$ -c_{24}^{1}$,
$ -c_{24}^{1}$,
$ 1+\sqrt{3}$,
$ 1+\sqrt{3}$,
$ \frac{-3-3\sqrt{3}}{\sqrt{6}}$,
$ \frac{-3-3\sqrt{3}}{\sqrt{6}}$,
$ 2+\sqrt{3}$,
$ 2+\sqrt{3}$,
$ -2c_{24}^{1}$;\ \ 
$(\frac{3+3\sqrt{3}}{\sqrt{6}})\mathrm{i}$,
$(\frac{-3-3\sqrt{3}}{\sqrt{6}})\mathrm{i}$,
$ -2c_{24}^{1}$,
$ 2c_{24}^{1}$,
$(\frac{3+3\sqrt{3}}{\sqrt{6}})\mathrm{i}$,
$(\frac{-3-3\sqrt{3}}{\sqrt{6}})\mathrm{i}$,
$ -c_{24}^{1}$,
$ c_{24}^{1}$,
$0$;\ \ 
$(\frac{3+3\sqrt{3}}{\sqrt{6}})\mathrm{i}$,
$ -2c_{24}^{1}$,
$ 2c_{24}^{1}$,
$(\frac{-3-3\sqrt{3}}{\sqrt{6}})\mathrm{i}$,
$(\frac{3+3\sqrt{3}}{\sqrt{6}})\mathrm{i}$,
$ -c_{24}^{1}$,
$ c_{24}^{1}$,
$0$;\ \ 
$ 1+\sqrt{3}$,
$ 1+\sqrt{3}$,
$0$,
$0$,
$ -1-\sqrt{3}$,
$ -1-\sqrt{3}$,
$ 2c_{24}^{1}$;\ \ 
$ 1+\sqrt{3}$,
$0$,
$0$,
$ -1-\sqrt{3}$,
$ -1-\sqrt{3}$,
$ -2c_{24}^{1}$;\ \ 
$(\frac{-3-3\sqrt{3}}{\sqrt{6}})\mathrm{i}$,
$(\frac{3+3\sqrt{3}}{\sqrt{6}})\mathrm{i}$,
$ \frac{3+3\sqrt{3}}{\sqrt{6}}$,
$ \frac{-3-3\sqrt{3}}{\sqrt{6}}$,
$0$;\ \ 
$(\frac{-3-3\sqrt{3}}{\sqrt{6}})\mathrm{i}$,
$ \frac{3+3\sqrt{3}}{\sqrt{6}}$,
$ \frac{-3-3\sqrt{3}}{\sqrt{6}}$,
$0$;\ \ 
$ 1$,
$ 1$,
$ -2c_{24}^{1}$;\ \ 
$ 1$,
$ 2c_{24}^{1}$;\ \ 
$0$)

\vskip 0.7ex
\hangindent=3em \hangafter=1
$\tau_n$ = ($3.62 + 8.75 i$, $15.69 + 37.9 i$, $-12.12 - 29.25 i$, $9.47 - 35.32 i$, $32.63 - 13.51 i$, $44.79 + 0.01 i$, $-13.51 + 32.63 i$, $0.$, $-29.25 - 12.12 i$, $47.37 - 19.63 i$, $8.75 + 3.62 i$, $44.78 + 44.77 i$, $-8.75 + 3.62 i$, $-2.58 + 6.23 i$, $29.25 - 12.12 i$, $44.78 - 25.85 i$, $13.51 + 32.63 i$, $44.78 + 0.01 i$, $-32.63 - 13.51 i$, $35.32 - 9.46 i$, $12.12 - 29.25 i$, $29.09 + 12.05 i$, $-3.62 + 8.75 i$, $0.01$, $-3.62 - 8.75 i$, $29.09 - 12.05 i$, $12.12 + 29.25 i$, $35.32 + 9.46 i$, $-32.63 + 13.51 i$, $44.78 - 0.01 i$, $13.51 - 32.63 i$, $44.78 + 25.85 i$, $29.25 + 12.12 i$, $-2.58 - 6.23 i$, $-8.75 - 3.62 i$, $44.78 - 44.77 i$, $8.75 - 3.62 i$, $47.37 + 19.63 i$, $-29.25 + 12.12 i$, $0.$, $-13.51 - 32.63 i$, $44.79 - 0.01 i$, $32.63 + 13.51 i$, $9.47 + 35.32 i$, $-12.12 + 29.25 i$, $15.69 - 37.9 i$, $3.62 - 8.75 i$, $89.56$)

\vskip 0.7ex
\hangindent=3em \hangafter=1
\textit{Intrinsic sign problem}

  \vskip 2ex

\noindent35. $11_{\frac{144}{23},310.1}^{23,306}$ \irep{1836}:\ \ 
$d_i$ = ($1.0$,
$1.981$,
$2.925$,
$3.815$,
$4.634$,
$5.367$,
$5.999$,
$6.520$,
$6.919$,
$7.190$,
$7.326$) 

\vskip 0.7ex
\hangindent=3em \hangafter=1
$D^2= 310.117 = 
66+55c^{1}_{23}
+45c^{2}_{23}
+36c^{3}_{23}
+28c^{4}_{23}
+21c^{5}_{23}
+15c^{6}_{23}
+10c^{7}_{23}
+6c^{8}_{23}
+3c^{9}_{23}
+c^{10}_{23}
$

\vskip 0.7ex
\hangindent=3em \hangafter=1
$T = ( 0,
\frac{5}{23},
\frac{21}{23},
\frac{2}{23},
\frac{17}{23},
\frac{20}{23},
\frac{11}{23},
\frac{13}{23},
\frac{3}{23},
\frac{4}{23},
\frac{16}{23} )
$,

\vskip 0.7ex
\hangindent=3em \hangafter=1
$S$ = ($ 1$,
$ -c_{23}^{11}$,
$ \xi_{23}^{3}$,
$ \xi_{23}^{19}$,
$ \xi_{23}^{5}$,
$ \xi_{23}^{17}$,
$ \xi_{23}^{7}$,
$ \xi_{23}^{15}$,
$ \xi_{23}^{9}$,
$ \xi_{23}^{13}$,
$ \xi_{23}^{11}$;\ \ 
$ -\xi_{23}^{19}$,
$ \xi_{23}^{17}$,
$ -\xi_{23}^{15}$,
$ \xi_{23}^{13}$,
$ -\xi_{23}^{11}$,
$ \xi_{23}^{9}$,
$ -\xi_{23}^{7}$,
$ \xi_{23}^{5}$,
$ -\xi_{23}^{3}$,
$ 1$;\ \ 
$ \xi_{23}^{9}$,
$ \xi_{23}^{11}$,
$ \xi_{23}^{15}$,
$ \xi_{23}^{5}$,
$ -c_{23}^{11}$,
$ -1$,
$ -\xi_{23}^{19}$,
$ -\xi_{23}^{7}$,
$ -\xi_{23}^{13}$;\ \ 
$ -\xi_{23}^{7}$,
$ \xi_{23}^{3}$,
$ 1$,
$ -\xi_{23}^{5}$,
$ \xi_{23}^{9}$,
$ -\xi_{23}^{13}$,
$ \xi_{23}^{17}$,
$ c_{23}^{11}$;\ \ 
$ c_{23}^{11}$,
$ -\xi_{23}^{7}$,
$ -\xi_{23}^{11}$,
$ -\xi_{23}^{17}$,
$ -1$,
$ \xi_{23}^{19}$,
$ \xi_{23}^{9}$;\ \ 
$ \xi_{23}^{13}$,
$ -\xi_{23}^{19}$,
$ c_{23}^{11}$,
$ \xi_{23}^{15}$,
$ -\xi_{23}^{9}$,
$ \xi_{23}^{3}$;\ \ 
$ \xi_{23}^{3}$,
$ \xi_{23}^{13}$,
$ \xi_{23}^{17}$,
$ -1$,
$ -\xi_{23}^{15}$;\ \ 
$ -\xi_{23}^{5}$,
$ -\xi_{23}^{3}$,
$ \xi_{23}^{11}$,
$ -\xi_{23}^{19}$;\ \ 
$ -\xi_{23}^{11}$,
$ c_{23}^{11}$,
$ \xi_{23}^{7}$;\ \ 
$ -\xi_{23}^{15}$,
$ \xi_{23}^{5}$;\ \ 
$ -\xi_{23}^{17}$)

\vskip 0.7ex
\hangindent=3em \hangafter=1
$\tau_n$ = ($3.58 - 17.24 i$, $-26.25 + 126.3 i$, $-110.55 + 30.97 i$, $-43.48 - 83.9 i$, $-94.5 - 76.89 i$, $38.75 + 54.89 i$, $121.91 + 34.16 i$, $-29.71 - 42.08 i$, $5.56 + 81.4 i$, $7.2 - 105.4 i$, $-16.05 + 30.97 i$, $-16.05 - 30.97 i$, $7.2 + 105.4 i$, $5.56 - 81.4 i$, $-29.71 + 42.08 i$, $121.91 - 34.16 i$, $38.75 - 54.89 i$, $-94.5 + 76.89 i$, $-43.48 + 83.9 i$, $-110.55 - 30.97 i$, $-26.25 - 126.3 i$, $3.58 + 17.24 i$, $310.05$)

\vskip 0.7ex
\hangindent=3em \hangafter=1
\textit{Intrinsic sign problem}

  \vskip 2ex

\noindent36. $11_{\frac{40}{23},310.1}^{23,508}$ \irep{1836}:\ \ 
$d_i$ = ($1.0$,
$1.981$,
$2.925$,
$3.815$,
$4.634$,
$5.367$,
$5.999$,
$6.520$,
$6.919$,
$7.190$,
$7.326$) 

\vskip 0.7ex
\hangindent=3em \hangafter=1
$D^2= 310.117 = 
66+55c^{1}_{23}
+45c^{2}_{23}
+36c^{3}_{23}
+28c^{4}_{23}
+21c^{5}_{23}
+15c^{6}_{23}
+10c^{7}_{23}
+6c^{8}_{23}
+3c^{9}_{23}
+c^{10}_{23}
$

\vskip 0.7ex
\hangindent=3em \hangafter=1
$T = ( 0,
\frac{18}{23},
\frac{2}{23},
\frac{21}{23},
\frac{6}{23},
\frac{3}{23},
\frac{12}{23},
\frac{10}{23},
\frac{20}{23},
\frac{19}{23},
\frac{7}{23} )
$,

\vskip 0.7ex
\hangindent=3em \hangafter=1
$S$ = ($ 1$,
$ -c_{23}^{11}$,
$ \xi_{23}^{3}$,
$ \xi_{23}^{19}$,
$ \xi_{23}^{5}$,
$ \xi_{23}^{17}$,
$ \xi_{23}^{7}$,
$ \xi_{23}^{15}$,
$ \xi_{23}^{9}$,
$ \xi_{23}^{13}$,
$ \xi_{23}^{11}$;\ \ 
$ -\xi_{23}^{19}$,
$ \xi_{23}^{17}$,
$ -\xi_{23}^{15}$,
$ \xi_{23}^{13}$,
$ -\xi_{23}^{11}$,
$ \xi_{23}^{9}$,
$ -\xi_{23}^{7}$,
$ \xi_{23}^{5}$,
$ -\xi_{23}^{3}$,
$ 1$;\ \ 
$ \xi_{23}^{9}$,
$ \xi_{23}^{11}$,
$ \xi_{23}^{15}$,
$ \xi_{23}^{5}$,
$ -c_{23}^{11}$,
$ -1$,
$ -\xi_{23}^{19}$,
$ -\xi_{23}^{7}$,
$ -\xi_{23}^{13}$;\ \ 
$ -\xi_{23}^{7}$,
$ \xi_{23}^{3}$,
$ 1$,
$ -\xi_{23}^{5}$,
$ \xi_{23}^{9}$,
$ -\xi_{23}^{13}$,
$ \xi_{23}^{17}$,
$ c_{23}^{11}$;\ \ 
$ c_{23}^{11}$,
$ -\xi_{23}^{7}$,
$ -\xi_{23}^{11}$,
$ -\xi_{23}^{17}$,
$ -1$,
$ \xi_{23}^{19}$,
$ \xi_{23}^{9}$;\ \ 
$ \xi_{23}^{13}$,
$ -\xi_{23}^{19}$,
$ c_{23}^{11}$,
$ \xi_{23}^{15}$,
$ -\xi_{23}^{9}$,
$ \xi_{23}^{3}$;\ \ 
$ \xi_{23}^{3}$,
$ \xi_{23}^{13}$,
$ \xi_{23}^{17}$,
$ -1$,
$ -\xi_{23}^{15}$;\ \ 
$ -\xi_{23}^{5}$,
$ -\xi_{23}^{3}$,
$ \xi_{23}^{11}$,
$ -\xi_{23}^{19}$;\ \ 
$ -\xi_{23}^{11}$,
$ c_{23}^{11}$,
$ \xi_{23}^{7}$;\ \ 
$ -\xi_{23}^{15}$,
$ \xi_{23}^{5}$;\ \ 
$ -\xi_{23}^{17}$)

\vskip 0.7ex
\hangindent=3em \hangafter=1
$\tau_n$ = ($3.58 + 17.24 i$, $-26.25 - 126.3 i$, $-110.55 - 30.97 i$, $-43.48 + 83.9 i$, $-94.5 + 76.89 i$, $38.75 - 54.89 i$, $121.91 - 34.16 i$, $-29.71 + 42.08 i$, $5.56 - 81.4 i$, $7.2 + 105.4 i$, $-16.05 - 30.97 i$, $-16.05 + 30.97 i$, $7.2 - 105.4 i$, $5.56 + 81.4 i$, $-29.71 - 42.08 i$, $121.91 + 34.16 i$, $38.75 + 54.89 i$, $-94.5 - 76.89 i$, $-43.48 - 83.9 i$, $-110.55 + 30.97 i$, $-26.25 + 126.3 i$, $3.58 - 17.24 i$, $310.05$)

\vskip 0.7ex
\hangindent=3em \hangafter=1
\textit{Intrinsic sign problem}

  \vskip 2ex

\noindent37. $11_{\frac{54}{19},696.5}^{19,306}$ \irep{1746}:\ \ 
$d_i$ = ($1.0$,
$2.972$,
$4.864$,
$6.54$,
$6.54$,
$6.623$,
$8.201$,
$9.556$,
$10.650$,
$11.453$,
$11.944$) 

\vskip 0.7ex
\hangindent=3em \hangafter=1
$D^2= 696.547 = 
171+152c^{1}_{19}
+133c^{2}_{19}
+114c^{3}_{19}
+95c^{4}_{19}
+76c^{5}_{19}
+57c^{6}_{19}
+38c^{7}_{19}
+19c^{8}_{19}
$

\vskip 0.7ex
\hangindent=3em \hangafter=1
$T = ( 0,
\frac{1}{19},
\frac{3}{19},
\frac{7}{19},
\frac{7}{19},
\frac{6}{19},
\frac{10}{19},
\frac{15}{19},
\frac{2}{19},
\frac{9}{19},
\frac{17}{19} )
$,

\vskip 0.7ex
\hangindent=3em \hangafter=1
$S$ = ($ 1$,
$ 2+c^{1}_{19}
+c^{2}_{19}
+c^{3}_{19}
+c^{4}_{19}
+c^{5}_{19}
+c^{6}_{19}
+c^{7}_{19}
+c^{8}_{19}
$,
$ 2+2c^{1}_{19}
+c^{2}_{19}
+c^{3}_{19}
+c^{4}_{19}
+c^{5}_{19}
+c^{6}_{19}
+c^{7}_{19}
+c^{8}_{19}
$,
$ \xi_{19}^{9}$,
$ \xi_{19}^{9}$,
$ 2+2c^{1}_{19}
+c^{2}_{19}
+c^{3}_{19}
+c^{4}_{19}
+c^{5}_{19}
+c^{6}_{19}
+c^{7}_{19}
$,
$ 2+2c^{1}_{19}
+2c^{2}_{19}
+c^{3}_{19}
+c^{4}_{19}
+c^{5}_{19}
+c^{6}_{19}
+c^{7}_{19}
$,
$ 2+2c^{1}_{19}
+2c^{2}_{19}
+c^{3}_{19}
+c^{4}_{19}
+c^{5}_{19}
+c^{6}_{19}
$,
$ 2+2c^{1}_{19}
+2c^{2}_{19}
+2c^{3}_{19}
+c^{4}_{19}
+c^{5}_{19}
+c^{6}_{19}
$,
$ 2+2c^{1}_{19}
+2c^{2}_{19}
+2c^{3}_{19}
+c^{4}_{19}
+c^{5}_{19}
$,
$ 2+2c^{1}_{19}
+2c^{2}_{19}
+2c^{3}_{19}
+2c^{4}_{19}
+c^{5}_{19}
$;\ \ 
$ 2+2c^{1}_{19}
+2c^{2}_{19}
+c^{3}_{19}
+c^{4}_{19}
+c^{5}_{19}
+c^{6}_{19}
+c^{7}_{19}
$,
$ 2+2c^{1}_{19}
+2c^{2}_{19}
+2c^{3}_{19}
+c^{4}_{19}
+c^{5}_{19}
$,
$ -\xi_{19}^{9}$,
$ -\xi_{19}^{9}$,
$ 2+2c^{1}_{19}
+2c^{2}_{19}
+2c^{3}_{19}
+2c^{4}_{19}
+c^{5}_{19}
$,
$ 2+2c^{1}_{19}
+2c^{2}_{19}
+c^{3}_{19}
+c^{4}_{19}
+c^{5}_{19}
+c^{6}_{19}
$,
$ 2+2c^{1}_{19}
+c^{2}_{19}
+c^{3}_{19}
+c^{4}_{19}
+c^{5}_{19}
+c^{6}_{19}
+c^{7}_{19}
+c^{8}_{19}
$,
$ -1$,
$ -2-2  c^{1}_{19}
-c^{2}_{19}
-c^{3}_{19}
-c^{4}_{19}
-c^{5}_{19}
-c^{6}_{19}
-c^{7}_{19}
$,
$ -2-2  c^{1}_{19}
-2  c^{2}_{19}
-2  c^{3}_{19}
-c^{4}_{19}
-c^{5}_{19}
-c^{6}_{19}
$;\ \ 
$ 2+2c^{1}_{19}
+2c^{2}_{19}
+2c^{3}_{19}
+c^{4}_{19}
+c^{5}_{19}
+c^{6}_{19}
$,
$ \xi_{19}^{9}$,
$ \xi_{19}^{9}$,
$ 2+c^{1}_{19}
+c^{2}_{19}
+c^{3}_{19}
+c^{4}_{19}
+c^{5}_{19}
+c^{6}_{19}
+c^{7}_{19}
+c^{8}_{19}
$,
$ -2-2  c^{1}_{19}
-c^{2}_{19}
-c^{3}_{19}
-c^{4}_{19}
-c^{5}_{19}
-c^{6}_{19}
-c^{7}_{19}
$,
$ -2-2  c^{1}_{19}
-2  c^{2}_{19}
-2  c^{3}_{19}
-2  c^{4}_{19}
-c^{5}_{19}
$,
$ -2-2  c^{1}_{19}
-2  c^{2}_{19}
-c^{3}_{19}
-c^{4}_{19}
-c^{5}_{19}
-c^{6}_{19}
$,
$ -1$,
$ 2+2c^{1}_{19}
+2c^{2}_{19}
+c^{3}_{19}
+c^{4}_{19}
+c^{5}_{19}
+c^{6}_{19}
+c^{7}_{19}
$;\ \ 
$ s^{1}_{19}
+s^{2}_{19}
+s^{3}_{19}
+s^{4}_{19}
+2\zeta^{5}_{19}
-\zeta^{-5}_{19}
+\zeta^{6}_{19}
+2\zeta^{7}_{19}
-\zeta^{-7}_{19}
+2\zeta^{8}_{19}
-\zeta^{-8}_{19}
+\zeta^{9}_{19}
$,
$ -1-2  \zeta^{1}_{19}
-2  \zeta^{2}_{19}
-2  \zeta^{3}_{19}
-2  \zeta^{4}_{19}
-2  \zeta^{5}_{19}
+\zeta^{-5}_{19}
-\zeta^{6}_{19}
-2  \zeta^{7}_{19}
+\zeta^{-7}_{19}
-2  \zeta^{8}_{19}
+\zeta^{-8}_{19}
-\zeta^{9}_{19}
$,
$ -\xi_{19}^{9}$,
$ \xi_{19}^{9}$,
$ -\xi_{19}^{9}$,
$ \xi_{19}^{9}$,
$ -\xi_{19}^{9}$,
$ \xi_{19}^{9}$;\ \ 
$ s^{1}_{19}
+s^{2}_{19}
+s^{3}_{19}
+s^{4}_{19}
+2\zeta^{5}_{19}
-\zeta^{-5}_{19}
+\zeta^{6}_{19}
+2\zeta^{7}_{19}
-\zeta^{-7}_{19}
+2\zeta^{8}_{19}
-\zeta^{-8}_{19}
+\zeta^{9}_{19}
$,
$ -\xi_{19}^{9}$,
$ \xi_{19}^{9}$,
$ -\xi_{19}^{9}$,
$ \xi_{19}^{9}$,
$ -\xi_{19}^{9}$,
$ \xi_{19}^{9}$;\ \ 
$ -2-2  c^{1}_{19}
-2  c^{2}_{19}
-c^{3}_{19}
-c^{4}_{19}
-c^{5}_{19}
-c^{6}_{19}
$,
$ -2-2  c^{1}_{19}
-2  c^{2}_{19}
-2  c^{3}_{19}
-c^{4}_{19}
-c^{5}_{19}
-c^{6}_{19}
$,
$ 1$,
$ 2+2c^{1}_{19}
+2c^{2}_{19}
+2c^{3}_{19}
+c^{4}_{19}
+c^{5}_{19}
$,
$ 2+2c^{1}_{19}
+2c^{2}_{19}
+c^{3}_{19}
+c^{4}_{19}
+c^{5}_{19}
+c^{6}_{19}
+c^{7}_{19}
$,
$ -2-2  c^{1}_{19}
-c^{2}_{19}
-c^{3}_{19}
-c^{4}_{19}
-c^{5}_{19}
-c^{6}_{19}
-c^{7}_{19}
-c^{8}_{19}
$;\ \ 
$ 2+2c^{1}_{19}
+c^{2}_{19}
+c^{3}_{19}
+c^{4}_{19}
+c^{5}_{19}
+c^{6}_{19}
+c^{7}_{19}
+c^{8}_{19}
$,
$ 2+2c^{1}_{19}
+2c^{2}_{19}
+2c^{3}_{19}
+c^{4}_{19}
+c^{5}_{19}
$,
$ -2-c^{1}_{19}
-c^{2}_{19}
-c^{3}_{19}
-c^{4}_{19}
-c^{5}_{19}
-c^{6}_{19}
-c^{7}_{19}
-c^{8}_{19}
$,
$ -2-2  c^{1}_{19}
-2  c^{2}_{19}
-2  c^{3}_{19}
-2  c^{4}_{19}
-c^{5}_{19}
$,
$ 1$;\ \ 
$ -2-2  c^{1}_{19}
-c^{2}_{19}
-c^{3}_{19}
-c^{4}_{19}
-c^{5}_{19}
-c^{6}_{19}
-c^{7}_{19}
$,
$ -2-2  c^{1}_{19}
-2  c^{2}_{19}
-c^{3}_{19}
-c^{4}_{19}
-c^{5}_{19}
-c^{6}_{19}
-c^{7}_{19}
$,
$ 2+2c^{1}_{19}
+2c^{2}_{19}
+2c^{3}_{19}
+c^{4}_{19}
+c^{5}_{19}
+c^{6}_{19}
$,
$ 2+c^{1}_{19}
+c^{2}_{19}
+c^{3}_{19}
+c^{4}_{19}
+c^{5}_{19}
+c^{6}_{19}
+c^{7}_{19}
+c^{8}_{19}
$;\ \ 
$ 2+2c^{1}_{19}
+2c^{2}_{19}
+2c^{3}_{19}
+2c^{4}_{19}
+c^{5}_{19}
$,
$ -2-2  c^{1}_{19}
-c^{2}_{19}
-c^{3}_{19}
-c^{4}_{19}
-c^{5}_{19}
-c^{6}_{19}
-c^{7}_{19}
-c^{8}_{19}
$,
$ -2-2  c^{1}_{19}
-c^{2}_{19}
-c^{3}_{19}
-c^{4}_{19}
-c^{5}_{19}
-c^{6}_{19}
-c^{7}_{19}
$;\ \ 
$ -2-c^{1}_{19}
-c^{2}_{19}
-c^{3}_{19}
-c^{4}_{19}
-c^{5}_{19}
-c^{6}_{19}
-c^{7}_{19}
-c^{8}_{19}
$,
$ 2+2c^{1}_{19}
+2c^{2}_{19}
+c^{3}_{19}
+c^{4}_{19}
+c^{5}_{19}
+c^{6}_{19}
$;\ \ 
$ -2-2  c^{1}_{19}
-2  c^{2}_{19}
-2  c^{3}_{19}
-c^{4}_{19}
-c^{5}_{19}
$)

\vskip 0.7ex
\hangindent=3em \hangafter=1
$\tau_n$ = ($-24.48 + 29.82 i$, $131.92 - 182.99 i$, $-247.74 + 120.42 i$, $-119.54 + 72.23 i$, $-286.32 - 84.44 i$, $2.99 - 66.61 i$, $-241.7 + 95.48 i$, $181.01 + 38.93 i$, $258.99 + 183.61 i$, $258.99 - 183.61 i$, $181.01 - 38.93 i$, $-241.7 - 95.48 i$, $2.99 + 66.61 i$, $-286.32 + 84.44 i$, $-119.54 - 72.23 i$, $-247.74 - 120.42 i$, $131.92 + 182.99 i$, $-24.48 - 29.82 i$, $708.72$)

\vskip 0.7ex
\hangindent=3em \hangafter=1
\textit{Intrinsic sign problem}

  \vskip 2ex

\noindent38. $11_{\frac{98}{19},696.5}^{19,829}$ \irep{1746}:\ \ 
$d_i$ = ($1.0$,
$2.972$,
$4.864$,
$6.54$,
$6.54$,
$6.623$,
$8.201$,
$9.556$,
$10.650$,
$11.453$,
$11.944$) 

\vskip 0.7ex
\hangindent=3em \hangafter=1
$D^2= 696.547 = 
171+152c^{1}_{19}
+133c^{2}_{19}
+114c^{3}_{19}
+95c^{4}_{19}
+76c^{5}_{19}
+57c^{6}_{19}
+38c^{7}_{19}
+19c^{8}_{19}
$

\vskip 0.7ex
\hangindent=3em \hangafter=1
$T = ( 0,
\frac{18}{19},
\frac{16}{19},
\frac{12}{19},
\frac{12}{19},
\frac{13}{19},
\frac{9}{19},
\frac{4}{19},
\frac{17}{19},
\frac{10}{19},
\frac{2}{19} )
$,

\vskip 0.7ex
\hangindent=3em \hangafter=1
$S$ = ($ 1$,
$ 2+c^{1}_{19}
+c^{2}_{19}
+c^{3}_{19}
+c^{4}_{19}
+c^{5}_{19}
+c^{6}_{19}
+c^{7}_{19}
+c^{8}_{19}
$,
$ 2+2c^{1}_{19}
+c^{2}_{19}
+c^{3}_{19}
+c^{4}_{19}
+c^{5}_{19}
+c^{6}_{19}
+c^{7}_{19}
+c^{8}_{19}
$,
$ \xi_{19}^{9}$,
$ \xi_{19}^{9}$,
$ 2+2c^{1}_{19}
+c^{2}_{19}
+c^{3}_{19}
+c^{4}_{19}
+c^{5}_{19}
+c^{6}_{19}
+c^{7}_{19}
$,
$ 2+2c^{1}_{19}
+2c^{2}_{19}
+c^{3}_{19}
+c^{4}_{19}
+c^{5}_{19}
+c^{6}_{19}
+c^{7}_{19}
$,
$ 2+2c^{1}_{19}
+2c^{2}_{19}
+c^{3}_{19}
+c^{4}_{19}
+c^{5}_{19}
+c^{6}_{19}
$,
$ 2+2c^{1}_{19}
+2c^{2}_{19}
+2c^{3}_{19}
+c^{4}_{19}
+c^{5}_{19}
+c^{6}_{19}
$,
$ 2+2c^{1}_{19}
+2c^{2}_{19}
+2c^{3}_{19}
+c^{4}_{19}
+c^{5}_{19}
$,
$ 2+2c^{1}_{19}
+2c^{2}_{19}
+2c^{3}_{19}
+2c^{4}_{19}
+c^{5}_{19}
$;\ \ 
$ 2+2c^{1}_{19}
+2c^{2}_{19}
+c^{3}_{19}
+c^{4}_{19}
+c^{5}_{19}
+c^{6}_{19}
+c^{7}_{19}
$,
$ 2+2c^{1}_{19}
+2c^{2}_{19}
+2c^{3}_{19}
+c^{4}_{19}
+c^{5}_{19}
$,
$ -\xi_{19}^{9}$,
$ -\xi_{19}^{9}$,
$ 2+2c^{1}_{19}
+2c^{2}_{19}
+2c^{3}_{19}
+2c^{4}_{19}
+c^{5}_{19}
$,
$ 2+2c^{1}_{19}
+2c^{2}_{19}
+c^{3}_{19}
+c^{4}_{19}
+c^{5}_{19}
+c^{6}_{19}
$,
$ 2+2c^{1}_{19}
+c^{2}_{19}
+c^{3}_{19}
+c^{4}_{19}
+c^{5}_{19}
+c^{6}_{19}
+c^{7}_{19}
+c^{8}_{19}
$,
$ -1$,
$ -2-2  c^{1}_{19}
-c^{2}_{19}
-c^{3}_{19}
-c^{4}_{19}
-c^{5}_{19}
-c^{6}_{19}
-c^{7}_{19}
$,
$ -2-2  c^{1}_{19}
-2  c^{2}_{19}
-2  c^{3}_{19}
-c^{4}_{19}
-c^{5}_{19}
-c^{6}_{19}
$;\ \ 
$ 2+2c^{1}_{19}
+2c^{2}_{19}
+2c^{3}_{19}
+c^{4}_{19}
+c^{5}_{19}
+c^{6}_{19}
$,
$ \xi_{19}^{9}$,
$ \xi_{19}^{9}$,
$ 2+c^{1}_{19}
+c^{2}_{19}
+c^{3}_{19}
+c^{4}_{19}
+c^{5}_{19}
+c^{6}_{19}
+c^{7}_{19}
+c^{8}_{19}
$,
$ -2-2  c^{1}_{19}
-c^{2}_{19}
-c^{3}_{19}
-c^{4}_{19}
-c^{5}_{19}
-c^{6}_{19}
-c^{7}_{19}
$,
$ -2-2  c^{1}_{19}
-2  c^{2}_{19}
-2  c^{3}_{19}
-2  c^{4}_{19}
-c^{5}_{19}
$,
$ -2-2  c^{1}_{19}
-2  c^{2}_{19}
-c^{3}_{19}
-c^{4}_{19}
-c^{5}_{19}
-c^{6}_{19}
$,
$ -1$,
$ 2+2c^{1}_{19}
+2c^{2}_{19}
+c^{3}_{19}
+c^{4}_{19}
+c^{5}_{19}
+c^{6}_{19}
+c^{7}_{19}
$;\ \ 
$ -1-2  \zeta^{1}_{19}
-2  \zeta^{2}_{19}
-2  \zeta^{3}_{19}
-2  \zeta^{4}_{19}
-2  \zeta^{5}_{19}
+\zeta^{-5}_{19}
-\zeta^{6}_{19}
-2  \zeta^{7}_{19}
+\zeta^{-7}_{19}
-2  \zeta^{8}_{19}
+\zeta^{-8}_{19}
-\zeta^{9}_{19}
$,
$ s^{1}_{19}
+s^{2}_{19}
+s^{3}_{19}
+s^{4}_{19}
+2\zeta^{5}_{19}
-\zeta^{-5}_{19}
+\zeta^{6}_{19}
+2\zeta^{7}_{19}
-\zeta^{-7}_{19}
+2\zeta^{8}_{19}
-\zeta^{-8}_{19}
+\zeta^{9}_{19}
$,
$ -\xi_{19}^{9}$,
$ \xi_{19}^{9}$,
$ -\xi_{19}^{9}$,
$ \xi_{19}^{9}$,
$ -\xi_{19}^{9}$,
$ \xi_{19}^{9}$;\ \ 
$ -1-2  \zeta^{1}_{19}
-2  \zeta^{2}_{19}
-2  \zeta^{3}_{19}
-2  \zeta^{4}_{19}
-2  \zeta^{5}_{19}
+\zeta^{-5}_{19}
-\zeta^{6}_{19}
-2  \zeta^{7}_{19}
+\zeta^{-7}_{19}
-2  \zeta^{8}_{19}
+\zeta^{-8}_{19}
-\zeta^{9}_{19}
$,
$ -\xi_{19}^{9}$,
$ \xi_{19}^{9}$,
$ -\xi_{19}^{9}$,
$ \xi_{19}^{9}$,
$ -\xi_{19}^{9}$,
$ \xi_{19}^{9}$;\ \ 
$ -2-2  c^{1}_{19}
-2  c^{2}_{19}
-c^{3}_{19}
-c^{4}_{19}
-c^{5}_{19}
-c^{6}_{19}
$,
$ -2-2  c^{1}_{19}
-2  c^{2}_{19}
-2  c^{3}_{19}
-c^{4}_{19}
-c^{5}_{19}
-c^{6}_{19}
$,
$ 1$,
$ 2+2c^{1}_{19}
+2c^{2}_{19}
+2c^{3}_{19}
+c^{4}_{19}
+c^{5}_{19}
$,
$ 2+2c^{1}_{19}
+2c^{2}_{19}
+c^{3}_{19}
+c^{4}_{19}
+c^{5}_{19}
+c^{6}_{19}
+c^{7}_{19}
$,
$ -2-2  c^{1}_{19}
-c^{2}_{19}
-c^{3}_{19}
-c^{4}_{19}
-c^{5}_{19}
-c^{6}_{19}
-c^{7}_{19}
-c^{8}_{19}
$;\ \ 
$ 2+2c^{1}_{19}
+c^{2}_{19}
+c^{3}_{19}
+c^{4}_{19}
+c^{5}_{19}
+c^{6}_{19}
+c^{7}_{19}
+c^{8}_{19}
$,
$ 2+2c^{1}_{19}
+2c^{2}_{19}
+2c^{3}_{19}
+c^{4}_{19}
+c^{5}_{19}
$,
$ -2-c^{1}_{19}
-c^{2}_{19}
-c^{3}_{19}
-c^{4}_{19}
-c^{5}_{19}
-c^{6}_{19}
-c^{7}_{19}
-c^{8}_{19}
$,
$ -2-2  c^{1}_{19}
-2  c^{2}_{19}
-2  c^{3}_{19}
-2  c^{4}_{19}
-c^{5}_{19}
$,
$ 1$;\ \ 
$ -2-2  c^{1}_{19}
-c^{2}_{19}
-c^{3}_{19}
-c^{4}_{19}
-c^{5}_{19}
-c^{6}_{19}
-c^{7}_{19}
$,
$ -2-2  c^{1}_{19}
-2  c^{2}_{19}
-c^{3}_{19}
-c^{4}_{19}
-c^{5}_{19}
-c^{6}_{19}
-c^{7}_{19}
$,
$ 2+2c^{1}_{19}
+2c^{2}_{19}
+2c^{3}_{19}
+c^{4}_{19}
+c^{5}_{19}
+c^{6}_{19}
$,
$ 2+c^{1}_{19}
+c^{2}_{19}
+c^{3}_{19}
+c^{4}_{19}
+c^{5}_{19}
+c^{6}_{19}
+c^{7}_{19}
+c^{8}_{19}
$;\ \ 
$ 2+2c^{1}_{19}
+2c^{2}_{19}
+2c^{3}_{19}
+2c^{4}_{19}
+c^{5}_{19}
$,
$ -2-2  c^{1}_{19}
-c^{2}_{19}
-c^{3}_{19}
-c^{4}_{19}
-c^{5}_{19}
-c^{6}_{19}
-c^{7}_{19}
-c^{8}_{19}
$,
$ -2-2  c^{1}_{19}
-c^{2}_{19}
-c^{3}_{19}
-c^{4}_{19}
-c^{5}_{19}
-c^{6}_{19}
-c^{7}_{19}
$;\ \ 
$ -2-c^{1}_{19}
-c^{2}_{19}
-c^{3}_{19}
-c^{4}_{19}
-c^{5}_{19}
-c^{6}_{19}
-c^{7}_{19}
-c^{8}_{19}
$,
$ 2+2c^{1}_{19}
+2c^{2}_{19}
+c^{3}_{19}
+c^{4}_{19}
+c^{5}_{19}
+c^{6}_{19}
$;\ \ 
$ -2-2  c^{1}_{19}
-2  c^{2}_{19}
-2  c^{3}_{19}
-c^{4}_{19}
-c^{5}_{19}
$)

\vskip 0.7ex
\hangindent=3em \hangafter=1
$\tau_n$ = ($-24.48 - 29.82 i$, $131.92 + 182.99 i$, $-247.74 - 120.42 i$, $-119.54 - 72.23 i$, $-286.32 + 84.44 i$, $2.99 + 66.61 i$, $-241.7 - 95.48 i$, $181.01 - 38.93 i$, $258.99 - 183.61 i$, $258.99 + 183.61 i$, $181.01 + 38.93 i$, $-241.7 + 95.48 i$, $2.99 - 66.61 i$, $-286.32 - 84.44 i$, $-119.54 + 72.23 i$, $-247.74 + 120.42 i$, $131.92 - 182.99 i$, $-24.48 + 29.82 i$, $708.72$)

\vskip 0.7ex
\hangindent=3em \hangafter=1
\textit{Intrinsic sign problem}

  \vskip 2ex

\noindent39. $11_{3,1337.}^{48,634}$ \irep{2187}:\ \ 
$d_i$ = ($1.0$,
$6.464$,
$6.464$,
$7.464$,
$7.464$,
$12.928$,
$12.928$,
$12.928$,
$13.928$,
$14.928$,
$14.928$) 

\vskip 0.7ex
\hangindent=3em \hangafter=1
$D^2= 1337.107 = 
672+384\sqrt{3}$

\vskip 0.7ex
\hangindent=3em \hangafter=1
$T = ( 0,
0,
0,
\frac{1}{4},
\frac{1}{4},
\frac{3}{4},
\frac{3}{16},
\frac{11}{16},
0,
\frac{1}{3},
\frac{7}{12} )
$,

\vskip 0.7ex
\hangindent=3em \hangafter=1
$S$ = ($ 1$,
$ 3+2\sqrt{3}$,
$ 3+2\sqrt{3}$,
$ 4+2\sqrt{3}$,
$ 4+2\sqrt{3}$,
$ 6+4\sqrt{3}$,
$ 6+4\sqrt{3}$,
$ 6+4\sqrt{3}$,
$ 7+4\sqrt{3}$,
$ 8+4\sqrt{3}$,
$ 8+4\sqrt{3}$;\ \ 
$ -7+4\zeta^{1}_{12}
-8  \zeta^{-1}_{12}
+8\zeta^{2}_{12}
$,
$ 1-8  \zeta^{1}_{12}
+4\zeta^{-1}_{12}
-8  \zeta^{2}_{12}
$,
$(-6-4\sqrt{3})\mathrm{i}$,
$(6+4\sqrt{3})\mathrm{i}$,
$ -6-4\sqrt{3}$,
$ 6+4\sqrt{3}$,
$ 6+4\sqrt{3}$,
$ -3-2\sqrt{3}$,
$0$,
$0$;\ \ 
$ -7+4\zeta^{1}_{12}
-8  \zeta^{-1}_{12}
+8\zeta^{2}_{12}
$,
$(6+4\sqrt{3})\mathrm{i}$,
$(-6-4\sqrt{3})\mathrm{i}$,
$ -6-4\sqrt{3}$,
$ 6+4\sqrt{3}$,
$ 6+4\sqrt{3}$,
$ -3-2\sqrt{3}$,
$0$,
$0$;\ \ 
$ (-8-4\sqrt{3})\zeta_{6}^{1}$,
$ (8+4\sqrt{3})\zeta_{3}^{1}$,
$0$,
$0$,
$0$,
$ 4+2\sqrt{3}$,
$ 8+4\sqrt{3}$,
$ -8-4\sqrt{3}$;\ \ 
$ (-8-4\sqrt{3})\zeta_{6}^{1}$,
$0$,
$0$,
$0$,
$ 4+2\sqrt{3}$,
$ 8+4\sqrt{3}$,
$ -8-4\sqrt{3}$;\ \ 
$ 12+8\sqrt{3}$,
$0$,
$0$,
$ -6-4\sqrt{3}$,
$0$,
$0$;\ \ 
$ \frac{24+12\sqrt{3}}{\sqrt{6}}$,
$ \frac{-24-12\sqrt{3}}{\sqrt{6}}$,
$ -6-4\sqrt{3}$,
$0$,
$0$;\ \ 
$ \frac{24+12\sqrt{3}}{\sqrt{6}}$,
$ -6-4\sqrt{3}$,
$0$,
$0$;\ \ 
$ 1$,
$ 8+4\sqrt{3}$,
$ 8+4\sqrt{3}$;\ \ 
$ -8-4\sqrt{3}$,
$ -8-4\sqrt{3}$;\ \ 
$ 8+4\sqrt{3}$)

\vskip 0.7ex
\hangindent=3em \hangafter=1
$\tau_n$ = ($-25.86 + 25.86 i$, $-236.36 + 236.36 i$, $501.4 - 167.13 i$, $334.27 + 51.71 i$, $360.12 - 360.12 i$, $236.36 + 236.36 i$, $360.12 + 360.12 i$, $0. - 385.98 i$, $501.4 + 167.13 i$, $236.36 - 236.36 i$, $-25.86 - 25.86 i$, $1002.8 + 334.27 i$, $-25.86 + 25.86 i$, $-236.36 - 236.36 i$, $501.4 - 167.13 i$, $668.53 + 385.98 i$, $360.12 - 360.12 i$, $-236.36 + 236.36 i$, $360.12 + 360.12 i$, $334.27 - 720.25 i$, $501.4 + 167.13 i$, $236.36 + 236.36 i$, $-25.86 - 25.86 i$, $668.54$, $-25.86 + 25.86 i$, $236.36 - 236.36 i$, $501.4 - 167.13 i$, $334.27 + 720.25 i$, $360.12 - 360.12 i$, $-236.36 - 236.36 i$, $360.12 + 360.12 i$, $668.53 - 385.98 i$, $501.4 + 167.13 i$, $-236.36 + 236.36 i$, $-25.86 - 25.86 i$, $1002.8 - 334.27 i$, $-25.86 + 25.86 i$, $236.36 + 236.36 i$, $501.4 - 167.13 i$, $0. + 385.98 i$, $360.12 - 360.12 i$, $236.36 - 236.36 i$, $360.12 + 360.12 i$, $334.27 - 51.71 i$, $501.4 + 167.13 i$, $-236.36 - 236.36 i$, $-25.86 - 25.86 i$, $1337.07$)

\vskip 0.7ex
\hangindent=3em \hangafter=1
\textit{Intrinsic sign problem}

  \vskip 2ex

\noindent40. $11_{3,1337.}^{48,924}$ \irep{2187}:\ \ 
$d_i$ = ($1.0$,
$6.464$,
$6.464$,
$7.464$,
$7.464$,
$12.928$,
$12.928$,
$12.928$,
$13.928$,
$14.928$,
$14.928$) 

\vskip 0.7ex
\hangindent=3em \hangafter=1
$D^2= 1337.107 = 
672+384\sqrt{3}$

\vskip 0.7ex
\hangindent=3em \hangafter=1
$T = ( 0,
0,
0,
\frac{1}{4},
\frac{1}{4},
\frac{3}{4},
\frac{7}{16},
\frac{15}{16},
0,
\frac{1}{3},
\frac{7}{12} )
$,

\vskip 0.7ex
\hangindent=3em \hangafter=1
$S$ = ($ 1$,
$ 3+2\sqrt{3}$,
$ 3+2\sqrt{3}$,
$ 4+2\sqrt{3}$,
$ 4+2\sqrt{3}$,
$ 6+4\sqrt{3}$,
$ 6+4\sqrt{3}$,
$ 6+4\sqrt{3}$,
$ 7+4\sqrt{3}$,
$ 8+4\sqrt{3}$,
$ 8+4\sqrt{3}$;\ \ 
$ -7+4\zeta^{1}_{12}
-8  \zeta^{-1}_{12}
+8\zeta^{2}_{12}
$,
$ 1-8  \zeta^{1}_{12}
+4\zeta^{-1}_{12}
-8  \zeta^{2}_{12}
$,
$(-6-4\sqrt{3})\mathrm{i}$,
$(6+4\sqrt{3})\mathrm{i}$,
$ -6-4\sqrt{3}$,
$ 6+4\sqrt{3}$,
$ 6+4\sqrt{3}$,
$ -3-2\sqrt{3}$,
$0$,
$0$;\ \ 
$ -7+4\zeta^{1}_{12}
-8  \zeta^{-1}_{12}
+8\zeta^{2}_{12}
$,
$(6+4\sqrt{3})\mathrm{i}$,
$(-6-4\sqrt{3})\mathrm{i}$,
$ -6-4\sqrt{3}$,
$ 6+4\sqrt{3}$,
$ 6+4\sqrt{3}$,
$ -3-2\sqrt{3}$,
$0$,
$0$;\ \ 
$ (-8-4\sqrt{3})\zeta_{6}^{1}$,
$ (8+4\sqrt{3})\zeta_{3}^{1}$,
$0$,
$0$,
$0$,
$ 4+2\sqrt{3}$,
$ 8+4\sqrt{3}$,
$ -8-4\sqrt{3}$;\ \ 
$ (-8-4\sqrt{3})\zeta_{6}^{1}$,
$0$,
$0$,
$0$,
$ 4+2\sqrt{3}$,
$ 8+4\sqrt{3}$,
$ -8-4\sqrt{3}$;\ \ 
$ 12+8\sqrt{3}$,
$0$,
$0$,
$ -6-4\sqrt{3}$,
$0$,
$0$;\ \ 
$ \frac{-24-12\sqrt{3}}{\sqrt{6}}$,
$ \frac{24+12\sqrt{3}}{\sqrt{6}}$,
$ -6-4\sqrt{3}$,
$0$,
$0$;\ \ 
$ \frac{-24-12\sqrt{3}}{\sqrt{6}}$,
$ -6-4\sqrt{3}$,
$0$,
$0$;\ \ 
$ 1$,
$ 8+4\sqrt{3}$,
$ 8+4\sqrt{3}$;\ \ 
$ -8-4\sqrt{3}$,
$ -8-4\sqrt{3}$;\ \ 
$ 8+4\sqrt{3}$)

\vskip 0.7ex
\hangindent=3em \hangafter=1
$\tau_n$ = ($-25.86 + 25.86 i$, $236.36 - 236.36 i$, $501.4 - 167.13 i$, $334.27 + 51.71 i$, $360.12 - 360.12 i$, $-236.36 - 236.36 i$, $360.12 + 360.12 i$, $0. - 385.98 i$, $501.4 + 167.13 i$, $-236.36 + 236.36 i$, $-25.86 - 25.86 i$, $1002.8 + 334.27 i$, $-25.86 + 25.86 i$, $236.36 + 236.36 i$, $501.4 - 167.13 i$, $668.53 + 385.98 i$, $360.12 - 360.12 i$, $236.36 - 236.36 i$, $360.12 + 360.12 i$, $334.27 - 720.25 i$, $501.4 + 167.13 i$, $-236.36 - 236.36 i$, $-25.86 - 25.86 i$, $668.54$, $-25.86 + 25.86 i$, $-236.36 + 236.36 i$, $501.4 - 167.13 i$, $334.27 + 720.25 i$, $360.12 - 360.12 i$, $236.36 + 236.36 i$, $360.12 + 360.12 i$, $668.53 - 385.98 i$, $501.4 + 167.13 i$, $236.36 - 236.36 i$, $-25.86 - 25.86 i$, $1002.8 - 334.27 i$, $-25.86 + 25.86 i$, $-236.36 - 236.36 i$, $501.4 - 167.13 i$, $0. + 385.98 i$, $360.12 - 360.12 i$, $-236.36 + 236.36 i$, $360.12 + 360.12 i$, $334.27 - 51.71 i$, $501.4 + 167.13 i$, $236.36 + 236.36 i$, $-25.86 - 25.86 i$, $1337.07$)

\vskip 0.7ex
\hangindent=3em \hangafter=1
\textit{Intrinsic sign problem}

  \vskip 2ex

\noindent41. $11_{5,1337.}^{48,528}$ \irep{2187}:\ \ 
$d_i$ = ($1.0$,
$6.464$,
$6.464$,
$7.464$,
$7.464$,
$12.928$,
$12.928$,
$12.928$,
$13.928$,
$14.928$,
$14.928$) 

\vskip 0.7ex
\hangindent=3em \hangafter=1
$D^2= 1337.107 = 
672+384\sqrt{3}$

\vskip 0.7ex
\hangindent=3em \hangafter=1
$T = ( 0,
0,
0,
\frac{3}{4},
\frac{3}{4},
\frac{1}{4},
\frac{1}{16},
\frac{9}{16},
0,
\frac{2}{3},
\frac{5}{12} )
$,

\vskip 0.7ex
\hangindent=3em \hangafter=1
$S$ = ($ 1$,
$ 3+2\sqrt{3}$,
$ 3+2\sqrt{3}$,
$ 4+2\sqrt{3}$,
$ 4+2\sqrt{3}$,
$ 6+4\sqrt{3}$,
$ 6+4\sqrt{3}$,
$ 6+4\sqrt{3}$,
$ 7+4\sqrt{3}$,
$ 8+4\sqrt{3}$,
$ 8+4\sqrt{3}$;\ \ 
$ 1-8  \zeta^{1}_{12}
+4\zeta^{-1}_{12}
-8  \zeta^{2}_{12}
$,
$ -7+4\zeta^{1}_{12}
-8  \zeta^{-1}_{12}
+8\zeta^{2}_{12}
$,
$(-6-4\sqrt{3})\mathrm{i}$,
$(6+4\sqrt{3})\mathrm{i}$,
$ -6-4\sqrt{3}$,
$ 6+4\sqrt{3}$,
$ 6+4\sqrt{3}$,
$ -3-2\sqrt{3}$,
$0$,
$0$;\ \ 
$ 1-8  \zeta^{1}_{12}
+4\zeta^{-1}_{12}
-8  \zeta^{2}_{12}
$,
$(6+4\sqrt{3})\mathrm{i}$,
$(-6-4\sqrt{3})\mathrm{i}$,
$ -6-4\sqrt{3}$,
$ 6+4\sqrt{3}$,
$ 6+4\sqrt{3}$,
$ -3-2\sqrt{3}$,
$0$,
$0$;\ \ 
$ (8+4\sqrt{3})\zeta_{3}^{1}$,
$ (-8-4\sqrt{3})\zeta_{6}^{1}$,
$0$,
$0$,
$0$,
$ 4+2\sqrt{3}$,
$ 8+4\sqrt{3}$,
$ -8-4\sqrt{3}$;\ \ 
$ (8+4\sqrt{3})\zeta_{3}^{1}$,
$0$,
$0$,
$0$,
$ 4+2\sqrt{3}$,
$ 8+4\sqrt{3}$,
$ -8-4\sqrt{3}$;\ \ 
$ 12+8\sqrt{3}$,
$0$,
$0$,
$ -6-4\sqrt{3}$,
$0$,
$0$;\ \ 
$ \frac{-24-12\sqrt{3}}{\sqrt{6}}$,
$ \frac{24+12\sqrt{3}}{\sqrt{6}}$,
$ -6-4\sqrt{3}$,
$0$,
$0$;\ \ 
$ \frac{-24-12\sqrt{3}}{\sqrt{6}}$,
$ -6-4\sqrt{3}$,
$0$,
$0$;\ \ 
$ 1$,
$ 8+4\sqrt{3}$,
$ 8+4\sqrt{3}$;\ \ 
$ -8-4\sqrt{3}$,
$ -8-4\sqrt{3}$;\ \ 
$ 8+4\sqrt{3}$)

\vskip 0.7ex
\hangindent=3em \hangafter=1
$\tau_n$ = ($-25.86 - 25.86 i$, $236.36 + 236.36 i$, $501.4 + 167.13 i$, $334.27 - 51.71 i$, $360.12 + 360.12 i$, $-236.36 + 236.36 i$, $360.12 - 360.12 i$, $0. + 385.98 i$, $501.4 - 167.13 i$, $-236.36 - 236.36 i$, $-25.86 + 25.86 i$, $1002.8 - 334.27 i$, $-25.86 - 25.86 i$, $236.36 - 236.36 i$, $501.4 + 167.13 i$, $668.53 - 385.98 i$, $360.12 + 360.12 i$, $236.36 + 236.36 i$, $360.12 - 360.12 i$, $334.27 + 720.25 i$, $501.4 - 167.13 i$, $-236.36 + 236.36 i$, $-25.86 + 25.86 i$, $668.54$, $-25.86 - 25.86 i$, $-236.36 - 236.36 i$, $501.4 + 167.13 i$, $334.27 - 720.25 i$, $360.12 + 360.12 i$, $236.36 - 236.36 i$, $360.12 - 360.12 i$, $668.53 + 385.98 i$, $501.4 - 167.13 i$, $236.36 + 236.36 i$, $-25.86 + 25.86 i$, $1002.8 + 334.27 i$, $-25.86 - 25.86 i$, $-236.36 + 236.36 i$, $501.4 + 167.13 i$, $0. - 385.98 i$, $360.12 + 360.12 i$, $-236.36 - 236.36 i$, $360.12 - 360.12 i$, $334.27 + 51.71 i$, $501.4 - 167.13 i$, $236.36 - 236.36 i$, $-25.86 + 25.86 i$, $1337.07$)

\vskip 0.7ex
\hangindent=3em \hangafter=1
\textit{Intrinsic sign problem}

  \vskip 2ex

\noindent42. $11_{5,1337.}^{48,372}$ \irep{2187}:\ \ 
$d_i$ = ($1.0$,
$6.464$,
$6.464$,
$7.464$,
$7.464$,
$12.928$,
$12.928$,
$12.928$,
$13.928$,
$14.928$,
$14.928$) 

\vskip 0.7ex
\hangindent=3em \hangafter=1
$D^2= 1337.107 = 
672+384\sqrt{3}$

\vskip 0.7ex
\hangindent=3em \hangafter=1
$T = ( 0,
0,
0,
\frac{3}{4},
\frac{3}{4},
\frac{1}{4},
\frac{5}{16},
\frac{13}{16},
0,
\frac{2}{3},
\frac{5}{12} )
$,

\vskip 0.7ex
\hangindent=3em \hangafter=1
$S$ = ($ 1$,
$ 3+2\sqrt{3}$,
$ 3+2\sqrt{3}$,
$ 4+2\sqrt{3}$,
$ 4+2\sqrt{3}$,
$ 6+4\sqrt{3}$,
$ 6+4\sqrt{3}$,
$ 6+4\sqrt{3}$,
$ 7+4\sqrt{3}$,
$ 8+4\sqrt{3}$,
$ 8+4\sqrt{3}$;\ \ 
$ 1-8  \zeta^{1}_{12}
+4\zeta^{-1}_{12}
-8  \zeta^{2}_{12}
$,
$ -7+4\zeta^{1}_{12}
-8  \zeta^{-1}_{12}
+8\zeta^{2}_{12}
$,
$(-6-4\sqrt{3})\mathrm{i}$,
$(6+4\sqrt{3})\mathrm{i}$,
$ -6-4\sqrt{3}$,
$ 6+4\sqrt{3}$,
$ 6+4\sqrt{3}$,
$ -3-2\sqrt{3}$,
$0$,
$0$;\ \ 
$ 1-8  \zeta^{1}_{12}
+4\zeta^{-1}_{12}
-8  \zeta^{2}_{12}
$,
$(6+4\sqrt{3})\mathrm{i}$,
$(-6-4\sqrt{3})\mathrm{i}$,
$ -6-4\sqrt{3}$,
$ 6+4\sqrt{3}$,
$ 6+4\sqrt{3}$,
$ -3-2\sqrt{3}$,
$0$,
$0$;\ \ 
$ (8+4\sqrt{3})\zeta_{3}^{1}$,
$ (-8-4\sqrt{3})\zeta_{6}^{1}$,
$0$,
$0$,
$0$,
$ 4+2\sqrt{3}$,
$ 8+4\sqrt{3}$,
$ -8-4\sqrt{3}$;\ \ 
$ (8+4\sqrt{3})\zeta_{3}^{1}$,
$0$,
$0$,
$0$,
$ 4+2\sqrt{3}$,
$ 8+4\sqrt{3}$,
$ -8-4\sqrt{3}$;\ \ 
$ 12+8\sqrt{3}$,
$0$,
$0$,
$ -6-4\sqrt{3}$,
$0$,
$0$;\ \ 
$ \frac{24+12\sqrt{3}}{\sqrt{6}}$,
$ \frac{-24-12\sqrt{3}}{\sqrt{6}}$,
$ -6-4\sqrt{3}$,
$0$,
$0$;\ \ 
$ \frac{24+12\sqrt{3}}{\sqrt{6}}$,
$ -6-4\sqrt{3}$,
$0$,
$0$;\ \ 
$ 1$,
$ 8+4\sqrt{3}$,
$ 8+4\sqrt{3}$;\ \ 
$ -8-4\sqrt{3}$,
$ -8-4\sqrt{3}$;\ \ 
$ 8+4\sqrt{3}$)

\vskip 0.7ex
\hangindent=3em \hangafter=1
$\tau_n$ = ($-25.86 - 25.86 i$, $-236.36 - 236.36 i$, $501.4 + 167.13 i$, $334.27 - 51.71 i$, $360.12 + 360.12 i$, $236.36 - 236.36 i$, $360.12 - 360.12 i$, $0. + 385.98 i$, $501.4 - 167.13 i$, $236.36 + 236.36 i$, $-25.86 + 25.86 i$, $1002.8 - 334.27 i$, $-25.86 - 25.86 i$, $-236.36 + 236.36 i$, $501.4 + 167.13 i$, $668.53 - 385.98 i$, $360.12 + 360.12 i$, $-236.36 - 236.36 i$, $360.12 - 360.12 i$, $334.27 + 720.25 i$, $501.4 - 167.13 i$, $236.36 - 236.36 i$, $-25.86 + 25.86 i$, $668.54$, $-25.86 - 25.86 i$, $236.36 + 236.36 i$, $501.4 + 167.13 i$, $334.27 - 720.25 i$, $360.12 + 360.12 i$, $-236.36 + 236.36 i$, $360.12 - 360.12 i$, $668.53 + 385.98 i$, $501.4 - 167.13 i$, $-236.36 - 236.36 i$, $-25.86 + 25.86 i$, $1002.8 + 334.27 i$, $-25.86 - 25.86 i$, $236.36 - 236.36 i$, $501.4 + 167.13 i$, $0. - 385.98 i$, $360.12 + 360.12 i$, $236.36 + 236.36 i$, $360.12 - 360.12 i$, $334.27 + 51.71 i$, $501.4 - 167.13 i$, $-236.36 + 236.36 i$, $-25.86 + 25.86 i$, $1337.07$)

\vskip 0.7ex
\hangindent=3em \hangafter=1
\textit{Intrinsic sign problem}

  \vskip 2ex

\noindent43. $11_{\frac{32}{5},1964.}^{35,581}$ \irep{2077}:\ \ 
$d_i$ = ($1.0$,
$8.807$,
$8.807$,
$8.807$,
$11.632$,
$13.250$,
$14.250$,
$14.250$,
$14.250$,
$19.822$,
$20.440$) 

\vskip 0.7ex
\hangindent=3em \hangafter=1
$D^2= 1964.590 = 
910-280  c^{1}_{35}
+280c^{2}_{35}
+280c^{3}_{35}
+175c^{4}_{35}
+280c^{5}_{35}
-105  c^{6}_{35}
+490c^{7}_{35}
-280  c^{8}_{35}
+175c^{9}_{35}
+280c^{10}_{35}
$

\vskip 0.7ex
\hangindent=3em \hangafter=1
$T = ( 0,
\frac{2}{35},
\frac{22}{35},
\frac{32}{35},
\frac{1}{5},
0,
\frac{3}{7},
\frac{5}{7},
\frac{6}{7},
\frac{3}{5},
\frac{1}{5} )
$,

\vskip 0.7ex
\hangindent=3em \hangafter=1
$S$ = ($ 1$,
$ 4-c^{1}_{35}
+c^{2}_{35}
+c^{3}_{35}
+c^{4}_{35}
+c^{5}_{35}
+2c^{7}_{35}
-c^{8}_{35}
+c^{9}_{35}
+c^{10}_{35}
$,
$ 4-c^{1}_{35}
+c^{2}_{35}
+c^{3}_{35}
+c^{4}_{35}
+c^{5}_{35}
+2c^{7}_{35}
-c^{8}_{35}
+c^{9}_{35}
+c^{10}_{35}
$,
$ 4-c^{1}_{35}
+c^{2}_{35}
+c^{3}_{35}
+c^{4}_{35}
+c^{5}_{35}
+2c^{7}_{35}
-c^{8}_{35}
+c^{9}_{35}
+c^{10}_{35}
$,
$ 5-2  c^{1}_{35}
+2c^{2}_{35}
+2c^{3}_{35}
+c^{4}_{35}
+2c^{5}_{35}
-c^{6}_{35}
+3c^{7}_{35}
-2  c^{8}_{35}
+c^{9}_{35}
+2c^{10}_{35}
$,
$ 6-2  c^{1}_{35}
+2c^{2}_{35}
+2c^{3}_{35}
+c^{4}_{35}
+2c^{5}_{35}
-c^{6}_{35}
+4c^{7}_{35}
-2  c^{8}_{35}
+c^{9}_{35}
+2c^{10}_{35}
$,
$ 7-2  c^{1}_{35}
+2c^{2}_{35}
+2c^{3}_{35}
+c^{4}_{35}
+2c^{5}_{35}
-c^{6}_{35}
+4c^{7}_{35}
-2  c^{8}_{35}
+c^{9}_{35}
+2c^{10}_{35}
$,
$ 7-2  c^{1}_{35}
+2c^{2}_{35}
+2c^{3}_{35}
+c^{4}_{35}
+2c^{5}_{35}
-c^{6}_{35}
+4c^{7}_{35}
-2  c^{8}_{35}
+c^{9}_{35}
+2c^{10}_{35}
$,
$ 7-2  c^{1}_{35}
+2c^{2}_{35}
+2c^{3}_{35}
+c^{4}_{35}
+2c^{5}_{35}
-c^{6}_{35}
+4c^{7}_{35}
-2  c^{8}_{35}
+c^{9}_{35}
+2c^{10}_{35}
$,
$ 9-3  c^{1}_{35}
+3c^{2}_{35}
+3c^{3}_{35}
+2c^{4}_{35}
+3c^{5}_{35}
-c^{6}_{35}
+4c^{7}_{35}
-3  c^{8}_{35}
+2c^{9}_{35}
+3c^{10}_{35}
$,
$ 9-3  c^{1}_{35}
+3c^{2}_{35}
+3c^{3}_{35}
+2c^{4}_{35}
+3c^{5}_{35}
-c^{6}_{35}
+5c^{7}_{35}
-3  c^{8}_{35}
+2c^{9}_{35}
+3c^{10}_{35}
$;\ \ 
$ 1-c^{1}_{35}
+4c^{2}_{35}
-2  c^{3}_{35}
+c^{4}_{35}
+2c^{5}_{35}
-c^{6}_{35}
+c^{7}_{35}
+2c^{9}_{35}
-3  c^{10}_{35}
+2c^{11}_{35}
$,
$ 1+5c^{1}_{35}
+4c^{3}_{35}
+3c^{4}_{35}
+c^{5}_{35}
+4c^{6}_{35}
+2c^{8}_{35}
+3c^{10}_{35}
+c^{11}_{35}
$,
$ 5-6  c^{1}_{35}
-2  c^{2}_{35}
-3  c^{4}_{35}
-c^{5}_{35}
-4  c^{6}_{35}
+3c^{7}_{35}
-4  c^{8}_{35}
-c^{9}_{35}
+2c^{10}_{35}
-3  c^{11}_{35}
$,
$ 7-2  c^{1}_{35}
+2c^{2}_{35}
+2c^{3}_{35}
+c^{4}_{35}
+2c^{5}_{35}
-c^{6}_{35}
+4c^{7}_{35}
-2  c^{8}_{35}
+c^{9}_{35}
+2c^{10}_{35}
$,
$ 4-c^{1}_{35}
+c^{2}_{35}
+c^{3}_{35}
+c^{4}_{35}
+c^{5}_{35}
+2c^{7}_{35}
-c^{8}_{35}
+c^{9}_{35}
+c^{10}_{35}
$,
$ -1-3  c^{1}_{35}
-2  c^{3}_{35}
-2  c^{4}_{35}
-c^{5}_{35}
-2  c^{6}_{35}
-c^{8}_{35}
-2  c^{10}_{35}
$,
$ -3+4c^{1}_{35}
+c^{2}_{35}
+2c^{4}_{35}
+c^{5}_{35}
+2c^{6}_{35}
-2  c^{7}_{35}
+2c^{8}_{35}
+c^{9}_{35}
-c^{10}_{35}
+2c^{11}_{35}
$,
$ -2  c^{2}_{35}
+c^{3}_{35}
-c^{4}_{35}
-c^{5}_{35}
-2  c^{9}_{35}
+2c^{10}_{35}
-2  c^{11}_{35}
$,
$0$,
$ -7+2c^{1}_{35}
-2  c^{2}_{35}
-2  c^{3}_{35}
-c^{4}_{35}
-2  c^{5}_{35}
+c^{6}_{35}
-4  c^{7}_{35}
+2c^{8}_{35}
-c^{9}_{35}
-2  c^{10}_{35}
$;\ \ 
$ 5-6  c^{1}_{35}
-2  c^{2}_{35}
-3  c^{4}_{35}
-c^{5}_{35}
-4  c^{6}_{35}
+3c^{7}_{35}
-4  c^{8}_{35}
-c^{9}_{35}
+2c^{10}_{35}
-3  c^{11}_{35}
$,
$ 1-c^{1}_{35}
+4c^{2}_{35}
-2  c^{3}_{35}
+c^{4}_{35}
+2c^{5}_{35}
-c^{6}_{35}
+c^{7}_{35}
+2c^{9}_{35}
-3  c^{10}_{35}
+2c^{11}_{35}
$,
$ 7-2  c^{1}_{35}
+2c^{2}_{35}
+2c^{3}_{35}
+c^{4}_{35}
+2c^{5}_{35}
-c^{6}_{35}
+4c^{7}_{35}
-2  c^{8}_{35}
+c^{9}_{35}
+2c^{10}_{35}
$,
$ 4-c^{1}_{35}
+c^{2}_{35}
+c^{3}_{35}
+c^{4}_{35}
+c^{5}_{35}
+2c^{7}_{35}
-c^{8}_{35}
+c^{9}_{35}
+c^{10}_{35}
$,
$ -3+4c^{1}_{35}
+c^{2}_{35}
+2c^{4}_{35}
+c^{5}_{35}
+2c^{6}_{35}
-2  c^{7}_{35}
+2c^{8}_{35}
+c^{9}_{35}
-c^{10}_{35}
+2c^{11}_{35}
$,
$ -2  c^{2}_{35}
+c^{3}_{35}
-c^{4}_{35}
-c^{5}_{35}
-2  c^{9}_{35}
+2c^{10}_{35}
-2  c^{11}_{35}
$,
$ -1-3  c^{1}_{35}
-2  c^{3}_{35}
-2  c^{4}_{35}
-c^{5}_{35}
-2  c^{6}_{35}
-c^{8}_{35}
-2  c^{10}_{35}
$,
$0$,
$ -7+2c^{1}_{35}
-2  c^{2}_{35}
-2  c^{3}_{35}
-c^{4}_{35}
-2  c^{5}_{35}
+c^{6}_{35}
-4  c^{7}_{35}
+2c^{8}_{35}
-c^{9}_{35}
-2  c^{10}_{35}
$;\ \ 
$ 1+5c^{1}_{35}
+4c^{3}_{35}
+3c^{4}_{35}
+c^{5}_{35}
+4c^{6}_{35}
+2c^{8}_{35}
+3c^{10}_{35}
+c^{11}_{35}
$,
$ 7-2  c^{1}_{35}
+2c^{2}_{35}
+2c^{3}_{35}
+c^{4}_{35}
+2c^{5}_{35}
-c^{6}_{35}
+4c^{7}_{35}
-2  c^{8}_{35}
+c^{9}_{35}
+2c^{10}_{35}
$,
$ 4-c^{1}_{35}
+c^{2}_{35}
+c^{3}_{35}
+c^{4}_{35}
+c^{5}_{35}
+2c^{7}_{35}
-c^{8}_{35}
+c^{9}_{35}
+c^{10}_{35}
$,
$ -2  c^{2}_{35}
+c^{3}_{35}
-c^{4}_{35}
-c^{5}_{35}
-2  c^{9}_{35}
+2c^{10}_{35}
-2  c^{11}_{35}
$,
$ -1-3  c^{1}_{35}
-2  c^{3}_{35}
-2  c^{4}_{35}
-c^{5}_{35}
-2  c^{6}_{35}
-c^{8}_{35}
-2  c^{10}_{35}
$,
$ -3+4c^{1}_{35}
+c^{2}_{35}
+2c^{4}_{35}
+c^{5}_{35}
+2c^{6}_{35}
-2  c^{7}_{35}
+2c^{8}_{35}
+c^{9}_{35}
-c^{10}_{35}
+2c^{11}_{35}
$,
$0$,
$ -7+2c^{1}_{35}
-2  c^{2}_{35}
-2  c^{3}_{35}
-c^{4}_{35}
-2  c^{5}_{35}
+c^{6}_{35}
-4  c^{7}_{35}
+2c^{8}_{35}
-c^{9}_{35}
-2  c^{10}_{35}
$;\ \ 
$ -6+2c^{1}_{35}
-2  c^{2}_{35}
-2  c^{3}_{35}
-c^{4}_{35}
-2  c^{5}_{35}
+c^{6}_{35}
-4  c^{7}_{35}
+2c^{8}_{35}
-c^{9}_{35}
-2  c^{10}_{35}
$,
$ -9+3c^{1}_{35}
-3  c^{2}_{35}
-3  c^{3}_{35}
-2  c^{4}_{35}
-3  c^{5}_{35}
+c^{6}_{35}
-5  c^{7}_{35}
+3c^{8}_{35}
-2  c^{9}_{35}
-3  c^{10}_{35}
$,
$ -4+c^{1}_{35}
-c^{2}_{35}
-c^{3}_{35}
-c^{4}_{35}
-c^{5}_{35}
-2  c^{7}_{35}
+c^{8}_{35}
-c^{9}_{35}
-c^{10}_{35}
$,
$ -4+c^{1}_{35}
-c^{2}_{35}
-c^{3}_{35}
-c^{4}_{35}
-c^{5}_{35}
-2  c^{7}_{35}
+c^{8}_{35}
-c^{9}_{35}
-c^{10}_{35}
$,
$ -4+c^{1}_{35}
-c^{2}_{35}
-c^{3}_{35}
-c^{4}_{35}
-c^{5}_{35}
-2  c^{7}_{35}
+c^{8}_{35}
-c^{9}_{35}
-c^{10}_{35}
$,
$ 9-3  c^{1}_{35}
+3c^{2}_{35}
+3c^{3}_{35}
+2c^{4}_{35}
+3c^{5}_{35}
-c^{6}_{35}
+4c^{7}_{35}
-3  c^{8}_{35}
+2c^{9}_{35}
+3c^{10}_{35}
$,
$ 1$;\ \ 
$ 1$,
$ 7-2  c^{1}_{35}
+2c^{2}_{35}
+2c^{3}_{35}
+c^{4}_{35}
+2c^{5}_{35}
-c^{6}_{35}
+4c^{7}_{35}
-2  c^{8}_{35}
+c^{9}_{35}
+2c^{10}_{35}
$,
$ 7-2  c^{1}_{35}
+2c^{2}_{35}
+2c^{3}_{35}
+c^{4}_{35}
+2c^{5}_{35}
-c^{6}_{35}
+4c^{7}_{35}
-2  c^{8}_{35}
+c^{9}_{35}
+2c^{10}_{35}
$,
$ 7-2  c^{1}_{35}
+2c^{2}_{35}
+2c^{3}_{35}
+c^{4}_{35}
+2c^{5}_{35}
-c^{6}_{35}
+4c^{7}_{35}
-2  c^{8}_{35}
+c^{9}_{35}
+2c^{10}_{35}
$,
$ -9+3c^{1}_{35}
-3  c^{2}_{35}
-3  c^{3}_{35}
-2  c^{4}_{35}
-3  c^{5}_{35}
+c^{6}_{35}
-4  c^{7}_{35}
+3c^{8}_{35}
-2  c^{9}_{35}
-3  c^{10}_{35}
$,
$ -5+2c^{1}_{35}
-2  c^{2}_{35}
-2  c^{3}_{35}
-c^{4}_{35}
-2  c^{5}_{35}
+c^{6}_{35}
-3  c^{7}_{35}
+2c^{8}_{35}
-c^{9}_{35}
-2  c^{10}_{35}
$;\ \ 
$ -5+6c^{1}_{35}
+2c^{2}_{35}
+3c^{4}_{35}
+c^{5}_{35}
+4c^{6}_{35}
-3  c^{7}_{35}
+4c^{8}_{35}
+c^{9}_{35}
-2  c^{10}_{35}
+3c^{11}_{35}
$,
$ -1+c^{1}_{35}
-4  c^{2}_{35}
+2c^{3}_{35}
-c^{4}_{35}
-2  c^{5}_{35}
+c^{6}_{35}
-c^{7}_{35}
-2  c^{9}_{35}
+3c^{10}_{35}
-2  c^{11}_{35}
$,
$ -1-5  c^{1}_{35}
-4  c^{3}_{35}
-3  c^{4}_{35}
-c^{5}_{35}
-4  c^{6}_{35}
-2  c^{8}_{35}
-3  c^{10}_{35}
-c^{11}_{35}
$,
$0$,
$ 4-c^{1}_{35}
+c^{2}_{35}
+c^{3}_{35}
+c^{4}_{35}
+c^{5}_{35}
+2c^{7}_{35}
-c^{8}_{35}
+c^{9}_{35}
+c^{10}_{35}
$;\ \ 
$ -1-5  c^{1}_{35}
-4  c^{3}_{35}
-3  c^{4}_{35}
-c^{5}_{35}
-4  c^{6}_{35}
-2  c^{8}_{35}
-3  c^{10}_{35}
-c^{11}_{35}
$,
$ -5+6c^{1}_{35}
+2c^{2}_{35}
+3c^{4}_{35}
+c^{5}_{35}
+4c^{6}_{35}
-3  c^{7}_{35}
+4c^{8}_{35}
+c^{9}_{35}
-2  c^{10}_{35}
+3c^{11}_{35}
$,
$0$,
$ 4-c^{1}_{35}
+c^{2}_{35}
+c^{3}_{35}
+c^{4}_{35}
+c^{5}_{35}
+2c^{7}_{35}
-c^{8}_{35}
+c^{9}_{35}
+c^{10}_{35}
$;\ \ 
$ -1+c^{1}_{35}
-4  c^{2}_{35}
+2c^{3}_{35}
-c^{4}_{35}
-2  c^{5}_{35}
+c^{6}_{35}
-c^{7}_{35}
-2  c^{9}_{35}
+3c^{10}_{35}
-2  c^{11}_{35}
$,
$0$,
$ 4-c^{1}_{35}
+c^{2}_{35}
+c^{3}_{35}
+c^{4}_{35}
+c^{5}_{35}
+2c^{7}_{35}
-c^{8}_{35}
+c^{9}_{35}
+c^{10}_{35}
$;\ \ 
$ -9+3c^{1}_{35}
-3  c^{2}_{35}
-3  c^{3}_{35}
-2  c^{4}_{35}
-3  c^{5}_{35}
+c^{6}_{35}
-4  c^{7}_{35}
+3c^{8}_{35}
-2  c^{9}_{35}
-3  c^{10}_{35}
$,
$ 9-3  c^{1}_{35}
+3c^{2}_{35}
+3c^{3}_{35}
+2c^{4}_{35}
+3c^{5}_{35}
-c^{6}_{35}
+4c^{7}_{35}
-3  c^{8}_{35}
+2c^{9}_{35}
+3c^{10}_{35}
$;\ \ 
$ -6+2c^{1}_{35}
-2  c^{2}_{35}
-2  c^{3}_{35}
-c^{4}_{35}
-2  c^{5}_{35}
+c^{6}_{35}
-4  c^{7}_{35}
+2c^{8}_{35}
-c^{9}_{35}
-2  c^{10}_{35}
$)

\vskip 0.7ex
\hangindent=3em \hangafter=1
$\tau_n$ = ($13.68 - 42.14 i$, $-159.33 + 490.37 i$, $-159.33 - 490.37 i$, $-181.49 - 558.53 i$, $982.26 + 371.23 i$, $-181.49 + 558.53 i$, $271.45 + 835.56 i$, $-279.95 - 861.6 i$, $-181.49 - 558.53 i$, $982.26 + 371.23 i$, $13.68 - 42.14 i$, $-279.95 + 861.6 i$, $-159.33 - 490.37 i$, $710.7 - 516.38 i$, $982.26 - 371.23 i$, $13.68 - 42.14 i$, $-279.95 + 861.6 i$, $-279.95 - 861.6 i$, $13.68 + 42.14 i$, $982.26 + 371.23 i$, $710.7 + 516.38 i$, $-159.33 + 490.37 i$, $-279.95 - 861.6 i$, $13.68 + 42.14 i$, $982.26 - 371.23 i$, $-181.49 + 558.53 i$, $-279.95 + 861.6 i$, $271.45 - 835.56 i$, $-181.49 - 558.53 i$, $982.26 - 371.23 i$, $-181.49 + 558.53 i$, $-159.33 + 490.37 i$, $-159.33 - 490.37 i$, $13.68 + 42.14 i$, $1964.45$)

\vskip 0.7ex
\hangindent=3em \hangafter=1
\textit{Intrinsic sign problem}

  \vskip 2ex

\noindent44. $11_{\frac{8}{5},1964.}^{35,508}$ \irep{2077}:\ \ 
$d_i$ = ($1.0$,
$8.807$,
$8.807$,
$8.807$,
$11.632$,
$13.250$,
$14.250$,
$14.250$,
$14.250$,
$19.822$,
$20.440$) 

\vskip 0.7ex
\hangindent=3em \hangafter=1
$D^2= 1964.590 = 
910-280  c^{1}_{35}
+280c^{2}_{35}
+280c^{3}_{35}
+175c^{4}_{35}
+280c^{5}_{35}
-105  c^{6}_{35}
+490c^{7}_{35}
-280  c^{8}_{35}
+175c^{9}_{35}
+280c^{10}_{35}
$

\vskip 0.7ex
\hangindent=3em \hangafter=1
$T = ( 0,
\frac{3}{35},
\frac{13}{35},
\frac{33}{35},
\frac{4}{5},
0,
\frac{1}{7},
\frac{2}{7},
\frac{4}{7},
\frac{2}{5},
\frac{4}{5} )
$,

\vskip 0.7ex
\hangindent=3em \hangafter=1
$S$ = ($ 1$,
$ 4-c^{1}_{35}
+c^{2}_{35}
+c^{3}_{35}
+c^{4}_{35}
+c^{5}_{35}
+2c^{7}_{35}
-c^{8}_{35}
+c^{9}_{35}
+c^{10}_{35}
$,
$ 4-c^{1}_{35}
+c^{2}_{35}
+c^{3}_{35}
+c^{4}_{35}
+c^{5}_{35}
+2c^{7}_{35}
-c^{8}_{35}
+c^{9}_{35}
+c^{10}_{35}
$,
$ 4-c^{1}_{35}
+c^{2}_{35}
+c^{3}_{35}
+c^{4}_{35}
+c^{5}_{35}
+2c^{7}_{35}
-c^{8}_{35}
+c^{9}_{35}
+c^{10}_{35}
$,
$ 5-2  c^{1}_{35}
+2c^{2}_{35}
+2c^{3}_{35}
+c^{4}_{35}
+2c^{5}_{35}
-c^{6}_{35}
+3c^{7}_{35}
-2  c^{8}_{35}
+c^{9}_{35}
+2c^{10}_{35}
$,
$ 6-2  c^{1}_{35}
+2c^{2}_{35}
+2c^{3}_{35}
+c^{4}_{35}
+2c^{5}_{35}
-c^{6}_{35}
+4c^{7}_{35}
-2  c^{8}_{35}
+c^{9}_{35}
+2c^{10}_{35}
$,
$ 7-2  c^{1}_{35}
+2c^{2}_{35}
+2c^{3}_{35}
+c^{4}_{35}
+2c^{5}_{35}
-c^{6}_{35}
+4c^{7}_{35}
-2  c^{8}_{35}
+c^{9}_{35}
+2c^{10}_{35}
$,
$ 7-2  c^{1}_{35}
+2c^{2}_{35}
+2c^{3}_{35}
+c^{4}_{35}
+2c^{5}_{35}
-c^{6}_{35}
+4c^{7}_{35}
-2  c^{8}_{35}
+c^{9}_{35}
+2c^{10}_{35}
$,
$ 7-2  c^{1}_{35}
+2c^{2}_{35}
+2c^{3}_{35}
+c^{4}_{35}
+2c^{5}_{35}
-c^{6}_{35}
+4c^{7}_{35}
-2  c^{8}_{35}
+c^{9}_{35}
+2c^{10}_{35}
$,
$ 9-3  c^{1}_{35}
+3c^{2}_{35}
+3c^{3}_{35}
+2c^{4}_{35}
+3c^{5}_{35}
-c^{6}_{35}
+4c^{7}_{35}
-3  c^{8}_{35}
+2c^{9}_{35}
+3c^{10}_{35}
$,
$ 9-3  c^{1}_{35}
+3c^{2}_{35}
+3c^{3}_{35}
+2c^{4}_{35}
+3c^{5}_{35}
-c^{6}_{35}
+5c^{7}_{35}
-3  c^{8}_{35}
+2c^{9}_{35}
+3c^{10}_{35}
$;\ \ 
$ 1+5c^{1}_{35}
+4c^{3}_{35}
+3c^{4}_{35}
+c^{5}_{35}
+4c^{6}_{35}
+2c^{8}_{35}
+3c^{10}_{35}
+c^{11}_{35}
$,
$ 1-c^{1}_{35}
+4c^{2}_{35}
-2  c^{3}_{35}
+c^{4}_{35}
+2c^{5}_{35}
-c^{6}_{35}
+c^{7}_{35}
+2c^{9}_{35}
-3  c^{10}_{35}
+2c^{11}_{35}
$,
$ 5-6  c^{1}_{35}
-2  c^{2}_{35}
-3  c^{4}_{35}
-c^{5}_{35}
-4  c^{6}_{35}
+3c^{7}_{35}
-4  c^{8}_{35}
-c^{9}_{35}
+2c^{10}_{35}
-3  c^{11}_{35}
$,
$ 7-2  c^{1}_{35}
+2c^{2}_{35}
+2c^{3}_{35}
+c^{4}_{35}
+2c^{5}_{35}
-c^{6}_{35}
+4c^{7}_{35}
-2  c^{8}_{35}
+c^{9}_{35}
+2c^{10}_{35}
$,
$ 4-c^{1}_{35}
+c^{2}_{35}
+c^{3}_{35}
+c^{4}_{35}
+c^{5}_{35}
+2c^{7}_{35}
-c^{8}_{35}
+c^{9}_{35}
+c^{10}_{35}
$,
$ -3+4c^{1}_{35}
+c^{2}_{35}
+2c^{4}_{35}
+c^{5}_{35}
+2c^{6}_{35}
-2  c^{7}_{35}
+2c^{8}_{35}
+c^{9}_{35}
-c^{10}_{35}
+2c^{11}_{35}
$,
$ -1-3  c^{1}_{35}
-2  c^{3}_{35}
-2  c^{4}_{35}
-c^{5}_{35}
-2  c^{6}_{35}
-c^{8}_{35}
-2  c^{10}_{35}
$,
$ -2  c^{2}_{35}
+c^{3}_{35}
-c^{4}_{35}
-c^{5}_{35}
-2  c^{9}_{35}
+2c^{10}_{35}
-2  c^{11}_{35}
$,
$0$,
$ -7+2c^{1}_{35}
-2  c^{2}_{35}
-2  c^{3}_{35}
-c^{4}_{35}
-2  c^{5}_{35}
+c^{6}_{35}
-4  c^{7}_{35}
+2c^{8}_{35}
-c^{9}_{35}
-2  c^{10}_{35}
$;\ \ 
$ 5-6  c^{1}_{35}
-2  c^{2}_{35}
-3  c^{4}_{35}
-c^{5}_{35}
-4  c^{6}_{35}
+3c^{7}_{35}
-4  c^{8}_{35}
-c^{9}_{35}
+2c^{10}_{35}
-3  c^{11}_{35}
$,
$ 1+5c^{1}_{35}
+4c^{3}_{35}
+3c^{4}_{35}
+c^{5}_{35}
+4c^{6}_{35}
+2c^{8}_{35}
+3c^{10}_{35}
+c^{11}_{35}
$,
$ 7-2  c^{1}_{35}
+2c^{2}_{35}
+2c^{3}_{35}
+c^{4}_{35}
+2c^{5}_{35}
-c^{6}_{35}
+4c^{7}_{35}
-2  c^{8}_{35}
+c^{9}_{35}
+2c^{10}_{35}
$,
$ 4-c^{1}_{35}
+c^{2}_{35}
+c^{3}_{35}
+c^{4}_{35}
+c^{5}_{35}
+2c^{7}_{35}
-c^{8}_{35}
+c^{9}_{35}
+c^{10}_{35}
$,
$ -1-3  c^{1}_{35}
-2  c^{3}_{35}
-2  c^{4}_{35}
-c^{5}_{35}
-2  c^{6}_{35}
-c^{8}_{35}
-2  c^{10}_{35}
$,
$ -2  c^{2}_{35}
+c^{3}_{35}
-c^{4}_{35}
-c^{5}_{35}
-2  c^{9}_{35}
+2c^{10}_{35}
-2  c^{11}_{35}
$,
$ -3+4c^{1}_{35}
+c^{2}_{35}
+2c^{4}_{35}
+c^{5}_{35}
+2c^{6}_{35}
-2  c^{7}_{35}
+2c^{8}_{35}
+c^{9}_{35}
-c^{10}_{35}
+2c^{11}_{35}
$,
$0$,
$ -7+2c^{1}_{35}
-2  c^{2}_{35}
-2  c^{3}_{35}
-c^{4}_{35}
-2  c^{5}_{35}
+c^{6}_{35}
-4  c^{7}_{35}
+2c^{8}_{35}
-c^{9}_{35}
-2  c^{10}_{35}
$;\ \ 
$ 1-c^{1}_{35}
+4c^{2}_{35}
-2  c^{3}_{35}
+c^{4}_{35}
+2c^{5}_{35}
-c^{6}_{35}
+c^{7}_{35}
+2c^{9}_{35}
-3  c^{10}_{35}
+2c^{11}_{35}
$,
$ 7-2  c^{1}_{35}
+2c^{2}_{35}
+2c^{3}_{35}
+c^{4}_{35}
+2c^{5}_{35}
-c^{6}_{35}
+4c^{7}_{35}
-2  c^{8}_{35}
+c^{9}_{35}
+2c^{10}_{35}
$,
$ 4-c^{1}_{35}
+c^{2}_{35}
+c^{3}_{35}
+c^{4}_{35}
+c^{5}_{35}
+2c^{7}_{35}
-c^{8}_{35}
+c^{9}_{35}
+c^{10}_{35}
$,
$ -2  c^{2}_{35}
+c^{3}_{35}
-c^{4}_{35}
-c^{5}_{35}
-2  c^{9}_{35}
+2c^{10}_{35}
-2  c^{11}_{35}
$,
$ -3+4c^{1}_{35}
+c^{2}_{35}
+2c^{4}_{35}
+c^{5}_{35}
+2c^{6}_{35}
-2  c^{7}_{35}
+2c^{8}_{35}
+c^{9}_{35}
-c^{10}_{35}
+2c^{11}_{35}
$,
$ -1-3  c^{1}_{35}
-2  c^{3}_{35}
-2  c^{4}_{35}
-c^{5}_{35}
-2  c^{6}_{35}
-c^{8}_{35}
-2  c^{10}_{35}
$,
$0$,
$ -7+2c^{1}_{35}
-2  c^{2}_{35}
-2  c^{3}_{35}
-c^{4}_{35}
-2  c^{5}_{35}
+c^{6}_{35}
-4  c^{7}_{35}
+2c^{8}_{35}
-c^{9}_{35}
-2  c^{10}_{35}
$;\ \ 
$ -6+2c^{1}_{35}
-2  c^{2}_{35}
-2  c^{3}_{35}
-c^{4}_{35}
-2  c^{5}_{35}
+c^{6}_{35}
-4  c^{7}_{35}
+2c^{8}_{35}
-c^{9}_{35}
-2  c^{10}_{35}
$,
$ -9+3c^{1}_{35}
-3  c^{2}_{35}
-3  c^{3}_{35}
-2  c^{4}_{35}
-3  c^{5}_{35}
+c^{6}_{35}
-5  c^{7}_{35}
+3c^{8}_{35}
-2  c^{9}_{35}
-3  c^{10}_{35}
$,
$ -4+c^{1}_{35}
-c^{2}_{35}
-c^{3}_{35}
-c^{4}_{35}
-c^{5}_{35}
-2  c^{7}_{35}
+c^{8}_{35}
-c^{9}_{35}
-c^{10}_{35}
$,
$ -4+c^{1}_{35}
-c^{2}_{35}
-c^{3}_{35}
-c^{4}_{35}
-c^{5}_{35}
-2  c^{7}_{35}
+c^{8}_{35}
-c^{9}_{35}
-c^{10}_{35}
$,
$ -4+c^{1}_{35}
-c^{2}_{35}
-c^{3}_{35}
-c^{4}_{35}
-c^{5}_{35}
-2  c^{7}_{35}
+c^{8}_{35}
-c^{9}_{35}
-c^{10}_{35}
$,
$ 9-3  c^{1}_{35}
+3c^{2}_{35}
+3c^{3}_{35}
+2c^{4}_{35}
+3c^{5}_{35}
-c^{6}_{35}
+4c^{7}_{35}
-3  c^{8}_{35}
+2c^{9}_{35}
+3c^{10}_{35}
$,
$ 1$;\ \ 
$ 1$,
$ 7-2  c^{1}_{35}
+2c^{2}_{35}
+2c^{3}_{35}
+c^{4}_{35}
+2c^{5}_{35}
-c^{6}_{35}
+4c^{7}_{35}
-2  c^{8}_{35}
+c^{9}_{35}
+2c^{10}_{35}
$,
$ 7-2  c^{1}_{35}
+2c^{2}_{35}
+2c^{3}_{35}
+c^{4}_{35}
+2c^{5}_{35}
-c^{6}_{35}
+4c^{7}_{35}
-2  c^{8}_{35}
+c^{9}_{35}
+2c^{10}_{35}
$,
$ 7-2  c^{1}_{35}
+2c^{2}_{35}
+2c^{3}_{35}
+c^{4}_{35}
+2c^{5}_{35}
-c^{6}_{35}
+4c^{7}_{35}
-2  c^{8}_{35}
+c^{9}_{35}
+2c^{10}_{35}
$,
$ -9+3c^{1}_{35}
-3  c^{2}_{35}
-3  c^{3}_{35}
-2  c^{4}_{35}
-3  c^{5}_{35}
+c^{6}_{35}
-4  c^{7}_{35}
+3c^{8}_{35}
-2  c^{9}_{35}
-3  c^{10}_{35}
$,
$ -5+2c^{1}_{35}
-2  c^{2}_{35}
-2  c^{3}_{35}
-c^{4}_{35}
-2  c^{5}_{35}
+c^{6}_{35}
-3  c^{7}_{35}
+2c^{8}_{35}
-c^{9}_{35}
-2  c^{10}_{35}
$;\ \ 
$ -1+c^{1}_{35}
-4  c^{2}_{35}
+2c^{3}_{35}
-c^{4}_{35}
-2  c^{5}_{35}
+c^{6}_{35}
-c^{7}_{35}
-2  c^{9}_{35}
+3c^{10}_{35}
-2  c^{11}_{35}
$,
$ -5+6c^{1}_{35}
+2c^{2}_{35}
+3c^{4}_{35}
+c^{5}_{35}
+4c^{6}_{35}
-3  c^{7}_{35}
+4c^{8}_{35}
+c^{9}_{35}
-2  c^{10}_{35}
+3c^{11}_{35}
$,
$ -1-5  c^{1}_{35}
-4  c^{3}_{35}
-3  c^{4}_{35}
-c^{5}_{35}
-4  c^{6}_{35}
-2  c^{8}_{35}
-3  c^{10}_{35}
-c^{11}_{35}
$,
$0$,
$ 4-c^{1}_{35}
+c^{2}_{35}
+c^{3}_{35}
+c^{4}_{35}
+c^{5}_{35}
+2c^{7}_{35}
-c^{8}_{35}
+c^{9}_{35}
+c^{10}_{35}
$;\ \ 
$ -1-5  c^{1}_{35}
-4  c^{3}_{35}
-3  c^{4}_{35}
-c^{5}_{35}
-4  c^{6}_{35}
-2  c^{8}_{35}
-3  c^{10}_{35}
-c^{11}_{35}
$,
$ -1+c^{1}_{35}
-4  c^{2}_{35}
+2c^{3}_{35}
-c^{4}_{35}
-2  c^{5}_{35}
+c^{6}_{35}
-c^{7}_{35}
-2  c^{9}_{35}
+3c^{10}_{35}
-2  c^{11}_{35}
$,
$0$,
$ 4-c^{1}_{35}
+c^{2}_{35}
+c^{3}_{35}
+c^{4}_{35}
+c^{5}_{35}
+2c^{7}_{35}
-c^{8}_{35}
+c^{9}_{35}
+c^{10}_{35}
$;\ \ 
$ -5+6c^{1}_{35}
+2c^{2}_{35}
+3c^{4}_{35}
+c^{5}_{35}
+4c^{6}_{35}
-3  c^{7}_{35}
+4c^{8}_{35}
+c^{9}_{35}
-2  c^{10}_{35}
+3c^{11}_{35}
$,
$0$,
$ 4-c^{1}_{35}
+c^{2}_{35}
+c^{3}_{35}
+c^{4}_{35}
+c^{5}_{35}
+2c^{7}_{35}
-c^{8}_{35}
+c^{9}_{35}
+c^{10}_{35}
$;\ \ 
$ -9+3c^{1}_{35}
-3  c^{2}_{35}
-3  c^{3}_{35}
-2  c^{4}_{35}
-3  c^{5}_{35}
+c^{6}_{35}
-4  c^{7}_{35}
+3c^{8}_{35}
-2  c^{9}_{35}
-3  c^{10}_{35}
$,
$ 9-3  c^{1}_{35}
+3c^{2}_{35}
+3c^{3}_{35}
+2c^{4}_{35}
+3c^{5}_{35}
-c^{6}_{35}
+4c^{7}_{35}
-3  c^{8}_{35}
+2c^{9}_{35}
+3c^{10}_{35}
$;\ \ 
$ -6+2c^{1}_{35}
-2  c^{2}_{35}
-2  c^{3}_{35}
-c^{4}_{35}
-2  c^{5}_{35}
+c^{6}_{35}
-4  c^{7}_{35}
+2c^{8}_{35}
-c^{9}_{35}
-2  c^{10}_{35}
$)

\vskip 0.7ex
\hangindent=3em \hangafter=1
$\tau_n$ = ($13.68 + 42.14 i$, $-159.33 - 490.37 i$, $-159.33 + 490.37 i$, $-181.49 + 558.53 i$, $982.26 - 371.23 i$, $-181.49 - 558.53 i$, $271.45 - 835.56 i$, $-279.95 + 861.6 i$, $-181.49 + 558.53 i$, $982.26 - 371.23 i$, $13.68 + 42.14 i$, $-279.95 - 861.6 i$, $-159.33 + 490.37 i$, $710.7 + 516.38 i$, $982.26 + 371.23 i$, $13.68 + 42.14 i$, $-279.95 - 861.6 i$, $-279.95 + 861.6 i$, $13.68 - 42.14 i$, $982.26 - 371.23 i$, $710.7 - 516.38 i$, $-159.33 - 490.37 i$, $-279.95 + 861.6 i$, $13.68 - 42.14 i$, $982.26 + 371.23 i$, $-181.49 - 558.53 i$, $-279.95 - 861.6 i$, $271.45 + 835.56 i$, $-181.49 + 558.53 i$, $982.26 + 371.23 i$, $-181.49 - 558.53 i$, $-159.33 - 490.37 i$, $-159.33 + 490.37 i$, $13.68 - 42.14 i$, $1964.45$)

\vskip 0.7ex
\hangindent=3em \hangafter=1
\textit{Intrinsic sign problem}

  \vskip 2ex 

%% file: modular_data/SsL12U_.tex
\noindent1. $12_{2,12.}^{6,694}$ \irep{0}:\ \ 
$d_i$ = ($1.0$,
$1.0$,
$1.0$,
$1.0$,
$1.0$,
$1.0$,
$1.0$,
$1.0$,
$1.0$,
$1.0$,
$1.0$,
$1.0$) 

\vskip 0.7ex
\hangindent=3em \hangafter=1
$D^2= 12.0 = 
12$

\vskip 0.7ex
\hangindent=3em \hangafter=1
$T = ( 0,
0,
0,
\frac{1}{2},
\frac{1}{3},
\frac{1}{3},
\frac{1}{3},
\frac{1}{3},
\frac{1}{3},
\frac{1}{3},
\frac{5}{6},
\frac{5}{6} )
$,

\vskip 0.7ex
\hangindent=3em \hangafter=1
$S$ = ($ 1$,
$ 1$,
$ 1$,
$ 1$,
$ 1$,
$ 1$,
$ 1$,
$ 1$,
$ 1$,
$ 1$,
$ 1$,
$ 1$;\ \ 
$ 1$,
$ -1$,
$ -1$,
$ 1$,
$ -1$,
$ 1$,
$ -1$,
$ 1$,
$ 1$,
$ -1$,
$ -1$;\ \ 
$ 1$,
$ -1$,
$ -1$,
$ 1$,
$ -1$,
$ 1$,
$ 1$,
$ 1$,
$ -1$,
$ -1$;\ \ 
$ 1$,
$ -1$,
$ -1$,
$ -1$,
$ -1$,
$ 1$,
$ 1$,
$ 1$,
$ 1$;\ \ 
$ \zeta_{3}^{1}$,
$ -\zeta_{3}^{1}$,
$ -\zeta_{6}^{1}$,
$ \zeta_{6}^{1}$,
$ -\zeta_{6}^{1}$,
$ \zeta_{3}^{1}$,
$ \zeta_{6}^{1}$,
$ -\zeta_{3}^{1}$;\ \ 
$ \zeta_{3}^{1}$,
$ \zeta_{6}^{1}$,
$ -\zeta_{6}^{1}$,
$ -\zeta_{6}^{1}$,
$ \zeta_{3}^{1}$,
$ \zeta_{6}^{1}$,
$ -\zeta_{3}^{1}$;\ \ 
$ \zeta_{3}^{1}$,
$ -\zeta_{3}^{1}$,
$ \zeta_{3}^{1}$,
$ -\zeta_{6}^{1}$,
$ -\zeta_{3}^{1}$,
$ \zeta_{6}^{1}$;\ \ 
$ \zeta_{3}^{1}$,
$ \zeta_{3}^{1}$,
$ -\zeta_{6}^{1}$,
$ -\zeta_{3}^{1}$,
$ \zeta_{6}^{1}$;\ \ 
$ \zeta_{3}^{1}$,
$ -\zeta_{6}^{1}$,
$ \zeta_{3}^{1}$,
$ -\zeta_{6}^{1}$;\ \ 
$ \zeta_{3}^{1}$,
$ -\zeta_{6}^{1}$,
$ \zeta_{3}^{1}$;\ \ 
$ \zeta_{3}^{1}$,
$ -\zeta_{6}^{1}$;\ \ 
$ \zeta_{3}^{1}$)

Factors = $3_{2,3.}^{3,527}\boxtimes 4_{0,4.}^{2,750}$

\vskip 0.7ex
\hangindent=3em \hangafter=1
$\tau_n$ = ($0. + 3.46 i$, $0. - 6.93 i$, $6.$, $0. + 6.93 i$, $0. - 3.46 i$, $12.$)

\vskip 0.7ex
\hangindent=3em \hangafter=1
\textit{Intrinsic sign problem}

  \vskip 2ex

\noindent2. $12_{6,12.}^{6,277}$ \irep{0}:\ \ 
$d_i$ = ($1.0$,
$1.0$,
$1.0$,
$1.0$,
$1.0$,
$1.0$,
$1.0$,
$1.0$,
$1.0$,
$1.0$,
$1.0$,
$1.0$) 

\vskip 0.7ex
\hangindent=3em \hangafter=1
$D^2= 12.0 = 
12$

\vskip 0.7ex
\hangindent=3em \hangafter=1
$T = ( 0,
0,
0,
\frac{1}{2},
\frac{2}{3},
\frac{2}{3},
\frac{2}{3},
\frac{2}{3},
\frac{2}{3},
\frac{2}{3},
\frac{1}{6},
\frac{1}{6} )
$,

\vskip 0.7ex
\hangindent=3em \hangafter=1
$S$ = ($ 1$,
$ 1$,
$ 1$,
$ 1$,
$ 1$,
$ 1$,
$ 1$,
$ 1$,
$ 1$,
$ 1$,
$ 1$,
$ 1$;\ \ 
$ 1$,
$ -1$,
$ -1$,
$ 1$,
$ -1$,
$ 1$,
$ -1$,
$ 1$,
$ 1$,
$ -1$,
$ -1$;\ \ 
$ 1$,
$ -1$,
$ -1$,
$ 1$,
$ -1$,
$ 1$,
$ 1$,
$ 1$,
$ -1$,
$ -1$;\ \ 
$ 1$,
$ -1$,
$ -1$,
$ -1$,
$ -1$,
$ 1$,
$ 1$,
$ 1$,
$ 1$;\ \ 
$ -\zeta_{6}^{1}$,
$ \zeta_{6}^{1}$,
$ \zeta_{3}^{1}$,
$ -\zeta_{3}^{1}$,
$ \zeta_{3}^{1}$,
$ -\zeta_{6}^{1}$,
$ -\zeta_{3}^{1}$,
$ \zeta_{6}^{1}$;\ \ 
$ -\zeta_{6}^{1}$,
$ -\zeta_{3}^{1}$,
$ \zeta_{3}^{1}$,
$ \zeta_{3}^{1}$,
$ -\zeta_{6}^{1}$,
$ -\zeta_{3}^{1}$,
$ \zeta_{6}^{1}$;\ \ 
$ -\zeta_{6}^{1}$,
$ \zeta_{6}^{1}$,
$ -\zeta_{6}^{1}$,
$ \zeta_{3}^{1}$,
$ \zeta_{6}^{1}$,
$ -\zeta_{3}^{1}$;\ \ 
$ -\zeta_{6}^{1}$,
$ -\zeta_{6}^{1}$,
$ \zeta_{3}^{1}$,
$ \zeta_{6}^{1}$,
$ -\zeta_{3}^{1}$;\ \ 
$ -\zeta_{6}^{1}$,
$ \zeta_{3}^{1}$,
$ -\zeta_{6}^{1}$,
$ \zeta_{3}^{1}$;\ \ 
$ -\zeta_{6}^{1}$,
$ \zeta_{3}^{1}$,
$ -\zeta_{6}^{1}$;\ \ 
$ -\zeta_{6}^{1}$,
$ \zeta_{3}^{1}$;\ \ 
$ -\zeta_{6}^{1}$)

Factors = $3_{6,3.}^{3,138}\boxtimes 4_{0,4.}^{2,750}$

\vskip 0.7ex
\hangindent=3em \hangafter=1
$\tau_n$ = ($0. - 3.46 i$, $0. + 6.93 i$, $6.$, $0. - 6.93 i$, $0. + 3.46 i$, $12.$)

\vskip 0.7ex
\hangindent=3em \hangafter=1
\textit{Intrinsic sign problem}

  \vskip 2ex

\noindent3. $12_{6,12.}^{6,213}$ \irep{0}:\ \ 
$d_i$ = ($1.0$,
$1.0$,
$1.0$,
$1.0$,
$1.0$,
$1.0$,
$1.0$,
$1.0$,
$1.0$,
$1.0$,
$1.0$,
$1.0$) 

\vskip 0.7ex
\hangindent=3em \hangafter=1
$D^2= 12.0 = 
12$

\vskip 0.7ex
\hangindent=3em \hangafter=1
$T = ( 0,
\frac{1}{2},
\frac{1}{2},
\frac{1}{2},
\frac{1}{3},
\frac{1}{3},
\frac{5}{6},
\frac{5}{6},
\frac{5}{6},
\frac{5}{6},
\frac{5}{6},
\frac{5}{6} )
$,

\vskip 0.7ex
\hangindent=3em \hangafter=1
$S$ = ($ 1$,
$ 1$,
$ 1$,
$ 1$,
$ 1$,
$ 1$,
$ 1$,
$ 1$,
$ 1$,
$ 1$,
$ 1$,
$ 1$;\ \ 
$ 1$,
$ -1$,
$ -1$,
$ 1$,
$ 1$,
$ 1$,
$ -1$,
$ -1$,
$ 1$,
$ -1$,
$ -1$;\ \ 
$ 1$,
$ -1$,
$ 1$,
$ 1$,
$ -1$,
$ 1$,
$ -1$,
$ -1$,
$ 1$,
$ -1$;\ \ 
$ 1$,
$ 1$,
$ 1$,
$ -1$,
$ -1$,
$ 1$,
$ -1$,
$ -1$,
$ 1$;\ \ 
$ \zeta_{3}^{1}$,
$ -\zeta_{6}^{1}$,
$ \zeta_{3}^{1}$,
$ \zeta_{3}^{1}$,
$ \zeta_{3}^{1}$,
$ -\zeta_{6}^{1}$,
$ -\zeta_{6}^{1}$,
$ -\zeta_{6}^{1}$;\ \ 
$ \zeta_{3}^{1}$,
$ -\zeta_{6}^{1}$,
$ -\zeta_{6}^{1}$,
$ -\zeta_{6}^{1}$,
$ \zeta_{3}^{1}$,
$ \zeta_{3}^{1}$,
$ \zeta_{3}^{1}$;\ \ 
$ \zeta_{3}^{1}$,
$ -\zeta_{3}^{1}$,
$ -\zeta_{3}^{1}$,
$ -\zeta_{6}^{1}$,
$ \zeta_{6}^{1}$,
$ \zeta_{6}^{1}$;\ \ 
$ \zeta_{3}^{1}$,
$ -\zeta_{3}^{1}$,
$ \zeta_{6}^{1}$,
$ -\zeta_{6}^{1}$,
$ \zeta_{6}^{1}$;\ \ 
$ \zeta_{3}^{1}$,
$ \zeta_{6}^{1}$,
$ \zeta_{6}^{1}$,
$ -\zeta_{6}^{1}$;\ \ 
$ \zeta_{3}^{1}$,
$ -\zeta_{3}^{1}$,
$ -\zeta_{3}^{1}$;\ \ 
$ \zeta_{3}^{1}$,
$ -\zeta_{3}^{1}$;\ \ 
$ \zeta_{3}^{1}$)

Factors = $3_{2,3.}^{3,527}\boxtimes 4_{4,4.}^{2,250}$

\vskip 0.7ex
\hangindent=3em \hangafter=1
$\tau_n$ = ($0. - 3.46 i$, $0. - 6.93 i$, $-6.$, $0. + 6.93 i$, $0. + 3.46 i$, $12.$)

\vskip 0.7ex
\hangindent=3em \hangafter=1
\textit{Intrinsic sign problem}

  \vskip 2ex

\noindent4. $12_{2,12.}^{6,119}$ \irep{0}:\ \ 
$d_i$ = ($1.0$,
$1.0$,
$1.0$,
$1.0$,
$1.0$,
$1.0$,
$1.0$,
$1.0$,
$1.0$,
$1.0$,
$1.0$,
$1.0$) 

\vskip 0.7ex
\hangindent=3em \hangafter=1
$D^2= 12.0 = 
12$

\vskip 0.7ex
\hangindent=3em \hangafter=1
$T = ( 0,
\frac{1}{2},
\frac{1}{2},
\frac{1}{2},
\frac{2}{3},
\frac{2}{3},
\frac{1}{6},
\frac{1}{6},
\frac{1}{6},
\frac{1}{6},
\frac{1}{6},
\frac{1}{6} )
$,

\vskip 0.7ex
\hangindent=3em \hangafter=1
$S$ = ($ 1$,
$ 1$,
$ 1$,
$ 1$,
$ 1$,
$ 1$,
$ 1$,
$ 1$,
$ 1$,
$ 1$,
$ 1$,
$ 1$;\ \ 
$ 1$,
$ -1$,
$ -1$,
$ 1$,
$ 1$,
$ 1$,
$ -1$,
$ -1$,
$ 1$,
$ -1$,
$ -1$;\ \ 
$ 1$,
$ -1$,
$ 1$,
$ 1$,
$ -1$,
$ 1$,
$ -1$,
$ -1$,
$ 1$,
$ -1$;\ \ 
$ 1$,
$ 1$,
$ 1$,
$ -1$,
$ -1$,
$ 1$,
$ -1$,
$ -1$,
$ 1$;\ \ 
$ -\zeta_{6}^{1}$,
$ \zeta_{3}^{1}$,
$ -\zeta_{6}^{1}$,
$ -\zeta_{6}^{1}$,
$ -\zeta_{6}^{1}$,
$ \zeta_{3}^{1}$,
$ \zeta_{3}^{1}$,
$ \zeta_{3}^{1}$;\ \ 
$ -\zeta_{6}^{1}$,
$ \zeta_{3}^{1}$,
$ \zeta_{3}^{1}$,
$ \zeta_{3}^{1}$,
$ -\zeta_{6}^{1}$,
$ -\zeta_{6}^{1}$,
$ -\zeta_{6}^{1}$;\ \ 
$ -\zeta_{6}^{1}$,
$ \zeta_{6}^{1}$,
$ \zeta_{6}^{1}$,
$ \zeta_{3}^{1}$,
$ -\zeta_{3}^{1}$,
$ -\zeta_{3}^{1}$;\ \ 
$ -\zeta_{6}^{1}$,
$ \zeta_{6}^{1}$,
$ -\zeta_{3}^{1}$,
$ \zeta_{3}^{1}$,
$ -\zeta_{3}^{1}$;\ \ 
$ -\zeta_{6}^{1}$,
$ -\zeta_{3}^{1}$,
$ -\zeta_{3}^{1}$,
$ \zeta_{3}^{1}$;\ \ 
$ -\zeta_{6}^{1}$,
$ \zeta_{6}^{1}$,
$ \zeta_{6}^{1}$;\ \ 
$ -\zeta_{6}^{1}$,
$ \zeta_{6}^{1}$;\ \ 
$ -\zeta_{6}^{1}$)

Factors = $3_{6,3.}^{3,138}\boxtimes 4_{4,4.}^{2,250}$

\vskip 0.7ex
\hangindent=3em \hangafter=1
$\tau_n$ = ($0. + 3.46 i$, $0. + 6.93 i$, $-6.$, $0. - 6.93 i$, $0. - 3.46 i$, $12.$)

\vskip 0.7ex
\hangindent=3em \hangafter=1
\textit{Intrinsic sign problem}

  \vskip 2ex

\noindent5. $12_{2,12.}^{12,123}$ \irep{0}:\ \ 
$d_i$ = ($1.0$,
$1.0$,
$1.0$,
$1.0$,
$1.0$,
$1.0$,
$1.0$,
$1.0$,
$1.0$,
$1.0$,
$1.0$,
$1.0$) 

\vskip 0.7ex
\hangindent=3em \hangafter=1
$D^2= 12.0 = 
12$

\vskip 0.7ex
\hangindent=3em \hangafter=1
$T = ( 0,
0,
\frac{1}{3},
\frac{1}{3},
\frac{1}{3},
\frac{1}{3},
\frac{1}{4},
\frac{3}{4},
\frac{1}{12},
\frac{1}{12},
\frac{7}{12},
\frac{7}{12} )
$,

\vskip 0.7ex
\hangindent=3em \hangafter=1
$S$ = ($ 1$,
$ 1$,
$ 1$,
$ 1$,
$ 1$,
$ 1$,
$ 1$,
$ 1$,
$ 1$,
$ 1$,
$ 1$,
$ 1$;\ \ 
$ 1$,
$ 1$,
$ 1$,
$ 1$,
$ 1$,
$ -1$,
$ -1$,
$ -1$,
$ -1$,
$ -1$,
$ -1$;\ \ 
$ \zeta_{3}^{1}$,
$ -\zeta_{6}^{1}$,
$ -\zeta_{6}^{1}$,
$ \zeta_{3}^{1}$,
$ -1$,
$ -1$,
$ -\zeta_{3}^{1}$,
$ \zeta_{6}^{1}$,
$ \zeta_{6}^{1}$,
$ -\zeta_{3}^{1}$;\ \ 
$ \zeta_{3}^{1}$,
$ \zeta_{3}^{1}$,
$ -\zeta_{6}^{1}$,
$ -1$,
$ -1$,
$ \zeta_{6}^{1}$,
$ -\zeta_{3}^{1}$,
$ -\zeta_{3}^{1}$,
$ \zeta_{6}^{1}$;\ \ 
$ \zeta_{3}^{1}$,
$ -\zeta_{6}^{1}$,
$ 1$,
$ 1$,
$ -\zeta_{6}^{1}$,
$ \zeta_{3}^{1}$,
$ \zeta_{3}^{1}$,
$ -\zeta_{6}^{1}$;\ \ 
$ \zeta_{3}^{1}$,
$ 1$,
$ 1$,
$ \zeta_{3}^{1}$,
$ -\zeta_{6}^{1}$,
$ -\zeta_{6}^{1}$,
$ \zeta_{3}^{1}$;\ \ 
$ -1$,
$ 1$,
$ 1$,
$ 1$,
$ -1$,
$ -1$;\ \ 
$ -1$,
$ -1$,
$ -1$,
$ 1$,
$ 1$;\ \ 
$ -\zeta_{3}^{1}$,
$ \zeta_{6}^{1}$,
$ -\zeta_{6}^{1}$,
$ \zeta_{3}^{1}$;\ \ 
$ -\zeta_{3}^{1}$,
$ \zeta_{3}^{1}$,
$ -\zeta_{6}^{1}$;\ \ 
$ -\zeta_{3}^{1}$,
$ \zeta_{6}^{1}$;\ \ 
$ -\zeta_{3}^{1}$)

Factors = $2_{1,2.}^{4,437}\boxtimes 6_{1,6.}^{12,701}$

\vskip 0.7ex
\hangindent=3em \hangafter=1
$\tau_n$ = ($0. + 3.46 i$, $0.$, $6.$, $0. + 6.93 i$, $0. - 3.46 i$, $0.$, $0. + 3.46 i$, $0. - 6.93 i$, $6.$, $0.$, $0. - 3.46 i$, $12.$)

\vskip 0.7ex
\hangindent=3em \hangafter=1
\textit{Intrinsic sign problem}

  \vskip 2ex

\noindent6. $12_{6,12.}^{12,143}$ \irep{0}:\ \ 
$d_i$ = ($1.0$,
$1.0$,
$1.0$,
$1.0$,
$1.0$,
$1.0$,
$1.0$,
$1.0$,
$1.0$,
$1.0$,
$1.0$,
$1.0$) 

\vskip 0.7ex
\hangindent=3em \hangafter=1
$D^2= 12.0 = 
12$

\vskip 0.7ex
\hangindent=3em \hangafter=1
$T = ( 0,
0,
\frac{2}{3},
\frac{2}{3},
\frac{2}{3},
\frac{2}{3},
\frac{1}{4},
\frac{3}{4},
\frac{5}{12},
\frac{5}{12},
\frac{11}{12},
\frac{11}{12} )
$,

\vskip 0.7ex
\hangindent=3em \hangafter=1
$S$ = ($ 1$,
$ 1$,
$ 1$,
$ 1$,
$ 1$,
$ 1$,
$ 1$,
$ 1$,
$ 1$,
$ 1$,
$ 1$,
$ 1$;\ \ 
$ 1$,
$ 1$,
$ 1$,
$ 1$,
$ 1$,
$ -1$,
$ -1$,
$ -1$,
$ -1$,
$ -1$,
$ -1$;\ \ 
$ -\zeta_{6}^{1}$,
$ -\zeta_{6}^{1}$,
$ \zeta_{3}^{1}$,
$ \zeta_{3}^{1}$,
$ -1$,
$ -1$,
$ -\zeta_{3}^{1}$,
$ \zeta_{6}^{1}$,
$ -\zeta_{3}^{1}$,
$ \zeta_{6}^{1}$;\ \ 
$ -\zeta_{6}^{1}$,
$ \zeta_{3}^{1}$,
$ \zeta_{3}^{1}$,
$ 1$,
$ 1$,
$ \zeta_{3}^{1}$,
$ -\zeta_{6}^{1}$,
$ \zeta_{3}^{1}$,
$ -\zeta_{6}^{1}$;\ \ 
$ -\zeta_{6}^{1}$,
$ -\zeta_{6}^{1}$,
$ -1$,
$ -1$,
$ \zeta_{6}^{1}$,
$ -\zeta_{3}^{1}$,
$ \zeta_{6}^{1}$,
$ -\zeta_{3}^{1}$;\ \ 
$ -\zeta_{6}^{1}$,
$ 1$,
$ 1$,
$ -\zeta_{6}^{1}$,
$ \zeta_{3}^{1}$,
$ -\zeta_{6}^{1}$,
$ \zeta_{3}^{1}$;\ \ 
$ -1$,
$ 1$,
$ 1$,
$ 1$,
$ -1$,
$ -1$;\ \ 
$ -1$,
$ -1$,
$ -1$,
$ 1$,
$ 1$;\ \ 
$ \zeta_{6}^{1}$,
$ -\zeta_{3}^{1}$,
$ -\zeta_{6}^{1}$,
$ \zeta_{3}^{1}$;\ \ 
$ \zeta_{6}^{1}$,
$ \zeta_{3}^{1}$,
$ -\zeta_{6}^{1}$;\ \ 
$ \zeta_{6}^{1}$,
$ -\zeta_{3}^{1}$;\ \ 
$ \zeta_{6}^{1}$)

Factors = $2_{1,2.}^{4,437}\boxtimes 6_{5,6.}^{12,298}$

\vskip 0.7ex
\hangindent=3em \hangafter=1
$\tau_n$ = ($0. - 3.46 i$, $0.$, $6.$, $0. - 6.93 i$, $0. + 3.46 i$, $0.$, $0. - 3.46 i$, $0. + 6.93 i$, $6.$, $0.$, $0. + 3.46 i$, $12.$)

\vskip 0.7ex
\hangindent=3em \hangafter=1
\textit{Intrinsic sign problem}

  \vskip 2ex

\noindent7. $12_{4,12.}^{12,347}$ \irep{0}:\ \ 
$d_i$ = ($1.0$,
$1.0$,
$1.0$,
$1.0$,
$1.0$,
$1.0$,
$1.0$,
$1.0$,
$1.0$,
$1.0$,
$1.0$,
$1.0$) 

\vskip 0.7ex
\hangindent=3em \hangafter=1
$D^2= 12.0 = 
12$

\vskip 0.7ex
\hangindent=3em \hangafter=1
$T = ( 0,
\frac{1}{2},
\frac{1}{3},
\frac{1}{3},
\frac{1}{4},
\frac{1}{4},
\frac{5}{6},
\frac{5}{6},
\frac{7}{12},
\frac{7}{12},
\frac{7}{12},
\frac{7}{12} )
$,

\vskip 0.7ex
\hangindent=3em \hangafter=1
$S$ = ($ 1$,
$ 1$,
$ 1$,
$ 1$,
$ 1$,
$ 1$,
$ 1$,
$ 1$,
$ 1$,
$ 1$,
$ 1$,
$ 1$;\ \ 
$ 1$,
$ 1$,
$ 1$,
$ -1$,
$ -1$,
$ 1$,
$ 1$,
$ -1$,
$ -1$,
$ -1$,
$ -1$;\ \ 
$ \zeta_{3}^{1}$,
$ -\zeta_{6}^{1}$,
$ 1$,
$ 1$,
$ -\zeta_{6}^{1}$,
$ \zeta_{3}^{1}$,
$ -\zeta_{6}^{1}$,
$ \zeta_{3}^{1}$,
$ \zeta_{3}^{1}$,
$ -\zeta_{6}^{1}$;\ \ 
$ \zeta_{3}^{1}$,
$ 1$,
$ 1$,
$ \zeta_{3}^{1}$,
$ -\zeta_{6}^{1}$,
$ \zeta_{3}^{1}$,
$ -\zeta_{6}^{1}$,
$ -\zeta_{6}^{1}$,
$ \zeta_{3}^{1}$;\ \ 
$ -1$,
$ 1$,
$ -1$,
$ -1$,
$ -1$,
$ -1$,
$ 1$,
$ 1$;\ \ 
$ -1$,
$ -1$,
$ -1$,
$ 1$,
$ 1$,
$ -1$,
$ -1$;\ \ 
$ \zeta_{3}^{1}$,
$ -\zeta_{6}^{1}$,
$ -\zeta_{3}^{1}$,
$ \zeta_{6}^{1}$,
$ \zeta_{6}^{1}$,
$ -\zeta_{3}^{1}$;\ \ 
$ \zeta_{3}^{1}$,
$ \zeta_{6}^{1}$,
$ -\zeta_{3}^{1}$,
$ -\zeta_{3}^{1}$,
$ \zeta_{6}^{1}$;\ \ 
$ -\zeta_{3}^{1}$,
$ \zeta_{6}^{1}$,
$ -\zeta_{6}^{1}$,
$ \zeta_{3}^{1}$;\ \ 
$ -\zeta_{3}^{1}$,
$ \zeta_{3}^{1}$,
$ -\zeta_{6}^{1}$;\ \ 
$ -\zeta_{3}^{1}$,
$ \zeta_{6}^{1}$;\ \ 
$ -\zeta_{3}^{1}$)

Factors = $2_{1,2.}^{4,437}\boxtimes 6_{3,6.}^{12,534}$

\vskip 0.7ex
\hangindent=3em \hangafter=1
$\tau_n$ = ($-3.46$, $0.$, $0. - 6. i$, $0. + 6.93 i$, $3.46$, $0.$, $3.46$, $0. - 6.93 i$, $0. + 6. i$, $0.$, $-3.46$, $12.$)

\vskip 0.7ex
\hangindent=3em \hangafter=1
\textit{Intrinsic sign problem}

  \vskip 2ex

\noindent8. $12_{0,12.}^{12,138}$ \irep{0}:\ \ 
$d_i$ = ($1.0$,
$1.0$,
$1.0$,
$1.0$,
$1.0$,
$1.0$,
$1.0$,
$1.0$,
$1.0$,
$1.0$,
$1.0$,
$1.0$) 

\vskip 0.7ex
\hangindent=3em \hangafter=1
$D^2= 12.0 = 
12$

\vskip 0.7ex
\hangindent=3em \hangafter=1
$T = ( 0,
\frac{1}{2},
\frac{1}{3},
\frac{1}{3},
\frac{3}{4},
\frac{3}{4},
\frac{5}{6},
\frac{5}{6},
\frac{1}{12},
\frac{1}{12},
\frac{1}{12},
\frac{1}{12} )
$,

\vskip 0.7ex
\hangindent=3em \hangafter=1
$S$ = ($ 1$,
$ 1$,
$ 1$,
$ 1$,
$ 1$,
$ 1$,
$ 1$,
$ 1$,
$ 1$,
$ 1$,
$ 1$,
$ 1$;\ \ 
$ 1$,
$ 1$,
$ 1$,
$ -1$,
$ -1$,
$ 1$,
$ 1$,
$ -1$,
$ -1$,
$ -1$,
$ -1$;\ \ 
$ \zeta_{3}^{1}$,
$ -\zeta_{6}^{1}$,
$ 1$,
$ 1$,
$ -\zeta_{6}^{1}$,
$ \zeta_{3}^{1}$,
$ -\zeta_{6}^{1}$,
$ \zeta_{3}^{1}$,
$ \zeta_{3}^{1}$,
$ -\zeta_{6}^{1}$;\ \ 
$ \zeta_{3}^{1}$,
$ 1$,
$ 1$,
$ \zeta_{3}^{1}$,
$ -\zeta_{6}^{1}$,
$ \zeta_{3}^{1}$,
$ -\zeta_{6}^{1}$,
$ -\zeta_{6}^{1}$,
$ \zeta_{3}^{1}$;\ \ 
$ -1$,
$ 1$,
$ -1$,
$ -1$,
$ -1$,
$ -1$,
$ 1$,
$ 1$;\ \ 
$ -1$,
$ -1$,
$ -1$,
$ 1$,
$ 1$,
$ -1$,
$ -1$;\ \ 
$ \zeta_{3}^{1}$,
$ -\zeta_{6}^{1}$,
$ -\zeta_{3}^{1}$,
$ \zeta_{6}^{1}$,
$ \zeta_{6}^{1}$,
$ -\zeta_{3}^{1}$;\ \ 
$ \zeta_{3}^{1}$,
$ \zeta_{6}^{1}$,
$ -\zeta_{3}^{1}$,
$ -\zeta_{3}^{1}$,
$ \zeta_{6}^{1}$;\ \ 
$ -\zeta_{3}^{1}$,
$ \zeta_{6}^{1}$,
$ -\zeta_{6}^{1}$,
$ \zeta_{3}^{1}$;\ \ 
$ -\zeta_{3}^{1}$,
$ \zeta_{3}^{1}$,
$ -\zeta_{6}^{1}$;\ \ 
$ -\zeta_{3}^{1}$,
$ \zeta_{6}^{1}$;\ \ 
$ -\zeta_{3}^{1}$)

Factors = $2_{7,2.}^{4,625}\boxtimes 6_{1,6.}^{12,701}$

\vskip 0.7ex
\hangindent=3em \hangafter=1
$\tau_n$ = ($3.46$, $0.$, $0. + 6. i$, $0. + 6.93 i$, $-3.46$, $0.$, $-3.46$, $0. - 6.93 i$, $0. - 6. i$, $0.$, $3.46$, $12.$)

\vskip 0.7ex
\hangindent=3em \hangafter=1
\textit{Intrinsic sign problem}

  \vskip 2ex

\noindent9. $12_{0,12.}^{12,168}$ \irep{0}:\ \ 
$d_i$ = ($1.0$,
$1.0$,
$1.0$,
$1.0$,
$1.0$,
$1.0$,
$1.0$,
$1.0$,
$1.0$,
$1.0$,
$1.0$,
$1.0$) 

\vskip 0.7ex
\hangindent=3em \hangafter=1
$D^2= 12.0 = 
12$

\vskip 0.7ex
\hangindent=3em \hangafter=1
$T = ( 0,
\frac{1}{2},
\frac{2}{3},
\frac{2}{3},
\frac{1}{4},
\frac{1}{4},
\frac{1}{6},
\frac{1}{6},
\frac{11}{12},
\frac{11}{12},
\frac{11}{12},
\frac{11}{12} )
$,

\vskip 0.7ex
\hangindent=3em \hangafter=1
$S$ = ($ 1$,
$ 1$,
$ 1$,
$ 1$,
$ 1$,
$ 1$,
$ 1$,
$ 1$,
$ 1$,
$ 1$,
$ 1$,
$ 1$;\ \ 
$ 1$,
$ 1$,
$ 1$,
$ -1$,
$ -1$,
$ 1$,
$ 1$,
$ -1$,
$ -1$,
$ -1$,
$ -1$;\ \ 
$ -\zeta_{6}^{1}$,
$ \zeta_{3}^{1}$,
$ 1$,
$ 1$,
$ -\zeta_{6}^{1}$,
$ \zeta_{3}^{1}$,
$ -\zeta_{6}^{1}$,
$ -\zeta_{6}^{1}$,
$ \zeta_{3}^{1}$,
$ \zeta_{3}^{1}$;\ \ 
$ -\zeta_{6}^{1}$,
$ 1$,
$ 1$,
$ \zeta_{3}^{1}$,
$ -\zeta_{6}^{1}$,
$ \zeta_{3}^{1}$,
$ \zeta_{3}^{1}$,
$ -\zeta_{6}^{1}$,
$ -\zeta_{6}^{1}$;\ \ 
$ -1$,
$ 1$,
$ -1$,
$ -1$,
$ -1$,
$ 1$,
$ -1$,
$ 1$;\ \ 
$ -1$,
$ -1$,
$ -1$,
$ 1$,
$ -1$,
$ 1$,
$ -1$;\ \ 
$ -\zeta_{6}^{1}$,
$ \zeta_{3}^{1}$,
$ \zeta_{6}^{1}$,
$ \zeta_{6}^{1}$,
$ -\zeta_{3}^{1}$,
$ -\zeta_{3}^{1}$;\ \ 
$ -\zeta_{6}^{1}$,
$ -\zeta_{3}^{1}$,
$ -\zeta_{3}^{1}$,
$ \zeta_{6}^{1}$,
$ \zeta_{6}^{1}$;\ \ 
$ \zeta_{6}^{1}$,
$ -\zeta_{6}^{1}$,
$ -\zeta_{3}^{1}$,
$ \zeta_{3}^{1}$;\ \ 
$ \zeta_{6}^{1}$,
$ \zeta_{3}^{1}$,
$ -\zeta_{3}^{1}$;\ \ 
$ \zeta_{6}^{1}$,
$ -\zeta_{6}^{1}$;\ \ 
$ \zeta_{6}^{1}$)

Factors = $2_{1,2.}^{4,437}\boxtimes 6_{7,6.}^{12,113}$

\vskip 0.7ex
\hangindent=3em \hangafter=1
$\tau_n$ = ($3.46$, $0.$, $0. - 6. i$, $0. - 6.93 i$, $-3.46$, $0.$, $-3.46$, $0. + 6.93 i$, $0. + 6. i$, $0.$, $3.46$, $12.$)

\vskip 0.7ex
\hangindent=3em \hangafter=1
\textit{Intrinsic sign problem}

  \vskip 2ex

\noindent10. $12_{4,12.}^{12,138}$ \irep{0}:\ \ 
$d_i$ = ($1.0$,
$1.0$,
$1.0$,
$1.0$,
$1.0$,
$1.0$,
$1.0$,
$1.0$,
$1.0$,
$1.0$,
$1.0$,
$1.0$) 

\vskip 0.7ex
\hangindent=3em \hangafter=1
$D^2= 12.0 = 
12$

\vskip 0.7ex
\hangindent=3em \hangafter=1
$T = ( 0,
\frac{1}{2},
\frac{2}{3},
\frac{2}{3},
\frac{3}{4},
\frac{3}{4},
\frac{1}{6},
\frac{1}{6},
\frac{5}{12},
\frac{5}{12},
\frac{5}{12},
\frac{5}{12} )
$,

\vskip 0.7ex
\hangindent=3em \hangafter=1
$S$ = ($ 1$,
$ 1$,
$ 1$,
$ 1$,
$ 1$,
$ 1$,
$ 1$,
$ 1$,
$ 1$,
$ 1$,
$ 1$,
$ 1$;\ \ 
$ 1$,
$ 1$,
$ 1$,
$ -1$,
$ -1$,
$ 1$,
$ 1$,
$ -1$,
$ -1$,
$ -1$,
$ -1$;\ \ 
$ -\zeta_{6}^{1}$,
$ \zeta_{3}^{1}$,
$ 1$,
$ 1$,
$ -\zeta_{6}^{1}$,
$ \zeta_{3}^{1}$,
$ -\zeta_{6}^{1}$,
$ -\zeta_{6}^{1}$,
$ \zeta_{3}^{1}$,
$ \zeta_{3}^{1}$;\ \ 
$ -\zeta_{6}^{1}$,
$ 1$,
$ 1$,
$ \zeta_{3}^{1}$,
$ -\zeta_{6}^{1}$,
$ \zeta_{3}^{1}$,
$ \zeta_{3}^{1}$,
$ -\zeta_{6}^{1}$,
$ -\zeta_{6}^{1}$;\ \ 
$ -1$,
$ 1$,
$ -1$,
$ -1$,
$ -1$,
$ 1$,
$ -1$,
$ 1$;\ \ 
$ -1$,
$ -1$,
$ -1$,
$ 1$,
$ -1$,
$ 1$,
$ -1$;\ \ 
$ -\zeta_{6}^{1}$,
$ \zeta_{3}^{1}$,
$ \zeta_{6}^{1}$,
$ \zeta_{6}^{1}$,
$ -\zeta_{3}^{1}$,
$ -\zeta_{3}^{1}$;\ \ 
$ -\zeta_{6}^{1}$,
$ -\zeta_{3}^{1}$,
$ -\zeta_{3}^{1}$,
$ \zeta_{6}^{1}$,
$ \zeta_{6}^{1}$;\ \ 
$ \zeta_{6}^{1}$,
$ -\zeta_{6}^{1}$,
$ -\zeta_{3}^{1}$,
$ \zeta_{3}^{1}$;\ \ 
$ \zeta_{6}^{1}$,
$ \zeta_{3}^{1}$,
$ -\zeta_{3}^{1}$;\ \ 
$ \zeta_{6}^{1}$,
$ -\zeta_{6}^{1}$;\ \ 
$ \zeta_{6}^{1}$)

Factors = $2_{7,2.}^{4,625}\boxtimes 6_{5,6.}^{12,298}$

\vskip 0.7ex
\hangindent=3em \hangafter=1
$\tau_n$ = ($-3.46$, $0.$, $0. + 6. i$, $0. - 6.93 i$, $3.46$, $0.$, $3.46$, $0. + 6.93 i$, $0. - 6. i$, $0.$, $-3.46$, $12.$)

\vskip 0.7ex
\hangindent=3em \hangafter=1
\textit{Intrinsic sign problem}

  \vskip 2ex

\noindent11. $12_{3,12.}^{24,267}$ \irep{0}:\ \ 
$d_i$ = ($1.0$,
$1.0$,
$1.0$,
$1.0$,
$1.0$,
$1.0$,
$1.0$,
$1.0$,
$1.0$,
$1.0$,
$1.0$,
$1.0$) 

\vskip 0.7ex
\hangindent=3em \hangafter=1
$D^2= 12.0 = 
12$

\vskip 0.7ex
\hangindent=3em \hangafter=1
$T = ( 0,
\frac{1}{2},
\frac{1}{3},
\frac{1}{3},
\frac{5}{6},
\frac{5}{6},
\frac{1}{8},
\frac{1}{8},
\frac{11}{24},
\frac{11}{24},
\frac{11}{24},
\frac{11}{24} )
$,

\vskip 0.7ex
\hangindent=3em \hangafter=1
$S$ = ($ 1$,
$ 1$,
$ 1$,
$ 1$,
$ 1$,
$ 1$,
$ 1$,
$ 1$,
$ 1$,
$ 1$,
$ 1$,
$ 1$;\ \ 
$ 1$,
$ 1$,
$ 1$,
$ 1$,
$ 1$,
$ -1$,
$ -1$,
$ -1$,
$ -1$,
$ -1$,
$ -1$;\ \ 
$ \zeta_{3}^{1}$,
$ -\zeta_{6}^{1}$,
$ -\zeta_{6}^{1}$,
$ \zeta_{3}^{1}$,
$ 1$,
$ 1$,
$ -\zeta_{6}^{1}$,
$ \zeta_{3}^{1}$,
$ \zeta_{3}^{1}$,
$ -\zeta_{6}^{1}$;\ \ 
$ \zeta_{3}^{1}$,
$ \zeta_{3}^{1}$,
$ -\zeta_{6}^{1}$,
$ 1$,
$ 1$,
$ \zeta_{3}^{1}$,
$ -\zeta_{6}^{1}$,
$ -\zeta_{6}^{1}$,
$ \zeta_{3}^{1}$;\ \ 
$ \zeta_{3}^{1}$,
$ -\zeta_{6}^{1}$,
$ -1$,
$ -1$,
$ -\zeta_{3}^{1}$,
$ \zeta_{6}^{1}$,
$ \zeta_{6}^{1}$,
$ -\zeta_{3}^{1}$;\ \ 
$ \zeta_{3}^{1}$,
$ -1$,
$ -1$,
$ \zeta_{6}^{1}$,
$ -\zeta_{3}^{1}$,
$ -\zeta_{3}^{1}$,
$ \zeta_{6}^{1}$;\ \ 
$-\mathrm{i}$,
$\mathrm{i}$,
$-\mathrm{i}$,
$-\mathrm{i}$,
$\mathrm{i}$,
$\mathrm{i}$;\ \ 
$-\mathrm{i}$,
$\mathrm{i}$,
$\mathrm{i}$,
$-\mathrm{i}$,
$-\mathrm{i}$;\ \ 
$ \zeta_{12}^{1}$,
$ \zeta_{12}^{5}$,
$ -\zeta_{12}^{5}$,
$ -\zeta_{12}^{1}$;\ \ 
$ \zeta_{12}^{1}$,
$ -\zeta_{12}^{1}$,
$ -\zeta_{12}^{5}$;\ \ 
$ \zeta_{12}^{1}$,
$ \zeta_{12}^{5}$;\ \ 
$ \zeta_{12}^{1}$)

Factors = $3_{2,3.}^{3,527}\boxtimes 4_{1,4.}^{8,718}$

\vskip 0.7ex
\hangindent=3em \hangafter=1
$\tau_n$ = ($-2.45 + 2.45 i$, $3.46 - 3.46 i$, $-4.24 + 4.24 i$, $0.$, $-2.45 + 2.45 i$, $6. - 6. i$, $2.45 + 2.45 i$, $0. - 6.93 i$, $4.24 + 4.24 i$, $-3.46 + 3.46 i$, $2.45 + 2.45 i$, $0.$, $2.45 - 2.45 i$, $-3.46 - 3.46 i$, $4.24 - 4.24 i$, $0. + 6.93 i$, $2.45 - 2.45 i$, $6. + 6. i$, $-2.45 - 2.45 i$, $0.$, $-4.24 - 4.24 i$, $3.46 + 3.46 i$, $-2.45 - 2.45 i$, $12.$)

\vskip 0.7ex
\hangindent=3em \hangafter=1
\textit{Intrinsic sign problem}

  \vskip 2ex

\noindent12. $12_{5,12.}^{24,114}$ \irep{0}:\ \ 
$d_i$ = ($1.0$,
$1.0$,
$1.0$,
$1.0$,
$1.0$,
$1.0$,
$1.0$,
$1.0$,
$1.0$,
$1.0$,
$1.0$,
$1.0$) 

\vskip 0.7ex
\hangindent=3em \hangafter=1
$D^2= 12.0 = 
12$

\vskip 0.7ex
\hangindent=3em \hangafter=1
$T = ( 0,
\frac{1}{2},
\frac{1}{3},
\frac{1}{3},
\frac{5}{6},
\frac{5}{6},
\frac{3}{8},
\frac{3}{8},
\frac{17}{24},
\frac{17}{24},
\frac{17}{24},
\frac{17}{24} )
$,

\vskip 0.7ex
\hangindent=3em \hangafter=1
$S$ = ($ 1$,
$ 1$,
$ 1$,
$ 1$,
$ 1$,
$ 1$,
$ 1$,
$ 1$,
$ 1$,
$ 1$,
$ 1$,
$ 1$;\ \ 
$ 1$,
$ 1$,
$ 1$,
$ 1$,
$ 1$,
$ -1$,
$ -1$,
$ -1$,
$ -1$,
$ -1$,
$ -1$;\ \ 
$ \zeta_{3}^{1}$,
$ -\zeta_{6}^{1}$,
$ -\zeta_{6}^{1}$,
$ \zeta_{3}^{1}$,
$ 1$,
$ 1$,
$ -\zeta_{6}^{1}$,
$ \zeta_{3}^{1}$,
$ \zeta_{3}^{1}$,
$ -\zeta_{6}^{1}$;\ \ 
$ \zeta_{3}^{1}$,
$ \zeta_{3}^{1}$,
$ -\zeta_{6}^{1}$,
$ 1$,
$ 1$,
$ \zeta_{3}^{1}$,
$ -\zeta_{6}^{1}$,
$ -\zeta_{6}^{1}$,
$ \zeta_{3}^{1}$;\ \ 
$ \zeta_{3}^{1}$,
$ -\zeta_{6}^{1}$,
$ -1$,
$ -1$,
$ -\zeta_{3}^{1}$,
$ \zeta_{6}^{1}$,
$ \zeta_{6}^{1}$,
$ -\zeta_{3}^{1}$;\ \ 
$ \zeta_{3}^{1}$,
$ -1$,
$ -1$,
$ \zeta_{6}^{1}$,
$ -\zeta_{3}^{1}$,
$ -\zeta_{3}^{1}$,
$ \zeta_{6}^{1}$;\ \ 
$\mathrm{i}$,
$-\mathrm{i}$,
$\mathrm{i}$,
$\mathrm{i}$,
$-\mathrm{i}$,
$-\mathrm{i}$;\ \ 
$\mathrm{i}$,
$-\mathrm{i}$,
$-\mathrm{i}$,
$\mathrm{i}$,
$\mathrm{i}$;\ \ 
$ -\zeta_{12}^{1}$,
$ -\zeta_{12}^{5}$,
$ \zeta_{12}^{5}$,
$ \zeta_{12}^{1}$;\ \ 
$ -\zeta_{12}^{1}$,
$ \zeta_{12}^{1}$,
$ \zeta_{12}^{5}$;\ \ 
$ -\zeta_{12}^{1}$,
$ -\zeta_{12}^{5}$;\ \ 
$ -\zeta_{12}^{1}$)

Factors = $3_{2,3.}^{3,527}\boxtimes 4_{3,4.}^{8,468}$

\vskip 0.7ex
\hangindent=3em \hangafter=1
$\tau_n$ = ($-2.45 - 2.45 i$, $-3.46 - 3.46 i$, $4.24 + 4.24 i$, $0.$, $-2.45 - 2.45 i$, $6. + 6. i$, $2.45 - 2.45 i$, $0. - 6.93 i$, $-4.24 + 4.24 i$, $3.46 + 3.46 i$, $2.45 - 2.45 i$, $0.$, $2.45 + 2.45 i$, $3.46 - 3.46 i$, $-4.24 - 4.24 i$, $0. + 6.93 i$, $2.45 + 2.45 i$, $6. - 6. i$, $-2.45 + 2.45 i$, $0.$, $4.24 - 4.24 i$, $-3.46 + 3.46 i$, $-2.45 + 2.45 i$, $12.$)

\vskip 0.7ex
\hangindent=3em \hangafter=1
\textit{Intrinsic sign problem}

  \vskip 2ex

\noindent13. $12_{7,12.}^{24,331}$ \irep{0}:\ \ 
$d_i$ = ($1.0$,
$1.0$,
$1.0$,
$1.0$,
$1.0$,
$1.0$,
$1.0$,
$1.0$,
$1.0$,
$1.0$,
$1.0$,
$1.0$) 

\vskip 0.7ex
\hangindent=3em \hangafter=1
$D^2= 12.0 = 
12$

\vskip 0.7ex
\hangindent=3em \hangafter=1
$T = ( 0,
\frac{1}{2},
\frac{1}{3},
\frac{1}{3},
\frac{5}{6},
\frac{5}{6},
\frac{5}{8},
\frac{5}{8},
\frac{23}{24},
\frac{23}{24},
\frac{23}{24},
\frac{23}{24} )
$,

\vskip 0.7ex
\hangindent=3em \hangafter=1
$S$ = ($ 1$,
$ 1$,
$ 1$,
$ 1$,
$ 1$,
$ 1$,
$ 1$,
$ 1$,
$ 1$,
$ 1$,
$ 1$,
$ 1$;\ \ 
$ 1$,
$ 1$,
$ 1$,
$ 1$,
$ 1$,
$ -1$,
$ -1$,
$ -1$,
$ -1$,
$ -1$,
$ -1$;\ \ 
$ \zeta_{3}^{1}$,
$ -\zeta_{6}^{1}$,
$ -\zeta_{6}^{1}$,
$ \zeta_{3}^{1}$,
$ 1$,
$ 1$,
$ -\zeta_{6}^{1}$,
$ \zeta_{3}^{1}$,
$ \zeta_{3}^{1}$,
$ -\zeta_{6}^{1}$;\ \ 
$ \zeta_{3}^{1}$,
$ \zeta_{3}^{1}$,
$ -\zeta_{6}^{1}$,
$ 1$,
$ 1$,
$ \zeta_{3}^{1}$,
$ -\zeta_{6}^{1}$,
$ -\zeta_{6}^{1}$,
$ \zeta_{3}^{1}$;\ \ 
$ \zeta_{3}^{1}$,
$ -\zeta_{6}^{1}$,
$ -1$,
$ -1$,
$ -\zeta_{3}^{1}$,
$ \zeta_{6}^{1}$,
$ \zeta_{6}^{1}$,
$ -\zeta_{3}^{1}$;\ \ 
$ \zeta_{3}^{1}$,
$ -1$,
$ -1$,
$ \zeta_{6}^{1}$,
$ -\zeta_{3}^{1}$,
$ -\zeta_{3}^{1}$,
$ \zeta_{6}^{1}$;\ \ 
$-\mathrm{i}$,
$\mathrm{i}$,
$-\mathrm{i}$,
$-\mathrm{i}$,
$\mathrm{i}$,
$\mathrm{i}$;\ \ 
$-\mathrm{i}$,
$\mathrm{i}$,
$\mathrm{i}$,
$-\mathrm{i}$,
$-\mathrm{i}$;\ \ 
$ \zeta_{12}^{1}$,
$ \zeta_{12}^{5}$,
$ -\zeta_{12}^{5}$,
$ -\zeta_{12}^{1}$;\ \ 
$ \zeta_{12}^{1}$,
$ -\zeta_{12}^{1}$,
$ -\zeta_{12}^{5}$;\ \ 
$ \zeta_{12}^{1}$,
$ \zeta_{12}^{5}$;\ \ 
$ \zeta_{12}^{1}$)

Factors = $3_{2,3.}^{3,527}\boxtimes 4_{5,4.}^{8,312}$

\vskip 0.7ex
\hangindent=3em \hangafter=1
$\tau_n$ = ($2.45 - 2.45 i$, $3.46 - 3.46 i$, $4.24 - 4.24 i$, $0.$, $2.45 - 2.45 i$, $6. - 6. i$, $-2.45 - 2.45 i$, $0. - 6.93 i$, $-4.24 - 4.24 i$, $-3.46 + 3.46 i$, $-2.45 - 2.45 i$, $0.$, $-2.45 + 2.45 i$, $-3.46 - 3.46 i$, $-4.24 + 4.24 i$, $0. + 6.93 i$, $-2.45 + 2.45 i$, $6. + 6. i$, $2.45 + 2.45 i$, $0.$, $4.24 + 4.24 i$, $3.46 + 3.46 i$, $2.45 + 2.45 i$, $12.$)

\vskip 0.7ex
\hangindent=3em \hangafter=1
\textit{Intrinsic sign problem}

  \vskip 2ex

\noindent14. $12_{1,12.}^{24,565}$ \irep{0}:\ \ 
$d_i$ = ($1.0$,
$1.0$,
$1.0$,
$1.0$,
$1.0$,
$1.0$,
$1.0$,
$1.0$,
$1.0$,
$1.0$,
$1.0$,
$1.0$) 

\vskip 0.7ex
\hangindent=3em \hangafter=1
$D^2= 12.0 = 
12$

\vskip 0.7ex
\hangindent=3em \hangafter=1
$T = ( 0,
\frac{1}{2},
\frac{1}{3},
\frac{1}{3},
\frac{5}{6},
\frac{5}{6},
\frac{7}{8},
\frac{7}{8},
\frac{5}{24},
\frac{5}{24},
\frac{5}{24},
\frac{5}{24} )
$,

\vskip 0.7ex
\hangindent=3em \hangafter=1
$S$ = ($ 1$,
$ 1$,
$ 1$,
$ 1$,
$ 1$,
$ 1$,
$ 1$,
$ 1$,
$ 1$,
$ 1$,
$ 1$,
$ 1$;\ \ 
$ 1$,
$ 1$,
$ 1$,
$ 1$,
$ 1$,
$ -1$,
$ -1$,
$ -1$,
$ -1$,
$ -1$,
$ -1$;\ \ 
$ \zeta_{3}^{1}$,
$ -\zeta_{6}^{1}$,
$ -\zeta_{6}^{1}$,
$ \zeta_{3}^{1}$,
$ 1$,
$ 1$,
$ -\zeta_{6}^{1}$,
$ \zeta_{3}^{1}$,
$ \zeta_{3}^{1}$,
$ -\zeta_{6}^{1}$;\ \ 
$ \zeta_{3}^{1}$,
$ \zeta_{3}^{1}$,
$ -\zeta_{6}^{1}$,
$ 1$,
$ 1$,
$ \zeta_{3}^{1}$,
$ -\zeta_{6}^{1}$,
$ -\zeta_{6}^{1}$,
$ \zeta_{3}^{1}$;\ \ 
$ \zeta_{3}^{1}$,
$ -\zeta_{6}^{1}$,
$ -1$,
$ -1$,
$ -\zeta_{3}^{1}$,
$ \zeta_{6}^{1}$,
$ \zeta_{6}^{1}$,
$ -\zeta_{3}^{1}$;\ \ 
$ \zeta_{3}^{1}$,
$ -1$,
$ -1$,
$ \zeta_{6}^{1}$,
$ -\zeta_{3}^{1}$,
$ -\zeta_{3}^{1}$,
$ \zeta_{6}^{1}$;\ \ 
$\mathrm{i}$,
$-\mathrm{i}$,
$\mathrm{i}$,
$\mathrm{i}$,
$-\mathrm{i}$,
$-\mathrm{i}$;\ \ 
$\mathrm{i}$,
$-\mathrm{i}$,
$-\mathrm{i}$,
$\mathrm{i}$,
$\mathrm{i}$;\ \ 
$ -\zeta_{12}^{1}$,
$ -\zeta_{12}^{5}$,
$ \zeta_{12}^{5}$,
$ \zeta_{12}^{1}$;\ \ 
$ -\zeta_{12}^{1}$,
$ \zeta_{12}^{1}$,
$ \zeta_{12}^{5}$;\ \ 
$ -\zeta_{12}^{1}$,
$ -\zeta_{12}^{5}$;\ \ 
$ -\zeta_{12}^{1}$)

Factors = $3_{2,3.}^{3,527}\boxtimes 4_{7,4.}^{8,781}$

\vskip 0.7ex
\hangindent=3em \hangafter=1
$\tau_n$ = ($2.45 + 2.45 i$, $-3.46 - 3.46 i$, $-4.24 - 4.24 i$, $0.$, $2.45 + 2.45 i$, $6. + 6. i$, $-2.45 + 2.45 i$, $0. - 6.93 i$, $4.24 - 4.24 i$, $3.46 + 3.46 i$, $-2.45 + 2.45 i$, $0.$, $-2.45 - 2.45 i$, $3.46 - 3.46 i$, $4.24 + 4.24 i$, $0. + 6.93 i$, $-2.45 - 2.45 i$, $6. - 6. i$, $2.45 - 2.45 i$, $0.$, $-4.24 + 4.24 i$, $-3.46 + 3.46 i$, $2.45 - 2.45 i$, $12.$)

\vskip 0.7ex
\hangindent=3em \hangafter=1
\textit{Intrinsic sign problem}

  \vskip 2ex

\noindent15. $12_{7,12.}^{24,732}$ \irep{0}:\ \ 
$d_i$ = ($1.0$,
$1.0$,
$1.0$,
$1.0$,
$1.0$,
$1.0$,
$1.0$,
$1.0$,
$1.0$,
$1.0$,
$1.0$,
$1.0$) 

\vskip 0.7ex
\hangindent=3em \hangafter=1
$D^2= 12.0 = 
12$

\vskip 0.7ex
\hangindent=3em \hangafter=1
$T = ( 0,
\frac{1}{2},
\frac{2}{3},
\frac{2}{3},
\frac{1}{6},
\frac{1}{6},
\frac{1}{8},
\frac{1}{8},
\frac{19}{24},
\frac{19}{24},
\frac{19}{24},
\frac{19}{24} )
$,

\vskip 0.7ex
\hangindent=3em \hangafter=1
$S$ = ($ 1$,
$ 1$,
$ 1$,
$ 1$,
$ 1$,
$ 1$,
$ 1$,
$ 1$,
$ 1$,
$ 1$,
$ 1$,
$ 1$;\ \ 
$ 1$,
$ 1$,
$ 1$,
$ 1$,
$ 1$,
$ -1$,
$ -1$,
$ -1$,
$ -1$,
$ -1$,
$ -1$;\ \ 
$ -\zeta_{6}^{1}$,
$ \zeta_{3}^{1}$,
$ -\zeta_{6}^{1}$,
$ \zeta_{3}^{1}$,
$ 1$,
$ 1$,
$ -\zeta_{6}^{1}$,
$ -\zeta_{6}^{1}$,
$ \zeta_{3}^{1}$,
$ \zeta_{3}^{1}$;\ \ 
$ -\zeta_{6}^{1}$,
$ \zeta_{3}^{1}$,
$ -\zeta_{6}^{1}$,
$ 1$,
$ 1$,
$ \zeta_{3}^{1}$,
$ \zeta_{3}^{1}$,
$ -\zeta_{6}^{1}$,
$ -\zeta_{6}^{1}$;\ \ 
$ -\zeta_{6}^{1}$,
$ \zeta_{3}^{1}$,
$ -1$,
$ -1$,
$ \zeta_{6}^{1}$,
$ \zeta_{6}^{1}$,
$ -\zeta_{3}^{1}$,
$ -\zeta_{3}^{1}$;\ \ 
$ -\zeta_{6}^{1}$,
$ -1$,
$ -1$,
$ -\zeta_{3}^{1}$,
$ -\zeta_{3}^{1}$,
$ \zeta_{6}^{1}$,
$ \zeta_{6}^{1}$;\ \ 
$-\mathrm{i}$,
$\mathrm{i}$,
$-\mathrm{i}$,
$\mathrm{i}$,
$-\mathrm{i}$,
$\mathrm{i}$;\ \ 
$-\mathrm{i}$,
$\mathrm{i}$,
$-\mathrm{i}$,
$\mathrm{i}$,
$-\mathrm{i}$;\ \ 
$ \zeta_{12}^{5}$,
$ -\zeta_{12}^{5}$,
$ \zeta_{12}^{1}$,
$ -\zeta_{12}^{1}$;\ \ 
$ \zeta_{12}^{5}$,
$ -\zeta_{12}^{1}$,
$ \zeta_{12}^{1}$;\ \ 
$ \zeta_{12}^{5}$,
$ -\zeta_{12}^{5}$;\ \ 
$ \zeta_{12}^{5}$)

Factors = $3_{6,3.}^{3,138}\boxtimes 4_{1,4.}^{8,718}$

\vskip 0.7ex
\hangindent=3em \hangafter=1
$\tau_n$ = ($2.45 - 2.45 i$, $-3.46 + 3.46 i$, $-4.24 + 4.24 i$, $0.$, $2.45 - 2.45 i$, $6. - 6. i$, $-2.45 - 2.45 i$, $0. + 6.93 i$, $4.24 + 4.24 i$, $3.46 - 3.46 i$, $-2.45 - 2.45 i$, $0.$, $-2.45 + 2.45 i$, $3.46 + 3.46 i$, $4.24 - 4.24 i$, $0. - 6.93 i$, $-2.45 + 2.45 i$, $6. + 6. i$, $2.45 + 2.45 i$, $0.$, $-4.24 - 4.24 i$, $-3.46 - 3.46 i$, $2.45 + 2.45 i$, $12.$)

\vskip 0.7ex
\hangindent=3em \hangafter=1
\textit{Intrinsic sign problem}

  \vskip 2ex

\noindent16. $12_{1,12.}^{24,151}$ \irep{0}:\ \ 
$d_i$ = ($1.0$,
$1.0$,
$1.0$,
$1.0$,
$1.0$,
$1.0$,
$1.0$,
$1.0$,
$1.0$,
$1.0$,
$1.0$,
$1.0$) 

\vskip 0.7ex
\hangindent=3em \hangafter=1
$D^2= 12.0 = 
12$

\vskip 0.7ex
\hangindent=3em \hangafter=1
$T = ( 0,
\frac{1}{2},
\frac{2}{3},
\frac{2}{3},
\frac{1}{6},
\frac{1}{6},
\frac{3}{8},
\frac{3}{8},
\frac{1}{24},
\frac{1}{24},
\frac{1}{24},
\frac{1}{24} )
$,

\vskip 0.7ex
\hangindent=3em \hangafter=1
$S$ = ($ 1$,
$ 1$,
$ 1$,
$ 1$,
$ 1$,
$ 1$,
$ 1$,
$ 1$,
$ 1$,
$ 1$,
$ 1$,
$ 1$;\ \ 
$ 1$,
$ 1$,
$ 1$,
$ 1$,
$ 1$,
$ -1$,
$ -1$,
$ -1$,
$ -1$,
$ -1$,
$ -1$;\ \ 
$ -\zeta_{6}^{1}$,
$ \zeta_{3}^{1}$,
$ -\zeta_{6}^{1}$,
$ \zeta_{3}^{1}$,
$ 1$,
$ 1$,
$ -\zeta_{6}^{1}$,
$ -\zeta_{6}^{1}$,
$ \zeta_{3}^{1}$,
$ \zeta_{3}^{1}$;\ \ 
$ -\zeta_{6}^{1}$,
$ \zeta_{3}^{1}$,
$ -\zeta_{6}^{1}$,
$ 1$,
$ 1$,
$ \zeta_{3}^{1}$,
$ \zeta_{3}^{1}$,
$ -\zeta_{6}^{1}$,
$ -\zeta_{6}^{1}$;\ \ 
$ -\zeta_{6}^{1}$,
$ \zeta_{3}^{1}$,
$ -1$,
$ -1$,
$ \zeta_{6}^{1}$,
$ \zeta_{6}^{1}$,
$ -\zeta_{3}^{1}$,
$ -\zeta_{3}^{1}$;\ \ 
$ -\zeta_{6}^{1}$,
$ -1$,
$ -1$,
$ -\zeta_{3}^{1}$,
$ -\zeta_{3}^{1}$,
$ \zeta_{6}^{1}$,
$ \zeta_{6}^{1}$;\ \ 
$\mathrm{i}$,
$-\mathrm{i}$,
$\mathrm{i}$,
$-\mathrm{i}$,
$\mathrm{i}$,
$-\mathrm{i}$;\ \ 
$\mathrm{i}$,
$-\mathrm{i}$,
$\mathrm{i}$,
$-\mathrm{i}$,
$\mathrm{i}$;\ \ 
$ -\zeta_{12}^{5}$,
$ \zeta_{12}^{5}$,
$ -\zeta_{12}^{1}$,
$ \zeta_{12}^{1}$;\ \ 
$ -\zeta_{12}^{5}$,
$ \zeta_{12}^{1}$,
$ -\zeta_{12}^{1}$;\ \ 
$ -\zeta_{12}^{5}$,
$ \zeta_{12}^{5}$;\ \ 
$ -\zeta_{12}^{5}$)

Factors = $3_{6,3.}^{3,138}\boxtimes 4_{3,4.}^{8,468}$

\vskip 0.7ex
\hangindent=3em \hangafter=1
$\tau_n$ = ($2.45 + 2.45 i$, $3.46 + 3.46 i$, $4.24 + 4.24 i$, $0.$, $2.45 + 2.45 i$, $6. + 6. i$, $-2.45 + 2.45 i$, $0. + 6.93 i$, $-4.24 + 4.24 i$, $-3.46 - 3.46 i$, $-2.45 + 2.45 i$, $0.$, $-2.45 - 2.45 i$, $-3.46 + 3.46 i$, $-4.24 - 4.24 i$, $0. - 6.93 i$, $-2.45 - 2.45 i$, $6. - 6. i$, $2.45 - 2.45 i$, $0.$, $4.24 - 4.24 i$, $3.46 - 3.46 i$, $2.45 - 2.45 i$, $12.$)

\vskip 0.7ex
\hangindent=3em \hangafter=1
\textit{Intrinsic sign problem}

  \vskip 2ex

\noindent17. $12_{3,12.}^{24,684}$ \irep{0}:\ \ 
$d_i$ = ($1.0$,
$1.0$,
$1.0$,
$1.0$,
$1.0$,
$1.0$,
$1.0$,
$1.0$,
$1.0$,
$1.0$,
$1.0$,
$1.0$) 

\vskip 0.7ex
\hangindent=3em \hangafter=1
$D^2= 12.0 = 
12$

\vskip 0.7ex
\hangindent=3em \hangafter=1
$T = ( 0,
\frac{1}{2},
\frac{2}{3},
\frac{2}{3},
\frac{1}{6},
\frac{1}{6},
\frac{5}{8},
\frac{5}{8},
\frac{7}{24},
\frac{7}{24},
\frac{7}{24},
\frac{7}{24} )
$,

\vskip 0.7ex
\hangindent=3em \hangafter=1
$S$ = ($ 1$,
$ 1$,
$ 1$,
$ 1$,
$ 1$,
$ 1$,
$ 1$,
$ 1$,
$ 1$,
$ 1$,
$ 1$,
$ 1$;\ \ 
$ 1$,
$ 1$,
$ 1$,
$ 1$,
$ 1$,
$ -1$,
$ -1$,
$ -1$,
$ -1$,
$ -1$,
$ -1$;\ \ 
$ -\zeta_{6}^{1}$,
$ \zeta_{3}^{1}$,
$ -\zeta_{6}^{1}$,
$ \zeta_{3}^{1}$,
$ 1$,
$ 1$,
$ -\zeta_{6}^{1}$,
$ -\zeta_{6}^{1}$,
$ \zeta_{3}^{1}$,
$ \zeta_{3}^{1}$;\ \ 
$ -\zeta_{6}^{1}$,
$ \zeta_{3}^{1}$,
$ -\zeta_{6}^{1}$,
$ 1$,
$ 1$,
$ \zeta_{3}^{1}$,
$ \zeta_{3}^{1}$,
$ -\zeta_{6}^{1}$,
$ -\zeta_{6}^{1}$;\ \ 
$ -\zeta_{6}^{1}$,
$ \zeta_{3}^{1}$,
$ -1$,
$ -1$,
$ \zeta_{6}^{1}$,
$ \zeta_{6}^{1}$,
$ -\zeta_{3}^{1}$,
$ -\zeta_{3}^{1}$;\ \ 
$ -\zeta_{6}^{1}$,
$ -1$,
$ -1$,
$ -\zeta_{3}^{1}$,
$ -\zeta_{3}^{1}$,
$ \zeta_{6}^{1}$,
$ \zeta_{6}^{1}$;\ \ 
$-\mathrm{i}$,
$\mathrm{i}$,
$-\mathrm{i}$,
$\mathrm{i}$,
$-\mathrm{i}$,
$\mathrm{i}$;\ \ 
$-\mathrm{i}$,
$\mathrm{i}$,
$-\mathrm{i}$,
$\mathrm{i}$,
$-\mathrm{i}$;\ \ 
$ \zeta_{12}^{5}$,
$ -\zeta_{12}^{5}$,
$ \zeta_{12}^{1}$,
$ -\zeta_{12}^{1}$;\ \ 
$ \zeta_{12}^{5}$,
$ -\zeta_{12}^{1}$,
$ \zeta_{12}^{1}$;\ \ 
$ \zeta_{12}^{5}$,
$ -\zeta_{12}^{5}$;\ \ 
$ \zeta_{12}^{5}$)

Factors = $3_{6,3.}^{3,138}\boxtimes 4_{5,4.}^{8,312}$

\vskip 0.7ex
\hangindent=3em \hangafter=1
$\tau_n$ = ($-2.45 + 2.45 i$, $-3.46 + 3.46 i$, $4.24 - 4.24 i$, $0.$, $-2.45 + 2.45 i$, $6. - 6. i$, $2.45 + 2.45 i$, $0. + 6.93 i$, $-4.24 - 4.24 i$, $3.46 - 3.46 i$, $2.45 + 2.45 i$, $0.$, $2.45 - 2.45 i$, $3.46 + 3.46 i$, $-4.24 + 4.24 i$, $0. - 6.93 i$, $2.45 - 2.45 i$, $6. + 6. i$, $-2.45 - 2.45 i$, $0.$, $4.24 + 4.24 i$, $-3.46 - 3.46 i$, $-2.45 - 2.45 i$, $12.$)

\vskip 0.7ex
\hangindent=3em \hangafter=1
\textit{Intrinsic sign problem}

  \vskip 2ex

\noindent18. $12_{5,12.}^{24,899}$ \irep{0}:\ \ 
$d_i$ = ($1.0$,
$1.0$,
$1.0$,
$1.0$,
$1.0$,
$1.0$,
$1.0$,
$1.0$,
$1.0$,
$1.0$,
$1.0$,
$1.0$) 

\vskip 0.7ex
\hangindent=3em \hangafter=1
$D^2= 12.0 = 
12$

\vskip 0.7ex
\hangindent=3em \hangafter=1
$T = ( 0,
\frac{1}{2},
\frac{2}{3},
\frac{2}{3},
\frac{1}{6},
\frac{1}{6},
\frac{7}{8},
\frac{7}{8},
\frac{13}{24},
\frac{13}{24},
\frac{13}{24},
\frac{13}{24} )
$,

\vskip 0.7ex
\hangindent=3em \hangafter=1
$S$ = ($ 1$,
$ 1$,
$ 1$,
$ 1$,
$ 1$,
$ 1$,
$ 1$,
$ 1$,
$ 1$,
$ 1$,
$ 1$,
$ 1$;\ \ 
$ 1$,
$ 1$,
$ 1$,
$ 1$,
$ 1$,
$ -1$,
$ -1$,
$ -1$,
$ -1$,
$ -1$,
$ -1$;\ \ 
$ -\zeta_{6}^{1}$,
$ \zeta_{3}^{1}$,
$ -\zeta_{6}^{1}$,
$ \zeta_{3}^{1}$,
$ 1$,
$ 1$,
$ -\zeta_{6}^{1}$,
$ -\zeta_{6}^{1}$,
$ \zeta_{3}^{1}$,
$ \zeta_{3}^{1}$;\ \ 
$ -\zeta_{6}^{1}$,
$ \zeta_{3}^{1}$,
$ -\zeta_{6}^{1}$,
$ 1$,
$ 1$,
$ \zeta_{3}^{1}$,
$ \zeta_{3}^{1}$,
$ -\zeta_{6}^{1}$,
$ -\zeta_{6}^{1}$;\ \ 
$ -\zeta_{6}^{1}$,
$ \zeta_{3}^{1}$,
$ -1$,
$ -1$,
$ \zeta_{6}^{1}$,
$ \zeta_{6}^{1}$,
$ -\zeta_{3}^{1}$,
$ -\zeta_{3}^{1}$;\ \ 
$ -\zeta_{6}^{1}$,
$ -1$,
$ -1$,
$ -\zeta_{3}^{1}$,
$ -\zeta_{3}^{1}$,
$ \zeta_{6}^{1}$,
$ \zeta_{6}^{1}$;\ \ 
$\mathrm{i}$,
$-\mathrm{i}$,
$\mathrm{i}$,
$-\mathrm{i}$,
$\mathrm{i}$,
$-\mathrm{i}$;\ \ 
$\mathrm{i}$,
$-\mathrm{i}$,
$\mathrm{i}$,
$-\mathrm{i}$,
$\mathrm{i}$;\ \ 
$ -\zeta_{12}^{5}$,
$ \zeta_{12}^{5}$,
$ -\zeta_{12}^{1}$,
$ \zeta_{12}^{1}$;\ \ 
$ -\zeta_{12}^{5}$,
$ \zeta_{12}^{1}$,
$ -\zeta_{12}^{1}$;\ \ 
$ -\zeta_{12}^{5}$,
$ \zeta_{12}^{5}$;\ \ 
$ -\zeta_{12}^{5}$)

Factors = $3_{6,3.}^{3,138}\boxtimes 4_{7,4.}^{8,781}$

\vskip 0.7ex
\hangindent=3em \hangafter=1
$\tau_n$ = ($-2.45 - 2.45 i$, $3.46 + 3.46 i$, $-4.24 - 4.24 i$, $0.$, $-2.45 - 2.45 i$, $6. + 6. i$, $2.45 - 2.45 i$, $0. + 6.93 i$, $4.24 - 4.24 i$, $-3.46 - 3.46 i$, $2.45 - 2.45 i$, $0.$, $2.45 + 2.45 i$, $-3.46 + 3.46 i$, $4.24 + 4.24 i$, $0. - 6.93 i$, $2.45 + 2.45 i$, $6. - 6. i$, $-2.45 + 2.45 i$, $0.$, $-4.24 + 4.24 i$, $3.46 - 3.46 i$, $-2.45 + 2.45 i$, $12.$)

\vskip 0.7ex
\hangindent=3em \hangafter=1
\textit{Intrinsic sign problem}

  \vskip 2ex 